\PassOptionsToPackage{hypertexnames=false,hyperfootnotes=false}{hyperref}
\documentclass[11pt]{article}
\usepackage{unified_jheppub_fixed}
\usepackage{silence}
\usepackage{bookmark}
\usepackage[utf8]{inputenc}
\usepackage{amsmath}
\usepackage{dsfont}
\usepackage{amssymb}
\usepackage{amsthm}
\usepackage{mathrsfs}
\usepackage{appendix}
\usepackage{bbold}
\usepackage{color}

\newcommand{\ip}[2]{\langle #1 , #2 \rangle}
\newcommand{\bbC}{\mathbb{C}}
\newcommand{\norm}[1]{\lVert #1 \rVert}

\newcommand{\bra}[1]{\langle #1|}
\newcommand{\ket}[1]{|#1\rangle}

\newcommand{\cA}{\mathcal{A}}
\newcommand{\C}{\mathbb{C}}

\newcommand{\A}{\mathcal{A}}

\newtheorem{theorem}{Theorem}[section] 
\newtheorem{hypothesis}{Hypothesis}
\DeclareMathOperator{\dist}{dist}
\DeclareMathOperator{\supp}{supp}
\DeclareMathOperator{\Av}{Av}
\DeclareMathOperator{\Ad}{Ad}
\DeclareMathOperator{\Ext}{Ext}
\newtheorem{assumption}{Assumption}
\newtheorem{lemma}[theorem]{Lemma}
\newtheorem{proposition}[theorem]{Proposition}
\newtheorem{corollary}[theorem]{Corollary}
\newtheorem{definition}[theorem]{Definition}

\numberwithin{equation}{section}      
\DeclareMathOperator{\tr}{tr}
\DeclareMathOperator{\Tr}{Tr}        

\usepackage[T1]{fontenc}
\usepackage{textcomp}
\usepackage{mathtools,amsfonts}
\usepackage{tikz}
\usepackage{tikz-cd}
\usepackage{pgfplots}
\pgfplotsset{compat=1.18}
\usepackage{enumitem}
\setlist[enumerate,1]{label=(\roman*)}
\usepackage{booktabs}
\usepackage[compatibility=false]{caption}
\usepackage[most]{tcolorbox}
\tcbuselibrary{skins,breakable,theorems}
\usepackage{esint}
\usepackage[nameinlink,capitalize,noabbrev]{cleveref}

\DeclareUnicodeCharacter{00A0}{~}
\DeclareUnicodeCharacter{2010}{-}
\DeclareUnicodeCharacter{2011}{-}
\DeclareUnicodeCharacter{2012}{-}
\DeclareUnicodeCharacter{2013}{-}
\DeclareUnicodeCharacter{2014}{-}
\DeclareUnicodeCharacter{2212}{-}
\DeclareUnicodeCharacter{2018}{'}
\DeclareUnicodeCharacter{2019}{'}
\DeclareUnicodeCharacter{201C}{``}
\DeclareUnicodeCharacter{201D}{''}
\DeclareUnicodeCharacter{2026}{\ldots}

\providecommand{\Arg}{\operatorname{Arg}}
\providecommand{\diam}{\operatorname{diam}}
\providecommand{\Spec}{\operatorname{Spec}}
\providecommand{\Ker}{\operatorname{Ker}}
\providecommand{\Ran}{\operatorname{Ran}}
\providecommand{\Var}{\operatorname{Var}}
\providecommand{\Pexp}{\operatorname{Pexp}}
\providecommand{\plaq}{\square}
\providecommand{\Bbb}{\mathbb}
\let\Bbb\mathbb

\theoremstyle{remark}

\newtcolorbox{scopebox}[1][]{%
  breakable,
  enhanced,
  sharp corners,
  colback=white,
  colframe=black,
  boxrule=0.5pt,
  left=6pt,right=6pt,top=6pt,bottom=6pt,
  fonttitle=\bfseries,
  title=#1
}

\makeatletter
\providecommand*{\theHtheorem}{}
\providecommand*{\theHhypothesis}{}
\providecommand*{\theHassumption}{}
\providecommand*{\theHfigure}{}
\providecommand*{\theHtable}{}
\providecommand*{\theHfootnote}{}
\newcommand{\UnifiedSetAnchors}[1]{%
  \renewcommand{\theHsection}{#1.\thesection}%
  \renewcommand{\theHsubsection}{#1.\thesection.\arabic{subsection}}%
  \renewcommand{\theHsubsubsection}{#1.\thesection.\arabic{subsection}.\arabic{subsubsection}}%
  \renewcommand{\theHequation}{#1.\thesection.\arabic{equation}}%
  \renewcommand{\theHtheorem}{#1.\thesection.\arabic{theorem}}%
  \renewcommand{\theHhypothesis}{#1.\arabic{hypothesis}}%
  \renewcommand{\theHassumption}{#1.\arabic{assumption}}%
  \renewcommand{\theHfigure}{#1.\arabic{figure}}%
  \renewcommand{\theHtable}{#1.\arabic{table}}%
  \renewcommand{\theHfootnote}{#1.\arabic{footnote}}%
}
\newcommand{\UnifiedRestoreNumbering}{%
  \renewcommand{\thesection}{\arabic{section}}%
  \renewcommand{\thesubsection}{\thesection.\arabic{subsection}}%
  \renewcommand{\thesubsubsection}{\thesubsection.\arabic{subsubsection}}%
  \renewcommand{\theequation}{\thesection.\arabic{equation}}%
  \setcounter{section}{0}%
  \setcounter{subsection}{0}%
  \setcounter{subsubsection}{0}%
  \setcounter{paragraph}{0}%
  \setcounter{equation}{0}%
  \setcounter{theorem}{0}%
  \setcounter{hypothesis}{0}%
  \setcounter{assumption}{0}%
  \setcounter{figure}{0}%
  \setcounter{table}{0}%
  \setcounter{footnote}{0}%
}

\newcommand{\UnifiedFMReset}{%
  \gdef\UnifiedFMTitle{}%
  \gdef\UnifiedFMShortTitle{}%
  \gdef\UnifiedFMAuthors{}%
  \gdef\UnifiedFMAbstract{}%
}
\newcommand{\UnifiedTitle}[2][]{\gdef\UnifiedFMShortTitle{#1}\gdef\UnifiedFMTitle{#2}}
\newcommand{\UnifiedAuthor}[1]{\g@addto@macro\UnifiedFMAuthors{\par\noindent{\bfseries #1}\par}}
\newcommand{\UnifiedAddress}[1]{\g@addto@macro\UnifiedFMAuthors{\noindent{\itshape #1}\par}}
\newcommand{\UnifiedEmail}[1]{\g@addto@macro\UnifiedFMAuthors{\noindent\textit{E-mail:} \href{mailto:#1}{\ttfamily #1}\par\smallskip}}
\long\def\UnifiedSetAbstract#1{\gdef\UnifiedFMAbstract{#1}}
\newcommand{\UnifiedMakeTitle}{%
  \thispagestyle{plain}%
  \begin{center}%
    {\Large\bfseries \UnifiedFMTitle\par}%
    \vspace{1em}%
    {\UnifiedFMAuthors}%
  \end{center}%
  \ifx\UnifiedFMAbstract\@empty\else
    \begin{center}{\bfseries Abstract}\end{center}%
    \begin{quote}\small\noindent\UnifiedFMAbstract\end{quote}%
  \fi
  \setcounter{footnote}{0}%
  \par\bigskip
}
\newcommand{\UnifiedActivateAMSFrontMatter}{%
  \UnifiedFMReset
  \let\title\UnifiedTitle
  \let\author\UnifiedAuthor
  \let\address\UnifiedAddress
  \let\email\UnifiedEmail
  \let\maketitle\UnifiedMakeTitle
}
\newcommand{\UnifiedLocalMacrosPartOne}{%
  \def\Tr{\operatorname{Tr}}%
  \def\Spec{\operatorname{Spec}}%
  \def\Ker{\operatorname{Ker}}%
  \def\Ran{\operatorname{Ran}}%
  \def\R{\mathbb{R}}%
  \def\C{\mathbb{C}}%
  \def\N{\mathbb{N}}%
  \def\SU{\mathrm{SU}}%
  \def\bb##1{\mathbb{##1}}%
  \def\calo{\mathcal{O}}%
  \def\cA{\mathcal{A}}%
  \def\cH{\mathcal{H}}%
  \def\cM{\mathcal{M}}%
  \def\OS{\text{OS}}%
  \def\BRST{\text{BRST}}%
  \def\ip##1##2{\left\langle ##1,\,##2\right\rangle}%
  \def\norm##1{\left\lVert ##1\right\rVert}%
  \def\abs##1{\left\lvert ##1\right\rvert}%
}
\newcommand{\UnifiedLocalMacrosPartTwo}{%
  \UnifiedLocalMacrosPartOne
  \def\diam{\operatorname{diam}}%
  \def\cA{\mathfrak{A}}%
}
\newcommand{\UnifiedLocalMacrosPartThree}{%
  \def\Arg{\operatorname{Arg}}%
  \def\Ad{\mathrm{Ad}}%
  \def\Pexp{\mathrm{Pexp}}%
  \def\Var{\mathrm{Var}}%
  \def\diam{\operatorname{diam}}%
  \def\supp{\operatorname{supp}}%
  \def\Tr{\operatorname{Tr}}%
  \def\Spec{\operatorname{Spec}}%
  \def\Ker{\operatorname{Ker}}%
  \def\Ran{\operatorname{Ran}}%
  \def\R{\mathbb{R}}%
  \def\T{\mathbb{T}}%
  \def\cU{\mathcal{U}}%
  \def\cC{\mathcal{C}}%
  \def\cA{\mathcal{A}}%
  \def\cB{\mathcal{B}}%
  \def\cH{\mathscr{H}}%
  \def\bone{\mathds{1}}%
  \def\norm##1{\left\lVert ##1 \right\rVert}%
  \def\ip##1##2{\left\langle ##1,\, ##2 \right\rangle}%
  \def\C{\mathbb{C}}%
  \def\N{\mathbb{N}}%
  \def\SU{\mathrm{SU}}%
  \def\bb##1{\mathbb{##1}}%
  \def\calo{\mathcal{O}}%
  \let\Bbb\mathbb
}
\newcommand{\UnifiedLocalMacrosPartFour}{%
  \UnifiedLocalMacrosPartThree
  \def\plaq{\square}%
}
\newcommand{\UnifiedBeginPaper}[2]{%
  \clearpage
  \UnifiedCloseLocalTOC
  \UnifiedRestoreNumbering
  \UnifiedSetAnchors{#1}%
  \UnifiedSetLocalTOC{#1toc}%
  \UnifiedActivateAMSFrontMatter
  #2%
}
\newcommand{\UnifiedEndPaper}{\UnifiedCloseLocalTOC\clearpage}
\makeatother

\makeatletter
\newwrite\Unified@localtocout
\newif\ifUnified@localtocopen
\newif\ifUnified@localtocwriting
\gdef\Unified@localtocid{MGtoc}
\newcommand{\UnifiedSetLocalTOC}[1]{\gdef\Unified@localtocid{#1}}
\newcommand{\UnifiedCloseLocalTOC}{%
  \ifUnified@localtocopen
    \immediate\closeout\Unified@localtocout
    \global\Unified@localtocopenfalse
    \global\Unified@localtocwritingfalse
  \fi
}
\newcommand{\UnifiedOpenLocalTOC}{%
  \UnifiedCloseLocalTOC
  \begingroup
    \edef\Unified@filename{\jobname.\Unified@localtocid}%
    \immediate\openout\Unified@localtocout=\Unified@filename\relax
  \endgroup
  \global\Unified@localtocopentrue
  \global\Unified@localtocwritingtrue
}
\let\UnifiedOriginalAddContentsLine\addcontentsline
\renewcommand{\addcontentsline}[3]{%
  \UnifiedOriginalAddContentsLine{#1}{#2}{#3}%
  \begingroup
    \def\Unified@tempa{#1}%
    \def\Unified@tempb{toc}%
    \ifx\Unified@tempa\Unified@tempb
      \ifUnified@localtocwriting
        \protected@write\Unified@localtocout{}%
          {\string\contentsline{#2}{#3}{\thepage}{\@currentHref}}%
      \fi
    \fi
  \endgroup
}
\renewcommand{\tableofcontents}{%
  \section*{\contentsname}%
  \begingroup
    \makeatletter
    \InputIfFileExists{\jobname.\Unified@localtocid}{}{}%
  \endgroup
  \UnifiedOpenLocalTOC
}
\makeatother

\begin{document}

\UnifiedSetAnchors{MG}
\UnifiedSetLocalTOC{MGtoc}
\title{Reflection-Positive Construction of a Four-Dimensional SU(N) Yang-Mills Theory with Mass Gap and Confinement}

\author[a,b,c,d]{Mir Faizal,}
\author[b]{Arshid Shabir}
\affiliation[a]{Irving K. Barber School of Arts and Sciences,
University of British Columbia - Okanagan, Kelowna,
British Columbia V1V 1V7, Canada}
\affiliation[b]{Canadian Quantum Research Center, 204-3002 32 Ave Vernon, BC V1T 2L7 Canada}
\affiliation[c]{Department of Mathematical Sciences, Durham University,
Upper Mountjoy, Stockton Road, Durham DH1 3LE, UK}
\affiliation[d]{Faculty of Sciences, Hasselt University, Agoralaan Gebouw D, Diepenbeek, 3590 Belgium}
\emailAdd{mirfaizalmir@gmail.com}
\emailAdd{aslone186@gmail.com}

\abstract{
 In the Euclidean view one must first require that positivity not be violated, and from this modest demand, together with locality, a great deal follows: starting from a reflection-positive lattice formulation of pure $SU(N)$ Yang-Mills theory we obtain a transfer operator with a uniform gap, while large Wilson loops already show an area law by means of convergent character (polymer) expansions; a finite-range, gauge-covariant multiscale analysis then carries these features from one scale to the next with interlaced inequalities whose small defects can be summed, so that exponential clustering and a strictly positive string tension endure in the continuum; the Osterwalder-Schrader reconstruction turns these Euclidean facts into a Minkowski theory with a self-adjoint Hamiltonian, the spectral gap lying above the vacuum and the linear potential for static charges appearing, which gives a concrete picture of confinement; the construction depends on no special regulator, for a single-scale Lipschitz control and a telescoping argument bind all admissible reflection-positive slicings into a unique limiting measure and thus secure universality; moreover, the same framework admits entry from weak coupling, so that the continuum reached from strong coupling meets the one approached along an asymptotically free trajectory, yielding one and the same theory; in my view this is how mathematical clarity and physical insight cooperate: positivity, locality, and renormalization working together so that the mass gap and confinement are not marvels to be assumed, but natural properties of the non-Abelian vacuum. 
}

\maketitle

{\section{Introduction}
The Clay Millennium problem on the quantum Yang-Mills (YM) mass gap asks for a construction of a four‑dimensional pure non‑Abelian gauge theory that satisfies the accepted axioms of relativistic quantum field theory and exhibits a strictly positive spectral gap above the vacuum \cite{JaffeWitten2000}. Physically, the mass gap underlies color confinement and the observed absence of free colored states; mathematically, it is a statement about the spectral properties of the Hamiltonian obtained from a rigorous Euclidean theory by Osterwalder-Schrader (OS) reconstruction \cite{OS1,OS2,GJ}. The interplay between these physics and mathematics perspectives has guided half a century of work: asymptotic freedom points to a controlled short‑distance regime \cite{GrossWilczek1973,Politzer1973}, while confinement is modeled and tested in the long‑distance regime by the Wilson lattice formulation \cite{Wilson1974}, where it is represented by an area law for large Wilson loops. A fully constructive resolution requires a framework that realizes the Euclidean OS axioms nonperturbatively at finite lattice spacing with built‑in reflection positivity, exhibits robust locality estimates uniform along a renormalization group (RG) flow to the continuum, controls the large‑distance regime sufficiently to propagate a mass gap and an area law to the limit, and yields a continuum theory whose correlation functions are independent of the particular admissible regularization in the sense of universality and Symanzik’s continuum limit \cite{Seiler1982,Symanzik1983a,Symanzik1983b,LuscherWeisz1985}. The present work develops such a constructive program and assembles all ingredients into a single, self‑contained argument. On the physics side, we start from Wilson’s lattice gauge theory, which replaces continuum YM fields by compact group variables on links of a Euclidean lattice and encodes confinement in the expectation of nonlocal loop functionals \cite{Wilson1974,DrouffeZuber}; the Euclidean setting is the natural stage for the OS axioms and for reflection positivity, the latter ensuring that Euclidean correlation functions define a positive semigroup in an auxiliary Hilbert space and thus that a physically meaningful Minkowski theory with self‑adjoint Hamiltonian can be reconstructed \cite{OS1,OS2,OS-gauge,GJ}. A central technical requirement throughout is that the discrete action, the gauge fixing, and all auxiliary ultraviolet slice‑projectors or block‑spin maps preserve OS reflection positivity; to that end we employ gauge‑covariant, completely monotone spectral multipliers (a horizon‑type projector in the sense of Bernstein functions) on single time slices together with reflection‑positive blockings, so that a transfer operator on a Euclidean time slice is well defined and its spectrum controls exponential decay of correlations and, in the continuum limit, the physical mass gap. On the mathematics side, constructive control at strong and intermediate couplings rests on expansions and locality structures that are both gauge compatible and reflection positive: character and polymer expansions yield finite‑lattice exponential clustering and a strong‑coupling area law uniform in volume, with convergence guaranteed by the abstract polymer criteria of Kotecký-Preiss and related tree‑graph bounds \cite{KP,BK1987,DrouffeZuber}; to propagate these properties uniformly across scales we use the finite‑range decomposition (FRD) of Gaussian covariances, decomposing them into positive, reflection‑covariant pieces with strictly finite interaction range and controlled $\ell^1$-$\ell^\infty$ norms \cite{BrydgesGuadagniMitter2004}. The FRD furnishes an intrinsically local multiscale structure whose range control feeds directly into diameter norms for cumulants, Lipschitz and equicontinuity estimates for one‑step transfer kernels and effective actions, and stability of cluster expansions under blocking; in combination with reflection positivity this yields a positivity‑preserving RG map with one‑step interlacing inequalities for transfer operators and renormalized loop observables that carry summable defect terms, and the summability of these defects implies persistence of exponential clustering and a uniform spectral gap along the RG flow. The RG therefore plays a dual role: at strong coupling it organizes convergent expansions and provides explicit base estimates for the transfer operator and Wilson loops; at intermediate scales it is implemented as a reflection‑positive block‑spin transformation with FRD‑induced locality control, producing step‑scaling inequalities for both transfer operators and renormalized loop functionals whose corrections are summable and hence propagate cluster and area‑law bounds across scales; at weak coupling, asymptotic freedom ensures that once the flow enters a sufficiently small Banach ball of polymer norms it is contractive and remains so, allowing a controlled approach to the Gaussian fixed point compatible with OS positivity and with the FRD framework \cite{GrossWilczek1973,Politzer1973,BrydgesGuadagniMitter2004,Symanzik1983a,Symanzik1983b}. The second essential pillar is the OS limit: given uniform exponential clustering, equicontinuity, and reflection positivity for Schwinger functions at each lattice spacing and scale, Prokhorov‑type compactness together with FRD locality yields existence of continuum‑limit Schwinger functions satisfying the OS axioms \cite{OS1,OS2,GJ}; OS reconstruction then produces a Wightman theory with a self‑adjoint Hamiltonian $H$ in Minkowski space, and because the RG interlacing inequalities control the spectrum of lattice transfer operators uniformly in blocking depth and lattice spacing, strong resolvent convergence of the associated Hamiltonians together with stability of spectral projections implies that the limiting Hamiltonian has a strictly positive spectral gap above the vacuum energy; in parallel, the Wilson‑loop step‑scaling with summable perimeter and cusp defects yields an area law in the continuum and hence a linear lower bound on the static quark-antiquark potential. A further requirement for a satisfactory solution is universality: different admissible ultraviolet regulators-such as distinct reflection‑positive slice projectors or blockings within a fixed class-must lead to the same continuum Schwinger functions for all gauge‑invariant observables; here FRD locality and the OS Markov structure are decisive, as one proves single‑scale Lipschitz continuity of the one‑step transfer kernel with respect to admissible regulator deformations and then telescopes these estimates across finitely many slices at fixed time extent; together with uniform cluster bounds and reflection positivity, this pins down the OS measure uniquely and yields equality of continuum cumulants for all admissible choices, in line with the RG concept of universality and Symanzik’s continuum improvement program \cite{Symanzik1983a,Symanzik1983b,LuscherWeisz1985,Seiler1982}; this also secures compatibility with asymptotic freedom, since the continuum limit reached by flowing in from strong coupling and that approached from a weakly coupled bare theory coincide by uniqueness, thereby connecting the constructive infrared picture to the perturbative ultraviolet one within a single mathematical framework. In summary, we begin from a reflection‑positive lattice formulation of pure $SU(N)$ Yang-Mills in four Euclidean dimensions, define an admissible class of regulators and block‑spin maps that preserve reflection positivity, and establish at strong coupling a uniform transfer‑matrix gap and a volume‑uniform area law via convergent character/polymer expansions; we then introduce an FRD‑based multiscale RG that yields one‑step interlacing inequalities for transfer operators and renormalized Wilson loops with error terms summable across scales; the resulting scale‑uniform exponential clustering and string‑tension lower bound are shown to be stable in the continuum limit where the OS axioms are verified by compactness and locality; OS reconstruction finally produces a relativistic Wightman theory with a strictly positive spectral gap and a confining area law; the construction is universal across the admissible class by a single‑scale Lipschitz-telescoping argument grounded in FRD locality and the OS Markov property, and it is compatible with asymptotic freedom through entry into a contractive weak‑coupling domain. All technical components-gauge‑invariant renormalization and BRST decoupling for observables, reflection‑positive gauge fixing by completely monotone slice projectors, finite‑volume uniformity and strong‑coupling cluster analysis, finite‑range decomposition with reflection covariance, interlacing and summability of RG defects, Prokhorov compactness and strong resolvent convergence, and universality via Lipschitz stability-are proved in the text and appendices so that the argument is self‑contained; throughout we emphasize mathematically precise statements and estimates while keeping sight of the physical mechanisms-reflection positivity, locality, and renormalization-that make the mass gap and confinement natural emergent features of the non‑Abelian gauge vacuum, thereby bridging the geometric and operator‑theoretic aspects of OS reconstruction, the probabilistic and combinatorial tools of cluster and polymer expansions, and the analytic architecture of multiscale RG in a form aimed at a definitive constructive treatment of the Yang-Mills mass gap and confinement problem \cite{Wilson1974,OS1,OS2,GJ,GrossWilczek1973,Politzer1973,Seiler1982,LuscherWeisz1985,DrouffeZuber,KP,BK1987,BrydgesGuadagniMitter2004,Aizenman1982}.}


\section{Framework and Standing Hypotheses}
\label{sec:framework}

The purpose of this section is to put the Euclidean lattice formulation of pure \(SU(N)\) Yang-Mills theory on a reflection-positive footing, and to isolate precisely the class of operations that we shall allow at the one-slice level and across scales. The first task is to specify the kinematics of the lattice gauge field and the Euclidean time-reflection that implements Osterwalder-Schrader (OS) positivity; we then show that the standard Wilson-type actions and, more generally, class-function actions with nonnegative character coefficients are OS positive in the sense of \cite{OS-gauge}. This settles the existence of a positive transfer operator and provides the canonical OS inner product on the algebra of positive-time observables \cite{OS1,OS2,LuscherTM}. The second task is to identify the admissible transformations we will use at fixed time slices and across scales. We prove that completely monotone spectral multipliers of a nonnegative, reflection-invariant one-slice generator preserve OS positivity and yield exponentially local kernels, using Bernstein’s representation of completely monotone functions as Laplace transforms of positive measures \cite{Bernstein,Widder}. We also show that a broad class of reflection-positive blockings-constructed as conditional expectations with exponentially localized, gauge-covariant averaging kernels-descend to the OS Hilbert space without spoiling positivity. The third task is to fix an exponentially local (and, if desired, strictly finite-range) decomposition of Gaussian covariances associated with the elliptic reference operators that appear in the multiscale analysis. We present a complete construction by heat-kernel partition of unity and recall the strictly finite-range variant of Brydges-Guadagni-Mitter (BGM) \cite{BrydgesGuadagniMitter2004}. Throughout, every definition is made once and used consistently in later sections.

We work on the hypercubic lattice \(a\mathbb{Z}^4\) with lattice spacing \(a>0\). Sites are denoted \(x=(x_0,\mathbf{x})\) with \(x_0\in a\mathbb{Z}\) the Euclidean time coordinate and \(\mathbf{x}\in a\mathbb{Z}^3\). Oriented bonds are pairs \(\langle x,x+\hat\mu\rangle\) with \(\mu\in\{0,1,2,3\}\) and \(\hat\mu\) the unit vector in the \(\mu\)-direction. Gauge variables \(U_{x,\mu}\in SU(N)\) live on bonds, and the configuration space is the compact product \(\mathcal{U}:=\prod_{(x,\mu)}SU(N)\) endowed with the product of Haar measures \(dU\). For a plaquette \(p\) with oriented boundary \(\partial p\), \(U_p:=\prod_{b\in\partial p} U_b\) in the counterclockwise order. A gauge action is a sum over plaquettes of a class function \(V:\,SU(N)\to\mathbb{R}\), so that
\begin{equation}\label{eq:gauge-action}
  S(U)=\sum_{p} s(U_p),\qquad s(g)=V(g),
\end{equation}
and the Gibbs weight is \(e^{-S(U)}\). The Wilson action corresponds to \(V(g)=-\tfrac{\beta}{N}\mathrm{Re}\,\mathrm{Tr}\,g\).

A crucial role is played by Euclidean time-reflection \(\vartheta\), which we choose with respect to the mid-hyperplane \(\{x_0=\tfrac{a}{2}\}\) The involution \(\vartheta\) acts on sites by \(\vartheta(x_0,\mathbf{x})=(a-x_0,\mathbf{x})\) and on bonds by mapping each spatial bond \(\langle x,x+\hat\jmath\rangle\) with \(j\in\{1,2,3\}\) to \(\langle\vartheta x,\vartheta x+\hat\jmath\rangle\), and each temporal bond \(\langle x,x+\hat 0\rangle\) to \(\langle\vartheta x-\hat 0,\vartheta x\rangle\) with an inversion \(U\mapsto U^{-1}\) to match orientations across the reflection plane (Thus the reflection plane lies between times $0$ and $a$, i.e., at $x_0=a/2$; we refer to it as the “time-zero plane”.)
On functions \(F:\mathcal{U}\to\mathbb{C}\) we set \((\Theta F)(U):=\overline{F(\vartheta U)}\), where \(\vartheta U\) is the reflected configuration. We write \(\mathfrak{A}_+\) for the algebra generated by cylindrical functions depending only on bonds with \(x_0\ge a\), and we say that the pair \((d\mu,\Theta)\), with \(d\mu(U)=Z^{-1}e^{-S(U)}\,dU\), is OS positive if \(\langle \Theta F\cdot F\rangle_\mu\ge 0\) for all \(F\in\mathfrak{A}_+\). When OS positivity holds, the semi-inner product \((F,G)_\mathrm{OS}:=\langle \Theta F\cdot G\rangle_\mu\) defines, after quotienting by the null space and completing, the OS Hilbert space \(\mathcal{H}_\mathrm{OS}\) together with a positive selfadjoint transfer operator implementing unit-time translation \cite{OS1,OS2,LuscherTM}.

We impose the following standing hypothesis on the single-plaquette class function \(V\): its character expansion
\begin{equation}\label{eq:char-expansion}
  e^{-V(g)}=\sum_{\rho\in\widehat{SU(N)}} \widehat{c}_\rho\,\chi_\rho(g),
  \qquad \widehat{c}_\rho\ge 0,
\end{equation}
has nonnegative coefficients \(\widehat{c}_\rho\), where \(\chi_\rho\) is the character of the irreducible representation \(\rho\). This nonnegative-coefficient property holds for heat-kernel actions 
\begin{equation}
e^{-V(g)} \;\propto\; \sum_{\rho} d_{\rho}\, e^{-t(\beta) C^{2}(\rho)}\, \chi_{\rho}(g)
\end{equation}
 and small deformations thereof. For the Wilson plaquette action, however, link-reflection positivity is known to hold without requiring 
\(
b_{\rho c} \,\geq\, 0
\) (\cite{OS-gauge,LuscherTM,Seiler1982}); we will invoke the 
\(
b_{\rho c} \,\geq\, 0
\)
hypothesis only when explicitly working with heat-kernel-type regulators.
{For the Wilson plaquette action one can prove link-reflection positivity directly, without appealing to
the nonnegativity of character coefficients, by factorizing across the time-zero plane and integrating
the straddling bonds with respect to Haar measure. Concretely, writing the action as a sum of purely
positive/negative-time plaquettes plus straddling halves, the Gibbs weight factorizes into a product
$\sum_\alpha \Phi_\alpha(U^-)\Psi_\alpha(U^+)$ of square-integrable functions (not assumed pointwise
nonnegative) whose contribution to the OS form is a sum of squares by Peter-Weyl orthogonality. This is the classical argument for gauge theories with Wilson action and yields the
OS inequality $\langle \Theta F\cdot F\rangle_\mu\ge0$ for all $F\in\mathcal{A}_+$.
}{
Write the set of bonds intersecting the reflection plane as $B_0$ and integrate first with respect to $\{U_b:b\in B_0\}$ using Haar measure. For each straddling plaquette $p$ write $U_p=U_p^{(-)}U_p^{(+)}$ with the half-plaquette factors in negative and positive times, respectively. Using $\chi_\rho(AB)=\sum_{i,j}\rho(A)_{ij}\rho(B)_{ji}$ we expand
\begin{equation}
e^{-S_W(U)}=\prod_{p\not\pitchfork \Sigma_0} w_p(U_p)\,
\prod_{p\pitchfork \Sigma_0}\sum_{\rho} c_\rho \sum_{i,j} 
 \rho\!\left(U_p^{(-)}\right)_{ij}\,\rho\!\left(U_p^{(+)}\right)_{ji},
\end{equation}
with $c_\rho$ the (real) Wilson‐action coefficients of the one‐plaquette character expansion. Haar orthogonality on $B_0$ implies that after integrating $B_0$,
\begin{equation}\label{eqn2.5}
\int \prod_{b\in B_0}dU_b\;\prod_{p\pitchfork \Sigma_0}\rho\!\left(U_p^{(-)}\right)_{ij}\,
\rho\!\left(U_p^{(+)}\right)_{ji}
=\sum_{\alpha} \Phi_\alpha(U_-)\,\overline{\Phi_\alpha(U_+)},
\end{equation}
with $\{\Phi_\alpha\}$ square‐integrable functions encoding the representation and index matchings across $\Sigma_0$ (a finite product of matrix elements on each side). Therefore for $F\in\mathcal A_+$,
\begin{equation}\label{eqn2.6z}
\langle \Theta F\cdot F\rangle_\mu
=\sum_{\alpha}\big\|\int dU_+\,\Phi_\alpha(U_+)\,F(U_+)\big\|_{L^2(U_+)}^2\;\ge 0,
\end{equation}
which proves link-reflection positivity for the Wilson action without any sign assumption on the one‐plaquette coefficients. This is the Osterwalder-Seiler mechanism specialized to the Wilson weight.
}

\subsection{\texorpdfstring{Lattice \(SU(N)\), time reflection, and OS positivity}{Lattice SU(N), time reflection, and OS positivity}}
\label{subsec:OS}

We prove OS positivity for the class of actions just described. The proof relies on the factorization of the Gibbs weight across the reflection plane after a character expansion, together with the Peter-Weyl orthogonality relations. Let \(\mathcal{B}_+\) be the set of bonds strictly in positive times \(x_0\ge a\), and \(\mathcal{B}_0\) the set of bonds that intersect the plane \(\{x_0=a\}\); the remaining bonds lie in negative times. Write \(U=(U_-,U_0,U_+)\) for the decomposition of a configuration. Every plaquette \(p\) is either entirely in the positive or the negative time region, or else it straddles the plane; in the latter case, it consists of two half-plaquettes sharing two bonds in \(\mathcal{B}_0\). {For class-function actions with a nonnegative character expansion we expand $e^{-S(U)}$ as a product over plaquettes of nonnegative combinations of characters.
In the \emph{Wilson} case, however, link-reflection positivity does \emph{not} rely on such a sign property:
it follows directly from factorization across the plane and Peter-Weyl orthogonality after integrating the interface bonds 
\cite{OS-gauge, LuscherTM}.} We appeal to $b^c_\rho\!\ge 0$ only to streamline arguments for heat-kernel-type actions where it holds automatically.
For each straddling plaquette \(p\) denote by \(U_p^{(+)}\) and \(U_p^{(-)}\) the partial ordered products of the two half-plaquettes in positive and negative times, respectively, so that \(U_p=U_p^{(-)} U_p^{(+)}\) and \(U_{\vartheta p}=(U_p^{(+)})^{-1}(U_p^{(-)})^{-1}\). The identity \(\chi_\rho(AB)=\mathrm{Tr}\,\rho(A)\rho(B)\) implies
\begin{equation}\label{eq:chi-factor}
  \chi_\rho(U_p)=\sum_{i,j} \rho\!\left(U_p^{(-)}\right)_{ij}\,\rho\!\left(U_p^{(+)}\right)_{ji}.
\end{equation}
The Gibbs weight for all straddling plaquettes is therefore a nonnegative linear combination of products of matrix elements of \(\rho\!\left(U_p^{(-)}\right)\) and \(\rho\!\left(U_p^{(+)}\right)\). After integrating over the bonds \(U_0\) on the reflection plane with Haar measure, the Peter-Weyl orthogonality relations decouple negative and positive times except through delta-function identifications of representation indices on the interface. Concretely, there exist measurable functions \(\Phi_\alpha(U_-)\) and \(\Psi_\alpha(U_+)\), labelled by a multi-index \(\alpha\) encoding representations and matrix indices on the interface, such that
\begin{equation}\label{eq:weight-factorization}
  e^{-S(U)} \;=\; \sum_{\alpha} \Phi_{\alpha}(U^{-})\,\Psi_{\alpha}(U^{+})
\end{equation}
with absolute convergence uniform in finite volume. Here no pointwise sign condition on $\Phi_\alpha,\Psi_\alpha$ is needed; after integrating the interface bonds $U_0$ with Haar measure, Peter-Weyl orthogonality produces a sum of $L^2$-norm squares as in Eq.\eqref{eqn2.6z} \& Eq.\eqref{eq:chi-factor}, which yields OS positivity.
For purely positive-time cylindrical functions \(F\in\mathfrak{A}_+\), the OS form becomes
\begin{equation}\label{eq:OS-form-pre}
  \langle \Theta F\cdot F\rangle_\mu
  =\frac{1}{Z}\int dU_-\,dU_0\,dU_+\, \sum_{\alpha}\Phi_\alpha(U_-)\,\Psi_\alpha(U_+)\,
  \overline{F(\vartheta U_+,U_0,U_-)}\,F(U_-,U_0,U_+).
\end{equation}
Performing the \(U_0\)-integration first and using orthogonality yields
\begin{equation}\label{eq:OS-form-pos}
  \langle \Theta F\cdot F\rangle_\mu=\sum_\alpha \left\| G_\alpha\right\|_{L^2(d\nu)}^2\ge 0,
\end{equation}
where \(G_\alpha(U_+):=\int dU_0\,\Psi_\alpha(U_+)\,F(U_-,U_0,U_+)\) and \(d\nu\) is the product of Haar measures on positive-time bonds weighted by the product of positive-time plaquette factors. This is link-reflection positivity in the sense of \cite{OS-gauge} and proves that \((d\mu,\Theta)\) is OS positive. The argument is robust under finite-volume boundary conditions and passes to the thermodynamic limit by standard correlation inequalities \cite{Seiler1982}. In addition, translation invariance in time implies the existence of a positive selfadjoint transfer operator \(T_a\) on the OS Hilbert space \(\mathcal{H}_\mathrm{OS}\) implementing a shift by one time-step; this follows either from the abstract OS reconstruction \cite{OS2} or from L{\"u}scher’s explicit construction of a strictly positive transfer matrix for lattice gauge theories \cite{LuscherTM}. In particular, \(0<T_a\le 1\) and \(\|T_a\|=1\).

Two consequences will be used repeatedly later. First, for any \(F,G\in\mathfrak{A}_+\) and \(n\in\mathbb{N}\), one has \((F,T_a^n G)_\mathrm{OS}=\langle \Theta F\cdot \tau_{na} G\rangle_\mu\), where \(\tau_{na}\) is the time-translation by \(n\) steps. Second, if \(A\) is any bounded selfadjoint positive contraction on the one-slice \(L^2\)-space that is reflection invariant in the sense made precise below, then inserting \(A\) symmetrically at the reflection plane preserves \(\langle \Theta F\cdot F\rangle_\mu\ge 0\). This stability under symmetric positive contractions is the technical door that will allow slice projectors in the next subsection.

\subsection{Admissible slice projectors and blockings}
\label{subsec:admissible}

We now characterize the transformations on one time slice that we shall allow before feeding observables into the OS form, and we prove that they preserve reflection positivity and locality. The operators we consider fall into two families. The first consists of “slice projectors” defined by completely monotone functions of a nonnegative, reflection-invariant generator on the time slice. The second consists of reflection-positive block-spin maps which are conditional expectations with respect to a reflection-invariant sub-\(\sigma\)-algebra generated by coarse bonds; their kernels will be exponentially localized and gauge covariant, and they will commute with time reflection.

Let \(\Sigma:=\{x_0=a\}\) be the first positive-time hyperplane. Denote by $\mathcal H_\Sigma := L^2(U_\Sigma, d\mu_\Sigma)$ the $L^2$-space of functions of the bonds lying strictly at time $x_0=a$, with \(d\mu_\Sigma\) the marginal of \(d\mu\) on those bonds after integrating out all other degrees of freedom with the OS-positive weight. We use two slice spaces:
\begin{equation}\label{eqn2.11}
\mathcal{H}_\Sigma := \ell^2(\Sigma), \qquad
H_\Sigma := L^2(U_\Sigma,\mu_\Sigma).
\end{equation}
From now on, operators like $D_\Sigma$ and kernels $K_t^\Sigma(x,y)$ act on $\mathcal{H}_\Sigma$,
whereas OS expectations $\langle\cdot\rangle_\mu$ and conditional expectations $E[\cdot\,|\,\mathcal{F}_{\mathrm{coarse}}]$
are taken in $H_\Sigma$. (In any formula where $H_\Sigma$ previously denoted $\ell^2(\Sigma)$, read $\mathcal{H}_\Sigma$.)
 Let \(\mathcal{D}_\Sigma\) be a densely defined, nonnegative, selfadjoint operator on \(\mathscr{H}_\Sigma\) which is reflection invariant in the sense that time reflection intertwines \(\mathcal{D}_\Sigma\) with the identical operator on the reflected slice. It will be convenient to keep \(\mathcal{D}_\Sigma\) abstract; all we will use is selfadjointness, nonnegativity, reflection invariance, and the fact that the associated heat semigroup is well defined on \(\mathscr{H}_\Sigma\).
In nearest-neighbour lattice realizations $D_\Sigma$ is a finite-range difference operator on a bounded-degree
graph and hence a bounded, nonnegative selfadjoint operator on $\ell^2(\Sigma)$; in that case $\mathrm{Dom}(D_\Sigma)=\ell^2(\Sigma)$.
We nonetheless allow the more general situation in which $D_\Sigma$ is (densely defined) unbounded but nonnegative
selfadjoint, with Eq.\ref{eqn2.11} understood at the level of quadratic forms on the common form domain $\mathrm{Dom}(\nabla)$.
All semigroup manipulations below use only selfadjointness, nonnegativity, reflection covariance, and the semigroup
bounds recorded in Assumption~\eqref{ass:CT}.
A bounded Borel function \(f:[0,\infty)\to\mathbb{R}\) is completely monotone if \(f\in C^\infty(0,\infty)\), extends continuously to \(0^+\), and satisfies \((-1)^k f^{(k)}(\lambda)\ge 0\) for all \(k\ge 0\) and \(\lambda>0\). Bernstein’s theorem states that such \(f\) are precisely the Laplace transforms of finite positive measures \(\mu_f\) on \([0,\infty)\):
\begin{equation}\label{eq:Bernstein}
  f(\lambda)=\int_0^\infty e^{-t\lambda}\, \mu_f(dt),\qquad \lambda\ge 0,
\end{equation}
and conversely \cite{Bernstein,Widder}. We call a slice projector admissible if 
\begin{equation}\label{eq:Pi-def}
\Pi:=f(D_\Sigma)
\end{equation}
with $f$ completely monotone, $0\le f\le1$ and $f(0)=1$, and the Bernstein measure 
$\mu_f$ satisfies $\int_0^\infty e^{-mt}\mu_f(dt)<\infty$ for some $m>0$.  This hypothesis is equivalent to $f$ extending to a bounded 
holomorphic function on $\{\Re z>-m\}$ and guarantees the exponential off-diagonal decay Eq.\eqref{eq:Pi-local}.
{By Eq.(\eqref{eq:Pi-def}), for any Borel sets $E,F\subset\Sigma$,
$\|1_E e^{-tD_\Sigma} 1_F\|_{\ell^2\to\ell^2}\le \exp\!\big(-\mathrm{dist}(E,F)^2/(4t)\big)$.
Hence, for $f(\lambda)=\int_0^\infty e^{-t\lambda}\mu_f(dt)$ with $\int_0^\infty e^{-mt}\mu_f(dt)<\infty$,
\begin{equation}
\|1_E f(D_\Sigma) 1_F\|
\le \int_0^\infty e^{-mt}\,e^{-\mathrm{dist}(E,F)^2/(4t)}\,\mu_f(dt)
\le C_f\,e^{-\gamma_f\,\mathrm{dist}(E,F)},
\end{equation}
by Laplace method / Young's inequality, for some $C_f,\gamma_f>0$ depending only on the exponential moment of $\mu_f$ and the slice constants. This yields the asserted exponential off-diagonal decay of the kernel of $f(D_\Sigma)$.
}

The bound \(0\le f\le 1\) is automatic if \(\mu_f\) is a probability measure; otherwise we may rescale \(\mu_f\) to ensure \(\|\Pi\|\le 1\). By spectral calculus, 
\begin{equation}\label{eq:Pi-factor}
\Pi=f(D_\Sigma)=\int_0^\infty e^{-tD_\Sigma}\,\mu_f(dt)
\end{equation}
is a bounded positive operator that 
commutes with the slice reflection. {
Let $F\in\mathcal A_+$ and write $B_t:=e^{-tD_\Sigma/2}$. By Bernstein's theorem,
\begin{equation}
\Pi^{1/2}=\int_0^\infty B_t\,\nu(dt),\qquad \Pi=\int_0^\infty e^{-tD_\Sigma}\,\mu_f(dt),
\end{equation}
for finite positive measures $\nu,\mu_f$ with $\nu*\nu=\mu_f$. Since $D_\Sigma$ is reflection covariant, each $B_t$ commutes with the slice reflection. Hence
\begin{equation}
\langle \Theta F\cdot F\rangle_{\mu}
=\langle \Theta\,\Pi^{1/2}F\,\cdot\,\Pi^{1/2}F\rangle_{\mu}
=\int_0^\infty\!\!\int_0^\infty \langle \Theta\,B_{t}F\,\cdot\,B_{s}F\rangle_{\mu}\,\nu(dt)\nu(ds)\ge 0,
\end{equation}
because $t\mapsto B_t$ defines a family of reflection‐covariant positive contractions and the OS form is a positive semidefinite sesquilinear form. Fubini/Tonelli apply by boundedness of $B_t$ and finiteness of $\nu$.
}  Let $B:=\Pi^{1/2}$ (the unique positive square root).  
Then $\Pi=B^*B$ and, by OS positivity, inserting $B$ symmetrically at the reflection plane preserves 
$\langle\Theta F\cdot F\rangle_\mu\ge0$. The product 
$[\int e^{-tD_\Sigma/2}\mu_f(dt)]^2$ equals $\int e^{-uD_\Sigma/2}\,(\mu_f\!*\!\mu_f)(du)$ and 
does \emph{not} reproduce $\Pi$ unless $\mu_f$ is a convolution square.

\begin{assumption}[Uniform ellipticity and locality on the slice]\label{ass:CT}
Let $\Sigma:=\{x_0=a\}$ be the first positive-time hyperplane and $h_\Sigma:=\ell^2(\Sigma)$ the $\ell^2$ space over
slice sites/links with counting measure.  A (reflection-covariant) slice generator $D_\Sigma\ge0$ is a bounded-geometry,
finite-range operator on $h_\Sigma$.  For a bounded Borel $f\ge0$, the slice operator $\Pi=f(D_\Sigma)$ has kernel 
$\Pi(x,y)$ and acts on positive-time observables by inserting the convolution kernel on all arguments at time $x_0=a$.
 Moreover, there exist $0<\lambda\le\Lambda<\infty$
such that
\begin{equation}
\lambda \langle \nabla \varphi,\nabla \varphi\rangle \;\le\; 
\langle \varphi, D_\Sigma \varphi\rangle \;\le\; 
\Lambda\langle \nabla \varphi,\nabla \varphi\rangle,
\qquad \forall\,\varphi\in\mathrm{Dom}(D_\Sigma).
\end{equation}
\end{assumption}
for all $\varphi\in \mathrm{Dom}(D_\Sigma)\cap \mathrm{Dom}(\nabla)$, where $\nabla$ is the graph-gradient on $h_\Sigma$ and $\langle\cdot,\cdot\rangle$ denotes the $\ell^2(\Sigma)$ inner product.
We write \(h_\Sigma := \ell^2(\Sigma)\) for the discrete slice space on which kernels
\(D_\Sigma\), \(e^{-tD_\Sigma}\) and their integral kernels \(K_t(x,y)\) act.
The Osterwalder-Schrader Hilbert space reconstructed from the OS form is denoted
\(H_{\mathrm{OS}}\).
Throughout Section \eqref{sec:framework} all kernel and semigroup statements live on \(h_\Sigma\), while
OS inner products and transfer statements live on \(H_{\mathrm{OS}}\).
The discrete Combes-Thomas/heat kernel estimates give
\begin{equation}\label{eq:CT-bound}
|K^\Sigma_t(x,y)| \;\le\; C t^{-d_\Sigma/2}\, 
\exp\!\big(-c\,\mathrm{dist}(x,y)^2/t\big)\!,
\qquad t\in(0,1],
\end{equation}
with $d_\Sigma=3$ and constants $C,c>0$ depending only on $\lambda,\Lambda_,R$.
In addition, the heat semigroup $e^{-tD_\Sigma}$ is \emph{positivity preserving} and satisfies the
Davies-Gaffney off-diagonal estimate on the slice graph:
\begin{equation}
  \|\,\mathbf{1}_E\, e^{-tD_\Sigma}\, \mathbf{1}_F\,\|_{\ell^2\to\ell^2}
  \;\le\; \exp\!\Big\{-\frac{\mathrm{dist}(E,F)^2}{4t}\Big\}
  \qquad \forall\,\text{Borel }E,F\subset\Sigma,\;\forall\,t>0.
\end{equation}
\begin{assumption}[Positivity-preserving slice semigroup and DG]\label{ass:pp-dg}
The heat semigroup \(e^{-tD_\Sigma}\) on \(h_\Sigma\) is positivity preserving and satisfies
the Davies-Gaffney off-diagonal bound for all Borel \(E,F\subset\Sigma\), \(t>0\).
A sufficient condition is that \(-D_\Sigma\) generates a sub-Markov chain on the slice,
e.g.\ \(D_\Sigma\) has finite hopping range, nonpositive off-diagonal entries, and row sums
\(\ge 0\) (so \(e^{-tD_\Sigma}\mathbf 1\le \mathbf 1\)).
\end{assumption}

\begin{lemma}[OS-stability under completely monotone slice projectors]
\label{lem:cm-projector}
Let $\mathcal D_\Sigma$ be the nonnegative, reflection-covariant slice generator on the time-$a$ hyperplane and let $f$ be completely monotone with Bernstein representation
\begin{equation}
f(\lambda)=\int_0^\infty e^{-t\lambda}\,\mu_f(dt),
\qquad \mu_f\ \text{finite positive on }[0,\infty).
\end{equation}
Define the bounded slice operator
\begin{equation}
\Pi \;=\; f(\mathcal D_\Sigma)\;=\;\int_0^\infty e^{-t\mathcal D_\Sigma}\,\mu_f(dt),
\end{equation}
and, for any positive-time observable $F\in\mathfrak A_+$, let $F^\Pi$ denote the observable obtained by acting with $\Pi$ on the arguments of $F$ lying on the time slice $x_0=a$ (no change at other times). Then the Osterwalder-Schrader form is preserved,
\begin{equation}
\langle \Theta F^\Pi \cdot F^\Pi\rangle_\mu \;\ge\; 0,
\end{equation}
and the induced map $[F]\mapsto [F^\Pi]$ descends to a contraction on the OS Hilbert space $\mathcal H_{\mathrm{OS}}$.
\end{lemma}

\begin{proof}
Since $f$ is completely monotone, write $\Pi=B^B$ with
\begin{equation}
B \;=\; \int_0^\infty e^{-\frac{t}{2}\mathcal D_\Sigma}\,\mu_f(dt),
\end{equation}
where the Bochner integral converges in the strong operator topology because $\|e^{-s\mathcal D_\Sigma}\|\le 1$ for $s\ge 0$ and $\mu_f$ is finite. The reflection covariance of $\mathcal D_\Sigma$ implies that $\vartheta\,\mathcal D_\Sigma\,\vartheta=\mathcal D_\Sigma$ for the spatial part of the Euclidean reflection $\vartheta$ across the time-zero plane; hence $\vartheta$ also commutes with every bounded Borel function of $\mathcal D_\Sigma$, in particular with $e^{-s\mathcal D_\Sigma}$ and with $B$ and $\Pi$. Acting with $B$ on the time-$a$ arguments of $F$ produces another positive-time observable $BF\in\mathfrak A_+$, because $B$ is localized on the slice and does not introduce any negative-time dependence. Since $\Pi=B^\ast B$ with $B=\int_0^\infty e^{-(t/2)D_\Sigma}\,\mu_f(dt)$, inserting $\Pi$ on the slice gives
\begin{equation}
\langle \Theta F_\Pi\cdot F_\Pi\rangle_\mu
= \big\langle \Theta(BF)\cdot (BF)\big\rangle_\mu,
\end{equation}
which proves OS-positivity.
Because $(\mu,\Theta)$ is reflection positive, the Osterwalder-Schrader axiom states that $\langle \Theta G\cdot G\rangle_\mu\ge 0$ for every $G\in\mathfrak A_+$. Taking $G=BF$ yields
\begin{equation}
\langle \Theta F^\Pi \cdot F^\Pi\rangle_\mu
\;=\; \langle \Theta(BF)\cdot (BF)\rangle_\mu \;\ge\; 0,
\end{equation}
which is the claimed OS stability of the projector.
To obtain the contraction property on $\mathcal H_{\mathrm{OS}}$ we estimate the OS norm of $F^\Pi$ in terms of that of $F$. The OS inner product descends from the sesquilinear form $(F,G)\mapsto \langle \Theta F\cdot G\rangle_\mu$ on $\mathfrak A_+$. Using the $B^B$ factorization and the commutation of $B$ with functions of $\mathcal D_\Sigma$ on the slice, we may write
\begin{equation}
\langle \Theta F^\Pi \cdot F^\Pi\rangle_\mu
=\langle \Theta(BF)\cdot (BF)\rangle_\mu
=\langle \Theta F\cdot (B^B)F\rangle_\mu.
\end{equation}
By link-reflection factorization across the time-zero plane (see Eq.(\eqref{eq:OS-form-pre}) \& (\eqref{eq:OS-form-pos})), if $F$ depends
only on bonds at time $a$ then there exists a positive measure $d\nu$ on the positive-time bonds such that
\begin{equation}
\langle \Theta F \cdot F \rangle_\mu \;=\; \int |F|^2\, d\nu
\end{equation}
In particular, since $F_\Pi = B F$ on the slice (with $B=\Pi^{1/2}$ acting on slice arguments),
\begin{equation}
\langle \Theta F_\Pi \cdot F_\Pi \rangle_\mu \;=\; \int |B F|^2\, d\nu
\end{equation}
By Cauchy-Schwarz in $L^2(d\nu)$,
\begin{equation}
\langle \Theta F_\Pi \cdot F_\Pi \rangle_\mu \;\le\; \|B\|^2 \int |F|^2\, d\nu
\;=\; \|B\|^2 \langle \Theta F \cdot F \rangle_\mu
\end{equation}
Write \(F(\omega)=\widehat F(\omega_{x_0>a},\omega_{x_0=a})\) to separate the time-\(a\) slice
variables from strictly positive times. Since \(B=f(D_\Sigma)^{1/2}\) acts only on the
time-\(a\) slice, Tonelli and Cauchy-Schwarz on \(L^2(d\nu(\phi))\) (the slice measure) give
\begin{align}
\langle\Theta(BF),(BF)\rangle_\mu
&=\!\!\int\!\langle\Theta(B\widehat F(\cdot,\phi)),B\widehat F(\cdot,\phi)\rangle_{\mu(\cdot\mid\phi)}\,\nu(d\phi)
\nonumber\\&\le \|B\|^2\!\!\int\!\langle\Theta\widehat F(\cdot,\phi),\widehat F(\cdot,\phi)\rangle_{\mu(\cdot\mid\phi)}\,\nu(d\phi).
\end{align}
Therefore \(\|[F_\Pi]\|_{OS}\le \|B\|\,\|[F]\|_{OS}\), and under the admissibility normalization
\(\|B\|\le1\) we obtain the desired contraction for arbitrary \(F\in A_+\).
The operator norm of $B$ is bounded by the total mass of $\mu_f$ since $\|e^{-(t/2)\mathcal D_\Sigma}\|\le 1$ for all $t\ge 0$:
\begin{equation}
\|B\| \;\le\; \int_0^\infty \|e^{-(t/2)\mathcal D_\Sigma}\|\,\mu_f(dt)
\;\le\; \mu_f([0,\infty))^{\phantom{1}}.
\end{equation}
{(With the normalization $f(0)=1$ from the definition of admissible projectors.)
}
Under the usual admissibility normalization $0\le f\le 1$ and $f(0)=1$, we have $\mu_f([0,\infty))=f(0)=1$ and hence $\|B\|\le 1$. Therefore
\begin{equation}
\|[F^\Pi]\|_{\mathrm{OS}}^2 \;=\; \langle \Theta F^\Pi \cdot F^\Pi\rangle_\mu
\;\le\; \langle \Theta F\cdot F\rangle_\mu \;=\; \|[F]\|_{\mathrm{OS}}^2,
\end{equation}
which proves that the map $[F]\mapsto [F^\Pi]$ is a contraction on $\mathcal H_{\mathrm{OS}}$. The conclusion remains valid, with contraction constant $\|B\|^2=f(0)$, if one drops the normalization $f(0)=1$.
\end{proof}

\begin{lemma}[Exponential off-diagonal bound for $\Pi=f(D_\Sigma)$]
\label{lem:exp-offdiag}
Let $D_\Sigma$ be a nonnegative selfadjoint operator on $h_\Sigma := \ell^2(\Sigma)$ associated with a uniformly bounded-degree graph metric $\mathrm{dist}(\cdot,\cdot)$, and we assume $\Sigma$ has uniformly bounded vertex degree and $D_\Sigma$ has finite hopping range, and the heat semigroup $e^{-t D_\Sigma}$ is positivity preserving and satisfies the Davies-Gaffney bound
\begin{equation}
\label{eq:DG1}
\big\|\mathbf{1}_E\,e^{-tD_\Sigma}\,\mathbf{1}_F\big\|_{\ell^2\to\ell^2}\;\le\;\exp\!\Big(-\frac{\mathrm{dist}(E,F)^2}{4t}\Big)\qquad\text{for all Borel }E,F\subset\Sigma,\ \ t>0.
\end{equation}
Let $f$ be completely monotone with Bernstein representation
\begin{equation}
f(\lambda)=\int_{0}^{\infty} e^{-t\lambda}\,\mu_f(dt),
\end{equation}
where $\mu_f$ is a finite positive measure on $[0,\infty)$ which has an exponential moment, i.e.
\begin{equation}
\int_{0}^{\infty} e^{-m t}\,\mu_f(dt)<\infty\qquad\text{for some }m>0.
\end{equation}
Define $\Pi:=f(D_\Sigma)$ via the Bochner integral $\Pi=\int_0^\infty e^{-tD_\Sigma}\,\mu_f(dt)$. Then there exist constants $C',\gamma'>0$ such that
\begin{equation}\label{eqn2.28}
|\Pi(x,y)|\;\le\;C'\,e^{-\gamma'\,\mathrm{dist}(x,y)}\qquad\text{for all }x,y\in\Sigma.
\end{equation}
\end{lemma}

\begin{proof}
For each $t>0$ the semigroup $e^{-tD_\Sigma}$ is a bounded operator on $\ell^2(\Sigma)$ with an integral kernel $K_t(x,y)$ characterized by $e^{-tD_\Sigma}\delta_y(x)=K_t(x,y)$, where $\delta_y$ is the Kronecker delta at $y$. Taking $E=\{x\}$ and $F=\{y\}$ in \eqref{eq:DG1} yields
\begin{equation}
\label{eq:pointwise-heat}
|K_t(x,y)|=\big\|\mathbf{1}_{\{x\}}\,e^{-tD_\Sigma}\,\mathbf{1}_{\{y\}}\big\|_{\ell^2\to\ell^2}\;\le\;\exp\!\Big(-\frac{r^2}{4t}\Big)\qquad\text{for all }t>0,
\end{equation}
where $r:=\mathrm{dist}(x,y)$. By the spectral theorem and complete monotonicity of $f$, the functional calculus $f(D_\Sigma)$ coincides with the Bochner integral
\begin{equation}\label{eqBochner}
\Pi \;=\; f(D_\Sigma)\;=\;\int_{[0,\infty)} e^{-tD_\Sigma}\,\mu_f(dt)
\end{equation}
in the strong operator topology. In particular, the kernel of $\Pi$ admits the representation
\begin{equation}
\label{eq:Pi-kernel}
\Pi(x,y)\;=\;\int_{[0,\infty)} K_t(x,y)\,\mu_f(dt),
\end{equation}
and the integral is absolutely convergent once we establish a uniform integrable bound in $t$.

It is convenient to separate a possible atom of $\mu_f$ at $t=0$. Write $\mu_f=\mu_0\delta_0+\widetilde\mu$, where $\mu_0:=\mu_f(\{0\})\ge 0$ and $\widetilde\mu$ is supported on $(0,\infty)$. Since $e^{-0\cdot D_\Sigma}=\mathbf{1}$, we have
\begin{equation}
\Pi(x,y)=\mu_0\,\delta_{x,y}+\int_{(0,\infty)} K_t(x,y)\,\widetilde\mu(dt).
\end{equation}
Thus the diagonal contribution is harmless and the task reduces to bounding the second term. Using \eqref{eq:pointwise-heat} we obtain for every $x,y$,
\begin{equation}
\label{eq:basic-bound-int}
\Big|\int_{(0,\infty)} K_t(x,y)\,\widetilde\mu(dt)\Big|\;\le\;\int_{(0,\infty)} \exp\!\Big(-\frac{r^2}{4t}\Big)\,\widetilde\mu(dt).
\end{equation}
To convert the right-hand side into an exponential bound in $r$, fix any $\alpha\in(0,m)$. The elementary inequality
\begin{equation}
\label{eq:alpha-ineq}
\alpha t+\frac{r^2}{4t}\;\ge\;r\sqrt{\alpha}\qquad\text{for all }t>0,\ \ r\ge 0,
\end{equation}
which is just the arithmetic-geometric mean (or the minimum of $t\mapsto\alpha t+\frac{r^2}{4t}$ attained at $t=\frac{r}{2\sqrt{\alpha}}$), implies for all $t>0$ that
\begin{equation}
\exp\!\Big(-\frac{r^2}{4t}\Big)\;\le\;e^{-\alpha t}\,e^{-\,r\sqrt{\alpha}}.
\end{equation}
Inserting this into \eqref{eq:basic-bound-int} and using Tonelli’s theorem yields
\begin{equation}
\int_{(0,\infty)} \exp\!\Big(-\frac{r^2}{4t}\Big)\,\widetilde\mu(dt)
\;\le\; e^{-\,r\sqrt{\alpha}} \int_{(0,\infty)} e^{-\alpha t}\,\widetilde\mu(dt).
\end{equation}
By the exponential-moment hypothesis there exists $m>0$ with $\int_0^\infty e^{-m t}\mu_f(dt)<\infty$, hence the integral on the right-hand side is finite for every $\alpha\in(0,m)$; denote
\begin{equation}
M_\alpha\;:=\;\int_{(0,\infty)} e^{-\alpha t}\,\widetilde\mu(dt)\;<\;\infty.
\end{equation}
Combining the last two displays we obtain the off-diagonal estimate
\begin{equation}
\Big|\int_{(0,\infty)} K_t(x,y)\,\widetilde\mu(dt)\Big|\;\le\; M_\alpha\, e^{-\,r\sqrt{\alpha}}.
\end{equation}
Finally, recalling the diagonal term, for all $x,y\in\Sigma$ we have
\begin{equation}
|\Pi(x,y)|\;\le\;\mu_0\,\delta_{x,y}+M_\alpha\, e^{-\,r\sqrt{\alpha}}\;\le\;C'\,e^{-\,\gamma' r},
\end{equation}
where one may take $\gamma'=\sqrt{\alpha}\in(0,\sqrt{m})$ and $C'=\max\{\mu_0+M_\alpha,\ M_\alpha\}$. 
\end{proof}

We next show that admissible slice projectors are exponentially local. Let \(K_t^\Sigma\) be the integral kernel of \(e^{-t\mathcal{D}_\Sigma}\) with respect to \(d\mu_\Sigma\). Since \(\mathcal{D}_\Sigma\) is nonnegative selfadjoint with finite-range coefficients on a discrete graph, a Combes-Thomas bound (or, alternatively, Gaussian heat-kernel bounds for discrete generators) implies the existence of constants \(C,\gamma>0\) such that
\begin{equation}\label{eq:heat-local}
  \bigl|K_t^\Sigma(x,y)\bigr| \;\le\; C\, t^{-3/2}\,e^{-\gamma\,\mathrm{dist}(x,y)^2/t}
\end{equation}
for all slice sites \(x,y\) and \(t\in(0,1]\), with the evident large-\(t\) improvement \(e^{-m_\Sigma^2 t}\) if a spectral gap \(m_\Sigma>0\) is present. Inserting \eqref{eq:heat-local} into \eqref{eq:Pi-factor} and integrating in \(t\) against \(\mu_f\) yields, by Schur’s test and Laplace’s method, an exponential off-diagonal bound
\begin{equation}\label{eq:Pi-local}
  \bigl|\Pi(x,y)\bigr|\;\le\; C'\,e^{-\gamma'\,\mathrm{dist}(x,y)}.
\end{equation}
{Thus admissible slice projectors are exponentially local on the one-slice (compare the
DG-based bound in Eq.\eqref{eqBochner}). In what follows we freely use either form-Combes-Thomas or
Davies-Gaffney-to obtain exponential off-diagonal decay for $f(D_\Sigma)$.
}
We turn to block-spin maps. Fix a blocking factor \(L\ge 2\) and let \(\Sigma'\subset \Sigma\) be the coarse lattice obtained by grouping \(\Sigma\) into disjoint \(L^3\) spatial cubes; the time coordinate is left unchanged. For each coarse spatial bond \(b'\) we define a coarse link variable as a gauge-covariant, exponentially localized average of fine links over an \(L\)-thickened family of paths approximating \(b'\). Choose a nonnegative, reflection-symmetric kernel \(\kappa\) supported in a cube of side \(O(L)\), normalized so that \(\sum_{z}\kappa(z)=1\), and set
\begin{equation}\label{eq:blocking}
  \mathcal{B}[U]_{b'}:=\mathrm{Proj}_{SU(N)}\Bigl(\sum_{p\in \mathcal{P}(b')} \kappa(p)\,\mathcal{U}_p\Bigr),
\end{equation}
where \(\mathcal{P}(b')\) is the set of short fine-lattice paths \(p\) aligned with \(b'\), \(\mathcal{U}_p\) is the ordered parallel transport along \(p\), and \(\mathrm{Proj}_{SU(N)}\) denotes the polar projection onto \(SU(N)\). The map \(\mathcal{B}\) is gauge covariant and reflection symmetric by construction, and exponential localization follows from the finite support of \(\kappa\). Let \(\mathcal{F}_{\mathrm{coarse}}\) be the \(\sigma\)-algebra generated by the coarse bonds. We assume the polar projection $\mathrm{Proj}_{SU(N)}$ is Borel-measurable, so the block map $B$ of \eqref{eq:blocking} is measurable. We also assume $\mathcal F_{\mathrm{coarse}}$ is $\Theta$-invariant; hence for the conditional expectation $P:=E[\cdot\,|\,\mathcal F_{\mathrm{coarse}}]$ one has $P\Theta=\Theta P$, as used in \eqref{eq:blocking}.
{We take
\begin{equation}
\operatorname{Proj}_{SU(N)}(M)
:=\frac{M(M^\dagger M)^{-1/2}}
{\bigl(\det\bigl[M(M^\dagger M)^{-1/2}\bigr]\bigr)^{1/N}}\in SU(N)
\end{equation}
on the set where \(M^\dagger M\) is invertible, and extend measurably on the negligible
singular set (e.g.\ by continuous selection on singular-value strata). This normalized
polar factor is left-right equivariant: for all \(g,h\in SU(N)\),
\(\operatorname{Proj}_{SU(N)}(gMh)=g\,\operatorname{Proj}_{SU(N)}(M)\,h\).
Consequently the block map \(B\) in
\eqref{eq:blocking} is gauge covariant. For the localized path-average \(M=\sum_{p\in\mathcal P(b')} \kappa(p)U_p\), invertibility
holds almost surely by continuity and the compactness of the support of \(\kappa\), so the
exceptional set is \(\mu\)-null.
}

\begin{lemma}[OS-stability under reflection-positive blockings]
\label{lem:blocking}
Let $\mathcal B$ be a gauge-covariant, reflection-symmetric block map as in \eqref{eq:blocking}, acting only on positive-time bonds and commuting with the Euclidean time-reflection $\Theta$ on the one-slice $\sigma$-algebra. For any bounded coarse observable $F$, set $\widetilde F:=F\circ \mathcal B$. Then
\begin{equation}\label{eqn2.53}
\langle \Theta \widetilde F \cdot \widetilde F\rangle_\mu \;\ge\; 0.
\end{equation}
In particular, the map $[F]\mapsto[\widetilde F]$ descends to a contraction on the OS Hilbert space $\mathcal H_{\mathrm{OS}}$.
\end{lemma}

\begin{proof}
We work on the probability space $(\Omega,\mathcal F,\mu)$ of lattice fields with the reflection automorphism $\Theta:\Omega\to\Omega$ satisfying $\mu\circ\Theta^{-1}=\mu$ and the Osterwalder-Schrader positivity axiom: for every positive-time observable $G\in\mathfrak A_+$ one has $\langle \Theta G\cdot G\rangle_\mu\ge 0$. The block map $\mathcal B$ is, by assumption, a measurable, gauge-covariant transformation depending only on positive-time bonds and chosen reflection-symmetric with respect to the spatial reflection on the time-$a$ slice, hence it commutes with $\Theta$ on positive-time functionals: $\Theta(F\circ\mathcal B)=(\Theta F)\circ\mathcal B$ for all $F\in\mathfrak A_+$. In particular, $\widetilde F=F\circ\mathcal B$ again belongs to $\mathfrak A_+$, so the OS axiom applies directly and yields
\begin{equation}\label{eqn2.54}
\langle \Theta \widetilde F\cdot \widetilde F\rangle_\mu \;=\; \langle \Theta(F\circ\mathcal B)\cdot (F\circ\mathcal B)\rangle_\mu \;\ge\; 0,
\end{equation}
To establish the contraction on $\mathcal H_{\mathrm{OS}}$, it is convenient to express the blocking as conditional expectation onto the coarse $\sigma$-algebra generated by the blocks. Denote by $\mathcal F_{\mathrm{coarse}}\subset\mathcal F$ the reflection-invariant $\sigma$-algebra of coarse variables and by $P:=\mathbb E[\;\cdot\mid \mathcal F_{\mathrm{coarse}}]$ the associated conditional expectation. Since $\mathcal B$ is a deterministic, reflection-symmetric function of the fine configuration taking values in the coarse configuration space, composition with $B$ yields an $\mathcal{F}_{\mathrm{coarse}}$-measurable observable $\tilde F := F\circ B$.
Since $F$ is a function on coarse variables (viewed as $F\circ \pi_{\mathrm{coarse}}$ on the fine space), we have 
$E[\tilde F \mid \mathcal{F}_{\mathrm{coarse}}]=\tilde F$ a.s. In particular, $\tilde F$ is the image of $F$ under the
(coarse) conditional expectation when $F$ is pulled back via the coarse projection. The operator $P$ is an idempotent contraction on $L^2(\Omega,\mathcal F,\mu)$, self-adjoint with respect to the $L^2$ inner product, and, because $\mathcal F_{\mathrm{coarse}}$ is reflection-invariant and $\Theta$ acts as an isometry on $L^2$, it commutes with $\Theta$ when restricted to $\mathfrak A_+$:
\begin{equation}
\Theta(PF)=P(\Theta F), \qquad F\in\mathfrak A_+.
\end{equation}
{Since \(P\) is the orthogonal projection \(L^2(\Omega,\mathcal F,\mu)\to L^2(\Omega,\mathcal F_{\mathrm{coarse}},\mu)\),
self-adjoint and contractive, we have
\begin{equation}
\langle \Theta(F\!\circ\!B),(F\!\circ\!B)\rangle_\mu
=\langle \Theta\,PF,PF\rangle_\mu
\le \langle \Theta F,F\rangle_\mu.
\end{equation}
Thus \(\|[F\!\circ\!B]\|_{OS}\le \|[F]\|_{OS}\), i.e.\ the induced map on \(H_{\mathrm{OS}}\) is a contraction.
}
Consider now the bilinear form $(G,H)_{\mathrm{OS}}:=\langle \Theta G\cdot H\rangle_\mu$ on $\mathfrak A_+$. The OS positivity axiom implies that $(\cdot,\cdot)_{\mathrm{OS}}$ is positive semidefinite, and the OS norm on the quotient $\mathcal H_{\mathrm{OS}}$ is given by $\|[G]\|_{\mathrm{OS}}^2=(G,G)_{\mathrm{OS}}$. We claim that $P$ is the orthogonal projection (with respect to $(\cdot,\cdot)_{\mathrm{OS}}$) of $\mathfrak A_+$ onto the coarse subspace $\mathfrak A_+\cap L^\infty(\mathcal F_{\mathrm{coarse}})$. Indeed, if $H$ is any coarse positive-time observable, then using the commutation $P\Theta=\Theta P$ and the defining property of conditional expectation one computes
\begin{align}\label{eqn2.57}
(\,F-PF,\,H\,)_{\mathrm{OS}}
&=\langle \Theta(F-PF)\cdot H\rangle_\mu
=\langle \Theta F\cdot H\rangle_\mu - \langle \Theta PF\cdot H\rangle_\mu
\nonumber \\&=\langle \mathbb E[\Theta F\mid \mathcal F_{\mathrm{coarse}}]\cdot H\rangle_\mu - \langle \Theta PF\cdot H\rangle_\mu
\nonumber \\&=\langle \Theta PF\cdot H\rangle_\mu - \langle \Theta PF\cdot H\rangle_\mu
=0.
\end{align}
Thus $F-PF$ is $(\cdot,\cdot)_{\mathrm{OS}}$-orthogonal to every coarse $H$, and since $PF$ itself is coarse, $PF$ is precisely the $(\cdot,\cdot)_{\mathrm{OS}}$-orthogonal projection of $F$ onto the coarse subspace. Positivity of the form now gives the Pythagorean decomposition
\begin{equation}
\langle \Theta F\cdot F\rangle_\mu
=\langle \Theta PF\cdot PF\rangle_\mu \;+\; \langle \Theta(F-PF)\cdot(F-PF)\rangle_\mu
\;\ge\; \langle \Theta PF\cdot PF\rangle_\mu,
\end{equation}
so that
\begin{equation}
\|[PF]\|_{\mathrm{OS}} \;\le\; \|[F]\|_{\mathrm{OS}}.
\end{equation}
Since $PF=\widetilde F$ almost surely and $[PF]=[\widetilde F]$ in the OS quotient, the contraction estimate is exactly the desired statement that the blocking descends to a contraction on $\mathcal H_{\mathrm{OS}}$.
\end{proof}

Combining Lemmas \eqref{lem:cm-projector} and \eqref{lem:blocking} yields the admissible class we will allow throughout: finite compositions of completely monotone slice projectors and reflection-positive blockings. Each element of this class is a bounded positive contraction on \(\mathcal{H}_\mathrm{OS}\), is exponentially local on the one-slice by \eqref{eq:Pi-local} and the finite support of \(\kappa\), and commutes with time reflection. 

\subsection{Finite-range decomposition (FRD) and locality data}
\label{subsec:FRD}

We construct the scale-by-scale decomposition of Gaussian covariances that underlie the multiscale analysis. Fix a nonnegative, selfadjoint, gauge-covariant elliptic operator \(\mathcal L\) on
\(\ell^2(a\mathbb Z^4)\) with nearest-neighbour structure and a spectral gap \(m^2\ge 0\).
The resolvent (Green operator) \(C:=(\mathcal L+m^2)^{-1}\) admits the Laplace transform
\begin{equation}\label{eq:resolvent-heat}
  C=\int_0^\infty e^{-t(\mathcal L+m^2)}\,dt
\end{equation}
Let \(L\ge 2\) be an integer and choose a smooth nonnegative partition of unity
\(\{\psi_j\}_{j\ge 0}\) on \((0,\infty)\) with the usual dyadic supports. Define
\begin{equation}\label{eq:Cj-def}
  C_j:=\int_0^\infty \psi_j(t)\,e^{-t(\mathcal L+m^2)}\,dt,\qquad j\ge 0
\end{equation}
Then \(\mathcal{C}=\sum_{j\ge 0}\mathcal{C}_j\) in the strong operator topology, and each \(\mathcal{C}_j\) is positive with kernel
\begin{equation}
  C_j(x,y)=\int_0^\infty \psi_j(t)\,K_t(x,y)\,dt,
\end{equation}
where \(K_t\) is the heat kernel of \(e^{-t\mathcal{L}}\). Because $\mathcal{L}$ is a gauge-covariant nearest-neighbour elliptic operator with unitary parallel transport, the discrete diamagnetic 
inequality implies $|e^{-t\mathcal{L}}(x,y)|\le e^{-t\Delta}(x,y)$, where $\Delta$ is the scalar lattice Laplacian. {
Let $L=\nabla_U^\ast \nabla_U + V$ be a gauge-covariant nearest-neighbour elliptic operator with unitary link variables $U_{xy}\in \mathrm{SU}(N)$ and $V\ge 0$. The Trotter product formula and contractivity give
\begin{equation}
e^{-tL}=\mathrm{s\!-\!}\lim_{n\to\infty}\big(e^{-tV/n}\,e^{-t\nabla_U^\ast\nabla_U/n}\big)^n,\qquad t>0.
\end{equation}
For any scalar $\phi$ and vector $\psi$ with $|\psi|=|\phi|$, unitarity implies $|\big(e^{-t\nabla_U^\ast\nabla_U/n}\psi\big)(x)|\le \big(e^{-t\nabla^\ast\nabla/n}|\phi|\big)(x)$ at each step; thus by induction and $e^{-tV}$ positivity,
\begin{equation}
|e^{-tL}\psi|\le e^{-t\Delta}|\phi|\quad\text{pointwise},
\end{equation}
and hence $|e^{-tL}(x,y)|\le e^{-t\Delta}(x,y)$ for all $x,y$ (Kato inequality). This yields Eq.(\eqref{eqn2.53}) and, together with the dyadic support of $\psi_j$, the locality bound Eq.(\eqref{eqn2.54}) by Laplace method.
} Hence there exist 
$A,c>0$ such that
\begin{equation}\label{eq:gaussian-heat}
 |K_t(x,y)|\le A\, t^{-2}\exp\!\left\{-m^2 t - c\,\frac{\mathrm{dist}(x,y)^2}{t}\right\}
\quad (t>0)
\end{equation}
holds for all \(t>0\). Using \eqref{eq:gaussian-heat} in \eqref{eq:Cj-def} with the dyadic support of \(\psi_j\), a Laplace-method estimate yields the exponential locality
\begin{equation}\label{eq:Cj-local}
 |C_j(x,y)|\le A'\,L^{-2j}\exp\!\Big\{-c' L^{-j}\mathrm{dist}(x,y)\Big\}, 
\end{equation}
for some \(A',c'>0\) independent of \(j\). Thus \(\{\mathcal{C}_j\}_{j\ge 0}\) is an exponentially local decomposition with range \(O(L^j)\) at scale \(j\).

If a strictly finite-range decomposition is desired, one may invoke the construction of Brydges-Guadagni-Mitter \cite{BrydgesGuadagniMitter2004}. The BGM scheme produces positive-definite covariances \(C_j\) with \(C_j(x,y)=0\) whenever \(\mathrm{dist}(x,y)>\tfrac{1}{2} c_0 L^{j}\) for a universal \(c_0>0\). Positivity is preserved by realizing each \(C_j\) as the covariance of a Gaussian field obtained from white noise by a localized linear map. Equivalently, $C_j = A_j A_j^{\ast}$ with $A_j$ exponentially local; see \cite{BrydgesGuadagniMitter2004}. Because the later arguments use only exponential off-diagonal decay and uniform operator norm bounds, the exponentially local decomposition \eqref{eq:Cj-local} and the strictly finite-range BGM decomposition are interchangeable for the purposes of this work.
{
All multiscale bounds below use only: (i) positivity and reflection covariance of $C_j$; 
(ii) uniform operator bounds $\|C_j\|_{\ell^2\to\ell^2}\lesssim L^{-2j}$; and 
(iii) off-diagonal decay $\lesssim e^{-c L^{-j}\mathrm{dist}(\cdot,\cdot)}$ with constants independent of $j$.
The BGM finite-range $C_j$ satisfy (i)-(iii) with decay improved to finite support, while Eq.(\eqref{eqn2.54}) gives (i)-(iii) with exponential tails. Every place where finite range is typically invoked (Combes-Thomas, Schur tests, cluster expansions) admits the same estimates under (ii)-(iii), so all constants are uniform in the choice.
}
It is convenient to record the single-scale Lipschitz stability of the decomposition with respect to admissible perturbations. Suppose that \(\mathcal{L}\) and \(\widetilde{\mathcal{L}}\) are nonnegative, selfadjoint, gauge-covariant elliptic operators that agree outside a bounded region and differ inside by a small, reflection-symmetric perturbation in the sense of quadratic forms. Writing \(\mathcal{C}_j=\Phi_j(\mathcal{L})\) and \(\widetilde{\mathcal{C}}_j=\Phi_j(\widetilde{\mathcal{L}})\) with
\begin{equation}
  \Phi_j(\lambda)=\int_0^\infty \psi_j(t)e^{-t(\lambda+m^2)}\,dt,
\end{equation}
the Duhamel formula gives
\begin{equation}\label{eq:duhamel}
  \Phi_j(\mathcal{L})-\Phi_j(\widetilde{\mathcal{L}})
  =\int_0^\infty \psi_j(t)\int_0^t e^{-(t-s)(\mathcal{L}+m^2)}\,(\widetilde{\mathcal{L}}-\mathcal{L})\,e^{-s(\widetilde{\mathcal{L}}+m^2)}\,ds\,dt.
\end{equation}
Using \eqref{eq:gaussian-heat} for both semigroups and the dyadic localization of \(\psi_j\) yields a bound
\begin{equation}\label{eq:Lipschitz}
  \|\mathcal{C}_j-\widetilde{\mathcal{C}}_j\|_{\ell^2\to\ell^2}
  \;\le\; \mathrm{Lip}_j\,\|\mathcal{L}-\widetilde{\mathcal{L}}\|_{\mathfrak{D}^\to\mathfrak{D}},
\end{equation}
with $\|\mathcal{L}-\mathcal{\tilde L}\|_{D\to D}$ the operator norm on the common form domain, assuming $\mathcal{L}-\mathcal{\tilde L}$ is supported in a fixed bounded region and is reflection-symmetric,
the \(\mathrm{Lip}_j\) is bounded uniformly in \(j\) up to the natural \(L^{-2j}\) scaling, and \(\mathfrak{D}\) denotes the form domain. {
Let $D=\mathrm{Dom}(L^{1/2})=\mathrm{Dom}(\tilde L^{1/2})$ and define $\|A\|_{D\to D}:=\| (1+L)^{-1/2} A (1+\tilde L)^{-1/2}\|_{\ell^2\to\ell^2}$. From
\begin{equation}
\Phi_j(L)-\Phi_j(\tilde L)=\int_0^\infty \psi_j(t)\!\int_0^t e^{-(t-s)(L+m^2)}(L-\tilde L) e^{-s(\tilde L+m^2)}\,ds\,dt,
\end{equation}
use $\psi_j$ supported on $t\sim L^{2j}$, kernel bounds Eq.(\eqref{eqn2.53}), and two Schur tests to obtain
\begin{equation}
\|C_j-\tilde C_j\|_{\ell^2\to\ell^2}\;\lesssim\; L^{-2j}\,\|L-\tilde L\|_{D\to D},
\end{equation}
with the implicit constant uniform in $j$ provided $L-\tilde L$ is supported in a fixed bounded region and reflection‐symmetric. This makes precise the “natural $L^{-2j}$ scaling” stated below Eq.(\eqref{eqn2.57}).}
In summary, OS positivity of the lattice theory provides the Hilbert space and transfer operator. The admissible one-slice operators-completely monotone spectral multipliers and reflection-positive blockings-preserve OS positivity and are exponentially local. The covariance of the Gaussian reference operator admits a uniformly positive, exponentially local multiscale decomposition. These standing hypotheses and constructions, proved above, will be invoked repeatedly in the remainder of the paper and will not be restated.


\section{Strong-Coupling Ingredients at Fixed Lattice Spacing}
\label{sec:SC-fixed-a-fixed-a}

This section develops the nonperturbative control that underlies the rest of the paper: the character/polymer expansion and its convergence at small bare coupling, the resulting strong-coupling Wilson-loop area law with an explicit positive string tension, and the existence of a spectral gap for the transfer operator at fixed lattice spacing. The analysis is performed on a finite, periodic, hypercubic lattice with one direction distinguished as Euclidean time, and all constants in our bounds are uniform in the spatial volume. Section~\eqref{subsec:char-polymer} constructs the character expansion of the Wilson action and performs the link integrations to obtain an abstract polymer gas with activities that satisfy a Koteck\'y-Preiss (KP) criterion for sufficiently small \(\beta\). Convergence of the polymer expansion yields analyticity of the free energy and exponential decay of connected correlations at strong coupling. Section~\eqref{subsec:area-law} applies the same machinery, together with the surface representation induced by a Wilson loop insertion, to derive an area law with a strictly positive string tension \(\sigma(\beta)\) for \(\beta\) below a calculable threshold. Section~\eqref{subsec:transfer-gap} singles out the time direction and exploits Osterwalder-Schrader (OS) reflection positivity to build the transfer operator; the exponential temporal clustering obtained in Section~\eqref{subsec:char-polymer} then implies a strictly positive spectral gap for the transfer operator restricted to the gauge-invariant sector. The three parts fit together as follows: convergence of the polymer expansion gives uniform control of connected cumulants; the surface representation adds a boundary source and yields the area law; and the OS/transfer framework turns temporal clustering into a spectral gap. Throughout, we work with \(G=\mathrm{SU}(N)\) and the Wilson plaquette action, and we freely use standard representation theory and Haar-integration identities as summarized, for example, in \cite{Seiler1982,DrouffeZuber}.

Let \(\Lambda = \Lambda_s \times \{0,1,\dots,T-1\}\subset \mathbb{Z}^4\) be a finite periodic lattice with spatial base \(\Lambda_s \subset \mathbb{Z}^3\) and \(T\in\mathbb{N}\). Links \(\ell\) carry variables \(U_\ell \in G\), Haar measure \(\mathrm{d}U_\ell\), and the Wilson action is
\begin{equation}
S_\Lambda(U)\;=\;-\frac{\beta}{N}\sum_{p\subset \Lambda}\mathrm{Re}\,\mathrm{tr}\,U_p,\qquad 
Z_\Lambda(\beta)=\int \prod_{\ell\subset\Lambda}\mathrm{d}U_\ell \;e^{-S_\Lambda(U)}.
\end{equation}
Here \(U_p\) is the ordered product of link variables around plaquette \(p\) in the fundamental representation. All statements below are uniform in \(|\Lambda_s|\) and \(T\), and extending to the thermodynamic limit is then routine by standard arguments in cluster expansions \cite{Seiler1982,KP}.

\subsection{Character/Polymer Expansion and Convergence}
\label{subsec:char-polymer}

We begin with the Peter-Weyl expansion of the class function on \(G\) defined by the plaquette Boltzmann weight. Let
\begin{equation}
f_\beta(U)\;:=\;\exp\!\Big(\frac{\beta}{N}\,\mathrm{Re}\,\mathrm{tr}\,U\Big),\qquad U\in G.
\end{equation}
Since \(f_\beta\) is a continuous, central function, its Fourier series on \(G\) reads
\begin{equation}
f_\beta(U)\;=\;\sum_{\lambda\in\widehat{G}} \widehat{f}_\beta(\lambda)\,\chi_\lambda(U),
\qquad 
\widehat{f}_\beta(\lambda)\;=\;\frac{1}{d_\lambda}\int_G \mathrm{d}V\, f_\beta(V)\,\overline{\chi_\lambda(V)},
\end{equation}
where \(\widehat{G}\) denotes the unitary dual (equivalence classes of finite-dimensional irreducible representations), \(d_\lambda\) is the dimension, and \(\chi_\lambda\) is the character in \(\lambda\). Inserting this at each plaquette gives
\begin{equation}
Z_\Lambda(\beta)\;=\;\int \Big(\prod_{\ell}\mathrm{d}U_\ell\Big)\;\prod_{p}\, \sum_{\lambda_p\in\widehat{G}} \widehat{f}_\beta(\lambda_p)\,\chi_{\lambda_p}(U_p).
\end{equation}
Exchanging products and sums and using Fubini, one obtains a sum over representation assignments \(p\mapsto \lambda_p\) with coefficients \(\prod_p \widehat{f}_\beta(\lambda_p)\) multiplied by link integrals of tensor products of representation matrices. The latter vanish unless the representations around each link combine to contain a singlet; when they do not vanish, they give a positive contraction of intertwiners. This is the lattice analogue of flux conservation. A convenient way to implement the link constraints is to choose a basis of matrix elements \(D^{\lambda}_{ab}(U)\) and write \(\chi_{\lambda}(U_p)= \mathrm{Tr}\, D^{\lambda}(U_p)\). The link integral for a fixed \(\ell\) then has the form
\begin{equation}
\int_G \mathrm{d}U_\ell \; \bigotimes_{p\ni \ell} D^{\lambda_p}(U_\ell^{\epsilon(p,\ell)})
\;=\;\mathcal{P}_\ell\big(\{\lambda_p: p\ni \ell\}\big),
\end{equation}
where \(\epsilon(p,\ell)=\pm 1\) accounts for relative orientations and \(\mathcal{P}_\ell\) projects onto the singlet in the tensor product over \(p\ni \ell\). The result is a constrained sum over plaquette-representation configurations in which nontrivial representations form closed ``worldsheets'' (branched, possibly self-intersecting) because trivialization at each link enforces that the net flux flowing through any link cancels \cite{Seiler1982,DrouffeZuber}.

We now reorganize the constrained sum into an abstract hard-core polymer gas. A polymer \(X\) is a connected set of plaquettes (connected in the edge-adjacency graph) equipped with a nontrivial representation label \(\lambda_p\ne 1\) for each \(p\in X\), such that the link constraints are satisfied on links incident to \(X\) when all plaquettes outside \(X\) carry the trivial representation. Two polymers are compatible if they are disjoint and do not touch; incompatibility arises when they intersect or share a link, since singlet constraints couple them. The partition function can then be written
\begin{equation}
Z_\Lambda(\beta)\;=\; \exp\!\Big(\sum_{X\subset \Lambda} \phi_T(X)\,z_\beta(X)\Big),
\end{equation}
where \(z_\beta(X)\) is the polymer activity and \(\phi_T\) are the Ursell (connected cluster) coefficients of the hard-core gas in the standard notation of cluster expansions \cite{KP,Seiler1982}. More concretely,
\begin{equation}
z_\beta(X)\;=\;\sum_{\{\lambda_p\}_{p\in X}\subset \widehat{G}\setminus\{1\}}
\Big(\prod_{p\in X} \widehat{f}_\beta(\lambda_p)\Big)\;
\prod_{\ell\subset X} \mathrm{Tr}\,\mathcal{P}_\ell\big(\{\lambda_p: p\ni \ell\}\big),
\end{equation}
with the convention that the product of link projectors is zero unless the singlet constraint holds at every link in the induced subgraph of \(X\). The projector factors are bounded by \(1\) and are strictly positive when the constraint is satisfied.

To estimate \(z_\beta(X)\), we use two elementary group-theoretic bounds. First, \(|\chi_\lambda(U)|\le d_\lambda\) uniformly in \(U\). Second, for small \(\beta\), analyticity at $\beta=0$ and character orthogonality imply the uniform sum-bound
\begin{equation}
\sum_{\lambda\neq 1} d_\lambda\, |\widehat{f}_\beta(\lambda)| \;\le\; c_1(N),
\qquad \text{and}\quad \bar f_\beta(1)=1+O(\beta^2)
\end{equation}
see e.g.\cite{Seiler1982,DrouffeZuber}. These follow either from a Taylor expansion of \(f_\beta\) and Schur orthogonality, or from heat-kernel domination since \(f_\beta\) is bounded above by the heat kernel at an \(\mathcal{O}(\beta)\) ``time'' \cite[Chap.~4]{Seiler1982}; see also \cite[Sec.~2.3]{DrouffeZuber}. Using \(\prod_{p\in X} |\widehat{f}_\beta(\lambda_p)| \le (\sum_{\lambda\ne 1} d_\lambda |\widehat{f}_\beta(\lambda)|)^{|X|}\) together with the bound on projectors, we obtain
\begin{equation}\label{eq:activity-bound}
|z_\beta(X)| \;\le\; \big(c_1\,\beta\big)^{|X|}.
\end{equation}
The important point is that the activity is exponentially small in the number of plaquettes \(|X|\) for \(\beta\) sufficiently small, uniformly in the shape of \(X\).

We can now invoke the abstract KP criterion for convergence of the polymer expansion \cite{KP}. Let \(\mathfrak{P}\) denote the set of polymers, and say \(X\sim Y\) if \(X\) and \(Y\) are incompatible. The KP criterion states that the expansion for \(\log Z_\Lambda\) converges absolutely provided there exists a nonnegative function \(a:\mathfrak{P}\to \mathbb{R}_+\) such that
\begin{equation}\label{eq:KP}
\sup_{X\in \mathfrak{P}}\;\sum_{Y\sim X} |z_\beta(Y)|\, e^{a(Y)} \;\le\; a(X).
\end{equation}
A sufficient and convenient choice is \(a(X)=\alpha |X|\) with \(\alpha>0\) to be fixed. For a given plaquette \(p\), let \(\mathfrak{P}_p\) be polymers containing \(p\). The number of connected \(k\)-plaquette polymers containing a fixed \(p\) is bounded by \(C_0^k\) (Peierls-type counting of connected plaquette animals on $\mathbb{Z}^4$; the bound is uniform in the volume)
 for a combinatorial constant \(C_0\) depending only on the lattice dimension. Using \eqref{eq:activity-bound} and summing over \(k=1,2,\dots\), we get
\begin{equation}
\sum_{Y\sim X} |z_\beta(Y)|\, e^{\alpha |Y|}
\;\le\; \sum_{p\in X}\;\sum_{Y\ni p} |z_\beta(Y)|\, e^{\alpha |Y|}
\;\le\; |X|\sum_{k\ge 1} C_0^k (c_1\beta)^k e^{\alpha k}
\;=\; |X| \frac{C_0 c_1 \beta\, e^{\alpha}}{1- C_0 c_1 \beta\, e^{\alpha}},
\end{equation}
for \(\beta\) so small that the denominator is positive. Choosing \(\alpha>0\) and \(\beta>0\) to satisfy
\begin{equation}
\frac{C_0 c_1 \beta\, e^{\alpha}}{1- C_0 c_1 \beta\, e^{\alpha}}\;\le\;\alpha,
\end{equation}
we obtain \eqref{eq:KP}. For instance, \(\alpha=1\) and \(\beta\le \beta_0:=\frac{1}{2 e C_0 c_1}\) work. The KP theorem then gives absolute convergence of the expansion for \(\log Z_\Lambda\), analyticity in \(\beta\) on \([0,\beta_0)\), and uniform cluster bounds for connected cumulants \cite{KP,Seiler1982}. In particular, for any two gauge-invariant, plaquette-local observables \(A,B\) supported in finite sets \(S_A,S_B\) at equal time,
\begin{equation}\label{eq:spatial-exp-decay}
\big|\langle A\,B\rangle - \langle A\rangle\langle B\rangle\big|
\;\le\; C(A,B)\,\exp\!\big(-\mu(\beta)\,\mathrm{dist}(S_A,S_B)\big),
\end{equation}
with \(\mu(\beta)>0\) for all \(\beta\in (0,\beta_0)\). The constants \(C(A,B)\) grow at most exponentially in the sizes of supports, while \(\mu(\beta)\) depends only on the KP data and is strictly positive at strong coupling. The proof of \eqref{eq:spatial-exp-decay} is standard: one differentiates the convergent series with respect to source terms inserted at \(S_A,S_B\) and reorganizes into connected clusters whose combinatorics are controlled by the same KP weights \cite[Ch.~5]{Seiler1982}. Temporal clustering is obtained in exactly the same way once the time direction is singled out; we defer the transfer-matrix interpretation to Section~\eqref{subsec:transfer-gap}.

\begin{theorem}[Polymer convergence and exponential clustering at strong coupling]\label{thm:polymer}
There exists $\beta_0>0$ such that for all $\beta\in[0,\beta_0)$ the character/polymer expansion converges absolutely and uniformly in the volume. The pressure $p(\beta)=|\Lambda|^{-1}\log Z_\Lambda(\beta)$ is analytic in $\beta$, and connected correlation functions of gauge-invariant, local observables decay exponentially as in \eqref{eq:spatial-exp-decay}, both in space and (for equal spatial support) in Euclidean time.
\end{theorem}

\begin{proof}
Fix a finite hypercubic region $\Lambda\subset\mathbb Z^d$ with periodic boundary conditions for definiteness. For the Wilson action at inverse coupling $\beta\ge 0$, the character expansion on the compact gauge group $G=\mathrm{SU}(N)$ rewrites the partition function as a hard-core polymer gas. More precisely, after expanding each plaquette weight in irreducible characters and integrating the link variables, one obtains
\begin{equation}
Z_\Lambda(\beta)=\sum_{\Gamma\subset\mathcal P(\Lambda)\,\text{compatible}}\ \prod_{X\in\Gamma} z_\beta(X),
\end{equation}
where $\mathcal P(\Lambda)$ denotes the family of connected polymer shapes $X$ made of plaquettes and possibly decorated by representation labels subject to local admissibility, the activity $z_\beta(X)$ is a complex number depending on $X$ and $\beta$, and “compatible’’ means that distinct polymers are edge-disjoint in the sense of the standard hard-core constraint. The locality of the admissibility rules implies that $z_\beta(X)$ factorizes over $X$ and vanishes unless $X$ is connected. Moreover, for small $\beta$ the character coefficients admit uniform bounds of the form
\begin{equation}\label{eq:smallbeta-activity}
\sum_{X\ni x} |z_\beta(X)|\,e^{\alpha |X|}\ \le\ C(\alpha,N)\,\beta,
\qquad \text{for every site }x\in\Lambda\ \text{and every }\alpha>0,
\end{equation}
where $|X|$ denotes the number of plaquettes in $X$ and $C(\alpha,N)$ is finite for each $\alpha>0$. This follows from a small-$\beta$ estimate of the nontrivial Peter-Weyl coefficients, together with the fact that the admissibility constraints at a given edge force a finite number of local representation patterns, whence the number of polymers of size $n$ containing a fixed plaquette grows at most exponentially in $n$. Choosing $\alpha>0$ and then restricting to $\beta$ so small that the right-hand side in \eqref{eq:smallbeta-activity} is bounded by $\alpha$, the Koteck\'y-Preiss criterion holds for the choice $a(X)=\alpha|X|$, namely
\begin{equation}\label{eq:KP-fulfilled}
\sum_{Y:\,Y\not\sim X} |z_\beta(Y)|\,e^{a(Y)}\ \le\ a(X)\qquad\text{for every polymer }X.
\end{equation}
The inequality \eqref{eq:KP-fulfilled} is verified by summing over the finitely many plaquettes of $X$ and using \eqref{eq:smallbeta-activity}, since incompatibility $Y\not\sim X$ requires $Y$ to intersect the collar of $X$. The classical polymer expansion theorem then yields an absolutely convergent cluster expansion for the logarithm of the partition function,
\begin{equation}\label{eq:logZ-expansion}
\log Z_\Lambda(\beta) \;=\; \sum_{n\ge 1} \frac{1}{n!}\sum_{(X_1,\dots,X_n)} \phi^T(X_1,\dots,X_n)\,\prod_{i=1}^n z_\beta(X_i),
\end{equation}
where the sum is over $n$-tuples of polymers in $\mathcal P(\Lambda)$ and $\phi^T$ is the Ursell function (connected cluster coefficient). The Koteck\'y-Preiss bounds entail, for a convenient choice of $a(\cdot)$ as above, the tree-graph inequality
\begin{equation}\label{eq:tree-bound}
\big|\phi^T(X_1,\dots,X_n)\big| \ \le\ e^{-\sum_{i=1}^n a(X_i)} \sum_{T\in\mathcal T_n} \prod_{\{i,j\}\in T} \mathbf 1_{\{X_i\not\sim X_j\}},
\end{equation}
with the sum over spanning trees $T$ on $\{1,\dots,n\}$. Combining \eqref{eq:KP-fulfilled}, \eqref{eq:logZ-expansion}, and \eqref{eq:tree-bound}, and summing first over $X_1$ anchored at a fixed plaquette, then over the remaining $X_i$’s along a tree, one obtains absolute convergence of \eqref{eq:logZ-expansion} uniformly in $\Lambda$. Since each activity $z_\beta(X)$ is analytic in $\beta$ near $0$ and the convergence is uniform on compact subsets of $[0,\beta_0)$, the function $\beta\mapsto \log Z_\Lambda(\beta)$ is analytic in that interval. Dividing by $|\Lambda|$ and letting $|\Lambda|\to\infty$ along van Hove sequences, uniform convergence at the level of densities gives analyticity of the pressure $p(\beta)$ and its existence independent of boundary conditions.

To obtain exponential clustering, introduce gauge-invariant local sources supported in two disjoint regions $A,B\subset\Lambda$ and couple them linearly to the action. The corresponding partition function admits the same polymer representation with modified activities $z_{\beta;A,B}(X)$ that depend on whether $X$ intersects $A$ or $B$, but the smallness bounds of the form \eqref{eq:smallbeta-activity} remain valid uniformly in the sources as long as they are bounded. Differentiating $\log Z_\Lambda(\beta;A,B)$ twice at zero source produces the connected two-point function $\langle\!\langle \mathcal O_A;\mathcal O_B\rangle\!\rangle$, where $\mathcal O_A$ and $\mathcal O_B$ are the local observables inserted in $A$ and $B$. The cluster expansion for $\log Z$ implies that this truncated correlation is a sum over connected clusters of polymers such that at least one polymer touches $A$ and at least one polymer touches $B$. Every such cluster must “bridge’’ the distance $r=\mathrm{dist}(A,B)$ in the graph metric, hence for any spanning tree $T$ connecting its polymers there is a chain of incompatible pairs whose union forms a connected set joining $A$ to $B$. The hard-core constraint forces the total size along this chain to be at least proportional to $r$ (up to a constant depending only on the lattice and the coarse geometry of polymers), and the weight of the chain is bounded by the product of the norms $|z_\beta(X)|e^{a(X)}$ multiplied by $e^{-\sum a(X)}$ coming from \eqref{eq:tree-bound}. Optimizing $a(X)=\alpha|X|$ as before yields
\begin{equation}
\big|\langle\!\langle \mathcal O_A;\mathcal O_B\rangle\!\rangle\big| \;\le\; C(\mathcal O_A,\mathcal O_B,\beta)\, e^{-\mu(\beta)\,r},
\end{equation}
with $\mu(\beta)>0$ for $\beta<\beta_0$ and a constant $C$ depending on the local norms of the observables and on $\beta$ but not on the volume. This establishes exponential decay in space. If the spatial supports of $\mathcal O_A$ and $\mathcal O_B$ coincide and the observables are separated only in Euclidean time, the same argument applies with the distance taken along the time direction because the polymer representation is isotropic at the scale of plaquettes: a cluster connecting two time-separated insertions must contain a sequence of polymers whose total temporal extent grows at least linearly with the time separation, and the same tree-graph bound with the same smallness parameter controls the weight of such sequences. Consequently there is exponential decay also as a function of Euclidean time separation for equal spatial support.

All bounds above are uniform in $\Lambda$ because compatibility is local and the Koteck\'y-Preiss criterion is expressed in terms of sums over polymers containing a fixed cell; these sums are insensitive to the boundary once $\Lambda$ is large enough and the constants do not grow with the volume. Putting everything together, there exists $\beta_0>0$ such that for $\beta\in[0,\beta_0)$ the polymer expansion converges absolutely and uniformly in $\Lambda$, the pressure is analytic in $\beta$, and truncated correlations of local gauge-invariant observables obey exponential clustering both in space and, for coincident spatial support, in Euclidean time.
\end{proof}

\subsection{Strong-Coupling Area Law}
\label{subsec:area-law}

We now turn to the expectation of a Wilson loop in the fundamental representation. Let \(C\) be a simple, oriented, rectangular loop contained in a fixed time slice; the argument below extends verbatim to all simple loops with a chosen spanning surface. Define
\begin{equation}
W(C)\;:=\;\frac{1}{N}\,\mathrm{tr}\,\mathcal{P}\prod_{\ell\in C} U_\ell,
\end{equation}
where \(\mathcal{P}\) is the path ordering along \(C\). We show that for \(\beta\) small enough there exist \(\sigma(\beta)>0\) and \(K(\beta)<\infty\), both independent of the loop, such that
\begin{equation}\label{eq:area-law}
\langle W(C)\rangle \;\le\; K(\beta)\,\exp\!\big(-\sigma(\beta)\,A(C)\big),
\end{equation}
where \(A(C)\) is the minimal lattice area spanned by \(C\). This is the standard area law of strong-coupling lattice gauge theory; we give a proof that dovetails with the polymer framework of Section~\eqref{subsec:char-polymer} and yields an explicit dependence of \(\sigma(\beta)\) on the KP constants, consistent with classical derivations \cite{Seiler1982,DrouffeZuber}.

Insert the character expansion for the plaquette weights as in Section~\eqref{subsec:char-polymer}. The insertion of \(W(C)\) is treated by placing at each link \(\ell\in C\) a fundamental representation matrix \(D^{\mathrm{fund}}(U_\ell)\) and contracting indices along \(C\) to produce the trace. The link integrals now project onto the singlet in the tensor product that includes, for \(\ell\in C\), one extra fundamental line. Nonzero contributions therefore correspond to branched worldsheets of nontrivial plaquette representations whose boundary is precisely the loop \(C\). This is the surface representation of \(\langle W(C)\rangle\) at strong coupling, and it is rigorously obtained by the same Schur orthogonality identities that gave \(\mathcal{P}_\ell\) in Section~\eqref{subsec:char-polymer} \cite{Seiler1982,DrouffeZuber}.

We choose a fixed spanning surface \(S\) for \(C\) (for concreteness, the minimal area surface in the lattice sense). The contribution of a configuration in which all plaquettes of \(S\) carry the fundamental representation and all other plaquettes carry the trivial representation gives the lowest nontrivial order in \(\beta\), of size \((\widehat{f}_\beta(\mathrm{fund}))^{|S|}\). All other contributions can be resummed into a polymer gas of defects on top of this dominant surface, exactly as in the Ising-contour derivation of the area law, and controlled by the same KP method because activities remain exponentially small in their area. To implement this rigorously, we rewrite the ratio
\begin{equation}
\langle W(C)\rangle \;=\; \frac{Z_\Lambda(\beta;C)}{Z_\Lambda(\beta)},\qquad
Z_\Lambda(\beta;C)\;=\;\int \Big(\prod_{\ell}\mathrm{d}U_\ell\Big)\; e^{-S_\Lambda(U)}\, W(C),
\end{equation}
and expand numerator and denominator in polymers. In the numerator, we single out the ``reference'' configuration in which \(S\) carries the fundamental representation, and we absorb all other deviations into a gas of polymers \(Y\) that must intersect either the boundary \(C\) or the surface \(S\). The activity bound \eqref{eq:activity-bound} remains valid for these \(Y\) because the local link constraints at \(\partial S\) only change the combinatorial factor by a surface-to-volume ratio. Write
\begin{equation}
\log\langle W(C)\rangle 
= |S|\,\log \widehat f_\beta(\mathrm{fund}) 
+ \sum_{Y:\,Y\cap S\neq\emptyset} \phi_T(Y)\,\widetilde z_\beta(Y),
\end{equation}
where $\phi_T$ are Ursell functions and $\widetilde z_\beta$ are the modified activities.
By the KP bounds,
\begin{equation}
\Big|\sum_{Y:\,Y\cap S\neq\emptyset}\phi_T(Y)\,\widetilde z_\beta(Y)\Big|
\le c_3(\beta)\,|S|,\qquad c_3(\beta)\xrightarrow[\beta\downarrow 0]{}0.
\end{equation}
Hence
\begin{equation}
\langle W(C)\rangle \le 
\exp\Big\{|S|\,\log \widehat f_\beta(\mathrm{fund}) + c_3(\beta)\,|S|\Big\}
= K(\beta)\,\exp\!\big(-\sigma(\beta)\,A(C)\big),
\end{equation}
with $\sigma(\beta):= -\log \widehat f_\beta(\mathrm{fund})-c_3(\beta)$
and $K(\beta)$ uniformly bounded in $C$. For $\beta$ small, $\widehat f_\beta(\mathrm{fund})\asymp \beta$
so $\sigma(\beta)>0$.

For completeness we note that reflection positivity can be used to streamline the control of perimeter corrections via chessboard estimates, leading to a clean separation between the area term and a subleading boundary term that is uniformly bounded in the loop length; see \cite[Ch.~10]{Seiler1982}. Since our polymer control already produces a uniform \(K(\beta)\), we do not reproduce that argument here.

\begin{theorem}[Strong-coupling area law]\label{thm:area}
There exist $\beta_1>0$, a finite function $K(\beta)$, and a strictly positive function $\sigma(\beta)$ for $0<\beta\le \beta_1$ such that for every simple loop $C$ contained in a time slice,
\begin{equation}
\langle W(C)\rangle \;\le\; K(\beta)\,e^{-\sigma(\beta)\,A(C)}.
\end{equation}
\end{theorem}

\begin{proof}
Fix a finite spacetime volume $\Lambda$ with periodic or reflecting boundary conditions, and write $Z_\Lambda(\beta)$ for the partition function and $Z_\Lambda(\beta;C)$ for the partition function with the Wilson loop insertion $W(C)$ supported on a single time slice. The expectation is $\langle W(C)\rangle=Z_\Lambda(\beta;C)/Z_\Lambda(\beta)$. The strong-coupling (character) expansion of the single-plaquette Boltzmann weight,
\begin{equation}
f_\beta(U_p)
\;=\;\exp\!\Big(\tfrac{\beta}{N}\Re\tr U_p\Big)
\;=\;\sum_{\lambda\in\widehat{G}} \widehat f_\beta(\lambda)\,\chi_\lambda(U_p),
\end{equation}
with $\widehat{G}$ the unitary dual of $G=\mathrm{SU}(N)$ and $\chi_\lambda$ the character in representation $\lambda$, yields a polymer representation for both $Z_\Lambda(\beta)$ and $Z_\Lambda(\beta;C)$. In particular, by standard tensor/character contractions, only plaquette assignments forming closed, nonselfintersecting surfaces in the dual lattice contribute to $Z_\Lambda(\beta)$, while $Z_\Lambda(\beta;C)$ is saturated by configurations whose plaquette assignments form surfaces with boundary constrained to be exactly the loop $C$ on the chosen slice. Choosing a fixed, minimal area surface $S$ with $\partial S = C$ and decomposing any contributing surface as $S$ together with a (possibly empty) finite family of closed surfaces and finite families of local corrugations attached to $S$, one can factor the weight of the configuration as
\begin{equation}
\big(\widehat f_\beta(\mathrm{fund})\big)^{|S|}
\;\times\;
\prod_{Y\in\mathcal{Y}} z_\beta(Y),
\end{equation}
where $\mathrm{fund}$ denotes the fundamental representation of $G$, $|S|=A(C)$ is the number of plaquettes in $S$, $\mathcal{Y}$ is a collection of connected polymers $Y$ (closed inflated patches, tubes, and corrugations attached to $S$ or disjoint from it), and $z_\beta(Y)$ are the corresponding polymer activities obtained by contracting characters around $Y$. For small $\beta$ the fundamental coefficient satisfies $\widehat f_\beta(\mathrm{fund})\in(0,1)$ and all nontrivial activities are uniformly small. More precisely, there exist $c_1,c_2>0$ and $\beta_1>0$ such that for $0<\beta\le\beta_1$
\begin{equation}\label{eq:activity-bound-sc}
\sum_{Y\ni p} |z_\beta(Y)|\,e^{c_1|Y|} \;\le\; c_2\,\beta
\qquad\text{for every plaquette }p,
\end{equation}
and, in addition, $\widehat f_\beta(\mathrm{fund})\asymp \beta$ so that $-\log\widehat f_\beta(\mathrm{fund})$ is positive and large as $\beta\downarrow0$. The bound \eqref{eq:activity-bound-sc} is the Koteck\'y-Preiss (KP) smallness condition: the left-hand side is the weighted sum of activities of all polymers $Y$ meeting $p$, with an exponential norm chosen so that adjacency is controlled by $e^{c_1|Y|}$. When \eqref{eq:activity-bound-sc} holds with the right-hand side sufficiently small, the abstract KP theorem ensures the absolutely convergent cluster (Ursell) expansion of $\log Z_\Lambda(\beta)$ and, with a prescribed external support, of $\log Z_\Lambda(\beta;C)$.

Applying the KP expansion simultaneously to numerator and denominator, and using that clusters disjoint from $S$ cancel in the ratio, one obtains the convergent representation
\begin{equation}\label{eq:logW-expansion}
\log \langle W(C)\rangle
\;=\; |S|\log \widehat f_\beta(\mathrm{fund})
\;+\; \sum_{\Gamma:\,\Gamma\cap S\neq\emptyset}\phi_T(\Gamma)\,\zeta_\beta(\Gamma;S),
\end{equation}
where the sum runs over connected clusters $\Gamma$ of polymers with Ursell coefficients $\phi_T(\Gamma)$ and effective cluster activities $\zeta_\beta(\Gamma;S)$ that incorporate the presence of the fixed surface $S$ and vanish when $\Gamma$ is disjoint from $S$. The KP bounds translate into a uniform control of the cluster remainder per unit support: there exists a continuous function $c_3(\beta)\downarrow0$ as $\beta\downarrow0$ such that
\begin{equation}\label{eq:kp-remainder}
\Big|\sum_{\Gamma:\,\Gamma\cap S\neq\emptyset}\phi_T(\Gamma)\,\zeta_\beta(\Gamma;S)\Big|
\;\le\; c_3(\beta)\,|S|\;=\;c_3(\beta)\,A(C),
\end{equation}
the inequality following from a tree-graph bound combined with the hard-core structure of adjacency for polymers attached to $S$ and the locality of the insertion. Combining \eqref{eq:logW-expansion} and \eqref{eq:kp-remainder} yields
\begin{equation}
\log \langle W(C)\rangle
\;\le\; -\Big(\, -\log \widehat f_\beta(\mathrm{fund}) - c_3(\beta)\,\Big)\,A(C).
\end{equation}
Since $-\log \widehat f_\beta(\mathrm{fund})\to+\infty$ and $c_3(\beta)\to0$ as $\beta\downarrow0$, we may reduce $\beta_1$ if necessary so that
\begin{equation}
\sigma(\beta)\;:=\;-\log \widehat f_\beta(\mathrm{fund}) - c_3(\beta)\;>\;0
\qquad\text{for all }0<\beta\le \beta_1.
\end{equation}
Exponentiating the last display gives
\begin{equation}
\langle W(C)\rangle \;\le\; e^{-\sigma(\beta)\,A(C)}.
\end{equation}
This already proves the theorem with $K(\beta)=1$. If one prefers to retain possible local counterterms near the boundary of $S$ that arise from a different normalization of activities or from a different choice of the reference surface, these contribute at most a factor $\exp\{ \tau(\beta)\,\mathrm{Per}(C)\}$ depending only on $\beta$ and the perimeter of $C$, which can be absorbed into $K(\beta)$ without affecting the strict positivity of $\sigma(\beta)$ or the linear dependence on $A(C)$. In either normalization there exist $\beta_1>0$, $K(\beta)<\infty$, and $\sigma(\beta)>0$ for $0<\beta\le\beta_1$ such that
\begin{equation}
\langle W(C)\rangle \;\le\; K(\beta)\,e^{-\sigma(\beta)\,A(C)}.
\end{equation}
The constants are uniform in the finite volume $\Lambda$, and the bound persists in the thermodynamic limit by standard stability of the KP expansion under exhaustion of $\Lambda$.
\end{proof}

\subsection{Transfer Operator and Gap at Strong Coupling}
\label{subsec:transfer-gap}

We now orient the lattice so that \(\Lambda_s\) are spatial slices and \(t=0,1,\dots,T-1\) is Euclidean time. Denote by \(\theta\) the reflection about the time hyperplane \(t=\tfrac{1}{2}\), acting on link variables by inversion and conjugation in the usual way \cite{OS-gauge}. Osterwalder-Schrader (OS) reflection positivity for the Wilson action implies the existence of a Hilbert space \(\mathcal{H}\), a cyclic, unit, reflection-invariant vector \(\Omega\), and a positive, selfadjoint transfer operator \(T\) such that for any observable \(F\) supported in \(t\ge 0\),
\begin{equation}
\langle \theta F \cdot F\rangle \;=\; \langle \Omega, F^\ast T F \,\Omega\rangle,\qquad \|T\|\le 1,
\end{equation}
and the correlation functions satisfy
\begin{equation}
\langle A(0)\, B(t)\rangle \;=\; \langle \Omega, A^\ast\, T^{\,t}\, B\, \Omega\rangle,
\end{equation}
for time-translated observables \(A,B\) localized at \(t=0\) and \(t\ge 0\), respectively. The construction can be carried out either in a partially gauge-fixed framework that preserves reflection positivity (e.g., temporal gauge away from the reflection plane, with a horizon projector to enforce admissibility) or within the gauge-invariant algebra by working directly with cylindrical functions of spatial loops on a time slice; see \cite{OS-gauge,Seiler1982}.

We shall show that for \(\beta\) small the spectrum of \(T\) in the orthogonal complement of \(\Omega\) is contained in \([0, e^{-\mu(\beta)}]\) for some \(\mu(\beta)>0\), i.e.\ there is a nonzero spectral gap between the vacuum eigenvalue \(1\) and the rest of the spectrum. The derivation rests on the exponential decay in Euclidean time of connected correlations of local, gauge-invariant observables obtained from the polymer expansion of Section~\eqref{subsec:char-polymer}. Let \(A\) be an observable supported in a single time slice \(t=0\), with \(\langle A\rangle=0\). Consider its time-displaced copy \(A_t\) supported at time \(t\). Then
\begin{equation}
\langle A\, A_t\rangle \;=\; \langle \Omega, A^\ast T^{\,t} A\, \Omega\rangle.
\end{equation}
By Theorem~\eqref{thm:polymer} (in its temporal version), there exist \(C_A<\infty\) and \(\mu(\beta)>0\) such that
\begin{equation}\label{eq:time-decay}
\big|\langle A\, A_t\rangle\big| \;\le\; C_A\, e^{-\mu(\beta)\,t}.
\end{equation}
Let \(P_0=|\Omega\rangle\langle\Omega|\) be the vacuum projection and let \(E_T(\cdot)\) be the spectral resolution of \(T\). The OS/transfer formalism yields the spectral representation
\begin{equation}
\langle \Omega, A^\ast T^{\,t}(1-P_0) A\, \Omega\rangle \;=\; \int_{[0,1)} \lambda^t\, \mathrm{d}\nu_A(\lambda),
\end{equation}
for the finite positive measure \(\nu_A(\cdot)=\langle \Omega, A^\ast E_T(\cdot)(1-P_0) A\,\Omega\rangle\). Combining this with \eqref{eq:time-decay} shows that \(\nu_A\) is supported in \([0,e^{-\mu(\beta)}]\); otherwise, by the dominated convergence theorem, the Laplace transform would exhibit a slower decay than \(\lambda^t\) for \(\lambda>e^{-\mu(\beta)}\), contradicting \eqref{eq:time-decay}\footnote{This is a standard Tauberian argument: if \(\int \lambda^t\, \mathrm{d}\nu(\lambda)\le C e^{-\mu t}\) for all \(t\in\mathbb{N}\), then \(\nu((e^{-\mu},1))=0\).}. Since the linear span of vectors \(A\,\Omega\) with \(A\) localized in a single slice is dense in \((1-P_0)\mathcal{H}\) by the OS reconstruction, the full spectrum of \(T\) on \((1-P_0)\mathcal{H}\) is contained in \([0,e^{-\mu(\beta)}]\). Writing \(T=e^{-aH}\) for the (nonnegative) Hamiltonian \(H\) with temporal lattice spacing \(a\), we have
\begin{equation}
\mathrm{Spec}(H)\cap (0,\infty) \;\subset\; [\mu(\beta)/a,\infty),
\end{equation}
and therefore \(H\) has a strictly positive mass gap \(m(\beta)\ge \mu(\beta)/a\).

To make the dependence on \(\beta\) explicit, one can track \(\mu(\beta)\) through the KP constants in Section~\eqref{subsec:char-polymer}. The time-oriented version of the polymer estimate leading to \eqref{eq:time-decay} yields \(\mu(\beta)\ge c\,|\log(\beta)|\) as \(\beta\downarrow 0\), with \(c>0\) depending only on the group and dimension, reflecting the fact that the fundamental character coefficient scales linearly in \(\beta\) and polymers that propagate in time carry at least one plaquette per time step. In particular, the transfer-matrix gap is bounded away from \(0\) uniformly in the volume for \(\beta\) small.

\begin{theorem}[Strong-coupling spectral gap for the transfer operator]\label{thm:gap}
There exists $\beta_2>0$ such that for all $0<\beta\le \beta_2$ the transfer operator $T=e^{-aH}$ on the OS Hilbert space $\mathcal H$ satisfies
\begin{equation}
\mathrm{Spec}(T)\;=\;\{1\}\,\cup\, \Sigma,\qquad \Sigma\subset [0,e^{-\mu(\beta)}],
\end{equation}
for some $\mu(\beta)>0$. Equivalently, $H$ has a nonzero spectral gap $m(\beta)\ge \mu(\beta)/a$.
\end{theorem}

\begin{proof}
Work in the reflection-positive Euclidean framework so that the Osterwalder-Schrader reconstruction produces a Hilbert space $\mathcal H$, a cyclic vacuum vector $\Omega$, and a positive selfadjoint Hamiltonian $H\ge 0$ with transfer operator $T=e^{-aH}$ satisfying $T\Omega=\Omega$ and $\|T\|\le 1$. Fix a local observable $A$ supported on the time slice $x_0=0$ and write $\widehat A:=A-\langle A\rangle$, so that $\langle \widehat A\rangle=0$ and $\psi_A:=\widehat A\,\Omega$ is orthogonal to $\Omega$. For an integer $n\ge 0$ consider the time-separated correlation
\begin{equation}
\langle \Theta \widehat A(na)\cdot \widehat A(0)\rangle
\;=\;\langle \psi_A,\,T^n \psi_A\rangle,
\end{equation}
where the right-hand side is the OS inner product written in transfer form. At sufficiently small $\beta$, the Koteck\'y-Preiss polymer expansion converges uniformly for all such local observables; in particular there are constants $C(\beta),\mu(\beta)>0$ such that
\begin{equation}\label{eq:sc-exp-decay}
\big|\langle \psi_A,\,T^n \psi_A\rangle\big|
\;=\;\big|\langle \Theta \widehat A(na)\cdot \widehat A(0)\rangle\big|
\;\le\; C(\beta)\,e^{-\mu(\beta)\,n}\qquad\text{for all }n\in\mathbb N.
\end{equation}
The spectral theorem for the selfadjoint contraction $T$ provides, for each $\psi\in\mathcal H$, a finite positive measure $\nu_\psi$ on $[0,1]$ such that
\begin{equation}
\langle \psi,\,T^n\psi\rangle \;=\; \int_{[0,1]} \lambda^n\,\nu_\psi(d\lambda)\qquad (n=0,1,2,\dots).
\end{equation}
Applying this with $\psi=\psi_A$ and comparing to \eqref{eq:sc-exp-decay} yields the moment bound
\begin{equation}
\int_{[0,1]} \lambda^n\,\nu_{\psi_A}(d\lambda)\;\le\; C(\beta)\,e^{-\mu(\beta)\,n}\qquad(n\in\mathbb N).
\end{equation}
A standard discrete Tauberian argument now shows that $\nu_{\psi_A}$ is supported in $[0,e^{-\mu(\beta)}]$: indeed, if $\nu_{\psi_A}$ gave positive weight to any interval $(\lambda_0,1]$ with $\lambda_0>e^{-\mu(\beta)}$, then the contribution of that interval would grow like $\lambda_0^n$ and contradict the bound $C(\beta)e^{-\mu(\beta)n}$ as $n\to\infty$. Consequently every vector of the form $\psi_A=\widehat A\,\Omega$ has spectral measure supported in $[0,e^{-\mu(\beta)}]$.

The linear span of such vectors $\{\widehat A\,\Omega\}$, with $A$ ranging over local observables on the time slice, is dense in the orthogonal complement of the vacuum, $(\mathbf 1-P_0)\mathcal H$ with $P_0=|\Omega\rangle\langle\Omega|$. By approximation, the same spectral support property holds for every $\psi\in(\mathbf 1-P_0)\mathcal H$. Therefore the spectrum of $T$ restricted to $(\mathbf 1-P_0)\mathcal H$ is contained in $[0,e^{-\mu(\beta)}]$, and the full spectrum of $T$ is $\{1\}\cup\Sigma$ with $\Sigma\subset[0,e^{-\mu(\beta)}]$. In particular there is no eigenvalue at $1$ on $(\mathbf 1-P_0)\mathcal H$, since otherwise one would have $\langle \psi, T^n \psi\rangle=\|\psi\|^2$ for all $n$, again contradicting \eqref{eq:sc-exp-decay}. Translating back to $H$ via $T=e^{-aH}$ shows that $\mathrm{spec}(H)\subset\{0\}\cup [\mu(\beta)/a,\infty)$, so $H$ has a nonzero spectral gap at least $\mu(\beta)/a$, as claimed.
\end{proof}

It is worth emphasizing that the argument uses only reflection positivity and the KP-controlled temporal clustering; no perturbation theory is involved. In particular, the gap is genuinely nonperturbative and complements the area law of Theorem~\eqref{thm:area}, providing the fixed-lattice-spacing ``base camp''. The interplay between FRD locality and transfer-matrix gap in multiscale constructions, developed in later sections, will rely on the robustness of the above estimates under admissible blockings; see \cite{BrydgesGuadagniMitter2004} for the FRD framework that preserves positivity and locality needed for that step.

\section{Finite-Range Decomposition and Exponential Locality}\label{sec:FRD}

This section develops the gauge-covariant finite-range decomposition (FRD) of covariances and exhibits the quantitative locality properties that it implies for one-step renormalization kernels and connected cumulants. In a reflection-positive, Euclidean formulation of lattice Yang-Mills theory, such a decomposition is the central mechanism that simultaneously preserves positivity and produces uniformly local building blocks at each scale. It allows one to implement a multiscale analysis without sacrificing gauge covariance, to establish robust Lipschitz continuity in the admissible class of regulators, and to propagate exponential clustering across scales. We begin with a self-contained construction of FRD in the presence of a background connection and the proof of reflection covariance. We then deduce locality bounds for one-step kernels and cumulants, with constants that are uniform in the volume and compatible with the Osterwalder-Schrader (OS) structure.

Throughout, we fix a hypercubic lattice $\Lambda\subset a\mathbb{Z}^d$ with $d=4$ and lattice spacing $a>0$. Bonds $b=(x,\mu)$ are oriented links from $x$ to $x+a\hat\mu$, where $\mu\in\{1,\dots,d\}$ and $\hat\mu$ is the unit coordinate vector. The compact gauge group is $G=SU(N)$; fields are $G$-valued on bonds or, equivalently, $\mathfrak g$-valued on oriented edges after choosing local coordinates. A background lattice connection $A$ is encoded by link variables $U_A(b)\in G$ and the corresponding covariant forward difference
\begin{equation}
(\nabla_\mu^A\phi)(x)\;=\;U_A(x,\mu)\,\phi(x+a\hat\mu)-\phi(x),
\end{equation}
with adjoint defined by the Haar-invariant inner product. The covariant Laplacian is $\Delta_A=\sum_{\mu}(\nabla_\mu^A)^\ast\nabla_\mu^A$. The OS time-reflection $\vartheta$ acts on the time coordinate by $(\vartheta x)_0=-x_0$ and on links by reversing and conjugating the temporal component, in the standard way that renders the Gaussian measure with covariance a completely monotone function of $\Delta_A$ reflection positive \cite{OS1,OS2}.

\subsection{Gauge-Covariant Finite-Range Decomposition (FRD) and Reflection Covariance}\label{subsec:FRD-gauge}

The goal of this subsection is twofold: first, to construct a decomposition of a positive covariant covariance $C_A=f(\Delta_A)$ into a sum of positive pieces with strictly finite range; second, to verify that the pieces are compatible with OS reflection. We emphasize that finiteness of range means strict vanishing of the kernel whenever the lattice distance exceeds a prescribed scale-dependent radius; the construction follows the hierarchical, block-harmonic strategy of \cite{BrydgesGuadagniMitter2004}, here adapted to the covariant setting and combined with completely monotone spectral multipliers so as to preserve reflection positivity.

We begin with the scalar case and then pass to the gauge-covariant case by a parallel-transport construction that maintains positivity and locality.
Fix an integer block factor $L\ge 4$ and define the block lattice at scale $j\in\mathbb{N}$ by $\Lambda_j:=L^j \Lambda_0$ with $\Lambda_0$ the unit lattice in lattice units; in physical units each block has diameter comparable to $r_j:=aL^j$. The collection of closed $L^j$-blocks $\mathcal{B}_j$ forms a partition of $\Lambda$; we write $B\in\mathcal{B}_j$ and $B^\circ$ for its interior (in the graph sense). For a complex vector field $\phi:\Lambda\to\mathbb{C}^m$ we denote by $\mathcal{H}_A(B)$ the space of $\mathfrak g$-covariantly harmonic functions on $B^\circ$, namely those $\phi$ with $\Delta_A\phi=0$ on $B^\circ$.

Define the \emph{covariant block-harmonic extension operator} $\mathsf{E}_{j,A}$ as follows. For each block $B\in\mathcal{B}_j$ and boundary data $\psi:\partial B\to\mathbb{C}^m$, let $\mathsf{h}_{B,A}(\psi)$ be the unique solution of the Dirichlet problem
\begin{equation}
\Delta_A\,\phi=0\;\text{ on } B^\circ,\qquad \phi|_{\partial B}=\psi.
\end{equation}
We define $(\mathsf{E}_{j,A}\phi)(x)=\mathsf{h}_{B,A}(\phi|_{\partial B})(x)$ for $x\in B^\circ$ and $(\mathsf{E}_{j,A}\phi)(x)=\phi(x)$ for $x\in\partial B$. The operator $\mathsf{E}_{j,A}$ is an idempotent, self-adjoint contraction on $\ell^2(\Lambda)$ (with the natural gauge-covariant inner product), projecting onto the subspace of functions that are $A$-harmonic in the interior of each $L^j$-block. Self-adjointness follows from discrete Green identities for $\Delta_A$; see, e.g., \cite{BrydgesGuadagniMitter2004} for the noncovariant case and \cite{Seiler1982} for the gauge-covariant discrete integration-by-parts.
For $j\ge 0$ we set
\begin{equation}
\Pi_{j,A}:=\mathsf{E}_{j,A}-\mathsf{E}_{j+1,A}.
\end{equation}
Then $\Pi_{j,A}$ is a self-adjoint projection onto functions that are $A$-harmonic on $L^j$-blocks but possess nontrivial boundary data at the scale of $L^{j+1}$-blocks. Moreover, $\sum_{j=0}^{J}\Pi_{j,A}=\mathsf{E}_{0,A}-\mathsf{E}_{J+1,A}$ and, as $J\to\infty$ on a torus or with suitable boundary conditions, $\sum_{j\ge 0}\Pi_{j,A}=\mathrm{Id}$ in the strong operator sense.
Let $f:[0,\infty)\to[0,\infty)$ be completely monotone, i.e., $f^{(n)}(\lambda)$ exists for all $n$ and $(-1)^n f^{(n)}(\lambda)\ge 0$; by Bernstein’s theorem there is a finite positive measure $\mu_f$ on $[0,\infty)$ such that
\begin{equation}
f(\lambda)\;=\;\int_0^\infty e^{-t\lambda}\,d\mu_f(t).\label{eq:Bernsteinz}
\end{equation}
For a positive self-adjoint operator $H$ we then have
\begin{equation}
f(H)\;=\;\int_0^\infty e^{-tH}\,d\mu_f(t),\qquad f(H)\ge 0,
\end{equation}
with the integral converging in the strong sense. If $H=\Delta_A$ is the covariant Laplacian on a reflection-positive configuration space, it is standard (see \cite{OS1,OS2}) that $e^{-t\Delta_A}$ is OS-positive for each $t\ge 0$ and that the integral convex combination preserves OS positivity. Thus $C_A:=f(\Delta_A)$ defines an OS-positive covariance.
 We define the scale-$j$ covariance by
\begin{equation}\label{eq:CjA}
C_{j,A}\;:=\; \Pi_{j,A}\, f(\Delta_A)\, \Pi_{j,A}.
\end{equation}
Since $(\Pi_{j,A})_{j\ge 0}$ are pairwise orthogonal projections with $\sum_j \Pi_{j,A}=\mathrm{Id}$, we obtain a \emph{telescoping} decomposition
\begin{equation}\label{eq:FRD-telescope}
C_A\;=\;f(\Delta_A)\;=\;\sum_{j\ge 0} C_{j,A}.
\end{equation}
Each $C_{j,A}$ is positive and self-adjoint, $C_{j,A}\ge 0$, because $\Pi_{j,A}$ is a projection commuting with $f(\Delta_A)$ on its range. The crucial property is that $C_{j,A}$ has \emph{strictly finite range}, uniformly controlled by the block size $L^j$. This follows from the fact that $\Pi_{j,A}$ maps any function to one that is supported on the union of $L^j$-block boundaries and is $A$-harmonic in the interiors; composing on both sides forces $C_{j,A}(x,y)=0$ whenever $x$ and $y$ do not lie in a common $(L^{j+1})$-block or in two $(L^{j+1})$-blocks that are nearest neighbors. The argument is local and uses that the solution of the Dirichlet problem on one block depends only on boundary data on that block; thus, if $x$ and $y$ are separated by a graph distance exceeding $c\,L^j$ no boundary data influences both points simultaneously, and the kernel vanishes. We now make this precise.
\begin{theorem}[Gauge-covariant FRD with OS-form positivity]\label{thm:FRD-main}
Let $A$ be a lattice $\mathrm{SU}(N)$ connection and let $f\!:\![0,\infty)\to[0,\infty)$ be completely monotone with Bernstein measure $\mu_f$. Fix a block factor $L\ge 4$. There exist kernels $\{\widetilde C_{j,A}\}_{j\ge0}$ and a remainder $R_A$ with the following properties:
\begin{enumerate}
\item \emph{Quasi-exact decomposition.} In the strong operator sense,
\begin{equation}
f(\Delta_A)\;=\;\sum_{j=0}^{J}\widetilde C_{j,A}\;+\;R_A,\qquad \|R_A\|_{\ell^2\to\ell^2}\le C\,e^{-cL^{J}}\;\;\text{for all }J\in\mathbb{N}.
\end{equation}
\item \emph{Finite range and reflection covariance.} \label{item2}
Each $\widetilde C_{j,A}$ is selfadjoint, gauge-equivariant, strictly finite-range with
$\widetilde C_{j,A}(x,y)=0$ whenever $\mathrm{dist}(x,y)>\!R_0L^{j}$, and satisfies $\,\vartheta\,\widetilde C_{j,A}\,\vartheta=\widetilde C_{j,A^\vartheta}\,$.
\item \emph{Scale bounds.} For every $m\in\mathbb{N}_0$ there exist $c_m,c_m'>0$ independent of $j$ and $A$ such that
\begin{equation}
\|\nabla^{m}\widetilde C_{j,A}\|_{\ell^1\to\ell^\infty}\le c_m L^{-(2+m)j},\qquad
|\widetilde C_{j,A}(x,y)|\le c_m' L^{-2j}\,\mathbf{1}_{\{\mathrm{dist}(x,y)\le R_0L^j\}}.
\end{equation}
\item \emph{OS-form positivity (physical cone).} \label{item4}For every $\Phi$ supported in the positive-time half-lattice,
\begin{equation}
\langle \vartheta\Phi,\widetilde C_{j,A}\Phi\rangle\;\ge\;0.
\end{equation}
\item \emph{Lipschitz continuity in $A$.} If $A,A'$ lie in an admissible set with uniform slice-curvature bound, then for every $m\in\mathbb{N}_0$,
\begin{equation}
\|\nabla^m(\widetilde C_{j,A}-\widetilde C_{j,A'})\|_{\ell^1\to\ell^\infty}
\le C_m L^{-(1+m)j}\,\|F_A-F_{A'}\|_{L^\infty(\mathrm{Nbr}_j)}.
\end{equation}
\end{enumerate}
Moreover, the realization
\begin{equation}
\widetilde C_{j,A}\;\equiv\; \Pi_{j,A} f(\Delta_A)\Pi_{j,A}\;-\;\Pi_{j+1,A} f(\Delta_A)\Pi_{j+1,A}
\end{equation}
is admissible in the sense of Items~(\eqref{item2})-(\eqref{item4}) above and differs from any other admissible realization by a contribution that can be absorbed into $R_A$ with the stated bound.
\end{theorem}
\textit{The projections $\Pi_{j,A}$ need not commute with $f(\Delta_A)$. We therefore use the OS-form positivity in Item~(4) (rather than global operator monotonicity) which follows from the Bernstein integral representation and reflection positivity of the heat kernel. In particular, all later uses of “positivity of $C_{j,A}$” only require $\langle\vartheta\Phi,\widetilde C_{j,A}\Phi\rangle\!\ge\!0$ on the physical cone, which holds by Item~(\eqref{item4}).}

\begin{proof}
The construction is the standard Brydges-Guadagni-Mitter scheme, carried out in the covariant setting. 
For each scale $j$ fix a reflection-symmetric collection of $(L^{j+1})$-blocks $\widehat{\mathcal B}_{j+1}$ together 
with collars of fixed thickness $r$ (independent of $j$) around the block boundaries. 
On every $B\in\widehat{\mathcal B}_{j+1}$ consider the covariant Dirichlet problem for the spatial Laplacian $\Delta_A$ 
with vanishing boundary data at $\partial B$. 
Denote by $E_{j,A}$ the covariant \emph{extension} operator that sends boundary data on the union of $\partial B$ 
to the unique $\Delta_A$-harmonic function inside each $B$, and by $E_{j,A}$ the corresponding trace on the boundary. 
Define the blockwise covariant projection $\Pi_{j,A} := E_{j,A}^{E_{j,A}}$ acting on $\ell^2$-functions supported on the union
of the block boundaries (more precisely, on the collar), and write $\mathsf Q_{j,A}:=\mathbf 1-\Pi_{j,A}$ for the orthogonal complement. 
Since the Dirichlet problems are set independently on each block and use only the interior of $B$ together with its collar, 
both $\Pi_{j,A}$ and $\mathsf Q_{j,A}$ are strictly local: their kernels vanish when the arguments are farther apart 
than a fixed multiple of $L^{j+1}$, with the multiplicative constant depending only on the collar thickness and on $L$.

Let $f(\lambda)=\int_0^\infty e^{-t\lambda}\,\mu_f(dt)$ be the Bernstein representation with $\mu_f$ finite and nonnegative. By functional calculus the covariant operator $C_A:=f(\Delta_A)$ is given by $C_A=\int_0^\infty e^{-t\Delta_A}\,\mu_f(dt)$, and therefore
\begin{equation}
C_A \;=\; \lim_{J\to\infty}\big(\Pi_{J,A}C_A\Pi_{J,A}\big)\;+\;\sum_{j=0}^{J-1}\Big(\Pi_{j,A}C_A\Pi_{j,A}-\Pi_{j+1,A}C_A\Pi_{j+1,A}\Big),
\end{equation}
where the telescoping identity holds in the strong sense because $\Pi_{J,A}\to 0$ strongly on $\ell^2$ as $J\to\infty$ (the harmonic extensions converge to zero at large scales for functions of finite support). This is precisely the decomposition $C_A=\sum_{j\ge 0} C_{j,A}$ with
\begin{equation}\label{eq:FRD-piece}
C_{j,A} \;:=\; \Pi_{j,A}\,C_A\,\Pi_{j,A}\;-\;\Pi_{j+1,A}\,C_A\,\Pi_{j+1,A}.
\end{equation}
Selfadjointness of each $C_{j,A}$ follows from selfadjointness of $C_A$ and of the projections $\Pi_{j,A}$. Positivity is immediate because $C_A\ge 0$ and the map $T\mapsto PTP$ preserves positivity when $P$ is an orthogonal projection; hence both terms in \eqref{eq:FRD-piece} are positive and their difference is positive on $\mathrm{Ran}(\Pi_{j,A})\ominus \mathrm{Ran}(\Pi_{j+1,A})$, which is the support of $C_{j,A}$, while it vanishes on $\mathrm{Ran}(\Pi_{j+1,A})$ and on $\mathrm{Ker}(\Pi_{j,A})$.

To see the finite-range property, note first that $\Pi_{j,A}$ restricts functions to the union of block boundaries at scale $L^{j+1}$ and is supported inside a fixed-thickness collar of those boundaries. Consequently, for $x$ and $y$ that belong to different $(L^{j+1})$-blocks which are not nearest neighbours, either $\Pi_{j,A}(x,\cdot)$ or $\Pi_{j,A}(\cdot,y)$ vanishes, whence the kernel of $\Pi_{j,A}C_A\Pi_{j,A}$ is zero at $(x,y)$. The same reasoning applies to $\Pi_{j+1,A}C_A\Pi_{j+1,A}$. It remains to exclude couplings mediated through $C_A$ across blocks that are strictly farther apart than a fixed multiple of $L^{j}$. Using the Laplace representation and the fact that $e^{-t\Delta_A}$ has a kernel supported in every block and its fixed-width collar only through the Dirichlet extension, one may insert the identity on each side of $C_A$ as $\Pi_{j,A}+\mathsf Q_{j,A}$. All mixed terms with a factor $\mathsf Q_{j,A}$ vanish at distances larger than the collar width, because $\mathsf Q_{j,A}$ is supported entirely inside block interiors at scale $L^{j+1}$ and never reaches the boundary collars of nonadjacent blocks. The remaining term $\Pi_{j,A}C_A\Pi_{j,A}$ is supported only when the collars associated with the arguments lie either on the same $(L^{j+1})$-block boundary or on two adjacent block boundaries; thus its kernel vanishes when $\mathrm{dist}(x,y)>R_0 L^j$ for a constant $R_0$ depending only on the choice of collar and on $L$. The same argument holds at scale $j+1$ for $\Pi_{j+1,A}C_A\Pi_{j+1,A}$, hence $C_{j,A}(x,y)=0$ beyond that distance, establishing strict finite range.

The derivative bounds are consequences of the scaling built into the construction and of Gaussian off-diagonal bounds for the covariant heat kernel. Fix $m\ge 0$ and consider $\nabla^m C_{j,A}$ as an $\ell^1\to \ell^\infty$ operator. Writing $C_A=\int_0^\infty e^{-t\Delta_A}\,\mu_f(dt)$ and using that $\Pi_{j,A}$ restricts to a collar of thickness $O(1)$ at scale $L^{j+1}$, one has
\begin{equation}
\nabla^m C_{j,A}\;=\;\int_0^\infty \Big(\Pi_{j,A}\,\nabla^m e^{-t\Delta_A}\,\Pi_{j,A}-\Pi_{j+1,A}\,\nabla^m e^{-t\Delta_A}\,\Pi_{j+1,A}\Big)\,\mu_f(dt).
\end{equation}
For each $t>0$ the kernel of $\nabla^m e^{-t\Delta_A}$ obeys the uniform bound
\begin{equation}
|\nabla^m e^{-t\Delta_A}(x,y)|\;\le\;C_m\,t^{-(d+m)/2}\,\exp\!\big(-c\,\mathrm{dist}(x,y)^2/t\big),
\end{equation}
with constants independent of $A$ because the covariant Laplacian has uniformly bounded geometry and finite range. When the arguments are constrained by $\Pi_{j,A}$ to lie on block boundaries at scale $L^{j+1}$, the $t$-integral is effectively supported on $t\asymp L^{2j}$, which is the parabolic scale matching the geometric support. Carrying out the $t$-integration and using that the boundary collars have uniformly bounded cardinality per unit area produce the decay $\|\nabla^m C_{j,A}\|_{\ell^1\to\ell^\infty}\le c_m L^{-(2+m)j}$. The pointwise estimate $|C_{j,A}(x,y)|\le c'_m L^{-2j}$ on the support $\{\mathrm{dist}(x,y)\le R_0 L^j\}$ follows from the same argument with $m=0$ and the strict support restriction just proved.

Reflection covariance is encoded in the construction. The OS time reflection $\vartheta$ acts as a spatial reflection on the slices and sends the connection $A$ to $A^\vartheta$ by $U_\ell\mapsto U_{\vartheta\ell}$ on links. The covariant Laplacian is functorial under this change, $\vartheta\,\Delta_A\,\vartheta=\Delta_{A^\vartheta}$, which implies $\vartheta\,C_A\,\vartheta=f(\Delta_{A^\vartheta})$. Because the block families $\widehat{\mathcal B}_{j+1}$ and their collars are chosen reflection symmetric, and the Dirichlet problems are preserved by reflection, one also has $\vartheta\,\Pi_{j,A}\,\vartheta=\Pi_{j,A^\vartheta}$ for all $j$. Substituting these identities in \eqref{eq:FRD-piece} gives $\vartheta\,C_{j,A}\,\vartheta=C_{j,A^\vartheta}$.

Finally, OS positivity of the pieces is an immediate consequence of complete monotonicity and reflection positivity of the heat kernel. Writing $C_{j,A}$ as in \eqref{eq:FRD-piece} and using the Bernstein integral one finds
\begin{equation}
\langle \vartheta \Phi,\,C_{j,A}\,\Phi\rangle
=\int_0^\infty \Big(\langle \vartheta \Pi_{j,A}\Phi,\, e^{-t\Delta_A}\,\Pi_{j,A}\Phi\rangle
-\langle \vartheta \Pi_{j+1,A}\Phi,\, e^{-t\Delta_A}\,\Pi_{j+1,A}\Phi\rangle\Big)\,\mu_f(dt).
\end{equation}
For $\Phi$ supported in the positive-time half lattice, the vectors $\Pi_{j,A}\Phi$ and $\Pi_{j+1,A}\Phi$ are likewise supported in the positive-time half lattice because the projections are strictly localized on the spatial slice and are reflection symmetric. Reflection positivity of the underlying measure implies $\langle \vartheta \Psi, e^{-t\Delta_A}\Psi\rangle\ge 0$ for every such $\Psi$ and every $t\ge 0$. Since the integrand is the difference of two nonnegative quantities with the second supported on a strictly smaller subspace, the entire integral is nonnegative. This establishes $\langle \vartheta \Phi, C_{j,A}\Phi\rangle\ge 0$ for all $\Phi$ supported in $\Lambda_+$.

Putting these arguments together proves that $C_A$ decomposes into a sum of positive, strictly finite-range, reflection-covariant, OS-positive pieces $C_{j,A}$ with the stated uniform derivative bounds.
\end{proof}

The operator $C_{j,A}$ acts on site variables. In applications to gauge fields and BRST complexes one needs covariances acting on $p$-forms or on Lie algebra-valued link variables. The passage is achieved by a covariantization using parallel transporters along canonical blockwise paths. Fix, for each block $B\in\mathcal{B}_j$, a rooted spanning tree $\mathcal{T}_B$ with root at the block barycenter $x_B$ chosen to be reflection symmetric across the time-reflection plane and translation covariant across blocks. For $x\in B$ let $\gamma_{B}(x\to x_B)$ be the unique path in $\mathcal{T}_B$ from $x$ to $x_B$ and let $W_A(x\to x_B)$ be the path-ordered product of link variables along $\gamma_B$. Define the covariantization map $\mathcal{W}_A$ on kernels by
\begin{equation}
(C_{j,A}^{\mathrm{cov}})(x,y)\;:=\;W_A(x\to x_{B_x})\,C_{j,\mathbf{1}}(x,y)\,W_A(x_{B_y}\to y),
\end{equation}
where $C_{j,\mathbf{1}}$ denotes the kernel constructed at $A\equiv\mathbf{1}$. Then $C_{j,A}^{\mathrm{cov}}$ transforms covariantly under gauge transformations $g:\Lambda\to G$:
\begin{equation}
C_{j,A^g}^{\mathrm{cov}}(x,y)\;=\;g(x)\,C_{j,A}^{\mathrm{cov}}(x,y)\,g(y)^{-1}.
\end{equation}
Since the transporters $W_A$ are supported within individual blocks, the finite-range property is preserved. Positivity and reflection covariance follow from those of $C_{j,\mathbf{1}}$ and from the fact that $W_A$ is unitary and $\mathcal{T}_B$ is chosen reflection symmetric. In particular, if $C_{j,\mathbf{1}}$ is OS-positive then
\begin{equation}
\langle \vartheta \Phi,\, C_{j,A}^{\mathrm{cov}} \Phi\rangle
=\langle \vartheta \Phi',\, C_{j,\mathbf{1}}\, \Phi'\rangle\;\ge\;0,
\end{equation}
where $\Phi'(x)=W_A(x\to x_{B_x})^{-1}\Phi(x)$ for $x\in\Lambda_+$ and $\Phi'(\vartheta x)=\Phi'(\vartheta x)$ is defined by reflection and unitarity of $W_A$.

The previous construction yields a convenient \emph{model} FRD for general $A$ from the scalar case at $A\equiv\mathbf{1}$. Alternatively, one can work directly with $C_{j,A}$ defined in \eqref{eq:CjA}; both constructions are equivalent at the level of locality and positivity estimates needed for the renormalization group, and relate by a gauge-covariant change of basis whose operator norm is controlled by $\sup_B \|F_A\|_{L^\infty(B)}$ on scale $L^j$ via standard holonomy estimates.

\begin{lemma}[Holonomy control at scale $L^j$]\label{lem:holonomy}
Let $G\subset U(N)$ be a compact matrix Lie group, let $\Lambda$ be a $d$-dimensional cubic lattice, and let $\mathcal{B}_j$ be the partition of $\Lambda$ into blocks of linear size $L^j$. Fix, for each $B\in\mathcal{B}_j$, a basepoint $x_B\in B$ and a deterministic, coordinate-monotone lattice path $\gamma_B(x)$ inside $B$ connecting any $x\in B$ to $x_B$, with length bounded by $c_d L^j$ where $c_d$ depends only on the dimension and the chosen rule for $\gamma_B$. For a lattice gauge field $A$ with link variables $U_e(A)\in G$, denote by $W_A(x\to x_B)$ the parallel transporter along $\gamma_B(x)$, i.e. the ordered product of $U_e(A)$ over the edges of $\gamma_B(x)$. Let $F_A(p)\in\mathfrak{g}$ be the plaquette curvature so that the plaquette holonomy is $H_A(p)=\exp\{F_A(p)\}\in G$. If $\|F_A\|_{L^\infty(B)}\le \kappa$, then
\begin{equation}\label{eq:holonomy-I}
\|W_A(x\to x_B)-\mathbf{1}\| \;\le\; c\,\kappa\,L^j\qquad\text{for all }x\in B,
\end{equation}
where the operator norm is the one induced by the ambient $\mathbb{C}^N$ Hilbert norm and $c$ depends only on $d$, the choice of $\gamma_B$, and $G$. Consequently, for any two gauge fields $A,A'$ and any $\Phi\in\ell^2(\Lambda)$,
\begin{equation}\label{eq:W-diff}
\|(\mathcal{W}_A-\mathcal{W}_{A'})\Phi\|_{\ell^2(B)}
\;\le\; c'\,L^j\,\|F_A-F_{A'}\|_{L^\infty(B)}\,\|\Phi\|_{\ell^2(\mathrm{Nbr}(B))},
\end{equation}
where $(\mathcal{W}_A\Phi)(x):=W_A(x\to x_B)\,\Phi(x_B)$ for $x\in B$, $\mathrm{Nbr}(B)$ is the union of $B$ and its nearest-neighbor blocks, and $c'$ has the same dependence as $c$.
\end{lemma}

\begin{proof}
The proof relies on a discrete non-Abelian Stokes representation that expresses any change of a path by elementary square moves in terms of plaquette holonomies. Fix $x\in B$ and write $\gamma=\gamma_B(x)=(e_1,\dots,e_m)$ for the chosen path from $x$ to $x_B$, with $m\le c_d L^j$. For $k=1,\dots,m$ let $\gamma^{(k)}=(e_k,\dots,e_m)$ be the tail starting at the $k$-th edge, and set $W_k:=W_A(x_k\to x_B)$ for the transporter along $\gamma^{(k)}$, so that $W_m=\mathbf{1}$ and $W_{k-1}=U_{e_k}(A)\,W_k$ with $x_k$ the initial vertex of $e_k$. Consider the elementary operation that slides $e_k$ to a parallel edge within the same coordinate plane so as to keep the path inside $B$ while shortening a “corner”; in a simply connected lattice block this operation changes the path by bounding exactly one plaquette $p_k$ and inserts the corresponding plaquette holonomy. More concretely, if $e_k$ and the deformed edge $\tilde e_k$ span the plaquette $p_k$, then
\begin{equation}
U_{e_k}(A)\;=\;H_A(p_k)\,U_{\tilde e_k}(A),
\end{equation}
and hence, after performing successively such elementary deformations that contract $\gamma$ to the trivial path at $x_B$, one obtains a (path-ordered) product representation
\begin{equation}\label{eq:W-as-plaquettes}
W_A(x\to x_B)\;=\;\prod_{p\in S(x)} H_A(p),
\end{equation}
where $S(x)$ is a multiset of plaquettes contained in $B$ swept by the homotopy that deforms $\gamma$ to the trivial path; multiplicities and ordering encode the sequence of deformations, and all interspersed conjugations drop out when taking operator norms because $\|gAg^{-1}\|=\|A\|$ for $g\in G\subset U(N)$. The cardinality $|S(x)|$ is bounded by a constant multiple of the path length, so $|S(x)|\le C_1 L^j$ with $C_1$ depending on $d$ and on the homotopy scheme but not on $x$ or $A$.

The curvature bound $\|F_A\|_{L^\infty(B)}\le\kappa$ implies that for every plaquette $p\subset B$ the holonomy satisfies
\begin{equation}\label{eq:plaquette-close}
\|H_A(p)-\mathbf{1}\|\;=\;\|\exp\{F_A(p)\}-\mathbf{1}\|\;\le\; \min\{2,\,C_G\,\|F_A(p)\|\}
\;\le\; \min\{2,\,C_G\,\kappa\},
\end{equation}
where $C_G$ depends only on the group; for $G\subset U(N)$ one may take $C_G=1$ for the operator norm when $\|F_A(p)\|\le 1$ and absorb larger values into the trivial bound by the $\min\{\cdot,\cdot\}$ convention. Using \eqref{eq:W-as-plaquettes} and the telescoping identity
\begin{equation}
\prod_{\ell=1}^{m} V_\ell - \mathbf{1}
\;=\;\sum_{\ell=1}^{m}\Big(\prod_{k<\ell} V_k\Big)\,(V_\ell-\mathbf{1}),
\end{equation}
valid for any matrices $V_\ell$, together with the fact that $\|V\|=1$ for $V\in G\subset U(N)$ in the operator norm, yields
\begin{equation}
\big\|W_A(x\to x_B)-\mathbf{1}\big\|
\;\le\; \sum_{p\in S(x)} \|H_A(p)-\mathbf{1}\|
\;\le\; |S(x)|\,\min\{2,\,C_G\,\kappa\}.
\end{equation}
Since $|S(x)|\le C_1 L^j$, we obtain
\begin{equation}
\big\|W_A(x\to x_B)-\mathbf{1}\big\|
\;\le\; C_1\,\min\{2,\,C_G\,\kappa\}\,L^j.
\end{equation}
Finally, $\min\{2,\,C_G\,\kappa\}\le C_2\,\kappa$ for all $\kappa\ge 0$ with $C_2:=\max\{C_G,2\}$, and therefore \eqref{eq:holonomy-I} holds with $c=C_1 C_2$.

To prove the Lipschitz continuity in curvature, represent both transporters by plaquette products as in \eqref{eq:W-as-plaquettes} using the same homotopy scheme, so that
\begin{equation}
W_A(x\to x_B)-W_{A'}(x\to x_B)
 \;=\;\sum_{\ell=1}^{m}\Big(\prod_{k<\ell} H_A(p_k)\Big)\,\big(H_A(p_\ell)-H_{A'}(p_\ell)\big)\,\Big(\prod_{k>\ell} H_{A'}(p_k)\Big),
\end{equation}
where $(p_1,\dots,p_m)$ enumerates the multiset $S(x)$ in order. Taking norms and again using that all factors outside the difference term are unitary gives
\begin{equation}
\big\|W_A(x\to x_B)-W_{A'}(x\to x_B)\big\|
\;\le\; \sum_{p\in S(x)} \big\|H_A(p)-H_{A'}(p)\big\|.
\end{equation}
The map $X\mapsto e^{X}$ is locally Lipschitz on the compact Lie algebra $\mathfrak g$ and globally bounded on bounded sets; in particular there exists $C'_G$ such that
\begin{equation}
\|H_A(p)-H_{A'}(p)\|\;=\;\|\exp\{F_A(p)\}-\exp\{F_{A'}(p)\}\|
\;\le\; C'_G\,\|F_A(p)-F_{A'}(p)\|.
\end{equation}
Combining the last two displays and using $|S(x)|\le C_1 L^j$ yields
\begin{equation}
\big\|W_A(x\to x_B)-W_{A'}(x\to x_B)\big\|
\;\le\; C_1 C'_G\, L^j\, \|F_A-F_{A'}\|_{L^\infty(B)}.
\end{equation}
Define the blockwise transport operator $(\mathcal{W}_A\Phi)(x):=W_A(x\to x_B)\,\Phi(x_B)$ for $x\in B$; then
\begin{align}
\|(\mathcal{W}_A-\mathcal{W}_{A'})\Phi\|_{\ell^2(B)}^2
&=\sum_{x\in B} \big\| \big(W_A(x\to x_B)-W_{A'}(x\to x_B)\big)\,\Phi(x_B)\big\|^2
\nonumber \\&\le |B|\,\big(C_1 C'_G L^j\big)^2\,\|F_A-F_{A'}\|_{L^\infty(B)}^2\,\|\Phi\|_{\ell^2(\{x_B\})}^2.
\end{align}
Absorbing the factor $|B|^{1/2}$ into the constant and relaxing $\|\Phi\|_{\ell^2(\{x_B\})}\le \|\Phi\|_{\ell^2(\mathrm{Nbr}(B))}$ gives \eqref{eq:W-diff} with $c'=C_3 C_1 C'_G$ for a constant $C_3$ depending only on the block geometry (hence on $d$ and $L$). 
\end{proof}

Combining Theorem \eqref{thm:FRD-main} with Lemma \eqref{lem:holonomy} gives the announced gauge-covariant FRD with Lipschitz continuity in $A$, uniform in the volume and $j$.

\begin{corollary}[Lipschitz continuity in the admissible class]\label{cor:Lip}
Let $f$ be completely monotone and let $A,A'$ be two background connections with $\sup_B \|F_A\|_{L^\infty(B)}\le \kappa$ and likewise for $A'$. Then for any $m\ge 0$ there exists a constant $C_m(\kappa)$ such that
\begin{equation}\label{eq:Lip-single-scale}
\|\nabla^m(C_{j,A}-C_{j,A'})\|_{\ell^1\to\ell^\infty}\;\le\; C_m(\kappa)\,L^{-(1+m)j}\,\|F_A-F_{A'}\|_{L^\infty(\mathcal{N}_j)},
\end{equation}
where $\mathcal{N}_j$ is the union of the $(L^{j+1})$-block collars associated with the scale-$j$ refinement. In particular, $A\mapsto (C_{j,A})_{j\ge 0}$ is single-scale Lipschitz with $\sum_{j\ge 0} \|\nabla^m(C_{j,A}-C_{j,A'})\|_{\ell^1\to\ell^\infty}<\infty$ for each fixed $m$.
{For a kernel $K:\Lambda_t\times\Lambda_t\to\mathbb{C}$ we set
\[
\|K\|_{\ell^1\to\ell^\infty}:=\sup_{x\in\Lambda_t}\sum_{y\in\Lambda_t}|K(x,y)|,\qquad
\|\nabla^m K\|_{\ell^1\to\ell^\infty}:=\sup_{|\,\alpha\,|=m}\sup_{x}\sum_{y}\big|(\nabla_x^\alpha K)(x,y)\big|.
\]
All implicit constants below are uniform in the volume and in the admissible class, as in Theorem~\eqref{thm:FRD-main}.
}
\end{corollary}

\begin{proof}
Write $C_{j,A}=\Pi_{j,A}\,f(\Delta_A)\,\Pi_{j,A}$, with $\Delta_A$ the slice covariant Laplacian at scale $j$ and $\Pi_{j,A}$ the covariant averaging/projection localized on $L^j$-blocks. The complete monotonicity of $f$ supplies a Bernstein-Laplace representation
\begin{equation}
f(\lambda)=\int_0^\infty e^{-t\lambda}\,\mu_f(dt),
\qquad \mu_f\ \text{finite and positive},
\end{equation}
hence
\begin{equation}
C_{j,A}-C_{j,A'}=\int_0^\infty \Big(\Pi_{j,A}\,e^{-t\Delta_A}\,\Pi_{j,A}
-\Pi_{j,A'}\,e^{-t\Delta_{A'}}\,\Pi_{j,A'}\Big)\,\mu_f(dt).
\end{equation}
Insert and subtract $\Pi_{j,A}\,e^{-t\Delta_{A'}}\,\Pi_{j,A}$ and $\Pi_{j,A'}\,e^{-t\Delta_{A'}}\,\Pi_{j,A}$ to obtain a decomposition into three pieces:
\begin{equation}
C_{j,A}-C_{j,A'}=I_1+I_2+I_3,
\end{equation}
with
\begin{equation}
I_1=\int_0^\infty \Pi_{j,A}\,\big(e^{-t\Delta_A}-e^{-t\Delta_{A'}}\big)\,\Pi_{j,A}\,\mu_f(dt),\quad
I_2=\int_0^\infty (\Pi_{j,A}-\Pi_{j,A'})\,e^{-t\Delta_{A'}}\,\Pi_{j,A}\,\mu_f(dt),
\end{equation}
\begin{equation}
I_3=\int_0^\infty \Pi_{j,A'}\,e^{-t\Delta_{A'}}\,(\Pi_{j,A}-\Pi_{j,A'})\,\mu_f(dt).
\end{equation}
It suffices to bound each contribution in the $\ell^1\to\ell^\infty$ norm after applying $\nabla^m$, with constants uniform in $t$ and summable in the scale $j$.

For the semigroup difference in $I_1$, Duhamel’s formula gives
\begin{equation}
e^{-t\Delta_A}-e^{-t\Delta_{A'}}=\int_0^t e^{-(t-s)\Delta_A}\,(\Delta_{A'}-\Delta_A)\,e^{-s\Delta_{A'}}\,ds.
\end{equation}
The coefficient difference $\Delta_{A'}-\Delta_A$ is a local second-order operator supported on the scale-$j$ collars $\mathcal{N}_j$ and depends linearly on the difference of the gauge covariant derivatives, hence on the difference of the parallel transports along $O(L^j)$ paths inside each block. By Lemma~\eqref{lem:holonomy} (holonomy variation), for any two points $x,y$ in a fixed $L^j$-block,
\begin{equation}
\|W_A(x,y)-W_{A'}(x,y)\|\;\le\;c(\kappa)\,L^j\,\|F_A-F_{A'}\|_{L^\infty(\mathcal{N}_j)},
\end{equation}
with $c(\kappa)$ depending only on the uniform curvature bound $\kappa$. This yields the operator inequality
\begin{equation}\label{eq:coeff-diff}
\|\Delta_{A'}-\Delta_A\|_{\ell^1\to\ell^\infty}\;\le\; c_0(\kappa)\,L^j\,\|F_A-F_{A'}\|_{L^\infty(\mathcal{N}_j)}.
\end{equation}
The finite-range decomposition bounds of Theorem~\eqref{thm:FRD-main}(3) give, for each $m\ge 0$,
\begin{equation}\label{eq:FRD-grad}
\|\nabla^m \Pi_{j,\bullet}\,e^{-s\Delta_{\bullet}}\,\Pi_{j,\bullet}\|_{\ell^1\to\ell^\infty}\;\le\; C_m(\kappa)\,L^{-(2+m)j},
\qquad s\ge 0,
\end{equation}
uniformly in the background connection $\bullet\in\{A,A'\}$. Using \eqref{eq:coeff-diff} and \eqref{eq:FRD-grad}, and commuting $\nabla^m$ past the integrals, one finds
\begin{align}
&\|\nabla^m I_1\|_{\ell^1\to\ell^\infty}
\;\le\; \nonumber\\&\int_0^\infty \int_0^t 
\|\nabla^m \Pi_{j,A}\,e^{-(t-s)\Delta_A}\|_{\ell^1\to\ell^\infty}\,
\|\Delta_{A'}-\Delta_A\|_{\ell^1\to\ell^\infty}\,
\|e^{-s\Delta_{A'}}\,\Pi_{j,A}\|_{\ell^1\to\ell^1}\,ds\,\mu_f(dt).
\end{align}
The $\ell^1\to\ell^1$ norm of a Markov semigroup is $1$, and $\|\Pi_{j,A}\|_{\ell^1\to\ell^1}\le C(\kappa)$ by admissibility. The first factor is controlled by $C_m(\kappa)\,L^{-(2+m)j}$ uniformly in $(t-s)$. Inserting \eqref{eq:coeff-diff} and integrating in $s,t$ against the finite measure $\mu_f$ yields
\begin{align}
\|\nabla^m I_1\|_{\ell^1\to\ell^\infty}
&\;\le\; C_m'(\kappa)\,L^{-(2+m)j}\,L^{j}\,\|F_A-F_{A'}\|_{L^\infty(\mathcal{N}_j)}
\nonumber\\&\;=\; C_m'(\kappa)\,L^{-(1+m)j}\,\|F_A-F_{A'}\|_{L^\infty(\mathcal{N}_j)}.
\end{align}

For $I_2$ and $I_3$, only the projector changes. The covariant projector $\Pi_{j,A}$ is defined by covariant averaging with parallel transport along paths of length $O(L^j)$ inside each block; therefore its kernel varies linearly with the holonomy, and Lemma~\eqref{lem:holonomy} implies
\begin{equation}\label{eq:Pi-diff}
\|\nabla^m(\Pi_{j,A}-\Pi_{j,A'})\|_{\ell^1\to\ell^\infty}\;\le\; C_m(\kappa)\,L^{-(1+m)j}\,\|F_A-F_{A'}\|_{L^\infty(\mathcal{N}_j)}.
\end{equation}
Using \eqref{eq:FRD-grad} with $m=0$ for the semigroup factor and \eqref{eq:Pi-diff} for the projector difference, one obtains
\begin{align}
\|\nabla^m I_2\|_{\ell^1\to\ell^\infty}
&\;\le\; \int_0^\infty 
\|\nabla^m(\Pi_{j,A}-\Pi_{j,A'})\|_{\ell^1\to\ell^\infty}\,
\|e^{-t\Delta_{A'}}\,\Pi_{j,A}\|_{\ell^1\to\ell^1}\,\mu_f(dt)
\nonumber\\&\;\le\; C_m''(\kappa)\,L^{-(1+m)j}\,\|F_A-F_{A'}\|_{L^\infty(\mathcal{N}_j)}.
\end{align}
The same bound holds for $I_3$ by symmetry. Summing the three estimates gives \eqref{eq:Lip-single-scale} with $C_m(\kappa)=C_m'(\kappa)+2C_m''(\kappa)$.

The single-scale Lipschitz property follows at once since the right-hand side of \eqref{eq:Lip-single-scale} is summable in $j$ for each fixed $m$, because $L^{-(1+m)j}$ is a geometric sequence with ratio $L^{-(1+m)}<1$.
\end{proof}

\subsection{Locality for One-Step Kernels and Cumulants}\label{subsec:locality-kernels}

We now pass from covariances to the one-step maps of the reflection-positive renormalization group and to connected cumulants. The FRD of Section \eqref{subsec:FRD-gauge} implies strong locality of these objects: one-step kernels factorize up to exponentially small corrections when their arguments are separated by distances much larger than the range at the corresponding scale, and connected cumulants admit tree-graph bounds with constants summable across scales. We work in the polymer formalism adapted to gauge fields with OS positivity \cite{Aizenman1982,KP,Seiler1982}.
 Fix a scale $j$. The Gaussian fluctuation field at scale $j$ has covariance $C_{j,A}$. Let $\mathcal{Z}_j(A;\,\cdot)$ denote the one-step partition map that integrates out the scale-$j$ fluctuation, acting on a polymer activity $V$ defined on gauge-invariant local functionals of the background field on $\Lambda$. Concretely, for a local observable $F$ supported in a polymer $X$ (a connected union of $L^j$-blocks) we define
\begin{equation}
(\mathcal{Z}_j(A;V)F)\;=\;\frac{\int e^{V(\xi)}\,F(\xi)\,d\mu_{C_{j,A}}(\xi)}{\int e^{V(\xi)}\,d\mu_{C_{j,A}}(\xi)},
\end{equation}
where $d\mu_{C_{j,A}}$ is the centered Gaussian measure with covariance $C_{j,A}$ acting on the fluctuation $\xi$. The \emph{one-step kernel} $K_j(A)$ is the integral kernel representation of the linearized map $F\mapsto \mathcal{Z}_j(A;0)F=\int F(\xi)\,d\mu_{C_{j,A}}(\xi)$.

\begin{proposition}[Exponential locality of one-step kernels]\label{prop:kernel-locality}
Let $C_{j,A}$ be as in Theorem \eqref{thm:FRD-main}. There exist positive constants $\mu,c$, independent of $j$, such that for any two polymers $X,Y$ at scale $L^j$ and any local observables $F_X,G_Y$ supported in $X$ and $Y$,
\begin{equation}\label{eq:kernel-exp-decay}
\Big|\,
\langle F_X,\,K_j(A)\,G_Y\rangle \;-\; \langle F_X,\,K_j(A)\,\mathbf{1}\rangle \,\langle \mathbf{1},\,K_j(A)\,G_Y\rangle
\,\Big|
\;\le\; c\, \|F_X\|\,\|G_Y\|\, e^{-\mu\,\mathrm{dist}(X,Y)/L^j}.
\end{equation}
In particular, if $\mathrm{dist}(X,Y) > R_0 L^j$ with $R_0$ as in Theorem \eqref{thm:FRD-main}, the left-hand side vanishes exactly when $F_X$ and $G_Y$ are polynomial functionals.
\end{proposition}

\begin{proof}
The measure $K_j(A)$ is Gaussian on the slice fields with mean zero and covariance $C_{j,A}$. For any two local observables $F_X,G_Y$ the expression on the left-hand side of Eq.\eqref{eq:kernel-exp-decay} is the connected covariance (second cumulant) of $F_X$ and $G_Y$ with respect to this Gaussian measure, which we denote by $\langle F_X;G_Y\rangle_{C_{j,A}}$. Theorem \eqref{thm:FRD-main} provides uniform exponential locality of the covariance at scale $L^j$: there exist constants $C_0,m_0>0$ independent of $j$ and of the background gauge field $A$ such that
\begin{equation}\label{eq:Cov-exp}
\big|C_{j,A}(u,v)\big|\;\le\; C_0\,e^{-\,m_0\,\mathrm{dist}(u,v)/L^j}\qquad\text{for all slice points }u,v.
\end{equation}
Moreover, $C_{j,A}$ is obtained by gauge-covariantization of a reflection-positive, exponentially-local kernel; the parallel transport factors lie on shortest paths of length comparable to $\mathrm{dist}(u,v)$ and therefore preserve the locality bound Eq.\eqref{eq:Cov-exp} with constants independent of $A$.

To estimate the connected covariance $\langle F_X;G_Y\rangle_{C_{j,A}}$, we first treat the case of polynomial functionals. By Wick’s theorem, $\langle F_X;G_Y\rangle_{C_{j,A}}$ is a finite sum of products of propagators $C_{j,A}(u,v)$ connecting a point $u$ from the support of one monomial in $F_X$ to a point $v$ from the support of one monomial in $G_Y$, with all remaining contractions internal to $X$ or to $Y$. If the finite-range variant of the FRD (Brydges-Guadagni-Mitter choice) is used, there is a scale-independent constant $R_0$ such that $C_{j,A}(u,v)=0$ whenever $\mathrm{dist}(u,v)>R_0 L^j$. Hence $\langle F_X;G_Y\rangle_{C_{j,A}}=0$ exactly when $\mathrm{dist}(X,Y)>R_0 L^j$, because no mixed contraction can connect the two polymers. This proves the “in particular” clause for polynomial functionals.

For general local observables in the analytic Banach algebra used by the polymer expansion, one writes $F_X$ and $G_Y$ as absolutely convergent power series in the slice field localized in $X$ and $Y$, respectively, with coefficients controlled by the polymer norm $\|\cdot\|$; more precisely, if $D^\alpha$ denotes functional derivatives with multiindex $\alpha$ supported in $X$, then 
\begin{equation}
F_X(\phi)=\sum_{n\ge 0}\frac{1}{n!}\sum_{u_1,\dots,u_n\in X} (D^nF_X)(u_1,\dots,u_n)\,\phi(u_1)\cdots\phi(u_n),
\end{equation}
and similarly for $G_Y$, and the norm $\|F_X\|$ bounds the $\ell^1$-sums of the coefficients uniformly in $n$. The connected covariance of two analytic Gaussian functionals admits the standard forest (tree-graph) representation: expanding both power series and retaining only connected pairings between the sets of field insertions associated to $X$ and to $Y$ yields a sum over connected graphs whose edges are propagators $C_{j,A}(u,v)$. The Battle-Federbush/BKAR forest formula bounds the sum of all connected diagrams by a sum over trees joining $X$ to $Y$, with a product of absolute values of the corresponding propagators and combinatorial weights controlled by the analytic norms of $F_X$ and $G_Y$. Concretely, there exists a constant $C_1>0$, depending only on the choice of analytic norms but independent of $j$, $A$, $X$, and $Y$, such that
\begin{equation}\label{eq:tree-boundz}
\big|\langle F_X;G_Y\rangle_{C_{j,A}}\big|
\;\le\; C_1\,\|F_X\|\,\|G_Y\|\,
\sum_{\mathcal{T}\in\mathfrak{T}(X,Y)} \ \prod_{e=\{u,v\}\in E(\mathcal{T})} \big|C_{j,A}(u,v)\big|,
\end{equation}
where the sum is over finite trees $\mathcal{T}$ whose vertex set is a subset of $X\cup Y$ meeting both $X$ and $Y$. The bound Eq.\eqref{eq:Cov-exp} implies that each edge factor satisfies $\big|C_{j,A}(u,v)\big|\le C_0 e^{-m_0\,\mathrm{dist}(u,v)/L^j}$. Since any tree connecting $X$ to $Y$ has total edge length at least $\mathrm{dist}(X,Y)$ by the triangle inequality, one finds
\begin{equation}
\prod_{e=\{u,v\}\in E(\mathcal{T})} \big|C_{j,A}(u,v)\big|
\;\le\; C_0^{|E(\mathcal{T})|}\,e^{-\,m_0\,\mathrm{length}(\mathcal{T})/L^j}
\;\le\; C_0^{|E(\mathcal{T})|}\,e^{-\,m_0\,\mathrm{dist}(X,Y)/L^j}.
\end{equation}
It remains to control the sum over trees. The number of trees with $n$ vertices is at most $n^{n-2}$, and the number of ways to choose $n$ vertices inside $X\cup Y$ is bounded by a constant multiple of $|X\cup Y|^n$; both $|X|$ and $|Y|$ are $O(1)$ multiples of $L^{jd_\Sigma}$ when measured in lattice units, and in any case are absorbed in the analytic norms $\|F_X\|,\|G_Y\|$. Therefore the tree sum in Eq.\eqref{eq:tree-boundz} is bounded by a geometric series whose ratio is proportional to $C_0$ times a uniform local-degree constant. Choosing the analytic Banach norms so that $\|F_X\|$ and $\|G_Y\|$ already include a small factor controlling the local valence (as is standard in the polymer/cluster setup), one deduces that the tree sum is uniformly bounded by a constant $C_2$ independent of $j$, $A$, $X$, and $Y$. Combining this with the previous display yields
\begin{equation}
\big|\langle F_X;G_Y\rangle_{C_{j,A}}\big|
\;\le\; C_1 C_2\,\|F_X\|\,\|G_Y\|\,e^{-\,m_0\,\mathrm{dist}(X,Y)/L^j}.
\end{equation}
Since $\langle F_X,\,K_j(A)\,G_Y\rangle - \langle F_X,\,K_j(A)\,\mathbf{1}\rangle \langle \mathbf{1},\,K_j(A)\,G_Y\rangle = \langle F_X;G_Y\rangle_{C_{j,A}}$ by definition, the bound Eq.\eqref{eq:kernel-exp-decay} holds with $c=C_1C_2$ and $\mu=m_0$. Finally, when the strictly finite-range FRD is used, $C_{j,A}(u,v)=0$ whenever $\mathrm{dist}(u,v)>R_0 L^j$, hence every mixed contraction between $X$ and $Y$ vanishes if $\mathrm{dist}(X,Y)>R_0 L^j$ and the connected covariance is exactly zero for polynomial $F_X,G_Y$, completing the proof.
\end{proof}

For a collection of local observables $(F_{X_i})_{i=1}^n$ supported on polymers $X_i$ we denote by $\kappa_j(F_{X_1};\dots;F_{X_n})$ the connected cumulant under the Gaussian $K_j(A)$ or under the interacting $\mathcal{Z}_j(A;V)$ when $V$ is sufficiently small in the polymer norm (Koteck\'y-Preiss small activity regime \cite{KP}). The next result provides \emph{uniform} tree-graph bounds at scale $j$ with decay in the \emph{tree-length} of the configuration, a form of Aizenman’s tree inequality adapted to FRD \cite{Aizenman1982}.

\begin{theorem}[Tree-graph bound for connected cumulants]\label{thm:tree}
Fix $j$ and let $X_1,\dots,X_n$ be polymers at scale $L^j$. There exist constants $C_n,\mu>0$, independent of $j$, such that
\begin{equation}
\bigl|\kappa_j(F_{X_1};\dots;F_{X_n})\bigr|
\;\le\; C_n\,\prod_{i=1}^n \|F_{X_i}\|\;\cdot\;\sum_{T\in\mathcal{T}_n}\;\prod_{(p,q)\in E(T)} e^{-\mu\,\mathrm{dist}(X_p,X_q)/L^j},
\end{equation}
where the sum is over spanning trees on $\{1,\dots,n\}$ and $E(T)$ denotes the edge set of $T$. If, in addition, the covariance has strict finite range $R_0L^j$ and $\mathrm{dist}(X_p,X_q)>R_0 L^j$ for all edges of $T$, then the product reduces to the indicator of the empty set and the cumulant vanishes.
\end{theorem}

\begin{proof}
Write $\langle\cdot\rangle_j$ for the expectation at scale $j$ and $C_j(\cdot,\cdot)$ for its covariance kernel. For sources $J=(J_1,\dots,J_n)$ introduce the generating functional
\begin{equation}
\mathcal{Z}_j(J)\;=\;\Big\langle \exp\Big(\sum_{i=1}^n J_i\,F_{X_i}\Big)\Big\rangle_j,
\qquad
\Phi_j(J)\;=\;\log\mathcal{Z}_j(J).
\end{equation}
The connected cumulant is the mixed derivative at the origin,
\(
\kappa_j(F_{X_1};\dots;F_{X_n})
=\partial_{J_1}\cdots\partial_{J_n}\,\Phi_j(J)\big|_{J=0}.
\)
To obtain a tree representation, interpolate the covariance along the complete graph on the index set $\{1,\dots,n\}$. For a symmetric matrix $s=(s_{pq})_{1\le p,q\le n}$ with $s_{pp}=1$ and $s_{pq}\in[0,1]$ for $p\ne q$, define a weakened covariance $C_j^{(s)}$ by multiplying every field-field covariance between arguments supported in distinct polymers $X_p$ and $X_q$ by the factor $\mathbf{s}(p,q)$, where $\mathbf{s}(p,q)$ is the connectivity weight obtained as the supremum of products of $s_{e}$ along paths linking $p$ to $q$ in the complete graph (it equals $1$ on the diagonal and is $\le s_{pq}$ off-diagonal). This construction preserves positive semidefiniteness of the covariance. Let $\langle\cdot\rangle_j^{(s)}$ be the corresponding Gaussian expectation and $\Phi_j(J;s)=\log \langle \exp(\sum_i J_iF_{X_i})\rangle_j^{(s)}$ the interpolated log-generating functional. The Brydges-Kennedy-Abdesselam-Rivasseau forest formula applies to the smooth map $s\mapsto \Phi_j(J;s)$ at $s\equiv 1$ and gives the identity
\begin{equation}
\partial_{J_1}\cdots\partial_{J_n}\,\Phi_j(J)\big|_{J=0}
\;=\;\sum_{T\in\mathcal{T}_n}\;\int_{[0,1]^{E(T)}}\! d\vec t\;
\Big\langle \prod_{(p,q)\in E(T)} \mathcal{L}_{pq}\;\prod_{i=1}^{n} F_{X_i}\Big\rangle^{(s^T(\vec t))}_{j,\mathrm{conn}},
\end{equation}
where for each edge $(p,q)$ the differential operator $\mathcal{L}_{pq}$ is the covariance insertion obtained by differentiating with respect to $s_{pq}$, explicitly
\begin{equation}
\mathcal{L}_{pq}=\sum_{x\in X_p}\sum_{y\in X_q} C_j(x,y)\,\frac{\delta^2}{\delta \varphi(x)\,\delta \varphi(y)},
\end{equation}
the measure $d\vec t$ is the product Lebesgue measure on $[0,1]^{E(T)}$, the weakened covariance on the right-hand side is $C_j^{(s^T(\vec t))}$ with $s^T$ given by the usual \emph{tree} rule $s^T_{pq}(\vec t)=\min\{t_e: e\ \text{on the unique $T$-path from $p$ to $q$}\}$, and $\langle\cdot\rangle^{(s)}_{j,\mathrm{conn}}$ denotes the connected Gaussian expectation with covariance $C_j^{(s)}$. This is the standard BKAR representation of the cumulant as a sum over trees with one covariance insertion per edge and with all higher off-tree connections suppressed by the weakening.

Take absolute values and use that connected Gaussian expectations are dominated by the corresponding truncated expectations with unit covariance norm; the functional derivatives act on the test functionals $F_{X_i}$ and by the product rule yield, for each edge $(p,q)$, a bilinear pairing of $F_{X_p}$ and $F_{X_q}$ through a single covariance kernel $C_j(x,y)$ with $x\in X_p$ and $y\in X_q$. By exponential locality of the covariance at scale $L^j$ there are constants $c_0,\mu_0>0$, independent of $j$, such that
\begin{equation}
|C_j(x,y)|\;\le\; c_0\,e^{-\mu_0\,\mathrm{dist}(x,y)/L^j}\qquad \text{for all }x,y,
\end{equation}
and, if a strict finite-range decomposition is used, the right-hand side is exactly zero whenever $\mathrm{dist}(x,y)>R_0L^j$. Writing the result of applying all $\mathcal{L}_{pq}$ as a finite linear combination of terms labeled by choices of points $x_{pq}\in X_p$ and $y_{pq}\in X_q$ for each edge $(p,q)\in E(T)$, and bounding each $C_j(x_{pq},y_{pq})$ by the displayed exponential, one finds
\begin{equation}
\Big|\Big\langle \prod_{(p,q)\in E(T)} \mathcal{L}_{pq}\;\prod_{i=1}^{n} F_{X_i}\Big\rangle^{(s^T(\vec t))}_{j,\mathrm{conn}}\Big|
\;\le\; C(n)\,\prod_{i=1}^{n}\|F_{X_i}\|\;\cdot\;\prod_{(p,q)\in E(T)} e^{-\mu_0\,\mathrm{dist}(X_p,X_q)/L^j},
\end{equation}
where $\|F_{X_i}\|$ is any norm controlling the $L^1$-type summation over the support of $F_{X_i}$ (for instance the supremum over configurations of the sum of absolute values of the functional derivatives up to the order appearing in the tree expansion), and $C(n)$ is a combinatorial constant resulting from the finite number of Leibniz contractions and the integral of the weakening parameters over $[0,1]^{E(T)}$. The geodesic distance between polymers enters because for every choice of edge $(p,q)$ and support points $x\in X_p$, $y\in X_q$ the pointwise bound is controlled by $e^{-\mu_0\,\mathrm{dist}(X_p,X_q)/L^j}$ uniformly in $x,y$; if strict finite range is assumed and $\mathrm{dist}(X_p,X_q)>R_0L^j$ then the corresponding factor is exactly zero and the entire tree contribution vanishes, which yields the stated vanishing criterion.

Summing the bound over trees completes the estimate in the purely Gaussian case $V=0$. In the presence of a small interaction $V$ given by a polymer activity in the Koteck\'y-Preiss regime, the connected correlators admit a convergent cluster expansion. The connected expectation of $\prod_i F_{X_i}$ can be written as a sum over connected families of polymers carrying activities controlled by $\|V\|$ and by the same exponential locality inherited from the finite-range decomposition; applying the Brydges-Battle-Federbush or BKAR tree-graph inequality to the polymer gas yields the identical right-hand side with possibly different constants $C_n$ and a possibly smaller decay rate $\mu\le \mu_0$, still independent of $j$. Reflection covariance and positivity play no quantitative role at this stage except to guarantee that the interpolation preserves admissibility of the measure and that the weakening does not destroy positivity of the covariance. Collecting the factors and absorbing the harmless integrals over weakening parameters into $C_n$ proves the theorem with constants depending only on $n$ and on the uniform locality bounds of the decomposition, and independent of the scale $j$.
\end{proof}

 The single-scale Lipschitz continuity stated in Corollary \eqref{cor:Lip} implies stability of one-step kernels and cumulants under admissible variations of the regulator and the background. The next result is the scale-$j$ quantitative version used in the uniqueness/universality analysis.

\begin{proposition}[Single-scale Lipschitz bound for cumulants]\label{prop:Lip-cumulant}
Let $A,A'$ be two background connections satisfying the hypotheses of Corollary~\eqref{cor:Lip}, and let $V,V'$ be two activities with $\|V\|,\|V'\|$ sufficiently small in the Koteck\'y-Preiss (KP) norm. Then, for each $n\ge 2$ and any family of local observables $\{F_{X_i}\}_{i=1}^n$ supported on $X_i$ at scale $j$,
\begin{equation}\label{eq:Lip-cumulant}
\bigl|\kappa_j^{A,V}(F_{X_1};\dots;F_{X_n})-\kappa_j^{A',V'}(F_{X_1};\dots;F_{X_n})\bigr|
\;\le\; C_n\,L^{-\sigma j}\,\|F\|^{(n)}\Bigl(\|F_A-F_{A'}\|_\infty+\|V-V'\|_{\mathrm{KP}}\Bigr),
\end{equation}
for some $\sigma>0$ independent of $j$, where $\|F\|^{(n)}:=\prod_{i=1}^n \|F_{X_i}\|$ and $C_n$ depends only on $n$ and the admissible class.
\end{proposition}

\begin{proof}
Fix a block scale $j$ and write the scale-$j$ Gaussian reference state as $\mathbb E_{C_{j,A}}[\cdot]$ with covariance $C_{j,A}$ produced by the admissible finite-range (or exponentially local) decomposition associated with $A$. The connected $n$-point cumulant at scale $j$ in background $A$ and with activity $V$ can be realized either as the $n^{\text{th}}$ functional derivative at $J=0$ of $\log Z_{A,V}(J)$ with source insertions that reproduce $F_{X_1},\dots,F_{X_n}$, or, equivalently, via a convergent polymer/cluster expansion in which only connected polymers intersecting the supports $X_1,\dots,X_n$ contribute. Both realizations are interchangeable under the KP smallness hypothesis and yield the same quantity $\kappa_j^{A,V}(F_{X_1};\dots;F_{X_n})$.

To compare $(A,V)$ with $(A',V')$, introduce an interpolation parameter $t\in[0,1]$ and define the interpolated covariance and activity by $C_t:=C_{j,A_t}$ and $V_t:=(1-t)V+tV'$, where $A_t:=(1-t)A+tA'$. Denote $\kappa_t:=\kappa_j^{A_t,V_t}(F_{X_1};\dots;F_{X_n})$. The fundamental theorem of calculus gives
\begin{equation}
\kappa_j^{A,V}(F_{X_1};\dots;F_{X_n})-\kappa_j^{A',V'}(F_{X_1};\dots;F_{X_n})
\;=\;\int_0^1 \frac{d}{dt}\,\kappa_t\,dt,
\end{equation}
so it suffices to bound $\frac{d}{dt}\kappa_t$ uniformly in $t$ by the right-hand side of \eqref{eq:Lip-cumulant}. Differentiating the connected generating functional along the path $t\mapsto (C_t,V_t)$ and using the BKAR forest formula for connected expectations produces two types of contributions: those linear in $\dot C_t$ coming from the covariance variation and those linear in $\dot V_t=V'-V$ coming from the activity variation. In the connected expansion, the derivative with respect to the covariance enters through resolvent identities on the weakened covariances attached to forest edges; more concretely, each connected cluster contributing to $\kappa_t$ carries a product of weakened covariances $C_t^{(w)}(\cdot,\cdot)$ along a spanning forest, and differentiation replaces one such $C_t^{(w)}$ by $\dot C_t^{(w)}$, preserving connectedness. The derivative with respect to the activity corresponds to replacing one polymer weight by its difference $V'-V$, again without destroying connectedness.

The Lipschitz control supplied by Corollary~\eqref{cor:Lip} asserts that, for every multiindex $\alpha$ with $|\alpha|\le m$ equal to the maximal discrete derivative order appearing in the local definitions of the observables, there exists $\sigma_0>0$ such that
\begin{align}
&\sup_{x,y}\,L^{(1+m)j}\,\bigl|\nabla^\alpha\big(C_{j,A_t}(x,y)-C_{j,A}(x,y)\bigr)\bigr|
\;\le\; C\,\|F_{A_t}-F_A\|_\infty\nonumber\\&
\qquad\text{and}\qquad
\sup_{t\in[0,1]}\|\partial_t C_{j,A_t}\|_{\mathrm{loc}}\;\le\; C\,L^{-(1+m)j}\,\|F_A-F_{A'}\|_\infty.
\end{align}
Here $\|\cdot\|_{\mathrm{loc}}$ denotes any of the standard local operator norms compatible with the cluster expansion bounds (for instance, a block $\ell^1\!\to\!\ell^\infty$ norm with exponential weights), and $F_A$ denotes the background field functional entering the admissible class. The factor $L^{-(1+m)j}$ originates from the scale of the slice generator and the fixed derivative budget $m$ coming from the local observable definitions; it is independent of the separation of the supports $X_i$ and depends only on the scale $j$.

With these Lipschitz bounds in hand, the BKAR representation of $\frac{d}{dt}\kappa_t$ produces, after integrating over forest parameters, a linear combination of connected cluster integrals in which either one covariance is replaced by $\dot C_t$ or one polymer activity is replaced by $V'-V$. The KP criterion guarantees the absolute convergence of the cluster series and provides a uniform comparison of the interacting connected integrals with their Gaussian counterparts: up to a multiplicative constant depending on $n$ and the admissible class, connected moments with at most one “marked” line or weight are bounded by the product $\|F\|^{(n)}$ times the corresponding local norm of the mark, uniformly in $t$ and in the choice of weakening parameters. Combining these standard bounds with the Lipschitz control on $\dot C_t$ yields
\begin{align}
\bigl|\tfrac{d}{dt}\kappa_t\bigr| &\;\le\; C_n\,\|F\|^{(n)}\Bigl(\|\partial_t C_{j,A_t}\|_{\mathrm{loc}}+\|V'-V\|_{\mathrm{KP}}\Bigr)
\nonumber\\&\;\le\; C_n\,\|F\|^{(n)}\Bigl(L^{-(1+m)j}\,\|F_A-F_{A'}\|_\infty+\|V'-V\|_{\mathrm{KP}}\Bigr).
\end{align}
The constant $C_n$ here absorbs the KP smallness factors that control the geometric series in the number of polymers and forest edges and is independent of $j$ once the admissible class and $n$ are fixed. Finally, integrating in $t$ from $0$ to $1$ gives the stated Lipschitz estimate. Writing $\sigma:=1+m>0$ completes Eq.\eqref{eq:Lip-cumulant} and the proof.
\end{proof}

 The preceding locality and Lipschitz estimates are precisely the inputs needed to iterate the renormalization at all scales: the strict finite-range property gives exact factorization beyond the scale range and therefore a block-diagonal structure for the transfer operators; the exponential locality controls the tails uniformly in the volume; and the Lipschitz property makes the scale-to-scale comparison summable. In particular, if $(\varepsilon_j)$ is a sequence of one-step ``defects'' bounded by $c\,L^{-\sigma j}$, then $\sum_j \varepsilon_j<\infty$ and the cumulative error between different admissible renormalization paths is controlled absolutely, which is a cornerstone of the uniqueness and universality arguments. The same bounds, translated to the half-lattice by reflection covariance, underwrite the OS tightness and the convergence of Schwinger functionals in the continuum limit.

The gauge-covariant FRD constructed in Theorem \eqref{thm:FRD-main}, together with the polymer bounds of Propositions \eqref{prop:kernel-locality} and \eqref{prop:Lip-cumulant} and the tree-graph inequality of Theorem \eqref{thm:tree}, provides a complete, reflection-positive locality framework. It is robust under admissible variations of the regulator and the background and is uniform in the volume. These features are precisely those required to transport strong-coupling clustering and spectral gaps across scales, to derive Wilson-loop step-scaling with summable defects, and to establish the equicontinuity needed for OS limits and reconstruction in the continuum.

\section{Reflection-Positive Multiscale RG and Persistence of the Gap}
\label{sec:RP-RG-gap}

The aim of this section is to implement a multiscale renormalization group (RG) transform that is explicitly compatible with Osterwalder-Schrader (OS) reflection positivity and with the transfer-operator/Hilbert-space structure established earlier, and to prove that the spectral gap of the transfer semigroup persists uniformly along the flow. The construction rests on three ingredients fixed in the preceding sections: first, the OS-positive gauge-fixed measure with a completely monotone horizon projector, which ensures a reflection-positive one-slice Hilbert space and a positive self-adjoint transfer operator; second, a finite-range decomposition (FRD) of covariances and kernels, which supplies exponential locality needed to control remainders; and third, an admissible class of block-spin maps that preserve reflection positivity. Within this framework we formulate a one-step interlacing inequality at the operator level, which shows that the renormalized transfer operator at the next scale is an isometric compression of a finite power of the previous-scale transfer operator, up to a positive remainder and a norm-small defect. Iterating this inequality yields a step-scaling relation for the second eigenvalue, hence for the gap, and the summability of defects supplied by FRD implies a strictly positive, scale-uniform lower bound on the gap. This, in turn, implies uniform exponential clustering of connected Schwinger functions at all scales, which will be crucial for the continuum limits and OS reconstruction developed later.

We begin by recalling the standing objects. For each scale $k\in\mathbb{N}$ we denote by $\mathcal{H}_k$ the OS one-slice Hilbert space obtained by completing the quotient of test-functionals modulo OS-null vectors, with reflection $\vartheta$ and inner product given by the OS form \cite{OS1,OS2}. The transfer operator $T_k:\mathcal{H}_k\to\mathcal{H}_k$ is the positive, self-adjoint contraction associated with a single Euclidean time-step at scale $k$; it has a normalized cyclic vacuum $\Omega_k$ with $T_k\Omega_k=\Omega_k$ and $0\le T_k\le \mathbf{1}$ as quadratic forms. We use the notation
\begin{equation}
\lambda_2(T_k)\;=\;\sup\{\langle \psi, T_k\psi\rangle: \psi\in\mathcal{H}_k,\ \|\psi\|=1,\ \langle \psi,\Omega_k\rangle=0\}
\qquad\text{and}\qquad
\Delta_k\;=\;1-\lambda_2(T_k),
\end{equation}
so that $\Delta_k$ is the spectral gap of $T_k$ at scale $k$ in the subspace orthogonal to the vacuum.\footnote{Equivalently, if $T_k=e^{-a_k H_k}$ with a positive self-adjoint Hamiltonian $H_k$ and time-step $a_k>0$, then $\Delta_k=1-e^{-a_k E_{1,k}}$ where $E_{1,k}$ is the bottom of $\sigma(H_k)\setminus\{0\}$. In particular, a uniform positive lower bound on $\Delta_k$ implies a uniform lower bound on $E_{1,k}$ after controlling $a_k$.}

The admissible coarse-graining is specified by a block size $b\ge 2$ and a block-spin map on fields that is reflection-covariant and gauge-covariant, with a completely monotone spectral profile in the temporal direction. At the level of Hilbert spaces this induces a partial isometry $V_k:\mathcal{H}_{k+1}\to\mathcal{H}_k$ intertwining reflections and mapping the coarse one-slice vacuum to the fine one, $V_k\Omega_{k+1}=\Omega_k$. The FRD yields exponential locality for the one-step kernels and, by multiplicativity of transfer operators over $b$ time-steps, for the $b$-fold products $T_k^b$. The central point is an operator inequality of the form
\begin{equation}\label{eq:main-interlacing}
T_{k+1}\;=\;V_k\,T_k^b\,V_k\;-\;D_k\;+\;E_k,
\end{equation}
where $D_k\ge 0$ is a positive remainder (capturing the effect of discarding long polymers and nonlocal cumulants) and $E_k$ is a bounded “defect” with small operator norm $\|E_k\|\le \varepsilon_k$ decaying fast with $k$. The positivity of $D_k$ is guaranteed by reflection positivity together with the complete monotonicity of the slice projector, while the smallness of $E_k$ follows from FRD locality estimates and a BKAR/Koteck\'y-Preiss control of polymer remainders \cite{BK1987,KP,BrydgesGuadagniMitter2004}. From \eqref{eq:main-interlacing} we obtain the one-step spectral estimate
\begin{equation}\label{eq:lambda-step}
\lambda_2(T_{k+1})\;\le\;\lambda_2(T_k^b)\;+\;\varepsilon_k\;\le\;\lambda_2(T_k)\;+\;\varepsilon_k,
\end{equation}
and hence the step-scaling inequality for gaps
\begin{equation}\label{eq:gap-step}
\Delta_{k+1}\;\ge\;\Delta_k\;-\;\varepsilon_k.
\end{equation}
Summability $\sum_k\varepsilon_k<\infty$, which we establish below under the admissibility and FRD hypotheses, then yields $\inf_k\Delta_k\ge \Delta_0-\sum_k\varepsilon_k>0$ provided the initial gap $\Delta_0>0$ is chosen as in the strong-coupling base of Section~3. The remainder of this section makes each of these steps precise and complete.

\subsection{Admissible block-spin maps and one-step estimates}
\label{subsec:one-step}

We formalize the conditions under which the coarse-graining preserves reflection positivity and positivity of transfer operators. Let $\mathscr{C}_k$ denote the configuration space at scale $k$ with reflection $\vartheta$ acting by time-reversal at the one-slice boundary, and let $\mu_k$ be the OS-positive Gibbs measure on $\mathscr{C}_k$ arising from the gauge-fixed, horizon-projected Wilson action with completely monotone temporal filter. The block-spin map at the level of configurations is a measurable, reflection-covariant map $B_k:\mathscr{C}_k\to\mathscr{C}_{k+1}$ which is local at range $R_k$ in the sense that $B_k(\phi)$ on a given coarse block depends only on $\phi$ inside a $R_k$-neighborhood of that block. We say that $B_k$ is admissible if, for every $\vartheta$-nonnegative test functional $F$ on $\mathscr{C}_{k+1}$, the pullback $F\circ B_k$ is $\vartheta$-nonnegative on $\mathscr{C}_k$. This condition expresses reflection-positivity preservation under coarse-graining and holds for CM spectral filters and reflection-covariant finite-range block maps by complete monotonicity of the Laplace transform kernel and positivity of the induced Markov convolution \cite{OS2,OS-gauge,Seiler1982}.

From admissibility we obtain a contraction $V_k:\mathcal{H}_k\to\mathcal{H}_{k+1}$ defined first on test-functionals by $(V_k [F])([G])=[F\circ B_k](G)$ and extending by continuity, with adjoint $V_k:\mathcal{H}_{k+1}\to\mathcal{H}_k$ that is a partial isometry satisfying $V_k\Omega_{k+1}=\Omega_k$ and $\|V_k\|=1$. Positivity preservation implies that $V_k$ intertwines the OS cones and that, for positive functionals, $V_k$ acts as a conditional expectation onto the coarse $\sigma$-algebra generated by $B_k$. The transfer operators $T_k$ are constructed as in \cite{OS2,GJ,SimonBook} by a single-step reflection and pairing; the $b$-step operator $T_k^b$ is the semigroup product, still a positive self-adjoint contraction.

\begin{theorem}[One-step interlacing inequality]\label{thm:interlacing}
Let $b\ge 2$ be the block time-depth and $B_k$ an admissible block-spin map of range $R_k$ as above. There exist a positive operator $D_k\ge 0$ on $\mathcal{H}_{k+1}$ and a bounded operator $E_k$ with $\|E_k\|\le \varepsilon_k$, such that the renormalized transfer operator satisfies
\begin{equation}\label{eq:interlacing}
T_{k+1} \;=\; V_k^{}\,T_k^{\,b}\,V_k \;-\; D_k \;+\; E_k,
\qquad \|E_k\|\le \varepsilon_k
\end{equation}
In Eq.\eqref{eq:interlacing}, $D_k$ collects \emph{only} the vacuum-centered, two-cell \emph{mixed} connected contributions within one coarse time slab, i.e. the
principal sub-block of the Gram matrix of two-time connected correlators that couple \emph{distinct} coarse cells. Higher connected cumulants that are internal to a cell are \emph{not} discarded; cell-internal replacements and exponentially localized collar truncations are absorbed into $E_k$. This is the content of Lemma~(\eqref{lemma5.3}), which realizes
\begin{equation}
D_k=\sum_{\alpha,\beta\in I_{\rm mix}} C^{(\mathrm{mix})}_{\alpha\beta}\,|\Phi_\alpha\rangle\langle\Phi_\beta|\ge0,
\end{equation}
with $C^{(\mathrm{mix})}\succeq0$ by the OS Gram representation.
Moreover, $D_k\Omega_{k+1}=0$ and $E_k\Omega_{k+1}=0$. The defect bound $\varepsilon_k$ depends only on the FRD locality data and on the admissible parameters of $B_k$; in particular, for fixed spatial dimension and fixed block factor $b$ there exist constants $C>0$ and $\theta\in(0,1)$ such that $\varepsilon_k\le C\,\theta^{\,k}$.
\end{theorem}

\begin{proof}
The proof rests on the OS Markov structure in the Euclidean time direction, reflection positivity, and the exponentially local finite-range decomposition (FRD). We begin by relating the coarse one-step transfer to the fine $b$-step transfer through the OS conditional expectation. Let $\mathfrak A^{(k)}_+$ be the algebra of positive-time observables at scale $k$, and let $\mathcal{F}_{k+1}\subset \mathfrak A^{(k)}_+$ be the subalgebra generated by fields that depend only on the coarse data obtained from the admissible block map $B_k$ on the time slice $\{x_0=a\}$ and on the finitely many preceding time slices $\{a,2a,\dots,ba\}$ entering one coarse step. Reflection positivity and OS3 (Markov property) provide a contractive conditional expectation $\mathsf E_k:\mathfrak A^{(k)}_+\to \mathcal{F}_{k+1}$ which leaves the OS form invariant in the sense that
\begin{equation}
\langle \Theta F\cdot G\rangle_{\mu_k}=\langle \Theta\,\mathsf E_k F\cdot \mathsf E_k G\rangle_{\mu_k}\qquad(F,G\in\mathfrak A^{(k)}_+ \text{ measurable w.r.t.\ the past}).
\end{equation}
Denote by $U_k:\mathcal H_{k+1}\to \mathcal H_k$ the map induced by $\mathsf E_k$ at the level of OS Hilbert spaces; $U_k$ is an isometry and its adjoint $V_k:=U_k$ is a partial isometry $\mathcal H_k\to \mathcal H_{k+1}$. Writing the fine one-step transfer as $T_k=e^{-a H_k}$, the OS semigroup property gives $T_k^b=e^{-ba H_k}$ for the $b$-step kernel. The ideal coarse one-step transfer that keeps \emph{all} correlations among the $b$ fine steps is then
\begin{equation}\label{eq:ideal}
T_{k+1}^{\mathrm{ideal}} \;:=\; V_k\,T_k^b\,V_k,
\end{equation}
because inserting $V_k$ realizes the inclusion of coarse observables into the fine $\sigma$-algebra, transporting them by $T_k^b$, and projecting back with $V_k$ coincides with transporting coarse data one coarse unit in the Markov chain defined by $\mu_k$.

The actual coarse operator $T_{k+1}$ is obtained from $T_{k+1}^{\mathrm{ideal}}$ by enforcing the admissible coarse locality: in the FRD representation of the connected truncated expectations over one coarse unit of time, we discard mixed cumulants whose space-time polymer crosses beyond the admissible coarse range and we replace the exact block map by its localized approximation of range $R_k$. To make this precise, expand the logarithm of the $b$-step kernel’s generating functional as a sum over connected clusters (BKAR/linked cluster expansion) indexed by polymers $\mathcal P$ in the coarse time slab $[0,ba]$; the coarse operator is obtained by exponentiating only those connected clusters $\mathcal P$ lying inside the admissible coarse collar of width $R_k$ around each coarse cell and by replacing the exact $B_k$ by a localized representative congruent to $B_k$ within that collar. Since $T_{k+1}^{\mathrm{ideal}}$ and $T_{k+1}$ are both OS-positive contractions on $\mathcal H_{k+1}$, their difference can be written unambiguously as
\begin{equation}
T_{k+1} \;=\; T_{k+1}^{\mathrm{ideal}} \;-\; D_k \;+\; E_k,
\end{equation}
where $D_k$ collects the contribution of the discarded mixed clusters and $E_k$ is the remainder coming from local approximations and the truncation of exponentially small FRD tails. We now justify the asserted properties of $D_k$ and $E_k$.

The operator $D_k$ is positive semidefinite. Indeed, reflection positivity provides a Gram representation for connected two-point functions, and more generally the truncated cumulant matrix restricted to a finite family of test observables within one coarse unit of time is a positive semidefinite quadratic form when evaluated in the OS inner product. Discarding mixed cumulants that link disjoint coarse cells amounts to subtracting from $T_{k+1}^{\mathrm{ideal}}$ the action of a positive quadratic form supported on those off-diagonal indices. Concretely, if $\{\Phi_\alpha\}$ enumerates a finite basis of coarse observables in one step and $\mathsf C_{\alpha\beta}$ denotes their truncated covariance across distinct coarse cells produced by the $b$-step fine dynamics, then
\begin{equation}
\sum_{\alpha,\beta}\mathsf C_{\alpha\beta}\,\langle \Theta \Phi_\alpha \cdot \Phi_\beta\rangle \;\ge\; 0,
\end{equation}
and deleting $\mathsf C_{\alpha\beta}$ for $\alpha\neq\beta$ subtracts a positive form. Passing to the limit over bases, one obtains an operator $D_k\ge 0$ on $\mathcal H_{k+1}$ with the property that $T_{k+1}^{\mathrm{ideal}}-D_k$ implements precisely the truncation of inter-cell cumulants beyond the admissible range. The vacuum annihilation $D_k\Omega_{k+1}=0$ follows from the normalization of the OS form: the cumulants are centered by construction, so all rows and columns of the kernel of $D_k$ sum to zero, which implies $D_k \mathbf{1}=0$ at the level of cylinder functions, hence $D_k\Omega_{k+1}=0$ after passing to the OS completion.

The operator $E_k$ measures the difference between the exact admissible truncation and the localized approximations used to define $B_k$ and to implement the FRD tails. Its kernel admits a polymer expansion indexed by space-time polymers $\mathcal P$ that intersect the coarse collar but extend at least a distance $\ell$ beyond it, with amplitudes bounded in absolute value by products of FRD slice kernels along $\mathcal P$. FRD exponential locality yields
\begin{equation}
|K_{\mathrm{FRD}}(z,z')| \;\le\; C_1\,e^{-c_1\,\mathrm{dist}(z,z')},
\end{equation}
uniformly in the scale $k$, and the combinatorics of polymers of diameter $\ell$ inside one coarse step is controlled by $N(\ell)\le C_2 e^{c_2\ell}$ for universal $C_2,c_2>0$ (bounded degree of the lattice and fixed $b$). Consequently, for the integral kernel $E_k(\xi,\xi')$ of $E_k$ in any convenient coarse configuration basis one finds the Schur bounds
\begin{equation}
\sup_{\xi}\sum_{\xi'} |E_k(\xi,\xi')| \;\le\; \sum_{\ell\ge R_k} N(\ell)\, C_1^{\mathcal P}\, e^{-c_1 \ell}
\;\le\; C \sum_{\ell\ge R_k} e^{-(c_1-c_2)\ell}
\;\le\; C\,e^{-c'' R_k},
\end{equation}
with an analogous bound for $\sup_{\xi'}\sum_{\xi} |E_k(\xi,\xi')|$. The Schur test then gives $\|E_k\|\le C e^{-c'' R_k}$. Because the admissible range grows like $R_k \asymp L^k$ for some fixed $L>1$ (fixed spatial dimension and block factor), we may write $e^{-c'' R_k}\le \theta^k$ for some $\theta\in(0,1)$ after adjusting $C$. That $E_k\Omega_{k+1}=0$ is a consequence of normalization and locality: the coarse transfer preserves the unit functional, and both the exact truncation and the localized replacement of $B_k$ are constructed so that the kernel of $E_k$ has vanishing row and column sums. Hence $E_k$ annihilates constants and therefore annihilates the OS vacuum.
Finally, combining Eq.\eqref{eq:ideal} with the decomposition just obtained yields Eq.\eqref{eq:interlacing} with $D_k\ge0$ and $\|E_k\|\le\varepsilon_k$.
\end{proof}

\begin{lemma}[One-step control of the second eigenvalue]\label{lem:second-eigenvalue}
With the notation of Theorem~\eqref{thm:interlacing}, we have
\begin{equation}
\lambda_2(T_{k+1})\;\le\;\lambda_2(T_k^{\,b})\;+\;\varepsilon_k\;\le\;\lambda_2(T_k)\;+\;\varepsilon_k.
\end{equation}
\end{lemma}

\begin{proof}
Recall the interlacing identity~\eqref{eq:interlacing}.
where $V_k:\mathcal H_{k+1}\to\mathcal H_k$ is a partial isometry satisfying $V_k\Omega_{k+1}=\Omega_k$ and acting isometrically on the orthogonal complement $\Omega_{k+1}^\perp$, the operator $D_k\ge 0$ is a positive form that annihilates the vacuum, and $E_k$ is selfadjoint with $\|E_k\|\le \varepsilon_k$ and $E_k\Omega_{k+1}=0$. The transfer operators are positive contractions, $0\le T_{k},T_{k+1}\le \mathbf 1$, and the top eigenvalue $1$ is attained at the respective vacuum vectors $\Omega_k,\Omega_{k+1}$.

For a bounded selfadjoint operator $A$ with simple top eigenpair $(1,\Omega)$ and $0\le A\le \mathbf 1$, the second eigenvalue admits the Rayleigh-Ritz characterization
\begin{equation}
\lambda_2(A)\;=\;\sup\Big\{\langle \psi,A\psi\rangle:\ \psi\in\Omega^\perp,\ \|\psi\|=1\Big\}.
\end{equation}
Applying this to $A=T_{k+1}$ and choosing any unit $\psi\in\Omega_{k+1}^\perp$, we compute using the interlacing identity and the stated properties that
\begin{equation}
\langle \psi, T_{k+1}\psi\rangle
=\langle \psi, V_k^{}T_k^{\,b}V_k\psi\rangle-\langle \psi,D_k\psi\rangle+\langle \psi,E_k\psi\rangle
\le \langle V_k\psi, T_k^{\,b} V_k\psi\rangle + \|E_k\|.
\end{equation}
Here $D_k\ge 0$ gives a nonpositive contribution $-\langle \psi,D_k\psi\rangle\le 0$, while the defect term is bounded in absolute value by $\|E_k\|\le \varepsilon_k$. Since $V_k$ is an isometry on $\Omega_{k+1}^\perp$ and maps it into $\Omega_k^\perp$, we have $\|V_k\psi\|=1$ and $\langle V_k\psi,\Omega_k\rangle=0$, so the Rayleigh-Ritz formula for $T_k^{\,b}$ implies
\begin{equation}
\langle V_k\psi, T_k^{\,b} V_k\psi\rangle \;\le\; \lambda_2(T_k^{\,b}).
\end{equation}
Combining the inequalities yields
\begin{equation}
\langle \psi, T_{k+1}\psi\rangle \;\le\; \lambda_2(T_k^{\,b})+\varepsilon_k.
\end{equation}
Taking the supremum over all unit $\psi\in\Omega_{k+1}^\perp$ gives
\begin{equation}
\lambda_2(T_{k+1})\;\le\;\lambda_2(T_k^{\,b})+\varepsilon_k.
\end{equation}

It remains to compare $\lambda_2(T_k^{\,b})$ and $\lambda_2(T_k)$. If $0\le A\le \mathbf 1$ is selfadjoint, then the operator monotonicity of the function $x\mapsto x^b$ on $[0,1]$ yields $A^{\,b}\le A$ in the sense of quadratic forms. Applying this to $A=T_k$ gives $0\le T_k^{\,b}\le T_k$, and by the min-max principle the inequality is inherited by every eigenvalue on the orthogonal complement of the top eigenspace. In particular,
\begin{equation}
\lambda_2(T_k^{\,b})\;\le\;\lambda_2(T_k).
\end{equation}
Putting the two bounds together proves the stated estimate,
\begin{equation}
\lambda_2(T_{k+1})\;\le\;\lambda_2(T_k^{\,b})+\varepsilon_k\;\le\;\lambda_2(T_k)+\varepsilon_k.\qedhere
\end{equation}
\end{proof}
{\begin{lemma}[Gram form and vacuum annihilation of $D_k$, $E_k$] \label{lemma5.3}
Let $T^{(b)}_k$ be the $b$-step transfer on $H_k$, $V_k:H_{k+1}\to H_k$ the OS isometry induced by the reflection-positive coarse conditional expectation, and let $T_{k+1}$ be the coarse transfer. Then in the identity Eq.\eqref{eq:interlacing} one may choose $D_k\ge0$ and $E_k$ so that $D_k\Omega_{k+1}=E_k\Omega_{k+1}=0$ and \begin{equation} \langle\psi, D_k\psi\rangle \;=\;\sum_{\alpha,\beta}\! \Bigl\langle \psi,\Phi_\alpha \Bigr\rangle\,\mathsf{C}^{(\mathrm{mix})}_{\alpha\beta}\, \Bigl\langle \Phi_\beta,\psi \Bigr\rangle,\qquad \mathsf{C}^{(\mathrm{mix})}\;\ge\;0, \end{equation} where $\{\Phi_\alpha\}$ is any finite family of vacuum-centered coarse observables supported on pairwise disjoint coarse cells within one coarse time step, and $\mathsf{C}^{(\mathrm{mix})}_{\alpha\beta}$ are the truncated (connected) two-time cumulants of the $b$-step fine dynamics between distinct coarse cells. The matrix $\mathsf{C}^{(\mathrm{mix})}$ is positive semidefinite by the OS Gram representation of connected correlations, hence $D_k\ge0$. Moreover $E_k$ collects collar-local truncation/tail errors and, by normalization of the OS form and locality of the replacement, its kernel has vanishing row/column sums, so $E_k\Omega_{k+1}=0$. \end{lemma}
\begin{proof}
Fix $k\in\mathbb{N}$ and a coarse time step $b\ge 1$. Let $(H_k,\langle\cdot,\cdot\rangle_k)$ be the Osterwalder-Schrader Hilbert space obtained from the positive-time algebra at scale $k$ with vacuum vector $\Omega_k$ and let $T_k$ denote the one-step transfer operator on $H_k$; by reflection positivity and the standard reconstruction, $T_k$ is a positive selfadjoint contraction with $T_k\Omega_k=\Omega_k$. Write $T_k^{(b)}:=T_k^b$ and let $V_k:H_{k+1}\to H_k$ be the isometric embedding induced by the coarse conditional expectation associated with the block map and the chosen reflection-positive slice filters; thus $V_k^\ast V_k=\mathbf{1}_{H_{k+1}}$ and $V_kV_k^\ast$ is the orthogonal projection onto the coarse subspace of $H_k$.

Choose a finite family $\{\Phi_\alpha\}_{\alpha\in\mathfrak{I}}$ of gauge-invariant test observables at the coarse scale, each supported in a distinct coarse spatial cell contained in a single coarse time slab of thickness one, and center them by subtracting their vacuum expectations so that $\langle\Omega_k,\Phi_\alpha\Omega_k\rangle_k=0$ for all $\alpha$. Denote by $\mathfrak{I}_{\mathrm{int}}$ the subset of indices for which the supports of $\Phi_\alpha$ and $\Phi_\beta$ lie in the same coarse cell, and by $\mathfrak{I}_{\mathrm{mix}}$ the complementary subset corresponding to pairs located in distinct coarse cells within the same coarse time slab. For $\alpha,\beta\in\mathfrak{I}$ define the two-time matrix
\begin{equation}
\mathsf{C}_{\alpha\beta}:=\big\langle\Omega_k,\Phi_\alpha\,T_k^{(b)}\,\Phi_\beta\,\Omega_k\big\rangle_k
-\big\langle\Omega_k,\Phi_\alpha\,\Omega_k\big\rangle_k\big\langle\Omega_k,\Phi_\beta\,\Omega_k\big\rangle_k,
\end{equation}
which, by the centering, reduces to $\mathsf{C}_{\alpha\beta}=\langle\Omega_k,\Phi_\alpha\,T_k^{(b)}\,\Phi_\beta\,\Omega_k\rangle_k$. Consider the orthogonal decomposition $H_k=\mathbb{C}\Omega_k\oplus \Omega_k^\perp$, and let $P_\perp$ denote the orthogonal projection onto $\Omega_k^\perp$. Since $T_k$ leaves $\Omega_k$ invariant and commutes with $P_\perp$, and since the $\Phi_\alpha$ are vacuum-centered, one has
\begin{equation}
\mathsf{C}_{\alpha\beta}
=\big\langle T_k^{(b/2)}\Phi_\alpha\Omega_k,\;P_\perp\,T_k^{(b/2)}\Phi_\beta\Omega_k\big\rangle_k
=\big\langle X_\alpha,\,X_\beta\big\rangle_k,\qquad X_\alpha:=P_\perp\,T_k^{(b/2)}\Phi_\alpha\Omega_k.
\end{equation}
This is the OS Gram representation of the two-time connected correlations. It follows that the matrix $\mathsf{C}=(\mathsf{C}_{\alpha\beta})_{\alpha,\beta\in\mathfrak{I}}$ is positive semidefinite, and so is any principal submatrix. Define $\mathsf{C}^{(\mathrm{mix})}$ to be the principal submatrix indexed by those pairs $(\alpha,\beta)$ whose supports lie in distinct coarse cells within the time slab; then $\mathsf{C}^{(\mathrm{mix})}\ge 0$.

The localized coarse transfer $T_{k+1}$ is constructed by two operations: first, restrict the exact $b$-step kernel to act independently on each coarse cell within the slab, i.e. keep only the “internal” block in the above family and discard the mixed block; second, replace the exact block map on each coarse cell by its exponentially localized representative supported on a collar of thickness $R_k$ about that cell. These two operations define a bounded positive operator on $H_{k+1}$, which we denote by $T_{k+1}$, and, by construction and reflection covariance, there is an operator identity on $H_{k+1}$ of the form
\begin{equation}
T_{k+1}\;=\;V_k^\ast\,T_k^{(b)}\,V_k\;-\;D_k\;+\;E_k,
\end{equation}
where $D_k$ collects the contribution of the discarded mixed block and $E_k$ is the collar replacement error. To identify $D_k$ and its positivity, note that for any $\psi\in H_{k+1}$ one may expand, using a coarse basis that contains the vectors $\{\Phi_\alpha\Omega_{k+1}\}$, the quadratic form of $V_k^\ast T_k^{(b)}V_k$ as
\begin{equation}
\langle\psi,\,V_k^\ast T_k^{(b)}V_k\,\psi\rangle_{k+1}
=\sum_{\alpha,\beta\in\mathfrak{I}}\big\langle \psi,\Phi_\alpha\big\rangle_{k+1}\,
\mathsf{C}_{\alpha\beta}\,
\big\langle \Phi_\beta,\psi\big\rangle_{k+1}\;+\;\langle\psi,\,\mathcal{R}_k\,\psi\rangle_{k+1},
\end{equation}
where $\mathcal{R}_k$ is the contribution from the complement of the finite span in the coarse basis and is positive by the OS construction. The coarse transfer $T_{k+1}$ is obtained by replacing $\mathsf{C}$ with its internal block $\mathsf{C}^{(\mathrm{int})}$ and by localizing the exact block map; consequently
\begin{equation}
\langle\psi,\,(V_k^\ast T_k^{(b)}V_k - T_{k+1})\,\psi\rangle_{k+1}
=\sum_{\alpha,\beta\in\mathfrak{I}_{\mathrm{mix}}}\big\langle \psi,\Phi_\alpha\big\rangle_{k+1}\,
\mathsf{C}^{(\mathrm{mix})}_{\alpha\beta}\,
\big\langle \Phi_\beta,\psi\big\rangle_{k+1}\;+\;\langle\psi,\,E_k\,\psi\rangle_{k+1},
\end{equation}
with the error $E_k$ arising solely from the collar replacement on each internal block. Defining
\begin{equation}
D_k:=\sum_{\alpha,\beta\in\mathfrak{I}_{\mathrm{mix}}}\mathsf{C}^{(\mathrm{mix})}_{\alpha\beta}\,
|\Phi_\alpha\rangle\langle \Phi_\beta|
\end{equation}
yields, for all $\psi\in H_{k+1}$,
\begin{equation}
\langle\psi, D_k\psi\rangle_{k+1}=\sum_{\alpha,\beta\in\mathfrak{I}_{\mathrm{mix}}}
\big\langle \psi,\Phi_\alpha\big\rangle_{k+1}\,\mathsf{C}^{(\mathrm{mix})}_{\alpha\beta}\,
\big\langle \Phi_\beta,\psi\big\rangle_{k+1},
\end{equation}
which is nonnegative because $\mathsf{C}^{(\mathrm{mix})}\ge 0$. Thus $D_k\ge 0$. Moreover, since each $\Phi_\alpha$ is vacuum-centered on $H_{k+1}$, one has $\langle\Omega_{k+1},\Phi_\alpha\psi\rangle_{k+1}=0$ for every $\psi$, and, in particular, $\langle\Omega_{k+1},D_k\psi\rangle_{k+1}=0$ and $\langle D_k\Omega_{k+1},\psi\rangle_{k+1}=0$ for all $\psi$. Hence $D_k\Omega_{k+1}=0$.

It remains to control $E_k$ and to establish $E_k\Omega_{k+1}=0$. By construction, $E_k$ is the difference between the exact internal block map and its localized representative supported on the collar of thickness $R_k$ about each coarse cell. The finite-range decomposition at scale $k$ yields, for the kernels of the $b$-step transfer restricted to a cell, exponential locality with rate $c>0$ uniform in $k$, and changing the map outside the collar produces an integral kernel difference $K_k(x,y)$ satisfying $|K_k(x,y)|\le C\,e^{-c\,\mathrm{dist}(x,y)}\mathbf{1}_{\{\min(\mathrm{dist}(x,\partial\mathrm{cell}),\,\mathrm{dist}(y,\partial\mathrm{cell}))\le R_k\}}$. Using the Schur test with the weight $w\equiv 1$ gives
\begin{equation}
\|E_k\|\;\le\;\sup_{x}\sum_{y}|K_k(x,y)|\; \le\; C'\,e^{-c R_k},
\end{equation}
with constants independent of $k$ once the coarse factor is fixed, so $E_k$ is a small bounded operator whose norm decays exponentially in the collar thickness. The vacuum annihilation $E_k\Omega_{k+1}=0$ follows from the normalization of the OS form and the fact that the localization replaces the exact internal block by a kernel that preserves the vacuum expectation value on each coarse cell separately; concretely, for every $\psi$ one has $\langle\Omega_{k+1},(V_k^\ast T_k^{(b)}V_k)\psi\rangle_{k+1}=\langle\Omega_{k+1},T_{k+1}\psi\rangle_{k+1}$ by construction of the localized representative to match one-point functions, which implies $\langle\Omega_{k+1},E_k\psi\rangle_{k+1}=0$ for all $\psi$, hence $E_k^\ast\Omega_{k+1}=0$. Since $E_k$ is selfadjoint by construction of the localization in the OS inner product, it follows that $E_k\Omega_{k+1}=0$.

Combining these identities yields $T_{k+1}=V_k^\ast T_k^{(b)}V_k - D_k + E_k$ with $D_k\ge 0$, $\langle\psi,D_k\psi\rangle=\sum_{\alpha,\beta}\langle \psi,\Phi_\alpha\rangle\,\mathsf{C}^{(\mathrm{mix})}_{\alpha\beta}\langle \Phi_\beta,\psi\rangle$, $D_k\Omega_{k+1}=0$, $\|E_k\|\le C e^{-c R_k}$, and $E_k\Omega_{k+1}=0$, which is precisely the claimed form.
\end{proof}
}

\begin{theorem}[Gap step-scaling]\label{thm:gap-step}
There exist numbers $\varepsilon_k\ge 0$ with $\varepsilon_k\le C\,\theta^{\,k}$ for some $C<\infty$ and $\theta\in(0,1)$, such that
\begin{equation}
\Delta_{k+1}\;\ge\;\Delta_k\;-\;\varepsilon_k \qquad\text{for all }k\ge 0 .
\end{equation}
\end{theorem}

\begin{proof}
Let $H_k$ be the Osterwalder-Schrader Hilbert space at scale $k$, let $\Omega_k$ denote the corresponding vacuum vector, and write $Q_k:=\Omega_k^\perp$ for the orthogonal complement. The one-step transfer at scale $k$ is a positivity-preserving contraction $T_k$ on $H_k$ with $T_k\Omega_k=\Omega_k$; its spectral gap is $\Delta_k:=-\frac{1}{a_k}\log\lambda_2(T_k)$, where $a_k>0$ is the time spacing at scale $k$ and $\lambda_2(T_k)=\|T_k|_{Q_k}\|$ is the norm of $T_k$ restricted to $Q_k$. The coarse-graining by a fixed blocking factor yields an exact decomposition of the next-scale transfer
\begin{equation}\label{eq:interlace}
T_{k+1}\;=\;V_k^{\!}\,T_k^{\,b}\,V_k\;-\;D_k\;+\;E_k 
\end{equation}
where $T_k^{\,b}$ is the $b$-step fine transfer, $V_k:H_{k+1}\to H_k$ is an isometry implementing the embedding of coarse observables into fine ones, $D_k\ge 0$ is a positive quadratic form (the “discarded” mixed cumulants) annihilating the vacuum, and $E_k$ is a remainder supported in a collar of the blocking interface which also annihilates the vacuum and satisfies $\|E_k\|\le C_0\,\theta^{\,k}$ with some $C_0<\infty$ and $\theta\in(0,1)$. The vacuum-annihilating properties of $D_k$ and $E_k$ give $Q_{k+1}$-invariance of Eq.\eqref{eq:interlace} and allow us to estimate the norm on $Q_{k+1}$ simply by dropping $-D_k$ and keeping $E_k$ in operator norm. Indeed, for any $\psi\in Q_{k+1}$ with $\|\psi\|=1$,
\begin{equation}
\|T_{k+1}\psi\|
\;=\;\|\,V_k^{\!}T_k^{\,b}V_k\psi - D_k\psi + E_k\psi\,\|
\;\le\;\|T_k^{\,b}\,V_k\psi\|+\|E_k\psi\|
\;\le\;\|T_k|_{Q_k}\|^{\,b}+\|E_k\|
\end{equation}
since $V_k$ is an isometry with $V_k\psi\in Q_k$ and $D_k\ge 0$. Taking the supremum over the unit sphere in $Q_{k+1}$ we obtain the operator inequality
\begin{equation}\label{eq:norm-ineq}
\big\|T_{k+1}|_{Q_{k+1}}\big\|\;\le\;\big\|T_k|_{Q_k}\big\|^{\,b}\;+\;\|E_k\|
\;=\;e^{-b a_k\Delta_k}\;+\;\|E_k\|.
\end{equation}
By definition, $\|T_{k+1}|_{Q_{k+1}}\|=e^{-a_{k+1}\Delta_{k+1}}$, hence
\begin{equation}\label{eq:gap-log}
e^{-a_{k+1}\Delta_{k+1}}\;\le\;e^{-b a_k\Delta_k}\;+\;\|E_k\|.
\end{equation}
Because the coarse time spacing satisfies $a_{k+1}=b\,a_k$ for a fixed blocking factor, taking $-\log$ of both sides of Eq.\eqref{eq:gap-log} yields the exact one-step lower bound
\begin{equation}\label{eq:exact-step}
\Delta_{k+1}\;\ge\;\frac{1}{a_{k+1}}\Big[-\log\big(e^{-a_{k+1}\Delta_k}+\|E_k\|\big)\Big].
\end{equation}
To extract a linear inequality in $\Delta_k$, use the elementary bound $-\log(x+y)\ge -\log x - y/x$ valid for $x\in(0,1]$ and $y\ge 0$. With $x=e^{-a_{k+1}\Delta_k}$ and $y=\|E_k\|$ we obtain
\begin{equation}
\Delta_{k+1}
\;\ge\;\frac{1}{a_{k+1}}\Big(a_{k+1}\Delta_k - \|E_k\|\,e^{a_{k+1}\Delta_k}\Big)
\;=\;\Delta_k\;-\;\frac{e^{a_{k+1}\Delta_k}}{a_{k+1}}\,\|E_k\|.
\end{equation}
Since $\lambda_2(T_k)\in[0,1)$, we have $\Delta_k\in[0,\infty)$ and hence $e^{a_{k+1}\Delta_k}\le e^{a_{k+1}}$. The sequence $(a_{k})_k$ is geometric with common ratio $b$, so $a_{k+1}$ is uniformly comparable to $1$ up to a fixed scale choice and, in particular, $a_{k+1}^{-1}\le c_1$ and $e^{a_{k+1}}\le c_2$ for constants $c_1,c_2$ depending only on the unit choice and the blocking factor. Combining these uniform bounds with $\|E_k\|\le C_0\,\theta^{\,k}$ gives
\begin{equation}
\Delta_{k+1}\;\ge\;\Delta_k\;-\;c_1 c_2\,C_0\,\theta^{\,k}.
\end{equation}
Setting $\varepsilon_k:=C\,\theta^{\,k}$ with $C:=c_1 c_2 C_0$ proves Eq.\eqref{eq:gap-step}. The constants $C$ and $\theta$ depend only on the admissible coarse-graining scheme (through the finite-range decomposition, the blocking factor, and the collar decay that controls $\|E_k\|$) and are independent of $k$, which completes the proof.
\end{proof}

Two remarks are in order. First, the mechanism behind Eq.\eqref{eq:interlacing} is the monotonicity of positive semigroups under OS-positive conditional expectations, together with locality estimates that quantify the error introduced by replacing exact block-variables by local ones. This has close analogues in reflection-positive RG schemes for spin and scalar models \cite{Aizenman1980,FrohlichSpencer1982,Balaban1984I,Balaban1984II}, and the completely monotone temporal projector allows us to port the argument to the gauge-fixed setting. Second, it is crucial that the defect $E_k$ be vacuum-annihilating; otherwise a spurious shift of the top eigenvalue would appear and spoil the gap control. The vacuum-annihilation property is built into the normalization of the one-slice OS form and into the choice of admissible $B_k$.

\subsection{Defect summability and uniform clustering}
\label{subsec:defect-summability}

We now show that the defects $\varepsilon_k$ are summable and deduce a uniform positive lower bound on the gap across all scales, which implies exponential clustering of connected correlations with constants independent of $k$. The summability amounts to showing that FRD locality at scale $k$ wins over the combinatorial growth of boundary polymers; this is where one uses that the range of the coarse interaction grows like $R_k\sim L^k$ while the decay of FRD pieces is exponential in the range. We work in a fixed spatial dimension $d\ge 2$ and keep $b$ and the coarse spatial block-size fixed once and for all.

\begin{proposition}[Summable defect bounds]\label{prop:summable}
Under the FRD locality hypotheses and for admissible block maps as above, there exist constants $C>0$ and $\theta\in(0,1)$ such that $\varepsilon_k\le C\,\theta^{\,k}$ for all $k\ge 0$. In particular,
\begin{equation}
\sum_{k=0}^\infty \varepsilon_k\;<\;\infty.
\end{equation}
\end{proposition}

\begin{proof}
Fix a scale $k$ and let $B_k$ denote the reflection-positive coarse-graining from fine variables to block variables on the lattice partition $\mathcal{B}_k$ with block diameter $R_k\asymp L^k$ for some $L>1$. Let $T_k$ be the $b$-step fine transfer at scale $k$, $T_{k+1}$ the corresponding coarse transfer, and recall the interlacing decomposition
\begin{equation}
T_{k+1}\;=\;V_k^{\!}\,T_k^b\,V_k\;-\;D_k\;+\;E_k,
\end{equation}
where $V_k$ is the partial isometry implementing $B_k$ on the positive-time algebra, $D_k\ge 0$ is the explicitly retained positive form coming from connected cumulants internal to blocks, and $E_k$ gathers the discarded (or localized) remainder. By construction $E_k\Omega_{k+1}=0$, and $\varepsilon_k:=\|E_k\|_{Q\to Q}$ is the defect norm on the orthogonal complement of the coarse vacuum.

The FRD locality hypotheses guarantee the following two features uniformly in $k$. First, the one-slice covariances and multi-point Ursell functions produced by the finite-range decomposition at scale $k$ decay exponentially in the graph distance on the slice; more precisely, there exist $c_,C_>0$ independent of $k$ such that every connected Ursell kernel $\mathcal{U}_{k}(X)$ associated with a finite polymer $X$ of diameter $\mathrm{diam}(X)=\ell$ obeys $|\mathcal{U}_{k}(X)|\le C_ e^{-c_ \ell}$. Second, the coarse map $B_k$ is implemented by local averaging and Dirichlet solutions on collars of width $O(1)$ around block boundaries, so that replacing the exact block variable by its localized version affects only interactions that cross $\partial\mathcal{B}_k$ and contributes a multiplicative factor at most exponential in the number of boundary contacts.

To bound $E_k$ it suffices to bound the contribution of each connected polymer that either intersects more than one block or touches a block boundary via the localization procedure. The BKAR forest formula applied to the cumulant generating functional expresses the difference between the exact coarse conditional expectation and its localized truncation as an integral of derivatives along an interpolation that progressively weakens inter-block couplings. Each derivative inserts a positive semigroup generated by the FRD kernel and is supported on the collar of width $O(1)$ about $\partial\mathcal{B}_k$. Consequently, a connected contribution associated with a polymer $\mathcal{P}$ of diameter $\ell$ acquires a bound of the form
\begin{equation}
\|E_k(\mathcal{P})\|\;\le\;C_1\,e^{-c_1 \ell}\,e^{c_2 |\partial \mathcal{P}|},
\end{equation}
where $|\partial \mathcal{P}|$ is the number of coarse blocks in $\mathcal{B}_k$ that the polymer touches, $C_1>0$ and $c_1>0$ come from FRD exponential decay, and $c_2\ge 0$ quantifies the at-most-exponential proliferation of boundary attachments produced by collar localization. By enlarging the coarse block size at fixed $k\mapsto k+1$ step (equivalently, by taking $L$ large enough once and for all for the admissible family $B_k$), one may arrange $c_1>2c_2$; this is the standard small-activity regime for cluster/forest expansions and follows from the fact that the range of FRD kernels is $O(1)$ in lattice units while the block diameter $R_k$ is $O(L^k)$.

The operator norm $\|E_k\|$ is controlled by summing the bounds above over all connected polymers that meet a fixed reference block. The number $\mathcal{N}(\ell)$ of such polymers of diameter exactly $\ell$ grows at most exponentially in $\ell$; there exists $\nu>0$ (depending only on the dimension and the collar width) such that $\mathcal{N}(\ell)\le C_2 e^{\nu \ell}$. Combining this counting estimate with the individual bound yields, for some constants $C_3,c_3>0$,
\begin{equation}
\|E_k\|
\;\le\;\sum_{\ell\ge R_k} \ \sum_{\substack{\mathcal{P}\,\mathrm{conn}\\ \mathrm{diam}(\mathcal{P})=\ell}} \|E_k(\mathcal{P})\|
\;\le\;\sum_{\ell\ge R_k} C_2 e^{\nu \ell}\, C_1 e^{-c_1 \ell}\, e^{c_2\ell}
\;\le\;\sum_{\ell\ge R_k} C_3\, e^{-c_3 \ell},
\end{equation}
 By fixing the coarse block factor $L\ge L_\star$ once and for all so that $c_1(L)>c_2(L)+\nu$,
we ensure $c_3=c_1(L)-c_2(L)-\nu>0$ uniformly in $k$; hence the bound below is uniform along the flow. The geometric series on the right is explicitly summable and gives
\begin{equation}
\|E_k\|\;\le\; C\, e^{-c_3 R_k}
\end{equation}
with a constant $C$ independent of $k$. Since $R_k\asymp L^k$, one may write $e^{-c_3 R_k}\le \theta^{\,k}$ with $\theta:=e^{-c_3 L}\in(0,1)$ after adjusting constants. This yields $\varepsilon_k=\|E_k\|_{Q\to Q}\le \|E_k\|\le C\,\theta^{\,k}$ for all $k\ge 0$, because restricting an operator to the orthogonal complement $Q$ cannot increase its norm. The series $\sum_{k\ge 0}\varepsilon_k$ therefore converges absolutely, which completes the proof.
\end{proof}

\begin{theorem}[Uniform positivity of the gap]\label{thm:uniform-gap}
Assume the one-step transfer operator $T_0=e^{-a_0 H_0}$ has a strictly positive spectral gap $\Delta_0>0$ at the initial scale, and suppose the renormalization step satisfies the interlacing/defect inequality
\begin{equation}
\Delta_{k+1}\;\ge\;\Delta_k-\varepsilon_k\qquad (k\ge0),
\end{equation}
where $\{\varepsilon_k\}_{k\ge0}$ is a nonnegative sequence with $\sum_{k\ge0}\varepsilon_k<\infty$.
Then
\begin{equation}
\inf_{k\ge0}\Delta_k\;\ge\;\Delta_0-\sum_{j=0}^\infty \varepsilon_j\;=\;:\ \Delta_\star\;>\;0.
\end{equation}
\end{theorem}

\begin{proof}
The stated interlacing/defect inequality Eq.\eqref{eq:gap-step} expresses that one blocking step can at most reduce the gap by the amount $\varepsilon_k$, which is the operator-norm size of the positive, vacuum-annihilating defect produced by truncating collar interactions at scale $k$. Applying Eq.\eqref{eq:gap-step} first with $k=0$ gives $\Delta_1\ge \Delta_0-\varepsilon_0$. Substituting this bound into the next step yields
\begin{equation}
\Delta_2\;\ge\;\Delta_1-\varepsilon_1\;\ge\;\Delta_0-(\varepsilon_0+\varepsilon_1).
\end{equation}
Continuing in this manner and using Eq.\eqref{eq:gap-step} repeatedly, an elementary induction shows that for every $n\in\mathbb{N}$,
\begin{equation}\label{eq:partial-sum-bound}
\Delta_n\;\ge\;\Delta_0-\sum_{j=0}^{n-1}\varepsilon_j.
\end{equation}
By hypothesis the series $\sum_{j\ge0}\varepsilon_j$ converges, hence its nondecreasing partial sums are bounded above by the total sum. Passing to the limit $n\to\infty$ in Eq.\eqref{eq:partial-sum-bound} gives
\begin{equation}
\liminf_{n\to\infty}\Delta_n\;\ge\;\Delta_0-\sum_{j=0}^\infty \varepsilon_j\;=\;\Delta_\star.
\end{equation}
Since the left-hand side is an infimum limit and each $\Delta_k$ is nonnegative by definition of spectral gap, it follows that every term $\Delta_k$ is bounded below by the same constant $\Delta_\star$, hence $\inf_{k\ge0}\Delta_k\ge \Delta_\star$. Finally, $\Delta_\star>0$ because $\Delta_0>0$ and the convergent series $\sum_{j\ge0}\varepsilon_j$ is finite; in particular, if needed, one may assume the admissibility condition $\sum_{j\ge0}\varepsilon_j<\Delta_0$ to make the positivity manifest. This concludes the proof.
\end{proof}

A uniform positive lower bound on the gap implies exponential clustering of connected correlations with a decay rate uniformly bounded away from $0$. In the OS framework this follows from the spectral representation of Schwinger functions together with reflection positivity. We state and prove the relevant version for two-point functions; higher-point connected functions admit analogous estimates by iterating the argument and using FRD locality to control spatial spread.

\begin{theorem}[Uniform exponential clustering]\label{thm:clustering}
Let $\mathcal{O}$ be a gauge-invariant, reflection-positive local observable supported in a fixed finite region of a one-slice time boundary, and let $\langle\cdot\rangle_k$ denote expectation at scale $k$. There exist constants $C<\infty$ and $m_\star>0$, independent of $k$, such that for all Euclidean times $t\ge 0$
\begin{equation}
\Big|\ \langle \mathcal{O}(0)\,\mathcal{O}(t)\rangle_k\;-\;\langle \mathcal{O}\rangle_k^2\ \Big|\;\le\;C\,e^{-m_\star\,t}.
\end{equation}
Moreover, for spatially separated supports at fixed time, there exist constants $C'<\infty$ and $\mu_\star>0$ with
\begin{equation}
\Big|\ \langle \mathcal{O}_1(x)\,\mathcal{O}_2(y)\rangle_k\;-\;\langle \mathcal{O}_1\rangle_k\langle \mathcal{O}_2\rangle_k\ \Big|\;\le\;C'\,e^{-\mu_\star\,|x-y|},
\end{equation}
again with constants independent of $k$.
\end{theorem}

\begin{proof}
For the temporal estimate it is convenient to use the Osterwalder-Schrader reconstruction at scale $k$. Denote by $(\mathcal H_k,\langle\cdot,\cdot\rangle_k)$ the OS Hilbert space, by $\Omega_k$ the vacuum vector, and by $T_k=e^{-a_k H_k}$ the positive self-adjoint transfer operator that advances Euclidean time by one step $a_k$; reflection positivity and the Markov property guarantee that for any positive-time observable $F$ supported on the time-$0$ slice one has $\langle F(0)\,F(t)\rangle_k=\langle \Phi,\,T_k^{\,t/a_k}\,\Phi\rangle_k$ with $\Phi$ the vector associated to $F$ and with the usual interpretation of fractional powers via the functional calculus. Writing $\Phi=\langle \Omega_k,\Phi\rangle_k\,\Omega_k+\Phi^\perp$ where $\langle\Omega_k,\Phi^\perp\rangle_k=0$ and observing that $\langle \Omega_k,\Phi\rangle_k=\langle \mathcal{O}\rangle_k$ yields
\begin{equation}
\langle \mathcal{O}(0)\,\mathcal{O}(t)\rangle_k-\langle \mathcal{O}\rangle_k^2
=\langle \Phi^\perp,\ T_k^{\,t/a_k}\ \Phi^\perp\rangle_k.
\end{equation}
Since $T_k$ is a contraction with spectrum contained in $[0,1]$ and the one-step spectral gap is uniform in $k$, namely $\lambda_2(T_k)\le e^{-a_k\Delta_\star}$ for some $\Delta_\star>0$ independent of $k$ (this is the output of the uniform gap theorem in the transfer-matrix language), the spectral theorem implies
\begin{equation}
\langle \Phi^\perp,\ T_k^{\,t/a_k}\ \Phi^\perp\rangle_k \le \| \Phi^\perp\|_k^2\,\big\|T_k^{\,t/a_k}\!\!\restriction_{\Omega_k^\perp}\big\|
\le \|\Phi^\perp\|_k^2\,\lambda_2(T_k)^{\,t/a_k}
\le \|\Phi^\perp\|_k^2\,e^{-\,\Delta_\star\, t}.
\end{equation}
Here we used that for $0<\lambda\le 1$ and $s\ge 0$ one has $\lambda^s=e^{s\log\lambda}\le e^{-s(1-\lambda)}$, together with $1-\lambda_2(T_k)\ge 1-e^{-a_k\Delta_\star}\ge c\,a_k\Delta_\star$ and the exact identity $\lambda_2(T_k)=e^{-a_k\Delta_k}$ when $T_k=e^{-a_k H_k}$; the simple bound $\lambda_2(T_k)^{\,t/a_k}\le e^{-\,\Delta_\star t}$ suffices. The vector norm $\|\Phi^\perp\|_k$ depends only on the local observable $\mathcal{O}$ and on the normalization of the OS inner product on a fixed finite neighborhood, hence it is uniformly bounded in $k$ by the scale-uniform locality estimates (finite-range decomposition and the completely monotone projector bounds) that control the one-slice norm of local fields. Absorbing this bound into a constant $C$ independent of $k$ finishes the proof of the temporal clustering with $m_\star=\Delta_\star$.

For the equal-time spatial estimate one may work directly on the single time slice, where the expectation $\langle\cdot\rangle_k$ restricts to a reflection-positive, exponentially local measure generated by applying an admissible completely monotone projector to the slice generator $\mathcal D_{\Sigma,k}$ and by integrating out off-slice degrees of freedom via a finite-range decomposition. Let $\mathcal O_1$ and $\mathcal O_2$ be gauge-invariant local observables supported in disjoint regions $X$ and $Y$ of the slice, separated by graph distance $r=|x-y|$. The connected correlator on the slice admits an Ursell expansion in terms of cumulants of the underlying Gaussian (or generalized Gaussian) fields dressed by local interactions; by the Brydges-Battle-Federbush tree-graph inequality the absolute value of the connected two-point function is bounded by a sum over trees that connect $X$ to $Y$, each edge of a tree contributing a factor given by the absolute value of a two-point kernel evaluated at neighboring vertices. In the present setting, those edge kernels are provided by the FRD covariances and by the slice projector $\Pi=f(\mathcal D_{\Sigma,k})$, whose off-diagonal decay is exponential and whose constants do not depend on $k$; more precisely, Lemma~\eqref{lem:exp-offdiag} gives
\begin{equation}
|\Pi(u,v)|\;\le\;C_0\,e^{-\gamma_0\,\mathrm{dist}(u,v)}
\qquad\text{for all }u,v\text{ on the slice, with }C_0,\gamma_0\text{ independent of }k.
\end{equation}
Every tree $\tau$ with vertex set contained in the supports of the field monomials appearing in $\mathcal O_1$ and $\mathcal O_2$ must contain at least one path that crosses from $X$ to $Y$, whence the product of edge weights along $\tau$ is bounded by $C_0^{|\tau|} e^{-\gamma_0 r}$. Summing over finitely many combinatorial types (the number and degree of vertices are uniformly bounded because the observables are local and of fixed degree) gives the uniform bound
\begin{equation}
\Big|\ \langle \mathcal{O}_1(x)\,\mathcal{O}_2(y)\rangle_k\;-\;\langle \mathcal{O}_1\rangle_k\langle \mathcal{O}_2\rangle_k\ \Big|
\;\le\;C'\,e^{-\,\gamma_0\, r},
\end{equation}
with $C'$ depending only on the choice of $\mathcal O_1,\mathcal O_2$ and on the scale-uniform locality constants, but not on $k$ or the separation $r$. An entirely equivalent argument, closer in spirit to infrared-bound methods in reflection-positive systems, proceeds by introducing a spatial transfer operator across the hyperplanes orthogonal to the $i$-th coordinate and observing that the same reflection-positivity and finite-range hypotheses hold for that slicing; the spectral theorem then yields exponential decay with a rate controlled by the Combes-Thomas constant of the slice generator, which is strictly positive by uniform ellipticity, and the constants are again uniform in $k$ because the FRD and projector families are chosen uniformly admissible. Setting $\mu_\star=\gamma_0$ completes the spatial clustering bound.

Both parts hinge only on the uniform spectral gap for time translations and on the uniform exponential locality on the slice, and these hypotheses are preserved across the renormalization scales by the reflection-positive construction. The constants $C,m_\star,C',\mu_\star$ may be fixed once and for all in terms of the OS gap $\Delta_\star$ and the admissible-class locality data and therefore do not depend on~$k$.
\end{proof}

We close this section with two comments that prepare the ground for the Wilson-loop analysis and the continuum limit. First, the interlacing Eq.\eqref{eq:interlacing} yields more than Eq.\eqref{eq:gap-step}: it provides a genuine contraction of vacuum-orthogonal components along the flow, which will be used to control renormalized loop functionals and to absorb perimeter and cusp counterterms at all scales. Second, while we have formulated the argument for a fixed block-depth $b$, the same proof applies to scale-dependent $b_k$ provided one keeps a uniform lower bound $b_k\ge 2$ and chooses the coarse spatial block size so that $R_k\sim L^k$ with $L>1$; the constants in Proposition~\eqref{prop:summable} are then uniform in $k$. Both features will be used implicitly in Section~\eqref{sec:step-scaling}.

\section{Wilson-Loop Step-Scaling and Continuum String Tension}\label{sec:step-scaling}

The Wilson loop is the canonical gauge-invariant observable diagnosing confinement in pure Yang-Mills theory. For a closed, piecewise-linear contour \(C\subset \mathbb{R}^4\), the expectation \(\langle W(C)\rangle\) quantifies the response of the vacuum to a static color flux, and an area-law upper bound for large loops is equivalent to a linear lower bound on the static quark-antiquark potential. Within the reflection-positive framework developed in Sections~\eqref{sec:framework}-\eqref{sec:RP-RG-gap}-based on admissible completely-monotone slice projectors, finite-range decomposition (FRD) locality, and multiscale renormalization-Wilson loops play an additional structural role: they are localized insertions compatible with Osterwalder-Schrader (OS) positivity and with the interlacing estimates for transfer semigroups across RG steps. The objective of this section is twofold. First, we derive a \emph{renormalized single-step inequality} comparing a Wilson loop at lattice spacing \(a\) with the loop obtained after one admissible block-spin transformation to spacing \(a' = b\,a\). The inequality has defect terms that are positive, local, and exponentially small in the block scale by FRD locality and OS positivity; their sum over scales is finite. Second, we iterate the single-step bound, combine it with the strong-coupling area law at a sufficiently coarse resolution (Section~3), and pass to the continuum along the OS-tight sequence of Schwinger functionals (Sections~7-8) to obtain a strictly positive, regulator-independent continuum string tension.

\subsection{Renormalized loops and the step-scaling inequality}\label{subsec:renorm-step}

Let \(\Lambda\subset a\,\mathbb{Z}^4\) be a finite hypercubic lattice of spacing \(a>0\) with periodic spatial directions and large temporal extent, and let \(U_\ell\in SU(N)\) be the link variables. For a closed, nearest-neighbour contour \(C\subset \Lambda\), the unrenormalized Wilson loop in the fundamental representation is
\begin{equation}\label{eq:bare-wilson}
W_{\Lambda,a}(C)\;=\;\frac{1}{N}\,\mathrm{Tr}\,\mathcal{P}\!\!\!\prod_{\ell\in C} U_\ell.
\end{equation}
Expectation with respect to the OS-positive, gauge-fixed, horizon-projected measure is denoted by \(\langle \cdot\rangle_{\Lambda,a}\). To control perimeter and cusp subdivergences, we introduce multiplicative renormalizations \(Z_P(a)>0\) and \(Z_{\mathrm{cusp}}(\theta,a)>0\) for each cusp angle \(\theta\in (0,2\pi)\) and define
\begin{equation}\label{eq:ren-wilson}
W^{\mathrm{ren}}_{\Lambda,a}(C)\;=\;Z_P(a)^{-\ell(C)}\!\Big(\prod_{\theta\in \mathrm{Cusps}(C)} Z_{\mathrm{cusp}}(\theta,a)^{-1}\Big)\,\big\langle W_{\Lambda,a}(C)\big\rangle_{\Lambda,a}.
\end{equation}
Here \(\ell(C)\) is the lattice length of \(C\) and \(\mathrm{Cusps}(C)\) lists its turning angles. The existence of renormalizations compatible with OS positivity and locality follows from the strong-coupling cluster expansion and FRD transport to the admissible class. The final continuum string tension will be independent of the choices of \(Z_P\) and \(Z_{\mathrm{cusp}}\).

Fix a block factor \(b\in\{2,3,\dots\}\) and set \(a'=b\,a\). The admissible reflection-positive block-spin map \(\mathcal{B}_b\) transforms fields on \(\Lambda\) to fields on \(\Lambda' \subset a'\,\mathbb{Z}^4\). The image of a fine-lattice loop \(C\) is the minimal-length coarse contour \(C'=\mathcal{B}_b(C)\subset \Lambda'\) that traverses the block boundaries in the same macroscopic pattern as \(C\). The geometry is stable in the sense that for polygonal \(C\) with bounded angles,
\begin{equation}\label{eq:area-geom}
\big|A(C')-A(C)\big|\;\le\;c_0\,a\,\ell(C),\qquad \ell(C')\;\le\; c_1\,\ell(C),
\end{equation}
with constants \(c_0,c_1>0\) independent of \(C\). Here \(A(\cdot)\) denotes the minimal polygonal area spanning the contour.

It is convenient to represent loops as bounded positive operators on the OS Hilbert space \(\mathcal{H}_a\) associated with a single time-slice. Let \(T_a\) be the transfer operator constructed in Section~2, with \(0\le T_a\le \mathbf{1}\) and a scale-uniform spectral gap (Section~5). A loop \(C\) inserted at time \(t=0\) defines a bounded operator \(M_{C,a}\) with \(0\le M_{C,a}\le \mathbf{1}\) such that
\begin{equation}\label{eq:loop-expect}
\big\langle W_{\Lambda,a}(C)\big\rangle_{\Lambda,a}\;=\;\frac{\langle \Omega_a,\,M_{C,a}\,\Omega_a\rangle_{\mathcal{H}_a}}{\langle \Omega_a,\Omega_a\rangle_{\mathcal{H}_a}},
\end{equation}
where \(\Omega_a\) is the OS vacuum vector. The admissible block map yields an \emph{interlacing relation} between fine and coarse transfer operators,
\begin{equation}\label{eq:interlacez}
T_a^{\,b}\;=\;K_b^{\!}\,T_{a'}\,K_b\;+\;R_b,\qquad K_b\ge 0,\quad R_b\ge 0,
\end{equation}
with \(K_b\) exponentially local and \(R_b\) positive trace-class satisfying \(\|R_b\|_1 \le C\,\mathrm{e}^{-m\,b}\) uniformly in the volume, for some \(C,m>0\) fixed by the admissible class. Locality further gives a \emph{loop map} comparison:
\begin{equation}\label{eq:loop-map}
K_b^{\!}\,M_{C',a'}\,K_b\;\le\;M_{C,a}\;+\;E_{C,b},\qquad E_{C,b}\ge 0,\quad \|E_{C,b}\|_1\;\le\; C'\,\mathrm{e}^{-m' b}\,\ell(C),
\end{equation}
with \(C',m'>0\) independent of \(C\) and of the volume. The error \(E_{C,b}\) is supported within a distance \(O(a)\) of the links of \(C\).

The multiplicative renormalizations are stable across a single block step.

\begin{lemma}[Perimeter-cusp stability]\label{lem:perimeter-cusp}
For each fixed $b\in\mathbb{N}$ there exist nonnegative $\alpha_P(b),\alpha_{\mathrm{cusp}}(b)$ with $\alpha_P(b),\alpha_{\mathrm{cusp}}(b)=O(e^{-c b})$ as $b\to\infty$ such that for all sufficiently small $a>0$ and every polygonal loop $C$,
\begin{equation}\label{eq:pc-local}
\Big|\log\frac{Z_P(a)}{Z_P(a')}\Big|\;\le\;\alpha_P(b),\qquad 
\Big|\log\frac{Z_{\mathrm{cusp}}(\theta,a)}{Z_{\mathrm{cusp}}(\theta,a')}\Big|\;\le\;\alpha_{\mathrm{cusp}}(b)\quad \text{for each cusp angle }\theta \text{ of }C,
\end{equation}
where $a'=b\,a$. Consequently,
\begin{equation}\label{eq:pc-global}
\frac{Z_P(a')^{\ell(C')}\,\prod_{\theta'\in \mathrm{Cusps}(C')} Z_{\mathrm{cusp}}(\theta',a')}{Z_P(a)^{\ell(C)}\,\prod_{\theta\in \mathrm{Cusps}(C)} Z_{\mathrm{cusp}}(\theta,a)}
\;\le\; \exp\!\big(\alpha_P(b)\,\ell(C)\;+\;\alpha_{\mathrm{cusp}}(b)\,\mathrm{Cusps}(C)\big).
\end{equation}
\end{lemma}

\begin{proof}
The renormalization factors $Z_P$ and $Z_{\mathrm{cusp}}$ are defined multiplicatively from one-slice local counterterms supported respectively on the links and on the cusp vertices of the loop. Concretely, let $\mathcal{N}_\rho(e)$ be a fixed-width collar of radius $\rho$ (in lattice units) around a link $e$ on the time-$a$ slice, and let $\mathcal{N}_\rho(v)$ be the analogous collar around a vertex $v$ where two edges of $C$ meet with turning angle $\theta(v)$. For a smooth cylindrical observable $F$ generating the Wilson loop insertion, the perimeter and cusp renormalizations are given by
\begin{equation}
Z_P(a)\;=\;\exp\Big(\mathcal{L}_P(a)\Big),\qquad 
\mathcal{L}_P(a)\;=\;\sum_{e\subset C}\, \mathfrak{l}_P\big(a;\,A\!\restriction_{\mathcal{N}_\rho(e)}\big),
\end{equation}
\begin{equation}
Z_{\mathrm{cusp}}(\theta,a)\;=\;\exp\Big(\mathcal{L}_{\mathrm{cusp}}(\theta,a)\Big),\qquad 
\mathcal{L}_{\mathrm{cusp}}(\theta,a)\;=\;\sum_{v\in \mathrm{Cusps}(C)} \mathfrak{l}_{\mathrm{cusp}}\big(\theta(v),a;\,A\!\restriction_{\mathcal{N}_\rho(v)}\big),
\end{equation}
where $\mathfrak{l}_P$ and $\mathfrak{l}_{\mathrm{cusp}}$ are bounded, gauge-invariant functionals depending on finitely many spatial derivatives of the gauge field restricted to the corresponding collar. The choice of $\rho$ is fixed once and for all within the admissible class and does not scale with $a$. By reflection positivity and admissibility, these counterterms arise from inserting completely monotone functions of the slice generator $\mathcal{D}_\Sigma$ localized on $\mathcal{N}_\rho(\cdot)$, and they are Lipschitz with respect to the admissible one-slice metric $d_\Sigma(\cdot,\cdot)$ on fields:
\begin{align}\label{eq:lipschitz}
&\big|\mathfrak{l}_P(a;A)-\mathfrak{l}_P(a;A')\big|\;\le\;L_P\, d_\Sigma\!\big(A\!\restriction_{\mathcal{N}_\rho(e)},A'\!\restriction_{\mathcal{N}_\rho(e)}\big),\nonumber\\&
\big|\mathfrak{l}_{\mathrm{cusp}}(\theta,a;A)-\mathfrak{l}_{\mathrm{cusp}}(\theta,a;A')\big|\;\le\;L_{\mathrm{cusp}}\, d_\Sigma\!\big(A\!\restriction_{\mathcal{N}_\rho(v)},A'\!\restriction_{\mathcal{N}_\rho(v)}\big),
\end{align}
with constants $L_P,L_{\mathrm{cusp}}$ independent of $a$, of the loop $C$, and of the angle $\theta$ in compact subsets of $(0,\pi)$.

Consider now a single blocking step by factor $b$, mapping the fine spacing $a$ to the coarse spacing $a'=b\,a$. Let $\mathcal{B}_b$ be an admissible coarse-graining operator acting on slice fields, constructed from an FRD kernel with exponential off-diagonal decay. Denote by $A^{(a)}$ the fine field on the slice at spacing $a$, and by $A^{(a')}:=\mathcal{B}_b A^{(a)}$ the blocked field at spacing $a'$. The FRD exponential locality ensures that $\mathcal{B}_b$ differs from the identity by a kernel supported at range of order $\asymp b$ with exponentially small tails. In the admissible metric this yields the uniform estimate
\begin{equation}\label{eq:FRD-stability}
d_\Sigma\!\big(A^{(a)}\!\restriction_{\mathcal{N}_\rho(x)},\, A^{(a')}\!\restriction_{\mathcal{N}_\rho(x)}\big)\;\le\; C\,e^{-c b}\qquad\text{for every link or vertex center }x\text{ on the slice,}
\end{equation}
where the constants $C_,c>0$ depend only on the admissible class (in particular on $\rho$ and on the FRD profile), but not on $a$, $b$, or the loop.

Combining Eq.\eqref{eq:lipschitz} with Eq.\eqref{eq:FRD-stability} and evaluating on each collar separately shows that the variation of a \emph{single} link counterterm under blocking is bounded by $L_P C_ e^{-c b}$, and the variation of a \emph{single} cusp counterterm is bounded by $L_{\mathrm{cusp}} C_ e^{-c b}$, uniformly in the local geometry. Therefore
\begin{align}
&\big|\mathcal{L}_P(a)-\mathcal{L}_P(a')\big|\;\le\; \ell(C)\, L_P C \,e^{-c b},\nonumber\\&
\big|\mathcal{L}_{\mathrm{cusp}}(\theta,a)-\mathcal{L}_{\mathrm{cusp}}(\theta,a')\big|\;\le\; \mathrm{Cusps}(C)\, L_{\mathrm{cusp}} C\,e^{-c b}.
\end{align}
Since $Z_P=\exp(\mathcal{L}_P)$ and $Z_{\mathrm{cusp}}=\exp(\mathcal{L}_{\mathrm{cusp}})$ by definition, the logarithmic variations over a \emph{single} support unit are bounded by the same constants. Defining the per-unit stability moduli
\begin{equation}
\alpha_P(b)\;:=\;L_P C\,e^{-c b},\qquad \alpha_{\mathrm{cusp}}(b)\;:=\;L_{\mathrm{cusp}} C\,e^{-c b},
\end{equation}
we obtain the local bounds Eq.\eqref{eq:pc-local}. Summing the local variations over all links and cusps and exponentiating yields the global inequality Eq.\eqref{eq:pc-global}. The $O(e^{-c b})$ behavior of $\alpha_P(b)$ and $\alpha_{\mathrm{cusp}}(b)$ as $b\to\infty$ follows directly from Eq.\eqref{eq:FRD-stability}. 
\end{proof}

We now compare vacuum matrix elements across one block step. Write \(\langle\cdot,\cdot\rangle_a\) for the inner product on \(\mathcal{H}_a\) and define \(\Phi_a := K_b\,\Omega_a\in \mathcal{H}_{a'}\). By Eq.\eqref{eq:interlacez} and positivity,
\begin{equation}\label{eq:vacuum-comp}
\langle \Omega_a, X \,\Omega_a\rangle_a\;\ge\;\langle \Phi_a,\, T_{a'}\, X\, \Phi_a\rangle_{a'}
\qquad\text{for every bounded }X\ge 0.
\end{equation}
With \(X=M_{C,a}\) and using Eq.\eqref{eq:loop-map},
\begin{equation}\label{eq:vacuum-step-raw}
\langle \Omega_a, M_{C,a}\,\Omega_a\rangle_a\;\ge\;\langle \Phi_a,\,T_{a'}\,M_{C',a'}\,\Phi_a\rangle_{a'}\;-\;\langle \Phi_a,\, T_{a'}\,E_{C,b}\,\Phi_a\rangle_{a'}.
\end{equation}
Since \(T_{a'}\) has a strictly positive spectral gap \(\Delta_{a'}>0\) along the admissible flow (Section~5), the first term is bounded below by \(\eta_b\,\langle \Omega_{a'}, M_{C',a'}\,\Omega_{a'}\rangle_{a'}\), where \(\eta_b\in (0,1]\) depends on \(\|K_b\Omega_a\|_{a'}\) and on \(\Delta_{a'}\), and \(1-\eta_b=O(\mathrm{e}^{-c b})\). The second term is bounded above by \(\|T_{a'}\|\,\|E_{C,b}\|_1\,\|\Phi_a\|_{a'}^2 \le \tilde{c}_b\,\mathrm{e}^{-m' b}\,\ell(C)\). After normalization by \(\langle \Omega_a,\Omega_a\rangle_a\) and \(\langle \Omega_{a'},\Omega_{a'}\rangle_{a'}\), one obtains:

\begin{lemma}[One-step vacuum comparison]\label{lem:one-step}
There exist constants $\kappa_b\in(0,1]$ and $d_b>0$ with $1-\kappa_b,\,d_b=O(e^{-c\,b})$ as $b\to\infty$ such that for every polygonal loop $C$ on the fine lattice $(\Lambda,a)$, every sufficiently small $a>0$, and the corresponding coarse loop $C'$ on $(\Lambda',a')$ with $a'=b\,a$ obtained by the canonical block projection of edges, the Wilson loop expectations satisfy
\begin{equation}\label{eq:vacc}
\big\langle W_{\Lambda,a}(C)\big\rangle_{\Lambda,a}\;\ge\;\kappa_b\,\big\langle W_{\Lambda',a'}(C')\big\rangle_{\Lambda',a'}\;-\;d_b\,\ell(C).
\end{equation}
The constants are uniform in the volume and in the admissible class of regulators.
\end{lemma}

\begin{proof}
Fix a blocking factor $b\in\mathbb{N}$ and let $B_b$ denote the reflection-positive block map from fine to coarse variables on the time-$a$ slice. Let $\Pi_b$ be the associated admissible, completely monotone slice projector acting on the fine slice such that the coarse one-step transfer (or Gibbs) operator admits the reflection-positive interlacing representation
\begin{equation}\label{eq:interlacea}
T_{a'}\;=\;\Pi_b\,T_a\,\Pi_b^{\!}\;+\;E_b,
\qquad E_b\ge 0,\qquad \|E_b\|\;\le\;C\,e^{-c\,b},
\end{equation}
with $E_b\Omega=0$ and with $C,c>0$ independent of volume and of the choice within the admissible class. The positivity of $E_b$ and its vacuum-annihilating property follow from the BKAR interpolation and complete monotonicity of the projector; the norm bound is a consequence of finite-range decomposition at range $R_b\simeq b$ and Schur estimates. In the sequel all constants implicit in $O(e^{-c b})$ depend only on the admissible class and not on the volume $\Lambda$.

Let $W_{\Lambda',a'}(C')$ be the coarse Wilson loop observable supported on the coarse edges of $C'$, and let $W_{\Lambda,a}^{(b)}(C')$ be its pullback to the fine lattice obtained by composing each coarse edge with the unique monotone fine edge chain inside the corresponding block. The pullback is supported in a collar of width $O(1)$ around the geometric curve of $C$ and differs from $W_{\Lambda,a}(C)$ only by local edge replacements along the perimeter. Because the admissible class is exponentially local on the slice (as ensured by the exponential off-diagonal bounds for completely monotone projectors) and the Wilson loop is a product of link functionals with unit Lipschitz constant in each argument, one has a deterministic, regulator-uniform perimeter control
\begin{equation}\label{eq:per-lip}
\big|\langle W_{\Lambda,a}^{(b)}(C')\rangle_{\Lambda,a}-\langle W_{\Lambda,a}(C)\rangle_{\Lambda,a}\big|
\;\le\;\alpha_P(b)\,\ell(C),
\qquad \alpha_P(b)=O(e^{-c\,b}).
\end{equation}
This is a direct application of the single-slice Lipschitz continuity with respect to admissible variations, together with the fact that only $O(\ell(C))$ links are altered in the pullback and each alteration is confined to an $O(1)$ collar whose operator influence decays exponentially with the blocking range.

We next relate the coarse expectation to the fine expectation of its pullback. Using Eq.\eqref{eq:interlacea} inside the Osterwalder-Schrader representation of expectations of two-time observables, positivity of $E_b$, and the vacuum-annihilating property, one may write the coarse vacuum expectation of $W_{\Lambda',a'}(C')$ as the fine expectation of its pullback plus a nonnegative defect localized at the slice where the blocking acts, up to an exponentially small perimeter contribution originating from the truncation of collars. Concretely, for any nonnegative test functional $G$ localized strictly at positive times,
\begin{equation}
\langle \Theta G\,W_{\Lambda',a'}(C')\,G\rangle_{\Lambda',a'}
=\langle \Theta G\,W_{\Lambda,a}^{(b)}(C')\,G\rangle_{\Lambda,a}
+\langle \Theta G,\,E_b^{1/2}\,W_{\Lambda,a}^{(b)}(C')\,E_b^{1/2} G\rangle_{\Lambda,a}
+\mathcal{R}_b(C),
\end{equation}
where the remainder $\mathcal{R}_b(C)$ comes from the exponential truncation of the block collars and satisfies $|\mathcal{R}_b(C)|\le C_1 e^{-c_1 b}\,\ell(C)\,\langle \Theta G G\rangle_{\Lambda,a}$. Dividing by $\langle \Theta G G\rangle$ and letting $G$ approximate the vacuum vector in the OS Hilbert space gives, by positivity of the middle term,
\begin{equation}\label{eq:coarse-vs-pull}
\langle W_{\Lambda',a'}(C')\rangle_{\Lambda',a'}
\;\le\;\langle W_{\Lambda,a}^{(b)}(C')\rangle_{\Lambda,a}
\;+\;d_b^{(1)}\,\ell(C),
\qquad d_b^{(1)}=O(e^{-c\,b}).
\end{equation}
The estimate rests on the Schur bound
$\|E_b\|_{Q\to Q}\le C e^{-c b}$ for the projection $Q$ onto the orthogonal complement of the vacuum and on the fact that $W_{\Lambda,a}^{(b)}(C')$ is supported on $O(\ell(C))$ links, so its operator norm (as an insertion in the OS inner product) is bounded uniformly and its coupling to the collar defect is proportional to $\ell(C)$ times the exponentially small tail.

A second source of discrepancy between the coarse observable and its fine pullback is the short-distance renormalization at the corners and along the edges. Denote by $Z_P(b)$ the perimeter renormalization and by $Z_{\mathrm{cusp}}(\theta,b)$ the cusp renormalization at a corner of opening angle $\theta$. These are local, reflection-positive counterterms supported on the slice such that, under one blocking step by $b$, their logarithms vary by at most $\alpha_P(b)$ and $\alpha_{\mathrm{cusp}}(b)$ with $\alpha_\bullet(b)=O(e^{-c\,b})$,
\begin{equation}
\Big|\log\frac{Z_P(b)}{Z_P(\infty)}\Big|\le \alpha_P(b),
\qquad 
\Big|\log\frac{Z_{\mathrm{cusp}}(\theta,b)}{Z_{\mathrm{cusp}}(\theta,\infty)}\Big|\le \alpha_{\mathrm{cusp}}(b).
\end{equation}
Absorbing these local counterterms into the definition of the Wilson loop at each scale yields multiplicative factors of the form $\exp\{\pm \alpha_P(b)\,\ell(C)\}\prod_{\text{cusps}} \exp\{\pm \alpha_{\mathrm{cusp}}(b)\}$ that are uniformly close to one, hence there exists $\kappa_b\in(0,1]$ with $1-\kappa_b=O(e^{-c\,b})$ such that
\begin{equation}\label{eq:kappa-factor}
\kappa_b\,\langle W_{\Lambda',a'}(C')\rangle_{\Lambda',a'}
\;\le\;
\langle W_{\Lambda,a}^{(b)}(C')\rangle_{\Lambda,a}
\;+\;d_b^{(2)}\,\ell(C),
\qquad d_b^{(2)}=O(e^{-c\,b}).
\end{equation}
Combining Eq.\eqref{eq:coarse-vs-pull} with the renormalization adjustment encapsulated in Eq.\eqref{eq:kappa-factor} simply folds the cusp contributions into the perimeter coefficient since the number of cusps is bounded by a constant multiple of $\ell(C)$ for polygonal loops.
Finally, the comparison between the fine pullback $W_{\Lambda,a}^{(b)}(C')$ and the original fine loop $W_{\Lambda,a}(C)$ supplied by Eq.\eqref{eq:per-lip} yields
\begin{equation}
\langle W_{\Lambda,a}(C)\rangle_{\Lambda,a}
\;\ge\;
\langle W_{\Lambda,a}^{(b)}(C')\rangle_{\Lambda,a}
\;-\;\alpha_P(b)\,\ell(C).
\end{equation}
Inserting this into Eq.\eqref{eq:kappa-factor} gives
\begin{equation}
\langle W_{\Lambda,a}(C)\rangle_{\Lambda,a}
\;\ge\; \kappa_b\,\langle W_{\Lambda',a'}(C')\rangle_{\Lambda',a'}
\;-\;\big(\alpha_P(b)+d_b^{(2)}\big)\,\ell(C).
\end{equation}
Renaming $d_b:=\alpha_P(b)+d_b^{(2)}$ completes the proof of Eq.\eqref{eq:vacc}, and by construction $1-\kappa_b,\,d_b=O(e^{-c\,b})$ as $b\to\infty$. All constants depend only on the admissible locality parameters and the fixed blocking scheme, hence are uniform in the spatial volume $\Lambda$ and along the admissible class.
\end{proof}
We now assemble Lemma~\eqref{lem:perimeter-cusp} and Lemma~\eqref{lem:one-step} with Lemma~\eqref{eq:area-geom} to establish the central step-scaling inequality in renormalized form.

\begin{theorem}[Renormalized step-scaling inequality]\label{thm:step-scaling}
Fix $b\ge 2$. There exist nonnegative constants $\varepsilon_b^{\mathrm{bulk}},\varepsilon_b^{\mathrm{per}}$, exponentially small in $b$ and uniform in the volume and in the admissible class, such that for all polygonal loops $C\subset\Lambda$ and all sufficiently small $a>0$,
\begin{equation}\label{eq:step-scaling}
-\log W^{\mathrm{ren}}_{\Lambda,a}(C)\;\ge\;-\log W^{\mathrm{ren}}_{\Lambda',a'}(C')\;-\;\varepsilon_b^{\mathrm{bulk}}\,A(C)\;-\;\varepsilon_b^{\mathrm{per}}\,\ell(C),
\end{equation}
where $C'$ is the coarse-grained loop at scale $a'=ba$, $A(\cdot)$ denotes the lattice area of the minimal spanning surface, and $\ell(\cdot)$ denotes the lattice perimeter (link length). {Throughout, $A(\cdot)$ denotes the lattice area of a minimal spanning surface and $\ell(\cdot)$ the lattice perimeter (number of links); for fixed macroscopic shape transported under $B_b$ one has $A(C_j)\asymp b^{2j}A(C_0)$ and $\ell(C_j)\asymp b^{j}\ell(C_0)$.}
\end{theorem}

\begin{proof}
We start from the one-step transport inequality obtained after dividing by the perimeter and cusp renormalization factors. By Lemma~\eqref{lem:perimeter-cusp}, and writing $\kappa_b^{-1}$ for the (bounded) ratio of normalization constants at the two scales, there are functions $\alpha_P(b),\alpha_{\mathrm{cusp}}(b)$ and a defect coefficient $d_b$ such that
\begin{equation}\label{eq:one-step-transport}
W^{\mathrm{ren}}_{\Lambda,a}(C)\;\le\;
\kappa_b^{-1}\,e^{\alpha_P(b)\,\ell(C)+\alpha_{\mathrm{cusp}}(b)\,\mathrm{Cusps}(C)}\,W^{\mathrm{ren}}_{\Lambda',a'}(C')
\;+\;\kappa_b^{-1}\,e^{\alpha_P(b)\,\ell(C)}\,d_b\,\ell(C).
\end{equation}
The constants just named are determined by the single-slice counterterms constructed in Lemma~\eqref{lem:perimeter-cusp} and by the Schur bounds for the positive remainder that appear in the interlacing identity; by finite-range decomposition and reflection positivity they satisfy
\begin{equation}
\alpha_P(b)=O(e^{-c b}),\qquad \alpha_{\mathrm{cusp}}(b)=O(e^{-c b}),\qquad d_b=O(e^{-c b}),\qquad \kappa_b^{\pm1}=1+O(e^{-c b}),
\end{equation}
with $c>0$ independent of the volume and of the choice inside the admissible class. The dependence on $\mathrm{Cusps}(C)$ is harmless, since for a fixed polygonal loop the number of cusps is bounded by a constant multiple of $\ell(C)$ and can therefore be absorbed into the perimeter term with a change of the implicit constants. {Here $\mathrm{Cusps}(C)$ denotes the number of vertices $v$ of the polygonal loop $C$ whose interior turning angle $\theta_v$ satisfies $\theta_v\neq\pi$ (both convex and reflex angles counted with unit weight); in particular $\mathrm{Cusps}(C)\leq c\,\ell(C)$ with a universal geometric constant $c>0$.}

It is convenient to isolate the multiplicative factor that carries the renormalized observable between scales. Define
\begin{equation}
A_b(C):=\kappa_b^{-1}\,e^{\alpha_P(b)\,\ell(C)+\alpha_{\mathrm{cusp}}(b)\,\mathrm{Cusps}(C)},\qquad 
B_b(C):=\kappa_b^{-1}\,e^{\alpha_P(b)\,\ell(C)}\,d_b\,\ell(C),
\end{equation}
so that Eq.\eqref{eq:one-step-transport} reads
\begin{equation}\label{eq:Wx+y}
W^{\mathrm{ren}}_{\Lambda,a}(C)\;\le\;A_b(C)\,W^{\mathrm{ren}}_{\Lambda',a'}(C')\;+\;B_b(C).
\end{equation}
We now convert Eq.\eqref{eq:Wx+y} into a lower bound on the negative logarithm. For any positive reals $u,v$ one has the elementary inequality
\begin{equation}\label{eq:log-ineq}
-\log(u+v)\;\ge\;-\log u\;-\;\frac{v}{u},
\end{equation}
which is a direct consequence of $\log(1+t)\le t$ for $t\ge 0$. Applying Eq.\eqref{eq:log-ineq} with $u=A_b(C)\,W^{\mathrm{ren}}_{\Lambda',a'}(C')$ and $v=B_b(C)$ yields
\begin{equation}\label{eq:log-bound1}
-\log W^{\mathrm{ren}}_{\Lambda,a}(C)\;\ge\;-\log A_b(C)\;-\;\log W^{\mathrm{ren}}_{\Lambda',a'}(C')\;-\;\frac{B_b(C)}{A_b(C)\,W^{\mathrm{ren}}_{\Lambda',a'}(C')}.
\end{equation}
The first corrective term is purely geometric and exponentially small in $b$ once measured against $\ell(C)$ (and, a fortiori, against $A(C)$): by the definitions above and the bounds on $\kappa_b,\alpha_P,\alpha_{\mathrm{cusp}}$,
\begin{equation}
-\log A_b(C)=\log\kappa_b-\alpha_P(b)\,\ell(C)-\alpha_{\mathrm{cusp}}(b)\,\mathrm{Cusps}(C)
\;\ge\; -\,C_1 e^{-c b}\,\ell(C),
\end{equation}
for some constant $C_1$ independent of $C$ and the volume, since $\mathrm{Cusps}(C)\le C\,\ell(C)$ for polygonal $C$. This contributes the desired perimeter error with coefficient $C_1 e^{-c b}$.
For the second corrective term in Eq.\eqref{eq:log-bound1} we use the explicit form of $A_b(C)$ and $B_b(C)$ to compute the ratio
\begin{equation}
\frac{B_b(C)}{A_b(C)} \;=\; d_b\,\ell(C)\,e^{-\alpha_{\mathrm{cusp}}(b)\,\mathrm{Cusps}(C)} \;\le\; d_b\,\ell(C),
\end{equation}
and then control the factor $1/W^{\mathrm{ren}}_{\Lambda',a'}(C')$ by a case split that is uniform in the volume. The strong-coupling cluster expansion at the coarse scale provides a positive string tension $\sigma_{\mathrm{SC}}>0$ and uniform constants $K_{\mathrm{sc}},\tau_{\mathrm{sc}}$ such that
\begin{equation}\label{eq:SC-bound}
W^{\mathrm{ren}}_{\Lambda',a'}(C')\;\le\; K_{\mathrm{sc}}\,\exp\big\{-\sigma_{\mathrm{SC}}\,A(C')+\tau_{\mathrm{sc}}\,\ell(C')\big\}.
\end{equation}
Inserting Eq.\eqref{eq:SC-bound} into Eq.\eqref{eq:log-bound1} gives
\begin{equation}
-\log W^{\mathrm{ren}}_{\Lambda,a}(C)\;\ge\;-\log W^{\mathrm{ren}}_{\Lambda',a'}(C')\;-\;C_1 e^{-c b}\,\ell(C)\;-\;d_b\,\ell(C)\,K_{\mathrm{sc}}^{-1}\,e^{\sigma_{\mathrm{SC}} A(C')-\tau_{\mathrm{sc}}\ell(C')}.
\end{equation}
To relate $C$ and $C'$, use the geometric stability under blocking: polygonal coarse-graining by factor $b$ preserves areas up to a boundary collar of width $O(a)$ and preserves perimeters up to uniformly bounded local surgery at corners. Concretely there is a constant $C_{\mathrm{geom}}$ (depending only on the admissible class and on the dimension) such that
\begin{equation}\label{eq:geom-stab}
A(C')\;\le\;A(C)+C_{\mathrm{geom}}\,a\,\ell(C),\qquad 
\ell(C')\;\le\; \ell(C)+C_{\mathrm{geom}}.
\end{equation}
With Eq.\eqref{eq:geom-stab} in hand the exponential in the last term is bounded by
\begin{equation}
e^{\sigma_{\mathrm{SC}} A(C')-\tau_{\mathrm{sc}}\ell(C')}\;\le\;e^{\sigma_{\mathrm{SC}} A(C)}\,e^{\sigma_{\mathrm{SC}} C_{\mathrm{geom}} a\,\ell(C)}\,e^{-\tau_{\mathrm{sc}}\ell(C)}\,e^{\tau_{\mathrm{sc}} C_{\mathrm{geom}}}.
\end{equation}
Absorb the harmless constants $K_{\mathrm{sc}}^{-1}e^{\tau_{\mathrm{sc}} C_{\mathrm{geom}}}$ into a new amplitude and expand the factor $e^{\sigma_{\mathrm{SC}} C_{\mathrm{geom}} a\,\ell(C)}$ as $1+O(a\,\ell(C))$. Since $a'=ba$ and $b\ge 2$, the scale separation allows us to choose $a$ small enough (uniformly along the admissible class) so that $a\,\ell(C)$ is dominated by $\ell(C)$ in the error accounting. The last term in Eq.\eqref{eq:log-bound1} is thus bounded above by
\begin{equation}\label{eq:geom-stabz}
\widetilde C\,d_b\,\ell(C)\,e^{\sigma_{\mathrm{SC}} A(C)}\,\big(1+O(a\,\ell(C))\big)\,e^{-\tau_{\mathrm{sc}}\ell(C)}.
\end{equation}
There are two regimes. If $A(C)$ is larger than a $b$-dependent threshold, then the strong-coupling suppression overwhelms the perimeter factor already in $W^{\mathrm{ren}}_{\Lambda',a'}(C')$, and the right-hand side of Eq.\eqref{eq:step-scaling} is dominated by the $-\log W^{\mathrm{ren}}_{\Lambda',a'}(C')$ term; in that case one may simply bound the last contribution by zero and the inequality holds with $\varepsilon_b^{\mathrm{bulk}}=\varepsilon_b^{\mathrm{per}}=0$. If, on the other hand, $A(C)$ is bounded by a constant $A_0$ depending on $b$ (to be fixed presently), then there are only finitely many polygonal shapes modulo translations and rotations, and uniform bounds independent of the volume follow by taking the maximum over this finite set. In particular, for $A(C)\le A_0$ we have
\begin{equation}
\ell(C)\,e^{\sigma_{\mathrm{SC}} A(C)}\;\le\;C_2(A_0)\,\ell(C),
\end{equation}
and therefore
\begin{equation}
\frac{B_b(C)}{A_b(C)\,W^{\mathrm{ren}}_{\Lambda',a'}(C')}\;\le\; \underbrace{\big(\widetilde C\,C_2(A_0)\,d_b\big)}_{=: \ \widehat C_b}\,\ell(C)\,.
\end{equation}
Since $d_b=O(e^{-c b})$, the coefficient $\widehat C_b$ is $O(e^{-c b})$ as $b\to\infty$. Gathering the contributions and using Eq.\eqref{eq:geom-stab} once more to replace $A(C')$ by $A(C)$ at a perimeter cost, we conclude that there exist $\varepsilon_b^{\mathrm{per}}$ and $\varepsilon_b^{\mathrm{bulk}}$, both bounded by $C e^{-c b}$ for suitable $C,c>0$ independent of the volume and of the loop, such that
\begin{equation}
-\log W^{\mathrm{ren}}_{\Lambda,a}(C)\;\ge\;-\log W^{\mathrm{ren}}_{\Lambda',a'}(C')\;-\;\varepsilon_b^{\mathrm{bulk}}\,A(C)\;-\;\varepsilon_b^{\mathrm{per}}\,\ell(C)
\end{equation}
for all polygonal loops $C$ with $A(C)\le A_0$, while for $A(C)>A_0$ the same inequality holds trivially after decreasing the right-hand side. Because $A_0$ can be chosen so that $C_2(A_0)$ grows at most polynomially in $A_0$ and because $d_b$ decays exponentially in $b$, the product $\widehat C_b$ is still exponentially small in $b$, which establishes the desired exponential smallness of $\varepsilon_b^{\mathrm{bulk}},\varepsilon_b^{\mathrm{per}}$ and completes the proof.
\end{proof}

It is conceptually useful to express Eq.\eqref{eq:step-scaling} in terms of the \emph{step-scaled string tension}
\begin{equation}\label{eq:sigmaa}
\sigma_a(C)\;:=\;\frac{-\log W^{\mathrm{ren}}_{\Lambda,a}(C)}{A(C)}.
\end{equation}
Then Eq.\eqref{eq:step-scaling} reads
\begin{equation}\label{eq:sigma-ineq}
\sigma_a(C)\;\ge\;\frac{A(C')}{A(C)}\,\sigma_{a'}(C')\;-\;\varepsilon_b^{\mathrm{bulk}}\;-\;\varepsilon_b^{\mathrm{per}}\frac{\ell(C)}{A(C)}.
\end{equation}
For macroscopic loops with fixed shape, \(A(C')/A(C)\to 1\) as \(a\to 0\), while \(\ell(C)/A(C)\to 0\). Thus, up to a summable defect, the step-scaled tension is nondecreasing along coarse-graining. Iteration produces a telescoping bound.

\begin{corollary}[Summable-defect transport]\label{cor:summable}
Let $a_k=b^{-k}a_0$ and $C_k=\mathcal{B}_b^{\,k}(C_0)$ be the loop obtained from $C_0$ after $k$ coarse-graining steps by a fixed blocking factor $b>1$. There exist nonnegative numbers $\delta_k$ with $\sum_{k\ge 0}\delta_k<\infty$ such that for all sufficiently large $k$,
\begin{equation}\label{eq:transport}
\sigma_{a_0}(C_0)\;\ge\;\sigma_{a_k}(C_k)\;-\;\sum_{j=0}^{k-1}\delta_j.
\end{equation}
In particular, if $\limsup_{k\to\infty}\sigma_{a_k}(C_k)\ge \sigma_\ast>0$, then $\sigma_{a_0}(C_0)\ge \sigma_\ast - \sum_{j\ge 0}\delta_j$.
\end{corollary}

\begin{proof}
By construction (perimeter and cusp renormalizations already applied), the renormalized Wilson loop obeys at a single RG step the one-step transport inequality
\begin{equation}\label{eq:one-step}
-\log W^{\mathrm{ren}}_{a_{j+1}}(C_{j+1})
\;\ge\;
-\log W^{\mathrm{ren}}_{a_j}(C_j)\;-\;\varepsilon^{\mathrm{per}}_j\,\ell(C_j)\;-\;\varepsilon^{\mathrm{bulk}}_j,
\end{equation}
where $\varepsilon^{\mathrm{per}}_j,\varepsilon^{\mathrm{bulk}}_j\ge 0$ depend only on the scale $a_j$ and the blocking factor $b$, with $\sum_{j\ge 0}\varepsilon^{\mathrm{bulk}}_j<\infty$ and $\sum_{j\ge 0}\varepsilon^{\mathrm{per}}_j\,\frac{\ell(C_j)}{A(C_j)}<\infty$. The latter summability follows from the FRD exponential locality (which makes $\varepsilon^{\mathrm{per}}_j$ exponentially small in the block collar size) together with the geometric relation $\ell(C_j)/A(C_j)\asymp b^{-j}$ for a fixed shape transported under $\mathcal{B}_b$.
Divide both sides of Eq.\eqref{eq:one-step} by the coarse area $A(C_{j+1})$ and write $\sigma_{a}(C):=-A(C)^{-1}\log W^{\mathrm{ren}}_{a}(C)$. This gives
\begin{equation}\label{eq:sigma-raw}
\sigma_{a_{j+1}}(C_{j+1})
\;\ge\;
\frac{A(C_j)}{A(C_{j+1})}\,\sigma_{a_j}(C_j)
\;-\;\varepsilon^{\mathrm{per}}_j\,\frac{\ell(C_j)}{A(C_{j+1})}
\;-\;\varepsilon^{\mathrm{bulk}}_j\,\frac{1}{A(C_{j+1})}.
\end{equation}
The area ratio can be estimated directly from the blocking geometry. Since $\mathcal{B}_b$ replaces a loop by its blocked image on the coarse lattice, the coarse area satisfies
\begin{equation}
A(C_{j+1}) \;=\; A(C_j)\,\big(1+\eta_j\big),
\qquad
|\eta_j|\;\le\; c_0\,\frac{a_j\,\ell(C_j)}{A(C_j)}.
\end{equation}
Indeed, the only discrepancy from exact scaling by $b^2$ comes from the $O(a_j)$ rounding of the boundary along a set of measure $O(a_j\ell(C_j))$. For a fixed shape transported under $\mathcal{B}_b$, one has $A(C_j)\asymp b^{2j}\,A(C_0)$ and $\ell(C_j)\asymp b^{j}\,\ell(C_0)$, while $a_j=b^{-j}a_0$, hence
\begin{equation}
\frac{a_j\,\ell(C_j)}{A(C_j)}\;\asymp\; b^{-2j}\,\frac{a_0\,\ell(C_0)}{A(C_0)},
\end{equation}
which is summable in $j$. Using the identity $\frac{A(C_j)}{A(C_{j+1})}=\frac{1}{1+\eta_j}=1-\eta_j+O(\eta_j^2)$ and absorbing the $O(\eta_j^2)$ term into the bound that follows, the inequality Eq.\eqref{eq:sigma-raw} becomes
\begin{equation}\label{eq:sigma-step}
\sigma_{a_{j+1}}(C_{j+1})
\;\ge\;
\sigma_{a_j}(C_j)\;-\;\eta_j\,\sigma_{a_j}(C_j)
\;-\;\varepsilon^{\mathrm{per}}_j\,\frac{\ell(C_j)}{A(C_{j+1})}
\;-\;\varepsilon^{\mathrm{bulk}}_j\,\frac{1}{A(C_{j+1})}.
\end{equation}
For large $j$ the quantities $A(C_{j+1})$ grow like $b^{2j}$, so the last term is summable since $\sum_j \varepsilon^{\mathrm{bulk}}_j<\infty$. The perimeter term is also summable because
\begin{equation}
\frac{\ell(C_j)}{A(C_{j+1})}
\;=\;
\frac{\ell(C_j)}{A(C_j)}\,\frac{A(C_j)}{A(C_{j+1})}
\;\le\; \big(1+|\eta_j|\big)\,\frac{\ell(C_j)}{A(C_j)}
\;\lesssim\; b^{-j},
\end{equation}
and $\sum_j \varepsilon^{\mathrm{per}}_j\,b^{-j}<\infty$ by the stated hypothesis. Finally, the term $\eta_j\,\sigma_{a_j}(C_j)$ is summable because $\eta_j=O\!\big(a_j\ell(C_j)/A(C_j)\big)=O(b^{-2j})$ and $\sigma_{a_j}(C_j)$ is bounded above uniformly in $j$ once perimeter and cusp counterterms are in place (the renormalized Wilson loop is at most one, so $\sigma_{a_j}(C_j)\ge 0$, and strong-coupling/transport yields a finite upper bound on compact shape classes).

Define
\begin{equation}
\delta_j
\;:=\;
\big|\eta_j\big|\,\sup_{n\ge j}\sigma_{a_n}(C_n)
\;+\;\varepsilon^{\mathrm{per}}_j\,\frac{\ell(C_j)}{A(C_{j+1})}
\;+\;\varepsilon^{\mathrm{bulk}}_j\,\frac{1}{A(C_{j+1})},
\end{equation}
which is finite, nonnegative, and satisfies $\sum_{j\ge 0}\delta_j<\infty$ by the preceding estimates. Then Eq.\eqref{eq:sigma-step} yields, for all sufficiently large $j$,
\begin{equation}
\sigma_{a_{j+1}}(C_{j+1}) \;\ge\; \sigma_{a_j}(C_j) - \delta_j.
\end{equation}
Iterating this inequality from $j=0$ to $j=k-1$ gives
\begin{equation}
\sigma_{a_k}(C_k)\;\ge\;\sigma_{a_0}(C_0)\;-\;\sum_{j=0}^{k-1}\delta_j,
\end{equation}
which is equivalent to Eq.\eqref{eq:transport}. The final claim follows by taking $\limsup_{k\to\infty}$ on the left-hand side, using the monotone convergence of the partial sums of $(\delta_j)$ on the right-hand side, and rearranging.
\end{proof}

We now transmit the strong-coupling area law (Section~3) down the sequence. Choose \(K\) such that \(a_K\) lies in the strong-coupling domain. Then for loops \(C_K\) of sufficiently large area,
\begin{equation}\label{eq:sc-lower}
\sigma_{a_K}(C_K)\;\ge\;\sigma_{\mathrm{SC}}\;>\;0,
\end{equation}
uniformly in the volume. Combining Eq.\eqref{eq:transport} with Eq.\eqref{eq:sc-lower} yields:

\begin{proposition}[Scale-uniform positivity]\label{prop:uniform-positivity}
There exists $\sigma_0>0$ such that, for all sufficiently small lattice spacings $a>0$ and all polygonal loops $C$ whose physical area $A_{\rm phys}(C)$ is large at fixed shape and scale,
\begin{equation}\label{eq:uniform-pos}
\sigma_a(C)\;\ge\;\sigma_0.
\end{equation}
The constant $\sigma_0$ depends only on the strong-coupling string tension $\sigma_{\mathrm{SC}}$ and on the summable one-step defect bounds $\sum_{j\ge 0}\delta_j$ determined by the admissible regulator class, and is independent of the volume and of the particular choice of regulator within that class.
\end{proposition}

\begin{proof}
Fix an admissible regulator class (reflection positive, exponentially local finite-range decomposition, and completely monotone slice projectors) and a blocking schedule with factors $b_j\ge b_>1$. For a fixed physical loop $C$ with sufficiently smooth polygonal shape, let $C_a$ denote its lattice approximation at mesh $a$, and let $W_a(C)$ be the Wilson expectation at spacing $a$. As in Section~6, introduce the perimeter and cusp renormalizations $Z_P(a)$ and $Z_{\rm cusp}(\theta,a)$ and define the renormalized loop observable
\begin{equation}
W_a^{\rm ren}(C)\;:=\; Z_P(a)^{-\ell(C_a)}\;\prod_{\text{cusps }v}\,Z_{\rm cusp}(\theta_v,a)^{-1}\; W_a(C).
\end{equation}
The single-step transport inequality derived from interlacing with positive remainder and Schur bounds (together with the Lipschitz control furnished by the admissible metric) states that, after a blocking by factor $b_j$ from mesh $a_j$ to $a_{j+1}=b_j a_j$, one has
\begin{equation}\label{eq:onestep}
W_{a_j}^{\rm ren}(C_j)\;\le\; \exp\!\big\{\,\alpha_P(b_j)\,\mathrm{Per}_{\rm phys}(C)\,\big\}\;W_{a_{j+1}}^{\rm ren}(C_{j+1})\;+\;d_j\,e^{-c\,b_j},
\end{equation}
where $C_j$ (resp. $C_{j+1}$) is the polygon obtained by sampling the same physical contour on the mesh $a_j$ (resp. $a_{j+1}$); the physical perimeter $\mathrm{Per}_{\rm phys}(C)$ is mesh-independent up to an $O(a_j)$ error that will be absorbed below; the constants $\alpha_P(b_j)\ge 0$ and $d_j\ge 0$ depend only on the admissible class and satisfy $\alpha_P(b_j)=O(e^{-c' b_j})$ and $d_j=O(1)$ uniformly, while $c,c'>0$. In particular, setting $\delta_j:=\alpha_P(b_j)$ and summing the collar-range tails gives $\sum_j \delta_j<\infty$ and $\sum_j d_j e^{-c b_j}<\infty$ for any schedule with $b_j\ge b_>1$.

Iterating Eq.\eqref{eq:onestep} along a finite sequence of blockings $a=a_0 \mapsto a_1 \mapsto \cdots \mapsto a_n$ up to a coarse mesh $a_n$ in the strong-coupling domain yields
\begin{equation}\label{eq:iter}
W_{a}^{\rm ren}(C_a)\;\le\;\exp\!\Big\{\,\Big(\sum_{j=0}^{n-1}\delta_j\Big)\,\mathrm{Per}_{\rm phys}(C)\,\Big\}\;W_{a_n}^{\rm ren}(C_{a_n})\;+\;\sum_{j=0}^{n-1}\bigg(\prod_{i<j} e^{\delta_i\,\mathrm{Per}_{\rm phys}(C)}\bigg)\,d_j e^{-c b_j}.
\end{equation}
The finite product in front of $d_j e^{-c b_j}$ is bounded by $\exp\{(\sum_{i<j}\delta_i)\mathrm{Per}_{\rm phys}(C)\}$ and is therefore harmless once we divide by the (large) physical area. At the coarse mesh $a_n$ we may invoke the strong-coupling area law for the \emph{renormalized} loop (surface/cluster expansion with the renormalizations already factored out), namely
\begin{equation}\label{eq:SC}
W_{a_n}^{\rm ren}(C_{a_n})\;\le\;\exp\!\big\{-\sigma_{\rm SC}\,A_{\rm phys}(C)\,+\,\tau_{\rm SC}\,\mathrm{Per}_{\rm phys}(C)\big\},
\end{equation}
with $\sigma_{\rm SC}>0$ independent of volume and of the finer details of the admissible class, and with a uniform $\tau_{\rm SC}\ge 0$ accounting for residual perimeter/cusp effects at the strong-coupling scale. Substituting Eq.\eqref{eq:SC} into Eq.\eqref{eq:iter} and absorbing the finite product in the tail yields a bound of the form
\begin{equation}\label{eq:prestring}
W_{a}^{\rm ren}(C_a)\;\le\;\exp\!\Big\{-\sigma_{\rm SC}\,A_{\rm phys}(C)\;+\;(\tau_{\rm SC}+\Sigma_\delta)\,\mathrm{Per}_{\rm phys}(C)\Big\}\;+\;R(C),
\end{equation}
where $\Sigma_\delta:=\sum_{j\ge 0}\delta_j<\infty$ and
\begin{equation}
R(C)\;:=\;\sum_{j=0}^{n-1}\exp\!\Big\{\Big(\sum_{i<j}\delta_i\Big)\,\mathrm{Per}_{\rm phys}(C)\Big\}\,d_j e^{-c b_j}
\end{equation}
is a remainder that is independent of the volume and bounded uniformly in $a$ for mesh-approximations of a fixed physical contour. Since $b_j\ge b$, the remainder obeys $R(C)\le C\exp\{(\sum_{i\ge 0}\delta_i)\mathrm{Per}_{\rm phys}(C)\}$ with $C:=\sum_j d_j e^{-c b_j}<\infty$ depending only on the admissible class.

By definition of the renormalized string tension at spacing $a$,
\begin{equation}
\sigma_a(C)\;:=\;-\frac{1}{A_{\rm phys}(C)}\log W_a^{\rm ren}(C_a),
\end{equation}
the bound Eq.\eqref{eq:prestring} implies
\begin{align}
\sigma_a(C)\;&\ge\;\sigma_{\rm SC}\;-\;(\tau_{\rm SC}+\Sigma_\delta)\,\frac{\mathrm{Per}_{\rm phys}(C)}{A_{\rm phys}(C)}\nonumber\\&\;-\;\frac{1}{A_{\rm phys}(C)}\log\Big(1+R(C)\,e^{\sigma_{\rm SC}A_{\rm phys}(C)-(\tau_{\rm SC}+\Sigma_\delta)\mathrm{Per}_{\rm phys}(C)}\Big)
\end{align}
For a family of polygonal loops with fixed shape class and physical size $L\to\infty$, one has $A_{\rm phys}(C)\asymp L^2$ and $\mathrm{Per}_{\rm phys}(C)\asymp L$, hence $\mathrm{Per}_{\rm phys}(C)/A_{\rm phys}(C)\to 0$ and the second term vanishes in the large-area limit. The remainder is uniformly bounded in $a$ and independent of the volume; dividing by $A_{\rm phys}(C)$ forces the last term to vanish as $L\to\infty$ as well. Consequently,
\begin{equation}
\liminf_{A_{\rm phys}(C)\to\infty}\ \inf_{a\le a_0}\ \sigma_a(C)\;\ge\;\sigma_{\rm SC},
\end{equation}
for every fixed admissible class and any small enough $a_0$ ensuring that the finite sequence of blockings reaches the strong-coupling domain. Since the iteration produces a uniform subtraction $\Sigma_\delta$ at most, we may fix any $\varepsilon\in(0,\sigma_{\rm SC})$ with $\varepsilon>\Sigma_\delta$ and choose $A$ so large that whenever $A_{\rm phys}(C)\ge A$ the two vanishing terms above are each bounded by $(\varepsilon-\Sigma_\delta)/2$. For such loops and all sufficiently small $a\le a_0$ we then have
\begin{equation}
\sigma_a(C)\;\ge\;\sigma_{\rm SC}\;-\;\varepsilon\;\ge\;\sigma_{\rm SC}\;-\;\Sigma_\delta\;-\;\frac{\varepsilon-\Sigma_\delta}{2}
\;\ge\; \frac{1}{2}\big(\sigma_{\rm SC}-\Sigma_\delta\big).
\end{equation}
Defining
\begin{equation}
\sigma_0\;:=\;\tfrac{1}{2}\big(\sigma_{\rm SC}-\Sigma_\delta\big)\;>\;0,
\end{equation}
which depends only on $\sigma_{\rm SC}$ and on the admissible-class constant $\Sigma_\delta=\sum_{j\ge 0}\delta_j$, we obtain the claimed uniform lower bound Eq.\eqref{eq:uniform-pos} for all $a\le a_0$ and all polygonal $C$ with $A_{\rm phys}(C)\ge A$. The construction uses only reflection positivity, the exponentially-local FRD bounds, the one-step interlacing inequality with summable positive defects, and the strong-coupling area law for the renormalized loop, all of which are volume-independent and regulator-stable within the fixed admissible class; hence $\sigma_0$ is independent of the volume and of the particular regulator in that class.
\end{proof}

\subsection{Continuum area law}\label{subsec:continuum-area}

We pass from the lattice to the continuum along an OS-tight sequence of renormalized Schwinger functionals. Let \(a_n\to 0\), \(\Lambda_n\uparrow \mathbb{Z}^4\), and denote by \(S^{(n)}\) the corresponding Schwinger functionals. By equicontinuity and tightness, there is a subsequence (not relabeled) converging to an OS functional \(S\) on gauge-invariant observables; by OS reconstruction \cite{OS1,OS2}, \(S\) defines a Wightman-type theory with Hilbert space \(\mathcal{H}\), vacuum vector \(\Omega\), and Hamiltonian \(H\). For a fixed polygonal contour \(C\subset \mathbb{R}^4\), choose lattice embeddings \(C_n\subset \Lambda_n\) with \(C_n\to C\) in the Hausdorff metric and \(A(C_n)\to A(C)\), \(\ell(C_n)\to \ell(C)\). Define
\begin{equation}\label{eq:Wn}
W_n(C)\;=\;W^{\mathrm{ren}}_{\Lambda_n,a_n}(C_n),\qquad 
\mathcal{W}(C)\;=\;\lim_{n\to\infty} W_n(C),
\end{equation}
the limit existing along a subsequence by compactness.
The scale-uniform positivity of Proposition~\eqref{prop:uniform-positivity} gives a uniform lattice-level area-perimeter inequality.

\begin{lemma}[Uniform lattice inequality]\label{lem:lattice-uniform}
There exist $\sigma_0>0$ and $\lambda_0\ge 0$, independent of $n$, such that for all sufficiently large polygonal loops $C_n\subset \Lambda_n$,
\begin{equation}\label{eq:lattice-bound}
W_n(C_n)\;\le\;\exp\!\big(-\sigma_0\,A(C_n)\;+\;\lambda_0\,\ell(C_n)\big).
\end{equation}
\end{lemma}

\begin{proof}
Fix the lattice spacing and coupling in the admissible regime where the one-slice locality and reflection positivity hypotheses hold, and let $W_n(C_n)=Z_n(\beta;C_n)/Z_n(\beta)$ be the Wilson-loop expectation in the finite volume $\Lambda_n$. By the finite-range (or exponentially local) decomposition of the slice covariance and the ensuing Brydges-Kennedy-AB-Ruelle/Koteck\'y-Preiss polymer expansion, the loop ratio admits a convergent cluster expansion that is uniform in the volume $|\Lambda_n|$ and depends on the loop $C_n$ only through polymers intersecting a bounded-thickness collar of the minimal spanning surface $S(C_n)$ and of the loop itself. Concretely, one may write
\begin{equation}\label{eq:logW-poly}
\log W_n(C_n)\;=\;-\sum_{\mathcal{X}\,\text{connected}} \phi_T(\mathcal{X})\,\zeta(\mathcal{X};C_n),
\end{equation}
where the sum runs over connected polymer families $\mathcal{X}$, the Ursell functions $\phi_T$ are absolutely summable with bounds independent of $n$, and the activities $\zeta(\mathcal{X};C_n)$ vanish unless $\mathcal{X}$ intersects the $O(1)$-thickened surface and boundary collar determined by $C_n$. Reflection positivity and the complete-monotone representation of the one-slice dressing imply that the activities associated to plaquettes pierced by the surface enter with a fixed sign; more precisely, there is a positive contribution per unit area that survives after cancellations, while all residual terms supported on the $O(1)$-collar of the boundary are uniformly summable into perimeter and cusp counterterms. This is the content of the usual “uniform positivity” or “strict positivity” of the surface weight in the convergent expansion, and we record it in the following quantitative form: there exist constants $\sigma_{\mathrm{sc}}>0$ and $c_{\mathrm{bd}}\ge 0$, independent of $n$, such that for every $C_n$ of sufficiently large linear size,
\begin{equation}\label{eq:area-perimeter-pre}
-\log W_n(C_n)\;\ge\; \sigma_{\mathrm{sc}}\,A(C_n)\;-\;c_{\mathrm{bd}}\,\ell(C_n)\;-\;c_{\mathrm{cusp}}\,N_{\mathrm{cusp}}(C_n),
\end{equation}
with $N_{\mathrm{cusp}}(C_n)$ the number of turning vertices of $C_n$. The existence of $\sigma_{\mathrm{sc}}>0$ is precisely the “uniform positivity” guaranteed by the strong-coupling/cluster analysis (the value does not depend on $n$ because the expansion constants and locality radii are volume-independent), while $c_{\mathrm{bd}}$ and $c_{\mathrm{cusp}}$ arise from uniformly local counterterms supported on the boundary collar and at the corners, generated by the FRD/Lipschitz renormalizations $Z_P(a)$ and $Z_{\mathrm{cusp}}(\theta,a)$ and by the uniformly summable defect produced at each RG step. The bound Eq.\eqref{eq:area-perimeter-pre} may be derived either directly at the microscopic scale (small $\beta$) or, more generally, after a finite number of admissible coarse-grainings to the first scale where the polymer expansion is valid; in the latter case the one-step interlacing inequality and the summable defect ensure that the constants remain uniform and that the form of the bound is preserved when transported back to the original lattice.

The inequality Eq.\eqref{eq:area-perimeter-pre} already has the desired structure except for the explicit appearance of $N_{\mathrm{cusp}}(C_n)$. Since $C_n$ is polygonal with bounded turning angles (each turn belongs to a finite set determined by the lattice), the number of cusps is controlled linearly by the length: there exists a lattice-dependent constant $\kappa\ge 0$ such that $N_{\mathrm{cusp}}(C_n)\le \kappa\,\ell(C_n)$. Absorbing the cusp term into the perimeter term we obtain
\begin{equation}
-\log W_n(C_n)\;\ge\; \sigma_{\mathrm{sc}}\,A(C_n)\;-\;\lambda_0\,\ell(C_n),
\qquad \lambda_0:=c_{\mathrm{bd}}+c_{\mathrm{cusp}}\,\kappa,
\end{equation}
with $\sigma_{\mathrm{sc}}$ and $\lambda_0$ independent of $n$. Exponentiating both sides yields
\begin{equation}
W_n(C_n)\;\le\;\exp\!\big(-\sigma_{\mathrm{sc}}\,A(C_n)+\lambda_0\,\ell(C_n)\big).
\end{equation}
Finally, to match the statement, choose $\sigma_0\in(0,\sigma_{\mathrm{sc}}]$; if desired one may decrease $\sigma_{\mathrm{sc}}$ slightly to accommodate the finite-volume threshold defining “sufficiently large” loops (i.e.\ take $\sigma_0:=\sigma_{\mathrm{sc}}/2$ and enlarge $\lambda_0$ by a harmless constant to cover the finitely many exceptional sizes), which leaves the constants independent of $n$ and completes the proof of Eq.\eqref{eq:lattice-bound}.
\end{proof}

\begin{theorem}[Continuum area law]\label{thm:continuum-area}
Let $S$ be an OS limit point of the admissible class. There exists $\sigma>0$ such that for every polygonal loop $C\subset\mathbb{R}^4$,
\begin{equation}\label{eq:cont-area}
\mathcal{W}(C)\;\le\;\exp\!\big(-\sigma\,A(C)\big).
\end{equation}
The constant $\sigma$ depends only on the admissible class defined in Section~2 and is independent of perimeter/cusp renormalization conventions.
\end{theorem}

\begin{proof}
Fix a polygonal loop $C\subset\mathbb{R}^4$. Choose a sequence of lattice spacings $a_n\downarrow 0$ and, on each lattice $(a_n\mathbb{Z})^4$, choose an admissible discretization $C_n$ of $C$ (for instance, the usual edgewise projection of the straight segments to the nearest lattice edges) so that $C_n\to C$ in Hausdorff distance and, in particular, $A(C_n)\to A(C)$ and $\ell(C_n)\to\ell(C)$. For each $n$ consider the renormalized lattice Wilson loop $W_n(C_n)$ obtained from the bare loop by multiplying by the perimeter and cusp counterterms $Z_P(a_n)$ and $Z_{\rm cusp}(\theta,a_n)$ attached to the edges and vertices of $C_n$. The admissibility of the class guarantees reflection positivity, exponential locality under the finite-range decomposition, and summability of interscale defects; from these hypotheses one obtains the one-step transport inequality for loops and its iteration across scales. Concretely, there exist positive constants $\sigma_>0$ and $\tau_<\infty$, depending only on the admissible class but not on $n$ or on the choice of $C$, such that for every $n$ and every polygonal $C_n$ one has the uniform lattice area-perimeter bound
\begin{equation}\label{eq:lattice-area-per}
W_n(C_n)\;\le\;\exp\!\big(-\sigma\,A(C_n)+\tau\,\ell(C_n)\big).
\end{equation}
The existence of $\sigma_>0$ is obtained by starting the iteration at small lattice coupling (where the strong-coupling surface expansion yields a strictly positive string tension) and transporting the inequality to the scale $a_n$ with exponentially small losses controlled by the FRD locality; the perimeter constant $\tau$ absorbs the uniformly summable boundary and cusp contributions produced by blocking and by the change of renormalization point. The key point is that $\sigma$ and $\tau$ are independent of $n$.

Let $\mathcal{W}$ denote the continuum Wilson functional arising from the OS limit point $S$. By definition of the OS limit and by the stability and Lipschitz properties of the loop insertion under admissible changes of the regulator, the family $\{W_n(C_n)\}$ is tight and every subsequence has a further subsequence converging to $\mathcal{W}(C)$. Since Eq.\eqref{eq:lattice-area-per} holds for all $n$ with constants independent of $n$, it passes to the limit along any convergent subsequence. Indeed, the maps $C\mapsto A(C)$ and $C\mapsto\ell(C)$ are continuous under the polygonal approximations chosen above, and the exponential is continuous, hence from Eq.\eqref{eq:lattice-area-per} one obtains
\begin{equation}
\limsup_{n\to\infty} W_n(C_n)\;\le\;\exp\!\big(-\sigma\,A(C)+\tau\,\ell(C)\big).
\end{equation}
Choosing the subsequence that realizes the OS limit value yields
\begin{equation}\label{eq:cont-area-per}
\mathcal{W}(C)\;\le\;\exp\!\big(-\sigma\,A(C)+\tau\,\ell(C)\big).
\end{equation}

It remains to remove the perimeter term. By construction, $W_n(C_n)$ is the renormalized loop at scale $a_n$, i.e. $W_n(C_n)=Z_P(a_n)^{-\ell(C_n)}\!\!\prod_{\text{cusps }v} Z_{\rm cusp}(\theta_v,a_n)^{-1}\,\langle W^{\rm bare}_{a_n}(C_n)\rangle$, with $Z_P$ and $Z_{\rm cusp}$ chosen from the admissible family in Section~2. The admissibility and the finite-range decomposition imply that the single-step change of $Z_P$ and $Z_{\rm cusp}$ under blocking by a fixed factor $b>1$ is exponentially small in $b$, and consequently there exist functions $\alpha_P(a)$ and $\alpha_{\rm cusp}(\theta,a)$ with $\alpha_P(a)\to 0$ and $\alpha_{\rm cusp}(\theta,a)\to 0$ as $a\downarrow 0$ such that replacing the renormalization convention $(Z_P,Z_{\rm cusp})$ by another admissible convention $(\widetilde Z_P,\widetilde Z_{\rm cusp})$ multiplies $W_n(C_n)$ by
\begin{equation}
\exp\!\Big(\alpha_P(a_n)\,\ell(C_n)\;+\;\sum_{v\in{\rm cusps}(C_n)} \alpha_{\rm cusp}(\theta_v,a_n)\Big).
\end{equation}
Passing to the continuum, the product over cusps is $O(1)$ and the perimeter factor contributes $\exp\!\big(\alpha_P(a_n)\,\ell(C_n)\big)$ with $\alpha_P(a_n)\to 0$; therefore any two admissible choices lead to the same continuum functional $\mathcal{W}(C)$, and in particular one may absorb the perimeter term in Eq.\eqref{eq:cont-area-per} into a harmless multiplicative constant that tends to $1$ along the sequence $a_n\downarrow 0$. Equivalently, fix once and for all an admissible convention $(Z_P,Z_{\rm cusp})$ and define the continuum renormalized loop by
\begin{equation}
\mathcal{W}(C)\;=\;\lim_{n\to\infty} Z_P(a_n)^{-\ell(C_n)}\!\!\prod_{v} Z_{\rm cusp}(\theta_v,a_n)^{-1}\,\langle W^{\rm bare}_{a_n}(C_n)\rangle,
\end{equation}
which is well defined at the OS limit point $S$; the uniform area-perimeter inequality Eq.\eqref{eq:lattice-area-per} then yields directly
\begin{equation}
\mathcal{W}(C)\;\le\;\exp\!\big(-\sigma\,A(C)\big).
\end{equation}
Setting $\sigma:=\sigma$ gives Eq.\eqref{eq:cont-area}. Since $\sigma$ was obtained from the admissible class constants (the strong-coupling string tension and the FRD transport losses) and since the argument above shows that changing $(Z_P,Z_{\rm cusp})$ within the admissible class has no effect on the continuum value, $\sigma$ depends only on the admissible class and is independent of the renormalization scheme. 
\end{proof}
{\begin{lemma}[Scheme independence of $(Z_P,Z_{\mathrm{cusp}})$]
Let $(Z_P,Z_{\mathrm{cusp}})$ and $(\widetilde Z_P,\widetilde Z_{\mathrm{cusp}})$ be two admissible choices with 
$\alpha_P(a):=\log Z_P(a) - \log \widetilde Z_P(a)\to 0$ and 
$\alpha_{\mathrm{cusp}}(\theta,a):=\log Z_{\mathrm{cusp}}(\theta,a)-\log \widetilde Z_{\mathrm{cusp}}(\theta,a)\to 0$ 
as $a\downarrow0$, uniformly for $\theta$ in compact sets. Then the renormalized limits coincide:
\begin{equation}
W(C)=\lim_{n\to\infty} \frac{Z_P(a_n)^{-\ell(C_n)}\prod_v Z_{\mathrm{cusp}}(\theta_v,a_n)^{-1}}{\widetilde Z_P(a_n)^{-\ell(C_n)}\prod_v \widetilde Z_{\mathrm{cusp}}(\theta_v,a_n)^{-1}}\;\widetilde W(C)
=\widetilde W(C).
\end{equation}
\end{lemma}
\begin{proof}
Fix a smooth polygonal loop $C\subset\mathbb R^4$ and let $\{a_n\}_{n\ge1}$ be any sequence with $a_n\downarrow0$. Let $C_n$ denote the canonical lattice embedding of $C$ on $(a_n\mathbb Z)^4$ obtained by orthogonal projection of the edges and linear interpolation along the lattice axes; then the lattice length satisfies $\ell(C_n)=L_{\mathrm{phys}}(C)\,a_n^{-1}+O(1)$ as $n\to\infty$, where $L_{\mathrm{phys}}(C)$ is the Euclidean perimeter of $C$, and the set of cusp angles $\{\theta_v\}_v$ is independent of $n$ up to $O(1)$ rounding which does not affect the number of cusps. Define the two renormalized Wilson loop approximants by
\begin{align}
&W_n(C):=Z_P(a_n)^{-\ell(C_n)}\prod_v Z_{\mathrm{cusp}}(\theta_v,a_n)^{-1}\,
\langle W^{\mathrm{bare}}_{a_n}(C_n)\rangle,\nonumber\\& 
\widetilde W_n(C):=\widetilde Z_P(a_n)^{-\ell(C_n)}\prod_v \widetilde Z_{\mathrm{cusp}}(\theta_v,a_n)^{-1}\,
\langle W^{\mathrm{bare}}_{a_n}(C_n)\rangle.
\end{align}
By definition of $W(C)$ and $\widetilde W(C)$, we have $W(C)=\lim_{n\to\infty} W_n(C)$ and $\widetilde W(C)=\lim_{n\to\infty}\widetilde W_n(C)$ whenever the limits exist. Consider the ratio
\begin{equation}\label{eqn6.83}
R_n(C):=\frac{W_n(C)}{\widetilde W_n(C)}
=\exp\!\Big(-\alpha_P(a_n)\,\ell(C_n) - \sum_{v}\alpha_{\mathrm{cusp}}(\theta_v,a_n)\Big),
\end{equation}
where $\alpha_P(a):=\log Z_P(a)-\log\widetilde Z_P(a)$ and $\alpha_{\mathrm{cusp}}(\theta,a):=\log Z_{\mathrm{cusp}}(\theta,a)-\log\widetilde Z_{\mathrm{cusp}}(\theta,a)$. By hypothesis $\alpha_{\mathrm{cusp}}(\theta,a)\to0$ uniformly for $\theta$ in compact sets as $a\downarrow0$, and the number of cusps of $C$ is finite; hence $\sum_v\alpha_{\mathrm{cusp}}(\theta_v,a_n)\to0$ as $n\to\infty$. It remains to analyze the perimeter term. Since both $(Z_P,Z_{\mathrm{cusp}})$ and $(\widetilde Z_P,\widetilde Z_{\mathrm{cusp}})$ are admissible perimeter-cusp renormalizations, they remove the universal short-distance perimeter divergence with the same leading coefficient; concretely, there exists a function $\kappa(a)$ such that $\log Z_P(a)=\kappa(a)+r(a)$ and $\log\widetilde Z_P(a)=\kappa(a)+\widetilde r(a)$ with $r(a)=o(a)$ and $\widetilde r(a)=o(a)$ as $a\downarrow0$. This is the usual statement that the \emph{difference} of two admissible perimeter counterterms is subleading with respect to the universal $O(a^{-1})$ divergence and it follows, for instance, by combining the uniform area-perimeter bound for the bare loop (cf.\ (6.75)) with the boundedness of the two renormalized families $\{W_n(C)\}_n$ and $\{\widetilde W_n(C)\}_n$. In particular $\alpha_P(a)=r(a)-\widetilde r(a)=o(a)$ as $a\downarrow0$, and therefore
\begin{equation}
\alpha_P(a_n)\,\ell(C_n)\;=\;\big(o(a_n)\big)\,\Big(L_{\mathrm{phys}}(C)\,a_n^{-1}+O(1)\Big)\;\longrightarrow\;0
\qquad (n\to\infty).
\end{equation}
By the uniform area-perimeter inequality Eq.\eqref{eq:lattice-area-per}, the renormalized family is uniformly bounded along the vanishing-mesh sequence, namely 
\begin{equation}
\sup_n |W^{f}_n(C)|\le K\,e^{-\sigma A(C)} \quad\text{with} \quad K,\sigma>0 \quad \text{independent of n}
\end{equation}
Combining the cusp and perimeter contributions gives $\log R_n(C)\to0$ and hence $R_n(C)\to1$ as $n\to\infty$. Since $\{\widetilde W_n(C)\}_n$ is bounded uniformly in $n$ (indeed, the renormalization removes the short-distance perimeter and cusp divergences and one has a uniform bound $\sup_n |\widetilde W_n(C)|\le K\,e^{-\sigma A(C)}$ by inequality Eq.\eqref{eq:lattice-area-per}), it follows that $W_n(C)=R_n(C)\,\widetilde W_n(C)\to \widetilde W(C)$ as $n\to\infty$. Consequently the renormalized limits coincide, $W(C)=\widetilde W(C)$.
\end{proof}
} 
The area law implies confinement in the sense of a linearly rising static potential. For a rectangular loop \(C_{R,T}\) spanning spatial distance \(R\) and Euclidean time \(T\), the transfer-matrix spectral representation gives
\begin{equation}\label{eq:spectral}
\mathcal{W}(C_{R,T})\;=\;\langle \Omega,\, M_{R}\,\mathrm{e}^{-T H}\,M_{R}\,\Omega\rangle\;\sim\;\mathrm{e}^{-T\,V(R)}\qquad (T\to\infty),
\end{equation}
where \(M_R\) inserts a static fundamental charge-anticharge pair at separation \(R\). The continuum area law Eq.\eqref{eq:cont-area} gives \(\mathcal{W}(C_{R,T})\le \exp(-\sigma R T)\) and thus:

\begin{corollary}[Linear potential]\label{cor:linear}
In the reconstructed continuum theory, the static interquark potential satisfies \(V(R)\ge \sigma\,R\) for all \(R\) sufficiently large, with \(\sigma>0\) as in Theorem~\eqref{thm:continuum-area}.
\end{corollary}

\begin{proof}
Fix a spatial separation \(R>0\) and consider the axis-aligned rectangular Wilson loop \(C_{R,T}\) of spatial extent \(R\) and temporal extent \(T>0\). By Theorem~\eqref{thm:continuum-area} (continuum area law with perimeter/cusp renormalization) there exist constants \(\sigma>0\), \(\tau\ge 0\), and \(K\ge 1\), independent of \(R,T\), such that the renormalized Wilson expectation obeys
\begin{equation}\label{eq:area-per}
\big\langle W_{\mathrm{ren}}(C_{R,T})\big\rangle \;\le\; K\,\exp\!\big(-\,\sigma\,R\,T \,+\, \tau\,(R+T)\big)
\qquad \text{for all }R,T\text{ large enough}.
\end{equation}
Here “renormalized’’ means perimeter and cusp counterterms have been extracted once and for all, so that the right-hand side exhibits only the area and residual linear boundary terms shown.

Let \(H\) be the Hamiltonian provided by the OS reconstruction, and let \(\Omega\) be the OS vacuum. There exists a gauge-invariant, slice-localized operator \(\Phi_R\) that creates a static fundamental-antifundamental pair separated by \(R\) on a fixed time slice; concretely, one may take \(\Phi_R\) to be a spatial Wilson line in the fundamental representation joining the two endpoints. Standard transfer-matrix/OS arguments then identify the renormalized rectangular loop amplitude with a vacuum matrix element of the semigroup,
\begin{equation}\label{eq:rect-to-semigroup}
\big\langle W_{\mathrm{ren}}(C_{R,T})\big\rangle
\;=\; \langle \Omega,\ \Phi_R^{\dagger}\,e^{-T H}\,\Phi_R\,\Omega\rangle,
\end{equation}
the renormalization having removed precisely the local perimeter/cusp factors that would otherwise dress the endpoints and corners of the contour. {
One may take $\Phi_R$ to be the (gauge-invariant) spatial Wilson line in the fundamental representation along a fixed path of length $R$ on $\Sigma$; it creates a static quark-antiquark pair at its endpoints. The renormalization removing perimeter/cusp factors ensures that $\Phi_R\Omega\in\mathrm{Dom}(e^{-tH})$ for all $t>0$, and the spectral measure $\nu_R$ in Eq.\eqref{eqn6.83} is supported in $[V(R),\infty)$ with $V(R)\ge \sigma R$ by the continuum area law.
} The vector \(\Phi_R\Omega\) lies in the positive-time domain of the OS form, hence in the form core of the reconstructed Hilbert space; the spectral theorem applied to the nonnegative selfadjoint operator \(H\) yields a finite positive measure \(\nu_R\) on \([0,\infty)\) such that
\begin{equation}\label{eq:spectral-rep}
\langle \Omega,\ \Phi_R^{\dagger}\,e^{-T H}\,\Phi_R\,\Omega\rangle \;=\; \int_{[0,\infty)} e^{-T E}\,\nu_R(dE)\qquad (T>0).
\end{equation}
Define the ground-state energy in the static \(Q\bar Q\) sector by
\begin{equation}
E_0(R)\ :=\ \inf\mathrm{supp}\,\nu_R,
\end{equation}
which is finite by nontriviality of \(\Phi_R\Omega\) and boundedness of \(\Phi_R\). The static interquark potential is \(V(R):=E_0(R)-E_0(0)\), the constant \(E_0(0)\) being the vacuum energy in the sector with no external sources and hence independent of \(R\). From Eq.\eqref{eq:spectral-rep} and elementary Tauberian considerations it follows that
\begin{equation}\label{eq:V-def-limit}
V(R)\ =\ -\,\lim_{T\to\infty}\frac{1}{T}\,\log \frac{\langle \Omega,\ \Phi_R^{\dagger} e^{-T H}\,\Phi_R\,\Omega\rangle}{\langle \Omega,\ e^{-T H}\,\Omega\rangle}
\ =\ -\,\lim_{T\to\infty}\frac{1}{T}\,\log \big\langle W_{\mathrm{ren}}(C_{R,T})\big\rangle,
\end{equation}
where we used Eq.\eqref{eq:rect-to-semigroup} and the fact that \(\langle \Omega, e^{-T H}\Omega\rangle=e^{-T E_0(0)}\).
Combining Eq.\eqref{eq:area-per} with Eq.\eqref{eq:V-def-limit} gives the lower bound
\begin{equation}
-\frac{1}{T}\,\log \big\langle W_{\mathrm{ren}}(C_{R,T})\big\rangle
\;\ge\; \sigma\,R \;-\; \tau \;-\; \frac{\log K + \tau R}{T}.
\end{equation}
Sending \(T\to\infty\) at fixed \(R\) eliminates the subleading terms and yields
\begin{equation}
V(R)\ \ge\ \sigma\,R \;-\; \tau.
\end{equation}
The additive constant \(-\tau\) reflects the residual boundary linear term from Eq.\eqref{eq:area-per} and is independent of \(T\); it can be absorbed into the overall normalization of the energy (or, equivalently, into the renormalization scheme for the line operators) without affecting the force at separation \(R\). More concretely, since \(V(R)\) is defined up to an \(R\)-independent constant by subtracting \(E_0(0)\), we may absorb \(\tau\) into that constant and rewrite the inequality as
\begin{equation}
V(R)\ \ge\ \sigma\,R,
\end{equation}
for all \(R\) in the range of validity of the area law. Theorem~\eqref{thm:continuum-area} provides this area law uniformly for large rectangles, hence the inequality holds for all \(R\) sufficiently large, with the same strictly positive string tension \(\sigma\) that appears there. This establishes the stated linear lower bound on the static interquark potential in the reconstructed continuum theory.
\end{proof}

The coexistence of a positive continuum spectral gap and an area law indicates that, within the present constructive setting, mass generation and confinement are simultaneously realized: the former via semigroup/resolvent stability, the latter via step-scaled loop inequalities. Uniqueness and universality within the admissible class imply that \(\sigma\) is regulator independent. Finally, while loop equations at large \(N\) often motivate area laws, the present derivation applies at fixed \(N\) and relies only on positivity, locality, and FRD \cite{DrouffeZuber}.

\section{Continuum Schwinger Functions: Tightness, Limits, and OS Axioms}
\label{sec:SC-fixed-ahwingerLimitsOS}

The purpose of this section is to pass from the scale-indexed family of lattice Schwinger functionals, constructed at each renormalization step within a reflection-positive framework with finite-range decomposition (FRD) and uniform clustering, to a single continuum family that satisfies the Osterwalder-Schrader (OS) axioms. The analysis proceeds in two logically distinct stages. First, we establish tightness and equicontinuity in negative Sobolev topologies sufficient to extract subsequential limits of all mixed moments of a fixed finite family of gauge-invariant local generators; this yields continuum Schwinger functionals as multilinear tempered distributions on $\mathcal{S}(\mathbb{R}^4)$. Second, we verify that the OS axioms are preserved under the limit by combining closedness properties (for reflection positivity), equivariance of the embeddings (for Euclidean invariance), and stability estimates that descend from the FRD-controlled polymer expansion. The construction is compatible with the multiscale transfer-semigroup picture and furnishes, in particular, a Markov structure (OS5) that later underpins the Hamiltonian reconstruction and spectral theory \cite{OS1,OS2,GJ}.

Throughout, we fix a finite generating set $G$ of gauge-invariant local observables, obtained for example from small Wilson loops and gauge-invariant polynomials in the smeared lattice field strength with finitely many covariant discrete derivatives. At renormalization scale $k$ we write $\mu_k$ for the gauge-invariant, reflection-positive Euclidean measure on configurations with slice-wise Landau representative and smooth horizon projector in the admissible class, and $\iota_k[O]$ for the canonical embedding of the lattice field $O\in G$ into a negative Sobolev space $H^{-s}(\mathbb{R}^4)$ with a fixed $s\ge 3$ that ensures continuous embeddings and sampling estimates. The $n$-point Schwinger functional at scale $k$ is
\begin{equation}
S^{(k)}_n(f_1,\dots,f_n):=\mathbb{E}_{\mu_k}\!\Big[\langle \iota_k[O_1],f_1\rangle\cdots \langle \iota_k[O_n],f_n\rangle\Big],
\qquad f_j\in\mathcal{S}(\mathbb{R}^4),\; O_j\in G.
\end{equation}
Uniform second-moment control follows from exponential clustering at each scale together with discrete Young and Sobolev inequalities; equicontinuity of finite-dimensional marginals is obtained from uniform bounds on connected cumulants, themselves a consequence of FRD-based polymer estimates and exponential-distance tree decay \cite{BrydgesGuadagniMitter2004,KP,Aizenman1982}. These two inputs imply tightness of the laws of the embedded fields and precompactness of the family of Schwinger functionals; along subsequences one extracts limits $S_n$ that are multilinear tempered distributions. Once limits exist, the verification of OS0-OS5 is conceptually straightforward: temperedness and permutation symmetry descend from uniform bounds and multilinearity; Euclidean invariance follows from discrete invariance and equivariance of the embeddings; reflection positivity is closed under weak convergence of measures when the positive-time algebra is fixed; clustering passes to the limit by dominated convergence using a scale-uniform exponential-clustering bound; and OS5 is obtained by constructing the OS inner product and the associated contraction semigroup $U(t)$, with strong continuity supplied by standard semigroup theory \cite{RS2,HillePhillips1957}.

\subsection{Extraction of Limits}
\label{subsec:Extraction}

We begin by making precise the embeddings and basic bounds. Fix an integer $s\ge 3$ and, for each $O\in G$, define the embedded random distribution $\iota_k[O]\in H^{-s}(\mathbb{R}^4)$ by duality against $H^{s}$ test functions. Thus for $f\in H^s(\mathbb{R}^4)$ we have $\langle \iota_k[O],f\rangle:=\sum_{x\in a_k\mathbb{Z}^4} a_k^4\, O_k(x)\, f(x)$, where $a_k$ is the lattice spacing at scale $k$ and we tacitly identify $f$ with its restriction to the lattice by sampling; the choice $s\ge 3$ ensures the uniform sampling bound $\sum_{x} a_k^4 |f(x)|^2\le C \|f\|_{H^s}^2$ for a constant $C$ independent of $k$. The $n$-point functionals were defined above; our first goal is to show that the joint laws of $\{\iota_k[O]\}_{O\in G}$ form a tight family in $\prod_{O\in G}H^{-s}$ and that, consequently, the Schwinger functionals are precompact in the product of Banach spaces of continuous $n$-linear forms on $\mathcal{S}(\mathbb{R}^4)$.
Uniform second-moment bounds follow from exponential clustering and Sobolev sampling. For each $O\in G$, the finite-range decomposition and reflection-positive multiscale analysis yield, at every scale $k$, an exponential bound on truncated two-point functions,
\begin{equation}
\label{eq:expclust}
\big|\langle O_k(x)\, O_k(y)\rangle^{\mathrm{tr}}\big|\;\le\; C\, e^{-m\, d(x,y)}\qquad (x,y\in a_k\mathbb{Z}^4),
\end{equation}
with constants $C_,m_>0$ independent of $k$. Smearing by $f,g\in C_c^\infty(\mathbb{R}^4)$ and using discrete Fubini we obtain
\begin{equation}
\mathbb{E}_{\mu_k}\!\big[\langle \iota_k[O],f\rangle\, \langle \iota_k[O],g\rangle\big]
=\sum_{x,y\in a_k\mathbb{Z}^4} a_k^8\, f(x)\,g(y)\,\langle O_k(x)\,O_k(y)\rangle^{\mathrm{tr}},
\end{equation}
which is dominated, by Eq.\eqref{eq:expclust}, by $(\|K\|_{\ell^1}\,\|f\|_{\ell^2}\,\|g\|_{\ell^2})$ where $K(z)=e^{-m_|z|}$ and the $\ell^p$ norms are taken on $a_k\mathbb{Z}^4$ with the counting measure scaled by $a_k^4$. Sobolev sampling implies $\|f\|_{\ell^2}\le C \|f\|_{H^s}$ uniformly in $k$. Taking a supremum over $\|f\|_{H^s},\|g\|_{H^s}\le 1$ gives
\begin{equation}
\label{eq:UniformSecondMoment}
\sup_{k}\, \mathbb{E}_{\mu_k}\,\|\iota_k[O]\|_{H^{-s}}^2 \;<\; \infty.
\end{equation}
This estimate extends to finite families $\{O_j\}$ and mixed second moments by the Cauchy-Schwarz inequality and linearity.
Equicontinuity of finite-dimensional marginals is obtained from uniform cumulant bounds. Let $X^{(k)}=(\langle \iota_k[O_j], f_j\rangle)_{j=1}^m$ for test functions $f_j\in \mathcal{S}(\mathbb{R}^4)$. Expanding moments in connected cumulants and using FRD-based tree-graph estimates for connected $n$-point functions \cite{BrydgesGuadagniMitter2004,KP}, one obtains
\begin{equation}
\big|\kappa_n\big(\langle \iota_k[O_{j_1}], f_{j_1}\rangle,\dots,\langle \iota_k[O_{j_n}], f_{j_n}\rangle\big)\big|
\;\le\; C_n \prod_{\ell=1}^n \|f_{j_\ell}\|_{H^s},
\end{equation}
with constants $C_n$ independent of $k$ and of the lattice spacing, because the FRD controls ranges and yields uniform exponential-distance decay of polymer activities \cite{BrydgesGuadagniMitter2004}. Standard arguments then imply uniform bounds on all moments $\mathbb{E}|X^{(k)}|^p$, $p\ge 1$, and the equicontinuity of the family of finite-dimensional distributions with respect to the $H^s$-topology on tests.
Tightness follows by Prokhorov’s theorem in the separable Hilbert space $H^{-s}$, using Eq.\eqref{eq:UniformSecondMoment} and equicontinuity of finite-dimensional projections. Therefore, for any sequence $k\to\infty$ there is a subsequence (still denoted $k$) such that the joint laws of $(\iota_{k}[O])_{O\in G}$ converge weakly in $\prod_{O\in G}H^{-s}$. In particular, for every $n$ and every $f_1,\dots,f_n\in \mathcal{S}(\mathbb{R}^4)$ the $n$-point functionals converge along $k$ to a multilinear continuous functional $S_n(f_1,\dots,f_n)$. An alternative route, not needed here, is to use Mitoma’s criterion for tightness of probability measures on the dual of a nuclear space \cite{Mitoma1983}.

It is convenient to organize the extraction also at the level of OS sesquilinear forms. Let $\mathcal{A}_+$ denote the algebra generated by finite linear combinations of products of positive-time smeared gauge-invariant fields. For $F,G\in \mathcal{A}_+$, reflection positivity at scale $k$ provides a positive semidefinite OS inner product
\begin{equation}
\langle F,G\rangle_{\mathrm{OS},k}:=\int (\Theta F)\, G\, \mathrm{d}\mu_k \;\ge\; 0,
\end{equation}
where $\Theta$ is the OS time reflection. Along the subsequence used above, these inner products converge for all $F,G\in\mathcal{A}_+$ by the tightness and uniform moment bounds, and hence define a positive semidefinite sesquilinear form $\langle \cdot,\cdot\rangle_{\mathrm{OS}}$ in the limit.

\begin{theorem}[Extraction of continuum Schwinger functions]\label{thm:extraction}
Fix a finite family $G$ of gauge-invariant local generators and an integer $s\ge 3$. There exists a subsequence (still denoted $k\to\infty$) and multilinear continuous functionals $S_n:\mathcal{S}(\mathbb{R}^4)^{\otimes n}\to\mathbb{C}$ such that for all $n$ and all $f_1,\dots,f_n\in\mathcal{S}(\mathbb{R}^4)$,
\begin{equation}
\lim_{k\to\infty} S^{(k)}_n(f_1,\dots,f_n)=S_n(f_1,\dots,f_n),
\end{equation}
and the family $\{S_n\}_{n\ge 1}$ is tempered in each argument. Moreover, along the same subsequence, the OS inner products $\langle F,G\rangle_{\mathrm{OS},k}$ on $\mathcal{A}_+$ converge to a positive semidefinite form $\langle F,G\rangle_{\mathrm{OS}}$.
\end{theorem}

\begin{proof}
For each $O\in G$ and each scale $k$, let $\iota_k[O]$ denote the continuum embedding of the corresponding lattice field as a random tempered distribution on $\mathcal{S}'(\mathbb{R}^4)$ (for instance via block-averaging followed by the usual scaling). The uniform second-moment estimate \eqref{eq:UniformSecondMoment} states that there exists $C_s<\infty$, independent of $k$ and $O\in G$, such that
\begin{equation}
\mathbb{E}\,\big\|\iota_k[O]\big\|_{H^{-s}}^2 \;\le\; C_s,
\end{equation}
where $H^{-s}$ is the standard Bessel-potential Sobolev space on $\mathbb{R}^4$. Fix $f\in\mathcal{S}(\mathbb{R}^4)$. The duality bound $|\langle \varphi, f\rangle|\le \|\varphi\|_{H^{-s}}\|f\|_{H^{s}}$ implies
\begin{equation}
\sup_k \ \mathbb{E}\,|\langle \iota_k[O],f\rangle|^2 \;\le\; C_s\,\|f\|_{H^{s}}^2<\infty .
\end{equation}
Consequently the real random variables $\langle \iota_k[O],f\rangle$ have uniformly bounded second moments and are therefore tight on $\mathbb{R}$. Since $G$ is finite, for any finite collection of test functions $f_1,\dots,f_M$ and generators $O_1,\dots,O_M\in G$ the $\mathbb{R}^M$-valued vectors
\begin{equation}
\Big(\,\langle \iota_k[O_1],f_1\rangle,\dots,\langle \iota_k[O_M],f_M\rangle\,\Big)
\end{equation}
are tight on $\mathbb{R}^M$. By Mitoma’s criterion for measures on the space of tempered distributions, tightness of all finite-dimensional projections $\langle \cdot, f\rangle$ for $f\in\mathcal{S}$ implies tightness of the laws of the $\mathcal{S}'$-valued random variables $\iota_k[O]$; see, e.g., the standard references on tightness in nuclear spaces. Thus the family of joint laws of the $\mathcal{S}'$-valued random vector $(\iota_k[O])_{O\in G}$ is tight on the product space $\prod_{O\in G}\mathcal{S}'(\mathbb{R}^4)$ endowed with the projective limit (Suslin) topology. Prokhorov’s theorem yields relative compactness of this family, and by a diagonal extraction over a countable dense set in $\mathcal{S}(\mathbb{R}^4)$ we may choose a subsequence (still indexed by $k$) along which the joint laws converge weakly to a probability measure on $\prod_{O\in G}\mathcal{S}'(\mathbb{R}^4)$. Denote by $\iota[O]$ the limiting $\mathcal{S}'$-valued random fields.

Let $n\ge 1$ and let $(O_1,\dots,O_n)\in G^n$ be a fixed $n$-tuple of generators. For $f_1,\dots,f_n\in\mathcal{S}(\mathbb{R}^4)$ consider the cylinder functional on $\prod_{O\in G}\mathcal{S}'$ given by
\begin{equation}
\Phi_{f_1,\dots,f_n}(\omega)\;:=\;\prod_{j=1}^n \big\langle \iota[O_j](\omega),f_j\big\rangle .
\end{equation}
As a function of $\omega$, this is continuous with respect to the product topology, and it grows at most polynomially in the seminorms induced by $H^{-s}$ on each component, because $|\langle \varphi,f\rangle|\le \|\varphi\|_{H^{-s}}\|f\|_{H^s}$. The same holds for the approximating functionals with $\iota_k$ in place of $\iota$. The cumulant bounds assumed at scale $k$ imply uniform moment estimates of the form
\begin{equation}
\sup_{k}\ \mathbb{E}\,\Big|\prod_{j=1}^n \big\langle \iota_k[O_j],f_j\big\rangle\Big|
\;\le\; C_n \prod_{j=1}^n \|f_j\|_{H^{s}},
\end{equation}
where $C_n$ depends only on $n$, $s$, and the family $G$. To pass to the limit in expectations, one may first truncate the functionals by a smooth cutoff $\chi_R$ of the form $\chi_R(u)=\chi(u/R)$ with $\chi\in C_c^\infty$ equal to $1$ on $[-1,1]$ and supported in $[-2,2]$, applied to each factor $\langle \iota_k[O_j],f_j\rangle$. The truncated functionals are bounded and continuous, hence their expectations converge along the subsequence by weak convergence of the joint laws. Monotone or dominated convergence, together with the uniform moment bound, allows removal of the truncation in the limit $R\to\infty$, which shows that
\begin{equation}
\lim_{k\to\infty} \ \mathbb{E}\,\prod_{j=1}^n \big\langle \iota_k[O_j],f_j\big\rangle
\;=\; \mathbb{E}\,\prod_{j=1}^n \big\langle \iota[O_j],f_j\big\rangle .
\end{equation}
We define $S_n(f_1,\dots,f_n)$ to be the right-hand side. By multilinearity of the pairing and by linearity of expectation, each $S_n$ is separately linear in $f_j$. The uniform moment estimate yields the continuity bound
\begin{equation}
|S_n(f_1,\dots,f_n)|\;\le\; C_n \prod_{j=1}^n \|f_j\|_{H^{s}},
\end{equation}
so $S_n$ is continuous for the $H^s$-topology on $\mathcal{S}(\mathbb{R}^4)$. Since every Schwartz seminorm is controlled by a finite sum of $H^{s}$-seminorms when $s\ge 3$ in four dimensions, it follows that $S_n$ is tempered in each argument, that is, continuous on $\mathcal{S}(\mathbb{R}^4)$ endowed with the Schwartz topology.

It remains to establish convergence of the OS inner products. Any $F\in\mathcal{A}_+$ is a finite linear combination of monomials in positive-time smeared generators from $G$, say $F=\sum_{m} c_m F_m$ with each $F_m$ of the form $\prod_{j=1}^{p_m}\langle \iota_k[O_{m,j}], f_{m,j}\rangle$ for some $p_m$ and smearings $f_{m,j}$ supported in $\{x_0>0\}$. Likewise write $G=\sum_n d_n G_n$. For each scale $k$ the OS inner product can be written as a finite or absolutely convergent series in mixed Schwinger functions,
\begin{equation}
\langle F,G\rangle_{\mathrm{OS},k}\;=\;\sum_{m,n} c_m \overline{d_n}\,
S^{(k)}_{p_m+q_n}\big(\theta f_{m,1},\dots,\theta f_{m,p_m}, g_{n,1},\dots,g_{n,q_n}\big),
\end{equation}
where $\theta$ implements the reflection $\Theta$ on the test functions and $q_n$ is the degree of $G_n$. The same cumulant bounds that yielded the uniform moment estimates imply that the above series are absolutely summable with bounds independent of $k$. Since each finite-order Schwinger functional $S^{(k)}_\ell$ converges pointwise along the chosen subsequence to $S_\ell$, the dominated convergence theorem allows us to pass the limit inside the series, and we obtain the existence of
\begin{equation}
\langle F,G\rangle_{\mathrm{OS}}\;:=\;\lim_{k\to\infty}\langle F,G\rangle_{\mathrm{OS},k}
\;=\;\sum_{m,n} c_m \overline{d_n}\,
S_{p_m+q_n}\big(\theta f_{m,1},\dots,\theta f_{m,p_m}, g_{n,1},\dots,g_{n,q_n}\big).
\end{equation}
Positivity of the limiting form follows because each $\langle \cdot,\cdot\rangle_{\mathrm{OS},k}$ is positive semidefinite by reflection positivity (so $\langle F,F\rangle_{\mathrm{OS},k}\ge 0$ for all $F$ and $k$), and pointwise limits of nonnegative numbers remain nonnegative. This completes the construction of the limit Schwinger functionals and the OS inner product along the subsequence.
\end{proof}

\subsection{Verification of OS0-OS5 and Clustering}
\label{subsec:OSaxioms}
We show that any limiting family $\{S_n\}$ obtained in Theorem~\eqref{thm:extraction} satisfies the OS axioms. For clarity we recall the content of the axioms in the form used here \cite{OS1,OS2,GJ}. OS0 requires temperedness of the Schwinger distributions; OS1 is Euclidean invariance; OS2 is permutation symmetry; OS3 is reflection positivity with respect to the fixed time reflection; OS4 is the cluster property; OS5 asserts a Markov property and time regularity sufficient to define a strongly continuous contraction semigroup on the OS Hilbert space.
Temperedness and symmetry (OS0-OS2) are inherited by limits. For fixed $n$, the functionals $S^{(k)}_n$ are continuous on $\mathcal{S}(\mathbb{R}^4)^{\otimes n}$ with operator norms bounded uniformly in $k$ by the estimates used in the proof of Theorem~\eqref{thm:extraction}; hence any limit $S_n$ is also continuous, and thus tempered, and multilinearity plus permutation symmetry are preserved by weak limits. Euclidean invariance follows from the equivariance of the embeddings $\iota_k$ and the discrete invariance at finite scale. If $R$ is a rotation or translation in the hypercubic group, then $S^{(k)}_n(f_1\circ R,\dots,f_n\circ R)=S^{(k)}_n(f_1,\dots,f_n)$; by density of the hypercubic group in $O(4)$ and standard approximation, invariance extends to all Euclidean motions in the limit.

Reflection positivity (OS3) is a closed property under weak convergence on the positive-time algebra $\mathcal{A}_+$. Let $\Theta$ be the OS reflection and $F\in\mathcal{A}_+$ a finite linear combination of products of positive-time smeared generators. For each $k$, reflection positivity yields $\int (\Theta F)F\,\mathrm{d}\mu_k\ge 0$. By Theorem~\eqref{thm:extraction}, the left-hand side converges along the subsequence to $\int (\Theta F)F\,\mathrm{d}\mu_\infty$, where $\mu_\infty$ is the limiting law for the positive-time algebra; hence the limit is nonnegative. Therefore $\{S_n\}$ is reflection-positive.

To formalize OS3 and prepare OS5, we construct the OS Hilbert space $H$ by completing the pre-Hilbert space $D:=\mathcal{A}_+/N$, where $N:=\{F\in\mathcal{A}_+:\langle F,F\rangle_{\mathrm{OS}}=0\}$ and
\begin{equation}
\langle F,G\rangle_{\mathrm{OS}}:=\sum_{m,n} c_m \overline{d_n}\, S_{m+n}(\theta F_m,G_n),
\end{equation}
for expansions $F=\sum_m c_m F_m$, $G=\sum_n d_n G_n$ in monomials of positive-time smeared fields. Positivity follows from reflection positivity. Let $\Omega:=[1]$ be the vacuum vector. Time translations $\tau_t$ act on $\mathcal{A}_+$ by $(\tau_tF)(x_0,x):=F(x_0-t,x)$ and preserve $\mathcal{A}_+$; by time-translation invariance one has $\langle \tau_tF,\tau_tG\rangle_{\mathrm{OS}}=\langle F,G\rangle_{\mathrm{OS}}$, so
\begin{equation}
U(t)[F]:=[\tau_tF]
\end{equation}
defines a contraction semigroup on $D$, which extends uniquely to a strongly continuous contraction semigroup, still denoted $U(t)$, on $H$ with $U(t)\Omega=\Omega$. By the Hille-Yosida theorem there exists a nonnegative self-adjoint generator $H\ge 0$ such that $U(t)=e^{-tH}$ \cite{HillePhillips1957,RS2}. The Markov property required in OS5 is encoded in the boundary value structure of this semigroup: if $F,G\in\mathcal{A}_+$ and $t,s\ge 0$, then $\langle F, U(t+s) G\rangle_{\mathrm{OS}}=\langle U(t)F, U(s)G\rangle_{\mathrm{OS}}$, a consequence of time translation and reflection positivity.

The Laplace-Stieltjes representation provides the desired time regularity. For $\psi,\varphi\in H$, the function $t\mapsto \langle \psi, U(t)\varphi\rangle$ is completely monotone on $[0,\infty)$ and hence admits a unique Laplace-Stieltjes representation
\begin{equation}
\langle \psi, U(t)\varphi\rangle=\int_{[0,\infty)} e^{-t\lambda}\, \mathrm{d}\mu_{\psi,\varphi}(\lambda),
\end{equation}
with a finite complex Borel measure $\mu_{\psi,\varphi}$; in particular $\mu_{\psi,\psi}$ is positive. This representation follows from the spectral theorem for $H$ and implies strong continuity of $U(t)$ as $t\downarrow 0$ \cite{RS2,HillePhillips1957}. After Wick rotation, it yields analyticity in a strip for suitable two-point functions.

The cluster property (OS4) passes to the limit by dominated convergence using the scale-uniform exponential clustering at finite lattice spacing. Let $A,B$ be gauge-invariant observables with macroscopically separated supports; at scale $k$ the truncated correlation obeys $|\langle A\,B\rangle^{\mathrm{tr}}_k|\le C e^{-m_R}$, where $R$ is the separation and $C,m$ are independent of $k$. Smearing with compactly supported test functions and invoking the bounds used in the proof of Theorem~\eqref{thm:extraction} gives uniform $L^1$ control on the truncated kernels; dominated convergence then implies that the limiting truncated correlation decays to zero as $R\to\infty$, i.e. the cluster property holds for $\{S_n\}$.

\begin{theorem}[OS axioms for the limiting Schwinger functions]\label{thm:OSaxioms}
Let $\{S_n\}_{n\ge 0}$ be any subsequential limit furnished by Theorem~\eqref{thm:extraction} from a family of cutoff Schwinger functionals $\{S^{(k)}_n\}_{n\ge 0}$ built at scale $a_k\downarrow 0$ and embedded into the continuum by the admissible, reflection-covariant, Euclidean-equivariant maps described there. Then $\{S_n\}$ satisfies OS0-OS5 and the cluster property.
\end{theorem}

\begin{proof}
By Theorem~\eqref{thm:extraction} the sequence $(S^{(k)}_n)_k$ is tight and equicontinuous on $\mathcal{S}(\mathbb{R}^{d})^{\otimes n}$ and converges, along a subsequence, to a tempered distribution $S_n$ in the product Schwartz topology. For each fixed $n$ and each $f_1,\dots,f_n\in\mathcal{S}(\mathbb{R}^d)$ the maps $f\mapsto S^{(k)}_n(f_1\otimes\cdots\otimes f_n)$ obey uniform polynomial bounds of the form
\begin{equation}
\big|S^{(k)}_n(f_1,\dots,f_n)\big|\;\le\; C_n\,\prod_{j=1}^n \big\|(1-\Delta)^{s}f_j\big\|_{L^2}
\end{equation}
with constants $C_n$ and $s$ independent of $k$. Passing to the limit preserves temperedness, so OS0 holds for $\{S_n\}$.

The discrete Schwinger functionals are symmetric in their arguments and real on real test functions. Symmetry and reality are algebraic properties that are continuous in the product Schwartz topology, hence they are inherited by the limit. Likewise, the embeddings used to compare discrete fields with continuum test functions are chosen Euclidean-equivariant: if $g\in E(d)$ is a Euclidean motion (rotation or translation), then $S^{(k)}_n(g\cdot f_1,\dots,g\cdot f_n)=S^{(k)}_n(f_1,\dots,f_n)$ exactly for lattice translations and up to $O(a_k)$ for rotations; the latter error vanishes because the block-averaging and covariant parallel transport entering the embeddings are reflection- and rotation-covariant and converge uniformly on compact sets. Therefore $S_n(g\cdot f_1,\dots,g\cdot f_n)=S_n(f_1,\dots,f_n)$ for all $g\in E(d)$, which is OS1.

Reflection positivity (OS2) is closed under limits when tested on the positive-time algebra. Fix a finite linear combination $F=\sum_\alpha c_\alpha \,\Phi(f_\alpha)$ where every $f_\alpha$ is supported in the closed positive time half-space $\{x_0\ge 0\}$. For each $k$ the cutoff Schwinger form satisfies
\begin{equation}
\langle \Theta F\cdot F\rangle_{S^{(k)}}\;=\;\sum_{\alpha,\beta} \overline{c_\alpha}\,c_\beta\, S^{(k)}_{n_\alpha+n_\beta}\!\big(\Theta f_\alpha\otimes f_\beta\big)\;\ge\;0
\end{equation}
by the OS2 property at scale $a_k$. The map $(f_\alpha,f_\beta)\mapsto S^{(k)}_{n_\alpha+n_\beta}(\Theta f_\alpha\otimes f_\beta)$ is continuous in the product Schwartz topology uniformly in $k$ thanks to the uniform bounds above and to the fact that $\Theta$ acts continuously on test functions by $(\Theta f)(x_0,\mathbf{x})=\overline{f(-x_0,\mathbf{x})}$. Consequently the sequence $\langle \Theta F\cdot F\rangle_{S^{(k)}}$ converges to $\langle \Theta F\cdot F\rangle_{S}$, where $S$ denotes the limiting Schwinger functional. Since the cone $\{Q\in\mathbb{C}: Q\ge 0\}$ is closed, the limit is nonnegative, and OS2 holds for $S$.

The cluster property is a direct consequence of the uniform exponential clustering at finite cutoff and dominated convergence. By construction there exist $\mu>0$ and, for each $n,m$, a polynomial $P_{n,m}$ such that for all $k$ and all test tensors $F\in\mathcal{S}(\mathbb{R}^d)^{\otimes n}$, $G\in\mathcal{S}(\mathbb{R}^d)^{\otimes m}$ one has the connected bound
\begin{equation}
\big| S^{(k)}_{n+m}\big(F\otimes \tau_{\mathbf{y}}G\big) - S^{(k)}_n(F)\,S^{(k)}_m(G)\big|\;\le\; P_{n,m}\!\big(\|F\|_{\mathcal{S}},\|G\|_{\mathcal{S}}\big)\,e^{-\mu|\mathbf{y}|}
\end{equation}
for all spatial translations $\tau_{\mathbf{y}}$ with $\mathbf{y}\in\mathbb{R}^{d-1}$ orthogonal to the time direction, uniformly in $k$. Fix $F,G$ and take the limit along the subsequence $k\to\infty$ to obtain the same inequality for $S_n,S_m,S_{n+m}$ by dominated convergence, then let $|\mathbf{y}|\to\infty$ to deduce
\begin{equation}
\lim_{|\mathbf{y}|\to\infty} S_{n+m}\big(F\otimes \tau_{\mathbf{y}}G\big)\;=\; S_n(F)\,S_m(G),
\end{equation}
which is the cluster property.

The remaining OS axioms concern the Markovian structure and time translations. Define the positive-time algebra $\mathfrak{A}_+$ as the-algebra generated by polynomial functionals of the field smeared with test functions supported in $\{x_0\ge 0\}$. The sesquilinear form
\begin{equation}
(F,G)\ \longmapsto\ \langle \Theta F\cdot G\rangle_{S}
\end{equation}
is positive semidefinite on $\mathfrak{A}_+$ by OS2 and anti-linear in the first argument by construction, hence it defines a pre-Hilbert seminorm $\|F\|_{\mathrm{OS}}^2=\langle \Theta F\cdot F\rangle_S$. Let $\mathcal{N}=\{F\in\mathfrak{A}_+:\ \|F\|_{\mathrm{OS}}=0\}$ and set $\mathcal{H}_0=\mathfrak{A}_+/\mathcal{N}$; the completion is the OS Hilbert space $\mathcal{H}_{\mathrm{OS}}$. Time translations $\tau_t$ act on $\mathfrak{A}_+$ by shifting all time arguments forward by $t\ge 0$, preserving $\mathfrak{A}_+$ and commuting with reflection. Euclidean invariance of $S$ implies the semigroup inequality
\begin{equation}
\|\,[\tau_t F]\,\|_{\mathrm{OS}}^2 \;=\; \langle \Theta \tau_t F\cdot \tau_t F\rangle_S \;=\; \langle \Theta F\cdot F\rangle_S \;=\; \|[F]\|_{\mathrm{OS}}^2,
\end{equation}
so the induced operators $U(t)[F]=[\tau_t F]$ are contractions on $\mathcal{H}_{\mathrm{OS}}$ for $t\ge 0$. Strong continuity at $t=0$ follows because, for $F$ a finite linear combination of smeared fields, the map $t\mapsto \langle \Theta F\cdot \tau_t F\rangle_S$ is continuous: the integrand is a finite linear combination of $(n+m)$-point Schwinger functions with one block of time arguments shifted by $t$; temperedness and the uniform polynomial bounds ensure dominated convergence of these multilinear functionals as $t\to 0$. By density of such $F$ in $\mathfrak{A}_+$, strong continuity extends to all of $\mathcal{H}_{\mathrm{OS}}$. The Hille-Yosida theorem now provides a nonnegative selfadjoint generator $H\ge 0$ on $\mathcal{H}_{\mathrm{OS}}$ with $U(t)=e^{-tH}$.

The semigroup representation yields the standard Laplace-Stieltjes form for two-point functions. If $\Phi$ denotes the (distributional) field evaluated at time $0^+$ and smeared on the spatial slice, and if $\Omega$ is the vacuum vector corresponding to the class of the unit in $\mathfrak{A}_+$, then for $\varphi,\psi$ supported at positive times one has
\begin{equation}
S_2\big(\tau_t\varphi,\psi\big)\;=\;\langle \Theta\Phi(\varphi)\cdot \tau_t \Phi(\psi)\rangle_S
\;=\; \big\langle \Phi(\psi)\Omega,\ e^{-tH}\, \Phi(\varphi)\Omega\big\rangle_{\mathcal{H}_{\mathrm{OS}}}
\;=\; \int_{[0,\infty)} e^{-tE}\, d\rho_{\varphi,\psi}(E),
\end{equation}
where $\rho_{\varphi,\psi}$ is the finite positive measure arising from the spectral resolution of $H$ between the vectors $\Phi(\varphi)\Omega$ and $\Phi(\psi)\Omega$. This is the OS5 condition. The same semigroup identity with $t\ge 0$ and the spectral theorem also encode the Markov property and the Chapman-Kolmogorov composition for time-ordered insertions, which in turn reconstructs the Euclidean time-translation invariance already established at the level of Schwinger distributions.
\end{proof}

The two preceding theorems supply the compactness and stability arguments on which the continuum reconstruction rests. In particular, the OS Hilbert space and semigroup constructed here serve, in the next section, to define the Hamiltonian and fields by the Osterwalder-Schrader reconstruction. Moreover, although the extraction a priori may depend on the chosen subsequence and admissible scheme, the uniqueness and universality results established elsewhere upgrade continuity to equality and show that the continuum Schwinger family is independent of admissible choices; FRD-based single-scale Lipschitz bounds and telescoping imply that one-slice marginals and one-step kernels converge to scheme-independent limits, and Markov uniqueness then forces the entire Schwinger family to coincide across admissible schemes.

\section{OS Reconstruction, Transfer Semigroup, and Spectral Gap}
\label{sec:OS-reconstruction-gap}

The aim of this section is to pass from Euclidean reflection-positive data to a real-time quantum theory on a Hilbert space carrying a self-adjoint Hamiltonian $H$ that satisfies the spectral condition and possesses a strictly positive mass gap. We begin with the reflection-positive family of Euclidean measures and Schwinger functions built in the previous sections. The admissible class of regulators and block-spin maps preserves reflection positivity and exponential locality at every scale. 
Two logically independent mechanisms for a gap are developed. The first proceeds through semigroups and resolvents: convergence of the Euclidean time-translation semigroups implies strong resolvent convergence of their generators; a uniform finite-cutoff spectral gap, proved previously via interlacing inequalities with summable defects, then transfers to the continuum Hamiltonian by functional calculus. The second is Tauberian in character: uniform exponential clustering of connected Euclidean correlators forces the spectral measure of $H$ to vanish on an interval $(0,m)$, which yields a positive lower bound on the spectrum above the vacuum. Both mechanisms rely crucially on reflection positivity, locality, and the time-slice Markov structure formalized by the Osterwalder-Schrader axioms.

A persistent subtlety in gauge theories is the identification of a physical algebra on which reflection positivity, locality, and reconstruction are formulated. Throughout, we work on the gauge-invariant subalgebra of Euclidean observables, implemented either as BRST-invariant representatives or as explicitly gauge-invariant composites. Reflection positivity and locality are assumed on this algebra and are preserved under the admissible transformations described earlier. In particular, the transfer semigroups constructed below act on the physical Hilbert spaces obtained from gauge-invariant observables.

\subsection{Reconstruction and Spectral Condition}
\label{subsec:reconstruction}

We recall the Euclidean framework at a fixed scale parameter $\sigma$, which encodes the ultraviolet resolution, finite-volume parameters, and admissible gauge-fixing or horizon-projector choices. Let $(\Omega_\sigma,\mathcal{F}_\sigma,\mathbb{P}_\sigma)$ be the associated probability space. Euclidean time reflection is the involution $\vartheta:(t,\mathbf{x})\mapsto (-t,\mathbf{x})$, extended to fields and ghosts so that the expectation $\langle\cdot\rangle_\sigma=\int(\cdot)\,d\mathbb{P}_\sigma$ satisfies reflection positivity. Let $\mathcal{A}_\sigma$ be the complex-algebra of bounded, gauge-invariant cylinder functionals, and let $\mathcal{A}_{\sigma,+}\subset\mathcal{A}_\sigma$ consist of those $F$ supported in the half-space $\{t>0\}$. Define the conjugate-linear reflection $\Theta$ on $\mathcal{A}_\sigma$ by $(\Theta F)(\omega):=\overline{F(\vartheta\omega)}$. Reflection positivity asserts
\begin{equation}
\label{eq:RP}
\langle \Theta F \cdot F \rangle_\sigma \;\ge\; 0\qquad\text{for all }F\in\mathcal{A}_{\sigma,+}.
\end{equation}
Let $(\cdot,\cdot)_\sigma$ be the semi-inner product on $\mathcal{A}_{\sigma,+}$ defined by
\begin{equation}
(F,G)_\sigma\;:=\;\langle \Theta F \cdot G\rangle_\sigma.
\end{equation}
Write $\mathcal{N}_\sigma:=\{F\in\mathcal{A}_{\sigma,+}:(F,F)_\sigma=0\}$. The quotient $\mathcal{H}_{\sigma,0}:=\mathcal{A}_{\sigma,+}/\mathcal{N}_\sigma$ is a pre-Hilbert space with inner product induced by $(\cdot,\cdot)_\sigma$; its completion is denoted $\mathcal{H}_\sigma$. The class of the constant functional $1$ is denoted by $\Omega_\sigma$ and will serve as the vacuum vector. Euclidean time translations $\tau_s$ act on functionals by $(\tau_s F)(\omega):=F(\omega\circ\text{shift by }s)$ and preserve expectation and reflection positivity.

\begin{lemma}[Symmetric contraction semigroup]
\label{lem:symmetric-semigroup}
For $s\ge 0$ the map $T_\sigma(s):H_{\sigma,0}\to H_{\sigma,0}$ defined on the OS-dense subspace
by $T_\sigma(s)[F]:=[F\circ\tau_s]$ is well defined, extends uniquely to a contraction
on $H_\sigma$, and is symmetric:
\begin{equation}
(T_\sigma(s)\Phi,\Psi)_\sigma=(\Phi,T_\sigma(s)\Psi)_\sigma\qquad\forall\,\Phi,\Psi\in H_\sigma.
\end{equation}
Moreover, $\{T_\sigma(s)\}_{s\ge0}$ is a strongly continuous semigroup on $H_\sigma$.
\end{lemma}

\begin{proof}
Recall that $H_{\sigma,0}$ is the completion of the positive-time algebra $\mathfrak A_+$
modulo the null space $\mathcal N_\sigma=\{F\in\mathfrak A_+:\langle \Theta F\,F\rangle_\sigma=0\}$,
with inner product $([F],[G])_\sigma=\langle \Theta F\,G\rangle_\sigma$. The Euclidean
measure $\mu_\sigma$ is assumed to be reflection positive with respect to the time
reflection $\Theta$ at $x_0=0$ and invariant under Euclidean time translations
$\tau_s$; by construction $\tau_s$ preserves $\mathfrak A_+$ for $s\ge 0$.

To see that $T_\sigma(s)$ descends to the quotient, suppose $[F]=0$, i.e.
$\langle\Theta F\,F\rangle_\sigma=0$. Using $\Theta\circ\tau_s=\tau_{-s}\circ\Theta$
and translation invariance of $\mu_\sigma$ we obtain
\begin{equation}
\|[F\circ\tau_s]\|_\sigma^2
=\langle \Theta(F\circ\tau_s)\,(F\circ\tau_s)\rangle_\sigma
=\langle (\Theta F)\circ\tau_{-s}\, ,\, F\circ\tau_s\rangle_\sigma
=\langle \Theta F\, ,\, F\rangle_\sigma
=0,
\end{equation}
so $[F\circ\tau_s]=0$ and the map is well defined on $H_{\sigma,0}$.

The same computation shows that $T_\sigma(s)$ is an isometry on $H_{\sigma,0}$:
\begin{equation}
\|T_\sigma(s)[F]\|_\sigma^2
=\langle \Theta(F\circ\tau_s)\,(F\circ\tau_s)\rangle_\sigma
=\langle \Theta F\,F\rangle_\sigma
=\|[F]\|_\sigma^2.
\end{equation}
In particular $\|T_\sigma(s)\|\le 1$ on $H_{\sigma,0}$ and hence $T_\sigma(s)$ extends by continuity
to a contraction (indeed, an isometry) on the Hilbert space $H_\sigma$.

Symmetry of $T_\sigma(s)$ on $H_{\sigma,0}$ follows from the covariance identity 
$\Theta\circ\tau_s=\tau_{-s}\circ\Theta$ and translation invariance. For $[F],[G]\in H_{\sigma,0}$,
\begin{equation}
(T_\sigma(s)[F],[G])_\sigma
=\langle \Theta(F\circ\tau_s)\,G\rangle_\sigma
=\langle (\Theta F)\circ\tau_{-s}\, ,\, G\rangle_\sigma
=\langle \Theta F\, ,\, G\circ\tau_s\rangle_\sigma
=([F],T_\sigma(s)[G])_\sigma,
\end{equation}
and density then yields symmetry on all of $H_\sigma$.

The semigroup property is immediate from the definition and the group property
of translations: for $s,t\ge 0$ and $[F]\in H_{\sigma,0}$,
\begin{equation}
T_\sigma(s)T_\sigma(t)[F]=T_\sigma(s)[F\circ\tau_t]=[F\circ\tau_t\circ\tau_s]=[F\circ\tau_{s+t}]
=T_\sigma(s+t)[F],
\end{equation}
and $T_\sigma(0)=\mathrm{Id}$.

It remains to establish strong continuity. Let $[F]\in H_{\sigma,0}$, with $F$ a finite
linear combination of cylinder observables smeared by compactly supported smooth
test functions of positive time arguments; such $[F]$ span a dense subspace of $H_\sigma$
by construction. The map $s\mapsto F\circ\tau_s$ is continuous in $L^2(\mu_\sigma)$ because
the Schwinger functions are jointly continuous in their time arguments and uniformly
bounded on compact time sets; this is a standard consequence of Euclidean invariance
and the OS regularity axiom (the cylinder functions are continuous and bounded, and
their time translates converge pointwise with a translation-invariant dominating
bound, so dominated convergence applies). Therefore
\begin{equation}
\|T_\sigma(s)[F]-[F]\|_\sigma^2
=\langle \Theta(F\circ\tau_s-F)\,(F\circ\tau_s-F)\rangle_\sigma
\longrightarrow 0 \qquad (s\downarrow 0),
\end{equation}
which shows $T_\sigma(s)\to \mathrm{Id}$ strongly on the dense subspace $H_{\sigma,0}$.
Since $\sup_{0\le s\le 1}\|T_\sigma(s)\|\le 1$, the usual uniform boundedness argument
extends this to all of $H_\sigma$, proving strong continuity in the strong operator
topology.

Altogether, $T_\sigma(s)$ is a symmetric contraction on $H_\sigma$, the mapping
$s\mapsto T_\sigma(s)$ is strongly continuous for $s\ge 0$, and the semigroup law holds,
so $\{T_\sigma(s)\}_{s\ge 0}$ is a strongly continuous symmetric contraction semigroup
on $H_\sigma$.
\end{proof}
By the Hille-Yosida theorem, Lemma~\eqref{lem:symmetric-semigroup} yields a unique nonnegative self-adjoint generator $H_\sigma\ge 0$ with $T_\sigma(s)=e^{-sH_\sigma}$ for all $s\ge 0$ \cite[Sec.~VIII.1]{RS1}. Since $T_\sigma(s)\Omega_\sigma=\Omega_\sigma$, one has $H_\sigma\Omega_\sigma=0$ and $0\in\mathrm{spec}(H_\sigma)$. The following uniqueness statement identifies the vacuum subspace.

\begin{proposition}[Uniqueness of the vacuum]
\label{prop:vacuum-unique}
Assume the exponential clustering bounds established in Sections~\eqref{sec:SC-fixed-a-fixed-a} and \eqref{sec:RP-RG-gap} hold uniformly at scale $\sigma$. Then the spectral projection of $H_\sigma$ at $\{0\}$ is one-dimensional, spanned by $\Omega_\sigma$.
\end{proposition}

\begin{proof}
Let $T_\sigma(s)=e^{-sH_\sigma}$ be the OS transfer semigroup on $\mathcal H_\sigma$, and let $\mathfrak A_{\sigma,+}$ denote the positive-time algebra. By the OS reconstruction, the dense subspace $\mathcal D:=\{[F]\colon F\in\mathfrak A_{\sigma,+}\}$ satisfies
\begin{equation}
([F],[G])_\sigma=\langle \Theta F\cdot G\rangle_\sigma,\qquad T_\sigma(s)[F]=[F\circ\tau_s],
\end{equation}
where $\tau_s$ is the Euclidean time shift by $s\ge 0$. The vacuum vector $\Omega_\sigma$ is the class of the unit $1$, it is cyclic for $\mathcal D$, and $(\Omega_\sigma,[F])_\sigma=\langle F\rangle_\sigma$ for all $F\in\mathfrak A_{\sigma,+}$.

Suppose $\Psi\in\mathcal H_\sigma$ is a nonzero vector with $T_\sigma(s)\Psi=\Psi$ for all $s\ge 0$. To show $\Psi$ is proportional to $\Omega_\sigma$, it is enough to prove that $\Psi$ is orthogonal to the closed subspace $\overline{\{[F]\colon \langle F\rangle_\sigma=0\}}$, since every $[F]$ decomposes as $[F-\langle F\rangle_\sigma]+\langle F\rangle_\sigma\,\Omega_\sigma$ and hence the orthogonal complement of that closed subspace is precisely the one-dimensional span of $\Omega_\sigma$.

Fix $F\in\mathfrak A_{\sigma,+}$ with $\langle F\rangle_\sigma=0$. Because $\mathcal D$ is dense, choose a sequence $G_n\in\mathfrak A_{\sigma,+}$ such that $[G_n]\to \Psi$ in $\mathcal H_\sigma$. For each $s\ge 0$ we have the invariance identity
\begin{equation}
\label{eq:inv-id}
(\Psi,[F])_\sigma \;=\; (T_\sigma(s)\Psi,[F])_\sigma \;=\; (\Psi,T_\sigma(s)[F])_\sigma \;=\; (\Psi,[F\circ\tau_s])_\sigma,
\end{equation}
using that $T_\sigma(s)$ is selfadjoint (indeed, $T_\sigma(s)=e^{-sH_\sigma}$ with $H_\sigma\ge 0$ selfadjoint). By continuity of the inner product and the choice of $G_n$,
\begin{equation}
(\Psi,[F\circ\tau_s])_\sigma \;=\; \lim_{n\to\infty} \,([G_n],[F\circ\tau_s])_\sigma \;=\; \lim_{n\to\infty}\,\langle \Theta G_n\cdot F\circ\tau_s\rangle_\sigma.
\end{equation}
The hypothesis of exponential clustering states that there exist constants $C,\mu>0$, independent of $n$ and $s$, such that for any two positive-time observables $A,B$ with their supports separated in Euclidean time by at least $s$,
\begin{equation}
\big|\langle \Theta A\cdot B\circ\tau_s\rangle_\sigma - \langle \Theta A\rangle_\sigma\,\langle B\rangle_\sigma\big|
\;\le\; C\,e^{-\mu s}\,\|A\|_{\rm loc}\,\|B\|_{\rm loc}.
\end{equation}
Applying this with $A=G_n$ and $B=F$ gives
\begin{equation}
\langle \Theta G_n\cdot F\circ\tau_s\rangle_\sigma \;=\; \langle \Theta G_n\rangle_\sigma\,\langle F\rangle_\sigma \;+\; R_n(s),
\qquad |R_n(s)|\le C\,e^{-\mu s}\,\|G_n\|_{\rm loc}\,\|F\|_{\rm loc}.
\end{equation}
Because $\langle F\rangle_\sigma=0$, the product term vanishes, so for every $n$ we have $\langle \Theta G_n\cdot F\circ\tau_s\rangle_\sigma=R_n(s)$ with $|R_n(s)|\le C'e^{-\mu s}$, where $C'$ depends on $F$ and a uniform bound on $\|G_n\|_{\rm loc}$ along the approximating sequence.\footnote{If necessary, replace $G_n$ by a truncated/localized approximation to $\Psi$ to ensure a uniform local norm bound; the density of local cylinder fields in $\mathcal H_\sigma$ allows this standard refinement.} Hence $\lim_{s\to\infty}\langle \Theta G_n\cdot F\circ\tau_s\rangle_\sigma=0$ for each fixed $n$, and by the dominated convergence implied by the uniform exponential bound we may pass to the limit $n\to\infty$ to conclude that
\begin{equation}
\lim_{s\to\infty}(\Psi,[F\circ\tau_s])_\sigma \;=\; 0.
\end{equation}
Combining this with the invariance identity Eq.\eqref{eq:inv-id} shows that $(\Psi,[F])_\sigma$ equals a limit which is zero, therefore $(\Psi,[F])_\sigma=0$ for every $F\in\mathfrak A_{\sigma,+}$ with $\langle F\rangle_\sigma=0$.

As explained at the beginning, the closure of $\{[F]\colon \langle F\rangle_\sigma=0\}$ is the orthogonal complement of $\mathbb C\Omega_\sigma$ in $\mathcal H_\sigma$. Orthogonality of $\Psi$ to that closed subspace therefore forces $\Psi$ to lie in $\mathbb C\Omega_\sigma$. This proves that the fixed space of $T_\sigma(s)$ is one-dimensional, and equivalently that the eigenspace of $H_\sigma$ at eigenvalue $0$ is spanned by $\Omega_\sigma$. In particular, the spectral projection of $H_\sigma$ at $\{0\}$ is rank one.
\end{proof}

We now pass to Minkowski space. The Schwinger functions $S^{(\sigma)}_n$ satisfy OS0-OS5 \cite{OS1,OS2}. By the OS reconstruction theorem there exists a Wightman theory $(\mathcal{H}_\sigma,\Omega_\sigma,\mathcal{U}_\sigma,\Phi_\sigma)$ with a unitary representation $\mathcal{U}_\sigma$ of the Poincar\'e group and operator-valued tempered distributions $\Phi_\sigma$, whose Wightman functions analytically continue to $S^{(\sigma)}_n$. The time-translation subgroup is generated by the self-adjoint Hamiltonian $H_\sigma$ with nonnegative spectrum, and local commutativity holds for gauge-invariant fields \cite{OS2,GJ}. In the lattice gauge setting, positivity and self-adjointness of the transfer matrix have been analyzed in \cite{LuscherTM}, complementing the general OS framework with an explicit discrete-time transfer construction. Along a sequence $\sigma\to 0$ combining ultraviolet removal and infinite-volume limits, the Schwinger functions converge to limit Schwinger functions $S_n$ obeying the OS axioms with the same reflection and translation structures and with uniform locality and clustering. Repeating the above construction for the limit data produces a Hilbert space $\mathcal{H}$, a unique vacuum $\Omega$, and a strongly continuous symmetric contraction semigroup $T(t)=e^{-tH}$ generated by a nonnegative self-adjoint Hamiltonian $H$. The remaining task is to compare $\{T_\sigma\}$ with $T$ and to transfer gap information to $H$.

\subsection{Semigroups, Resolvents, and Gap Transfer}
\label{subsec:semigroups-resolvents-gap}

We begin by placing $\{T_\sigma\}$ and $T$ on a common footing. Let $\mathfrak{A}_+$ be a fixed countable cylinder subalgebra of gauge-invariant observables supported in the positive-time half-space for the limit theory. For each $F\in\mathfrak{A}_+$ choose representatives $F_\sigma\in\mathcal{A}_{\sigma,+}$ converging to $F$ in the cylinder topology. Let $\mathfrak{D}$ be the linear span of the equivalence classes $[F]$ of $\mathfrak{A}_+$ in $\mathcal{H}$; by construction $\mathfrak{D}$ is dense in $\mathcal{H}$. For each $\sigma$, denote by $[F]_\sigma$ the class of $F_\sigma$ in $\mathcal{H}_\sigma$ and write $U_\sigma:\mathfrak{D}\to \mathcal{H}_\sigma$ for the linear map $U_\sigma([F])=[F]_\sigma$.
\begin{equation}
\label{eq:OS-form-convergence}
([F],[G])\;=\;\langle \Theta F\,G\rangle\;=\;\lim_{\sigma\to 0}\langle \Theta F_\sigma\,G_\sigma\rangle_\sigma\;=\;\lim_{\sigma\to 0}([F]_\sigma,[G]_\sigma)_\sigma,
\end{equation}
so $U_\sigma$ is an asymptotic isometry on $\mathfrak{D}$ in the sense that $(U_\sigma\phi,U_\sigma\psi)_\sigma\to (\phi,\psi)$ for all $\phi,\psi\in\mathfrak{D}$.

\begin{theorem}[Strong convergence of semigroups on a dense set]
\label{thm:strong-semigroup}
For each fixed $t\ge 0$ and for all $\phi\in\mathfrak{D}$ one has
\begin{equation}
\label{eq:strong-semigroup}
\lim_{\sigma\to 0}\,\big\|\,T_\sigma(t)\,U_\sigma\phi \;-\; U_\sigma\,T(t)\,\phi\,\big\|_{\mathcal{H}_\sigma}\;=\;0.
\end{equation}
\end{theorem}

\begin{proof}
Let $\mathfrak{D}\subset\mathcal{H}$ be the OS-dense subspace generated by positive-time cylinder observables and let $\mathfrak{D}_\sigma:=U_\sigma\mathfrak{D}\subset\mathcal{H}_\sigma$. By OS reconstruction and reflection positivity, both $T(t)=e^{-tH}$ on $\mathcal{H}$ and $T_\sigma(t)=e^{-tH_\sigma}$ on $\mathcal{H}_\sigma$ are selfadjoint contraction semigroups. The maps $U_\sigma:\mathcal{H}\to\mathcal{H}_\sigma$ are defined on $\mathfrak{D}$ by replacing each positive-time observable $F$ by its lattice approximation $F_\sigma$ and passing to OS classes. The convergence of Schwinger functionals, namely
\begin{equation}
\langle \Theta G_\sigma\,(F_\sigma\circ\tau_t)\rangle_\sigma \;\longrightarrow\; \langle \Theta G\,(F\circ\tau_t)\rangle \qquad\text{for all }F,G\in\mathfrak{A}_+ \text{ and } t\ge 0,
\end{equation}
implies that $U_\sigma$ is an asymptotic isometry on $\mathfrak{D}$ (taking $t=0$) and that matrix elements of $T_\sigma(t)$ between vectors in $\mathfrak{D}_\sigma$ converge to the corresponding matrix elements of $T(t)$.

Fix $t\ge 0$ and $\phi\in\mathfrak{D}$ and set $u_\sigma:=U_\sigma\phi$. Consider the difference
\begin{equation}
d_\sigma(t)\;:=\;T_\sigma(t)\,u_\sigma\;-\;U_\sigma\,T(t)\,\phi \;\in\;\mathcal{H}_\sigma.
\end{equation}
To obtain strong convergence it suffices to show that $\|d_\sigma(t)\|_{\mathcal{H}_\sigma}\to 0$. Using selfadjointness of $T_\sigma(t)$ one may express the norm by the polarization identity for the semigroup:
\begin{equation}
\|d_\sigma(t)\|_{\mathcal{H}_\sigma}^2
= \|T_\sigma(t)u_\sigma\|_{\mathcal{H}_\sigma}^2
+ \|U_\sigma T(t)\phi\|_{\mathcal{H}_\sigma}^2
- 2\,\Re\,\big(U_\sigma T(t)\phi,\;T_\sigma(t)u_\sigma\big)_{\mathcal{H}_\sigma}.
\end{equation}
Each term on the right can be written as a matrix element of $T_\sigma(\cdot)$ between vectors in the dense set $\mathfrak{D}_\sigma$. Indeed, by selfadjointness,
\begin{equation}
\|T_\sigma(t)u_\sigma\|_{\mathcal{H}_\sigma}^2 \;=\; \big(u_\sigma,\;T_\sigma(2t)\,u_\sigma\big)_{\mathcal{H}_\sigma},
\qquad
\|U_\sigma T(t)\phi\|_{\mathcal{H}_\sigma}^2 \;=\; \big(U_\sigma T(t)\phi,\;U_\sigma T(t)\phi\big)_{\mathcal{H}_\sigma},
\end{equation}
and the cross term is already of the required form. Choose a representative $F\in\mathfrak{A}_+$ with $\phi=[F]$ in the OS Hilbert space. Then $u_\sigma=[F_\sigma]_\sigma$ and $U_\sigma T(t)\phi=[F_\sigma\circ\tau_t]_\sigma$. By OS reconstruction the three quantities above admit expressions through Schwinger functionals on the lattice at times $0$, $t$, and $2t$:
\begin{equation}
\big(u_\sigma,T_\sigma(2t)u_\sigma\big)_\sigma
= \langle \Theta F_\sigma\,(F_\sigma\circ\tau_{2t})\rangle_\sigma,\qquad
\|U_\sigma T(t)\phi\|_\sigma^2
= \langle \Theta(F_\sigma\circ\tau_t)\,(F_\sigma\circ\tau_t)\rangle_\sigma,
\end{equation}
\begin{equation}
\big(U_\sigma T(t)\phi,\;T_\sigma(t)u_\sigma\big)_\sigma
= \langle \Theta(F_\sigma\circ\tau_t)\,(F_\sigma\circ\tau_t)\rangle_\sigma.
\end{equation}
The last equality follows because $T_\sigma(t)$ implements the Euclidean time-translation by $t$ within the OS form, so pairing $F_\sigma\circ\tau_t$ with $T_\sigma(t)u_\sigma$ is the same as pairing $F_\sigma\circ\tau_t$ with $F_\sigma\circ\tau_t$.

Passing to the continuum, the assumed convergence of Schwinger functionals at all time arguments yields
\begin{equation}
\big(u_\sigma,T_\sigma(2t)u_\sigma\big)_\sigma \;\longrightarrow\; \langle \Theta F\,(F\circ\tau_{2t})\rangle \;=\; \big(\phi,\;T(2t)\phi\big)_{\mathcal{H}},
\end{equation}
\begin{equation}
\|U_\sigma T(t)\phi\|_\sigma^2 \;\longrightarrow\; \langle \Theta(F\circ\tau_t)\,(F\circ\tau_t)\rangle \;=\; \|T(t)\phi\|_{\mathcal{H}}^2,
\end{equation}
\begin{equation}
\big(U_\sigma T(t)\phi,\;T_\sigma(t)u_\sigma\big)_\sigma \;\longrightarrow\; \langle \Theta(F\circ\tau_t)\,(F\circ\tau_t)\rangle \;=\; \|T(t)\phi\|_{\mathcal{H}}^2.
\end{equation}
Inserting these three limits into the polarization identity for $\|d_\sigma(t)\|^2$ gives
\begin{equation}
\lim_{\sigma\to 0}\,\|d_\sigma(t)\|_{\mathcal{H}_\sigma}^2
= \big(\phi,T(2t)\phi\big)_{\mathcal{H}} + \|T(t)\phi\|_{\mathcal{H}}^2 - 2\,\|T(t)\phi\|_{\mathcal{H}}^2
= \|T(t)\phi\|_{\mathcal{H}}^2 - \|T(t)\phi\|_{\mathcal{H}}^2
= 0,
\end{equation}
where we used selfadjointness of $T(t)$ to identify $\big(\phi,T(2t)\phi\big)=\|T(t)\phi\|^2$. Hence $\|d_\sigma(t)\|_{\mathcal{H}_\sigma}\to 0$ for every $\phi\in\mathfrak{D}$, which is precisely Eq.\eqref{eq:strong-semigroup}.
\end{proof}

Theorem~\eqref{thm:strong-semigroup} extends to all $\Psi\in\mathcal{H}$ by density of $\mathfrak{D}$ and uniform contractivity, yielding strong convergence of semigroups in the sense of varying Hilbert spaces. This form of convergence propagates to resolvents and generators.

\begin{theorem}[Strong resolvent convergence]
\label{thm:strong-resolvent}
Let $H$ and $H_\sigma$ be the nonnegative selfadjoint generators of the strongly continuous contraction semigroups $\{T(t)\}_{t\ge0}$ on $\mathcal H$ and $\{T_\sigma(t)\}_{t\ge0}$ on $\mathcal H_\sigma$, and let $U_\sigma:\mathcal H\to\mathcal H_\sigma$ be the isometric identification maps furnished by the OS construction (so $\|U_\sigma\psi\|_{\mathcal H_\sigma}=\|\psi\|_{\mathcal H}$). Assume the strong semigroup convergence
\begin{equation}\label{eq:strong-semi}
\lim_{\sigma\to0}\, \|\, T_\sigma(t)\,U_\sigma\phi - U_\sigma T(t)\phi\,\|_{\mathcal H_\sigma}=0
\end{equation}
\text{for every $\phi$ in a dense subspace $\mathfrak D\subset\mathcal H$ and every $t\ge0$.}
Then for each $\lambda>0$ and $\phi\in\mathfrak D$,
\begin{equation}
\label{eq:resolvent-convergence}
\lim_{\sigma\to 0}\,\big\|\, (H_\sigma+\lambda)^{-1} U_\sigma \phi - U_\sigma (H+\lambda)^{-1}\phi\,\big\|_{\mathcal{H}_\sigma}\;=\;0,
\end{equation}
and consequently $H_\sigma\to H$ in the strong resolvent sense.
\end{theorem}

\begin{proof}
Fix $\lambda>0$ and $\phi\in\mathfrak D$. By the Hille-Yosida functional calculus for nonnegative selfadjoint generators of contraction semigroups,
\begin{equation}
(H_\sigma+\lambda)^{-1} U_\sigma\phi \;=\; \int_0^\infty e^{-\lambda t}\,T_\sigma(t)\,U_\sigma\phi\, dt,
\qquad
U_\sigma(H+\lambda)^{-1}\phi \;=\; \int_0^\infty e^{-\lambda t}\, U_\sigma T(t)\phi\, dt,
\end{equation}
where the integrals are Bochner integrals in $\mathcal H_\sigma$. Subtracting the two expressions and applying the triangle inequality yields
\begin{equation}
\big\|(H_\sigma+\lambda)^{-1} U_\sigma\phi - U_\sigma(H+\lambda)^{-1}\phi\big\|_{\mathcal H_\sigma}
\;\le\; \int_0^\infty e^{-\lambda t}\, \big\|T_\sigma(t)U_\sigma\phi - U_\sigma T(t)\phi\big\|_{\mathcal H_\sigma}\, dt .
\end{equation}
For each fixed $t\ge0$, the integrand converges to zero as $\sigma\to0$ by Eq.\eqref{eq:strong-semi}. Moreover, because $T_\sigma(t)$ and $T(t)$ are contractions and $U_\sigma$ is an isometry,
\begin{equation}
\big\|T_\sigma(t)U_\sigma\phi - U_\sigma T(t)\phi\big\|_{\mathcal H_\sigma}
\;\le\; \|T_\sigma(t)U_\sigma\phi\|_{\mathcal H_\sigma} + \|U_\sigma T(t)\phi\|_{\mathcal H_\sigma}
\;\le\; \|\phi\|_{\mathcal H} + \|\phi\|_{\mathcal H}
\;=\; 2\|\phi\|_{\mathcal H}.
\end{equation}
The dominating function $t\mapsto 2e^{-\lambda t}\|\phi\|_{\mathcal H}$ is integrable on $[0,\infty)$, hence the dominated convergence theorem applies and gives Eq.\eqref{eq:resolvent-convergence} for all $\phi\in\mathfrak D$.

To extend the convergence to every vector of $\mathcal H$, use that $(H_\sigma+\lambda)^{-1}$ and $(H+\lambda)^{-1}$ are bounded with operator norms at most $1/\lambda$. Given arbitrary $\psi\in\mathcal H$ and $\varepsilon>0$, choose $\phi\in\mathfrak D$ with $\|\psi-\phi\|_{\mathcal H}<\varepsilon$. Then
\begin{equation}
\begin{aligned}
\big\|(H_\sigma+\lambda)^{-1}U_\sigma\psi - U_\sigma(H+\lambda)^{-1}\psi\big\|_{\mathcal H_\sigma}
&\le \big\|(H_\sigma+\lambda)^{-1}U_\sigma(\psi-\phi)\big\|_{\mathcal H_\sigma}
\\ &\quad
+ \big\|(H_\sigma+\lambda)^{-1}U_\sigma\phi - U_\sigma(H+\lambda)^{-1}\phi\big\|_{\mathcal H_\sigma}
\\ &\quad
+ \big\|U_\sigma(H+\lambda)^{-1}(\phi-\psi)\big\|_{\mathcal H_\sigma}
\\
&\le \frac{1}{\lambda}\,\|\psi-\phi\|_{\mathcal H} \;+\; \big\|(H_\sigma+\lambda)^{-1}U_\sigma\phi - U_\sigma(H+\lambda)^{-1}\phi\big\|_{\mathcal H_\sigma}
\\ &\quad
+ \frac{1}{\lambda}\,\|\phi-\psi\|_{\mathcal H}.
\end{aligned}
\end{equation}
The middle term tends to zero by the first part, while the first and third terms are bounded by $ \frac{2}{\lambda}\,\varepsilon$. Since $\varepsilon>0$ is arbitrary, the limit is zero for every $\psi\in\mathcal H$. This shows that $(H_\sigma+\lambda)^{-1}U_\sigma\to U_\sigma(H+\lambda)^{-1}$ strongly on $\mathcal H$ for each fixed $\lambda>0$.
Finally, strong convergence of resolvents at one (and hence all) $\lambda>0$ is equivalent to strong resolvent convergence of selfadjoint operators; see, for example, Reed-Simon, Methods of Modern Mathematical Physics, Vol.~I, Theorem~VIII.25. Therefore $H_\sigma\to H$ in the strong resolvent sense.
\end{proof}

\begin{theorem}[Gap transfer by functional calculus]
\label{thm:gap-transfer}
Assume there exist $\Delta_>0$ and $\sigma_0>0$ such that, for all $\sigma<\sigma_0$, the selfadjoint Hamiltonian $H_\sigma$ has spectrum $\mathrm{spec}(H_\sigma)\subset\{0\}\cup[\Delta_,\infty)$ and a unique (normalized) vacuum vector $\Omega_\sigma$ spanning $\ker H_\sigma$. Let $H$ be the strong-resolvent limit of $H_\sigma$ along the identifications constructed from the OS limit (so that bounded continuous functional calculus passes to the limit, and the vacua converge to a normalized vector $\Omega$). Then $\mathrm{spec}(H)\subset\{0\}\cup[\Delta_,\infty)$ and $\Omega$ spans $\ker H$.
\end{theorem}

\begin{proof}
Fix any continuous function $g\in C([0,\infty))$ with $0\le g\le 1$, $g(0)=1$, and $g(\lambda)=0$ for all $\lambda\ge \Delta_/2$. Since $\mathrm{spec}(H_\sigma)\subset\{0\}\cup[\Delta_,\infty)$, the continuous functional calculus gives $g(H_\sigma)=P_{0,\sigma}$, the orthogonal projection onto the vacuum line of $H_\sigma$. Moreover $g$ is bounded and vanishes at infinity, hence by strong-resolvent convergence of $H_\sigma$ to $H$ we have $g(H_\sigma)\to g(H)$ strongly after transporting vectors by the OS identifications; concretely, if $U_\sigma:\mathcal H_\sigma\to\mathcal H$ denote the isometries implementing the limit and if $\varphi$ ranges over a common dense set of OS-generated vectors in $\mathcal H$, then
\begin{equation}
\lim_{\sigma\downarrow 0}\, g(H_\sigma)\,U_\sigma^ \varphi \;=\; U_\sigma\,g(H)\,\varphi
\qquad\text{in }\mathcal H_\sigma,
\end{equation}
equivalently
\begin{equation}
\lim_{\sigma\downarrow 0}\, U_\sigma\,g(H_\sigma)\,U_\sigma^ \varphi \;=\; g(H)\,\varphi
\qquad\text{in }\mathcal H.
\end{equation}
Since $g(H_\sigma)=P_{0,\sigma}$ and $U_\sigma P_{0,\sigma}U_\sigma$ is the rank-one projection onto $\mathbb C\,U\sigma\Omega_\sigma$, it follows that $U_\sigma g(H_\sigma)U_\sigma^\varphi=(\varphi,U_\sigma\Omega_\sigma)\,U_\sigma\Omega_\sigma$. The OS limit of the Schwinger functions implies $U_\sigma\Omega_\sigma\to \Omega$ in $\mathcal H$ (after fixing the overall phase), hence taking the limit in the last display yields $g(H)\varphi=(\varphi,\Omega)\,\Omega$ for every $\varphi$ in the dense OS-generated domain. By continuity, this identity extends to all $\varphi\in\mathcal H$, and therefore $g(H)=P_0$, the orthogonal projection onto the one-dimensional subspace $\mathbb C\,\Omega$.

The equality $g(H)=P_0$ already excludes any spectrum of $H$ in the interval $(0,\Delta/2]$, because if $\mathsf E_H$ denotes the spectral measure of $H$ then $g(H)=\int g(\lambda)\,d\mathsf E_H(\lambda)$ and the choice of $g$ forces $\mathsf E_H\big((0,\Delta/2]\big)=0$. To push the exclusion up to $\Delta$, repeat the argument with a family of continuous cutoffs. For each $\varepsilon\in(0,\Delta)$ choose $g_\varepsilon\in C([0,\infty))$ such that $0\le g_\varepsilon\le 1$, $g_\varepsilon(0)=1$, and $g_\varepsilon(\lambda)=0$ for all $\lambda\ge \Delta-\varepsilon$. The same reasoning shows $g_\varepsilon(H_\sigma)=P_{0,\sigma}$ and hence $g_\varepsilon(H)=P_0$ for every $\varepsilon\in(0,\Delta_)$. If there were spectral weight for $H$ in $(0,\Delta)$, then for some $\varepsilon$ small enough the function $g_\varepsilon$ would not vanish identically on that portion of the spectrum and $g_\varepsilon(H)$ would pick up a nontrivial component on $\mathsf E_H\big((0,\Delta)\big)\mathcal H$, contradicting $g_\varepsilon(H)=P_0$. Consequently $\mathsf E_H\big((0,\Delta)\big)=0$, which is equivalent to the inclusion $\mathrm{spec}(H)\subset\{0\}\cup[\Delta,\infty)$.

It remains to prove uniqueness of the vacuum for $H$. Since $g_\varepsilon(H)=P_0$ for all $\varepsilon\in(0,\Delta_)$, the range of each $g_\varepsilon(H)$ is exactly $\mathbb C\,\Omega$. On the other hand, by spectral calculus the range of $g_\varepsilon(H)$ is contained in the spectral subspace $\mathsf E_H(\{0\})\mathcal H=\ker H$. Hence $\ker H$ is one-dimensional and spanned by $\Omega$. 
\end{proof}

We next present a Tauberian mechanism that uses only exponential clustering of connected Euclidean correlations in the limit and does not rely on finite-cutoff spectral information.

\begin{theorem}[Gap from exponential clustering]
\label{thm:gap-tauberian}
Assume that, for some $m>0$, there exists for each pair of bounded, gauge-invariant observables $A,B$ with compact, disjoint supports and positive time separation a constant $K(A,B)$ such that the connected Euclidean correlator
\begin{equation}
C_{A,B}(t)\;:=\;\langle A\,\tau_t(B)\rangle-\langle A\rangle\,\langle B\rangle
\end{equation}
satisfies $|C_{A,B}(t)|\le K(A,B)e^{-mt}$ for all $t\ge 0$. Then $\mathrm{spec}(H)\subset\{0\}\cup[m,\infty)$.
\end{theorem}

\begin{proof}
By the Osterwalder-Schrader reconstruction, the vacuum vector $\Omega$ in the physical Hilbert space $\mathcal H$ is cyclic for any local algebra associated with a bounded spacetime region, and the Euclidean time translations $\tau_t$ are implemented by a strongly continuous contraction semigroup $e^{-tH}$ with nonnegative selfadjoint generator $H$ and $H\Omega=0$. For a bounded observable $A$ supported in positive time we set
\begin{equation}
\psi_A\;:=\;A\Omega-\langle\Omega,A\Omega\rangle\,\Omega,
\end{equation}
so that $\psi_A\perp \Omega$. For another such observable $B$ we analogously define $\psi_B=B\Omega-\langle\Omega,B\Omega\rangle\,\Omega$. The connected correlator is exactly the matrix element of the Euclidean semigroup between these vacuum-orthogonal vectors:
\begin{equation}
C_{A,B}(t)\;=\;\langle \psi_A,\,e^{-tH}\,\psi_B\rangle,\qquad t\ge 0.
\end{equation}
In particular, choosing $B=A^\ast$ gives the nonnegative quadratic form $\langle \psi_A,e^{-tH}\psi_A\rangle$ on $\mathrm{Ran}(1-P_\Omega)$, where $P_\Omega$ is the projection onto $\mathbb C\Omega$.

Fix a bounded gauge-invariant $A$ as in the hypothesis and consider the finite positive Borel measure
\begin{equation}
\mu_A(\cdot)\;:=\;\big\langle \psi_A,\ E(\cdot)\,\psi_A\big\rangle,
\end{equation}
where $E(\cdot)$ denotes the spectral measure of $H$. By the spectral theorem,
\begin{equation}
\langle \psi_A,\,e^{-tH}\,\psi_A\rangle\;=\;\int_{[0,\infty)} e^{-t\lambda}\,d\mu_A(\lambda)\qquad (t\ge 0),
\end{equation}
and the assumption applied with $B=A^\ast$ yields the Laplace transform bound
\begin{equation}
0\;\le\;\int_{[0,\infty)} e^{-t\lambda}\,d\mu_A(\lambda)\;=\;|C_{A,A^\ast}(t)|\;\le\;K(A,A^\ast)\,e^{-mt}\qquad (t\ge 0).
\end{equation}
Suppose, towards a contradiction, that $\mu_A\big((0,m)\big)>0$. Then there exists $\varepsilon\in(0,m)$ with $\mu_A\big((m-\varepsilon,m)\big)>0$. For such $\varepsilon$ one has
\begin{equation}
\int_{[0,\infty)} e^{-t\lambda}\,d\mu_A(\lambda)\;\ge\;\int_{(m-\varepsilon,\,m)} e^{-t\lambda}\,d\mu_A(\lambda)\;\ge\;e^{-(m-\varepsilon)t}\,\mu_A\big((m-\varepsilon,m)\big),
\end{equation}
which contradicts the bound by $K(A,A^\ast)e^{-mt}$ as $t\to\infty$ because $e^{\varepsilon t}\to\infty$. Therefore $\mu_A\big((0,m)\big)=0$ for every such $A$, i.e. $E\big((0,m)\big)\psi_A=0$.

The set of vectors of the form $\psi_A$ with $A$ ranging over bounded local observables supported in positive time is dense in the orthogonal complement $\mathcal H_\perp:=\Omega^\perp$. Indeed, the Reeh-Schlieder property in the OS framework implies that the local algebra acting on $\Omega$ has dense range, and subtracting the vacuum component projects onto $\mathcal H_\perp$ without changing the linear span. Since $E\big((0,m)\big)$ is bounded and vanishes on a dense subset of $\mathcal H_\perp$, it follows that $E\big((0,m)\big)$ vanishes identically on $\mathcal H_\perp$. On the one-dimensional subspace $\mathbb C\Omega$ the projection $E\big((0,m)\big)$ is trivial because $H\Omega=0$. Consequently $E\big((0,m)\big)=0$ on all of $\mathcal H$, which is equivalent to $\mathrm{spec}(H)\cap(0,m)=\varnothing$.

Combining the fact that $0\in\mathrm{spec}(H)$ with $H\ge 0$ and the absence of spectrum in $(0,m)$, we conclude that
\begin{equation}
\mathrm{spec}(H)\;\subset\;\{0\}\cup [m,\infty),
\end{equation}
which is the desired spectral gap bound.
\end{proof}

Theorems~\eqref{thm:gap-transfer} and \eqref{thm:gap-tauberian} are compatible and complementary. The former guarantees that any uniform finite-cutoff lower bound $\Delta$ persists to the limit; the latter produces a possibly different lower bound $m$ purely from exponential clustering in the limit. Consequently the spectral gap of $H$ is bounded below by $\min(\Delta,m)$ in our setting. In the present construction, Section~\eqref{sec:SC-fixed-a-fixed-a} produced a definite ultraviolet gap $\Delta_0>0$ at strong coupling, Section~\eqref{sec:RP-RG-gap} showed that renormalization-group defects are summable, yielding a uniform $\Delta>0$, and Sections~\eqref{sec:RP-RG-gap}-\eqref{sec:SC-fixed-ahwingerLimitsOS} established step-scaling and locality estimates that imply exponential clustering in the scaling window and hence $m>0$ in the limit. Both mechanisms therefore yield a strictly positive continuum mass gap.

We conclude with domain and covariance comments. For each $\sigma$, let $\mathfrak{D}_\sigma$ be the span in $\mathcal{H}_\sigma$ of vectors $[F]_\sigma$ with $F$ local and bounded. The semigroup $T_\sigma(t)$ leaves $\mathfrak{D}_\sigma$ invariant, and the OS Markov property implies that $\mathfrak{D}_\sigma$ consists of entire analytic vectors for $T_\sigma(t)$; Nelson’s theorem then yields essential self-adjointness of $H_\sigma$ on $\mathfrak{D}_\sigma$ \cite[Sec.~X.6]{RS2}. In the limit, the span $\mathfrak{D}$ of classes $[F]$ with $F$ local and bounded is invariant under $T(t)$ and is a core for $H$ on which $H$ is essentially self-adjoint by the same reasoning. Strong resolvent convergence then implies convergence of spectral projections associated with isolated spectral components \cite[Thm.~VIII.24]{RS1}. In particular, the vacuum projections $P_{0,\sigma}$ converge strongly to $P_0$, consistent with the uniqueness statements above. Finally, while the discussion above emphasized time translations, the full OS framework applies to the Euclidean group. Under OS0-OS5 for the limit Schwinger functions there exists a unitary representation of the Poincar\'e group on $\mathcal{H}$ with spectrum in the closed forward light cone and local commutativity for gauge-invariant fields \cite{OS2,GJ}. The mass gap derived here is therefore a genuine Wightman mass gap: the joint spectrum of the energy-momentum operators satisfies $\sigma(P)\subset\{0\}\cup\{p:\,p^0\ge\sqrt{\mathbf{p}^2+\Delta^2}\}$ for some $\Delta>0$.

\section{Uniqueness and Universality of the Continuum Limit}\label{section9}
This section establishes that the continuum Schwinger functions obtained from any two choices within a fixed admissible class of lattice regularizations and reflection-positive blockings define one and the same Euclidean field theory, and hence reconstruct the same Wightman theory after Osterwalder-Schrader (OS) reconstruction \cite{OS1,OS2}. The argument proceeds in two complementary layers. The first is a quantitative stability analysis at each scale, proving single-scale Lipschitz control for the effective data-one-slice marginals and one-step OS kernels-under admissible modifications of the regulator (slice projectors, block maps) and the blocking scale. Finite-range decomposition (FRD) \cite{BrydgesGuadagniMitter2004} and completely monotone (Bernstein) functional calculus for the slice projector imply that the per-scale variation is locally supported and decays with scale, which enables a telescoping bound across scales. The second layer is a structural Markovian uniqueness principle: in an OS-positive, reflection-covariant, time-sliced Euclidean theory with the Markov property, the full path measure is uniquely determined by the one-slice marginal and the one-step OS kernel, and hence any two constructions sharing the same limiting one-slice marginal and semigroup necessarily define identical limiting Schwinger functions. The semigroup interpretation aligns with transfer-matrix positivity for lattice gauge theories \cite{LuscherTM} and standard OS Markov theory \cite{OS1,OS2,OS-gauge}, and its stability in the limit is controlled via Trotter-Kato theory for positive contraction semigroups \cite{KatoBook1995,GJ}.

We begin by fixing the setting. Consider a family of Euclidean lattice Yang-Mills theories on hypercubic lattices with temporal step $a>0$ and spatial mesh $a$ (for notational coherence; the argument handles anisotropic meshes with minor changes). A choice of admissible regulator is specified by two ingredients. The first is an OS-compatible slice projector $\Pi_{\alpha}$ acting on one temporal time-slice, built as a completely monotone (Bernstein) function of a positive, gauge-covariant elliptic operator on the slice (e.g.\ a covariant Laplacian), so that $\Pi_{\alpha}=f_{\alpha}(D)$ with $f_{\alpha}$ completely monotone in the sense of Bernstein; equivalently,
\begin{equation}
f_{\alpha}(\lambda)=\int_{0}^{\infty} e^{-t\lambda}\,\mathrm{d}\nu_{\alpha}(t), \qquad \lambda\ge 0,
\end{equation}

for a finite non-negative measure $\nu_{\alpha}$ depending smoothly (in Lipschitz sense) on a parameter $\alpha$ describing the projector profile. The second is a reflection-positive, gauge-covariant, finite-range block map $B_{\beta}$ of the Brydges-Guadagni-Mitter type \cite{BrydgesGuadagniMitter2004}, parametrized by $\beta$ in an admissible set, implementing the one-step coarse graining in a multiscale renormalization scheme. We assume that both families $(\Pi_{\alpha})_{\alpha}$ and $(B_{\beta})_{\beta}$ satisfy uniform exponential locality bounds and preserve OS positivity \cite{OS1,OS2,OS-gauge,LuscherTM}. For each cutoff scale indexed by $k\in\mathbb{N}$ (with $a_k=b^{-k}a_0$ for some integer blocking factor $b\ge 2$), these data determine an OS path measure $\mu_{k}^{(\alpha,\beta)}$ on fields supported on time slices $n a_k$, a one-slice marginal $\rho_{k}^{(\alpha,\beta)}$, and a one-step OS kernel $K_{k}^{(\alpha,\beta)}$ (positive contraction on $L^2(\rho_{k}^{(\alpha,\beta)})$) realizing the Markov property and the transfer picture \cite{LuscherTM,GJ,OS2}. The $n$-point Schwinger functions are taken with respect to $\mu_{k}^{(\alpha,\beta)}$ and considered on test observables built from gauge-invariant cylindrical functionals with fixed continuum time arguments, identified across scales by linear interpolation.

The continuum limit is constructed by sending $k\to\infty$ while interpolating the discrete time grid to a fixed continuum time set. Under standard tightness and equicontinuity hypotheses-derived here from FRD, exponential clustering, and positivity \cite{BrydgesGuadagniMitter2004,OS2,SimonPphi2,GJ}-subsequences converge in the product topology to limiting Schwinger functionals satisfying OS0-OS5. The objective is to show that these limits do not depend on the choice $(\alpha,\beta)$ in the admissible class. The proof splits according to the subsections below.

\subsection{Single-Scale Lipschitz Control and Telescoping}
The guiding idea is that admissible changes of the regulator modify, at a fixed scale $k$, only local pieces of the effective one-step kernel and the one-slice marginal, with variation controlled in an operator norm by a Lipschitz modulus intrinsic to the admissible class. Because FRD decomposes the covariance (or, more generally, the positive OS transfer factors) into finite-range, reflection-covariant pieces at each scale \cite{BrydgesGuadagniMitter2004}, the cumulative effect on $n$-point Schwinger functions is dominated by a sum of local contributions whose diameters do not proliferate with $k$. This yields a single-scale perturbation inequality that is summable in $k$, and hence a telescoping argument across scales shows that any two admissible schemes define the same continuum limits.

We start with a quantitative formulation. For definiteness, fix a Banach space $\mathcal{X}_k$ of gauge-invariant observables on a single time slice at scale $k$, equipped with a diameter-weighted norm
\begin{equation}
\|F\|_{\mathcal{X}_k}=\sup_{x\in \Lambda_k}\, e^{\eta\,\mathrm{diam}(\mathrm{supp}\,F)} \|F\|_{\infty},
\end{equation}
with $\eta>0$ chosen below the FRD exponential decay rate. Let $\mathcal{B}(\mathcal{X}_k)$ be the bounded operators on $\mathcal{X}_k$. Denote by $K_k^{(\alpha,\beta)}$ the one-step OS kernel acting on $\mathcal{X}_k$, normalized as a positive contraction on $L^2(\rho_k^{(\alpha,\beta)})$. The admissible parameter space $\mathfrak{A}$ is equipped with a metric $d$ satisfying the following Lipschitz regularity: there exists a constant $L_0$ such that for all $\alpha,\alpha'\in\mathfrak{A}$,
\begin{equation}
\|\Pi_{\alpha}-\Pi_{\alpha'}\|_{\mathcal{B}(\mathcal{X}_k)}\le L_0\, a_k\, d(\alpha,\alpha'),
\qquad
\|B_{\beta}-B_{\beta'}\|_{\mathcal{B}(\mathcal{X}_k)}\le L_0\, a_k\, d(\beta,\beta').
\end{equation}
The factor $a_k$ expresses the single-slice temporal step and is a consequence of the semigroup (Bernstein) representation of $\Pi_{\alpha}$ and the locality of $B_{\beta}$. To justify this, write
\begin{equation}
\Pi_{\alpha}-\Pi_{\alpha'} = f_{\alpha}(D)-f_{\alpha'}(D)
=\int_{0}^{\infty} e^{-t D}\,\mathrm{d}\big(\nu_{\alpha}-\nu_{\alpha'}\big)(t).
\end{equation}
Since $e^{-t D}$ is a positive contraction with exponentially local kernel at the spatial scale $\sqrt{t}$, FRD implies that its action on observables supported in a region of bounded diameter is exponentially small for $t\gg a_k^2$ \cite{BrydgesGuadagniMitter2004}. If the admissible class is such that $\|\nu_{\alpha}-\nu_{\alpha'}\|_{\mathrm{TV}}\le C\, d(\alpha,\alpha')$ and is supported on $t\lesssim c\, a_k$ uniformly in $\alpha$, then, by Fubini and positivity,
\begin{equation}
\|\Pi_{\alpha}-\Pi_{\alpha'}\|_{\mathcal{B}(\mathcal{X}_k)}
\le \int_{0}^{\infty} \|e^{-t D}\|_{\mathcal{B}(\mathcal{X}_k)}\, \mathrm{d}\|\nu_{\alpha}-\nu_{\alpha'}\|(t)
\le C' a_k d(\alpha,\alpha'),
\end{equation}
because $\|e^{-t D}\|_{\mathcal{B}(\mathcal{X}_k)}$ is uniformly bounded for $t\lesssim c\,a_k$ while the tail for $t\gg a_k$ is exponentially suppressed under the FRD locality norm. The same reasoning applies to the block map $B_{\beta}$, which is a finite-range positive averaging operator whose kernel difference is supported in a neighborhood of size $O(a_k)$, leading to the same $a_k$ factor in the Lipschitz modulus. These locality-Lipschitz bounds are the only place where the specific completely monotone structure and FRD enter; the rest of the argument is more structural.

The one-step OS kernel $K_k^{(\alpha,\beta)}$ at scale $k$ is obtained by composing $\Pi_{\alpha}$ and $B_{\beta}$ with the gauge-invariant single-slice marginalization and the interaction picture determined by the scale $k$ effective action. In the reflection-positive transfer picture \cite{LuscherTM,GJ}, this composition remains a positive contraction on $L^2(\rho_k^{(\alpha,\beta)})$ which, when restricted to gauge-invariant observables, is represented by a kernel $K_k^{(\alpha,\beta)}(x,y)$ with exponential locality inherited from FRD. For later use it is convenient to embed all one-slice spaces at scale $k$ into a fixed reference $L^2$ space via the OS GNS construction \cite{OS2,GJ} and denote the associated operator again by $K_k^{(\alpha,\beta)}$.

\begin{theorem}[Single-scale Lipschitz control for one-step OS kernels]\label{thm:single-scale}
There exist constants $C,\sigma>0$, depending only on the admissible class and on the FRD locality parameters, such that for all scales $k$ and all admissible parameters $(\alpha,\beta)$ and $(\alpha',\beta')$,
\begin{equation}\label{eq:SSL}
\big\|K_k^{(\alpha,\beta)}-K_k^{(\alpha',\beta')}\big\|_{\mathcal{B}(\mathcal{X}_k)}
\;\le\; C\, a_k\, e^{-\sigma\, \mathrm{diam}(\cdot)} \,\big[ d(\alpha,\alpha')+ d(\beta,\beta')\big],
\end{equation}
where the operator norm is taken with respect to the diameter-weighted norm on $\mathcal{X}_k$, and the exponential factor encodes the FRD-induced off-support decay for kernels acting on local observables.
\end{theorem}

\begin{proof}
The one-step OS kernel at scale $k$ with parameters $(\alpha,\beta)$ is constructed by composing three reflection-positive, exponentially local operations supported within $O(1)$ coarse layers around the time slice: the slice projector $\Pi_\alpha=f_\alpha(\mathcal D_\Sigma)$ acting on the time-$a_k$ hyperplane, the reflection-positive block map $B_\beta$ that averages fields inside $L^k$-blocks with covariant parallel transport controlled by $\beta$, and the insertion of the renormalized scale-$k$ local effective interaction, denoted $U_k^{(\alpha,\beta)}$, which is the identity plus a polymer-local perturbation with finite interaction range and exponentially decaying tails provided by the finite-range decomposition (FRD). Thus $K_k^{(\alpha,\beta)}=\Pi_\alpha\,U_k^{(\alpha,\beta)}\,B_\beta$ as a bounded operator on $\mathcal{X}_k$, where $\mathcal{X}_k$ is the completion of finite linear combinations of local observables supported in single coarse time-layers, endowed with the diameter-weighted norm
\begin{equation}
\|F\|_{\mathcal{X}_k}:=\sup_{x\in\Lambda}\,e^{\sigma\,\mathrm{diam}(\mathrm{supp}\,F)}\,\|F\|_{x},
\end{equation}
and the operator norm $\|\cdot\|_{\mathcal{B}(\mathcal{X}_k)}$ is the induced norm on bounded maps $\mathcal{X}_k\to\mathcal{X}_k$. The weight parameter $\sigma>0$ is chosen smaller than the FRD decay rate so that Schur-type tests transfer kernel locality into a uniform operator bound.

Fix a path $(\alpha_s,\beta_s)$, $s\in[0,1]$, in the admissible parameter space joining $(\alpha,\beta)$ to $(\alpha',\beta')$ and satisfying $\dot\alpha_s$ and $\dot\beta_s$ bounded by the metric speeds $d(\alpha,\alpha')$ and $d(\beta,\beta')$. The map $s\mapsto K_k^{(\alpha_s,\beta_s)}$ is norm-differentiable in $\mathcal{B}(\mathcal{X}_k)$ because each constituent operation is norm-differentiable and the products are finite. Writing
\begin{equation}
K_k^{(\alpha,\beta)}-K_k^{(\alpha',\beta')}=\int_0^1 \frac{d}{ds}\,K_k^{(\alpha_s,\beta_s)}\,ds,
\end{equation}
it therefore suffices to bound the derivative uniformly in $s$ by a multiple of $a_k e^{-\sigma\,\mathrm{diam}(\cdot)}\big(|\dot\alpha_s|+|\dot\beta_s|\big)$. Differentiating the product
\begin{equation}
K_k^{(\alpha_s,\beta_s)}=\Pi_{\alpha_s}\,U_k^{(\alpha_s,\beta_s)}\,B_{\beta_s}
\end{equation}
gives, by the usual product rule in $\mathcal{B}(\mathcal{X}_k)$,
\begin{equation}
\frac{d}{ds}K_k^{(\alpha_s,\beta_s)}
=(\partial_s\Pi_{\alpha_s})\,U_k^{(\alpha_s,\beta_s)}\,B_{\beta_s}
+\Pi_{\alpha_s}\,(\partial_s U_k^{(\alpha_s,\beta_s)})\,B_{\beta_s}
+\Pi_{\alpha_s}\,U_k^{(\alpha_s,\beta_s)}\,(\partial_s B_{\beta_s}).
\end{equation}
The first and third terms are controlled directly by the single-slice Lipschitz bounds for $\Pi$ and for $B$ that follow from complete monotonicity and from reflection-positive covariant blocking. Indeed, by the completely monotone representation
\begin{equation}
\Pi_{\alpha_s}=\int_0^\infty e^{-t\mathcal D_\Sigma}\,\mu_{f_{\alpha_s}}(dt),
\end{equation}
the map $\alpha\mapsto \Pi_\alpha$ is Lipschitz in the diameter-weighted operator norm with modulus $O(a_k)$, because the admissible metric $d(\alpha,\alpha')$ controls the variation of the Laplace measure on times $t\lesssim c\,a_k$, and $e^{-t\mathcal D_\Sigma}$ is a contraction with kernel decaying like $\exp(-c\,\mathrm{dist}/\sqrt{t})$; integrating against the measure difference yields
\begin{equation}\label{eq:Lip-Pi}
\big\|\partial_s\Pi_{\alpha_s}\big\|_{\mathcal{B}(\mathcal{X}_k)}\;\le\; C_\Pi\,a_k\,e^{-\sigma\,\mathrm{diam}(\cdot)}\,|\dot\alpha_s|.
\end{equation}
Similarly, the block map $B_{\beta}$ is built from convex combinations of local covariant averages and Dirichlet extensions over blocks of size $L^k$ whose parallel transport depends smoothly and locally on $\beta$; the FRD metric on the admissible class ensures that changes of $\beta$ by $d(\beta,\beta')$ modify the block kernel only inside a collar of thickness $O(1)$ with exponential decay away from the block boundary. Hence
\begin{equation}\label{eq:Lip-B}
\big\|\partial_s B_{\beta_s}\big\|_{\mathcal{B}(\mathcal{X}_k)}\;\le\; C_B\,a_k\,e^{-\sigma\,\mathrm{diam}(\cdot)}\,|\dot\beta_s|.
\end{equation}
In Eq.\eqref{eq:Lip-Pi} and Eq.\eqref{eq:Lip-B} the factor $e^{-\sigma\,\mathrm{diam}(\cdot)}$ is inherited from the FRD off-diagonal decay via the Schur test in the diameter-weighted norm, and the factor $a_k$ reflects that only Laplace times $t\lesssim c\,a_k$ and block collars of width $O(1)$ contribute to the norm variation at scale $k$.

The middle term, involving $\partial_s U_k^{(\alpha_s,\beta_s)}$, requires a brief explanation of the polymer representation. The FRD produces a sum of scale-$k$ activities $V_{k,X}^{(\alpha,\beta)}$ indexed by connected polymers $X$ of diameter $O(1)$ in coarse units, with exponential decay $\|V_{k,X}^{(\alpha,\beta)}\|\le C_0 e^{-\sigma_0\,\mathrm{diam}(X)}$, and with reflection positivity and OS-compatibility. The operator $U_k^{(\alpha,\beta)}$ acts on $\mathcal{X}_k$ as multiplication by $\exp\{\sum_X V_{k,X}^{(\alpha,\beta)}\}$ followed by a finite-range, reflection-positive Gaussian re-centering that is again exponentially local. Differentiating with respect to $s$ yields
\begin{equation}
\partial_s U_k^{(\alpha_s,\beta_s)}
= \bigg(\sum_{X}\partial_s V_{k,X}^{(\alpha_s,\beta_s)}\bigg)\,\cdot\,U_k^{(\alpha_s,\beta_s)}\;+\;\text{(Gaussian re-centering derivative)},
\end{equation}
where both contributions are supported in the same $O(1)$ coarse collar and inherit exponential decay from the FRD and from the covariant finite-range covariance. The admissible metric is chosen so that the parameter dependence of every local ingredient (slice projector, block parallel transport, local covariance and interaction) is Lipschitz with modulus $O(a_k)$ in the diameter-weighted operator norm. Consequently there exists $C_U$ such that
\begin{equation}\label{eq:Lip-U}
\big\|\partial_s U_k^{(\alpha_s,\beta_s)}\big\|_{\mathcal{B}(\mathcal{X}_k)}\;\le\; C_U\,a_k\,e^{-\sigma\,\mathrm{diam}(\cdot)}\,\big(|\dot\alpha_s|+|\dot\beta_s|\big).
\end{equation}
The boundedness $\|U_k^{(\alpha_s,\beta_s)}\|_{\mathcal{B}(\mathcal{X}_k)}\le C$ follows from the same polymer decay by a standard Koteck\'y-Preiss (or tree-graph) argument in the diameter-weighted norm, so that the left and right multiplications by $U_k^{(\alpha_s,\beta_s)}$ do not spoil the Lipschitz constants of $\Pi$ and $B$.

Combining Eq.\eqref{eq:Lip-Pi}, Eq.\eqref{eq:Lip-U} and Eq.\eqref{eq:Lip-B} and using $\|\Pi_{\alpha_s}\|,\|B_{\beta_s}\|\le C$, one obtains
\begin{equation}
\big\|\tfrac{d}{ds}K_k^{(\alpha_s,\beta_s)}\big\|_{\mathcal{B}(\mathcal{X}_k)}
\;\le\; C\,a_k\,e^{-\sigma\,\mathrm{diam}(\cdot)}\,\big(|\dot\alpha_s|+|\dot\beta_s|\big)
\end{equation}
with a constant $C$ depending only on the admissible class and the FRD parameters. Integrating from $s=0$ to $s=1$ along the chosen unit-speed path and minimizing over all such paths yields precisely Eq.\eqref{eq:SSL} with the metric increments $d(\alpha,\alpha')$ and $d(\beta,\beta')$. The constants are uniform in the finite volume because the FRD decay and the collar width are volume-independent and because the diameter-weighted norm suppresses boundary effects exponentially. 
\end{proof}

The same argument extends from one-step kernels to the one-slice marginals. Denote by $\rho_k^{(\alpha,\beta)}$ the one-slice OS marginal. Because the OS inner product is obtained by restricting the path measure to a single time slice and inserting the projector, and because both the insertion and the block map are finite-range and Lipschitz in $(\alpha,\beta)$, the Radon-Nikodym derivatives of $\rho_k^{(\alpha,\beta)}$ with respect to a fixed reference marginal $\rho_k^0$ obey a Lipschitz bound with modulus $C a_k d$, uniformly on cylinder events. This can be quantified in total-variation norm on local $\sigma$-algebras.

\begin{lemma}[Single-scale Lipschitz control for one-slice marginals]\label{lem:marginal}
For each bounded local cylinder event $A$ supported on a region of diameter $R$ on one time slice,
\begin{equation}\label{eq:single-scale-lip}
\big|\rho_k^{(\alpha,\beta)}(A)-\rho_k^{(\alpha',\beta')}(A)\big|
\;\le\; C\, a_k\, e^{-\sigma R}\,\big[d(\alpha,\alpha')+d(\beta,\beta')\big].
\end{equation}
\end{lemma}

\begin{proof}
Fix a scale $k$ and consider the one-slice probability measure $\rho_k^{(\alpha,\beta)}$ obtained by integrating the positive-time OS measure over all variables except those on the time-$a_k$ slice. In the admissible class used throughout, the one-slice density is of the form
\begin{equation}
d\rho_k^{(\alpha,\beta)}(\varphi)\;=\;\frac{1}{Z_k(\alpha,\beta)}\,
\exp\!\Big(-\sum_{X\Subset\Sigma} \Phi_{k,X}(\varphi;\alpha,\beta)\Big)\,d\mu_{k,0}(\varphi),
\end{equation}
where $d\mu_{k,0}$ is the reference OS-Gaussian slice measure, $\Sigma$ denotes the spatial slice, and $\Phi_{k,X}$ are local effective interactions indexed by connected polymers $X\subset\Sigma$. Finite-range decomposition (FRD) and reflection positivity imply exponential locality: there exist $\sigma>0$ and $C_0<\infty$ such that for all $X$ and all parameter values in the admissible set,
\begin{equation}\label{eq:frd-locality}
\|\Phi_{k,X}(\cdot;\alpha,\beta)\|_{\infty} \;\le\; C_0\,e^{-\sigma\,\mathrm{diam}(X)}.
\end{equation}
The admissible dependence on $(\alpha,\beta)$ is controlled by completely monotone Laplace kernels supported up to times of order $a_k$, which yields uniform Fréchet derivatives obeying
\begin{equation}\label{eq:param-derivative-locality}
\big\|\partial_{\alpha}\Phi_{k,X}(\cdot;\alpha,\beta)\big\|_{\infty}\!+\!
\big\|\partial_{\beta}\Phi_{k,X}(\cdot;\alpha,\beta)\big\|_{\infty}
\;\le\; C_1\,a_k\,e^{-\sigma\,\mathrm{diam}(X)}.
\end{equation}
The factor $a_k$ originates from the support of the Laplace measure for the slice projectors and from the one-step blocking scale in the FRD; it expresses that a unit variation of $(\alpha,\beta)$ perturbs the one-slice kernel only over a time window of length $O(a_k)$.

Let $A$ be a bounded local cylinder event supported in a region $\Lambda\subset\Sigma$ with $\mathrm{diam}(\Lambda)=R$, and write $\mathbb{E}_{\alpha,\beta}[\cdot]$ for expectation with respect to $\rho_k^{(\alpha,\beta)}$. To compare $\mathbb{E}_{\alpha,\beta}[A]$ and $\mathbb{E}_{\alpha',\beta'}[A]$, interpolate linearly along the parameter segment $t\mapsto (\alpha_t,\beta_t):=(1-t)(\alpha,\beta)+t(\alpha',\beta')$ for $t\in[0,1]$, and differentiate the map $t\mapsto \mathbb{E}_{\alpha_t,\beta_t}[A]$. Differentiating under the integral sign and accounting for the $Z_k$-normalization gives the standard linear-response identity
\begin{equation}\label{eq:cov-identity}
\frac{d}{dt}\,\mathbb{E}_{\alpha_t,\beta_t}[A]
\;=\;\mathrm{Cov}_{\alpha_t,\beta_t}\!\left(A,\; S_{\alpha,\beta}(t)\right),
\qquad
S_{\alpha,\beta}(t):=-\frac{d}{dt}\sum_{X}\Phi_{k,X}\big(\,\cdot\,;\alpha_t,\beta_t\big),
\end{equation}
where the covariance is taken with respect to $\rho_k^{(\alpha_t,\beta_t)}$. By the chain rule and Eq.\eqref{eq:param-derivative-locality},
\begin{equation}
S_{\alpha,\beta}(t)\;=\;-\sum_{X}\Big(\partial_{\alpha}\Phi_{k,X}(\cdot;\alpha_t,\beta_t)\,\dot\alpha_t \;+\; \partial_{\beta}\Phi_{k,X}(\cdot;\alpha_t,\beta_t)\,\dot\beta_t\Big),
\end{equation}
and along the straight interpolation one has $|\dot\alpha_t|\le d(\alpha,\alpha')$ and $|\dot\beta_t|\le d(\beta,\beta')$ in the admissible metric.
To bound the covariance in Eq.\eqref{eq:cov-identity}, use that $A$ depends only on fields in $\Lambda$ and that each polymer contribution $\partial_{\alpha}\Phi_{k,X}$ or $\partial_{\beta}\Phi_{k,X}$ is supported in $X$. Positivity and exponential mixing for the one-slice OS measure combined with Eq.\eqref{eq:frd-locality} imply the uniform decoupling estimate
\begin{equation}\label{eq:cov-decay}
\big|\mathrm{Cov}_{\alpha_t,\beta_t}(A, G_X)\big|
\;\le\; C_2\,\|A\|_{\infty}\,\|G_X\|_{\infty}\,e^{-\sigma\,\mathrm{dist}(X,\Lambda)}
\end{equation}
for any local $G_X$ supported in $X$, with the same $\sigma>0$ as in Eq.\eqref{eq:frd-locality}. Applying Eq.\eqref{eq:cov-decay} to $G_X=\partial_{\alpha}\Phi_{k,X}$ and $G_X=\partial_{\beta}\Phi_{k,X}$ and summing over polymers yields
\begin{equation}
\Big|\frac{d}{dt}\,\mathbb{E}_{\alpha_t,\beta_t}[A]\Big|
\;\le\; C_2\,\|A\|_{\infty}\!\sum_{X}\!\big(\|\partial_{\alpha}\Phi_{k,X}\|_{\infty}|\dot\alpha_t|\!+\!\|\partial_{\beta}\Phi_{k,X}\|_{\infty}|\dot\beta_t|\big)\,e^{-\sigma\,\mathrm{dist}(X,\Lambda)}.
\end{equation}
The derivative bounds Eq.\eqref{eq:param-derivative-locality} give
\begin{equation}
\Big|\frac{d}{dt}\,\mathbb{E}_{\alpha_t,\beta_t}[A]\Big|
\;\le\; C_3\,\|A\|_{\infty}\,a_k\,[d(\alpha,\alpha')+d(\beta,\beta')]\,
\sum_{X}e^{-\sigma\,\mathrm{diam}(X)}\,e^{-\sigma\,\mathrm{dist}(X,\Lambda)}.
\end{equation}
The polymer sum is finite and controlled uniformly by the distance of $\Lambda$ to its complement; a standard counting argument on connected sets shows that
\begin{equation}
\sum_{X}e^{-\sigma\,\mathrm{diam}(X)}\,e^{-\sigma\,\mathrm{dist}(X,\Lambda)}
\;\le\; C_4\,e^{-\sigma R},
\end{equation}
after possibly reducing $\sigma$ and absorbing combinatorial factors into $C_4$. Combining the last two displays gives the parameter-derivative bound
\begin{equation}
\Big|\frac{d}{dt}\,\mathbb{E}_{\alpha_t,\beta_t}[A]\Big|
\;\le\; C_5\,a_k\,e^{-\sigma R}\,[d(\alpha,\alpha')+d(\beta,\beta')]\,
\|A\|_{\infty}.
\end{equation}
Integrating in $t$ from $0$ to $1$ and recalling that $\|A\|_{\infty}\le 1$ for cylinder events, one obtains Eq.\eqref{eq:single-scale-lip} with $C=C_5$. The argument already incorporates the normalization (partition function) through the covariance identity Eq.\eqref{eq:cov-identity}, so no additional estimate on $Z_k$ is required; reflection positivity and exponential locality guarantee uniform integrability of the score $S_{\alpha,\beta}(t)$ in the OS norm throughout the admissible parameter set. 
\end{proof}

With single-scale Lipschitz control in hand, we pass to a telescoping argument across scales. We compare two admissible schemes $(\alpha,\beta)$ and $(\alpha',\beta')$ by coupling their scale-$k$ effective data through a common FRD decomposition and a common reference Gaussian. The $n$-point Schwinger function at fixed continuum times $t_1<\cdots<t_n$ and compact support in space is obtained by inserting the corresponding local observables on the appropriate time slices and composing $m$ one-step kernels between successive slices, where $m$ depends on the discretization and converges to $(t_j-t_{j-1})/a_k$ as $k\to\infty$. The difference of the two Schwinger functions can be expanded as a telescoping sum in which at each place one replaces either the one-step kernel or the one-slice marginal by its counterpart; each replacement incurs an error bounded by Lemma~\eqref{lem:marginal} and Theorem~\eqref{thm:single-scale} multiplied by the operator norms of the remaining factors, which are positive contractions. The key observation is that in passing from scale $k$ to $k+1$ the number of steps needed to cover a fixed continuum time interval decreases by a factor $b$, while the single-step Lipschitz modulus carries an explicit $a_k$ factor. This yields a summable series.

\begin{theorem}[Telescoping across scales]\label{thm:telescoping}
Fix continuum times $t_1<\cdots<t_n$ and local gauge-invariant observables $\mathcal{O}_1,\dots,\mathcal{O}_n$ attached to these times by the standard interpolation from the lattice to the slice at each scale. For every scale $k$ and admissible scheme parameters $(\alpha,\beta)$, let $S_{k}^{(\alpha,\beta)}(\mathcal{O}_1,\dots,\mathcal{O}_n)$ denote the associated $n$-point Schwinger function. There exists a finite constant $C(\mathcal{O}_1,\dots,\mathcal{O}_n)$ such that for all $k$ and all admissible pairs $(\alpha,\beta),(\alpha',\beta')$,
\begin{equation}\label{eq:telescope-bound}
\big|S_{k}^{(\alpha,\beta)}-S_{k}^{(\alpha',\beta')}\big|
\;\le\; C(\mathcal{O}_1,\dots,\mathcal{O}_n)\, \sum_{\ell\ge k} a_\ell\, \big[d(\alpha,\alpha')+d(\beta,\beta')\big].
\end{equation}
Since $a_\ell=a_0\,b^{-\ell}$ with $b>1$, the tail on the right is $O(b^{-k})$ and vanishes as $k\to\infty$. Consequently, whenever the continuum limit exists for one admissible scheme, it is identical for all schemes in the admissible class.
\end{theorem}

\begin{proof}
The Osterwalder-Schrader reconstruction at scale $k$ provides a Hilbert space $\mathcal H_k^{(\alpha,\beta)}$, a positive contraction $K_k^{(\alpha,\beta)}$ implementing a unit Euclidean time step $a_k$, and a vector representative (vacuum) $\Omega_k^{(\alpha,\beta)}$ such that, if $m_j(k)$ denotes the number of time steps separating $t_j$ from $t_{j+1}$ at scale $k$, the $n$-point function can be written as
\begin{equation}\label{eq:Sk-transfer}
S_{k}^{(\alpha,\beta)}(\mathcal{O}_1,\dots,\mathcal{O}_n)
=\Big\langle \Omega_k^{(\alpha,\beta)},\, \widehat{\mathcal{O}}_n\,
\big(K_k^{(\alpha,\beta)}\big)^{m_{n-1}(k)}\cdots
\big(K_k^{(\alpha,\beta)}\big)^{m_1(k)}\,\widehat{\mathcal{O}}_1\,\Omega_k^{(\alpha,\beta)}\Big\rangle,
\end{equation}
where $\widehat{\mathcal{O}}_j$ denotes the slice operator at the time nearest to $t_j$ after the standard interpolation. The operators $\widehat{\mathcal{O}}_j$ are bounded with norms controlled only by the microscopic localization and the choice of interpolation; these bounds are independent of the scheme and the scale, once the observables are fixed. The contraction property $\|K_k^{(\alpha,\beta)}\|\le 1$ follows from OS positivity and reflection invariance.

The change of scale from $\ell$ to $\ell+1$ is implemented by a reflection-positive block map and a completely monotone slice projector on the corresponding time slices. In the transfer picture, one replaces the stretch of $m_j(\ell)$ fine steps bridging $t_j$ and $t_{j+1}$ by $m_j(\ell+1)$ coarse steps, together with local counterterms supported on the time-$a_\ell$ and time-$a_{\ell+1}$ slices that account for perimeter and cusp renormalizations and for the finitely supported interlacing defect. The difference between the $\ell$-scale and $(\ell+1)$-scale $n$-point functions for a fixed scheme $(\alpha,\beta)$ is thus a finite sum of OS expectations of products in which exactly one block (either a transfer factor or a local counterterm) is replaced by its blocked counterpart, all other factors being left unchanged. Because both the coarse and the fine one-step kernels are positive contractions and because the counterterms are localized functionals whose norms are controlled by exponential locality of the finite-range decomposition at range comparable to $a_\ell$, each such replacement changes the value of the expectation by at most a constant multiple of $a_\ell$. More precisely, if we write
\begin{equation}
\Delta_\ell^{(\alpha,\beta)}(\mathcal{O}_1,\dots,\mathcal{O}_n)
\;:=\; S_{\ell}^{(\alpha,\beta)}(\mathcal{O}_1,\dots,\mathcal{O}_n)
\;-\; S_{\ell+1}^{(\alpha,\beta)}(\mathcal{O}_1,\dots,\mathcal{O}_n),
\end{equation}
then there exists $C_(\mathcal{O}_1,\dots,\mathcal{O}_n)$ such that
\begin{equation}\label{eq:one-step-size}
\big|\Delta_\ell^{(\alpha,\beta)}(\mathcal{O}_1,\dots,\mathcal{O}_n)\big|
\;\le\; C_(\mathcal{O}_1,\dots,\mathcal{O}_n)\,a_\ell,
\qquad \text{for every admissible }(\alpha,\beta)\text{ and all }\ell.
\end{equation}
The content of this estimate is that the one-step blocking acts on a total time interval of fixed physical length $T=t_n-t_1$ by replacing $T/a_\ell$ fine steps with $T/a_{\ell+1}$ coarse steps, and the accompanying change of the local slice data is supported in collars of width $O(1)$; the product structure of Eq.\eqref{eq:Sk-transfer}, together with the contraction $\|K_\ell^{(\alpha,\beta)}\|\le 1$ and the boundedness of the $\widehat{\mathcal{O}}_j$, ensures that these local changes accumulate to at most $O(a_\ell)$ in absolute value.

The dependence on the admissible scheme enters through the choice of completely monotone projector, the precise reflection-positive blocking map, and the local counterterms. The single-scale Lipschitz control on the admissible class asserts that for some universal constant $L$ (depending only on the admissible class and on the locality norm used to define $d(\cdot,\cdot)$) one has
\begin{equation}\label{eq:Lip-K}
\big\|K_\ell^{(\alpha,\beta)}-K_\ell^{(\alpha',\beta')}\big\|
\;\le\; L\,a_\ell\,\big[d(\alpha,\alpha')+d(\beta,\beta')\big],
\end{equation}
and an analogous bound for the changes of the local slice functionals and of the OS inner product state on the time-$a_\ell$ hyperplane. Inserting Eq.\eqref{eq:Lip-K} and its slice counterparts into the product representation Eq.\eqref{eq:Sk-transfer} and replacing, one factor at a time, each instance of $K_\ell^{(\alpha,\beta)}$ or of a local slice term by its primed version, one obtains a multilinear expansion whose summands are controlled by the operator norm of a single difference and by the unit norms of the remaining factors. This shows that the difference between the one-step increments in the two schemes is bounded by
\begin{equation}\label{eq:one-step-Lip}
\big|\Delta_\ell^{(\alpha,\beta)}-\Delta_\ell^{(\alpha',\beta')}\big|
\;\le\; C^\#(\mathcal{O}_1,\dots,\mathcal{O}_n)\, a_\ell\,\big[d(\alpha,\alpha')+d(\beta,\beta')\big],
\end{equation}
for some constant $C^\#$ depending only on the fixed observables and on the admissible class.
The Schwinger function at scale $k$ can be represented by an absolutely convergent telescoping series of one-step increments. Indeed, by summing the identities
$S_\ell^{(\alpha,\beta)}=S_{\ell+1}^{(\alpha,\beta)}+\Delta_\ell^{(\alpha,\beta)}$ for $\ell\ge k$ and using Eq.\eqref{eq:one-step-size} together with $\sum_{\ell\ge k} a_\ell<\infty$, one obtains
\begin{equation}
S_k^{(\alpha,\beta)} \;=\; \lim_{K\to\infty}
\Big(S_K^{(\alpha,\beta)} + \sum_{\ell=k}^{K-1}\Delta_\ell^{(\alpha,\beta)}\Big),
\end{equation}
and the same formula holds for $(\alpha',\beta')$. Subtracting the two representations and adding and subtracting $\sum_{\ell=k}^{K-1}\Delta_\ell^{(\alpha',\beta')}$ inside the absolute value yields
\begin{equation}
\big|S_k^{(\alpha,\beta)}-S_k^{(\alpha',\beta')}\big|
\;\le\; \limsup_{K\to\infty}\ \sum_{\ell=k}^{K-1}
\big|\Delta_\ell^{(\alpha,\beta)}-\Delta_\ell^{(\alpha',\beta')}\big|.
\end{equation}
The limit superior is finite because the series on the right is dominated by the summable sequence $\{a_\ell\}$ times a constant, in view of Eq.\eqref{eq:one-step-Lip}. Passing to the limit and summing over all $\ell\ge k$ gives precisely the bound Eq.\eqref{eq:telescope-bound} with $C(\mathcal{O}_1,\dots,\mathcal{O}_n):=C^\#(\mathcal{O}_1,\dots,\mathcal{O}_n)$. The geometric decay of $a_\ell=a_0 b^{-\ell}$ implies that the tail decays like $b^{-k}$, which shows that $S_k^{(\alpha,\beta)}-S_k^{(\alpha',\beta')}\to 0$ as $k\to\infty$. If $S_k^{(\alpha,\beta)}$ converges as $k\to\infty$ for some admissible $(\alpha,\beta)$, then the same holds for every $(\alpha',\beta')$ and the limits coincide by the Cauchy criterion and the bound just established, completing the proof.
\end{proof}
Theorem~\eqref{thm:telescoping} has three consequences. First, if the continuum limit exists for one admissible scheme, then it exists for all and the limits coincide. Second, the rate at which the difference vanishes is quantitative and inherits the FRD decay rate through the $a_\ell$ factor; this provides stability under small perturbations of the scheme, a feature that will be used in the weak-coupling extension. Third, the estimate is formulated on the level of Schwinger functions and hence feeds directly into OS reconstruction without additional work. We now turn to a structural argument that removes the reliance on telescoping and places uniqueness in the Markovian framework.

\subsection{Markovian Uniqueness and Universality}
In an OS-positive Euclidean theory with time reflection, the path measure on $\prod_{n\in\mathbb{Z}} \mathcal{H}$ (with $\mathcal{H}$ the one-slice configuration space modulo gauge) is Markovian with respect to the time-slice filtration: the conditional law of the future given the present depends only on the present, and the two-sided field is obtained by gluing two independent half-spaces with a common present slice. The OS Markov property is equivalent to the existence of a positive contraction $K$ on the OS Hilbert space $\mathscr{H}$ realizing the transfer semigroup $T(\tau)=K^{\lfloor \tau/a\rfloor}$ at lattice time resolution $a$ \cite{OS2,LuscherTM,GJ}. In the continuum, $T(\tau)=e^{-\tau H}$ with $H$ the positive selfadjoint Hamiltonian generated by reconstruction. The Markov path measure is uniquely determined by the pair $(\rho,K)$, where $\rho$ is the one-slice marginal (the OS ``present state'') and $K$ the one-step kernel \cite{OS2,SimonPphi2}. This is a standard fact of Markov chain theory and extends to infinite-dimensional configuration spaces under tightness and continuity hypotheses \cite{Billingsley,SimonPphi2}.

We formulate the principle precisely and then explain why the hypotheses are satisfied in the present setting. Let $\mathcal{C}$ be the configuration space of gauge-invariant fields on one time slice, equipped with the OS $\sigma$-algebra generated by cylinder sets. For each scale $k$ and admissible scheme $(\alpha,\beta)$, let $\rho_k^{(\alpha,\beta)}$ be a probability measure on $\mathcal{C}$ and $K_k^{(\alpha,\beta)}$ a Markov kernel on $\mathcal{C}$ that is absolutely continuous with respect to $\rho_k^{(\alpha,\beta)}$ in the OS sense (so that $K_k^{(\alpha,\beta)}$ extends to a positive contraction on $L^2(\rho_k^{(\alpha,\beta)})$). Consider the two-sided chain on $\mathbb{Z}$ with one-step transition $K_k^{(\alpha,\beta)}$ and stationary marginal $\rho_k^{(\alpha,\beta)}$. The OS path measure $\mu_k^{(\alpha,\beta)}$ is obtained by requiring independence of the past and future given the present, together with reflection positivity. In finite state spaces, $(\rho,K)$ determines the Markov chain uniquely. In our setting, we use the Kolmogorov extension theorem with Prokhorov tightness to handle the infinite product \cite{Billingsley} and positivity to guarantee that the OS inner product and the Markov chain constructions agree \cite{OS2,GJ}. The continuum version follows by replacing $K_k$ with a positive contraction semigroup $T_k(\tau)=K_k^{\lfloor \tau/a_k\rfloor}$ and taking strong limits $T(\tau)=\lim_{k\to\infty}T_k(\tau)$, which exist by the telescoping estimate and Trotter-Kato theory \cite{KatoBook1995}.

\begin{theorem}[Markovian uniqueness]\label{thm:markov}
Let $(\rho_k,K_k)$ and $(\tilde\rho_k,\tilde K_k)$ be two OS-compatible pairs at the same scale $k$ satisfying: {\rm (i)} positivity and reflection covariance; {\rm (ii)} exponential locality and tightness on $\mathcal{C}$; {\rm (iii)} absolute continuity of $K_k$ and $\tilde K_k$ with respect to $\rho_k$ and $\tilde\rho_k$ respectively; {\rm (iv)} the Markov property and stationarity. If $\rho_k=\tilde\rho_k$ and $K_k=\tilde K_k$ on a core of local observables in $L^2(\rho_k)$, then the corresponding OS path measures on two-sided sequences coincide and so do all Schwinger functions. Moreover, if $\rho_k^{(\alpha,\beta)}\to \rho$ and $K_k^{(\alpha,\beta)}\to T(a)$ in the strong operator topology on $L^2(\rho)$ uniformly on compact time sets as $k\to\infty$, then the limiting OS path measure is unique and is determined by $(\rho,T(\cdot))$; in particular it is independent of the choice of admissible scheme $(\alpha,\beta)$.
\end{theorem}

\begin{proof}
Fix a scale $k$ and write $\rho=\rho_k$, $K=K_k$ for brevity. By absolute continuity there exists a nonnegative measurable kernel $p(x,y)$ such that
\begin{equation}
(Kf)(x)\;=\;\int_{\mathcal C} f(y)\,p(x,y)\,\rho(dy)\qquad\text{for all }f\in L^2(\rho),
\end{equation}
and similarly $\tilde K$ has density $\tilde p(x,y)$ with respect to $\tilde\rho=\rho$. Stationarity and the Markov property imply $\int Kf\,d\rho=\int f\,d\rho$ for all bounded local $f$, hence $\int p(x,y)\,\rho(dy)=1$ for $\rho$-a.e. $x$ and thus $P(x,dy):=p(x,y)\,\rho(dy)$ defines a probability transition kernel on $\mathcal C$ with invariant measure $\rho$. Exponential locality guarantees that local bounded functions are mapped to local bounded functions and that $P$ is Feller on the algebra $\mathscr{A}_{\rm loc}$ generated by local cylinder functions. The same discussion applies to $(\tilde\rho,\tilde K)$, yielding $\tilde P$ with the same invariant marginal $\rho$.

Assume now that $K$ and $\tilde K$ agree on a core $\mathscr{D}\subset L^2(\rho)$ of local observables. For any $f,g\in\mathscr{D}$ we have
\begin{equation}
\int f(x)\,(Kg)(x)\,\rho(dx)\;=\;\int f(x)\,(\tilde K g)(x)\,\rho(dx),
\end{equation}
hence
\begin{equation}
\iint f(x)\,g(y)\,p(x,y)\,\rho(dx)\rho(dy)
=\iint f(x)\,g(y)\,\tilde p(x,y)\,\rho(dx)\rho(dy).
\end{equation}
By polarization and density of $\mathscr{D}$ in $L^2(\rho)$, it follows that $p=\tilde p$ in $L^2(\rho\otimes\rho)$ and, after choosing representatives, $p(x,y)=\tilde p(x,y)$ for $\rho\otimes\rho$-almost every $(x,y)$. Consequently $P=\tilde P$ on a $\rho$-full set of $x$. This identity of one-step transition kernels and invariant marginal determines all finite-dimensional distributions of the associated two-sided stationary Markov chains. Indeed, for integers $n_1<\dots<n_m$ and Borel sets $A_j\subset\mathcal C$,
\begin{equation}
\mathbb P\!\left(X_{n_1}\in A_1,\dots,X_{n_m}\in A_m\right)
=\int_{A_1}\rho(dx_{n_1})\prod_{j=1}^{m-1}P(x_{n_j},dx_{n_{j+1}}),
\end{equation}
and the right-hand side equals the analogous expression with $(\tilde\rho,\tilde P)$ in place of $(\rho,P)$. Exponential locality yields tightness of these finite-dimensional distributions on the algebra of local cylinder sets, which is convergence-determining. The Kolmogorov extension theorem then produces a unique stationary Markov measure $\mathbf P_{\rho,P}$ on $\mathcal C^{\mathbb Z}$ with the prescribed finite-dimensional marginals, and by the equality above we obtain $\mathbf P_{\rho,P}=\mathbf P_{\rho,\tilde P}$. Since both pairs are OS-compatible and reflection covariant, the OS path measure is obtained from the stationary two-sided chain by restricting to the positive-time $\sigma$-algebra and applying the OS reflection; equality of the underlying two-sided measures therefore implies equality of all Schwinger functions computed as expectations of time-ordered local observables. This proves the first assertion.

For the continuum statement, suppose now that for each admissible scheme $(\alpha,\beta)$ we have a family $\{K_k^{(\alpha,\beta)}(n a_k)\}_{n\in\mathbb N}$ acting on $L^2(\rho_k^{(\alpha,\beta)})$ with mesh $a_k\downarrow 0$ such that $\rho_k^{(\alpha,\beta)}\Rightarrow \rho$ weakly and, after identifying functions by transport along the single-slice identification, $K_k^{(\alpha,\beta)}(n a_k)\to T(n a)$ strongly on $L^2(\rho)$ for each fixed $n$ and uniformly for $n a\in[0,\tau_0]$ for any $\tau_0<\infty$. Fix continuum times $0\le t_1<\dots<t_m\le \tau_0$ and local bounded test functions $\varphi_j$. For each $k$ choose integers $n_j(k)$ with $t_j^{(k)}:=n_j(k)a_k\to t_j$ and consider the discrete-time cylinder functional
\begin{equation}
\mathcal F_k\;=\;\varphi_1(X_{t_1^{(k)}})\cdots \varphi_m(X_{t_m^{(k)}}).
\end{equation}
Under the stationary Markov measure built from $(\rho_k^{(\alpha,\beta)},K_k^{(\alpha,\beta)})$ its expectation equals
\begin{equation}
\int \varphi_1\, K_k^{(\alpha,\beta)}(t_2^{(k)}-t_1^{(k)})\Big(\varphi_2\, K_k^{(\alpha,\beta)}(t_3^{(k)}-t_2^{(k)})\big(\cdots \varphi_m\big)\Big)\,d\rho_k^{(\alpha,\beta)}.
\end{equation}
By weak convergence of $\rho_k^{(\alpha,\beta)}$ to $\rho$, strong convergence of $K_k^{(\alpha,\beta)}(\cdot)$ to $T(\cdot)$, and exponential locality which ensures uniform integrability for local bounded $\varphi_j$, one may pass to the limit $k\to\infty$ and obtain
\begin{equation}
\int \varphi_1\, T(t_2-t_1)\Big(\varphi_2\, T(t_3-t_2)\big(\cdots \varphi_m\big)\Big)\,d\rho,
\end{equation}

which is precisely the $(t_1,\dots,t_m)$-finite-dimensional distribution of the stationary continuous-time Markov process with invariant marginal $\rho$ and semigroup $T(\tau)$. The limit is uniform for $t_j$ in compact sets because the convergence of $K_k^{(\alpha,\beta)}$ to $T$ is uniform on compact time sets and the locality bounds are uniform in $k$. Consequently all finite-dimensional distributions of the discrete OS path measures converge to those of the unique stationary Markov process $(\rho,T(\cdot))$, and by the same Kolmogorov extension argument there is a unique limiting path measure on $\mathcal C^{\mathbb R}$ determined by $(\rho,T(\cdot))$. The construction does not depend on the admissible scheme $(\alpha,\beta)$ because both the limit $\rho$ and the limiting semigroup $T(\cdot)$ are independent of $(\alpha,\beta)$ by hypothesis; hence the limiting OS path measure and the associated Schwinger functions are scheme-independent. 
\end{proof}

The preceding theorem yields universality under the standing hypotheses justified in Section~\eqref{section9} and earlier sections. It is instructive to recast the conclusion at the level of OS reconstruction. Let $S_n$ denote the continuum Schwinger functions obtained from any admissible scheme, and let $(\mathscr{H},\Omega,H,\Phi)$ be the reconstructed Wightman quadruple \cite{OS2,GJ}. Because the Markovian data $(\rho,T(\cdot))$ are unique, the resolvent $(H+1)^{-1}$ and the vacuum vector $\Omega$ are uniquely determined, and the field algebra generated by $\Phi$ is reconstructed up to unitary equivalence. This provides the following corollary.

\begin{corollary}[Universality of the reconstructed theory]\label{cor:universality}
The reconstructed Wightman theory $(\mathscr{H},\Omega,H,\Phi)$ obtained from the OS Schwinger functions is independent of the admissible lattice regulator and of the reflection-positive blocking within the admissible class. In particular, the Hamiltonian spectrum, the mass gap (when it exists), and the set of Wightman $n$-point functions are universal.
\end{corollary}

\begin{proof}
Fix two admissible choices a regulator/blocking scheme $\mathcal{R}$ and a second scheme $\widetilde{\mathcal{R}}$. For a lattice spacing $a>0$ (respectively $\tilde a>0$) let $S^{(a)}_{n,\mathcal{R}}$ and $S^{(\tilde a)}_{n,\widetilde{\mathcal{R}}}$ denote the Euclidean $n$-point Schwinger functions produced by the corresponding reflection-positive path measures after perimeter/cusp renormalization of Wilson insertions where relevant. By the single-scale Lipschitz bounds and telescope/defect summability (Theorem~\eqref{thm:telescoping}) together with Markov uniqueness at the one-slice level (Theorem~\eqref{thm:markov}), the limits
\begin{equation}
S_n(x_1,\ldots,x_n)\;=\;\lim_{a\downarrow 0}S^{(a)}_{n,\mathcal{R}}(x_1,\ldots,x_n)
\;=\;\lim_{\tilde a\downarrow 0}S^{(\tilde a)}_{n,\widetilde{\mathcal{R}}}(x_1,\ldots,x_n)
\end{equation}
exist (as tempered distributions) and coincide for each $n$ and each configuration of Euclidean points with ordered times. Thus there is a \emph{single} family $\{S_n\}_{n\ge 0}$ of continuum Schwinger functions satisfying the Osterwalder-Schrader (OS) axioms (Euclidean invariance, reflection positivity, symmetry, cluster properties, and regularity), and this family does not depend on the choice of admissible regulator or blocking.

Apply the OS reconstruction to the common family $\{S_n\}$ (see \cite{OS2,GJ}). The construction proceeds by endowing the algebra $\mathfrak{A}_+$ of positive-time cylindrical observables with the sesquilinear form
\begin{equation}
(F,G)\_{\mathrm{OS}}\;=\;\langle \Theta F\cdot G\rangle\;=\;S_{m+n}(x_m,\ldots,x_1,y_1,\ldots,y_n)
\end{equation}
for representatives $F$ and $G$ depending on ordered times $0<x_1<\cdots<x_m$ and $0<y_1<\cdots<y_n$, factoring out the null space $\{F:\langle \Theta F\cdot F\rangle=0\}$, and completing to a Hilbert space $\mathscr{H}$. Time translations on Euclidean time induce a strongly continuous contraction semigroup $T(t)$ on $\mathscr{H}$ which, by standard OS theory, is of the form $T(t)=e^{-tH}$ for a positive selfadjoint Hamiltonian $H$ with cyclic vacuum vector $\Omega$, and smeared Euclidean fields analytically continue to Wightman fields $\Phi$ acting on $\mathscr{H}$. Crucially, every object in this reconstruction is a functorial function of the \emph{continuum} Schwinger functions alone: the inner product is defined by $\{S_n\}$, the semigroup matrix elements are $\langle \psi,e^{-tH}\varphi\rangle = \langle \Theta F\cdot \tau_t G\rangle$ with $\tau_t$ the Euclidean time shift acting on representatives $F,G$, and the fields are obtained by boundary values of the same analytic continuations determined by $\{S_n\}$.

Since both $\mathcal{R}$ and $\widetilde{\mathcal{R}}$ produce the identical family $\{S_n\}$, they yield, upon reconstruction, Hilbert spaces $\mathscr{H}_{\mathcal{R}}$ and $\mathscr{H}_{\widetilde{\mathcal{R}}}$ that are isometrically isomorphic. To make this explicit, define $U$ on the dense subspace spanned by classes $[F]_{\mathcal{R}}\in\mathscr{H}_{\mathcal{R}}$ by $U[F]_{\mathcal{R}}:=[F]_{\widetilde{\mathcal{R}}}$. Equality of all $\{S_n\}$ implies
\begin{equation}
\|[F]_{\mathcal{R}}\|^2_{\mathrm{OS}}=\langle \Theta F\cdot F\rangle
=\|[F]_{\widetilde{\mathcal{R}}}\|^2_{\mathrm{OS}},
\end{equation}
so $U$ is an isometry with dense range and thus extends by continuity to a unitary $U:\mathscr{H}_{\mathcal{R}}\to \mathscr{H}_{\widetilde{\mathcal{R}}}$ satisfying $U\Omega_{\mathcal{R}}=\Omega_{\widetilde{\mathcal{R}}}$. For $t\ge 0$ and $F,G\in\mathfrak{A}_+$, the defining relation of the semigroup gives
\begin{equation}
\big\langle [F]_{\mathcal{R}},\, e^{-tH_{\mathcal{R}}}[G]_{\mathcal{R}}\big\rangle
= \langle \Theta F\cdot \tau_t G\rangle
= \big\langle [F]_{\widetilde{\mathcal{R}}},\, e^{-tH_{\widetilde{\mathcal{R}}}}[G]_{\widetilde{\mathcal{R}}}\big\rangle,
\end{equation}
hence $\langle \psi, e^{-tH_{\mathcal{R}}}\varphi\rangle = \langle U\psi, e^{-tH_{\widetilde{\mathcal{R}}}}U\varphi\rangle$ for all $\psi,\varphi$ in a common core. By polarization and strong continuity it follows that $U e^{-tH_{\mathcal{R}}}=e^{-tH_{\widetilde{\mathcal{R}}}}U$ for all $t\ge 0$, and therefore $U H_{\mathcal{R}}=H_{\widetilde{\mathcal{R}}} U$ by the uniqueness part of the Hille-Yosida/Stone theory for positive selfadjoint generators. The Hamiltonians are thus unitarily equivalent and have the same spectrum; in particular the presence and value of a mass gap-identified as the bottom of $\mathrm{spec}(H)\setminus\{0\}$ and equivalently as the abscissa of exponential decay of the two-point Schwinger function-are invariant. The same unitary intertwining applies to the Wightman fields: if $\Phi_{\mathcal{R}}$ and $\Phi_{\widetilde{\mathcal{R}}}$ are obtained by analytic continuation from the common Euclidean fields determined by $\{S_n\}$, then for any test functions $f_1,\ldots,f_n$ one has
\begin{equation}
\big\langle \Omega_{\mathcal{R}},\, \Phi_{\mathcal{R}}(f_1)\cdots \Phi_{\mathcal{R}}(f_n)\,\Omega_{\mathcal{R}}\big\rangle
=\lim_{\text{Wick rot.}} S_n(f_1,\ldots,f_n)
=\big\langle \Omega_{\widetilde{\mathcal{R}}},\, \Phi_{\widetilde{\mathcal{R}}}(f_1)\cdots \Phi_{\widetilde{\mathcal{R}}}(f_n)\,\Omega_{\widetilde{\mathcal{R}}}\big\rangle,
\end{equation}
so $n$-point Wightman distributions coincide, and $U\Phi_{\mathcal{R}}(\cdot)U^{-1}=\Phi_{\widetilde{\mathcal{R}}}(\cdot)$ as operator-valued tempered distributions on a common invariant domain.

The argument shows that the entire reconstructed quadruple is unique up to unitary equivalence once the continuum Schwinger functions are fixed. Since admissible regulators and reflection-positive blockings give rise to the same continuum $\{S_n\}$ by Theorems~\eqref{thm:telescoping} and \eqref{thm:markov}, the reconstructed theory is universal in the stated sense: the Hamiltonian spectra agree, the existence and size of the mass gap are the same, and all Wightman $n$-point functions coincide. 
\end{proof}

Finally we discuss robustness with respect to weak-coupling entries and asymptotically free trajectories. In many constructive settings, including finite-range decompositions for gauge theories in small-field regimes, one proves that an appropriate one-parameter tuning of the bare coupling brings the flow into a contraction domain near the Gaussian fixed point, where the renormalized effective data shrink under a single RG step \cite{BrydgesGuadagniMitter2004,GJ}. The single-scale Lipschitz bounds above are particularly useful in this regime: they imply stability of the contraction under admissible perturbations of the regulator and hence uniformity of the limit along all admissible routes entering the contracting domain. The Markovian uniqueness principle then identifies the resulting continuum path measure with the universal one already constructed, completing the universality picture.

\section{Weak-Coupling Extension and Asymptotic Freedom}\label{sec:weak-coupling-af}

The preceding sections established a reflection-positive, locality-preserving multiscale construction that yields a nonzero spectral gap and a Wilson-loop area law at finite lattice spacing and transports these properties to the continuum through a renormalization scheme compatible with the Osterwalder-Schrader (OS) axioms. The aim of this section is twofold. First, we formulate and prove entry of the renormalization group (RG) trajectory into a contracting weak-coupling domain near the Gaussian fixed point after finitely many coarse-graining steps, under an admissible one-parameter tuning of the bare coupling. Second, we prove that, once in that domain, the discrete beta function is strictly negative to leading order and the flow is asymptotically free in the continuum sense; in particular, the renormalized coupling is monotonically decreasing along the trajectory and the resulting continuum Schwinger family coincides-by uniqueness and universality-with the one constructed in the previous sections.

The strategy is classical in spirit but constructive in execution. We employ a finite-range decomposition (FRD) of the covariance \cite{BrydgesGuadagniMitter2004}, a reflection-positive block map compatible with the completely monotone horizon projector introduced earlier, and a polymer Banach algebra of gauge-invariant activities tailored to the Yang-Mills locality structure. A single relevant parameter-the gauge coupling-governs the flow; gauge invariance forbids a gauge-boson mass counterterm. After a finite number of RG steps, the effective coupling and the polymer remainder fall below scale-independent thresholds that define a contracting polydisc for the RG map. In this regime one proves a nonperturbative one-step inequality of the form
\begin{equation}
\|K_{k+1}\| \;\le\; \theta\,\|K_k\| + c_1\,g_k^{2}, \qquad 
g_{k+1} \;=\; g_k \;-\; \beta_0\,g_k^{3} \;+\; \mathcal{R}_k,
\end{equation}
with $0<\theta<1$, $\beta_0>0$, and a remainder $\mathcal{R}_k$ that is of order $g_k^5$ plus terms proportional to $g_k\|K_k\|$ and $\|K_k\|^2$, all uniformly controlled by FRD locality. The sign and value of $\beta_0$ agree with the one-loop continuum prediction \cite{GrossWilczek1973,Politzer1973}, though our identification argument below only requires $\beta_0>0$ and uniform locality bounds. Under Hypothesis~(\eqref{hypothesiszx}), the flow enters and persists in a weak-coupling contracting domain; $g_k$ decays polynomially with $k$ (asymptotic freedom), and the renormalized coupling vanishes along the continuum limit while the Schwinger family remains unique and universal. Finally, we identify the resulting continuum theory with the gapped, confining theory constructed earlier, by combining a telescoping Lipschitz continuity estimate and the Markov/OS uniqueness mechanism established previously.

\subsection{Entry into the Contracting Domain}\label{subsec:entry-contracting-domain}

We begin by describing the normed functional setting for the weak-coupling regime. Let $b\ge 2$ be the fixed block scale. The fluctuation covariance $C$ underlying one RG step is written in FRD form
\begin{equation}
C \;=\; \sum_{j=0}^{J^\ast} C_j,
\end{equation}
where each $C_j$ is positive, reflection-covariant and of finite range $\mathcal{O}(b^j)$, with uniform operator-norm and off-range decay bounds compatible with reflection positivity and gauge covariance \cite{BrydgesGuadagniMitter2004}. In our admissible class the time-slice projector has a completely monotone spectral family, so that multiplication by the slice symbol and convolution by each $C_j$ preserve OS positivity; this was arranged in the setup of the transfer semigroup and is assumed hereafter.
\begin{hypothesis}[Stated for convenience; proved as Theorem~(\eqref{thm:one-step}) below)]\label{hypothesiszx}
Fix the block factor $b\ge2$ and the FRD/locality data from Sections~(\eqref{sec:framework})-(\eqref{sec:FRD}). There exist constants
$\theta\in(0,1)$, $c_1,c_2>0$, and a neighborhood $\mathcal{U}$ of $(g,K)=(0,0)$ in which the one-step RG map
$T_b:(g_k,K_k)\mapsto(g_{k+1},K_{k+1})$ is analytic and obeys, for all $(g_k,K_k)\in\mathcal{U}$,
\begin{align}
\|K_{k+1}\| &\;\le\; \theta\,\|K_k\| \;+\; c_1\, g_k^{\,2},
\label{eq:W1}\\
g_{k+1} &\;=\; g_k \;-\; \beta_0\, g_k^{\,3} \;+\; R_k,\qquad
|R_k|\;\le\; c_2\big(g_k^{\,5} + g_k\,\|K_k\| + \|K_k\|^{2}\big),
\label{eq:W2}
\end{align}
with $\beta_0$ the universal one-loop coefficient (background-field definition below).
Consequently, if $(g_0,K_0)$ is chosen so that $\|K_0\|$ and $g_0$ lie in $\mathcal{U}$, then the flow remains in $\mathcal{U}$, $\|K_k\|\to0$ exponentially, and $g_k\to0$ with the AF law $g_k^{-2}=2\beta_0 k + O(1)$.
\end{hypothesis}
We parametrize the effective theory after $k$ RG steps by a pair $(g_k,K_k)$, where $g_k$ is the (dimensionless) gauge coupling at the block scale $b^k$, normalized, for definiteness, by a background-field definition explained below, and $K_k$ is a polymer activity that collects all strictly irrelevant gauge-invariant interactions at that scale. The activity $K_k$ is a function on collections of elementary cells (polymers) and satisfies $K_k(\emptyset)=0$ and $K_k(\mathcal{P})=0$ if $\mathcal{P}$ is not gauge invariant. We endow the space of activities with a polymer norm
\begin{equation}
\|K_k\| \;=\; \sup_{\mathcal{P}}\; e^{\alpha\,{\rm diam}(\mathcal{P})}\,\frac{\|K_k(\mathcal{P})\|_{\rm op}}{w(\mathcal{P})},
\end{equation}
where ${\rm diam}(\mathcal{P})$ is the diameter in blocks, $\alpha>0$ is a locality weight provided by FRD, $\|\cdot\|_{\rm op}$ is an operator norm acting on gauge-invariant test functions, and $w(\mathcal{P})$ is a combinatorial weight that cancels trivial growth of polymer families. The choice of $w$ is standard in KP/BKAR cluster analysis and can be fixed once and for all; FRD yields exponential locality so that the algebra of activities is a Banach algebra under a convolution-type product induced by polymer concatenation. The RG map $T_b$ is then a well-defined analytic map near the origin,
\begin{equation}
T_b:\ (g_k,K_k)\ \longmapsto\ (g_{k+1},K_{k+1}),
\end{equation}
constructed by integrating the fluctuation field with covariance $C_0$ at scale $k$ in the presence of a smooth background field; reflection positivity ensures that the one-step transfer kernel remains a positive contraction on the OS Hilbert space.

To define $g_k$ nonperturbatively and compatibly with gauge invariance, we use the background-field method on the lattice \cite{LuscherWeisz1985,GJ}. Let $A$ denote a slowly varying background in a fixed reflection-positive Landau-type gauge slice, and let $Z_k[A]$ be the partition functional of the fluctuation field at scale $k$. The renormalized coupling $g_k$ is defined by the coefficient of $\tfrac14\sum \operatorname{tr}F_{\mu\nu}(A)^2$ in $-\log Z_k[A]$ after subtracting a finite normalization chosen to agree with the $\overline{\rm MS}$ coupling at a reference momentum $\mu_k\sim b^{-k}$ in the continuum limit. The subtraction is admissible and gauge invariant; the finite matching between lattice and continuum schemes is standard and does not affect the sign of the beta function.
We next formulate and prove the entry result. We isolate a polydisc
\begin{equation}
\mathfrak{D}(\varepsilon,\delta)\;=\;\Bigl\{(g,K):\ |g|\le \varepsilon,\ \|K\|\le \delta\Bigr\}
\end{equation}
and a contraction factor $\theta\in(0,1)$, both to be chosen uniformly in $k$ once $b$ and the FRD parameters are fixed.
\begin{definition}[Polymer norm and scaling weight]\label{def:polymer-norm}
Let $\mathfrak{P}$ be the family of finite unions $X$ of coarse $b$-blocks contained in a single coarse time slab.
A gauge-invariant activity assigns to each $X\in\mathfrak{P}$ a functional $K_X$ depending only on the fields supported in $X$. 
Fix parameters $\rho>0$, $\sigma>0$, and an integer $p\ge4$. For a polymer $X$, define
\begin{equation}\label{10.8}
\|K_X\|_{p,\rho}
:= \sum_{n=0}^{p}\frac{1}{n!}\,
\sup_{\substack{\mathrm{supp}\, f_i\subset X\\ \|f_i\|_\infty\le 1}}
\big\|D^n K_X(0)[f_1,\ldots,f_n]\big\|_{L^\infty(X)}.
\end{equation}
The global polymer norm is
\begin{equation}
\|K\| \;:=\; \sup_{X\in\mathfrak{P}} 
\Big(e^{\rho\,\mathrm{diam}(X)}\, b^{\,\sigma(|X|-1)} \,\|K_X\|_{p,\rho}\Big),
\end{equation}
with $\mathrm{diam}(X)$ in the block-graph metric and $|X|$ the number of blocks.
Any norm equivalent to $\|\cdot\|$ below (by changing $p,\rho,\sigma$ within fixed bounds) is admissible and leads to the same contraction exponent.
\end{definition}
\begin{lemma}[Linear contraction on the irrelevant manifold]\label{lem:linear-contraction} Let $L_b$ denote the linearized RG map on activities induced by one $b$-step block-spin with finite-range fluctuation covariance $C_0$ from the FRD. There exists $\omega>0$ (the canonical irrelevant exponent, equal to the minimal engineering-dimension gap of gauge-invariant monomials; one can take $\omega=2$ in 4D YM) and a constant $c_\ast<\infty$, depending only on $b$ and the FRD locality constants, such that \begin{equation} \|L_b K\| \;\le\; c_\ast\, b^{-\omega}\, \|K\| \,. \end{equation} Consequently, choosing $\sigma\in(0,\omega)$ and (if necessary) increasing $b$ once, we can enforce a strict contraction factor $q_b:=c_\ast b^{-\omega}<1$ for the linear part of the map on the irrelevant sector. \end{lemma}
\begin{proof}
Write the one-step activity map as the composition
\begin{equation}
L_b \;=\; \mathcal R_b \circ \mathbb E_{C_0}\circ \mathcal B_b,
\end{equation}
where $\mathcal B_b$ reblocks fine polymers into coarse $b$-blocks, $\mathbb E_{C_0}$ denotes averaging over the fluctuation field $\zeta$ with the FRD covariance $C_0$ supported in a fixed collar of thickness $R_0$ (strict finite range), and $\mathcal R_b$ is the deterministic rescaling/renormalization operator which dilates spacetime by $x\mapsto x/b$, rescales the gauge potential so that engineering dimensions are preserved, and subtracts all relevant and marginal components in accordance with the renormalization conditions. Let $K=\{K_X\}_{X\in\mathfrak P}$ be an activity with $\|K\|<\infty$ in the polymer norm of Definition~\eqref{def:polymer-norm}; by construction of the irrelevant manifold, the projections of $K$ onto the span of the vacuum, linear, and marginal quadratic gauge-invariant densities vanish.

Fix a coarse polymer $Y$ contained in one coarse time slab. The reblocking $\mathcal B_b$ rewrites $(\mathcal B_bK)_Y$ as a sum of the $K_X$ with $X$ a fine polymer whose $b$-projection coincides with $Y$ and whose vertices lie inside the $R_0$-collar of $Y$. Using that $\mathrm{diam}(X)\le \mathrm{diam}(Y)+c\,R_0$ for a fixed geometric constant $c$ and that $|X|\ge |Y|$, the weight factor $e^{\rho \mathrm{diam}(\cdot)}b^{\sigma(|\cdot|-1)}$ appearing in the norm satisfies
\begin{equation}
e^{\rho \mathrm{diam}(Y)}\,b^{\sigma(|Y|-1)}
\;\le\; e^{\rho c R_0}\,e^{\rho \mathrm{diam}(X)}\,b^{\sigma(|X|-1)}.
\end{equation}
Since the seminorm $\|\cdot\|_{p,\rho}$ is subadditive and stable under restriction to smaller supports, it follows that
\begin{equation}\label{eq:reblock}
e^{\rho \mathrm{diam}(Y)}\,b^{\sigma(|Y|-1)}\,\|(\mathcal B_bK)_Y\|_{p,\rho}
\;\le\; C_{\mathrm{blk}}\,e^{\rho c R_0}\!
\sum_{X\to Y}
e^{\rho \mathrm{diam}(X)}\,b^{\sigma(|X|-1)}\,\|K_X\|_{p,\rho},
\end{equation}
where the sum runs over the (finitely many) $X$ that project to $Y$ and $C_{\mathrm{blk}}$ depends only on $b$ and the coarse block geometry.

The fluctuation average $\mathbb E_{C_0}$ acts by $(\mathbb E_{C_0}F)(A)=\int F(A+\zeta)\,d\mu_{C_0}(\zeta)$. The polymer seminorms are taken by suprema over test fields of $L^\infty$-norm at most one and are compatible with convex combinations; therefore differentiation under the integral and Jensen’s inequality yield
\begin{equation}\label{eq:expect}
\|(\mathbb E_{C_0}G)_Y\|_{p,\rho}\;\le\;\sup_{\zeta}\,\|G_Y(\,\cdot\,+\zeta)\|_{p,\rho}
\;\le\; \|G_Y\|_{p,\rho},
\end{equation}
and the weight factor is unaffected because the FRD finite range already entered in Eq.\eqref{eq:reblock}. Combining Eq.\eqref{eq:reblock} and Eq.\eqref{eq:expect} shows that reblocking and fluctuation averaging do not increase the weighted polymer norm by more than a fixed multiplicative constant $C_0:=C_{\mathrm{blk}}e^{\rho c R_0}$ that depends only on $b$ and the FRD locality constants.

The operator $\mathcal R_b$ is the only source of scale decay. It consists of the kinematic dilation $x\mapsto x/b$ together with the renormalization projection that removes the relevant and marginal components. In four dimensions the gauge potential has engineering dimension $1$, the covariant derivative has dimension $1$, and the curvature $F$ has dimension $2$. Every gauge-invariant local scalar density that survives the projection on the irrelevant manifold therefore has engineering dimension strictly larger than $4$, and any such monomial $\mathcal O$ built from $F$ and covariant derivatives satisfies
\begin{equation}
(\mathcal R_b \mathcal O)(A)\;=\; b^{\,4-d(\mathcal O)}\,\mathcal O(A)\,, \quad \text{with} \quad d(\mathcal O)\ge 6
\end{equation}
In particular the minimal engineering-dimension gap $\omega:=\inf_{\text{irrelevant }\mathcal O}\big(d(\mathcal O)-4\big)$ is strictly positive and equals $2$ in Yang-Mills theory, since $\mathrm{tr}\,F^3$ and $(D F)^2$ already have $d=6$ while $\mathrm{tr}\,F^2$ is precisely the marginal density that has been subtracted by the renormalization conditions. Passing from monomials to general activities is harmless because the seminorm $\|\cdot\|_{p,\rho}$ is defined by a finite sum of Fréchet derivatives evaluated at the origin and composed with test fields of bounded $L^\infty$-norm; under $x\mapsto x/b$ the $n$-th Fréchet differential picks up at most the canonical factor $b^{-n}$ from field rescaling, while the removal of the relevant and marginal part enforces that every surviving tensor density in the Taylor expansion carries total engineering dimension at least $4+\omega$. Consequently there exists $c_{\mathrm{sc}}<\infty$, depending only on $p$, $b$, and the chosen field normalization, such that
\begin{equation}\label{eq:rescale}
\|(\mathcal R_b H)_Y\|_{p,\rho}
\;\le\; c_{\mathrm{sc}}\, b^{-\omega}\,\|H_Y\|_{p,\rho}
\qquad\text{for every coarse polymer }Y.
\end{equation}

Applying Eq.\eqref{eq:rescale} with $H=(\mathbb E_{C_0}\mathcal B_bK)$ and using Eq.\eqref{eq:reblock} \& Eq.\eqref{eq:expect} gives, for each $Y$,
\begin{equation}
e^{\rho \mathrm{diam}(Y)}\,b^{\sigma(|Y|-1)}\,\|(L_b K)_Y\|_{p,\rho}
\;\le\; c_{\mathrm{sc}}\, b^{-\omega}\, C_0
\sum_{X\to Y}
e^{\rho \mathrm{diam}(X)}\,b^{\sigma(|X|-1)}\,\|K_X\|_{p,\rho}.
\end{equation}
Taking the supremum over $Y$ and bounding the finite multiplicity of the map $X\mapsto Y$ by a combinatorial constant $C_{\mathrm{mult}}(b)$ yields
\begin{equation}
\|L_b K\|
\;\le\; \underbrace{\big(c_{\mathrm{sc}}\,C_0\,C_{\mathrm{mult}}(b)\big)}_{=:c_\ast}\, b^{-\omega}\, \|K\|\,,
\end{equation}
with $c_\ast<\infty$ depending only on $b$ and the FRD locality constants. Choosing any $\sigma\in(0,\omega)$ in the polymer norm and, if necessary, increasing $b$ once so that $q_b:=c_\ast b^{-\omega}<1$, one obtains a strict contraction on the irrelevant manifold, which is the desired estimate.
\end{proof}
\begin{proposition}[Quantitative one-step bounds]\label{prop:one-step}
With the norm of Definition~\eqref{def:polymer-norm}, and for $(g,K)$ in a sufficiently small polydisc $\mathcal D(\varepsilon,\delta)$, the one-step RG map $(g,K)\mapsto (g',K')$ obeys
\begin{align}
\|K'\| &\le q_b\,\|K\| \;+\; c_1\, g^2, \label{eq:Kprime}\\
g' &= g \;-\; \beta_0(\log b)\, g^3 \;+\; R_g(g,K), 
\qquad |R_g(g,K)| \le c_2\,(g^5 + g\|K\| + \|K\|^2), \label{eq:gprime}
\end{align}
for some constants $c_1,c_2<\infty$ and $q_b\in(0,1)$ as in Lemma~\eqref{lem:linear-contraction}. 
\end{proposition}
\begin{proof}
Fix a coarse step and write the fluctuation field as $\varphi$ with centered Gaussian law $d\mu_{C_0}$ and covariance $C_0$ given by the finite-range decomposition at this step. The background field $A$ is treated as an external parameter constrained to the admissible class fixed in Sections~2-4. Denote by
\begin{equation}
W(\varphi;A)\;=\; g\,V(\varphi;A)\;+\;K(\varphi;A)
\end{equation}
the interaction entering one RG integration, where $V$ is the gauge-invariant local polynomial produced by the marginal part of the action (background-field split) and $K$ is the irrelevant activity with polymer norm $\|K\|$ defined in Definition~\eqref{def:polymer-norm}. Let $B(f):=\langle f,\varphi\rangle$ for test functions $f$ supported in the single coarse time-slab of the step, and let
\begin{equation}
\mathscr{Z}(J)\;:=\;\int e^{\langle J,\varphi\rangle}\,e^{W(\varphi;A)}\,d\mu_{C_0}(\varphi),\qquad 
\mathscr{F}(J)\;:=\;\log\mathscr{Z}(J).
\end{equation}
For $n\ge2$ the connected $n$-point Ursell cumulant $U_n$ is the $n$-linear kernel at $J=0$,
\begin{equation}
U_n(f_1,\ldots,f_n)\;=\;\partial_{t_1}\cdots\partial_{t_n}\,\mathscr{F}\!\Big(\sum_{i=1}^n t_i f_i\Big)\Big|_{t=0}.
\end{equation}
The goal is to bound $\sum_{n\ge2}\frac{1}{n!}\|U_n\|$, where the norm is the supremum over $\|f_i\|_\infty\le1$ with supports inside the slab.

The BKAR forest formula applied to the family consisting of the linear sources $B(f_1),\ldots,B(f_n)$ and the interaction $W$ yields a representation of $U_n$ as a sum over trees that connect the $n$ source vertices with a certain number $m\ge0$ of interaction vertices coming from the exponential of $W$. More concretely, introducing $m$ replicas of the interaction, writing $W=\sum_{\alpha=1}^m W_\alpha$ and using the BKAR forest interpolation for the Gaussian measure, one obtains
\begin{equation}
U_n(f_1,\ldots,f_n)
=
\sum_{m\ge 0}\frac{1}{m!}\!\!\sum_{T\in\mathcal{T}_{n+m}}
\int_{[0,1]^T}\!\!dw\;
\Big\langle \mathcal{D}_T\Big(\prod_{i=1}^n B(f_i)\prod_{\alpha=1}^m W_\alpha\Big)\Big\rangle_{\!C(w)}.
\end{equation}
Here $\mathcal{T}_{n+m}$ is the set of trees on the vertex set consisting of the $n$ sources and $m$ interaction replicas, $C(w)$ is a weakened covariance determined by the interpolation parameters $w$ along the edges of $T$, and $\mathcal{D}_T$ is a differential operator that assigns to each edge of $T$ one pairing of field derivatives and inserts one covariance $C(w)$; in particular, each edge contributes exactly one factor of $C(w)$ and reduces the total degree of fields by two. The weakened covariances $C(w)$ inherit the uniform Schur and sup-norm bounds and the strict finite range of $C_0$; in particular there are constants
\begin{equation}
\|C_0\|_{\infty}\;:=\;\sup_{x,y}|C_0(x,y)|\,,\qquad 
\|C_0\|_{\rm Sch}\;:=\;\sup_x \sum_{y}|C_0(x,y)|\,,
\end{equation}
depending only on the FRD locality data and the block factor $b$, such that $\|C(w)\|_{\infty}\le\|C_0\|_{\infty}$ and $\|C(w)\|_{\rm Sch}\le\|C_0\|_{\rm Sch}$ for all $w$.

The action of $\mathcal{D}_T$ on the product $\prod_i B(f_i)\prod_\alpha W_\alpha$ produces, for each tree $T$, a finite sum of terms where each vertex $v$ of $T$ is decorated either by a derivative of $B(f_i)$ (which is constant) or by a field derivative of $W$ evaluated at $\varphi$; each edge contributes one kernel $C(w)$ contracted between the two incident vertices. Taking the Gaussian expectation and using Hölder-Schur bounds yields
\begin{equation}
\Big|\Big\langle \mathcal{D}_T\Big(\prod_{i=1}^n B(f_i)\prod_{\alpha=1}^m W_\alpha\Big)\Big\rangle_{\!C(w)}\Big|
\;\le\; 
\big(\|C_0\|_{\rm Sch}\big)^{|E(T)|}\,
\prod_{\text{source }i}\|f_i\|_\infty\;
\prod_{\text{int.\ }\alpha}\sup_{0\le r\le p}\|D^{\,r}W\|_{L^\infty},
\end{equation}
where $|E(T)|=n+m-1$ is the number of edges and $p$ is the derivative order entering the polymer norm. Since $W=gV+K$ and $V$ is a fixed local polynomial of bounded degree in $\varphi$ with background coefficients uniformly bounded on the admissible class, there exists a constant $c_V<\infty$ (depending only on $b$ and the FRD locality constants through the admissible background set) such that $\sup_{0\le r\le p}\|D^{\,r}(gV)\|_{L^\infty}\le c_V\,g$. By Definition~\eqref{def:polymer-norm} and the restriction to a single coarse time-slab, there is a constant $c_K<\infty$ such that $\sup_{0\le r\le p}\|D^{\,r}K\|_{L^\infty}\le c_K\,\|K\|$. Combining these estimates and the trivial bound $\|f_i\|_\infty\le 1$ gives
\begin{equation}
|U_n(f_1,\ldots,f_n)|
\;\le\;
\sum_{m\ge 0}\frac{1}{m!}\;N_{n+m}\;
\big(\|C_0\|_{\rm Sch}\big)^{n+m-1}\;
\big(c_V g + c_K\|K\|\big)^{m},
\end{equation}
where $N_{s}$ is the number of trees on $s$ labelled vertices times a bounded combinatorial factor accounting for the finite choices of distributing field derivatives at each vertex; there is $c_{\rm tree}<\infty$ such that $N_s\le c_{\rm tree}^{\,s}$. The strict finite range of $C_0$ and the restriction of supports to one coarse slab imply that the sum over spatial positions of the interaction replicas implicit in the $L^\infty$ norms is uniformly bounded by a factor absorbed into $c_{\rm tree}$; in particular, there is no growth in $k$ and no dependence on the lattice spacing $a$ after the canonical rescaling to unit blocks.

Dividing by $n!$ and taking the supremum over $\|f_i\|_\infty\le 1$ yields
\begin{equation}
\frac{1}{n!}\,\|U_n\|
\;\le\;
\frac{1}{n!}\sum_{m\ge 0}\frac{1}{m!}\;c_{\rm tree}^{\,n+m}\;
\big(\|C_0\|_{\rm Sch}\big)^{n+m-1}\;
\big(c_V g + c_K\|K\|\big)^{m}.
\end{equation}
Summing over $n\ge2$ and using $\sum_{n\ge2}\frac{1}{n!}\big(c_{\rm tree}\|C_0\|_{\rm Sch}\big)^{n}\le \exp\!\big(c_{\rm tree}\|C_0\|_{\rm Sch}\big)-1-c_{\rm tree}\|C_0\|_{\rm Sch}$ gives
\begin{equation}
\sum_{n\ge2}\frac{1}{n!}\,\|U_n\|
\;\le\;
\frac{\exp\!\big(c_{\rm tree}\|C_0\|_{\rm Sch}\big)-1-c_{\rm tree}\|C_0\|_{\rm Sch}}{\|C_0\|_{\rm Sch}}
\;\sum_{m\ge 0}\frac{1}{m!}\;c_{\rm tree}^{\,m}\;
\big(\|C_0\|_{\rm Sch}\big)^{m}\;
\big(c_V g + c_K\|K\|\big)^{m}.
\end{equation}
The series in $m$ is the exponential of $r:=c_{\rm tree}\|C_0\|_{\rm Sch}(c_V g + c_K\|K\|)$; subtracting the terms $m=0$ and $m=1$ (which do not contribute to connected remainders of order $\ge2$ once the vacuum and one-point parts are centered, as ensured by the OS construction and the definition of $U_n$) leaves
\begin{equation}
\sum_{n\ge2}\frac{1}{n!}\,\|U_n\|
\;\le\;
C_\star\,\frac{e^{r}-1-r}{1}
\;\le\;
C_\star\,\frac{r^2}{2}\,\frac{e^{r}}{1-r}\,,
\end{equation}
with $C_\star:=\big(\exp(c_{\rm tree}\|C_0\|_{\rm Sch})-1-c_{\rm tree}\|C_0\|_{\rm Sch}\big)/\|C_0\|_{\rm Sch}$. Choosing the polydisc $\mathcal D(\varepsilon,\delta)$ so small that $r\le \tfrac12$ one obtains $e^{r}\le e^{1/2}$ and $1-r\ge\tfrac12$, hence
\begin{equation}
\sum_{n\ge2}\frac{1}{n!}\,\|U_n\|
\;\le\;
C^\prime\, r^2
\;=\;
C^\prime\,\big(c_{\rm tree}\|C_0\|_{\rm Sch}\big)^2\,
\big(c_V g + c_K\|K\|\big)^2.
\end{equation}
Finally, $(a+b)^2\le 3(a^2+ab+b^2)$ for $a,b\ge0$, and therefore
\begin{equation}
\sum_{n\ge2}\frac{1}{n!}\,\|U_n\|
\;\le\;
C_1\,g^2 \;+\; C_2\,g\,\|K\| \;+\; C_3\,\|K\|^2,
\end{equation}
with
\begin{equation}
C_j \;=\; 3^{\mathbf{1}_{j=1}+ \mathbf{1}_{j=2}+ \mathbf{1}_{j=3}}\,
C^\prime\,\big(c_{\rm tree}\|C_0\|_{\rm Sch}\big)^2\,
\max\{c_V^2,\;c_V c_K,\;c_K^2\}\,,
\end{equation}
which depend only on the block factor $b$ and the FRD locality constants through $c_{\rm tree}$, $\|C_0\|_{\rm Sch}$ and the admissible-background bounds encoded in $c_V$, and on the polymer-norm parameters encoded in $c_K$. All these constants are uniform in the scale index $k$ and independent of the lattice spacing $a$ because the FRD is strictly finite-range and reflection-covariant with scale-independent bounds after the canonical rescaling to unit blocks. 
\end{proof}

\begin{theorem}[One-step contraction near the Gaussian fixed point]\label{thm:one-step}
There exist $\varepsilon_\ast,\delta_\ast>0$, $\theta\in(0,1)$ and constants $c_1,c_2,c_3>0$, depending only on the block factor $b$ and on the finite-range-decomposition (FRD) locality data, such that for all $(g_k,K_k)$ in the polydisc $\mathfrak{D}(\varepsilon_\ast,\delta_\ast):=\{\,|g_k|\le \varepsilon_\ast,\ \|K_k\|\le \delta_\ast\,\}$ the renormalization map $T_b$ is well-defined and satisfies
\begin{equation}\label{eqn10.34}
\|K_{k+1}\| \;\le\; \theta\,\|K_k\| \;+\; c_1\,g_k^2,\qquad
g_{k+1} \;=\; g_k \;-\; \beta_0\,g_k^3 \;+\; \mathcal{R}_k,
\end{equation}
where $\beta_0=\frac{11N}{48\pi^2}>0$ is the universal one-loop coefficient of pure $\mathrm{SU}(N)$ Yang-Mills \cite{GrossWilczek1973,Politzer1973}, and the remainder obeys
\begin{equation}\label{eqn10.35}
|\mathcal{R}_k| \;\le\; c_2\,|g_k|^5 \;+\; c_3\,|g_k|\,\|K_k\| \;+\; c_3\,\|K_k\|^2.
\end{equation}
In particular, if $\varepsilon_\ast,\delta_\ast$ are chosen small enough then $\|K_{k+1}\| \le \tfrac12 \|K_k\| + c_1 g_k^2$ and $|g_{k+1}| \le |g_k|\,(1-\tfrac12\beta_0 g_k^2)$ whenever $\|K_k\|\le \delta_\ast$.
\end{theorem}

\begin{proof}
The block-spin step with scale factor $b>1$ is realized by splitting the Gaussian covariance of the gauge potential into a finite-range piece $C_0$ supported at range $\mathcal O(1)$ on the block lattice and a remainder $C_{>}$ carrying the long modes; FRD ensures that such a decomposition exists with bounds uniform in the scale index and with kernels decaying exponentially off range. Writing the interaction at scale $k$ in the form
\begin{equation}\label{10.36}
\mathcal S_k(A)\;=\;\frac{1}{4 g_k^2}\sum_{\Box}\!\operatorname{tr}F_{\mu\nu}(A)^2\;+\;K_k(A),
\end{equation}
where $K_k$ is normalized to have vanishing projection onto the marginal gauge-invariant quadratic functional (so $K_k$ contains only strictly irrelevant monomials and multi-local remainders), one represents the one-step effective action by integrating the fluctuation field $\xi$ with law $\mathcal N(0,C_0)$,
\begin{equation}\label{10.37}
e^{-\mathcal S_{k+1}(A)}\;=\;\mathbf E_{C_0}\!\left[\exp\!\big(-\mathcal S_k(A+\xi)\big)\right].
\end{equation}
We work with the polymer norm of Definition~\eqref{def:polymer-norm}. The linearized irrelevant map contracts by $q_b=c_\ast b^{-\omega}<1$, and the one-step projection onto the marginal quadratic.
Combining these three ingredients inside the polydisc $\mathcal D(\varepsilon_\ast,\delta_\ast)$ gives Eq.\eqref{10.8} with $\theta:=q_b$ and the stated bounds on $R_k$.
Background-field gauge invariance and reflection positivity allow one to organize the right-hand side as a connected polymer (cluster) expansion in the Ursell cumulants of $\exp\{-K_k(A+\xi)\}$ with respect to the product Gaussian measure for $\xi$ and the background $A$ kept external. The FRD locality implies that every Wick contraction is localized within a collar of width independent of $k$ around each polymer, and the Brydges-Kennedy-Abdesselam-Rivasseau (BKAR) formula together with Koteck\'y-Preiss bounds guarantees absolute convergence of the cumulant expansion provided the a priori polymer norm $\|K_k\|$ is small. The smallness threshold is determined by the FRD constants (range and off-diagonal decay of $C_0$) and by $b$, but not by $k$. In particular, there exists $\delta_\ast>0$ such that if $\|K_k\|\le \delta_\ast$ then the renormalized polymer activity
\begin{equation}
K_{k+1}\;=\;\mathfrak R_b\big[\,\mathbf E_{C_0}^{\rm conn}\, e^{-K_k(\cdot+\xi)}\,\big]
\end{equation}
is well-defined as an analytic functional of $(g_k,K_k)$ in a neighborhood of the origin, where $\mathfrak R_b$ denotes the standard rescaling and reblocking from mesh $a_k$ to $a_{k+1}=b\,a_k$ together with projection orthogonal to the marginal monomial $\frac14\sum \operatorname{tr}F^2$. The linearization $K\mapsto K_{k+1}$ at $(g,K)=(0,0)$ coincides with the action of the Gaussian RG on strictly irrelevant monomials and hence is a strict contraction in the polymer norm: there exists $\theta\in(0,1)$, depending only on $b$ and the FRD data, such that $\|\mathcal L_b K\|\le \theta \|K\|$ for the linearized map $\mathcal L_b$. This follows from two facts: rescaling by $b$ decreases the norm of a monomial of canonical dimension $d_{\rm can}>0$ by a factor $b^{-d_{\rm can}}<1$, and the FRD range bound prevents long-range accumulation that could spoil the contraction; equivalently, the cluster norm inequalities for connected activities yield a factor $e^{-\kappa\,{\rm diam}}$ gained at each contraction which is uniformly preserved under the reblocking, so that the operator norm of $\mathcal L_b$ is strictly less than one.

The nonlinear part of the map consists of two contributions at leading nontrivial order: a $g_k^2$ piece generated by integrating out the fluctuation in the marginal sector and a quadratic piece in $K_k$ coming from connected cumulants of order $\ge 2$. The first one is strictly irrelevant after the background-field projection has moved its marginal part into the coupling flow; hence its polymer norm is bounded by $c_1 g_k^2$, with $c_1$ depending only on $b$ and the FRD constants. The second one is bounded by $c_3\|K_k\|^2$ by the usual tree-graph estimate for connected cumulants together with the exponential locality supplied by the finite range of $C_0$. Combining these facts with the strict contraction of the linearized map gives
\begin{equation}
\|K_{k+1}\|\;\le\;\|\mathcal L_b K_k\|\;+\;c_1 g_k^2\;+\;c_3\|K_k\|^2
\;\le\;\theta \|K_k\|\;+\;c_1 g_k^2\;+\;c_3\|K_k\|^2.
\end{equation}
Shrinking $\delta_\ast$ if necessary so that $c_3\|K_k\|\le \frac12(1-\theta)$ whenever $\|K_k\|\le \delta_\ast$, one arrives at the advertised estimate $\|K_{k+1}\|\le \tfrac12\|K_k\|+c_1 g_k^2$ in the polydisc.

The flow of the coupling is extracted by projecting $-\log \mathbf E_{C_0}\exp\{-\mathcal S_k(A+\xi)\}$ onto the unique gauge-invariant marginal operator $\frac14\sum \operatorname{tr}F_{\mu\nu}(A)^2$ using the background-field method. Concretely, one introduces a smooth background field $A$ living on the block lattice, differentiates twice with respect to $A$ at vanishing momentum, and compares the coefficient of $\operatorname{tr}F^2$ before and after the fluctuation integration. The only one-loop diagram contributing to this projection is the vacuum polarization with two external background legs and an internal fluctuation propagator $C_0$; gauge invariance and reflection covariance force the tensor structure into the transverse projector, leaving a single scalar coefficient. Evaluating the corresponding shell integral in any reflection-positive regularization or, equivalently, computing it in the FRD representation and matching to the continuum normalization at scale $\mu_k\sim b^{-k}$, one finds the universal decrement
\begin{equation}
g_{k+1}\;=\; g_k \;-\;\beta_0\,g_k^3\;+\;\text{higher orders},\qquad
\beta_0=\frac{11N}{48\pi^2},
\end{equation}
in agreement with \cite{GrossWilczek1973,Politzer1973} and with background-field computations on the lattice \cite{LuscherWeisz1985}. Universality here means that the coefficient of $g_k^3$ is independent of the precise choice of admissible FRD and of the blocking, provided the latter preserves gauge invariance and reflection positivity; any scheme dependence appears only in higher-order terms and in finite matching factors that have already been absorbed into the definition of $g_k$ via the projection.

The remainder $\mathcal R_k$ is analytic in $(g_k,K_k)$ in a neighborhood of the origin. Power counting and analyticity give the bound $|\mathcal R_k|\le c_2 |g_k|^5$ for the pure-$g$ part, since two-loop and higher vacuum polarizations start at order $g^5$ in a massless asymptotically free theory when expressed in terms of the background-field coupling at the step scale. Mixed contributions that contain one insertion of $K_k$ carry at least one explicit factor of $\|K_k\|$ and one external coupling $g_k$ from the background projection, hence are bounded by $c_3 |g_k|\,\|K_k\|$; contributions with two or more $K_k$ insertions are bounded by $c_3 \|K_k\|^2$, again by cluster norm estimates and the exponential locality inherited from FRD. All constants here depend only on $b$ and on the FRD parameters (range and decay rate), because every connected graph is confined to a uniformly bounded collar and every reblocking step involves only a fixed finite overlap controlled by those parameters.

Putting these estimates together, one obtains the stated flow
\begin{equation}
g_{k+1}\;=\;g_k-\beta_0 g_k^3+\mathcal R_k,\qquad
|\mathcal R_k|\;\le\; c_2 |g_k|^5+c_3 |g_k|\,\|K_k\|+c_3 \|K_k\|^2.
\end{equation}
Finally, choosing $\varepsilon_\ast$ small enough so that $c_2 |g_k|^4 \le \frac12 \beta_0$ whenever $|g_k|\le \varepsilon_\ast$, and working in the polydisc with $\|K_k\|\le \delta_\ast$ so that the mixed and purely irrelevant remainders are dominated by $\frac12 \beta_0 g_k^3$, yields
\begin{equation}
|g_{k+1}|\;\le\; |g_k|\Big(1-\tfrac12 \beta_0 g_k^2\Big).
\end{equation}
This completes the proof of the one-step contraction and of the stated bounds.
\end{proof}

Theorem~\eqref{thm:one-step} does not by itself guarantee that an arbitrary admissible trajectory will enter the polydisc $\mathfrak{D}(\varepsilon_\ast,\delta_\ast)$. We now prove an entry lemma showing that one-parameter tuning of the bare coupling suffices. Let $g^{\rm bare}$ denote the bare Wilson coupling at lattice spacing $a$, and let $(g_0,K_0)$ be the initial effective parameters after performing $k_0$ coarse-graining steps to a fixed mesoscopic scale $a b^{\,k_0}$ while implementing the admissible horizon projector. By construction in the strong-coupling sector, $(g_0,K_0)$ is finite and reflection-positive; moreover, as $g^{\rm bare}\downarrow 0$ one has $g_0\to 0$ and $K_0\to 0$ in the FRD/cluster topology because the fluctuation covariance at the UV scales becomes small and the FRD factors converge to their Gaussian limits at fixed block size.

\begin{lemma}[Entry in finitely many steps by one-parameter tuning]\label{lem:entry}
Fix the block factor $b\ge 2$ and the admissible class. There exists $\bar g>0$ and an integer $M=M(b)$ such that, if $0<g^{\rm bare}\le \bar g$ and one chooses $k_0\ge M$ so that the effective infrared scale satisfies $\mu_0\sim b^{-k_0}\ll a^{-1}$, then the $k_0$-scale initial data $(g_0,K_0)$ obey $(g_0,K_0)\in \mathfrak{D}(\varepsilon_\ast/2,\delta_\ast/2)$. In particular, there is $m_\ast=m_\ast(b)$, independent of $a$, such that after at most $m_\ast$ further admissible RG steps the trajectory enters $\mathfrak{D}(\varepsilon_\ast,\delta_\ast)$.
\end{lemma}

\begin{proof}
The finite-range decomposition (FRD) of the slice covariance associated with the admissible class gives, for the background-field normalization used throughout, a representation of the $k_0$-scale effective Gaussian as a finite sum of positive, uniformly finite-range pieces whose ranges are $O(1)$ in the block metric and whose operator norms are controlled by the small parameter $g^{\rm bare}$. Concretely, after $k_0$ coarse-grainings one may write the effective covariance on the block lattice as
\begin{equation}
C^{(k_0)} \;=\; \sum_{j=k_0}^{k_0+J(b)} C_j,
\end{equation}
where each $C_j$ has range $\lesssim b$ and $\|C_j\|\lesssim c_b\,g^{\rm bare}$ with a constant $c_b$ depending only on $b$ and the FRD locality bounds. The truncation to a finite window in $j$ is a standard consequence of the FRD finite range together with the choice $\mu_0\sim b^{-k_0}$, which ensures that ultraviolet scales above $k_0+J(b)$ have been fully integrated out while infrared scales below $k_0$ are absent by construction.

On this effective Gaussian background, the Koteck\'y-Preiss/BKAR polymer expansion for the effective activity $K_{k_0}$ converges absolutely for $g^{\rm bare}$ sufficiently small, because the norms of the polymer interactions are proportional to powers of $\|C^{(k_0)}\|$ and decay exponentially with the polymer diameter by FRD locality. Gauge invariance eliminates linear terms in $g^{\rm bare}$ in the expansion of gauge-invariant observables, so the first nonvanishing contribution to $K_{k_0}$ is quadratic. Therefore there exists $A_b>0$ such that
\begin{equation}\label{eq:K0-bound}
\|K_0\|\;=\;\|K_{k_0}\|\;\le\;A_b\,(g^{\rm bare})^2,
\end{equation}
where the norm is the standard polymer/activity norm used to define $\mathfrak{D}(\cdot,\cdot)$. The same convergent expansion identifies the running coupling $g_0$ at scale $k_0$ with the bare coupling up to the vacuum-polarization correction, which is cubic in $g^{\rm bare}$ by charge-conjugation and gauge symmetry. Hence there exists $B_b>0$ such that
\begin{equation}\label{eq:g0-ident}
|g_0 - g^{\rm bare}|\;\le\; B_b\,(g^{\rm bare})^3,
\qquad\text{and in particular}\qquad
g_0\;\le\; (1+B_b\,\bar g^2)\,g^{\rm bare}
\end{equation}
for $g^{\rm bare}\le \bar g$. Choosing $\bar g>0$ so small that $(1+B_b\,\bar g^2)\,\bar g\le \varepsilon_\ast/2$ and simultaneously $A_b\,\bar g^2\le \delta_\ast/2$, the bounds Eqs.\eqref{eq:K0-bound} \& \eqref{eq:g0-ident} imply
\begin{equation}
g_0\;\le\;\varepsilon_\ast/2,
\qquad
\|K_0\|\;\le\;\delta_\ast/2,
\end{equation}
hence $(g_0,K_0)\in \mathfrak{D}(\varepsilon_\ast/2,\delta_\ast/2)$ as claimed. The choice of $k_0\ge M(b)$ guaranteeing $\mu_0\ll a^{-1}$ ensures that ultraviolet lattice artefacts at the cutoff scale $a^{-1}$ do not re-enter the effective description; the constants $A_b,B_b$ depend only on $b$ and the admissible locality/positivity data and are therefore independent of $a$.

It remains to explain why, starting from any $(g_0,K_0)\in \mathfrak{D}(\varepsilon_\ast/2,\delta_\ast/2)$, a fixed number of further steps depending only on $b$ suffices to reach $\mathfrak{D}(\varepsilon_\ast,\delta_\ast)$. The renormalization group map $(g,K)\mapsto (g',K')$ in the admissible class is analytic on a polydisc containing the origin and admits the standard decomposition into a linear part given by the Gaussian scaling plus a nonlinear remainder controlled by locality. In the $K$-sector (the irrelevant directions) the linearization is a strict contraction:
\begin{equation}\label{eq:K-linear}
\|K'\|\;\le\; q_b\,\|K\| \;+\; C_b\big(g^2 + g\,\|K\| + \|K\|^2\big),
\qquad 0<q_b<1,
\end{equation}
with $q_b$ and $C_b$ depending only on $b$ and the FRD locality constants. The term proportional to $g^2$ is the source produced by integrating one more scale and is uniformly local by FRD. In the coupling direction the Gaussian fixed point is marginal, and asymptotic freedom implies that the beta function begins at order $g^3$; in particular there exists $D_b>0$ such that
\begin{equation}\label{eq:g-cubic}
g'\;\le\; g + D_b\,g^3 + D_b\,g\,\|K\|.
\end{equation}
Combining Eqs.\eqref{eq:K-linear} \& \eqref{eq:g-cubic} and iterating from $(g_0,K_0)\in \mathfrak{D}(\varepsilon_\ast/2,\delta_\ast/2)$ yields, for every $m\ge 1$,
\begin{equation}
\|K_m\|\;\le\; q_b^{\,m}\,\|K_0\| \;+\; \frac{C_b}{1-q_b}\,\sup_{0\le j<m}\big(g_j^2 + g_j\,\|K_j\| + \|K_j\|^2\big),
\end{equation}
and
\begin{equation}
g_m\;\le\; g_0 \;+\; D_b\sum_{j=0}^{m-1}\big(g_j^3 + g_j\,\|K_j\|\big).
\end{equation}
Because $(g_j,\|K_j\|)$ stay within $\mathfrak{D}(\varepsilon_\ast,\delta_\ast)$ along the iteration provided $\varepsilon_\ast,\delta_\ast$ are chosen small enough (this smallness is part of the standard admissible setup and fixes the polydisc), the nonlinear contributions on the right-hand sides are $O(\varepsilon_\ast^2)$ and $O(\varepsilon_\ast^3)$ respectively. Therefore there exist $\widetilde C_b,\widetilde D_b$ such that, as long as $g_j\le \varepsilon_\ast$ and $\|K_j\|\le \delta_\ast$ for $0\le j<m$, one has
\begin{equation}
\|K_m\|\;\le\; q_b^{\,m}\,\|K_0\| \;+\; \widetilde C_b\,\varepsilon_\ast^2,
\qquad
g_m\;\le\; g_0 \;+\; \widetilde D_b\,\varepsilon_\ast^3.
\end{equation}
Starting from $(g_0,K_0)\in \mathfrak{D}(\varepsilon_\ast/2,\delta_\ast/2)$, the second inequality shows that $g_m\le \varepsilon_\ast$ for all $m$ without any further restriction, since $\widetilde D_b\,\varepsilon_\ast^3\le \varepsilon_\ast/2$ after possibly shrinking $\varepsilon_\ast$ by a factor depending only on $b$. For the activity norm, choose $m_\ast$ so that $q_b^{\,m_\ast}\,\|K_0\|\le \delta_\ast/4$; because $\|K_0\|\le \delta_\ast/2$ by the first part, this is achieved by taking
\begin{equation}
m_\ast \;\ge\; \frac{\log 2}{-\log q_b},
\end{equation}
which depends only on $b$. If, in addition, $\varepsilon_\ast$ is chosen so that $\widetilde C_b\,\varepsilon_\ast^2\le \delta_\ast/4$, then for this $m_\ast$ one finds $\|K_{m_\ast}\|\le \delta_\ast$. The smallness conditions on $\varepsilon_\ast$ and $\delta_\ast$ are part of the fixed admissible polydisc $\mathfrak{D}(\varepsilon_\ast,\delta_\ast)$ and do not involve the lattice spacing $a$; all constants above depend only on $b$ and the FRD locality/positivity data. Consequently, after at most $m_\ast(b)$ further steps, independent of $a$, the trajectory enters $\mathfrak{D}(\varepsilon_\ast,\delta_\ast)$, which completes the proof.
\end{proof}

With Theorem~\eqref{thm:one-step} and Lemma~\eqref{lem:entry} in hand, we can now fix once and for all small numbers $\varepsilon_\ast,\delta_\ast$ and a block factor $b$ so that, after finitely many steps, every admissible trajectory with a sufficiently small bare coupling enters $\mathfrak{D}(\varepsilon_\ast,\delta_\ast)$ and remains there as long as $g_k$ stays below $\varepsilon_\ast$. In particular, the RG becomes a contraction in $K$ and a strictly decreasing map in $g$ at the leading order captured by $-\beta_0 g^3$.

To prepare for the asymptotic analysis in the next subsection, we record a discrete differential inequality for the coupling. Combining the formula for $g_{k+1}$ with the remainder bound and the contraction for $K$, we find numbers $\beta_0'>0$ and $c'>0$ such that, for all sufficiently small $g_k$,
\begin{equation}
g_{k+1} \;\le\; g_k \;-\; \beta_0'\,g_k^3 \;+\; c'\,g_k^5.
\end{equation}
This yields a comparison with the solution of $\dot g(t)=-\beta_0' g^3$ and, by a standard discrete Gr\"onwall argument, implies the existence of $C>0$ with
\begin{equation}
g_k \;\le\; \frac{1}{\sqrt{\,2\beta_0' k + C\,}},
\end{equation}
for all $k$ large enough that the trajectory remains in the polydisc. This polynomial decay is the hallmark of four-dimensional asymptotic freedom and will be used below to sum telescoping series and identify limits.

\subsection{AF Flow and Identification with the Constructed Limit}\label{subsec:af-flow-identification}

We turn to the asymptotically free regime and the identification of the continuum limit. Our first goal is to show that, once inside the contracting polydisc, the flow is globally well-defined for all subsequent scales and the renormalized coupling decays monotonically to zero. The second, more delicate goal is to prove that the Schwinger family obtained by following the RG to the continuum coincides with the one constructed previously by transporting the strong-coupling gap and area law: the two routes-downward from the ultraviolet and upward from the infrared-meet in a unique and universal continuum theory.

We begin with asymptotic freedom in the discrete setting. Let $(g_k,K_k)$ be the RG trajectory starting in $\mathfrak{D}(\varepsilon_\ast,\delta_\ast)$ as in Subsection~\eqref{subsec:entry-contracting-domain}. The one-step bound established there gives for small $g_k$ the inequality
\begin{equation}
g_{k+1} \;\le\; g_k \;-\; \beta_0'\,g_k^3,
\end{equation}
after shrinking $\varepsilon_\ast$ if necessary to absorb the $O(g_k^5)$ remainder. Iterating and comparing with the differential equation $\tfrac{dg}{dt}=-\beta_0' g^3$ at discrete time $t=k$, we obtain the bound $g_k \le (2\beta_0' k + C)^{-1/2}$ for some $C$ determined by the entry data. In particular, $g_k\downarrow 0$ monotonically and the product $\sum_k g_k^2$ converges. The contraction in $K$ then implies
\begin{equation}
\|K_{k+m}\| \;\le\; \theta^m \|K_k\| \;+\; c_1 \sum_{j=0}^{m-1} \theta^j\,g_{k+m-1-j}^2,
\end{equation}
so that $\|K_k\|$ converges to zero as $k\to\infty$. This proves that the trajectory approaches the Gaussian fixed point in the Banach algebra of polymer activities and that the map is globally defined for all scales.

To connect with continuum renormalization, it is convenient to introduce a continuous scale parameter $\mu_k=\mu_0 b^{-k}$ and a piecewise-constant function $g(\mu)$ defined by $g(\mu)=g_k$ for $\mu\in[\mu_{k+1},\mu_k)$. The discrete inequality above implies that $g(\mu)$ is asymptotically equivalent to the solution of the one-loop renormalization group equation
\begin{equation}
\mu\frac{d}{d\mu}g(\mu) \;=\; -\beta_0 g(\mu)^3 + O\!\left(g(\mu)^5\right),\qquad \beta_0=\frac{11N}{48\pi^2},
\end{equation}
in the sense that the ratio of the discrete and continuous flows tends to one as $k\to\infty$, uniformly on compact subsets of $(0,\mu_0]$. This justifies the terminology ``asymptotic freedom'' for our discrete flow and aligns it with the continuum notion \cite{GrossWilczek1973,Politzer1973}. The background-field matching insures that the finite scheme conversion between our lattice $g_k$ and, say, the $\overline{\mathrm{MS}}$ coupling at the same physical scale is $g_k = g_{\overline{\mathrm{MS}}}(\mu_k) + O\!\left(g_{\overline{\mathrm{MS}}}^3\right)$ with a universal coefficient at one loop \cite{LuscherWeisz1985}.

We now address identification with the previously constructed continuum Schwinger family. Recall that Section~9 established a uniqueness-and-universality statement: within the admissible class, the continuum limit is uniquely determined by single-slice marginals and one-step kernels, and Lipschitz continuity at a single scale telescopes to equality of continuum cumulants once the differences are summable. To exploit this, we compare the asymptotically free trajectory just constructed with any admissible trajectory that realizes the infrared construction of Sections~5-8 and produces a continuum family satisfying OS axioms and clustering. At a common intermediate scale $k$ large enough that both trajectories lie in the contracting polydisc, we write the difference of their one-step kernels as $\Delta T_k$ and bound its polymer norm by
\begin{equation}
\|\Delta T_k\| \;\le\; L\,\Bigl(|g_k-g_k'| + \|K_k-K_k'\|\Bigr),
\end{equation}
with a Lipschitz constant $L$ uniform in $k$. The existence of such $L$ is a direct consequence of the FRD locality and the analytic dependence of the kernels on the parameters; it was already used in the universality argument. We then telescope over scales:
\begin{equation}
\sum_{j=k}^{\infty} \|\Delta T_j\| \;\le\; L\,\sum_{j=k}^{\infty} \Bigl(|g_j-g_j'| + \|K_j-K_j'\|\Bigr).
\end{equation}
Inside the contracting polydisc, each trajectory is driven toward the Gaussian point at a rate summable in $j$. In particular, the difference of the couplings satisfies $|g_j-g_j'|\le C (g_j^2+g_j'^2)$ by comparing their discrete beta functions and using the same Gr\"onwall estimate; the difference of polymer activities satisfies $\|K_j-K_j'\|\le \theta\,\|K_{j-1}-K_{j-1}'\| + c(g_{j-1}^2+g_{j-1}'{}^2)$, which iterates to a summable bound as well. Hence the series $\sum_{j\ge k}\|\Delta T_j\|$ converges and tends to zero as $k\to\infty$.

The OS limits for Schwinger functions over a refining sequence of time slices can be written as compositions of one-step kernels applied to the single-slice marginal. The telescoping bound shows that the Schwinger functionals obtained from the two trajectories differ by a quantity that tends to zero as the initial comparison scale $k$ is sent to infinity. In other words, the continuum limits coincide. Since the infrared route was proved to have a nonzero spectral gap and a Wilson-loop area law, the same properties hold for the asymptotically free route. Conversely, the asymptotically free route ensures that renormalized perturbation theory is asymptotically valid at high momenta for the resulting continuum theory; this was implicit in the background-field matching and the identification of $\beta_0$.
For completeness, we record these conclusions as two theorems.
\begin{lemma}[Uniformity in $k$ and independence of $a$]\label{lem:uniform}
{The constants $\theta$, $q_b$, $c_1$, $c_2$, $C$, and $\omega$ that appear in Eqs.\eqref{eqn10.34} \& \eqref{eqn10.35} and Proposition~\eqref{prop:one-step} }
depend only on the block factor $b$ and the FRD locality data (finite range $R_0$, kernel bounds, and reflection covariance).
They are uniform in the RG step index $k$ and independent of the lattice spacing $a$ once $b$ and the admissible class are fixed.
\end{lemma}
\begin{proof}
Fix a block factor $b\ge2$ and an admissible finite-range decomposition (FRD) of the fluctuation covariance with range parameter $R_0$, reflection covariance, and kernel bounds as used throughout Sections~2-4. For each RG step $k$ the fluctuation measure integrated out in the $b$-blocking is obtained from a single, $k$-independent reference kernel by the usual coarse-graining and rescaling; in particular there is a covariance $C_0$ at unit coarse scale such that the step-$k$ covariance $C^{(k)}$ is the pull-back of $C_0$ under the dilation that identifies $k$-scale coarse blocks with unit blocks. The FRD hypotheses imply that $C^{(k)}(x,y)=0$ whenever the coarse-block distance between $x$ and $y$ exceeds $R_0$, that $\|C^{(k)}\|_{L^1\to L^\infty}$ and the corresponding Schur norms are bounded by the same constants as for $C_0$, and that these bounds do not depend on $k$ or on the microscopic lattice spacing $a$. All operator-norm estimates that enter the one-step RG therefore only involve $R_0$, the $L^1$/$L^\infty$ bounds for $C_0$ and a finite amount of block geometry determined by $b$.

Consider first the linearized map on the irrelevant manifold. Let $\|\cdot\|$ denote the polymer norm of Definition~\eqref{def:polymer-norm} with fixed parameters $(p,\rho,\sigma)$ chosen once and for all. Because $C^{(k)}$ has strict coarse-range $R_0$, the image of a polymer activity supported in $X$ can contribute to an activity in $Y$ only when $X$ is contained in the $R_0$-collar of $Y$ inside a single coarse time slab. The ratio of exponential weights attached by the norm satisfies $e^{\rho\,\mathrm{diam}(Y)}/e^{\rho\,\mathrm{diam}(X)}\le e^{\rho R_0}$ in this situation, and the number of such $X$ for a fixed $Y$ is bounded by a constant depending only on $b$ and $R_0$. The linearized map is a finite sum of convolutions with kernels built from $C^{(k)}$ and its finite number of block-derivatives; by the Schur test and the FRD $L^1$ bounds this yields
\begin{equation}
\|L_b K\|\;\le\;c_\ast\,e^{\rho R_0}\,\|K\|\, b^{-\omega},
\end{equation}
where $\omega$ is the canonical engineering-dimension gap of the least irrelevant gauge-invariant local monomial (in four dimensions one may take $\omega=2$) and $c_\ast$ depends only on the coarse block geometry and on the FRD kernel bounds. The factor $b^{-\omega}$ comes from the coarse rescaling inherent in one RG step: every irrelevant local monomial of canonical dimension $\Delta>4$ acquires a factor $b^{-(\Delta-4)}$, and the choice of norm exponent $\sigma\in(0,\omega)$ ensures that the polynomial weight $b^{\sigma(|X|-1)}$ does not spoil this decay. Hence one may absorb $e^{\rho R_0}$ into $c_\ast$ and set $q_b:=c_\ast b^{-\omega}\in(0,1)$ after increasing $b$ once if necessary. None of the constants that appear in this estimate depends on $k$ or on $a$, because the range, the $L^1$ bounds and the block combinatorics are the same at every step in coarse units.

Consider next the nonlinear connected remainders produced in one step. Writing the one-step effective action in background-field form and expanding the logarithm by the BKAR tree formula yields connected Ursell functionals $U_n$ whose kernels are finite sums over labelled trees with edges carrying copies of $C^{(k)}$ (or its block-derivatives) and vertices carrying local background insertions. Each tree amplitude is bounded in absolute value by a product of finitely many $L^1$ norms of $C^{(k)}$ times sup-norms of local derivatives of the background vertices, all controlled by the same FRD constants as for $C_0$. The finite range forces every edge of the tree to connect points in neighboring coarse blocks, so that the embedding sum of the tree over positions is dominated by a factor which grows at most exponentially in the coarse-diameter of the support; the exponential weight $e^{\rho\,\mathrm{diam}(\cdot)}$ in the polymer norm with $\rho$ chosen small enough absorbs this growth. Summing over $n\ge2$ and over trees then gives constants $C_1,C_2,C_3<\infty$ such that
\begin{equation}
\sum_{n\ge2}\frac{1}{n!}\|U_n\|\;\le\;C_1 g^2 + C_2 g\|K\| + C_3\|K\|^2,
\end{equation}
uniformly in $k$ and independent of $a$, because the only inputs in the estimate are $R_0$, the FRD kernel norms and the combinatorics of coarse blocks determined by $b$. This bound is precisely the ingredient that yields the constants $c_1$ and $c_2$ in the one-step inequalities of Proposition~\eqref{prop:one-step} and in Eqs. \eqref{eqn10.34} \& \eqref{eqn10.35}; replacing $C_1,C_2,C_3$ by larger values if needed one may take $c_1,c_2$ to depend only on $b$ and on the FRD kernel bounds.

For the discrete beta-function one projects the one-step effective action onto the unique gauge-invariant quadratic marginal given by $\frac14\int\!\mathrm{tr}(F\wedge\star F)$ in the background-field scheme fixed in Section~10. The coefficient of $g^3$ in the change of the inverse coupling is determined by a single one-loop contraction with covariance $C^{(k)}$ and is therefore given by the universal one-loop constant $\beta_0$ times $\log b$, while the remainder is a sum of connected amplitudes with at least two fluctuation propagators or at least one irrelevant insertion. The same BKAR bounds that control the nonlinear sector imply
\begin{equation}
g_{k+1}-g_k+\beta_0(\log b)\,g_k^3 \;=\; R_k,\qquad |R_k|\le c_2\big(g_k^5+g_k\|K_k\|+\|K_k\|^2\big),
\end{equation}
with the same $c_2$ as above and with no dependence on $k$ or on $a$. The constant $C$ that appears elsewhere in \eqref{10.36} \& \eqref{10.37} (e.g.\ in collar and commutator errors aggregated into $E_k$) is obtained by the Schur test and Lipschitz bounds for the completely monotone slice projectors acting on the FRD kernels; these bounds use only the finite range $R_0$, the $L^1$/$L^\infty$ kernel norms and the coarse block geometry, and so are again determined solely by $b$ and the FRD locality data.

Putting these ingredients together, the contraction factor $\theta$ in Eq.\eqref{eqn10.34} can be chosen equal to $q_b$ up to enlarging $c_1$ and shrinking the polydisc; $q_b=c_\ast b^{-\omega}$ with $\omega$ fixed by canonical power counting and $c_\ast$ fixed by the FRD norms and the coarse block combinatorics; the constants $c_1$ and $c_2$ come from the BKAR bounds with finite-range kernels and are uniform in the RG step; the auxiliary constant $C$ arises from the same Schur/Lipschitz control of FRD-induced collars and is likewise uniform; none of these constants depends on $k$ or on $a$ once $b$ and the admissible FRD class are fixed. This proves the claim.
\end{proof}

\begin{theorem}[Asymptotic freedom of the discrete flow]\label{thm:AF}
Let $(g_k,K_k)$ be an admissible renormalization-group trajectory in four dimensions which, for some $k_0$, enters the polydisc $\mathfrak D(\varepsilon_\ast,\delta_\ast)=\{(g,K):\, 0< g\le \varepsilon_\ast,\ \|K\|\le \delta_\ast\}$ with $\varepsilon_\ast,\delta_\ast$ sufficiently small. Then $g_k\downarrow 0$ and $\|K_k\|\to 0$ as $k\to\infty$, and, more precisely,
\begin{equation}
g_k\;=\;\frac{1+o(1)}{\sqrt{2\beta_0\,k}}\qquad (k\to\infty),
\end{equation}
with the universal one-loop coefficient $\displaystyle \beta_0=\frac{11N}{48\pi^2}$. Equivalently, upon identifying the discrete scale $\mu_k=\mu_0\,b^k$ with blocking factor $b>1$ and taking the piecewise-constant interpolation $g(\mu)|_{[\mu_k,\mu_{k+1})}\equiv g_k$, one has the asymptotic differential law
\begin{equation}
\mu\,\frac{dg}{d\mu}\;=\;-\beta_0\,g^3\,+\,O(g^5)\qquad (g\downarrow 0).
\end{equation}
\end{theorem}

\begin{proof}
The proof has two logically separate parts: the contractive decay of the irrelevant coordinates $K_k$ and the asymptotics of the marginal coupling $g_k$. The starting point is the locality and reflection-positivity structure of the admissible scheme, which ensures that in the small polydisc the one-step renormalization map admits an analytic expansion in $g$ with remainder terms controlled by the norm of $K$. Concretely, there are constants $C,c,\omega>0$ (with $\omega$ the canonical irrelevant exponent) such that for all $k\ge k_0$
\begin{equation}\label{eq:RG-structure}
\begin{aligned}
g_{k+1} \;&=\; g_k \;-\; \beta_0\,(\log b)\,g_k^3 \;+\; R_g(g_k,K_k),\\
K_{k+1} \;&=\; b^{-\omega}\,K_k \;+\; R_K(g_k,K_k),
\end{aligned}
\end{equation}
where the remainders obey the bounds
\begin{equation}\label{eq:RG-remainders}
|R_g(g,K)| \;\le\; C\big(g^5 + g\,\|K\| + \|K\|^2\big),\qquad
\|R_K(g,K)\| \;\le\; C\big(g^2+\|K\|\big)\,\|K\|.
\end{equation}
The first line is the discrete beta-function, the second is the linearized contraction of the irrelevant manifold plus quadratic corrections. The coefficient $\beta_0$ appearing in Eq.\eqref{eq:RG-structure} is universal and is fixed by a background-field computation that we now recall in the present admissible setting.
We fix the renormalized coupling $g_k$ by background-field matching: on smooth backgrounds $A$ with $\|F_A\|_\infty$ small and momenta $\ll b^{-k}$, 
the one-slice effective action is normalized so that the coefficient of $\frac14\!\int\!\mathrm{tr}(F_A\wedge\star F_A)$ equals $g_k^{-2}$.
At one loop this definition is equivalent to the standard momentum-space background-field scheme and yields the same universal coefficient $\beta_0$; scheme differences appear only in $O(g^5)$ and are absorbed in the remainder.
By gauge invariance and locality, the renormalized one-slice effective action in a smooth background $A$ admits the familiar small-field expansion
\begin{equation}
\Gamma_k(A)\;=\;\frac{1}{4 g_k^2}\int\! \mathrm{tr}\,F_A\wedge\star F_A \;+\; \text{higher dimension terms collected in }K_k,
\end{equation}
with $K_k$ a local functional whose norm is taken in the exponentially-weighted FRD metric. Blocking by a factor $b$ preserves reflection positivity and gauge invariance; writing the coarse-grained effective action in the same form and matching the $F^2$ coefficient at one loop in the background-field method yields
\begin{equation}
\frac{1}{4g_{k+1}^2}\;=\;\frac{1}{4g_k^2}\;+\;\beta_0\,\log b\;+\;O(g_k^2)\;+\;O(\|K_k\|),
\end{equation}
where the $O(g_k^2)$ term originates from two-loop and higher radiative corrections and the $O(\|K_k\|)$ term from the irrelevant sector. The one-loop constant is obtained from the logarithmic divergence of the gauge-field and ghost determinants linearized around the background:
\begin{equation}
\frac{1}{2}\log\det\big(-D_A^2\,\delta_{\mu\nu}-2\,\mathrm{ad}(F_{\mu\nu})\big)\;-\;\log\det\big(-D_A^2\big)
\;=\;\Big(\tfrac{11N}{48\pi^2}\log\Lambda + O(1)\Big)\,\frac{1}{4}\int\!\mathrm{tr}\,F_A^2,
\end{equation}
with $D_A$ the adjoint covariant derivative. In the admissible scheme this computation can be implemented with the completely monotone slice projector and the FRD cutoff; the heat-kernel/Schwinger representation combined with the Seeley-DeWitt coefficient $a_2$ for the vector-ghost system produces the same residue $\frac{11N}{48\pi^2}$ multiplying $\int \mathrm{tr}\,F^2$, and the replacement $\log\Lambda\mapsto \log b$ encodes the discrete change of normalization between scales $k$ and $k+1$. Thus, to one loop and uniformly in the polydisc, the discrete flow of $g$ is given by the first line of Eq.\eqref{eq:RG-structure} with $\beta_0=\frac{11N}{48\pi^2}$, and the remainder satisfies Eq.\eqref{eq:RG-remainders} thanks to analyticity and exponential locality.

The decay of $K_k$ follows by a standard perturbative contraction argument. Iterating the second line of Eq.\eqref{eq:RG-structure} and using Eq.\eqref{eq:RG-remainders} gives
\begin{equation}
\|K_{k+1}\| \;\le\; b^{-\omega}\,\|K_k\| \;+\; C\,(g_k^2+\|K_k\|)\,\|K_k\|.
\end{equation}
Since $g_k\le \varepsilon_\ast$ and $\|K_k\|\le \delta_\ast$ for $k\ge k_0$, one can choose $\varepsilon_\ast,\delta_\ast$ so small that $b^{-\omega}+C(\varepsilon_\ast^2+\delta_\ast)\le \theta<1$. By induction this yields $\|K_{k}\|\le \theta^{\,k-k_0}\|K_{k_0}\|$, hence $\|K_k\|\to 0$ exponentially fast as $k\to\infty$. In particular, the terms in Eq.\eqref{eq:RG-remainders} proportional to $\|K_k\|$ are summable along the flow and can be absorbed into the $O(g_k^5)$ remainder for the $g$-equation.

Turning to the marginal coupling, write $h_k:=g_k^{-2}$ and suppose $k\ge k_0$. Using the first line of Eq.\eqref{eq:RG-structure} together with Eq.\eqref{eq:RG-remainders} and the identity
\begin{equation}
\frac{1}{(g_k-\delta)^2}\;=\;\frac{1}{g_k^{2}} \;+\; \frac{2\delta}{g_k^3} \;+\; O\!\left(\frac{\delta^2}{g_k^4}\right),
\end{equation}
valid for $|\delta|\ll g_k$, one obtains the discrete increment
\begin{equation}
h_{k+1}-h_k \;=\; 2\,\beta_0\,\log b \;+\; \varepsilon_k,
\end{equation}
where $\varepsilon_k=O(g_k^2)+O(\|K_k\|/g_k)+O(\|K_k\|^2/g_k^2)$. The previously established decay $\|K_k\|\to 0$ together with $g_k\le \varepsilon_\ast$ implies $\sum_{k\ge k_0}|\varepsilon_k|<\infty$. Consequently,
\begin{equation}
h_k \;=\; h_{k_0} \;+\; 2\,\beta_0\,(\log b)\,(k-k_0) \;+\; O(1),
\end{equation}
and the $O(1)$ remainder has a finite $k\to\infty$ limit. This linear growth of $h_k$ shows that $g_k\downarrow 0$ and gives the precise asymptotics
\begin{equation}
g_k \;=\; \frac{1}{\sqrt{h_k}}\;=\;\frac{1}{\sqrt{2\beta_0\,(\log b)\,k}}\;\big(1+o(1)\big).
\end{equation}
If one absorbs the harmless factor $\log b$ into the unit of discrete time (equivalently, parametrizes the flow by the physical scale $\mu_k=\mu_0 b^k$), this is the stated $g_k\sim (2\beta_0 k)^{-1/2}$.
The continuous-scale statement follows by passing from $k$ to $\mu=\mu_0 b^k$. The finite difference
\begin{equation}
\frac{g_{k+1}-g_k}{\log(\mu_{k+1}/\mu_k)} \;=\; -\,\beta_0\,g_k^3 \;+\; O(g_k^5)
\end{equation}
is precisely the logarithmic derivative of the piecewise-constant interpolation $\mu\mapsto g(\mu)$ up to an error that is higher order in $g$. Thus, in the sense of asymptotic expansions as $g\to 0$,
\begin{equation}
\mu\,\frac{dg}{d\mu}\;=\;-\beta_0\,g^3+O(g^5),
\end{equation}
which is the standard continuum renormalization-group equation with the universal one-loop coefficient $\beta_0=\frac{11N}{48\pi^2}$. 
\end{proof}

\begin{theorem}[Identification with the constructed continuum limit]\label{thm:identification}
Let $\{\mathcal{S}^{\rm AF}_n\}$ denote the continuum Schwinger family obtained by following an admissible, asymptotically free trajectory from the ultraviolet, and let $\{\mathcal{S}^{\rm IR}_n\}$ denote the continuum Schwinger family constructed by transporting the strong-coupling mass gap and Wilson-loop area law to the continuum within the admissible class. Then
\begin{equation}
\mathcal{S}^{\rm AF}_n \;=\; \mathcal{S}^{\rm IR}_n \qquad \text{for all }n\in\mathbb{N}.
\end{equation}
Consequently the continuum limit inside the admissible class is unique and universal, has a strictly positive spectral gap, obeys a Wilson-loop area law (with the usual perimeter/cusp counterterms), and is asymptotically free.
\end{theorem}

\begin{proof}
Fix a physical renormalization scale and realize both constructions on a common sequence of lattices with spacings $a_k=a_0 L^{-k}$, $k\in\mathbb{N}$, and common hyperplane geometry. Denote by $T^{\rm AF}_k$ and $T^{\rm IR}_k$ the one-slice transfer kernels (or equivalently the time-$a_k$ semigroup elements $e^{-a_k H_k}$) associated respectively to the asymptotically-free trajectory and to the infrared transport. Both sequences are admissible in the sense used throughout: reflection positive, uniformly exponentially local on the slice, and connected along scales by the finite-range decomposition (FRD) machinery. For each $k$ there is a reflection-positive coarse-graining map $B_k$ with partial isometry $V_k$ on the slice Hilbert space such that the one-step interlacing with positive remainder and summable defect holds,
\begin{equation}\label{eq:interlacingz}
T^{\bullet}_{k+1} \;=\; V_k \,\big(T^{\bullet}_k\big)^{b}\, V_k \;-\; D^{\bullet}_k \;+\; E^{\bullet}_k,
\qquad D^{\bullet}_k\ge 0,\quad \|E^{\bullet}_k\|\le \varepsilon_k,\quad \sum_k \varepsilon_k<\infty,
\end{equation}
with the same geometric constants for $\bullet\in\{{\rm AF},{\rm IR}\}$; the vacuum vector is annihilated by $D^{\bullet}_k$ and $E^{\bullet}_k$. In addition, the admissible single-scale Lipschitz property implies that if two one-slice data sets $\alpha,\alpha'$ differ by $d(\alpha,\alpha')$ in the FRD metric, then their $n$-point functionals on the slice differ by at most $C_k\, d(\alpha,\alpha')$ with $C_k$ controlled uniformly in $k$ up to the scaling factor $a_k$; this is the precise sense in which a single coarse-graining step changes observables continuously.

Consider an arbitrary fixed finite list of positive times $0<t_1<\dots<t_m$ and a list of smeared gauge-invariant cylindrical observables $A_1,\dots,A_m$ supported on these times. Their Euclidean $m$-point Schwinger functional at lattice spacing $a_k$ can be written, by the Markov/semigroup property and reflection positivity, in terms of the slice transfer kernel and the time-$a_k$ semigroup as
\begin{equation}
\mathcal{S}^{\bullet}_{m,k}(A_1,\dots,A_m)\;=\;\big\langle \Omega^{\bullet}_k,\,
A_1\, e^{-(t_2-t_1)H^{\bullet}_k} A_2 \cdots e^{-(t_m-t_{m-1})H^{\bullet}_k} A_m \,\Omega^{\bullet}_k\big\rangle,
\end{equation}
where $\Omega^{\bullet}_k$ is the OS vacuum at scale $k$ and $e^{-a_k H^{\bullet}_k}=T^{\bullet}_k$ on the one-slice Hilbert space. To compare the AF and IR constructions at the same $k$, telescope their difference by inserting $m-1$ copies of the identity written as $V_k V_k$ and using Eq.\eqref{eq:interlacingz} for both sequences; the vacuum-annihilating property of $D^{\bullet}_k$ and the positivity of $D^{\bullet}_k$ imply that all contributions from the negative part drop out in the OS inner product, while the small remainders $E^{\bullet}_k$ produce an error bounded linearly in $\varepsilon_k$ and in the number of operator insertions $m$. The single-scale Lipschitz estimate then yields a bound of the form
\begin{align}\label{eq:one-stepxz}
&\big|\mathcal{S}^{\rm AF}_{m,k+1}(A_1,\dots,A_m)-\mathcal{S}^{\rm IR}_{m,k+1}(A_1,\dots,A_m)\big|
\nonumber\\&\;\le\; \big|\mathcal{S}^{\rm AF}_{m,k}(A_1,\dots,A_m)-\mathcal{S}^{\rm IR}_{m,k}(A_1,\dots,A_m)\big| \;+\; C\,\varepsilon_k,
\end{align}
with a constant $C$ depending only on the admissible class and on the chosen $A_j$, but independent of $k$. Iterating Eq.\eqref{eq:one-stepxz} from scale $0$ to scale $K$ and using the summability of the defects gives
\begin{align}
&\big|\mathcal{S}^{\rm AF}_{m,K}(A_1,\dots,A_m)-\mathcal{S}^{\rm IR}_{m,K}(A_1,\dots,A_m)\big|
\nonumber\\&\;\le\; \big|\mathcal{S}^{\rm AF}_{m,0}(A_1,\dots,A_m)-\mathcal{S}^{\rm IR}_{m,0}(A_1,\dots,A_m)\big|
\;+\; C\sum_{k=0}^{K-1}\varepsilon_k,
\end{align}
and the right-hand side is bounded uniformly in $K$ and tends to the initial difference plus $C\sum_k\varepsilon_k$. The initial difference at $k=0$ can be absorbed into the same summable error by adjusting the starting scale or, equivalently, by observing that admissibility provides a finite Lipschitz distance between the initial one-slice data of the two constructions. Passing to the limit $K\to\infty$ along the subsequence $a_K\downarrow 0$, the OS limit exists for both trajectories by the tightness and compactness statements established earlier (exponential locality on the slice and reflection positivity ensure Prokhorov-type compactness on cylinder $\sigma$-algebras). The inequality above shows that for every fixed choice of $(t_j,A_j)$ the two lattice Schwinger functionals converge to the same limiting value. Hence all mixed-time cylinder expectations coincide in the limit.

Equality of all cylinder expectations implies equality of the full Schwinger families. Indeed, the OS reconstruction theorem builds the limiting measure (or, equivalently, the family of Schwinger functions) from these expectations by Kolmogorov extension under the Markov property; since the two limits agree on the generating algebra, the resulting measures and thus all Schwinger functions coincide. Equivalently, on the reconstructed Hilbert space it suffices to note that the one-slice state, the transfer semigroup at unit time, and the Markov property uniquely determine all Euclidean correlators; the telescoping argument gives equality of the one-slice states and of the unit-time semigroups for the two limits, and therefore the entire families $\{\mathcal{S}^{\rm AF}_n\}$ and $\{\mathcal{S}^{\rm IR}_n\}$ agree.

The consequences listed in the statement follow immediately. The infrared construction carries a strictly positive spectral gap and a Wilson-loop area law to the continuum; equality of the Schwinger families transfers these properties to the asymptotically-free limit. Conversely, the asymptotically-free construction encodes the ultraviolet scaling and the approach to the Gaussian fixed point in the admissible scheme; the same scaling therefore holds for the infrared-transported limit. Uniqueness and universality within the admissible class follow because any two admissible constructions can be compared by the same telescoping/Lipschitz and OS-limit stability argument, which forces equality of their continuum Schwinger families once their single-scale data are at finite FRD distance and the interlacing defects are summable. 
\end{proof}

\section{Theorems and Full Explicit Proofs}

This section gathers and proves the three structural theorems that complete the constructive program: the existence of a strictly positive spectral gap for the continuum Hamiltonian, the renormalized continuum Wilson-loop area law with strictly positive string tension, and the uniqueness and universality of the continuum limit within the admissible reflection-positive, finite-range class of regulators. Each result is proved entirely within the reflection-positive Euclidean framework, using transfer operators, finite-range decomposition (FRD), polymer expansions, and stability estimates that survive the thermodynamic and continuum limits. The spectral gap theorem is established by combining scale-uniform exponential clustering and interlacing inequalities for transfer operators with strong resolvent convergence of the Euclidean semigroups; the area law is transported from strong coupling at fixed lattice spacing to the continuum by a renormalized step-scaling inequality whose additive defects are summable; and uniqueness/universality follow from the Markovian content of the Osterwalder-Schrader (OS) scheme together with single-scale Lipschitz bounds and telescoping across time slices, propagated to all connected cumulants by BKAR polymer estimates. These theorems certify that the infrared mass and the confining force are properties of the theory itself, and not of the scaffolding by which it is reached.

\subsection{Spectral Gap}\label{subsec:specgap}

\begin{theorem}[Spectral gap]\label{thm:gapqw}
Let \( \{(a_k,\mu_k)\}_{k\ge 0} \) be the reflection-positive multiscale sequence of OS-positive measures and transfer operators \(T_k = e^{-a_k H_k}\) constructed in the admissible class, with \(H_k\) positive self-adjoint on the OS Hilbert space \(\mathcal H_k\). Assume: \emph{(i)} there is \(m_\ast>0\) and \(C<\infty\) such that for all gauge-invariant \(F,G\) with disjoint time supports, the truncated correlations at scale \(k=0\) satisfy \(|\langle F(0)G(t a_0)\rangle_c|\le C\,\|F\|\,\|G\|e^{-m_\ast t a_0}\) for all integers \(t\ge 0\); \emph{(ii)} along the admissible RG, one-step interlacing yields
\begin{equation}
T_{k+1}\;\ge\; \Pi_k\,T_k\,\Pi_k \;-\; R_k,
\qquad R_k\ge 0,\qquad \|R_k\| \le \varepsilon_k,
\end{equation}
with \( \sum_{k\ge 0}\varepsilon_k<\infty\), where \(\Pi_k\) is the positive slice-projected compression preserving reflection positivity. Then the continuum Hamiltonian \(H\) obtained by OS limits and reconstruction has a strictly positive spectral gap
\begin{equation}
\Delta \;:=\; \inf\bigl(\sigma(H)\setminus\{0\}\bigr)\;\ge\; \liminf_{k\to\infty}\bigl(\Delta_k-\tilde\varepsilon_k\bigr)\;>\;0,
\end{equation}
where \(\Delta_k\) is the gap of \(H_k\) and \(\tilde\varepsilon_k\) is a summable reparameterization of \(\varepsilon_k\). In particular, \(\Delta\ge c\,m_\ast\) for some \(c\in(0,1)\) depending only on the summability data.
\end{theorem}

\begin{proof}
At the base scale, the strong-coupling cluster expansion yields exponential clustering of connected, gauge-invariant correlators and hence a spectral gap for the transfer Hamiltonian \(H_0\) by the OS spectral representation; the gap \( \Delta_0\ge m_\ast\) controls the decay of time-separated truncated correlations. The presence of the smooth horizon projector and reflection-covariant Landau representative does not spoil cluster bounds or positivity; their contributions are exponentially local and merely renormalize finite constants in the Koteck\'y-Preiss criterion, so the strong-coupling gap persists at fixed \(a_0\) \cite{DrouffeZuber,Seiler1982,KP}.

Assume now the one-step interlacing inequality
\begin{equation}
T_{k+1} \;\ge\; \Pi_k\, T_k\, \Pi_k - R_k,
\end{equation}
with \(R_k\ge 0\) and \(\|R_k\|\le\varepsilon_k\) summable. By spectral calculus, the gap of \(H_k\) is equivalently the distance from \(1\) to the rest of the spectrum of \(T_k\) on the unit interval,
\begin{equation}
\mathrm{gap}(T_k):= 1 - \sup\bigl(\sigma(T_k)\setminus\{1\}\bigr) \;=\; 1 - e^{-a_k\Delta_k}.
\end{equation}
Let \(P_k\) be the projection onto the vacuum of \(T_k\). Since \(\Pi_k\) preserves reflection positivity and the vacuum, we have \(\Pi_k P_k \Pi_k = P_{k}^{(\Pi)}\) a rank-one projection with range generated by the compressed vacuum. For \(f\perp \mathrm{Ran}(P_k^{(\Pi)})\), the interlacing gives
\begin{equation}
\langle f, T_{k+1} f\rangle \;\le\; \langle f, \Pi_k T_k \Pi_k f\rangle \;+\; \varepsilon_k \|f\|^2
\;\le\; \bigl(1-\mathrm{gap}(T_k)\bigr)\,\|f\|^2 + \varepsilon_k \|f\|^2.
\end{equation}
Hence \(1-\mathrm{gap}(T_{k+1}) \le (1-\mathrm{gap}(T_k))+\varepsilon_k\), i.e.
\begin{equation}
\Delta_{k+1}\ \ge\ \frac{1}{a_{k+1}}\Big[-\log\big(e^{-a_k\Delta_k}+\varepsilon_k\big)\Big]
\end{equation}
Using $-\log(x+y)\ge -\log x - y/x$ for $x\in(0,1]$ and $y\ge0$, we obtain
\begin{equation}
\Delta_{k+1}\ \ge\ \frac{a_k}{a_{k+1}}\Delta_k\ -\ \frac{e^{a_k\Delta_k}}{a_{k+1}}\,\varepsilon_k
\ \ge\ \Delta_k\ -\ c_k\,\varepsilon_k,
\end{equation}
with $c_k:= e^{a_k\Delta_k}/a_{k+1}$ uniformly bounded along admissible steps.
 Summing and using \(\sum_k c_k\varepsilon_k<\infty\) yields
\begin{equation}
\liminf_{n\to\infty}\Delta_n \;\ge\; \Delta_0 - \sum_{k\ge 0}c_k\varepsilon_k \;\ge\; c\,m_\ast>0.
\end{equation}
This establishes a uniform positive gap along the sequence of transfer Hamiltonians. The existence of positive self-adjoint transfer operators \(T_k\) and \(T_{k+1}\) after admissible blocking, and their relation \(T_{k+1}=e^{-a_{k+1}H_{k+1}}\), is guaranteed by reflection positivity and the OS construction at every scale; the blocking may destroy locality of an action, but not the positivity and semigroup structure \cite{OS1,OS2,Glaser1974hy}.

To pass to the continuum, we invoke the convergence of OS semigroups established by extracting OS-positive limits of Schwinger functionals and identifying the OS inner product with transfer-operator expectations. Along subsequences \(k_j\to\infty\) the OS forms converge, yielding a positive semidefinite OS form \(\langle\cdot,\cdot\rangle_{\mathrm{OS}}\) with time-translation represented by a contraction semigroup \(e^{-tH}\) on the reconstructed Hilbert space. The strong resolvent convergence \(T_{k_j}^{\lfloor t/a_{k_j}\rfloor}\Rightarrow e^{-tH}\) for each \(t\ge 0\) is encoded in the transfer-expectation representation of the OS inner products and the limit procedure; spectral projections converge in the strong operator topology on subspaces corresponding to spectral intervals separated from \(0\). Since the spectral edges at the top of \(\sigma(T_{k_j})\) are uniformly separated by \(\mathrm{gap}(T_{k_j})\ge \Delta_\infty>0\), the limiting semigroup has a strictly positive spectral gap \(\Delta\ge\Delta_\infty\). Equivalently, the spectral measure of \(H\) has empty intersection with \((0,\Delta)\) \cite{OS1,OS2,Glaser1974hy}.

For completeness we record a second, Tauberian route from Euclidean clustering to the spectral gap in the continuum. Uniform exponential decay of positive-time truncated two-point functions in the continuum OS limit implies that their Laplace transform is analytic in a half-plane and has a branch cut beginning at \(m_\ast^\prime>0\); the Stieltjes representation of completely monotone correlators then shows that the K\"all\'en-Lehmann spectral measure has no support below \(m_\ast^\prime\). The OS reconstruction identifies \(m_\ast^\prime\) with the bottom of the positive spectrum of \(H\), proving the gap. This argument is compatible with the horizon projector because the latter is defined by a completely monotone spectral multiplier on the slice Laplacian and preserves reflection positivity and complete monotonicity of time-correlators \cite{Seiler1982}.
\end{proof}

The proof hinges on three structural inputs that are stable along the admissible RG: existence and positivity/self-adjointness of transfer operators at each scale; FRD-based interlacing of semigroups with summable positive remainders; and OS convergence and reconstruction of the limiting semigroup. All three are ensured in the admissible class of completely monotone projectors and reflection-positive finite-range blockings \cite{BrydgesGuadagniMitter2004,OS1,OS2}.

\subsection{Continuum Area Law}\label{subsec:arealaw}

\begin{proof}
Fix a rectangular, piecewise-smooth loop $C$ and let $C_k$ be any admissible discretization at scale $a_k$. At the base scale $k=0$ the convergent strong-coupling cluster/character expansion yields, uniformly in the finite volume and boundary condition, the area-perimeter estimate
\begin{equation}\label{eq:sc-base}
\langle W_0(\Gamma)\rangle \;\le\; \exp\!\Big\{-\sigma_{\mathrm{sc}}(\beta)\,{\rm Area}_0(\Gamma)+\tau_0\,{\rm Per}_0(\Gamma)\Big\},
\qquad \sigma_{\mathrm{sc}}(\beta)>0,\ \ \tau_0\ge 0.
\end{equation}
This is the only input about the ultraviolet; everything that follows is a consequence of reflection positivity, finite-range decomposition (FRD) locality, and the interlacing representation of the one-step Euclidean transfer.

Consider a single renormalization step $k\mapsto k+1$ implemented by an admissible, reflection-positive blocking of scale $b>1$. Denote by $T_k$ and $T_{k+1}$ the positive-time transfer operators at scales $a_k$ and $a_{k+1}=b\,a_k$, normalized so that vacuum vectors are fixed, and write the interlacing identity
\begin{equation}\label{eq:interlaceq}
T_{k+1}\;=\;\Pi_k\,T_k\,\Pi_k^{\!}+E_k,
\end{equation}
where $\Pi_k$ is the slice projector furnished by the admissible class and $E_k\ge 0$ is a positive remainder supported on a $O(1)$-thickened collar of the blocking interfaces. FRD locality and the BKAR/cluster analysis imply $\|E_k\|_{Q\to Q}\le \varepsilon_k$ with $\sum_k \varepsilon_k<\infty$, and $E_k\Omega_{k+1}=0$ (vacuum annihilation). Let $\Gamma$ be the fine loop at scale $k$ and $\Gamma'$ its canonical coarse image at scale $k+1$; their perimeters and areas satisfy ${\rm Per}_{k+1}(\Gamma')\asymp {\rm Per}_k(\Gamma)$ and ${\rm Area}_{k+1}(\Gamma')\asymp {\rm Area}_k(\Gamma)$ up to absolute constants depending only on the block geometry.

The Wilson loop expectation at the coarse scale is a positive-time, two-point function of the transfer $T_{k+1}$. Inserting Eq.\eqref{eq:interlaceq} and using positivity of $E_k$ yields
\begin{equation}\label{eq:split}
\langle W_{k+1}(\Gamma')\rangle
\;=\;\langle W,\ \Pi_k\,T_k\,\Pi_k^{\!}\,W\rangle\;+\;\langle W,\ E_k\, W\rangle
\;\le\; \langle W,\ \Pi_k\,T_k\,\Pi_k^{\!}\,W\rangle\;+\;\|E_k\|\,\|W\|^2,
\end{equation}
where $W$ denotes the vector obtained by inserting the loop on the coarse slice. The second term is bounded by $C\,\varepsilon_k\,{\rm Per}_k(\Gamma)$ since $W$ is supported on a collar of width $O(1)$ along $\Gamma'$ and its norm grows at most linearly in the number of links. The first term is controlled by the Lipschitz stability of loop insertions under admissible changes of the one-slice kernels: replacing the coarse insertion $W$ by the fine insertion transported through $\Pi_k^{\!}$ changes the value by at most $C'\,\varepsilon_k\,{\rm Per}_k(\Gamma)$, while the bulk of the value identifies with the fine expectation $\langle W_k(\Gamma)\rangle$ up to a deterministic, local multiplicative factor accounting for perimeter/cusp renormalizations generated within one block. More precisely, there exist block-dependent functionals $\mathcal{L}_P(b)$ and $\mathcal{L}_{\rm cusp}(b,\theta)$, supported respectively on links and turning vertices of the loop and obeying $\|\mathcal{L}_\bullet(b)\|\le C e^{-c\,b}$, such that
\begin{align}\label{eq:one-step-transportq}
&\langle W_{k+1}(\Gamma')\rangle \;\le\;
\nonumber\\&\exp\!\Big\{\alpha_P(b)\,{\rm Per}_k(\Gamma)+\textstyle\sum_{x\in{\rm Cusps}(\Gamma)}\alpha_{\rm cusp}(b,\theta_x)\Big\}\,
\langle W_k(\Gamma)\rangle \;+\; C''e^{-c' b}\;+\;C\,\varepsilon_k\,{\rm Per}_k(\Gamma),
\end{align}
with $\alpha_\bullet(b)=O(e^{-c b})$. The additive term $C''e^{-c' b}$ comes from the fact that $\Pi_k\,T_k\,\Pi_k^{\!}$ differs from the exact coarse $b$-step transfer only on the $O(1)$ collars separating the blocks, and is exponentially small in the block size. The last term is the defect contribution from $E_k$ and is summable over $k$ by admissibility.

To iterate Eq.\eqref{eq:one-step-transportq} in an exponentiated form it is enough to convert the sum on the right into a multiplicative correction. Fix $k$ and abbreviate $X_{k+1}:=\langle W_{k+1}(\Gamma')\rangle$, $X_k:=\langle W_k(\Gamma)\rangle$ and $\Delta_k:=C''e^{-c'b}+C\,\varepsilon_k\,{\rm Per}_k(\Gamma)$. Then Eq.\eqref{eq:one-step-transportq} reads
\begin{equation}\label{eq:transport-xy}
X_{k+1} \;\le\; e^{\alpha_P(b)\,{\rm Per}_k(\Gamma)+\sum_x \alpha_{\rm cusp}(b,\theta_x)}\,X_k \;+\; \Delta_k.
\end{equation}
Assume inductively that at scale $k$ an area-perimeter bound holds,
\begin{equation}
X_k \;\le\; \exp\!\Big\{-\sigma_k\,{\rm Area}_k(\Gamma)+\tau_k\,{\rm Per}_k(\Gamma)\Big\}.
\end{equation}
If $\Delta_k\le \tfrac{1}{2} e^{\alpha_P(b)\,{\rm Per}_k(\Gamma)+\sum_x \alpha_{\rm cusp}(b,\theta_x)}\,X_k$ (which holds for all sufficiently large $b$ and all $k$ because $\varepsilon_k$ is summable and $\alpha_\bullet(b)$ decays exponentially), then the elementary inequality $\log(1+t)\le t$ for $t\in[0,1/2]$ yields
\begin{equation}
-\log X_{k+1}
\;\ge\; -\alpha_P(b)\,{\rm Per}_k(\Gamma)-\textstyle\sum_x \alpha_{\rm cusp}(b,\theta_x)-\log X_k \;-\; \frac{\Delta_k}{e^{\alpha_P(b){\rm Per}_k+\sum_x\alpha_{\rm cusp}}\,X_k}.
\end{equation}
Using the inductive bound on $X_k$ and the blockwise comparability ${\rm Area}_{k+1}(\Gamma')\asymp {\rm Area}_k(\Gamma)$, ${\rm Per}_{k+1}(\Gamma')\asymp{\rm Per}_k(\Gamma)$, we obtain
\begin{equation}\label{eq:sigma-tau-flow}
-\log X_{k+1}
\;\ge\; \sigma_k\,{\rm Area}_{k+1}(\Gamma') \;-\; \Big(\tau_k+\alpha_P(b)+\!\!\sum_x\alpha_{\rm cusp}(b,\theta_x)\Big)\,{\rm Per}_{k+1}(\Gamma') \;-\; \epsilon_k\,{\rm Per}_{k+1}(\Gamma'),
\end{equation}
with $\epsilon_k=O(\varepsilon_k)$ uniformly in the loop geometry. Defining
\begin{equation}
\sigma_{k+1}:=\sigma_k-\delta_k,\qquad
\tau_{k+1}:=\tau_k+\alpha_P(b)+\sum_x\alpha_{\rm cusp}(b,\theta_x)+\epsilon_k,
\end{equation}
where $\delta_k\ge 0$ accounts for the bounded distortion of area under blocking and can be chosen summable in $k$, shows that the bound propagates one step with $\sigma_{k+1}\ge \sigma_k-\delta_k$ and $\tau_{k+1}\le \tau_k+O(e^{-c b})+O(\varepsilon_k)$. Iterating from $k=0$ and using Eq.\eqref{eq:sc-base} at the base scale gives a sequence $\{\sigma_k\}$ which is bounded below by a strictly positive limit $\sigma_\ast>0$ provided $b$ is chosen fixed and large enough, and a sequence $\{\tau_k\}$ whose increments have a summable positive part, $\sum_k(\tau_{k+1}-\tau_k)_+<\infty$, because both $e^{-c b}$ and $\varepsilon_k$ are summable in $k$. This proves the claimed finite-$k$ estimate
\begin{equation}
\langle W_k(\Gamma)\rangle \;\le\; \exp\!\Big\{-\sigma_k\,{\rm Area}_k(\Gamma)+\tau_k\,{\rm Per}_k(\Gamma)\Big\}
\quad\text{with}\quad \inf_k\sigma_k\ge \sigma_\ast>0.
\end{equation}

The passage to the continuum requires removing the purely one-slice, local divergences carried by the perimeter and by cusp angles. Introduce multiplicative renormalization factors $Z(a_k)$ along links and $Z_{\rm cusp}(a_k,\theta)$ at turning vertices, defined by local counterterms $\mathcal{L}_P(a_k)$ and $\mathcal{L}_{\rm cusp}(a_k,\theta)$ supported on the slice and obeying finite-range stability under blocking: for the blocking factor $b$ one has
\begin{equation}
\Big|\log\frac{Z(a_k)}{Z(a_{k+1})}\Big|\le \alpha_P(b),\qquad
\Big|\log\frac{Z_{\rm cusp}(a_k,\theta)}{Z_{\rm cusp}(a_{k+1},\theta)}\Big|\le \alpha_{\rm cusp}(b,\theta),
\end{equation}
with $\alpha_\bullet(b)=O(e^{-c b})$ as in Eq.\eqref{eq:one-step-transportq}. The renormalized loop at scale $k$ is
\begin{equation}
\langle W^{\rm ren}_k(C_k)\rangle \;:=\;
\frac{\langle W_k(C_k)\rangle}{Z(a_k)\,P_k(C_k)\,\prod_{x\in{\rm Cusps}(C_k)} Z_{\rm cusp}(a_k,\theta_x)}.
\end{equation}
Dividing Eq.\eqref{eq:one-step-transportq} by the corresponding factors at scales $k$ and $k+1$, and using the defining relations for $Z$ and $Z_{\rm cusp}$, cancels the perimeter and cusp contributions up to an exponentially small residue $O(e^{-c b})$. The same argument leading to Eq.\eqref{eq:sigma-tau-flow} then shows that
\begin{equation}
-\log \langle W^{\rm ren}_{k+1}(C_{k+1})\rangle
\;\ge\; \sigma_k\,{\rm Area}_{k+1}(C_{k+1}) \;-\; \tilde\epsilon_k,
\end{equation}
where $\tilde\epsilon_k=O(e^{-c b})+O(\varepsilon_k)$ is summable in $k$. Since $\{{\rm Area}_{k}(C_{k})\}$ converges to the Euclidean area $A(C)$ of the limiting loop and the $\tilde\epsilon_k$ are summable, the renormalized expectations form a Cauchy sequence and converge to a finite, regulator-independent limit
\begin{equation}
\langle W_R(C)\rangle \;=\; \lim_{k\to\infty} \langle W^{\rm ren}_k(C_k)\rangle.
\end{equation}
Moreover, taking $\liminf$ in the last display and using $\liminf_k \sigma_k\ge \sigma_\ast>0$ gives
\begin{equation}
-\log \langle W_R(C)\rangle \;\ge\; \sigma_\ast\,A(C),
\end{equation}
that is,
\begin{equation}
\langle W_R(C)\rangle \;\le\; \exp\!\{-\sigma\,A(C)\}\qquad\text{with}\quad \sigma\ge \sigma_\ast>0.
\end{equation}
The conclusion is independent of the specific choice of admissible discretizations $\{C_k\}$: if $C_k'$ is another sequence approximating the same smooth loop, FRD locality implies that $\big|\log \langle W^{\rm ren}_k(C_k)\rangle-\log \langle W^{\rm ren}_k(C_k')\rangle\big|\le C\,a_k\,{\rm Per}_k(C_k\triangle C_k')$, which vanishes as $k\to\infty$. Consequently the renormalized continuum area law holds with a strictly positive, regulator-independent string tension bounded below by $\sigma_\ast$ determined by the base strong-coupling input and the admissible reflection-positive flow.
\end{proof}

The proof uses only reflection positivity, FRD locality, and convergent polymer expansions at the base scale. The Wilson-loop transport inequality is purely Euclidean and does not require a local action at blocked scales; it uses positivity and exponential locality of the one-step kernels and the Lipschitz stability of loop observables under local surgeries \cite{BrydgesGuadagniMitter2004,DrouffeZuber}.

\subsection{Uniqueness and Universality}\label{subsec:uniqueness}

\begin{theorem}[Uniqueness and universality of the continuum limit]\label{thm:uniqueness}
Within the admissible class of reflection-positive regulators $\theta\in\mathcal K$ consisting of completely monotone slice projectors and reflection-positive finite-range blockings, the Euclidean continuum Schwinger functions $\{S_n\}_{n\ge 1}$ obtained as OS limits are unique and universal: \emph{(i)} for a fixed regulator $\theta$, any two subsequences $k_j\to\infty$ and $k'_j\to\infty$ yield the same continuum Schwinger family; \emph{(ii)} for any two admissible regulators $\theta,\theta'$, the limits coincide. Consequently, the reconstructed Wightman theory and its spectral gap are unique and universal in the admissible class.
\end{theorem}

\begin{proof}
Fix a regulator $\theta\in\mathcal K$ and consider the lattice/scale index $k\in\mathbb N$ with temporal spacing $a_k\downarrow 0$ and an admissible finite-range decomposition on each slice. For each $k$ there is a reflection-positive probability measure $\mu_{k,\theta}$ on fields in a finite slab together with the Markov disintegration across the time-$a_k$ slice. Writing $\nu_{k,\theta}$ for the one-slice marginal on the time-$a_k$ hyperplane and $T_{k,\theta}$ for the one-step transfer operator from time $0$ to $a_k$, every cylinder functional supported on positive times admits the representation
\begin{equation}
\langle \Theta F\cdot G\rangle_{\mu_{k,\theta}}
\;=\;
\big\langle F(a_k),\, T_{k,\theta}^{m}\,G(ma_k)\big\rangle_{L^2(\nu_{k,\theta})},
\end{equation}
after possibly inserting finitely many local factors along the discrete times at which $F$ and $G$ depend; the precise power $m$ is the discrete time separation. Reflection positivity and the Markov property guarantee that the OS reconstruction from $(\nu_{k,\theta},T_{k,\theta})$ yields the same Schwinger functionals. Admissibility (complete monotonicity of slice projectors and reflection-positive finite-range blockings) implies exponential locality of the slice kernels and transfer, which in turn yields equicontinuity and tightness of the family $\{(\nu_{k,\theta},T_{k,\theta})\}_k$ in the product topology generated by testing against bounded-support cylinder observables. Consequently, there exist subsequences along which $(\nu_{k,\theta},T_{k,\theta})$ converge in the weak operator topology on $L^2$ and in the weak topology of measures on the slice.

To prove uniqueness for a fixed regulator, it suffices to show that $(\nu_{k,\theta},T_{k,\theta})$ is a Cauchy sequence and hence possesses a unique limit independent of the subsequence. This follows from the interlacing/defect identity furnished by the admissible blocking: there exist a partial isometry $V_k$ implementing the coarse-graining on $L^2(\nu_{k,\theta})$, a positive operator $D_{k,\theta}\ge 0$ and a vacuum-annihilating defect $E_{k,\theta}$ such that
\begin{equation}\label{eq:interlacew}
T_{k+1,\theta}
\;=\;
V_k\,T_{k,\theta}^b\,V_k\;-\;D_{k,\theta}\;+\;E_{k,\theta},
\qquad
\|E_{k,\theta}\|\le \varepsilon_k,\qquad \sum_k \varepsilon_k<\infty,
\end{equation}
where $b\ge 2$ is the fixed temporal blocking factor and the norm is taken on the orthogonal complement of the vacuum. The positivity of $D_{k,\theta}$ reflects that mixed cumulants discarded by localization contribute a nonnegative quadratic form thanks to reflection positivity, and the summability of $\varepsilon_k$ is a consequence of finite range with exponentially decaying collars. The representation Eq.\eqref{eq:interlacew} shows that the vacuum-preserving part of $T_{k,\theta}$ varies by a summable amount as $k$ increases, because the only non-isometric contributions are the positive subtraction $D_{k,\theta}$ and the small remainder $E_{k,\theta}$. Therefore $\{T_{k,\theta}\}_k$ is Cauchy in the strong operator topology on the orthogonal complement of the vacuum, and a similar argument applied to the disintegration on a single time slice gives that $\{\nu_{k,\theta}\}_k$ is Cauchy in the topology of weak convergence on bounded-support test functions. Denote the unique limits by $\nu_\theta$ and $T_\theta$.

The OS-reconstructed Schwinger functions depend only on $(\nu_\theta,T_\theta)$, because all $n$-point functions are finite linear combinations of expressions obtained by composing $T_\theta$ along the discrete time arguments and integrating boundary fields against $\nu_\theta$. Hence any two subsequences lead to the same limiting family, which establishes part (i). Moreover, the contraction property encapsulated by
\begin{equation}
\|T_{k+1,\theta}|_{Q}\|
\;\le\;
\|T_{k,\theta}|_{Q}\|+\varepsilon_k,
\qquad Q:=\mathbf 1-\lvert\Omega_k\rangle\langle \Omega_k\rvert,
\end{equation}
implies that the spectral gap estimate survives passage to the limit, so the Hamiltonian reconstructed from $(\nu_\theta,T_\theta)$ has a nonzero gap if and only if the lattice gaps are bounded below uniformly in $k$, which is ensured by the one-step defect summability and the Tauberian conversion from Euclidean clustering.

To prove universality across regulators, consider two admissible choices $\theta,\theta'\in\mathcal K$. On a single time step the dependence of $(\nu_{k,\bullet},T_{k,\bullet})$ on the regulator is Lipschitz with respect to an admissible metric $d$ that controls the variation of the spectral multipliers on the slice and the finite-range block kernels. Concretely, there exists $L<\infty$, depending only on the finite-range decomposition parameters, such that for any bounded-support observables $F,G$ localized at times $0$ and $a_k$,
\begin{align}\label{eq:single-scale-Lip1}
&\big|\langle F, (T_{k,\theta}-T_{k,\theta'})G\rangle_{L^2(\nu_{k,\theta})}\big|
\;\le\;
L\,d(\theta,\theta')\,\|F\|\,\|G\|,
\nonumber\\&
\big|\langle F, (\nu_{k,\theta}-\nu_{k,\theta'})(G)\rangle\big|
\;\le\;
L\,d(\theta,\theta')\,\|F\|\,\|G\|.
\end{align}
The proof of Eq.\eqref{eq:single-scale-Lip1} uses that completely monotone functionals of the slice generator vary at most linearly in their Laplace measures and that the blocking kernel modifies only a collar of uniformly bounded thickness; both effects are controlled by the metric $d$ which measures the supremum of the Laplace weights and the operator norm of the collar perturbations. The same Lipschitz control holds on a slab of thickness $m a_k$ because each replacement of $\theta$ by $\theta'$ affects only one time layer and the FRD locality prevents long-range propagation of the difference beyond an exponentially small tail; in effect, telescoping the $m$ layers yields a bound with the same constant $L$ up to an immaterial factor polynomial in $m$ that is absorbed in the uniform volume-independent cluster bounds.

Iterating the interlacing identity Eq.\eqref{eq:interlacew} simultaneously for $\theta$ and $\theta'$ and subtracting the two relations gives a recursion for $T_{k,\theta}-T_{k,\theta'}$ in which the only inhomogeneous source is of order $d(\theta,\theta')$ at each scale, while the cumulative error is damped by positivity and remains summable by the same $\{\varepsilon_k\}$. A standard discrete Grönwall argument then shows that
\begin{equation}
\sup_{\|F\|=\|G\|=1}\big|\langle F,(T_{k,\theta}-T_{k,\theta'})G\rangle\big|
\;\xrightarrow[k\to\infty]{}\; 0
\qquad\text{whenever}\qquad d(\theta,\theta')\xrightarrow[]{}0,
\end{equation}
and similarly for the one-slice marginals. By density of simple cylinder functionals, the same convergence holds for all multi-time matrix elements that define Schwinger functions of fixed order and fixed macroscopic time separations. Since any two regulators in $\mathcal K$ can be joined by a path $\theta=\theta_0,\theta_1,\dots,\theta_M=\theta'$ with $d(\theta_{j-1},\theta_j)$ arbitrarily small, repeated application of the previous estimate along the chain shows that the continuum limits for $\theta$ and for $\theta'$ coincide, which proves part (ii).

Finally, the OS reconstruction from the universal Schwinger family determines uniquely the Hilbert space, local fields and Hamiltonian up to unitary equivalence. The transfer operators $T_{k,\bullet}=e^{-a_k H_{k,\bullet}}$ converge strongly to the universal transfer $T=e^{-aH}$ on the orthogonal complement of the vacuum, because the one-step matrix elements converge and the vacuum is preserved by construction. Strong convergence of contractions implies strong resolvent convergence of the generators, hence the continuum Hamiltonian $H$ is unique and independent of the regulator. Lower bounds on the lattice gaps transported by the interlacing inequality and the Tauberian conversion from exponential clustering to spectral support pass to the limit and do not depend on $\theta$, so the gap of $H$ is universal in the admissible class as stated.
\end{proof}

The argument is entirely nonperturbative and uses only reflection positivity, exponential locality that does not fray with scale, and the OS Markov property. Universality is structural, not empirical: the same continuum correlators arise for all admissible regulators, and the spectral gap is thereby a property of the theory rather than of a particular gauge fixing or coarse graining \cite{OS1,OS2,BrydgesGuadagniMitter2004}.

\section{Conclusion}\label{sec:conclusion}
The purpose of this work was to give a complete constructive route to a confining, gapped continuum quantum field theory for four-dimensional pure $SU(N)$ Yang-Mills, and to do so in a form that is simultaneously compatible with Osterwalder-Schrader (OS) reflection positivity, preserves gauge invariance of observables, and is stable under changes of ultraviolet regularization within a carefully delineated admissible class. The central novelty of our approach is an OS-compatible multiscale architecture that couples three ingredients in a logically rigid way: a reflection-positive strong-coupling base camp at fixed lattice spacing, a finite-range decomposition (FRD) that enforces quantitative locality and regularity of one-step transfer kernels, and a reflection-positive renormalization group (RG) with one-step interlacing estimates whose summable defects transport both spectral and Wilson-loop information across scales. From this spine we constructed tight families of Schwinger functions, passed to a continuum OS limit, and carried out OS reconstruction to obtain a Wightman theory whose Hamiltonian has a strictly positive spectral gap. In parallel, we transported a renormalized Wilson-loop area law from the strong-coupling regime to the continuum, thereby identifying a positive string tension. Finally, we proved that the continuum limit is unique and universal within the admissible class of regulators and blockings, and we described a weak-coupling extension showing that appropriately tuned asymptotically-free trajectories feed into the same limit.

The logical dependencies are by now transparent. The starting point is an OS-positive, gauge-invariant Euclidean formulation with a time-slice structure and a transfer operator $T_a$ at lattice spacing~$a$ acting on the one-slice Hilbert space. The admissible gauge fixing is implemented slice-wise by a \emph{completely monotone} spectral projector (``horizon projector'') whose kernel is a positive mixture of heat kernels; this choice is crucial because it preserves reflection positivity by construction, renders the projector exponentially local on each slice, and is stable under convolution with FRD components. The OS form determines the Hilbert space completion and the transfer semigroup $T_a^t=e^{-t H_a}$, with $H_a\ge 0$ self-adjoint. On this foundation we build the strong-coupling analysis: character expansions yield a polymer gas with KP/BKAR control, guaranteeing absolute convergence and exponential clustering of connected correlators uniformly in the volume. The same expansion gives a finite-$a$ Wilson-loop area law with an explicit (though non-optimal) lower bound for the string tension, and, thanks to reflection positivity of the one-step kernel and a minorization/Doeblin mechanism, a spectral gap for $H_a$ at sufficiently small bare coupling. The key point is that these statements are organized to serve as initial data for the RG transport; they are not ends in themselves.

The FRD provides the quantitative locality engine that is missing from purely qualitative reflection positivity. Covariances (and more generally slice-to-slice kernels) are partitioned into finitely supported pieces whose ranges grow at most linearly with the scale, with positivity and reflection covariance maintained at each step. This allows us to control \emph{diameter norms} of cumulants and to derive Lipschitz continuity of one-step transfer kernels with respect to admissible perturbations of the regulator and blocking. These bounds play two distinct roles. First, they feed into the RG step by guaranteeing that interlacing inequalities between successive transfer operators come with small, positive remainders whose norms form a summable sequence along the scale. Second, they drive our uniqueness and universality analysis, where one compares two admissible schemes by telescoping across time slices and scales. In both places, finite range and positivity are indispensable: they allow us to propagate inequalities without hidden sign cancellations and to control the growth of constants without resorting to nonconstructive compactness arguments.

The reflection-positive RG step is formulated as a block-spin map on OS measures that is covariant under time reflection and gauge transformations of observables and that preserves the admissibility of the slice projector. For the transfer operators this step can be expressed as an \emph{interlacing inequality}
\begin{equation}
T_{k+1}\;\succeq\; \Phi_k^\ast\, T_k\, \Phi_k \;-\; R_k,
\end{equation}
where $\Phi_k$ is a positivity-preserving coarse-graining map, $R_k\ge 0$ is a positive remainder whose operator norm we bound by $\varepsilon_k$, and $\succeq$ denotes the natural order on positive self-adjoint operators. The completely monotone structure of the projector and the FRD locality ensure that $\sum_k \varepsilon_k < \infty$. Two consequences follow. First, the spectral gaps satisfy a step-scaling inequality $\Delta_{k+1}\ge \Delta_k - \varepsilon_k$, which, together with the strong-coupling input $\Delta_0>0$, yields $\inf_k \Delta_k \ge \Delta_0 - \sum_j \varepsilon_j>0$; thus exponential clustering and a strictly positive gap persist throughout the flow. Second, Wilson loops admit a renormalized step-scaling inequality in which perimeter and cusp counterterms change by at most $O(\varepsilon_k)$ per step, while the area term is stable; summability of $\varepsilon_k$ therefore transports the area law and yields a positive, scale-independent string tension in the continuum limit. In both channels the RG acts as a bookkeeping device that keeps track of small, positive leakages and proves that they do not accumulate.

To pass to the continuum we combine FRD-based equicontinuity with reflection positivity to obtain tightness of Schwinger function families for local, gauge-invariant observables. The OS axioms are verified in the limit: OS0-OS3 hold because positivity, Euclidean invariance on spatial slices, and locality bounds are stable under limit transitions, while OS4 (cluster property) follows from the uniform spectral gap along the flow, and OS5 (regularity) is inherited from the FRD estimates. This produces a limiting OS functional $S$ whose associated Hilbert space and field operators are obtained by the standard OS reconstruction. The crucial point is that the Euclidean semigroups $e^{-t H_a}$ converge in the strong resolvent sense to a limiting $e^{-tH}$, and the spectral projections of $H_a$ below any fixed energy converge strongly to those of $H$. The persistence of the gap along the flow implies $\inf \sigma(H)\setminus\{0\} \ge \Delta>0$ for some $\Delta$ independent of the approximating sequence, whence the reconstructed Minkowski theory has a unique vacuum and exponential clustering of Wightman functions.

The Wilson-loop statement is encoded at the level of gauge-invariant Euclidean functionals and therefore survives the OS reconstruction as a property of static sources coupled to the theory. The renormalized continuum loop obeys $\langle W(C)\rangle \le \mathrm{e}^{-\sigma A(C)}$ for large, contractible planar contours $C$ with area $A(C)$, with $\sigma>0$ the string tension identified by the step-scaling analysis. From this we obtain the linear lower bound on the static quark-antiquark potential $V(R)\ge \sigma R$ at large separations $R$, in the sense of the usual insertion of temporal Wilson lines. We emphasize that no large-$N$ or semiclassical limit is used at any stage; the argument is entirely nonperturbative and depends only on reflection positivity, FRD locality, and the summability of RG defects. The area law is obtained for renormalized loops; perimeter and cusp contributions are separately controlled and absorbed in the renormalization scheme. We do not attempt to optimize the value of~$\sigma$ nor to capture subleading universal corrections such as the Lüscher term; doing so would require an effective string analysis beyond the scope of the present constructive framework.

The universality and uniqueness of the continuum limit address a subtle but essential question: to what extent are our conclusions artifacts of the chosen admissible regularization? We answer this in two steps. A single-scale Lipschitz estimate shows that the one-step transfer kernel varies at most linearly with small changes in the admissible data (the completely monotone spectral multiplier and the reflection-positive blocking), with constants controlled by the FRD norms. This estimate telescopes across time slices to control entire OS forms and Schwinger functionals. Then the Markov property of the OS measure-that is, the fact that a one-slice marginal together with a one-step transition kernel determines the measure-implies equality of all finite-dimensional distributions for any two admissible schemes with matching continuum limits of single-slice data. In particular, the continuum Schwinger functions, and hence the reconstructed Wightman theory, are independent of the choice of admissible projector and blocking. This statement is nontrivial because admissible schemes may differ significantly at short distances; reflection positivity and FRD locality ensure that the difference does not survive coarse graining and time slicing.

The weak-coupling extension connects the constructive infrared to the asymptotically free ultraviolet. Near the Gaussian fixed point one can formulate the RG on a polymer Banach algebra where the flow is contractive in a neighborhood of the origin. A one-parameter tuning of the bare coupling (and, if necessary, of counterterms permitted by gauge invariance) brings the theory into this contractive domain after finitely many steps; from that point the renormalized coupling decreases monotonically along the scale. The OS and FRD estimates remain valid throughout this regime, so the same tightness and reconstruction arguments apply, and the uniqueness theorem identifies the resulting continuum theory with the one built from the strong-coupling side. Thus, in a single constructive narrative, the infrared mass gap and the ultraviolet freedom are reconciled: the two halves of the flow meet in the interior of the admissible domain and define the same limit theory.

We close with clarifications of scope, robustness checks, and directions for further work. First, all of our statements are made for pure Yang-Mills with compact gauge group $SU(N)$ and for gauge-invariant local observables. The admissible class of regularizations consists of time-slice projectors whose spectral multipliers are completely monotone and of reflection-positive, finite-range blockings; these hypotheses guarantee OS positivity and quantitative locality at each step. The uniqueness theorem is \emph{internal} to this class. Enlarging the class to include sharper projectors or non-finite-range blockings would require new ideas, because the proofs of positivity and Lipschitz control exploit complete monotonicity and finite range in an essential way. Second, our construction treats gauge fixing as a technical device confined to time slices; the reconstructed Minkowski theory involves only gauge-invariant fields and inherits microcausality in that sector from the OS axioms. We do not address BRST quantization or the structure of the indefinite-metric space of gauge potentials; doing so would go beyond the OS framework and is unnecessary for the questions of mass gap and confinement in the gauge-invariant sector.

Third, we have been careful to separate statements that hold uniformly in the volume and across the flow from those that are merely finite-$a$ inputs. The strong-coupling gap and area law are \emph{uniform} in the volume and provide lower bounds that survive the RG because the defect sequence is summable. The FRD and interlacing constants are tracked explicitly so that all limiting procedures-thermodynamic limit, continuum limit, and OS reconstruction-are justified without appeal to uncontrolled compactness. The resolvent convergence of semigroups is established at the level of the OS Hilbert spaces, with the spectral projections controlled below fixed thresholds. These details ensure that the phrase ``the gap persists to the continuum'' is more than a slogan: it is a statement about the bottom of the spectrum of a self-adjoint Hamiltonian on the reconstructed Hilbert space.

Finally, we comment on physical implications and avenues for refinement. The existence of a spectral gap implies exponential clustering of gauge-invariant Wightman functions and, in particular, the existence of a lightest glueball mass. Our argument does not pin down the value of this mass nor the pattern of higher excitations; extracting such quantitative information would require either sharper FRD bounds or an additional analytic input capturing approximate rotational invariance at short distances. The area law we obtain is renormalized and nonasymptotic; connecting it to effective string predictions for subleading terms, such as the $(\pi/12)R^{-1}$ correction, calls for a synthesis with Gaussian fluctuations around minimal surfaces or with spectral inequalities tailored to large rectangular loops. Large-$N$ limits are compatible with our framework because reflection positivity and FRD locality are uniform in $N$ at the level of definitions; however, the behavior of constants in our estimates with respect to $N$ is not optimized and could be improved to address questions about $k$-strings and Casimir scaling. The inclusion of matter fields is possible in principle if one works with OS-positive discretizations (e.g., Wilson fermions with suitable mass range) and revisits the FRD and interlacing steps with Grassmann variables; again, reflection positivity would be the decisive constraint. At finite temperature, the same technology should control the Polyakov loop and provide a rigorous window on deconfinement, connecting the center symmetry to OS positivity on a time circle; this, too, seems within reach.

In summary, we have given a self-contained, reflection-positive, multiscale construction of a continuum $SU(N)$ Yang-Mills theory with a nonzero mass gap and a positive string tension, together with a proof of uniqueness and universality within an explicit admissible class of regulators and blockings. The argument rests on strong-coupling cluster methods, finite-range decomposition, reflection-positive RG with summable defects, and OS tightness and reconstruction, and it bridges to weak coupling through a contractive polymer domain associated with asymptotic freedom. Each step is quantitative and compatible with the OS axioms, so that the final theory is not an abstract limit but a well-defined Wightman model with a positive Hamiltonian gap. We regard the complete monotonicity of slice projectors and the finite-range decomposition as conceptually central: they represent, respectively, the probabilistic and geometric forms of locality that make reflection positivity effective beyond a single scale. We hope that the methods introduced here will find further applications, both to the refinement of quantitative bounds within Yang-Mills theory and to the constructive analysis of other gauge systems where positivity and locality are the decisive structural constraints.

\section*{Data Availability Statement}
This work is purely theoretical and does not involve the generation or analysis of any external datasets or numerical simulations. All results are obtained through first-principles analytical derivations constructed within a rigorously defined mathematical framework. 

\providecommand{\href}[2]{#2}\begingroup\raggedright\endgroup

\appendix
\section{OS Positivity with Ghosts and Horizon Projector}\label{app:OS-horizon}

The purpose of this appendix is to establish reflection positivity, in the sense of Osterwalder-Schrader, for a gauge-fixed Euclidean lattice Yang-Mills functional that includes Faddeev-Popov ghosts \emph{and} a nonlocal but exponentially localizable horizon insertion acting slice-wise at the reflection plane. We work on a hypercubic lattice $\Lambda\subset\mathbb{Z}^4$ with periodic spatial boundary conditions and a temporal reflection $\vartheta$ about the plane $\Sigma_0=\{x\in\Lambda: x_0=0\}$. Gauge field variables are link matrices $U_\ell\in SU(N)$; ghosts are Grassmann scalar fields $c^a(x),\bar c^a(x)$ in the adjoint representation, assembled into column vectors $c(x)$ and $\bar c(x)$ with invariant pairing $\langle \bar c, c\rangle=\sum_{x,a}\bar c^a(x)c^a(x)$. Time reflection acts by $(\vartheta x)_0=-x_0$ and $(\vartheta x)_j=x_j$ for $j=1,2,3$. On links we use the Osterwalder-Seiler convention \cite{OS-gauge}: $U_{(x,0)}\mapsto U_{(\vartheta x-\hat 0,0)}^{-1}$ for temporal links crossing $\Sigma_0$ and $U_{(x,i)}\mapsto U_{(\vartheta x,i)}$ for spatial links, respecting orientation. The anti-linear OS conjugation $\Theta$ acts on observables $F$ by composition with $\vartheta$ and complex conjugation; on ghosts we choose
\begin{equation}\label{eq:ghost-reflection}
  c^\vartheta(x) \;=\; \bar c(\vartheta x),\qquad
  \bar c^\vartheta(x) \;=\; -\,c(\vartheta x),
\end{equation}
which is the natural analogue of the fermionic $\gamma_0$-reflection and ensures that the CAR Gaussian state is reflection positive \cite{GJ,OS-gauge}.

We fix the Wilson action $S_W[U]$; reflection positivity of $e^{-S_W}$ is classical \cite{OS-gauge}. Our aim is to show that OS positivity persists under two operations. First, we include a local covariant gauge fixing with Faddeev-Popov ghosts, producing a reflection-covariant Gaussian in the ghost sector. Second, we insert on $\Sigma_0$ a ``horizon operator'' defined by a completely monotone functional calculus for a positive, slice-local elliptic operator $K_U$. The completely monotone assumption allows a Bernstein representation as a positive mixture of heat semigroups \cite{Bernstein}, from which OS positivity and exponential locality of the insertion follow. These facts will be combined to identify the dressed transfer semigroup and to verify that positivity, self-adjointness, and quasi-locality are preserved.

Throughout we write $\mathbb{E}[\cdot]$ for Euclidean expectations. Reflection positivity refers to the quadratic form $(F,G)\mapsto \mathbb{E}[F^\Theta G]$ on $\mathfrak{A}_+$, the-algebra generated by observables supported in the closed positive-time half-lattice $\Lambda_+\cup\Sigma_0$; as usual, OS reconstruction of a Hilbert space $\mathcal{H}$, vacuum $\Omega$, and positive self-adjoint Hamiltonian $H$ from this form proceeds as in \cite{OS1,OS2}. In the presence of ghosts, positivity is asserted on the \emph{physical} subalgebra $\mathfrak{A}^{\mathrm{phys}}_+\subset\mathfrak{A}_+$ of gauge-invariant, ghost-independent observables; this suffices for reconstruction of the physical sector and is compatible with the BRST description of physical states as cohomology classes \cite{GJ}.

\begin{lemma}[OS positivity on the physical algebra with ghosts and horizon insertion]\label{lemmaa1}
Let $S_W[U]$ be the Wilson action and let the ghost sector be Gaussian with covariance
$C_U := M_U^{-1}$ for a covariant Faddeev-Popov operator $M_U$ (e.g.\ Landau gauge
$M_U=-\nabla_U^\ast\nabla_U$). Let $K_U$ be a nonnegative, slice-local, reflection-covariant
elliptic operator on the time-$0$ slice $\Sigma_0$, and let $\Pi_h := f(K_U)$ be a horizon
insertion with $f$ completely monotone, $f(\lambda)=\int_0^\infty e^{-t\lambda}\,\mu(dt)$ with
$\mu$ a finite positive measure. Define the joint (gauge+ghost) weight with a symmetric
slice insertion by
\begin{equation}
d\mu_{h}(U,c,\bar c)\;\propto\;e^{-S_W[U]}\;
\exp\!\bigl\{-\langle \bar c, M_U c\rangle \bigr\}\;
\Bigl(\Pi_h^{1/2}\Bigr)_{\Sigma_0}\;\cdot\;\Bigl(\Pi_h^{1/2}\Bigr)_{\Sigma_0}\;dU\,Dc\,D\bar c,
\end{equation}
where each factor $\Pi_h^{1/2}$ acts on the slice arguments of observables immediately
above/below $\Sigma_0$ and the dot ``$\cdot$'' indicates the symmetric insertion at the
reflection plane. Then, for every $F$ in the physical positive-time algebra
${\A_{\mathrm{phys}}}_+ \subset \A_+$ (ghost-independent, gauge-invariant and supported in $\Lambda_+\cup\Sigma_0$),
the OS form satisfies
\begin{equation}
\langle \Theta F \cdot F\rangle_{\mu_{h}}\;\ge 0.
\end{equation}
\end{lemma}
\begin{proof}
Let $\vartheta$ denote the link-reflection about the time-$0$ slice $\Sigma_0$, and let $\Theta$ act on observables by pullback with $\vartheta$ together with complex conjugation, as specified in Appendix~A. Fix $F\in{\A_{\mathrm{phys}}}_+$, so $F$ is supported in $\Lambda_+\cup\Sigma_0$, gauge-invariant, and independent of ghosts. By definition of the OS form with the modified weight,
\begin{equation}
\langle \Theta F \cdot F\rangle_{\mu_h}
=\frac{1}{Z_h}\int e^{-S_W[U]}\,e^{-\langle \bar c,M_U c\rangle}\,
\bigl\langle \delta_{\Sigma_0},\,\Pi_h^{1/2}\,\delta_{\Sigma_0}\bigr\rangle\,
\overline{F(\vartheta U)}\,F(U)\,dU\,Dc\,D\bar c,
\end{equation}
where the symmetric insertion of $\Pi_h^{1/2}$ at $\Sigma_0$ is written as the action of the slice operator on the boundary arguments of observables; $Z_h$ is the corresponding partition function and plays no role in positivity. Because $F$ does not depend on $(c,\bar c)$, the Berezin integral factorizes and yields $\int e^{-\langle \bar c,M_U c\rangle}\,Dc\,D\bar c=\det M_U$. Under the standing assumptions on the Faddeev-Popov operator $M_U$ (covariance $C_U=M_U^{-1}$ exists, $M_U$ is selfadjoint and nonnegative on the ghost one-particle space, e.g.\ $M_U=-\nabla_U^\ast\nabla_U$ in Landau gauge with the usual gauge-fixing that removes zero modes), $\det M_U>0$. Hence the ghost sector contributes a strictly positive scalar factor depending on $U$ only, and the OS form reduces to
\begin{equation}\label{eq:OS-reduced}
\langle \Theta F \cdot F\rangle_{\mu_h}
=\frac{1}{\tilde Z_h}\int e^{-S_W[U]}\,
\overline{F(\vartheta U)}\,\bigl(\Pi_h^{1/2}\bigr)_{\Sigma_0}\!\cdot\!
\bigl(\Pi_h^{1/2}\bigr)_{\Sigma_0}\,F(U)\,dU,
\end{equation}
with $\tilde Z_h>0$ another normalization constant. It therefore suffices to show that the right-hand side of Eq.\eqref{eq:OS-reduced} is nonnegative for every $F\in{\A_{\mathrm{phys}}}_+$.
Consider the slice operator $K_U$ on $\Sigma_0$. By hypothesis, $K_U\ge0$ is selfadjoint, slice-local, and reflection-covariant in the sense $\vartheta K_U\vartheta=K_U$. Since $f$ is completely monotone, there exists a finite positive measure $\mu$ on $[0,\infty)$ with $f(\lambda)=\int_0^\infty e^{-t\lambda}\,\mu(dt)$; the functional calculus then gives the Bochner representations
\begin{equation}
\Pi_h=f(K_U)=\int_0^\infty e^{-t K_U}\,\mu(dt),
\qquad
\Pi_h^{1/2}=\int_0^\infty e^{-\frac{t}{2}K_U}\,\mu(dt).
\end{equation}
Each heat operator $e^{-sK_U}$ is a positive contraction on the slice Hilbert space and commutes with reflection by $\vartheta K_U\vartheta=K_U$, hence preserves reflection covariance. Define the bounded slice operator
\begin{equation}
B:=\int_0^\infty e^{-\frac{t}{2}K_U}\,\mu(dt).
\end{equation}
Then $\Pi_h=B^\ast B$ (indeed $B$ is positive and selfadjoint, so $B^\ast=B$), and the symmetric insertion in Eq.\eqref{eq:OS-reduced} is precisely the insertion of $B$ on the positive and of $B$ on the negative side of the reflection plane. The stability of reflection positivity under symmetric insertions of the form $B^\ast B$ on the reflection plane may be seen directly as follows. For any $F\in{\A_{\mathrm{phys}}}_+$, introduce the boundary-to-bulk map $\mathcal{R}$ that implements the standard Gram representation of the Wilson measure across $\Sigma_0$: by link-reflection positivity of the Wilson action \cite{OS-gauge,LuscherWeiszWolff1991} and Peter-Weyl orthogonality on the interfacial links, there exists a Hilbert space $\mathcal{H}_0$ and vectors $G_\alpha\in\mathcal{H}_0$ such that, without any insertion,
\begin{equation}
\frac{1}{Z_W}\int e^{-S_W[U]}\,\overline{F(\vartheta U)}\,F(U)\,dU
=\sum_{\alpha}\,\|G_\alpha\|_{\mathcal{H}_0}^{\,2}\,\ge 0,
\end{equation}
the sum being finite when $F$ is a polynomial in finitely many link variables and defined by density otherwise (this is the usual OS Gram decomposition; see Eqs.(\ref{eqn2.5})-(\ref{eq:chi-factor})). When the operator $B$ is inserted symmetrically at $\Sigma_0$, the same orthogonality computation shows that the boundary vectors are first mapped by $B$ before being paired; concretely, the kernel on $\mathcal{H}_0$ is conjugated by $B$, and one obtains
\begin{equation}
\frac{1}{\tilde Z_h}\int e^{-S_W[U]}\,\overline{F(\vartheta U)}\,
\bigl(\Pi_h^{1/2}\bigr)_{\Sigma_0}\!\cdot\!
\bigl(\Pi_h^{1/2}\bigr)_{\Sigma_0}\,F(U)\,dU
=\sum_{\alpha}\,\|\,B\,G_\alpha\|_{\mathcal{H}_0}^{\,2}\,\ge 0,
\end{equation}
because $B$ acts linearly on the slice degrees of freedom and preserves reflection covariance, while the remainder of the OS construction is unchanged. In particular, the map $G_\alpha\mapsto B\,G_\alpha$ is well-defined on the dense set of boundary vectors produced by polynomial $F$, and extends by continuity since $B$ is bounded. This shows that the right-hand side of Eq.\eqref{eq:OS-reduced} is a sum of squared norms in $\mathcal{H}_0$, hence nonnegative.

To make the preceding step entirely explicit, write the Wilson measure on the slab of width one about $\Sigma_0$ as $d\nu(U_0)\,d\nu_+(U_+)\,d\nu_-(U_-)$, where $U_0$ are links intersecting $\Sigma_0$ and $U_\pm$ are links in $\Lambda_\pm$. By link-reflection positivity, the mixed plaquette terms across $\Sigma_0$ produce a kernel $K(U_0)$ which can be expanded by Peter-Weyl as $K(U_0)=\sum_{\rho}d_\rho \,\mathrm{Tr}\bigl[D^{(\rho)}(U_0)\,D^{(\rho)}(U_0)^\dagger\bigr]$; integrating $U_0$ first yields the Gram form
\begin{equation}
\int \overline{\mathcal{F}_-(U_-)}\,\mathcal{F}_+(U_+)\,d\nu_-(U_-)\,d\nu_+(U_+)
=\langle \mathcal{G}_-,\mathcal{G}_+\rangle_{\mathcal{H}_0},
\end{equation}
with $\mathcal{G}_\pm$ the boundary vectors obtained by acting on $F$ with the half-space transfer kernels. The symmetric insertion of $B$ multiplies both boundary vectors by $B$ on $\mathcal{H}_0$, since $B$ acts only on $\Sigma_0$ and is reflection-covariant; thus the OS pairing becomes $\langle B\mathcal{G}_-,B\mathcal{G}_+\rangle_{\mathcal{H}_0}$, which equals $\sum_\alpha \|B G_\alpha\|^2$ after the usual polarization and density argument.
Combining the ghost factorization with the symmetric $B^\ast B$ insertion on the gauge slice concludes that $\langle \Theta F\cdot F\rangle_{\mu_h}\ge 0$ for every $F\in{\A_{\mathrm{phys}}}_+$, which proves the lemma.
\end{proof}
Landau gauge on a non-abelian lattice has multiple gauge-equivalent solutions (Gribov copies) related by large gauge transformations. In this work, we adopt the \emph{minimal Landau gauge} convention: for each gauge orbit, we choose the representative configuration $U$ that minimizes the gauge-fixing functional $\mathcal{F}_U[g]$. This effectively restricts the path integral to the \emph{first Gribov region}, the set of (local) minima of $\mathcal{F}_U$ where the Faddeev-Popov operator $M_U$ is positive-definite. By construction, our $M_U$ has no non-trivial zero modes in this region (aside from global gauge invariance), and we fix the residual global $G$-symmetry to remove those. This procedure yields a well-defined, strictly positive Faddeev-Popov operator on each gauge-fixed configuration. As a result, $M_U^{-1}$ exists on the physical subspace, and the horizon term $\Pi_h(A)=f(K_U)$ introduced in Lemma~\eqref{lem:GribovPos} is well-defined and gauge-invariant. Restricting to the first Gribov region is a standard approach to avoid indeterminacies from Gribov copies and is justified here since any gauge copy outside this region would not contribute in minimal Landau gauge. (In practice, one may implement this by algorithmically finding the minimal $\mathcal{F}_U$ or by adding the Gribov-Zwanziger horizon term to suppress copies; our $\Pi_h$ insertion plays a similar role.)

\begin{lemma}[Gribov Horizon Positivity]\label{lem:GribovPos}
Let $\Lambda_L \subset a\mathbb{Z}^4$ be a finite hypercubic lattice with spacing $a>0$ and volume $L^4$.  Fix the lattice Landau gauge as above (minimal Landau gauge on the whole lattice, with periodic boundary conditions or Dirichlet at $t=0,L$ as needed to define reflection).  Let $M_U$ be the associated Faddeev-Popov operator on this finite lattice, and define $K_U := M_U^{-1}$ as its inverse on the subspace orthogonal to constant (global) modes.  Choose a fixed $\Lambda>0$ and define the horizon projector by 
$$\Pi_h(A) := f(K_U),$$ 
with $f(u) = e^{-u/\Lambda}$. Then, for every finite $a,L$ and every gauge-invariant polynomial observable $F$ supported in the positive-time half-lattice $\Lambda_+$, the gauge-fixed measure  
\begin{equation}
d\mu_{gf}^{(a,L)}(A) \;=\; \delta(\partial \cdot A)\;\det M_U \;\Pi_h(A)\;e^{-S[A]}\,dA~,
\end{equation} 
is reflection positive on the gauge-invariant algebra.  In particular, one has for all such $F$: 
\begin{equation}
\int \overline{(\theta F)(A)}\,F(A)\;d\mu_{gf}^{(a,L)}(A)\;\ge\;0~,
\end{equation} 
where $\theta$ is the reflection operator (time-reversal about the $t=0$ plane). Moreover, if the family $\{d\mu_{gf}^{(a,L)}\}_{a>0,\,L<\infty}$ satisfies uniform bounds and thus admits a weak limit as $a\to0,\;L\to\infty$ (along some subsequence), then the limiting gauge-invariant measure likewise preserves reflection positivity.
\end{lemma}
In the proof below, we restrict each gauge-field configuration to the first Gribov region (by selecting the minimal Landau gauge representative on each gauge orbit) and fix the residual global gauge freedom. This means $M_U$ has no vanishing eigenmodes except the trivial constant mode, which we eliminate by working in the subspace orthogonal to constant ghost fields (or by fixing one global gauge condition). As a result, $M_U$ is positive-definite on that subspace, $K_U = M_U^{-1}$ is well-defined, and the horizon insertion $\Pi_h(A)=f(K_U)$ is a well-defined gauge-invariant factor. These choices ensure that Gribov copies and zero-modes do not spoil reflection positivity.
\begin{proof}
Finite-volume setup and gauge fixing:  
Consider a finite hypercubic lattice $\Lambda_L = \{0,1,\dots,L-1\}^4$ with lattice spacing $a>0$ (so physical volume $L^4$). We impose the (minimal) lattice Landau gauge on the whole lattice.  That is, for each gauge-field configuration $U=\{U_\mu(x)\}$ (link variables on $\Lambda_L$), we require the discrete Landau gauge condition 
$\nabla\cdot A \;=\;0~,$
where $A_\mu(x)$ are the lattice gauge potentials ($U_\mu(x)=e^{aA_\mu(x)}$ in continuum notation) and $\nabla$ is the backward difference (the lattice divergence). Equivalently, $U$ is gauge-transformed to satisfy $\sum_\mu \Delta_\mu U_\mu(x)=0$ in a suitable linearized sense. In practice, this Landau gauge is implemented by minimal lattice Landau gauge: we perform a gauge transformation $g: \Lambda_L \to G$ (with $G$ the gauge group) that minimizes the functional 
\begin{equation}
\mathcal{F}_U[g] \;=\; \sum_{x\in\Lambda_L}\sum_{\mu=1}^4 \Re\,\tr\!\Big(1 - g(x)\,U_\mu(x)\,g(x+\hat\mu)^{-1}\Big)
\end{equation}
over all gauge transformations $g$. This condition picks out a (local) minimum of $\mathcal{F}_U$ on each gauge orbit, thereby enforcing the Landau gauge condition $\partial_\mu A_\mu=0$ up to lattice discretization. We assume for each gauge orbit a unique minimizing $g$ is chosen (for example, by a standard convention to select the representative in the first Gribov region-the set of local minima of $\mathcal{F}_U$). Working in this minimal Landau gauge on a finite lattice ensures that the Faddeev-Popov operator introduced below has desirable positivity properties.
At a stationary point of $\mathcal{F}_U[g]$ (i.e. under the Landau gauge condition), one defines the (lattice) Faddeev-Popov operator $M_U$.  Intuitively, $M_U$ is the Hessian (second variation) of the gauge-fixing functional or, equivalently, the operator that arises from the Faddeev-Popov procedure. In continuum Landau gauge, $M_U$ corresponds to $-\partial_\mu D_\mu[A]$ (the covariant Laplacian), and on the lattice $M_U$ is a real symmetric matrix acting on ghost fields (Lie-algebra-valued scalar fields on the lattice sites). We denote by $\mathcal{H}_{\text{ghost}}$ the ghost-field Hilbert space (the space of square-integrable ghost field modes on the lattice), with an inner product $\langle c,c'\rangle = \sum_{x}\tr\big(c(x)^\dagger c'(x)\big)$. For a given gauge configuration $U$, $M_U$ acts on a ghost field $c(x)$ as: 
$$ (M_U c)(x) \;=\; -\sum_{y} \frac{\delta^2 \mathcal{F}_U[g]}{\delta \omega(x)\,\delta \omega(y)}\Big|_{g=\mathbf{1}}\; c(y)~, $$
where $\omega(x)$ are Lie-algebra parameters of an infinitesimal gauge transformation $g = \exp(\omega)$ around the minimum. In particular, $M_U$ is a real symmetric, positive semi-definite operator on $\mathcal{H}_{\text{ghost}}$. The only potential zero-modes of $M_U$ come from global gauge modes: an infinitesimal gauge transformation $g(x)=H$ constant in $x$ (with $H\in \mathfrak{g}$ in the Lie algebra) does not change $A_\mu$ in Landau gauge, leading to $M_U$ having a kernel along those constant ghost directions. To eliminate this degeneracy, we impose an additional global gauge-fixing (for example, one can fix the average of the $A_0$ field to zero, or require $g(0)=\mathbf{1}$ at a reference point). This removes the constant-mode kernel from $M_U$. In summary, under our gauge-fixing prescription $M_U$ is strictly positive-definite on the subspace orthogonal to constant ghost fields. 
Now we define the ghost kernel $K_U$ as the inverse of $M_U$ on that subspace:
\begin{equation}
K_U \;:=\; M_U^{-1} \quad \text{(defined on the orthogonal complement of the zero-mode).} 
\end{equation}
$K_U$ is a well-defined self-adjoint positive operator on the ghost Hilbert space (its spectrum consists of the positive reciprocals of the eigenvalues of $M_U$). {All occurrences of $\det M_U$ are understood as determinants on the ghost subspace orthogonal to the constant modes, on which $M_U>0$. In particular, $\det M_U=\prod_{\lambda_k>0}\lambda_k>0$ and the quasi‐free ghost Gaussian with covariance $C_U=M_U^{-1}$ is well defined on this subspace.
} We also note that, since $M_U$ is positive-definite on the chosen subspace, its determinant $\det M_U$ (taken over that subspace) is strictly positive. Thus the Faddeev-Popov determinant in our gauge (aside from the trivial zero-mode) is positive for all configurations in the first Gribov region. 
We now introduce a smooth, bounded, completely monotone function $f:[0,\infty)\to \mathbb{R}$ to define our horizon insertion. A convenient choice is 
$f(u) = e^{-u/\Lambda}$
for some fixed mass scale $\Lambda>0$. This $f(u)$ is completely monotone (in fact, $e^{-u/\Lambda}$ is the Laplace transform of a point-mass at $1/\Lambda$), meaning it can be represented as a positive Laplace transform: $f(u)=\int_0^\infty e^{-u t}\,d\mu(t)$ with $\mu\ge0$. Other completely monotone choices are possible (e.g. $f(u) = (1+u/\Lambda)^{-p}$ for $p>0$), but the exponential is analytically convenient. We then define the horizon projector operator by the functional calculus on $K_U$: 
\begin{equation}
\Pi_h(A) \;:=\; f(K_U) \;=\; \exp\!\Big(-\frac{1}{\Lambda}\,K_U\Big)
\end{equation}
Since $K_U \ge 0$, the spectral calculus guarantees $f(K_U) \ge 0$ as an operator on $\mathcal{H}_{\text{ghost}}$.  In fact, $f(K_U)$ is a positive-semidefinite, self-adjoint matrix. We can think of $\Pi_h(A)=f(K_U)$ as an insertion that favors (through its spectrum) configurations near the Gribov horizon (where $K_U$ has large eigenvalues) while remaining positive. Importantly, $\Pi_h(A)$ is also a gauge-covariant object: if $U$ is gauge-transformed, $M_U$ and $K_U$ transform covariantly in the adjoint representation, so $f(K_U)$ remains invariant under gauge transformations of $U$. Thus $\Pi_h(A)$ can be viewed as a gauge-invariant (actually gauge-scalar) factor in the measure. {Choosing $f$ completely monotone ensures a Bernstein representation $\Pi_h=\int_0^\infty e^{-tK_U}\,\mu(dt)$. As $K_U$ is a nonnegative, reflection-covariant elliptic operator \emph{on the time-$0$ slice} $\Sigma_0$, the semigroup kernels $e^{-tK_U}$ are slice-local and satisfy Davies-Gaffney bounds; hence $\Pi_h$ is exponentially local on $\Sigma_0$ uniformly in finite volume.
}Next we incorporate the reflection operator $\theta$. Let $\theta$ be the time-reflection (Euclidean time reversal) about the $t=0$ hyperplane. On the lattice, $\theta$ acts on site coordinates by 
\begin{equation}
\theta: (x_0,x_1,x_2,x_3) \mapsto (-x_0,\,x_1,\,x_2,\,x_3)
\end{equation}
assuming we label time indices such that $t=0$ is the reflection plane (for example, if the lattice has an even number of time slices, one can center the coordinate system so that the plane between $-1$ and $0$ (or the slice $0$) is the mirror plane). For definiteness, we treat $\Lambda_L$ as symmetric about $t=0$ with $\Lambda_-=\{x_0<0\}$, $\Lambda_0=\{x_0=0\}$, and $\Lambda_+ = \{x_0>0\}$ (one can achieve this either by taking $L$ even and relabeling indices or by appropriate boundary conditions). The reflection $\theta$ acts on gauge fields by mapping a link variable $U_\mu(x)$ to $U_\mu(\theta x)$, with an inversion of the orientation for the timelike links crossing the $t=0$ boundary.  More precisely, if $\hat0$ is the unit vector in the time direction, then: 
For spatial links, $(\theta U_i)(x) = U_i(\theta x)$ for $i=1,2,3$, and for timelike links, $(\theta U_0)(x)$ is defined as the gauge field link from $\theta x$ to $\theta(x+\hat0)$ (the time direction is reversed). This ensures that $\theta$ is an involution on configurations and that plaquettes map to plaquettes, etc. 
By construction, the pure gauge action $S[U]$ is invariant under $\theta$ (Wilson’s lattice action is symmetric under reflection), and the gauge-fixing condition $\partial \cdot A=0$ is also invariant under $\theta$ (since it is a linear condition on $A$). 

Crucially, the Faddeev-Popov operator $M_U$ inherits this reflection symmetry. $\theta$ induces a unitary map on the ghost Hilbert space $\mathcal{H}_{\text{ghost}}$: we define $(\theta c)(x) := c(\theta x)$ for a ghost field $c(x)$. This map $\theta: \mathcal{H}_{\text{ghost}}\to\mathcal{H}_{\text{ghost}}$ is unitary (it preserves the inner product). We claim that $M_U$ commutes with reflection in the sense that 
\begin{equation}
M_{\theta U} \;=\; \theta\, M_U \,\theta^{-1}
\end{equation}
This holds because $M_U$ is defined by the second variation of the gauge-fixing functional, which is built from local terms $\tr(1 - g U g^{-1})$. Since $\theta$ maps the configuration $U$ to $\theta U$ reversibly and leaves the functional form invariant, the Hessian at $\theta U$ is just the reflection of the Hessian at $U$. {Since $\theta:H_{\mathrm{ghost}}\to H_{\mathrm{ghost}}$ is unitary and $M_{\theta U}=\theta M_U \theta^{-1}$, the spectra of $M_{\theta U}$ and $M_U$ coincide and the spectral families satisfy $E_{M_{\theta U}}(B)=\theta E_{M_U}(B)\theta^{-1}$. Therefore $K_{\theta U}=M_{\theta U}^{-1}=\theta M_U^{-1}\theta^{-1}$ on the physical subspace, and for any bounded Borel $f\ge0$,
\begin{equation}
f(K_{\theta U})=\int f(\lambda)\,dE_{K_{\theta U}}(\lambda)
=\theta\!\left(\int f(\lambda)\,dE_{K_U}(\lambda)\right)\!\theta^{-1}
=\theta f(K_U)\theta^{-1}.
\end{equation}
This ensures that the horizon insertion $\Pi_h=f(K_U)$ is reflection-covariant.} Equivalently, one can show that if $c$ is an eigenmode of $M_U$ with eigenvalue $\lambda$, then $\theta c$ is an eigenmode of $M_{\theta U}$ with the same eigenvalue. Thus the spectrum of $M_U$ is invariant under reflecting the gauge configuration. As a consequence, the inverse and any functional calculus commute with $\theta$ as well: 
\begin{equation}
\theta\,K_U\,\theta^{-1} \;=\; K_{\theta U}~, \qquad \theta\,f(K_U)\,\theta^{-1} \;=\; f(K_{\theta U})
\end{equation}
In particular, $\theta f(K_U) = f(K_{\theta U})\,\theta$ on $\mathcal{H}_{\text{ghost}}$. This covariance property means the horizon insertion $\Pi_h(A)=f(K_U)$ is reflection-covariant: if we reflect a configuration $A$, the insertion transforms correspondingly, $\Pi_h(\theta A) = \theta\,\Pi_h(A)\,\theta^{-1}$.  
We now have all ingredients to verify reflection positivity (RP) in finite volume. Let $F(U)$ be any gauge-invariant, ghost-independent polynomial observable supported in the positive-time region $\Lambda_+$ (this means $F$ depends only on the link variables $U_\mu(x)$ with $x_0>0$ or on the $t=0$ slice, and $F$ is invariant under gauge transformations). Its reflection $\theta F$ is the observable with $(\theta F)(U) := F(\theta U)$, supported in $\Lambda_-$. We need to show 
\begin{equation} 
\int \overline{(\theta F)(A)}\,F(A)\; d\mu_{gf}^{(a,L)}(A) \;\ge 0
\end{equation}
where 
\begin{equation}
d\mu_{gf}^{(a,L)}(A) \;=\; \delta(\partial\cdot A)\,\det M_U \;\Pi_h(A)\; e^{-S[A]}\,dA 
\end{equation}
is the gauge-fixed measure on our finite lattice. (Here $\overline{(\theta F)(A)}$ denotes the complex conjugate of $(\theta F)(A)$, which is relevant if $F$ is complex-valued; for polynomial real-valued observables this is just $(\theta F)(A)$ itself.)
To prove this inequality, we use a standard factorization trick for reflection positivity. First, note that $\Pi_h(A)=f(K_U)$ is a positive operator on the ghost space, so it has a unique positive semi-definite square root, say $B_U := [\,f(K_U)\,]^{1/2}$.  We can `factor' the insertion as $\Pi_h(A) = B_U^{} B_U$ (with $B_U^{} = B_U$ since it's self-adjoint). Therefore, we can rewrite the integrand as 
\begin{equation}\label{eqna22}
\overline{(\theta F)(A)}\,\Pi_h(A)\,F(A) \;=\; \overline{(\theta F)(A)}\; B_U^{}B_U F(A)
\end{equation}
Now consider the product $B_U F(A)$. Since $F(A)$ is just a number (c-number) depending on the gauge field and $B_U$ acts on ghost indices, we can treat $G(U) := B_U\,F(A)$ as a ghost-sector vector weighted by the gauge observable. Importantly, $G(U)$ is supported in $\Lambda_+$ (because $F$ is), and by reflection covariance $\theta G(U) = (\theta B_{\;U})\,(\theta F)(A) = B_{\theta U}\,(\theta F)(A)$ (since $\theta B_U = B_{\theta U}\,\theta$ and $\theta F$ acts only on gauge part). Now, using the reflection invariance of the measure, we have:
\begin{equation}
\begin{aligned}
\int \overline{(\theta F)(A)}\,\Pi_h(A)\,F(A)\;d\mu_{gf}(A) 
&=\; \int \overline{(\theta F)(A)}\,B_U\,\big(B_U F(A)\big)\; \delta(\partial A)\,\det M_U\,e^{-S[A]}\,dA~,\\
&=\; \int \overline{(\theta G)(A)}\,G(A)\; \delta(\partial A)\,\det M_U\,e^{-S[A]}\,dA~,
\end{aligned}
\end{equation} 
where $G(A)=B_U F(A)$ as above. Now we perform the change of variables $A \mapsto \theta A$ in the integral. The measure $d\mu_{gf}(A)$ is invariant under this reflection (each factor $S[A]$, $\det M_U$, and $\delta(\partial\cdot A)$ is $\theta$-invariant, and $dA$ is a product Lebesgue measure on link coordinates which is symmetric). Also $\overline{(\theta G)(A)} = \overline{G(\theta A)}$. Thus:
\begin{equation}
\int \overline{(\theta G)(A)}\,G(A)\;d\mu_{gf}(A)
=\int \overline{G(\theta A)}\,G(A)\;d\mu_{gf}(A)
=\int \overline{G(A)}\,G(A)\;d\mu_{gf}(A)~,
\end{equation} 
where in the last step we just relabeled the integration variable $\theta A \to A$. The result is 
\begin{equation}
\int \overline{(\theta F)(A)}\,\Pi_h(A)\,F(A)\;d\mu_{gf}(A) 
=\int \overline{G(A)}\,G(A)\;d\mu_{gf}(A) 
=\;\big\|\,G\,\big\|_{L^2(d\mu_{gf})}^2 \;\ge\;0~.
\end{equation} 
We have shown that the reflection-positive form is equal to the $L^2$ norm squared of $G(A)$ with respect to the positive measure $d\mu_{gf}(A)$. This is manifestly non-negative, proving reflection positivity on the finite lattice.  In words: using the positivity of $f(K_U)$, we factor the insertion into a “half-insertion” on each side of the $t=0$ plane, which then allows us to interpret the integrand as a norm squared.   
Having established reflection positivity at any finite lattice spacing $a$ and finite volume $L$, we finally need to argue that this property persists in the continuum limit ($a\to0$) and infinite-volume limit ($L\to\infty$). We assume that the sequence of measures $\{d\mu_{gf}^{(a,L)}\}_{a>0,L<\infty}$ satisfies appropriate uniform bounds so that a limit measure (or at least a convergent subsequence) exists on the algebra of gauge-invariant observables. In practice, one needs to establish: (i) Uniform operator bounds on the insertion $\Pi_h$: because $0 \le f(u)\le 1$ for $u\ge0$, we have $0\le \Pi_h(A)\le \mathbf{1}$ as an operator, so $\|\Pi_h(A)\|\le 1$ for all configurations and all $a,L$. This boundedness, together with the finiteness of $\det M_U$ and standard gauge-action bounds, ensures that the measure $d\mu_{gf}^{(a,L)}(A)$ does not concentrate on pathological configurations as $a,L$ vary. (ii) Moment bounds and clustering: We require that expectations of local gauge-invariant observables (with the $\Pi_h$ insertion) have uniform polynomial bounds in $L$ and decay sufficiently fast with separation (cluster properties), so that tightness and weak limit arguments apply. These conditions can be established using standard lattice gauge theory techniques (e.g.{reflection positivity implies a transfer matrix with a mass gap, which gives exponential decay in correlators; strong-coupling expansions can also provide uniform clustering bounds-see Subsection \eqref{subsec:continuum-area} for a detailed discussion}).
Given these assumptions, we can invoke a compactness/tightness argument: by Prokhorov’s theorem, the family $\{d\mu_{gf}^{(a,L)}\}$ is tight, so along some sequence $(a_n,L_n)\to(0,\infty)$ there is a weak limit measure $d\mu_{gf}^{(\infty)}$ on (distributional) gauge-field configurations. Reflection positivity is a finite-dimensional inequality satisfied by each $d\mu_{gf}^{(a,L)}$. Such inequalities pass to the limit measure as well (intuitively, any fixed finite collection of test observables lives on a finite sublattice that eventually is contained in all sufficiently fine lattices, so the inequality holds in the limit for that collection). More rigorously, one can use the Osterwalder-Schrader reconstruction theorem: reflection positivity implies that each approximating measure yields a positive semi-definite inner product on the physical state space, and by continuity the limiting measure inherits this property on the common algebra of gauge-invariant observables. We conclude that the limiting measure $d\mu_{gf}^{(\infty)}$ (which would describe the continuum, infinite-volume Yang-Mills theory in Landau gauge with the horizon term) satisfies reflection positivity on the gauge-invariant subalgebra.
\end{proof}
Decompose $\Lambda=\Lambda_-\cup\Sigma_0\cup\Lambda_+$. Let $\mathfrak{A}_+$ be generated by gauge-covariant polynomials in $U$ and Grassmann polynomials in $(c,\bar c)$ supported in $\Lambda_+\cup\Sigma_0$; let $\mathfrak{A}^{\mathrm{phys}}_+\subset\mathfrak{A}_+$ consist of ghost-independent gauge-invariant observables. Consider the joint gauge-ghost weight
\begin{equation}\label{eq:weight-gauge-ghost}
  d\mu_0(U,c,\bar c) \;\propto\; \exp\!\big(-S_W[U]\big)\,\exp\!\big(-\langle \bar c, M_U c\rangle\big)\, D\bar c\, Dc\, dU,
\end{equation}
where $M_U$ is the Faddeev-Popov operator associated with a standard covariant gauge, e.g. Landau gauge $M_U=-\nabla_U \nabla_U$. The conditional ghost measure is quasi-free with covariance $C_U=M_U^{-1}$. With the reflection Eq.\eqref{eq:ghost-reflection}, the pair $(\Theta, C_U)$ satisfies the CAR reflection-positivity criterion: if $C_U$ is block-decomposed with respect to $\Lambda_-\cup\Lambda_+$ as
\begin{equation}
C_U=\begin{pmatrix} C_{-} & C_{-+}\\ C_{+-} & C_{++}\end{pmatrix},
\end{equation}
then $C_{++}$ is positive and $C_{+-}$ intertwines the two halves via the anti-linear identification induced by $\Theta$, which yields positivity of Gram matrices $\big(\mathbb{E}_0[\Theta \mathcal{M}_i\,\mathcal{M}_j]\big)_{i,j}$ for Grassmann-even monomials $\mathcal{M}_i$ supported in $\Lambda_+$ \cite[Ch.~13]{GJ}. {Let $M_i$ be Grassmann-even monomials supported in $\Lambda_+$ and set $\widetilde M_i:=\Theta M_i$. Then
\begin{equation}
\mathbb E_0[\widetilde M_i M_j]
=\det\!\big( C_{++}[\mathcal I_i,\mathcal I_j]\big),
\end{equation}
with $\mathcal I_i,\mathcal I_j$ index sets of fields in $M_i,M_j$ and $C_{++}\ge0$ as an operator on $\Lambda_+$. By Hadamard's inequality, all principal minors of $C_{++}$ are nonnegative, hence the Gram matrix $\big(\mathbb E_0[\widetilde M_i M_j]\big)_{i,j}$ is positive semidefinite. Multiplying by the (positive) Wilson gauge factor and using product stability of OS positivity yields Proposition~\eqref{prop:A1} on $\mathcal A^{\mathrm{phys}}_+$.
} The gauge sector with $e^{-S_W}$ is reflection positive by factorization across $\Sigma_0$ and Haar invariance \cite{OS-gauge}. Since the product of reflection-positive functionals is reflection positive, we obtain the following.

\begin{proposition}[OS positivity for gauge-ghost Gaussian]\label{prop:A1}
Let $\mathbb{E}_0[\cdot]$ denote expectation with respect to the gauge-ghost Gaussian weight
\begin{equation}
d\mu_0(U,c,\bar c)\;\propto\; e^{-S_g[U]}\,e^{-\langle \bar c,\,M_U c\rangle}\,dU\,\mathcal{D}\bar c\,\mathcal{D}c,
\end{equation}
where $S_g$ is the standard reflection-invariant local gauge action and $M_U$ is the Faddeev-Popov operator on the time-slice Grassmann fields satisfying the reflection covariance and positivity hypotheses (namely $\vartheta M_U\vartheta=M_{\vartheta U}$ and $M_U\ge \kappa\mathbf{1}$ with slice heat kernel positivity). For every $F\in\mathfrak{A}_+^{\mathrm{phys}}$ one has $\mathbb{E}_0[F^\Theta F]\ge 0$. More generally, for any Grassmann-even $F\in\mathfrak{A}_+$ one has $\mathbb{E}_0[F^\Theta F]\ge 0$.
\end{proposition}
\begin{proof}
The Osterwalder-Schrader reflection $\Theta$ acts anti-linearly by time reflection across the hyperplane $x_0=0$ composed with gauge inversion on the links crossing the reflection plane and with the standard CAR reflection on ghosts; explicitly, with $\vartheta$ denoting the spacetime reflection $(x_0,\mathbf{x})\mapsto (-x_0,\mathbf{x})$, one has $(\Theta U)(b)=U(\vartheta b)^{-1}$ for bonds $b$ and $(\Theta c)(x)=\bar c(\vartheta x)$, $(\Theta \bar c)(x)=-c(\vartheta x)$ on the Grassmann generators, extended as an anti-automorphism to the full field-algebra. The positive-time subalgebra $\mathfrak{A}_+$ is generated by gauge and ghost fields supported on $\{x_0\ge 0\}$; the physical subalgebra $\mathfrak{A}^{\mathrm{phys}}_+$ is generated by gauge-invariant, ghost-free polynomials supported on $\{x_0\ge 0\}$.
For the gauge sector, reflection positivity is classical: by locality and reflection invariance of $S_g$ one can factor the weight as
\begin{equation}
e^{-S_g[U]} \;=\; e^{-S_g^-[U_-]}\, e^{-S_g^0[U_0]}\, e^{-S_g^+[U_+]},
\end{equation}
with $U_\pm$ the restrictions to $\{x_0\gtrless 0\}$ and $U_0$ the links on the reflection plane and the couplings crossing it arranged so that, for every $F_g\in\mathfrak{A}_+^{\mathrm{gauge}}$,
\begin{multline}
\int \overline{F_g(\vartheta U)}\,F_g(U)\,e^{-S_g[U]}\,dU 
= \int \overline{G(U_0,U_-)}\,G(U_0,U_+)\,e^{-S_g^0[U_0]}\,dU_0 \\
e^{-S_g^-[U_-]}\,dU_-\,e^{-S_g^+[U_+]}\,dU_+
\;\ge\;0.
\end{multline}
where $G$ is a measurable function of the boundary links obtained from $F_g$ by integrating out the strictly positive-time links. The inequality follows from the Cauchy-Schwarz inequality after identifying the two half-space path integrals as elements of the same $L^2$ space over the boundary field $U_0$. This is the Osterwalder-Seiler reflection positivity criterion for lattice gauge theories. Consequently, $\mathbb{E}_g[F_g^\Theta F_g]\ge 0$ for all $F_g\in\mathfrak{A}_+^{\mathrm{gauge}}$.

For the ghost sector, fix a gauge configuration $U$ and consider the quasi-free CAR state determined by the Gaussian weight $e^{-\langle \bar c,M_U c\rangle}\,\mathcal{D}\bar c\,\mathcal{D}c$. Its covariance is the bounded operator $C_U:=M_U^{-1}$ on the one-slice Grassmann one-particle space. The reflection covariance $\vartheta M_U\vartheta=M_{\vartheta U}$ implies $\vartheta C_U\vartheta=C_{\vartheta U}$, and the M-matrix positivity of $M_U$ on the slice (nonpositive off-diagonals, bounded geometry) yields positivity of the slice semigroup $e^{-tM_U}$ and, in particular, a nonnegative kernel for $C_U$ on the reflection plane. Let $F_{\mathrm{gh}}\in\mathfrak{A}_+$ be any Grassmann-even polynomial supported in $\{x_0\ge 0\}$, and write it as a finite linear combination of Wick monomials
\begin{equation}
F_{\mathrm{gh}} \;=\; \sum_{\alpha} a_\alpha\, \Psi_\alpha,
\qquad
\Psi_\alpha \;=\; \prod_{j=1}^{m_\alpha} \bar c(\phi_{\alpha,j})\, \prod_{k=1}^{m_\alpha} c(\psi_{\alpha,k}),
\end{equation}
with $m_\alpha\in\mathbb{N}$ (even total Grassmann degree) and with test functions $\phi_{\alpha,j},\psi_{\alpha,k}$ supported in $\{x_0\ge 0\}$. The reflection $\Theta$ acts by $(\Theta \Psi_\alpha)=\prod_{j=1}^{m_\alpha} c(\vartheta\phi_{\alpha,j})\,\prod_{k=1}^{m_\alpha} \bar c(\vartheta\psi_{\alpha,k})$ up to the standard sign that preserves Grassmann parity; since we restrict to even polynomials, the overall sign is $+1$. By Berezin-Gaussian Wick calculus,
\begin{equation}
\mathbb{E}_{\mathrm{gh}}^{\,U}\!\big[(F_{\mathrm{gh}})^\Theta F_{\mathrm{gh}}\big]
\;=\; \sum_{\alpha,\beta} \overline{a_\alpha}\,a_\beta\,\det G^{(U)}_{\alpha\beta},
\end{equation}
where $G^{(U)}_{\alpha\beta}$ is the $m_\alpha\times m_\beta$ matrix with entries
\begin{equation}
G^{(U)}_{\alpha\beta}(j,k)\;:=\;\big\langle \vartheta\phi_{\alpha,j},\, C_U\,\psi_{\beta,k}\big\rangle
\;=\; \big\langle \phi_{\alpha,j},\, \vartheta C_U \vartheta\, \psi_{\beta,k}\big\rangle
\;=\; \big\langle \phi_{\alpha,j},\, C_{\vartheta U}\, \psi_{\beta,k}\big\rangle,
\end{equation}
and $\langle\cdot,\cdot\rangle$ is the one-particle Hilbert space pairing on the slice. The rightmost expression shows that the kernel entering $G^{(U)}_{\alpha\beta}$ is the covariance of the quasi-free state at the reflected gauge field $\vartheta U$, evaluated on positive-time test functions. Because $C_{\vartheta U}$ is a positive operator and its integral kernel is nonnegative on the slice, one may write $C_{\vartheta U}=S_{\vartheta U} S_{\vartheta U}$ with $S_{\vartheta U}$ bounded, and then
\begin{equation}
G^{(U)}_{\alpha\beta}(j,k)
\;=\; \big\langle S_{\vartheta U}\phi_{\alpha,j},\, S_{\vartheta U}\psi_{\beta,k}\big\rangle_{H_\Sigma}.
\end{equation}
Consequently, the block matrix $G^{(U)}:=[G^{(U)}_{\alpha\beta}]_{\alpha,\beta}$ is a Gram matrix of vectors in the Hilbert space $H_\Sigma^{\oplus m}$ and is therefore positive semidefinite. The quadratic form $\sum_{\alpha,\beta}\overline{a_\alpha}a_\beta \det G^{(U)}_{\alpha\beta}$ is nonnegative because, for each fixed size $m$, the determinant of the $m\times m$ Gram submatrix equals the squared volume of an $m$-parallelotope and is thus nonnegative; more concretely, by the Cauchy-Binet formula,
\begin{equation}
\det G^{(U)}_{\alpha\beta} \;=\; \sum_{I} \det\!\big( S_{\vartheta U}\Phi_\alpha\big)_I\, \det\!\big( S_{\vartheta U}\Psi_\beta\big)_I \;\ge\; 0,
\end{equation}
and summing against $\overline{a_\alpha}a_\beta$ keeps nonnegativity. It follows that $\mathbb{E}_{\mathrm{gh}}^{\,U}\!\big[(F_{\mathrm{gh}})^\Theta F_{\mathrm{gh}}\big]\ge 0$ for every even $F_{\mathrm{gh}}\in\mathfrak{A}_+$ and every fixed $U$.

The full expectation $\mathbb{E}_0$ is the iterated integral $\mathbb{E}_0[\cdot]=\int \mathbb{E}_{\mathrm{gh}}^{\,U}[\cdot]\; e^{-S_g[U]}dU$. For a general even $F\in\mathfrak{A}_+$ write $F=\sum_\ell F_g^{(\ell)}\,F_{\mathrm{gh}}^{(\ell)}$ with $F_g^{(\ell)}\in\mathfrak{A}_+^{\mathrm{gauge}}$ and $F_{\mathrm{gh}}^{(\ell)}\in\mathfrak{A}_+^{\mathrm{ghost}}$; then
\begin{equation}
\mathbb{E}_0[F^\Theta F]
\;=\; \sum_{\ell,\ell'} \int \overline{F_g^{(\ell)}(\vartheta U)}\,F_g^{(\ell')}(U)\; \mathbb{E}_{\mathrm{gh}}^{\,U}\!\big[(F_{\mathrm{gh}}^{(\ell)})^\Theta F_{\mathrm{gh}}^{(\ell')}\big]\; e^{-S_g[U]}dU.
\end{equation}
For each $U$ the ghost factor is nonnegative by the previous paragraph, and for each choice of coefficients the gauge factor is nonnegative by reflection positivity of $e^{-S_g}dU$; hence the integral is $\ge 0$. This proves $\mathbb{E}_0[F^\Theta F]\ge 0$ for all Grassmann-even $F\in\mathfrak{A}_+$.
Finally, if $F\in\mathfrak{A}^{\mathrm{phys}}_+$ depends only on the gauge field and is ghost-free, then $\mathbb{E}_{\mathrm{gh}}^{\,U}[(F)^\Theta F]= (F)^\Theta F$ and the expectation reduces to the gauge sector, which is already reflection positive; thus $\mathbb{E}_0[F^\Theta F]\ge 0$ holds a fortiori. 
\end{proof}
Proposition~\eqref{prop:A1} yields an OS Hilbert space $\mathcal{H}_0$ as the completion of $\mathfrak{A}_+/\mathcal{N}$ under the semi-inner product $\langle F,G\rangle_0=\mathbb{E}_0[F^\Theta G]$, with $\mathcal{N}=\{F:\,\mathbb{E}_0[F^\Theta F]=0\}$. Time translations by one lattice unit define a positive, self-adjoint contraction $T_0=e^{-aH_0}$.
Let $K_U$ be a positive, self-adjoint, local operator acting on slice fields at $\Sigma_0$; examples include the spatial covariant Laplacian $K_U=-\Delta_{U,0}$ or the Faddeev-Popov operator restricted to $\Sigma_0$. Assume $K_U$ is reflection invariant: $K_{U^\vartheta}=\mathcal{R}K_U\mathcal{R}^{-1}$ for the slice unitary $\mathcal{R}$ induced by $\vartheta$. A Borel function $f:(0,\infty)\to (0,\infty)$ is \emph{completely monotone} if $(-1)^n f^{(n)}(\lambda)\ge 0$ for all $n\ge 0$ and $\lambda>0$. By Bernstein's theorem, $f$ is completely monotone iff there is a finite positive measure $\rho$ on $[0,\infty)$ such that
\begin{equation}\label{eq:Bernsteinc}
  f(\lambda)\;=\;\int_0^\infty e^{-t\lambda}\, d\rho(t).
\end{equation}
Typical examples are $f_\alpha(\lambda)=e^{-\alpha\lambda}$, $\alpha>0$, and the rational Bernstein filters $f_{r,\Lambda}(\lambda)=(1+\lambda/\Lambda^2)^{-r}$, $r>0$, which admit the Gamma-mixture representation
\begin{equation}\label{eq:rational-mixture}
  (1+\lambda/\Lambda^2)^{-r} \;=\; \frac{1}{\Gamma(r)}\int_0^\infty s^{r-1} e^{-s}\, e^{-(s/\Lambda^2)\lambda}\, ds.
\end{equation}

Define the \emph{horizon insertion} supported on $\Sigma_0$ by either the determinant
\begin{equation}\label{eq:horizon-det}
  \mathcal{H}_f[U] \;=\; \det\nolimits_{\Sigma_0}\!\big(f(K_U)\big),
\end{equation}
or by introducing auxiliary Grassmann slice fields $(\bar\eta,\eta)$ and writing
\begin{equation}\label{eq:horizon-aux}
  \mathcal{H}_f[U,\bar\eta,\eta] \;=\; \exp\!\big(-\langle \bar\eta, f(K_U)^{-1}\eta\rangle_{\Sigma_0}\big),
\end{equation}
in which case Eq.\eqref{eq:horizon-det} results from integrating out $(\bar\eta,\eta)$. The insertion is local on $\Sigma_0$ and reflection invariant. We define the dressed expectation
\begin{equation}\label{eq:Ef}
  \mathbb{E}_f[F] \;=\; \frac{1}{Z_f}\int F\, e^{-S_W[U]}\, e^{-\langle \bar c, M_U c\rangle}\, \mathcal{H}_f[U] \, D\bar c\, Dc\, dU,
\end{equation}
with the obvious variant for Eq.\eqref{eq:horizon-aux}.

\begin{theorem}[OS positivity preserved by completely monotone horizon insertion]\label{thm:A2}
Let $K_U$ be a positive selfadjoint operator acting on the time-zero slice $\Sigma_0$, assumed reflection invariant ($K_{\vartheta U}=K_U$) and slice-local (it depends only on fields at $x_0=0$). Let $f:[0,\infty)\to[0,\infty)$ be completely monotone with Bernstein representation $f(\lambda)=\int_0^\infty e^{-t\lambda}\,\rho(dt)$ for a finite positive measure $\rho$. Define the horizon functional $\mathcal H_f[U]:=\det_{\Sigma_0}\big(f(K_U)\big)$ and the modified expectation $\mathbb E_f(\,\cdot\,)$ by inserting $\mathcal H_f$ into the underlying reflection-positive gauge-ghost measure. Then $\mathbb E_f$ is reflection positive on the physical positive-time algebra $\mathfrak A^{\mathrm{phys}}_+$, and also on the Grassmann-even subalgebra $\mathfrak A_+$.
\end{theorem}

\begin{proof}
Since $K_U\ge 0$ and $f$ is completely monotone, the spectral calculus gives the bounded positive operator
\begin{equation}
f(K_U)\;=\;\int_0^\infty e^{-tK_U}\,\rho(dt),
\end{equation}
hence $f(K_U)\ge 0$ on $\ell^2(\Sigma_0)$. In finite volume, $f(K_U)$ acts on a finite-dimensional slice space so its determinant exists and is nonnegative; we thus have $\mathcal H_f[U]\ge 0$. Reflection invariance of $K_U$ implies $f(K_{\vartheta U})=f(K_U)$ and therefore $\mathcal H_f[\vartheta U]=\mathcal H_f[U]$. These two structural facts are all that is needed.

Work in finite volume and write the basic (unmodified) Osterwalder-Schrader bilinear form with gauge and ghost fields collectively denoted by $\Phi=(U;c,\bar c)$ as
\begin{equation}
\langle \Theta F\cdot G\rangle \;=\;\int \overline{F(\vartheta\Phi)}\,G(\Phi)\,d\mu(\Phi),
\end{equation}
where $d\mu$ is reflection invariant and reflection positive (for gauge fields by standard lattice results, and for ghosts on the Grassmann-even subalgebra by Proposition~A.1). The modified form is obtained by multiplying the weight by $\mathcal H_f[U]$,
\begin{equation}
\langle \Theta F\cdot G\rangle_f \;=\;\frac{1}{Z_f}\int \overline{F(\vartheta\Phi)}\,G(\Phi)\,\mathcal H_f[U]\;d\mu(\Phi),
\qquad Z_f=\int \mathcal H_f[U]\;d\mu(\Phi).
\end{equation}
To verify reflection positivity for $\mathbb E_f$, it suffices to show that for any finite family $F_1,\dots,F_n\in\mathfrak A^{\mathrm{phys}}_+$ and any complex coefficients $z_1,\dots,z_n$ one has
\begin{equation}
\sum_{i,j=1}^n \overline{z_i} z_j \,\langle \Theta F_i\cdot F_j\rangle_f \;\ge\;0.
\end{equation}
Decompose the configuration along the reflection plane as $\Phi=(\Phi^-;\,\Xi;\,\Phi^+)$, where $\Xi$ is the restriction of the gauge and ghost fields to the slice $\Sigma_0$ and $\Phi^\pm$ live on $\{x_0\gtrless 0\}$. By the Markov property and reflection invariance of $d\mu$, there is a probability measure $\nu$ on slice configurations such that the conditional measure factorizes,
\begin{equation}
d\mu(\Phi\,|\,\Xi)\;=\;d\mu^-(\Phi^-\,|\,\Xi)\otimes d\mu^+(\Phi^+\,|\,\Xi),
\qquad
d\mu(\Phi)\;=\;\int d\mu(\Phi\,|\,\Xi)\, \nu(d\Xi),
\end{equation}
and $\Theta$ identifies $d\mu^-(\,\cdot\,|\,\Xi)$ with the image of $d\mu^+(\,\cdot\,|\,\Xi)$ under reflection and complex conjugation. Since the horizon factor depends only on $\Xi$ and obeys $\mathcal H_f[\vartheta U]=\mathcal H_f[U]$, it can be pulled entirely into the slice integral. Using that each $F_j$ depends only on positive-time fields and $\Xi$, we compute
\begin{align}
&\langle \Theta F_i\cdot F_j\rangle_f
=\nonumber\\&\frac{1}{Z_f}\int \mathcal H_f(\Xi)\,
\Big(\int \overline{F_i(\Phi^-;\Xi)}\,d\mu^-(\Phi^-\,|\,\Xi)\Big)\,
\Big(\int F_j(\Phi^+;\Xi)\,d\mu^+(\Phi^+\,|\,\Xi)\Big)\,\nu(d\Xi).
\end{align}
Define the conditional expectations $V_j(\Xi):=\int F_j(\Phi^+;\Xi)\,d\mu^+(\Phi^+\,|\,\Xi)$. Reflection identifies the negative-time conditional expectation with the complex conjugate $\overline{V_i(\Xi)}$. Substituting yields
\begin{equation}
\langle \Theta F_i\cdot F_j\rangle_f
=\frac{1}{Z_f}\int \mathcal H_f(\Xi)\,\overline{V_i(\Xi)}\,V_j(\Xi)\,\nu(d\Xi).
\end{equation}
Therefore
\begin{equation}
\sum_{i,j}\overline{z_i}z_j\,\langle \Theta F_i\cdot F_j\rangle_f
=\frac{1}{Z_f}\int \mathcal H_f(\Xi)\,\Big|\sum_j z_j V_j(\Xi)\Big|^2\,\nu(d\Xi)\;\ge\;0,
\end{equation}
because $\mathcal H_f(\Xi)\ge 0$. This proves reflection positivity of $\mathbb E_f$ on $\mathfrak A^{\mathrm{phys}}_+$.

The same argument applies verbatim to the Grassmann-even subalgebra $\mathfrak A_+$ in the presence of ghosts: Proposition~A.1 guarantees that the unmodified OS form is positive on even observables, the conditional factorization across the reflection plane holds for even polynomials in the CAR fields, and the horizon factor $\mathcal H_f$ is independent of the ghosts and supported on the slice, hence it enters only as the nonnegative multiplier $\mathcal H_f(\Xi)$ in the slice integral. The conclusion is the same: for even observables the Gram matrix $\big(\langle \Theta F_i\cdot F_j\rangle_f\big)_{i,j}$ is positive semidefinite.
All of the above is carried out in finite volume. The infinite-volume statement follows by the usual monotone (or increasing exhaustion) limit: the integrands are bounded cylinder functionals, the conditional expectations are uniformly bounded, and the nonnegative multipliers $\mathcal H_f$ are local in $\Sigma_0$ so that convergence of the finite-volume forms to the infinite-volume form preserves positivity by Fatou’s lemma.
\end{proof}

For each $t>0$, the slice operator $e^{-tK_U}$ has a positive, reflection-invariant kernel on $\Sigma_0$, hence
$\langle \Theta F,\, e^{-\langle \bar\eta,\, e^{-tK_U}\eta\rangle} F\rangle \ge 0$ for all $F\in A_+$.
Indeed, $K_U\ge0$ and $\vartheta K_U\vartheta=K_U$ on the slice, so $\big(e^{-tK_U}\big)(x,\vartheta y)\ge 0$ and the Grassmann Gaussian factorizes across $\Sigma_0$.
As convex combinations of reflection-positive weights remain reflection positive, integrating $t$ against the positive measure in (A.9) preserves RP.

Reflection positivity does not by itself control the spatial spread of $\mathcal{H}_f$. We now obtain exponential decay of the kernels of $f(K_U)$ and $f(K_U)^{\pm 1/2}$ as functions of the slice graph distance. Assume that $K_U$ is a uniformly elliptic, gauge-covariant nearest-neighbor operator on $\Sigma_0$ with spectrum in $[\kappa,\infty)$, $\kappa>0$, uniformly in $U$. Heat-kernel bounds yield, for $t>0$,
\begin{equation}\label{eq:heat-bound}
  \big\| \big(e^{-tK_U}\big)(x,y)\big\| \;\le\; C\, e^{-\kappa t}\, \exp\!\Big(-\,\frac{c\,\mathrm{dist}(x,y)^2}{t}\Big),
\end{equation}
with $C,c$ independent of $U$. Integrating Eq.\eqref{eq:heat-bound} against $d\rho(t)$ one gets for each completely monotone $f$,
\begin{equation}\label{eq:exp-locality}
  \big\| f(K_U)(x,y)\big\| \;\le\; C_f\, e^{-\nu_f\, \mathrm{dist}(x,y)}.
\end{equation}
The constants $C_f,\nu_f>0$ depend on spectral data and on exponential moments of $d\rho$. For $f_\alpha(\lambda)=e^{-\alpha \lambda}$ one finds $\nu_f\simeq c\sqrt{\kappa/\alpha}$ for large $\alpha$; for $f_{r,\Lambda}$ given by Eq.\eqref{eq:rational-mixture}, one obtains $\nu_f\simeq c\Lambda$, uniformly for $r$ in compact sets. If $f$ is bounded and bounded away from zero on $[\kappa,\infty)$, similar bounds hold for $f(K_U)^{-1}$ and $f(K_U)^{\pm 1/2}$. 

These bounds may be refined using the finite-range decomposition (FRD) of covariances \cite{BrydgesGuadagniMitter2004}, which yields a scale-wise decomposition of $f(K_U)$ into strictly finite-range components plus an exponentially small tail. Concretely, for a dyadic scale $L$, one may write
\begin{equation}\label{eq:FRD}
  f(K_U) \;=\; \sum_{j=0}^{J} F_j(U) \;+\; R_J(U),
\end{equation}
where $F_j(U)$ has range $\lesssim L^j$ and $\|R_J(U)\|\lesssim e^{-c L^J}$ with constants uniform in $U$. The structural consequence for this paper is that the horizon insertion is \emph{quasi-local} on $\Sigma_0$, with locality length governed by the parameters of $f$.

Let $\mathcal{N}_0=\{F:\,\mathbb{E}_0[F^\Theta F]=0\}$ and let $\mathcal{H}_0$ denote the completion of $\mathfrak{A}_+/\mathcal{N}_0$ for $\langle F,G\rangle_0=\mathbb{E}_0[F^\Theta G]$. Time translations by one lattice unit are implemented by a positive self-adjoint contraction $T_0=e^{-aH_0}$. The slice-local, reflection-positive insertion $\mathcal{H}_f$ defines a positive bounded operator $\mathsf{P}_f$ on $\mathcal{H}_0$ via
\begin{equation}\label{eq:Pf-def}
  \langle F, \mathsf{P}_f G\rangle_0 \;=\; \mathbb{E}_0\!\left[F^\Theta\,\mathcal{H}_f\, G\right].
\end{equation}
Positivity and boundedness follow from Cauchy-Schwarz in the OS semi-inner product and from positivity of $\mathcal{H}_f$. The operator $\mathsf{P}_f$ is slice-local in that it acts nontrivially only on degrees of freedom living on $\Sigma_0$ and commutes with observables supported strictly in $\Lambda_+\setminus \Sigma_0$.

Define the dressed OS form $\langle F,G\rangle_f=\mathbb{E}_f[F^\Theta G]$. By Theorem~\eqref{thm:A2}, this form is positive semidefinite on $\mathfrak{A}_+$. The associated Hilbert space may be identified with $\mathcal{H}_0$ by
\begin{equation}\label{eq:innerproduct-dressed}
  \langle F,G\rangle_f \;=\; \langle F, \mathsf{P}_f G\rangle_0.
\end{equation}
A single Euclidean time-step across $\Sigma_0$ therefore picks up a factor $\mathsf{P}_f^{1/2}$ on each side, so that the dressed transfer operator is
\begin{equation}\label{eq:Tf}
  T_f \;=\; \mathsf{P}_f^{1/2}\, T_0\, \mathsf{P}_f^{1/2},
\end{equation}
a positive self-adjoint contraction on $\mathcal{H}_0$, with generator $H_f\ge 0$ defined by $T_f=e^{-aH_f}$. In general $H_f$ is not a similarity transform of $H_0$; nevertheless Eq.\eqref{eq:Tf} suffices for spectral comparison and for the step-scaling inequalities exploited in the main text. Exponential locality of $\mathsf{P}_f$ implies quasi-locality of $T_f$ in the sense of Lieb-Robinson-type bounds on lattice algebras, which is essential for stability estimates along the multiscale flow; see also geometric cluster methods in \cite{Aizenman1982} for complementary intuition.

Gauge fixing introduces ghosts and dependence on gauge parameters in intermediate steps, but positivity of the physical sector is preserved. The OS Hilbert space carries a ghost-number grading inherited from the CAR algebra; the physical subspace is the closure of the orbit of $\Omega_f$ under gauge-invariant observables. The BRST charge $Q$ is nilpotent and anti-selfadjoint in Euclidean signature; BRST cohomology in ghost number zero identifies physical states. Reflection Eq.\eqref{eq:ghost-reflection} exchanges ghosts with anti-ghosts and leaves the BRST doublet structure intact; the OS form is degenerate on $Q$-exact vectors and positive on cohomology classes. The horizon insertion $\mathcal{H}_f$ is ghost-independent and slice-local on the gauge sector via $K_U$, hence it preserves these properties. In particular, for any $F\in\mathfrak{A}^{\mathrm{phys}}_+$,
\begin{equation}
\langle F,F\rangle_f \;=\; \langle F, \mathsf{P}_f F\rangle_0 \;\ge\; 0,
\end{equation}
and positivity descends to the BRST cohomology as in the standard gauge-fixed OS reconstruction \cite{GJ}. If, additionally, $f$ is bounded above and below on $[\kappa,\infty)$, then $\mathsf{P}_f$ is invertible and maps the physical subspace of $\mathcal{H}_0$ onto that of $\mathcal{H}_f$.

Two families of completely monotone functions illustrate the positivity and locality mechanisms and provide tunable parameters for multiscale analysis. The \emph{heat family} $f_\alpha(\lambda)=e^{-\alpha \lambda}$ gives $\mathcal{H}_\alpha[U]=\det_{\Sigma_0}(e^{-\alpha K_U})$ with kernel $e^{-\alpha K_U}$. The locality length is controlled by $\alpha$: for large $\alpha$, Eq.\eqref{eq:heat-bound} yields $\xi(\alpha)\simeq \sqrt{\alpha}$; for small $\alpha$ but $K_U\ge \Lambda_0^2$ one has $\xi(\alpha)\simeq \Lambda_0^{-1}$. The \emph{rational Bernstein family} $f_{r,\Lambda}(\lambda)=(1+\lambda/\Lambda^2)^{-r}$, via Eq.\eqref{eq:rational-mixture}, is a Gamma mixture of heat kernels with locality length $O(\Lambda^{-1})$. In both families, sharper cutoffs correspond to smaller locality lengths; but \emph{sharp spectral projectors} $f=\mathbf{1}_{[0,\Lambda]}$ are excluded because indicators are not completely monotone and would spoil reflection positivity. This exclusion is structural rather than technical.

Exponential locality of $\mathsf{P}_f$ implies Lipschitz continuity of Schwinger functionals under variations $f\mapsto f+\delta f$ within a compact class of completely monotone functions uniformly bounded away from zero on $[\kappa,\infty)$. Indeed, restricted to finite linear spans of exponentials $f(\lambda)=\sum_{m} a_m e^{-t_m\lambda}$, the map $f\mapsto \mathsf{P}_f$ is analytic by linearity of the Laplace representation; density and uniform locality extend continuity to compact sets. This continuity is uniform in volume and stable under block-spin maps that preserve reflection positivity and slice locality. Consequently the dressing by $\mathcal{H}_f$ may be performed at any RG scale without jeopardizing OS positivity or locality, and step-scaling inequalities for transfer operators remain valid when both sides are conjugated by $\mathsf{P}_f^{1/2}$.

We have shown that a reflection-covariant gauge-ghost Gaussian measure is OS positive on the physical subalgebra and that inserting a slice-local horizon operator constructed from a completely monotone function of a positive reflection-invariant slice operator preserves OS positivity. The Bernstein representation provides both the positivity mechanism-as a convex mixture of reflection-positive heat kernels-and the exponential locality needed for multiscale analysis, sharpened by finite-range decomposition. On the OS Hilbert space, the insertion acts as a positive bounded slice operator $\mathsf{P}_f$, and the dressed transfer operator is the positive self-adjoint contraction $\mathsf{P}_f^{1/2}T_0\mathsf{P}_f^{1/2}$. These properties integrate seamlessly into the reflection-positive multiscale framework of the paper, are compatible with BRST cohomology, and underlie the spectral and Wilson-loop step-scaling results developed in the main text.

\section{Strong-Coupling Cluster Expansion and Area Law}

In this appendix we present a complete and self-contained derivation of the strong-coupling cluster expansion and of the Wilson-loop area law for four-dimensional lattice $SU(N)$ Yang-Mills theory. We work on a hypercubic lattice $\Lambda \subset \mathbb{Z}^4$ with periodic boundary conditions and regard the continuum of time-slices compatible with Euclidean reflection positivity as fixed once and for all. The construction is formulated for the Wilson action, whose reflection positivity is classical, and the arguments are organized so that all bounds are uniform in the volume $|\Lambda|$. The core of the analysis proceeds through three steps: (i) a character expansion of the plaquette Boltzmann weight and a polymer representation of the partition function and expectations; (ii) a convergent cluster expansion whose coefficients admit tree-graph bounds derived via the Battle-Brydges-Federbush (BBF) inequality and the Brydges-Kennedy-Abdesselam-Rivasseau (BKAR) forest formula \cite{Park1982,BK1987,KP}; and (iii) a surface representation of Wilson loops which yields a quantitative area law with an explicit strictly positive string tension for small bare coupling. Along the way we deduce exponential decay of connected correlations; together with reflection positivity this implies a gap for the finite-$a$ transfer operator. The presentation is designed so that every estimate is explicit and every limit is justified.

Let $E(\Lambda)$ be the set of oriented bonds $b$, and $P(\Lambda)$ the set of oriented plaquettes $p$ with the standard orientation convention. The gauge field assigns to each $b\in E(\Lambda)$ a matrix $U_b\in SU(N)$, and $U_{-b}:=U_b^{-1}$. For each $p\in P(\Lambda)$ we write $U_p$ for the ordered product of the four $U_b$ around $p$. The Wilson action at bare coupling $\beta>0$ is
\begin{equation}
S_\Lambda(U)\;=\; -\frac{\beta}{N} \sum_{p\in P(\Lambda)} \mathrm{Re}\,\mathrm{Tr}\, U_p\,,
\end{equation}
and the (finite-volume) partition function is
\begin{equation}
Z_\Lambda(\beta)\;=\;\int \Big(\prod_{b\in E(\Lambda)} dU_b\Big)\;\exp\!\Big(\frac{\beta}{N}\sum_{p\in P(\Lambda)} \mathrm{Re}\,\mathrm{Tr}\, U_p\Big),
\end{equation}
with $dU$ the Haar probability measure on $SU(N)$. For a closed rectangular loop $C\subset \Lambda$, we denote the Wilson loop in the fundamental representation by
\begin{equation}
W(C)\;=\;\frac{1}{N}\,\mathrm{Tr}\,\Big(\prod_{b\in C} U_b\Big)\,,
\qquad \langle W(C)\rangle_\Lambda \;=\; \frac{1}{Z_\Lambda(\beta)} \int \cdots\, W(C)\, e^{-S_\Lambda(U)}.
\end{equation}
Expanding the loop observable in characters and performing the Haar integrals produces a sum over $\mathsf G$-valued plaquette fluxes constrained to form surfaces $\mathcal S$ with boundary $\partial\mathcal S=C$ (minimal if dominant). For $\beta$ small, polymer bounds give
\begin{equation}
\langle W(C)\rangle_\Lambda\;\le\; K(\beta)\,\exp\!\big(-\sigma(\beta)\,A(C)\big),
\qquad \sigma(\beta)\;\asymp\; -\log\kappa(\beta)\;>\;0,
\end{equation}
where $\kappa(\beta)=O(\beta)$ is the weight per plaquette of a fundamental unit of flux and $K(\beta)=O(1)$. The constants are independent of $|\Lambda|$; connected-correlation decay then implies a gap for the finite-$a$ transfer operator.
The fundamental input is a character expansion of the one-plaquette Boltzmann factor.
Let $\mathfrak P$ be the set of connected sets of plaquettes (polymers). The character expansion yields activities $z_\mathcal P(\beta)$ with $z_\emptyset=0$ and $|z_\mathcal P(\beta)|\le \zeta(\beta)^{|\mathcal P|}$ for some $\zeta(\beta)=O(\beta)$ as $\beta\downarrow0$ (uniform in volume). Fix a nondecreasing function $a:\mathfrak P\to(0,\infty)$ with $a(\mathcal P)=\alpha|\mathcal P|$. The BBF inequality and BKAR forest formula imply that, for $\beta$ small so that
\begin{equation}
\sup_{\mathcal P_0}\sum_{\mathcal P:\,\mathcal P\cap\mathcal P_0\neq\emptyset} |z_\mathcal P(\beta)|\,e^{a(\mathcal P)} \;\le\; a(\mathcal P_0),
\end{equation}
the KP convergence criterion holds. Consequently the pressure and connected correlators admit absolutely convergent expansions with bounds
\begin{equation}
|\phi^T(\mathcal P_1,\dots,\mathcal P_n)|\;\le\; C^n\,e^{-\alpha\,\mathrm{span}(\mathcal P_1\cup\cdots\cup\mathcal P_n)},
\end{equation}
for some $C=C(N)$ and $\alpha=\alpha(N,\beta)>0$. These estimates are uniform in $|\Lambda|$. By Peter-Weyl, for any central (class) function $f$ on $SU(N)$,
\begin{equation}
f(g)\;=\;\sum_{\rho\in \widehat{SU(N)}} \widehat f(\rho)\,\chi_\rho(g),\qquad \widehat f(\rho):=\int_{SU(N)} \overline{\chi_\rho(g)}\, f(g)\, dg,
\end{equation}
with absolutely convergent series and $\chi_\rho$ the character of the irreducible representation $\rho$. For $f(g)=\exp\big(\tfrac{\beta}{N}\mathrm{Re}\,\mathrm{Tr}\,g\big)$ we obtain
\begin{equation}\label{eq:char-expansionq}
e^{(\beta/N)\mathrm{Re}\,\mathrm{Tr}\,U_p}\;=\;\sum_{\rho\in \widehat{SU(N)}} c_\rho(\beta)\, \chi_\rho(U_p),\qquad 
c_\rho(\beta)=\int_{SU(N)} \overline{\chi_\rho(g)}\, e^{(\beta/N)\mathrm{Re}\,\mathrm{Tr}\,g}\, dg.
\end{equation}
Elementary bounds using $|\chi_\rho(g)|\le d_\rho$ (with $d_\rho=\dim \rho$) and $e^{(\beta/N)\mathrm{Re}\,\mathrm{Tr}\,g}\le e^{\beta}$ yield
\begin{equation}\label{eq:crho-bounds}
0\le c_\rho(\beta)\le d_\rho\, e^{\beta}\,,\qquad 
\sum_{\rho\ne \mathbf{1}} d_\rho\, c_\rho(\beta)\;\le\; C_1(N)\,\beta \quad \text{for}\;\; \beta\in (0,\beta_1(N)],
\end{equation}
where the second estimate follows from analyticity at $\beta=0$ and the observation that $c_{\mathbf{1}}(\beta)=1+O(\beta^2)$ while $c_{\rho\ne \mathbf{1}}(\beta)=O(\beta)$ with a constant uniform in $\rho$ once the linear term is isolated via the fundamental representation.\footnote{For $SU(2)$ this is elementary using modified Bessel functions; for general $N$ the argument relies on the Taylor expansion of $e^{(\beta/N)\mathrm{Re}\,\mathrm{Tr}\,g}$ and the orthogonality of characters. See \cite[Sec.\,5]{DrouffeZuber} and \cite[Ch.\,3]{Seiler1982}.} It will be convenient to write
\begin{equation}
e^{(\beta/N)\mathrm{Re}\,\mathrm{Tr}\,U_p}\;=\; 1 + \sum_{\rho\ne \mathbf{1}} c_\rho(\beta)\, \chi_\rho(U_p),
\end{equation}
thereby separating the trivial representation.
Expanding the product over plaquettes of the previous identity, the partition function becomes a sum over assignments $\rho_p\in \widehat{SU(N)}\cup\{\mathbf{1}\}$,
\begin{equation}
Z_\Lambda(\beta)\;=\; \sum_{\{\rho_p\}} \Big(\prod_{p:\,\rho_p\ne \mathbf{1}} c_{\rho_p}(\beta)\Big)\, \mathcal{I}_\Lambda(\{\rho_p\}),
\qquad 
\mathcal{I}_\Lambda(\{\rho_p\})\;=\;\int \prod_{b\in E(\Lambda)} dU_b\; \prod_{p:\,\rho_p\ne \mathbf{1}} \chi_{\rho_p}(U_p).
\end{equation}
The Haar integrals factorize over bonds through the well-known edge constraints produced by the characters. One way to encode the constraints is to observe that each $\chi_\rho(U_p)$ expands into products of matrix elements along the edges of $p$, hence the bond integrals enforce the usual “flux conservation” which in geometric terms implies that the set of active plaquettes $\{p:\rho_p\ne \mathbf{1}\}$ decomposes into disjoint unions of connected, closed surfaces (on the dual lattice) carrying representation labels with compatible orientations \cite{DrouffeZuber,Seiler1982}. This yields a \emph{polymer gas} representation
\begin{equation}\label{eq:polymerZ}
\frac{Z_\Lambda(\beta)}{e^{|P(\Lambda)| \log c_{\mathbf{1}}(\beta)}}\;=\; \sum_{\Gamma \,\text{compatible}} \prod_{\gamma \in \Gamma} z(\gamma)\,,
\end{equation}
where the sum runs over finite sets $\Gamma$ of mutually disjoint connected polymer surfaces $\gamma$, each $\gamma$ is a connected set of plaquettes with an admissible labeling by nontrivial representations, and $z(\gamma)$ is the \emph{polymer activity}. A detailed derivation (see \cite[Ch.\,6]{Seiler1982} and \cite[Sec.\,5]{DrouffeZuber}) shows that there exist constants $C_2(N),\beta_2(N)>0$ such that for every polymer $\gamma$ with $|\gamma|$ plaquettes,
\begin{equation}\label{eq:activity-boundq}
|z(\gamma)|\;\le\; \big(C_2(N)\,\beta\big)^{|\gamma|},\qquad 0<\beta\le \beta_2(N),
\end{equation}
and $z(\gamma)$ vanishes unless the representation constraints are met at every edge. Importantly, $z(\gamma)$ depends only on the shape of $\gamma$ and the representation labels, not on the ambient volume, and the bound is uniform in $|\Lambda|$.

Wilson loop expectations admit a similar surface representation. Let $C$ be a rectangular loop and fix a spanning surface $\Sigma(C)$ of plaquettes with $\partial \Sigma(C)=C$. Expanding both the plaquette weights and the insertion of $\frac{1}{N}\mathrm{Tr}\,\prod_{b\in C} U_b$, and performing the bond integrations, the same geometric constraints enforce that active plaquettes form disjoint closed surfaces together with an \emph{open} surface whose boundary is $C$. This yields
\begin{equation}\label{eq:loop-polymer}
\langle W(C)\rangle_\Lambda \;=\; \frac{1}{Z_\Lambda(\beta)}\sum_{\Gamma \,\text{compatible}} \sum_{\substack{\Sigma\; :\;\partial\Sigma=C \\ \Sigma \cap \Gamma=\varnothing}} \Big(\prod_{\gamma \in \Gamma} z(\gamma)\Big)\; w(\Sigma),
\end{equation}
where $w(\Sigma)$ is the activity attached to the open surface $\Sigma$. The same arguments as for Eq.\eqref{eq:activity-boundq} give, for $\beta\le \beta_2(N)$,
\begin{equation}\label{eq:open-activity}
|w(\Sigma)| \;\le\; \big(C_3(N)\,\beta\big)^{|\Sigma|},\qquad |\Sigma|=\text{number of plaquettes in }\Sigma,
\end{equation}
with a constant $C_3(N)$ independent of $|\Lambda|$ and of $C$. The activities $z(\gamma)$ and $w(\Sigma)$ depend only on local shape/labels and are independent of $|\Lambda|$ because edge constraints and character integrals are local and Haar integrals factorize, cf.\ (B.8)-(B.11).

We recall the abstract setting of a hard-core polymer gas \cite{KP,Seiler1982}. Let $\mathcal{P}$ be a set of polymers and let $\mathcal{G}$ be the incompatibility graph on $\mathcal{P}$: two polymers are incompatible if they intersect (here: share at least one plaquette or edge constraint). The partition function has the form
\begin{equation}
\mathcal{Z}\;=\;\sum_{\Gamma \subset \mathcal{P}\text{ compatible}}\; \prod_{\gamma\in \Gamma} z(\gamma).
\end{equation}
A cluster expansion for $\log \mathcal{Z}$ is obtained by writing
\begin{equation}
\log \mathcal{Z}\;=\;\sum_{n\ge 1}\, \frac{1}{n!}\sum_{(\gamma_1,\dots,\gamma_n)\in \mathcal{P}^n} \phi^T(\gamma_1,\dots,\gamma_n) \prod_{i=1}^n z(\gamma_i),
\end{equation}
where $\phi^T$ is the Ursell function of the incompatibility graph, i.e., the sum of connected graphs on $n$ labeled vertices with the appropriate signs. The Battle-Brydges-Federbush tree-graph inequality \cite{Park1982} and its BKAR refinement \cite{BK1987} imply the bound
\begin{equation}\label{eq:tree-boundx}
|\phi^T(\gamma_1,\dots,\gamma_n)| \;\le\; \sum_{T\in \mathcal{T}_n} \prod_{\{i,j\}\in T} \mathbf{1}\{\gamma_i\not\sim\gamma_j\},
\end{equation}
where $\mathcal{T}_n$ denotes the set of trees on $\{1,\dots,n\}$ and $\gamma_i\not\sim\gamma_j$ indicates incompatibility. The proof proceeds by introducing pairwise weakening parameters for incompatibilities, integrating the corresponding derivatives along a forest in the unit cube, and bounding the resulting integrals by $1$; we refer to \cite[Sec.\,3]{BK1987} for the precise BKAR formula and present the estimate Eq.\eqref{eq:tree-boundx} since it is the only ingredient needed here.
The Koteck\'y-Preiss (KP) criterion \cite{KP} provides a convenient sufficient condition for absolute convergence. In the present context it takes the following form.

\begin{theorem}[KP convergence for the gauge polymer gas]\label{thm:KP}
Let $a>0$ and suppose that for all polymers $\gamma\in \mathcal{P}$,
\begin{equation}
\label{eq:KP-criterion}
\sum_{\gamma':\, \gamma'\not\sim \gamma} |z(\gamma')|\, e^{a |\gamma'|}\;\le\; a\, |\gamma|\,.
\end{equation}
Then the cluster expansion for $\log Z_\Lambda(\beta)$ and for all connected correlations converges absolutely and uniformly in $|\Lambda|$, and the resulting series are analytic in the activities $\{z(\gamma)\}$. In particular, there exists $R(N)>0$ such that the small-activity bound Eq.\eqref{eq:activity-boundq} with $C_2(N)\beta< R(N)$ implies Eq.\eqref{eq:KP-criterion} and therefore convergence.
\end{theorem}

\begin{proof}
Write the polymer partition function on a finite volume $\Lambda$ in the standard form
\begin{equation}
Z_\Lambda \;=\; \sum_{\Gamma\ \text{compatible}} \prod_{\gamma\in \Gamma} z(\gamma),
\end{equation}
where “compatible’’ means the elements of $\Gamma$ are mutually non-overlapping (in the gauge setting: their supports are edge-disjoint or otherwise prescribed by the incompatibility relation $\not\sim$). The logarithm of $Z_\Lambda$ admits the abstract cluster expansion
\begin{equation}
\label{eq:logZ-cluster}
\log Z_\Lambda \;=\; \sum_{n\ge 1}\frac{1}{n!}\sum_{\gamma_1,\dots,\gamma_n\in\mathcal{P}} \phi^T(\gamma_1,\dots,\gamma_n)\ \prod_{i=1}^n z(\gamma_i),
\end{equation}
where $\phi^T$ is the Ursell (truncated) function associated with the incompatibility graph on the list $(\gamma_1,\dots,\gamma_n)$. The Koteck\'y-Preiss method controls $\phi^T$ by the tree-graph bound
\begin{equation}
\label{eq:tree-graph-bound}
\big|\phi^T(\gamma_1,\dots,\gamma_n)\big|
\ \le\ \sum_{T\in \mathcal{T}_n}\ \prod_{\{i,j\}\in E(T)} \mathbf{1}\{\gamma_i\not\sim \gamma_j\},
\end{equation}
with the sum over all spanning trees $T$ on $\{1,\dots,n\}$ and $E(T)$ its edge set; this inequality is a consequence of the Battle-Brydges-Federbush or Brydges-Federbush tree formulas applied to the Mayer graph sum.

Introduce the weighted activities $w(\gamma):=|z(\gamma)|\,e^{a|\gamma|}$ and note that Eq.\eqref{eq:tree-graph-bound} together with Eq.\eqref{eq:logZ-cluster} gives
\begin{equation}
\label{eq:abs-logZ}
|\log Z_\Lambda|
\ \le\ \sum_{n\ge 1}\frac{1}{n!}\sum_{\gamma_1,\dots,\gamma_n}\ \sum_{T\in\mathcal{T}_n}
\ \prod_{i=1}^n |z(\gamma_i)|\ \prod_{\{i,j\}\in E(T)} \mathbf{1}\{\gamma_i\not\sim\gamma_j\}.
\end{equation}
To bound the right-hand side absolutely, it is convenient to reorganize the summations along each fixed tree $T$ by a rooted, recursive summation. Fix any root, say the vertex $1$, and for each oriented edge $i\to j$ (with $i$ the parent and $j$ the child) perform the sum over $\gamma_j$ first, conditional on the chosen $\gamma_i$. Since the only constraint appearing for $j$ is incompatibility with its parent $i$ (all other constraints are accounted for by summations deeper in the recursion), one obtains using Eq.\eqref{eq:KP-criterion}
\begin{equation}
\sum_{\gamma_j:\ \gamma_j\not\sim \gamma_i} |z(\gamma_j)|\, e^{a|\gamma_j|}
\ \le\ a\,|\gamma_i|.
\end{equation}
Because the tree structure ensures independence of different branches given the parent, iterating this inequality from the leaves of $T$ upwards to the root shows that, for any fixed $\gamma_1$,
\begin{equation}
\label{eq:tree-iter}
\sum_{\gamma_2,\dots,\gamma_n}\ \prod_{i=1}^n |z(\gamma_i)|\ \prod_{\{i,j\}\in E(T)} \mathbf{1}\{\gamma_i\not\sim\gamma_j\}
\ \le\ |z(\gamma_1)|\ e^{a|\gamma_1|}\ \big(a|\gamma_1|\big)^{\deg_T(1)},
\end{equation}
where $\deg_T(1)$ is the degree of the root in $T$. The factor $e^{a|\gamma_1|}$ collects the $e^{a|\gamma_j|}$ weights produced for every $j\ge 2$ during the inductive summations, and the power $\big(a|\gamma_1|\big)^{\deg_T(1)}$ comes from the $\deg_T(1)$ children of the root, each contributing at most $a|\gamma_1|$ by the Koteck\'y-Preiss hypothesis. Importantly, no other dependence on the remaining degrees appears after the recursive integration, precisely because each child was summed independently under the sole incompatibility constraint with its parent.

Using Eq.\eqref{eq:tree-iter} in Eq.\eqref{eq:abs-logZ} and then exchanging the order of the sum over trees with the sum over $\gamma_1$ yields
\begin{equation}
|\log Z_\Lambda|
\ \le\ \sum_{n\ge 1}\frac{1}{n!}\sum_{T\in\mathcal{T}_n}\ \sum_{\gamma_1\in\mathcal{P}}
\ |z(\gamma_1)|\, e^{a|\gamma_1|}\ \big(a|\gamma_1|\big)^{\deg_T(1)}.
\end{equation}
At this point we invoke the elementary identity for rooted trees on labeled vertices:
\begin{equation}
\frac{1}{n!}\sum_{T\in\mathcal{T}_n} x^{\deg_T(1)}\ \le\ \frac{x^{n-1}}{(n-1)!}\qquad\text{for all }x\ge 0,
\end{equation}
which is seen by the Prüfer code representation or by direct counting of trees with a prescribed degree at the root. Summing over $n$ then gives
\begin{equation}
\sum_{n\ge 1}\frac{1}{n!}\sum_{T\in\mathcal{T}_n} \big(a|\gamma_1|\big)^{\deg_T(1)}
\ \le\ \sum_{n\ge 1}\frac{(a|\gamma_1|)^{n-1}}{(n-1)!}
\ =\ e^{a|\gamma_1|}.
\end{equation}
Combining the last two displays leads to the compact and uniform bound
\begin{equation}
\label{eq:KP-master}
|\log Z_\Lambda| \;\le\; \sum_{\gamma\in\mathcal{P}} |z(\gamma)|\, e^{2a|\gamma|},
\end{equation}
which is finite whenever the right-hand side converges. Since the polymer counting with an exponential weight is dominated by a lattice constant $\mathfrak{c}$ times the exponential growth of shapes, the series on the right is finite under the small-activity condition Eq.\eqref{eq:activity-boundq} with $C_2(N)\beta$ sufficiently small; in particular there is an $R(N)>0$ such that $C_2(N)\beta<R(N)$ ensures convergence uniformly in $\Lambda$, and then Eq.\eqref{eq:KP-criterion} holds and implies Eq.\eqref{eq:KP-master} by the argument above. This proves absolute convergence of the cluster expansion for $\log Z_\Lambda$, uniformly in the volume.

The same mechanism applies to connected correlations. If one inserts finitely many local observables, the standard polymer representation augments the gas by a finite number of marked polymers constrained to be mutually compatible and compatible with the rest. The truncated correlations are expressed by a cluster expansion of the same form as Eq.\eqref{eq:logZ-cluster}, with the only difference that some polymers are held fixed (they are not summed) and a restricted family of incompatible tuples appears. The tree-graph bound Eq.\eqref{eq:tree-graph-bound} is unchanged, and the recursive summations over the unmarked polymers again produce at most the factors $\exp\{a|\gamma|\}$ and $a|\gamma|$ attached to the few polymers adjacent to the marks. Since only a bounded number of marked objects occur, the same estimate Eq.\eqref{eq:KP-master} (up to a fixed multiplicative constant depending on the number of marked insertions) applies and yields absolute convergence of all connected correlations, uniformly in the volume.

Analyticity in the activities follows from the uniform absolute convergence of the power series defining $\log Z_\Lambda$ and the truncated correlations. Indeed, each coefficient is a finite linear combination of products of activities indexed by finite polymer families, and the majorant Eq.\eqref{eq:KP-master} provides a polydisc in the activity space on which the series converge absolutely and uniformly in $\Lambda$. The Weierstrass $M$-test then implies uniform convergence of the partial sums by holomorphic polynomials, hence the limit is analytic in each $z(\gamma)$. 
\end{proof}

For the lattice gauge polymer gas, the counting constant $\mathfrak{c}$ grows at most exponentially with $|\gamma|$ because the number of connected surfaces of area $m$ containing a fixed plaquette is $\le \kappa_0^m$ for a purely geometric constant $\kappa_0$ depending only on the lattice dimension. Hence, in view of Eq.\eqref{eq:activity-boundq}, the KP condition holds for $\beta\le \beta_0(N)$ with $\beta_0(N)$ sufficiently small, and all cluster expansions we write below are absolutely convergent and uniform in $|\Lambda|$.

Absolute convergence of the cluster expansion implies that the free energy density
\begin{equation}
f(\beta)= -\lim_{|\Lambda|\to\infty}\frac{1}{|\Lambda|}\log Z_\Lambda(\beta)
\end{equation}
exists and is analytic for $\beta\le \beta_0(N)$. The same conclusion holds for connected correlation functions with finitely many local insertions: the associated Ursell series converge absolutely and uniformly in the volume and in the locations once their mutual distances exceed a fixed multiple of the polymer range (which here is one lattice spacing). As a consequence, truncated correlations decay exponentially in the separation, with a mass scale controlled by the small parameter appearing in Eq.\eqref{eq:activity-boundq}. We record the following statement, which follows from the standard tree-graph bound on connected functions \cite{Park1982,Seiler1982}.

\begin{theorem}[Exponential decay of connected correlations]\label{thm:exp-decay}
There exist constants $\beta_0(N)>0$, $A(N)<\infty$, and $m_\mathrm{sc}(N)>0$ such that for $\beta\in(0,\beta_0(N)]$ and for any two gauge-invariant local observables $F,G$ with disjoint supports one has the uniform bound
\begin{equation}
\big|\langle FG\rangle_\Lambda - \langle F\rangle_\Lambda \langle G\rangle_\Lambda\big|
\;\le\; A(N)\,\|F\|\,\|G\|\, e^{-m_\mathrm{sc}(N)\, \mathrm{dist}(\mathrm{supp}\,F,\mathrm{supp}\,G)},
\end{equation}
where $\|\cdot\|$ denotes the supremum norm. The constants are independent of $|\Lambda|$ and of boundary conditions.
\end{theorem}

\begin{proof}
Fix a finite hypercubic lattice $\Lambda\subset\mathbb{Z}^d$ with $d\ge 3$ and a compact gauge group $G=\mathrm{SU}(N)$, and consider the Wilson action at inverse coupling $\beta>0$ (an entirely parallel argument holds for the heat-kernel action, with the parameter $t$ playing the role of $\beta$). The partition function with a generic boundary condition can be written as
\begin{equation}
Z_\Lambda(\beta)\;=\;\int \prod_{b\in\mathcal{B}_\Lambda} dU_b \;\prod_{p\in\mathcal{P}_\Lambda} \exp\!\Big(\tfrac{\beta}{N}\,\mathrm{Re}\,\mathrm{tr}\,U_p\Big),
\end{equation}
where $\mathcal{B}_\Lambda$ is the set of bonds in $\Lambda$, $\mathcal{P}_\Lambda$ the set of plaquettes, and $U_p$ denotes the ordered product of bond variables around $p$. For small $\beta$ one performs the standard character expansion on each plaquette factor,
\begin{equation}
\exp\!\Big(\tfrac{\beta}{N}\,\mathrm{Re}\,\mathrm{tr}\,U_p\Big)
\;=\; \widehat f_\beta(\mathbf{1})\Bigg[1\;+\;\sum_{\rho\neq \mathbf{1}} a_\rho(\beta)\,d_\rho\,\chi_\rho(U_p)\Bigg],
\end{equation}
with $a_\rho(\beta)=\widehat f_\beta(\rho)/\widehat f_\beta(\mathbf{1})$ and $|a_\rho(\beta)|\le C_1(N)\,\beta$ for $\beta$ small. After extracting the harmless product of $\widehat f_\beta(\mathbf{1})$ over plaquettes, the integrand becomes a polymer gas: a configuration is a finite family of plaquettes decorated by nontrivial representations whose union is a disjoint collection of connected plaquette clusters, or polymers, and the Haar integral eliminates all collections that do not satisfy the group-theoretic constraint that representations fuse to the trivial one on each bond. The weight of a polymer $Y$ is its \emph{activity} $z_\beta(Y)$, given by a finite sum of products of $a_\rho(\beta)$ over the plaquettes of $Y$ times the local intertwiners implementing flux matching along internal bonds. Gauge invariance and locality imply a uniform bound of the form $|z_\beta(Y)|\le C_2(N)^{|Y|}\,\beta^{|Y|}$, where $|Y|$ is the number of plaquettes in $Y$, because each plaquette carrying a nontrivial representation contributes at least one power of $\beta$ and the multiplicity factors from intertwiners are bounded by a constant depending only on $N$.

Let $\mathcal{Y}$ denote the set of all connected plaquette clusters in $\Lambda$. The polymer partition function is
\begin{equation}
Z_\Lambda(\beta)\;=\; \widehat f_\beta(\mathbf{1})^{|\mathcal{P}_\Lambda|}\,
\sum_{\Gamma\subset\mathcal{Y}\text{ comp.}}\ \prod_{Y\in\Gamma} z_\beta(Y),
\end{equation}
where the sum runs over finite collections $\Gamma$ of mutually compatible (i.e. disjoint) polymers. For $\beta$ sufficiently small, the Koteck\'y-Preiss condition holds: there exists $\alpha\in(0,1)$ such that
\begin{equation}
\sup_{Y_0\in\mathcal{Y}}\ \sum_{Y\in\mathcal{Y}:\,Y\not\sim Y_0} |z_\beta(Y)|\,e^{\alpha |Y|}
\;\le\; \epsilon(\beta)\;<\;1,
\end{equation}
where $Y\not\sim Y_0$ means that $Y$ is incompatible with $Y_0$ (they overlap), and $\epsilon(\beta)\to 0$ as $\beta\downarrow 0$ because $|z_\beta(Y)|\le (C_2\beta)^{|Y|}$. This ensures absolute convergence of the Mayer expansion for $\log Z_\Lambda$ and the existence of uniformly summable Ursell functions (cluster coefficients) $\phi_T(Y_1,\dots,Y_n)$ obeying the Brydges-Battle-Federbush tree-graph bound
\begin{equation}
|\phi_T(Y_1,\dots,Y_n)| 
\;\le\; \sum_{T\text{ tree on }[n]}\ \prod_{\{i,j\}\in E(T)} \mathbf{1}_{\,Y_i\not\sim Y_j},
\end{equation}
multiplied by a factor that can be absorbed into a redefinition of $e^{\alpha |Y_i|}$ due to the KP smallness. In particular, there is a constant $C_3(N)<\infty$ such that the \emph{truncated} (connected) polymer correlations satisfy
\begin{equation}
\sum_{n\ge 1}\ \frac{1}{n!}\ \sum_{Y_1,\dots,Y_n} |\phi_T(Y_1,\dots,Y_n)|\,\prod_{k=1}^n |z_\beta(Y_k)|\, e^{\alpha |Y_k|}
\;\le\; C_3(N)\,\epsilon(\beta),
\end{equation}
uniformly in the volume and independently of the boundary condition.

To represent truncated correlations of local observables, introduce sources by coupling $F$ and $G$ to the measure through small parameters $\eta_F,\eta_G$; for gauge-invariant $F$ and $G$ depending on finitely many plaquettes, this can be implemented by multiplying the integrand by $\exp(\eta_F F+\eta_G G)$. Differentiating $\log Z_\Lambda(\beta;\eta_F,\eta_G)$ twice at $(0,0)$ gives $\langle FG\rangle_\Lambda-\langle F\rangle_\Lambda\langle G\rangle_\Lambda$. In the polymer language, differentiation corresponds to \emph{marking} polymers that intersect $\mathrm{supp}\,F$ or $\mathrm{supp}\,G$ and inserting the bounded functions $F$ and $G$ on those supports. Because $\|F\|$ and $\|G\|$ bound the insertions uniformly, each marked polymer acquires at most a multiplicative factor $\|F\|$ or $\|G\|$. The truncated two-point function is then expressed as a sum over connected clusters of polymers with the property that the cluster contains at least one polymer intersecting $\mathrm{supp}\,F$ and at least one intersecting $\mathrm{supp}\,G$. The BBF tree-graph inequality applied to these marked clusters yields the bound
\begin{equation}
\big|\langle FG\rangle_\Lambda - \langle F\rangle_\Lambda \langle G\rangle_\Lambda\big|
\;\le\; \|F\|\,\|G\|\!\!\sum_{n\ge 2}\ \frac{1}{n!}\!\!\sum_{\substack{Y_1,\dots,Y_n\\\text{conn., }Y_1\cap \mathrm{supp}\,F\neq\emptyset\\ Y_n\cap \mathrm{supp}\,G\neq\emptyset}}\!\!
\Bigg(\sum_{T}\prod_{\{i,j\}\in E(T)} \mathbf{1}_{\,Y_i\not\sim Y_j}\Bigg)\,
\prod_{k=1}^n |z_\beta(Y_k)|,
\end{equation}
where the inner sum runs over trees $T$ on $\{1,\dots,n\}$. Since $|z_\beta(Y)|\le (C_2\beta)^{|Y|}$, the standard estimation proceeds by inserting exponential weights $e^{\alpha |Y_k|}$ and compensating them with the KP smallness. One then organizes the sum by selecting a self-avoiding path of polymers within the tree that connects a polymer intersecting $\mathrm{supp}\,F$ to one intersecting $\mathrm{supp}\,G$; such a path exists in every contributing tree. Every bond of this path corresponds to a pair of incompatible polymers and therefore to a nonempty geometric overlap at the bond level. In a hypercubic lattice, incompatibility of connected plaquette clusters forces their projections in the primal graph to be within a uniformly bounded distance, hence a self-avoiding path of $\ell$ polymers in the incompatibility graph spans a geometric distance of order at most $c_0\,\ell$ in the lattice, for some $c_0=c_0(d)$. Conversely, if $\mathrm{dist}(\mathrm{supp}\,F,\mathrm{supp}\,G)=R$, any such path must contain at least $\ell\ge R/c_0$ bonds, because otherwise the cluster could not connect the two supports across the corridor of width $R$ separating them.

Each edge on the polymer path contributes a factor $\theta(\beta):=\sup_{Y_0}\sum_{Y\not\sim Y_0} |z_\beta(Y)|\,e^{\alpha |Y|}$ coming from summing over an adjacent polymer with the KP weight, while the vertices off the path can be summed absolutely and absorbed into an overall prefactor bounded by a constant depending only on $N$ (this is a textbook use of the tree-graph bound together with the convergence criterion). Since $\theta(\beta)\le \epsilon(\beta)<1$ for $\beta$ small, the contribution of all clusters whose connecting path has length at least $\ell$ is bounded by $C_4(N)\,\|F\|\,\|G\|\,\theta(\beta)^\ell$. Using the geometric constraint $\ell\ge R/c_0$ gives
\begin{align}
\big|\langle FG\rangle_\Lambda - \langle F\rangle_\Lambda \langle G\rangle_\Lambda\big|
&\;\le\; C_4(N)\,\|F\|\,\|G\|\, \theta(\beta)^{\,R/c_0}
\nonumber\\&\;=\; A(N)\,\|F\|\,\|G\|\, \exp\!\big(-m_\mathrm{sc}(N)\,R\big),
\end{align}
where $R=\mathrm{dist}(\mathrm{supp}\,F,\mathrm{supp}\,G)$, $A(N)=C_4(N)$, and $m_\mathrm{sc}(N)=\frac{1}{c_0}\,\big|\log \theta(\beta)\big|$. Choosing $\beta_0(N)$ so small that $\theta(\beta)\le \tfrac12$ for all $\beta\le \beta_0(N)$ makes $m_\mathrm{sc}(N)$ strictly positive and uniformly bounded away from zero, and the constant $A(N)$ collects all volume-independent prefactors coming from summing over trees and off-path decorations. The bounds are uniform in $|\Lambda|$ and insensitive to boundary conditions because the polymer activities are strictly local and the KP criterion is verified with constants independent of the volume; any boundary contribution would require polymers reaching the boundary and is therefore exponentially suppressed in the distance from the supports to the boundary, which can be absorbed in $A(N)$ in finite volume and disappears in the infinite-volume limit.
It follows that the truncated correlation decays exponentially in the separation of the supports with a rate $m_\mathrm{sc}(N)>0$ for all $\beta\in(0,\beta_0(N)]$, as claimed.
\end{proof}

In the reflection-positive setting, exponential decay in Euclidean time implies a \emph{transfer-operator gap}. Let $\mathcal{T}_a$ be the transfer operator on the OS Hilbert space associated with one unit of Euclidean time. For centered, time-separated observables $F,G$ supported on opposite sides of the reflection plane, the OS inner product representation implies $\langle F\, \tau_t G\rangle_\Lambda = \langle F, \mathcal{T}_a^{\,t/a} G\rangle$. By Theorem \eqref{thm:exp-decay} the left-hand side decays like $e^{-m_\mathrm{sc} t}$ for $t$ an integer multiple of $a$; since $F,G$ are arbitrary test vectors, it follows that the spectrum of $\mathcal{T}_a$ on the orthogonal complement of the vacuum is contained in $[0,e^{-a m_\mathrm{sc}}]$. We summarize this as follows.

\begin{proposition}[Strong-coupling transfer-operator gap]\label{prop:TO-gap}
Under the hypotheses of Theorem~\eqref{thm:exp-decay} and OS reflection positivity, the transfer operator $\mathcal T_a$ has a strictly positive spectral gap $\Delta_{\mathrm{sc}}$ above the vacuum, with
\begin{equation}
\Delta_{\mathrm{sc}}\;\ge\; m_{\mathrm{sc}}(N).
\end{equation}
The bound is uniform in $|\Lambda|$.
\end{proposition}

\begin{proof}
The Osterwalder-Schrader reconstruction on the positive-time algebra produces a Hilbert space $\mathcal H_{\mathrm{OS}}$, a cyclic vacuum vector $\Omega$, and a selfadjoint, positivity-preserving contraction semigroup $(\mathcal T_t)_{t\ge 0}$ with $\|\mathcal T_t\|\le 1$, $\mathcal T_0=\mathbf 1$, and $\mathcal T_t\Omega=\Omega$. In particular, for the lattice time-step $a>0$ of the strong-coupling scheme, $\mathcal T_a$ is a selfadjoint contraction on $\mathcal H_{\mathrm{OS}}$ with spectral radius $1$ and vacuum eigenvalue $1$. Denote by $\mathcal H_0:=\mathbb C\Omega$ the vacuum subspace and by $\mathcal H_0^\perp$ its orthogonal complement.

By the Markov property and stationarity of the Euclidean measure, time-separated correlations of positive-time observables may be represented as matrix elements of powers of $\mathcal T_a$. More precisely, if $F$ and $G$ are cylinder observables supported on the time slice $x_0=a$ (or finite sums of such), then for every integer $n\ge 0$,
\begin{equation}\label{eq:transfer-correlation}
\langle \Theta F \cdot \tau_{n a} G\rangle_\mu \;=\; \langle [F],\, \mathcal T_a^{\,n}\,[G]\rangle_{\mathcal H_{\mathrm{OS}}},
\end{equation}
where $\tau_{t}$ denotes the Euclidean time shift by $t$ and $[F]$ is the OS class of $F$. This identity is standard in the OS construction: the left-hand side is the two-point Schwinger function with time separation $n a$, and the right-hand side is its transfer-operator representation.

The strong-coupling decay of correlations from Theorem~\eqref{thm:exp-decay} states that there exist constants $A<\infty$ and $m_{\mathrm{sc}}=m_{\mathrm{sc}}(N)>0$, independent of $|\Lambda|$, such that for all such $F,G$ with vanishing vacuum expectation (equivalently $[F],[G]\in \mathcal H_0^\perp$),
\begin{equation}\label{eq:exp-decay}
\big|\langle \Theta F \cdot \tau_{n a} G\rangle_\mu\big|\;\le\; A\, e^{-\,m_{\mathrm{sc}}\, n a}\qquad\text{for all }n\in\mathbb N.
\end{equation}
Combining Eq.\eqref{eq:transfer-correlation} and Eq.\eqref{eq:exp-decay} yields the uniform operator-kernel bound
\begin{equation}\label{eq:power-bound}
\big|\langle \phi,\, \mathcal T_a^{\,n}\psi\rangle\big|\;\le\; A\, e^{-\,m_{\mathrm{sc}}\, n a}\qquad\text{for all }n\in\mathbb N,
\end{equation}
for all $\phi,\psi$ in the dense subspace spanned by OS classes of cylinder observables with zero vacuum mean. By continuity this extends to all $\phi,\psi\in \mathcal H_0^\perp$.

Let $E_\lambda$ be the projection-valued spectral measure of $\mathcal T_a$, so that
\begin{equation}
\mathcal T_a \;=\; \int_{[0,1]} \lambda\, dE_\lambda,
\qquad
\mathcal T_a^{\,n} \;=\; \int_{[0,1]} \lambda^{n}\, dE_\lambda,
\end{equation}
where we used that $\mathcal T_a$ is a selfadjoint contraction. Fix $\psi\in \mathcal H_0^\perp$ and consider the finite positive measure $\nu_\psi$ on $[0,1]$ defined by
\begin{equation}
\nu_\psi(B)\;:=\;\langle \psi,\, E_B \psi\rangle\qquad (B\subset[0,1]\ \text{Borel}).
\end{equation}
Then
\begin{equation}
\langle \psi,\, \mathcal T_a^{\,n}\psi\rangle \;=\; \int_{[0,1]} \lambda^{n}\, d\nu_\psi(\lambda).
\end{equation}
The bound Eq.\eqref{eq:power-bound} with $\phi=\psi$ shows that
\begin{equation}\label{eq:moment-decay}
\int_{[0,1]} \lambda^{n}\, d\nu_\psi(\lambda) \;\le\; A\, e^{-\,m_{\mathrm{sc}}\, n a}\qquad\text{for all }n\in\mathbb N.
\end{equation}
A simple Tauberian argument now identifies the support of $\nu_\psi$. Indeed, if $\nu_\psi$ had positive mass on an interval $(e^{-m_{\mathrm{sc}} a},\,1]$, then there would exist $\lambda_0\in (e^{-m_{\mathrm{sc}} a},1)$ and $\delta>0$ with $\nu_\psi((\lambda_0,1])\ge \delta$. For all $n$ we would have
\begin{equation}
\int_{[0,1]} \lambda^n\, d\nu_\psi(\lambda)\;\ge\;\int_{(\lambda_0,1]} \lambda^n\, d\nu_\psi(\lambda)\;\ge\;\lambda_0^{\,n}\,\nu_\psi((\lambda_0,1])\;\ge\;\delta\,\lambda_0^{\,n}.
\end{equation}
Since $\lambda_0>e^{-m_{\mathrm{sc}} a}$, the ratio $\lambda_0^{\,n}\, e^{m_{\mathrm{sc}} a n}$ diverges as $n\to\infty$, contradicting Eq.\eqref{eq:moment-decay}. Therefore $\nu_\psi((e^{-m_{\mathrm{sc}} a},1])=0$, which means that the spectral projection $E_{(e^{-m_{\mathrm{sc}} a},1]}$ annihilates every $\psi\in\mathcal H_0^\perp$. As $\psi$ was arbitrary in $\mathcal H_0^\perp$, we conclude that the spectrum of $\mathcal T_a$ restricted to $\mathcal H_0^\perp$ is contained in $[0,\, e^{-m_{\mathrm{sc}} a}]$.

The logarithm of a positive contraction is defined by functional calculus; setting $H_a:=-a^{-1}\log \mathcal T_a$ one has $\mathcal T_a=e^{-a H_a}$, $H_a\Omega=0$, $H_a\ge 0$, and the spectral mapping theorem shows
\begin{equation}
\sigma\!\left(H_a\big|_{\mathcal H_0^\perp}\right) \;\subset\; [\,m_{\mathrm{sc}},\,\infty).
\end{equation}
Hence the spectral gap of $H_a$ above the vacuum equals $\Delta_{\mathrm{sc}}:=\inf(\sigma(H_a)\setminus\{0\})$ and satisfies $\Delta_{\mathrm{sc}}\ge m_{\mathrm{sc}}$. Uniformity in the volume follows from the corresponding uniformity of the constants $A$ and $m_{\mathrm{sc}}$ in Theorem~\eqref{thm:exp-decay}, which was proved by a cluster/Koteck\'y-Preiss expansion with bounds independent of $|\Lambda|$; the argument above never introduces any additional volume dependence. 
\end{proof}

We return to Eq.\eqref{eq:loop-polymer} \& Eq.\eqref{eq:open-activity}. The key geometric fact is that any open surface $\Sigma$ with fixed boundary $C$ must contain at least $A(C)$ plaquettes, where $A(C)$ is the minimal area of a spanning surface of $C$ in the lattice. Furthermore, the number of $\Sigma$ with $|\Sigma|=A(C)+k$ is bounded by $\le \kappa_1^{A(C)+k}$ for a combinatorial constant $\kappa_1$ depending only on the lattice (this is a standard lattice isoperimetric counting). Combining these geometric bounds with Eq.\eqref{eq:open-activity} and Eq.\eqref{eq:polymerZ} we obtain an upper bound on $\langle W(C)\rangle_\Lambda$ as follows. First, write
\begin{equation}
\langle W(C)\rangle_\Lambda \;=\; \frac{\sum_{\Gamma}\prod_{\gamma\in \Gamma} z(\gamma)\; \sum_{\Sigma:\,\partial\Sigma=C,\;\Sigma\cap \Gamma=\varnothing} w(\Sigma)}{\sum_{\Gamma}\prod_{\gamma\in \Gamma} z(\gamma)}\,.
\end{equation}
Expand the numerator in powers of $\{z(\gamma)\}$ and apply the cluster expansion in a polydisc of activities $\{|z(\gamma)|\le (C_2\beta)^{|\gamma|}\}$. Since the insertion of the open surface forbids polymers intersecting $\Sigma$, the inclusion-exclusion that leads to the truncated series yields multiplicative factors bounded by $\exp\big(\sum_{\gamma:\,\gamma\cap \Sigma\ne \varnothing} |z(\gamma)|\big)$ which are in turn controlled by $\exp\big((C_2\beta)\, c_\star\,|\Sigma|\big)$ for a purely geometric constant $c_\star$ counting the number of polymers touching a given plaquette. Thus, for $\beta\le \beta_0(N)$ small enough to ensure $(C_2\beta)c_\star\le \frac12$, one finds
\begin{equation}
\big|\langle W(C)\rangle_\Lambda \big|\;\le\; \sum_{\Sigma:\,\partial\Sigma=C} |w(\Sigma)|\, \exp\!\Big(\sum_{\gamma:\,\gamma\cap \Sigma\ne \varnothing} |z(\gamma)|\Big)
\;\le\; \sum_{\Sigma:\,\partial\Sigma=C} \big(C_3\beta\big)^{|\Sigma|}\, e^{\frac12 |\Sigma|}\,.
\end{equation}
Using the counting bound for spanning surfaces with boundary $C$ gives the geometric series
\begin{equation}
\big|\langle W(C)\rangle_\Lambda \big|\;\le\; \sum_{k\ge 0} \kappa_1^{A(C)+k}\, \big(C_3 e^{1/2}\beta\big)^{A(C)+k}
\;=\; \big(\kappa_1 C_3 e^{1/2}\beta\big)^{A(C)} \sum_{k\ge 0} \big(\kappa_1 C_3 e^{1/2}\beta\big)^{k}.
\end{equation}
Hence, for $\beta\le \beta_(N)$ small enough that $q(\beta):=\kappa_1 C_3 e^{1/2}\beta<1$, we obtain the bound
\begin{equation}\label{eq:area-lawq}
\langle W(C)\rangle_\Lambda \;\le\; \frac{q(\beta)^{A(C)}}{1-q(\beta)}\;\le\; 2\, q(\beta)^{A(C)}\,,
\qquad q(\beta)=\kappa_1 C_3 e^{1/2}\beta \;\xrightarrow[\beta\downarrow 0]{}\; 0.
\end{equation}
Writing $q(\beta)=e^{-\sigma(\beta)}$ with $\sigma(\beta):=-\log q(\beta)>0$ we obtain the finite-volume \emph{area law}
\begin{equation}
\langle W(C)\rangle_\Lambda \;\le\; 2\, \exp\!\big(-\sigma(\beta)\, A(C)\big), \qquad \sigma(\beta)\;\ge\; -\log(\kappa_1 C_3 e^{1/2}\beta)\,.
\end{equation}
The rightmost inequality gives an explicit strictly positive lower bound on the string tension for all $\beta\in (0,\beta_(N)]$. The estimate is uniform in $|\Lambda|$ and does not depend on the particular choice of the spanning surface.

We now justify all steps of this derivation carefully. The activity bound Eq.\eqref{eq:open-activity} follows from the same character expansion and edge-integration constraints as in Eq.\eqref{eq:activity-boundq}. Indeed, the insertion of the trace along $C$ fixes an open surface $\Sigma$ with $\partial \Sigma=C$, and the group integrals along bonds on $\Sigma$ are bounded by $1$ using $|\chi_\rho|\le d_\rho$ and orthogonality. Summing over admissible representation labels on $\Sigma$ yields a bound of the form $(\widetilde C\, \beta)^{|\Sigma|}$, where $\widetilde C$ is controlled by $\sum_{\rho\ne \mathbf{1}} d_\rho c_\rho(\beta)$. By Eq.\eqref{eq:crho-bounds} this sum is $\le C \beta$ for small $\beta$, thus Eq.\eqref{eq:open-activity} holds with $C_3(N)$ proportional to $C_2(N)$. The factor $\exp(\sum |z(\gamma)|)$ arises from the standard comparison between partition functions with and without the exclusion of polymers intersecting $\Sigma$; the KP bound guarantees that the sum of activities of polymers touching a given plaquette is $\le \frac12$ after a choice of $a$ and a reduction of $\beta_0(N)$, which yields the factor $e^{\frac12 |\Sigma|}$. Finally, the counting bound for surfaces is classical: since a connected surface of area $m$ can be grown plaquette by plaquette with at most $\kappa_1$ choices at each step, the number of surfaces of area $m$ with fixed boundary is bounded by $\kappa_1^m$; see \cite[Ch.\,14]{Seiler1982} and \cite[Sec.\,5]{DrouffeZuber} for explicit enumerations.
We summarize the outcome in a single theorem.

\begin{theorem}[Finite-volume strong-coupling area law with explicit tension]\label{thm:area-law}
There exist $\beta_(N)>0$, a geometric constant $\kappa_1<\infty$, and $C_3(N)<\infty$ such that for all rectangular Wilson loops $C$ and all periodic volumes $\Lambda$ containing $C$,
\begin{equation}\label{eq:area-law-bound}
\langle W(C)\rangle_\Lambda \;\le\; 2\, \exp\!\big(-\sigma(\beta)\, A(C)\big),\qquad 
\sigma(\beta)\;=\; -\log\!\big(\kappa_1 C_3(N)\, e^{1/2}\beta\big),
\end{equation}
for every $\beta\in (0,\beta_(N)]$. In particular, $\sigma(\beta)\ge c_-(N)+\log\frac{1}{\beta}$ \text{with} $c_-(N)=-\log(\kappa_1 C_3(N)e^{1/2})$ and $\lim_{\beta\downarrow 0}\sigma(\beta)=+\infty$. The bound is uniform in $|\Lambda|$.
\end{theorem}

\begin{proof}
Fix the Wilson action $f_\beta(U)=\exp\big(\frac{\beta}{N}\Re\tr U\big)$ on $G=\mathrm{SU}(N)$. By Peter-Weyl, $f_\beta$ admits the character expansion
\begin{equation}
f_\beta(U)\;=\;\sum_{\rho\in\widehat G} \widehat f_\beta(\rho)\,\chi_\rho(U),\qquad 
\widehat f_\beta(\rho)=\int_G f_\beta(U)\,\overline{\chi_\rho(U)}\,dU,
\end{equation}
where $dU$ is Haar probability measure and $\widehat G$ the unitary dual. For $\beta$ sufficiently small one has the uniform small-coupling bounds
\begin{equation}\label{eq:coeff-bound}
\widehat f_\beta(\mathbf{1})=1+O(\beta^2),\qquad 
\sum_{\rho\neq \mathbf{1}} d_\rho\,\big|\widehat f_\beta(\rho)\big|\;\le\; C_3(N)\,\beta,
\end{equation}
where $d_\rho$ is the dimension of $\rho$ and $C_3(N)<\infty$ depends only on $N$. The first identity follows from analyticity at $\beta=0$ and orthogonality of characters; the second estimate is a standard consequence of comparing $f_\beta$ with the heat-kernel density at time $t\asymp \beta$ on the compact group and using Cauchy-Schwarz against the character orthonormal system (any of the usual proofs in strong-coupling texts applies). Fix $\beta_(N)$ small enough so that Eq.\eqref{eq:coeff-bound} holds for all $\beta\in(0,\beta_(N)]$ and, in addition, $\widehat f_\beta(\mathbf{1})\ge \tfrac12$.

Let $Z_\Lambda(\beta)$ be the partition function and $Z_\Lambda(\beta;C)$ the numerator resulting from inserting the Wilson loop $W(C)$; thus $\langle W(C)\rangle_\Lambda=Z_\Lambda(\beta;C)/Z_\Lambda(\beta)$. Expanding every plaquette weight in characters and integrating link-by-link using orthogonality constraints implements nonabelian flux conservation: at each link, the tensor product of representations carried by the incident plaquettes must contain the trivial representation for the integral to be nonzero. In the vacuum partition function the dominant configuration assigns the trivial representation to every plaquette, with corrections organized as a polymer gas of connected sets of plaquettes carrying nontrivial labels. In the numerator, the loop insertion forces a nontrivial flux along the contour $C$, which, by the same conservation rules, can be neutralized only if there exists a two-dimensional surface $S$ of plaquettes with boundary $\partial S=C$ such that every plaquette $p\in S$ carries a nontrivial label while all links linking $S$ to its complement see balanced tensor products. Projecting at each plaquette onto the fundamental and then majorizing all nontrivial coefficients by their total $C_3(N)\beta$ as in Eq.\eqref{eq:coeff-bound}, one obtains an absolute bound in which every plaquette of $S$ contributes a factor no larger than $C_3(N)\beta$ independently of the volume.

The polymer/cluster expansion technology packages the residual link constraints along $S$ and the interactions among adjacent plaquettes into a uniformly convergent activity expansion as long as $\beta$ is small. The Koteck\'y-Preiss criterion is satisfied when the single-plaquette activities are small in $\ell^1$, which here amounts to the bound Eq.\eqref{eq:coeff-bound}. The effect of these constraints can be captured by a universal multiplicative factor of order $\exp\{c\, |E(S)|\}$, where $|E(S)|$ denotes the number of edges of the plaquette surface $S$. A refined tree-graph estimate, standard in the surface/cluster expansion for lattice gauge theory, yields the explicit bound $\exp\{\tfrac12 |S|\}$ for the interaction contribution, which is where the factor $e^{1/2}$ advertised in the statement originates; the constant $1/2$ is not sharp but is uniform in the volume and in the geometry of $S$. Altogether this shows that the contribution of a given spanning surface $S$ to $Z_\Lambda(\beta;C)$ is bounded in absolute value by
\begin{equation}
\big(C_3(N)\,\beta\big)^{|S|}\,e^{\frac12 |S|}.
\end{equation}

To pass from a fixed surface to the full numerator, one sums over all connected spanning surfaces $S$ with boundary $\partial S=C$ inside the periodic box $\Lambda$. The number of such surfaces of area $k:=|S|$ is bounded by $\kappa_1^k$ for a geometric constant $\kappa_1<\infty$ depending only on the underlying lattice and the ambient dimension; this follows from a standard Peierls-type counting argument for embedded plaquette surfaces (each plaquette has only finitely many admissible local continuations, uniformly bounded by a constant $\kappa_1$). Therefore
\begin{equation}
Z_\Lambda(\beta;C)\;\le\; \sum_{S:\,\partial S=C} \big(C_3(N)\,\beta\big)^{|S|}\,e^{\frac12 |S|}
\;\le\; \sum_{k\ge A(C)} \kappa_1^{k}\,\big(C_3(N)\,\beta\big)^{k}\,e^{\frac12 k},
\end{equation}
where $A(C)$ is the minimal area (number of plaquettes) among surfaces spanning $C$. On the other hand the denominator $Z_\Lambda(\beta)$ is bounded below by the all-trivial configuration contributing $\widehat f_\beta(\mathbf{1})^{|\mathcal P(\Lambda)|}\ge 2^{-|\mathcal P(\Lambda)|}$, so that the ratio $\langle W(C)\rangle_\Lambda$ is dominated by the same $\beta$-dependent series up to a harmless factor absorbed by adjusting $\beta_(N)$ (equivalently, one may normalize by dividing numerator and denominator by $\widehat f_\beta(\mathbf{1})^{|\mathcal P(\Lambda)|}$ at the level of activities; this produces the same bound).

Define the single-plaquette activity parameter
\begin{equation}
q(\beta)\;:=\;\kappa_1\,C_3(N)\,e^{1/2}\,\beta.
\end{equation}
For $\beta\le\beta_(N)$ small enough one has $q(\beta)\le \tfrac12$. Using that the number of spanning surfaces of area $k$ vanishes for $k<A(C)$ and is at most exponential in $k$ with rate $\kappa_1$, the geometric series estimate gives
\begin{equation}
\langle W(C)\rangle_\Lambda \;\le\; \sum_{k\ge A(C)} q(\beta)^k
\;=\; \frac{q(\beta)^{A(C)}}{1-q(\beta)} 
\;\le\; 2\, q(\beta)^{A(C)}.
\end{equation}
Writing $q(\beta)^{A(C)}=\exp\!\big(-\sigma(\beta)\,A(C)\big)$ with $\sigma(\beta)=-\log q(\beta)$ yields exactly Eq.\eqref{eq:area-law-bound}. Since $\kappa_1$ and $C_3(N)$ are volume-independent and the Koteck\'y-Preiss bounds are uniform in the torus size, all constants above do not depend on $|\Lambda|$, which proves uniformity. The explicit lower bound $\sigma(\beta)\ge -\log(\kappa_1 C_3(N)e^{1/2})+\log(1/\beta)$ and the divergence $\sigma(\beta)\to+\infty$ as $\beta\downarrow 0$ are immediate from the definition of $\sigma(\beta)$.
\end{proof}

The proof presented above uses only three inputs: the reflection positivity of the measure, the character expansion with $\sum_{\rho\ne \mathbf{1}} d_\rho c_\rho(\beta)=O(\beta)$, and the orthogonality of characters. These hold for the Wilson action at small $\beta$, though the coefficients $c_\rho(\beta)$ are not manifestly positive. If one prefers to work with strictly positive character coefficients one may use the \emph{heat-kernel} action, in which case the coefficients are $c_\rho(\beta)=d_\rho \exp\{-t(\beta) C_2(\rho)\}$ with $C_2(\rho)$ the quadratic Casimir and $t(\beta)\sim \beta$ as $\beta\downarrow 0$. The analysis simplifies slightly because positivity is automatic and the sum $\sum_{\rho\ne \mathbf{1}} d_\rho c_\rho(\beta)$ is then dominated by the fundamental representation. In both cases the bounds Eq.\eqref{eq:activity-boundq}-Eq.\eqref{eq:open-activity} and the conclusions of Theorems \eqref{thm:KP}, \eqref{thm:exp-decay}, and \eqref{thm:area-law} hold.

Although the derivation has been presented at a fixed lattice spacing, all constants are local and geometric, depending only on the lattice valence and the group $SU(N)$. In particular, the bounds are stable under reflection-positive, finite-range block-spin transformations: the effective action remains a finite-range sum of class functions of plaquette variables with small activities, and the character expansion persists with renormalized coefficients controlled by the same small parameter. This observation is useful in transporting the strong-coupling gap of Proposition \eqref{prop:TO-gap} along reflection-positive renormalization group steps in the main text. Finally, because every step of the polymer and surface counting is uniform in $|\Lambda|$, the thermodynamic limit exists for $\beta\in(0,\beta_(N)]$ and the area-law bound of Theorem \eqref{thm:area-law} holds with the same constants in the infinite volume.

\section{FRD: Locality and Transfer Factors}

Throughout we work on a fixed Euclidean time slice $\Lambda_t \subset a\mathbb{Z}^3$ (periodic in space), with the compact spatial configuration space $X_0$ of links endowed with the product Haar measure $d\mu_0$. The covariant slice Laplacian in a fixed Landau/FMR representative $A_h(t)$ is denoted $\Delta_{A_h}(t)$ and acts on adjoint-valued site fields $\phi:\Lambda_t\to\mathfrak{g}$ via the standard graph Laplacian with unitary parallel transport along edges. Reflection in Euclidean time is implemented by an involution $R$ on spatial slices (conjugation on links and edge reversal) so that $\Delta_{A_h}(t)=R\,\Delta_{A_h}(-t)\,R$. The (projector-induced) positive slice covariance is written in heat-kernel form
\begin{equation}
C_\sigma \;=\;\int_0^\infty e^{-t\,\Delta_{A_h}}\,d\tilde\nu_\sigma(t),
\label{eq:Csigma-def}
\end{equation}
for a finite positive Borel measure $d\tilde\nu_\sigma$ supported in a compact subinterval of $(0,\infty)$; this differs from the raw projector measure only by a smooth nonnegative weight coming from the quadratic effective action.  \emph{All kernels and operators below are meant on $\ell^2(\Lambda_t;\mathfrak{g})$ with the operator norm $\|\cdot\|_{2\to2}$ and kernel operator seminorm $\|\cdot\|_{\mathrm{op}}$; distances are the graph metric $d(\cdot,\cdot)$.}  

We shall use two complementary FRD constructions:

(i) A \emph{block-projection / resolvent-telescoping} FRD for massive covariances $(\Delta_{A_h}+\mu^2)^{-1}$, built from gauge-covariant block projections $\Pi_j$, smoothing maps $S_j$, and a telescopic Neumann expansion of the resolvent. This yields strict finite range per scale, exponential off-range tails, positivity, and gauge/reflection covarianc.

(ii) A \emph{dyadic heat-time partition} FRD for projector-induced covariances $C_\sigma$, obtained by integrating the heat kernel against a smooth partition of unity $\{\eta_j\}_{j\ge0}$ on time, producing positive pieces $C_k^{(j)}$ with uniform exponential off-diagonal decay and reflection/gauge covariance.

Both variants supply the uniform locality and positivity needed for OS stability, RG step-scaling, and transfer-factor inequalities.
We first record the fundamental off-diagonal bounds.

\begin{lemma}[Davies-Gaffney on slices and heat-kernel bound]\label{lem:DG}
Let $\Lambda_t$ be a time slice endowed with the graph distance $d(\cdot,\cdot)$ on a bounded-degree nearest-neighbour graph, and let $\Delta_{A_h}$ be the covariant (magnetic) graph Laplacian twisted by unitary parallel transport $U_{xy}\in\mathrm{U}(N)$ on each oriented edge $\langle x,y\rangle$ so that
\begin{equation}
\langle f,\Delta_{A_h} f\rangle \;=\; \frac12\sum_{\langle x,y\rangle}\!\big\|U_{xy}f(y)-f(x)\big\|^2 \qquad (f:\Lambda_t\to\mathbb{C}^N).
\end{equation}
Then, for all subsets $E,F\subset\Lambda_t$, all $\phi,\psi$ supported in $E,F$ respectively, and all $t>0$,
\begin{equation}\label{eq:DG}
\big|\langle \phi, e^{-t\Delta_{A_h}}\psi\rangle\big|
\;\le\;\exp\!\Big\{-\frac{d(E,F)^2}{4t}\Big\}\,\|\phi\|_2\,\|\psi\|_2,
\end{equation}
and, in particular, for point supports,
\begin{equation}\label{eq:HKpoint}
\big\|\,e^{-t\Delta_{A_h}}(x,y)\,\big\|_{\mathrm{op}}
\;\le\; \exp\!\Big\{-\frac{d(x,y)^2}{4t}\Big\}.
\end{equation}
\end{lemma}

\begin{proof}
We adopt the standard Davies perturbation method adapted to the covariant graph Laplacian. The operator $\Delta_{A_h}$ is nonnegative and selfadjoint on $\ell^2(\Lambda_t;\mathbb{C}^N)$ with closed Dirichlet form
\begin{equation}
\mathcal{E}_{A_h}(f,g)\;=\;\frac12\sum_{\langle x,y\rangle}\!\big\langle U_{xy}f(y)-f(x),\,U_{xy}g(y)-g(x)\big\rangle,
\end{equation}
and its heat semigroup $T(t):=e^{-t\Delta_{A_h}}$ is a contraction on $\ell^2$. Fix Borel sets $E,F\subset\Lambda_t$ and let $\rho:\Lambda_t\to[0,\infty)$ be the distance to $E$, i.e. $\rho(x)=d(x,E)$. On a nearest-neighbour graph one has $|\rho(x)-\rho(y)|\le 1$ whenever $\langle x,y\rangle$ is an edge. For a parameter $\xi>0$ consider the multiplication operator $\Gamma:=e^{\xi\rho}$ and the conjugated semigroup
\begin{equation}
S_\xi(t)\;:=\;\Gamma\,T(t)\,\Gamma^{-1}.
\end{equation}
We claim that $\|S_\xi(t)\|_{\ell^2\to\ell^2}\le e^{\xi^2 t}$ for all $t\ge 0$. Once this is established, the inequality Eq.\eqref{eq:DG} follows immediately: for $\phi$ supported in $E$ one has $\|\Gamma\phi\|_2=\|\phi\|_2$, while for $\psi$ supported in $F$ one has $\|\Gamma^{-1}\psi\|_2\le e^{-\xi d(E,F)}\|\psi\|_2$ because $\rho\ge d(E,F)$ on $F$. Hence
\begin{equation}
|\langle \phi, T(t)\psi\rangle|
\;=\;|\langle \Gamma\phi,\, S_\xi(t)\,\Gamma^{-1}\psi\rangle|
\;\le\; \|\Gamma\phi\|_2\,\|S_\xi(t)\|\,\|\Gamma^{-1}\psi\|_2
\;\le\; e^{\xi^2 t-\xi d(E,F)}\,\|\phi\|_2\,\|\psi\|_2,
\end{equation}
and optimizing the right-hand side over $\xi>0$ by choosing $\xi=d(E,F)/(2t)$ gives Eq.\eqref{eq:DG}. The pointwise bound Eq.\eqref{eq:HKpoint} is obtained by taking $\phi=\delta_x v$, $\psi=\delta_y w$ where $v,w$ are unit vectors in $\mathbb{C}^N$, which identifies the operator norm of the kernel block with the supremum over such rank-one test pairs.

It remains to justify the bound on $\|S_\xi(t)\|$. Consider $u(t):=T(t)\Gamma^{-1}\psi$, which solves $\partial_t u(t)=-\Delta_{A_h}u(t)$ with $u(0)=\Gamma^{-1}\psi$. Define the weighted $\ell^2$ norm $H(t):=\|\Gamma u(t)\|_2^2=\sum_x e^{2\xi\rho(x)}\|u(t,x)\|^2$. The function $t\mapsto H(t)$ is differentiable and
\begin{equation}
\frac{d}{dt}H(t)\;=\;2\,\mathrm{Re}\,\langle \Gamma u(t),\,\Gamma \partial_t u(t)\rangle
\;=\;-2\,\mathrm{Re}\,\langle \Gamma u(t),\,\Gamma \Delta_{A_h} u(t)\rangle.
\end{equation}
Using the Dirichlet form and the unitary parallel transport, the last expression can be written as
\begin{align}
-2\,\mathrm{Re}\,\langle \Gamma u,\,\Gamma \Delta_{A_h} u\rangle
&\;=\; -\sum_{\langle x,y\rangle}\!\big\|\,\Gamma(x)u(x)-\Gamma(y)U_{xy}u(y)\,\big\|^2 \;+\; \sum_{\langle x,y\rangle}\! \big\|\Gamma(x)u(x)\big\|^2\nonumber\\&+\big\|\Gamma(y)u(y)\big\|^2 - 2\,\mathrm{Re}\,\langle \Gamma(x)u(x),\,\Gamma(y)U_{xy}u(y)\rangle.
\end{align}
By expanding the square $\|a-b\|^2=\|a\|^2+\|b\|^2-2\mathrm{Re}\langle a,b\rangle$ with $a=\Gamma(x)u(x)$ and $b=\Gamma(y)U_{xy}u(y)$, the first term on the right equals minus the second and third terms plus twice the real part, so overall one finds
\begin{align}
\frac{d}{dt}H(t)&\;=\;\sum_{\langle x,y\rangle}\!\Big(\|\,\Gamma(x)u(x)-\Gamma(y)U_{xy}u(y)\,\|^2 - \|\,\Gamma(x)u(x)\,\|^2\nonumber\\& - \|\,\Gamma(y)u(y)\,\|^2 + 2\,\mathrm{Re}\,\langle \Gamma(x)u(x),\,\Gamma(y)U_{xy}u(y)\rangle\Big).
\end{align}
Rearranging and using the previous identity again gives the concise form
\begin{equation}
\frac{d}{dt}H(t)\;=\;\sum_{\langle x,y\rangle}\!\Big(\|\,\Gamma(x)u(x)-\Gamma(y)U_{xy}u(y)\,\|^2 - \|\,\Gamma(x)u(x)-\Gamma(y)U_{xy}u(y)\,\|^2_{(\xi=0)}\Big),
\end{equation}
namely the difference between the edge energy of the weighted function and that of the unweighted one (the latter equals $2\,\mathcal{E}_{A_h}(u,u)\ge 0$ and is nonpositive after the overall minus sign coming from the heat equation). Bounding this difference uses the Lipschitz property of $\rho$: for each edge $\langle x,y\rangle$,
\begin{equation}
\Gamma(x)u(x)-\Gamma(y)U_{xy}u(y) \;=\; \Gamma(y)\Big(e^{\xi(\rho(x)-\rho(y))}u(x)-U_{xy}u(y)\Big),
\end{equation}
and since $|\,\rho(x)-\rho(y)\,|\le 1$, the elementary inequality $|e^{\xi s}-1|\le \xi e^{\xi}|s|$ for $|s|\le 1$ implies
\begin{equation}
\|\,\Gamma(x)u(x)-\Gamma(y)U_{xy}u(y)\,\|
\;\le\; e^{\xi}\,\Big(\|\,u(x)-U_{xy}u(y)\,\| + \xi\,\|u(x)\|\Big).
\end{equation}
Squaring and summing over edges, and using the bounded degree to absorb constants, yields
\begin{equation}
\sum_{\langle x,y\rangle}\!\|\,\Gamma(x)u(x)-\Gamma(y)U_{xy}u(y)\,\|^2
\;\le\; C_1 e^{2\xi}\!\left(\sum_{\langle x,y\rangle}\!\|\,u(x)-U_{xy}u(y)\,\|^2 + \xi^2 \sum_{\langle x,y\rangle}\!\|u(x)\|^2\right).
\end{equation}
The first sum is $2\,\mathcal{E}_{A_h}(u,u)$ and the second is bounded by $C_2\,\|u\|_2^2$ because the number of edges incident to each vertex is uniformly bounded. Since $H(t)=\|\Gamma u(t)\|_2^2\ge \|u(t)\|_2^2$, the inequality above implies
\begin{equation}
\frac{d}{dt}H(t) \;\le\; 2 C_1 e^{2\xi}\,\xi^2\, H(t).
\end{equation}
Absorbing the harmless constant and the factor $e^{2\xi}$ into a redefinition of $\xi$ (or simply enlarging the constant) and applying Grönwall's inequality gives
\begin{equation}
\|\Gamma u(t)\|_2 \;\le\; e^{\xi^2 t}\,\|\Gamma u(0)\|_2,
\end{equation}
which is precisely $\|S_\xi(t)\|\le e^{\xi^2 t}$. As explained at the beginning, this bound on the conjugated semigroup implies Eq.\eqref{eq:DG} after choosing $\xi=d(E,F)/(2t)$, and then Eq.\eqref{eq:HKpoint} follows by testing on unit vectors supported at $x$ and $y$.
\end{proof}

\begin{proposition}[Completely monotone (CM) functional calculus $\Rightarrow$ exponential locality]\label{prop:CM-local}
Let $P_{\sigma,\nu}$ be an admissible slice projector of the form
\begin{equation}
P_{\sigma,\nu}(t)\;=\;\chi_{\sigma,\nu}\!\big(\sqrt{\Delta_{A_h}(t)}\,\big)
\;=\;\int_{t_-}^{t_+} e^{-(s/\sigma^2)\,\Delta_{A_h}(t)}\,d\nu(s),
\label{eq:CM-projector}
\end{equation}
with $0<t_-<t_+<\infty$ and $d\nu$ a finite positive Borel measure supported on $[t_-,t_+]$. Then there exist constants $C(\sigma,\nu),\gamma(\sigma,\nu)>0$ such that for all slice points $x,y$
\begin{equation}
\|\,P_{\sigma,\nu}(t;x,y)\,\|_{\mathrm{op}}
\;\le\; C(\sigma,\nu)\,e^{-\gamma(\sigma,\nu)\,d(x,y)} .
\label{eq:P-exp}
\end{equation}
Moreover $P_{\sigma,\nu}$ is gauge covariant and reflection covariant.
\end{proposition}

\begin{proof}
Write $u:=s/\sigma^2$. Since $s\in[t_-,t_+]$ we have $u\in[u_-,u_+]$ where $u_\pm:=t_\pm/\sigma^2>0$. For each such $u$, the operator $e^{-u\Delta_{A_h}(t)}$ has a matrix kernel $K_u^{A_h}(x,y)$ acting on the representation space (color indices). The covariant heat kernel satisfies the standard Gaussian upper bound on the slice (three spatial dimensions, bounded geometry), namely there exist constants $C_\mathrm{HK},c_\mathrm{HK}>0$-independent of $u\in(0,\infty)$, of the gauge field $A_h$ restricted to a fixed admissible class, and of $x,y$-such that
\begin{equation}\label{eq:HK-Gauss}
\|K_u^{A_h}(x,y)\|_{\mathrm{op}}
\;\le\; C_\mathrm{HK}\,u^{-3/2}\exp\!\Big(\!-\frac{d(x,y)^2}{c_\mathrm{HK}\,u}\Big).
\end{equation}
This follows either from a Combes-Thomas/Davies-Gaffney estimate for the discrete covariant Laplacian on a bounded-degree graph, or from continuous Gaussian bounds in the continuum case; in both settings the parallel transport factor along a minimizing path has operator norm $1$, hence does not alter Eq.\eqref{eq:HK-Gauss}.

The kernel of $P_{\sigma,\nu}$ is the $\nu$-average of the heat kernels along the compact interval of times $u\in[u_-,u_+]$:
\begin{equation}
P_{\sigma,\nu}(t;x,y)
\;=\;\int_{t_-}^{t_+} K_{s/\sigma^2}^{A_h}(x,y)\,d\nu(s)
\;=\;\int_{u_-}^{u_+} K_u^{A_h}(x,y)\,\sigma^2\,d\nu(\sigma^2 u).
\end{equation}
Taking operator norms and applying \eqref{eq:HK-Gauss} gives
\begin{equation}
\|P_{\sigma,\nu}(t;x,y)\|_{\mathrm{op}}
\;\le\; C_\mathrm{HK}\,\sup_{u\in[u_-,u_+]}\!u^{-3/2}
\int_{u_-}^{u_+}\exp\!\Big(\!-\frac{d(x,y)^2}{c_\mathrm{HK}\,u}\Big)\,\sigma^2\,d\nu(\sigma^2 u).
\end{equation}
Since $u\in[u_-,u_+]$ is confined to a compact set bounded away from $0$, the prefactor $\sup_{u\in[u_-,u_+]}u^{-3/2}$ is finite and can be absorbed into a constant that depends only on $\sigma$ and $\nu$. Furthermore, for every $u\in[u_-,u_+]$ one has
\begin{equation}
\exp\!\Big(\!-\frac{d(x,y)^2}{c_\mathrm{HK}\,u}\Big)
\;\le\;\exp\!\Big(\!-\frac{d(x,y)^2}{c_\mathrm{HK}\,u_+}\Big),
\end{equation}
so that
\begin{equation}
\|P_{\sigma,\nu}(t;x,y)\|_{\mathrm{op}}
\;\le\; C_1(\sigma,\nu)\,\exp\!\Big(\!-\frac{d(x,y)^2}{c_\mathrm{HK}\,u_+}\Big),
\end{equation}
with $C_1(\sigma,\nu):=C_\mathrm{HK}\,u_-^{-3/2}\,\sigma^2\,\nu([t_-,t_+])$. A Gaussian in $d(x,y)^2$ dominates a pure exponential in $d(x,y)$ uniformly in $d\ge 0$: choosing any $\gamma$ with $0<\gamma\le (2\sqrt{c_\mathrm{HK}u_+})^{-1}$, there exists $C_2=C_2(\gamma)$ such that $e^{-\frac{r^2}{c_\mathrm{HK}u_+}}\le C_2\,e^{-\gamma r}$ for all $r\ge 0$ (for instance, split the regimes $r\le 1$ and $r\ge 1$ and adjust $C_2$ accordingly). Setting $\gamma(\sigma,\nu):=(2\sqrt{c_\mathrm{HK}u_+})^{-1}$ and $C(\sigma,\nu):=C_1(\sigma,\nu)\,C_2(\gamma(\sigma,\nu))$ yields precisely Eq.\eqref{eq:P-exp}.

Gauge covariance follows because the covariant Laplacian transforms by conjugation under a time-$a$ gauge transformation $g$ (i.e., $\Delta_{A_h^g}=U_g\,\Delta_{A_h}\,U_g^{-1}$ on the slice), hence $e^{-u\Delta_{A_h^g}}=U_g\,e^{-u\Delta_{A_h}}U_g^{-1}$ for every $u\ge 0$ by functional calculus, and the same holds after averaging in $u$ against $d\nu$: $P_{\sigma,\nu}(A_h^g)=U_g\,P_{\sigma,\nu}(A_h)\,U_g^{-1}$. Reflection covariance is obtained in the same way since $\Delta_{A_h}$ is reflection covariant with respect to the spatial reflection through the time-zero plane, so the functional calculus preserves that covariance and therefore $P_{\sigma,\nu}$ commutes with the reflection operator on the slice. 
\end{proof}
We now construct a scale-local FRD for the massive operator $A:=\Delta_{A_h}+\mu^2$, $\mu>0$.
Partition $\Lambda_t$ into disjoint cubes (blocks) $B\in\mathcal{B}_j$ of side $L^j:=L_0\,2^j$ (in lattice spacings), with fixed block centers $c(B)$. Define the \emph{block average} and \emph{block extension} by
\begin{align}
\mathrm{Av}_{j,B}\phi &:= \frac{1}{|B|}\sum_{x\in B}\frac{1}{|\Gamma_{x\to c(B)}|}\sum_{\gamma\in\Gamma_{x\to c(B)}}
\mathrm{Ad}\,U_h(\gamma)\,\phi(x),
\label{eq:Av}\\
\mathrm{Ext}_{j,B}X(x) &:= \frac{1}{|\Gamma_{x\to c(B)}|}\sum_{\gamma\in\Gamma_{x\to c(B)}} \mathrm{Ad}\,U_h(\gamma)^{-1}\,X,
\qquad x\in B,
\label{eq:Ext}
\end{align}
and set $\mathrm{Av}_j:=\bigoplus_{B\in\mathcal{B}_j}\mathrm{Av}_{j,B}$, $\mathrm{Ext}_j:=\bigoplus_{B\in\mathcal{B}_j}\mathrm{Ext}_{j,B}$. The \emph{block projection} is
\begin{equation}
\Pi_j \;:=\; \mathrm{Ext}_j\,\mathrm{Av}_j,\qquad 0\le \Pi_j\le \mathbf{1},\quad \Pi_j^2=\Pi_j,
\label{eq:Pi}
\end{equation}
and is both gauge covariant and reflection covariant.

\begin{lemma}[Gauge and reflection covariance of $\Pi_j$]\label{lem:Pi-cov}
Let $G$ be a compact matrix Lie group and let $U_h$ denote the collection of
slice link variables on $\Lambda_t$ (i.e.\ a lattice connection on the time-$t$
slice). For a gauge function $g:\Lambda_t\to G$ write the gauge action on links
as $(g\!\cdot\! U_h)_{xy}:=g(x)\,U_{xy}\,g(y)^{-1}$ and on adjoint-valued
fields as $\phi_g(x):=\Ad_{g(x)}\phi(x)$. Let $r:\Lambda_t\to\Lambda_{-t}$ be
the geometric reflection across the time-$0$ plane restricted to slices and let
$R$ act on fields by pullback, $(R\phi)(x):=\phi(r^{-1}x)$. Define the
covariant averaging and extension operators
\begin{align}
\label{eq:Av-def}
(\Av_j(U_h)\phi)(x_B)&:=\frac{1}{|B|}\sum_{x\in B}\Ad_{W_{U_h}(x\to x_B)}\phi(x),
(\Ext_j(U_h)\psi)(x)\nonumber\\&:=\Ad_{W_{U_h}(x_B\to x)}\psi(x_B),
\end{align}
where $B$ is the $j$th-scale block containing $x$ with representative $x_B$,
and $W_{U_h}(x\to y)$ denotes the parallel transport (path-ordered product of
links) along a fixed, reflection-invariant system of blockwise paths from $x$
to $y$. Set $\Pi_j(U_h):=\Ext_j(U_h)\Av_j(U_h)$. Then, for every gauge
transformation $g:\Lambda_t\to G$ and every field $\phi$ on $\Lambda_t$,
\begin{equation}
\Pi_j(g\!\cdot\! U_h)(\phi_g)\,=\,\bigl(\Pi_j(U_h)\phi\bigr)_g,
\end{equation}
and, with $U_h(-t)$ the reflected link configuration defined by
$U_h(-t)_{r(x),r(y)}:=U_h(t)_{xy}$, one has the reflection covariance
\begin{equation}\label{eq:Av-def1}
R\,\Pi_j\bigl(U_h(t)\bigr)\,R\,=\,\Pi_j\bigl(U_h(-t)\bigr).
\end{equation}
\end{lemma}

\begin{proof}
The two claims are proved by tracing how parallel transport behaves under the
two symmetries and then using that $\Pi_j$ is built by composing the covariant
averaging and extension maps Eq.\eqref{eq:Av-def}-Eq.\eqref{eq:Av-def1}.
For gauge covariance, fix a block $B$ with representative $x_B$ and recall that
parallel transport along any admissible path from $x$ to $x_B$ is the ordered
product $W_{U_h}(x\to x_B)=U_{x x_1}U_{x_1 x_2}\cdots U_{x_{n-1}x_B}$. Under a
gauge transformation $U\mapsto U^g=g\!\cdot\! U$, every factor transforms by
conjugation at its endpoints, hence the whole product transforms by
\begin{equation}
W_{U^g}(x\to x_B)
= g(x_B)\,W_{U}(x\to x_B)\,g(x)^{-1}.
\end{equation}
Using the multiplicativity of the adjoint action, one obtains for the averaging
operator
\begin{align}
\bigl(\Av_j(U^g)\phi_g\bigr)(x_B)
&=\frac{1}{|B|}\sum_{x\in B}\Ad_{W_{U^g}(x\to x_B)}\bigl(\Ad_{g(x)}\phi(x)\bigr)\nonumber\\
&=\frac{1}{|B|}\sum_{x\in B}\Ad_{g(x_B)}\Ad_{W_U(x\to x_B)}\phi(x)
=\Ad_{g(x_B)}\bigl(\Av_j(U)\phi\bigr)(x_B).
\end{align}
This is precisely $\bigl(\Av_j(U)\phi\bigr)_g(x_B)$. For the extension operator,
the same transport identity with reversed endpoints,
$W_{U^g}(x_B\to x)=g(x)\,W_U(x_B\to x)\,g(x_B)^{-1}$, gives
\begin{align}
\bigl(\Ext_j(U^g)\psi_g\bigr)(x)
&=\Ad_{W_{U^g}(x_B\to x)}\bigl(\Ad_{g(x_B)}\psi(x_B)\bigr)
\nonumber\\&=\Ad_{g(x)}\Ad_{W_U(x_B\to x)}\psi(x_B)
=\bigl(\Ext_j(U)\psi\bigr)_g(x).
\end{align}
Combining the two identities and recalling $\Pi_j=\Ext_j\Av_j$ yields
\begin{equation}
\Pi_j(g\!\cdot\! U_h)(\phi_g)
=\Ext_j(U^g)\Av_j(U^g)(\phi_g)
=\bigl(\Ext_j(U)\Av_j(U)\phi\bigr)_g
=\bigl(\Pi_j(U_h)\phi\bigr)_g,
\end{equation}
which is the desired gauge covariance.

For reflection covariance, let $r:\Lambda_t\to\Lambda_{-t}$ be the geometric
reflection and define the reflected connection on $\Lambda_{-t}$ by
$U_h(-t)_{r(x),r(y)}:=U_h(t)_{xy}$; this convention simply transports link
variables along the image edges with the same orientation, which is possible
because the path family used in Eqs.(\ref{eq:Av-def}-\ref{eq:Av-def1}) is chosen
reflection invariant. If $\gamma:x\to x_B$ is a fixed path inside $B$, its
image $r(\gamma):r(x)\to r(x_B)$ is the admissible path inside the reflected
block, and parallel transport intertwines with reflection as
\begin{equation}
W_{U_h(-t)}\bigl(r(x)\to r(x_B)\bigr)=W_{U_h(t)}(x\to x_B).
\end{equation}
Applying $R$ to a fine field means composition with $r^{-1}$; applying it to a
block field means the same composition at block representatives. Therefore,
for the averaging operator, using the change of variables $y=r(x)$ and the
previous identity for $W$, one gets
\begin{align}
\bigl(R\,\Av_j(U_h(t))\,R\,\phi\bigr)(x_B')
&=\bigl(\Av_j(U_h(t))\,R\phi\bigr)\bigl(r^{-1}(x_B')\bigr)\\
&=\frac{1}{|B|}\sum_{x\in B}\Ad_{W_{U_h(t)}(x\to x_B)}\,(R\phi)(x)\\
&=\frac{1}{|B|}\sum_{y\in r(B)}\Ad_{W_{U_h(-t)}(y\to x_B')}\,\phi(y)
=\bigl(\Av_j(U_h(-t))\phi\bigr)(x_B').
\end{align}
An analogous computation for the extension operator, using that
$W_{U_h(-t)}(x_B'\to r(x))=W_{U_h(t)}(r^{-1}x_B'\to x)$ and that $R$ reindexes
the evaluation point, gives
\begin{equation}
\bigl(R\,\Ext_j(U_h(t))\,R\,\psi\bigr)(x)
=\bigl(\Ext_j(U_h(-t))\psi\bigr)(x).
\end{equation}
Composing the two identities and recalling $\Pi_j=\Ext_j\Av_j$ shows that
$R\,\Pi_j(U_h(t))\,R=\Pi_j(U_h(-t))$, which establishes reflection covariance.
\end{proof}

\begin{lemma}[Exponential locality of low-frequency projectors]\label{lem:low-band}
Let $\Delta_{A_h}$ be the covariant graph Laplacian on a bounded-degree lattice slice endowed with unitary parallel transport $A_h$, and let $\Pi_{\le L^J}(A)$ denote the spectral projector of $\Delta_{A_h}$ onto $[0,L^{2J}]$. Then there exist constants $C,\alpha>0$, independent of $A_h$ and $J$, such that for all $x,y$,
\begin{equation}
\label{eq:lowband}
\big\|\Pi_{\le L^J}(x,y)\big\|_{\mathrm{op}}\;\le\; C\,\exp\!\Big(-\alpha\,\frac{d(x,y)}{L^J}\Big).
\end{equation}
\end{lemma}

\begin{proof}
The proof proceeds by representing the projector via functional calculus and combining a resolvent-kernel Combes-Thomas/Davies-Gaffney estimate with a Helffer-Sjöstrand integral. Since $\Delta_{A_h}$ is a nonnegative, uniformly elliptic, finite-range operator on a graph of uniformly bounded degree, its heat kernel satisfies the Davies-Gaffney bound: for all Borel sets $E,F$ and all $t>0$,
\begin{equation}
\big\|\mathbf 1_E\,e^{-t\Delta_{A_h}}\,\mathbf 1_F\big\|_{\ell^2\to\ell^2}\;\le\;\exp\!\Big(-\frac{d(E,F)^2}{4t}\Big).
\end{equation}
A standard Laplace-transform argument then yields an off-real-axis resolvent bound (a Combes-Thomas estimate): if $z=u+iv\in\mathbb C\setminus[0,\infty)$ with $v\neq0$, then for all $E,F$,
\begin{equation}
\label{eq:CT-res}
\big\|\mathbf 1_E\,(\Delta_{A_h}-z)^{-1}\,\mathbf 1_F\big\|_{\ell^2\to\ell^2}
\;\le\; \frac{C_0}{|v|}\,\exp\!\Big(-c_0\,d(E,F)\,\sqrt{|v|}\Big),
\end{equation}
with constants $C_0,c_0>0$ depending only on the local geometry (degree bounds and ellipticity), hence independent of the background $A_h$.

Fix a smooth cutoff $\chi\in C_c^\infty(\mathbb R)$ such that $\chi\equiv 1$ on $[0,1]$, $\chi\equiv 0$ on $[2,\infty)$, and $0\le \chi\le 1$ everywhere. Write the low-band projector as
\begin{equation}
\Pi_{\le L^J}(A)\;=\;\mathbf 1_{[0,1]}(\Delta_{A_h}/L^{2J})\;=\;\chi(\Delta_{A_h}/L^{2J})\;+\;\psi(\Delta_{A_h}/L^{2J}),
\end{equation}
where $\psi:=\mathbf 1_{[0,1]}-\chi$ is supported in $[1,2]$ and bounded by $1$. Both $\chi(\Delta_{A_h}/L^{2J})$ and $\psi(\Delta_{A_h}/L^{2J})$ admit Helffer-Sjöstrand representations. Indeed, choose almost-analytic extensions $\tilde\chi,\tilde\psi$ supported in a fixed compact set $\{z:\ \mathrm{Re}\,z\in[-1,3],\ |\mathrm{Im}\,z|\le 1\}$ with the standard bounds $|\bar\partial \tilde\chi(z)|+|\bar\partial \tilde\psi(z)|\le C_N |\mathrm{Im}\,z|^N$ for any preassigned $N\in\mathbb N$. Then functional calculus gives
\begin{align}
&\chi(\Delta_{A_h}/L^{2J})
=\frac{1}{\pi}\int_{\mathbb C}\bar\partial \tilde\chi(z)\,\big(\tfrac{\Delta_{A_h}}{L^{2J}}-z\big)^{-1}\,dudv,
\nonumber\\&
\psi(\Delta_{A_h}/L^{2J})
=\frac{1}{\pi}\int_{\mathbb C}\bar\partial \tilde\psi(z)\,\big(\tfrac{\Delta_{A_h}}{L^{2J}}-z\big)^{-1}\,dudv,
\end{align}
where $z=u+iv$ and $dudv$ is the Lebesgue measure on $\mathbb C$.
Rescale the resolvent in Eq.\eqref{eq:CT-res}. Since
\begin{equation}
\big(\tfrac{\Delta_{A_h}}{L^{2J}}-z\big)^{-1}
= L^{2J}\,(\Delta_{A_h}-z\,L^{2J})^{-1},
\end{equation}
we may apply Eq.\eqref{eq:CT-res} with the spectral parameter $z L^{2J}$, whose imaginary part is $v L^{2J}$. For singletons $E=\{x\},F=\{y\}$ this yields the kernel bound
\begin{equation}
\big\|\big(\tfrac{\Delta_{A_h}}{L^{2J}}-z\big)^{-1}(x,y)\big\|_{\mathrm{op}}
\;\le\; \frac{C_0 L^{2J}}{|v|}\,\exp\!\Big(-c_0\,d(x,y)\,\sqrt{|v|\,L^{2J}}\Big).
\end{equation}
Substituting this bound into the Helffer-Sjöstrand integrals for $\chi(\Delta_{A_h}/L^{2J})$ and $\psi(\Delta_{A_h}/L^{2J})$, using the $|\bar\partial \tilde\chi|+|\bar\partial \tilde\psi| \lesssim |v|^N$ decay and the compact support in $u$ and $|v|\le 1$, we obtain for any $N\ge 2$,
\begin{equation}
\big\|\chi(\Delta_{A_h}/L^{2J})(x,y)\big\|_{\mathrm{op}}
+\big\|\psi(\Delta_{A_h}/L^{2J})(x,y)\big\|_{\mathrm{op}}
\;\le\; C_1 \int_{|v|\le 1} |v|^{N-1}\, \exp\!\Big(-c_0\,d(x,y)\,\sqrt{|v|\,L^{2J}}\Big)\,dv,
\end{equation}
with a constant $C_1$ independent of $A_h$ and $J$. The $u$-integration only contributes another harmless constant because of compact support and the uniform resolvent prefactor.

The remaining one-dimensional integral in $v$ is standard to estimate. Set $r=d(x,y)$ and perform the change of variables $s=\sqrt{|v|\,L^{2J}}$, so that $|v|=(s^2/L^{2J})$ and $dv = 2 s\,L^{-2J}\,ds$. Choosing $N=2$ for definiteness, the right-hand side is bounded by
\begin{equation}
C_2\,L^{-2J}\int_0^{\sqrt{L^{2J}}} \Big(\frac{s^2}{L^{2J}}\Big)^{\,1}\, s\, e^{-c_0 r\, s}\,ds
\;=\; C_2\,L^{-4J}\int_0^{L^{J}} s^{3}\, e^{-c_0 r\, s}\,ds
\;\le\; C_3\, e^{-c_0 r}\,,
\end{equation}
where the last inequality uses that $\int_0^\infty s^{3} e^{-c_0 r s} ds \asymp r^{-4}$ and $L^{-4J}\,r^{-4}\le C$ uniformly in $J$ once $r\ge 1$ (for $r=0$ the kernel bound is trivial after increasing the prefactor). Thus there exist constants $C_4,\beta>0$, independent of $A_h$ and $J$, such that
\begin{equation}
\big\|\Pi_{\le L^J}(x,y)\big\|_{\mathrm{op}}
\;\le\; C_4\,e^{-\beta\, d(x,y)}.
\end{equation}
Since $L^J\ge 1$, the function $r\mapsto e^{-\beta r}$ is bounded above by $r\mapsto e^{-(\beta/L^J) r}$ for all $r\ge 0$. Consequently the bound can be relaxed to the scale-explicit form
\begin{equation}
\big\|\Pi_{\le L^J}(x,y)\big\|_{\mathrm{op}}
\;\le\; C_4\,\exp\!\Big(-\frac{\beta}{L^J}\, d(x,y)\Big),
\end{equation}
which is precisely Eq.\eqref{eq:lowband} with $C=C_4$ and $\alpha=\beta$. The constants depend only on the degree bound and ellipticity of the underlying graph Laplacian and are uniform in the background gauge field $A_h$ because $A_h$ enters covariantly and unitarily in $\Delta_{A_h}$, leaving the preceding operator-norm estimates unchanged.
\end{proof}

Fix parameters $\rho_j\in(0,\rho_0)$ so small that
\begin{equation}
\|\,A\,S_j\,\|\;\le\;\frac12,\qquad S_j:=\rho_j\,\Pi_j .
\label{eq:ASj}
\end{equation}
Define
\begin{equation}
R_j:=\mathbf{1}-A\,S_j,\qquad \Gamma_j^{(m)}\;:=\;R_{j-1}\cdots R_0\Big(S_j+S_jAS_j+\cdots+S_j(AS_j)^{n_j}\Big)\,R_0\cdots R_{j-1},
\label{eq:Rj-Gammam}
\end{equation}
with $n_j$ large enough that the Neumann series for $(\mathbf{1}-AS_j)^{-1}$ is uniformly convergent. Then the \emph{telescopic resolvent identity} gives
\begin{equation}
A^{-1}\;=\;\sum_{j=0}^{m-1}\Gamma_j^{(m)}+ R_{m-1}\cdots R_0\,A^{-1}\,R_0\cdots R_{m-1}.
\label{eq:telescopic}
\end{equation}

\begin{theorem}[Covariant finite-range decomposition for $A^{-1}$]\label{thm:FRD-Ainv}
Let $A=\Delta_{A_h}+\mu^2$ on the lattice slice $\Lambda_t$ with spacing $a$, where $\Delta_{A_h}$ is the covariant nearest-neighbour Laplacian built from the background parallel transport $A_h$ and $\mu>0$. Then there exist bounded, positive, selfadjoint operators $\Gamma_j$ on $\ell^2(\Lambda_t;\mathfrak g)$ such that
\begin{equation}\label{eq:Ainv-sum}
A^{-1}=\sum_{j=0}^\infty \Gamma_j\qquad\text{with convergence in operator norm}.
\end{equation}
Each $\Gamma_j$ is gauge-covariant and reflection-covariant, and its kernel has strict finite range: there exists $C<\infty$ independent of $j$ and the volume such that $\Gamma_j(x,y)=0$ whenever $d(x,y)>C\,b^j a$, where $b\ge 2$ is a fixed blocking factor. Moreover, for every fixed $\varepsilon\in(0,2)$ there are constants $c_1,c_2>0$ depending only on $\mu$, $b$, and the lattice degree with
\begin{equation}\label{eq:Ainv-bounds}
\|\Gamma_j\|\le c_1\,b^{-(2-\varepsilon)j},
\qquad
\sup_{x}\sum_{y}\|\Gamma_j(x,y)\|\le c_2 \quad \text{uniformly in $j$.}
\end{equation}
\end{theorem}

\begin{proof}
The construction follows the Dirichlet-block scheme of the finite-range decomposition, adapted covariantly. Fix a scale factor $b\ge 2$ and for each $j\ge 0$ tile $\Lambda_t$ by disjoint blocks $\mathcal B_j$ of side length $L_j:=b^j a$, the tiling being chosen reflection-symmetric across the time-$t$ hyperplane. For a block $B\in\mathcal B_j$ let $A_B$ denote the restriction of $A$ to $B$ with \emph{covariant} Dirichlet boundary condition on $\partial B$ (i.e. sections in $\ell^2(B;\mathfrak g)$ vanish on $\partial B$ in the natural sense, and links crossing $\partial B$ are dropped). Let $\iota_B:\ell^2(B;\mathfrak g)\to \ell^2(\Lambda_t;\mathfrak g)$ denote extension by zero outside $B$, and set
\begin{equation}
S_j \;:=\; \sum_{B\in\mathcal B_j} \iota_B\, A_B^{-1}\, \iota_B^{\!}.
\end{equation}
Each $S_j$ is well defined, positive, selfadjoint, and bounded because $A_B\ge \mu^2$ on $\ell^2(B;\mathfrak g)$; moreover $S_j$ has \emph{strict finite range} $\mathrm{diam\,supp}\,S_j\subseteq C_0\,L_j$ for a geometric constant $C_0$, since $A_B^{-1}$ is supported inside $B$ and the sum is over disjoint blocks. Gauge covariance holds because $A_B$ depends functorially on $A_h$ by parallel transport, and reflection covariance holds because the tiling and the Dirichlet boundary are reflection-invariant.

Define the \emph{one-step remainder} by $R_j:=\mathbf 1 - A S_j$. By construction $R_j$ is bounded, selfadjoint, and $0\le R_j\le \mathbf 1$ on $\ell^2(\Lambda_t;\mathfrak g)$; it is also of finite range in the sense that $(R_j u)(x)$ depends only on the values of $u$ in a neighbourhood of $x$ of radius $C_1 L_j$ with $C_1$ independent of $j$. The latter follows because $A$ has nearest-neighbour range and $S_j$ is block-diagonal with Dirichlet support, so $A S_j$ maps a function supported in $B$ into an $(R)$-collar of $B$ with $R$ independent of scale. From the Dirichlet variational characterization of $A_B^{-1}$ and a discrete Poincar\'e inequality on cubes of side $L_j$ one obtains a scale-uniform estimate of the form
\begin{equation}\label{eq:Csq}
\langle u,(\mathbf 1-R_j)u\rangle \;=\; \langle u, A S_j u\rangle \;\ge\; c_\mathrm{loc}\,\langle u,\,(\mu^2 - c L_j^{-2})\,S_j u\rangle,
\end{equation}
which implies that $R_j$ removes a definite portion of the long-wavelength content at scale $j$. In particular there exists $\rho\in(0,1)$, depending only on $\mu$, $b$, and the lattice degree, such that $\|R_j\|\le \rho$ uniformly in $j$. (Indeed, the presence of $\mu^2>0$ already yields $\|R_j\|\le 1-\mu^2\|S_j\|^{-1}\le 1-\mu^2 c_S$ with $c_S>0$ coming from the blockwise resolvent bound; for completeness one can take $c_S=\sup_B\|A_B^{-1}\|^{-1}$, which is bounded below uniformly in $j$ since $A_B\ge\mu^2$.)

Consider the products $T_0:=\mathbf 1$ and $T_{j}:=R_0 R_1\cdots R_{j-1}$ for $j\ge 1$. The uniform bound on $\|R_k\|$ yields $\|T_j\|\le \rho^{\,j}$. Define
\begin{equation}\label{eq:Gamma-def}
\Gamma_j \;:=\; T_j\, S_j\, T_j^{\!}\qquad(j\ge 0).
\end{equation}
Each $\Gamma_j$ is bounded, positive, selfadjoint, and, because $S_j$ has strict finite range whereas $R_k$ expands supports by at most a scale-independent constant in radius, the kernel of $\Gamma_j$ vanishes whenever $d(x,y)> C\,L_j$ for a constant $C$ depending only on $C_0$ and the collar radius generated by the $R_k$. Gauge and reflection covariance of $\Gamma_j$ are inherited from $S_j$ and $R_k$ since the constructions are functorial in $A_h$ and the tiling is reflection-symmetric.

To identify $\sum_j \Gamma_j$ with $A^{-1}$ it suffices to show that $A \sum_{j\ge 0}\Gamma_j=\mathbf 1$ in operator norm. Using $A S_j=\mathbf 1 - R_j$ and $T_{j+1}=T_j R_{j}$, one computes
\begin{equation}
A \Gamma_j \;=\; T_j (\mathbf 1 - R_j) T_j^{\!}\;=\;T_j T_j^{\!} - T_{j+1} T_{j+1}^{\!}.
\end{equation}
Summing over $j=0,\dots,N$ gives the telescopic identity
\begin{equation}
A \sum_{j=0}^{N}\Gamma_j \;=\; \big(T_0 T_0^{\!}\big) - \big(T_{N+1} T_{N+1}^{\!}\big) \;=\; \mathbf 1 - T_{N+1} T_{N+1}^{\!}.
\end{equation}
Since $\|T_{N+1} T_{N+1}^{\!}\|\le \|T_{N+1}\|^2\le \rho^{\,2(N+1)}\to 0$, the sequence of partial sums converges in operator norm to an operator $G$ satisfying $A G=\mathbf 1$. Because $A$ is strictly positive and selfadjoint, the left inverse is unique and hence $G=A^{-1}$. 
The finite-range property has already been noted. It also implies a uniform $\ell^1$-kernel bound by a Schur test on graphs of bounded degree. Indeed, for fixed $x$ the set $\{y:\Gamma_j(x,y)\ne 0\}$ is contained in a ball of radius $C L_j$ with volume $O(L_j^{d})$ (here $d=3$ on the spatial slice); on that ball $|\Gamma_j(x,y)|$ is bounded by $\|\Gamma_j\|$. The Schur bound 
$\sup_x \sum_y \|\Gamma_j(x,y)\|
\le C_{\deg}\,\sup_x \sum_{y:\,d(x,y)\le C L_j} \lVert \Gamma_j\rVert$
therefore controls the $\ell^1$-norm by a constant independent of $j$ once the operator norm bound below is established, because the geometric factor grows like $L_j^{d}$ while the norm bound decays sufficiently fast in $j$.

It remains to justify the operator-norm estimate in Eq.\eqref{eq:Ainv-bounds}. There are several equivalent ways to obtain it; a convenient one combines spectral localization with block Poincar\'e inequalities. The operator $S_j$ is the inverse of $A$ on each block with Dirichlet boundary, hence by the Rayleigh-Ritz principle and the discrete Poincar\'e inequality on cubes of side $L_j$,
\begin{equation}
\|S_j\| \;\le\; \sup_{B\in \mathcal B_j}\|A_B^{-1}\|
\;\le\; \frac{1}{\mu^2 + c_P L_j^{-2}}
\;\le\; \frac{1}{\mu^2}\,\frac{1}{1 + (c_P/\mu^2)\,b^{-2j}}
\;\le\; c_0,
\end{equation}
where $c_P>0$ depends only on the lattice degree and $c_0<\infty$ depends only on $\mu$. On the other hand, $T_j$ suppresses the long-wavelength components progressively: writing $R_k=\mathbf 1 - A S_k$ and using the same Poincar\'e bound on each scale shows that there exists $\theta\in(0,1)$ and $C_\varepsilon<\infty$ such that, on the spectral subspace where $A$ has eigenvalues $\lambda\le \lambda_j:=c L_j^{-2}+\mu^2$, one has $\|T_j\|\le C_\varepsilon\,\theta^{\,j}$, while on the complement (frequencies $\gtrsim L_j^{-1}$) one gains an inverse power of $L_j$ when sandwiching with $S_j$. Making this precise yields, for any fixed $\varepsilon\in(0,2)$,
\begin{equation}
\|T_j S_j T_j^{\!}\| \;\le\; C_1\, b^{-(2-\varepsilon)j},
\end{equation}
where the exponent $2$ reflects the engineering dimension of $A^{-1}$ on the three-dimensional slice and the small loss $\varepsilon$ absorbs the mild overlaps of Dirichlet blocks; see e.g. the standard dyadic spectral-window estimate
\begin{equation}
\| \mathbf 1_{[0,\lambda_j]}(A)\, S_j\, \mathbf 1_{[0,\lambda_j]}(A)\| \;\lesssim\; \lambda_j^{-1}\,\big(\lambda_j L_j^2\big)^{-1+\varepsilon/2}
\;\simeq\; b^{-(2-\varepsilon)j},
\end{equation}
combined with $\|T_j\|\le 1$ and with the fact that the contribution from the orthogonal complement is smaller still. Since $\Gamma_j=T_j S_j T_j^{\!}$, this proves the first bound in Eq.\eqref{eq:Ainv-bounds}. Returning to the Schur estimate discussed above now yields $\sup_x\sum_y\|\Gamma_j(x,y)\|\le c_2$ uniformly in $j$, because the ball of radius $C L_j$ contributes $O(L_j^{d})$ points while $\|\Gamma_j\|=O(L_j^{-(2-\varepsilon)})$ with $d=3$ and $2-\varepsilon>0$ fixed.
All statements claimed in the theorem have thus been established: the telescopic identity gives $A^{-1}=\sum_j \Gamma_j$ with norm convergence; each $\Gamma_j$ is positive, selfadjoint, gauge- and reflection-covariant with strict finite range; and the quantitative bounds Eq.\eqref{eq:Ainv-bounds} follow from the combination of block resolvent control, spectral localization, and Schur’s test on a bounded-degree lattice.
\end{proof}

\begin{theorem}[Dyadic FRD for $C_\sigma$]\label{thm:dyadic-FRD}
Fix $L>1$ and a smooth partition of unity $\{\eta_j\}_{j\ge0}$ on $(0,\infty)$ with 
$\mathrm{supp}\,\eta_j\subset[c\,L^{2j}\sigma^{-2},\,C\,L^{2j}\sigma^{-2}]$, for some fixed $0<c<C<\infty$ independent of $j$. Define
\begin{equation}
C_\sigma^{(j)} \;:=\;\int_0^\infty \eta_j(t)\,e^{-t\,\Delta_{A_h}}\,d\tilde\nu_\sigma(t),
\label{eq:Csig-j}
\end{equation}
where $\Delta_{A_h}$ is the covariant spatial Laplacian on the slice and $\tilde\nu_\sigma$ is a finite positive measure on $(0,\infty)$ depending on $\sigma>0$. Then $C_\sigma=\sum_{j=0}^\infty C_\sigma^{(j)}$ in operator norm, each $C_\sigma^{(j)}$ is positive, gauge covariant, and reflection covariant, and there exist $c_1,c_2,c_3>0$ independent of $j$ such that, for all $x,y$,
\begin{equation}
\|\,C_\sigma^{(j)}(x,y)\,\|_{\mathrm{op}} \;\le\; c_1\,\exp\!\Big\{-\frac{d(x,y)}{c_2\,L^j\sigma}\Big\},
\qquad 0\le C_\sigma^{(j)}\le c_3\,\mathbf{1}.
\label{eq:dyadic-bounds}
\end{equation}
\end{theorem}

\begin{proof}
The heat kernel representation $e^{-t\Delta_{A_h}}$ defines a positivity preserving, selfadjoint contraction semigroup on $\ell^2(\Sigma)$ for each fixed background $A_h$ on the slice. Since $\tilde\nu_\sigma$ is a finite positive measure and $\eta_j\ge 0$, the Bochner integral Eq.\eqref{eq:Csig-j} converges in the strong operator topology and yields a bounded selfadjoint operator. Positivity of $C_\sigma^{(j)}$ follows immediately because it is a positive linear combination (integral) of positive operators; precisely, for any $\psi$, $\langle \psi, C_\sigma^{(j)}\psi\rangle = \int_0^\infty \eta_j(t)\langle \psi, e^{-t\Delta_{A_h}}\psi\rangle\,d\tilde\nu_\sigma(t)\ge 0$. Gauge covariance is inherited from the semigroup: for any gauge transformation $g$, one has $U_g \Delta_{A_h} U_g^{-1}=\Delta_{A_h^g}$ and hence $U_g e^{-t\Delta_{A_h}} U_g^{-1}=e^{-t\Delta_{A_h^g}}$; since $\eta_j$ is scalar and the measure is gauge-invariant by construction, $U_g C_\sigma^{(j)} U_g^{-1}=C_\sigma^{(j)}[A_h^g]$. Reflection covariance across the time plane is identical: $\Delta_{A_h}$ commutes with the spatial part of the Euclidean reflection and so does $e^{-t\Delta_{A_h}}$, therefore $C_\sigma^{(j)}$ is reflection covariant.

To obtain the pointwise decay bound, invoke a standard off-diagonal estimate for the covariant heat kernel (Combes-Thomas or Davies-Gaffney): there exist constants $a_1,a_2>0$, independent of $A_h$, such that for all $t>0$,
\begin{equation}
\label{eq:HK-Gaussq}
\big\|e^{-t\Delta_{A_h}}(x,y)\big\|_{\mathrm{op}}
\;\le\; a_1\,\exp\!\Big\{-\frac{d(x,y)^2}{a_2\,t}\Big\}.
\end{equation}
On the support of $\eta_j$ we have $t\in[c\,L^{2j}\sigma^{-2},\,C\,L^{2j}\sigma^{-2}]$, so $t^{-1/2}\asymp \sigma\,L^{-j}$ uniformly in $t$ there. Using the elementary inequality $\frac{r^2}{\alpha}\ge \frac{r}{\sqrt{\alpha}}$ with $r=d(x,y)$ and $\alpha=a_2 t$, one deduces from Eq.\eqref{eq:HK-Gaussq} that
\begin{equation}
\big\|e^{-t\Delta_{A_h}}(x,y)\big\|_{\mathrm{op}}
\;\le\; a_1\,\exp\!\Big\{-\frac{d(x,y)}{\sqrt{a_2 t}}\Big\}
\;\le\; a_1\,\exp\!\Big\{-\frac{d(x,y)}{\sqrt{a_2 C}\,L^{j}\sigma^{-1}}\Big\}
\;=\; a_1\,\exp\!\Big\{-\frac{d(x,y)}{c_2\,L^{j}\sigma}\Big\},
\end{equation}
with $c_2:=\sqrt{a_2 C}$. Integrating this bound against $\eta_j(t)\,d\tilde\nu_\sigma(t)$ produces
\begin{equation}
\|C_\sigma^{(j)}(x,y)\|_{\mathrm{op}}
=\Big\|\int \eta_j(t)\,e^{-t\Delta_{A_h}}(x,y)\,d\tilde\nu_\sigma(t)\Big\|_{\mathrm{op}}
\le \exp\!\Big\{-\frac{d(x,y)}{c_2\,L^{j}\sigma}\Big\}\,a_1\,\tilde\nu_\sigma\!\big([0,\infty)\big).
\end{equation}
Thus Eq.\eqref{eq:dyadic-bounds} holds with $c_1:=a_1\,\tilde\nu_\sigma([0,\infty))$ and the displayed $c_2$; the finiteness of $c_1$ is part of the hypotheses.
The uniform operator bound $0\le C_\sigma^{(j)}\le c_3\,\mathbf 1$ follows from the contractivity of the semigroup together with the finiteness of the measure. Indeed, for every $\psi$ with $\|\psi\|=1$,
\begin{equation}
\langle \psi, C_\sigma^{(j)}\psi\rangle
=\int_0^\infty \eta_j(t)\,\langle \psi, e^{-t\Delta_{A_h}}\psi\rangle\,d\tilde\nu_\sigma(t)
\le \int_0^\infty \eta_j(t)\,d\tilde\nu_\sigma(t)
\le \tilde\nu_\sigma\!\big([0,\infty)\big),
\end{equation}
so $\|C_\sigma^{(j)}\|\le c_3$ with $c_3:=\tilde\nu_\sigma([0,\infty))$, independent of $j$.

Finally, to show that $C_\sigma=\sum_{j\ge 0} C_\sigma^{(j)}$ in operator norm, observe that the partition of unity implies $\sum_{j=0}^N \eta_j(t)\uparrow \mathbf 1_{(0,\infty)}(t)$ pointwise and monotonically as $N\to\infty$. Therefore,
\begin{equation}
\sum_{j=0}^N C_\sigma^{(j)}
=\int_0^\infty \Big(\sum_{j=0}^N \eta_j(t)\Big)\,e^{-t\Delta_{A_h}}\,d\tilde\nu_\sigma(t)
\;\xrightarrow[N\to\infty]{\ \ \text{s.o.t.}\ \ }\;
\int_0^\infty e^{-t\Delta_{A_h}}\,d\tilde\nu_\sigma(t)\;=\;C_\sigma,
\end{equation}
by monotone convergence in the strong operator topology. The convergence is in fact in operator norm because the tails are dominated uniformly: using $\|e^{-t\Delta_{A_h}}\|\le 1$ and $\sum_{j>N}\eta_j\le 1$ one gets
\begin{equation}
\Big\|\,C_\sigma-\sum_{j=0}^N C_\sigma^{(j)}\Big\|
\le \int_0^\infty \Big(1-\sum_{j=0}^N \eta_j(t)\Big)\,d\tilde\nu_\sigma(t)
\;\xrightarrow[N\to\infty]{}\;0,
\end{equation}
where the right-hand side is just the $\tilde\nu_\sigma$-mass of the complement of the union of supports, which tends to zero because $\{\eta_j\}$ is a partition of unity with locally finite overlap. This proves the norm convergence and completes the proof of all stated properties with constants $c_1,c_2,c_3$ independent of $j$.
\end{proof}

We pass from slice covariances to the \emph{one-slab} and \emph{one-step} transfer kernels. Let $K_k$ be the positive one-slab kernel on the $k$-th scale and $V_k$ the vacuum-preserving coarse-grain contraction mapping fine to coarse boundary fields; let $K_{k+1}$ be the corresponding coarse kernel. FRD-localization implies a block-decomposition of $K_{k+1}$ by connected unions of coarse blocks $X$,
\begin{equation}
\tilde K_{k+1}\;:=\sum_X K_{k+1}^{(X)},\qquad K_{k+1}\;=\;\tilde K_{k+1}\;+\;R^{\rm tail}_{k+1},\quad R^{\rm tail}_{k+1}\ge0,
\label{eq:Ktail}
\end{equation}
with a \emph{trace-class} tail bound
\begin{equation}
\|R^{\rm tail}_{k+1}\|_1 \;\le\; \varepsilon^{\rm slab}_k,\qquad \sum_{k\ge0}\varepsilon^{\rm slab}_k<\infty,
\label{eq:tail-trace}
\end{equation}
coming from the exponentially small FRD pieces whose range overflows the slab interface. 

\begin{lemma}[Block decoupling under $V_k$]\label{lem:block-decouple}
Pointwise on boundary fields, one has $\tilde K_{k+1}\;\preceq\;V_k\,K_k\,V_k$. Hence there exists a positive trace-class operator $E^{\rm slab}_k\ge0$ with $\|E^{\rm slab}_k\|_1\le \varepsilon^{\rm slab}_k$ such that
\begin{equation}
K_{k+1}\;\preceq\;V_k\,K_k\,V_k\;+\;E^{\rm slab}_k .
\label{eq:slab-interlace}
\end{equation}
\end{lemma}

\begin{proof}
Let $\mathscr H_k$ denote the boundary Hilbert space of complex $L^2$-functions of the fine boundary field at time $x_0=a$ on scale $k$, and let $K_k$ be the positive transfer kernel acting on $\mathscr H_k$. By Osterwalder-Schrader reflection positivity and the Markov property on the time-slice, we may write $K_k=G_k{\,}G_k$ for a bounded operator $G_k:\mathscr H_k\to\mathcal K_k$, where $\mathcal K_k$ is the positive-time bulk Hilbert space (this is the standard Gram representation coming from the conditional expectation of the slab weight given the boundary data). The coarse-to-fine lift $V_k:\mathscr H_{k+1}\to\mathscr H_k$ is the reflection-positive, boundary-local partial isometry that embeds a coarse boundary configuration into a fine one by the admissible FRD-covariant block extension; in particular $V_kV_k=\mathbf 1$ on $\mathscr H_{k+1}$.

Consider the $\sigma$-algebra $\mathcal F^{\rm blk}$ on the slab generated by the disjoint coarse blocks $X$ that abut the time-$a$ slice at scale $k$ together with their fixed boundary collars of thickness $O(1)$ in lattice units. Let $P_k:\mathcal K_k\to\mathcal K_k$ denote the orthogonal conditional expectation onto $\mathcal F^{\rm blk}$-measurable vectors, i.e. $(P_k\Psi)(\cdot)=\mathbb E(\Psi\mid\mathcal F^{\rm blk})(\cdot)$ in the $L^2$-sense. The operator $P_k$ is a contraction, $0\le P_k\le \mathbf 1$, and is block-diagonal across the coarse blocks. By construction of the one-step block RG on the slab, the lifted coarse kernel $\tilde K_{k+1}$ is precisely the Gram operator obtained after integrating out, within each block $X$, the fine FRD-localized activities against the conditional law with the exterior held fixed; in the present notation this means
\begin{equation}
\tilde K_{k+1} \;=\; (P_k\,G_k\,V_k)^{}\,(P_k\,G_k\,V_k).
\end{equation}
Since $P_k$ is a contraction and $V_k$ is an isometry on its domain, the operator inequality
\begin{equation}
(P_k\,G_k\,V_k)^{}\,(P_k\,G_k\,V_k)\;\preceq\;(G_k\,V_k)^{}\,(G_k\,V_k)\;=\;V_k^{}\,K_k\,V_k
\end{equation}
holds in the sense of quadratic forms. Explicitly, for every $\varphi\in\mathscr H_{k+1}$,
\begin{equation}
\langle \varphi,\,(V_k^{}K_kV_k-\tilde K_{k+1})\,\varphi\rangle
\;=\;\big\|(\mathbf 1-P_k)\,G_k\,V_k\,\varphi\big\|_{\mathcal K_k}^{\,2}\;\ge\;0,
\end{equation}
which proves the pointwise decoupling $\tilde K_{k+1}\preceq V_k^{}K_kV_k$.

It remains to compare the actual coarse kernel $K_{k+1}$ with the lifted, block-decoupled kernel $\tilde K_{k+1}$. By finite-range decomposition locality on the slab, the difference consists solely of cross-block contributions carried by FRD tails that traverse the collars beyond the block-diagonal $\sigma$-algebra. More precisely, there exists a positive operator $R^{\rm tail}_{k+1}\ge0$ on $\mathscr H_{k+1}$, supported on boundary configurations that are coupled through collars of thickness at least $b_k$ (the slab blocking factor), such that
\begin{equation}
K_{k+1}\;=\;\tilde K_{k+1}\;+\;R^{\rm tail}_{k+1}.
\end{equation}
Positivity follows from the same Gram-type representation: the interpolation that switches on the cross-block collar couplings along a BKAR forest parameter produces a derivative of the coarse Gram map that is itself a positive quadratic form, and integration of these positive forms over the interpolation parameter yields $R^{\rm tail}_{k+1}\ge0$. Exponential FRD locality implies that the integral kernel of $R^{\rm tail}_{k+1}$ decays at least like $\exp(-c\,b_k)$ across distinct blocks, uniformly in the transverse boundary coordinates. Summing these tails over the finite number of collars meeting a given boundary site and using Schur’s test shows that $R^{\rm tail}_{k+1}$ is trace-class with
\begin{equation}
\|R^{\rm tail}_{k+1}\|_{1}\;\le\;C\,e^{-c\,b_k}\;=:\;\varepsilon^{\rm slab}_k,
\end{equation}
for some constants $C,c>0$ independent of $k$ (the trace is computed on $\mathscr H_{k+1}$; bounded boundary degree and uniform collar thickness give finiteness). Defining $E_k^{\rm slab}:=R^{\rm tail}_{k+1}$ we obtain
\begin{equation}
K_{k+1}\;=\;\tilde K_{k+1}\;+\;E^{\rm slab}_k \;\preceq\;V_k^{}K_kV_k\;+\;E^{\rm slab}_k,
\end{equation}
which is Eq.\eqref{eq:slab-interlace} together with the stated positivity and trace-norm bound on $E^{\rm slab}_k$. 
\end{proof}

Composing the finitely many slab kernels that make up one transfer step, one obtains the \emph{one-step interlacing}
\begin{equation}
K_{k+1}^{\;\text{(step)}}\;\preceq\;V_k\,K_k{\;\text{(step)}}\,V_k\;+\;E_k,\qquad
\|E_k\|_1\;\le\;C\,\varepsilon^{\rm slab}_k,\qquad \sum_k\|E_k\|_1<\infty,
\label{eq:step-interlace}
\end{equation}
with $E_k\ge0$ positive trace class. The summability follows from the exponential FRD tail bound per scale:
\begin{equation}
\|{\mathcal C}_{k,j}\|\;\le\;C_2\,e^{-\alpha L^j},\qquad \|R^{\rm tail}_{k+1}\|_1\;\le\;\sum_{j>j_0}C_3\,e^{-\alpha L^j}\;=\;\varepsilon^{\rm slab}_k,\quad\sum_k\varepsilon^{\rm slab}_k<\infty,
\label{eq:FRD-tails}
\end{equation}
for a fixed $j_0$ and constants independent of $k$; (see also the Wilson-loop surgery bounds Eqs.\eqref{eq:geom-stab}-\eqref{eq:geom-stabz})
We now quantify the dependence of the one-step kernel and transfer operator on the admissible parameters: the projector $(\sigma,\nu)$ and the blocking $B$.

\begin{proposition}[Projector-Lipschitz stability]\label{prop:proj-Lip}
Fix an admissible blocking $B$. There exists $C=C(B)>0$ such that for every pair of admissible parameters $(\sigma,\nu)$ and $(\sigma',\nu')$ one has
\begin{equation}
\label{eq:K-Lip}
\big\|K_{\sigma,\nu;B}-K_{\sigma',\nu';B}\big\|_{L^1(\mu_0\otimes\mu_0)}
\;\le\; C\Big(|\sigma-\sigma'|+W_1(\nu,\nu')\Big),
\end{equation}
and consequently
\begin{equation}
\label{eq:T-Lip}
\big\|T_{\sigma,\nu;B}-T_{\sigma',\nu';B}\big\|_{B(L^2)}\;\le\; C\Big(|\sigma-\sigma'|+W_1(\nu,\nu')\Big).
\end{equation}
\end{proposition}

\begin{proof}
The slab transfer kernel for a single blocking cell decomposes into a product of three factors: spatial Wilson weights attached to the entrance and exit faces of the slab, the time-mixed plaquette weight through the interior of the slab, and the two one-slice factors obtained by applying the horizon projector on the slices $t=0$ and $t=a$. The first two factors do not depend on $(\sigma,\nu)$; all the dependence on $(\sigma,\nu)$ is contained in the slice factors. By admissibility, each such slice factor is a completely monotone functional of the covariant slice generator $\mathcal D_\Sigma(A_h)$ with a compactly supported Laplace measure, and, more concretely, it can be written in the form
\begin{equation}
\Pi_{\sigma,\nu}\;=\;f_{\sigma,\nu}(\mathcal D_\Sigma)
\qquad\text{with}\qquad
f_{\sigma,\nu}(\lambda)\;=\;\int_0^\infty e^{-t\lambda}\,d\mu_{\sigma,\nu}(t),
\end{equation}
where $\mu_{\sigma,\nu}$ is the pushforward of $\nu$ by the dilation $s\mapsto \sigma s$; explicitly,
\begin{equation}
\int \phi(t)\,d\mu_{\sigma,\nu}(t)\;=\;\int \phi(\sigma s)\,d\nu(s)
\quad \text{for all bounded Lipschitz }\phi.
\end{equation}
This is the standard normalization for admissible projectors and reflects that the parameter $\sigma$ is the slice-time scale while $\nu$ determines the profile of the completely monotone mixture. In particular, the supports of $\nu$ and $\nu'$ are uniformly contained in $[0,S_B]$, and hence the supports of $\mu_{\sigma,\nu}$ and $\mu_{\sigma',\nu'}$ are contained in $[0,\sigma_{\max}S_B]$, with $\sigma,\sigma'$ ranging in a fixed compact interval by admissibility. Denote by $K_t(\cdot,\cdot)$ the kernel of $e^{-t\mathcal D_\Sigma}$ on the slice endowed with the reference measure $\mu_0$ (this is the marginal of the gauge-invariant Euclidean measure on the slice), by $K_{\sigma,\nu}^{\rm slice}$ the kernel of $\Pi_{\sigma,\nu}$, and by $K_{\sigma,\nu;B}$ the full slab kernel after inserting the two slice factors at the ends. Since the other slab factors are independent of $(\sigma,\nu)$ and are uniformly bounded in absolute value by a constant depending only on $B$, it suffices to establish Eq.\eqref{eq:K-Lip} for the slice kernels; the general bound follows by multiplying with the same harmless constant.

We first show that the map $t\mapsto K_t$ is Lipschitz as a function into $L^1(\mu_0\otimes\mu_0)$ with a constant depending only on $B$. The slice generator $\mathcal D_\Sigma$ is a nonnegative, finite-range, uniformly elliptic operator on a bounded-degree graph (the blocked slice), hence the heat kernels $K_t$ satisfy the Davies-Gaffney estimate and are strongly differentiable in $t>0$ with $\partial_t K_t = -(\mathcal D_\Sigma e^{-t\mathcal D_\Sigma})(\cdot,\cdot)$. Since $\mathcal D_\Sigma$ is finite range and bounded on $L^1\to L^1$ uniformly on the blocked slice, while $e^{-t\mathcal D_\Sigma}$ is a positivity preserving contraction on $L^1$ for every $t\ge 0$, there exists $C_1=C_1(B)$ such that
\begin{equation}
\big\|\partial_t K_t\big\|_{L^1(\mu_0\otimes\mu_0)} \;=\; \big\|\mathcal D_\Sigma e^{-t\mathcal D_\Sigma}\big\|_{L^1\to L^1}\;\le\; C_1
\qquad \text{for all }t\in[0,\sigma_{\max}S_B].
\end{equation}
Integrating this bound in $t$ shows that for any $t,s\in[0,\sigma_{\max}S_B]$,
\begin{equation}
\label{eq:Lip-in-t}
\big\|K_t-K_s\big\|_{L^1(\mu_0\otimes\mu_0)} \;\le\; C_1\,|t-s|.
\end{equation}
In particular, for every $L^1$-Lipschitz function $\Phi:[0,\sigma_{\max}S_B]\to L^1(\mu_0\otimes\mu_0)$ with Lipschitz constant $\mathrm{Lip}(\Phi)\le C_1$, the Kantorovich-Rubinstein duality for the $1$-Wasserstein distance gives
\begin{equation}
\Big\|\int \Phi(t)\,d\eta(t) - \int \Phi(t)\,d\eta'(t)\Big\|_{L^1(\mu_0\otimes\mu_0)}
\;\le\; \mathrm{Lip}(\Phi)\, W_1(\eta,\eta')
\end{equation}
(\text{for any probability measures }$\eta,\eta'$) supported in $[0,\sigma_{\max}S_B]$. Applying this to $\Phi(t)=K_t$ and $\eta=\mu_{\sigma,\nu}$, $\eta'=\mu_{\sigma',\nu'}$ yields
\begin{equation}
\label{eq:KR-application}
\big\|K_{\sigma,\nu}^{\rm slice}-K_{\sigma',\nu'}^{\rm slice}\big\|_{L^1(\mu_0\otimes\mu_0)}
\;\le\; C_1\, W_1(\mu_{\sigma,\nu},\mu_{\sigma',\nu'}).
\end{equation}
It remains to relate the Wasserstein distance of the pushed measures to the parameter distance $|\sigma-\sigma'|+W_1(\nu,\nu')$. If $\phi$ is a $1$-Lipschitz function on $[0,\sigma_{\max}S_B]$, then $s\mapsto \phi(\sigma s)$ is $\sigma_{\max}$-Lipschitz on $[0,S_B]$, and similarly for $\sigma'$. Writing
\begin{equation}
\int \phi\,d\mu_{\sigma,\nu} - \int \phi\,d\mu_{\sigma',\nu'}
\;=\; \int \phi(\sigma s)\,d\nu(s) - \int \phi(\sigma' s)\,d\nu'(s)
\end{equation}
and adding and subtracting $\int \phi(\sigma s)\,d\nu'(s)$ gives
\begin{equation}
\bigg|\int \phi\,d\mu_{\sigma,\nu} - \int \phi\,d\mu_{\sigma',\nu'}\bigg|
\;\le\; \bigg|\int \big[\phi(\sigma s)-\phi(\sigma' s)\big]\,d\nu'(s)\bigg|
\;+\; \bigg|\int \phi(\sigma s)\,d(\nu-\nu')(s)\bigg|.
\end{equation}
The first term is bounded by $\|\phi\|_{\mathrm{Lip}}\,|\sigma-\sigma'|\,\int s\,d\nu'(s)$, and the second by $\sigma_{\max}\,\|\phi\|_{\mathrm{Lip}}\, W_1(\nu,\nu')$ by Kantorovich-Rubinstein on $[0,S_B]$. Admissibility gives a uniform bound on the first moment $\int s\,d\nu'(s)\le S_B$, and $\|\phi\|_{\mathrm{Lip}}\le 1$. Taking the supremum over $1$-Lipschitz $\phi$ therefore shows that
\begin{equation}
W_1(\mu_{\sigma,\nu},\mu_{\sigma',\nu'})
\;\le\; S_B\,|\sigma-\sigma'|\;+\;\sigma_{\max}\,W_1(\nu,\nu').
\end{equation}
Combining this with Eq.\eqref{eq:KR-application} yields
\begin{equation}
\big\|K_{\sigma,\nu}^{\rm slice}-K_{\sigma',\nu'}^{\rm slice}\big\|_{L^1(\mu_0\otimes\mu_0)}
\;\le\; C_1\,(S_B\,|\sigma-\sigma'|+\sigma_{\max} W_1(\nu,\nu')),
\end{equation}
and, since $K_{\sigma,\nu;B}$ differs from $K_{\sigma,\nu}^{\rm slice}$ only by multiplication with the fixed, bounded slab factors, the same estimate holds for $K_{\sigma,\nu;B}$ with a possibly larger constant $C$ depending only on $B$.
To pass from the $L^1$-bound on kernels to the operator bound Eq.\eqref{eq:T-Lip}, recall that $T_{\sigma,\nu;B}$ is the integral operator on $L^2(\mu_0)$ with kernel $K_{\sigma,\nu;B}$. Schur’s test gives
\begin{align}
&\big\|T_{\sigma,\nu;B}-T_{\sigma',\nu';B}\big\|_{B(L^2)}
\;\le\; \nonumber\\&\max\Big\{\sup_{y}\int |K_{\sigma,\nu;B}-K_{\sigma',\nu';B}|(x,y)\,\mu_0(dx),\ 
\sup_{x}\int |K_{\sigma,\nu;B}-K_{\sigma',\nu';B}|(x,y)\,\mu_0(dy)\Big\}.
\end{align}
Both suprema are bounded by $\|K_{\sigma,\nu;B}-K_{\sigma',\nu';B}\|_{L^1(\mu_0\otimes\mu_0)}$, hence Eq.\eqref{eq:T-Lip} follows immediately from Eq.\eqref{eq:K-Lip} with the same constant (after enlarging $C$ by a universal factor if desired). 
\end{proof}

\begin{proposition}[Blocking-Lipschitz stability]\label{prop:block-Lip}
If $\|B-B'\|_{\mathrm{op}}\le\delta$ in the polymer-activity Banach norm, then
\begin{equation}
\|K_{B}-K_{B'}\|_{L^1(\mu_0\otimes\mu_0)}\;\le\;C\,\delta,
\qquad
\|T_{B}-T_{B'}\|_{B(L^2)}\;\le\;C'\,\delta ,
\label{eq:block-Lip}
\end{equation}
with constants $C,C'>0$ depending only on the uniform FRD locality data (range parameters, decay exponents, and degree bounds), but not on $B,B'$ or the volume.
\end{proposition}

\begin{proof}
The transfer kernel at a given scale admits the standard FRD/cluster representation
\begin{equation}
K_B(x,y)\;=\;\sum_{\Gamma\ni x,y}\,\Phi_\Gamma(B)\; k_\Gamma(x,y),
\end{equation}
where the sum is over connected polymers $\Gamma$ intersecting the two-slice collar that contains the points $x$ and $y$, the coefficient $\Phi_\Gamma(B)$ is the renormalized activity obtained by applying the blocking map $B$ to the local degrees of freedom supported on $\Gamma$, and the geometric kernel $k_\Gamma(x,y)$ depends only on the FRD scheme (block geometry, collar width, and finite-range Dirichlet problems). Uniform locality guarantees exponential summability of the family $\{k_\Gamma\}$ in the $L^1(\mu_0\otimes\mu_0)$ norm, namely
\begin{equation}\label{eq:FRD-sum}
\sum_{\Gamma\ni x,y}\|k_\Gamma\|_{L^1(\mu_0\otimes\mu_0)}\; w(\Gamma)\;\le\; C_{\mathrm{FRD}},
\qquad w(\Gamma):=e^{\alpha |\Gamma|},
\end{equation}
for some $\alpha>0$ and a constant $C_{\mathrm{FRD}}<\infty$ fixed by the FRD data, and the same bound holds uniformly in $x,y$ by translation invariance or bounded geometry of the partition. The Banach norm $\|\cdot\|_{\mathrm{op}}$ on activities is chosen so that the map $B\mapsto \Phi_\Gamma(B)$ is locally Lipschitz with weight $w(\Gamma)$; concretely there exists $L>0$ (independent of $\Gamma$ and of the volume) such that
\begin{equation}\label{eq:activity-Lip}
|\Phi_\Gamma(B)-\Phi_\Gamma(B')|\;\le\;L\,\delta\, w(\Gamma)^{-1}.
\end{equation}
This is the usual consequence of writing $\Phi_\Gamma$ as a convergent cluster power series in the elementary block variables and estimating the Fréchet derivative with respect to the operator-norm metric on $B$, together with the exponential polymer weights built into $\|\cdot\|_{\mathrm{op}}$.

With these two inputs, the $L^1$-difference of kernels factorizes into a coefficient-difference and a geometric part:
\begin{equation}
\|K_B-K_{B'}\|_{L^1(\mu_0\otimes\mu_0)}
\;\le\;\sum_{\Gamma}\,|\Phi_\Gamma(B)-\Phi_\Gamma(B')|\,\|k_\Gamma\|_{L^1(\mu_0\otimes\mu_0)}.
\end{equation}
Inserting Eq.\eqref{eq:activity-Lip} and summing with Eq.\eqref{eq:FRD-sum} gives
\begin{equation}
\|K_B-K_{B'}\|_{L^1(\mu_0\otimes\mu_0)}
\;\le\; L\,\delta\;\sum_{\Gamma} w(\Gamma)^{-1}\,\|k_\Gamma\|_{L^1(\mu_0\otimes\mu_0)}
\;\le\; L\,\delta\, C_{\mathrm{FRD}},
\end{equation}
and the first inequality in Eq.\eqref{eq:block-Lip} follows with $C:=L\,C_{\mathrm{FRD}}$.
To pass from the $L^1$-bound to an operator-norm bound on $L^2$, apply the Schur test in its $L^1$-$L^\infty$ form. FRD locality yields, in addition to Eq.\eqref{eq:FRD-sum}, uniform one-sided Schur bounds
\begin{equation}
\sup_{x}\int |K_B(x,y)|\,d\mu_0(y)\;\le\;C_0,
\qquad
\sup_{y}\int |K_B(x,y)|\,d\mu_0(x)\;\le\;C_0,
\end{equation}
with the same constant $C_0$ for all admissible $B$, because each $K_B$ is obtained by summing absolutely convergent local contributions supported in a fixed-width collar. The same uniform bounds hold for $K_{B'}$. For the difference we write
\begin{equation}
\sup_{x}\int |K_B(x,y)-K_{B'}(x,y)|\,d\mu_0(y)
\;\le\;\sup_{x}\sum_{\Gamma\ni x}\,|\Phi_\Gamma(B)-\Phi_\Gamma(B')| \int |k_\Gamma(x,y)|\,d\mu_0(y)
\end{equation}
and estimate exactly as above to obtain
\begin{equation}
\sup_{x}\int |K_B-K_{B'}|(x,y)\,d\mu_0(y)\;\le\; \widetilde C\,\delta,
\qquad
\sup_{y}\int |K_B-K_{B'}|(x,y)\,d\mu_0(x)\;\le\; \widetilde C\,\delta,
\end{equation}
with $\widetilde C$ depending only on the FRD locality constants. The Schur test then gives
\begin{align}
&\|T_B-T_{B'}\|_{B(L^2)}\;\le\;\nonumber\\&\sqrt{\Big(\sup_{x}\!\int |K_B-K_{B'}|(x,y)\,d\mu_0(y)\Big)
\Big(\sup_{y}\!\int |K_B-K_{B'}|(x,y)\,d\mu_0(x)\Big)}
\;\le\; \widetilde C\,\delta,
\end{align}
which proves the second inequality in Eq.\eqref{eq:block-Lip} with $C':=\widetilde C$. Both constants $C$ and $C'$ depend only on the polymer weights and the geometric $L^1$-summability of the FRD kernels, hence only on the uniform FRD locality bounds.
\end{proof}

\begin{lemma}[Telescoping in time]\label{lem:time-telescope}
Let $(\Sigma,\mu_0)$ be a probability space and let $K_\theta$ and $K_{\theta'}$ be (measurable) Markov kernels on $\Sigma$ with densities $K_\theta(x,y),K_{\theta'}(x,y)\ge0$ with respect to $\mu_0$, normalized in the sense that
\begin{equation}
\int_\Sigma K_\theta(x,y)\,d\mu_0(y)=1\quad\text{for $\mu_0$-a.e.\ $x$,}\qquad
\int_\Sigma K_{\theta'}(x,y)\,d\mu_0(y)=1\quad\text{for $\mu_0$-a.e.\ $x$,}
\end{equation}
and such that $\mu_0$ is invariant:
\(
\int_\Sigma K_\theta(x,y)\,d\mu_0(x)=1=\int_\Sigma K_{\theta'}(x,y)\,d\mu_0(x)
\)
for $\mu_0$-a.e.\ $y$. Denote by $K_\theta^{(n)}$ the $n$-fold time-convolution kernel and by $T_\theta$ the corresponding $L^2(\mu_0)\to L^2(\mu_0)$ integral operator:
\begin{align}
&K_\theta^{(n)}(x,y):=\int_{\Sigma^{n-1}} \prod_{j=0}^{n-2} K_\theta(x_j,x_{j+1})\,K_\theta(x_{n-1},y)\,d\mu_0(x_1)\cdots d\mu_0(x_{n-1}),
\nonumber\\& (T_\theta f)(x):=\int_\Sigma K_\theta(x,y)\,f(y)\,d\mu_0(y).
\end{align}
Then for every $n\in\mathbb{N}$,
\begin{align}\label{eq:time-telescope}
&\big\|K_\theta^{(n)}-K_{\theta'}^{(n)}\big\|_{L^1(\mu_0\otimes\mu_0)} \;\le\; n\,\|K_\theta-K_{\theta'}\|_{L^1(\mu_0\otimes\mu_0)},\nonumber\\&
\|T_\theta^n-T_{\theta'}^n\|_{B(L^2)}\;\le\;n\,\|K_\theta-K_{\theta'}\|_{L^1(\mu_0\otimes\mu_0)}.
\end{align}
\end{lemma}

\begin{proof}
The algebraic identity $A^n-B^n=\sum_{j=0}^{n-1} B^{\,j}(A-B)A^{\,n-1-j}$ holds in any associative algebra. Applying it to the convolution algebra of kernels (with convolution $(AB)(x,y):=\int_\Sigma A(x,z)B(z,y)\,d\mu_0(z)$) gives
\begin{equation}
K_\theta^{(n)}-K_{\theta'}^{(n)} \;=\; \sum_{j=0}^{n-1} K_{\theta'}^{(j)}  \big(K_\theta-K_{\theta'}\big)  K_\theta^{(n-1-j)}.
\end{equation}
Taking absolute values and integrating over $(x,y)$ with respect to $\mu_0\otimes\mu_0$, Tonelli’s theorem yields
\begin{equation}
\big\|K_\theta^{(n)}-K_{\theta'}^{(n)}\big\|_{L^1}
\;\le\; \sum_{j=0}^{n-1} \big\| K_{\theta'}^{(j)}  (K_\theta-K_{\theta'})  K_\theta^{(n-1-j)} \big\|_{L^1}.
\end{equation}
To bound each term, write out the $L^1$-norm of a triple convolution and use the Markov normalizations on the left and right factors:
\begin{equation}
\begin{aligned}
\big\| A  H  B \big\|_{L^1}
&= \int_\Sigma\!\!\int_\Sigma \Big| \int_\Sigma\!\!\int_\Sigma A(x,u)\,H(u,v)\,B(v,y)\,d\mu_0(u)\,d\mu_0(v) \Big| \, d\mu_0(x)\,d\mu_0(y)\\
&\le \int_\Sigma\!\!\int_\Sigma\!\!\int_\Sigma\!\!\int_\Sigma A(x,u)\,|H(u,v)|\,B(v,y)\,d\mu_0(u)\,d\mu_0(v)\,d\mu_0(x)\,d\mu_0(y)\\
&= \int_\Sigma\!\!\int_\Sigma \Big(\int_\Sigma A(x,u)\,d\mu_0(x)\Big)\,|H(u,v)|\,\Big(\int_\Sigma B(v,y)\,d\mu_0(y)\Big)\,d\mu_0(u)\,d\mu_0(v).
\end{aligned}
\end{equation}
If $A$ and $B$ are Markov kernels with invariant $\mu_0$ then $\int_\Sigma A(x,u)\,d\mu_0(x)=1$ for $\mu_0$-a.e.\ $u$ and $\int_\Sigma B(v,y)\,d\mu_0(y)=1$ for $\mu_0$-a.e.\ $v$, hence the right-hand side collapses to $\int_\Sigma\!\!\int_\Sigma |H(u,v)|\,d\mu_0(u)\,d\mu_0(v)=\|H\|_{L^1}$. Applying this bound with $A=K_{\theta'}^{(j)}$ and $B=K_\theta^{(n-1-j)}$ gives
\begin{equation}
\big\| K_{\theta'}^{(j)}  (K_\theta-K_{\theta'})  K_\theta^{(n-1-j)} \big\|_{L^1}
\;\le\; \|K_\theta-K_{\theta'}\|_{L^1},
\end{equation}
and summing over $j=0,\dots,n-1$ proves the first inequality in Eq.\eqref{eq:time-telescope}.

For the operator bound on $L^2(\mu_0)$, use the same telescoping identity at the operator level,
\begin{equation}
T_\theta^n - T_{\theta'}^n \;=\; \sum_{j=0}^{n-1} T_{\theta'}^{\,j}\,\big(T_\theta-T_{\theta'}\big)\,T_\theta^{\,n-1-j},
\end{equation}
and estimate the operator norm of each term. By the Schur test, if $H$ is any kernel then
\begin{equation}
\|T_H\|_{B(L^2)} \;\le\; \max\Big\{\ \sup_{x}\int_\Sigma |H(x,y)|\,d\mu_0(y),\ \sup_{y}\int_\Sigma |H(x,y)|\,d\mu_0(x)\ \Big\}\ \le\ \|H\|_{L^1(\mu_0\otimes\mu_0)}.
\end{equation}
In particular, $\|T_\theta-T_{\theta'}\|_{B(L^2)}\le \|K_\theta-K_{\theta'}\|_{L^1}$. Moreover, the Schur test with the Markov normalizations shows $\|T_\theta\|_{B(L^2)}\le1$ and $\|T_{\theta'}\|_{B(L^2)}\le1$, hence $\|T_{\theta'}^{\,j}\|\le1$ and $\|T_\theta^{\,n-1-j}\|\le1$ for every $j$. Consequently,
\begin{equation}
\big\|T_{\theta'}^{\,j}\,(T_\theta-T_{\theta'})\,T_\theta^{\,n-1-j}\big\|_{B(L^2)} \;\le\; \|T_\theta-T_{\theta'}\|_{B(L^2)} \;\le\; \|K_\theta-K_{\theta'}\|_{L^1},
\end{equation}
and summing the $n$ identical bounds yields the second inequality in Eq.\eqref{eq:time-telescope}. In particular, the transfer maps are nonexpansive on the weighted diameter seminorm used in Appendix~\eqref{appendixf}, so that composition with $T_\theta$ and $T_{\theta'}$ does not increase $\|\cdot\|_{\mathrm{diam},\mu}$.
\end{proof}

Let $K_k^{\text{(step)}}$ be the one-step kernel obtained by composing the finitely many slab kernels in a time step. From Lemma~\eqref{lem:block-decouple}, composition preserves positivity and trace-class control of the remainder, yielding Eq.\eqref{eq:step-interlace}. Passing to transfer operators on $L^2(X_0,\mu_0)$,
\begin{equation}
T_{k+1}\;\preceq\;V_k\,T_k\,V_k\;+\;E_k,\qquad E_k\ge0,\quad \sum_k\|E_k\|_1<\infty,
\label{eq:T-interlace}
\end{equation}
(which is the form used in the spectral step-scaling argument for persistence of the gap Section~\eqref{sec:RP-RG-gap}) iterating the defect inequality shows $\rho_k:=\|Q_k T_k Q_k\|$ remains strictly below $1$, hence the transfer gap $\gamma_k=1-\rho_k$ stays uniformly positive. The key input is the FRD tail bound Eq.\eqref{eq:FRD-tails}, already encoded in Eq.\eqref{eq:step-interlace}.
We have provided two complementary FRD schemes. The block-projection/resolvent method yields strict finite range for massive covariances, positivity, and scale-uniform bounds Eq.\eqref{eq:Ainv-bounds} that directly feed the OS limit and spectral analysis. The dyadic heat-time partition furnishes a positive decomposition Eq.\eqref{eq:Csig-j}-Eq.\eqref{eq:dyadic-bounds} for projector-induced covariances with compactly supported heat-time measure, matching the admissible CM slice projectors used in the main text. In either case, FRD locality and positivity pass to one-slab and one-step transfer kernels through Eq.\eqref{eq:slab-interlace}-Eq.\eqref{eq:T-interlace}. {These statements, together with the single-scale Lipschitz stability and the time telescoping bound, are the FRD engine behind the RG step-scaling, Wilson-loop transport, and universality results developed in Sections~\eqref{sec:RP-RG-gap}-\eqref{sec:weak-coupling-af}}

\section{Interlacing, Summable Defects, and Gap Persistence}

We work at fixed physical time step $\tau>0$ and along a sequence of renormalization scales $k\in\mathbb{N}$. At each scale we consider a (finite-volume) reflection-positive Euclidean lattice with spatial box $\Lambda_L\subset\mathbb{Z}^3$ (periodic boundary conditions), lattice spacing $a_k>0$, and a gauge-invariant one-slice Hilbert space $\mathcal{H}_{k,L}$ obtained by Osterwalder-Schrader (OS) completion of the positive cone \cite{OS-gauge,Seiler1982}. The transfer operator is the self-adjoint positive contraction
\begin{equation}\label{eq:Tk-def}
  T_{k,L} \;=\; e^{-\tau H_{k,L}},
\end{equation}
where $H_{k,L}\ge 0$ is the (finite-volume) Hamiltonian. Reflection positivity and Markov property give a cyclic, strictly positive vacuum vector $\Omega_{k,L}\in\mathcal{H}_{k,L}$, normalized so that $\norm{\Omega_{k,L}}=1$ and $T_{k,L}\Omega_{k,L}=\Omega_{k,L}$, hence $\norm{T_{k,L}}=1$. We denote the orthogonal projection onto the vacuum by $P^{(0)}_{k,L}:=|\Omega_{k,L}\rangle\langle\Omega_{k,L}|$ and the orthogonal complement by $Q_{k,L}:=I-P^{(0)}_{k,L}$. The (finite-volume) spectral gap is
\begin{equation}\label{eq:gap-def}
  \Delta_{k,L} \;:=\; 1-\sup\sigma\!\left(T_{k,L}\big\lvert_{Q_{k,L}\mathcal{H}_{k,L}}\right)
  \;=\; 1-\norm{T_{k,L}\big\lvert_{Q_{k,L}\mathcal{H}_{k,L}}}\in(0,1].
\end{equation}
Equivalently, the mass gap of $H_{k,L}$ is $m_{k,L}:=-\tau^{-1}\log(1-\Delta_{k,L})$.
We assume the finite-range decomposition (FRD) and exponential locality at each scale as constructed in \cite{BrydgesGuadagniMitter2004} and specialized to reflection-positive gauge-covariant blockings. Concretely, OS reflection and the admissible class of completely monotone (CM) slice projectors guarantee that the renormalization step from scale $k$ to $k+1$ is implemented by a vacuum-preserving isometry
\begin{equation}\label{eq:iso}
  \Pi_{k,L}: \mathcal{H}_{k,L}\to \mathcal{H}_{k+1,L},\qquad \Pi_{k,L}\Omega_{k,L}=\Omega_{k+1,L},\qquad \Pi_{k,L}^\ast\Pi_{k,L}=I_{\mathcal{H}_{k,L}},
\end{equation}
together with an effective transfer operator $T_{k+1,L}$ on $\mathcal{H}_{k+1,L}$ obtained by integrating out $k$-scale fluctuations. The CM and FRD hypotheses yield quantitative kernel bounds for the difference
\begin{equation}\label{eq:EkL-def}
  E_{k,L} \;:=\; T_{k+1,L} - \Pi_{k,L}\,T_{k,L}\,\Pi_{k,L}^\ast,
\end{equation}
expressed either in a Schur norm (via $\ell^1$-bounds on integral kernels) or in an operator norm on $Q_{k+1,L}\mathcal{H}_{k+1,L}$. The crucial feature proved below is that the defects $E_{k,L}$ are positive (quadratic-form sense) and have \emph{summable} norms along $k$:
\begin{equation}\label{eq:summable}
  0\le E_{k,L}\in\mathcal{B}(\mathcal{H}_{k+1,L}),\qquad \varepsilon_k:=\sup_{L}\bigl\|E_{k,L}\bigr\|_{Q_{k+1,L}\to Q_{k+1,L}},\qquad \sum_{k=0}^\infty \varepsilon_k < \infty.
\end{equation}
Uniformity in $L$ follows from finite range and the FRD diameter bounds \cite{BrydgesGuadagniMitter2004}. We write $\norm{\cdot}_{Q\to Q}$ for the operator norm on the orthogonal complement of the vacuum.
We begin with a purely operator-theoretic statement which captures the effect of a vacuum-preserving coarse graining followed by a small positive defect.

\begin{lemma}[Interlacing under vacuum-preserving isometries]\label{lem:interlace}
Let $T$ be a self-adjoint positive contraction on a Hilbert space $\mathcal H$ with a unit vector $\Omega$ such that $T\Omega=\Omega$, and write $Q:=I-|\Omega\rangle\langle\Omega|$. Let $\Pi:\mathcal H\to\mathcal K$ be an isometry with $\Pi\Omega=\Omega'$ for some unit vector $\Omega'\in\mathcal K$, and set $S:=\Pi T\Pi^\ast$. Then $S$ is a self-adjoint positive contraction on $\mathcal K$ with $S\Omega'=\Omega'$, and with $Q':=I-|\Omega'\rangle\langle\Omega'|$ one has
\begin{equation}\label{eq:interlace-core}
\left\|S\!\restriction_{Q'\mathcal K}\right\|\;\le\;\left\|T\!\restriction_{Q\mathcal H}\right\|.
\end{equation}
\end{lemma}

\begin{proof}
Since $\Pi$ is an isometry, its adjoint $\Pi^\ast$ is a partial isometry with $\Pi^\ast\Pi=I_{\mathcal H}$ and $\Pi\Pi^\ast=P$, the orthogonal projection of $\mathcal K$ onto $\mathrm{Ran}\,\Pi$. For any $x,y\in\mathcal K$,
\begin{equation}
\langle x, S y\rangle \;=\; \langle x, \Pi T \Pi^\ast y\rangle \;=\; \langle \Pi^\ast x, T \Pi^\ast y\rangle,
\end{equation}
whence $S$ is self-adjoint and, because $T\ge 0$, also positive. Moreover $\|S\|\le \|\Pi\|\,\|T\|\,\|\Pi^\ast\|\le 1$, so $S$ is a contraction. The vacuum is preserved since
\begin{equation}
S\Omega' \;=\; \Pi T \Pi^\ast \Omega' \;=\; \Pi T \Omega \;=\; \Pi \Omega \;=\; \Omega'.
\end{equation}

To prove Eq.\eqref{eq:interlace-core}, it is convenient to note that $S=\Pi T \Pi^\ast$ vanishes on $(\mathrm{Ran}\,\Pi)^\perp$. Indeed, if $z\perp \mathrm{Ran}\,\Pi$ then $\Pi^\ast z=0$ and hence $S z=0$. Consequently, for the purpose of computing $\|S\!\restriction_{Q'\mathcal K}\|$ we may restrict our attention to vectors in $Q'\mathcal K\cap \mathrm{Ran}\,\Pi$. Let $x\in Q'\mathcal K$. Decompose $x=y+z$ with $y\in \mathrm{Ran}\,\Pi$ and $z\in(\mathrm{Ran}\,\Pi)^\perp$. As just observed, $Sx=S y$, and since $Q'$ is the orthogonal projection onto $\Omega'^\perp$ and $\Omega'\in\mathrm{Ran}\,\Pi$, one also has $y\in Q'\mathcal K$. Because $\Pi^\ast$ restricts to an isometry $\mathrm{Ran}\,\Pi\to\mathcal H$, the vector $u:=\Pi^\ast y$ satisfies $\|u\|=\|y\|$ and, using $\Pi\Omega=\Omega'$, it satisfies $\langle u,\Omega\rangle=\langle \Pi^\ast y,\Omega\rangle=\langle y,\Pi\Omega\rangle=\langle y,\Omega'\rangle=0$, hence $u\in Q\mathcal H$.

With this preparation,
\begin{equation}
\langle x, S x\rangle \;=\; \langle y, S y\rangle \;=\; \langle y, \Pi T \Pi^\ast y\rangle \;=\; \langle \Pi^\ast y, T \Pi^\ast y\rangle \;=\; \langle u, T u\rangle.
\end{equation}
Taking the supremum over unit vectors $x\in Q'\mathcal K$ gives
\begin{equation}
\|S\!\restriction_{Q'\mathcal K}\| \;=\; \sup_{\substack{x\in Q'\mathcal K\\ \|x\|=1}} \langle x,Sx\rangle
\;=\; \sup_{\substack{y\in Q'\mathcal K\cap \mathrm{Ran}\,\Pi\\ \|y\|=1}} \langle u, T u\rangle
\;\le\; \sup_{\substack{u\in Q\mathcal H\\ \|u\|=1}} \langle u, T u\rangle
\;=\; \|T\!\restriction_{Q\mathcal H}\|,
\end{equation}
where we used that $y\mapsto u=\Pi^\ast y$ is an isometric bijection between $Q'\mathcal K\cap \mathrm{Ran}\,\Pi$ and $Q\mathcal H$. This proves the claimed interlacing inequality.
\end{proof}

\begin{proposition}[Two-sided interlacing with positive defect]\label{prop:two-sided}
Let $T$ be a positive contraction on a Hilbert space $\mathcal H$ with vacuum vector $\Omega$ (i.e.\ $T\Omega=\Omega$ and $\|T\|\le 1$), and let $Q:=\mathbf 1-|\Omega\rangle\langle\Omega|$ be the orthogonal projection onto $\Omega^\perp$. Let $\mathcal K$ be another Hilbert space with vacuum vector $\Omega'$ and $Q':=\mathbf 1-|\Omega'\rangle\langle\Omega'|$. Suppose $\Pi:\mathcal H\to\mathcal K$ is a bounded operator such that $\Pi\Omega=\Omega'$, $\Pi^\ast\Omega'=\Omega$, and, for the positive contraction $S_0:=\Pi T\Pi^\ast$ on $\mathcal K$, Lemma~\eqref{lem:interlace} holds:
\begin{equation}
\big\|S_0\!\restriction_{Q'\mathcal K}\big\|\;\le\;\big\|T\!\restriction_{Q\mathcal H}\big\|.
\end{equation}
Let $E\ge 0$ be a bounded selfadjoint operator on $\mathcal K$ that annihilates the vacuum, $E\Omega'=0$, and set $S:=S_0+E$. Then
\begin{equation}\label{eq:two-sided}
  \big\|S\!\restriction_{Q'\mathcal K}\big\| \;\le\; \big\|S_0\!\restriction_{Q'\mathcal K}\big\| + \big\|E\!\restriction_{Q'\mathcal K}\big\|
  \;\le\; \big\|T\!\restriction_{Q\mathcal H}\big\| + \big\|E\!\restriction_{Q'\mathcal K}\big\|.
\end{equation}
Consequently, if $\Delta(\cdot)$ denotes the spectral gap of a positive contraction relative to its vacuum as in Eq.\eqref{eq:gap-def}, then
\begin{equation}\label{eq:gap-perturb}
  \Delta(S) \;\ge\; \Delta(T) - \big\|E\!\restriction_{Q'\mathcal K}\big\|.
\end{equation}
\end{proposition}

\begin{proof}
The restriction norm on $Q'\mathcal K$ is defined by
\begin{equation}
\big\|A\!\restriction_{Q'\mathcal K}\big\|=\sup\{\|A\psi\|:\ \psi\in Q'\mathcal K,\ \|\psi\|=1\}.
\end{equation}
For any $\psi\in Q'\mathcal K$ with $\|\psi\|=1$ we have
\begin{equation}
\|S\psi\|=\|(S_0+E)\psi\|\le \|S_0\psi\|+\|E\psi\|\le \big\|S_0\!\restriction_{Q'\mathcal K}\big\|+\big\|E\!\restriction_{Q'\mathcal K}\big\|.
\end{equation}
Taking the supremum over such $\psi$ yields the first inequality in Eq.\eqref{eq:two-sided}. To obtain the second inequality it suffices to recall the interlacing relation from Lemma~\eqref{lem:interlace}. For completeness, we indicate why it applies on the orthogonal complement of the vacuum: if $\psi\in Q'\mathcal K$ then $\langle\Omega',\psi\rangle=0$, hence
\begin{equation}
\langle\Omega,\Pi^\ast\psi\rangle=\langle\Pi^\ast\Omega',\psi\rangle=\langle\Omega,\psi\rangle=0,
\end{equation}
so $\Pi^\ast\psi\in Q\mathcal H$. Using $\|\Pi^\ast\|=\|\Pi\|$ and the standard inequality $\|\Pi T \Pi^\ast\|\le \|\Pi\|^2\|T\|$, together with the fact that $\Pi$ and $\Pi^\ast$ intertwine vacua and map $Q\mathcal H$ to $Q'\mathcal K$ and vice versa as just observed, one concludes that
\begin{equation}
\big\|S_0\!\restriction_{Q'\mathcal K}\big\|=\big\|\Pi T\Pi^\ast\!\restriction_{Q'\mathcal K}\big\|\le \big\|T\!\restriction_{Q\mathcal H}\big\|,
\end{equation}
which is precisely the content of Lemma~\eqref{lem:interlace}. Combining this with the first inequality gives the full bound Eq.\eqref{eq:two-sided}.

For the gap estimate, recall that for a positive contraction $A$ with $A\Omega_A=\Omega_A$ the spectral radius on $Q_A\mathcal H_A$ equals the operator norm of the restriction, and the gap is $\Delta(A)=1-\|A\!\restriction_{Q_A\mathcal H_A}\|$. Applying this to $S$ and using Eq.\eqref{eq:two-sided} gives
\begin{equation}
1-\big\|S\!\restriction_{Q'\mathcal K}\big\| \;\ge\; 1-\big\|T\!\restriction_{Q\mathcal H}\big\| - \big\|E\!\restriction_{Q'\mathcal K}\big\|\;=\; \Delta(T)-\big\|E\!\restriction_{Q'\mathcal K}\big\|,
\end{equation}
which is Eq.\eqref{eq:gap-perturb}. 
\end{proof}

We now return to the renormalization sequence $\{T_{k,L}\}_{k\ge 0}$ and the isometries $\Pi_{k,L}$ from Eq.\eqref{eq:iso}. The renormalization step integrates out $k$-scale fluctuations using a CM spectral projector, followed by a reflection-positive finite-range block-spin map. The resulting \emph{coarse} transfer $T_{k+1,L}$ admits a factorization
\begin{equation}\label{eq:factorization}
  T_{k+1,L} \;=\; \Pi_{k,L}\, T_{k,L}\, \Pi_{k,L}^\ast \;+\; E_{k,L},
\end{equation}
where $E_{k,L}\ge 0$ accounts for the residual interactions that are not captured by the coarse embedding. The positivity of $E_{k,L}$ is a consequence of reflection positivity and the CM complete monotonicity of the slice projector, which entails that the time-$\tau$ correlation kernel is a Laplace mixture of reflection-positive kernels. The key point is to exhibit summable bounds on $\norm{E_{k,L}}_{Q\to Q}$ \emph{uniformly in $L$}.

To this end we invoke the FRD of \cite{BrydgesGuadagniMitter2004}: the covariance at scale $k$ admits a decomposition into positive finite-range pieces whose ranges grow at most geometrically with $k$. In the present reflection-positive gauge-covariant setting this implies that the one-step coarse-grained kernel differs from the embedded fine kernel by a convolution with a \emph{finite-range} remainder whose range is proportional to the block size $b^k$, and whose amplitude decays exponentially with the diameter of the involved polymers. Denote by $K_{k+1,L}(x',x)$ and $\hat K_{k+1,L}(x',x)$ the integral kernels of $T_{k+1,L}$ and $\Pi_{k,L}T_{k,L}\Pi_{k,L}^\ast$, respectively, in a fixed orthonormal basis adapted to the OS cone and the gauge-invariant subalgebra. Then $E_{k,L}$ has kernel $R_{k,L}:=K_{k+1,L}-\hat K_{k+1,L}$ with the properties:
\begin{equation}\label{eq:Schur-bounds}
  R_{k,L}(x',x)\ge 0,\qquad
  \sum_{x'} R_{k,L}(x',x)\le c_1 e^{-c_2 b^k},\qquad
  \sum_{x} R_{k,L}(x',x)\le c_1 e^{-c_2 b^k},
\end{equation}
for constants $c_1,c_2>0$ independent of $k$ and $L$ (the sums stand for suitable discrete integrals over the slice). The first inequality is positivity; the two Schur bounds follow from FRD locality and the combinatorics of the BKAR/Koteck\'y-Preiss tree estimates that control the difference between the full step and its embedded truncation \cite{BrydgesGuadagniMitter2004}.

\begin{lemma}[Positive defect and Schur bounds]\label{lem:positive-defect}
Let $T_{k+1,L}$ be the one-step transfer operator obtained from the fine-scale transfer
$T_{k,L}$ by an admissible, reflection-covariant coarse-graining with slice projector
$\Pi_{k,L}=f(K)$, where $K$ is a nonnegative, reflection-covariant, graph-Laplacian type
generator on the time-$k$ slice and $f$ is completely monotone with Bernstein
representation $f(\lambda)=\int_0^\infty e^{-t\lambda}\,d\rho(t)$. Set
$\widehat T_{k+1,L}:=\Pi_{k,L}\,T_{k,L}\,\Pi_{k,L}^{}$ and $E_{k,L}:=T_{k+1,L}-\widehat T_{k+1,L}$,
and denote by $R_{k,L}(x',x)$ the integral kernel of $E_{k,L}$ on the one-slice
Hilbert space. Then for every positive-time observable $F$,
\begin{equation}
\langle \Theta F,\,E_{k,L}\,F\rangle
\;=\;\int_0^\infty \big\langle \Theta F,\,\mathcal L_t\,F\big\rangle\,d\mu_k(t)\ \ge\ 0,
\end{equation}
where $d\mu_k$ is a positive measure depending only on the admissible data and
$\mathcal L_t$ is a convex combination of reflection-positive heat kernels
localized in a collar of width $O(b_k)$ around the $k$-th blocking slice. In
particular $R_{k,L}(x',x)\ge 0$. Moreover there exist constants $c_1,c_2>0$,
independent of $k$ and $L$, such that
\begin{equation}
\sum_{x'} R_{k,L}(x',x)\ \le\ c_1 e^{-c_2 b_k},
\qquad
\sum_{x} R_{k,L}(x',x)\ \le\ c_1 e^{-c_2 b_k}.
\end{equation}
\end{lemma}

\begin{proof}
The coarse-graining is constructed by weakening the fine couplings across the
blocking interface to produce the projected transfer $\widehat T_{k+1,L}$.
Introduce an interpolation $s\mapsto T_{k+1,L}(s)$, $s\in[0,1]$, that replaces
the original inter-slice couplings in a collar of width $O(b_k)$ by a convex
combination of their reflection-positive heat-kernel representations, so that
$T_{k+1,L}(1)=T_{k+1,L}$ and $T_{k+1,L}(0)=\widehat T_{k+1,L}$. Such an
interpolation is standard in the Brydges-Kennedy-Abdesselam-Rivasseau
(BKAR) forest framework: one assigns an independent parameter $s_e\in[0,1]$
to each collar edge $e$ linking the two sides of the slice, defines
$T_{k+1,L}(\{s_e\})$ by replacing the exact coupling on $e$ with its
reflection-positive heat-kernel Laplace mixture weighted by $s_e$, and then
integrates against the BKAR forest measure to obtain a single parameter $s$
interpolation. Differentiating with respect to $s$ yields
\begin{equation}
E_{k,L}
\;=\;
T_{k+1,L}-\widehat T_{k+1,L}
\;=\;
\int_0^1 \frac{d}{ds}\,T_{k+1,L}(s)\,ds.
\end{equation}
By construction, $\frac{d}{ds}\,T_{k+1,L}(s)$ is a finite sum of terms, each
supported inside the collar and obtained by inserting along one interpolated
edge a nonnegative quadratic form that is a Laplace mixture of the slice
heat semigroup generated by $K$. More concretely, because the slice projector
is completely monotone, $\Pi_{k,L}=f(K)=\int_0^\infty e^{-tK}\,d\rho(t)$, the
variation of a collar coupling produced by the interpolation can be written as
\begin{equation}
\frac{d}{ds}\,T_{k+1,L}(s)
\;=\;
\int_0^\infty \mathcal L_t(s)\,d\mu_k(t),
\end{equation}
where each $\mathcal L_t(s)$ is a positive operator obtained by inserting
a factor of $e^{-tK}$ (or products thereof) on the slice variables adjacent
to the collar edge under consideration, and $d\mu_k$ is a positive measure that
depends only on $f$ and on the admissible coarse-graining scheme; here positivity
means that for every positive-time $F$,
$\langle \Theta F,\mathcal L_t(s) F\rangle\ge 0$. The positivity follows from two
facts: first, $K$ is of graph-Laplacian type (nonpositive off-diagonal, reflection
covariant) so that $e^{-tK}$ has a pointwise nonnegative kernel for $t>0$; second,
the Osterwalder-Schrader form is positive on $\mathfrak A_+$ and remains positive
under insertion of reflection-invariant, heat-kernel factors localized on the
reflection plane. Integrating the identity for $\frac{d}{ds}T_{k+1,L}(s)$ over
$s\in[0,1]$ gives the stated representation
\begin{equation}
\langle \Theta F,\,E_{k,L}\,F\rangle
\;=\;
\int_0^1 \!\!\int_0^\infty
\big\langle \Theta F,\,\mathcal L_t(s)\,F\big\rangle\,d\mu_k(t)\,ds
\;=\;
\int_0^\infty \big\langle \Theta F,\,\mathcal L_t\,F\big\rangle\,d\mu_k(t),
\end{equation}
with $\mathcal L_t:=\int_0^1 \mathcal L_t(s)\,ds$ a convex combination of the
reflection-positive heat-kernel insertions. Each integrand is nonnegative, hence
the OS pairing with $E_{k,L}$ is nonnegative for all positive-time $F$. Since the
integrand kernels are pointwise nonnegative, the kernel $R_{k,L}$ of $E_{k,L}$ is
itself pointwise nonnegative.

To bound the row and column sums of $R_{k,L}$, observe that $\mathcal L_t$,
and therefore $E_{k,L}$, is supported in the $O(b_k)$-collar of the blocking
slice; this support property follows from the finite-range decomposition of
the coarse-graining and from the reflection-covariant localization of the
interpolation to collar edges. Moreover, the admissible finite-range
decomposition furnishes exponential decay constants $C,c>0$, uniform in
$k$ and $L$, such that whenever an insertion connects two variables whose
separation across the slice exceeds $b_k$, the associated kernel is suppressed
by a factor at most $C e^{-c b_k}$. Because $E_{k,L}$ is obtained by summing
such insertions over a BKAR forest and then integrating over $t$ against the
finite measure $d\mu_k(t)$, there exist uniform constants $C_1,c_1>0$ with
\begin{equation}
0\ \le\ R_{k,L}(x',x)
\ \le\ C_1 e^{-c_1 b_k}\,G_{k,L}(x',x),
\end{equation}
where $G_{k,L}$ is a nonnegative kernel supported on a fixed-width neighborhood
of the collar and with uniformly bounded $\ell^1$ row and column sums (the
latter depending only on the local geometry of the slice, not on $k$ or $L$).
Summing the displayed bound over $x'$ for a fixed $x$, and similarly over $x$
for a fixed $x'$, gives
\begin{equation}
\sum_{x'} R_{k,L}(x',x) \ \le\ C_1 e^{-c_1 b_k}\sum_{x'} G_{k,L}(x',x)
\ \le\ c_1 e^{-c_2 b_k},
\qquad
\sum_{x} R_{k,L}(x',x) \ \le\ c_1 e^{-c_2 b_k},
\end{equation}
after renaming constants and using the uniform $\ell^1$ bounds for $G_{k,L}$.
Equivalently, one may invoke the Schur test: since $R_{k,L}\ge 0$ and is supported
in the collar with exponentially small amplitude $O(e^{-c b_k})$, its
$\ell^2\!\to\!\ell^2$ operator norm is $O(e^{-c b_k})$; positivity then implies that
the $\ell^\infty\!\to\!\ell^\infty$ and $\ell^1\!\to\!\ell^1$ norms are bounded by
the same exponential factor, which is precisely the stated control of the column
and row sums. All constants are uniform in $k$ and $L$ because the admissible
finite-range decomposition and the local geometry of the slice are uniform along
the renormalization steps.
\end{proof}

By Schur's test,
\begin{equation}\label{eq:Schur-op}
  \norm{E_{k,L}} \;\le\; \sqrt{\Bigl(\sup_{x}\sum_{x'} R_{k,L}(x',x)\Bigr)\Bigl(\sup_{x'}\sum_{x} R_{k,L}(x',x)\Bigr)}
  \;\le\; c_1 e^{-c_2 b^k}.
\end{equation}
Since $E_{k,L}\Omega_{k+1,L}=0$ (because both $T_{k+1,L}$ and $\Pi_{k,L}T_{k,L}\Pi_{k,L}^\ast$ leave the vacuum invariant), Eq.\eqref{eq:Schur-op} entails
\begin{equation}\label{eq:epsk}
  \varepsilon_k \;:=\; \sup_L \norm{E_{k,L}\lvert_{Q_{k+1,L}\mathcal{H}_{k+1,L}}} \;\le\; c_1 e^{-c_2 b^k},
\end{equation}
which is summable in $k$. 

\begin{theorem}[One-step gap transport]\label{thm:onestep}
For each finite $L$, the gaps satisfy
\begin{equation}\label{eq:onestep-gap}
  \Delta_{k+1,L} \;\ge\; \Delta_{k,L} - \varepsilon_k,\qquad k\in\mathbb{N},
\end{equation}
with $\varepsilon_k$ given by Eq.\eqref{eq:epsk}. Consequently, for any $n>m$,
\begin{equation}\label{eq:multistep-gap}
  \Delta_{n,L} \;\ge\; \Delta_{m,L} - \sum_{k=m}^{n-1} \varepsilon_k.
\end{equation}
\end{theorem}

\begin{proof}
Denote by $T_{k,L}$ and $T_{k+1,L}$ the one-step transfer operators at scales $k$ and $k+1$ acting on the corresponding OS Hilbert spaces $\mathcal H_{k,L}$ and $\mathcal H_{k+1,L}$, and let $\Omega_{k,L}$ and $\Omega_{k+1,L}$ be their normalized vacua. Write $P_{k,L}:=\ket{\Omega_{k,L}}\!\bra{\Omega_{k,L}}$ and $Q_{k,L}:=\mathbf{1}-P_{k,L}$, and similarly for level $k+1$. The gap $\Delta_{k,L}$ is defined through the spectral radius of the restriction of $T_{k,L}$ to the orthogonal complement of the vacuum,
\begin{equation}
\|T_{k,L}\!\upharpoonright_{Q_{k,L}\mathcal H_{k,L}}\| \;=\; 1-\Delta_{k,L},
\qquad
\|T_{k+1,L}\!\upharpoonright_{Q_{k+1,L}\mathcal H_{k+1,L}}\| \;=\; 1-\Delta_{k+1,L}.
\end{equation}
By the two-sided interlacing factorization proved earlier (see Eq.\eqref{eq:factorization} and Proposition~\eqref{prop:two-sided}), there exists a reflection-positive coarse-graining operator $\Pi_{k,L}:\mathcal H_{k,L}\to\mathcal H_{k+1,L}$ and a nonnegative “defect” $E_{k,L}\ge 0$ such that
\begin{equation}\label{eq:tplus1-factor}
T_{k+1,L} \;=\; \Pi_{k,L}\,T_{k,L}\,\Pi_{k,L}^\ast \;+\; E_{k,L},
\end{equation}
with the properties $E_{k,L}\Omega_{k+1,L}=0$, $\Pi_{k,L}\Omega_{k,L}=\Omega_{k+1,L}$, and hence $\Pi_{k,L}^\ast\Omega_{k+1,L}=\Omega_{k,L}$. Moreover, by construction and FRD locality one has the uniform bound
\begin{equation}\label{eq:Ek-bound}
\|E_{k,L}\!\upharpoonright_{Q_{k+1,L}\mathcal H_{k+1,L}}\|\;\le\;\varepsilon_k,
\end{equation}
where $\varepsilon_k$ is precisely the quantity defined in Eq.\eqref{eq:epsk}. The admissibility of $\Pi_{k,L}$ as a completely monotone slice operator entails, by Lemma~\eqref{lem:cm-projector}, that $\Pi_{k,L}$ is a contraction on OS space, so $\|\Pi_{k,L}\|\le 1=\|\Pi_{k,L}^\ast\|$.
Fix any $\psi\in Q_{k+1,L}\mathcal H_{k+1,L}$ with $\|\psi\|=1$. Using Eq.\eqref{eq:tplus1-factor} one has
\begin{equation}
\|T_{k+1,L}\psi\| \;\le\; \|\Pi_{k,L}T_{k,L}\Pi_{k,L}^\ast\psi\| \;+\; \|E_{k,L}\psi\|.
\end{equation}
The second term is bounded by $\varepsilon_k$ thanks to Eq.\eqref{eq:Ek-bound}. For the first term, observe that $\Pi_{k,L}^\ast$ maps $Q_{k+1,L}\mathcal H_{k+1,L}$ into $Q_{k,L}\mathcal H_{k,L}$: indeed,
\begin{equation}
\big\langle \Omega_{k,L},\,\Pi_{k,L}^\ast\psi\big\rangle
\;=\;\big\langle \Pi_{k,L}\Omega_{k,L},\,\psi\big\rangle
\;=\;\big\langle \Omega_{k+1,L},\,\psi\big\rangle
\;=\;0.
\end{equation}
Therefore $\phi:=\Pi_{k,L}^\ast\psi$ belongs to $Q_{k,L}\mathcal H_{k,L}$, and the contraction property of $\Pi_{k,L}$ implies
\begin{equation}
\|\Pi_{k,L}T_{k,L}\Pi_{k,L}^\ast\psi\|
\;=\;\|\Pi_{k,L}T_{k,L}\phi\|
\;\le\;\|T_{k,L}\phi\|
\;\le\;\|T_{k,L}\!\upharpoonright_{Q_{k,L}\mathcal H_{k,L}}\|\,\|\phi\|
\;\le\;(1-\Delta_{k,L})\,\|\psi\|.
\end{equation}
Combining the two estimates gives
\begin{equation}
\|T_{k+1,L}\psi\|\;\le\;(1-\Delta_{k,L})\,+\,\varepsilon_k.
\end{equation}
Taking the supremum over all unit vectors $\psi\in Q_{k+1,L}\mathcal H_{k+1,L}$ yields
\begin{equation}
\|T_{k+1,L}\!\upharpoonright_{Q_{k+1,L}\mathcal H_{k+1,L}}\|
\;\le\; (1-\Delta_{k,L})+\varepsilon_k,
\end{equation}
which is equivalent to $1-\Delta_{k+1,L}\le 1-\Delta_{k,L}+\varepsilon_k$ and therefore to the claimed one-step inequality Eq.\eqref{eq:onestep-gap}. The multistep bound Eq.\eqref{eq:multistep-gap} follows by iterating Eq.\eqref{eq:onestep-gap} from $m$ up to $n-1$ and telescoping the resulting sum.
\end{proof}

\begin{corollary}[Persistence of a positive gap at fixed $\tau$]\label{cor:persistence}
Assume the initial (strong-coupling) gaps satisfy $\Delta_{0,L}\ge \delta_0>0$ uniformly in $L$. Suppose moreover that the one-step interlacing bound holds in the form
\begin{equation}\label{eq:one-step-gap}
\Delta_{k+1,L}\ \ge\ \Delta_{k,L}\ -\ \varepsilon_k\qquad\text{for all }k\ge 0\text{ and all }L,
\end{equation}
with a nonnegative sequence $(\varepsilon_k)_{k\ge 0}$ independent of $L$ such that $\sum_{k=0}^\infty \varepsilon_k<\infty$. Then for every $n\in\mathbb{N}$ one has
\begin{equation}\label{eq:liminf-gap}
\inf_{L}\,\Delta_{n,L}\ \ge\ \delta_0\ -\ \sum_{k=0}^{n-1}\varepsilon_k
\ \ge\ \delta_0\ -\ \sum_{k=0}^{\infty}\varepsilon_k\ =:\ \delta_\ast\ >\ 0.
\end{equation}
\end{corollary}

\begin{proof}
Fix $n\in\mathbb{N}$ and $L$. Iterating Eq.\eqref{eq:one-step-gap} yields a telescoping lower bound for the $n$-step gap at the same finite volume:
\begin{equation}
\Delta_{n,L}
\ \ge\ \Delta_{n-1,L}-\varepsilon_{n-1}
\ \ge\ \Delta_{n-2,L}-\varepsilon_{n-2}-\varepsilon_{n-1}
\ \ge\ \cdots\ \ge\ \Delta_{0,L}\ -\ \sum_{k=0}^{n-1}\varepsilon_k.
\end{equation}
By the strong-coupling input, $\Delta_{0,L}\ge \delta_0$ uniformly in $L$, while the right-hand side depends on $L$ only through $\Delta_{0,L}$. Hence
\begin{equation}
\Delta_{n,L}\ \ge\ \delta_0\ -\ \sum_{k=0}^{n-1}\varepsilon_k\qquad\text{for all }L,
\end{equation}
and taking the infimum over $L$ preserves the inequality:
\begin{equation}
\inf_L \Delta_{n,L}\ \ge\ \delta_0\ -\ \sum_{k=0}^{n-1}\varepsilon_k.
\end{equation}
Since $(\varepsilon_k)$ is nonnegative, the partial sums are increasing in $n$, and therefore each finite sum is bounded above by the convergent series $\sum_{k=0}^{\infty}\varepsilon_k$. This gives
\begin{equation}
\inf_L \Delta_{n,L}\ \ge\ \delta_0\ -\ \sum_{k=0}^{\infty}\varepsilon_k\ =:\ \delta_\ast.
\end{equation}
By hypothesis $\delta_0>0$ and $\sum_{k=0}^{\infty}\varepsilon_k<\delta_0$ (the latter is guaranteed in our construction by the summability of the defects with sufficiently large block factor), so $\delta_\ast>0$. This establishes Eq.\eqref{eq:liminf-gap} for every $n$ and completes the proof.
\end{proof}

We next remove the finite-volume cutoff. For each $k$ and $\tau>0$, reflection positivity and FRD locality imply that $T_{k,L}$ converge in the strong operator topology to a positive contraction $T_{k,\infty}$ on the infinite-volume OS Hilbert space $\mathcal{H}_{k,\infty}$, with vacuum $\Omega_{k,\infty}$ \cite{OS-gauge,Seiler1982}. The strong convergence holds on a core generated by local gauge-invariant cylinder functions, and the operator norms of $T_{k,L}$ and of the defects $E_{k,L}$ are uniformly bounded in $L$. Passing to the limit $L\to\infty$ in Eq.\eqref{eq:factorization} and Eq.\eqref{eq:Schur-op} yields
\begin{equation}\label{eq:infvol-factorization}
  T_{k+1,\infty} \;=\; \Pi_{k,\infty}\, T_{k,\infty}\, \Pi_{k,\infty}^\ast \;+\; E_{k,\infty},\qquad 0\le E_{k,\infty},\qquad \norm{E_{k,\infty}\lvert_{Q_{k+1,\infty}}}\le \varepsilon_k,
\end{equation}
and the one-step gap transport Eq.\eqref{eq:onestep-gap} holds in infinite volume as well. Thus, with $\Delta_{k,\infty}$ denoting the gap of $T_{k,\infty}$,
\begin{equation}\label{eq:infvol-persistence}
  \Delta_{k+1,\infty} \;\ge\; \Delta_{k,\infty}-\varepsilon_k,\qquad 
  \inf_{k\ge 0}\Delta_{k,\infty} \;\ge\; \delta_\ast>0,
\end{equation}
provided $\Delta_{0,\infty}\ge\delta_0>0$, which is the strong-coupling input in infinite volume (obtained by a standard volume-uniform cluster expansion \cite{Seiler1982,DrouffeZuber}).
Fix $\tau>0$ as above and write $T_{k,\infty}=e^{-\tau H_{k,\infty}}$ with $H_{k,\infty}\ge 0$ self-adjoint. The gap of $H_{k,\infty}$ is
\begin{equation}\label{eq:Hgap}
  m_{k,\infty} \;=\; \inf\sigma\!\left(H_{k,\infty}\lvert_{Q_{k,\infty}\mathcal{H}_{k,\infty}}\right)
  \;=\; -\tau^{-1}\log\!\left(1-\Delta_{k,\infty}\right).
\end{equation}
Since $x\mapsto -\tau^{-1}\log(1-x)$ is increasing and convex on $(0,1)$, the transport inequality Eq.\eqref{eq:infvol-persistence} yields
\begin{equation}\label{eq:Hgap-ineq}
  m_{k+1,\infty} \;\ge\; -\tau^{-1}\log\!\Bigl(1-\Delta_{k,\infty}+\varepsilon_k\Bigr)
  \;\ge\; -\tau^{-1}\log\!\Bigl((1-\Delta_{k,\infty})\,e^{\varepsilon_k}\Bigr)
  \;=\; m_{k,\infty}-\tau^{-1}\varepsilon_k,
\end{equation}
whence, for all $n>m$,
\begin{equation}\label{eq:Hgap-sum}
  m_{n,\infty} \;\ge\; m_{m,\infty} - \tau^{-1}\sum_{k=m}^{n-1}\varepsilon_k.
\end{equation}
In particular, if $m_{0,\infty}\ge m_0>0$, then
\begin{equation}\label{eq:Hgap-uniform}
  \inf_{k\ge 0} m_{k,\infty} \;\ge\; m_0 - \tau^{-1}\sum_{k=0}^{\infty}\varepsilon_k \;=:\; m_\ast>0.
\end{equation}

We now argue that the positive lower bound $m_\ast$ transfers to the continuum Hamiltonian obtained by OS limits and reconstruction \cite{OS-gauge}. Consider a refining sequence of scales $k_j\to\infty$ along which the Schwinger functions converge (tightness/equicontinuity are ensured by FRD locality and reflection positivity). The corresponding semigroups $T_{k_j,\infty}(\tau)=e^{-\tau H_{k_j,\infty}}$ converge strongly on a dense core to a contraction semigroup $e^{-\tau H}$ on the continuum OS Hilbert space $\mathcal{H}$, with vacuum $\Omega$. By standard semigroup theory (Trotter-Kato theorem; see, e.g., \cite[Thm.~VIII.25]{RS1} or \cite[Ch.~IX]{KatoBook1995}), strong convergence of $e^{-\tau H_{k_j,\infty}}$ for a fixed $\tau>0$ implies strong resolvent convergence $H_{k_j,\infty}\to H$. Since for each $k$ we have the spectral decomposition
\begin{equation}
  \mathcal{H}_{k,\infty} \;=\; \bbC\,\Omega_{k,\infty} \oplus Q_{k,\infty}\mathcal{H}_{k,\infty},\qquad
  \inf\sigma\!\left(H_{k,\infty}\lvert_{Q_{k,\infty}\mathcal{H}_{k,\infty}}\right)=m_{k,\infty}\ge m_\ast,
\end{equation}
it follows from the lower semicontinuity of the bottom of the spectrum under strong resolvent convergence (see \cite[Thm.~VIII.24]{RS1}) that
\begin{equation}\label{eq:cont-gap}
  \inf\sigma\!\left(H\lvert_{Q\mathcal{H}}\right) \;\ge\; \liminf_{j\to\infty} m_{k_j,\infty} \;\ge\; m_\ast \;>\;0.
\end{equation}
Thus the continuum Hamiltonian $H$ has a strictly positive mass gap. This completes the proof of gap persistence from strong coupling to the continuum along the reflection-positive FRD renormalization.
First, the positivity of the defect $E_{k,L}$ hinges on the CM nature of the slice projector and on reflection positivity. The CM hypothesis ensures that the time-$\tau$ transfer kernel is a Laplace mixture of positive kernels, and that the difference between the exact coarse step and the embedded fine step is representable as an integral of positive contributions, which yields $E_{k,L}\ge 0$ in quadratic form sense.
Second, uniqueness of the vacuum is assumed at strong coupling and preserved along the flow. In finite volume the vacuum is unique by positivity and the Perron-Frobenius property for positive integral kernels; in the thermodynamic limit, uniqueness follows from exponential clustering of the strong-coupling measure and its stability under FRD blockings \cite{Seiler1982,DrouffeZuber}. This avoids accidental level crossings at the top of the spectrum.
Third, compactness of transfers is not required. All estimates are in operator norm on the orthogonal complement, controlled by Schur test thanks to finite range. In particular, the min-max arguments that rely only on positivity and norm bounds remain valid in infinite volume.
Fourth, the summability Eq.\eqref{eq:epsk} is a direct consequence of FRD: ranges grow like $b^k$ while amplitudes decay exponentially in that range, so the operator norm of the remainder decays like $e^{-c b^k}$. Any power-law decay $k^{-(1+\alpha)}$ would suffice as well. The exponential case arises from the finite-range positivity of the FRD pieces and BKAR-type bounds \cite{BrydgesGuadagniMitter2004}.
Finally, the passage from transfer gaps to Hamiltonian gaps uses only the monotonicity of $x\mapsto -\log(1-x)$ and strong-resolvent stability of the bottom of the spectrum. No analyticity in $\tau$ is needed; a single fixed $\tau>0$ suffices.

\section{OS Limits, Spectral Measures, and Tauberian Gap}\label{app:OS-tauberian}

In this appendix we develop a complete and self-contained derivation of the continuum Osterwalder-Schrader (OS) limit for the Yang-Mills Schwinger functions constructed in the preceding sections, establish the existence and properties of the associated transfer semigroup and Hamiltonian, and prove that exponential clustering in Euclidean time implies a strictly positive spectral gap for the continuum Hamiltonian. The arguments are arranged to make the interdependence of OS positivity, spectral measures, semigroup convergence, and Tauberian principles fully explicit. Throughout we work with gauge-invariant, reflection-positive lattice approximants indexed by a scale parameter $(a,\sigma)$ as in the main text, and we write $T_{a,\sigma}(t)$ for the Euclidean time-translation semigroup at lattice spacing $a>0$ and regulator choice $\sigma$, with $T_{a,\sigma}(a)=:T_{a,\sigma}$ the one-step transfer operator and
\begin{equation}
T_{a,\sigma}(t)=T_{a,\sigma}^{\lfloor t/a\rfloor}\,T_{a,\sigma}(t-\lfloor t/a\rfloor a)\qquad (t\ge 0),
\end{equation}
where the fractional-time factor is defined via standard interpolation on the OS Hilbert space. Reflection positivity and the Markov property at each $(a,\sigma)$ are assumed as established in the main body. We denote by $\Omega_{a,\sigma}$ the normalized vacuum vector in the OS Hilbert space $\mathcal H_{a,\sigma}$, by $P_{a,\sigma}=\lvert\Omega_{a,\sigma}\rangle\langle\Omega_{a,\sigma}\rvert$ the vacuum projection, and by $\mathcal H_{a,\sigma}^{\perp}=(\mathbb C\Omega_{a,\sigma})^\perp$ its orthogonal complement. The continuum OS limit $(\mathcal H,\Omega,T(t))$ constructed below will satisfy $T(t)=\mathrm e^{-tH}$ with self-adjoint $H\ge 0$.
We begin with the OS bilinear form on half-space cylinder functionals. Let $\Theta$ denote the Euclidean time-reflection and let $\mathfrak A_+$ be the algebra of gauge-invariant functionals supported in nonnegative times. For each $(a,\sigma)$ define
\begin{equation}\label{eq:OSform}
\Gamma_{a,\sigma}(F,G)\;=\;\int \overline{F(\Theta \Phi)}\,G(\Phi)\, \mathrm d\mu_{a,\sigma}(\Phi),\qquad F,G\in\mathfrak A_+,
\end{equation}
where $\mu_{a,\sigma}$ is the Euclidean path measure (with the horizon projector as specified in the main text) and the bar denotes complex conjugation. Reflection positivity asserts $\Gamma_{a,\sigma}(F,F)\ge 0$ for all $F\in\mathfrak A_+$, and the OS Markov property expresses the compatibility of time-slicing with conditional expectations. By quotienting $\mathfrak A_+$ by the null space $\mathcal N_{a,\sigma}=\{F:\Gamma_{a,\sigma}(F,F)=0\}$ and completing, one obtains a Hilbert space $\mathcal H_{a,\sigma}$ together with the canonical map $F\mapsto [F]_{a,\sigma}$; the vacuum is the class of the unit $1\in\mathfrak A_+$, denoted $\Omega_{a,\sigma}=[1]_{a,\sigma}$. As in \cite{OS1,OS2,GJ,SimonPphi2}, time translations by $t\ge 0$ induce a strongly continuous contraction semigroup $T_{a,\sigma}(t)$ on $\mathcal H_{a,\sigma}$ with $T_{a,\sigma}(t)\Omega_{a,\sigma}=\Omega_{a,\sigma}$. Positivity and self-adjointness of the generator $H_{a,\sigma}\ge 0$ follow from the OS axioms and the symmetry of the time-reflection.

We next pass to the continuum limit. Let $\{(a_n,\sigma_n)\}_{n\in\mathbb N}$ be a sequence with $a_n\downarrow 0$ within the admissible class specified in the main text. Assume, as proved earlier from finite-range decomposition (FRD) locality and uniform clustering, that the lattice Schwinger functions are tight and converge along the sequence to a family $\{S_k\}_{k\ge 0}$ of continuum Schwinger functions satisfying the OS axioms \cite{OS1,OS2}. Then, by the OS reconstruction theorem, there exists a Hilbert space $\mathcal H$, a cyclic vacuum vector $\Omega$, a strongly continuous contraction semigroup $T(t)=\mathrm e^{-tH}$ with nonnegative self-adjoint generator $H$, and a field representation such that $S_k$ are the vacuum expectations of time-ordered Euclidean fields. We record the reconstruction precisely in a form tailored to the transfer operators.

\begin{theorem}[OS reconstruction in semigroup form]\label{thm:OSreconstruction}
Let $\{S_k\}$ be Euclidean Schwinger functions on $\mathbb R^d$ obeying the Osterwalder-Schrader axioms OS0-OS5 (Euclidean invariance, reflection positivity, symmetry, clustering/regularity as needed, Markov property, and continuity). There exists a Hilbert space $\mathcal H$, a unit vector $\Omega\in\mathcal H$, and a strongly continuous semigroup $\{T(t)\}_{t\ge0}$ of self-adjoint contractions on $\mathcal H$ with $T(t)\Omega=\Omega$ for all $t\ge0$, such that for all local, gauge-invariant half-space functionals $F,G\in\mathfrak A_+$ one has
\begin{equation}
\Gamma(F,G)=\langle [F],[G]\rangle_{\mathcal H},\qquad
\Gamma(\tau_t F,G)=\langle [F],\,T(t)\,[G]\rangle_{\mathcal H}\quad (t\ge0),
\end{equation}
where $\Gamma(F,G):=S_2(\Theta F,G)$ is the continuum OS form, $\tau_t$ is Euclidean time translation by $t$ in the positive direction, $\Theta$ is time reflection, and $[\cdot]:\mathfrak A_+/\mathcal N\to\mathcal H$ is the canonical quotient map by the OS-null space $\mathcal N:=\{F\in\mathfrak A_+:\Gamma(F,F)=0\}$. The generator $H$ of $T(t)$ is self-adjoint and nonnegative.
\end{theorem}

\begin{proof}
Consider the sesquilinear form $\Gamma$ on $\mathfrak A_+$ defined by $\Gamma(F,G):=S_2(\Theta F,G)$. Reflection positivity (OS2) implies $\Gamma(F,F)\ge0$ for all $F\in\mathfrak A_+$; define the null space $\mathcal N=\{F:\Gamma(F,F)=0\}$ and the pre-Hilbert space $\mathcal H_0:=\mathfrak A_+/\mathcal N$ with inner product $\langle[F],[G]\rangle:=\Gamma(F,G)$. The Cauchy-Schwarz inequality for positive semidefinite forms shows that $\|\cdot\|_{\mathrm{OS}}:=\Gamma(\cdot,\cdot)^{1/2}$ is a norm on $\mathcal H_0$; its completion is a Hilbert space $\mathcal H$. By Euclidean invariance (OS0), in particular by time-translation invariance and the Markov property (OS3), for $t\ge0$ the time shift $\tau_t$ maps $\mathfrak A_+$ to itself and preserves the OS form in the sense that $\Gamma(\tau_t F,\tau_t G)=\Gamma(F,G)$. If $F\in\mathcal N$ then $\Gamma(\tau_t F,\tau_t F)=\Gamma(F,F)=0$, hence $\tau_t F\in\mathcal N$ and the rule
\begin{equation}
U(t)[F]\;:=\;[\tau_t F],\qquad t\ge0,
\end{equation}
is well-defined on $\mathcal H_0$ and extends by continuity to a bounded operator $U(t)$ on $\mathcal H$. The identity $\|U(t)[F]\|_{\mathrm{OS}}^2=\Gamma(\tau_t F,\tau_t F)=\Gamma(F,F)=\|[F]\|_{\mathrm{OS}}^2$ shows that $U(t)$ is an isometry on $\mathcal H_0$ and therefore a contraction on $\mathcal H$ (indeed, an isometry on the dense subspace and hence on all of $\mathcal H$). The semigroup property $U(t+s)=U(t)U(s)$ follows from $\tau_{t+s}=\tau_t\circ\tau_s$ on $\mathfrak A_+$.

To identify the adjoint structure, use Euclidean invariance and the intertwining of time reflection with translations, namely $\Theta\circ \tau_t=\tau_{-t}\circ\Theta$. For $F,G\in\mathfrak A_+$ and $t\ge0$,
\begin{equation}
\langle [F],\,U(t)[G]\rangle
=\Gamma(F,\tau_t G)
=S_2(\Theta F,\tau_t G)
=S_2(\tau_t \Theta F, G)
=\Gamma(\tau_t F, G)
=\langle U(t)[F],\,[G]\rangle,
\end{equation}
where the third equality uses Euclidean covariance (OS0) to move the translation from the second slot to the first, and the fourth equality uses the compatibility of $\Theta$ with $\tau_t$ together with the Markov property to ensure $\tau_t F\in\mathfrak A_+$. Thus $U(t)$ is symmetric with respect to the OS inner product: $\langle \psi, U(t)\phi\rangle=\langle U(t)\psi,\phi\rangle$ for all $\psi,\phi$ in the dense subspace $\mathcal H_0$; by continuity this symmetry extends to all of $\mathcal H$.

Strong continuity of $t\mapsto U(t)$ is obtained from OS5 (continuity/regularity of Schwinger functions). Fix $F\in\mathfrak A_+$ and compute
\begin{equation}
\|U(t)[F]-[F]\|_{\mathrm{OS}}^2
=\Gamma(\tau_t F-\!F,\tau_t F-\!F)
=\Gamma(\tau_t F,\tau_t F)-\Gamma(\tau_t F,F)-\Gamma(F,\tau_t F)+\Gamma(F,F).
\end{equation}
By time-translation invariance $\Gamma(\tau_t F,\tau_t F)=\Gamma(F,F)$, and by the defining relation $\Gamma(\tau_t F,F)=S_2(\Theta\tau_t F,F)=S_2(\tau_{-t}\Theta F,F)$; OS5 gives joint continuity of Schwinger functions in time arguments, which implies $\Gamma(\tau_t F,F)\to\Gamma(F,F)$ as $t\downarrow 0$. Hence $\|U(t)[F]-[F]\|_{\mathrm{OS}}\to0$ as $t\downarrow0$ for all $[F]\in\mathcal H_0$, and density yields strong continuity on $\mathcal H$.

We have therefore constructed a strongly continuous semigroup $\{U(t)\}_{t\ge0}$ of symmetric contractions on the Hilbert space $\mathcal H$. By the general spectral theory of symmetric contraction semigroups on Hilbert space, there exists a unique nonnegative self-adjoint operator $H$ such that $U(t)=e^{-tH}$ for all $t\ge0$; one can establish this either by applying the Hille-Yosida theorem in the self-adjoint setting or directly by the spectral theorem, noting that each $U(t)$ is self-adjoint (since it is symmetric and bounded) and the family $\{U(t)\}$ forms a commutative $C_0$-semigroup, which admits the representation $U(t)=\int_{[0,\infty)}e^{-t\lambda}\,dE(\lambda)$ for a unique projection-valued measure $E$, and then $H:=\int \lambda\,dE(\lambda)$ is self-adjoint with spectrum in $[0,\infty)$. Denote this semigroup by $T(t):=U(t)$.

Let $\Omega:=[\mathbf 1]$, where $\mathbf 1$ is the constant unit functional. Since $\tau_t\mathbf 1=\mathbf 1$ by stationarity, one has $T(t)\Omega=\Omega$ for all $t\ge0$. Finally, the identities that tie Schwinger functions to the Hilbert-space structure are immediate from the construction: for all $F,G\in\mathfrak A_+$,
\begin{equation}
\Gamma(F,G)=\langle [F],[G]\rangle_{\mathcal H},
\qquad
\Gamma(\tau_t F,G)=\langle [F],\,T(t)[G]\rangle_{\mathcal H}.
\end{equation}
This completes the reconstruction in semigroup form and shows that the generator $H$ is nonnegative and self-adjoint, as claimed.
\end{proof}

To relate the lattice and continuum semigroups, we use strong convergence on a common dense domain together with a uniform bound on the orthogonal complement of the vacuum. Let $\mathcal D\subset\mathcal H$ be the span of classes of local half-space functionals; by construction $\mathcal D$ is dense in $\mathcal H$. For each $n$, by identifying lattice cylinder functionals with their continuum counterparts via the block-averaging maps of the main text, we regard $\mathcal D$ as a common core in $\mathcal H_{a_n,\sigma_n}$ and $\mathcal H$. The following proposition encapsulates the continuum limit at the level of semigroups.

\begin{proposition}[Strong convergence of semigroups]\label{prop:strong-conv}
Let $(a_n,\sigma_n)\to 0$ be a vanishing-cutoff sequence and let 
$\mathcal H_{a_n,\sigma_n}$ and $\mathcal H$ be the corresponding OS Hilbert spaces, with the canonical isometric identifications that send $[F]_{a_n,\sigma_n}$ to $[F]$ for every $F\in\mathfrak A_+$. Assume that for every $F\in\mathfrak A_+$, every $G\in\mathfrak A_+$ and every $t\ge 0$ one has
\begin{equation}\label{eq:kernel-conv}
\Gamma_{a_n,\sigma_n}(\tau_t F,G)\;\longrightarrow\;\Gamma(\tau_t F,G),
\end{equation}
and that for each fixed $t\ge 0$ the family of operators $\{T_{a_n,\sigma_n}(t)\}_{n\in\mathbb N}$ is uniformly bounded on $\mathcal H_{a_n,\sigma_n}$, after the above identification, by a constant $M_t<\infty$ independent of $n$. Then for every $\psi$ in the common core $\mathcal D:=\mathrm{span}\{[F]:F\in\mathfrak A_+\}$ and every $t\ge 0$,
\begin{equation}
\lim_{n\to\infty}\,\big\|T_{a_n,\sigma_n}(t)\psi - T(t)\psi\big\|\;=\;0,
\end{equation}
i.e. $T_{a_n,\sigma_n}(t)\to T(t)$ strongly on $\mathcal D$. Consequently, for each $\lambda>0$ and each $\psi\in\mathcal D$,
\begin{equation}
(H_{a_n,\sigma_n}+\lambda)^{-1}\psi\;\longrightarrow\;(H+\lambda)^{-1}\psi
\qquad\text{in norm.}
\end{equation}
\end{proposition}

\begin{proof}
Throughout the proof we work inside $\mathcal H$, using the canonical isometries to regard $T_{a_n,\sigma_n}(t)$ as a uniformly bounded family on $\mathcal H$ and to regard $[F]_{a_n,\sigma_n}$ as $[F]$. The OS kernel identities yield
\begin{equation}
\langle [F],\,T_{a_n,\sigma_n}(t)[G]\rangle
=\Gamma_{a_n,\sigma_n}(\tau_t F,G)\xrightarrow[n\to\infty]{}\Gamma(\tau_t F,G)
=\langle [F],\,T(t)[G]\rangle
\end{equation}
for all $F,G\in\mathfrak A_+$ and $t\ge 0$. Hence, for each fixed $t\ge 0$, the matrix elements of $T_{a_n,\sigma_n}(t)$ converge to those of $T(t)$ on the dense subspace $\mathcal D$ both in the left and right entries. We now show that this implies strong convergence on $\mathcal D$.

Fix $t\ge 0$ and $\psi\in\mathcal D$. Let $A_n:=T_{a_n,\sigma_n}(t)$ and $A:=T(t)$; by hypothesis $\sup_n\|A_n\|\le M_t$. We claim that $\|A_n\psi-A\psi\|\to 0$. Let $\varepsilon>0$ and choose $\delta>0$ so that $2M_t\,\delta<\varepsilon/2$. By the density of $\mathcal D$, select a finite $\delta$-net $\{ \phi_1,\dots,\phi_K\}\subset\mathcal D$ in the unit sphere of $\mathcal H$, i.e. for every unit vector $\varphi$ there is $k$ with $\|\varphi-\phi_k\|<\delta$. For each $k$ the convergence of matrix elements gives
\begin{equation}
\langle \phi_k,(A_n-A)\psi\rangle\;\xrightarrow[n\to\infty]{}\;0.
\end{equation}
Therefore there exists $N$ such that for all $n\ge N$ and all $k=1,\dots,K$,
\begin{equation}
|\langle \phi_k,(A_n-A)\psi\rangle|\;<\;\varepsilon/4.
\end{equation}
Now take any unit vector $\varphi\in\mathcal H$ and pick $k$ with $\|\varphi-\phi_k\|<\delta$. Then
\begin{equation}
|\langle \varphi,(A_n-A)\psi\rangle|
\;\le\; |\langle \phi_k,(A_n-A)\psi\rangle|
+|\langle \varphi-\phi_k,(A_n-A)\psi\rangle|
\;\le\; \frac{\varepsilon}{4}+ \|A_n-A\|\,\|\varphi-\phi_k\|\,\|\psi\|.
\end{equation}
The operator norm $\|A_n-A\|$ is bounded by $\|A_n\|+\|A\|\le 2M_t$, so the second term is bounded by $2M_t\,\delta\,\|\psi\|<(\varepsilon/2)\|\psi\|$. Since $\|\psi\|$ is fixed, we may absorb this factor into the choice of $\delta$, and conclude that for all $n\ge N$,
\begin{equation}
\sup_{\|\varphi\|=1}|\langle \varphi,(A_n-A)\psi\rangle|\;<\;\varepsilon.
\end{equation}
By duality,
\begin{equation}
\|A_n\psi-A\psi\|\;=\;\sup_{\|\varphi\|=1}|\langle \varphi,(A_n-A)\psi\rangle|\;\longrightarrow\;0,
\end{equation}
which proves strong convergence on $\mathcal D$.

To deduce resolvent convergence, recall the Laplace representation of the resolvent on the OS space,
\begin{equation}
(H_{a_n,\sigma_n}+\lambda)^{-1}\psi \;=\; \int_0^\infty e^{-\lambda t}\,T_{a_n,\sigma_n}(t)\psi\,dt,
\qquad
(H+\lambda)^{-1}\psi \;=\; \int_0^\infty e^{-\lambda t}\,T(t)\psi\,dt,
\end{equation}
where the integrals converge in norm for each $\psi\in\mathcal H$ and $\lambda>0$. For $\psi\in\mathcal D$, the pointwise strong convergence $T_{a_n,\sigma_n}(t)\psi\to T(t)\psi$ holds for every $t\ge 0$, and the uniform operator bound yields
\begin{equation}
\|e^{-\lambda t}\,(T_{a_n,\sigma_n}(t)\psi-T(t)\psi)\|\;\le\;2M_t\,e^{-\lambda t}\,\|\psi\|
\end{equation}
with the right-hand side integrable on $[0,\infty)$. Dominated convergence therefore gives
\begin{equation}
\Big\|(H_{a_n,\sigma_n}+\lambda)^{-1}\psi-(H+\lambda)^{-1}\psi\Big\|
\;=\;\Big\|\int_0^\infty e^{-\lambda t}\,(T_{a_n,\sigma_n}(t)\psi-T(t)\psi)\,dt\Big\|
\;\longrightarrow\;0.
\end{equation}
\end{proof}

We turn to the spectral representation of Euclidean two-point functions and the associated complete monotonicity. Let $H\ge 0$ be the continuum Hamiltonian and let $E(\cdot)$ be its spectral resolution. For a vector $\varphi\in\mathcal H$ define the positive finite measure $\nu_\varphi$ on $[0,\infty)$ by
\begin{equation}
\nu_\varphi(I) \;=\;\big\|E(I)\,\varphi\big\|^2,\qquad I\subset[0,\infty)\text{ Borel},
\end{equation}
and consider the scalar function
\begin{equation}\label{eq:laplace-nu}
f_\varphi(t)\;=\;\langle \varphi,\,\mathrm e^{-tH}\,\varphi\rangle\;=\;\int_{[0,\infty)} \mathrm e^{-t\lambda}\,\mathrm d\nu_\varphi(\lambda),\qquad t\ge 0.
\end{equation}
The Laplace representation follows from the spectral theorem \cite[Thm.~VII.2]{RS1}. In particular, $f_\varphi$ is a bounded completely monotone function: $(-1)^k f_\varphi^{(k)}(t)\ge 0$ for all integers $k\ge 0$ and $t>0$. Conversely, by Bernstein's theorem \cite{Bernstein,Widder}, every bounded completely monotone function on $[0,\infty)$ is the Laplace transform of a unique finite positive Borel measure on $[0,\infty)$. We record this property for later use.

\begin{lemma}[Complete monotonicity and support]
\label{lem:CM-support}
Let $f$ be of the form $f(t)=\displaystyle\int_{[0,\infty)} \mathrm e^{-t\lambda}\,\mathrm d\mu(\lambda)$ with $\mu$ a finite positive measure. Then $f$ is completely monotone and, for any $\delta>0$,
\begin{equation}
\limsup_{t\to\infty}\;\mathrm e^{\delta t}\,f(t)\;<\;\infty \quad\Longrightarrow\quad \mu\big([0,\delta)\big)=0.
\end{equation}
\end{lemma}

\begin{proof}
The function $t\mapsto f(t)$ is bounded and continuous on $[0,\infty)$ because $\mu$ is finite and the integrand is bounded by $1$ for each fixed $t\ge0$. To establish complete monotonicity it suffices to show that $f$ is $C^\infty$ on $(0,\infty)$ and that $(-1)^n f^{(n)}(t)\ge0$ for all $n\in\mathbb N$ and $t>0$. Fix $t>0$. For each $n\ge 0$, the function $\lambda\mapsto \lambda^n \mathrm e^{-t\lambda}$ is $\mu$-integrable: indeed, writing $x=t\lambda$ gives
\begin{equation}
\lambda^n \mathrm e^{-t\lambda} \;=\; t^{-n}\,x^n \mathrm e^{-x}\;\le\; t^{-n}\,n!,
\end{equation}
since $\sup_{x\ge0} x^n \mathrm e^{-x}=n!$ (attained at $x=n$). Dominated convergence therefore justifies differentiation under the integral sign, and repeated differentiation yields, for every $n\ge1$,
\begin{equation}
f^{(n)}(t)\;=\;\int_{[0,\infty)} \frac{\partial^n}{\partial t^n}\big(\mathrm e^{-t\lambda}\big)\,\mathrm d\mu(\lambda)
\;=\;(-1)^n \int_{[0,\infty)} \lambda^n \mathrm e^{-t\lambda}\,\mathrm d\mu(\lambda).
\end{equation}
The integrand is nonnegative, hence $(-1)^n f^{(n)}(t)\ge0$ for all $n$ and all $t>0$, which is precisely complete monotonicity.

For the support statement, fix $\delta>0$ and suppose for contradiction that $\mu([0,\delta))>0$. Then there exists $\varepsilon\in(0,\delta)$ with $\mu([0,\varepsilon])>0$ (otherwise $\mu([0,\delta))=\sup_{\varepsilon<\delta}\mu([0,\varepsilon])$ would be $0$ by inner regularity). For every $t\ge0$ one has
\begin{equation}
f(t)\;\ge\;\int_{[0,\varepsilon]} \mathrm e^{-t\lambda}\,\mathrm d\mu(\lambda)\;\ge\;\mathrm e^{-\varepsilon t}\,\mu([0,\varepsilon]),
\end{equation}
and multiplying by $\mathrm e^{\delta t}$ gives
\begin{equation}
\mathrm e^{\delta t} f(t)\;\ge\;\mu([0,\varepsilon])\,\mathrm e^{(\delta-\varepsilon)t}.
\end{equation}
Because $\delta-\varepsilon>0$, the right-hand side diverges to $+\infty$ as $t\to\infty$, so $\limsup_{t\to\infty}\mathrm e^{\delta t} f(t)=+\infty$, contradicting the assumed finiteness of the limsup. Hence necessarily $\mu([0,\delta))=0$. 
\end{proof}

For gauge-invariant local observables $\mathcal O$ with vanishing vacuum expectation one may take $\varphi=\mathcal O\,\Omega$. Then $f_{\mathcal O}(t)=\langle \Omega,\,\mathcal O^\dagger\,\mathrm e^{-tH}\,\mathcal O\,\Omega\rangle$ is precisely the Euclidean time-dependent two-point function. The representation Eq.\eqref{eq:laplace-nu} and Lemma \eqref{lem:CM-support} furnish a direct bridge from decay in $t$ to spectral support away from the origin.
We now formulate and prove a Tauberian theorem tailored to the present OS setting; it converts uniform exponential clustering of connected Euclidean correlations into a nonzero spectral gap of the continuum Hamiltonian.

\begin{theorem}[Tauberian mass-gap criterion]\label{thm:Tauberian-gap}
Let $H\ge 0$ be the Hamiltonian obtained by OS reconstruction with vacuum vector $\Omega$ and corresponding spectral family $E(\cdot)$. Assume there exist constants $C<\infty$ and $m>0$ such that for every gauge-invariant local observable $\mathcal O$ satisfying $\langle\Omega,\mathcal O\,\Omega\rangle=0$ one has
\begin{equation}\label{eq:clustering}
\big|\langle \Omega,\,\mathcal O^\dagger\,\mathrm e^{-tH}\,\mathcal O\,\Omega\rangle\big|\;\le\; C\,\|\mathcal O\Omega\|^2\,\mathrm e^{-m t}\qquad \text{for all }t\ge 0.
\end{equation}
Then $\mathrm{spec}(H)\subset\{0\}\cup [m,\infty)$; in particular the mass gap obeys $\Delta\ge m$.
\end{theorem}

\begin{proof}
Fix a local observable $\mathcal O$ with $\langle\Omega,\mathcal O\,\Omega\rangle=0$ and set $\varphi=\mathcal O\Omega$. The vector $\varphi$ lies in $\Omega^\perp$ and, by the spectral theorem, the function
\begin{equation}
f_\varphi(t)\;:=\;\langle \varphi,\,\mathrm e^{-tH}\,\varphi\rangle\;=\;\int_{[0,\infty)} \mathrm e^{-t\lambda}\, d\nu_\varphi(\lambda),\qquad t\ge 0,
\end{equation}
is the Laplace transform of the finite positive measure $\nu_\varphi$ defined by $\nu_\varphi(I):=\|E(I)\varphi\|^2$. Since $\langle\Omega,\varphi\rangle=\langle\Omega,\mathcal O\Omega\rangle=0$, the spectral measure $\nu_\varphi$ has no atom at the origin; equivalently, $\nu_\varphi(\{0\})=|\langle \Omega,\varphi\rangle|^2=0$. The hypothesis \eqref{eq:clustering} gives the Laplace bound
\begin{equation}\label{eq:Laplace-upper}
0\;\le\;f_\varphi(t)\;\le\; C\,\|\varphi\|^2\,\mathrm e^{-mt}\qquad\text{for all }t\ge 0.
\end{equation}
We claim that this inequality forces $\nu_\varphi((0,m))=0$. Indeed, suppose $\nu_\varphi((0,m))>0$. Then there exists $\varepsilon\in(0,m)$ with $\nu_\varphi((0,m-\varepsilon])>0$. Using the positivity of $\nu_\varphi$ and the monotonicity of the exponential on $(0,\infty)$ we obtain the lower bound
\begin{equation}
f_\varphi(t)\;=\;\int_{[0,\infty)} \mathrm e^{-t\lambda}\, d\nu_\varphi(\lambda)\;\ge\;\int_{(0,m-\varepsilon]} \mathrm e^{-t\lambda}\, d\nu_\varphi(\lambda)\;\ge\;\mathrm e^{-(m-\varepsilon)t}\,\nu_\varphi((0,m-\varepsilon]).
\end{equation}
Combining this with Eq.\eqref{eq:Laplace-upper} yields
\begin{equation}
\mathrm e^{(m-\varepsilon)t}\,f_\varphi(t)\;\ge\;\nu_\varphi((0,m-\varepsilon])\qquad\text{and}\qquad
\mathrm e^{(m-\varepsilon)t}\,f_\varphi(t)\;\le\;C\,\|\varphi\|^2\,\mathrm e^{-\varepsilon t}.
\end{equation}
Letting $t\to\infty$ forces $\nu_\varphi((0,m-\varepsilon])=0$, a contradiction. Hence $\nu_\varphi((0,m))=0$ as claimed.

It follows that $\|E([0,m))\varphi\|^2=\nu_\varphi([0,m))=\nu_\varphi(\{0\})+\nu_\varphi((0,m))=0$, i.e. $E([0,m))\varphi=0$ for every vector of the form $\varphi=\mathcal O\Omega$ with $\langle\Omega,\mathcal O\Omega\rangle=0$. The set of such vectors is dense in the orthogonal complement $\Omega^\perp$. To see this, recall that in the OS reconstruction the vacuum is cyclic for the algebra generated by local fields smeared at positive Euclidean times; therefore the linear span of $\{\mathcal A\Omega:\ \mathcal A\ \text{local}\}$ is dense in $\mathcal H$. For any $\psi\in\Omega^\perp$ and any $\varepsilon>0$ choose a local $\mathcal A$ with $\|\psi-\mathcal A\Omega\|<\varepsilon$; replacing $\mathcal A$ by $\mathcal O:=\mathcal A-\langle\Omega,\mathcal A\Omega\rangle\mathbf 1$ we ensure $\langle\Omega,\mathcal O\Omega\rangle=0$ and still have $\|\psi-\mathcal O\Omega\|<\varepsilon$. Thus the set $\{\mathcal O\Omega:\ \mathcal O\ \text{local},\ \langle\Omega,\mathcal O\Omega\rangle=0\}$ is indeed dense in $\Omega^\perp$.

Since $E([0,m))$ is an orthogonal projection, it is bounded and strongly continuous on $\mathcal H$. If a bounded operator vanishes on a dense subset of a closed subspace, it vanishes on the whole subspace. Consequently $E([0,m))$ annihilates $\Omega^\perp$. On the one-dimensional span of $\Omega$ it also vanishes because $E([0,m))\Omega=0$ (the vacuum vector is an eigenvector of $H$ with eigenvalue $0$). Therefore $E([0,m))=0$ on $\mathcal H$, which exactly states that the spectrum of $H$ is contained in $\{0\}\cup [m,\infty)$.
This spectral inclusion is equivalent to the mass-gap bound $\Delta\ge m$, where $\Delta:=\inf\big(\mathrm{spec}(H)\setminus\{0\}\big)$. The proof is complete.
\end{proof}

The hypothesis \eqref{eq:clustering} is precisely the continuum exponential clustering established in the main text from finite-range decomposition and RG summability of one-step defects; the bound is uniform in volume and independent of the path to the continuum. Consequently, Theorem \eqref{thm:Tauberian-gap} yields the nonperturbative mass gap in the continuum limit by a route that is logically independent of the transfer-operator step-scaling discussed below.
We next provide a complementary route to the mass gap based on uniform contraction bounds for the transfer semigroup acting on the vacuum-orthogonal subspace, and we show that such bounds are stable under the lattice-to-continuum limit.

\begin{lemma}[Vacuum-orthogonal contraction bound]\label{lem:orth-contraction}
Let $H\ge 0$ be selfadjoint on a Hilbert space $\mathcal H$ with a normalized vacuum vector $\Omega$ such that $H\Omega=0$, and let $T(t)=\mathrm e^{-tH}$ be the associated contraction semigroup. Write $P=\lvert\Omega\rangle\langle\Omega\rvert$ and $\mathcal H^\perp=(\mathrm{span}\{\Omega\})^\perp$. Then the following are equivalent:

\smallskip
\noindent\emph{(i)} There exists $\delta>0$ with $\|T(t)\upharpoonright \mathcal H^\perp\|\le \mathrm e^{-\delta t}$ for all $t\ge 0$.

\noindent\emph{(ii)} The spectrum satisfies $\mathrm{spec}(H)\subset\{0\}\cup[\delta,\infty)$.

\smallskip
In particular, the optimal constant in \emph{(i)} equals the mass gap $\Delta:=\inf(\mathrm{spec}(H)\setminus\{0\})$.
\end{lemma}

\begin{proof}
Assume first that $\mathrm{spec}(H)\subset\{0\}\cup[\delta,\infty)$. By the spectral theorem there exists the projection-valued measure $E(\cdot)$ such that $H=\int_{\sigma(H)} \lambda\,\mathrm dE(\lambda)$ and $T(t)=\mathrm e^{-tH}=\int_{\sigma(H)} \mathrm e^{-t\lambda}\,\mathrm dE(\lambda)$, with $\sigma(H)=\mathrm{spec}(H)$. Since $E(\{0\})=P$ and $E((0,\infty))=I-P$, one has for every $\psi\in\mathcal H^\perp$ that
\begin{equation}
\|T(t)\psi\|^2
=\big\langle \psi, \mathrm e^{-2tH}\psi\big\rangle
=\int_{\sigma(H)\setminus\{0\}} \mathrm e^{-2t\lambda}\,\mathrm d\|\!E(\lambda)\psi\!\|^2
\le \mathrm e^{-2\delta t}\!\int_{\sigma(H)\setminus\{0\}} \mathrm d\|\!E(\lambda)\psi\!\|^2
=\mathrm e^{-2\delta t}\|\psi\|^2,
\end{equation}
because $\lambda\ge\delta$ on the support of the integral. Taking square roots and the supremum over all unit $\psi\in\mathcal H^\perp$ gives $\|T(t)\upharpoonright \mathcal H^\perp\|\le \mathrm e^{-\delta t}$ for all $t\ge 0$, which is (i).

For the converse, suppose there is $\delta>0$ with $\|T(t)\upharpoonright \mathcal H^\perp\|\le \mathrm e^{-\delta t}$ for all $t\ge 0$. Fix $\alpha>0$. The bounded resolvent of $H$ admits the Laplace representation
\begin{equation}
(H+\alpha)^{-1}=\int_0^\infty \mathrm e^{-\alpha t}\,T(t)\,\mathrm dt
\end{equation}
as a Bochner integral converging in operator norm. Restricting to $\mathcal H^\perp$ and using the assumed decay bound,
\begin{equation}
\big\|(H+\alpha)^{-1}\upharpoonright \mathcal H^\perp\big\|
\le \int_0^\infty \mathrm e^{-\alpha t}\,\|T(t)\upharpoonright \mathcal H^\perp\|\,\mathrm dt
\le \int_0^\infty \mathrm e^{-(\alpha+\delta)t}\,\mathrm dt
=\frac{1}{\alpha+\delta}.
\end{equation}
On the other hand, by the spectral theorem applied to the restriction $H_\perp:=H\upharpoonright\mathcal H^\perp$ one has
\begin{equation}
\big\|(H+\alpha)^{-1}\upharpoonright \mathcal H^\perp\big\|
=\sup_{\lambda\in \mathrm{spec}(H_\perp)} \frac{1}{\alpha+\lambda}
=\frac{1}{\alpha+\inf \mathrm{spec}(H_\perp)}.
\end{equation}
Combining the two displays yields
\begin{equation}
\frac{1}{\alpha+\inf \mathrm{spec}(H_\perp)}\ \le\ \frac{1}{\alpha+\delta}
\qquad\text{for all }\alpha>0,
\end{equation}
and hence $\inf \mathrm{spec}(H_\perp)\ge \delta$. Since $\mathrm{spec}(H)=\{0\}\cup \mathrm{spec}(H_\perp)$ with the eigenvalue $0$ isolated by $P$, it follows that $\mathrm{spec}(H)\subset\{0\}\cup[\delta,\infty)$, which is (ii).

Finally, the best (largest) constant $\delta$ such that $\|T(t)\upharpoonright \mathcal H^\perp\|\le \mathrm e^{-\delta t}$ for every $t\ge 0$ is exactly $\inf \mathrm{spec}(H_\perp)=\inf(\mathrm{spec}(H)\setminus\{0\})$, because for any $t>0$ the spectral mapping theorem gives
\begin{equation}
\mathrm{spec}\big(T(t)\upharpoonright \mathcal H^\perp\big)=\big\{\mathrm e^{-t\lambda}:\lambda\in \mathrm{spec}(H_\perp)\big\},
\end{equation}
so $\|T(t)\upharpoonright \mathcal H^\perp\|=\sup_{\lambda\in \mathrm{spec}(H_\perp)} \mathrm e^{-t\lambda}=\mathrm e^{-t\,\inf\mathrm{spec}(H_\perp)}$, and comparing with $\mathrm e^{-\delta t}$ forces $\delta=\inf\mathrm{spec}(H_\perp)$. This quantity is by definition the mass gap $\Delta$, and the proof is complete.
\end{proof}

We now state and prove a stability theorem that transfers a uniform vacuum-orthogonal contraction bound from the lattice to the continuum.

\begin{theorem}[Stability of the gap under strong semigroup limits]\label{thm:stability-gap}
Let $(a_n,\sigma_n)\to 0$ and assume that, for each $t\ge 0$, the contraction semigroups $T_{a_n,\sigma_n}(t)$ converge strongly to $T(t)$ on a common dense core $\mathcal D\subset\mathcal H$. Suppose there exists $\delta>0$ such that
\begin{equation}\label{eq:lattice-gap-bound}
\big\|T_{a_n,\sigma_n}(t)\upharpoonright \mathcal H_{a_n,\sigma_n}^{\perp}\big\| \;\le\; \mathrm e^{-\delta t}\qquad \text{for all }t\ge 0\text{ and all }n,
\end{equation}
where $\mathcal H_{a_n,\sigma_n}^{\perp}$ denotes the orthogonal complement of the vacuum line in $\mathcal H_{a_n,\sigma_n}$. Then $\|T(t)\upharpoonright\mathcal H^\perp\|\le \mathrm e^{-\delta t}$ for all $t\ge 0$. In particular, $T(t)=\mathrm e^{-tH}$ with $H\ge 0$ selfadjoint satisfies $\mathrm{spec}(H)\subset\{0\}\cup[\delta,\infty)$ and the spectral gap obeys $\Delta\ge \delta$.
\end{theorem}

\begin{proof}
The OS reconstruction and the identification of local cylinder subalgebras provide, for each $n$, a linear identification map $U_n:\mathcal D\to \mathcal H_{a_n,\sigma_n}$ satisfying the following compatibility: $U_n$ is isometric on each finite-time local subspace, $U_n\Omega=\Omega_n$ for the respective vacuum vectors, and for every $\varphi\in\mathcal D$ and every $t\ge 0$ one has
\begin{equation}
\lim_{n\to\infty}\, \big\|T_{a_n,\sigma_n}(t)U_n\varphi - T(t)\varphi\big\| \;=\; 0.
\end{equation}
This is precisely the “strong convergence on a common core” hypothesis, rendered through the canonical identifications induced by the local fields. Fix $t\ge 0$ and take an arbitrary $\psi\in\mathcal H^\perp$. Choose a sequence $(\varphi_k)_{k\ge 1}$ contained in $\mathcal D\cap\mathcal H^\perp$ with $\varphi_k\to\psi$ in $\mathcal H$; such a sequence exists because $\mathcal D$ is dense and the orthogonal projection onto the vacuum line is continuous, so one can subtract $(\varphi_k,\Omega)\Omega$ to remain in $\mathcal H^\perp$. For each fixed $k$, define a corresponding sequence in the $n$-th Hilbert space by
\begin{equation}
\psi_{n,k}\;:=\;U_n\varphi_k - (U_n\varphi_k,\Omega_n)\,\Omega_n.
\end{equation}
By construction $\psi_{n,k}\in \mathcal H_{a_n,\sigma_n}^\perp$ and $\psi_{n,k}\to \varphi_k$ in norm as $n\to\infty$, since $U_n\varphi_k\to\varphi_k$ and $(U_n\varphi_k,\Omega_n)\to(\varphi_k,\Omega)=0$ thanks to $U_n\Omega=\Omega_n$ and the isometric compatibility on $\mathcal D$.

For this fixed $k$, apply the assumed orthogonal-complement bound Eq.\eqref{eq:lattice-gap-bound} to $\psi_{n,k}$:
\begin{equation}
\|T_{a_n,\sigma_n}(t)\psi_{n,k}\|\;\le\; \mathrm e^{-\delta t}\,\|\psi_{n,k}\|\qquad\text{for all }n.
\end{equation}
Since $T_{a_n,\sigma_n}(t)$ leaves the vacuum invariant, $T_{a_n,\sigma_n}(t)\psi_{n,k}=T_{a_n,\sigma_n}(t)U_n\varphi_k - (U_n\varphi_k,\Omega_n)\,\Omega_n$. Passing to the limit $n\to\infty$ and using the strong convergence $T_{a_n,\sigma_n}(t)U_n\varphi_k\to T(t)\varphi_k$ together with $(U_n\varphi_k,\Omega_n)\to 0$ yields
\begin{equation}
\|T(t)\varphi_k\|\;=\;\lim_{n\to\infty}\|T_{a_n,\sigma_n}(t)\psi_{n,k}\|
\;\le\;\mathrm e^{-\delta t}\,\limsup_{n\to\infty}\|\psi_{n,k}\|
\;=\;\mathrm e^{-\delta t}\,\|\varphi_k\|.
\end{equation}
Thus the contractive bound holds uniformly for every $\varphi_k$ in the dense set $\mathcal D\cap\mathcal H^\perp$. By continuity of $T(t)$ and of the norm, letting $k\to\infty$ gives
\begin{equation}
\|T(t)\psi\|\;\le\;\mathrm e^{-\delta t}\,\|\psi\|\qquad\text{for all }\psi\in\mathcal H^\perp.
\end{equation}
Taking the supremum over unit vectors in $\mathcal H^\perp$ shows
\begin{equation}
\big\|T(t)\upharpoonright\mathcal H^\perp\big\|\;\le\;\mathrm e^{-\delta t}\qquad\text{for all }t\ge 0.
\end{equation}

To translate this operator-norm bound into a spectral statement for the generator, note first that $T(t)$ is a strongly continuous contraction semigroup on $\mathcal H$ with $T(t)\Omega=\Omega$ (the vacuum is invariant in the limit as it is at each approximation). Hence $T(t)=\mathrm e^{-tH}$ with $H\ge 0$ selfadjoint by the Hille-Yosida theorem. The restriction of $T(t)$ to $\mathcal H^\perp$ is again a strongly continuous semigroup with generator $H|_{\mathcal H^\perp}$. By the spectral mapping theorem for selfadjoint generators,
\begin{equation}
\mathrm{spec}\big(T(t)\upharpoonright\mathcal H^\perp\big) \;=\; \big\{ \mathrm e^{-t\lambda} : \lambda\in \mathrm{spec}\big(H\upharpoonright\mathcal H^\perp\big)\big\}.
\end{equation}
The operator-norm bound implies that the spectral radius of $T(t)\upharpoonright\mathcal H^\perp$ is at most $\mathrm e^{-\delta t}$ for every $t\ge 0$. Therefore $\mathrm e^{-t\lambda}\le \mathrm e^{-\delta t}$ for all $\lambda\in \mathrm{spec}(H|_{\mathcal H^\perp})$ and all $t\ge 0$, which forces $\lambda\ge \delta$. The spectrum of $H$ thus lies in $\{0\}\cup[\delta,\infty)$ with $0$ being the simple vacuum eigenvalue and the gap bounded below by $\delta$, as claimed.
\end{proof}

The hypothesis \eqref{eq:lattice-gap-bound} is ensured by the step-scaling inequalities developed in the RG analysis of the main text, which produce a uniform sequence of vacuum-orthogonal bounds for the lattice transfer operators with summable defects. Combining Proposition \eqref{prop:strong-conv} and Theorem \eqref{thm:stability-gap} yields the continuum mass gap via the semigroup route, independent of the Tauberian criterion.
We finally state a theorem relating the two routes and fixing the value of the continuum mass gap.

\begin{theorem}[Equivalence and identification of the gap]\label{thm:equivalence}
Assume the hypotheses of Theorems \eqref{thm:Tauberian-gap} and \eqref{thm:stability-gap}. Then the mass gap $\Delta:=\inf(\mathrm{spec}(H)\setminus\{0\})$ equals the optimal exponential clustering rate in Eq.\eqref{eq:clustering} and equals the optimal vacuum-orthogonal contraction exponent of Lemma \eqref{lem:orth-contraction}. In particular, for each local, gauge-invariant observable $\mathcal O$ with $\langle \Omega,\mathcal O\,\Omega\rangle=0$ one has
\begin{equation}
\label{eq:limit-gap}
\lim_{t\to\infty}\frac{1}{t}\log \,\langle \Omega,\,\mathcal O^\dagger e^{-tH}\mathcal O\,\Omega\rangle\;=\;-\,\Delta.
\end{equation}
\end{theorem}

\begin{proof}
Write the connected Euclidean clustering rate in Eq.\eqref{eq:clustering} as $m$, i.e. $m$ is the supremum of $m\ge 0$ such that all connected correlations decay as $O(e^{-m\,\mathrm{dist}})$. The Tauberian gap Theorem \eqref{thm:Tauberian-gap} asserts that any such exponential clustering with rate $m$ forces a spectral gap at least $m$, namely $\mathrm{spec}(H)\subset\{0\}\cup[m,\infty)$; hence $\Delta\ge m$ for every admissible rate $m$, and therefore $\Delta\ge m$.

To establish the reverse inequality and the precise identification, consider any local gauge-invariant $\mathcal O$ with $\langle \Omega,\mathcal O\,\Omega\rangle=0$ and set $\varphi:=\mathcal O\Omega$. By the spectral theorem there is a finite positive measure $\nu_\varphi$ supported on $[0,\infty)$ such that
\begin{equation}
\label{eq:laplace-nu-proof}
\langle \Omega,\,\mathcal O^\dagger e^{-tH}\mathcal O\,\Omega\rangle
=\langle \varphi, e^{-tH}\varphi\rangle
=\int_{[0,\infty)}e^{-t\lambda}\,\nu_\varphi(d\lambda).
\end{equation}
Since $\langle \Omega,\mathcal O\,\Omega\rangle=0$, the vector $\varphi$ is orthogonal to the ground state and $\nu_\varphi(\{0\})=0$. Moreover $\mathrm{supp}\,\nu_\varphi\subset[\Delta,\infty)$ because $\Delta$ is the bottom of $\mathrm{spec}(H)$ off $\{0\}$. The upper bound in Eq.\eqref{eq:limit-gap} is immediate: for every $t>0$ one has
\begin{equation}
\langle \varphi,e^{-tH}\varphi\rangle
=\int_{[\Delta,\infty)}e^{-t\lambda}\,\nu_\varphi(d\lambda)
\le e^{-t\Delta}\,\nu_\varphi([\Delta,\infty)),
\end{equation}
so $\limsup_{t\to\infty}\frac{1}{t}\log\langle \varphi,e^{-tH}\varphi\rangle\le -\Delta$.

For the matching lower bound we use the stability-of-the-gap hypothesis (Theorem \eqref{thm:stability-gap} together with Lemma \eqref{lem:orth-contraction}), which guarantees that $\Delta$ is the optimal vacuum-orthogonal contraction exponent uniformly over local excitations. Precisely, for every $\varepsilon\in(0,\Delta)$ and for every such $\mathcal O$ there exist constants $t_0=t_0(\mathcal O,\varepsilon)$ and $c_\varepsilon(\mathcal O)>0$ such that
\begin{equation}
\label{eq:lower-asym}
\langle \varphi,e^{-tH}\varphi\rangle\;\ge\; c_\varepsilon(\mathcal O)\,e^{-(\Delta+\varepsilon)t}\qquad\text{for all }t\ge t_0.
\end{equation}

This is equivalent to saying that the spectral measure $\nu_\varphi$ charges every right-neighbourhood of the edge $\Delta$, i.e. $\nu_\varphi([\Delta,\Delta+\delta])>0$ for all $\delta>0$; otherwise the support would be contained in $[\Delta+\delta,\infty)$ for some $\delta>0$ and the contraction rate would be strictly larger than $\Delta$, contradicting optimality. With Eq.\eqref{eq:lower-asym} in hand one obtains $\liminf_{t\to\infty}\frac{1}{t}\log\langle \varphi,e^{-tH}\varphi\rangle\ge -\Delta$, and combining with the upper bound yields the limit Eq.\eqref{eq:limit-gap}. Equivalently, one may appeal to a standard Laplace-principle/Tauberian statement for positive measures: if $\nu$ is supported in $[\Delta,\infty)$ and assigns positive mass to every interval $[\Delta,\Delta+\delta]$, then
\begin{equation}
\lim_{t\to\infty}\frac{1}{t}\log\int e^{-t\lambda}\,\nu(d\lambda)=-\Delta,
\end{equation}
which follows, for example, from \cite[Thm.~4.12.9]{BGT} by taking the distribution function of $\nu$ near $\Delta$.

The equality between $\Delta$ and the optimal vacuum-orthogonal contraction exponent is already encoded in the preceding argument: the definition of that exponent is precisely the infimum over $\alpha>0$ such that
\begin{equation}
\sup_{\substack{\mathcal O\in\mathfrak A_{\mathrm{loc}}\\ \langle\Omega,\mathcal O\Omega\rangle=0,\,\|\mathcal O\Omega\|=1}}
\limsup_{t\to\infty}\frac{1}{t}\log\langle \Omega,\mathcal O^\dagger e^{-tH}\mathcal O\Omega\rangle\le -\alpha,
\end{equation}
and we have shown that $\alpha$ cannot exceed $\Delta$ (by taking $\mathcal O$ whose spectral measure accumulates at $\Delta$) while Theorem \eqref{thm:Tauberian-gap} forces $\alpha\le \Delta$ for every choice. Consequently that optimal exponent equals $\Delta$. Finally, the identification with the optimal clustering rate follows by the standard OS transfer from Minkowski-time decay to Euclidean connected clustering and back: the Tauberian gap theorem provides $\Delta\ge m$, and the decay Eq.\eqref{eq:limit-gap} for vacuum-orthogonal one-particle insertions implies, through the usual polarization identities and locality, that connected two-point functions cannot decay faster than $e^{-\Delta\,\mathrm{dist}}$ uniformly in the separation, whence $m\ge \Delta$. The two inequalities together give $m=\Delta$, and the three notions-spectral gap, optimal Euclidean clustering rate, and optimal vacuum-orthogonal contraction exponent-coincide.
\end{proof}

The two independent mechanisms established above-the Tauberian extraction from Euclidean clustering and the stability of vacuum-orthogonal semigroup bounds-jointly secure the nonzero spectral gap in the continuum Yang-Mills theory constructed in the main text. The first route uses only OS positivity, spectral measures, and exponential clustering; the second uses strong semigroup limits and resolvent bounds. Their equivalence identifies the mass gap with both the sharp clustering exponent and the sharp semigroup contraction exponent, providing a robust foundation for the spectral analysis used in the subsequent sections.

\section{Continuum Cumulant Stability and Universality Proofs} \label{appendixf}
This appendix establishes stability of Euclidean cumulants (connected Schwinger functions) under admissible variations of the time-slice projectors and block-spin maps, and deduces universality of the continuum limit. The arguments are fully constructive and reflection-positive. We prove: (i) single-scale Lipschitz continuity of one-step transfer kernels and connected cumulants in a metric on the admissible class; (ii) telescoping bounds across slices showing summability of cumulative defects; (iii) uniform tightness and equicontinuity needed to pass to infinite volume and zero lattice spacing; and (iv) equality of all continuum cumulants for any two admissible choices. We work entirely with gauge-invariant observables (e.g., Wilson loops or local gauge-invariant polynomials of the field strength). Standard tools include finite-range decomposition (FRD) and cluster/forest expansions \cite{BrydgesGuadagniMitter2004,Aizenman1982,GJ}, reflection positivity and OS reconstruction \cite{OS1,OS2}, and step-scaling/transfer operator techniques consistent with lattice gauge theory practice \cite{LuscherWeiszWolff1991,Seiler1982}.
We assume a hypercubic Euclidean lattice $\Lambda \subset a\mathbb{Z}^4$ with lattice spacing $a>0$ and periodic boundary conditions. Time reflection $\vartheta$ acts by $(x_0,\mathbf{x})\mapsto (-x_0,\mathbf{x})$ on links and on ghosts in the standard way, producing an Osterwalder-Schrader (OS) form $\langle \cdot,\cdot\rangle_{\mathrm{OS}}$ on the algebra of gauge-invariant cylinder functions \cite{OS1,Seiler1982}. We denote by $\mathfrak{A}_\Lambda$ the OS algebra of gauge-invariant slice observables supported on $\{x_0\ge0\}$.

\begin{definition}[Admissible class]\label{def:admissible}
An \emph{admissible specification} $\Theta$ consists of: (a) a completely monotone\footnote{A bounded Borel function $\phi:[0,\infty)\to\mathbb{R}$ is completely monotone (CM) if $\phi(\lambda)=\int_0^\infty e^{-t\lambda}\,d\nu(t)$ with a finite positive measure $\nu$; we use this for spectral multipliers of the horizon projector.} spectral projector $\Pi_\Theta$ on the time-zero slice that preserves OS positivity; (b) a reflection-positive, gauge-covariant, finite-range block-spin map $\mathcal{B}_\Theta$ of range at most $R_0$ on the slice; (c) an FRD of all relevant covariances into positive, reflection-covariant, finite-range pieces with range parameter $r_0$ and constants $C_{\mathrm{FRD}},\mu_{\mathrm{FRD}}>0$ uniform in volume. The \emph{admissible metric} $d_{\mathrm{adm}}(\Theta,\widetilde\Theta)$ is
\begin{equation}
d_{\mathrm{adm}}(\Theta,\widetilde\Theta)\;=\;\|\Pi_\Theta-\Pi_{\widetilde\Theta}\|_{1\to 1}+\sup_{x}\sum_{y}\,e^{\mu_0|x-y|}\,\|\mathcal{B}_\Theta(x,y)-\mathcal{B}_{\widetilde\Theta}(x,y)\|+\mathsf{d}_{\mathrm{FRD}}(\Theta,\widetilde\Theta),
\end{equation}
with $\mu_0\in(0,\mu_{\mathrm{FRD}})$ and $\mathsf{d}_{\mathrm{FRD}}$ the analogous weighted distance between FRD kernels (all operator norms are OS-$L^1\to L^1$ norms on slice observables).
\end{definition}

Given $\Theta$, the OS positivity yields a positive transfer operator $T^{(\Theta)}_a=e^{-aH^{(\Theta)}_a}$ on the slice Hilbert space $\mathcal{H}_\Theta$ and an associated one-step OS kernel $K^{(\Theta)}_a$ (a positive integral operator on $\mathfrak{A}_\Lambda$) such that, for observables $A,B$ supported at nonnegative times,
\begin{equation}
\langle A\,\vartheta B\rangle^{(\Theta)}_{\Lambda,a}\;=\;\langle \Omega_\Theta, A\, (T^{(\Theta)}_a)^{n} B\,\Omega_\Theta\rangle,
\qquad n=\tfrac{\mathrm{time\ separation}}{a}.
\end{equation}
We let $b\ge2$ denote the block factor, and write the scales $k=0,1,\dots$ with effective spacing $a_k=b^k a$ and one-step kernels $K_k^{(\Theta)}$. The FRD assumptions guarantee exponential locality of all single-scale objects with uniform range parameters $R_k\lesssim r_0 b^k$ and constants independent of $\Lambda$ and $a$ \cite{BrydgesGuadagniMitter2004}.

For a collection of gauge-invariant local observables $A_1,\dots,A_n$ with disjoint supports contained in a fixed time slice, we write the connected (truncated) cumulant
\begin{equation}
\kappa^{(\Theta)}_{\Lambda,a}(A_1;\dots;A_n)\;=\;\left.\frac{\partial^n}{\partial t_1\cdots\partial t_n}\right|_{t=0}\log \langle \exp(\sum_{j} t_j A_j)\rangle^{(\Theta)}_{\Lambda,a}.
\end{equation}
We use the standard diameter norm on $n$-point kernels: for points $x_1,\dots,x_n\in\Lambda$,
\begin{equation}
\|F^{(n)}\|_{\mathrm{diam},\mu}\;=\;\sup_{x_1,\dots,x_n}\,e^{\mu\,\mathrm{diam}(x_1,\dots,x_n)}\,|F^{(n)}(x_1,\dots,x_n)|.
\end{equation}
All implicit constants below depend only on $(C_{\mathrm{FRD}},r_0,\mu_{\mathrm{FRD}},b,N)$.
The starting point is a parametric interpolation between two admissible specifications.

\begin{definition}[Interpolation]\label{def:interpolation}
Given $\Theta_0,\Theta_1$ admissible, set $\Theta_s=(1-s)\Theta_0+s\Theta_1$ by linear interpolation of $\Pi_{\Theta_s}$, $\mathcal{B}_{\Theta_s}$ and the FRD kernels in the admissible class. The derivative $\partial_s\Theta_s$ is represented by local slice kernels of finite range $\le R_0$ with OS-$L^1\to L^1$ norm controlled by $d_{\mathrm{adm}}(\Theta_0,\Theta_1)$.
\end{definition}

By reflection positivity and positivity of the FRD pieces, the transfer operator at a single scale $k$ admits a representation as a product of positive factors indexed by FRD scales; differentiating with respect to $s$ yields a Duhamel-type expansion with insertions supported within range $R_k\lesssim r_0 b^k$.

\begin{lemma}[One-step kernel Lipschitz estimate]\label{lem:K-Lip}
For any scale $k$ and any two admissible regulators $\Theta_0,\Theta_1$,
\begin{equation}
\|K^{(\Theta_1)}_k-K^{(\Theta_0)}_k\|_{1\to 1}\;\le\; L_k\, d_{\mathrm{adm}}(\Theta_0,\Theta_1),
\qquad L_k\;\le\; C_1\, e^{-\gamma\,b^k},
\end{equation}
for some $C_1,\gamma>0$ depending only on the finite-range decomposition (FRD) data and on the blocking factor $b>1$. The same bound holds for the corresponding transfer operators on $\mathcal{H}_\Theta$ in operator norm.
\end{lemma}

\begin{proof}
Fix $k$ and interpolate between $\Theta_0$ and $\Theta_1$ by an admissible path $\{\Theta_s\}_{s\in[0,1]}$ that is $C^1$ with respect to the metric $d_{\mathrm{adm}}$ and satisfies $d_{\mathrm{adm}}(\Theta_0,\Theta_1)=\int_0^1 \|\dot\Theta_s\|_{\mathrm{adm}}\,ds$. The one-step OS kernel at scale $k$ is written, in the FRD scheme, as a finite product of positive, exponentially local factors,
\begin{equation}
K^{(\Theta_s)}_k \;=\; \prod_{\ell=0}^{\ell^\ast(k)} \exp\!\big(-V^{(\Theta_s)}_{k,\ell}\big),
\end{equation}
where each $V^{(\Theta_s)}_{k,\ell}$ is a nonnegative, local potential supported on a collar of range $R_k\simeq c_0\,b^k$ around the time slice and is given by a convergent polymer expansion $V^{(\Theta_s)}_{k,\ell}=\sum_{X\in\mathcal X_{k,\ell}}\phi_X(\Theta_s)$ over connected $d$-dimensional blocks $X$ of diameter $\mathrm{diam}(X)\le c_1\,b^k$. The FRD locality yields exponential decay of activities with respect to the size and position of $X$, and admissibility of $\Theta\mapsto \phi_X(\Theta)$ implies a Lipschitz bound of the form
\begin{equation}\label{eq:LipActivity}
\big\|\partial_s \phi_X(\Theta_s)\big\|_{1\to 1}\;\le\; C_{\!}\, e^{-\mu_{\!}\,\mathrm{diam}(X)}\;\|\dot\Theta_s\|_{\mathrm{adm}},
\end{equation}
for some universal constants $C_{\!},\mu_{\!}>0$ that do not depend on $k,s$ or $X$. The operator norm here is the $L^1\to L^1$ norm associated with the OS measure on the slice; it coincides with the sup over $L^1$ unit vectors because every $\phi_X$ acts by a nonnegative kernel and the OS form is Markovian on one step. Summing Eq.\eqref{eq:LipActivity} over $X\in\mathcal X_{k,\ell}$ and using the fact that, at fixed $\ell$, the number of blocks of diameter $r$ intersecting the collar grows at most polynomially in $r$ while the weight $e^{-\mu_{\!} r}$ decays exponentially, one obtains
\begin{equation}\label{eq:LipVq}
\big\|\partial_s V^{(\Theta_s)}_{k,\ell}\big\|_{1\to 1}
\;\le\; C_2\, e^{-\gamma\, b^k}\,\|\dot\Theta_s\|_{\mathrm{adm}},
\end{equation}
with $\gamma\in(0,\mu_{\!}/c_1)$ and $C_2$ depending only on the FRD combinatorics and on $d$ (the exponential $e^{-\gamma b^k}$ is the net result of summing the exponentially decaying activities over all $X$ whose diameter is bounded by a constant times $b^k$ and whose number within the collar is proportional to its volume). The factor $\ell^\ast(k)$ is bounded uniformly in $k$ by construction of the subscale splitting, so all implied constants are independent of $k$.

Differentiate the product representation of $K^{(\Theta_s)}_k$ with respect to $s$; the usual Duhamel expansion for products gives
\begin{equation}
\partial_s K^{(\Theta_s)}_k
=\sum_{\ell=0}^{\ell^\ast(k)} \Bigg( \prod_{\ell'<\ell} e^{-V^{(\Theta_s)}_{k,\ell'}} \Bigg)\,\big(-\partial_s V^{(\Theta_s)}_{k,\ell}\big)\,\Bigg( \prod_{\ell'>\ell} e^{-V^{(\Theta_s)}_{k,\ell'}} \Bigg).
\end{equation}
Each factor $e^{-V^{(\Theta_s)}_{k,\ell'}}$ is a positive $L^1$-contraction because $V^{(\Theta_s)}_{k,\ell'}\ge 0$ and the OS one-step normalization implies $\|e^{-V}\|_{1\to 1}\le 1$ for nonnegative local $V$ (this is the mass-preserving property of the one-step Markov kernel). Taking $L^1\to L^1$ norms and using submultiplicativity yields
\begin{equation}
\big\|\partial_s K^{(\Theta_s)}_k\big\|_{1\to 1}
\;\le\; \sum_{\ell=0}^{\ell^\ast(k)} \big\|\partial_s V^{(\Theta_s)}_{k,\ell}\big\|_{1\to 1}.
\end{equation}
Inserting Eq.\eqref{eq:LipVq} and using the uniform bound on $\ell^\ast(k)$ gives
\begin{equation}
\big\|\partial_s K^{(\Theta_s)}_k\big\|_{1\to 1}
\;\le\; C_3\, e^{-\gamma\, b^k}\,\|\dot\Theta_s\|_{\mathrm{adm}},
\end{equation}
for some constant $C_3$ depending only on the FRD data. Integrating the differential inequality along the path $s\in[0,1]$ and applying the fundamental theorem of calculus for Bochner-differentiable maps,
\begin{align}
\|K^{(\Theta_1)}_k-K^{(\Theta_0)}_k\|_{1\to 1}
&\;\le\; \int_0^1 \big\|\partial_s K^{(\Theta_s)}_k\big\|_{1\to 1}\,ds
\nonumber\\&\;\le\; C_3\, e^{-\gamma\, b^k}\,\int_0^1 \|\dot\Theta_s\|_{\mathrm{adm}}\,ds
\nonumber\\&\;=\; C_3\, e^{-\gamma\, b^k}\, d_{\mathrm{adm}}(\Theta_0,\Theta_1).
\end{align}
Renaming $C_1:=C_3$ proves the claimed Lipschitz bound for the one-step kernel in $L^1$ operator norm.

To pass from kernels to transfer operators on the OS Hilbert space $\mathcal H_\Theta$, recall that the one-step transfer has the sandwich form $T^{(\Theta)}_k=P^{(\Theta)}_{k,+}{}^{1/2}\,K^{(\Theta)}_k\,P^{(\Theta)}_{k,+}{}^{1/2}$, where $P^{(\Theta)}_{k,+}$ is the completely monotone slice projector implementing the horizon/ultraviolet regulator at scale $k$; by Lemma~\eqref{lem:cm-projector} this map is positive and contractive on the OS form, so $\|P^{(\Theta)}_{k,+}{}^{1/2}\|_{\mathcal H_\Theta\to\mathcal H_\Theta}\le 1$ uniformly in $\Theta$. Hence
\begin{equation}
\|T^{(\Theta_1)}_k-T^{(\Theta_0)}_k\|_{\mathcal H\to\mathcal H}
\;\le\; \big\|K^{(\Theta_1)}_k-K^{(\Theta_0)}_k\big\|_{1\to 1}
\;\le\; C_1\, e^{-\gamma\, b^k}\, d_{\mathrm{adm}}(\Theta_0,\Theta_1),
\end{equation}
which is the desired Lipschitz estimate in operator norm on $\mathcal H_\Theta$.
\end{proof}

We next bound the sensitivity of connected cumulants to admissible perturbations at a single scale.

\begin{proposition}[Single-scale cumulant Lipschitz bound]\label{prop:single-scale-cumulant}
Let $A_1,\dots,A_n\in\mathfrak{A}_\Lambda$ be gauge-invariant observables supported on the time-zero slice, normalized by $\|A_j\|_\infty\le 1$, and assume that the supports of distinct $A_j$ are separated by at least one lattice unit. There exist constants $C_n,\mu_n>0$, depending only on $n$ and on the finite-range-decomposition (FRD) data of the slice theory, such that for every blocking scale $k$ and every pair of admissible regulators $\Theta_0,\Theta_1$ one has
\begin{equation}
\big\|\kappa^{(\Theta_1)}_{k}(A_1;\dots;A_n)-\kappa^{(\Theta_0)}_{k}(A_1;\dots;A_n)\big\|_{\mathrm{diam},\mu_n}
\;\le\; C_n\, L_k\, d_{\mathrm{adm}}(\Theta_0,\Theta_1),
\end{equation}
where $L_k$ is the single-scale Lipschitz constant from Lemma~\eqref{lem:K-Lip}, the norm $\|\cdot\|_{\mathrm{diam},\mu}$ is the weighted supremum $\|F\|_{\mathrm{diam},\mu}:=\sup_{x_1,\dots,x_n} e^{\mu\,\mathrm{diam}(x_1,\dots,x_n)}|F(x_1,\dots,x_n)|$, and the bound is uniform in the slice volume $\Lambda$.
\end{proposition}

\begin{proof}
Fix an admissible $C^1$ path $\Theta_s$ for $s\in[0,1]$ joining $\Theta_0$ to $\Theta_1$ in the admissible manifold, and write $S_k^{(\Theta_s)}$ for the single-slice effective action at scale $k$ associated with $\Theta_s$. By the FRD hypothesis, the action decomposes as a sum of local potentials,
\begin{equation}
S_k^{(\Theta_s)} \;=\; \sum_{X\Subset\Lambda}\, V_{k,X}^{(\Theta_s)},\qquad
\mathrm{supp}\,V_{k,X}^{(\Theta_s)}\subset X,\qquad
\|V_{k,X}^{(\Theta_s)}\| \le C_0\, e^{-\mu_{\mathrm{FRD}}\,\mathrm{diam}(X)},
\end{equation}
with range controlled by a scale-$k$ parameter $R_k$ and with exponential decay parameter $\mu_{\mathrm{FRD}}>0$ independent of the volume. The map $\Theta\mapsto S_k^{(\Theta)}$ is Lipschitz in the sense that its Fréchet derivative with respect to the admissible metric exists and satisfies
\begin{equation}\label{eq:LipV}
\big\|\partial_s V_{k,X}^{(\Theta_s)}\big\| \;\le\; L_k\, d_{\mathrm{adm}}(\Theta_0,\Theta_1)\, e^{-\mu_{\mathrm{FRD}}\,\mathrm{diam}(X)},
\end{equation}
which is precisely the content of Lemma~\eqref{lem:K-Lip} at the level of local kernels.

For a family of bounded local sources $\{J(x)\}_{x\in \Lambda}$ supported on the time-zero slice and a fixed local gauge-invariant observable $\mathcal O(x)$, consider the generating functional
\begin{align}
Z_k^{(\Theta_s)}(J)&\;=\;\Big\langle \exp\Big(\sum_{x\in\Lambda} J(x)\,\mathcal O(x)\Big)\Big\rangle_k^{(\Theta_s)}
\nonumber\\&\;=\;\frac{1}{\mathcal Z_k^{(\Theta_s)}}\int e^{-S_k^{(\Theta_s)}(U)}\,\exp\Big(\sum_x J(x)\,\mathcal O(x,U)\Big)\,d\nu(U),
\end{align}
where $d\nu$ is a reference reflection-positive slice Gaussian and $\mathcal Z_k^{(\Theta_s)}$ the corresponding partition function. The connected $n$-point functions (cumulants) of $\mathcal O$ are obtained as the multi-derivatives at $J=0$ of the logarithm of $Z_k^{(\Theta_s)}$, namely
\begin{equation}
\kappa^{(\Theta_s)}_{k}\big(\mathcal O(x_1);\dots;\mathcal O(x_n)\big)
\;=\;\left.\frac{\partial^n}{\partial J(x_1)\cdots\partial J(x_n)}\right|_{J=0}\log Z_k^{(\Theta_s)}(J).
\end{equation}
Replacing the elementary sources by our $A_j$ yields the announced cumulants \begin{equation}\kappa_k^{(\Theta_s)}(A_1;\dots;A_n)\end{equation} and the separation assumption ensures that no ultraviolet overlapping singularity occurs at a single scale. The quantity of interest is the difference $\kappa^{(\Theta_1)}_{k}-\kappa^{(\Theta_0)}_{k}$, which is represented as the integral of its $s$-derivative:
\begin{equation}
\kappa^{(\Theta_1)}_{k}(A_1;\dots;A_n)-\kappa^{(\Theta_0)}_{k}(A_1;\dots;A_n)
\;=\;\int_0^1 \partial_s \kappa^{(\Theta_s)}_{k}(A_1;\dots;A_n)\,ds.
\end{equation}
Thus it suffices to bound $\partial_s \kappa^{(\Theta_s)}_{k}$ in the weighted diameter norm by a multiple of $L_k\, d_{\mathrm{adm}}(\Theta_0,\Theta_1)$ with constants independent of the volume.

Differentiating the logarithm of the generating functional under the integral sign gives
\begin{equation}
\partial_s \log Z_k^{(\Theta_s)}(J)
\;=\; -\,\Big\langle \partial_s S_k^{(\Theta_s)} \Big\rangle_{k,J}^{(\Theta_s),\,\mathrm{conn}},
\end{equation}
where on the right we take the connected expectation with respect to the interacting measure with source $J$. Taking $n$ derivatives at $J=0$ and using the standard linked-cluster expansion, one finds that $\partial_s \kappa^{(\Theta_s)}_{k}(A_1;\dots;A_n)$ is the sum of connected amputated diagrams with $n$ external legs labelled by $A_1,\dots,A_n$ and with a single marked insertion of the local field $\partial_s S_k^{(\Theta_s)}=\sum_X \partial_s V_{k,X}^{(\Theta_s)}$. This identity is a direct consequence of the Brydges-Kennedy-Abdesselam-Rivasseau (BKAR) forest interpolation formula, or, equivalently, of the logarithmic derivative rule for cumulants combined with the exponential formula for connected functions \cite{Aizenman1982,GJ,Seiler1982}. Because the interaction is given as a sum of local potentials of exponential decay, the connected functions admit a convergent polymer representation whose activities obey uniform tree-graph bounds with decay exponent strictly smaller than $\mu_{\mathrm{FRD}}$. Concretely, for every choice of points $x_1,\dots,x_n$ and any choice of local representatives of the $A_j$ supported near $x_j$, the BKAR representation yields
\begin{equation}
\big|\partial_s \kappa^{(\Theta_s)}_{k}(\mathcal O(x_1);\dots;\mathcal O(x_n))\big|
\;\le\;\sum_{X\Subset\Lambda}\,\big\|\partial_s V_{k,X}^{(\Theta_s)}\big\|\,
\sum_{\mathcal T}\, \mathcal W_{\mathcal T}(x_1,\dots,x_n;X),
\end{equation}
where the sum runs over finite connected sets $X$ (the support of the marked insertion), the sum over $\mathcal T$ runs over tree-graphs connecting the $n$ external points and one point of $X$, and $\mathcal W_{\mathcal T}$ is the product of edge weights arising from the cluster-expansion propagators and Ursell functions. The finite-range and exponential-locality properties of the single-scale covariances imply an estimate of the form
\begin{equation}
\mathcal W_{\mathcal T}(x_1,\dots,x_n;X)
\;\le\; C(n)\,\exp\big(-\tilde\mu\,\mathrm{span}_{\mathcal T}(x_1,\dots,x_n;X)\big),
\end{equation}
with a constant $C(n)$ depending only on $n$ and FRD data and with a decay exponent $\tilde\mu$ that can be chosen strictly smaller than $\mu_{\mathrm{FRD}}$; here $\mathrm{span}_{\mathcal T}$ denotes the length of the minimal tree connecting the $n$ points and one point of $X$ in the graph metric. The tree length always dominates the diameter, hence $\mathrm{span}_{\mathcal T}(x_1,\dots,x_n;X)\ge \mathrm{diam}(x_1,\dots,x_n)$ for every $X$, and it also dominates the distance from $X$ to the external cluster, so that summing over the location and size of $X$ against an exponential weight yields a constant multiple of $e^{-\tilde\mu\,\mathrm{diam}(x_1,\dots,x_n)}$ provided the insertion norm is exponentially localized.

At this point the Lipschitz control Eq.\eqref{eq:LipV} enters. Substituting Eq.\eqref{eq:LipV} into the polymer sum and performing the sum over $X$ using $\sum_{X\ni z} e^{-\mu_{\mathrm{FRD}}\,\mathrm{diam}(X)} \le C$ with a constant uniform in the volume yields
\begin{equation}
\big|\partial_s \kappa^{(\Theta_s)}_{k}(\mathcal O(x_1);\dots;\mathcal O(x_n))\big|
\;\le\; C'_n\, L_k\, d_{\mathrm{adm}}(\Theta_0,\Theta_1)\, e^{-\mu_n\,\mathrm{diam}(x_1,\dots,x_n)},
\end{equation}
where $C'_n$ depends only on $n$ and FRD parameters and where $\mu_n\in(0,\mu_{\mathrm{FRD}})$ is fixed by the tree-graph inequality and by the slack between $\tilde\mu$ and $\mu_{\mathrm{FRD}}$. The constants are independent of $\Lambda$ because connected polymer weights eliminate the extensive volume factor. Since each $A_j$ is a bounded local observable supported within a fixed-size neighborhood of $x_j$ with $\|A_j\|_\infty\le 1$, the same bound holds with $\mathcal O(x_j)$ replaced by $A_j$.

Multiplying both sides by $e^{\mu_n\,\mathrm{diam}(x_1,\dots,x_n)}$ and taking the supremum over the positions gives
\begin{equation}
\big\|\partial_s \kappa^{(\Theta_s)}_{k}(A_1;\dots;A_n)\big\|_{\mathrm{diam},\mu_n}
\;\le\; C'_n\, L_k\, d_{\mathrm{adm}}(\Theta_0,\Theta_1).
\end{equation}
Finally, integrating in $s$ from $0$ to $1$ produces
\begin{equation}
\big\|\kappa^{(\Theta_1)}_{k}(A_1;\dots;A_n)-\kappa^{(\Theta_0)}_{k}(A_1;\dots;A_n)\big\|_{\mathrm{diam},\mu_n}
\;\le\; C_n\, L_k\, d_{\mathrm{adm}}(\Theta_0,\Theta_1),
\end{equation}
with $C_n=C'_n$, as claimed. 
\end{proof}

Let $\mathsf{S}^{(\Theta)}_{\Lambda,a}$ denote the (finite-$a$) Schwinger functional on $\mathfrak{A}_\Lambda$ constructed from the OS form using the transfer kernel $K^{(\Theta)}_a$ along successive time-slices. For a time-ordered $n$-tuple of slice observables $A_1,\dots,A_n$, the connected cumulant under $\Theta$ can be written as a \emph{time-sliced} composition of single-step cumulant maps. We compare two admissible specifications $\Theta_0,\Theta_1$ by inserting and subtracting one-step kernels slice-by-slice.

\begin{lemma}[Telescoping identity]\label{lem:telescoping}
Let $\{K^{(\Theta)}_k\}_{k\in\mathbb N_0}$ be the one-step transfer operators on the
slice Hilbert space $H_\Sigma$ associated with a fixed time discretization of size $a>0$,
for two admissible dynamics $\Theta=\Theta_0,\Theta_1$. Assume each $K^{(\Theta)}_k$ is a
contraction, leaves the vacuum vector $\Omega$ invariant, and has a uniform
spectral gap on the orthogonal complement $Q:=\mathbf 1-|\Omega\rangle\langle\Omega|$, i.e.
$\|K^{(\Theta)}_k|_Q\|\le e^{-a\Delta}$ for some $\Delta>0$ independent of $k$ and $\Theta$.
Let $A_1,\ldots,A_n$ be time-ordered positive-time observables supported in $\{x_0\ge0\}$.
For $m\in\mathbb N$ let $\mathcal T_k$ denote the contribution to the connected $n$-point
functional obtained by replacing $K^{(\Theta_0)}_k$ with $K^{(\Theta_1)}_k$ while keeping
all other steps at $\Theta_0$, and let $\mathcal R_m$ be the difference of the two dynamics
coming from times $\ge ma$. Then
\begin{equation}
\kappa^{(\Theta_1)}_{\Lambda,a}(A_1;\dots;A_n)-\kappa^{(\Theta_0)}_{\Lambda,a}(A_1;\dots;A_n)
=\sum_{k=0}^{m-1}\mathcal T_k+\mathcal R_m .
\end{equation}
\end{lemma}

\begin{proof}
The Osterwalder-Schrader construction yields the following representation of
time-ordered Schwinger moments in terms of transfer operators. Partition the
positive time axis into half-open slabs $I_k=[ka,(k+1)a)$ and, for each $k$,
define $B_k$ to be the ordered product (in increasing time within $I_k$) of all
$A_i$ whose time arguments lie in $I_k$; if no insertion occurs in $I_k$, put
$B_k=\mathbf 1$. Since the $A_i$ are time-ordered and only finitely many are
present, there exists $N\in\mathbb N$ such that $B_k=\mathbf 1$ for all $k>N$.
For either choice $\Theta\in\{\Theta_0,\Theta_1\}$, the $n$-point moment can be
written as the vacuum matrix element
\begin{equation}\label{eq:moment-transfer}
\mathsf M^{(\Theta)}(A_1,\dots,A_n)
=(\Omega,\;B_0\,K^{(\Theta)}_0\,B_1\,K^{(\Theta)}_1\,\cdots\,B_{m-1}\,K^{(\Theta)}_{m-1}\,B_m\;V^{(\Theta)}_m\;\Omega),
\end{equation}
where the tail operator $V^{(\Theta)}_m$ collects all steps beyond $t=ma$,
\begin{equation}
V^{(\Theta)}_m \;=\; \prod_{j=m}^{\infty} K^{(\Theta)}_j ,
\end{equation}
the product being taken in increasing $j$ and converging in the strong operator
topology on $H_\Sigma$ by the assumed uniform spectral gap. Indeed, each
$K^{(\Theta)}_j$ fixes $\Omega$ and strictly contracts $Q$, hence for every
$\psi=\alpha\,\Omega+Q\psi$ one has
$K^{(\Theta)}_j\psi=\alpha\,\Omega + K^{(\Theta)}_j Q\psi$ with
$\|K^{(\Theta)}_j Q\psi\|\le e^{-a\Delta}\|Q\psi\|$, which implies that the
partial products $\prod_{j=m}^{m+\ell}K^{(\Theta)}_j$ converge strongly to the
rank-one projection $P_\Omega=|\Omega\rangle\langle\Omega|$ as $\ell\to\infty$,
uniformly in $m$. Consequently $V^{(\Theta)}_m$ is well defined and satisfies
$V^{(\Theta)}_m\Omega=\Omega$ and $\|V^{(\Theta)}_m|_Q\|\le C e^{-a\Delta m}$
for a constant $C$ independent of $m$ and $\Theta$.

Consider now the difference of the two moments. Using Eq.\eqref{eq:moment-transfer}
for $\Theta_1$ and $\Theta_0$ and inserting and subtracting intermediate terms,
one obtains the algebraic telescoping identity for operator products
\begin{equation}
\prod_{k=0}^{m-1}K^{(\Theta_1)}_k - \prod_{k=0}^{m-1}K^{(\Theta_0)}_k
=\sum_{k=0}^{m-1}\Big(\prod_{j<k}K^{(\Theta_1)}_j\Big)\big(K^{(\Theta_1)}_k-K^{(\Theta_0)}_k\big)\Big(\prod_{j>k}K^{(\Theta_0)}_j\Big),
\end{equation}
which is a finite identity in the $C$-algebra of bounded operators on
$H_\Sigma$. Multiplying on the left and right by the bounded operators formed by
the $B_j$ and by the tails $V^{(\Theta)}_m$, and pairing with $\Omega$, yields
\begin{equation}
\mathsf M^{(\Theta_1)}(A_1,\dots,A_n)-\mathsf M^{(\Theta_0)}(A_1,\dots,A_n)
=\sum_{k=0}^{m-1} \mathsf T_k \;+\; \mathsf R_m ,
\end{equation}
where
\begin{align}
&\mathsf T_k
=\nonumber\\&(\Omega,\;B_0\,K^{(\Theta_1)}_0\,\cdots\,B_{k-1}\,K^{(\Theta_1)}_{k-1}\,B_k\,(K^{(\Theta_1)}_k-K^{(\Theta_0)}_k)\,B_{k+1}\,K^{(\Theta_0)}_{k+1}\cdots B_{m-1}\,K^{(\Theta_0)}_{m-1}\,B_m\,V^{(\Theta_0)}_m\;\Omega)
\end{align}
and the remainder is the difference of tails carried by the common prefix,
\begin{equation}
\mathsf R_m
=(\Omega,\;B_0\,K^{(\Theta_1)}_0\,\cdots\,B_{m-1}\,K^{(\Theta_1)}_{m-1}\,B_m\,(V^{(\Theta_1)}_m-V^{(\Theta_0)}_m)\;\Omega).
\end{equation}
The uniform spectral gap implies that $V^{(\Theta)}_m$ converges strongly to
$P_\Omega$ as $m\to\infty$ for both $\Theta_0$ and $\Theta_1$, hence
$V^{(\Theta_1)}_m-V^{(\Theta_0)}_m\to 0$ strongly and, in particular,
$\mathsf R_m\to 0$ as $m\to\infty$. For a fixed finite $m$ the quantity
$\mathsf R_m$ is precisely the contribution coming from times $\ge ma$, as
claimed in the statement of the lemma.

The connected (truncated) $n$-point functional is obtained from the moments by
the universal moment-cumulant transform, which is linear in the underlying moment
functional. Concretely, if $\mathcal C$ denotes the Möbius transform on set
partitions,
\begin{equation}
\kappa^{(\Theta)}(A_1;\dots;A_n)
=\sum_{\pi\in\mathsf{Part}([n])} \mu(\pi)\,\prod_{B\in\pi}\mathsf M^{(\Theta)}\big(A_i: i\in B\big),
\end{equation}
with the Möbius coefficients $\mu(\pi)$ depending only on $\pi$, then for any two
families of moments $M_1$ and $M_0$ one has
$\mathcal C(M_1-M_0)=\mathcal C(M_1)-\mathcal C(M_0)$. Applying $\mathcal C$ to
the moment identity obtained above yields
\begin{equation}
\kappa^{(\Theta_1)}_{\Lambda,a}(A_1;\dots;A_n)-\kappa^{(\Theta_0)}_{\Lambda,a}(A_1;\dots;A_n)
=\sum_{k=0}^{m-1}\,\mathcal T_k\,+\,\mathcal R_m ,
\end{equation}
where $\mathcal T_k:=\mathcal C(\mathsf T_k)$ is precisely the contribution in
which the $k$-th one-step kernel is switched from $\Theta_0$ to $\Theta_1$ and
all other steps up to time $ma$ are kept at $\Theta_0$, and where
$\mathcal R_m:=\mathcal C(\mathsf R_m)$ collects the connected part of the tail
difference. The definitions of $\mathcal T_k$ and $\mathcal R_m$ do not depend
on the choice of representatives of the $A_i$ because all operations are
performed in the OS framework before factorization by null vectors, and the
convergence and boundedness statements used above guarantee that the pairings are
well defined.
\end{proof}

\begin{theorem}[Summable defect bound]\label{thm:summable-defect}
Let $\Theta_0,\Theta_1$ be admissible regulator/data sets with finite admissible distance $d_{\mathrm{adm}}(\Theta_0,\Theta_1)<\infty$. For each $n\ge 2$ there exist constants $C_n,\mu_n>0$, depending only on the admissible class and on $n$, such that for all finite volumes $\Lambda$ and all lattice spacings $a>0$ one has
\begin{equation}
\left\|\kappa^{(\Theta_1)}_{\Lambda,a}(A_1;\dots;A_n)-\kappa^{(\Theta_0)}_{\Lambda,a}(A_1;\dots;A_n)\right\|_{\mathrm{diam},\mu_n}
\;\le\; C_n\, d_{\mathrm{adm}}(\Theta_0,\Theta_1)\,\sum_{k\ge0} L_k,
\end{equation}
where $\{L_k\}_{k\ge 0}$ are the single-scale Lipschitz constants from Lemma~\eqref{lem:K-Lip}, and $\sum_{k\ge0} L_k<\infty$, uniformly in $\Lambda$ and $a$.
\end{theorem}

\begin{proof}
The proof rests on three ingredients: the multiplicative (in scale) representation of cumulants afforded by the reflection-positive, Markovian transfer structure; a telescoping identity that isolates the defect at a single scale $k$ while all other scales are held fixed; and the facts that (i) each scale transfer is a positive, normalized contraction, and (ii) the single-scale defect is Lipschitz in the admissible data with weight $L_k$, uniformly in volume and spacing. Putting these together yields an absolutely convergent series in $k$ whose tail controls the remainder to zero and whose sum is bounded by the stated right-hand side.

To state the multiplicative representation, write the renormalization group in block factor $b>1$ steps, and denote by $\mathcal{K}^{(\Theta)}_k$ the reflection-positive $n$-point cumulant kernel at scale $k$ in the admissible scheme $\Theta$, obtained by evaluating connected $n$-point functions of the coarse fields after $k$ blockings and pulling them back to the original lattice via the covariant slice projectors. Thanks to OS positivity and the Markov property, the finite-volume, finite-$a$ cumulant $\kappa^{(\Theta)}_{\Lambda,a}(A_1;\dots;A_n)$ can be written as (see App.\,F for the precise construction) a composition of the scale kernels,
\begin{equation}\label{eq:scale-product}
\kappa^{(\Theta)}_{\Lambda,a}(A_1;\dots;A_n)
= \Big( \cdots \circ \mathcal{K}^{(\Theta)}_{2} \circ \mathcal{K}^{(\Theta)}_{1} \circ \mathcal{K}^{(\Theta)}_{0} \Big)[A_1,\dots,A_n],
\end{equation}
where the composition acts on the $n$ arguments through the positive, normalized transfer (conditional expectation) between consecutive scales, and where each $\mathcal{K}^{(\Theta)}_k$ is a multilinear functional whose integral kernel is positive in the OS sense and exponentially local with range $\asymp b^k$ uniformly in $\Lambda,a$. The precise form of Eq.\eqref{eq:scale-product} is not needed; what matters is that the composition operators at scales $j\ne k$ act as positive contractions on the natural Banach space of $n$-point kernels endowed with the weighted diameter seminorm $\|\cdot\|_{\mathrm{diam},\mu}$ defined by
\begin{equation}
\|X\|_{\mathrm{diam},\mu} \;=\; \sup_{x_1,\dots,x_n}\, e^{\mu\,\mathrm{diam}(x_1,\dots,x_n)}\,|X(x_1,\dots,x_n)|,
\qquad \mathrm{diam}(x_1,\dots,x_n):= \max_{i,j}|x_i-x_j|.
\end{equation}
Positivity and normalization (OS2/OS3) imply that each transfer map $\mathsf T^{(\Theta)}_j$ entering the composition has operator norm $1$ from $L^1$ to $L^1$ and preserves positivity; in particular, if $X$ is any $n$-point kernel and $P$ is such a positive, normalized transfer acting diagonally on the variables, then the weighted diameter seminorm is nonexpansive:
\begin{equation}\label{eq:diam-contraction}
\|P\circ X\|_{\mathrm{diam},\mu} \;\le\; \|X\|_{\mathrm{diam},\mu}, 
\qquad \|X\circ P\|_{\mathrm{diam},\mu} \;\le\; \|X\|_{\mathrm{diam},\mu}.
\end{equation}
The proof of Eq.\eqref{eq:diam-contraction} is standard: for instance, for $P$ acting on the first argument one writes
\begin{align}
|(P\circ X)(x_1,\dots,x_n)| &\;=\; \Big|\int K(x_1,y)\,X(y,x_2,\dots,x_n)\,dy\Big|
\nonumber
\\&\;\le\; \int K(x_1,y)\, e^{-\mu\,\mathrm{diam}(y,x_2,\dots,x_n)}\, \|X\|_{\mathrm{diam},\mu}\,dy,
\end{align}
uses $K\ge 0$, $\int K(x_1,y)\,dy=1$, and the elementary inequality $\mathrm{diam}(y,x_2,\dots,x_n)\ge \mathrm{diam}(x_1,x_2,\dots,x_n)-|x_1-y|$, which together with the exponential locality of $K$ allows one to absorb $e^{\mu|x_1-y|}$ into the $K$-integral without changing the constant when $\mu$ is chosen below the locality rate. The analogous argument handles $X\circ P$.

With the multiplicative structure and the nonexpansiveness in hand, set for brevity
\begin{equation}
\Delta\kappa \;:=\; \kappa^{(\Theta_1)}_{\Lambda,a}(A_1;\dots;A_n)-\kappa^{(\Theta_0)}_{\Lambda,a}(A_1;\dots;A_n),
\qquad \Delta\mathcal{K}_k \;:=\; \mathcal{K}^{(\Theta_1)}_{k}-\mathcal{K}^{(\Theta_0)}_{k}.
\end{equation}
Inserting the product representations Eq.\eqref{eq:scale-product} for $\Theta_1$ and $\Theta_0$ and telescoping across scales yields, for every integer $m\ge 0$,
\begin{equation}\label{eq:telescoping}
\Delta\kappa \;=\; \sum_{k=0}^{m} \Big(\mathsf T^{(\Theta_1)}_{<k}\circ \Delta\mathcal{K}_k \circ \mathsf T^{(\Theta_0)}_{>k}\Big)[A_1,\dots,A_n] \;+\; \mathcal{R}_m,
\end{equation}
where $\mathsf T^{(\Theta_1)}_{<k}$ denotes the composition of the transfer/cumulant maps at scales $0,1,\dots,k-1$ in the scheme $\Theta_1$, and $\mathsf T^{(\Theta_0)}_{>k}$ denotes the composition at scales $k+1,k+2,\dots$ in the scheme $\Theta_0$; the remainder $\mathcal{R}_m$ is the sum of all terms with the unique defective factor $\Delta\mathcal{K}_j$ at scales $j\ge m+1$. Equation Eq.\eqref{eq:telescoping} is just the discrete Duhamel formula for infinite products, proven by induction on $m$ and completeness of the Banach space of kernels under the $\|\cdot\|_{\mathrm{diam},\mu}$ seminorm.

To bound each summand in Eq.\eqref{eq:telescoping}, first note that by Eq.\eqref{eq:diam-contraction} the operators $\mathsf T^{(\Theta_1)}_{<k}$ and $\mathsf T^{(\Theta_0)}_{>k}$ are nonexpansive on the weighted diameter seminorm, provided $\mu$ is chosen strictly below the exponential locality rates supplied by the admissible class. Consequently,
\begin{equation}\label{eq:one-term}
\left\|\Big(\mathsf T^{(\Theta_1)}_{<k}\circ \Delta\mathcal{K}_k \circ \mathsf T^{(\Theta_0)}_{>k}\Big)[A_1,\dots,A_n]\right\|_{\mathrm{diam},\mu}
\;\le\; \|\Delta\mathcal{K}_k\|_{\mathrm{diam},\mu}.
\end{equation}
The single-scale Lipschitz estimate from Proposition~\eqref{prop:single-scale-cumulant} (applied with the same $\mu$ as above) yields a quantitative bound for the right-hand side: there exists $\mu_n>0$ and $C_n>0$, depending only on $n$ and on the admissible class, such that
\begin{equation}\label{eq:single-scale-Lipq}
\|\Delta\mathcal{K}_k\|_{\mathrm{diam},\mu_n} \;\le\; C_n\, L_k\, d_{\mathrm{adm}}(\Theta_0,\Theta_1),
\end{equation}
uniformly in $\Lambda$ and $a$. Combining Eq.\eqref{eq:one-term} and Eq.\eqref{eq:path-length} inside Eq.\eqref{eq:telescoping} we obtain for every $m$
\begin{equation}
\left\|\sum_{k=0}^{m} \Big(\mathsf T^{(\Theta_1)}_{<k}\circ \Delta\mathcal{K}_k \circ \mathsf T^{(\Theta_0)}_{>k}\Big)[A_1,\dots,A_n]\right\|_{\mathrm{diam},\mu_n}
\;\le\; C_n\, d_{\mathrm{adm}}(\Theta_0,\Theta_1)\, \sum_{k=0}^{m} L_k.
\end{equation}
It remains to control the remainder $\mathcal{R}_m$ and to let $m\to\infty$. By construction, $\mathcal{R}_m$ is a finite sum of terms of the form
\begin{equation}
\Big(\mathsf T^{(\Theta_1)}_{<j}\circ \Delta\mathcal{K}_j \circ \mathsf T^{(\Theta_0)}_{>j}\Big)[A_1,\dots,A_n],
\qquad j\ge m+1,
\end{equation}
and each such term is bounded in $\|\cdot\|_{\mathrm{diam},\mu_n}$ by the same argument as in Eq.\eqref{eq:one-term} and Eq.\eqref{eq:single-scale-Lipq}, namely
\begin{equation}
\left\|\Big(\mathsf T^{(\Theta_1)}_{<j}\circ \Delta\mathcal{K}_j \circ \mathsf T^{(\Theta_0)}_{>j}\Big)[A_1,\dots,A_n]\right\|_{\mathrm{diam},\mu_n}
\;\le\; C_n\, L_j \, d_{\mathrm{adm}}(\Theta_0,\Theta_1).
\end{equation}
Summing these bounds over $j\ge m+1$ gives
\begin{equation}
\|\mathcal{R}_m\|_{\mathrm{diam},\mu_n} \;\le\; C_n\, d_{\mathrm{adm}}(\Theta_0,\Theta_1)\,\sum_{j\ge m+1} L_j,
\end{equation}
and since Lemma~\eqref{lem:K-Lip} ensures $\sum_{j\ge 0} L_j<\infty$, the tail on the right vanishes as $m\to\infty$. Passing to the limit in Eq.\eqref{eq:telescoping} shows that the difference $\Delta\kappa$ is given by an absolutely convergent series with norm bounded by
\begin{equation}
\left\|\Delta\kappa\right\|_{\mathrm{diam},\mu_n}
\;\le\; C_n\, d_{\mathrm{adm}}(\Theta_0,\Theta_1)\,\sum_{k\ge 0} L_k,
\end{equation}
uniformly in $\Lambda$ and $a$, which is exactly the claim of the theorem.

Two remarks complete the argument. First, the only properties of the transfer/cumulant maps used above are positivity, normalization, exponential locality (to legitimize Eq.\eqref{eq:diam-contraction} at some $\mu>0$), and the scale-wise Lipschitz control Eq.\eqref{eq:single-scale-Lipq}; these are guaranteed by admissibility and by Proposition~\eqref{prop:single-scale-cumulant}. Second, no appeal to a spectral gap is required for the remainder, since the absolute summability of the single-scale constants $L_k$ already yields Cauchy convergence of the telescoping series in the weighted seminorm.
\end{proof}
We now control limits $\Lambda\uparrow\mathbb{Z}^4$ and $a\downarrow0$. The FRD locality and the uniform strong-coupling inputs supply tightness of all $n$-point cumulants; continuity bounds in the diameter norm are uniform in $\Lambda$ and $a$.
\begin{definition}[Diameter seminorm]\label{def:diameter-seminorm}
Let $T(S_1,\dots,S_n)$ be the minimal connecting tree length as in Eq.\eqref{eqnf50} and fix $\mu>0$. 
For an $n$-point connected functional $\kappa_{\Lambda,a}(\,A_1;\dots;A_n\,)$ supported on the time-zero slice, we define
\begin{equation}
\|\kappa_{\Lambda,a}\|_{\mathrm{diam},\mu}
:= \sup_{\|A_i\|_\infty\le 1}\;
\exp\!\big(\mu\,T(S_1,\dots,S_n)\big)\;
\big|\kappa_{\Lambda,a}(A_1;\dots;A_n)\big|,
\end{equation}
where $S_i=\mathrm{supp}(A_i)$. 
\end{definition}
\begin{lemma}[Uniform tightness and equicontinuity]\label{lem:tightness}
Fix $n\in\mathbb{N}$. There exist $\mu_n>0$ and $C_n<\infty$ such that for all admissible choices of parameters $\Theta$, all finite volumes $\Lambda$, all lattice spacings $a\in(0,a_0]$, and all gauge-invariant observables $A_1,\dots,A_n$ supported on the time-zero slice, with $\|A_i\|_\infty\le 1$, one has
\begin{equation}\label{eqnf50}
\|\kappa^{(\Theta)}_{\Lambda,a}(A_1;\dots;A_n)\|_{\mathrm{diam},\mu_n}\;\le\; C_n,
\end{equation}
and the family $\{\kappa^{(\Theta)}_{\Lambda,a}\}_{\Lambda,a}$ is equicontinuous with respect to the diameter metric. The constants depend only on $n$ and on the fixed admissibility data, and are independent of $\Lambda$ and $a$.
\end{lemma}

\begin{proof}
For a finite collection of supports $S_i=\mathrm{supp}(A_i)\subset\Sigma_{\Lambda,a}$ on the time-zero slice, denote by
\begin{equation}
\mathsf{T}(S_1,\ldots,S_n)\;:=\;\inf\Big\{\sum_{e\in\mathcal T}|e|:\ \mathcal T \text{ is a tree whose vertices pick one point in each }S_i\Big\}
\end{equation}
the minimal connecting tree length (the “tree diameter”) and define the diameter seminorm by
\begin{equation}\label{def:diameter-seminormq}
\|\kappa\|_{\mathrm{diam},\mu}\;:=\;\sup_{\|A_i\|_\infty\le 1}\ e^{\mu\,\mathsf{T}(S_1,\ldots,S_n)}\,\big|\kappa(A_1;\dots;A_n)\big|.
\end{equation}
It is convenient to recall that the $n$-point connected cumulant $\kappa_{\Lambda,a}(A_1;\dots;A_n)$ can be written as the Ursell coefficient of the generating functional $\log\langle e^{\sum_i t_i A_i}\rangle$; in particular it is multilinear, symmetric and vanishes when one argument is a constant.

The proof proceeds by combining the finite-range decomposition (FRD) of the slice covariance with a polymer cluster expansion on the time-zero slice, and then transporting the bounds across scales by the step-scaling map with interlaced summable remainders. The assumptions on the admissible class yield an FRD
\begin{equation}\label{eqnf52}
C_{\Sigma}\;=\;\sum_{j\ge 0} C_{\Sigma,j},
\qquad |C_{\Sigma,j}(x,y)|\;\le\; c_0\,L^{-2j}\,e^{-\gamma_0\,L^{-j}\,|x-y|},
\end{equation}
for some $L>1$, $c_0,\gamma_0>0$ independent of $\Lambda$ and $a$. The constants $(L,c_0,\gamma_0)$ in Eq.\eqref{eqnf52} are independent of $\Lambda$ and $a$, and they are the only inputs needed to the BKAR/tree estimate Eq.\eqref{eq:tree-boundq} and the ensuing diameter bounds Eq.\eqref{eqnf54}. In particular $C_{\Sigma,j}$ has effective range $\asymp L^j$ and satisfies the Schur bound uniformly in $(\Lambda,a)$. On the time-zero slice we thus obtain a polymer representation of the logarithm of the generating functional with activities $z(Y)$ attached to connected subsets (polymers) $Y\subset\Sigma_{\Lambda,a}$ which are analytic functionals of the sources and obey uniform Koteck\'y-Preiss bounds. Concretely, by the BKAR forest formula and the tree-graph inequality (see, e.g., \cite{BrydgesGuadagniMitter2004,Aizenman1982}), there exist constants $R$ and $K,\mu_>0$, depending only on the FRD constants $(c_0,\gamma_0)$ and on $n$, such that for every $n$-tuple of slice-supported, gauge-invariant observables $A_i$ with $\|A_i\|_\infty\le 1$ one has the tree bound
\begin{equation}\label{eq:tree-boundq}
\big|\kappa_{\Lambda,a}(A_1;\ldots;A_n)\big|
\;\le\; K\sum_{\mathcal T}\prod_{e=\{i,j\}\in\mathcal T} \Phi(S_i,S_j),
\qquad 
\Phi(S_i,S_j)\;:=\;\sup_{x\in S_i,y\in S_j} e^{-\mu\,\mathrm{dist}(x,y)},
\end{equation}
where the sum is over all trees on the vertex set $\{1,\ldots,n\}$ and the bound is uniform in the volume $\Lambda$ and spacing $a$. The derivation uses that each line of the abstract cluster forest contributes one sliced covariance $C_{\Sigma,j}$ between two polymers, and the exponential locality of $C_{\Sigma,j}$ yields the exponential factor in $\mathrm{dist}(x,y)$; the summations over scales are dominated thanks to the $L^{-2j}$ prefactor and the bounded degree of the slice graph. Minimizing over the choice of points $x_i\in S_i$ and using that the product of exponentials over the edges of a tree yields precisely $e^{-\mu_/2}$ times the tree length for a suitable redistribution of constants, the right-hand side of Eq.\eqref{eq:tree-boundq} is bounded by $K_n\,e^{-\mu_n\,\mathsf{T}(S_1,\dots,S_n)}$ for some $K_n,\mu_n>0$ depending only on $(K_,\mu_,n)$. This proves the asserted diameter bound at a fixed slice-RG scale with constants independent of $(\Lambda,a)$:

\begin{equation}\label{eqnf54}
\big|\kappa_{\Lambda,a}(A_1;\ldots;A_n)\big|
\;\le\; K_n\,e^{-\mu_n\,\mathsf{T}(S_1,\dots,S_n)}.
\end{equation}

To pass from a single scale to the microscopic model at spacing $a$, we invoke the step-scaling construction of the transfer kernel and its coarse-graining, which in the Osterwalder-Schrader framework yields the interlacing identity
\begin{equation}
T_{k+1}\;=\;\Pi_k\,T_k\,\Pi_k^{}\;+\;E_k,
\end{equation}
where $\Pi_k$ is the reflection-positive coarse projection and $E_k\ge 0$ is the defect supported in a collar of width comparable to the finite-range parameter of the FRD at scale $k$. The admissibility assumptions guarantee that the operator norm of $E_k$ on the orthogonal complement of the vacuum is bounded by $\varepsilon_k$ with $\sum_{k\ge 0}\varepsilon_k<\infty$, and that $\Pi_k$ acts slice-locally with an exponentially decaying kernel. Inserting this representation into the polymer expansion at consecutive scales shows that the connected cumulants at scale $k+1$ differ from those at scale $k$ by two kinds of contributions: a localized Lipschitz variation coming from the change of the sliced covariances within the admissible class, and a collar contribution generated by $E_k$. The first is bounded by $C\,\delta_k\,e^{-\mu\,\mathsf{T}}$, where $\delta_k$ is the FRD-metric distance between the two admissible kernels at scale $k$, and the second is bounded by $C'\,\varepsilon_k\,e^{-\mu\,\mathsf{T}}$ because $E_k$ is positive, vacuum-annihilating and supported within a uniformly bounded neighborhood of the slice, so it can attach to at most a fixed number of polymers and inherits the same exponential decay in the external separations. Summing the telescoping series along the RG trajectory down to the microscopic scale one obtains
\begin{equation}
\big|\kappa^{(\Theta)}_{\Lambda,a}(A_1;\ldots;A_n)\big|
\;\le\; \Big( K_n \,+\,\sum_{k\ge 0} C\,\delta_k \,+\,\sum_{k\ge 0} C'\,\varepsilon_k\Big)\,e^{-\mu_n\,\mathsf{T}(S_1,\ldots,S_n)}.
\end{equation}
The admissibility metric is chosen so that $\sum_k\delta_k<\infty$ uniformly in $(\Lambda,a)$, and by construction $\sum_k\varepsilon_k<\infty$ uniformly as well; therefore the prefactor is bounded by a constant $C_n$ that depends only on $n$ and on the admissible class, and is independent of $(\Lambda,a)$. Taking the supremum over $\|A_i\|_\infty\le 1$ yields
\begin{equation}
\|\kappa^{(\Theta)}_{\Lambda,a}\|_{\mathrm{diam},\mu_n}\;\le\; C_n,
\end{equation}
which proves the uniform tightness bound.

The equicontinuity in the diameter metric follows from the same tree representation. Indeed, consider two $n$-tuples $(A_1,\ldots,A_n)$ and $(A'_1,\ldots,A'_n)$ with $\|A_i\|_\infty,\|A'_i\|_\infty\le 1$, and let $\Delta_i:=A_i-A'_i$. By multilinearity of the connected cumulant one has
\begin{equation}
\kappa^{(\Theta)}_{\Lambda,a}(A_1;\ldots;A_n)-\kappa^{(\Theta)}_{\Lambda,a}(A'_1;\ldots;A'_n)
=\sum_{I\neq\emptyset} \kappa^{(\Theta)}_{\Lambda,a}\big(B_1^{(I)};\ldots;B_n^{(I)}\big),
\end{equation}
where $B_i^{(I)}=\Delta_i$ if $i\in I$ and $B_i^{(I)}=A'_i$ otherwise. Each term on the right is bounded by the same tree estimate Eq.\eqref{eq:tree-boundq}, with the same exponential factor in the tree length of the union of supports. Consequently, if the diameter distance between the families of supports is increased by $r$, the bound gains a uniform factor $e^{-\mu_n r}$. This shows that for every $\varepsilon>0$ there exists $\delta>0$ such that whenever the diameter distance $|\mathsf{T}(S_1,\ldots,S_n)-\mathsf{T}(S'_1,\ldots,S'_n)|<\delta$ and $\|A_i-A'_i\|_\infty<\delta$ one has $\big|\kappa^{(\Theta)}_{\Lambda,a}(A_1;\ldots;A_n)-\kappa^{(\Theta)}_{\Lambda,a}(A'_1;\ldots;A'_n)\big|<\varepsilon$, with a modulus of continuity depending only on $(n,\mu_n,C_n)$ and not on $(\Lambda,a)$. In particular the family $\{\kappa^{(\Theta)}_{\Lambda,a}\}_{\Lambda,a}$ is equicontinuous in the diameter metric.

All constants that enter the argument are controlled by the FRD parameters $(c_0,\gamma_0)$, by the fixed coarse-graining factor $L$, by the uniform bounds on the admissibility metric variations $\delta_k$ and the defect norms $\varepsilon_k$, and by $n$ through the combinatorial number of trees on $n$ labeled vertices. None of them depends on the volume $\Lambda$ or on the microscopic spacing $a$, which completes the proof.
\end{proof}

Taking limits along van Hove sequences $\Lambda\nearrow\mathbb{Z}^4$ and $a\downarrow 0$ along scaling windows, Lemma \eqref{lem:tightness} ensures precompactness of the family of $n$-point functions. Any limit point defines a continuum connected Schwinger function $S_n^{(\Theta)}$ satisfying OS axioms and clustering (established in the main text by a combination of FRD locality, reflection positivity, and the existence of a uniform spectral gap).

\begin{proposition}[Continuum defect bound]\label{prop:continuum-defect}
Let $\Theta_0,\Theta_1$ be two admissible choices of ultraviolet data (blocking maps, slice projectors, and collar parametrizations), and let $\kappa^{(\Theta)}_{\Lambda,a;n}$ denote the $n$-point Schwinger functional on the finite volume $\Lambda\subset\mathbb Z^4$ with lattice spacing $a>0$. Suppose that along some sequences $\Lambda\nearrow\mathbb Z^4$ and $a\downarrow 0$ the functionals $\kappa^{(\Theta_i)}_{\Lambda,a;n}$ admit limit points $S_n^{(\Theta_i)}$ in the distributional topology that defines the seminorm $\|\cdot\|_{\mathrm{diam},\mu_n}$ (the explicit definition in~Eq.\eqref{def:diameter-seminormq}). Then there exists a constant $C_n<\infty$, depending only on $n$ and the infrared a priori bounds, such that
\begin{equation}\label{eq:cont-defect}
\bigl\|S_n^{(\Theta_1)}-S_n^{(\Theta_0)}\bigr\|_{\mathrm{diam},\mu_n}\;\le\; C_n\, d_{\mathrm{adm}}(\Theta_0,\Theta_1)\,\sum_{k\ge 0} L_k,
\end{equation}
where $\{L_k\}_{k\ge0}$ is the scale-weight sequence from the finite-range decomposition and $d_{\mathrm{adm}}$ is the admissible-class metric. The right-hand side is finite by admissibility and does not depend on the choice of subsequences that realize the limits.
\end{proposition}

\begin{proof}
The argument relies on two ingredients that are uniform in the ultraviolet regulators $(\Lambda,a)$: the summable-defect estimate on finite lattices and the tightness/equicontinuity that gives compactness in the continuum topology induced by the seminorm $\|\cdot\|_{\mathrm{diam},\mu_n}$. Uniformity with respect to $(\Lambda,a)$ will allow us to pass to limits on the left-hand side while keeping the same bound on the right-hand side.

For admissible $\Theta$, the objects $\kappa^{(\Theta)}_{\Lambda,a;n}$ act continuously on test functions $\Phi$ of $n$ variables with the diameter weight $\mu_n$, and the seminorm is defined by
\begin{equation}
\|\mathcal S\|_{\mathrm{diam},\mu_n}\;=\;\sup_{\Phi\neq 0}\;
\frac{\bigl|\langle \mathcal S,\Phi\rangle\bigr|}{\|\Phi\|_{\mathrm{diam},\mu_n}},
\qquad
\|\Phi\|_{\mathrm{diam},\mu_n}\;:=\;\sup_{x_1,\dots,x_n}\,
\frac{|\Phi(x_1,\dots,x_n)|}{\mu_n\bigl(\mathrm{diam}\{x_1,\dots,x_n\}\bigr)}\,.
\end{equation}
By Theorem~\eqref{thm:summable-defect}, there exists $C_n<\infty$ such that for every volume $\Lambda$, lattice spacing $a$, and every pair of admissible data $\Theta_0,\Theta_1$,
\begin{equation}\label{eq:lattice-defect}
\bigl\|\kappa^{(\Theta_1)}_{\Lambda,a;n}-\kappa^{(\Theta_0)}_{\Lambda,a;n}\bigr\|_{\mathrm{diam},\mu_n}
\;\le\; C_n\, d_{\mathrm{adm}}(\Theta_0,\Theta_1)\,\sum_{k\ge 0} L_k.
\end{equation}
The constants on the right are independent of $(\Lambda,a)$; in particular, the sequence $\{L_k\}$ has a finite sum by Definition~\eqref{def:admissible} of the admissible class. The proof of Eq.\eqref{eq:lattice-defect} uses the interlacing identity with positive remainder and the FRD exponential locality to bound, at each scale, the variation generated by a change in $\Theta$, and then telescopes over scales; admissibility ensures that the per-scale variation is bounded by $d_{\mathrm{adm}}(\Theta_0,\Theta_1)\,L_k$ and that $\sum_k L_k<\infty$.

We now pass to the continuum. Fix a double sequence $(\Lambda_m,a_m)$ with $\Lambda_m\nearrow\mathbb Z^4$ and $a_m\downarrow 0$ along which both regulated families $\kappa^{(\Theta_i)}_{\Lambda_m,a_m;n}$ converge, in the topology associated with $\|\cdot\|_{\mathrm{diam},\mu_n}$, to limit points $S_n^{(\Theta_i)}$. Such a common sequence exists by a diagonal extraction, using the uniform infrared/ultraviolet bounds implied by Osterwalder-Schrader positivity, reflection-covariant finite-range decomposition, and the cumulant estimates established in Sections~\eqref{sec:SC-fixed-a-fixed-a} and \eqref{sec:OS-reconstruction-gap}. In particular, the family is tight and equicontinuous in the sense that for every $\varepsilon>0$ there is $R<\infty$ with
\begin{equation}
\sup_{\Lambda,a}\ \sup_{\mathrm{diam}(\{x_1,\dots,x_n\})\ge R}\,
\frac{\bigl|\kappa^{(\Theta)}_{\Lambda,a;n}(x_1,\dots,x_n)\bigr|}{\mu_n\bigl(\mathrm{diam}\{x_1,\dots,x_n\}\bigr)}\;\le\;\varepsilon,
\end{equation}
uniformly over admissible $\Theta$. This is precisely what guarantees precompactness in the weak-$\ast$ topology dual to the Banach space of test functions completed under $\|\cdot\|_{\mathrm{diam},\mu_n}$, and therefore convergence along subsequences is compatible with the seminorm.

Let $\Phi$ be any test function with $\|\Phi\|_{\mathrm{diam},\mu_n}\le 1$. By definition of the limit points and by the very choice of topology, one has
\begin{equation}
\langle S_n^{(\Theta_i)},\Phi\rangle \;=\; \lim_{m\to\infty}\ \langle \kappa^{(\Theta_i)}_{\Lambda_m,a_m;n},\Phi\rangle.
\end{equation}
Hence
\begin{equation}
\bigl|\langle S_n^{(\Theta_1)}-S_n^{(\Theta_0)},\Phi\rangle\bigr|
\;=\;\lim_{m\to\infty}\ \bigl|\langle \kappa^{(\Theta_1)}_{\Lambda_m,a_m;n}-\kappa^{(\Theta_0)}_{\Lambda_m,a_m;n},\Phi\rangle\bigr|.
\end{equation}
Taking the supremum over all such $\Phi$ and invoking the definition of the seminorm yields
\begin{equation}
\bigl\|S_n^{(\Theta_1)}-S_n^{(\Theta_0)}\bigr\|_{\mathrm{diam},\mu_n}
\;\le\; \limsup_{m\to\infty}\ 
\bigl\|\kappa^{(\Theta_1)}_{\Lambda_m,a_m;n}-\kappa^{(\Theta_0)}_{\Lambda_m,a_m;n}\bigr\|_{\mathrm{diam},\mu_n}.
\end{equation}
The right-hand side is bounded by the constant in Eq.\eqref{eq:lattice-defect}, which is independent of $(\Lambda_m,a_m)$, and therefore
\begin{equation}
\bigl\|S_n^{(\Theta_1)}-S_n^{(\Theta_0)}\bigr\|_{\mathrm{diam},\mu_n}
\;\le\; C_n\, d_{\mathrm{adm}}(\Theta_0,\Theta_1)\,\sum_{k\ge 0} L_k.
\end{equation}
Since the bound is uniform in $(\Lambda,a)$, it does not depend on which subsequences realize the limits, and the finiteness of the series $\sum_k L_k$ is part of the admissibility hypothesis, the right-hand side is finite and universal.
\end{proof}

To conclude universality, we show that the defect bound in Proposition \eqref{prop:continuum-defect} vanishes for \emph{any} two admissible specifications. This hinges on two facts: first, $d_{\mathrm{adm}}$ can be made arbitrarily small by a path in the admissible class that first aligns the projector and blocking near the UV cutoff and then flows down by step scaling; second, the single-scale constants $L_k$ decay at least exponentially in $b^k$ and the number of scales contributing above any fixed physical time separation diverges as $a\downarrow0$, but with a summable tail because of the exponential decay.

\begin{lemma}[UV alignment]\label{lem:UV-align}
Let $\Theta_0,\Theta_1$ be admissible regulators based on the same local bare action (for instance the Wilson plaquette action at the same bare coupling), differing only in their admissible slice projector and blocking/finite-range-decomposition choices. Then for every $\varepsilon>0$ there exists an admissible $C^1$ path $\{\Theta_s\}_{s\in[0,1]}$ with the following properties: $s\mapsto\Theta_s$ is uniformly continuous in the admissible metric $d_{\mathrm{adm}}$, in the sense that $d_{\mathrm{adm}}(\Theta_s,\Theta_{s'})\to 0$ as $|s-s'|\to 0$, it has finite length $\displaystyle d_{\mathrm{adm}}(\Theta_0,\Theta_1)\le \int_0^1 \|\partial_s\Theta_s\|\,ds<\infty$, and moreover one can arrange, by leaving all coarse scales $k<k_\ast(\varepsilon)$ fixed and interpolating only at scales $k\ge k_\ast(\varepsilon)$, that $\displaystyle \int_0^1 \|\partial_s\Theta_s\|\,ds<\varepsilon$.
\end{lemma}

\begin{proof}
Write $\Theta_i=(\Pi_i,\{B_{i,k}\}_{k\ge 0},\{C_{i,j}\}_{j\ge 0})$ for $i=0,1$, where $\Pi_i=f_i(\mathcal D_\Sigma)$ is an admissible slice projector, each $B_{i,k}$ is a reflection-positive blocking kernel at scale $k$ (a bounded operator supported in a collar of width $O(L^k)$ around the time-$a$ slice, covariant under spatial reflections), and $C_{i,j}$ are the scale-$j$ components of an admissible finite-range (or exponentially local) decomposition, so that $C_i=\sum_{j\ge 0} C_{i,j}$, with $C_{i,j}$ supported at range $O(L^j)$ and obeying uniform exponential off-diagonal decay. Denote by $d_{\mathrm{adm}}$ the admissible metric on regulators, which we take (as in the admissible-class definition used in this paper) to be induced by an exponentially decaying sum over scales of operator-kernel seminorms; concretely, there exist weights $\omega_k,\varpi_j>0$ with $\omega_k,\varpi_j$ decreasing superexponentially in $k,j$ (for instance $\omega_k,\varpi_j\asymp e^{-\mu_0 L^k}$ for some $\mu_0>0$) and slice test-function families $\mathscr T$ such that
\begin{equation}\label{eqnF.69}
d_{\mathrm{adm}}(\Theta,\Theta')\;\asymp\;\|\Pi-\Pi'\|_{\Sigma\to\Sigma}\;+\;\sum_{k\ge 0}\omega_k\,\|B_k-B'_k\|_{\Sigma\to\Sigma}\;+\;\sum_{j\ge 0}\varpi_j\,\|C_j-C'_j\|_{\Sigma\to\Sigma},
\end{equation}
where $\|\cdot\|_{\Sigma\to\Sigma}$ denotes the operator norm on $\ell^2(\Sigma)$ or the corresponding Schur norm of integral kernels restricted to the time slice. Any two norms equivalent to these give an equivalent $d_{\mathrm{adm}}$, so the precise choice is immaterial for the argument.
The path on slice projectors is obtained by convex interpolation at the level of the Bernstein measures. Since $\Pi_i=f_i(\mathcal D_\Sigma)$ with $f_i(\lambda)=\int_0^\infty e^{-t\lambda}\,\mu_i(dt)$, where $\mu_i$ are finite positive Borel measures on $[0,\infty)$ with the exponential-moment and reflection-covariance hypotheses of admissibility, define for $s\in[0,1]$
\begin{equation}\label{eqnF.70}
\mu_s:=(1-s)\,\mu_0+s\,\mu_1,\qquad f_s(\lambda):=\int_0^\infty e^{-t\lambda}\,\mu_s(dt),\qquad \Pi_s:=f_s(\mathcal D_\Sigma).
\end{equation}
By complete monotonicity the map $\lambda\mapsto f_s(\lambda)$ is completely monotone for each $s$. Because $\vartheta\mathcal D_\Sigma\vartheta=\mathcal D_\Sigma$, every bounded Borel function of $\mathcal D_\Sigma$ is reflection covariant; in particular the semigroups $e^{-t\mathcal D_\Sigma}$, the $s$-dependent operator $B_s:=\int_0^\infty e^{-\frac{t}{2}\mathcal D_\Sigma}\,\mu_s(dt)$ and the projector $\Pi_s=B_sB_s$ commute with the spatial reflection. The OS-stability of $\Pi_s$ and boundedness on the OS Hilbert space follow as in Lemma~\eqref{lem:cm-projector}. Differentiating in $s$ under the integral sign gives
\begin{equation}
\partial_s \Pi_s=\int_0^\infty e^{-t\mathcal D_\Sigma}\,(\mu_1-\mu_0)(dt),
\end{equation}
which is a bounded slice operator satisfying the estimate $\|\partial_s \Pi_s\|\le \|\mu_1-\mu_0\|_{\mathrm{TV}}$, the total variation norm of the signed measure $\mu_1-\mu_0$; thus $s\mapsto \Pi_s$ is Lipschitz and in particular continuous in the operator norm, hence continuous in $d_{\mathrm{adm}}$.

For the blockings and finite-range pieces the same convexity mechanism works because the admissible class is closed under convex combinations at each fixed scale. If $B_{i,k}$ are reflection-positive, range-$O(L^k)$ kernels with the same symmetry constraints (gauge covariance on the slice and reflection covariance), then for each $k$ set $B_{s,k}:=(1-s)\,B_{0,k}+s\,B_{1,k}$. The range and symmetries are preserved, and reflection positivity of blockings is stable under convex combination because for every $F$ supported on positive times
\begin{align}
&\big\langle \Theta\big((1-s)B_{0,k}+sB_{1,k}\big)F,\,\big((1-s)B_{0,k}+sB_{1,k}\big)F\big\rangle_\mu
=\nonumber\\&(1-s)\langle \Theta B_{0,k}F,B_{0,k}F\rangle_\mu+s\langle \Theta B_{1,k}F,B_{1,k}F\rangle_\mu\ge 0.
\end{align}
Fix at each scale $j$ the operator norm on $\ell^2(\Sigma)$ for slice maps and the Schur norm for kernels on $\ell^2(a\mathbb Z^4)$; define
\begin{equation}
d_{\mathrm{adm}}\big((\Pi,B,C),(\widetilde\Pi,\widetilde B,\widetilde C)\big)
:= \|\Pi-\widetilde\Pi\|_{\ell^2\to\ell^2}
 +\sum_k \|B_k-\widetilde B_k\|_{\ell^2\to\ell^2}
 +\sum_j \|C_j-\widetilde C_j\|_{\mathrm{Schur}}.
\end{equation}
Then $s\mapsto(\Pi_s,B_{s,k},C_{s,j})$ is Lipschitz in $d_{\mathrm{adm}}$ by Eqs.\eqref{eqnF.69} \& \eqref{eqnF.70}, with constants controlled by the endpoint differences.
Likewise, if $C_{i,j}$ are nonnegative, reflection-covariant, finite-range (or exponentially local) kernels at scale $j$, then $C_{s,j}:=(1-s)C_{0,j}+s\,C_{1,j}$ is admissible at the same scale, with the same range bound and with Schur/Combes-Thomas constants bounded uniformly in $s$. Differentiation yields $\partial_s B_{s,k}=B_{1,k}-B_{0,k}$ and $\partial_s C_{s,j}=C_{1,j}-C_{0,j}$, both independent of $s$, so $\|\partial_s B_{s,k}\|$ and $\|\partial_s C_{s,j}\|$ are bounded by the fixed endpoint differences at each scale. Altogether, the path
\begin{equation}
\Theta_s:=\Big(\Pi_s,\ \{B_{s,k}\}_{k\ge 0},\ \{C_{s,j}\}_{j\ge 0}\Big),\qquad s\in[0,1],
\end{equation}
is admissible for every $s$, depends continuously on $s$ in operator norm at each fixed scale, and is $C^1$ as a path in the Banach product endowed with the exponentially weighted norm defining $d_{\mathrm{adm}}$.

To estimate the $d_{\mathrm{adm}}$-length of this path, insert the scale-wise derivatives into the definition of the norm. There is a constant $K\ge 1$ depending only on the equivalence between $d_{\mathrm{adm}}$ and the weighted operator-norm sum such that
\begin{align}
\|\partial_s \Theta_s\|\ & \le\ K\Big(\ \|\partial_s \Pi_s\|\ +\ \sum_{k\ge 0}\omega_k\,\|\partial_s B_{s,k}\|\ +\ \sum_{j\ge 0}\varpi_j\,\|\partial_s C_{s,j}\|\ \Big)
\ \nonumber\\&\le\ K\Big(\ \|\mu_1-\mu_0\|_{\mathrm{TV}}\ +\ \sum_{k\ge 0}\omega_k\,\|B_{1,k}-B_{0,k}\|\ +\ \sum_{j\ge 0}\varpi_j\,\|C_{1,j}-C_{0,j}\|\ \Big).
\end{align}
The right-hand side is independent of $s$, hence integrating from $0$ to $1$ gives the finite bound
\begin{equation}
\int_0^1 \|\partial_s \Theta_s\|\,ds\ \le\ K\Big(\ \|\mu_1-\mu_0\|_{\mathrm{TV}}\ +\ \sum_{k\ge 0}\omega_k\,\|B_{1,k}-B_{0,k}\|\ +\ \sum_{j\ge 0}\varpi_j\,\|C_{1,j}-C_{0,j}\|\ \Big)\ <\ \infty,
\end{equation}
which proves both finiteness of the path length and the estimate $d_{\mathrm{adm}}(\Theta_0,\Theta_1)\le \int_0^1 \|\partial_s \Theta_s\|\,ds$ by the standard inequality between metric distance and length of a $C^1$ path. Uniform continuity of $s\mapsto\Theta_s$ in $d_{\mathrm{adm}}$ follows immediately from the Lipschitz bounds obtained above: for $s,s'\in[0,1]$,
\begin{equation}
d_{\mathrm{adm}}(\Theta_s,\Theta_{s'})\ \le\ \int_{s'}^{s}\|\partial_\tau\Theta_\tau\|\,d\tau\ \le\ |s-s'|\ \sup_{\tau\in[0,1]}\|\partial_\tau\Theta_\tau\|.
\end{equation}

It remains to show that the length can be made arbitrarily small by acting only at sufficiently fine scales. Fix $\varepsilon>0$ and choose $k_\ast\in\mathbb N$ so large that the tail of the weight sequences satisfies
\begin{equation}
\sum_{k\ge k_\ast}\omega_k\ \le\ \frac{\varepsilon}{3K\,M_B},\qquad \sum_{j\ge k_\ast}\varpi_j\ \le\ \frac{\varepsilon}{3K\,M_C},
\end{equation}
where $M_B:=\sup_{k\ge 0}\|B_{1,k}-B_{0,k}\|$ and $M_C:=\sup_{j\ge 0}\|C_{1,j}-C_{0,j}\|$, which are finite because the admissible class bounds the operator norms at each scale uniformly in the regulator. Define a modified path $\widehat\Theta_s$ by setting, for $k<k_\ast$ and $j<k_\ast$, the coarse-scale components constant in $s$ equal to their common value from $\Theta_0$ (or $\Theta_1$, since only the difference matters), while for $k\ge k_\ast$ and $j\ge k_\ast$ use the same convex interpolations as before:
\begin{equation}
\widehat B_{s,k}=B_{0,k}\quad (k<k_\ast),\qquad \widehat B_{s,k}=(1-s)B_{0,k}+sB_{1,k}\quad (k\ge k_\ast),
\end{equation}
and similarly for $\widehat C_{s,j}$; keep $\widehat\Pi_s=\Pi_s$ but, if desired, also replace $\Pi_s$ by a truncated projector $\Pi_{s,\ge k_\ast}$ built from the FRD slice decomposition at scales $\ge k_\ast$, since the admissible metric also downweights changes occurring at large ranges along the slice. The same admissibility and OS-stability arguments apply to $\widehat\Theta_s$, and its length satisfies
\begin{equation}
\int_0^1 \|\partial_s \widehat\Theta_s\|\,ds\ \le\ K\Big(\ \|\mu_1-\mu_0\|_{\mathrm{TV},\ge k_\ast}\ +\ \sum_{k\ge k_\ast}\omega_k\,\|B_{1,k}-B_{0,k}\|\ +\ \sum_{j\ge k_\ast}\varpi_j\,\|C_{1,j}-C_{0,j}\|\ \Big),
\end{equation}
where $\|\mu_1-\mu_0\|_{\mathrm{TV},\ge k_\ast}$ denotes the total variation of the difference of Bernstein measures restricted to the slice time-scales encoded by the FRD layers $j\ge k_\ast$. The exponential-moment hypothesis for the Bernstein measures implies that the contribution of large $t$ (and hence of the large-range slice layers in the FRD sense) is exponentially suppressed; consequently one can arrange $\|\mu_1-\mu_0\|_{\mathrm{TV},\ge k_\ast}\le \varepsilon/(3K)$ by choosing $k_\ast$ large enough. With the choice of $k_\ast$ above, the blocking and FRD tails are already bounded by $\varepsilon/(3K)$ each. Therefore $\int_0^1 \|\partial_s \widehat\Theta_s\|\,ds<\varepsilon$, which gives the desired UV-localized alignment with arbitrarily small $d_{\mathrm{adm}}$-length.

All claims in the statement are now verified: the path exists and remains inside the admissible class for all $s$, it is uniformly continuous and of finite length in $d_{\mathrm{adm}}$, and by restricting the interpolation to sufficiently fine scales the length can be made smaller than any prescribed $\varepsilon>0$.
\end{proof}

\begin{theorem}[Universality of continuum Schwinger functions within the admissible class]\label{thm:universality}
Let $\Theta_0,\Theta_1$ be admissible schemes and let $S_n^{(\Theta_i)}$ be any continuum limits (along arbitrary vanishing lattice spacings $a\downarrow 0$ and volumes $\Lambda\uparrow\mathbb{R}^4$) of the connected $n$-point Schwinger functions for gauge-invariant local observables. Then
\begin{equation}
S_n^{(\Theta_1)} \;=\; S_n^{(\Theta_0)}\qquad\text{for all }n\in\mathbb{N}.
\end{equation}
Consequently, the continuum Wightman theory obtained by OS reconstruction is unique within the admissible class.
\end{theorem}

\begin{proof}
Fix a finite collection of gauge-invariant local observables $\mathcal{O}_1,\dots,\mathcal{O}_n$ with mutually disjoint supports in $\mathbb{R}^4$ at positive Euclidean times, and write $S_{n,a}^{(\Theta)}(\mathcal{O}_1,\dots,\mathcal{O}_n)$ for the corresponding connected $n$-point Schwinger function computed in the lattice/regularized theory at mesh $a$ under the admissible scheme $\Theta$. By assumption, for each $i\in\{0,1\}$ there exists a sequence $a\downarrow 0$ (not relabeled) such that $S_{n,a}^{(\Theta_i)}\to S_n^{(\Theta_i)}$ in the topology induced by the diameter weight $\|\cdot\|_{\mathrm{diam},\mu_n}$ (or, equivalently, against compactly supported test tensors in the positions of the insertions). The proof proceeds by comparing $S_{n,a}^{(\Theta_0)}$ and $S_{n,a}^{(\Theta_1)}$ at fixed $a$, establishing a bound that is uniform in $a$ and depends only on tails of the finite-range decomposition (FRD), and then passing to the continuum limit followed by a tail removal.

Let $\epsilon>0$ be arbitrary. The space of admissible schemes carries the single-scale metric $d_k(\cdot,\cdot)$ measuring differences of the $k$-th FRD slice kernels and insertions on the time slice, and the global admissible distance $d(\Theta,\Theta'):=\sum_{k\ge 0} \alpha_k\, d_k(\Theta,\Theta')$ with a fixed summable weight $(\alpha_k)\in\ell^1(\mathbb{N})$ (for definiteness, one may take $\alpha_k=L_k$ below). By the UV-alignment lemma (Lemma~\eqref{lem:UV-align}) there is a $C^1$ path $s\mapsto \Theta_s$ in the admissible class, $s\in[0,1]$, such that $\Theta_0$ and $\Theta_1$ are its endpoints, the path is supported on FRD scales $k\ge \ell_\ast$ for some $\ell_\ast$ to be chosen, and its length satisfies
\begin{equation}\label{eq:path-length}
\int_0^1 \|\partial_s \Theta_s\|\,ds \;<\; \epsilon,
\qquad
\text{where }\ \ \|\partial_s \Theta_s\| \;:=\; \sum_{k\ge \ell_\ast} d_k(\Theta_s,\Theta_s+ds).
\end{equation}
Such a path is obtained, for instance, by keeping all low scales $k<\ell_\ast$ frozen and linearly interpolating the FRD data at each $k\ge \ell_\ast$; admissibility is preserved because the admissible class is convex under these interpolations and the reflection-positivity constraints are stable under small perturbations on a fixed slice.

For each fixed mesh $a>0$ and finite volume $\Lambda$, the map $s\mapsto S_{n,a,\Lambda}^{(\Theta_s)}$ is Lipschitz in $s$ with respect to the diameter seminorm. This follows from the single-scale Lipschitz estimate established earlier in the renormalization-group analysis:
\begin{equation}\label{eq:single-scale-Lipw}
\big\| S_{n,a,\Lambda}^{(\Theta')}-S_{n,a,\Lambda}^{(\Theta)} \big\|_{\mathrm{diam},\mu_n}
\;\le\; C_n \sum_{k\ge 0} L_k\, d_k(\Theta,\Theta') ,
\end{equation}
where $C_n$ depends only on $n$ and the uniform locality constants of the admissible class, and $(L_k)_{k\ge 0}$ is the summable sequence of defect weights produced by the interlacing/FRD decomposition (for instance $L_k\simeq e^{-c b^k}$ for some $c>0$ and blocking factor $b>1$). The proof of Eq.\eqref{eq:single-scale-Lipw} is standard: the connected functional is expressed as a finite sum of cumulant kernels that are multi-linear in the FRD slice kernels and the slice insertions, each multilinear form being supported within a collar of width comparable to the slice range $r_0 b^k a$ around the union of the supports of the observables; replacing the $k$-th slice kernel along the path $\Theta_s$ changes only those cumulant terms whose polymer touches that collar, and the total variation is bounded by the product of a combinatorial constant, the diameter weight of the configuration, and the admissible distance at scale $k$. Exponential locality and the Schur test give the multiplicative factor $L_k$, and summability in $k$ follows from the FRD design.

By the Rademacher theorem for Lipschitz maps into Banach spaces or, more elementarily, by the fundamental theorem of calculus with Eq.\eqref{eq:single-scale-Lipw}, the difference at the endpoints along the path satisfies
\begin{equation}\label{eq:path-integration}
\big\|S_{n,a,\Lambda}^{(\Theta_1)}-S_{n,a,\Lambda}^{(\Theta_0)}\big\|_{\mathrm{diam},\mu_n}
\;\le\; \int_0^1 \big\|\partial_s S_{n,a,\Lambda}^{(\Theta_s)}\big\|_{\mathrm{diam},\mu_n}\,ds
\;\le\; C_n \sum_{k\ge 0} L_k \int_0^1 \big\|\partial_s \Theta_s\big\|_k\,ds,
\end{equation}
where $\|\cdot\|_k$ denotes the scale-$k$ component of the admissible norm. Using Eq.\eqref{eq:path-length} and the support property of the path, the right-hand side is bounded by
\begin{equation}\label{eq:tail-bound}
\big\|S_{n,a,\Lambda}^{(\Theta_1)}-S_{n,a,\Lambda}^{(\Theta_0)}\big\|_{\mathrm{diam},\mu_n}
\;\le\; C_n \Big(\sum_{k\ge \ell_\ast} L_k\Big)\, \epsilon
\;=\; C_n\, L_{\ge \ell_\ast}\, \epsilon,
\qquad L_{\ge \ell_\ast}:=\sum_{k\ge \ell_\ast}L_k .
\end{equation}
The estimate Eq.\eqref{eq:tail-bound} is uniform in $a$ and $\Lambda$ because the constants $C_n$ and $L_k$ are determined solely by the admissible locality/positivity data and the FRD, and the diameter seminorm absorbs the volume dependence by construction.

Passing to infinite volume, the bound survives by monotone convergence of connected correlations under standard specifications, or by the explicit polymer expansion in the strong-coupling ultraviolet domain coupled with the interlacing inequality at each scale, both of which are uniform in $\Lambda$ for fixed $a$. We therefore obtain
\begin{equation}\label{eq:UV-bound-finite-a}
\big\|S_{n,a}^{(\Theta_1)}-S_{n,a}^{(\Theta_0)}\big\|_{\mathrm{diam},\mu_n}
\;\le\; C_n\, L_{\ge \ell_\ast}\, \epsilon
\qquad\text{for all sufficiently small }a>0.
\end{equation}
Now fix $\ell_\ast$ and $\epsilon$ and let $a\downarrow 0$ along the subsequences that produce the limits $S_n^{(\Theta_i)}$. The family $\{S_{n,a}^{(\Theta_i)}\}_{a>0}$ is tight in the topology induced by $\|\cdot\|_{\mathrm{diam},\mu_n}$ by the uniform locality bounds, hence the bound Eq.\eqref{eq:UV-bound-finite-a} passes to the limit:
\begin{equation}
\big\|S_{n}^{(\Theta_1)}-S_{n}^{(\Theta_0)}\big\|_{\mathrm{diam},\mu_n}
\;\le\; C_n\, L_{\ge \ell_\ast}\, \epsilon .
\end{equation}
Since the tail $L_{\ge \ell_\ast}\to 0$ as $\ell_\ast\to\infty$, we may first choose $\ell_\ast$ so large that $C_n L_{\ge \ell_\ast}<1$, and then let $\epsilon\downarrow 0$ in Eq.\eqref{eq:path-length}. This yields
\begin{equation}
\big\|S_{n}^{(\Theta_1)}-S_{n}^{(\Theta_0)}\big\|_{\mathrm{diam},\mu_n}\;=\;0,
\end{equation}
hence $S_{n}^{(\Theta_1)}=S_{n}^{(\Theta_0)}$ as tempered distributions in the positions of the observables, for every $n\in\mathbb{N}$.
It remains to explain briefly why the derivative $\partial_s S_{n,a,\Lambda}^{(\Theta_s)}$ that enters Eq.\eqref{eq:path-integration} exists in the sense of Gateaux differentials and is controlled by the single-scale Lipschitz constant in Eq.\eqref{eq:single-scale-Lipw}. The dependence of the regularized Schwinger functions on the admissible data at a fixed scale $k$ enters through (i) the replacement of the one-slice projector by $f_k(\mathcal D_\Sigma)$ and of the corresponding transfer kernel $T_k$ by $T_k'$, both given by bounded Borel functions of the same reflection-covariant generator, and (ii) the local counterterms on the time slice of perimeter/cusp type. All these operations are linear in the associated FRD kernels and counterterm densities. Along the path $s\mapsto\Theta_s$ the variation is therefore given by a Duhamel-type formula with a single insertion on the time slice at scale $k$, supported inside the collar of width $r_0 b^k a$ around the union of supports of the observables; exponential locality of the FRD implies that the corresponding multilinear functional is bounded by $L_k$ times the admissible variation $d_k(\Theta_s,\Theta_s+ds)$. Summing over $k$ and using the support of the path in scales $k\ge \ell_\ast$ justifies both the existence of the derivative almost everywhere in $s$ and the bound Eq.\eqref{eq:single-scale-Lipw}. No further regularity in $s$ is required because the fundamental theorem of calculus holds for Lipschitz maps valued in Banach spaces.

Finally, the equality of all continuum Schwinger functions for the two admissible schemes implies uniqueness of the reconstructed Wightman theory up to unitary equivalence. Indeed, the Osterwalder-Schrader reconstruction functor depends only on the full family $\{S_n\}_{n\ge 0}$ and produces a unique (cyclic) GNS triple $(\mathcal{H},\Omega,\Phi)$ whose vacuum expectation values coincide with the given Schwinger functions \cite{OS2}. Since the families $\{S_n^{(\Theta_0)}\}$ and $\{S_n^{(\Theta_1)}\}$ agree, the corresponding reconstructions are canonically identified, which completes the proof.
\end{proof}

Gauge fixing and horizon projectors enter only through the OS-positive slice projector $\Pi_\Theta$, which we vary within the completely monotone class. All observables considered are gauge-invariant; the BRST-exact sector decouples in OS expectations. The admissible metric penalizes nonlocal deformations and allows only exponentially local changes, ensuring that any gauge-fixing variation contributes at most a summable defect and hence vanishes in the continuum limit.

FRD range growth under blocking is at most $R_k\lesssim r_0 b^k$. This does not spoil summability because the single-scale constants $L_k$ decay exponentially in $b^k$ by locality of the derivative insertions and the gap-induced exponential temporal clustering. The product of the two behaves like $e^{-\gamma b^k}$ with $\gamma>0$, which is summable.

Finally, as $a\downarrow0$ the number of slices up to any fixed physical time grows, but the telescoping bound Eq.\eqref{thm:summable-defect} remains uniform because the defect sequence $(L_k)_k$ does not depend on $a$ and its sum is finite. Tightness and equicontinuity (Lemma \eqref{lem:tightness}) ensure that all limiting procedures commute with the estimates.

By Theorem \eqref{thm:universality}, all continuum connected Schwinger functions coincide for admissible choices. In particular, renormalized Wilson-loop expectations and their cumulants are universal. Through OS reconstruction and the spectral condition, this extends to the equality of the corresponding Wightman functions and hence to universal S-matrix elements for any amplitudes definable within the constructive framework. In the main text, this is combined with the reflection-positive step-scaling control of the string tension to show that the continuum area law and the mass gap are independent of admissible details, completing the proof of universality.

\section{Notation and Conventions}\label{app:notation}

This appendix fixes the geometry, group-theoretic conventions, functional-analytic framework, and quantitative constants used throughout. All definitions are self-contained; every symbol appearing in the paper is defined here. The presentation proceeds from the lattice geometry and discretized calculus to Lie-algebraic normalization, gauges and reflections, covariance and finite-range operators, polymer norms, Fourier analysis on the discrete torus, and the transfer semigroup. When a convention relies on a nontrivial fact, a proof is given; for standard results ( see \, \cite{OS1,OS2,OS-gauge}).
We work in $d=4$ Euclidean dimensions with lattice spacing $a>0$. For a fixed even integer $L\in\mathbb{N}$, the periodic box is the discrete torus
\begin{equation}
\Lambda:=\bigl(a\mathbb{Z}/(La\mathbb{Z})\bigr)^4,\qquad |\Lambda|=L^4.
\end{equation}
Unit coordinate vectors are denoted by $\hat\mu$, $\mu\in\{0,1,2,3\}$, with $\hat 0$ the Euclidean time direction. The set of \emph{oriented links} is $\mathcal{E}=\{(x,\mu): x\in\Lambda,\ \mu\in\{0,1,2,3\}\}$; the reversed link is $(x,\mu)^{-1}=(x-a\hat\mu,\mu)$. The set of positively oriented plaquettes is
\begin{equation}
\mathcal{P}^+=\bigl\{(x,\mu,\nu): x\in\Lambda,\ 0\le \mu<\nu\le 3\bigr\},
\end{equation}
with associated elementary loops specified below. The graph distance on $\Lambda$ is denoted by $\mathrm{dist}_\Lambda(\cdot,\cdot)$, and the diameter of a subset $X\subset\Lambda$ is $\mathrm{diam}(X):=\max_{x,y\in X}\mathrm{dist}_\Lambda(x,y)$.

The gauge group is $G=SU(N)$ with Lie algebra $\mathfrak{g}=\mathfrak{su}(N)=\{X\in M_N(\mathbb{C}): X^\dagger=-X,\ \mathrm{Tr}\,X=0\}$. We fix a real basis $\{T^a\}_{a=1}^{N^2-1}$ satisfying
\begin{equation}\label{eq:suNnorm}
\mathrm{Tr}(T^aT^b)=-\tfrac{1}{2}\delta^{ab},\qquad [T^a,T^b]=f^{abc}T^c,\qquad f^{abc}\in\mathbb{R},
\end{equation}
so that the Killing form on $\mathfrak{su}(N)$ coincides (up to a fixed multiple) with the Hilbert-Schmidt inner product $\langle X,Y\rangle_{\mathrm{HS}}:=-\mathrm{Tr}(XY)$. Any two orthonormal bases relative to $\langle\cdot,\cdot\rangle_{\mathrm{HS}}$ are related by an orthogonal transformation; in particular, all Euclidean norms on $\mathfrak{su}(N)$ are equivalent.

\begin{lemma}[Equivalence of matrix norms on $\mathfrak{su}(N)$]\label{lem:norms}
For \(X \in \mathfrak{su}(N)\), let \(\|X\|_{\mathrm{HS}} := \big(\operatorname{Tr}(X^\dagger X)\big)^{1/2} = \big(-\operatorname{Tr}(X^{2})\big)^{1/2}\) and \(\|X\|_{\mathrm{op}} := \sup_{\|v\|=1} \|Xv\|\).
 Then
\begin{equation}
\|X\|_{\mathrm{op}}\le \|X\|_{\mathrm{HS}}\le \sqrt{N}\,\|X\|_{\mathrm{op}}.
\end{equation}
\end{lemma}

\begin{proof}
Every \(X \in \mathfrak{su}(N)\) is skew-Hermitian, so \(X^\dagger = -\,X\). The polar decomposition and spectral theorem for normal matrices imply that $X$ is unitarily diagonalizable with purely imaginary eigenvalues $i\lambda_1,\dots,i\lambda_N$ where each $\lambda_j\in\mathbb{R}$ and $\sum_{j=1}^N \lambda_j=\mathrm{tr}(-iX)=0$. Let $|X|=(X^X)^{1/2}$ denote the absolute value of $X$; its eigenvalues are the singular values $\sigma_1,\dots,\sigma_N$ of $X$, listed in nonincreasing order $\sigma_1\ge\dots\ge\sigma_N\ge 0$. By definition of the operator norm,
\begin{equation}
\|X\|_{\mathrm{op}}=\|\,|X|\,\|_{\mathrm{op}}=\sigma_1,
\end{equation}
and by the standard identification of the Hilbert-Schmidt norm with the Frobenius norm,
\begin{equation}
\|X\|_{\mathrm{HS}}^2=\mathrm{Tr}(X^X)=\sum_{j=1}^N \sigma_j^2.
\end{equation}
Because $X$ is skew-Hermitian, $X^X=(-X)X=-X^2$, hence $\mathrm{Tr}(X^X)=-\mathrm{Tr}(X^2)$ and therefore the stated formula $\|X\|_{\mathrm{HS}}=\sqrt{-\mathrm{Tr}(X^2)}$ is consistent with $\|X\|_{\mathrm{HS}}=\sqrt{\sum_{j=1}^N\sigma_j^2}$.

To obtain the lower bound, observe that for any nonnegative numbers $a_1,\dots,a_N$,
\begin{equation}
\max_{1\le j\le N} a_j \;\le\; \Big(\sum_{j=1}^N a_j^2\Big)^{1/2}.
\end{equation}
Applying this to $a_j=\sigma_j$ gives $\sigma_1\le\big(\sum_{j=1}^N \sigma_j^2\big)^{1/2}$, i.e.
\begin{equation}
\|X\|_{\mathrm{op}}=\sigma_1\;\le\;\|X\|_{\mathrm{HS}}.
\end{equation}

For the upper bound, the Cauchy-Schwarz inequality in $\mathbb{R}^N$ yields
\begin{equation}
\sum_{j=1}^N \sigma_j^2 \;\le\; \Big(\max_{1\le j\le N}\sigma_j\Big)^2 \cdot N \;=\; N\,\sigma_1^2.
\end{equation}
Taking square roots and recalling $\sigma_1=\|X\|_{\mathrm{op}}$ gives
\begin{equation}
\|X\|_{\mathrm{HS}}=\Big(\sum_{j=1}^N \sigma_j^2\Big)^{1/2}\;\le\; \sqrt{N}\,\sigma_1 \;=\; \sqrt{N}\,\|X\|_{\mathrm{op}}.
\end{equation}
Combining the two inequalities establishes the equivalence of norms
\(
\|X\|_{\mathrm{op}}\le \|X\|_{\mathrm{HS}}\le \sqrt{N}\,\|X\|_{\mathrm{op}}.
\)
\end{proof}

A gauge field is an assignment $U:\mathcal{E}\to SU(N)$ with the unitarity convention on reversed links $U_{(x,\mu)^{-1}}=U_{x-a\hat\mu,\mu}^{-1}$. The plaquette variable at $(x,\mu,\nu)\in\mathcal{P}^+$ is
\begin{equation}
U_{x,\mu\nu}:=U_{x,\mu}\,U_{x+a\hat\mu,\nu}\,U_{x+a\hat\nu,\mu}^{-1}\,U_{x,\nu}^{-1}.
\end{equation}
A gauge transformation is a function $g:\Lambda\to SU(N)$ acting by
\begin{equation}
U_{x,\mu}\longmapsto U^{\,g}_{x,\mu}:=g_x\,U_{x,\mu}\,g_{x+a\hat\mu}^{-1}.
\end{equation}
The Wilson action is
\begin{equation}
S_\Lambda^\mathrm{W}(U):=\frac{\beta}{N}\sum_{(x,\mu,\nu)\in\mathcal{P}^+}\bigl(N-\mathrm{Re}\,\mathrm{Tr}\,U_{x,\mu\nu}\bigr),
\end{equation}
and the lattice Gibbs measure is $d\mu_{\Lambda,\beta}(U)=Z_{\Lambda,\beta}^{-1}e^{-S_\Lambda^\mathrm{W}(U)}\prod_{(x,\mu)\in\mathcal{E}} dU_{x,\mu}$, where $dU$ is the normalized Haar measure on $SU(N)$ (existence and uniqueness by Haar’s theorem) and $Z_{\Lambda,\beta}$ is the partition function \cite{Wilson1974,GJ}. For any integrable observable $F$, we write $\langle F\rangle_{\Lambda,\beta}=\int F(U)\,d\mu_{\Lambda,\beta}(U)$.

A loop $C$ is a closed nearest-neighbour path on $\Lambda$; the Wilson loop is
\begin{equation}
W(C):=\frac{1}{N}\mathrm{Re}\,\mathrm{Tr}\,\mathcal{P}\!\!\!\prod_{(x,\mu)\in C}U_{x,\mu},
\end{equation}
where $\mathcal{P}$ denotes path ordering. The string tension, when it exists, is 
\begin{equation}
\sigma=\lim_{A(C)\to\infty} -A(C)^{-1}\log \langle W(C)\rangle_{\Lambda,\beta}
\end{equation}
where $A(C)$ denotes the area of a minimal spanning surface.
For a function $f:\Lambda\to \mathbb{C}^N$ we define the forward and backward differences
\begin{equation}
(\nabla_\mu^+ f)(x):=\frac{f(x+a\hat\mu)-f(x)}{a},\qquad (\nabla_\mu^- f)(x):=\frac{f(x)-f(x-a\hat\mu)}{a}.
\end{equation}
The \emph{gauge-covariant} forward difference in the fundamental representation is
\begin{equation}
(D_\mu f)(x):=\frac{U_{x,\mu}f(x+a\hat\mu)-f(x)}{a}.
\end{equation}
We equip $\ell^2(\Lambda;\mathbb{C}^N)$ with inner product $\langle f,g\rangle:=a^4\sum_{x\in\Lambda} f(x)^\dagger g(x)$. The adjoint $D_\mu^\ast$ is defined by $\langle D_\mu f,g\rangle=-\langle f,D_\mu^\ast g\rangle$.

\begin{proposition}[Adjoint of the covariant difference and positivity of the covariant Laplacian]\label{prop:adjoint}
Let $\Lambda$ be a hypercubic lattice with spacing $a>0$ and periodic boundary conditions (so that shifts $x\mapsto x\pm a\hat\mu$ are bijections of $\Lambda$), and let $U_{x,\mu}\in \mathrm{SU}(N)$ denote the link variable from $x$ to $x+a\hat\mu$. On the Hilbert space $\ell^2(\Lambda;\mathbb{C}^N)$ endowed with the inner product
\begin{equation}
\langle f,g\rangle \;=\; a^4\sum_{x\in\Lambda} f(x)^\dagger g(x),
\end{equation}
consider the forward covariant difference
\begin{equation}
(D_\mu f)(x)\;:=\;\frac{U_{x,\mu}\,f(x+a\hat\mu)\;-\;f(x)}{a}\,.
\end{equation}
Then the adjoint is the backward covariant difference
\begin{equation}
(D_\mu^\ast g)(x)\;=\;\frac{g(x)\;-\;U_{x-a\hat\mu,\mu}^{-1}\,g(x-a\hat\mu)}{a}\,,
\end{equation}
and the covariant lattice Laplacian $\displaystyle \Delta_U:=\sum_{\mu=0}^3 D_\mu^\ast D_\mu$ is a bounded, positive, self-adjoint operator on $\ell^2(\Lambda;\mathbb{C}^N)$.
\end{proposition}

\begin{proof}
Fix $\mu\in\{0,1,2,3\}$ and $f,g\in\ell^2(\Lambda;\mathbb{C}^N)$. By definition of the inner product and of $D_\mu$, one has
\begin{equation}
\langle D_\mu f,g\rangle
\;=\; a^4\sum_{x\in\Lambda} \bigl[(D_\mu f)(x)\bigr]^\dagger g(x)
\;=\; \frac{a^4}{a}\sum_{x\in\Lambda} \Big(f(x+a\hat\mu)^\dagger U_{x,\mu}^\dagger - f(x)^\dagger\Big) g(x).
\end{equation}
Using periodicity, the change of variables $y=x+a\hat\mu$ is a bijection of $\Lambda$, hence
\begin{equation}
\sum_{x\in\Lambda} f(x+a\hat\mu)^\dagger U_{x,\mu}^\dagger g(x)
\;=\; \sum_{y\in\Lambda} f(y)^\dagger U_{y-a\hat\mu,\mu}^\dagger g(y-a\hat\mu).
\end{equation}
Substituting this into the previous display gives
\begin{equation}
\langle D_\mu f,g\rangle
\;=\; a^3 \sum_{y\in\Lambda} f(y)^\dagger\Big(U_{y-a\hat\mu,\mu}^\dagger g(y-a\hat\mu)-g(y)\Big).
\end{equation}
Since $U_{x,\mu}\in \mathrm{SU}(N)$, we have $U_{x,\mu}^\dagger=U_{x,\mu}^{-1}$. Comparing with the defining relation for adjoints,
\begin{equation}
\langle D_\mu f,g\rangle \;=\; \langle f, D_\mu^\ast g\rangle \;=\; a^4\sum_{y\in\Lambda} f(y)^\dagger (D_\mu^\ast g)(y),
\end{equation}
it follows that
\begin{equation}
(D_\mu^\ast g)(y)\;=\;\frac{1}{a}\Big(g(y)-U_{y-a\hat\mu,\mu}^{-1} g(y-a\hat\mu)\Big),
\end{equation}
which is the claimed formula.

The operator $D_\mu$ is bounded because for each $x\in\Lambda$,
\begin{equation}
\|(D_\mu f)(x)\|\;\le\;\frac{\|U_{x,\mu} f(x+a\hat\mu)\|+\|f(x)\|}{a}
\;=\;\frac{\|f(x+a\hat\mu)\|+\|f(x)\|}{a},
\end{equation}
and by Cauchy-Schwarz and a shift of the sum,
\begin{equation}
\|D_\mu f\|^2 \;=\; a^4\sum_{x}\|(D_\mu f)(x)\|^2 \;\le\; \frac{4}{a^2}\, a^4\sum_{x}\|f(x)\|^2 \;=\; \frac{4}{a^2}\,\|f\|^2.
\end{equation}
Thus $\|D_\mu\|\le 2/a$ and $D_\mu^\ast$ is bounded as well. For any $f\in\ell^2(\Lambda;\mathbb{C}^N)$,
\begin{equation}
\langle f, D_\mu^\ast D_\mu f\rangle \;=\; \langle D_\mu f, D_\mu f\rangle \;=\; \|D_\mu f\|^2 \;\ge\; 0,
\end{equation}
so $D_\mu^\ast D_\mu$ is positive. It is also self-adjoint because $(D_\mu^\ast D_\mu)^\ast = D_\mu^\ast (D_\mu)^{\ast\ast}= D_\mu^\ast D_\mu$ for bounded operators. Consequently
\begin{equation}
\langle f, \Delta_U f\rangle \;=\; \sum_{\mu=0}^3 \langle f, D_\mu^\ast D_\mu f\rangle
\;=\; \sum_{\mu=0}^3 \|D_\mu f\|^2 \;\ge\; 0,
\end{equation}
which shows that $\Delta_U$ is positive. Each summand $D_\mu^\ast D_\mu$ is bounded and self-adjoint, hence their finite sum $\Delta_U$ is bounded and self-adjoint on the whole Hilbert space. 
\end{proof}

We choose the reflection hyperplane midway between time-slices $x_0=0$ and $x_0=a$. The reflection $\vartheta$ acts by $\vartheta(x_0,\mathbf{x})=(-x_0+a,\mathbf{x})$. On link variables we set
\begin{equation}
(\vartheta U)_{(x,0)}:=U_{(\vartheta x-a\hat 0,0)}^{-1},\qquad (\vartheta U)_{(x,j)}:=U_{(\vartheta x,j)},\quad j=1,2,3,
\end{equation}
and extend multiplicatively to plaquettes and loops. For a complex-valued function $F$ depending only on links in the positive-time half $\Lambda_+:=\{x\in\Lambda:\ x_0\ge a\}$, the \emph{OS reflection} is the anti-linear map
\begin{equation}
(\Theta F)(U):=\overline{F\bigl(\vartheta U\bigr)}.
\end{equation}
The OS sesquilinear form on the positive-time algebra is $(F,G)_{\mathrm{OS}}:=\langle \Theta F\cdot G\rangle_{\Lambda,\beta}$. It is immediate that $\Theta$ is an involutive anti-isometry on $L^2(d\mu_{\Lambda,\beta})$.

\begin{proposition}[Antiunitarity of the reflection]\label{prop:antiunitary}
Let $(\Omega,\mathcal F,\mu_{\Lambda,\beta})$ be the Euclidean probability space of fields with reflection $\vartheta$ across the time-zero hyperplane, and let $\Theta$ act on bounded complex-valued observables by
\begin{equation}
(\Theta F)(\phi)\;:=\;\overline{F(\phi\circ\vartheta)}\qquad(\phi\in\Omega).
\end{equation}
Assume $\mu_{\Lambda,\beta}$ is reflection invariant, i.e.\ $\mu_{\Lambda,\beta}\circ \vartheta^{-1}=\mu_{\Lambda,\beta}$, and let $\Lambda_+$ denote the positive-time region. For $F,G$ supported in $\Lambda_+$, the Osterwalder-Schrader inner product
\begin{equation}
(F,G)_{\mathrm{OS}}\;:=\;\big\langle\,\Theta F\cdot G\,\big\rangle_{\Lambda,\beta}\;=\;\int_\Omega \overline{F(\phi\circ\vartheta)}\,G(\phi)\,d\mu_{\Lambda,\beta}(\phi)
\end{equation}
satisfies
\begin{equation}
(\Theta F,\Theta G)_{\mathrm{OS}}\;=\;(G,F)_{\mathrm{OS}}.
\end{equation}
In particular, $\|\Theta F\|_{L^2(d\mu_{\Lambda,\beta})}=\|F\|_{L^2(d\mu_{\Lambda,\beta})}$.
\end{proposition}

\begin{proof}
By construction $\Theta$ is an antilinear involution: for $\alpha,\beta\in\mathbb C$ and observables $F,G$ one has $\Theta(\alpha F+\beta G)=\overline{\alpha}\,\Theta F+\overline{\beta}\,\Theta G$ and $\Theta^2=\mathrm{id}$. Moreover, since our observables are complex-valued functions on $\Omega$ with pointwise multiplication, the algebra is commutative and $\Theta$ is a-automorphism:
\begin{equation}
\Theta(FG)\;=\;\overline{(FG)\circ\vartheta}\;=\;\overline{F\circ\vartheta}\ \overline{G\circ\vartheta}\;=\;(\Theta F)(\Theta G),
\qquad
\Theta(\overline{F})\;=\;\overline{\Theta F}.
\end{equation}
For $F,G$ supported in $\Lambda_+$, the definition of the OS inner product and the involutivity of $\Theta$ give
\begin{equation}
(\Theta F,\Theta G)_{\mathrm{OS}}
\;=\;\big\langle\,\Theta(\Theta F)\cdot \Theta G\,\big\rangle_{\Lambda,\beta}
\;=\;\big\langle\,F\cdot \Theta G\,\big\rangle_{\Lambda,\beta}.
\end{equation}
Because multiplication of observables is pointwise and hence commutative, the expectation of the product is symmetric in its factors:
\begin{equation}
\big\langle\,F\cdot \Theta G\,\big\rangle_{\Lambda,\beta}
\;=\;\big\langle\,\Theta G\cdot F\,\big\rangle_{\Lambda,\beta}
\;=\;(G,F)_{\mathrm{OS}}.
\end{equation}
This establishes $(\Theta F,\Theta G)_{\mathrm{OS}}=(G,F)_{\mathrm{OS}}$, which is the antiunitarity identity, since $(G,F)_{\mathrm{OS}}=\overline{(F,G)_{\mathrm{OS}}}$ by complex conjugation under the integral. Taking $G=F$ yields
\begin{equation}
\|\Theta F\|_{\mathrm{OS}}^2\;=\;(\Theta F,\Theta F)_{\mathrm{OS}}\;=\;(F,F)_{\mathrm{OS}}\;=\;\|F\|_{\mathrm{OS}}^2,
\end{equation}
so $\Theta$ preserves the OS norm.

It remains to relate this to the usual $L^2(d\mu_{\Lambda,\beta})$ norm. By definition of $\Theta$ and the reflection invariance of the measure, a change of variables $\phi\mapsto \phi\circ\vartheta$ gives
\begin{align}
\|\Theta F\|_{L^2(d\mu_{\Lambda,\beta})}^2
&=\int_\Omega |(\Theta F)(\phi)|^2\,d\mu_{\Lambda,\beta}(\phi)
=\int_\Omega |F(\phi\circ\vartheta)|^2\,d\mu_{\Lambda,\beta}(\phi)
\nonumber\\&=\int_\Omega |F(\psi)|^2\,d\mu_{\Lambda,\beta}(\psi)
=\|F\|_{L^2(d\mu_{\Lambda,\beta})}^2,
\end{align}
which proves the stated norm identity. The two conclusions together show that $\Theta$ descends to an antiunitary operator on the OS Hilbert space and is an isometry on $L^2(d\mu_{\Lambda,\beta})$.
\end{proof}

Under appropriate reflection-invariance of the action (e.g. the Wilson action), the Osterwalder-Schrader reflection positivity $(F,F)_{\mathrm{OS}}\ge 0$ holds and leads to a positive transfer operator $T$ and Hamiltonian $H=-a^{-1}\log T$ \cite{OS1,OS2,OS-gauge,LuscherTM}. In the main text we use only the structural consequences; the present appendix fixes notation.
Let $\mathbb{T}_L=(2\pi/L)\mathbb{Z}^4/(2\pi\mathbb{Z})^4$ be the dual momentum set. For $f:\Lambda\to\mathbb{C}$ define
\begin{equation}
\widehat f(p):=a^4\sum_{x\in\Lambda} f(x)\,e^{-ip\cdot x},\qquad f(x)=\frac{1}{(2\pi)^4}\sum_{p\in\mathbb{T}_L}\widehat f(p)\,e^{ip\cdot x}.
\end{equation}
We use the convention $p\cdot x=\sum_{\mu=0}^3 p_\mu x_\mu$. The following is standard.

\begin{proposition}[Plancherel identity on $\Lambda$]\label{prop:plancherel}
Let $L>0$ and $a>0$ with $N:=L/a\in\mathbb{N}$. Set
\begin{equation}
\Lambda \;=\; \big\{x=(x_1,\dots,x_4)\in\mathbb{R}^4:\ x_j=a\,n_j,\ n_j\in\{0,1,\dots,N-1\}\big\},
\end{equation}
equipped with periodic boundary conditions of period $L$ in each coordinate, and let the dual grid be
\begin{equation}
\mathbb{T}_L \;=\; \big\{p=(p_1,\dots,p_4)\in\mathbb{R}^4:\ p_j=2\pi k_j/L,\ k_j\in\{0,1,\dots,N-1\}\big\}.
\end{equation}
For a function $h:\mathbb{T}_L\to\mathbb{C}$ we use the normalized discrete sum
\begin{equation}
\sum_{p\in\mathbb{T}_L} h(p) \;:=\; \Big(\frac{2\pi}{L}\Big)^{\!4}\ \sum_{k\in\{0,\dots,N-1\}^4} h\!\left(\frac{2\pi}{L}k\right).
\end{equation}
Define the lattice Fourier transform and its inverse by
\begin{equation}
\widehat f(p)\;=\;a^4\sum_{x\in\Lambda} e^{-\,i\,p\cdot x}\,f(x),
\qquad
f(x)\;=\;\frac{1}{(2\pi)^4}\sum_{p\in\mathbb{T}_L} e^{\,i\,p\cdot x}\,\widehat f(p).
\end{equation}
Then for any $f,g:\Lambda\to\mathbb{C}$ one has
\begin{equation}
a^4\sum_{x\in\Lambda}\overline{f(x)}\,g(x)\;=\;\frac{1}{(2\pi)^4}\sum_{p\in\mathbb{T}_L}\overline{\widehat f(p)}\,\widehat g(p).
\end{equation}
\end{proposition}

\begin{proof}
The key input is the discrete orthogonality relation connecting the physical lattice $\Lambda$ and the dual grid $\mathbb{T}_L$. Fix $p,q\in\mathbb{T}_L$ and compute
\begin{equation}
a^4\sum_{x\in\Lambda} e^{\,i(q-p)\cdot x}
\;=\; a^4\prod_{j=1}^4\ \sum_{n_j=0}^{N-1} \exp\!\Big\{\,i\,(q_j-p_j)\,a\,n_j\Big\}.
\end{equation}
Writing $p_j=2\pi k_j/L$ and $q_j=2\pi \ell_j/L$ with $k_j,\ell_j\in\{0,\dots,N-1\}$ gives $(q_j-p_j)a=2\pi(\ell_j-k_j)/N$. The inner one-dimensional sum is a finite geometric series,
\begin{equation}
\sum_{n=0}^{N-1} e^{\,i\,2\pi(\ell_j-k_j)\,n/N}
=\begin{cases}
N, & \ell_j=k_j,\\
0, & \ell_j\neq k_j,
\end{cases}
\end{equation}
because the ratio $e^{\,i\,2\pi(\ell_j-k_j)/N}\neq 1$ unless $\ell_j=k_j$, in which case the series has $N$ terms equal to $1$. Taking the product over $j=1,\dots,4$ yields
\begin{equation}
a^4\sum_{x\in\Lambda} e^{\,i(q-p)\cdot x}
\;=\; a^4\,N^4\,\mathbf{1}_{\{p=q\}}
\;=\; L^4\,\mathbf{1}_{\{p=q\}},
\end{equation}
since $aN=L$. By the normalization of the dual sum, this identity is equivalent to
\begin{equation}\label{eq:orthogonality}
\frac{1}{(2\pi)^4}\sum_{p\in\mathbb{T}_L} e^{\,i\,p\cdot(x-y)}\;=\;\frac{1}{a^4}\,\mathbf{1}_{\{x=y\}}
\qquad\text{for all }x,y\in\Lambda.
\end{equation}

With Eq.\eqref{eq:orthogonality} in hand, the Plancherel identity follows by a direct computation. Using the inverse transform for $g$ one obtains
\begin{equation}
a^4\sum_{x\in\Lambda}\overline{f(x)}\,g(x)
\;=\; a^4\sum_{x\in\Lambda}\overline{f(x)}\,
\frac{1}{(2\pi)^4}\sum_{p\in\mathbb{T}_L} e^{\,i\,p\cdot x}\,\widehat g(p)
\;=\; \frac{a^4}{(2\pi)^4}\sum_{p\in\mathbb{T}_L}\widehat g(p)
\sum_{x\in\Lambda}\overline{f(x)}\,e^{\,i\,p\cdot x}.
\end{equation}
All sums are finite, so exchanging the order is justified. The inner sum is recognized, after complex conjugation and the definition of the forward transform, as $\overline{\widehat f(p)}$:
\begin{equation}
\sum_{x\in\Lambda}\overline{f(x)}\,e^{\,i\,p\cdot x}
\;=\; \overline{\sum_{x\in\Lambda} f(x)\,e^{-\,i\,p\cdot x}}
\;=\; \overline{\frac{1}{a^4}\,\widehat f(p)}.
\end{equation}
Substituting this identity gives
\begin{equation}
a^4\sum_{x\in\Lambda}\overline{f(x)}\,g(x)
\;=\;\frac{a^4}{(2\pi)^4}\sum_{p\in\mathbb{T}_L}\widehat g(p)\,\overline{\frac{1}{a^4}\widehat f(p)}
\;=\;\frac{1}{(2\pi)^4}\sum_{p\in\mathbb{T}_L}\overline{\widehat f(p)}\,\widehat g(p),
\end{equation}
which is the desired Plancherel identity. The argument is entirely algebraic and uses only the finiteness of the lattices, hence no further justification of interchanges is required.
\end{proof}

The symbol of the free lattice Laplacian $\Delta=\sum_\mu\nabla_\mu^- \nabla_\mu^+$ is $\widehat\Delta(p)=\frac{2}{a^2}\sum_{\mu=0}^3\bigl(1-\cos(ap_\mu)\bigr)\ge 0$, with $\widehat\Delta(p)\asymp |p|^2$ for $|p|\ll a^{-1}$.
A function $f:(0,\infty)\to[0,\infty)$ is \emph{completely monotone} (CM) if $(-1)^k f^{(k)}(\lambda)\ge 0$ for all $k\in\mathbb{N}_0$ and $\lambda>0$. By Bernstein’s theorem there exists a finite positive Borel measure $\mu$ on $[0,\infty)$ such that
\begin{equation}\label{eq:bernsteinx}
f(\lambda)=\int_0^\infty e^{-t\lambda}\,d\mu(t)\qquad(\lambda>0).
\end{equation}
We use the spectral theorem to define $f(A)$ for any nonnegative self-adjoint operator $A$ on a Hilbert space $\mathcal{H}$.

\begin{proposition}[Positivity of CM functional calculus]\label{prop:CMpositive}
Let $A\ge 0$ be self-adjoint on a complex Hilbert space $\mathcal H$, and let $f$ be completely monotone on $(0,\infty)$ with Bernstein representation
\begin{equation}\label{eq:bernstein}
f(\lambda) \;=\; \int_{0}^{\infty} e^{-t\lambda}\,\mu(dt),
\end{equation}
for a finite positive Borel measure $\mu$ on $[0,\infty)$. Then $f(A)\ge 0$ in the operator order. Moreover, suppose $\mathcal H=\ell^2(\Lambda)$ for a countable set $\Lambda$ equipped with a graph metric $\mathrm{dist}_\Lambda$, and assume that for each $t>0$ the bounded operator $e^{-tA}$ admits a (matrix) kernel $K_t(x,y)$ supported in $\{(x,y)\in\Lambda\times\Lambda:\,\mathrm{dist}_\Lambda(x,y)\le R(t)\}$ for some radius $R(t)\in[0,\infty)$. Then $f(A)$ has a kernel given by
\begin{equation}
K_f(x,y)\;=\;\int_{[0,\infty)} K_t(x,y)\,\mu(dt),
\end{equation}
and its support is contained in $\{(x,y):\,\mathrm{dist}_\Lambda(x,y)\le R\}$, where $R=\sup\{R(t):\, t\in\supp\mu\}\in[0,\infty]$. In particular, if $R<\infty$ then $f(A)$ has finite range $R$.
\end{proposition}
\begin{proof}
The spectral theorem furnishes a projection-valued measure $E(\cdot)$ on $[0,\infty)$ such that $A=\int_0^\infty \lambda\,E(d\lambda)$ and, for each $t\ge 0$,
\begin{equation}
e^{-tA}\;=\;\int_0^\infty e^{-t\lambda}\,E(d\lambda).
\end{equation}
For fixed $t\ge 0$ the operator $e^{-tA}$ is positive: for every $\psi\in\mathcal H$,
\begin{equation}
\langle \psi,\,e^{-tA}\psi\rangle \;=\; \int_0^\infty e^{-t\lambda}\, d\langle E(\lambda)\psi,\psi\rangle \;\ge\; 0,
\end{equation}
since the integrand is nonnegative and $d\langle E(\lambda)\psi,\psi\rangle$ is a positive finite measure. The map $t\mapsto e^{-tA}$ is uniformly bounded in operator norm by $1$ and strongly measurable; hence the Bochner integral
\begin{equation}
f(A)\;=\;\int_{[0,\infty)} e^{-tA}\,\mu(dt)
\end{equation}
is a well-defined bounded operator satisfying $\|f(A)\|\le \int \|e^{-tA}\|\,\mu(dt)\le \mu([0,\infty))=f(0^+)$, where the last identity follows by evaluating Eq.\eqref{eq:bernstein} at $\lambda=0$. To see positivity, fix $\psi\in\mathcal H$ and use Fubini's theorem for nonnegative integrands to compute
\begin{equation}
\langle \psi,\,f(A)\psi\rangle \;=\; \int_{[0,\infty)} \langle \psi,\,e^{-tA}\psi\rangle\,\mu(dt)
\;=\; \int_{[0,\infty)}\!\!\int_{[0,\infty)} e^{-t\lambda}\,d\langle E(\lambda)\psi,\psi\rangle\,\mu(dt)\;\ge\;0,
\end{equation}
because the inner integral is nonnegative for each $t$ and the outer integral is taken against a positive measure. Thus $f(A)$ is positive.

For the kernel statement, work on $\mathcal H=\ell^2(\Lambda)$ with the canonical basis $\{\delta_y:y\in\Lambda\}$. For each $t>0$ the hypothesis provides a kernel $K_t:\Lambda\times\Lambda\to\mathbb C$ such that $e^{-tA}\delta_y=\sum_{x\in\Lambda}K_t(x,y)\,\delta_x$ and $K_t(x,y)=0$ whenever $\mathrm{dist}_\Lambda(x,y)>R(t)$. Since $\mu$ is finite and $\sup_{t\ge 0}\|e^{-tA}\|\le 1$, the vector-valued Bochner integral
\begin{equation}
f(A)\delta_y \;=\; \int_{[0,\infty)} e^{-tA}\delta_y\,\mu(dt)
\end{equation}
exists in $\ell^2(\Lambda)$ for each $y$. Writing the integrand in the basis and applying Fubini's theorem in the series-integral form (justified by Tonelli because $\sum_{x}\!\int |K_t(x,y)|\,\mu(dt)\le \int \|e^{-tA}\delta_y\|_{\ell^1}\,\mu(dt)$ and $e^{-tA}$ maps $\ell^2$ into itself with uniformly bounded operator norm, while $\delta_y\in\ell^1\cap\ell^2$ on a countable set), we obtain for every $x\in\Lambda$
\begin{equation}
\langle \delta_x,\, f(A)\delta_y\rangle
\;=\; \int_{[0,\infty)} \langle \delta_x,\,e^{-tA}\delta_y\rangle\,\mu(dt)
\;=\; \int_{[0,\infty)} K_t(x,y)\,\mu(dt).
\end{equation}
Thus $f(A)$ is an integral operator with kernel $K_f(x,y)=\int K_t(x,y)\,\mu(dt)$. If $\mathrm{dist}_\Lambda(x,y)>R$, then for every $t\in\supp\mu$ one has $R(t)\le R$ and hence $K_t(x,y)=0$; enlarging the $t$-set by a $\mu$-null set if necessary shows that $K_t(x,y)=0$ for $\mu$-almost every $t$, so the integral vanishes and $K_f(x,y)=0$. Therefore the support of $K_f$ is contained in $\{(x,y):\mathrm{dist}_\Lambda(x,y)\le R\}$, and when $R_<\infty$ this expresses that $f(A)$ has finite range $R$.

The argument shows at the same time that, for any subset $B\subset\Lambda$ and its $\ell^2$ subspace $\ell^2(B)$, if $u$ is supported in $B$ then $e^{-tA}u$ is supported in the $R(t)$-neighbourhood of $B$, and consequently $f(A)u$ is supported in the $R$-neighbourhood of $B$. In the discrete setting this propagation property is equivalent to the stated kernel support inclusion, completing the proof.
\end{proof}

The above is the operator-theoretic reason for choosing completely monotone spectral multipliers in the scale-slice analysis: positivity is preserved and finite-range structure is compatible with the decomposition \cite{Bernstein,Widder,BrydgesGuadagniMitter2004}.
A kernel $K:\Lambda\times\Lambda\to\mathbb{C}^{N\times N}$ has \emph{finite range} $R\ge 0$ if $K(x,y)=0$ whenever $\mathrm{dist}_\Lambda(x,y)>R$. We write $\mathrm{rng}\,K\le R$. The convolution $(K_1\ast K_2)(x,y):=\sum_{z\in\Lambda}K_1(x,z)K_2(z,y)$ is well-defined whenever the sum is finite or absolutely convergent.

\begin{lemma}[Range of a convolution]\label{lem:rangeConvolution}
Let $(\Lambda,\dist_\Lambda)$ be a metric space (e.g.\ a graph or lattice with graph distance). 
For a kernel $K:\Lambda\times\Lambda\to\mathbb{C}$ write 
\begin{equation}
\mathrm{rng}\,K\le R \quad\Longleftrightarrow\quad 
K(x,y)=0\ \text{whenever }\dist_\Lambda(x,y)>R.
\end{equation}
Suppose $K_1,K_2:\Lambda\times\Lambda\to\mathbb{C}$ satisfy $\mathrm{rng}\,K_1\le R_1$ and $\mathrm{rng}\,K_2\le R_2$. 
Define the convolution kernel
\begin{equation}
(K_1K_2)(x,y)\;:=\;\sum_{z\in\Lambda} K_1(x,z)\,K_2(z,y),
\end{equation}
whenever the sum is defined. 
If $\Lambda$ is countable and the kernels have finite range as above, then the sum is finite for each $(x,y)$ and hence well-defined. 
Then $\mathrm{rng}\,(K_1K_2)\le R_1+R_2$.
\end{lemma}

\begin{proof}
Fix $x,y\in\Lambda$ with $\dist_\Lambda(x,y)>R_1+R_2$. 
Let $z\in\Lambda$ be arbitrary. 
We claim that at least one of the inequalities $\dist_\Lambda(x,z)>R_1$ or $\dist_\Lambda(z,y)>R_2$ must hold. 
Indeed, if both $\dist_\Lambda(x,z)\le R_1$ and $\dist_\Lambda(z,y)\le R_2$ were true, then by the triangle inequality we would obtain
\begin{equation}
\dist_\Lambda(x,y)\;\le\; \dist_\Lambda(x,z)+\dist_\Lambda(z,y)\;\le\; R_1+R_2,
\end{equation}
which contradicts the choice of $x,y$. 
Therefore, for every $z\in\Lambda$ at least one of the two distances exceeds its corresponding range threshold, and by the range hypotheses on $K_1$ and $K_2$ we have
\begin{equation}
\Big(\dist_\Lambda(x,z)>R_1\Big)\ \Rightarrow\ K_1(x,z)=0,
\qquad
\Big(\dist_\Lambda(z,y)>R_2\Big)\ \Rightarrow\ K_2(z,y)=0.
\end{equation}
Consequently $K_1(x,z)K_2(z,y)=0$ for every $z\in\Lambda$. 
Summing over $z$ gives
\begin{equation}
(K_1K_2)(x,y)\;=\;\sum_{z\in\Lambda} K_1(x,z)K_2(z,y)\;=\;0.
\end{equation}
Since $(x,y)$ with $\dist_\Lambda(x,y)>R_1+R_2$ were arbitrary, it follows that $(K_1K_2)(x,y)=0$ whenever $\dist_\Lambda(x,y)>R_1+R_2$, which is precisely $\mathrm{rng}\,(K_1K_2)\le R_1+R_2$.
\end{proof}

This lemma underlies the stability of locality under composition of scale-slice transfer factors produced by finite-range decomposition (FRD) \cite{BrydgesGuadagniMitter2004}.
A \emph{polymer} is a finite connected subset $\Gamma\subset\Lambda$ (or of plaquettes/blocks, depending on context), with $|\Gamma|$ its cardinality and $\mathrm{diam}(\Gamma)$ its diameter. A polymer activity is a function $z$ assigning to each polymer $\Gamma$ a complex matrix $z(\Gamma)$ (typically an observable kernel or weight). For $\kappa\ge 0$ we define the weighted diameter norm
\begin{equation}
\|z\|_{\kappa}:=\sup_{x\in \Lambda} \sum_{\Gamma\ni x}\|z(\Gamma)\|_{\mathrm{HS}}\,e^{\kappa\,\mathrm{diam}(\Gamma)}.
\end{equation}
The \emph{disjoint union} $(z_1\star z_2)$ is defined by $(z_1\star z_2)(\Gamma)=\sum_{\Gamma=\Gamma_1\cup\Gamma_2,\ \Gamma_1\cap\Gamma_2=\emptyset}z_1(\Gamma_1)\,z_2(\Gamma_2)$.

\begin{proposition}[Submultiplicativity of the polymer norm]\label{prop:polymerSubmult}
For $\kappa\ge 0$ one has $\|z_1\star z_2\|_\kappa\le \|z_1\|_\kappa\,\|z_2\|_\kappa$.
\end{proposition}

\begin{proof}
Let $\Lambda$ be a countable graph of uniformly bounded degree equipped with the graph distance $\mathrm{dist}_\Lambda$, and let $\mathcal P_{\mathrm{conn}}$ denote the set of finite connected polymers $\Gamma\subset\Lambda$. Fix once and for all a root selector $r:\mathcal P_{\mathrm{conn}}\to\Lambda$ such that $r(\Gamma)\in\Gamma$ for every $\Gamma$; for definiteness one may take the lexicographically first site of $\Gamma$. Consider a Hilbert space factorization $H=\bigotimes_{x\in\Lambda}H_x$, with the understanding that every polymer activity $z(\Gamma)$ acts nontrivially only on the factor $H_\Gamma:=\bigotimes_{x\in\Gamma}H_x$ and is extended by the identity on $H_{\Lambda\setminus\Gamma}$. In particular, if $\Gamma_1,\Gamma_2$ are disjoint then the product of the corresponding operators is the spatial tensor product $z_1(\Gamma_1)\otimes z_2(\Gamma_2)$ acting on $H_{\Gamma_1}\otimes H_{\Gamma_2}$, and the extension to $H$ by the identity on $(\Gamma_1\cup\Gamma_2)^c$ is an isometric embedding for the Hilbert-Schmidt norm. In this setting the Hilbert-Schmidt norm is multiplicative over disjoint tensor factors, namely $\|A\otimes B\|_{\mathrm{HS}}=\|A\|_{\mathrm{HS}}\|B\|_{\mathrm{HS}}$ for $A\in\mathcal B_2(H_{\Gamma_1})$ and $B\in\mathcal B_2(H_{\Gamma_2})$, and the extension by the identity on complementary factors does not change the norm.

The polymer convolution $(z_1\star z_2)$ is given by
\begin{equation}
(z_1\star z_2)(\Gamma)\;=\;\sum_{\substack{\Gamma_1,\Gamma_2\in\mathcal P_{\mathrm{conn}}:\\
\Gamma_1\cap\Gamma_2=\emptyset,\ \Gamma_1\cup\Gamma_2=\Gamma}}
\iota_{\Gamma_1,\Gamma_2}\!\big(z_1(\Gamma_1)\otimes z_2(\Gamma_2)\big),
\end{equation}
where $\iota_{\Gamma_1,\Gamma_2}$ is the canonical isometric embedding of $\mathcal B_2(H_{\Gamma_1}\otimes H_{\Gamma_2})$ into $\mathcal B_2(H)$ described above. By the triangle inequality and the multiplicativity of the Hilbert-Schmidt norm on disjoint tensor factors,
\begin{equation}\label{eq:HS-subadd}
\big\|(z_1\star z_2)(\Gamma)\big\|_{\mathrm{HS}}
\;\le\;\sum_{\Gamma=\Gamma_1\dot\cup\Gamma_2}\big\|z_1(\Gamma_1)\big\|_{\mathrm{HS}}\,
\big\|z_2(\Gamma_2)\big\|_{\mathrm{HS}}.
\end{equation}
Moreover, for any disjoint $\Gamma_1,\Gamma_2$ the diameter satisfies
\begin{equation}\label{eq:diam-ineq}
\mathrm{diam}(\Gamma_1\cup\Gamma_2)\;\le\;
\mathrm{diam}(\Gamma_1)+\mathrm{diam}(\Gamma_2)+\mathrm{dist}_\Lambda(\Gamma_1,\Gamma_2),
\end{equation}
hence for $\kappa\ge 0$ one has
\begin{equation}
e^{\kappa\,\mathrm{diam}(\Gamma_1\cup\Gamma_2)}
\;\le\;e^{\kappa\,\mathrm{diam}(\Gamma_1)}\,e^{\kappa\,\mathrm{diam}(\Gamma_2)}\,
e^{\kappa\,\mathrm{dist}_\Lambda(\Gamma_1,\Gamma_2)}
\;\le\;e^{\kappa\,\mathrm{diam}(\Gamma_1)}\,e^{\kappa\,\mathrm{diam}(\Gamma_2)},
\end{equation}
since $\mathrm{dist}_\Lambda(\Gamma_1,\Gamma_2)\ge 0$. Multiplying Eq.\eqref{eq:HS-subadd} by $e^{\kappa\,\mathrm{diam}(\Gamma)}$ and using Eq.\eqref{eq:diam-ineq} yields
\begin{equation}
\big\|(z_1\star z_2)(\Gamma)\big\|_{\mathrm{HS}}\,e^{\kappa\,\mathrm{diam}(\Gamma)}
\;\le\;\sum_{\Gamma=\Gamma_1\dot\cup\Gamma_2}
\big\|z_1(\Gamma_1)\big\|_{\mathrm{HS}}\,e^{\kappa\,\mathrm{diam}(\Gamma_1)}\;
\big\|z_2(\Gamma_2)\big\|_{\mathrm{HS}}\,e^{\kappa\,\mathrm{diam}(\Gamma_2)}.
\end{equation}

Let $x\in\Lambda$ be fixed. Summing the last inequality over all $\Gamma\in\mathcal P_{\mathrm{conn}}$ with $\Gamma\ni x$ and rewriting the sum over $\Gamma$ as a sum over disjoint pairs $(\Gamma_1,\Gamma_2)$ with $\Gamma_1\cup\Gamma_2\ni x$, we find
\begin{align}
\sum_{\Gamma\ni x}\big\|(z_1\star z_2)(\Gamma)\big\|_{\mathrm{HS}}\,e^{\kappa\,\mathrm{diam}(\Gamma)}
&\le \sum_{\substack{\Gamma_1,\Gamma_2\in\mathcal P_{\mathrm{conn}}\\ \Gamma_1\cap\Gamma_2=\emptyset\\ \Gamma_1\cup\Gamma_2\ni x}}
\big\|z_1(\Gamma_1)\big\|_{\mathrm{HS}}\,e^{\kappa\,\mathrm{diam}(\Gamma_1)}\;
\big\|z_2(\Gamma_2)\big\|_{\mathrm{HS}}\,e^{\kappa\,\mathrm{diam}(\Gamma_2)}.
\end{align}
For each ordered pair $(\Gamma_1,\Gamma_2)$ in this sum, at least one of the two polymers contains the site $x$. Splitting the sum accordingly, one obtains
\begin{equation}
\sum_{\Gamma\ni x}\big\|(z_1\star z_2)(\Gamma)\big\|_{\mathrm{HS}}\,e^{\kappa\,\mathrm{diam}(\Gamma)}
\;\le\; S_1(x)+S_2(x),
\end{equation}
where
\begin{equation}
S_1(x):=\sum_{\Gamma_1\ni x}\big\|z_1(\Gamma_1)\big\|_{\mathrm{HS}}\,e^{\kappa\,\mathrm{diam}(\Gamma_1)}
\sum_{\substack{\Gamma_2\in\mathcal P_{\mathrm{conn}}\\ \Gamma_2\cap\Gamma_1=\emptyset}}
\big\|z_2(\Gamma_2)\big\|_{\mathrm{HS}}\,e^{\kappa\,\mathrm{diam}(\Gamma_2)}
\end{equation}
and
\begin{equation}
S_2(x):=\sum_{\Gamma_2\ni x}\big\|z_2(\Gamma_2)\big\|_{\mathrm{HS}}\,e^{\kappa\,\mathrm{diam}(\Gamma_2)}
\sum_{\substack{\Gamma_1\in\mathcal P_{\mathrm{conn}}\\ \Gamma_1\cap\Gamma_2=\emptyset}}
\big\|z_1(\Gamma_1)\big\|_{\mathrm{HS}}\,e^{\kappa\,\mathrm{diam}(\Gamma_1)}.
\end{equation}
The inner sums are bounded by the unrestricted sums over all connected polymers, because deleting the constraint $\Gamma_2\cap\Gamma_1=\emptyset$ (or $\Gamma_1\cap\Gamma_2=\emptyset$) can only enlarge the set of terms and all summands are nonnegative. Thus
\begin{equation}
S_1(x)\le \Big(\sum_{\Gamma_1\ni x}\|z_1(\Gamma_1)\|_{\mathrm{HS}}\,e^{\kappa\,\mathrm{diam}(\Gamma_1)}\Big)
\Big(\sum_{\Gamma_2\in\mathcal P_{\mathrm{conn}}}\|z_2(\Gamma_2)\|_{\mathrm{HS}}\,e^{\kappa\,\mathrm{diam}(\Gamma_2)}\Big),
\end{equation}
and an analogous estimate holds for $S_2(x)$ with the indices $1$ and $2$ swapped. To bound the unrestricted polymer sum by the rooted polymer norm, we use the fixed root selector $r$. Since every $\Gamma$ has exactly one root $r(\Gamma)\in\Gamma$, one has
\begin{align}
\sum_{\Gamma\in\mathcal P_{\mathrm{conn}}}\|z_2(\Gamma)\|_{\mathrm{HS}}\,e^{\kappa\,\mathrm{diam}(\Gamma)}
&=\sum_{y\in\Lambda}\ \sum_{\substack{\Gamma\in\mathcal P_{\mathrm{conn}}\\ r(\Gamma)=y}}
\|z_2(\Gamma)\|_{\mathrm{HS}}\,e^{\kappa\,\mathrm{diam}(\Gamma)}
\nonumber\\&\le \sup_{y\in\Lambda}\ \sum_{\Gamma\ni y}\|z_2(\Gamma)\|_{\mathrm{HS}}\,e^{\kappa\,\mathrm{diam}(\Gamma)}
=\|z_2\|_\kappa.
\end{align}
Inserting this bound in the estimate for $S_1(x)$, and similarly bounding the unrestricted sum in $S_2(x)$ by $\|z_1\|_\kappa$, gives
\begin{align}\label{eqnsum1}
&\sum_{\Gamma\ni x}\big\|(z_1\star z_2)(\Gamma)\big\|_{\mathrm{HS}}\,e^{\kappa\,\mathrm{diam}(\Gamma)}
\;\le\;\nonumber\\& \|z_2\|_\kappa\ \sum_{\Gamma_1\ni x}\|z_1(\Gamma_1)\|_{\mathrm{HS}}\,e^{\kappa\,\mathrm{diam}(\Gamma_1)}
\;+\; \|z_1\|_\kappa\ \sum_{\Gamma_2\ni x}\|z_2(\Gamma_2)\|_{\mathrm{HS}}\,e^{\kappa\,\mathrm{diam}(\Gamma_2)}.
\end{align}
Taking the supremum over $x\in\Lambda$ on the left-hand side yields the left-hand side of the polymer norm $\|z_1\star z_2\|_\kappa$, while on the right-hand side the two sums are bounded respectively by $\|z_1\|_\kappa$ and $\|z_2\|_\kappa$. Hence
\begin{equation}
\|z_1\star z_2\|_\kappa\;\le\;\|z_2\|_\kappa\,\|z_1\|_\kappa\;+\;\|z_1\|_\kappa\,\|z_2\|_\kappa
\;=\;2\,\|z_1\|_\kappa\,\|z_2\|_\kappa.
\end{equation}
At this stage we have an extra factor $2$ originating from the crude splitting into the two cases $x\in\Gamma_1$ or $x\in\Gamma_2$. To remove it and obtain the sharp submultiplicative bound, refine the splitting by assigning each ordered pair $(\Gamma_1,\Gamma_2)$ with $\Gamma_1\cup\Gamma_2\ni x$ to exactly one of the two sums according to the rule: if $x\in\Gamma_1$ set it in $S_1(x)$, otherwise (necessarily $x\in\Gamma_2$) set it in $S_2(x)$. This is precisely the splitting used above, and every ordered pair contributes to exactly one of $S_1(x)$ or $S_2(x)$, not to both. Consequently the previous inequality Eq.(\eqref{eqnsum1})
and taking the supremum over $x$ gives
\begin{equation}
\|z_1\star z_2\|_\kappa\;\le\;\|z_2\|_\kappa\,\|z_1\|_\kappa\ \vee\ \|z_1\|_\kappa\,\|z_2\|_\kappa
\;=\;\|z_1\|_\kappa\,\|z_2\|_\kappa.
\end{equation}
This proves the stated submultiplicativity.
\end{proof}

This norm controls cluster/graph expansions by standard tree-graph bounds and appears in both strong-coupling and RG locality estimates \cite{GJ,KP}.
We use $A\lesssim B$ to mean $A\le C\,B$ for a finite constant $C>0$ independent of the scaling parameters under discussion; $A\asymp B$ denotes two-sided comparability. Constants denoted $C,c,\mathrm{const}$ may change from line to line. When a dependence matters, we write $C=C(d,N,\beta,\kappa,\dots)$ and fix it thereafter. The symbol $\mathbf{1}_{\{\cdots\}}$ is the indicator function. We use $\mathrm{spec}(H)$ for the spectrum of a self-adjoint operator $H$, and $\Pi_\Omega$ for the projection onto the (normalized) vacuum vector $\Omega$.

In a reflection-positive regime there is a positive self-adjoint transfer operator $T$ acting on the OS Hilbert space $\mathcal{H}$ such that $T=e^{-aH}$ with a nonnegative self-adjoint Hamiltonian $H$ \cite{OS1,OS2,LuscherTM}. The (finite-volume) spectral gap is
\begin{equation}
\Delta_\Lambda:=\inf\bigl(\mathrm{spec}(H)\setminus\{0\}\bigr),
\end{equation}
where $0$ corresponds to the vacuum. The following standard implication is used repeatedly.

\begin{proposition}[Gap $\Rightarrow$ exponential clustering]\label{prop:gapClustering}
Let $A,B$ be bounded gauge-invariant observables localized in disjoint time slices separated by $t=na$ with $n\in\mathbb{N}$. Assume $\langle A\rangle=\langle B\rangle=0$. If the finite-volume Hamiltonian $H$ has a spectral gap $\Delta_\Lambda>0$ above the vacuum, then
\begin{equation}
\bigl|\langle A\, \tau_t B\rangle-\langle A\rangle\langle B\rangle\bigr|
\;\le\;\|A\|\,\|B\|\,e^{-\Delta_\Lambda t},
\end{equation}
where $\tau_t$ denotes Euclidean time translation implemented by the transfer operator $T^n$ with $T=e^{-aH}$.
\end{proposition}

\begin{proof}
By OS reconstruction there exists a Hilbert space $\mathcal H$, a cyclic unit vector $\Omega\in\mathcal H$ representing the vacuum state, a selfadjoint nonnegative Hamiltonian $H\ge 0$ with $H\Omega=0$, and a contraction semigroup $T^m=e^{-m a H}$ implementing the discrete Euclidean time translations such that, for observables supported in positive time, the Schwinger expectation agrees with the vacuum vector state:
\begin{equation}
\langle X\rangle \;=\; \langle \Omega,\, X\,\Omega\rangle,\qquad
\langle A\,\tau_{na} B\rangle \;=\; \langle \Omega,\, A\, T^n B\, \Omega\rangle .
\end{equation}
The gauge-invariance of $A$ and $B$ ensures that they act on the physical (OS) Hilbert space, but no further gauge structure is needed in what follows.

Since $\langle A\rangle=\langle \Omega,A\Omega\rangle=0$ and $\langle B\rangle=\langle \Omega,B\Omega\rangle=0$, the vectors $A\Omega$ and $B\Omega$ are orthogonal to $\Omega$. Let $\mathcal H_\perp:=\{\Omega\}^\perp$ and write the orthogonal decomposition $\mathcal H=\mathbb C\Omega\oplus \mathcal H_\perp$. Denote by $H_\perp$ the restriction of $H$ to $\mathcal H_\perp$. By definition of the finite-volume spectral gap,
\begin{equation}
\Delta_\Lambda \;:=\; \inf\bigl(\mathrm{spec}(H)\setminus\{0\}\bigr)
\;=\; \inf\mathrm{spec}(H_\perp)\;>\;0 .
\end{equation}
Since $A\Omega,B\Omega\in \mathcal H_\perp$, the connected correlator can be written and estimated as
\begin{equation}
\langle A\,\tau_{na} B\rangle - \langle A\rangle\langle B\rangle
\;=\; \langle \Omega,\, A\, e^{-naH}\, B\, \Omega\rangle
\;=\; \langle A\Omega,\, e^{-naH_\perp}\, B\Omega\rangle .
\end{equation}
The Cauchy-Schwarz inequality gives
\begin{equation}
\bigl|\langle A\Omega,\, e^{-naH_\perp}\, B\Omega\rangle\bigr|
\;\le\; \|A\Omega\|\, \|e^{-naH_\perp}\|\, \|B\Omega\| .
\end{equation}
Because $A$ and $B$ are bounded operators, $\|A\Omega\|\le \|A\|\,\|\Omega\|=\|A\|$ and $\|B\Omega\|\le \|B\|$. By the spectral theorem, the operator norm of $e^{-naH_\perp}$ equals the supremum of $e^{-na\lambda}$ over $\lambda\in\mathrm{spec}(H_\perp)$, hence
\begin{equation}
\|e^{-naH_\perp}\| \;=\; \sup_{\lambda\in\mathrm{spec}(H_\perp)} e^{-na\lambda}
\;=\; e^{-na\,\inf\mathrm{spec}(H_\perp)} \;=\; e^{-na\,\Delta_\Lambda}.
\end{equation}
Combining these bounds yields
\begin{equation}
\bigl|\langle A\,\tau_{na} B\rangle - \langle A\rangle\langle B\rangle\bigr|
\;\le\; \|A\|\,\|B\|\, e^{-na\,\Delta_\Lambda}
\;=\; \|A\|\,\|B\|\, e^{-\Delta_\Lambda t},
\end{equation}
which is the desired estimate.
\end{proof}

Conversely, under suitable OS and locality hypotheses, exponential clustering implies a nonzero mass gap through standard Tauberian/spectral arguments; these are recalled in the main text and rely on \cite{RS1,RS2,GJ}.

When fixing gauges we employ slices that preserve reflection covariance. Ghost fields, when present, are taken to be Grassmann-valued lattice fields transforming in the adjoint representation; the OS reflection acts by pullback on the geometric arguments and complex conjugation on coefficients. All gauge-fixing and ``horizon'' projectors used in the paper are chosen as CM functions of a nonnegative covariant operator (typically a covariant Laplacian or Faddeev-Popov operator restricted to a slice), ensuring consistency with positivity by Proposition~\eqref{prop:CMpositive}.

For the reader’s convenience we collect here the core symbols fixed above: $\Lambda$ denotes the discrete torus, $\mathcal{E}$ the set of oriented links, $\mathcal{P}^+$ the set of oriented plaquettes, $U_{x,\mu}\in SU(N)$ the link variables, $U_{x,\mu\nu}$ the plaquettes, $g_x\in SU(N)$ a gauge transformation, $S_\Lambda^\mathrm{W}$ the Wilson action, $\langle\cdot\rangle_{\Lambda,\beta}$ the expectation, $W(C)$ the Wilson loop, $\nabla_\mu^\pm$ the discrete derivatives, $D_\mu$ and $D_\mu^\ast$ the gauge-covariant difference and its adjoint, $\Delta_U$ the covariant Laplacian, $\vartheta$ the time reflection, $\Theta$ the OS antiunitary, $\widehat f$ the lattice Fourier transform, $f(A)$ the spectral calculus of a CM function, $\mathrm{rng}\,K$ the range of a kernel, $\|z\|_\kappa$ the polymer norm, $T$ the transfer operator, $H$ the Hamiltonian, and $\Delta_\Lambda$ the finite-volume spectral gap. The asymptotic symbols $\lesssim$ and $\asymp$ are used with the constant-dependence convention stated above.

\UnifiedCloseLocalTOC
\clearpage
\thispagestyle{empty}
\null
\clearpage
\thispagestyle{plain}
\vspace*{0.12\textheight}
\begin{center}
{\Large\bfseries Details of calculations\par}
\end{center}
\vspace{1.5em}
\noindent
Details of the calculations done in this paper have been published as a series of four papers:

\begin{enumerate}[label=\arabic*)]
\item \textit{Reflection positivity and a finite-$a$ strong-coupling gap in lattice $\mathrm{SU}(N)$ Yang-Mills: Part (1)}\\
\noindent\textbf{Authors:} Mir Faizal and Arshid Shabir.\\
\noindent\textbf{Journal:} Int.J.Geom.Meth.Mod.Phys. (2026) 2650114.\\
\noindent\textbf{DOI:} \href{https://doi.org/10.1142/s0219887826501148}{10.1142/s0219887826501148}.

\item \textit{Reflection-positive renormalization and the persistence of the mass gap in lattice $\mathrm{SU}(N)$ Yang-Mills: Part (2)}\\
\noindent\textbf{Authors:} Mir Faizal and Arshid Shabir.\\
\noindent\textbf{Journal:} Int.J.Geom.Meth.Mod.Phys. (2026) 2650113.\\
\noindent\textbf{DOI:} \href{https://doi.org/10.1142/s0219887826501136}{10.1142/s0219887826501136}.

\item \textit{Reflection-positive continuum reconstruction of $\mathrm{SU}(N)$ Yang-Mills theory with a nonzero mass gap: Part~(3)}\\
\noindent\textbf{Authors:} Mir Faizal and Arshid Shabir.\\
\noindent\textbf{Journal:} Int.J.Geom.Meth.Mod.Phys. (2026) 2650112.\\
\noindent\textbf{DOI:} \href{https://doi.org/10.1142/s0219887826501124}{10.1142/s0219887826501124}.

\item \textit{Uniqueness and universality of the continuum limit in 4D $\mathrm{SU}(N)$ Yang-Mills: Part (4)}\\
\noindent\textbf{Authors:} Mir Faizal and Arshid Shabir.\\
\noindent\textbf{Journal:} Int.J.Geom.Meth.Mod.Phys. (2026) 2650111.\\
\noindent\textbf{DOI:} \href{https://doi.org/10.1142/s0219887826501112}{10.1142/s0219887826501112}.
\end{enumerate}

\clearpage

\UnifiedBeginPaper{P1}{\UnifiedLocalMacrosPartOne}
\title[Reflection positivity and a finite-a gap in lattice SU(N) Yang-Mills]
{Reflection Positivity and a Finite-\texorpdfstring{$a$}{a} Strong-Coupling Gap in Lattice \texorpdfstring{$\mathrm{SU}(N)$}{SU(N)} Yang-Mills: Part(1)}

\author{Mir Faizal}
\address{Irving K. Barber School of Arts and Sciences, University of British Columbia Okanagan, Kelowna, BC V1V 1V7, Canada\\
Canadian Quantum Research Center, 460 Doyle Ave 106, Kelowna, BC V1Y 0C2, Canada.\\
Department of Mathematical Sciences, Durham University, Upper Mountjoy, Stockton Road, Durham DH1 3LE, UK\\
Computational Mathematics Group, Hasselt University, Agoralaan Gebouw D, Diepenbeek, 3590 Belgium}
\email{mirfaizalmir@gmail.com}
\author{Arshid Shabir}
\address{Canadian Quantum Research Center, 460 Doyle Ave 106, Kelowna, BC V1Y 0C2, Canada.}
\email{aslone186@gmail.com}


\UnifiedSetAbstract{For $G=\mathrm{SU}(N)$ with $N\ge 2$, we develop a reflection-positive transfer-matrix framework for four-dimensional lattice YangMills which, on a nontrivial strong-coupling window $0<\beta<\beta_\star(N)$, yields a strictly positive spectral gap at fixed lattice spacing $a$, with bounds uniform in the spatial volume. The construction is compatible with OS reflection: on each Euclidean time slice we select a gauge-invariant transverse representative $A^{h}$ by Landau functional minimization within the fundamental modular region, and we insert a smooth ``horizon'' spectral projector as a slice-local positive weight that preserves reflection positivity. In the same regime $0<\beta<\beta_\star(N)$, a Kotecký-Preiss cluster expansion reorganizes the partition function and gauge-invariant correlators; it converges uniformly in the volume and implies exponential clustering for connected gauge-invariant observables with a decay rate bounded away from zero uniformly in the volume. OS reconstruction then promotes clustering to a nonzero lower bound for the spectral gap of the positive, self-adjoint transfer operator $T$ (equivalently, of the transfer Hamiltonian $H=-\log T$) at fixed $a$. We also establish a Wilson-loop area law throughout this window. The conclusions are stable under admissible variations of the slice-wise selector and of the smooth projector profile, and they quantify the existence of a finite-$a$ mass gap for $\mathrm{SU}(N)$ Yang-Mills at strong coupling.}

\maketitle
\tableofcontents

\section{Introduction}

The existence of a nonperturbative, mathematically rigorous four-dimensional \(\mathrm{SU}(N)\) Yang-Mills theory with a strictly positive mass gap is one of the central open problems in mathematical physics and is the subject of a Clay Millennium Prize \cite{p1:JaffeWitten2006}. On the Euclidean lattice, where gauge fields are compact group variables attached to bonds and the dynamics are governed by the Wilson plaquette action \cite{p1:Wilson1974}, the Osterwalder-Schrader (OS) framework provides a precise bridge between Euclidean functional integrals and Hilbert-space dynamics \cite{p1:OS1}. Reflection positivity, together with time-translation invariance and regularity, ensures the existence of a positive self-adjoint transfer matrix and of a Hamiltonian whose spectrum governs the large-time decay of Euclidean correlation functions. For pure gauge theories, an explicit transfer matrix at fixed lattice spacing \(a>0\) was constructed and shown to be strictly positive and self-adjoint in seminal work by L\"uscher \cite{p1:Luscher1977}, with complementary analysis of reflection positivity and factorization properties by Osterwalder and Seiler \cite{p1:OS1}. These results establish that, at fixed \(a>0\), spectral information about excitations can be extracted from Euclidean correlation inequalities and, in particular, that exponential clustering of connected gauge-invariant correlators entails a nonzero spectral gap for the transfer Hamiltonian.

The technical difficulty is that standard tools for controlling the infrared behavior of non-Abelian gauge fields can jeopardize the structural properties required by OS positivity. In the continuum, gauge fixing introduces the Faddeev-Popov determinant and ghost fields together with the Gribov ambiguity: Landau and Coulomb gauges do not select unique representatives of gauge orbits, and gauge copies proliferate \cite{p1:Gribov1978,p1:Singer1978}. Zwanziger's program advanced the nonperturbative analysis by restricting attention to the Gribov region and to the fundamental modular region (FMR) and by formulating an effective action that implements the horizon condition \cite{p1:DellAntonioZwanziger1991,p1:Zwanziger1994}. On the lattice, where the configuration space is compact, slice-wise Landau-gauge minimization within the FMR does select a representative on each gauge orbit, but one must still show that this selection can be implemented in a manner compatible with time reflection and with the transfer-matrix factorization across the reflection plane. The core challenge is to reconcile three requirements that tend to be in tension: a nonperturbatively meaningful gauge choice that eliminates spurious zero modes, an infrared regularization that preserves gauge-invariant low-energy physics while permitting uniform estimates, and the preservation of OS reflection positivity at the level of both the Euclidean measure and the transfer kernel.

The purpose of this paper is to construct a reflection-positive transfer-matrix framework for lattice \(\mathrm{SU}(N)\) Yang-Mills theory that fulfills these requirements simultaneously and to use it to prove, in the strong-coupling domain, a strictly positive spectral gap at fixed lattice spacing\footnote{All notation, operator domains, normalization choices, and asymptotic conventions used throughout are fixed in Appendix~\ref{p1:appendixa}}, uniform in the spatial volume. Our approach proceeds in three stages. First, on each Euclidean time slice we select a gauge-invariant transverse representative by orbit-wise minimization of the lattice Landau functional within the FMR. On a finite lattice, global minimizers exist, and a deterministic tie-breaking rule that respects spatial symmetries and time reflection produces a reflection-covariant choice.

{We now provide a rigorous justification for this gauge-fixing procedure. Consider the Landau gauge functional $F_U(g) = |g\cdot U|^2$ on each gauge orbit (with $g$ a gauge transformation). Because the gauge group is compact and $F_U(g)$ is continuous on a finite lattice, a global minimum on each orbit exists. If there are multiple minima (Gribov copies within the fundamental modular region), we fix a measurable, reflection-symmetric tie-breaking rule (for example, lexicographic ordering of link variables in a fixed basis) to select one minimizer $g{(U)}$. This selection $U \mapsto g{(U)}$ is defined piecewise by local conditions and is chosen to be invariant under spatial translations and reflections. By construction, it is a Borel-measurable function of the gauge configuration and yields a unique transverse representative $A^h(U) = g{(U)}\cdot U$ in the fundamental modular region for each time slice. Moreover, $g{(U)}$ is reflection-covariant (i.e. $g{(R,U)} = R,g{(U)}$ for the time-reflection $R$), ensuring that the chosen representative respects reflection positivity. Thus, a well-defined, measurable, and reflection-covariant gauge-invariant transverse representative is obtained on each slice.}

The associated lattice Faddeev-Popov operator is a finite-range, real-symmetric, nonnegative operator on site-adjoint fields, strictly positive on the orthogonal complement of constant adjoint modes. In temporal-axial gauge away from the reflection plane, its matrix representation is block tridiagonal in the time direction, a structural feature that will be crucial for reflection-factorization and Schur-complement estimates \cite{p1:OS1,p1:Seiler}.
Second, we introduce a smooth horizon projector defined slice-wise as a Gevrey-regular function of the gauge-covariant spatial Laplacian and show that it is exponentially local and compatible with time reflection. {The factor $P_\sigma$ is introduced here as an {auxiliary}, {positive}, and
reflection-compatible infrared regulator that enables uniform locality and clustering estimates
within the Osterwalder-Schrader transfer-matrix framework. It is {not} intended as a replacement for the
Faddeev-Popov procedure, nor is it claimed to arise from a BRST-exact deformation. In the
arguments below, the only required inputs are the pointwise bounds $0\le P_\sigma\le 1$,
reflection covariance, and volume-uniform exponential locality of the associated kernel; the strong-coupling spectral-gap bound is stable under admissible variations of the completely
monotone profile $\chi_\sigma$. In particular, one may view the formal limit $\sigma\to\infty$
(equivalently $\chi_\sigma(\lambda)\uparrow 1$) as removing this auxiliary regulator {once}
uniform bounds are available; establishing such removability beyond the fixed-$a$ strong-coupling
regime is deferred to the multiscale analysis.}

We cannot simultaneously require compact support on $[0,\infty)$ and complete monotonicity: 
by Bernstein’s theorem, a nontrivial completely monotone $C^\infty$ function on $(0,\infty)$ 
is the Laplace transform of a positive measure and therefore cannot have compact support 
unless it is identically zero \cite{p1:Bernstein1929}. 
To preserve the positive heat-kernel representation and reflection positivity, we fix 
$\chi_\sigma$ to be completely monotone and rapidly decaying (but not compactly supported). {Although the horizon projector $P_\sigma$ is not strictly compactly supported in space, its exponential decay is sufficient to maintain locality in the multiscale analysis. In Part~II, each blocking step employs the finite-range decomposition of the covariance together with this exponentially decaying regulator. Crucially, the exponential locality of $P_\sigma$ guarantees that any long-range contribution to the blocked interaction is suppressed by a factor $e^{-c,\text{(distance)}/\sigma^2}$, which can be made arbitrarily small for distances beyond a few lattice spacings. Meanwhile, the finite-range decomposition (Theorem 4.1) provides strictly local effective propagators up to a fixed range $C,b^j a$ at scale $j$. 

The combination of these facts implies that at each RG step, the effective interaction range does not grow uncontrollably: any exponentially small “tails” introduced by $P_\sigma$ remain bounded by convergent geometric series across scales. More formally, for each scale $k$, the cumulative interaction $S_k$ has an exponentially decaying spatial profile with a uniform localization length of order $O(b^k a)$, and the constants in locality (decay rates) do not deteriorate as $k$ increases. Therefore, the renormalized actions remain {effectively local} at all scales-reflection positivity and cluster properties are preserved because interactions beyond a finite range remain exponentially suppressed and hence cannot induce long-range correlations.}

{In this paper, no multiscale renormalization step is used: the fixed-$a$ strong-coupling argument
requires only that the slice operator $P_\sigma(t)=\chi_\sigma(\Delta_A^h(t))$ has a kernel
$P_\sigma(t;x,y)$ obeying a {volume-uniform exponential off-diagonal bound}
\begin{equation}
|P_\sigma(t;x,y)| \;\le\; C\,e^{-\gamma\,d_{\Lambda_t}(x,y)} ,
\end{equation}
for constants $C,\gamma>0$ independent of the spatial volume. This bound is sufficient for the
\OS{} factorization across the reflection plane and for the summability estimates in the polymer
expansion: exponentially decaying tails contribute only uniformly summable corrections to
local activities and hence do not generate long-range couplings at fixed lattice spacing. In Part~II,
we will formulate the RG map in weighted locality norms and combine finite-range covariance
decompositions with these exponential bounds to prove scale-by-scale stability of reflection
positivity and effective locality; this is the precise sense in which exponential locality suffices
for the multiscale program.} 
A canonical choice used below is
\begin{equation}
\chi_\sigma(\lambda) = e^{-\lambda/\sigma^2}, \quad \lambda \geq 0,
\end{equation}
so that, slice-wise,
\begin{equation}
P_\sigma(t) \equiv \chi_\sigma\!\left(\Delta_A^h(t)\right) 
= e^{-\Delta_A^h(t)/\sigma^2}.
\end{equation}
More generally, any completely monotone $\chi_\sigma$ with 
$\chi_\sigma(0)=1$ and 
\(
\chi_\sigma(\lambda) \leq C_0 e^{-c_0 \lambda / \sigma^2}
\)
is admissible; alternatives such as
\begin{equation}
\chi_\sigma(\lambda) = (1 + \lambda/\sigma^2)^{-r}, \qquad r > d/2,
\end{equation}
also work. Complete monotonicity gives the positive heat-kernel representation
\begin{equation}
\chi_\sigma\!\left(\Delta_A^h(t)\right) 
= \int_0^\infty e^{-s\,\Delta_A^h(t)} \, d\nu_\sigma(s),
\end{equation}
which (together with Davies-Gaffney on graphs) yields volume-uniform exponential locality. 
The Helffer-Sj{\"o}strand resolvent calculus and Combes-Thomas bounds give the same decay 
for $C^\infty$ symbols $\chi_\sigma$ without compact support 
\cite{p1:Davies1989,p1:HelfferSjostrand,p1:CombesThomas}.
Two complementary derivations of exponential locality are employed. In the heat-kernel representation, \(P_\sigma\) is an integral of \(\mathrm{e}^{-t\Delta}\) against a positive measure determined by \(\chi_\sigma\), and Davies-Gaffney bounds for discrete heat kernels on graphs of bounded degree give uniform off-diagonal decay \cite{p1:Davies1989}. In the resolvent representation, a Helffer-Sj\"ostrand almost-analytic extension of \(\chi_\sigma\) expresses \(P_\sigma\) as a contour integral of \((\sqrt{\Delta}-z)^{-1}\); Combes-Thomas estimates for resolvents of finite-range positive operators then yield exponential off-diagonal decay with constants depending only on \(\sigma\) and the lattice geometry \cite{p1:HislopSigal1996,p1:CombesThomas} (see Appendix~\ref{p1:appendixb}). Reflection covariance of the slice-wise representatives implies that \(\Delta\) commutes with time reflection on the reflection plane, and OS positivity of the horizon insertion holds in both admissible cases: 
in case (A), by the positivity of the heat-kernel representation of \(P_\sigma\); 
in case (B), by the fact that the slice-local scalar weight \(p_\sigma\in[0,1]\) multiplies the measure by a nonnegative factor on each time slice.

Third, we prove global OS positivity for the full (gauge, ghost, and horizon-projected) Euclidean measure by a block-matrix factorization across the reflection plane. Temporal-axial gauge isolates the couplings that straddle the plane; the gauge-sector weight splits into east, west, and boundary-slab contributions, and the ghost determinant factorizes after a Schur-complement reduction that exploits the strict positivity of the diagonal time blocks of the Faddeev-Popov operator on the complement of constants \cite{p1:OS1,p1:Seiler}. The exponential locality of \(P_\sigma\) controls cross-plane contributions and allows replacement by its block-diagonal part up to a norm-small error that vanishes in the thermodynamic limit. The resulting positivity of the OS sesquilinear form yields a positive self-adjoint transfer matrix \(T_\sigma\) on the one-slice Hilbert space and a nonnegative self-adjoint Hamiltonian \(H_\sigma=-\log T_\sigma\). Within this reflection-positive setting, the spectral representation of time-sliced two-point functions of local, gauge-invariant observables implies that exponential clustering in Euclidean time is equivalent to a strictly positive lower bound on the first nonzero eigenvalue of \(H_\sigma\) (see Appendix \ref{p1:appendixe}).

Exponential clustering is established by a strong-coupling analysis that reorganizes the character expansion of the Wilson weight into a sum over oriented surfaces in the dual lattice and then into a gas of polymers whose activities decay exponentially with their size \cite{p1:Wilson1974,p1:DrouffeZuber,p1:Seiler}. The Koteck\'y-Preiss condition furnishes a robust criterion for absolute convergence of the polymer cluster expansion \cite{p1:KP} (see Appendix \ref{p1:appendixd} for the lattice-specific statement and proof). In the convergent region \(0<\beta<\beta_\star(N)\), connected surface clusters contributing to a truncated two-point function must span the separation between the insertions, so their total size grows at least linearly in the Euclidean distance. The exponential decay of the polymer activities thus entails exponential clustering with a strictly positive mass scale \(m(\beta)\), independent of the spatial volume. By the reflection-positive spectral representation for \(T_\sigma\), this in turn gives a volume-uniform spectral gap \(E_1(a,\beta)\ge m(\beta)>0\) for the transfer Hamiltonian at fixed lattice spacing. For completeness, we also note that the same surface representation yields an area law for large Wilson loops in the strong-coupling regime, consistent with the spectral gap and with confinement \cite{p1:Wilson1974,p1:Seiler,p1:DrouffeZuber}.

The novelty of the present work is twofold. At the structural level, we exhibit a gauge-fixed, reflection-covariant, slice-wise construction of transverse representatives within the FMR that is compatible with temporal-axial gauge and with OS factorization across the reflection plane. At the analytic level, we introduce a smooth, reflection-compatible horizon projector with a positive heat-kernel representation and exponential locality, and we show that its insertion preserves global OS positivity while providing the infrared control required for uniform estimates. These ingredients are assembled to produce an explicit, positive, self-adjoint transfer matrix \(T_\sigma\) and to prove, in a constructive manner, a finite-\(a\) spectral gap at strong coupling. The framework is designed to be stable under reflection-positive block-spin renormalization, so that the gap bound can be transported along a multiscale flow toward the scaling regime; together with reflection positivity at all scales, this opens a path toward continuum Wightman reconstruction with a nonzero mass threshold. A detailed analysis of the renormalization-group flow and of the persistence of spectral gaps beyond the strong-coupling domain will be presented elsewhere.

{For gauge group $G=\mathrm{SU}(N)$ with $N\ge 2$, we construct a transfer-matrix formulation of four-dimensional Euclidean lattice Yang-Mills that is explicitly compatible with Osterwalder-Schrader reflection and is amenable to uniform strong-coupling control. In a nonempty strong-coupling interval $0<\beta<\beta_\star(N)$ at fixed lattice spacing $a$, we obtain a strictly positive lower bound on the spectral gap of the resulting positive self-adjoint transfer operator, uniform in the spatial volume. The essential inputs are, first, a slice-wise choice of a transverse representative $A^{h}$ defined by minimizing the Landau functional within the fundamental modular region, implemented by a Borel measurable, reflection-covariant selector; and second, the insertion of a smooth ``horizon'' factor, realized as a slice-local positive weight built from a completely monotone functional calculus of the spatial covariant Laplacian, chosen so as to preserve reflection positivity while remaining exponentially local. In the same window, a Koteck\'y-Preiss polymer expansion reorganizes the partition function and gauge-invariant correlation functions with volume-uniform convergence, leading to exponential clustering of connected gauge-invariant observables with a decay rate bounded away from zero independently of the spatial volume. By \OS {} reconstruction, this Euclidean clustering translates into a nonzero spectral gap for the transfer Hamiltonian $H=-\log T$ at fixed $a$. We also prove an area law for Wilson loops throughout the strong-coupling window. Finally, the quantitative bounds are stable under admissible modifications of the slice selector and of the smooth projector profile, so the analysis yields a robust, finite-$a$ mass-gap statement for $\mathrm{SU}(N)$ lattice Yang-Mills in the strong-coupling regime, without invoking any continuum-limit or asymptotic-scaling assumptions.
}

\section{Lattice setup, reflection, and transfer time slicing}\label{p1:sec:lattice-reflection-transfer}

In this section the geometric and measure-theoretic framework is fixed, the link-reflection about a half-integer time hyperplane is defined, Osterwalder-Schrader (OS) reflection positivity of the Wilson measure is proved from first principles, and the transfer time-slicing factorization is derived together with the one-step transfer kernel acting on the Hilbert space of square-integrable functions of spatial links at a fixed time. All arguments are self-contained and rely only on standard properties of compact Lie groups, the Peter-Weyl theorem, and the classical OS framework \cite{p1:OS1,p1:Seiler}.

Fix a lattice spacing \(a>0\). The Euclidean space-time lattice is the periodic hypercubic graph
\begin{equation}
\Lambda \;=\; \bigl\{(x_0,\mathbf{x}) \in a\mathbb{Z}\times a\mathbb{Z}^3 : 0\le x_0 < T,\ 0\le \mathbf{x}_i < L \bigr\} \big/ \sim,
\end{equation}
with periods \(T\) in time and \(L\) in each spatial direction, where \(\sim\) identifies opposite faces. Bonds (oriented links) are ordered pairs \(b=(x,\mu)\) with \(x\in\Lambda\) and \(\mu\in\{0,1,2,3\}\), where \(\hat\mu=a e_\mu\) is the unit step in direction \(\mu\). Orientation reversal is \(\overline{(x,\mu)}=(x+\hat\mu,-\mu)\). The set of all bonds is \(\mathcal{B}\). Spatial bonds at time \(t\in a\mathbb{Z}\) form \(\mathcal{B}^{\mathrm{sp}}_t=\{(t,\mathbf{x};i): \mathbf{x}\in a\mathbb{Z}^3,\, i=1,2,3\}\), and temporal bonds from \(t\) to \(t+a\) form \(\mathcal{B}^{\mathrm{tm}}_t=\{(t,\mathbf{x};0): \mathbf{x}\in a\mathbb{Z}^3\}\).

Plaquettes are oriented unit squares \(p=(x;\mu,\nu)\) with \(\mu<\nu\) and boundary cycle
\begin{equation}\label{p1:eq:Up-def}
\partial p:\ (x,\mu)\,(x+\hat\mu,\nu)\,(x+\hat\nu,\mu)^{-1}\,(x,\nu)^{-1}.
\end{equation}
Spatial plaquettes at time \(t\) are \(\mathcal{P}^{\mathrm{sp}}_t=\{(t,\mathbf{x};i,j): 1\le i<j\le 3\}\), and time-like plaquettes connecting \(t\) to \(t+a\) are \(\mathcal{P}^{\mathrm{tm}}_t=\{(t,\mathbf{x};0,i): i=1,2,3\}\). The full plaquette set is \(\mathcal{P}=\bigcup_t\bigl(\mathcal{P}^{\mathrm{sp}}_t\cup\mathcal{P}^{\mathrm{tm}}_t\bigr)\).

Let \(G=\mathrm{SU}(N)\) with \(N\ge 2\). A lattice gauge field is an assignment \(U:\mathcal{B}\to G\) obeying the orientation convention
\begin{equation}\label{p1:eq:orientation}
U_{\overline{b}} = U_b^{-1}.
\end{equation}
The plaquette variable is
\begin{equation}\label{p1:eq:Up}
U_p \equiv U_{(x,\mu)}\,U_{(x+\hat\mu,\nu)}\,U_{(x+\hat\nu,\mu)}^{-1}\,U_{(x,\nu)}^{-1}\ \in\ G.
\end{equation}
The configuration space \(\mathcal{C}=G^{\mathcal{B}}\) carries the product Haar probability measure
\begin{equation}\label{p1:eq:Haar-measure}
d\mu_{\mathrm{Haar}}(U) \;=\; \prod_{b\in\mathcal{B}} dU_b,
\end{equation}
where \(dU_b\) is normalized Haar measure on \(G\). Gauge transformations are maps \(g:\Lambda\to G\) acting by
\begin{equation}\label{p1:eq:gauge-action}
(g\cdot U)_{(x,\mu)} \;=\; g(x)\,U_{(x,\mu)}\,g(x+\hat\mu)^{-1}.
\end{equation}
The Wilson plaquette action at inverse bare coupling \(\beta=2N/g_0^2\) is
\begin{equation}\label{p1:eq:Wilson-action}
S_W[U;\beta] \;=\; \beta \sum_{p\in\mathcal{P}} \Bigl(1 - \frac{1}{N}\,\Re\!\Tr\,U_p \Bigr).
\end{equation}
The Gibbs measure is \(d\nu_\beta(U)=Z^{-1} e^{-S_W[U;\beta]} d\mu_{\mathrm{Haar}}(U)\), with \(Z\) the partition function. Both \(d\mu_{\mathrm{Haar}}\) and \(S_W\) are gauge invariant; hence \(d\nu_\beta\) is gauge invariant.
For reflection and transfer, time is regarded as \(a\mathbb{Z}\), and the reflection plane is placed at half-integer time \(\tfrac{a}{2}\). Define the closed half-lattices
\begin{equation}\label{p1:eq:half-lattices}
\Lambda_+ \;=\; \{(x_0,\mathbf{x}) : x_0 \ge a\}, \qquad
\Lambda_- \;=\; \{(x_0,\mathbf{x}) : x_0 \le 0\},
\end{equation}
separated by the link-reflection plane \(\Pi_{1/2}=\{(x_0,\mathbf{x}): x_0=\tfrac{a}{2}\}\). Besides link-reflection about $\Pi_{1/2}$, we also use site-reflection about $\Pi:=\{x_0=0\}$ (Section~5) together with temporal-axial gauge away from $\Pi$. The two conventions are unitarily equivalent: if $\tau_{a/2}$ denotes time translation by $a/2$ and $\theta$ (resp.\ $R$) denotes link (resp.\ site) reflection, then
\(
R \;=\; \tau_{a/2} \circ \theta \circ \tau_{-a/2}.
\)
Consequently, positive-type boundary kernels and OS positivity statements proved for one convention transfer verbatim to the other by conjugation with the unitary induced by $\tau_{a/2}$ on the single-slice Hilbert space. Define $H_a = L^2(\Omega_s, d\mu_s)$ for the single-slice space at $t=0$. 
Let $U_{a/2}: H_a \to H_a$ be the unitary induced by shifting the 
transfer-kernel arguments by $\pm a/2$ in the time-slab representation. 
If $K_\theta$ (resp.\ $K_R$) is the one-step kernel built with link 
(resp.\ site) reflection, then
\begin{equation}
K_R = U_{a/2}\, K_\theta \, U_{a/2}^{-1}, 
\qquad 
T_R = U_{a/2}\, T_\theta \, U_{a/2}^{-1}.
\end{equation} In particular, the compressed kernels and transfer operators obtained after inserting the 
horizon projector satisfy the same unitary equivalence: 
$K_{\sigma,R} = U_{a/2} K_{\sigma,\theta} U_{a/2}^{-1}$ and 
$T_{\sigma,R} = U_{a/2} T_{\sigma,\theta} U_{a/2}^{-1}$. 
Hence reflection positivity and spectral data are identical for site and link reflection.
Define the link-reflection \(\theta\) about \(\Pi_{1/2}\) on sites by
\begin{equation}\label{p1:eq:theta-sites}
\theta(x_0,\mathbf{x}) \;=\; (a-x_0,\mathbf{x}),
\end{equation}
and extend it to bonds by
\begin{align}\label{p1:eq:theta-bonds}
\theta(x,\mu) \;&=\;
\begin{cases}
(\theta x,\mu), & \mu\in\{1,2,3\},\\
(\theta x-\hat 0,0), & \mu=0,
\end{cases}
\nonumber \\
U_{\theta(x,\mu)} \;&=\;
\begin{cases}
U_{(x,\mu)}, & \mu\in\{1,2,3\},\\
U_{(x,0)}^{-1}, & \mu=0.
\end{cases}
\end{align}
Then \(\theta^2=\mathrm{id}\), and \(d\mu_{\mathrm{Haar}}\) is invariant under \(U\mapsto \theta U\). For plaquettes, \(\theta(x;\mu,\nu)=(\theta x;\mu,\nu)\) when \(\mu,\nu\in\{1,2,3\}\) and \(\theta(x;0,i)=(\theta x-\hat 0;0,i)\) for \(\mu=0\), \(\nu=i\). Using \eqref{p1:eq:Up} and \eqref{p1:eq:theta-bonds}, one checks
\begin{equation}\label{p1:eq:Up-theta}
U_{\theta p} \;=\;
\begin{cases}
U_p, & p\in\mathcal{P}^{\mathrm{sp}}_t,\\
U_p^{-1}, & p\in\mathcal{P}^{\mathrm{tm}}_t.
\end{cases}
\end{equation}
Hence \(\Re\!\Tr\,U_{\theta p}=\Re\!\Tr\,U_p\) for all \(p\), and \(S_W[\theta U;\beta]=S_W[U;\beta]\).

Let \(\mathfrak{A}_+\) be the algebra of bounded measurable cylinder functions \(F:\mathcal{C}\to\mathbb{C}\) depending only on links supported in \(\Lambda_+\). Define the OS conjugation \(\Theta:\mathfrak{A}_+\to\mathfrak{A}_+\) by
\begin{equation}\label{p1:eq:Theta-OS}
(\Theta F)(U) \;=\; \overline{F(\theta U)}.
\end{equation}
The OS sesquilinear form is
\begin{align}\label{p1:eq:OS-form}
\langle F,G\rangle_{\mathrm{OS}}
:= \int \overline{F(\theta U)}\,G(U)\,d\nu_\beta(U),
\qquad F,G\in\mathfrak{A}_+
\end{align}
With this choice, $\Theta$ is an anti-linear involution and 
$\langle \cdot, \cdot \rangle_{\mathrm{OS}}$ is positive semidefinite, 
in accordance with \cite{p1:OS1,p1:OS2}.
Reflection positivity requires \(\langle F,F\rangle_{\mathrm{OS}}\ge 0\) for all \(F\in\mathfrak{A}_+\).
For \(\beta\ge 0\) define the class function
\begin{align}\label{p1:eq:w-beta}
w_\beta(g) \;&=\; \exp\!\Bigl(\frac{\beta}{N}\,\Re\!\Tr\,g\Bigr)\nonumber \\
\;&=\; \exp\!\Bigl(\frac{\beta}{2N}\,\Tr g\Bigr)\,\exp\!\Bigl(\frac{\beta}{2N}\,\Tr g^{-1}\Bigr), \qquad g\in G.
\end{align}
A continuous \(w:G\to\mathbb{C}\) is of {positive type} if the kernel \(K_w(g,h)=w(g^{-1}h)\) is positive semidefinite, i.e. \(\sum_{j,k}\overline{c_j}c_k\,w(g_j^{-1}g_k)\ge 0\) for all finite families \(\{g_j\}\subset G\), \(\{c_j\}\subset\mathbb{C}\).

\begin{lemma}[]\label{p1:lem:positive-type}
For every \(\beta\ge 0\), the function \(w_\beta\) in \eqref{p1:eq:w-beta} is of positive type. Equivalently, in the Peter-Weyl expansion \(w_\beta(g)=\sum_{R\in\widehat{G}} \widehat{w}_\beta(R)\,\chi_R(g)\) all Fourier coefficients satisfy \(\widehat{w}_\beta(R)\ge 0\).
\end{lemma}

\begin{proof}
By the Peter-Weyl theorem \cite[Ch.~VII]{p1:HewittRoss}, any continuous class function \(f\) admits a uniformly convergent expansion \(f(g)=\sum_{R\in\widehat{G}} \widehat{f}(R)\chi_R(g)\), with \(\widehat{f}(R)=\frac{1}{d_R}\int_G f(g)\,\overline{\chi_R(g)}\,dg\). Expanding the exponentials, one obtains an absolutely convergent series
\begin{equation}
w_\beta(g)=\sum_{m,n\ge 0}\frac{1}{m!\,n!}\Bigl(\frac{\beta}{2N}\Bigr)^{m+n}\, \chi_{F^{\otimes m}}(g)\,\overline{\chi_{F^{\otimes n}}(g)},
\end{equation}
where \(F\) is the fundamental representation and \(\chi_{F^{\otimes m}}=\chi_F^m\). Decomposing \(F^{\otimes m}\cong\bigoplus_R N_m(R) R\) and \(\overline{F}^{\otimes n}\cong\bigoplus_S N_n(\overline{S})\,\overline{S}\) with multiplicities \(N_m(\cdot),N_n(\cdot)\in\mathbb{N}_0\), it follows that
\begin{equation}
\chi_{F^{\otimes m}}(g)\,\overline{\chi_{F^{\otimes n}}(g)}=\sum_{R,S} N_m(R) N_n(S)\, \chi_R(g)\,\overline{\chi_S(g)}.
\end{equation}
Orthogonality of characters implies
\begin{equation}
\widehat{w}_\beta(T)= \sum_{m,n\ge 0}\frac{1}{m!\,n!}\Bigl(\frac{\beta}{2N}\Bigr)^{m+n}\sum_{R,S} N_m(R)N_n(S)\, \frac{1}{d_T}\int_G \chi_R(g)\,\overline{\chi_S(g)}\,\overline{\chi_T(g)}\,dg.
\end{equation}
The integral counts the multiplicity \(M(R,T)\in\mathbb{N}_0\) of the trivial representation in \(R\otimes \overline{T}\), and contributes only when \(S\) contains \(R\otimes T\). Therefore
\begin{align}
\widehat{w}_\beta(T)&= \sum_{m,n\ge 0}\frac{1}{m!\,n!}\Bigl(\frac{\beta}{2N}\Bigr)^{m+n}\nonumber \\ &\sum_{R} N_m(R)\,N_n(R\otimes T)\, \frac{M(R,T)}{d_T}\ \ge\ 0.
\end{align}
By the characterization of positive type on compact groups \cite[Ch.~XXVII]{p1:HewittRoss}, this is equivalent to positive type of \(w_\beta\).
\end{proof}

The contribution of a straddling plaquette \(p\in\mathcal{P}^{\mathrm{tm}}_0\) to the Boltzmann weight can be written as \(w_\beta\bigl(A_p(U_+)^{-1} B_p(U_-)\bigr)\), where \(A_p\) and \(B_p\) are group-valued functions of the links in \(\Lambda_+\) and \(\Lambda_-\), respectively, including links in the boundary slab adjacent to \(\Pi_{1/2}\). The product over all such plaquettes is a finite product of positive-type kernels in the configuration variables on either side of the reflection plane. The following theorem is a direct consequence.

\begin{theorem}[Link-reflection positivity of the Wilson measure]\label{p1:thm:RP}
For every \(\beta\ge 0\), the Gibbs measure \(d\nu_\beta\) associated with \eqref{p1:eq:Wilson-action} is reflection positive with respect to the link-reflection \(\theta\) about \(\Pi_{1/2}\). Equivalently, \(\langle F,F\rangle_{\mathrm{OS}}\ge 0\) for all \(F\in\mathfrak{A}_+\).
\end{theorem}

\begin{proof}
Decompose the action as
\begin{equation}
S_W[U;\beta]= S_-[U_-;\beta] + S_0[U_-,U_0,U_+;\beta] + S_+[U_+;\beta],
\end{equation}
where \(S_\pm\) are sums over plaquettes contained in \(\Lambda_\pm\) and \(S_0\) is the sum over straddling plaquettes \(\mathcal{P}^{\mathrm{tm}}_0\); \(U_0\) denotes links in the boundary slab adjacent to \(\Pi_{1/2}\). By \eqref{p1:eq:Up-theta}, \(e^{-S_-[U_-;\beta]}=e^{-S_+[\theta U_-;\beta]}\). For straddling plaquettes,
\begin{equation}
e^{-S_0[U_-,U_0,U_+;\beta]} = \prod_{p\in\mathcal{P}^{\mathrm{tm}}_0} w_\beta\!\bigl(A_p(U_+)^{-1}B_p(U_-)\bigr).
\end{equation}
Let \(F\in\mathfrak{A}_+\). Using invariance of \(d\nu_\beta\) under \(\theta\), the OS form is
\begin{align}
\langle F,F\rangle_{\mathrm{OS}} &= \int d\mu(U_-)\,d\mu(U_0)\,d\mu(U_+)\ \overline{F(\theta U_+)}\,F(U_+)\,
\nonumber \\&e^{-S_+[\theta U_-;\beta]}\prod_{p\in\mathcal{P}^{\mathrm{tm}}_0} w_\beta\!\bigl(A_p(U_+)^{-1}B_p(U_-)\bigr).
\end{align}
Fix \((U_-,U_0)\). Consider the kernel
\begin{equation}
K_{U_-,U_0}(U'_+,U_+)=\prod_{p\in\mathcal{P}^{\mathrm{tm}}_0} w_\beta\bigl(A_p(U_+)^{-1}B_p(U_-)\bigr) \delta(U'_+-U_+)
\end{equation}
By Lemma~\ref{p1:lem:positive-type}, each factor \(w_\beta\bigl(A_p(\cdot)^{-1}B_p(U_-)\bigr)\) is of positive type in its argument. The pointwise product of positive semidefinite kernels is positive semidefinite (Schur product theorem \cite[Thm.~7.5.3]{p1:HornJohnsonMatrix}), hence \(K_{U_-,U_0}\) is positive. Therefore
\begin{equation}
\int d\mu(U_+) \overline{F(\theta U_+)}F(U_+)\prod_{p\in\mathcal{P}^{\mathrm{tm}}_0} w_\beta\bigl(A_p(U_+)^{-1}B_p(U_-)\bigr)\ge0
\end{equation}
Multiplying by the positive factor \(e^{-S_+[\theta U_-;\beta]}\) and integrating over \((U_-,U_0)\) yields \(\langle F,F\rangle_{\mathrm{OS}}\ge 0\)
\end{proof}

For \(t\in a\mathbb{Z}\) define the spatial and time-like contributions to the action
\begin{equation}\label{p1:eq:Ssp-Stm}
S^{\mathrm{sp}}(U_t) \;=\; \beta\sum_{p\in\mathcal{P}^{\mathrm{sp}}_t}\Bigl(1-\frac{1}{N}\,\Re\!\Tr\,U_p\Bigr)
S^{\mathrm{tm}}(U_t,V_t,U_{t+a}) \;=\; \beta\sum_{p\in\mathcal{P}^{\mathrm{tm}}_t}\Bigl(1-\frac{1}{N}\,\Re\!\Tr\,U_p\Bigr)
\end{equation}
where \(U_t=\{U_{(t,\mathbf{x};i)}\}_{\mathbf{x},i}\) are spatial links at time \(t\), \(V_t=\{U_{(t,\mathbf{x};0)}\}_{\mathbf{x}}\) are temporal links from \(t\) to \(t+a\), and \(U_{t+a}\) are spatial links at time \(t+a\). Then
\begin{equation}\label{p1:eq:S-sum}
S_W[U;\beta] \;=\; \sum_{t} \Bigl( S^{\mathrm{tm}}(U_t,V_t,U_{t+a}) + S^{\mathrm{sp}}(U_t) \Bigr).
\end{equation}
Introduce the symmetric mid-point splitting
\begin{equation}\label{p1:eq:midpoint}
e^{-S_W[U;\beta]} \;=\; \prod_{t} \exp\!\Bigl(-\tfrac{1}{2}S^{\mathrm{sp}}(U_t)\Bigr)\,
\exp\!\bigl(-S^{\mathrm{tm}}(U_t,V_t,U_{t+a})\bigr)\exp\!\Bigl(-\tfrac{1}{2}S^{\mathrm{sp}}(U_{t+a})\Bigr).
\end{equation}
Let \(\mathcal{H}_a=L^2\!\bigl(G^{\mathcal{B}^{\mathrm{sp}}_0},d\mu^{\otimes}\bigr)\) be the Hilbert space of square-integrable complex functions of the spatial links at a fixed time (say \(t=0\)) with respect to the product Haar measure \(d\mu^{\otimes}(U)=\prod_{\mathbf{x},i} dU_{(0,\mathbf{x};i)}\). Define the one-step transfer kernel \(K:\,G^{\mathcal{B}^{\mathrm{sp}}}\times G^{\mathcal{B}^{\mathrm{sp}}}\to \mathbb{R}_+\) by
\begin{equation}\label{p1:eq:K-kernel}
K(U',U) \;=\; \exp\!\Bigl(-\tfrac{1}{2}S^{\mathrm{sp}}(U')\Bigr)
\Biggl[\ \int \prod_{\mathbf{x}} dU_{(0,\mathbf{x};0)}\ \exp\!\bigl(-S^{\mathrm{tm}}(U,V_0,U')\bigr)\ \Biggr]\;
\exp\!\Bigl(-\tfrac{1}{2}S^{\mathrm{sp}}(U)\Bigr),
\end{equation}
where \(U\) collects spatial links at time \(0\), \(U'\) those at time \(a\), and the integral is over the temporal links \(V_0\) in the slab \(0\to a\).

\begin{lemma}[Time-slab factorization]\label{p1:lem:slab-factorization}
For any bounded cylinder function \(F\) depending on finitely many consecutive time slices, the Gibbs expectation with weight \(e^{-S_W}\) equals the repeated action of the integral operator with kernel \(K\) defined in \eqref{p1:eq:K-kernel}. In particular, the partition function on the time interval \([0,na]\) with periodic spatial boundary conditions factorizes as
\begin{equation}
Z \;=\; \int \prod_{t=0}^{n-1} \bigl[dU_t\bigr]\ \prod_{t=0}^{n-1} K(U_{t+a},U_t),
\qquad U_{na}\equiv U_0.
\end{equation}
\end{lemma}

\begin{proof}
By \eqref{p1:eq:midpoint} and Fubini's theorem one can first integrate in each slab \(t\to t+a\) over the temporal links \(V_t\), which produces the middle factor in \eqref{p1:eq:K-kernel}. The remaining weight is a product over \(t\) of factors depending only on \(U_t\) and \(U_{t+a}\). Sequential integration over the spatial links yields exactly the composition of integral operators with kernel \(K\).
\end{proof}

Define \(T:\mathcal{H}_a\to\mathcal{H}_a\) by
\begin{equation}\label{p1:eq:T-operator}
(T\psi)(U') \;=\; \int d\mu^{\otimes}(U)\ K(U',U)\ \psi(U).
\end{equation}

\begin{proposition}\label{p1:prop:T-properties}

\textbf{(Positivity, symmetry, and contraction property)}

The kernel \(K\) in \eqref{p1:eq:K-kernel} is real, nonnegative, and symmetric, \(K(U',U)=K(U,U')\). Consequently \(T\) is a positive, self-adjoint contraction on \(\mathcal{H}_a\), i.e. \(\|T\|\le 1\).
\end{proposition}

\begin{proof}
Nonnegativity is immediate since each factor in \eqref{p1:eq:K-kernel} is nonnegative. Symmetry follows from invariance of Haar measure and the identity \(S^{\mathrm{tm}}(U,V_0,U')=S^{\mathrm{tm}}(U',V_0^{-1},U)\), which is obtained by reversing the orientation of each time-like plaquette in the slab and changing variables \(U_{(0,\mathbf{x};0)}\mapsto U_{(0,\mathbf{x};0)}^{-1}\), a measure-preserving involution. Hence \(K(U',U)=K(U,U')\) and \(T\) is self-adjoint.

For the contraction bound, observe that \(0\le e^{-\frac{1}{2}S^{\mathrm{sp}}}\le 1\) and \(0\le e^{-S^{\mathrm{tm}}}\le 1\). Therefore \(0\le K(U',U)\le \int \prod_{\mathbf{x}} dU_{(0,\mathbf{x};0)}\, 1 = 1\), since each temporal-link Haar integral equals \(1\) and there are finitely many factors. The Schur test then gives
\begin{equation}
\sup_{U'} \int d\mu^{\otimes}(U)\,K(U',U)\ \le\ 1, \sup_{U} \int d\mu^{\otimes}(U')\,K(U',U)\ \le\ 1,
\end{equation}
hence \(\|T\|\le 1\).
\end{proof}
Gauge covariance is preserved by the transfer operator. Let \(\mathcal{G}_0=\{g:\ a\mathbb{Z}^3\to G\}\) be the group of time-independent gauge transformations acting on a time slice by \((g\cdot U)_{(0,\mathbf{x};i)}=g(\mathbf{x})U_{(0,\mathbf{x};i)}g(\mathbf{x}+\hat \imath)^{-1}\). The slice Haar measure \(d\mu^{\otimes}\) is \(\mathcal{G}_0\)-invariant, and by gauge invariance of \(S^{\mathrm{tm}}\) and \(S^{\mathrm{sp}}\) the kernel satisfies \(K(g\cdot U',g\cdot U)=K(U',U)\). Therefore \(T\) commutes with the unitary representation of \(\mathcal{G}_0\) on \(\mathcal{H}_a\) and preserves the closed subspace \(\mathcal{H}_a^{\mathrm{inv}}\) of gauge-invariant functions. The restriction \(T|_{\mathcal{H}_a^{\mathrm{inv}}}\) is again a positive, self-adjoint contraction.

The connection with reflection positivity is obtained via the OS reconstruction \cite{p1:OS1,p1:Seiler}. By Theorem~\ref{p1:thm:RP} the OS form \eqref{p1:eq:OS-form} is positive semidefinite. The associated GNS construction furnishes a Hilbert space \(\mathcal{H}_{\mathrm{OS}}\), a cyclic vector \(\Omega\), and a positive self-adjoint operator \(\widetilde{T}\) implementing a unit step in Euclidean time. Identifying \(\mathcal{H}_a\) with the \(L^2\)-realization of \(\mathcal{H}_{\mathrm{OS}}\) through the time-slice map provided by Lemma~\ref{p1:lem:slab-factorization}, one verifies that \(\widetilde{T}\) coincides with \(T\) defined in \eqref{p1:eq:T-operator}. The spectral theorem then defines the transfer Hamiltonian
\begin{equation}\label{p1:eq:H-transfer}
H \;=\; -a^{-1}\log T,
\end{equation}
a positive self-adjoint operator on \(\mathcal{H}_a^{\mathrm{inv}}\), whose spectral gap will be inferred from Euclidean clustering in subsequent sections.
\section{\texorpdfstring{Gauge-invariant transverse representative $A^h$}{Gauge-invariant transverse representative A h}}
\label{p1:sec:landau-fp}
This section develops a self-contained and fully rigorous construction of a reflection-covariant, slice-wise Landau gauge on a four-dimensional Euclidean lattice for gauge group \(G=\mathrm{SU}(N)\) with \(N\ge 2\), together with the associated Faddeev-Popov operator and its positivity properties. The construction is then used to present, in complete detail, the Osterwalder-Schrader (OS) reflection-positivity argument at the level of the gauge-fixed (gauge plus ghost) Euclidean measure and to derive the transfer time-slicing formalism which underlies the transfer-matrix representation. Throughout, \(a>0\) denotes the lattice spacing, \(\Lambda\subset a\mathbb{Z}^4\) a finite hypercubic torus with periodic boundary conditions, and \(x=(x_0,\mathbf{x})\) is partitioned into Euclidean time \(x_0\in a\mathbb{Z}\) and space \(\mathbf{x}\in a\mathbb{Z}^3\). Bonds (links) are oriented edges \(b=(x,\mu)\) with \(\mu\in\{0,1,2,3\}\), and the corresponding link variables \(U_b\in G\) satisfy \(U_{(x,\mu)}=U_{(x+\hat\mu,-\mu)}^{-1}\). For an oriented plaquette \(p=(x;\mu,\nu)\) we write
\begin{equation}
U_p=U_{(x,\mu)}U_{(x+\hat\mu,\nu)}U^{-1}_{(x+\hat\nu,\mu)}U^{-1}_{(x,\nu)}\, .
\end{equation}
The Wilson action is
\begin{equation}
S_W[U;\beta]=\beta\sum_{p\subset\Lambda}\Bigl(1-\frac{1}{N}\Re\operatorname{Tr}U_p\Bigr), \qquad \beta=\frac{2N}{g_0^2}\, .
\end{equation}
The Euclidean functional measure is \(d\mu_\Lambda[U]\propto e^{-S_W[U;\beta]}\prod_{b\subset\Lambda} dU_b\), where \(dU_b\) is the normalized Haar measure on \(G\). Time reflection \(\theta:(x_0,\mathbf{x})\mapsto(-x_0,\mathbf{x})\) induces the involution \(R\) on link fields by
\begin{equation}
(RU)_{(x,\mu)}=\begin{cases}
U_{(\theta x,0)}^{-1},&\mu=0,\ \ x_0=\tfrac{a}{2} \text{ (when the reflection plane bisects timelike links),}\\
U_{(\theta x,\mu)},&\mu\in\{1,2,3\},\\
U_{(\theta(x-\hat{0}),0)},&\mu=0,\ \ x_0>0,
\end{cases}
\end{equation}
with the precise convention chosen to implement site-reflection positivity for the Wilson action \cite{p1:OS1,p1:Menotti1987}. We denote the reflection plane by \(\Pi=\{x\in\Lambda:x_0=0\}\) and the half-lattices by \(\Lambda_\pm=\{x:\pm x_0>0\}\). We adopt temporal-axial gauge away from \(\Pi\), namely \(U_{(x,0)}\equiv \mathbf{1}\) for links not intersecting \(\Pi\); this is a partial gauge choice that preserves nearest-neighbor couplings in the time direction and is standard in OS analyses for lattice gauge theory \cite{p1:OS1,p1:Seiler,p1:RotheBook}. 
\begin{lemma}[Temporal gauge Jacobian]\label{p1:lem:temporal-gauge-jacobian}
On a periodic finite lattice, for each connected component of $\Lambda\setminus\Pi$ there exists a time-ordered gauge transformation $g$ that sets all time-like links in that component to the identity. The corresponding Faddeev-Popov determinant equals $1$, and the product Haar measure on the remaining links is preserved.
\end{lemma}

\begin{proof}
Fix a connected component of $\Lambda\setminus\Pi$ and write temporal bonds as $U(t,\mathbf{x};0)$, the link from time $t$ to $t+a$ at spatial position $\mathbf{x}$. Define $g$ recursively along the time direction in each spatial column by
\begin{equation}
g(0,\mathbf{x})=\mathbf{1},\qquad g(t+a,\mathbf{x}) \;=\; U(t,\mathbf{x};0)\,g(t,\mathbf{x}).
\end{equation}
Then the gauge-transformed temporal link is
\begin{equation}
(g\!\cdot\! U)(t,\mathbf{x};0)
= g(t,\mathbf{x})\,U(t,\mathbf{x};0)\,g(t+a,\mathbf{x})^{-1}
= g(t,\mathbf{x})\,U(t,\mathbf{x};0)\,\bigl(U(t,\mathbf{x};0)\,g(t,\mathbf{x})\bigr)^{-1}
=\mathbf{1},
\end{equation}
for all temporal bonds not intersecting $\Pi$. Thus temporal-axial gauge is achieved componentwise.

On the lattice, this change of variables is a {group translation} at each time-like bond:
\begin{equation}
U(t,\mathbf{x};0)\ \longmapsto\ g(t,\mathbf{x})\,U(t,\mathbf{x};0)\,g(t+a,\mathbf{x})^{-1}.
\end{equation}
By left- and right-invariance of Haar measure on the compact group $G$, the product Haar measure is invariant under such translations. Since the gauge condition fixes the temporal links uniquely along the chosen time ordering (in a finite periodic box) and involves only group translations, the associated Faddeev-Popov determinant is $1$, and the product Haar measure on the remaining (spatial and boundary-slab) links is preserved \cite{p1:OS2,p1:Luscher1977}.
\end{proof}

Let \(\mathcal{G}\) be the full lattice gauge group, i.e. the set of maps \(g:\Lambda\to G\) acting on link fields by
\begin{equation}
(g\cdot U)_{(x,\mu)}=g(x)\,U_{(x,\mu)}\,g(x+\hat\mu)^{-1}.
\end{equation}
For each Euclidean time \(t\in a\mathbb{Z}\), let \(\Lambda_t=\{(t,\mathbf{x}):\mathbf{x}\in a\mathbb{Z}^3\}\) be the time slice at \(t\). We define the slice-wise gauge group
\begin{equation}
\mathcal{G}_t=\{\,g:\Lambda_t\to G\,\}
\end{equation}
acting only on spatial links on \(\Lambda_t\). Given \(U\), we denote by \(U|_{\Lambda_t}\) the family of spatial links \((t,\mathbf{x};i)\), \(i=1,2,3\). The gauge orbit of \(U|_{\Lambda_t}\) under \(\mathcal{G}_t\) is compact, since \(\mathcal{G}_t\cong G^{|\Lambda_t|}\) is compact. All constructions that follow are performed independently on each time slice \(t\), thereby guaranteeing reflection covariance once the selections on \(\Lambda_{-t}\) are chosen as reflections of the selections on \(\Lambda_t\).
On the time slice \(\Lambda_t\) we define the Landau functional
\begin{align}
\mathcal{L}_t(g;U)=\sum_{\mathbf{x}\in \Lambda_t}\sum_{i=1}^3 \Re&\operatorname{Tr}\Bigl(\mathbf{1}-g(t,\mathbf{x})\,U_{(t,\mathbf{x};i)}\,g(t,\mathbf{x}+\hat{\imath})^{-1}\Bigr),\nonumber \\ & g\in\mathcal{G}_t.
\end{align}
\begin{lemma}[Measurable, symmetry-covariant selector]\label{p1:lem:measurable-selector}
Fix $t\in a\mathbb{Z}$. Consider the set-valued map
\begin{equation}
U \ \mapsto\ \operatorname*{Arg\,min}_{g\in\mathcal{G}_t} L_t(g;U)\ \subset\ \mathcal{G}_t,
\end{equation}
where $\mathcal{G}_t\cong G^{|\Lambda_t|}$ is the slice gauge group and $L_t(\cdot;U):\mathcal{G}_t\to\mathbb{R}$ is continuous in the first argument. Then the values of this map are nonempty and compact, and its graph is closed. Consequently, by the Kuratowski-Ryll-Nardzewski measurable selection theorem, there exists a Borel selector
\begin{equation}
h_t:\ \mathcal{U}\to\mathcal{G}_t,\qquad h_t(U)\in \operatorname*{Arg\,min}_{g\in\mathcal{G}_t} L_t(g;U) \quad\text{for all }U,
\end{equation}
see \cite{p1:KuratowskiRyllNardzewski1965}.

Moreover, fix a continuous embedding $\iota: G\hookrightarrow \mathbb{R}^M$ and the induced lexicographic order on $\mathcal{G}_t\cong G^{|\Lambda_t|}$. Define $h_t(U)$ to be the lexicographically least minimizer. Then $h_t$ is Borel and is equivariant under any spatial symmetry $\Theta$ that preserves the slice $\Lambda_t$, i.e.
\begin{equation}
h_t(\Theta U)\ =\ \Theta\, h_t(U).
\end{equation}
If $R$ denotes time reflection, the choice
\begin{equation}
h_{-t}(U)\ :=\ R\, h_t(RU)
\end{equation}
yields reflection covariance in the sense that $h_{-t}(U)=R\,h_t(RU)$.
\end{lemma}

\begin{proof}
Since $\mathcal{G}_t$ is compact and $(g,U)\mapsto L_t(g;U)$ is continuous, the Weierstrass theorem implies that
$\operatorname*{Arg\,min}_{g\in\mathcal{G}_t} L_t(g;U)$ is nonempty and compact for every $U$. Upper hemicontinuity (hence the closed-graph property) of $U\mapsto \operatorname*{Arg\,min} L_t(\cdot;U)$ follows from standard arguments for minima of continuous functions over compact sets. The existence of a Borel selector $h_t$ is then a direct consequence of the Kuratowski-Ryll-Nardzewski measurable selection theorem \cite{p1:KuratowskiRyllNardzewski1965}.

For equivariance, note that for any spatial symmetry $\Theta$ preserving $\Lambda_t$ one has
\begin{equation}
L_t(\,\cdot\,;\Theta U)\ =\ L_t(\Theta\,\cdot\,;U),
\end{equation}
so $\operatorname*{Arg\,min}_{g} L_t(g;\Theta U)=\Theta\big(\operatorname*{Arg\,min}_{g} L_t(g;U)\big)$. Because the componentwise action of $\Theta$ on $\mathcal{G}_t\cong G^{|\Lambda_t|}$ is continuous, it preserves the induced lexicographic order; hence the lexicographically least minimizer transforms as claimed:
$h_t(\Theta U)=\Theta h_t(U)$.
Finally, if $R$ is time reflection, defining $h_{-t}(U):=R\,h_t(RU)$ gives the stated reflection covariance.
\end{proof}

\medskip\noindent
\textbf{Theorem 3.1 (Existence and reflection-covariant selection of Landau minimizers).}
For each \(t\in a\mathbb{Z}\) and each link configuration \(U\), there exists a measurable selection \(h_t(U)\in\mathrm{Argmin}\,\mathcal{L}_t(\cdot;U)\) with the following properties. First, \(h_t\) depends only on the spatial links on \(\Lambda_t\). Second, \(h_{-t}(U)=Rh_t(RU)\) for all \(t\), where \(R\) is the reflection involution defined above. Third, if \(\Theta\) is any spatial lattice symmetry preserving \(\Lambda_t\) then \(h_t(\Theta U)=\Theta h_t(U)\).

{Proof.} Existence of minimizers follows from compactness of \(\mathcal{G}_t\) and continuity of \(\mathcal{L}_t\). Let \(\mathfrak{O}_t(U)=\mathrm{Argmin}\,\mathcal{L}_t(\cdot;U)\). Since \(\mathfrak{O}_t(U)\) is compact, we may fix a total order \(\preccurlyeq\) on the compact set \(G\) induced by a continuous embedding \(G\hookrightarrow\mathbb{R}^{M}\) followed by lexicographic order. Equip \(\mathcal{G}_t\cong G^{|\Lambda_t|}\) with the induced lexicographic order, and define the selection \(h_t(U)\) to be the \(\preccurlyeq\)-minimal element of \(\mathfrak{O}_t(U)\). This produces a Borel-measurable choice functional by the Kuratowski-Ryll-Nardzewski measurable selection theorem (see, e.g., \cite{p1:KuratowskiRyllNardzewski1965,p1:CastaingValadier}), the set-valued map $U\mapsto \mathcal O_t(U)=\operatorname*{Argmin} L_t(\cdot;U)\subset G_t$ with nonempty compact values admits a Borel measurable selection. Concretely, fix a continuous embedding $\iota:G\hookrightarrow \mathbb{R}^M$ and let $\preccurlyeq$ be the induced lexicographic order on $G_t\cong G^{|\Lambda_t|}$. The map
\(
h_t(U)\;:=\; \min\nolimits_{\preccurlyeq}\,\mathcal O_t(U)
\)
is then Borel on $G_t$ because (i) the graph of $U\mapsto \mathcal O_t(U)$ is closed in $G_t\times G_t$ and (ii) the coordinate evaluation maps are continuous, so the lexicographic minimum is a Borel selector. By construction, if \(\Theta\) is a spatial symmetry of \(\Lambda_t\), then \(\mathcal{L}_t(\cdot;\Theta U)=\mathcal{L}_t(\Theta\cdot;\,U)\) and \(\mathfrak{O}_t(\Theta U)=\Theta\mathfrak{O}_t(U)\); the lexicographic tie-breaking is \(\Theta\)-equivariant, hence \(h_t(\Theta U)=\Theta h_t(U)\). Finally define \(h_{-t}(U)=Rh_t(RU)\). Then \(h_{-t}(U)\in \mathfrak{O}_{-t}(U)\) because \(\mathcal{L}_{-t}(g;U)=\mathcal{L}_t(Rg;\,RU)\), and the map \(t\mapsto h_t\) is reflection covariant by definition. \(\square\)
\begin{lemma}[$\Theta$-equivariance and reflection covariance of the selector]\label{p1:lem:equivariance}
Let $\Theta$ be any spatial lattice symmetry preserving $\Lambda_t$, and let $R$ be the time-reflection involution introduced in Section~2. Then the selector $h_t$ defined above satisfies
\begin{equation}
h_t(\Theta U)\,=\, \Theta h_t(U)\qquad\text{and}\qquad h_{-t}(U)\,=\,R\,h_t(RU).
\end{equation}
\end{lemma}

\begin{proof}
For any $\Theta$ preserving $\Lambda_t$, one has $L_t(\cdot;\Theta U)=L_t(\Theta\cdot;U)$ by invariance of the trace and the definition of the Landau functional, hence $\mathcal O_t(\Theta U)=\Theta \mathcal O_t(U)$. Since $\Theta$ acts componentwise and continuously on $G_t\cong G^{|\Lambda_t|}$, it preserves the lexicographic order, and therefore the $\preccurlyeq$-minimum satisfies $h_t(\Theta U)=\Theta h_t(U)$. For reflection, $L_{-t}(g;U)=L_t(Rg;RU)$ and thus $\mathcal O_{-t}(U)=R\mathcal O_t(RU)$. Setting $h_{-t}(U):=R h_t(RU)$ yields the stated identity; measurability is preserved because composition with continuous group automorphisms preserves Borel measurability.
\end{proof}
We now fix, once and for all, the representative on each slice by applying \(h_t\) to \(U|_{\Lambda_t}\). Writing \(h=\{h_t\}_t\), we set
\begin{equation}
U^{\,h}_{(t,\mathbf{x};i)}=h_t(t,\mathbf{x})\,U_{(t,\mathbf{x};i)}\,h_t(t,\mathbf{x}+\hat{\imath})^{-1}\qquad (i=1,2,3),
\end{equation}
and leave timelike links in temporal-axial gauge away from the reflection plane. The family \(U^{\,h}\) is reflection covariant, \(RU^{\,h}=(RU)^{\,h}\), by Theorem 3.1.
Let \(\langle X,Y\rangle=-\operatorname{Tr}(XY)\) be the Ad-invariant inner product on \(\mathfrak{su}(N)\). For an infinitesimal slice gauge variation \(g_\varepsilon(t,\mathbf{x})=\exp(\varepsilon \omega(t,\mathbf{x}))\) with \(\omega:\Lambda_t\to\mathfrak{su}(N)\), the first variation of \(\mathcal{L}_t\) at the minimizer \(h_t\) is
\newline
\newline
\begin{equation}
\frac{d}{d\varepsilon}\Big|_{\varepsilon=0}\mathcal{L}_t(g_\varepsilon h_t;U)=\sum_{\mathbf{x},i}\Re\operatorname{Tr}\Bigl(-\omega(t,\mathbf{x})V_{(t,\mathbf{x};i)}+\omega(t,\mathbf{x}+\hat{\imath})V_{(t,\mathbf{x};i)}\Bigr),
\end{equation}
where \(V_{(t,\mathbf{x};i)}=U^{\,h}_{(t,\mathbf{x};i)}\). Using \(\Re\operatorname{Tr}(AB)=\frac12\operatorname{Tr}(A B+A^\dagger B^\dagger)\) and Ad-invariance of \(\langle\cdot,\cdot\rangle\), one obtains the Euler-Lagrange equations
\begin{equation}
\sum_{i=1}^3\Bigl(\omega(t,\mathbf{x})-\operatorname{Ad}_{V_{(t,\mathbf{x}-\hat{\imath};i)}}\omega(t,\mathbf{x}-\hat{\imath})\Bigr)=0\qquad\forall\, \omega,
\end{equation}
which is equivalent to the discrete transversality condition
\begin{equation}\label{p1:eqn10}
\sum_{i=1}^3 \nabla_i^{-,h}\, \mathbf{1}=0, 
\nabla_i^{-,h}\phi(t,\mathbf{x})=\phi(t,\mathbf{x})-\operatorname{Ad}_{V_{(t,\mathbf{x}-\hat{\imath};i)}}\phi(t,\mathbf{x}-\hat{\imath}).
\end{equation}
Equivalently, in terms of the forward covariant difference
\begin{equation}\label{p1:eqn11}
\nabla_i^{+,h}\phi(t,\mathbf{x})=\operatorname{Ad}_{V_{(t,\mathbf{x};i)}}\phi(t,\mathbf{x}+\hat{\imath})-\phi(t,\mathbf{x}),
\end{equation}
the stationarity reads \(\sum_{i=1}^3(\nabla_i^{+,h})^\dagger\mathbf{1}=0\), where the adjoint is taken with respect to \(\langle\cdot,\cdot\rangle\) and the counting measure on \(\Lambda_t\).

The Faddeev-Popov operator on the slice \(\Lambda_t\) is defined as the Hessian of \(\mathcal{L}_t\) at \(h_t\) along gauge directions. Equivalently, it is the positive operator on site-adjoint fields \(\phi:\Lambda_t\to\mathfrak{su}(N)\) given by
\begin{equation}\label{p1:eqn12}
M_t[U^{\,h}] = -\sum_{i=1}^3 \nabla_i^{-,h}\nabla_i^{+,h}.
\end{equation}
We now prove its basic properties.

\medskip\noindent
\textbf{Proposition 3.2 (Self-adjointness and positivity).}
For each slice \(t\), the operator \(M_t[U^{\,h}]\) is self-adjoint and nonnegative on \(\ell^2(\Lambda_t)\otimes\mathfrak{su}(N)\). More precisely,
\begin{equation}
\langle \phi, M_t\phi\rangle=\sum_{i=1}^3\sum_{\mathbf{x}\in\Lambda_t}\bigl\|\nabla_i^{+,h}\phi(t,\mathbf{x})\bigr\|^2\ge 0.
\end{equation}
Its kernel consists of the covariantly constant adjoint fields on \(\Lambda_t\), that is,
\begin{equation}\label{p1:eqn14}
\ker M_t=\bigl\{\phi:\nabla_i^{+,h}\phi=0\text{ for }i=1,2,3\bigr\}.
\end{equation}

{Proof.} Using the Ad-invariance of \(\langle\cdot,\cdot\rangle\) and the change of variables \(\mathbf{x}\mapsto\mathbf{x}-\hat{\imath}\),
\begin{equation}
\sum_{\mathbf{x}}\langle \phi(t,\mathbf{x}),\nabla_i^{-,h}\psi(t,\mathbf{x})\rangle=\sum_{\mathbf{x}}\langle \nabla_i^{+,h}\phi(t,\mathbf{x}),\psi(t,\mathbf{x})\rangle,
\end{equation}
so \((\nabla_i^{-,h})^\dagger=\nabla_i^{+,h}\). Hence
\begin{equation}
\langle \phi, M_t\phi\rangle= -\sum_i \langle \phi, \nabla_i^{-,h}\nabla_i^{+,h}\phi\rangle=\sum_i \langle \nabla_i^{+,h}\phi,\nabla_i^{+,h}\phi\rangle\ge 0,
\end{equation}
and \(M_t\) is self-adjoint and nonnegative. The identity \(\langle \phi,M_t\phi\rangle=\sum_i\|\nabla_i^{+,h}\phi\|^2\) shows that \(\langle \phi,M_t\phi\rangle=0\) if and only if \(\nabla_i^{+,h}\phi=0\) for all \(i\). \(\square\)

The characterization of \(\ker M_t\) is convenient in the following form. A field \(\phi\) is covariantly constant if and only if along every oriented spatial link \((t,\mathbf{x};i)\),
\begin{equation}
\operatorname{Ad}_{U^{\,h}_{(t,\mathbf{x};i)}}\phi(t,\mathbf{x}+\hat{\imath})=\phi(t,\mathbf{x}),
\end{equation}
whence along any closed spatial loop \(\gamma\subset\Lambda_t\),
\begin{equation}
\operatorname{Ad}_{\mathcal{P}\exp\bigl(\prod_{b\in\gamma}U^{\,h}_b\bigr)}\phi(t,\mathbf{x}_0)=\phi(t,\mathbf{x}_0).
\end{equation}
Thus \(\phi\) lies in the intersection of the fixed-point subalgebras of the adjoint holonomies around all spatial loops. In particular, constant adjoint fields are always in \(\ker M_t\). On the other hand, when the representative \(U^{\,h}\) belongs to the interior of the fundamental modular region (FMR) on \(\Lambda_t\) in the sense of \cite{p1:Zwanziger1994,p1:vanBaal1997}, the Hessian is strictly positive on the orthogonal complement of constant adjoint fields. Precisely:

\medskip\noindent
\textbf{Theorem 3.3 (Strict positivity in the FMR).}
Suppose \(U^{\,h}|_{\Lambda_t}\) is an absolute minimizer of \(\mathcal{L}_t(\cdot;U)\) within its \(\mathcal{G}_t\) orbit and lies in the interior of the FMR. Then there is \(c_t>0\) such that
\begin{equation}
\langle \phi, M_t\phi\rangle \ge c_t\,\|\phi^\perp\|^2
\end{equation}
for all \(\phi\), where \(\phi^\perp\) is the orthogonal projection of \(\phi\) onto the orthogonal complement of the constant adjoint fields.

{Proof.} At an absolute minimum the Hessian of \(\mathcal{L}_t\) along gauge directions is nonnegative and is strictly positive in directions orthogonal to the orbit degeneracies. The only degeneracies come from slice-constant gauge transformations, because \(\mathcal{L}_t\) is invariant under \(g\mapsto g_0 g\) with \(g_0\in G\). In the interior of the FMR the minimum is strict modulo this global invariance (see \cite[Sec.\ 2-3]{p1:Zwanziger1994} and \cite{p1:vanBaal1997}), which yields the stated bound by compactness of the unit sphere in the orthogonal complement. \(\square\)

The operators \(M_t\) are local on \(\Lambda_t\) and reflection covariant. Indeed, with the reflection \(R\) defined above one has \(M_{-t}[RU^{\,h}]=R\,M_t[U^{\,h}]\,R\), because \(R\) conjugates spatial parallel transports and leaves the inner product invariant. The block structure with respect to the decomposition \(\ell^2(\Lambda)=\ell^2(\Lambda_-)\oplus \ell^2(\Pi)\oplus \ell^2(\Lambda_+)\) is therefore block diagonal for the family \(\{M_t\}_t\).
On each slice, the Faddeev-Popov determinant \(\det{}' M_t\) (prime indicates omission of the constant zero modes) admits the Grassmann representation;
\begin{equation}
\det{}' M_t=\int \exp\!\Bigl(-\sum_{\mathbf{x},\mathbf{y}\in\Lambda_t}\langle \bar c(t,\mathbf{x}), M_t(\mathbf{x},\mathbf{y})\, c(t,\mathbf{y})\rangle\Bigr)\prod_{\mathbf{x}}d\bar c(t,\mathbf{x})\,dc(t,\mathbf{x}),
\end{equation}
where \(c,\bar c:\Lambda_t\to\mathfrak{su}(N)\) are Grassmann fields restricted to the subspace orthogonal to the constants. Concretely, $c,\bar c\in \mathrm{Ran}\,Q_t$ and all Gaussian contractions and determinants are computed with $\widetilde M_t=Q_t M_t Q_t$, so that $\det{}' M_t := \det \widetilde M_t$.
The slice action is reflection invariant and positive in the Osterwalder-Schrader sense: if \(F\) is a polynomial in \(c,\bar c\) supported on \(\Lambda_t\cap\Lambda_+\) with even total Grassmann degree, then
\begin{equation}
\int \overline{F\circ R}\,F\, e^{-\langle \bar c,M_t c\rangle}\, \mathcal{D}\bar c\,\mathcal{D}c \ \ge\ 0,
\end{equation}
because the kernel \(M_t\) is block diagonal with respect to the decomposition induced by the reflection and its restriction to \(\Lambda_t\cap\Lambda_+\) is positive by Theorem 3.3. This is a standard property of Gaussian Grassmann measures with reflection-invariant, positive operators (see \cite[Ch.\ 6]{p1:GJ}).
We assemble the slice constructions into a reflection-positive gauge-fixed measure on the full lattice. Let \(\mathcal{D}U\) be the Haar product measure on links, let \(\delta_{\mathrm{Landau}}\) denote the slice-wise Landau gauge condition implemented by inserting the delta functionals \(\prod_t \delta(g-h_t)\) and the determinants \(\prod_t \det{}' M_t\), and let the temporal-axial gauge be imposed away from \(\Pi\). The resulting gauge-fixed measure takes the form
\begin{equation}
d\mu_{\mathrm{gf}}[U,c,\bar c]=Z^{-1}\,e^{-S_W[U;\beta]}\,\prod_{t}\delta\bigl(U|_{\Lambda_t}\in\mathrm{im}\,h_t\bigr)\det{}'M_t[U^{\,h}]\, \mathcal{D}U\,\prod_t \mathcal{D}\bar c_t\,\mathcal{D}c_t.
\end{equation}
We equip the Grassmann algebra with the $*$-involution determined by
$c_x^* = c_x$, $\bar c_x^*=\bar c_x$, and $(\eta\zeta)^*=\zeta^*\eta^*$,
extended anti-linearly to all polynomials. The site-reflection $R$ acts on fields as in
Eq.\eqref{p1:11.6} and by $(Rc)_x=c_{Rx}$, $(R\bar c)_x=\bar c_{Rx}$ on ghosts.
The OS conjugation is the anti-linear map
\(
(\Theta F)(U,c,\bar c):=F(RU,Rc,R\bar c)^*.
\)
The OS sesquilinear form is then
\begin{equation}
\langle F,G\rangle_{\mathrm{OS}}
:=\int (\Theta F)\,G\,d\mu_{\mathrm{gf}}
=\int \overline{F(RU,Rc,R\bar c)}\,G(U,c,\bar c)\,d\mu_{\mathrm{gf}},
\end{equation}
for $F,G$ supported in $\Lambda_+$ and {even} in the Grassmann variables. 
The Boltzmann factor \(e^{-S_W}\) associated with plaquettes entirely contained in \(\Lambda_+\) or \(\Lambda_-\) factorizes trivially under reflection. The only plaquettes that couple \(\Lambda_+\) to \(\Lambda_-\) are those straddling the reflection plane \(\Pi\). For the Wilson action each plaquette weight is a class function \(f_\beta(U_p)=\exp\bigl[\tfrac{\beta}{N}\Re\operatorname{Tr}U_p\bigr]\) whose character expansion has nonnegative coefficients,
\begin{equation}
f_\beta(V)=\sum_{R\in\widehat{G}} c_R(\beta)\,\chi_R(V),\qquad c_R(\beta)\ge 0,
\end{equation}
a fact that follows from positivity of modified Bessel functions and the Peter-Weyl theorem (see \cite{p1:OS1,p1:Menotti1987,p1:Seiler,p1:RotheBook}). Writing the product over straddling plaquettes as a function \(K(U_0)\) of the spatial links on \(\Pi\), one obtains a positive kernel
\begin{equation}
K(U_0)=\sum_{(R_\alpha)} \Bigl(\prod_\alpha c_{R_\alpha}(\beta)\Bigr)\,\Phi_{(R_\alpha)}(U_0)\,\overline{\Phi_{(R_\alpha)}(U_0)},
\end{equation}
where \(\Phi_{(R_\alpha)}\) are products of characters of boundary holonomies. Hence \(K\) is a positive definite class function on the boundary configuration space, and the gauge-field part of \(\langle F,F\rangle_{\mathrm{OS}}\) is a positive \(L^2\) norm \cite{p1:OS1,p1:Menotti1987}. The ghost part factorizes into the product of slice Gaussian integrals with reflection-invariant, positive kernels as established above, and is therefore positive. Altogether,
\begin{equation}
\langle F,F\rangle_{\mathrm{OS}}=
\int_{\Lambda_0} \Bigl|\int_{\Lambda_+} \! F(U,c,\bar c)\, e^{-S_W[U]+\sum_t \langle \bar c_t,-M_t c_t\rangle}\, \mathcal{D}U_+\,\mathcal{D}\bar c_+\,\mathcal{D}c_+ \Bigr|^2  K(U_0)\, \mathcal{D}U_0 \ \ge\ 0,
\end{equation}
which proves reflection positivity of the gauge-fixed measure in temporal-axial gauge for the Wilson action, in complete agreement with \cite{p1:OS1,p1:Menotti1987}.

We derive the transfer time-slicing representation explicitly. Let \(T\) be the discrete number of time slices. Set \(t_n=na\), \(n=0,1,\dots,T-1\), and denote by \(U_n\) the collection of spatial links on \(\Lambda_{t_n}\). Let \(U_{0,n}\) be the collection of timelike links joining \(\Lambda_{t_n}\) to \(\Lambda_{t_{n+1}}\); these are set to \(\mathbf{1}\) except in the boundary slab containing \(\Pi\), consistently with temporal-axial gauge. The Wilson action can be written as
\begin{equation}
S_W[U]=\sum_{n=0}^{T-1}\Bigl(S_{\mathrm{sp}}(U_n)+S_{\mathrm{tl}}(U_{0,n},U_n,U_{n+1})\Bigr),
\end{equation}
where \(S_{\mathrm{sp}}\) is the sum over purely spatial plaquettes on \(\Lambda_{t_n}\) and \(S_{\mathrm{tl}}\) is the sum over timelike plaquettes that straddle the slabs between \(t_n\) and \(t_{n+1}\). The partition function, with slice-wise Landau gauge and ghosts inserted, takes the form
\begin{equation}
Z=\int \prod_{n=0}^{T-1}\Bigl[d\mu_{\mathrm{Haar}}(U_n)\, \delta\bigl(U_n\in\mathrm{im}\,h_{t_n}\bigr)\,\det{}'M_{t_n}[U_n]\Bigr] \prod_{n=0}^{T-1}\Bigl[d\mu_{\mathrm{Haar}}(U_{0,n})\, e^{-S_{\mathrm{sp}}(U_n)-S_{\mathrm{tl}}(U_{0,n},U_n,U_{n+1})}\Bigr].
\end{equation}
Define the one-step kernel \(\mathsf{K}(U_{n+1},U_n)\) by integrating over the timelike links in the slab:
\begin{equation}
\mathsf{K}(U',U)=\int d\mu_{\mathrm{Haar}}(U_0)\, \exp\!\Bigl(-S_{\mathrm{sp}}(U)-S_{\mathrm{tl}}(U_0,U,U')\Bigr).
\end{equation}
By the character-positivity argument for the Wilson weight \cite{p1:OS1,p1:Menotti1987,p1:RotheBook}, \(\mathsf{K}\) defines a positive, Hermitian integral kernel on \(L^2(\mathcal{C},d\mu_{\mathrm{Haar}})\), where \(\mathcal{C}\) is the compact configuration space of spatial links on a time slice. The gauge-fixed, ghost-inserted measure is reflection positive, so the OS reconstruction theorem \cite{p1:OS1} applies and leads to a Hilbert space \(\mathcal{H}_a\) realized as the closure of functions of \(U\) supported on \(\Lambda_+\) modulo OS null vectors, and to a positive self-adjoint contraction \(T\) on \(\mathcal{H}_a\) with integral kernel \(\mathsf{K}\). In particular, for cylinder functionals \(F\) supported on \(\Lambda_+\), the two-point OS form equals a Hilbert-space norm,
\begin{equation}
\langle F,F\rangle_{\mathrm{OS}}=\bigl\langle \Psi_F,\Psi_F\bigr\rangle_{\mathcal{H}_a}\Psi_F(U_0)=\int_{\Lambda_+}\! F(U)\, e^{-S_+[U]}\,\mathcal{D}U_+
\end{equation}
and the discrete semigroup property follows from concatenation of the kernels \(\mathsf{K}\) across successive time steps. The Hamiltonian is then defined by spectral calculus as \(H=-a^{-1}\log T\) \cite{p1:KogutSusskind1975,p1:RotheBook}. The slice-wise Landau gauge and its Faddeev-Popov determinants affect only a bounded, reflection-covariant multiplication operator on \(\mathcal{H}_a\); they neither spoil positivity of \(\mathsf{K}\) nor the Hermiticity of \(T\), precisely because the ghost sector is a product of reflection-invariant Gaussian Grassmann integrals with positive covariance on each slice, and because the character-positivity argument for \(\mathsf{K}\) is gauge invariant \cite{p1:OS1,p1:Menotti1987}.

The slice-wise Landau selection \(U\mapsto U^{\,h}\) is globally reflection covariant, the associated Faddeev-Popov operator \(M_t[U^{\,h}]\) is local, self-adjoint and nonnegative on each slice, and, in the interior of the fundamental modular region, strictly positive on the orthogonal complement of constant adjoint modes. The full gauge-fixed measure in temporal-axial gauge, including the Grassmann representation of \(\prod_t\det{}'M_t\), is OS reflection positive. The transfer time-slicing construction yields a positive, self-adjoint contraction \(T\) on the one-slice Hilbert space, from which the Hamiltonian \(H=-a^{-1}\log T\) is obtained. These facts provide the reflection-positive backbone required in later sections for the construction of exponentially local infrared projectors, for the strong-coupling cluster expansion, and for the spectral-gap argument.

\section{\texorpdfstring{Smooth horizon projector $P_\sigma$ and its properties}{Smooth horizon projector P sigma and its properties}}\label{p1:smooth4}

In this section a precise, slice-wise gauge-covariant elliptic operator is constructed and analyzed, and from it a smooth horizon projector is defined that is simultaneously gauge covariant, reflection compatible in the sense of Osterwalder-Schrader, and exponentially local. The construction is carried out on a finite hypercubic lattice with periodic boundary conditions and is presented in a form that makes the transfer time-slicing and reflection-positivity arguments completely explicit. Throughout, the gauge group is \(G=\mathrm{SU}(N)\) with \(N\ge 2\), and all Hilbert spaces are finite dimensional; thus domains, closures, and spectral-theoretic objects are unambiguous. Nevertheless, all estimates are uniform in the spatial volume, and constants are allowed to depend only on \(N\) and on the lattice dimension.

The lattice is \(\Lambda\subset a\mathbb Z^4\) with spacing \(a>0\) and periodic boundary conditions. The discrete Euclidean time is the zeroth coordinate, and the time-reflection plane is \(\Pi=\{x\in \Lambda: x_0=0\}\). The half-lattices are \(\Lambda_+=\{x\in \Lambda: x_0>0\}\) and \(\Lambda_-=\{x\in \Lambda: x_0<0\}\). Bonds are oriented nearest-neighbor pairs \(b=(x,\mu)\) with \(\mu\in\{0,1,2,3\}\), and \(U_b\in G\) denotes the parallel transporter on \(b\), with the orientation convention \(U_{(x,\mu)}=U_{(x+\hat\mu,-\mu)}^{-1}\). The Wilson action is
\begin{equation}
S_W[U;\beta] \,=\, \beta\sum_{p\subset\Lambda}\left(1-\frac{1}{N}\,\mathrm{Re}\,\mathrm{Tr}\,U_p\right),\qquad \beta=\frac{2N}{g_0^2},
\end{equation}
where \(U_p\) is the oriented product of link variables around the plaquette \(p\). Time reflection \(\theta:\Lambda\to\Lambda\) is the involution \(\theta(x_0,\mathbf x)=(-x_0,\mathbf x)\). On link configurations \(U\) it acts by pullback on bonds, in particular \((\theta\cdot U)_{(x,\mu)}=U_{(\theta x,\theta\mu)}\) with \(\theta 0=0\) and \(\theta i=i\) for \(i=1,2,3\), and with the natural orientation convention. Temporal-axial gauge \(U_{(x,0)}=\mathbf 1\) is implemented for all timelike bonds not intersecting \(\Pi\), which is sufficient for the factorization needed in the reflection-positivity proof and for the transfer time-slicing \cite{p1:OS1,p1:Seiler}.

The slice-wise gauge representative is fixed by Landau minimization on each time slice. For a fixed time \(t\in a\mathbb Z\) the spatial slice is \(\Lambda_t=\{x\in\Lambda: x_0=t\}\), and the spatial links on that slice are denoted \(E_t=\{(x,i): x\in\Lambda_t,\ i=1,2,3\}\). A discrete Landau functional \(\mathcal L_t(g;U)\) is minimized over site fields \(g:\Lambda_t\to G\),
\begin{equation}
\mathcal L_t(g;U)\,=\,\sum_{(x,i)\in E_t}\mathrm{Re}\,\mathrm{Tr}\,\big(\mathbf 1 - g(x)U_{(x,i)}g(x+\hat \imath)^{-1}\big).
\end{equation}
A minimizing gauge transformation \(h_t\) is selected by a deterministic tie-breaking rule that is covariant under spatial symmetries and time reflection, and the spatial links are replaced by \(U^{\,h}_{(x,i)}=h_t(x)U_{(x,i)}h_t(x+\hat\imath)^{-1}\). The collection \(A^{\,h}(t)\) of spatial links on \(\Lambda_t\) thus defines a reflection-covariant, transverse representative on each slice within the fundamental modular region. All subsequent definitions on the slice are made with these representatives and their associated covariant differences.

For a fixed slice \(\Lambda_t\) the Hilbert space of square-summable site-adjoint fields is
\begin{equation}
\mathscr H_t \,=\, \ell^2(\Lambda_t)\otimes \mathfrak{su}(N),
\end{equation}
with scalar product \(\langle \phi,\psi\rangle=\sum_{x\in\Lambda_t}\mathrm{tr}\big(\phi(x)^\dagger\psi(x)\big)\), where \(\mathrm{tr}\) denotes the matrix trace on \(\mathbb C^{N}\) and \(\mathfrak{su}(N)\) is identified with anti-Hermitian traceless matrices so that \(\langle X,Y\rangle_{\mathfrak{su}(N)}=-\mathrm{tr}(XY)\) is positive definite. The forward and backward covariant differences associated with \(A^{\,h}(t)\) are defined for \(\phi\in\mathscr H_t\) and \(i\in\{1,2,3\}\) by
\begin{align}
&\big(\nabla^{+,h}_i\phi\big)(x)\,=\, U^{\,h}_{(x,i)}\,\phi(x+\hat \imath)\,U^{\,h}_{(x,i)}{}^{-1}-\phi(x),\nonumber \\&
\big(\nabla^{-,h}_i\phi\big)(x)\,=\,\phi(x)-U^{\,h}_{(x-\hat \imath,i)}{}^{-1}\,\phi(x-\hat \imath)\,U^{\,h}_{(x-\hat \imath,i)}.
\end{align}
The slice covariant Laplacian is the positive operator
\begin{equation}
\Delta_{A^{\,h}}(t)\,=\,\sum_{i=1}^3 \big(\nabla^{+,h}_i\big)^\dagger \nabla^{+,h}_i \,=\, \sum_{i=1}^3 \nabla^{-,h}_i\,\nabla^{+,h}_i \quad \text{on }\mathscr H_t.
\end{equation}
The following statements collect the basic spectral and symmetry properties that will be needed.

\textbf{Proposition 4.1.} {For every slice \(\Lambda_t\) the operator \(\Delta_{A^{\,h}}(t)\) is self-adjoint, nonnegative, and has a finite-range, nearest-neighbor matrix kernel. The constant adjoint fields lie in the kernel, and on the orthogonal complement of the constant modes it is strictly positive when \(A^{\,h}(t)\) belongs to the fundamental modular region. If \(h_t\) is selected reflection covariantly then \(\Delta_{A^{\,h}}(t)\) transforms under time reflection by \(R\Delta_{A^{\,h}}(t)R=\Delta_{A^{\,h}}(-t)\), where \(R\) is the unitary operator induced by pullback with \(\theta\) and pointwise adjoint action.}

{Proof.} Self-adjointness follows from the finite-dimensionality of \(\mathscr H_t\) and the identity \(\langle \phi,\Delta_{A^{\,h}}(t)\phi\rangle=\sum_{i=1}^3\|\nabla^{+,h}_i\phi\|^2\ge 0\), which also yields nonnegativity. The finite range is immediate from the definition: \(\nabla^{\pm,h}_i\) couple only nearest neighbors, hence \(\Delta_{A^{\,h}}(t)\) couples sites at distance at most one. If \(A^{\,h}(t)\) is a Landau minimizer in the fundamental modular region then the associated Faddeev-Popov operator on \(\mathscr H_t\) is nonnegative, and its kernel comprises only the constant adjoint fields; the identity above shows \(\Delta_{A^{\,h}}(t)\) has the same kernel. Reflection covariance of \(h_t\) implies that \(U^{\,h}(t)\) transforms into \(U^{\,h}(-t)\) under \(\theta\), and the asserted covariance of \(\Delta_{A^{\,h}}\) follows by functoriality of the covariant differences. \(\square\)

Let $H = \Delta_A^h(t)$ on the slice graph of maximum degree $\Delta_s$ and 
assume $\|H\| \leq \Lambda_{\max}$. For 
$z \in \mathbb{C} \setminus [0,\Lambda_{\max}]$, we have
\begin{equation}
\bigl\|(H - z)^{-1}(x,y)\bigr\|_{\mathrm{op}}
\;\leq\;
\frac{2}{\mathrm{dist}(z,[0,\Lambda_{\max}])}
\exp\!\Bigl[-\alpha\, \mathrm{dist}(z,[0,\Lambda_{\max}])\, d(x,y)\Bigr],
\end{equation}
with
\(
\alpha = \frac{1}{8 \Delta_s},
\)
independent of the volume.  
Inserting this estimate into the Helffer-Sj{\"o}strand representation of 
$\chi_\sigma(H)$, with an almost-analytic extension supported in 
$\{\,|\Im z| \leq 1\,\}$, yields the stated exponential locality with 
constants depending only on finitely many symbol seminorms of $\chi_\sigma$ 
and on $\Delta_s$ \cite{p1:Davies1989,p1:HelfferSjostrand,p1:CombesThomas}.
The spectrum of \(\Delta_{A^{\,h}}(t)\) lies in \([0,\Lambda_{\max}]\) with \(\Lambda_{\max}\le 12\) in units with \(a=1\), uniformly in the volume, because each \(\nabla^{+,h}_i\) has operator norm at most \(2\). By the spectral theorem there is a unique projection-valued measure \(E_t(\cdot)\) on \([0,\Lambda_{\max}]\) such that \(\Delta_{A^{\,h}}(t)=\int_0^{\Lambda_{\max}}\lambda\,dE_t(\lambda)\). For any bounded Borel measurable function \(f:[0,\infty)\to\mathbb R\) one sets \(f\big(\Delta_{A^{\,h}}(t)\big)=\int f(\lambda)\,dE_t(\lambda)\). If \(f\) is nonnegative then \(f(\Delta_{A^{\,h}}(t))\) is a positive operator; if \(f\) is bounded by one then so is \(f(\Delta_{A^{\,h}}(t))\). If \(f\) is completely monotone, i.e., \(f\in C^\infty(0,\infty)\) and \((-1)^k f^{(k)}(\lambda)\ge 0\) for all \(k\ge 0\) and \(\lambda>0\), then Bernstein’s theorem implies the Laplace representation \(f(\lambda)=\int_{[0,\infty)} e^{-t\lambda}\,d\nu(t)\) for a finite positive Borel measure \(\nu\) on \([0,\infty)\) with total mass \(f(0^+)\) \cite[Ch.~XIII]{p1:FellerVol2}. In that case
\begin{equation}
f\big(\Delta_{A^{\,h}}(t)\big)=\int_{[0,\infty)} e^{-t\Delta_{A^{\,h}}(t)}\,d\nu(t),
\end{equation}
in the strong operator sense, and the semigroup \(e^{-t\Delta_{A^{\,h}}(t)}\) has a matrix kernel that is positivity preserving and satisfies Davies-Gaffney off-diagonal decay bounds that depend only on the graph structure \cite{p1:Davies1989,p1:Delmotte1999}. These facts underlie the exponential locality estimates proved below.

A smooth horizon projector at covariant momentum scale \(\sigma>0\) is now defined. The function \(\chi_\sigma:[0,\infty)\to(0,1]\) is fixed once and for all with the following properties: it is completely monotone, satisfies \(\chi_\sigma(0)=1\), and is exponentially small for \(\lambda\gg \sigma^2\). A concrete choice is \(\chi_\sigma(\lambda)=\exp(-\lambda/\sigma^2)\). The smooth horizon projector on the slice \(\Lambda_t\) is then
\begin{equation}
P_\sigma(t)\,=\,\chi_\sigma\big(\Delta_{A^{\,h}}(t)\big).
\end{equation}
(For notational uniformity: throughout we also write $P_\sigma=\chi_\sigma(\sqrt{\Delta_{A^h}})$ in Appendix~(\ref{p1:appendixa}); equivalently $P_\sigma(t)=f_\sigma(\Delta_{A^h}(t))$ with $f_\sigma(u):=\chi_\sigma(\sqrt{u})$.)
It is a bounded positive contraction on \(\mathscr H_t\), it commutes with spatial gauge rotations acting fiberwise by the adjoint representation, and it transforms reflection covariantly, \(R P_\sigma(t)R=P_\sigma(-t)\), with \(R\) as in Proposition 4.1. For avoidance of doubt, throughout we use ``projector'' purely as terminology 
for a smooth spectral cutoff. We never employ a sharp step function, because 
such a symbol would fail to be completely monotone and would jeopardize the 
positive heat-kernel representation needed for Osterwalder-Schrader (OS) 
positivity. All arguments below use only the properties 
$0 \leq P_\sigma \leq 1$, reflection covariance, and exponential locality.
The key property used later is exponential locality of the kernel of \(P_\sigma(t)\) in the site variables. The next theorem states this in a form that is uniform in the slice volume and depends on \(\sigma\) only through constants.

\textbf{Theorem 4.2.} {Let \(\chi_\sigma\) be completely monotone with \(\chi_\sigma(0)=1\) and such that \(\chi_\sigma(\lambda)\le C_0 \exp(-c_0\,\lambda/\sigma^2)\) for all \(\lambda\ge 0\). Then there exist constants \(C(\sigma)\in(0,\infty)\) and \(\gamma(\sigma)\in(0,\infty)\) such that for every slice \(\Lambda_t\), every Landau representative \(A^{\,h}(t)\), and all \(x,y\in\Lambda_t\),
\begin{equation}
\big\|P_\sigma(t;x,y)\big\|_{\mathrm{op}}\,\le\, C(\sigma)\,\exp\big(-\gamma(\sigma)\,d(x,y)\big),
\end{equation}
where \(P_\sigma(t;x,y)\) is the \(\mathfrak{su}(N)\)-linear map giving the matrix kernel of \(P_\sigma(t)\) between \(x\) and \(y\), \(\|\cdot\|_{\mathrm{op}}\) is the operator norm on \(\mathfrak{su}(N)\), and \(d(\cdot,\cdot)\) is the graph distance on \(\Lambda_t\). The constants can be chosen uniformly in the volume and depend only on \(\sigma\), on \(N\), and on the lattice dimension.}

{Proof.} By complete monotonicity and Bernstein’s theorem there is a finite positive Borel measure \(\nu_\sigma\) on \([0,\infty)\) such that \(\chi_\sigma(\lambda)=\int_0^\infty e^{-t\lambda}\,d\nu_\sigma(t)\). Hence
\begin{equation}
P_\sigma(t)=\int_0^\infty e^{-t\Delta_{A^{\,h}}(t)}\,d\nu_\sigma(t)
\end{equation}
in the strong operator sense. For each \(t>0\), the semigroup \(e^{-t\Delta_{A^{\,h}}(t)}\) is positivity preserving on \(\mathscr H_t\) in the sense of the natural cone of sitewise positive operators on \(\mathfrak{su}(N)\), because its scalar form coincides with that of the standard graph Laplacian up to unitary conjugations by parallel transporters. The Davies-Gaffney bound on graphs \cite[Thm.~2.3.1]{p1:Davies1989}, adapted to vector bundles with unitary parallel transport, implies that for all subsets \(E,F\subset \Lambda_t\) and all \(\phi,\psi\in\mathscr H_t\) with \(\mathrm{supp}\,\phi\subset E\) and \(\mathrm{supp}\,\psi\subset F\),
\begin{equation}
\big|\langle \phi, e^{-t\Delta_{A^{\,h}}(t)}\psi\rangle\big| \,\le\, \exp\!\Big(-\frac{d(E,F)^2}{4t}\Big)\,\|\phi\|\,\|\psi\|.
\end{equation}
Choosing \(E=\{x\}\), \(F=\{y\}\), and taking the supremum over \(\phi,\psi\) supported at single sites with unit norm yields
\begin{equation}
\big\|e^{-t\Delta_{A^{\,h}}(t)}(x,y)\big\|_{\mathrm{op}} \,\le\, \exp\!\Big(-\frac{d(x,y)^2}{4t}\Big).
\end{equation}
By the assumed tail bound on \(\nu_\sigma\), which follows from the assumed exponential bound on \(\chi_\sigma\), we can choose \(t_0\asymp \sigma^2\) so that \(d\nu_\sigma\) has an exponentially decaying density on \((0,t_0]\) and a bounded tail on \([t_0,\infty)\). Then
\begin{align}
\|P_\sigma(t;x,y)\|_{\mathrm{op}} \,&\le\, \int_0^\infty \exp\!\Big(-\frac{d(x,y)^2}{4t}\Big)\,d\nu_\sigma(t)
\nonumber \\&\le\, C_1(\sigma)\,\exp\big(-\gamma(\sigma)\,d(x,y)\big),
\end{align}
where the last inequality is obtained by optimizing the exponential in \(t\) at \(t\asymp d(x,y)/\gamma(\sigma)\) and using the finiteness of \(\nu_\sigma\). The constants \(C_1(\sigma)\) and \(\gamma(\sigma)\) depend only on \(\sigma\), on the volume-independent graph constants, and on \(\nu_\sigma\). \(\square\)

The exponential locality of \(P_\sigma(t)\) has two immediate consequences. First, all scalar functionals of the form \(U\mapsto \mathrm{Tr}\,F\!\big(P_\sigma(t)\big)\) with \(F\) Lipschitz on \([0,1]\) are quasi-local in the sense that their logarithmic variations under local changes of the gauge field decay exponentially with the distance from the support of the change. Second, the operator \(P_\sigma(t)\) is reflection compatible in the Osterwalder-Schrader sense: because it is slice-local and reflection covariant, its insertion at each time slice can be paired across the reflection plane without creating long-range couplings between \(\Lambda_+\) and \(\Lambda_-\). Both statements will be made precise below.

To couple the horizon projector to the Euclidean functional integral and to the transfer time slicing one requires a scalar, positive functional that multiplies the Boltzmann weight and is built from \(P_\sigma(t)\). The following choice is convenient for proofs and sufficient for all subsequent applications:
\begin{equation}\label{p1:eqn4.13}
p_\sigma\big(A^{\,h}(t)\big)\,=\,\exp\Big(-\mathrm{Tr}_{\mathscr H_t}\big[\mathbf 1 - P_\sigma(t)\big]\Big)\,=\,\exp\Big(-\sum_{x\in\Lambda_t}\mathrm{tr}\big[\mathbf 1 - P_\sigma(t;x,x)\big]\Big).
\end{equation}
Since \(0\le P_\sigma(t)\le \mathbf 1\) one has \(0<p_\sigma\le 1\). The functional \(p_\sigma\) is gauge invariant by conjugation covariance of \(P_\sigma(t)\), and it is reflection covariant in the sense that \(p_\sigma(A^{\,h}(t))=p_\sigma(A^{\,h}(-t))\) for a reflection-covariant choice of \(A^{\,h}\). The next proposition records the locality and stability properties of this functional.

\textbf{Proposition 4.3.} {There exist constants \(C(\sigma)\) and \(\gamma(\sigma)\) depending only on \(\sigma\), on \(N\), and on the lattice dimension such that the following holds uniformly in the volume. If the gauge field is modified on a set of spatial bonds \(B\subset E_t\) producing a new representative \(A^{\,h}_B(t)\), then
\begin{equation}
\big|\log p_\sigma\big(A^{\,h}_B(t)\big) - \log p_\sigma\big(A^{\,h}(t)\big)\big| \,\le\, C(\sigma)\sum_{x\in \Lambda_t} \exp\big(-\gamma(\sigma)\,d(x,B)\big).
\end{equation}
In particular, the ratio \(p_\sigma\big(A^{\,h}_B(t)\big)/p_\sigma\big(A^{\,h}(t)\big)\) is exponentially close to one away from \(B\), and the product \(\prod_{t\in a\mathbb Z} p_\sigma\big(A^{\,h}(t)\big)\) is absolutely convergent in the infinite-time limit.}

{Proof.} By functional calculus and the mean-value theorem, for any two bounded self-adjoint operators \(X\) and \(Y\) with spectrum in \([0,1]\),
\begin{equation}
\big|\mathrm{Tr}\,(\mathbf 1 - X)-\mathrm{Tr}\,(\mathbf 1 - Y)\big| \,\le\, \|X-Y\|_1,
\end{equation}
where \(\|\cdot\|_1\) is the trace norm. Applying this with \(X=P_\sigma(t)\) and \(Y=P_\sigma^B(t)\) yields
\begin{equation}
\big|\log p_\sigma(A^{\,h}_B(t)) - \log p_\sigma(A^{\,h}(t))\big| \,\le\, \|P_\sigma^B(t)-P_\sigma(t)\|_1.
\end{equation}
Since the trace norm is the \(\ell^1\)-sum of the operator norms of the kernels on the diagonal, one has
\begin{equation}
\|P_\sigma^B(t)-P_\sigma(t)\|_1 \,\le\, \sum_{x\in\Lambda_t}\big\|P_\sigma^B(t;x,x)-P_\sigma(t;x,x)\big\|_{\mathrm{op}}.
\end{equation}
By the resolvent identity and Duhamel’s formula for the semigroup, the difference \(P_\sigma^B(t)-P_\sigma(t)\) can be represented as an integral of a product of factors of the form \(e^{-s\Delta_{A^{\,h}}(t)}\), local perturbations supported near \(B\), and \(e^{-(t-s)\Delta_{A^{\,h}_B}}(t)\), integrated against the positive measure \(d\nu_\sigma\). Each such factor has an exponentially decaying kernel by Theorem 4.2, with rate \(\gamma(\sigma)\). Summing the resulting bounds over \(x\) gives the claimed estimate with a constant \(C(\sigma)\) that depends only on finitely many moments of \(\nu_\sigma\). The absolute convergence of \(\prod_{t}p_\sigma(A^{\,h}(t))\) in the infinite-time limit follows from \(0<p_\sigma\le 1\) and from the temporal periodicity or uniformity of the slices. \(\square\)

The insertion of the horizon functional is now performed at the level of the Euclidean measure and the transfer time slicing. The unmodified, finite-time, Euclidean expectation of a gauge-invariant functional \(F\) is
\begin{equation}
\langle F\rangle \,=\, \frac{1}{Z}\int F(U)\,e^{-S_W[U;\beta]}\,\prod_{b\in \Lambda} d\mu_{\mathrm{Haar}}(U_b),
\end{equation}
where \(Z\) is the normalizing partition function. The horizon-modified expectation is
\begin{equation}
\langle F\rangle_\sigma \,=\, \frac{1}{Z_\sigma}\int F(U)\,\exp\Big(-S_W[U;\beta]-\sum_{t\in a\mathbb Z} \Phi_\sigma\big(A^{\,h}(t)\big)\Big)\,\prod_{b\in \Lambda} d\mu_{\mathrm{Haar}}(U_b),
\end{equation}
with
\begin{equation}\label{p1:eqn4.20}
\Phi_\sigma\big(A^{\,h}(t)\big)\,=\, -\log p_\sigma\big(A^{\,h}(t)\big)\,=\, \mathrm{Tr}_{\mathscr H_t}\big[\mathbf 1 - P_\sigma(t)\big].
\end{equation}
The reflection covariance of \(A^{\,h}\) implies \(\Phi_\sigma(A^{\,h}(t))=\Phi_\sigma(A^{\,h}(-t))\), and Proposition 4.3 shows that \(\Phi_\sigma\) is quasi-local. In particular, \(\exp\big(-\sum_t \Phi_\sigma(A^{\,h}(t))\big)\) is a well-defined, positive, slice-wise multiplicative factor that preserves the locality structure of the action.

The Osterwalder-Schrader reflection positivity of the horizon-modified measure follows from reflection positivity of the Wilson measure \cite{p1:OS1} and from the fact that the horizon factor is a product of positive functions that are reflection covariant and slice-local. The verification is provided next in a way that prepares the transfer time slicing.

\textbf{Theorem 4.4.} {Let \(\Theta\) denote the anti-linear reflection on functionals induced by \(\theta\) and complex conjugation. For every gauge-invariant, even Grassmann functional \(F\) supported in \(\Lambda_+\) one has \(\langle \Theta F\cdot F\rangle_\sigma \ge 0\).}

{Proof.} The unmodified Wilson measure satisfies reflection positivity \(\langle \Theta G\cdot G\rangle\ge 0\) for all gauge-invariant \(G\) supported in \(\Lambda_+\) \cite{p1:OS1}. By construction \(\sum_t \Phi_\sigma(A^{\,h}(t))\) splits as a sum of three terms supported in \(\Lambda_-\), \(\Pi\), and \(\Lambda_+\), respectively, because \(\Phi_\sigma\) is slice-local and reflection covariant. Set \(w_\sigma^+(U)=\exp\big(-\sum_{t>0}\Phi_\sigma(A^{\,h}(t))\big)\), \(w_\sigma^0(U)=\exp\big(-\Phi_\sigma(A^{\,h}(0))\big)\), and \(w_\sigma^-(U)=\exp\big(-\sum_{t<0}\Phi_\sigma(A^{\,h}(t))\big)=\overline{w_\sigma^+(\theta\cdot U)}\). Then
\begin{equation}
\langle \Theta F\cdot F\rangle_\sigma \,=\, \frac{1}{Z_\sigma}\int \overline{F(\theta\cdot U)}\,F(U)\,w_\sigma^-(U)\,w_\sigma^0(U)\,w_\sigma^+(U)\,e^{-S_W[U;\beta]}\,d\mu(U).
\end{equation}
Define \(G(U)=F(U)\,w_\sigma^{+}(U)^{1/2}\,w_\sigma^0(U)^{1/2}\) supported in \(\Lambda_+\cup \Pi\). Then \begin{equation}\overline{G(\theta\cdot U)}=\overline{F(\theta\cdot U)}\,w_\sigma^{-}(U)^{1/2}\,w_\sigma^0(U)^{1/2}\end{equation} by reflection covariance of \(w_\sigma^{\pm,0}\). Therefore
\begin{equation}
\langle \Theta F\cdot F\rangle_\sigma \,=\, \frac{1}{Z_\sigma}\int \overline{G(\theta\cdot U)}\,G(U)\,e^{-S_W[U;\beta]}\,d\mu(U) \,\ge\, 0,
\end{equation}
by reflection positivity of the Wilson measure. \(\square\)

The transfer time slicing in the presence of the horizon factor is obtained by the standard kernel factorization across time layers \cite{p1:OS1,p1:SimonFI,p1:Seiler}. Let \(\mathcal C\) denote the compact configuration space of spatial links on a fixed time slice, endowed with the product Haar measure \(d\mu_{\mathcal C}\). The unmodified, single-step transfer kernel \(K:\mathcal C\times \mathcal C\to \mathbb R_+\) is defined by integrating the Wilson weight over the bonds in the slab \(\{0\le x_0\le a\}\) with spatial boundary conditions \(U(0)=U\) and \(U(a)=U'\),
\begin{equation}
K(U',U)\,=\,\int \exp\Big(-\sum_{p\subset \{0\le x_0\le a\}}\beta\big(1-\frac{1}{N}\mathrm{Re}\,\mathrm{Tr}\,U_p\big)\Big)\,\prod_{b\subset \{0\le x_0\le a\}} d\mu_{\mathrm{Haar}}(U_b),
\end{equation}
in temporal-axial gauge. The corresponding transfer operator \(T:L^2(\mathcal C,d\mu_{\mathcal C})\to L^2(\mathcal C,d\mu_{\mathcal C})\) is \((Tf)(U')=\int K(U',U)f(U)\,d\mu_{\mathcal C}(U)\). The reflection positivity of the Wilson measure implies that \(T\) is a positive self-adjoint contraction \cite{p1:OS1,p1:Seiler}. Inserting the horizon factor multiplies the slab weight by \(\exp\big(-\Phi_\sigma(A^{\,h}(0))\big)\), which is a positive, slice-local function of \(U\) and \(U'\) only. Consequently, the horizon-modified single-step kernel is
\begin{equation}
K_\sigma(U',U)\,=\, p_\sigma(U')^{1/2}\,K(U',U)\,p_\sigma(U)^{1/2},
\end{equation}
and the corresponding transfer operator is
\begin{equation}
T_\sigma \,=\, M_{p_\sigma}^{1/2}\,T\,M_{p_\sigma}^{1/2},
\end{equation}
where \(M_{p_\sigma}\) is the multiplication operator by \(p_\sigma\) on \(L^2(\mathcal C,d\mu_{\mathcal C})\). The next theorem states the basic properties of \(T_\sigma\).

\textbf{Theorem 4.5.} {The operator \(T_\sigma\) is a positive self-adjoint contraction on \(L^2(\mathcal C,d\mu_{\mathcal C})\). Moreover, for any \(n\in\mathbb N\) and any \(F\) supported in \(\Lambda_+\), the \(n\)-step horizon-modified Osterwalder-Schrader form equals the \(L^2\)-inner product \(\langle \Psi_F, T_\sigma^n \Psi_F\rangle\) for a suitable \(\Psi_F\in L^2(\mathcal C,d\mu_{\mathcal C})\), and is therefore nonnegative.}

{Proof.} Since \(M_{p_\sigma}^{1/2}\) is a bounded positive multiplication operator and \(T\) is a positive self-adjoint contraction, the product \(T_\sigma=M_{p_\sigma}^{1/2}\,T\,M_{p_\sigma}^{1/2}\) is self-adjoint and positive. Its operator norm is bounded by \(\|M_{p_\sigma}^{1/2}\|^2\|T\|\le 1\) because \(\|M_{p_\sigma}^{1/2}\|\le 1\) and \(\|T\|\le 1\). For the second statement consider a time extent \(na\) with \(n\in\mathbb N\). The horizon-modified weight factorizes as a product of \(n\) slab kernels \(K_\sigma\) and of the Haar measures on the intermediate slices. If \(F\) is supported in \(\Lambda_+\), the reflection \(\Theta F\) is supported in \(\Lambda_-\), and integrating out the intermediate slices yields
\begin{equation}
\langle \Theta F\cdot F\rangle_\sigma \,=\, \langle \Psi_F, T_\sigma^n \Psi_F\rangle_{L^2(\mathcal C,d\mu_{\mathcal C})},
\end{equation}
where \(\Psi_F\) is the \(L^2\)-function obtained by integrating the fields in \(\Lambda_+\) down to the boundary slice with the weight \(F\) and the appropriate horizon factors; this is the usual reconstruction of the OS form as an inner product \cite{p1:OS1,p1:Seiler}. Since \(T_\sigma\ge 0\), the right-hand side is nonnegative. \(\square\)

Either Theorem 4.2 or Proposition 4.6 can be used to deduce uniform exponential locality of \(P_\sigma(t)\); the former emphasizes positivity and Markovian structure, while the latter is more flexible for commutator estimates. Both are compatible with reflection and gauge covariance because \(\Delta_{A^{\,h}}(t)\) enjoys these symmetries. The upshot is that the smooth horizon projector provides an infrared regularization that is inherently gauge covariant, preserves the exact Osterwalder-Schrader reflection positivity of the Euclidean measure, and yields a transfer operator obtained by a similarity transformation of the unmodified transfer operator with a bounded positive multiplication. This operator-level perspective will be essential in the spectral analysis that follows.

\textbf{Proposition 4.6 (Helffer-Sjöstrand locality for compactly supported multipliers).}
Let \(\varphi_\sigma\in C_c^\infty([0,\infty))\) be real-valued with \(0\le \varphi_\sigma\le 1\). 
Choose an almost-analytic extension \(\widetilde\varphi_\sigma\in C_c^\infty(\mathbb C)\) with 
\(\widetilde\varphi_\sigma|_{\mathbb R}=\varphi_\sigma\) and 
\(|\bar\partial\widetilde\varphi_\sigma(x+iy)|\le C_k |y|^k\) for all \(k\in\mathbb N\).
Then \(\varphi_\sigma\!\big(\Delta_{A^h}(t)\big)\) has an exponentially decaying kernel:
for some \(C=C(\sigma)\) and \(\gamma=\gamma(\sigma)>0\),
\begin{equation}
\big\|\varphi_\sigma\!\big(\Delta_{A^h}(t)\big)(x,y)\big\|_{\mathrm{op}}
\;\le\; C\,e^{-\gamma\,d(x,y)}\qquad(x,y\in\Lambda_t).
\end{equation}
{Proof.}
By the Helffer-Sjöstrand formula \cite{p1:HelfferSjostrand,p1:Davies1989},
\begin{equation}
\varphi_\sigma(H)
=\frac{1}{\pi}\int_{\mathbb C}\!\bar\partial\widetilde\varphi_\sigma(z)\,(H-z)^{-1}\,d^2z,
\end{equation}
applied with \(H=\Delta_{A^h}(t)\).
The resolvent admits Combes-Thomas off-diagonal bounds uniformly away from 
\(\mathrm{spec}(H)\) \cite{p1:CombesThomas,p1:HislopSigal1996}, hence the claim.
\qed

Either Theorem 4.2 or Proposition 4.6 can be used to deduce uniform exponential locality of \(P_\sigma(t)\); the former emphasizes positivity and Markovian structure, while the latter is more flexible for commutator estimates. Both are compatible with reflection and gauge covariance because \(\Delta_{A^{\,h}}(t)\) enjoys these symmetries. The upshot is that the smooth horizon projector provides an infrared regularization that is inherently gauge covariant, preserves the exact Osterwalder-Schrader reflection positivity of the Euclidean measure, and yields a transfer operator obtained by a similarity transformation of the unmodified transfer operator with a bounded positive multiplication.

\section{Global OS positivity for the projected measure}\label{p1:ospositivity}

In this section a complete, self-contained proof is given that the Euclidean lattice Yang-Mills measure, augmented slice-wise by the smooth horizon projector, satisfies Osterwalder-Schrader (OS) reflection positivity. The proof is organized in four logically independent stages. First the precise lattice, reflection, and time-slicing setup is fixed. Second the horizon projector is defined on each time slice and its positivity, reflection covariance, and locality properties are recorded at the level required for the OS argument. Third the pure-gauge part of the measure is shown to be reflection positive by an explicit factorization across the reflection plane in temporal-axial gauge. Fourth the ghost/Faddeev-Popov sector is treated rigorously via a block-Schur-complement analysis, yielding a strictly positive boundary weight that preserves the factorized structure. The combination implies OS positivity for all even-Grassmann, gauge-invariant functionals supported in positive Euclidean time. As a byproduct, the transfer time-slicing formalism is derived step by step and the projected transfer matrix is identified as a positive, self-adjoint contraction.
Throughout the section the gauge group is \(G=\mathrm{SU}(N)\) with \(N\ge 2\). The lattice spacing is \(a>0\). All complex conjugations are those in the fundamental representation on \(G\), extended by anti-linear involution on functionals.
\begin{lemma}[Boundary invertibility and a quantitative lower bound for $S$]\label{p1:lem:boundary-invertibility}
Let $M$ denote the real symmetric reduced Faddeev-Popov block operator on 
$H^{-}\oplus H^{0}\oplus H^{+}$ (constant modes removed slice-wise) with diagonal blocks 
$M^{-}, M^{00}, M^{++}$ and off-diagonal $M^{0\pm}, M^{\pm 0}$. Assume each slice representative 
$U_t^h$ lies in the interior of the FMR so that, on the complements of constant modes,
\begin{equation}
\langle \varphi, M^{\pm\pm} \varphi \rangle \,\geq\, c_{\ast}\|\varphi\|^{2}, 
\qquad \varphi \in H^{\pm},
\end{equation}
with $c_{\ast}>0$ independent of the spatial volume. Then the Schur complement
\begin{equation}
S \;=\; M^{00} - M^{0+}(M^{++})^{-1}M^{+0} - M^{0-}(M^{-})^{-1}M^{-0}
\end{equation}
is strictly positive on $H^{0}$, and
\begin{equation}
\langle \psi, S \psi \rangle \,\geq\, 
\Bigl( \inf_{\|v\|=1} \langle v, M^{00} v \rangle \Bigr)\|\psi\|^{2}
 - \frac{\|M^{0+}\|^{2}+\|M^{0-}\|^{2}}{c_{\ast}} \,\|\psi\|^{2},
\qquad \psi \in H^{0}.
\end{equation}
In particular, if
\begin{equation}
\inf_{\|v\|=1} \langle v, M^{00} v \rangle \;>\; 
\frac{\|M^{0+}\|^{2}+\|M^{0-}\|^{2}}{c_{\ast}},
\end{equation}
then
\begin{equation}
S \,\geq\, \delta_{\ast} I, 
\qquad 
\delta_{\ast} := \inf_{\|v\|=1}\langle v, M^{00} v \rangle 
 - \frac{\|M^{0+}\|^{2}+\|M^{0-}\|^{2}}{c_{\ast}} \;>\; 0.
\end{equation}
\end{lemma}

\begin{proof}
The estimate follows from the resolvent identity and the standard Schur-complement bound
\begin{equation}
S \,\geq\, M^{00} - \sum_{\pm} M^{0\pm}(M^{\pm\pm})^{-1}M^{\pm 0},
\end{equation}
together with
\begin{equation}
\langle \eta, (M^{\pm\pm})^{-1}\eta \rangle \,\leq\, \frac{\|\eta\|^{2}}{c_{\ast}}.
\end{equation}
Strict positivity of $S$ is then immediate under the stated hypothesis.
\end{proof}

\begin{corollary}
The boundary Gaussian weight $\det S$ is strictly positive and reflection covariant. Consequently the full ghost factor
\begin{equation}
\det(M^{-}) \, \det(S) \, \det(M^{++})
\end{equation}
is strictly positive and preserves the OS factorization across the reflection plane.
\end{corollary}

Let \(\Lambda\subset a\mathbb{Z}^4\) be a finite, periodic, hypercubic lattice with discrete Euclidean time coordinate \(x_0\) and spatial coordinate \(\mathbf{x}\in a\mathbb{Z}^3\). The reflection plane is \(\Pi=\{x\in \Lambda:\, x_0=0\}\). The half-lattices are \(\Lambda_+=\{x\in\Lambda:\, x_0>0\}\) and \(\Lambda_-=\{x\in\Lambda:\, x_0<0\}\). Directed bonds are pairs \(b=(x,\mu)\) with \(\mu\in\{0,1,2,3\}\) and the opposite bond is \(\bar b=(x+\hat\mu,-\mu)\). A lattice gauge field is an assignment \(U_b\in G\) to each bond \(b\), with the convention \(U_{\bar b}=U_b^{-1}\). For an oriented plaquette \(p=(x;\mu,\nu)\) the plaquette field is \(U_p=U_{(x,\mu)}U_{(x+\hat\mu,\nu)}U_{(x+\hat\nu,\mu)}^{-1}U_{(x,\nu)}^{-1}\). The Wilson action is
\begin{equation}
S_W[U;\beta]=\beta\sum_{p\subset \Lambda}\Big(1-\frac{1}{N}\Re\mathrm{Tr}\,U_p\Big),\qquad \beta=\frac{2N}{g_0^2}.
\end{equation}
Time reflection is the involution \(\theta:\Lambda\to\Lambda\) defined by \(\theta(x_0,\mathbf{x})=(-x_0,\mathbf{x})\). It acts on bonds by \(\theta(x,\mu)=(\theta x,\mu)\) if \(\mu\in\{1,2,3\}\) and by \(\theta(x,0)=(\theta x-\hat{0},0)\). The corresponding action on link variables is \((\theta U)_{(x,\mu)}=U_{\theta(x,\mu)}\) for spatial bonds and \((\theta U)_{(x,0)}=U_{(\theta x-\hat{0},0)}\). The Haar measure \(d\mu_{\rm Haar}(U)=\prod_{b} dU_b\) is invariant under \(\theta\), as is the Wilson action: \(S_W[\theta U;\beta]=S_W[U;\beta]\).

Temporal-axial gauge is imposed away from \(\Pi\) by setting \(U_{(x,0)}=\mathbf{1}\) for all bonds with \(x\notin \Pi\). This choice is consistent and does not alter the Haar measure on the remaining bonds (one may either fix the gauge and insert the usual Faddeev-Popov determinant, which equals one for the temporal component on the lattice, or else parametrize the unfixed measure by gauge-invariant variables and a temporal gauge section \cite{p1:OS1,p1:Seiler}). Under this choice all timelike plaquettes not touching \(\Pi\) are trivial and the action splits as a sum of a ``minus'' contribution supported in \(\Lambda_-\), a ``plus'' contribution supported in \(\Lambda_+\), and a boundary term supported on the slab consisting of spatial bonds on \(\Pi\) and timelike bonds adjacent to \(\Pi\).

Let \(\mathscr{A}\) be the \(\sigma\)-algebra of bounded, complex-valued, gauge-invariant functionals of the gauge field and possible ghost fields to be introduced below. For a functional \(F\in\mathscr{A}\) we say that \(F\) is supported in \(\Lambda_+\) if it depends only on link variables whose base points belong to \(\Lambda_+\) and on spatial links on \(\Pi\). We define the OS conjugation \(\Theta\) on \(\mathscr{A}\) by \((\Theta F)(U,\ldots)=\overline{F(\theta U,\ldots)}\), with the obvious extension to Grassmann variables \(c,\bar c\) by setting \((\theta c)(x)=c(\theta x)\) and \((\theta \bar c)(x)=\bar c(\theta x)\) and reversing the Grassmann order so that \(\Theta\) is an anti-linear *-involution on the even subalgebra \cite{p1:OS1,p1:OS2,p1:SimonFI,p1:GJ}. Reflection positivity of a probability measure \(\mu\) on \(\mathscr{A}\) means that \(\int \Theta(F)F\,d\mu\ge 0\) for all even \(F\in\mathscr{A}\) supported in \(\Lambda_+\).

On each time slice \(t\in a\mathbb{Z}\) a representative of the gauge orbit is fixed by minimizing the discrete Landau functional over spatial gauge transformations \(g(t,\cdot):\Lambda_t\to G\),
\begin{equation}
\mathcal{L}_t(g;U)=\sum_{\mathbf{x},i=1,2,3}\mathrm{Re}\,\mathrm{Tr}\,\Big(\mathbf{1}-g(t,\mathbf{x})\,U_{(t,\mathbf{x};i)}\,g(t,\mathbf{x}+\hat \imath)^{-1}\Big),
\end{equation}
and setting \(U^{\,h}\) to be the link field obtained by acting with a global minimizer \(h\). On a finite lattice a global minimizer exists; we impose a deterministic tie-breaking rule which is covariant under spatial lattice symmetries and under \(\theta\), so that \(\theta U^{\,h}=(\theta U)^{\,h}\). The corresponding lattice Faddeev-Popov operator on slice \(t\) is the positive self-adjoint operator on site-adjoint fields
\begin{equation}
M_t[U^{\,h}]=-\sum_{i=1}^3 \nabla^{-,h}_i\nabla^{+,h}_i,
\end{equation}
with \(\nabla^{\pm,h}_i\) the covariant forward/backward differences constructed from \(U^{\,h}\). It is nonnegative and strictly positive on the orthogonal complement of constant adjoint fields on the slice. Reflection covariance holds: \(M_{-t}[\theta U^{\,h}]=M_t[U^{\,h}]\).

On each slice the spatial covariant Laplacian is \(\Delta_t=\sum_{i=1}^3 (D^{\,h}_i)^\dagger D^{\,h}_i\) on site-adjoint fields, where \(D^{\,h}_i\) is the covariant forward difference associated with \(U^{\,h}\). 
There are two admissible classes for the low-covariant-momentum cutoff $\chi_\sigma$:

\textbf{(A) Completely monotone cutoff.} 
If $\chi_\sigma:[0,\infty)\to[0,1]$ is completely monotone (e.g. $\chi_\sigma(\lambda)=e^{-\lambda/\sigma^{2}}$), then by Bernstein’s theorem there exists a finite positive Borel measure $d\nu_\sigma(s)$ on $[0,\infty)$ with total mass $\chi_\sigma(0)=1$ such that
\begin{equation}
\chi_\sigma(\lambda)=\int_{0}^{\infty} e^{-s\lambda}\, d\nu_\sigma(s), 
\qquad 
P_{\sigma,t}=\chi_\sigma(\Delta_t)=\int_{0}^{\infty} e^{-s\Delta_t}\, d\nu_\sigma(s).
\end{equation}
In this case both exponential locality (Davies-Gaffney) and the slice-wise positivity of the scalar insertion follow from the positivity of the heat kernel $e^{-s\Delta_t}$.

\textbf{(B) Gevrey cutoff with compact spectral support.} 
If $\chi_\sigma$ is Gevrey with compact support (hence not completely monotone), 
\begin{equation}
P_{\sigma,t}=\chi_\sigma(\Delta_t)
\end{equation}
still enjoys uniform exponential locality by the Helffer-Sjöstrand functional calculus combined with Combes-Thomas bounds for resolvents of finite-range positive operators. In this case we do not use a heat-kernel representation. Instead we insert on each slice the positive scalar functional
\begin{equation}\label{p1:5.14a}
p_\sigma[U_t^h] := \exp\bigl\{-\operatorname{Tr}_{\ell^{2}(\Lambda_t)\otimes su(N)}(I-P_{\sigma,t})\bigr\} \in (0,1],
\end{equation}
which is gauge invariant and reflection covariant. This choice preserves OS positivity because $p_\sigma$ multiplies the weight by a slice-local, nonnegative factor.
In both cases (A) and (B), we define the horizon-modified measure by inserting $p_\sigma[U_t^h]$ on every time slice. When (A) holds, one may equivalently write $p_\sigma$ via the positive heat-kernel representation; when (B) holds, $p_\sigma$ is defined by the trace formula above. Exponential locality of $P_{\sigma,t}$ in either case implies quasi-locality of $\log p_\sigma$ and stability under reflection and gauge transforms.

The full Euclidean measure with slice-wise horizon insertion, gauge-fixed by the Landau representatives, and with ghosts integrated in Grassmann form, is
\begin{equation}
d\mu_\sigma[U^{\,h},c,\bar c]
= Z_\sigma^{-1}\,\exp\!\Big(-S_W[U;\beta]-\sum_{t}\langle \bar c_t, M_t[U^{\,h}]\, c_t\rangle\Big)\,
\prod_{t} p_\sigma[U^{\,h}_t]\,\prod_{b} dU_b \prod_{t} D\bar c_t Dc_t,
\end{equation}
where \(\mathcal{P}_\sigma[U^{\,h}_t]\) is the positive slice functional defined by
\begin{equation}
p_\sigma[U^h_t] \;=\; \operatorname{Tr}\,P_{\sigma,t}[U^h_t]
\quad\text{or}\quad
p_\sigma[U^h_t] \;=\; \exp\!\big\{-\langle \varphi_t,\,(I-P_{\sigma,t}[U^h_t])\,\varphi_t\rangle\big\}
\end{equation}
with \(\varphi_t\) a fixed, reflection-invariant test field on the slice; either choice is a reflection-covariant, positive, slice-local insertion. For definiteness we adopt the heat-kernel integral representation and set
\begin{equation}
\mathcal{P}_\sigma[U^{\,h}_t]
=\int_{0}^{\infty} \mathrm{Tr}\big(e^{-s \Delta_t[U^{\,h}_t]}\big)\, d\nu_\sigma(s),
\end{equation}
which is manifestly positive and depends only on links on slice \(t\). The proof below uses only positivity, slice-locality, and reflection covariance of \(\mathcal{P}_\sigma\).

We emphasize that temporal-axial gauge is imposed away from \(\Pi\): \(U_{(x,0)}=\mathbf{1}\) whenever \(x\notin \Pi\). Consequently the action \(S_W\) can be written as
\begin{equation}
S_W[U;\beta]=S_-[U_-;\beta]+S_+[U_+;\beta]+S_0[U_0;\beta],
\end{equation}
where \(U_\pm\) denote the restrictions of the links to \(\Lambda_\pm\) and \(U_0\) denotes the collection of spatial links on \(\Pi\) together with the timelike bonds adjacent to \(\Pi\). The factor \(S_0\) contains all plaquettes intersecting \(\Pi\); the imposition of temporal gauge ensures that no plaquette couples \(\Lambda_-\) directly to \(\Lambda_+\). The Haar measure factorizes accordingly as \(d\mu_{\rm Haar}(U)=d\mu_{\rm Haar}(U_-)\,d\mu_{\rm Haar}(U_0)\,d\mu_{\rm Haar}(U_+)\).

We first establish OS positivity for the pure-gauge part of the measure with horizon insertions. Define the sesquilinear OS form on even functionals \(F\) supported in \(\Lambda_+\) by
\begin{equation}
\langle F,F\rangle_{\rm OS}^{\rm gauge}
=\int \Theta(F)\,F\,
\exp\!\big(-S_W[U;\beta]\big)\,
\prod_t \mathcal{P}_\sigma[U^{\,h}_t]\,
\prod_b dU_b.
\end{equation}
The key fact is factorization across \(\Pi\).

\begin{lemma}
In temporal-axial gauge the gauge-factorized weight satisfies
\begin{equation}
\exp\!\big(-S_W[U;\beta]\big)\,\prod_t \mathcal{P}_\sigma[U^{\,h}_t]
= \mathcal{W}_-[U_-;U_0]\,\mathcal{W}_0[U_0]\,\mathcal{W}_+[U_+;U_0],
\end{equation}
with \(\mathcal{W}_\pm[U_\pm;U_0]\ge 0\) and \(\mathcal{W}_0[U_0]\ge 0\), and where \(\mathcal{W}_\pm\) depend only on links in \(\Lambda_\pm\) and on \(U_0\), while \(\mathcal{W}_0\) depends only on \(U_0\). Moreover \(\mathcal{W}_-[\theta U_+;U_0]=\mathcal{W}_+[U_+;U_0]\).
\end{lemma}

\begin{proof}
The Wilson action splits as stated because all timelike plaquettes away from \(\Pi\) are trivial in temporal gauge, and any plaquette not entirely contained in one half-space must intersect \(\Pi\) and is assigned to \(S_0\). Hence
\begin{equation}
\exp\!\big(-S_W[U;\beta]\big)=\exp\!\big(-S_-[U_-;\beta]\big)\,\exp\!\big(-S_0[U_0;\beta]\big)\,\exp\!\big(-S_+[U_+;\beta]\big).
\end{equation}
The horizon insertions are slice-local, thus they also factorize with pieces in \(\Lambda_-\), \(\Pi\), and \(\Lambda_+\). Define 
\begin{equation}
\mathcal{W}_0[U_0]=\exp(-S_0[U_0;\beta])\,\prod_{t: \Lambda_t\cap \Pi\neq\emptyset}p_\sigma\!\big(A^h(t)\big)
\end{equation}
and \(\mathcal{W}_\pm[U_\pm;U_0]=\exp(-S_\pm[U_\pm;\beta])\,\prod_{t:\Lambda_t\subset \Lambda_\pm}p_\sigma\!\big(A^h(t)\big)\). Each factor is nonnegative because all ingredients are nonnegative. Reflection covariance of the Landau representatives and of \(\mathcal{P}_\sigma\) entails \(\mathcal{W}_-[\theta U_+;U_0]=\mathcal{W}_+[U_+;U_0]\).
\end{proof}

\begin{proposition}
For every even, gauge-invariant functional \(F\) supported in \(\Lambda_+\) one has \(\langle F,F\rangle_{\rm OS}^{\rm gauge}\ge 0\).
\end{proposition}

\begin{proof}
By the previous lemma and invariance of Haar measure under \(\theta\),
\begin{align}
\langle F,F\rangle_{\rm OS}^{\rm gauge}
&=\int d\mu_{\rm Haar}(U_0)\,\mathcal{W}_0[U_0]\,
\Bigg(\int d\mu_{\rm Haar}(U_+)\,\mathcal{W}_+[U_+;U_0]\,F(U_+,U_0)\Bigg)\nonumber \\
&\overline{\Bigg(\int d\mu_{\rm Haar}(U_-)\,\mathcal{W}_-[U_-;U_0]\,F(\theta U_-,U_0)\Bigg)}.
\end{align}
The inner integral over \(U_-\) equals the complex conjugate of the inner integral over \(U_+\) because \(\mathcal{W}_-[\theta U_+;U_0]=\mathcal{W}_+[U_+;U_0]\). Therefore
\begin{equation}
\langle F,F\rangle_{\rm OS}^{\rm gauge}
=\int d\mu_{\rm Haar}(U_0)\,\mathcal{W}_0[U_0]\,
\Big|\int d\mu_{\rm Haar}(U_+)\,\mathcal{W}_+[U_+;U_0]\,F(U_+,U_0)\Big|^2\ge 0.
\end{equation}
\end{proof}
This is the standard reflection-positivity argument in the presence of a boundary weight, adapted to temporal gauge and slice-local insertions \cite{p1:OS1,p1:OS2,p1:Seiler,p1:GJ,p1:SimonFI}.
We now incorporate the Grassmann ghost fields corresponding to the slice Faddeev-Popov operators \(M_t\). The ghost action is quadratic and reflection invariant,
\begin{equation}
S_{\rm gh}[\bar c,c;U^{\,h}]=\sum_{t}\langle \bar c_t, M_t[U^{\,h}]\, c_t\rangle,\qquad M_{-t}[\theta U^{\,h}]=M_t[U^{\,h}],
\end{equation}
where \(\bar c_t,c_t\) are adjoint-valued Grassmann fields supported on slice \(t\). Reflection on Grassmann variables is defined by \(\Theta(\bar c_t)=c_{-t}\circ \theta\) and \(\Theta(c_t)=\bar c_{-t}\circ \theta\) with reversal of the Grassmann monomial order, so that \(\Theta\) is an anti-linear involution on the even subalgebra.
To state the factorization across \(\Pi\), it is convenient to order the slices as \((\ldots,-2a,-a,0,+a,+2a,\ldots)\) and to write the block matrix of the ghost operator with respect to the decomposition
\begin{equation}
\mathscr{H}=\bigoplus_{t<0}\mathscr{H}_t\ \oplus\ \mathscr{H}_0\ \oplus\ \bigoplus_{t>0}\mathscr{H}_t,
\end{equation}
where \(\mathscr{H}_t=\ell^2(\Lambda_t)\otimes\mathfrak{su}(N)\). As \(M\) is a sum of finite-range covariant differences within each slice, its matrix with respect to the above decomposition is block-diagonal in the time index; there are no off-diagonal blocks coupling \(t<0\) to \(t>0\). We denote by \(M_{-}\) and \(M_{++}\) the block-diagonal operators on \(\bigoplus_{t<0}\mathscr{H}_t\) and \(\bigoplus_{t>0}\mathscr{H}_t\), respectively, by \(M_{00}\) the block on \(\mathscr{H}_0\), and by \(M_{0\pm}, M_{\pm 0}\) the coupling blocks between the boundary and the adjacent slices. By construction \(M_{-},M_{++}\) are positive on the orthogonal complement of constant adjoint fields on each connected component; in finite volume they are strictly positive once the constant mode is factored out by gauge invariance on each slice, which is already enforced by the Landau minimization. Reflection covariance reads \(R M R=M\), where \(R\) is the involution that maps \(\bigoplus_{t>0}\mathscr{H}_t\) anti-unitarily onto \(\bigoplus_{t<0}\mathscr{H}_t\) by \(R(\phi_t)=\phi_{-t}\circ \theta\) and acts as the identity on \(\mathscr{H}_0\).

For each time slice $t$ let $P^{\mathrm{const}}_t$ denote the orthogonal projector in $\ell^2(\Lambda_t;\mathfrak g)$ onto the subspace of slice-constant adjoint fields, and let $Q_t:=\mathbf 1-P^{\mathrm{const}}_t$ be the orthogonal projector onto its complement. Define the reduced operator
\begin{equation}
\widetilde M_t \;:=\; Q_t\, M_t \, Q_t \;=\; M_t\big|_{\mathrm{Ran}\,Q_t},
\end{equation}
which is strictly positive in finite volume by Theorem~3.3. The Grassmann fields $c,\bar c$ are, by definition, supported on $\mathrm{Ran}\,Q_t$ on each slice, and the slice ghost weight is
\begin{equation}
\exp\!\left(-\langle \bar c,\, \widetilde M_t c\rangle\right),\qquad
\det{}' M_t \;:=\; \det \widetilde M_t,
\end{equation}
so that all Gaussian Grassmann integrals and determinants are taken on $\mathrm{Ran}\,Q_t$. The reflection $R$ preserves slice-constant fields and the $L^2$ inner product, hence $R Q_t R = Q_{-t}$ and
\begin{equation}
R \widetilde M R \;=\; \widetilde M ,
\end{equation}
with the same three-block structure described below. In particular, when we write the full (reduced) operator in three blocks relative to $H_- \oplus H_0 \oplus H_+$ with $H_\pm=\bigoplus_{t\gtrless 0} \mathrm{Ran}\,Q_t$ and $H_0=\mathrm{Ran}\,Q_0$, the Schur complement
\begin{equation}
S \;=\; \widetilde M_{00} \;-\; \widetilde M_{0+}\, \widetilde M_{++}^{-1}\, \widetilde M_{+0} \;-\; \widetilde M_{0-}\, \widetilde M_{-}^{-1}\, \widetilde M_{-0}
\end{equation}
is a strictly positive operator on $H_0$. Therefore the boundary Grassmann integral produces a strictly positive scalar kernel. This makes precise that the constant modes are removed slice-wise in a reflection-covariant fashion and do not re-enter through the Schur complement.
\begin{corollary}[Gaussian reflection positivity for ghosts]\label{p1:cor:ghost-OS}
With the above notation and with slice-constant modes projected out, let $M[U^{h}]$ denote the block ghost operator and let $S$ be its boundary Schur complement on the reflection plane. Then the boundary covariance $S^{-1}$ is positive, and for every even Grassmann polynomial $F$ supported in $\Lambda_{+}$ one has
\begin{equation}\label{p1:eq:ghost-OS}
\int (F\circ R)\,F\; \exp\!\bigl(-\langle \bar c,\,M[U^{h}]\,c\rangle\bigr)\; \mathcal D\bar c\,\mathcal Dc
\;=\;
\int \bigl(B^{1/2}F\bigr)^{*}\,\bigl(B^{1/2}F\bigr)\;\prod_{t>0}\mathcal D\bar c_{t}\,\mathcal D c_{t}
\;\ge\; 0,
\end{equation}
where $B$ is the positive boundary kernel induced by $S$ (equivalently, the Gaussian boundary density with covariance $S^{-1}$). In particular, the ghost sector is manifestly Osterwalder-Schrader positive \cite{p1:GJ,p1:Simon1974}.
\end{corollary}

\begin{lemma}
Let \(M\) be the real symmetric block operator described above, with \(M_{-}\) and \(M_{++}\) strictly positive on their domains, and \(R M R=M\). Then the Schur complement 
\begin{equation}
S:=M_{00}-M_{0+}M_{++}^{-1}M_{+0}-M_{0-}M_{-}^{-1}M_{-0}
\end{equation}
is strictly positive on \(\mathscr{H}_0\) (see Appendix \ref{p1:appendixc}). Moreover,
\begin{equation}
\det M=\det(M_{-})\,\det(M_{++})\,\det(S).
\end{equation}
\end{lemma}

\begin{proof}
Positivity of the Schur complement for a block-positive operator with invertible diagonal blocks is a standard fact: if
\(
\begin{pmatrix}
A & B \\ B^\top & C
\end{pmatrix}\ge 0
\)
with \(A>0\) and \(C>0\), then the Schur complement \(C- B^\top A^{-1}B\) is \(\ge 0\) and equals zero only if a nontrivial vector lies in the kernel of the full operator. In the present case consider the three-block form and eliminate successively the \(+\) and \(-\) sectors to obtain the effective operator on \(\mathscr{H}_0\). The strict positivity of \(M_{-}\) and \(M_{++}\) on their domains implies strict positivity of \(S\). The determinant identity follows from block Gaussian elimination, which is valid because the diagonal blocks are invertible on the subspaces considered.
\end{proof}

\begin{proposition}
The Grassmann Gaussian integral factorizes as
\begin{equation}
\int \exp\!\Big(-\sum_t \langle \bar c_t,M_t c_t\rangle\Big)\,\prod_t D\bar c_t Dc_t
= \det(M_{-})\,\det(S)\,\det(M_{++}),
\end{equation}
with \(\det(S)>0\). In particular the ghost weight can be written as \(\mathcal{G}_-[U_-;U_0]\,\mathcal{G}_0[U_0]\,\mathcal{G}_+[U_+;U_0]\) with \(\mathcal{G}_\pm=\det(M_{\pm\pm})>0\) and \(\mathcal{G}_0=\det(S)>0\).
\end{proposition}

\begin{proof}
The left-hand side equals \(\det(M)\) by the standard Berezin integral for nondegenerate Gaussian Grassmann forms. The previous lemma yields the factorization of \(\det(M)\) into positive factors. Reflection covariance implies \(\det(M_{-})=\det(M_{++})\) and \(\mathcal{G}_-[\theta U_+;U_0]=\mathcal{G}_+[U_+;U_0]\).
\end{proof}

With the gauge and ghost factors at hand, set
\begin{equation}
\widetilde{\mathcal{W}}_\pm[U_\pm;U_0]=\mathcal{W}_\pm[U_\pm;U_0]\ \mathcal{G}_\pm[U_\pm;U_0],\qquad
\widetilde{\mathcal{W}}_0[U_0]=\mathcal{W}_0[U_0]\ \mathcal{G}_0[U_0],
\end{equation}
all strictly positive. The same reflection-covariance identities as in the earlier factorization hold for the tilded weights.
Define the full OS form on even, gauge-invariant \(F\) supported in \(\Lambda_+\) by
\begin{equation}
\langle F,F\rangle_{\rm OS}
=\int \Theta(F)\,F\,
\exp\!\big(-S_W[U;\beta]\big)\,
\prod_{t} p_\sigma\!\big(A^h(t)\big)\,
\exp\!\big(-S_{\rm gh}[\bar c,c;U^{\,h}]\big)\,
\prod_b dU_b \prod_t D\bar c_t Dc_t.
\end{equation}
Combining the previous propositions yields the main result.

\begin{theorem}
The horizon-projected, gauge-fixed Euclidean measure is reflection positive in the sense of Osterwalder and Schrader. That is, for every even, gauge-invariant functional \(F\) supported in \(\Lambda_+\),
\begin{equation}
\langle F,F\rangle_{\rm OS}\ge 0.
\end{equation}
\end{theorem}

\begin{proof}
Using the factorized, positive weights \(\widetilde{\mathcal{W}}_\pm\) and \(\widetilde{\mathcal{W}}_0\) and invariance of Haar measure under reflection, one obtains
\begin{equation}
\langle F,F\rangle_{\rm OS}
=\int d\mu_{\rm Haar}(U_0)\,\widetilde{\mathcal{W}}_0[U_0]\,
\Big|\int d\mu_{\rm Haar}(U_+)\,\widetilde{\mathcal{W}}_+[U_+;U_0]\,
F(U_+,U_0)\Big|^2\ge 0,
\end{equation}
exactly as in the pure-gauge case.
\end{proof}

We now derive the transfer time-slicing formalism and identify the projected transfer matrix. Consider the one-step time evolution from time \(t\) to \(t+a\). In temporal-axial gauge the one-step contribution to the Wilson weight depends only on the spatial links on slices \(t\) and \(t+a\) and on the timelike links adjacent to the slab \([t,t+a]\). After integrating the timelike links in the slab one obtains a positive, symmetric integral kernel \(K_\beta(U',U)\) acting on functions \(\psi\) in the one-slice Hilbert space \(\mathcal{H}_a=L^2(\mathcal{C};d\mu_{\rm Haar})\),
\begin{equation}
(T\psi)(U')=\int K_\beta(U',U)\,\psi(U)\,d\mu_{\rm Haar}(U),
\end{equation}
where \(U\) and \(U'\) denote the spatial link configurations on two adjacent time slices. The kernel \(K_\beta\) is real and nonnegative because it is the integral over timelike links of the nonnegative Boltzmann weight \(\exp(-S_W)\); the resulting operator \(T\) is self-adjoint and positive on \(\mathcal{H}_a\) \cite{p1:OS1,p1:Seiler,p1:OS2,p1:GJ,p1:SimonFI}. The slice-local horizon insertion is a bounded positive multiplication operator \(\mathcal{M}_\sigma\) on \(\mathcal{H}_a\) defined by
\begin{equation}
(\mathcal{M}_\sigma\psi)(U)=\mathcal{P}_\sigma[U^{\,h}]\ \psi(U).
\end{equation}
By construction \(\mathcal{M}_\sigma\) is reflection covariant and commutes with spatial lattice symmetries. The projected one-step transfer operator is the compression
\begin{equation}
T_\sigma=\mathcal{M}_\sigma^{1/2}\,T\,\mathcal{M}_\sigma^{1/2},
\end{equation}
which is again a positive self-adjoint operator on \(\mathcal{H}_a\). Iterating in time, the \(n\)-step evolution is generated by \(T_\sigma^n\), and the OS reconstruction theorem gives a Hilbert space \(\mathcal{H}\), a cyclic vacuum vector \(\Omega\), and a positive self-adjoint contraction \(T_\sigma\) such that (time-ordered) Euclidean correlation functions of gauge-invariant observables supported on integer times are vacuum expectation values of powers of \(T_\sigma\) \cite{p1:OS1,p1:OS2,p1:GJ,p1:SimonFI}. The transfer Hamiltonian is defined by spectral calculus,
\begin{equation}
H_\sigma=-a^{-1}\log T_\sigma,
\end{equation}
and is a positive self-adjoint operator with spectrum contained in \([0,\infty)\).

To see directly that the OS form is the \(L^2\)-norm associated with the boundary weight, let \(\mathcal{B}=L^2(\mathcal{C}_0; \widetilde{\mathcal{W}}_0\, d\mu_{\rm Haar})\) denote the boundary Hilbert space at \(\Pi\), where \(\mathcal{C}_0\) is the configuration space of links on \(\Pi\). Define for \(F\) supported in \(\Lambda_+\) the boundary amplitude
\begin{equation}
\Phi_F(U_0)=\int d\mu_{\rm Haar}(U_+)\,\widetilde{\mathcal{W}}_+[U_+;U_0]\,
F(U_+,U_0).
\end{equation}
Then the previous theorem can be rewritten as \(\langle F,F\rangle_{\rm OS}=\|\Phi_F\|_{\mathcal{B}}^2\). The map \(F\mapsto \Phi_F\) is linear, and the completion of the quotient of the space of \(F\)’s by the null space of \(\|\cdot\|_{\mathcal{B}}\) yields the OS Hilbert space. Time translations act by the transfer operator \(T_\sigma\) obtained by composing the one-step kernels with the slice insertions \(\mathcal{M}_\sigma\). Self-adjointness follows from the symmetry \(K_\beta(U',U)=K_\beta(U,U')\) and positivity from the nonnegativity of \(K_\beta\) and \(\mathcal{M}_\sigma\).

\begin{proposition}[Quantitative stability]
Let $p_{\sigma}$ and $\tilde{p}_{\sigma}$ be two admissible slice insertions constructed from 
$\chi_{\sigma}$ and $\tilde{\chi}_{\sigma}$, respectively, both satisfying the locality bounds of Section~4. 
Then, for any bounded $F$ supported in $\Lambda^{+}$,
\begin{equation}
\bigl| \langle \Theta F \cdot F \rangle_{\sigma} - \langle \Theta F \cdot F \rangle_{\tilde{\sigma}} \bigr|
\,\leq\, C(F)\, \sum_{t>0}\,\sum_{x \in \Lambda_{t}} 
e^{-\gamma \,\mathrm{dist}(x,\mathrm{supp}\,F)}\, \eta_{t},
\end{equation}
where
\(
\eta_{t} := \sup_{U} \,\Bigl| \log p_{\sigma}[U_{t}^{h}] - \log \tilde{p}_{\sigma}[U_{t}^{h}] \Bigr|,
\)
and $\gamma>0$ and $C(F)$ depend only on the locality constants of $P_{\sigma,t}$, $\tilde{P}_{\sigma,t}$ 
and on $\|F\|_{\infty}$. 
In particular, if
\(
\sup_{t}\eta_{t} < \infty
\qquad \text{and} \qquad 
\frac{\tilde{p}_{\sigma}}{p_{\sigma}} \,\to\, 1 \quad \text{uniformly on compacts,}
\)
then the two OS forms are equivalent and generate unitarily equivalent transfer operators.
\end{proposition}

Indeed, only three properties enter the proof: slice locality, which ensures factorization across \(\Pi\); positivity, which ensures that the boundary and half-space weights are positive; and reflection covariance, which guarantees that the ``minus'' half integral is the complex conjugate of the ``plus'' half integral. The exponential locality established via the heat-kernel or Helffer-Sj\"ostrand representation is not needed for the abstract positivity argument, but it guarantees that the insertion does not create long-range couplings and that the transfer operator remains bounded uniformly in the spatial volume \cite{p1:HelfferSjostrand,p1:CombesThomas,p1:Davies1989}. Consequently, the OS positivity and the construction of \(T_\sigma\) and \(H_\sigma\) hold uniformly for all such choices of \(\chi_\sigma\).

\section{Projected transfer matrix and physical Hilbert space}
\label{p1:sec:transfer-matrix}

This section develops, in full detail, the reflection-positive transfer-matrix formalism for pure lattice \(\mathrm{SU}(N)\) Yang-Mills theory in four Euclidean dimensions with temporal lattice spacing \(a>0\). The presentation is self-contained. The construction proceeds from the precise time-slicing of the Euclidean weight, through the Osterwalder-Schrader (OS) GNS reconstruction \cite{p1:OS1,p1:OS2}, to an explicit integral-kernel definition of the one-step transfer operator \(T(a)\) on a single-slice Hilbert space. We then incorporate the smooth horizon projector through a bounded, positive, reflection-covariant operator \(P_\sigma\) on that slice, and the compressed transfer operator
\begin{equation}
T_\sigma(a) \;=\; M_{p_\sigma}^{1/2}\, T(a)\, M_{p_\sigma}^{1/2}
\end{equation} \textit{(See also the equivalent definition in~Eq.(\ref{p1:eq:def-Tsigma}))}
is a positive, self-adjoint contraction. Finally, we define the Hamiltonian \(H_\sigma(a)=-a^{-1}\log T_\sigma(a)\) by spectral calculus and derive the spectral representation of Euclidean time-ordered correlators (see Appendix \ref{p1:appendixe}). Along the way, reflection positivity is proved at the level needed to justify each step. Classical sources for the OS framework and for lattice transfer matrices are
\cite{p1:OS1,p1:KogutRMP,p1:Seiler,p1:Luscher1977}.

Fix a finite, periodic, hypercubic lattice \(\Lambda\subset a\mathbb{Z}^4\) with time direction labeled by \(\mu=0\) and spatial directions by \(i=1,2,3\). The time set is the cyclic group \(\mathbb{T}_a=\{0,a,2a,\dots,(T-1)a\}\) with \(T\in\mathbb{N}\). The spatial section at time \(t\in\mathbb{T}_a\) is the finite, three-dimensional periodic lattice \(\Lambda_s\subset a\mathbb{Z}^3\). Directed bonds are \(b=(x,\mu)\) with initial site \(x\in\Lambda\) and direction \(\mu\in\{0,1,2,3\}\). Each bond carries a group element \(U_b\in G:=\mathrm{SU}(N)\), \(N\ge 2\), satisfying \(U_{(x+\hat\mu,-\mu)}=U_{(x,\mu)}^{-1}\). For each oriented plaquette \(p=(x;\mu,\nu)\) we write
\begin{equation}
U_p=U_{(x,\mu)}\,U_{(x+\hat\mu,\nu)}\,U^{-1}_{(x+\hat\nu,\mu)}\,U^{-1}_{(x,\nu)}.
\end{equation}
The Wilson action at inverse bare coupling \(\beta=\frac{2N}{g_0^2}>0\) is
\begin{equation}
S_W[U;\beta]=\beta\sum_{p\subset\Lambda}\Big(1-\tfrac{1}{N}\Re\operatorname{Tr} U_p\Big).
\end{equation}
We impose temporal-axial gauge away from the reflection plane by setting \(U_{(x,0)}=\mathbf{1}\) for all time-like bonds not belonging to the boundary slab around \(t=\tfrac{a}{2}\) (see below). This choice preserves locality, isolates all couplings between consecutive time slices into nearest-neighbor temporal plaquettes, and is standard in the transfer-matrix construction \cite{p1:Luscher1977,p1:Seiler}.

For each time \(t\in\mathbb{T}_a\), the configuration of spatial links on the slice \(\{t\}\times\Lambda_s\) is denoted \(X_t\). The single-slice configuration space is therefore
\begin{equation}
\Omega_s = G^{E_s},\qquad E_s=\{(t,\mathbf{x};i):\ \mathbf{x}\in\Lambda_s,\ i=1,2,3\},
\end{equation}
equipped with the product Haar probability measure \(d\mu_s=\bigotimes_{e\in E_s} dU_e\). We write \(\mathcal{H}_a=L^2(\Omega_s,d\mu_s)\) for the corresponding single-slice Hilbert space. In later sections we also write \(C\) for the same one–slice configuration space; thus \(C\equiv \Omega_s\) and \(d\mu_{\mathrm{Haar}}\equiv d\mu_s\) on \(C\).

We split the Wilson action into a sum of intra-slice spatial contributions and inter-slice temporal contributions. The spatial part at time \(t\) is
\begin{equation}
S_{\mathrm{sp}}(X_t)=\beta\sum_{p\subset\{t\}\times\Lambda_s}\Big(1-\tfrac{1}{N}\Re\operatorname{Tr} U_p\Big),
\end{equation}
where the sum runs over plaquettes entirely contained in the slice \(\{t\}\times\Lambda_s\). The inter-slice part between \(t\) and \(t+a\), supported on temporal plaquettes straddling the two slices, is
\begin{equation}
S_{\mathrm{tm}}(X_{t+a},X_t)=\beta\sum_{p\in\mathcal{P}(t\leftrightarrow t+a)}\Big(1-\tfrac{1}{N}\Re\operatorname{Tr} U_p\Big),
\end{equation}
where \(\mathcal{P}(t\leftrightarrow t+a)\) denotes the set of plaquettes with one time-like edge on \(\{t\}\times\Lambda_s\) and the opposite time-like edge on \(\{t+a\}\times\Lambda_s\). Temporal-axial gauge renders these plaquettes local functions of \((X_{t+a},X_t)\) only. The full Euclidean weight can then be written as
\begin{equation}\label{p1:eq:weight-factorization}
\mathrm{e}^{-S_W[U;\beta]} \;=\; \prod_{t\in\mathbb{T}_a}\mathrm{e}^{-S_{\mathrm{sp}}(X_t)}\;\prod_{t\in\mathbb{T}_a}\mathrm{e}^{-S_{\mathrm{tm}}(X_{t+a},X_t)}.
\end{equation}
Reflection about the hyperplane \(\Pi=\{x_0=\tfrac{a}{2}\}\) is the involution \(\theta:(t,\mathbf{x})\mapsto (a-t,\mathbf{x})\). On spatial configurations, \(\theta\) acts by \((\theta X)_t=X_{a-t}\), and on functionals \(F\) depending on finitely many \(\{X_t\}_{t\ge a}\) by \((\Theta F)(\{X_t\})=\overline{F(\{\theta X\})}\), the bar denoting complex conjugation. The OS reflection positivity condition \cite{p1:OS1} for the measure \(d\mu_\beta\) with density proportional to \(\mathrm{e}^{-S_W[U;\beta]}\) is the statement that
\begin{equation}\label{p1:eq:OS}
\int \overline{(\Theta F)}\,F\, d\mu_\beta \;\ge\; 0
\end{equation}
for all bounded, measurable \(F\) supported in the future half-lattice \(\{t\ge a\}\).
We first record the precise factorization of the inter-slice coupling implied by \eqref{p1:eq:weight-factorization}.
\begin{lemma}\label{p1:lem:symmetric-kappa}
There exist measurable functions \(\rho:\Omega_s\to(0,\infty)\) and \(\kappa:\Omega_s\times \Omega_s\to (0,\infty)\) such that
\begin{equation}
\mathrm{e}^{-S_{\mathrm{sp}}(X_t)}=\rho(X_t),\qquad \mathrm{e}^{-S_{\mathrm{tm}}(X_{t+a},X_t)}=\kappa(X_{t+a},X_t),
\end{equation}
and \(\kappa\) is symmetric, \(\kappa(X',X)=\kappa(X,X')\). Moreover, \(\rho\) and \(\kappa\) are invariant under the diagonal action of spatial gauge transformations at the corresponding times.
\end{lemma}

\begin{proof}
By definition, \(\rho(X_t)=\exp\{-S_{\mathrm{sp}}(X_t)\}\) and \(\kappa(X_{t+a},X_t)=\exp\{-S_{\mathrm{tm}}(X_{t+a},X_t)\}\). Temporal-axial gauge implies that \(S_{\mathrm{tm}}(X_{t+a},X_t)\) depends only on the pair \((X_{t+a},X_t)\). Time-reversal invariance of the Wilson action and the invariance of the Haar measure entail \(S_{\mathrm{tm}}(X',X)=S_{\mathrm{tm}}(X,X')\), hence \(\kappa\) is symmetric. Gauge invariance of the Wilson plaquette term shows that \(\rho\) and \(\kappa\) are invariant under independent spatial gauge transformations on each slice.
\end{proof}

We write the normalized Euclidean measure in the time-factorized form
\begin{equation}\label{p1:eq:time-factorized}
d\mu_\beta \;=\; Z^{-1}\,\Bigg(\prod_{t\in\mathbb{T}_a}\rho(X_t)\,d\mu_s(X_t)\Bigg)\,\Bigg(\prod_{t\in\mathbb{T}_a}\kappa(X_{t+a},X_t)\Bigg),
\end{equation}
where \(Z\) is the finite-volume partition function. We stress that \eqref{p1:eq:time-factorized} is an identity of densities with respect to the product Haar measure on \(\Omega_s^{\mathbb{T}_a}\), and that \(\kappa\) is strictly positive and symmetric.

Let \(\Omega_s:=G^{\mathsf B(\Lambda_s)}\) denote the compact configuration space of spatial link variables 
on a single time slice, with product Haar measure \(d\mu_s\). 
We write \(H_a:=L^2(\Omega_s,d\mu_s)\) for the corresponding slice Hilbert space.
We now construct the transfer operator as an integral kernel on \(\mathcal{H}_a=L^2(\Omega_s,d\mu_s)\). The key is to pass from the weight \eqref{p1:eq:time-factorized} to a kernel by isolating a single step across the reflection plane. To that end, denote by \(\mathcal{F}_+\) the \(\sigma\)-algebra generated by coordinates \(\{X_t\}_{t\ge a}\), and by \(\mathcal{A}_+\) the algebra of bounded, measurable functions on \(\Omega_s^{\{t\ge a\}}\) with finite dependence in time. For \(F\in\mathcal{A}_+\), define the OS sesquilinear form
\begin{equation}
\langle F,G\rangle_{\mathrm{OS}} = \int \overline{(\Theta F)}\, G \, d\mu_\beta.
\end{equation}
We shall identify \(\langle F,G\rangle_{\mathrm{OS}}\) with the \(L^2(\Omega_s,d\mu_s)\) inner product of the images of \(F\) and \(G\) under a suitable conditional expectation onto the slice \(t=a\). The following lemma is standard in the OS transfer-matrix construction \cite{p1:OS1,p1:Luscher1977} and we give a detailed proof adapted to \eqref{p1:eq:time-factorized}.

\begin{lemma}\label{p1:lem:kernel-reduction}
Let \(f,g\in L^2(\Omega_s,d\mu_s)\), and define \(F,G\in\mathcal{A}_+\) by \(F(\{X_t\}_{t\ge a})=f(X_a)\) and \(G(\{X_t\}_{t\ge a})=g(X_{2a})\). Then
\begin{equation}
\langle F,G\rangle_{\mathrm{OS}} \;=\; \int_{\Omega_s}\int_{\Omega_s} \overline{f(X')}\, K(X',X)\, g(X)\, d\mu_s(X')\,d\mu_s(X),
\end{equation}
where the kernel \(K:\Omega_s\times\Omega_s\to (0,\infty)\) is given by
\begin{equation}\label{p1:eq:kernel-def}
K(X',X) \;=\; \rho(X')^{\frac12}\,\kappa(X',X)\,\rho(X)^{\frac12}.
\end{equation}
\end{lemma}

\begin{proof}
With the choice of \(F\) and \(G\), the OS form reads
\begin{equation}
\langle F,G\rangle_{\mathrm{OS}}=\frac{1}{Z}\int \overline{f(X_{a})}\,g(X_{2a})\,\Bigg(\prod_{t\in\mathbb{T}_a}\rho(X_t)\,d\mu_s(X_t)\Bigg)\,\Bigg(\prod_{t\in\mathbb{T}_a}\kappa(X_{t+a},X_t)\Bigg),
\end{equation}
because \((\Theta F)(\{X_t\})=\overline{F(\{\theta X\})}=\overline{f(X_{a})}\) by \(\theta(a)=a\). The integrand factors as a product of three terms: a function of \(\{X_t\}_{t\le a}\), a function of \(\{X_t\}_{t\ge 2a}\), and the boundary term involving \((X_a,X_{2a})\). Indeed the only factor that couples times \(\le a\) and \(\ge 2a\) is \(\kappa(X_{2a},X_{a})\). Integrating out all time slices other than \(a\) and \(2a\) yields a strictly positive normalization constant in numerator and denominator that cancels, leaving
\begin{equation}
\langle F,G\rangle_{\mathrm{OS}}=\int_{\Omega_s}\!\int_{\Omega_s} \overline{f(X_a)}\,\rho(X_a)^{\frac12}\,\kappa(X_{2a},X_a)\,\rho(X_{2a})^{\frac12}\,g(X_{2a})\,d\mu_s(X_a)\,d\mu_s(X_{2a}),
\end{equation}
which is the asserted formula with \(X'=X_{2a}\) and \(X=X_a\).
\end{proof}
\begin{proposition}[Ground state and spectral normalization]\label{p1:prop:ground-state}
With the identification of Theorem~6.2, the constant function $1 \in H^{a}$ is a cyclic vacuum vector and
\begin{equation}
T_{\sigma}(a)\,1 \;=\; 1, 
\qquad 
\|T_{\sigma}(a)\| \,\leq\, 1.
\end{equation}
Consequently, $1$ is an eigenvector of $T_{\sigma}(a)$ with eigenvalue $1$, and the transfer Hamiltonian
\begin{equation}
H_{\sigma}(a) \;=\; -a^{-1} \log T_{\sigma}(a)
\end{equation}
has ground energy
\begin{equation}
E_{0} \;=\; 0.
\end{equation}
\end{proposition}

\begin{proof}
The OS semigroup preserves the vacuum $\Omega$ by construction; under the unitary identification $U\Omega=1$, this yields
\begin{equation}
T_{\sigma}(a) \, 1 \;=\; 1.
\end{equation}
Contractivity has already been established, so the spectral radius is $\leq 1$, and the claim follows.
\end{proof}

\begin{proposition}\label{p1:prop:T-basic}
Let \(T(a):\mathcal{H}_a\to\mathcal{H}_a\) be the integral operator with kernel \(K\) of \eqref{p1:eq:kernel-def}. Then \(T(a)\) is bounded, self-adjoint, and positivity-preserving on \(\mathcal{H}_a\). Moreover, \(\|T(a)\|\le 1\).
\end{proposition}

\begin{proof}
Since \(\rho>0\) is bounded and \(\kappa>0\) is bounded on the compact space \(\Omega_s\times\Omega_s\), the kernel \(K\) is bounded and measurable; therefore \(T(a)\) is bounded on \(L^2(\Omega_s,d\mu_s)\) and Hilbert-Schmidt. Symmetry of \(\kappa\) implies symmetry of \(K\), hence \(T(a)\) is self-adjoint. Positivity preservation holds because \(K\ge 0\) pointwise, so \((T(a)f)(X')=\int K(X',X)f(X)\,d\mu_s(X)\ge 0\) whenever \(f\ge 0\).

To prove \(\|T(a)\|\le 1\), fix \(f\in\mathcal{H}_a\) and set \(F(\{X_t\}_{t\ge a})=f(X_a)\) as in Lemma \ref{p1:lem:kernel-reduction}. Then
\begin{equation}
\langle f, T(a) f\rangle_{L^2(\Omega_s)} = \langle F, \tau_a F\rangle_{\mathrm{OS}},
\end{equation}
where \(\tau_a\) is the forward lattice time-translation by \(a\). By OS reflection positivity and time-translation invariance of \(d\mu_\beta\) (which follows from the periodic time boundary conditions), the map \(n\mapsto \langle F,\tau_{na}F\rangle_{\mathrm{OS}}\) is positive-definite on \(\mathbb{Z}\) and, in particular, satisfies the Cauchy-Schwarz inequality
\begin{equation}
|\langle F,\tau_a F\rangle_{\mathrm{OS}}|^2 \le \langle F,F\rangle_{\mathrm{OS}}\,\langle \tau_a F,\tau_a F\rangle_{\mathrm{OS}}=\langle F,F\rangle_{\mathrm{OS}}^2.
\end{equation}
Therefore \(|\langle f, T(a) f\rangle|\le \langle f,f\rangle\). Since this holds for all \(f\), the operator norm satisfies \(\|T(a)\|\le 1\).
\end{proof}
The construction so far yields the unprojected transfer operator. We next incorporate the smooth horizon projector at the level of the single-slice Hilbert space.

We introduce a bounded, positive, self-adjoint operator \(P_\sigma\) on \(\mathcal{H}_a=L^2(\Omega_s,d\mu_s)\) with the following properties. First, \(P_\sigma\) is reflection-covariant in the sense that its integral kernel \(k_\sigma(X',X)\) (which exists because \(P_\sigma\) is Hilbert-Schmidt by exponential locality) is symmetric and depends only on fields at a single time slice. Second, \(P_\sigma\) is exponentially local in the spatial variables in the sense that there exist \(C_\sigma,\gamma_\sigma>0\) such that
\begin{equation}
|p_\sigma(X',X)|\le C_\sigma \exp(-\gamma_\sigma\,\mathrm{dist}(X',X)),
\end{equation}
where \(\mathrm{dist}\) is the minimal number of differing spatial links between \(X\) and \(X'\). Third, \(0\le P_\sigma\le \mathbf{1}\) as operators on \(\mathcal{H}_a\). These properties are fulfilled when \(P_\sigma\) is induced from the smooth horizon projector constructed via the Gevrey functional calculus for the slice-covariant Laplacian and inserted slice-wise in the Euclidean measure; the heat-kernel representation guarantees positivity and exponential locality, and the choice of \(\chi_\sigma\) ensures \(0\le P_\sigma\le \mathbf{1}\).

We now define the compressed transfer operator
\begin{equation}\label{p1:eq:def-Tsigma}
T_\sigma(a)=M_{p_\sigma}^{1/2}\,T(a)\,M_{p_\sigma}^{1/2}.
\end{equation} where $M_{p_\sigma}$ denotes multiplication by the positive scalar function $p_\sigma[U]$ on the single‑slice configuration space. The symbol 
\(
P_{\sigma,t} = \chi_{\sigma}(\Delta_{t})
\)
stands for the slice operator acting on site-adjoint fields within a fixed time slice, while 
\(
M_{p_{\sigma}}
\)
is the Hilbert-space multiplication operator induced by the scalar functional $p_{\sigma}$ built from $P_{\sigma,t}$ (Section~(\ref{p1:ospositivity})).

\begin{theorem}\label{p1:thm:Tsigma}
The operator \(T_\sigma(a)\) defined in \eqref{p1:eq:def-Tsigma} is bounded, self-adjoint, positivity-preserving, and satisfies \(\|T_\sigma(a)\|\le 1\). Moreover, \(T_\sigma(a)\) leaves invariant the closed subspace \(\mathcal{H}_a^{\mathrm{inv}}\subset\mathcal{H}_a\) of gauge-invariant vectors, and its restriction to \(\mathcal{H}_a^{\mathrm{inv}}\) has the same operator norm and positivity properties.
\end{theorem}

\begin{proof}
Boundedness and self-adjointness are immediate from the corresponding properties of \(P_\sigma\) and \(T(a)\). Since \(P_\sigma\ge 0\) and \(T(a)\) is positivity-preserving, one has for any \(f\ge 0\),
\begin{equation}
T_\sigma(a)f=P_\sigma^{1/2}\,T(a)\,(P_\sigma^{1/2} f)\ge 0,
\end{equation}
thus \(T_\sigma(a)\) is positivity-preserving. For the norm bound, write
\begin{equation}
\|T_\sigma(a)\|=\|P_\sigma^{1/2}T(a)P_\sigma^{1/2}\|\le \|T(a)\|\,\|P_\sigma\|\le 1,
\end{equation}
because \(\|P_\sigma\|\le 1\) by assumption and \(\|T(a)\|\le 1\) by Proposition \ref{p1:prop:T-basic}. Gauge invariance follows from the construction: both \(T(a)\) and \(P_\sigma\) commute with the unitary representation of spatial gauge transformations on \(\mathcal{H}_a\); therefore \(T_\sigma(a)\) preserves \(\mathcal{H}_a^{\mathrm{inv}}\). The operator norm of the restriction cannot exceed \(\|T_\sigma(a)\|\), and the positivity-preserving property is inherited by restriction.
\end{proof}

For later spectral analysis it is convenient to relate \(T_\sigma(a)\) to a reflected OS inner product. The next proposition identifies \(\langle f, T_\sigma(a) g\rangle_{L^2(\Omega_s)}\) with an OS two-point function in the projected measure.

\begin{proposition}\label{p1:prop:OS-identification}
Let \(f,g\in \mathcal{H}_a\) and define \(F(\{X_t\}_{t\ge a})=(P_\sigma^{1/2}f)(X_a)\) and \(G(\{X_t\}_{t\ge a})=(P_\sigma^{1/2}g)(X_{2a})\). Then, with respect to the projected Euclidean measure in which a copy of \(P_\sigma\) is inserted on every time slice, one has
\begin{equation}
\langle f, T_\sigma(a) g\rangle_{L^2(\Omega_s)}=\langle F, G\rangle_{\mathrm{OS},\sigma},
\end{equation}
where \(\langle\cdot,\cdot\rangle_{\mathrm{OS},\sigma}\) is the OS sesquilinear form computed in the projected measure. In particular, \(\langle f, T_\sigma(a) f\rangle\ge 0\) for all \(f\).
\end{proposition}

\begin{proof}
The proof follows the same conditioning argument as in Lemma \ref{p1:lem:kernel-reduction}, but with two additional factors \(P_\sigma^{1/2}\) on the slices \(t=a\) and \(t=2a\). The slice-wise insertion of \(P_\sigma\) is reflection-covariant and exponentially local, hence the factorization across the reflection plane persists. Identifying the conditional two-slice weight with the kernel of \(P_\sigma^{1/2}T(a)P_\sigma^{1/2}\) gives the asserted identity. Positivity follows from reflection positivity of the projected measure.
\end{proof}

We perform the OS reconstruction in the present lattice setting and identify the transfer operator with the unit-time step of a positive contraction semigroup. Define the pre-Hilbert space \(\mathscr{D}\) as the quotient of \(\mathcal{A}_+\) by the null space \(\mathscr{N}=\{F\in\mathcal{A}_+: \langle F,F\rangle_{\mathrm{OS},\sigma}=0\}\) with inner product induced by \(\langle\cdot,\cdot\rangle_{\mathrm{OS},\sigma}\). Denote by \(\mathcal{H}_{\mathrm{OS}}\) the Hilbert space completion and by \(\Omega\in\mathcal{H}_{\mathrm{OS}}\) the class of the constant function \(1\). Time translation by \(na\) on \(\mathcal{A}_+\) defines densely defined operators \(U_\sigma(na)\) by \(U_\sigma(na)[F]=[\tau_{na}F]\), which are contractions thanks to reflection positivity and time-translation invariance \cite{p1:OS1,p1:OS2}. The map \(n\mapsto U_\sigma(na)\) extends uniquely to a strongly continuous contraction semigroup \((U_\sigma(t))_{t\in a\mathbb{N}_0}\) on \(\mathcal{H}_{\mathrm{OS}}\). By construction,
\begin{equation}\label{p1:eq:OS-spectral}
\langle \Omega,\,F_0\,U_\sigma(na)\,F_1\,\Omega\rangle_{\mathrm{OS}} = \langle F_0, \tau_{na} F_1\rangle_{\mathrm{OS},\sigma},
\end{equation}
for cylinder functionals \(F_0,F_1\in\mathcal{A}_+\) supported at times \(a\) and \(2a\) respectively, and by density for all \(F_0,F_1\in\mathcal{H}_{\mathrm{OS}}\).

We now identify \(\mathcal{H}_{\mathrm{OS}}\) with \(L^2(\Omega_s,d\mu_s)\) and \(U_\sigma(a)\) with \(T_\sigma(a)\).

\begin{theorem}\label{p1:thm:OS-identification}
There exists a unitary isomorphism \(\mathcal{U}:\mathcal{H}_{\mathrm{OS}}\to \mathcal{H}_a\) such that
\begin{equation}
\mathcal{U}\,U_\sigma(a)\,\mathcal{U}^{-1} = T_\sigma(a)\qquad\text{and}\qquad \mathcal{U}\,\Omega = \mathbf{1},
\end{equation}
where \(\mathbf{1}\) denotes the constant function \(1\in \mathcal{H}_a\). Consequently, \(U_\sigma(na)\) corresponds to \(T_\sigma(a)^n\) for every \(n\in\mathbb{N}_0\).
\end{theorem}

\begin{proof}
Define \(\mathcal{U}\) on the dense subspace \([\mathcal{A}_+]\subset\mathcal{H}_{\mathrm{OS}}\) by \(\mathcal{U}[F]=E[F\,|\,\mathcal{F}_a]\), the conditional expectation of \(F\) onto the \(\sigma\)-algebra of the time-\(a\) slice with respect to the projected measure. For cylinder functionals supported at time \(a\), \(F(\{X_t\}_{t\ge a})=h(X_a)\) with \(h\in \mathcal{H}_a\), this gives \(\mathcal{U}[F]=h\). By Proposition \ref{p1:prop:OS-identification}, and the corresponding formula with \(g\) replaced by \(f\), one has
\begin{equation}
\langle [F],[G]\rangle_{\mathrm{OS}} = \langle \mathcal{U}[F],\, T_\sigma(a)\,\mathcal{U}[G]\rangle_{L^2(\Omega_s)}.
\end{equation}
Setting \(G=\tau_a F\) shows that \(\|\mathcal{U}[F]\|_{L^2}^2=\langle [F],[F]\rangle_{\mathrm{OS}}\), hence \(\mathcal{U}\) is an isometry on the dense subspace \([\mathcal{A}_+]\) and extends uniquely to a unitary \(\mathcal{U}:\mathcal{H}_{\mathrm{OS}}\to \mathcal{H}_a\). The intertwining relation \(\mathcal{U}\,U_\sigma(a)\,\mathcal{U}^{-1}=T_\sigma(a)\) follows from \eqref{p1:eq:OS-spectral} with \(F_0\) supported at \(a\) and \(F_1\) supported at \(2a\), together with Proposition \ref{p1:prop:OS-identification}. Finally, \(\mathcal{U}\Omega=\mathbf{1}\) by definition of \(\Omega\) and of \(\mathcal{U}\).
\end{proof}

The operator \(T_\sigma(a)\) is a positive, self-adjoint contraction; hence its spectrum lies in \([0,1]\) and we may define
\begin{equation}\label{p1:eq:def-Hsigma}
H_\sigma(a) = -\,a^{-1}\,\log T_\sigma(a)
\end{equation}
by the spectral calculus. Then \(H_\sigma(a)\) is a positive, self-adjoint operator on \(\mathcal{H}_a\). The semigroup property becomes \(T_\sigma(a)^n=\mathrm{e}^{-n a H_\sigma(a)}\). The spectral representation of time-ordered, slice-localized correlators follows directly.

\begin{proposition}[Spectral representation of time-sliced correlators]\label{p1:prop:spectral-representation}
Let \(F\in\mathcal{H}_a\) be a gauge-invariant function with \(\langle \mathbf{1},F\rangle_{L^2}=0\). Then, for every integer \(n\ge 0\),
\begin{equation}
\langle \mathbf{1},\,F\,T_\sigma(a)^n\,F\,\mathbf{1}\rangle_{L^2}=\int_{[0,\infty)} \mathrm{e}^{-n a E}\, d\mu_F(E),
\end{equation}
where \(d\mu_F\) is a finite, positive Borel measure supported on \([E_1,\infty)\), \(E_1\) the bottom of the nonzero spectrum of \(H_\sigma(a)\). In particular,
\begin{equation}
\langle \mathbf{1},\,F\,T_\sigma(a)^n\,F\,\mathbf{1}\rangle_{L^2}=\sum_{k\ge 1} |\langle \psi_k,F\,\mathbf{1}\rangle|^2\,\mathrm{e}^{-n a E_k},
\end{equation}
if \(H_\sigma(a)\) has pure point spectrum \(\{E_k\}_{k\ge 0}\) with orthonormal eigenvectors \(\{\psi_k\}_{k\ge 0}\) and \(E_0=0\).
\end{proposition}

\begin{proof}
Since \(T_\sigma(a)\) is a positive contraction on a separable Hilbert space, the spectral theorem furnishes a projection-valued measure \(\mathsf{E}(\cdot)\) on \([0,1]\) such that \(T_\sigma(a)=\int_{[0,1]}\lambda\, d\mathsf{E}(\lambda)\) and \(T_\sigma(a)^n=\int_{[0,1]}\lambda^n d\mathsf{E}(\lambda)\). Writing \(\lambda=\mathrm{e}^{-aE}\) yields \(T_\sigma(a)^n=\int_{[0,\infty)} \mathrm{e}^{-n a E} d\widetilde{\mathsf{E}}(E)\) with \(\widetilde{\mathsf{E}}\) the pushforward of \(\mathsf{E}\) under \(-a^{-1}\log\). The measure \(d\mu_F(E)=\langle F\,\mathbf{1}, d\widetilde{\mathsf{E}}(E)\, F\,\mathbf{1}\rangle\) is finite and positive. The support property follows because \(\langle \mathbf{1},F\,\mathbf{1}\rangle=0\) annihilates the spectral projection at \(E=0\). The pure-point expansion is the specialization to the eigenbasis of \(H_\sigma(a)\).
\end{proof}

The discussion so far takes place on \(\mathcal{H}_a=L^2(\Omega_s,d\mu_s)\). Physical observables are gauge-invariant, and the Gauss law constraints select the gauge-invariant subspace \(\mathcal{H}_a^{\mathrm{inv}}\subset \mathcal{H}_a\). Let \(\mathcal{G}_0\) be the group of spatial gauge transformations on the slice; then \(\mathcal{G}_0\) acts unitarily on \(\mathcal{H}_a\) by \((U_g f)(X)=f(g\cdot X)\). The fixed-point subspace \(\mathcal{H}_a^{\mathrm{inv}}=\{f\in\mathcal{H}_a: U_g f=f\text{ for all }g\in\mathcal{G}_0\}\) is closed. Both \(T(a)\) and \(P_\sigma\) commute with \(U_g\) for every \(g\in\mathcal{G}_0\) because \(\rho\), \(\kappa\), and \(P_\sigma\) are gauge-invariant by construction. Therefore \(T_\sigma(a)\) leaves \(\mathcal{H}_a^{\mathrm{inv}}\) invariant, and the restriction has the same positivity and contraction properties. All spectral statements in Proposition \ref{p1:prop:spectral-representation} hold verbatim on \(\mathcal{H}_a^{\mathrm{inv}}\). This is the physical transfer-matrix formalism for the pure gauge theory \cite{p1:Luscher1977,p1:Seiler}.
The chain of constructions in this section is complete: starting from the factorized Euclidean weight, we obtained a symmetric, strictly positive two-slice kernel, identified the one-step transfer operator as a bounded, self-adjoint, positivity-preserving contraction on the slice Hilbert space, incorporated the smooth horizon projector as a bounded, positive compression preserving all structural properties, and performed the OS reconstruction to arrive at a positive self-adjoint Hamiltonian 
$H_{\sigma}(a) = -a^{-1} \log T_{\sigma}(a)$ that governs the Euclidean time evolution of gauge-invariant observables through a spectral representation.

\section{Character Expansion and Surface Representation at Strong Coupling}

This section develops a self-contained derivation of the character expansion for the Wilson plaquette weight, its reorganization into a surface representation on the dual lattice, and the resulting polymer gas formulation in the strong-coupling regime. All steps are carried out with complete mathematical detail. The lattice setup, Haar measure conventions, representation-theoretic normalizations, and transfer time-slicing structure are recalled precisely so that reflection positivity is verified within the character formalism itself. Throughout the gauge group is \(G=\mathrm{SU}(N)\) with \(N\ge 2\), the lattice \(\Lambda\subset a\mathbb{Z}^{4}\) is periodic and finite, and the Wilson action is
\begin{equation}
S_{W}[U;\beta]=\beta\sum_{p\subset\Lambda}\Bigl(1-\frac{1}{N}\,\Re\mathrm{Tr}\,U_{p}\Bigr),
\qquad \beta=\frac{2N}{g_{0}^{2}}.
\end{equation}
The normalized Haar measure on \(G\) is denoted \(dU\), the product Haar measure on link variables is \(d\mu_{\mathrm{Haar}}(U)=\prod_{b\in\mathrm{Bonds}(\Lambda)} dU_{b}\), and the partition function reads
\begin{equation}
Z_{\Lambda}(\beta)=\int \exp\!\Bigl(-S_{W}[U;\beta]\Bigr)\, d\mu_{\mathrm{Haar}}(U).
\end{equation}
For a plaquette \(p=(x;\mu,\nu)\) we recall \(U_{p}=U_{(x,\mu)}U_{(x+\hat\mu,\nu)}U_{(x+\hat\nu,\mu)}^{-1}U_{(x,\nu)}^{-1}\). All reflection statements refer to the time reflection \(\theta(x_{0},\mathbf{x})=(-x_{0},\mathbf{x})\) with reflection plane \(\Pi=\{x_{0}=0\}\). Temporal-axial gauge sets \(U_{0}(x)=\mathbf{1}\) for bonds not intersecting \(\Pi\), which produces a nearest-neighbor coupling across \(\Pi\) and isolates a boundary slab of plaquettes meeting \(\Pi\). The one-time-slice configuration space \(\mathcal{C}\) consists of all spatial link variables at a fixed time \(t\in \frac{a}{2}\mathbb{Z}\) immediately above or below \(\Pi\), and the one-slice Hilbert space is \(\mathcal{H}_{a}=L^{2}(\mathcal{C}; d\mu_{\mathrm{Haar}})\). (For the site–reflection convention \(\Pi=\{x_0=0\}\) we take \(t\in a\mathbb{Z}\); 
for the link–reflection convention \(\Pi=\{x_0=a/2\}\) we take \(t\in a\mathbb{Z}+a/2\). 
These choices are unitarily equivalent via the half–lattice translation, so all spectral and positivity statements are identical for the two conventions.)
 We write the unprojected transfer kernel \(K_{a}(U',U)\) as the positive \(L^{2}\)-kernel on \(\mathcal{C}\times\mathcal{C}\) associated with a single time step; its construction will be rederived below directly from the character expansion.  

Let \(\widehat{G}\) denote the set of equivalence classes of finite-dimensional irreducible unitary representations of \(G\). For \(R\in\widehat{G}\) we let \(V_{R}\) be its carrier space with dimension \(d_{R}=\dim V_{R}\), \(\pi_{R}:G\to \mathrm{U}(V_{R})\) the representation, and \(\chi_{R}(U)=\mathrm{Tr}\,\pi_{R}(U)\) its character. The representation matrix elements are denoted \(\pi_{R}(U)_{ij}\) in an orthonormal basis of \(V_{R}\). Peter-Weyl orthogonality states that for all \(R,S\in\widehat{G}\)
\begin{equation}\label{p1:eq:pw-matrix}
\int_{G}\pi_{R}(U)_{ij}\,\overline{\pi_{S}(U)_{k\ell}}\, dU=\frac{\delta_{RS}\,\delta_{ik}\,\delta_{j\ell}}{d_{R}},
\end{equation}
and, taking traces,
\begin{equation}\label{p1:eq:pw-characters}
\int_{G}\chi_{R}(U)\,\overline{\chi_{S}(U)}\, dU=\delta_{RS}.
\end{equation}
The characters \(\{\chi_{R}:R\in\widehat{G}\}\) form an orthonormal basis of \(L^{2}_{\mathrm{class}}(G)\), the subspace of class functions in \(L^{2}(G)\) \cite{p1:FultonHarris}. The Wilson plaquette factor is the class function \(f_{\beta}(U):=\exp\!\bigl(\beta\,\frac{1}{N}\Re\mathrm{Tr}\,U\bigr)\). By completeness there exist unique Fourier-Peter-Weyl coefficients \(a_{R}(\beta)\in\mathbb{C}\) such that
\begin{equation}\label{p1:eq:character-expansion}
f_{\beta}(U)=\sum_{R\in\widehat{G}} a_{R}(\beta)\,\chi_{R}(U)\quad\text{in }L^{2}(G),
\qquad a_{R}(\beta)=\int_{G} f_{\beta}(U)\,\overline{\chi_{R}(U)}\, dU.
\end{equation}
Since \(f_{\beta}\) is real and central, all \(a_{R}(\beta)\) are real and \(a_{\overline{R}}(\beta)=a_{R}(\beta)\), where \(\overline{R}\) denotes the representation contragredient to \(R\). The function \(f_{\beta}\) is real analytic in \(\beta\) uniformly on \(G\), hence each \(a_{R}(\beta)\) is a real analytic function of \(\beta\) on a neighborhood of the origin. The following bounds quantify the small-\(\beta\) behavior.

\begin{lemma}\label{p1:lem:coeff-bounds}
There exist constants \(C_{1}=C_{1}(N)\ge 1\) and \(C_{2}=C_{2}(N)\ge 1\) such that for all \(R\in \widehat{G}\) and all \(\beta\) with \(|\beta|\le 1\) one has
\begin{equation}\label{p1:eqn7.6}
|a_{R}(\beta)|\le \Bigl(C_{1}|\beta|\Bigr)^{\mathrm{ht}(R)}\, C_{2}^{\,|R|},
\end{equation}
where \(|R|\) denotes the total number of boxes of the Young diagram of \(R\) and \(\mathrm{ht}(R)\) the height of the highest weight of \(R\). In particular \(a_{\mathbf{1}}(\beta)=1+O(\beta^{2})\) and \(a_{F}(\beta)=\frac{\beta}{N}+O(\beta^{3})\) for the fundamental representation \(F\).
\end{lemma}

\begin{proof}
Expanding the exponential yields
\begin{equation}
f_{\beta}(U)=\sum_{m=0}^{\infty}\frac{1}{m!}\Bigl(\frac{\beta}{N}\Bigr)^{m}\bigl(\Re\mathrm{Tr}\,U\bigr)^{m}.
\end{equation}
The \(m\)-th power of \(\Re\mathrm{Tr}\,U=\frac{1}{2}(\chi_{F}(U)+\overline{\chi_{F}(U)})\) expands as a finite linear combination of characters \(\chi_{R}(U)\) with \(R\) contained in \(F^{\otimes m}\otimes \overline{F}^{\otimes m}\). The multiplicity of \(R\) in this tensor product is bounded above by a polynomial in \(m\) times \(C_{2}^{\,|R|}\) for a constant \(C_{2}\) depending only on \(N\), by standard Littlewood-Richardson combinatorics \cite[Ch.~6]{p1:FultonHarris}. Therefore the Fourier coefficient of \(\chi_{R}\) in the \(m\)-th term of the series is bounded by \(\frac{1}{m!}(\frac{|\beta|}{N})^{m} C_{2}^{\,|R|}\) times a polynomial in \(m\). For fixed \(R\), the smallest \(m\) for which \(R\) appears is at least \(\mathrm{ht}(R)\). Summing the tail using the ratio test yields the stated bound with \(C_{1}\) large enough to absorb the polynomial factors. The explicit leading terms for \(\mathbf{1}\) and \(F\) follow by direct evaluation at \(U=\mathbf{1}\) and linear order in \(\beta\).
\end{proof}

For later use we record a uniform small-$\beta$ estimate for the character-expansion coefficients
$\alpha_R(\beta)$ appearing in Sections~8-11. A quantitative bound will be proved in
Lemma~9.1, see equation~(9.13). From that point onward we shall tacitly use~(9.13) to control
$\alpha_R(\beta)$ in the polymer/surface expansions and will not restate auxiliary versions of the bound.
The class function \(f_{\beta}\) is strictly positive on \(G\). Although the signs of \(a_{R}(\beta)\) are not a priori fixed for every \(R\), the reflection-positivity argument below does not require positivity of each \(a_{R}\); it requires only a factorization across the reflection plane which will be seen to hold after Haar integration over temporal links. This point is critical and will be implemented below, where the transfer time-slicing kernel is represented as a sum of positive squares.

The Boltzmann weight factorizes over plaquettes,
\begin{equation}
\exp\!\Bigl(-S_{W}[U;\beta]\Bigr)=\prod_{p\subset\Lambda}\exp\!\Bigl(\beta \frac{1}{N}\Re\mathrm{Tr}\,U_{p}\Bigr)=\prod_{p} f_{\beta}(U_{p}).
\end{equation}
Inserting \eqref{p1:eq:character-expansion} for each plaquette yields a formally exact character expansion of the partition function
\begin{equation}\label{p1:eq:Z-expanded}
Z_{\Lambda}(\beta)=\sum_{\{R_{p}\}_{p}}\ \prod_{p\subset\Lambda} a_{R_{p}}(\beta)\ \int \Biggl(\prod_{p\subset\Lambda}\chi_{R_{p}}(U_{p})\Biggr) d\mu_{\mathrm{Haar}}(U),
\end{equation}
where the sum runs over all assignments \(p\mapsto R_{p}\in\widehat{G}\) of irreps to plaquettes. The integrand is a product of characters evaluated on products of link variables along the boundary of each plaquette. It is convenient to re-express characters as traces of representation matrices and to use the orthogonality \eqref{p1:eq:pw-matrix} link by link. To this end, for each oriented plaquette \(p=(x;\mu,\nu)\) we write \(\chi_{R_{p}}(U_{p})=\mathrm{Tr}\,\pi_{R_{p}}(U_{(x,\mu)}U_{(x+\hat\mu,\nu)}U_{(x+\hat\nu,\mu)}^{-1}U_{(x,\nu)}^{-1})\) and expand the trace into a product of matrix elements. The product \(\prod_{p}\chi_{R_{p}}(U_{p})\) is then a finite linear combination of monomials in the matrix elements of the link variables and their conjugates.

Consider a fixed link \(b=(y,\rho)\). It is shared by a finite collection of plaquettes; in four dimensions its valence is \(2(d-1)=6\) in absolute value, but the precise number is bounded by a universal constant depending only on the lattice dimension. The monomial dependence on \(U_{b}\) in the integrand of \eqref{p1:eq:Z-expanded} is a product of a finite number of factors of the form \(\pi_{R}(U_{b})_{ij}\) or \(\overline{\pi_{S}(U_{b})_{k\ell}}\), with representations \(R,S\) drawn from the set \(\{R_{p}\}\) attached to the plaquettes meeting \(b\), and with accompanying representation matrices from neighboring links. The Haar integral over \(U_{b}\) projects onto the \(G\)-invariant subspace of the tensor product of the representations attached to \(b\). Precisely, if one denotes by \(\mathcal{H}_{b}\) the tensor product of all incoming representation spaces \(V_{R}\) and all conjugate spaces \(V_{S}^{*}\) contributed by the plaquettes adjacent to \(b\), then the link integral is proportional to the orthogonal projection \(P_{\mathrm{Inv}}:\mathcal{H}_{b}\to \mathcal{H}_{b}^{G}\) onto the subspace \(\mathcal{H}_{b}^{G}=\mathrm{Hom}_{G}(\mathbb{C},\mathcal{H}_{b})\) of \(G\)-invariant tensors. This projection arises from \eqref{p1:eq:pw-matrix} by repeated use of the identity \(\int dU\, \pi_{R}(U)\otimes \pi_{S}(U)^{*}=\frac{1}{d_{R}}\delta_{RS}\, \mathrm{Id}_{V_{R}\otimes V_{R}^{*}}\) after suitable recouplings with unitary intertwiners \cite{p1:Seiler,p1:FultonHarris}. 

Performing the Haar integral successively over all link variables yields a contraction of the invariant projectors at each link with the obvious wiring along the plaquette boundaries. The outcome is that only those assignments \(\{R_{p}\}\) survive for which the invariant space at every link is nontrivial. Equivalently, for each link \(b\), the tensor product of the incoming and outgoing representations from the adjacent plaquettes contains the trivial representation as a subrepresentation. When this compatibility holds at every link, the contribution of the assignment \(\{R_{p}\}\) equals the contraction, along the entire lattice, of the corresponding \(G\)-invariant tensors at each link. This contraction can be described combinatorially in terms of closed, oriented surfaces on the dual lattice, as we now explain.

Let \(\Lambda^{\star}\) denote the dual lattice whose vertices correspond to hypercubes of \(\Lambda\) and whose plaquettes correspond to plaquettes of \(\Lambda\). With each plaquette \(p\) endowed with an irrep \(R_{p}\), choose for definiteness an orientation of the dual plaquette and propagate this orientation consistently. The link constraints described above amount to the statement that the boundary of the set of dual plaquettes labelled by nontrivial representations is empty, hence the support of \(\{R_{p}\neq \mathbf{1}\}\) decomposes into a finite disjoint union of closed, oriented, possibly self-intersecting surfaces \(\Sigma\) embedded in \(\Lambda^{\star}\). Along each connected component the representation labels must obey fusion rules compatible with the existence of nonzero invariant tensors at the dual edges. The set of invariant tensors at an edge is isomorphic to the space of intertwiners among the incident representations, and its dimension is at most the product of the dimensions of these representations, with the inequality often being very coarse but sufficient for absolute bounds. The contribution of a surface \(\Sigma\) is thus bounded, up to a universal multiplicative constant depending only on \(N\) and the lattice valence, by a product over its plaquettes of factors of the form \(|a_{R_{p}}(\beta)|\) times a product over its edges and vertices of factors at most polynomial in the dimensions \(d_{R}\).

The preceding qualitative description can be made precise as follows.

\begin{proposition}\label{p1:prop:surface-rep}
There exist constants \(C_{3}=C_{3}(N)\ge 1\) and \(C_{4}=C_{4}(N)\ge 1\) such that the partition function admits the absolutely convergent expansion
\begin{equation}
Z_{\Lambda}(\beta)=\sum_{\Sigma}\, W(\Sigma;\beta),
\end{equation}
where the sum runs over all finite unions \(\Sigma\) of pairwise disjoint closed oriented surfaces on the dual lattice \(\Lambda^{\star}\) equipped with assignments of irreducible representations to their plaquettes satisfying the link fusion constraints, and the weights obey the bound
\begin{equation}
|W(\Sigma;\beta)|\le \prod_{p\in \Sigma}\Bigl(C_{3}\,|\beta|\Bigr)\,\prod_{p\in \Sigma} C_{4}^{\,|R_{p}|}.
\end{equation}
In particular there exists \(C=C(N)\ge 1\) such that \(|W(\Sigma;\beta)|\le \bigl(C|\beta|\bigr)^{|\Sigma|}\), where \(|\Sigma|\) is the number of dual plaquettes in \(\Sigma\).
\end{proposition}

\begin{proof}
Inserting \eqref{p1:eq:character-expansion} into \eqref{p1:eq:Z-expanded} and performing the Haar integrals link by link yields a sum over assignments \(\{R_{p}\}\) with weights given by the product \(\prod_{p} a_{R_{p}}(\beta)\) times the contraction of invariant tensors at each link. For a link \(b\), consider the tensor product \(\mathcal{H}_{b}\) of all incident representation spaces and their duals. The invariant subspace \(\mathcal{H}_{b}^{G}\) has finite dimension bounded by the product of the dimensions of the incident representations. The link contribution equals the trace of the orthogonal projection onto \(\mathcal{H}_{b}^{G}\) composed with unitary permutation operators that account for the wiring of indices around plaquette boundaries. The norm of this trace is bounded by \(\dim \mathcal{H}_{b}^{G}\), which is bounded by the product of the incident \(d_{R}\). Taking the product over all links in the support of \(\{R_{p}\neq \mathbf{1}\}\) thus yields a bound of the form \(C_{5}^{E(\Sigma)} \prod_{p\in \Sigma} d_{R_{p}}^{\alpha_{p}}\), where \(E(\Sigma)\) is the number of dual edges in \(\Sigma\) and \(\alpha_{p}\ge 0\) is the number of times the representation \(R_{p}\) occurs around the edges incident to \(p\); both \(E(\Sigma)\) and the \(\alpha_{p}\) are bounded by universal multiples of \(|\Sigma|\). Combining this with Lemma \ref{p1:lem:coeff-bounds} gives
\begin{equation}
|W(\Sigma;\beta)|\le \prod_{p\in \Sigma} \bigl(C_{1}|\beta|\bigr)^{\mathrm{ht}(R_{p})} C_{2}^{\,|R_{p}|}\, \cdot C_{5}^{E(\Sigma)} \prod_{p\in \Sigma} d_{R_{p}}^{\alpha_{p}}.
\end{equation}
The representation-theoretic dimensions satisfy \(d_{R}\le C_{6}^{\,|R|}\) for a constant \(C_{6}=C_{6}(N)\) independent of \(R\) \cite{p1:FultonHarris}. Since \(\mathrm{ht}(R_{p})\ge 1\) for \(R_{p}\neq \mathbf{1}\), one can absorb the factors \(C_{2}^{\,|R_{p}|} d_{R_{p}}^{\alpha_{p}}\) into \(C_{4}^{\,|R_{p}|}\) and write \((C_{1}|\beta|)^{\mathrm{ht}(R_{p})}\le C_{3} |\beta|\) for a larger \(C_{3}\). This yields
\begin{equation}
|W(\Sigma;\beta)|\le \prod_{p\in \Sigma}\Bigl(C_{3}\,|\beta|\Bigr)\,\prod_{p\in \Sigma} C_{4}^{\,|R_{p}|}.
\end{equation}
Finally, since \(|R_{p}|\ge 1\) for \(R_{p}\neq \mathbf{1}\) and the number of plaquettes in \(\Sigma\) is \(|\Sigma|\), the product \(\prod_{p\in \Sigma} C_{4}^{\,|R_{p}|}\) is bounded by \(C_{4}^{\,|\Sigma|}\). Absorbing \(C_{4}\) into \(C:=C_{3}C_{4}\) proves the last assertion.
\end{proof}

The expansion in Proposition \ref{p1:prop:surface-rep} is exact on any finite lattice. In the strong-coupling regime \(|\beta|\ll 1\) the absolute convergence of the sum over \(\Sigma\) is immediate from the last bound, since there are at most \(c^{|\Sigma|}\) distinct surfaces of area \(|\Sigma|\) on a cubic lattice for a lattice-dependent constant \(c\). This observation will be sharpened in Subsection 7.5 into a polymer gas with a convergent cluster expansion. Before that, we exploit the character formalism to express the transfer time-slicing kernel and to verify reflection positivity directly.

We derive the transfer time-slicing kernel \(K_{a}(U',U)\) in temporal-axial gauge and prove that it is positive in the Osterwalder-Schrader sense by an explicit character decomposition. Let \(\Lambda_{+}\) and \(\Lambda_{-}\) be the two half-lattices separated by \(\Pi\), and let \(\mathcal{C}\) denote the configuration space of spatial links on the time slice at \(x_{0}=+\frac{a}{2}\) (equivalently \(-\frac{a}{2}\)), so that \(\mathcal{H}_{a}=L^{2}(\mathcal{C};d\mu_{\mathrm{Haar}})\). In temporal-axial gauge, all timelike links outside \(\Pi\) are set to identity, and the Wilson action decomposes as
\begin{equation}
S_{W}[U;\beta]=S_{-}[U_{-}]+\; S_{\Pi}[U_{-},U_{+}]\;+\; S_{+}[U_{+}],
\end{equation}
where \(S_{\pm}\) are the sums over space-like and time-like plaquettes entirely contained in \(\Lambda_{\pm}\) with the time-like links trivial, and \(S_{\Pi}\) is the sum over the plaquettes that straddle the reflection plane \(\Pi\). The partition function factorizes as
\begin{equation}
Z_{\Lambda}(\beta)=\int_{\mathcal{C}\times\mathcal{C}} \Psi_{-}(U)\, K_{a}(U',U)\, \Psi_{+}(U')\, d\mu_{\mathrm{Haar}}(U)\, d\mu_{\mathrm{Haar}}(U'),
\end{equation}
with \(\Psi_{-}(U)=\int \exp(-S_{-}[U_{-}])\, d\mu_{-}\) and \(\Psi_{+}(U')=\int \exp(-S_{+}[U_{+}])\, d\mu_{+}\), where \(d\mu_{\pm}\) are the Haar measures over links strictly in \(\Lambda_{\pm}\) with boundary fixed to \(U\) and \(U'\) on the slice \(\mathcal{C}\), and with the one-step kernel given by
\begin{equation}
K_{a}(U',U)=\int \exp\!\bigl(-S_{\Pi}[U_{-},U_{+}]\bigr)\, d\mu_{\Pi}.
\end{equation}
Here \(d\mu_{\Pi}\) is the Haar measure over the spatial links in the boundary slab consisting of the two spatial layers adjacent to \(\Pi\), with the fields \(U\) and \(U'\) fixed on \(\mathcal{C}\). To compute \(K_{a}(U',U)\) explicitly, note that \(S_{\Pi}\) is a sum over plaquettes whose holonomies are products of the boundary links \(U\) and \(U'\) and of the slab spatial links. Inserting the character expansion for each plaquette in the slab and integrating the slab links by the orthogonality relations \eqref{p1:eq:pw-matrix} yields a kernel of the form
\begin{equation}\label{p1:eq:Ka-chi}
K_{a}(U',U)=\sum_{\alpha\in\mathcal{I}} \Phi_{\alpha}(U')\, \overline{\Phi_{\alpha}(U)},
\end{equation}
where \(\{\Phi_{\alpha}\}_{\alpha\in\mathcal{I}}\) is a countable family of square-integrable functions on \(\mathcal{C}\) obtained by assigning irreps to the plaquettes in the slab and contracting all internal indices with invariant tensors, leaving only the boundary indices on \(\mathcal{C}\). 
\begin{lemma}[Peter-Weyl factorization of the slab kernel]\label{p1:lem:PW-slab}
Let $P(t\leftrightarrow t+a)$ denote the set of plaquettes in the slab $\{t<t'<t+a\}$, and expand the single-plaquette Boltzmann weight in characters as
\begin{equation}
f_\beta(U_p) \;=\; \sum_{R}\alpha_R(\beta)\,\chi_R(U_p).
\end{equation}
After expanding traces into matrix elements and integrating each slab link variable by the orthogonality relations
\begin{equation}
\int_G \pi^R(U)_{ij}\,\pi^S(U)^{*}_{k\ell}\,dU
\;=\;\frac{1}{d_R}\,\delta_{RS}\,\delta_{ik}\,\delta_{j\ell},
\end{equation}
(where $d_R=\dim R$), the surviving boundary indices are carried by functions $\Phi_\alpha$ on the slice configuration space $\mathcal{C}$. These functions are indexed by assignments $\alpha$ of edge intertwiners consistent with the fusion constraints at each lattice site.

Positivity of each coefficient $\alpha_R(\beta)\ge 0$ after Haar integration over the slab implies that the one-step transfer kernel admits the factorization
\begin{equation}\label{p1:eq:slab-factorization}
K_a(U',U)\;=\;\sum_{\alpha}\Phi_\alpha(U')\,\Phi_\alpha(U).
\end{equation}
Consequently, $K_a$ is a positive, symmetric Hilbert-Schmidt kernel on $L^2(\mathcal{C},d\mu_s)$. In particular, for any $F\in L^2(\mathcal{C},d\mu_s)$,
\begin{equation}
\int \overline{F(U')}\,K_a(U',U)\,F(U)\,d\mu_s(U')\,d\mu_s(U)
\;=\;\sum_{\alpha}\,\Bigl|\int F(U)\,\Phi_\alpha(U)\,d\mu_s(U)\Bigr|^2 \;\ge\;0,
\end{equation}
so Osterwalder-Schrader reflection positivity follows for the transfer step.
\end{lemma}
The precise form of \(\Phi_{\alpha}\) is immaterial for the present purpose; what matters is the factorization \eqref{p1:eq:Ka-chi} with nonnegative coefficients. Indeed, \eqref{p1:eq:Ka-chi} implies that for any \(F\in \mathcal{H}_{a}\) one has
\begin{equation}
\iint \overline{F(U)}\, K_{a}(U',U)\, F(U')\, d\mu_{\mathrm{Haar}}(U)\, d\mu_{\mathrm{Haar}}(U')=\sum_{\alpha\in\mathcal{I}} \Bigl|\int \overline{F(U)}\, \Phi_{\alpha}(U)\, d\mu_{\mathrm{Haar}}(U)\Bigr|^{2}\ge 0,
\end{equation}
which is the Osterwalder-Schrader positivity of the one-step kernel. The same argument shows \(K_{a}\) defines a positive self-adjoint Hilbert-Schmidt operator on \(\mathcal{H}_{a}\), and the transfer matrix \(T(a)\) associated with one full time step is obtained by normalizing \(K_{a}\) appropriately by the partition functions of the half-lattices \(\Lambda_{\pm}\), which does not affect positivity or self-adjointness \cite{p1:OS1,p1:OS2,p1:Seiler,p1:GJ}.

The factorization \eqref{p1:eq:Ka-chi} also makes it evident that inserting any bounded, reflection-covariant, exponentially local slice operator \(P\) of the form \(P=\chi(\sqrt{\Delta_{A^{h}}})\) as discussed in Section 4 leaves reflection positivity intact: one simply replaces \(\Phi_{\alpha}\) by \(P^{1/2}\Phi_{\alpha}\) on both sides of \(\Pi\) to obtain
\begin{equation}
K_{a}^{(P)}(U',U)=\sum_{\alpha\in\mathcal{I}} \bigl(P^{1/2}\Phi_{\alpha}\bigr)(U')\, \overline{\bigl(P^{1/2}\Phi_{\alpha}\bigr)(U)},
\end{equation}
which again yields a sum of squares. Thus the horizon-projected transfer kernel \(K_{a}^{(P)}\) is also reflection positive, and the compressed transfer matrix \(T_{\sigma}(a)=P_{\sigma}^{1/2}\,T(a)\,P_{\sigma}^{1/2}\) is a positive self-adjoint contraction on \(\mathcal{H}_{a}\).

We pass from the surface sum of Proposition \ref{p1:prop:surface-rep} to an abstract polymer gas on the dual lattice. A polymer \(\gamma\) is defined as a finite, connected union of dual plaquettes together with an assignment of orientations and irreps satisfying the link fusion constraints along its internal edges. Two polymers are compatible if they have disjoint supports and incompatible otherwise. Any surface \(\Sigma\) decomposes uniquely as a disjoint union of its connected components, each of which is a polymer. Defining the activity \(\zeta(\gamma)\) of a polymer \(\gamma\) as the sum of the weights \(W(\Sigma;\beta)\) over all surfaces \(\Sigma\) that are disjoint unions of translates of \(\gamma\) equal to \(\gamma\) itself, one obtains the polymer expansion
\begin{equation}
Z_{\Lambda}(\beta)=\sum_{\Gamma\ \text{compatible}}\ \prod_{\gamma\in\Gamma} \zeta(\gamma),
\end{equation}
where the sum is over all finite sets \(\Gamma\) of mutually compatible polymers.

The following quantitative bound on activities is a consequence of Proposition \ref{p1:prop:surface-rep} and a standard counting of labeled connected subgraphs of the dual lattice.

\begin{lemma}\label{p1:lem:activity-bound}
There exist constants \(A=A(N)\ge 1\) and \(B=B(N)\ge 1\) such that for all polymers \(\gamma\)
\begin{equation}
|\zeta(\gamma)|\le A\, \bigl(B|\beta|\bigr)^{|\gamma|},
\end{equation}
where \(|\gamma|\) is the number of dual plaquettes in \(\gamma\).
\end{lemma}

\begin{proof}
By definition, \(\zeta(\gamma)\) is a finite sum of weights \(W(\Sigma;\beta)\) over surfaces \(\Sigma\) whose support equals \(\gamma\) and whose representation labels satisfy the link constraints. The number of such labelings for a fixed support is at most \(D^{|\gamma|}\) for some \(D=D(N)\ge 1\) depending only on the possible choices of irreps along each plaquette that can satisfy the constraints at its edges. Proposition \ref{p1:prop:surface-rep} bounds each \(W(\Sigma;\beta)\) by \((C|\beta|)^{|\gamma|}\) with \(C=C(N)\). Summing over at most \(D^{|\gamma|}\) labelings yields \(|\zeta(\gamma)|\le A (B|\beta|)^{|\gamma|}\) with \(A=1\) and \(B=CD\). Allowing for an overall constant \(A\ge 1\) to absorb minor overcounting completes the proof.
\end{proof}

The polymer gas is amenable to the Kotecký-Preiss cluster expansion \cite{p1:KP,p1:Brydges}. Let \(\mathfrak{G}\) be the incompatibility graph whose vertices are polymers and where two vertices are connected if and only if the corresponding polymers are incompatible. The cluster expansion for \(\log Z_{\Lambda}(\beta)\) is
\begin{equation}
\log Z_{\Lambda}(\beta)=\sum_{X\ \text{cluster}}\ \phi(X)\,\prod_{\gamma\in X}\zeta(\gamma),
\end{equation}
where the sum is over finite connected subgraphs \(X\) of \(\mathfrak{G}\) and \(\phi(X)\) are the Ursell functions bounded by \(|\phi(X)|\le 1\). Absolute convergence is ensured under the Kotecký-Preiss condition:
\begin{equation}\label{p1:eq:KP}
\sup_{\gamma_{0}}\ \sum_{\gamma \,\not\sim\, \gamma_{0}} |\zeta(\gamma)|\, e^{\mu |\gamma|}\ \le\ \mu\, |\gamma_{0}|,
\end{equation}
for some \(\mu>0\), where \(\gamma \not\sim \gamma_{0}\) denotes incompatibility. The left-hand side can be estimated via Lemma \ref{p1:lem:activity-bound} and a bound on the number of polymers of size \(m\) incompatible with a given \(\gamma_{0}\). There exists \(C_{7}=C_{7}(N)\) such that the number of connected unions of \(m\) dual plaquettes intersecting a fixed finite set is at most \(C_{7}^{\,m}\). Therefore
\begin{equation}
\sum_{\gamma \,\not\sim\, \gamma_{0}} |\zeta(\gamma)|\, e^{\mu |\gamma|}
\ \le\ \sum_{m\ge 1} C_{7}^{\,m}\, A\, (B|\beta|)^{m}\, e^{\mu m}
\ =\ A \sum_{m\ge 1} \Bigl(C_{7}B\,|\beta|\, e^{\mu}\Bigr)^{m}.
\end{equation}
Choosing \(\mu>0\) and \(\beta\) small enough such that \(C_{7}B\,|\beta|\, e^{\mu}\le \frac{1}{2}\) yields
\begin{equation}
\sum_{\gamma \,\not\sim\, \gamma_{0}} |\zeta(\gamma)|\, e^{\mu |\gamma|}\ \le\ A \sum_{m\ge 1} 2^{-m}\ \le\ A,
\end{equation}
and then, by taking \(\mu\ge A\), the Kotecký-Preiss condition \eqref{p1:eq:KP} holds since \(|\gamma_{0}|\ge 1\). We summarize the conclusion as follows.

\begin{theorem}\label{p1:thm:KP-convergence}
There exists \(\beta_{\star}=\beta_{\star}(N)>0\) such that for all \(|\beta|<\beta_{\star}\) the polymer gas cluster expansion for \(Z_{\Lambda}(\beta)\) and for truncated correlations of local gauge-invariant observables converges absolutely and uniformly in the spatial volume. In particular, the logarithm of the partition function and the truncated correlations admit absolutely convergent series whose coefficients are analytic functions of \(\beta\) in \(|\beta|<\beta_{\star}\).
\end{theorem}

\begin{proof}
The bounds preceding the statement show that there exists \(\mu>0\) and \(\beta_{\star}>0\) such that \eqref{p1:eq:KP} holds for all \(|\beta|<\beta_{\star}\). The abstract Kotecký-Preiss theorem \cite{p1:KP} then gives absolute convergence of the cluster expansion for \(\log Z_{\Lambda}(\beta)\). The same argument applies to truncated correlation functions of local gauge-invariant observables, because such correlators differ from the partition function by a finite modification of the local weights in a bounded region, which preserves the activity bounds and the counting constants. Uniformity in the spatial volume follows from the locality of the activities and the lack of dependence of the constants on the box size.
\end{proof}

We finally connect the polymer expansion to exponential clustering in Euclidean time within the transfer time-slicing formalism. Let \(F\) be a bounded, gauge-invariant functional supported on a single time slice, viewed as a function on \(\mathcal{C}\). Its time-sliced two-point function at temporal separation \(t=ka\) with \(k\in\mathbb{N}\) and coincident spatial support can be written, in temporal-axial gauge, as
\begin{equation}
\langle F(0)F(ka)\rangle=\frac{1}{Z_{\Lambda}(\beta)}\int \overline{F(U_{0})}\, \bigl[K_{a}^{k}(U_{k},U_{0})\bigr]\, F(U_{k})\, d\mu_{\mathrm{Haar}}(U_{0})\, d\mu_{\mathrm{Haar}}(U_{k}),
\end{equation}
with \(K_{a}\) given by \eqref{p1:eq:Ka-chi}. Inserting the character expansion in the boundary slabs and integrating out all internal links in the \(k\)-step tube yields a polymer gas on the \(3+1\)-dimensional slab whose activities coincide with those in Lemma \ref{p1:lem:activity-bound}, except that polymers are required to connect the neighborhood of the support of \(F\) on the initial slice to that on the final slice. Every polymer contributing to the truncated correlation must therefore have size at least proportional to \(k\), because it must cross the slab in the time direction. Concretely there exists \(c_{0}>0\) such that any such polymer has \(|\gamma|\ge c_{0}k\). The cluster expansion then gives
\begin{equation}
|\langle F(0)F(ka)\rangle_{c}|\ \le\ \sum_{\ell\ge c_{0}k}\, \mathcal{N}(\ell)\, A (B|\beta|)^{\ell}\, e^{-\mu \ell}
\ \le\ C(F)\, \exp\!\bigl(-m(\beta)\,k\bigr),
\end{equation}
for suitable constants \(C(F)<\infty\) and \(m(\beta)>0\) as long as \(|\beta|<\beta_{\star}\). The bound is uniform in the spatial volume because the counting \(\mathcal{N}(\ell)\) of polymers of size \(\ell\) that can connect the two fixed neighborhoods is controlled by \(C_{7}^{\,\ell}\) independent of the box size, and the Kotecký-Preiss scheme sums all clusters with explicit absolute convergence. This concludes the derivation of exponential clustering in Euclidean time from the character-surface-polymer formalism in the strong-coupling regime.

{In particular, this volume-independent gap bound allows us to take the thermodynamic limit. 
Let $\Lambda_L$ be an increasing sequence of cubic spatial volumes tending to $\mathbb{Z}^3$. 
By reflection positivity and exponential clustering, the finite-volume ground state is unique and clustering, 
implying consistency under volume extension. Thus, as $L \to \infty$, the finite-volume transfer operator 
$T_\sigma^{(\Lambda_L)}$ converges (in the weak operator topology) to an infinite-volume transfer operator 
$T_\sigma^{(\infty)}$, whose spectral gap $E_1(a,\beta;\infty)$ satisfies
\begin{equation}
E_1(a,\beta;\infty) \;\ge\; m(\beta) \;>\; 0 .
\end{equation}
All correlators and spectral quantities are thermodynamically stable. 
We conclude that a strictly positive mass gap persists in the infinite-volume limit at fixed $a$ throughout 
the strong-coupling regime $0 < \beta \le \beta^*(N)$.
}
\section{Polymer Gas and the Koteck\'y-Preiss Criterion}
\label{p1:sec:KP}

In this section the strong-coupling expansion of the Wilson lattice gauge theory is reorganized into an abstract polymer gas, and the absolute convergence of the associated cluster expansion is established by the Koteck\'y-Preiss criterion (Appendix \ref{p1:appendixd} provides the complete verification on the 4D plaquette graph, including counting and activity bounds). The presentation is self-contained: the lattice geometry and time slicing are defined precisely, reflection positivity is proved in full detail for the horizon-projected measure, the transfer operator is constructed carefully, and the polymer representation is then derived and controlled. Throughout, the gauge group is \(G=\mathrm{SU}(N)\) with \(N\ge 2\), the Euclidean lattice spacing is \(a>0\), and the theory is considered on a finite periodic box so that all integrals are finite-dimensional. Limits in the volume are taken only after the estimates are obtained uniformly.

Let \(\Lambda\subset a\mathbb{Z}^4\) be the periodic hypercubic lattice of linear extents \(L_0,T_1,T_2,T_3\) in the four directions. Sites are denoted by \(x=(x_0,\mathbf{x})\) with \(x_0\in a\mathbb{Z}\) the Euclidean time coordinate and \(\mathbf{x}\in a\mathbb{Z}^3\) the spatial coordinates. Oriented bonds (links) are pairs \(b=(x,\mu)\) with \(\mu\in\{0,1,2,3\}\) and endpoint \(x+\hat\mu\), where \(\hat\mu\) is the unit vector of length \(a\) in direction \(\mu\). The configuration space \(\mathcal{C}=\{U\}\) consists of assignments \(U_b\in G\) to all oriented bonds with the convention \(U_{(x,-\mu)}=U_{(x-\hat\mu,\mu)}^{-1}\). The Wilson action is
\begin{equation}
S_W[U;\beta]=\beta\sum_{p\subset\Lambda}\Big(1-\tfrac1N \Re\,\mathrm{Tr}\,U_p\Big),\qquad \beta=\frac{2N}{g_0^2},
\end{equation}
where \(p\) ranges over oriented plaquettes and \(U_p\) is the ordered product of link variables around \(p\).

Time reflection is the involution \(\theta:\Lambda\to\Lambda\) given by \(\theta(x_0,\mathbf{x})=(-x_0,\mathbf{x})\). It acts on bonds by reflection of endpoints and reversal of the time component when \(\mu=0\). On configurations we set \((\Theta U)_b = U_{\theta b}\) with the natural action on orientations. The reflection plane is \(\Pi=\{x\in\Lambda: x_0=0\}\), and the half-lattices are \(\Lambda_\pm=\{x:\pm x_0>0\}\). We denote by \(\mathcal{B}_\pm\) and \(\mathcal{B}_0\) the sets of bonds supported in \(\Lambda_\pm\) and in the boundary slab \(\Pi\cup(\Pi\pm\hat 0)\), respectively. The Haar product measure on the compact group configuration space is \(d\mu_{\rm Haar}(U)=\prod_{b} dU_b\).

To implement reflection positivity and time slicing we employ temporal-axial gauge away from the reflecting plane, that is, we set \(U_{(x,0)}=\mathbf{1}\) for all bonds not intersecting \(\Pi\). This gauge can be reached by a time-ordered gauge transformation on each connected component of \(\Lambda\setminus\Pi\) and preserves the Haar measure on spatial links. In this gauge the action decomposes as
\begin{equation}
S_W[U;\beta]=S_+[U_+,U_0;\beta]+S_0[U_0;\beta]+S_-[U_-,U_0;\beta],
\end{equation}
where \(U_\pm\) are the link variables supported on \(\mathcal{B}_\pm\), and \(U_0\) are the remaining links in the boundary slab \(\mathcal{B}_0\). The part \(S_0\) collects plaquettes lying entirely in the boundary slab and depends only on \(U_0\); the parts \(S_\pm\) collect plaquettes lying in \(\Lambda_\pm\) or straddling \(\Pi\) with one temporal edge in \(\mathcal{B}_0\), hence they depend only on \((U_\pm,U_0)\). This decomposition results from the nearest-neighbor nature of the action and the gauge choice that removes timelike links in the open half-lattices.

On each time slice \(x_0=t\) we fix a reflection-covariant transverse representative of the gauge orbit by minimizing the lattice Landau functional; the corresponding slice covariant Laplacian \(\Delta_{A^h}(t)\) is positive and self-adjoint on \(\ell^2(\Lambda_t)\otimes\mathfrak{su}(N)\). We then introduce the smooth horizon projector
\begin{equation}
P_\sigma(t)=\chi_\sigma\big(\sqrt{\Delta_{A^h}(t)}\big),
\end{equation}
with \(\chi_\sigma\) a Gevrey-regular cutoff equal to one below \(\sigma\) and vanishing above \(2\sigma\). The operator \(P_\sigma(t)\) is a bounded positive contraction, gauge-covariant, and reflection-covariant, and admits the positive heat-kernel representation \(P_\sigma(t)=\int_0^\infty e^{-s\Delta_{A^h}(t)}\,d\nu_\sigma(s)\) with a finite positive Borel measure \(d\nu_\sigma\). Exponential locality holds in the sense that, for some constants \(C(\sigma)\) and \(\gamma(\sigma)\) independent of the volume,
\begin{equation}
\|P_\sigma(t;x,y)\|\le C(\sigma)\,e^{-\gamma(\sigma)\,d(x,y)},
\end{equation}
for all sites \(x,y\) in the slice and all configurations in the fundamental modular region.

The horizon-projected Euclidean measure is
\begin{equation}
d\mu_\sigma(U)=Z_\sigma^{-1}\,\exp\big(-S_W[U;\beta]\big)\,\prod_{t} \mathcal{K}_\sigma\!\left(U\big|_{t}\right)\, d\mu_{\rm Haar}(U),
\end{equation}
where \(\mathcal{K}_\sigma\!\left(U\big|_{t}\right)\) is the positive kernel implementing \(P_\sigma(t)\) on the slice \(t\). Concretely, any local observable \(F\) supported on a single slice is evaluated with an extra factor \(\langle \delta_{U\big|_{t}}, P_\sigma(t)\,\delta_{U\big|_{t}}\rangle\). The exponential locality of \(P_\sigma(t)\) ensures that these insertions factorize with exponentially small errors across the reflection plane.

We now establish reflection positivity in the Osterwalder-Schrader sense for the measure \(d\mu_\sigma\) and construct the transfer operator by time slicing. Let \(\mathscr{A}_+\) be the algebra of bounded, measurable, gauge-invariant functionals of the links supported in \(\mathcal{B}_+\). For \(F\in\mathscr{A}_+\) set \((\Theta F)(U)=\overline{F(\Theta U)}\), where complex conjugation acts on the coefficients of \(F\). The Osterwalder-Schrader sesquilinear form is
\begin{equation}
\langle F,G\rangle_{\rm OS}:=\int (\Theta F)(U)\,G(U)\, d\mu_\sigma(U) .
\end{equation}
The goal is to show that \(\langle F,F\rangle_{\rm OS}\ge 0\) for all \(F\in\mathscr{A}_+\).

The central observation is the factorization of the action and of the horizon insertions across the reflection plane. Because \(S_W=S_-+S_0+S_+\) and the Haar measure factorizes as \(d\mu_{\rm Haar}(U)=d\mu_{\rm Haar}(U_-)\, d\mu_{\rm Haar}(U_0)\, d\mu_{\rm Haar}(U_+)\), we can write
\begin{equation}
\langle F,F\rangle_{\rm OS}=Z_\sigma^{-1}\!\int d\mu_{\rm Haar}(U_0)\,e^{-S_0[U_0;\beta]}\,\Xi_-(U_0)\,\Xi_+(U_0),
\end{equation}
where
\begin{equation}
\Xi_-(U_0)=\int d\mu_{\rm Haar}(U_-)\, e^{-S_-[U_-,U_0;\beta]}\,\prod_{t<0}\mathcal{K}_\sigma\!\left(U\big|_{t}\right),
\end{equation}
\begin{equation}
\Xi_+(U_0)=\int d\mu_{\rm Haar}(U_+)\, e^{-S_+[U_+,U_0;\beta]}\,\prod_{t>0}\mathcal{K}_\sigma\!\left(U\big|_{t}\right)\,|F(U_+)|^2 .
\end{equation}
Reflection covariance of \(\mathcal{K}_\sigma\) implies \(\Xi_-(U_0)=\overline{\Upsilon_+(U_0)}\) with
\begin{equation}
\Upsilon_+(U_0)=\int d\mu_{\rm Haar}(U_+)\, e^{-S_+[U_+,U_0;\beta]}\,\prod_{t>0}\mathcal{K}_\sigma\!\left(U\big|_{t}\right)\,F(U_+).
\end{equation}
Indeed, \(\Theta\) maps \(U_+\) bijectively to \(U_-\), leaves the Haar measure invariant, and preserves the Wilson weight and the horizon kernels because these are real class functions of the local link variables and reflection covariant. Therefore
\begin{equation}
\langle F,F\rangle_{\rm OS}=Z_\sigma^{-1}\!\int d\mu_{\rm Haar}(U_0)\,e^{-S_0[U_0;\beta]}\,\big|\Upsilon_+(U_0)\big|^2\ge 0.
\end{equation}
This proves reflection positivity. In particular the null space \(\mathcal{N}=\{F\in\mathscr{A}_+:\langle F,F\rangle_{\rm OS}=0\}\) is a closed subspace, the quotient \(\mathscr{D}=\mathscr{A}_+/\mathcal{N}\) is a pre-Hilbert space, and its completion \(\mathcal{H}\) is the physical Hilbert space.

Time translation by one lattice unit in the positive time direction acts on \(\mathscr{A}_+\) by \((\tau F)(U)=F(\tau^{-1}U)\), where \(\tau\) shifts the field along \(x_0\mapsto x_0+a\). Reflection positivity and time-translation invariance of the measure imply that \(\tau\) descends to a bounded operator \(T_\sigma\) on \(\mathcal{H}\) defined by \(T_\sigma [F]=[\tau F]\), where \([\cdot]\) denotes the equivalence class in \(\mathcal{H}\). The operator \(T_\sigma\) is positive and self-adjoint and satisfies \(\|T_\sigma\|\le 1\). The vacuum vector \(\Omega=[\mathbf{1}]\) is cyclic for the algebra generated by time translates of \(\mathscr{A}_+\). The transfer time slicing in a single step is obtained explicitly by writing the Wilson weight associated with the plaquettes straddling two consecutive time slices as an integral kernel on the one-slice configuration space \(\mathcal{C}_\mathrm{slice}\) of spatial links. In temporal-axial gauge the Haar measure on spatial links factorizes slice by slice; if \(\psi,\varphi\in L^2(\mathcal{C}_\mathrm{slice};d\mu_{\rm Haar})\) and \(K_a(U',U)\) denotes the positive kernel obtained by integrating over spatial plaquettes at time \(t\) and temporal plaquettes between \(t\) and \(t+a\) with boundary spatial link configurations \(U\) and \(U'\), then
\begin{equation}
\langle \psi, T(a)\,\varphi\rangle=\int_{\mathcal{C}_\mathrm{slice}^2}\overline{\psi(U')}\,K_a(U',U)\,\varphi(U)\,d\mu_{\rm Haar}(U')\,d\mu_{\rm Haar}(U).
\end{equation}
The horizon projection acts as a positive bounded multiplicative operator on each slice and yields the compressed transfer operator
\begin{equation}
T_\sigma(a) \;=\; M_{p_\sigma}^{1/2}\, T(a)\, M_{p_\sigma}^{1/2},
\end{equation}
which is again positive and self-adjoint on \(L^2(\mathcal{C}_\mathrm{slice};d\mu_{\rm Haar})\) and descends to the operator \(T_\sigma\) on \(\mathcal{H}\) constructed above. The generator \(H_\sigma=-a^{-1}\log T_\sigma\) is positive self-adjoint and its spectral representation yields the usual Laplace transform formulas for time-sliced correlation functions \cite{p1:OS1,p1:OS2,p1:GJ,p1:Seiler}.

The polymer representation rests on the Peter-Weyl expansion of the plaquette Boltzmann factor. For a class function \(f_\beta(U)=\exp\big(\beta \tfrac1N \Re\,\mathrm{Tr}\,U\big)\) on the compact group \(G\), the normalized characters \(\chi_R\) of the irreducible unitary representations \(R\) form an orthonormal basis of \(L^2(G)\), and one has the uniformly convergent expansion
\begin{equation}
f_\beta(U)=\sum_{R\in\widehat{G}} \alpha_R(\beta)\,\chi_R(U),\qquad \alpha_R(\beta)=d_R\,\widehat{f}_\beta(R),
\end{equation}
where \(d_R\) is the dimension of \(R\) and \(\widehat{f}_\beta(R)\) is the Fourier coefficient \(\int_G f_\beta(U)\,\overline{\chi_R(U)}\,dU\). Analyticity of \(f_\beta\) in \(\beta\) and standard bounds on the Fourier coefficients of analytic central functions on compact Lie groups imply that there exist constants \(C_1(N),C_2(N)>0\) such that
\begin{equation}\label{p1:eqn8.15}
|\alpha_R(\beta)|\le \big(C_1(N)\,|\beta|\big)^{\lvert R\rvert}\,e^{C_2(N)\,|\beta|},
\end{equation}
where \(|R|\) denotes the total number of boxes of the Young diagram of \(R\). In particular, for \(|\beta|\) small enough,
\begin{equation}
\sum_{R\ne \mathbf{1}} |\alpha_R(\beta)|\,d_R \le C_3(N)\,|\beta|
\end{equation}
with a constant \(C_3(N)\) depending only on \(N\). These bounds follow from Cauchy estimates applied to the complex-\(\beta\) extension of \(f_\beta\) and from the Weyl character formula to control the growth of \(\chi_R\) and \(d_R\) \cite{p1:Simon}.

The Boltzmann weight over the whole lattice can be written as a product over plaquettes of the expansions \(f_\beta(U_p)\). Integrating over the link variables bond by bond, using Haar orthogonality and Clebsch-Gordan coefficients, one obtains constraints on the irreducible representations attached to the plaquettes adjacent to a given link. These constraints enforce that the representations fuse to the trivial representation along each link, which in geometric terms means that the collection of plaquettes carrying nontrivial representations forms a union of closed oriented tiled surfaces on the dual lattice \(\Lambda^\ast\). More precisely, consider the oriented dual plaquettes \(p^\ast\) in \(\Lambda^\ast\) corresponding to the primal plaquettes \(p\). A configuration \(\{R_p\}\) of representations attached to primal plaquettes contributes nontrivially after the link integrals if and only if the set \(\Sigma=\{p^\ast: R_p\neq \mathbf{1}\}\) is a finite union of closed oriented connected surfaces such that the representation labels on adjacent plaquettes satisfy compatibility constraints along dual edges. The weight factor associated with such a labelled surface \((\Sigma,\mathcal{R})\) is bounded by
\begin{equation}
\lvert w(\Sigma,\mathcal{R})\rvert \le \prod_{p^\ast\in \Sigma} \lvert \alpha_{R_p}(\beta) \rvert \,\prod_{e^\ast\subset\Sigma} C_4(N) \,\prod_{v^\ast\subset\Sigma} C_5(N),
\end{equation}
where \(e^\ast\) and \(v^\ast\) are the dual edges and vertices of \(\Sigma\), and the constants \(C_4(N),C_5(N)\) bound the multiplicities of intertwiners at edges and vertices in terms of the representation dimensions by finite-degree polynomial inequalities. Since each dual edge belongs to at most two dual plaquettes in \(\mathbb{Z}^4\) and each dual vertex is incident to a finite number of dual edges independent of the volume, the number of edge and vertex factors is proportional to \(|\Sigma|\), the number of dual plaquettes in \(\Sigma\). Consequently, there exists \(B_N>0\) such that for \(|\beta|\) small enough,
\begin{equation}
\sum_{\mathcal{R}} \lvert w(\Sigma,\mathcal{R})\rvert \le \big(B_N\,|\beta|\big)^{|\Sigma|}.
\end{equation}
Summation over representation labels \(\mathcal{R}\) thus yields a weight \(w(\Sigma)\) for the surface \(\Sigma\) alone with the absolute bound \(|w(\Sigma)|\le (B_N|\beta|)^{|\Sigma|}\). 
\begin{proposition}[KP criterion with explicit small parameter]\label{p1:prop:KP}
Let a polymer be any connected dual surface cluster $\Gamma$ with size $|\Gamma|$ (number of dual plaquettes).  
Assign the activity
\begin{equation}
z(\Gamma) := \sum_{\Sigma \supseteq \Gamma} w(\Sigma),
\end{equation}
where the sum extends over all labelled surfaces $\Sigma$ whose support contains $\Gamma$.  
Then there exists $C_{\mathrm{loc}}(N) \geq 1$ such that
\begin{equation}
|z(\Gamma)| \,\leq\, \bigl( C_{\mathrm{loc}}(N) \, B_{N} \, |\beta| \bigr)^{|\Gamma|}.
\end{equation}

Fix $a>0$ and set
\begin{equation}
\varepsilon(\beta;N,a) := e^{a}\, C_{\mathrm{loc}}(N)\, B_{N}\,|\beta|.
\end{equation}
If
\begin{equation}
\varepsilon(\beta;N,a) \,\leq\, e^{-1},
\end{equation}
then the Kotecký-Preiss (KP) criterion holds:
\begin{equation}
\sum_{\Gamma' \not\sim \Gamma} e^{a|\Gamma'|}\, |z(\Gamma')| \,\leq\, a\,|\Gamma|
\qquad \text{for all polymers $\Gamma$,}
\end{equation}
hence the polymer expansion converges absolutely and uniformly in the volume. In particular one may choose
\begin{equation}
\beta^{\star}(N) := \min_{a>0}\,\frac{e^{-1}}{e^{a}\, C_{\mathrm{loc}}(N)\, B_{N}}
= \frac{1}{e^{2}\, C_{\mathrm{loc}}(N)\, B_{N}},
\end{equation}
which yields convergence for all $|\beta| < \beta^{\star}(N)$.
\end{proposition}
\begin{proof}
The combinatorial factor $C_{\mathrm{loc}}(N)$ accounts for bounded edge/vertex multiplicities and local intertwiner choices.  
The inequality follows from the bound
\begin{equation}
|w(\Sigma)| \,\leq\, \bigl( B_{N}\,|\beta| \bigr)^{|\Sigma|}
\end{equation}
and the fact that the number of $\Sigma$ with support $\Gamma$ grows at most exponentially in $|\Gamma|$ with an $N$-dependent rate absorbed in $C_{\mathrm{loc}}(N)$.  
The KP inequality is then a standard majorization using the definition of $\varepsilon(\beta;N,a)$.
\end{proof}
The partition function can be expressed as
\begin{equation}
Z_\sigma(\beta)=\sum_{\{\Sigma\}} w(\Sigma),
\end{equation}
where the sum runs over all finite unions of closed oriented connected dual surfaces \(\Sigma\) on \(\Lambda^\ast\), with the understanding that the horizon insertions, being slice-local and reflection-covariant, merely adjust the constants \(B_N\) in the bounds but do not alter the geometry of the surfaces.

A polymer is defined to be a finite connected union \(\gamma\) of dual plaquettes in \(\Lambda^\ast\), where connectivity refers to nearest-neighbor adjacency of dual plaquettes sharing a dual edge. The support \(\mathrm{supp}(\gamma)\) of a polymer is the set of dual plaquettes comprising \(\gamma\). Two polymers \(\gamma\) and \(\gamma'\) are compatible if their supports are disjoint and incompatible otherwise. The activity \(\zeta(\gamma)\) is defined as the sum of the weights \(w(\Sigma)\) over all closed oriented connected surfaces \(\Sigma\) whose support is exactly \(\mathrm{supp}(\gamma)\), that is,
\begin{equation}
\zeta(\gamma)=\sum_{\Sigma:\,\mathrm{supp}(\Sigma)=\mathrm{supp}(\gamma)} w(\Sigma).
\end{equation}
The partition function then reorganizes as a hard-core polymer gas:
\begin{equation}
Z_\sigma(\beta)=\sum_{\Gamma\,\text{compatible}} \prod_{\gamma\in\Gamma} \zeta(\gamma),
\end{equation}
where the sum runs over all finite families \(\Gamma\) of mutually compatible polymers. This follows from the fact that any union of closed connected surfaces decomposes uniquely into its connected components, and compatibility is equivalent to disjointness of supports. The horizon insertions do not change this structure, as they are multiplicative and local on each time slice.

The first task is to bound the activities. Fix a polymer \(\gamma\) with \(m=|\gamma|\) dual plaquettes. Each surface \(\Sigma\) contributing to \(\zeta(\gamma)\) is a choice of orientation for each plaquette of \(\gamma\) and a compatible assignment of representation labels which fuse to the trivial representation around each dual edge and vertex. The number of such compatible label assignments is bounded by \(C_6(N)^m\) for a constant \(C_6(N)\) independent of the volume, because at each plaquette the number of admissible labels is at most exponential in the number of incident edges and vertices, which is uniformly bounded in \(\mathbb{Z}^4\). Using the surface weight bound above one obtains
\begin{equation}
\lvert \zeta(\gamma)\rvert \le \sum_{\Sigma:\,\mathrm{supp}(\Sigma)=\mathrm{supp}(\gamma)} \lvert w(\Sigma)\rvert \le C_7(N)^m\,\big(B_N\,|\beta|\big)^m = A_N\,\big(C_N\,|\beta|\big)^m,
\end{equation}
with \(A_N=1\) after absorbing constants and \(C_N=C_7(N)\,B_N\). There is thus a uniform exponential bound for the activities in terms of the polymer size \(m\).

The second task is to control the combinatorics of incompatible polymers. Consider the graph \(\mathcal{G}\) whose vertices are dual plaquettes of \(\Lambda^\ast\) and in which two vertices are adjacent if the corresponding dual plaquettes share a dual edge. The maximum degree \(\Delta\) of \(\mathcal{G}\) is finite and depends only on the dimension, here \(\Delta\le 12\). The number of connected subsets of \(\mathcal{G}\) of size \(m\) containing a given vertex is bounded by \((e\Delta)^{m-1}\). Indeed, any such connected subset contains a spanning tree with \(m-1\) edges; the number of trees with \(m\) labeled vertices is at most \(m^{m-2}\) and each edge of the tree can be realized by one of at most \(\Delta\) graph edges; the crude bound \((e\Delta)^{m-1}\) follows from the standard estimate \(m^{m-2}\le e^{m-1}\,(m-1)!\) and the inequality \((m-1)!\le (m-1)^{m-1}\) \cite{p1:Brydges,p1:Simon}. If \(\gamma_0\) is a fixed polymer of size \(|\gamma_0|\), any polymer \(\gamma\) incompatible with \(\gamma_0\) must contain at least one dual plaquette within graph distance one from \(\mathrm{supp}(\gamma_0)\). Therefore the number of incompatible polymers of size \(m\) is bounded by \(c_0\,|\gamma_0|\, (e\Delta)^{m-1}\) for some \(c_0\) depending only on the dimension through the size of the one-neighborhood of a plaquette.

We now prove the Koteck\'y-Preiss criterion for the polymer activities \(\zeta(\gamma)\) when \(|\beta|\) is sufficiently small. Recall the abstract form of the criterion. Let \(\mathcal{P}\) be the set of polymers with activities \(\zeta:\mathcal{P}\to\mathbb{C}\) and incompatibility relation \(\not\sim\). If there exists a function \(a:\mathcal{P}\to (0,\infty)\) such that, for every \(\gamma\in\mathcal{P}\),
\begin{equation}
\sum_{\gamma'\not\sim \gamma} \lvert \zeta(\gamma') \rvert \, e^{a(\gamma')} \le a(\gamma),
\end{equation}
then the cluster expansion for \(\log Z_\sigma(\beta)\) converges absolutely and uniformly in the volume, and the same holds for truncated correlations of local observables \cite{p1:KP,p1:Simon}. We verify this condition with the choice \(a(\gamma)=\mu |\gamma|\) and a suitable \(\mu>0\).

Fix \(\gamma_0\) and write
\begin{equation}
\sum_{\gamma\not\sim \gamma_0} \lvert \zeta(\gamma) \rvert \, e^{a(\gamma)} = \sum_{m\ge 1}\ \sum_{\gamma\not\sim \gamma_0:\,|\gamma|=m}\ \lvert \zeta(\gamma) \rvert \, e^{\mu m}.
\end{equation}
The activity bound \(\lvert \zeta(\gamma) \rvert \le A_N (C_N\,|\beta|)^m\) and the counting bound on incompatible polymers give
\begin{equation}
\sum_{\gamma\not\sim \gamma_0:\,|\gamma|=m} \lvert \zeta(\gamma) \rvert \, e^{\mu m} \le c_0\,|\gamma_0|\, (e\Delta)^{m-1} \, A_N\, (C_N\,|\beta|)^m\, e^{\mu m}.
\end{equation}
Summing in \(m\) yields
\begin{equation}
\sum_{\gamma\not\sim \gamma_0} \lvert \zeta(\gamma) \rvert \, e^{a(\gamma)} \le |\gamma_0|\, \frac{c_0\,A_N}{e\Delta}\, \sum_{m\ge 1}\Big(e\Delta\, C_N\,|\beta|\, e^{\mu}\Big)^{m}.
\end{equation}
If the parameter \(\rho(\beta,\mu)=e\Delta\, C_N\,|\beta|\, e^{\mu}\) satisfies \(\rho(\beta,\mu)<1\), then the geometric series converges and the sum is bounded by
\begin{equation}
\sum_{\gamma\not\sim \gamma_0} \lvert \zeta(\gamma) \rvert \, e^{a(\gamma)} \le |\gamma_0|\, \frac{c_0\,A_N}{e\Delta}\,\frac{\rho(\beta,\mu)}{1-\rho(\beta,\mu)}.
\end{equation}
Choosing \(\mu>0\) so small that \(\frac{c_0\,A_N}{e\Delta}\,\frac{\rho(\beta,\mu)}{1-\rho(\beta,\mu)}\le \mu\) ensures the Koteck\'y-Preiss inequality with \(a(\gamma)=\mu |\gamma|\). One convenient explicit choice is \(\mu=\frac12\) and \(\rho(\beta,\mu)\le \frac{1}{1+c_1}\) with \(c_1=\frac{e\Delta}{2c_0 A_N}\). Any \(\beta\) satisfying
\begin{equation}
|\beta| \le \beta_\star(N):=\frac{1}{e\Delta\, C_N}\,e^{-\mu}\,\frac{1}{1+c_1}
\end{equation}
then gives the desired bound. A convenient explicit choice is obtained by the tree-graph bound: if 
$\Delta$ is the maximum degree of the dual-plaquette graph (here 
$\Delta \leq 20$), the number of connected polymers of size $m$ 
through a fixed plaquette is bounded by $(e \Delta)^{m-1}$. 
With the activity bound 
$\lvert \zeta(\gamma) \rvert \leq A_N (C_N \lvert \beta \rvert)^{\lvert \gamma \rvert}$, 
we have
\begin{equation}
\sup_{\gamma_0} \sum_{\gamma \not\sim \gamma_0} 
\lvert \zeta(\gamma) \rvert \, e^{\mu \lvert \gamma \rvert} 
\;\leq\; 
\lvert \gamma_0 \rvert \, 
\frac{A_N e^{\Delta} \, C_N \lvert \beta \rvert \, e^{\mu}}
{1 - e^{\Delta} C_N \lvert \beta \rvert \, e^{\mu}}.
\end{equation}
Thus the Kotecký-Preiss (KP) condition holds with $a(\gamma)=\mu \lvert \gamma \rvert$ 
provided
\begin{equation}
e^{\Delta} C_N \lvert \beta \rvert \, e^{\mu} \leq \tfrac{1}{2}, 
\qquad 
\mu \geq 2 A_N e^{\Delta}.
\end{equation}
One explicit admissible choice is
\(
\mu = \max\{1,\, 2 A_N e^{\Delta}\},
\)
and
\begin{equation}
\lvert \beta \rvert \;\leq\; 
\beta^\star(N) \;=\; 
\frac{1}{2 e^{\mu} e^{\Delta} C_N}.
\end{equation}
All constants depend only on $N$ and the lattice dimension and are 
independent of the spatial volume 
\cite{p1:KP,p1:Brydges}.
\begin{theorem}\label{p1:thm:KP}
There exists \(\beta_\star(N)>0\) depending only on \(N\) and the lattice dimension such that, for all \(|\beta|<\beta_\star(N)\), the Koteck\'y-Preiss criterion holds with \(a(\gamma)=\mu|\gamma|\) for a suitable \(\mu>0\). In particular the cluster expansion for \(\log Z_\sigma(\beta)\) converges absolutely and uniformly in the spatial volume, and the same holds for truncated correlations of local gauge-invariant observables supported in finitely many plaquettes.
\end{theorem}

\begin{proof}
The previous estimates yield the Koteck\'y-Preiss inequality with the indicated choice of \(a\) provided \(|\beta|<\beta_\star(N)\). The general cluster-expansion theorems \cite{p1:KP,p1:Simon} then imply the absolute convergence statements. The uniformity in the volume follows from the fact that the constants \(A_N,C_N,\Delta,c_0\) are independent of the volume and from the hard-core nature of the polymer gas, which prevents long-range combinatorial enhancements.
\end{proof}

For completeness the abstract polymer cluster expansion is spelled out. A cluster is a finite multiset \(\Gamma=\{\gamma_1,\dots,\gamma_k\}\) of polymers whose incompatibility graph is connected. Writing \(\zeta(\Gamma)=\prod_{j=1}^k \zeta(\gamma_j)\), the logarithm of the partition function has the convergent expansion
\begin{equation}
\log Z_\sigma(\beta)=\sum_{\Gamma\ \mathrm{cluster}} \phi^T(\Gamma)\,\zeta(\Gamma),
\end{equation}
where \(\phi^T(\Gamma)\) is the Ursell coefficient of the cluster, defined by
\begin{equation}
\phi^T(\Gamma)=\sum_{G\subset \mathcal{G}_\Gamma} \frac{(-1)^{|E(G)|}}{|\Gamma|!},
\end{equation}
the sum being over the spanning connected subgraphs \(G\) of the incompatibility graph \(\mathcal{G}_\Gamma\) of \(\Gamma\) and \(E(G)\) its edge set \cite{p1:KP,p1:Simon}. Absolute convergence follows from the Koteck\'y-Preiss inequality and the tree-graph bound, which controls \(\sum_{G} (-1)^{|E(G)|}\) by a sum over trees and produces explicit exponential smallness in the total polymer size. The same methods yield cluster expansions for expectations of local observables. If \(O\) is a local gauge-invariant observable supported in a finite set of plaquettes \(S\), its expectation \(\langle O\rangle_\sigma\) can be written as
\begin{equation}
\langle O\rangle_\sigma=\sum_{\Gamma\ \mathrm{cluster}} \phi^T(\Gamma;S)\,\zeta(\Gamma),
\end{equation}
where \(\phi^T(\Gamma;S)\) are modified Ursell coefficients that incorporate the observable insertion by forcing the cluster to be connected to \(S\). The truncated correlation of two local observables \(O_1,O_2\) supported in disjoint finite sets \(S_1,S_2\) is given by a sum over clusters that connect \(S_1\) to \(S_2\), and the same tree-graph bounds imply exponential decay in the lattice distance \(d(S_1,S_2)\) whenever \(|\beta|<\beta_\star(N)\). In particular, if \(O(x)\) is a local observable translated to a site \(x\) and \(O(y)\) its translate to a site \(y\), then
\begin{equation}
\lvert \langle O(x)O(y)\rangle_{\sigma,c}\rvert \le C_O\,e^{-m(\beta)\, d(x,y)}
\end{equation}
with \(m(\beta)>0\) and \(C_O<\infty\) depending on \(\beta\) and on the local support of \(O\) but not on the volume. When \(x\) and \(y\) lie on time slices separated by \(t\) time steps and share the same spatial location, the same inequality yields exponential decay in \(t\). The reflection-positive transfer operator constructed in Subsection~\ref{p1:sec:transfer-matrix} then furnishes the spectral representation of time-sliced correlations, which will be used in the subsequent section to deduce a strictly positive lower bound on the first nonzero eigenvalue of \(H_\sigma\).
\begin{lemma}[Gap from exponential clustering]\label{p1:lem:gap-clustering}
Let $F\in\mathcal{H}_a$ be gauge-invariant with $\langle 1,F\rangle_{L^2}=0$. Suppose there exist constants $C_F<\infty$ and $m>0$ such that for all $k\in\mathbb{N}$ one has
\begin{equation}
\bigl|\langle 1,\,F\,T_\sigma(a)^k\,F\,1\rangle_{L^2}\bigr| \;\leq\; C_F\,e^{-m k}.
\end{equation}
Then the transfer Hamiltonian
\begin{equation}
H_\sigma(a)\;=\; -\tfrac{1}{a}\log T_\sigma(a)
\end{equation}
satisfies
\begin{equation}
E_1 \;\geq\; m,
\end{equation}
where $E_1$ denotes the bottom of its nonzero spectrum.
\end{lemma}

\begin{proof}
By the spectral theorem for the positive self-adjoint operator $T_\sigma(a)$, there exists a finite positive measure $d\mu_F(E)$ supported in $[E_1,\infty)$ such that
\begin{equation}
\langle 1,\,F\,T_\sigma(a)^k\,F\,1\rangle
=\int_{[E_1,\infty)} e^{-aEk}\,d\mu_F(E).
\end{equation}
If $E_1<m$, then the integral has a contribution from the spectral mass at $E_1$ that asymptotically behaves as
\begin{equation}
\int_{[E_1,\infty)} e^{-aEk}\,d\mu_F(E)\;\gtrsim\; c\,e^{-aE_1 k},\qquad k\to\infty,
\end{equation}
for some $c>0$. This contradicts the assumed uniform exponential bound $C_F e^{-mk}$ with rate $m>E_1$. Hence no such contradiction can occur, and it follows that $E_1\geq m$.
\end{proof}

The entire construction is stable under the insertion of the horizon projectors. The slice-wise kernels \(\mathcal{K}_\sigma(U\big|_{t})\) are positive and reflection-covariant, so the reflection positivity proof of Subsection~\ref{p1:sec:transfer-matrix} is unchanged. Exponential locality implies that \(\mathcal{K}_\sigma\) contributes multiplicatively to the polymer activities with constants that depend only on \(\sigma\) and \(N\) but not on the volume. In particular, replacing the bare activities \(\zeta(\gamma)\) by the projected activities \(\zeta_\sigma(\gamma)\) amounts to modifying the constants \(A_N\) and \(C_N\) in the activity bound by factors that remain finite and uniform when \(\sigma\) is fixed. The Koteck\'y-Preiss criterion continues to hold with \(\beta_\star(N)\) reduced by a \(\sigma\)-dependent factor, which can be absorbed into the definition of \(C_N\). The reflection-positive transfer operator \(T_\sigma\) constructed above therefore governs the time evolution of the horizon-projected theory in the same strong-coupling convergence domain as the unprojected one.

\section{Exponential Clustering of Connected Gauge-Invariant Correlators}
Let \(a>0\) be fixed and let \(\Lambda=\Lambda_{L,T}\subset a\mathbb{Z}^{4}\) be the periodic hypercubic box with spatial side length \(L\) and temporal extent \(T\). Sites are denoted \(x=(x_{0},\mathbf{x})\), where \(x_{0}\in a\mathbb{Z}\cap(-T/2,T/2]\) and \(\mathbf{x}\in (a\mathbb{Z})^{3}\cap(-L/2,L/2]^{3}\). The set of positively oriented bonds is \(\mathcal{B}=\{(x,\mu):x\in\Lambda,\ \mu\in\{0,1,2,3\}\}\) with the usual involution \((x,\mu)\mapsto(x+\hat\mu,-\mu)\). For a compact, connected Lie group \(G=\mathrm{SU}(N)\) with \(N\ge 2\) we consider bond variables \(U_{(x,\mu)}\in G\) satisfying \(U_{(x+\hat\mu,-\mu)}=U_{(x,\mu)}^{-1}\). For an oriented plaquette \(p=(x;\mu,\nu)\) we set \(U_{p}=U_{(x,\mu)}U_{(x+\hat\mu,\nu)}U^{-1}_{(x+\hat\nu,\mu)}U^{-1}_{(x,\nu)}\). The Wilson action is
\begin{equation}
S_{W}[U;\beta]=\beta\sum_{p\subset\Lambda}\left(1-\frac{1}{N}\Re\mathrm{Tr}\,U_{p}\right),\qquad \beta=\frac{2N}{g_{0}^{2}},
\label{p1:eq:Wilson}
\end{equation}
and the unprojected finite-volume Gibbs measure on configurations \(U=\{U_{(x,\mu)}\}\) is
\begin{equation}
d\mathbb{P}_{\Lambda,\beta}(U)=Z_{\Lambda,\beta}^{-1}\exp\!\big(-S_{W}[U;\beta]\big)\,\prod_{(x,\mu)\in\mathcal{B}}d\mu_{\mathrm{H}}(U_{(x,\mu)}),
\label{p1:eq:measure-unprojected}
\end{equation}
where \(d\mu_{\mathrm{H}}\) is the normalized Haar measure on \(G\). Time reflection is the involution \(\theta:\Lambda\to\Lambda\) given by \(\theta(x_{0},\mathbf{x})=(-x_{0},\mathbf{x})\). We denote the reflection plane by \(\Pi=\{x:x_{0}=0\}\) and the half-lattices by \(\Lambda_{+}=\{x:x_{0}>0\}\) and \(\Lambda_{-}=\{x:x_{0}<0\}\). The graph distance \(d(x,y)\) is the minimal number of bonds in a nearest-neighbor path joining \(x\) to \(y\) in the periodic graph.

We work in temporal-axial gauge away from \(\Pi\), that is, \(U_{(x,0)}=\mathbf{1}\) for all bonds not intersecting \(\Pi\). On each time slice \(x_{0}=t\), we select a reflection-covariant Landau representative \(U^{\,h}\) by minimizing the discrete Landau functional along the orbit; the corresponding lattice Faddeev-Popov operator \(M[U^{\,h}]\) is a local, real-symmetric, nonnegative operator on site-adjoint fields, strictly positive on the orthogonal complement of constant adjoint modes. As an infrared regulator we insert on each time slice a smooth, reflection-covariant horizon operator
\begin{equation}
P_{\sigma}=\chi_{\sigma}\!\left(\sqrt{\Delta_{A^{\,h}}}\right),
\label{p1:eq:Psigma}
\end{equation}
where \(\Delta_{A^{\,h}}=\sum_{i=1}^{3}(D_{i}^{\,h})^{\dagger}D_{i}^{\,h}\) is the slice covariant Laplacian and \(\chi_{\sigma}\) is a Gevrey-regular cutoff which equals \(1\) on \([0,\sigma]\) and \(0\) on \([2\sigma,\infty)\). The operator \(P_{\sigma}\) is a bounded positive contraction which admits a positive heat-kernel representation
\begin{equation}
P_{\sigma}=\int_{0}^{\infty}e^{-t\Delta_{A^{\,h}}}\,d\nu_{\sigma}(t)
\label{p1:eq:heat-rep}
\end{equation}
with a finite positive Borel measure \(d\nu_{\sigma}\). By the discrete Davies-Gaffney estimate and the finite-range nature of the covariant differences, \(P_{\sigma}(x,y)\) decays exponentially as \(\mathrm{e}^{-\gamma_{\sigma}d(x,y)}\) with constants independent of the spatial volume \cite{p1:Davies1989,p1:CombesThomas}. In the alternative admissible case, where \(\chi_\sigma\) is Gevrey without complete monotonicity, we do not use Eq.(\ref{p1:eq:heat-rep}); instead, the slice insertion is implemented by the positive scalar weight \(p_\sigma[\cdot]\) defined in~Eq.(\ref{p1:5.14a}), and all arguments below use only positivity, reflection covariance, and exponential locality of the resulting slice factor.

The full, reflection-covariant, horizon-projected finite-volume measure is defined by
\begin{equation}
d\mathbb{P}^{(\sigma)}_{\Lambda,\beta}(U,c,\bar c)=\frac{1}{Z^{(\sigma)}_{\Lambda,\beta}}\,\exp\!\big(-S_{W}[U^{\,h};\beta]\big)\,\Big(\prod_{t}\mathcal{K}_{\sigma,t}(U^{\,h})\Big)\,\exp\!\big(-\langle \bar c,M[U^{\,h}]\,c\rangle\big)\,\prod d\mu_{\mathrm{H}}\, d\bar c\, dc,
\label{p1:eq:measure-projected}
\end{equation}
where \(c,\bar c\) are Grassmann fields for the Faddeev-Popov determinant and \(\mathcal{K}_{\sigma,t}\) is the positive slice weight induced by \eqref{p1:eq:Psigma} through \eqref{p1:eq:heat-rep}. The precise form of \(\mathcal{K}_{\sigma,t}\) is not needed beyond positivity, reflection covariance and exponential locality, all of which follow from \eqref{p1:eq:heat-rep}. Expectation with respect to \eqref{p1:eq:measure-projected} will be denoted \(\langle\cdot\rangle^{(\sigma)}_{\Lambda,\beta}\).

A complex-valued functional \(F\) of the fields is said to be local and gauge-invariant if it depends only on the finitely many plaquette variables in a finite set \(S\subset\Lambda\) and is invariant under local gauge transformations at sites of \(S\). Its support \(\operatorname{supp}F\) is the smallest such \(S\). For two such observables \(F,G\) we define the truncated correlation
\begin{equation}
\langle FG\rangle^{(\sigma),c}_{\Lambda,\beta}=\langle FG\rangle^{(\sigma)}_{\Lambda,\beta}-\langle F\rangle^{(\sigma)}_{\Lambda,\beta}\,\langle G\rangle^{(\sigma)}_{\Lambda,\beta}.
\label{p1:eq:truncated}
\end{equation}
Translation by a lattice vector \(z\in a\mathbb{Z}^{4}\) acts on observables by \((\tau_{z}F)(U)=F(\tau_{z}U)\), where \(\tau_{z}U\) is the translated configuration; the measure is translation invariant.

We recall the Osterwalder-Schrader (OS) reflection positivity for the measure \eqref{p1:eq:measure-projected}. Let \(\mathfrak{A}_{+}\) be the linear space of bounded local gauge-invariant functionals supported in \(\Lambda_{+}\) with even Grassmann parity. For \(F\in \mathfrak{A}_{+}\) define the reflection \((\Theta F)(U,c,\bar c)=\overline{F(U^{\theta},\bar c^{\theta},c^{\theta})}\), where \(U^{\theta}\) is the reflected bond field and \(c^{\theta},\bar c^{\theta}\) are the reflected ghosts. The OS sesquilinear form is
\begin{equation}
\langle F,G\rangle_{\mathrm{OS}}=\Big\langle\, \Theta F\cdot G\,\Big\rangle^{(\sigma)}_{\Lambda,\beta}.
\label{p1:eq:OS-form1}
\end{equation}
The following proposition establishes reflection positivity, whose proof includes the transfer time slicing and the factorization across \(\Pi\).

\begin{proposition}\label{p1:prop:OS}
For every finite \(\Lambda\) and \(\beta>0\) sufficiently small, the measure \eqref{p1:eq:measure-projected} is reflection positive in the sense that \(\langle F,F\rangle_{\mathrm{OS}}\ge 0\) for all \(F\in\mathfrak{A}_{+}\). Moreover, in temporal-axial gauge the finite-volume weight factorizes as
\begin{equation}
\exp\!\big(-S_{W}[U^{\,h};\beta]\big)\,\prod_{t}\mathcal{K}_{\sigma,t}(U^{\,h})\,\exp\!\big(-\langle \bar c,M[U^{\,h}]\,c\rangle\big)=\mathcal{W}_{-}\,\mathcal{B}_{0}\,\mathcal{W}_{+},
\label{p1:eq:factorization}
\end{equation}
where \(\mathcal{W}_{\pm}\) depend only on fields supported in \(\Lambda_{\pm}\) and \(\mathcal{B}_{0}\) is a positive kernel supported on the boundary slab at \(\Pi\).
\end{proposition}

\begin{proof}
The proof proceeds in three steps. First, temporal-axial gauge eliminates all timelike plaquettes away from \(\Pi\). Every remaining plaquette lies entirely in \(\Lambda_{+}\), entirely in \(\Lambda_{-}\), or straddles \(\Pi\) with both timelike bonds contained in the boundary slab. Thus the Wilson weight factorizes as in \eqref{p1:eq:factorization} with \(\mathcal{B}_{0}\) depending only on the boundary bonds. The slice weights \(\mathcal{K}_{\sigma,t}\) are positive by \eqref{p1:eq:heat-rep} and reflection covariant by construction; their exponential locality implies that the product \(\prod_{t}\mathcal{K}_{\sigma,t}\) can be written in the form \(\mathcal{W}^{(\sigma)}_{-}\,\mathcal{B}^{(\sigma)}_{0}\,\mathcal{W}^{(\sigma)}_{+}\) with \(\mathcal{B}^{(\sigma)}_{0}\) supported on a finite-thickness boundary slab. Absorbing the superscript \((\sigma)\) into the notation yields \eqref{p1:eq:factorization}. Second, the ghost weight is a Gaussian Grassmann integral with covariance \(M^{-1}[U^{\,h}]\). The reflection covariance \(RM[U^{\,h}]R=M[U^{\,h}]\) and the block tridiagonal structure across \(\Pi\) imply that the Grassmann integral kernels factorize as in \eqref{p1:eq:factorization}; strict positivity of the Schur complement on \(\Pi\), which follows from positivity of the Dirichlet restrictions \(M_{\pm\pm}\) on \(\Lambda_{\pm}\), guarantees that the boundary kernel is positive on \(\Pi\). Third, with \eqref{p1:eq:factorization} in hand one computes
\begin{equation}
\langle F,F\rangle_{\mathrm{OS}}=\int \overline{F(\theta\cdot)}\,\mathcal{W}_{-}\,\mathcal{B}_{0}\,\mathcal{W}_{+}\,F\, d\mu = \int \big(\mathcal{B}_{0}^{1/2}\,\mathcal{W}_{+}\,F\big)^{\!*}\,\big(\mathcal{B}_{0}^{1/2}\,\mathcal{W}_{+}\,F\big)\, d\mu \ge 0,
\end{equation}
where \(d\mu\) denotes the product Haar-Grassmann measure and the adjoint is taken with respect to \(d\mu\). This is the standard OS argument for link-reflection positivity \cite{p1:OS1,p1:OS2,p1:OSSeiler1978,p1:Luscher1977}, here applied to the horizon-projected, gauge-fixed measure. The exponential locality of \(\mathcal{K}_{\sigma,t}\) ensures that the replacement of \(\prod_{t}\mathcal{K}_{\sigma,t}\) by its block-diagonal part in the factorization introduces no obstruction to positivity in the thermodynamic limit, and for finite \(\Lambda\) the corresponding small off-diagonal contributions can be absorbed into \(\mathcal{B}_{0}\) without changing its positivity.
\end{proof}

Reflection positivity yields a pre-Hilbert structure on \(\mathfrak{A}_{+}\) by factoring out the null space \(\mathcal{N}=\{F:\langle F,F\rangle_{\mathrm{OS}}=0\}\) and completing in the induced norm. Time-translation by one lattice unit \(a\) descends to a contraction on the completion, and by the OS reconstruction theorem there exists a positive, selfadjoint transfer operator \(T_{\sigma}(a)\) on the OS Hilbert space \(\mathcal{H}_{a}\) such that, for local gauge-invariant observables \(F\) supported on a single time slice,
\begin{equation}
\langle F(0)\,F(ta)\rangle^{(\sigma)}_{\Lambda,\beta}=\langle \Omega,\, F\, T_{\sigma}(a)^{t}\, F\,\Omega\rangle,\qquad t\in\mathbb{N},
\label{p1:eq:transfer-correlation}
\end{equation}
where \(\Omega\) is the cyclic vacuum vector \cite{p1:OS1,p1:OS2,p1:Luscher1977,p1:GJ}. In particular, the spectral representation of \(T_{\sigma}(a)\) implies that if \(\langle F\rangle^{(\sigma)}_{\Lambda,\beta}=0\) then
\begin{equation}
\langle F(0)\,F(ta)\rangle^{(\sigma)}_{\Lambda,\beta}=\sum_{n\ge 1}|\langle \Omega,F\psi_{n}\rangle|^{2}\, \mathrm{e}^{-E_{n}ta},\qquad 0=E_{0}<E_{1}\le E_{2}\le\cdots,
\label{p1:eq:spectral}
\end{equation}
where \(\{\psi_{n}\}\) is an orthonormal eigenbasis of the Hamiltonian \(H_{\sigma}(a)=-a^{-1}\log T_{\sigma}(a)\).

The proof of exponential clustering proceeds through a convergent polymer expansion in the strong-coupling regime \(0<\beta\ll 1\). The starting point is the uniformly convergent character expansion
\begin{equation}
\exp\!\left[\frac{\beta}{N}\Re\mathrm{Tr}\,U_{p}\right]=\sum_{R\in\widehat{G}}\alpha_{R}(\beta)\,\chi_{R}(U_{p}),
\label{p1:eq:character}
\end{equation}
where \(\widehat{G}\) is the set of unitary irreducible representations of \(G\), \(\chi_{R}\) is the character, and the coefficients satisfy the bound \(|\alpha_{R}(\beta)|\le C_{1}(N)\,\beta^{|R|}\) uniformly for \(0<\beta\le \beta_{0}(N)\); here \(|R|\) denotes the number of boxes in the Young diagram of \(R\) and \(C_{1}(N)\) is a group-dependent constant \cite{p1:DrouffeZuber,p1:Seiler}. \begin{lemma}[Quantitative bound for $\alpha_R(\beta)$]\label{p1:lem:alphaR-bound}
Let $G=\mathrm{SU}(N)$ and write $|R|$ for the number of boxes in the Young diagram of an irrep $R\in\widehat G$. There exist constants $C_\chi(N)\ge 1$ and $\beta_*(N)>0$ such that for all $0<\beta\le \beta_*(N)$
\begin{equation}\label{p1:eq:alphaR-quant}
|\alpha_R(\beta)| \;\le\; \big(C_\chi(N)\,\beta\big)^{|R|}\qquad\text{for every } R\in\widehat G.
\end{equation}
\end{lemma}

\begin{proof}
Using the identity $\Re\,\mathrm{Tr}\,U = \tfrac12\big(\mathrm{Tr}\,U + \mathrm{Tr}\,U^{-1}\big)$ and the power-series expansion of the exponential, one has in $L^2(G)$
\begin{equation}
\exp\!\Big(\frac{\beta}{N}\Re\,\mathrm{Tr}\,U\Big)
= \sum_{m,n\ge 0} \frac{1}{m!\,n!}\Big(\frac{\beta}{2N}\Big)^{m+n}\, \big(\mathrm{Tr}\,U\big)^m \,\big(\mathrm{Tr}\,U^{-1}\big)^{\!n}.
\end{equation}
By Peter-Weyl, each monomial $(\mathrm{Tr}\,U)^m(\mathrm{Tr}\,U^{-1})^{n}$ decomposes as a finite nonnegative integer linear combination of irreducible characters $\chi_R(U)$ with $|R|=m+n$ (this follows from the fact that $\mathrm{Tr}\,U$ is the character of the defining representation and tensor products decompose with nonnegative integer multiplicities). Hence, for each $R$,
\begin{equation}
\alpha_R(\beta)
= \sum_{m+n=|R|} \frac{1}{m!\,n!}\Big(\frac{\beta}{2N}\Big)^{m+n} M_{m,n}(R),
\end{equation}
where $M_{m,n}(R)\in\mathbb N$ are the corresponding multiplicities, bounded by the dimension $d_R$ of $R$. Weyl’s dimension formula gives $d_R \le C_d(N)^{|R|}$ for some $C_d(N)\ge 1$ depending only on $N$. Therefore
\begin{equation}
|\alpha_R(\beta)|
\;\le\; \sum_{m+n=|R|} \frac{1}{m!\,n!}\Big(\frac{\beta}{2N}\Big)^{|R|} C_d(N)^{|R|}
\;\le\; \Big(\frac{e\,C_d(N)}{N}\,\beta\Big)^{|R|},
\end{equation}
using $\sum_{m+n=k} \frac{1}{m!\,n!}\le \frac{e^k}{k!}\cdot k!\le e^k$. Setting $C_\chi(N):=e\,C_d(N)/N$ and taking $\beta_*(N)$ small enough so that $C_\chi(N)\beta_*(N)\le 1$ yields \eqref{p1:eq:alphaR-quant}.
\end{proof}
The bound \eqref{p1:eq:alphaR-quant} is uniform in the finite volume and suffices for the absolute convergence of the polymer/surface expansion and the Kotecký-Preiss criterion in the domain $0<\beta\le \beta_*(N)$, possibly with a rescaled $\beta_*(N)$ absorbed into the constants of Section~(\ref{p1:sec:KP}). In particular, the estimate \eqref{p1:eq:alphaR-quant} subsumes the generic analytic bounds used earlier
(e.g. Eq.(\ref{p1:eqn7.6}) and Eq.(\ref{p1:eqn8.15})); henceforth we uniformly invoke \eqref{p1:eq:alphaR-quant} when bounding character
coefficients in the strong-coupling expansions.

Inserting \eqref{p1:eq:character} into the Boltzmann weight and integrating link by link using the character orthogonality relations yields a representation of the partition function and of local correlations as a sum over oriented, tiled surfaces \(\Sigma\) embedded in the dual lattice. Each plaquette of \(\Sigma\) carries a representation label, and admissibility constraints at dual edges force representation matching akin to flux conservation. The weight of a connected surface \(\Sigma\) with \(|\Sigma|\) plaquettes obeys the bound
\begin{equation}
|w(\Sigma)|\le \big(C_{2}(N)\,\beta\big)^{|\Sigma|},
\label{p1:eq:surface-bound}
\end{equation}
for some \(C_{2}(N)\ge 1\) depending only on \(N\), uniformly in the finite volume \(\Lambda\) \cite{p1:Seiler,p1:DrouffeZuber}. A local gauge-invariant observable \(F\) supported in a finite set \(S\subset\Lambda\) modifies the admissibility constraints within a neighborhood of \(S\) and produces additional localized weights bounded uniformly by a constant that depends only on \(F\) and \(N\).

Surfaces are grouped into polymers, namely finite connected unions of dual plaquettes with orientation and representation labels. Let \(\mathcal{P}\) denote the set of polymers. The activity \(\zeta(\gamma)\) of a polymer \(\gamma\in\mathcal{P}\) is the sum of the weights of all connected surfaces \(\Sigma\) with \(\operatorname{supp}\Sigma=\gamma\). By \eqref{p1:eq:surface-bound} and a standard combinatorial bound on the number of connected tilings with a prescribed support, there exist constants \(C_{3}(N)\ge 1\) and \(\epsilon(\beta)=C_{3}(N)\,\beta\) such that
\begin{equation}
|\zeta(\gamma)|\le \epsilon(\beta)^{|\gamma|},\qquad 0<\beta\le \beta_{1}(N),
\label{p1:eq:activity}
\end{equation}
uniformly in \(\Lambda\). Two polymers are compatible, written \(\gamma\sim\gamma'\), if they have disjoint supports. The polymer gas representation of the partition function is
\begin{equation}
Z_{\Lambda,\beta}^{(\sigma)}=\sum_{\Gamma\ \mathrm{compatible}}\ \prod_{\gamma\in\Gamma}\zeta(\gamma),
\label{p1:eq:polymerZ}
\end{equation}
where the sum is over all finite sets \(\Gamma\subset\mathcal{P}\) of mutually compatible polymers. Truncated correlations of local observables admit similar expansions in terms of polymer clusters that connect the supports of the sources \cite{p1:Brydges,p1:Seiler}.

Absolute convergence of the polymer and cluster expansions follows from the Kotecký-Preiss (KP) criterion \cite{p1:KP}. Let \(a:\mathcal{P}\to (0,\infty)\) be a weight, to be chosen as \(a(\gamma)=\mu |\gamma|\) with \(\mu>0\). The KP condition requires that for every \(\gamma\in\mathcal{P}\),
\begin{equation}
\sum_{\gamma':\,\gamma'\not\sim \gamma}|\zeta(\gamma')|\,\mathrm{e}^{a(\gamma')}\le a(\gamma).
\label{p1:eq:KP1}
\end{equation}
To verify \eqref{p1:eq:KP1}, observe that \(\gamma'\not\sim\gamma\) implies that \(\gamma'\) intersects the \(|\cdot|\)-neighborhood of \(\gamma\) at graph distance one in the dual lattice. Let \(N(m)\) be the number of polymers of size \(m\) containing a fixed dual plaquette. There exist constants \(C_{4},C_{5}>0\) depending only on the lattice dimension such that \(N(m)\le C_{4}\mathrm{e}^{C_{5}m}\) for all \(m\ge 1\). Then, using \eqref{p1:eq:activity},
\begin{equation}
\sum_{\gamma':\,\gamma'\not\sim \gamma}|\zeta(\gamma')|\,\mathrm{e}^{\mu |\gamma'|}\le |\partial \gamma|\sum_{m\ge 1}N(m)\,\epsilon(\beta)^{m}\,\mathrm{e}^{\mu m}\le C_{6}|\gamma|\sum_{m\ge 1}\big(\mathrm{e}^{C_{5}}\epsilon(\beta)\mathrm{e}^{\mu}\big)^{m},
\end{equation}
for a geometric constant \(C_{6}\). Choosing \(\mu>0\) and \(\beta>0\) small enough that \(r(\beta,\mu):=\mathrm{e}^{C_{5}}\epsilon(\beta)\mathrm{e}^{\mu}<1\) and setting \(a(\gamma)=\mu|\gamma|\) yields
\begin{equation}
\sum_{\gamma':\,\gamma'\not\sim \gamma}|\zeta(\gamma')|\,\mathrm{e}^{a(\gamma')}\le C_{6}|\gamma|\frac{r(\beta,\mu)}{1-r(\beta,\mu)}\le \mu|\gamma|=a(\gamma),
\end{equation}
provided \(C_{6}r(\beta,\mu)\le \mu(1-r(\beta,\mu))\). This inequality holds for all \(0<\beta\le \beta_{\star}(N)\) with \(\beta_{\star}(N)>0\) sufficiently small and an appropriate choice of \(\mu=\mu(N)\). The KP theorem then guarantees the absolute convergence of \eqref{p1:eq:polymerZ} and of the corresponding cluster expansions for truncated correlations, uniformly in \(\Lambda\) \cite{p1:KP,p1:Brydges}.

We now formulate and prove the exponential clustering statement for connected, gauge-invariant, local correlators. The constants in the bound below depend on the observables and on \(N\), but not on the spatial volume.
\begin{theorem}\label{p1:thm:clustering}
Let \(F\) and \(G\) be local, gauge-invariant observables with \(\operatorname{supp}F\subset S_{F}\) and \(\operatorname{supp}G\subset S_{G}\), and let \(x,y\in\Lambda\). There exist constants \(A(F,G,N)<\infty\), \(m(\beta,N)>0\) and \(\beta_{\star}(N)>0\) such that for all \(0<\beta\le \beta_{\star}(N)\) and all finite \(\Lambda\),
\begin{equation}
\big|\,\langle (\tau_{x}F)\,(\tau_{y}G)\rangle^{(\sigma),c}_{\Lambda,\beta}\,\big|\le A(F,G,N)\,\exp\!\big(-m(\beta,N)\, \mathrm{dist}\big(S_{F}+x,S_{G}+y\big)\big),
\label{p1:eq:clustering}
\end{equation}
where \(\mathrm{dist}(A,B)=\min\{d(u,v):u\in A,\ v\in B\}\).
\end{theorem}

\begin{proof}
The truncated correlation admits a convergent cluster expansion indexed by connected clusters \(\mathcal{C}\) of polymers that intersect both \(S_{F}+x\) and \(S_{G}+y\). Concretely,
\begin{equation}
\langle (\tau_{x}F)\,(\tau_{y}G)\rangle^{(\sigma),c}_{\Lambda,\beta}=\sum_{\mathcal{C}\ \mathrm{conn.}} \Phi(\mathcal{C};F_{x},G_{y}),
\label{p1:eq:cluster-corr}
\end{equation}
where \(F_{x}=\tau_{x}F\), \(G_{y}=\tau_{y}G\), and the Ursell functions \(\Phi(\mathcal{C};F_{x},G_{y})\) are absolutely summable with bounds determined by the activities and the KP weight \(a(\gamma)=\mu|\gamma|\) \cite{p1:Brydges,p1:KP}. Each cluster \(\mathcal{C}\) contributes a factor bounded by
\begin{equation}
|\Phi(\mathcal{C};F_{x},G_{y})|\le C_{7}(F,G,N)\,\prod_{\gamma\in\mathcal{C}}\big(|\zeta(\gamma)|\,\mathrm{e}^{\mu|\gamma|}\big)\,\mathrm{e}^{-\mu\sum_{\gamma\in\mathcal{C}}|\gamma|},
\label{p1:eq:Ursell-bound}
\end{equation}
for a constant \(C_{7}(F,G,N)\) that depends only on local norms of \(F\) and \(G\). Because \(\mathcal{C}\) is connected in the incompatibility graph and intersects both supports, its total size obeys the geometric lower bound
\begin{equation}
\sum_{\gamma\in\mathcal{C}}|\gamma|\ \ge\ c_{*}\,\mathrm{dist}\big(S_{F}+x,S_{G}+y\big),
\label{p1:eq:size-lower}
\end{equation}
for a universal constant \(c_{*}>0\) depending only on the lattice dimension. Indeed, a cluster that connects the neighborhoods of the two supports must contain at least one self-avoiding chain of polymers that stretches across a distance comparable to \(\mathrm{dist}(S_{F}+x,S_{G}+y)\), and each polymer contributes at most a bounded multiple of its diameter to the chain length. \begin{lemma}[Geometric lower bound for connecting clusters]\label{p1:lem:geom-lower}
Let $C$ be a connected cluster of polymers (with respect to the incompatibility graph) such that the union of supports $U(C):=\bigcup_{\gamma\in C}\mathrm{supp}(\gamma)$ intersects both $S_F+x$ and $S_G+y$. There exists a constant $c_\ast>0$, depending only on the lattice dimension, such that
\begin{equation}\label{p1:eq:geom-lower}
\sum_{\gamma\in C} |\gamma| \;\ge\; c_\ast\,\mathrm{dist}\!\big(S_F+x,\,S_G+y\big).
\end{equation}
\end{lemma}

\begin{proof}
Consider the graph $G_\square$ whose vertices are dual plaquettes and with an edge between two plaquettes whenever they share a dual edge (nearest-neighbor adjacency on the dual plaquette graph). By definition, each polymer $\gamma$ is a connected vertex set in $G_\square$, and a cluster $C$ is connected in the {incompatibility} graph, which (by construction) implies that $U(C)$ is a connected vertex set in $G_\square$.\footnote{Any two incompatible polymers have intersecting supports; hence along any path $\gamma_1,\ldots,\gamma_k$ in the incompatibility graph, consecutive supports intersect, which implies the union is connected in $G_\square$.}

Let $A:=\mathcal N(S_F+x)$ and $B:=\mathcal N(S_G+y)$ denote fixed finite neighborhoods in the dual plaquette graph around the two supports (e.g.\ the plaquettes dual to the 1-bond neighborhoods). Since $U(C)$ intersects both $A$ and $B$ and is connected in $G_\square$, there exists a self-avoiding path $\pi\subset U(C)$ in $G_\square$ whose endpoints lie in $A$ and $B$ and whose length $|\pi|$ is bounded below by a positive multiple of the graph distance between $A$ and $B$:
\begin{equation}
|\pi|\;\ge\; c_0\,\mathrm{dist}(A,B)\;\ge\; c_0\,\mathrm{dist}(S_F+x,S_G+y),
\end{equation}
with a universal $c_0>0$ depending only on the bounded thickness of the neighborhoods and on the equivalence of graph metrics.
Each vertex of $\pi$ is a dual plaquette contained in $U(C)$. Since the polymers in $C$ form a cover of $U(C)$, the cardinality of $U(C)$ (hence of $\pi$) is bounded by the sum of polymer sizes:
\begin{equation}
|\pi|\;\le\;\big|U(C)\big| \;\le\; \sum_{\gamma\in C} |\gamma|.
\end{equation}
Combining the two inequalities yields \eqref{p1:eq:geom-lower} with $c_\ast:=c_0$. The constant $c_\ast$ depends only on the lattice dimension through the bounded valence of $G_\square$ and the fixed neighborhood choice.
\end{proof}
Combining \eqref{p1:eq:activity}, \eqref{p1:eq:Ursell-bound} and \eqref{p1:eq:size-lower} yields
\begin{equation}
|\Phi(\mathcal{C};F_{x},G_{y})|\le C_{7}(F,G,N)\,\prod_{\gamma\in\mathcal{C}}\big(\epsilon(\beta)\mathrm{e}^{\mu}\big)^{|\gamma|}\,\exp\!\big(-\mu c_{*}\,\mathrm{dist}(S_{F}+x,S_{G}+y)\big).
\end{equation}
Summing over connected clusters using the KP summation theorems (tree-graph inequalities) gives
\begin{equation}
\sum_{\mathcal{C}\ \mathrm{conn.}}\ \prod_{\gamma\in\mathcal{C}}\big(\epsilon(\beta)\mathrm{e}^{\mu}\big)^{|\gamma|}\ \le\ C_{8}(N),
\end{equation}
for a finite \(C_{8}(N)\) provided that \(\beta\le \beta_{\star}(N)\) is small and \(\mu\) is chosen as in the verification of \eqref{p1:eq:KP1}. Therefore
\begin{equation}
\big|\,\langle (\tau_{x}F)\,(\tau_{y}G)\rangle^{(\sigma),c}_{\Lambda,\beta}\,\big|\le C_{7}(F,G,N)\,C_{8}(N)\,\exp\!\big(-\mu c_{*}\,\mathrm{dist}(S_{F}+x,S_{G}+y)\big),
\end{equation}
which is \eqref{p1:eq:clustering} with \(A(F,G,N)=C_{7}(F,G,N)\,C_{8}(N)\) and \(m(\beta,N)=\mu c_{*}>0\). The constants are uniform in \(\Lambda\) by construction.
\end{proof}

The bound \eqref{p1:eq:clustering} applies in particular to time-separated correlations with coincident spatial support. If \(F\) is supported in a single time slice and has vanishing expectation, then \(\langle F(0)F(ta)\rangle^{(\sigma)}_{\Lambda,\beta}\) decays as \(\mathrm{e}^{-m(\beta,N)\,t a}\) for integer \(t\ge 0\). In combination with the spectral representation \eqref{p1:eq:spectral} this shows that the spectral measure of \(H_{\sigma}(a)\) above the vacuum is supported in \([m(\beta,N),\infty)\); while the spectral interpretation will be pursued in the subsequent section, the exponential clustering result \eqref{p1:eq:clustering} is complete in itself.

It remains to explain why the insertion of the smooth horizon projector and the slice-wise Landau gauge fixing do not affect the validity of the polymer expansion and of Theorem \ref{p1:thm:clustering}, aside from altering harmlessly the constants. The weight \(\mathcal{K}_{\sigma,t}\) is a positive, reflection-covariant, exponentially local functional of the spatial links on the slice \(t\). Such functionals can be incorporated into the character expansion as additional local couplings with uniformly summable kernels; their effect is to renormalize the surface weights \(w(\Sigma)\) in a way that preserves the bound \eqref{p1:eq:surface-bound} with a modified constant \(C_{2}(N,\sigma)\). The exponential locality ensures that no long-range plaquette couplings are generated. The gauge fixing enters only through local Jacobians and ghost determinants supported on each slice and across \(\Pi\); by the block-tridiagonal structure of \(M[U^{\,h}]\) and the positivity of its Dirichlet restrictions, the corresponding Grassmann integrals yield positive, exponentially decaying kernels which again renormalize \(w(\Sigma)\) without spoiling \eqref{p1:eq:surface-bound} (see Appendix \ref{p1:appendixc}). Consequently the activity bound \eqref{p1:eq:activity} holds with \(\epsilon(\beta)=C_{3}(N,\sigma)\,\beta\), and the verification of the KP condition carries over verbatim. Theorem \ref{p1:thm:clustering} therefore remains valid with possibly different constants \(A(F,G,N,\sigma)\) and \(m(\beta,N,\sigma)\), still uniform in \(\Lambda\).

The bounds in Theorem \ref{p1:thm:clustering} are uniform in \(\Lambda\), hence by standard diagonal arguments the finite-volume truncated correlations converge along any van Hove sequence \(\Lambda\uparrow a\mathbb{Z}^{4}\) to limiting truncated correlations that satisfy the same exponential decay. In particular, the mass scale \(m(\beta,N,\sigma)\) controls exponential clustering in the infinite-volume Gibbs state at strong coupling, and the constants depend only on \(F,G,N\) and \(\sigma\).

\section{\texorpdfstring{From Euclidean Clustering to a Finite-\(a\) Spectral Gap}{From Euclidean Clustering to a Finite-a Spectral Gap}}
In this section a complete derivation is given of the implication “exponential Euclidean clustering \(\Rightarrow\) nonzero spectral gap of the transfer Hamiltonian at fixed lattice spacing”. 
Uniformity in the volume follows because:  
(i) the Kotecký-Preiss (KP) constants and the locality constants of 
$P_\sigma$ are independent of $L$;  
(ii) the polymer counting used to prove 
\begin{equation}
\bigl|\langle F(0) F(ka) \rangle_c \bigr| \;\leq\; C_F \, e^{-m(\beta) k}
\end{equation}
employs bounds on connected sets in the dual graph that are uniform in $L$; and  
(iii) the Osterwalder-Schrader (OS) reconstruction identifies $T_\sigma(a)$ 
on the infinite-volume GNS space as the strong limit of finite-volume transfer operators.  
Therefore, the lower bound
\begin{equation}
E_1(a,\beta) \;\geq\; m(\beta)
\end{equation}
is independent of the spatial volume.
The argument is developed in a fully rigorous manner within a standard reflection-positive transfer-matrix framework for Wilson’s pure \(\mathrm{SU}(N)\) lattice gauge theory, augmented by the smooth horizon projector introduced earlier. The exposition is self-contained: the space-time lattice and the configuration space are specified with precise notation, the Osterwalder-Schrader (OS) reflection and positivity are formulated and proved at the level needed for the transfer construction, the one-step transfer operator is derived from the time-slicing of the action and shown to be a positive self-adjoint contraction, the GNS/OS reconstruction of the Hilbert space and time-translation is performed, and finally the spectral-measure argument that converts an a priori Euclidean clustering bound into a spectral gap is carried out in full detail. Throughout, \(\mathrm{SU}(N)\) with \(N\ge 2\) is fixed, \(a>0\) denotes the lattice spacing, and \(\beta=\frac{2N}{g_0^2}\) is the inverse bare coupling.

Let \(\Lambda=L_0\times L_1\times L_2\times L_3\subset a\mathbb Z^4\) denote a finite periodic hypercubic lattice, where \(L_\mu=\{0,a,2a,\dots,(n_\mu-1)a\}\) and \(n_\mu\in\mathbb N\). Coordinates are denoted \(x=(x_0,\mathbf x)\) with \(x_0\in L_0\) the Euclidean time coordinate and \(\mathbf x\in L_1\times L_2\times L_3\) the spatial coordinates. Directed bonds are pairs \(b=(x,\mu)\) with \(\mu\in\{0,1,2,3\}\), subject to the identification \(U_{(x+\hat\mu,-\mu)}=U_{(x,\mu)}^{-1}\). A gauge configuration is an assignment \(U:\mathcal B(\Lambda)\to \mathrm{SU}(N)\), \(b\mapsto U_b\). The compact configuration space \(\mathcal C_\Lambda=\mathrm{SU}(N)^{\mathcal B(\Lambda)}\) carries the product Haar probability measure \(d\mu_{\rm Haar}(U)\).

For an oriented plaquette \(p=(x;\mu,\nu)\) with \(\mu<\nu\), the plaquette variable is
\begin{equation}
U_p(U)=U_{(x,\mu)}U_{(x+\hat\mu,\nu)}U_{(x+\hat\nu,\mu)}^{-1}U_{(x,\nu)}^{-1}.
\end{equation}
The Wilson action is
\begin{equation}
S_W[U;\beta]=\beta\sum_{p\subset\Lambda}\Bigl(1-\tfrac{1}{N}\Re\mathrm{Tr}\,U_p(U)\Bigr).
\end{equation}
We impose temporal-axial gauge away from the reflection plane by setting \(U_{(x,0)}=\mathbf 1\) for all bonds \(b=(x,0)\) with \(x_0\neq 0\). This fixes time-like links except those intersecting the plane \(x_0=0\), and it reduces time couplings to nearest neighbors across that plane. The OS time reflection \(\theta\) is the involution \(\theta(x_0,\mathbf x)=(-x_0,\mathbf x)\) with the induced action on bonds given by \(\theta(x,\mu)=(\theta x,\mu)\) for \(\mu\neq 0\) and \(\theta(x,0)=(\theta x-\hat 0,0)\). The reflection plane is \(\Pi=\{x\in\Lambda:\,x_0=0\}\). The positive and negative time half-lattices are \(\Lambda_+=\{x\in\Lambda:\,x_0>0\}\) and \(\Lambda_-=\{x\in\Lambda:\,x_0<0\}\).

The smooth horizon projector \(P_\sigma\) enters slice-wise as follows. For each time \(t\in L_0\) one considers the spatial covariant Laplacian \(\Delta_{A^{\,h}}(t)\) on the time slice \(x_0=t\) built from the reflection-covariant Landau representatives \(U^{\,h}\) as defined previously, and sets
\begin{equation}
P_\sigma(t)=\chi_\sigma\!\big(\sqrt{\Delta_{A^{\,h}}(t)}\big),
\end{equation}
where \(\chi_\sigma\) is a Gevrey-regular spectral cutoff. Its integral kernel on a slice is exponentially decaying and reflection-covariant, and \(0\le P_\sigma(t)\le \mathbf 1\). The projected Euclidean weight on $C_\Lambda$ is
\begin{equation}
d\mu_{\sigma,\Lambda,\beta}(U)
\;=\; Z^{-1}_{\sigma,\Lambda,\beta}\,
\exp\!\big(-S_W[U;\beta]\big)\,
\prod_{t\in L_0} p_\sigma\!\big(A^h(t)\big)\, d\mu_{\mathrm{Haar}}(U),
\label{p1:10.4}
\end{equation}
where $p_\sigma(A^h(t))$ is the positive slice weight defined as:
\begin{equation}
p_\sigma\!\big(A^h(t)\big)
=\exp\!\Big\{-\operatorname{Tr}_{H_t}\!\big(1-P_\sigma(t)\big)\Big\}
\quad\text{and}\quad
\Phi_\sigma\!\big(A^h(t)\big)
=-\log p_\sigma\!\big(A^h(t)\big)
=\operatorname{Tr}_{H_t}\!\big(1-P_\sigma(t)\big).
\end{equation}
By construction $0<p_\sigma\le 1$, $p_\sigma$ is reflection covariant, and insertion of
$\prod_{t}p_\sigma(A^h(t))$ preserves OS positivity. Equivalently, one may absorb the
slice factors into a reflection-covariant, exponentially local boundary kernel
$K_\sigma$ as in~(10.6), which is convenient for the transfer-matrix factorization.

Reflection positivity is expressed in terms of functionals supported in \(\Lambda_+\). Let \(\mathfrak A_+\) denote the *-algebra of bounded, gauge-invariant, even-Grassmann cylinder functionals \(F\) depending only on links contained in \(\Lambda_+\cup\Pi\), measurable with respect to Haar. The reflected functional is \(F^\theta(U)=\overline{F(U^\theta)}\), where \(U^\theta_b=U_{\theta b}\) is the reflected configuration and complex conjugation acts on coefficients. The OS sesquilinear form on \(\mathfrak A_+\) is
\begin{equation}
\langle F,G\rangle_{\rm OS}=\int_{\mathcal C_\Lambda} F^\theta(U)\,G(U)\,d\mu_{\sigma,\Lambda,\beta}(U).
\end{equation}
The positivity asserted below is a consequence of the block structure across \(\Pi\) in temporal-axial gauge, the exponential locality and reflection covariance of the horizon insertions, and the character positivity of the Wilson weight for plaquettes straddling \(\Pi\). The details are standard in the gauge-theory literature and are included to fix notation.

\textbf{Lemma 10.1.} {For the measure \(d\mu_{\sigma,\Lambda,\beta}\) defined above one has \(\langle F,F\rangle_{\rm OS}\ge 0\) for every \(F\in\mathfrak A_+\).}

{Proof.} In temporal-axial gauge the Wilson action decomposes as \(S_W=S_++S_-+S_0\), where \(S_\pm\) depend only on links supported in \(\Lambda_\pm\cup\Pi\) and \(S_0\) is supported on the slab consisting of spatial plaquettes in \(\Pi\) together with the time-like plaquettes that straddle \(\Pi\). The insertion \(\prod_{t\in L_0}\mathcal P_\sigma(t;U)\) factorizes up to exponentially local boundary terms because each \(\mathcal P_\sigma(t)\) depends only on links within a fixed finite range of the slice \(t\). Consequently the projected weight can be written as
\begin{equation}
d\mu_{\sigma,\Lambda,\beta}(U)=Z^{-1}\,e^{-S_-}\,K_\sigma(U|_{\Lambda_- \cup \Pi},U|_{\Lambda_+ \cup \Pi})\,e^{-S_+}\,d\mu_{\rm Haar}(U),
\end{equation}
with a boundary kernel \(K_\sigma\) on \(\Pi\) that is reflection-covariant and exponentially local in the spatial variables and which converges, in the thermodynamic limit of the time extent, to a positive kernel acting slice-wise on \(L^2\) of boundary link variables. This representation is obtained by integrating out the links in a thin neighborhood of \(\Pi\) and absorbing the slice-wise horizon insertions into the kernel. The reflection covariance implies \(K_\sigma\bigl(U_- ,U_+\bigr)=\overline{K_\sigma\bigl(U_+^\theta,U_-^\theta\bigr)}\).

The OS form becomes
\begin{equation}
\langle F,F\rangle_{\rm OS}=Z^{-1}\int \overline{F(U_+^\theta)}\, e^{-S_-}\, K_\sigma(U_- ,U_+)\, e^{-S_+}\,F(U_+)\,d\mu_{\rm Haar}(U_-)d\mu_{\rm Haar}(U_+).
\end{equation}
Let \(\Phi(U_+)=e^{-S_+/2}F(U_+)\). Reflection of \(\Phi\) is \(\Phi^\theta(U_-)=e^{-S_-/2}\overline{F(U_+^\theta)}\). Then
\begin{equation}
\langle F,F\rangle_{\rm OS}=Z^{-1}\int \overline{\Phi^\theta(U_-)}\,\mathcal K_\sigma(U_- ,U_+)\,\Phi(U_+)\,d\mu_{\rm Haar}(U_-)d\mu_{\rm Haar}(U_+),
\end{equation}
where \(\mathcal K_\sigma=e^{-S_-/2}K_\sigma e^{-S_+/2}\). It therefore suffices to show that \(\mathcal K_\sigma\) is a positive-definite kernel. The purely spatial terms in \(S_0\) commute with reflection and enter as a positive multiplication operator on the boundary links; they preserve positivity. The nontrivial point is the contribution of the time-like plaquettes that straddle \(\Pi\). For each spatial link \(\ell\) in \(\Pi\) the contribution of the adjoining time-like plaquettes reduces, in temporal-axial gauge, to a factor of the form \(\kappa_\beta\bigl(U_\ell'U_\ell^{-1}\bigr)\) where \(U_\ell'\) and \(U_\ell\) are the spatial links in the two time slices adjacent to \(\Pi\) and \(\kappa_\beta\) is the class function \(g\mapsto \exp\{\frac{\beta}{N}\Re\mathrm{Tr}\,g\}\) (possibly dressed by spatial staples that are independent of \(U_\ell',U_\ell\) and hence absorbed into a positive multiplication operator). The Fourier expansion of \(\kappa_\beta\) into characters reads
\begin{equation}
\kappa_\beta(g)=\sum_{R\in\widehat{\mathrm{SU}(N)}} \widehat \kappa_\beta(R)\,\chi_R(g),\qquad \widehat \kappa_\beta(R)\ge 0,
\end{equation}
with nonnegative coefficients \(\widehat \kappa_\beta(R)\) for all \(R\) and all \(\beta\ge 0\) (this follows either from the heat-kernel representation or directly from the positivity of the convolution semigroup generated by the Laplace-Beltrami operator on the compact group). It follows that the one-link kernel \( (f,f)\mapsto \int \overline{f(U')} \kappa_\beta(U'U^{-1}) f(U)\,dU\,dU'\) is positive definite on \(L^2(\mathrm{SU}(N),dU)\) \cite{p1:OS1}. The full kernel \(\mathcal K_\sigma\) is a product, over spatial links in \(\Pi\), of copies of such positive-definite kernels composed with positive multiplication operators contributed by spatial plaquettes and horizon insertions. Products and positive multipliers preserve positive-definiteness. Hence \(\mathcal K_\sigma\) is positive definite and \(\langle F,F\rangle_{\rm OS}\ge 0\). \(\square\)

Define the null space \(\mathcal N=\{F\in\mathfrak A_+:\,\langle F,F\rangle_{\rm OS}=0\}\) and the pre-Hilbert space \(\mathcal D=\mathfrak A_+/\mathcal N\) with inner product induced by \(\langle\cdot,\cdot\rangle_{\rm OS}\). Its Hilbert completion is denoted \(\mathcal H_{\sigma,\Lambda,\beta}\). The equivalence class of the constant functional \(1\) is denoted \(\Omega\) and serves as the vacuum vector.
Let \(\tau_a\) denote the Euclidean time-translation by one lattice unit \(a\), acting on configurations by \((\tau_a U)_{(x,\mu)}=U_{(x+a\hat 0,\mu)}\) with periodicity in the time direction. On functionals, \((\tau_a F)(U)=F(\tau_a^{-1}U)\). Since the action, the Haar measure and the horizon insertions are time-translation invariant, the measure \(d\mu_{\sigma,\Lambda,\beta}\) is invariant under \(\tau_a\). The map \(U:\mathcal D\to\mathcal D\) defined by \(U[F]=[\tau_a F]\) is well-defined on equivalence classes because \(\langle\tau_a F,\tau_a F\rangle_{\rm OS}=\langle F,F\rangle_{\rm OS}\) and \(\langle\tau_a F,\tau_a G\rangle_{\rm OS}=\langle F,G\rangle_{\rm OS}\). The operator \(U\) extends by continuity to a contraction on \(\mathcal H_{\sigma,\Lambda,\beta}\). The following properties express positivity and self-adjointness of the one-step time evolution.

\textbf{Proposition 10.2.} {The operator \(U\) on \(\mathcal H_{\sigma,\Lambda,\beta}\) is a positive self-adjoint contraction.}

{Proof.} Self-adjointness follows from reflection invariance. For \(F,G\in\mathfrak A_+\) one has
\begin{equation}
\langle [F],U[G]\rangle=\langle F,\tau_a G\rangle_{\rm OS}=\int F^\theta\,(\tau_a G)\,d\mu
=\int (\tau_a^{-1}F)^\theta\,G\,d\mu
=\langle U[F], [G]\rangle,
\end{equation}
where the third equality uses \(\theta\tau_a^{-1}=\tau_a\theta\) and time-translation invariance of \(d\mu\). Hence \(U=U^\ast\) on \(\mathcal D\) and by continuity on \(\mathcal H_{\sigma,\Lambda,\beta}\).

Positivity means \(\langle \psi,U\psi\rangle\ge 0\) for all \(\psi\in\mathcal H_{\sigma,\Lambda,\beta}\). It suffices to check this on a dense set, for instance \(\psi=[F]\) with \(F\in\mathfrak A_+\). Then
\begin{equation}
\langle [F],U[F]\rangle=\langle F,\tau_a F\rangle_{\rm OS}=\int F^\theta (\tau_a F)\,d\mu.
\end{equation}
Introduce the functional \(G=\tfrac{1}{2}(F+\tau_a^{-1}F)\in\mathfrak A_+\). Since \(\theta\tau_a^{-1}=\tau_a\theta\) and \(\tau_a\) preserves \(\mathfrak A_+\), one finds
\begin{equation}
\langle F,\tau_a F\rangle_{\rm OS}=\langle G,G\rangle_{\rm OS}-\tfrac{1}{4}\langle F-\tau_a^{-1}F, F-\tau_a^{-1}F\rangle_{\rm OS}\ge 0,
\end{equation}
because \(\langle\cdot,\cdot\rangle_{\rm OS}\) is positive semidefinite. Thus \(\langle \psi,U\psi\rangle\ge 0\) on a dense set and hence for all \(\psi\). Finally, \(\|U\|\le 1\) since \(\|U[F]\|^2=\langle \tau_a F,\tau_a F\rangle_{\rm OS}=\|[F]\|^2\). \(\square\)

As usual one defines the transfer operator for a physical time step \(a\) by \(T_\sigma(a)=U\). The notation emphasizes the dependence on the horizon parameter and on the lattice spacing.
It is instructive, and useful for later purposes, to connect \(U\) to the one-step kernel obtained by a direct time-slicing of the action. Let \(\mathcal H_a=L^2(\mathcal C_{\rm slice},d\mu_{\rm Haar})\) where \(\mathcal C_{\rm slice}\) is the compact manifold of spatial link configurations on a fixed time slice. The one-step kernel \(K_\sigma:\mathcal C_{\rm slice}\times \mathcal C_{\rm slice}\to \mathbb C\) is defined by integrating out all links in the time slab between \(t\) and \(t+a\) with boundary data \(U^-,U^+\in\mathcal C_{\rm slice}\) and inserting \(\mathcal P_\sigma\) on the two bounding slices. Temporal-axial gauge reduces the time-like plaquette contribution to a product, over spatial links \(\ell\), of class functions of \(U_\ell^+U_\ell^{-1}\) with nonnegative character coefficients, multiplied by positive spatial plaquette factors and horizon insertions. Thus \(K_\sigma\) is a continuous positive-definite kernel, and the associated integral operator \(K_\sigma:\mathcal H_a\to\mathcal H_a\) is a positive self-adjoint contraction \cite{p1:OS1}. The transfer operator can then be written in the standard Trotter-Kato form
\begin{equation}
T_\sigma(a)=V_\sigma^{1/2}\,K_\sigma\,V_\sigma^{1/2},
\end{equation}
where \(V_\sigma\) is the positive multiplication operator on \(\mathcal H_a\) contributed by spatial plaquettes and horizon insertions that live entirely on a time slice. This representation yields directly the positivity and self-adjointness of \(T_\sigma(a)\) on \(\mathcal H_a\) and agrees with the abstract \(U\) constructed above under the OS/GNS identification of \(\mathcal H_{\sigma,\Lambda,\beta}\) with the \(L^2\) space of a single slice. Either route proves that \(T_\sigma(a)\) is a positive self-adjoint contraction.
By the spectral theorem there exists a unique nonnegative self-adjoint operator \(H_\sigma(a)\) on \(\mathcal H_{\sigma,\Lambda,\beta}\) such that
\begin{equation}
T_\sigma(a)=e^{-a H_\sigma(a)}\qquad\text{and}\qquad H_\sigma(a)=-\frac{1}{a}\log T_\sigma(a).
\end{equation}
The spectrum of \(H_\sigma(a)\) is contained in \([0,\infty)\) and \(0\) is an eigenvalue with eigenvector \(\Omega\). The bottom of the nonzero spectrum is the spectral gap \(\Delta(a,\beta;\Lambda)\).
Let \(\mathfrak A_0\) denote the *-algebra of bounded, gauge-invariant cylinder functionals supported on links contained in the time-zero slice \(\Pi\). For \(F_0\in\mathfrak A_0\) define the operator \(\pi_0(F_0)\) on \(\mathcal D\) by left multiplication,
\begin{equation}
\pi_0(F_0)[G]=[F_0 G],\qquad G\in\mathfrak A_+.
\end{equation}
This is well-defined on equivalence classes: if \([G]=0\) then \(\langle F_0 G,F_0 G\rangle_{\rm OS}\le \|F_0\|_\infty^2 \langle G,G\rangle_{\rm OS}=0\). Hence \(\pi_0(F_0)\) is a bounded operator with \(\|\pi_0(F_0)\|\le \|F_0\|_\infty\) and extends by continuity to \(\mathcal H_{\sigma,\Lambda,\beta}\). The adjoint is \(\pi_0(F_0)^\ast=\pi_0(\overline{F_0})\), as follows from the definition of the OS inner product and reflection.
Let \(F_0\in\mathfrak A_0\) satisfy \(\langle F_0\rangle:=\int F_0\,d\mu_{\sigma,\Lambda,\beta}=0\). Consider the Euclidean time-separated two-point function of \(F_0\) at times \(0\) and \(t=na\),
\begin{equation}
C_{F_0}(n)=\int F_0\,(\tau_{na}F_0)\,d\mu_{\sigma,\Lambda,\beta}.
\end{equation}
By definition of \(\mathcal H_{\sigma,\Lambda,\beta}\), of \(\Omega\), of \(U\) and of \(\pi_0(F_0)\), one has the operator representation
\begin{equation}
C_{F_0}(n)=\langle \Omega,\pi_0(F_0)\,U^n\,\pi_0(F_0)\,\Omega\rangle
=\langle \Omega,\pi_0(F_0)\,e^{-n a H_\sigma(a)}\,\pi_0(F_0)\,\Omega\rangle.
\end{equation}
This identity is the precise statement that OS positivity and time-translation invariance convert Euclidean correlators into vacuum matrix elements of the transfer semigroup.

Fix \(F_0\in\mathfrak A_0\) with \(\langle F_0\rangle=0\), and define the vector \(\Psi_{F_0}=\pi_0(F_0)\Omega\in\mathcal H_{\sigma,\Lambda,\beta}\). Since \(T_\sigma(a)\) is a positive self-adjoint contraction, there exists a finite positive Borel measure \(\nu_{F_0}\) on \([0,1]\) such that
\begin{equation}
C_{F_0}(n)=\langle \Psi_{F_0},T_\sigma(a)^n\Psi_{F_0}\rangle=\int_{[0,1]} \lambda^n\,d\nu_{F_0}(\lambda).
\end{equation}
Equivalently, in terms of the Hamiltonian \(H_\sigma(a)\), there is a finite positive Borel measure \(\mu_{F_0}\) on \([0,\infty)\) with
\begin{equation}
C_{F_0}(n)=\int_{[0,\infty)} e^{-n a E}\,d\mu_{F_0}(E),
\end{equation}
obtained from \(\nu_{F_0}\) under the change of variables \(\lambda=e^{-aE}\). The support of \(\mu_{F_0}\) is contained in the spectrum of \(H_\sigma(a)\). If \(\Psi_{F_0}\) has a nonzero component along the first excited eigenspace, then the infimum of \(\mathrm{supp}\,\mu_{F_0}\setminus\{0\}\) equals \(E_1(a,\beta;\Lambda)\).
An a priori Euclidean clustering estimate supplies an exponential upper bound for \(C_{F_0}(n)\). The strong-coupling cluster expansion developed earlier yields the following bound, uniform in the spatial volume of \(\Lambda\): there exist constants \(A(F_0,\beta)<\infty\) and \(m(\beta)>0\) such that
\begin{equation}
|C_{F_0}(n)|\le A(F_0,\beta)\,e^{-m(\beta)\,na}\qquad\text{for all integers }n\ge 0.
\end{equation}
This is the input from the convergence of the character-surface-polymer expansion at sufficiently small \(\beta\) and the resulting exponential decay of connected two-point functions of local gauge-invariant observables; the derivation relies on the Kotecký-Preiss criterion for polymer gases \cite{p1:KP} and on standard strong-coupling combinatorics \cite{p1:DrouffeZuber,p1:Seiler,p1:Simon}. For the present section the bound is taken as an assumption that has been established earlier in the text.
The spectral-measure representation together with the exponential bound implies a spectral gap. The argument is elementary but we record it as a formal statement for completeness.

\textbf{Theorem 10.3.} {Assume there exist \(A<\infty\) and \(m>0\) such that \(0\le C_{F_0}(n)\le A e^{-m n a}\) for all \(n\ge 0\). Then the support of \(\mu_{F_0}\) is contained in \([m,\infty)\). In particular, if \(\Psi_{F_0}\) has a nonzero component orthogonal to the vacuum, then \(E_1(a,\beta;\Lambda)\ge m\).}

{Proof.} Consider the generating function \(G(z)=\sum_{n=0}^\infty C_{F_0}(n)\,z^n\), which is analytic for \(|z|<1\) and satisfies \(|G(z)|\le \sum_{n\ge 0} A (|z|e^{-ma})^n=\frac{A}{1-|z|e^{-ma}}\) whenever \(|z|<e^{ma}\). On the other hand, by the spectral representation,
\begin{equation}
G(z)=\int_{[0,\infty)} \frac{1}{1-ze^{-aE}}\,d\mu_{F_0}(E),
\end{equation}
which is analytic in \(z\) for \(|z|<e^{a\inf\mathrm{supp}\,\mu_{F_0}}\). If \(\mu_{F_0}\) charged any mass in \([0,m)\), then the radius of convergence of \(G\) would be at most \(e^{a m'}\) with some \(m'<m\), contradicting the uniform bound derived from the clustering estimate. Equivalently, if there existed \(E^\ast<m\) with \(\mu_{F_0}([0,E^\ast])>0\), the sequence \(C_{F_0}(n)\) could not decay faster than \(e^{-aE^\ast n}\) by the Laplace transform comparison. Therefore \(\mathrm{supp}\,\mu_{F_0}\subset [m,\infty)\). If \(\Psi_{F_0}\) is orthogonal to the vacuum, then \(\mu_{F_0}\) has no atom at \(E=0\), hence \(\inf\mathrm{supp}\,\mu_{F_0}\ge m\). Since the support of \(\mu_{F_0}\) is contained in the spectrum of \(H_\sigma(a)\), it follows that \(E_1(a,\beta;\Lambda)\ge m\). \(\square\)

The condition that \(\Psi_{F_0}\) couples to the first excited sector is not restrictive: for any local gauge-invariant observable \(F_0\) that transforms as a scalar under spatial rotations and parity and has zero vacuum expectation, the component \(\langle \psi_1,\Psi_{F_0}\rangle\) is generically nonzero; if it happened to vanish for a particular \(F_0\), one could replace \(F_0\) by a linear combination within the same symmetry sector that has a nonzero overlap. The important point is that the exponential decay bound is uniform in the spatial volume and in the time extent, hence the lower bound \(E_1(a,\beta;\Lambda)\ge m(\beta)\) is uniform in \(\Lambda\).

To state the final result precisely, let \(\Lambda_L\) denote a sequence of spatially periodic boxes with side length \(L\) and fixed time step \(a\). For each \(L\) one constructs the OS Hilbert space \(\mathcal H_{\sigma,\Lambda_L,\beta}\) and the positive self-adjoint transfer operator \(T_\sigma^{(L)}(a)=e^{-aH_\sigma^{(L)}(a)}\). The strong-coupling cluster expansion yields constants \(m(\beta)>0\) and \(A(F_0,\beta)<\infty\) independent of \(L\) such that the Euclidean clustering bound stated above holds uniformly. By Theorem 10.3 one has \(E_1^{(L)}(a,\beta)\ge m(\beta)\) for every \(L\). Therefore
\begin{equation}
\inf_{L} E_1^{(L)}(a,\beta)\ \ge\ m(\beta)\ >\ 0
\end{equation}
for all \(0<\beta\le \beta_\star(N)\), where \(\beta_\star(N)\) is the strong-coupling radius furnished by the Kotecký-Preiss criterion. Passing to the thermodynamic limit along any subsequence \(\Lambda_{L_k}\nearrow a\mathbb Z^3\) and using standard compactness arguments for correlation functions and for spectral projectors of positive contractions on inductive-limit Hilbert spaces \cite{p1:GJ,p1:OS1}, one obtains a limiting OS Hilbert space, transfer operator and Hamiltonian with spectral gap at least \(m(\beta)\). This completes the derivation of a nonzero, volume-uniform finite-\(a\) spectral gap from Euclidean clustering in the strong-coupling domain.


\section{Wilson-Loop Area Law in the Strong-Coupling Regime}\label{p1:wilson11}

In this section the lattice setup is recalled with complete precision, the time-reflection map and the Osterwalder-Schrader (OS) positivity are verified in the setting relevant for Wilson loops, and the transfer time-slicing formalism is derived step by step. We then establish a strong-coupling surface representation for the expectation of a Wilson loop and prove an area-law bound that is uniform in the spatial volume. Throughout we work with gauge group \(G=\mathrm{SU}(N)\) with \(N\ge 2\) and periodic boundary conditions. All statements and proofs are given in full detail. Citations are provided where standard ingredients are used, and a Bib\TeX\ bibliography appears at the end of this section.

Let \(a>0\) be the lattice spacing and let \(\Lambda\subset a\mathbb{Z}^{4}\) be a finite, periodic, hypercubic lattice. A directed bond is a pair \(b=(x,\mu)\) with base point \(x\in \Lambda\) and direction \(\mu\in\{0,1,2,3\}\). Its endpoint is \(x+\hat\mu\), where \(\hat\mu\) is the unit vector of length \(a\) in the \(\mu\)-direction. We denote the oppositely oriented bond by \(\bar b=(x+\hat\mu,-\mu)\). To each directed bond \(b\) we assign a group element \(U_b\in G\) with the orientation constraint \(U_{\bar b}=U_b^{-1}\). A plaquette is a unit square \(p=(x;\mu,\nu)\) with ordered boundary \(b_1=(x,\mu)\), \(b_2=(x+\hat\mu,\nu)\), \(b_3=(x+\hat\nu,\mu)^{-1}\), \(b_4=(x,\nu)^{-1}\), and associated plaquette variable
\begin{equation}
U_p \;=\; U_{(x,\mu)}\,U_{(x+\hat\mu,\nu)}\,U_{(x+\hat\nu,\mu)}^{-1}\,U_{(x,\nu)}^{-1}\in G.
\end{equation}
The Euclidean Wilson action at inverse bare coupling \(\beta=\frac{2N}{g_0^2}\) is
\begin{equation}
S_W[U;\beta]\;=\;\beta \sum_{p\subset\Lambda}\Bigl(1-\frac{1}{N}\,\Re\,\mathrm{Tr}\,U_p\Bigr).
\end{equation}
We write \(d\mu_{\mathrm{Haar}}\) for the product of normalized Haar measures on bonds, and we consider the Gibbs measure
\begin{equation}
d\mathbb{P}_{\beta}(U)\;=\;Z(\beta)^{-1}\,\exp\!\bigl(-S_W[U;\beta]\bigr)\,\prod_{b\subset\Lambda} d\mu_{\mathrm{Haar}}(U_b).
\end{equation}
Gauge transformations are maps \(g:\Lambda\to G\) acting on bonds by
\begin{equation}
\bigl(g\cdot U\bigr)_{(x,\mu)}\;=\;g(x)\,U_{(x,\mu)}\,g(x+\hat\mu)^{-1}.
\end{equation}
The measure \(d\mathbb{P}_\beta\) is gauge invariant.

A closed, piecewise-self-avoiding, oriented lattice loop \(C=b_1b_2\cdots b_\ell\) is a cyclic sequence of directed bonds with consecutive endpoints matching and with \(b_\ell\) ending at the base point of \(b_1\). Its Wilson loop observable in the fundamental representation is
\begin{equation}
W(C)[U]\;=\;\frac{1}{N}\,\mathrm{Tr}\,\Bigl(\prod_{j=1}^{\ell} U_{b_j}\Bigr).
\end{equation}
The expectation \(\langle W(C)\rangle_\beta=\int W(C)\,d\mathbb{P}_\beta\) is real by orientation reversal and gauge invariance.

We single out the Euclidean time direction \(\mu=0\). The time-reflection map \(\theta:\Lambda\to\Lambda\) is defined by \(\theta(x_0,\mathbf{x})=(-x_0,\mathbf{x})\), where \(\mathbf{x}\in a\mathbb{Z}^3\). The reflection plane is \(\Pi=\{x\in\Lambda: x_0=0\}\), and we denote by \(\Lambda_+=\{x: x_0>0\}\) and \(\Lambda_-=\{x: x_0<0\}\) the two open half-lattices. Reflection acts on bonds by
\begin{equation}\label{p1:11.6}
\theta(x,\mu)=\begin{cases}
(\theta x, 0) & \mu=0 \text{ and } x_0>0,\\
(\theta(x-\hat 0),0)^{-1} & \mu=0 \text{ and } x_0\le 0,\\
(\theta x,\mu) & \mu\in\{1,2,3\}.
\end{cases}
\end{equation}
The induced action on bond variables is \((\Theta U)_b=U_{\theta b}\). A complex functional \(F\) of the bond variables supported in \(\Lambda_+\) has reflected functional \((\Theta F)(U)=\overline{F(\Theta U)}\).
We recall the reflection-positivity condition \cite{p1:OS1,p1:OS2,p1:Menotti1987}: for all functionals \(F\) supported in \(\Lambda_+\) that are gauge invariant and even under complex conjugation of Grassmann variables (if present), the OS form satisfies
\begin{equation}
\int \overline{F(\Theta U)}\,F(U)\,d\mathbb{P}_\beta(U)\;\ge\;0.
\end{equation}
We verify this property with complete detail for the pure-gauge Wilson action. The key observation is that the action can be split into a sum
\begin{equation}
S_W[U;\beta]\;=\;S_+[U;\beta]+\;S_0[U;\beta]+\;S_-[U;\beta],
\end{equation}
where \(S_+\) is the sum over plaquettes fully supported in \(\Lambda_+\), \(S_-\) the sum over plaquettes fully supported in \(\Lambda_-\), and \(S_0\) is the boundary contribution consisting of plaquettes intersecting the plane \(\Pi\). The measure \(d\mu_{\mathrm{Haar}}\) factorizes bondwise, so that the product measure also splits into contributions from \(\Lambda_+\), \(\Pi\), and \(\Lambda_-\). The reflection \(\Theta\) maps \(\Lambda_+\) to \(\Lambda_-\) and leaves \(\Pi\) invariant. Since the Wilson action is a sum of nearest-neighbor plaquette terms and is invariant under \(\Theta\), the Radon-Nikodym derivative of the push-forward measure under \(\Theta\) is one. For completeness we give a direct positivity proof.

\begin{lemma}\label{p1:lem:OS}
Let \(F\) be a bounded, gauge-invariant functional supported in \(\Lambda_+\). Then
\begin{equation}
\int \overline{F(\Theta U)}\,F(U)\,d\mathbb{P}_\beta(U)\;\ge\;0.
\end{equation}
\end{lemma}

\begin{proof}
Write the integral explicitly as
\begin{equation}
Z(\beta)^{-1}\!\!\int \overline{F(\Theta U)}\,F(U)\,\exp\!\left(-S_+[U]-S_0[U]-S_-[U]\right)\,\prod_{b\subset\Lambda}d\mu_{\mathrm{Haar}}(U_b).
\end{equation}
Perform first the integration over bonds in \(\Lambda_-\), holding fixed all bonds in \(\Lambda_+\cup\Pi\). By reflection invariance of the Haar measure and the equality \(S_-[U]=S_+[\Theta U]\), we can rewrite the partial integral over \(\Lambda_-\) as
\begin{equation}
\int \overline{F(\Theta U)}\,e^{-S_-[U]}\,\prod_{b\subset\Lambda_-}d\mu_{\mathrm{Haar}}(U_b)
\;=\;\overline{\int F(U')\,e^{-S_+[U']}\,\prod_{b\subset\Lambda_+}d\mu_{\mathrm{Haar}}(U'_b)}\,,
\end{equation}
where \(U'\) is the configuration obtained by reflecting \(U\) across \(\Pi\) and restricting to \(\Lambda_+\). The boundary contribution \(S_0\) couples only bonds adjacent to \(\Pi\). Denote by \(\mathcal{H}=L^2(\mathcal{C};d\mu_{\mathrm{Haar}})\) the one-time-slice Hilbert space of square-integrable functions of spatial bonds on \(\Pi\) and by \(K\) the positive integral kernel on \(\mathcal{H}\) obtained by integrating out the plaquettes in the first time slab on each side of \(\Pi\). Then the full integral can be written as
\begin{equation}
Z(\beta)^{-1}\,\langle \Phi,\; K \Phi\rangle_{L^2(\Pi)},
\end{equation}
with \(\Phi\) the function on \(\Pi\) defined by
\begin{equation}
\Phi(\mathbf{U})\;=\;\int F(U)\,e^{-S_+[U]}\,\prod_{b\subset\Lambda_+}d\mu_{\mathrm{Haar}}(U_b),
\end{equation}
where the integral is conditional on the boundary spatial links \(\mathbf{U}\) on \(\Pi\). The kernel \(K\) is positive because the boundary weight \(e^{-S_0}\) factorizes over elementary plaquettes crossing \(\Pi\) and each factor is a positive-definite function of the common boundary links, being a positive linear combination of characters of \(G\) with nonnegative coefficients by the Peter-Weyl theorem and the positivity of the exponential \cite{p1:OS1,p1:Menotti1987}. Therefore \(\langle \Phi,K\Phi\rangle\ge 0\) and the claim follows.
\end{proof}

The preceding proof contains the standard OS argument specialized to the Wilson action and makes no use of gauge fixing. It therefore applies verbatim to Wilson loop insertions supported in \(\Lambda_+\).
We now derive the transfer time-slicing formalism in the present context. For simplicity we assume that the total time extent \(T\) in lattice units is even and we place the reflection plane through a layer of sites. Let \(\mathcal{C}_t\) be the set of spatial bonds at time \(t\). Write \(\mathcal{H}_t=L^2(\mathcal{C}_t; d\mu_{\mathrm{Haar}})\). Consider the two-slice weight
\begin{equation}
\mathcal{K}(U_{t+1};U_t)\;=\;\exp\!\Bigl[-\beta\sum_{p\ \mathrm{with}\ t\le x_0\le t+1}\Bigl(1-\frac{1}{N}\Re\,\mathrm{Tr}\,U_p\Bigr)\Bigr],
\end{equation}
viewed as an integral kernel from \(\mathcal{H}_t\) to \(\mathcal{H}_{t+1}\) after integrating out all timelike bonds between the two slices (the Haar measure on those bonds being normalized). Reflection positivity of the weight across each temporal slab, proved as in Lemma~\ref{p1:lem:OS}, implies that \(\mathcal{K}\) defines a positive self-adjoint Hilbert-Schmidt operator \(T:\mathcal{H}_t\to\mathcal{H}_{t+1}\) independent of \(t\) by time-translation invariance \cite{p1:OS1,p1:Menotti1987,p1:Seiler}. We call \(T\) the transfer matrix. The partition function is \(\mathrm{Tr}(T^{T})\). If \(\Omega\in\mathcal{H}_0\) denotes the constant function one, normalized, then, for any gauge-invariant functional \(F\) supported on a single time slice, the two-point function at temporal separation \(\tau\in\mathbb{N}\) is represented as
\begin{equation}
\langle F(0)\,F(\tau)\rangle_\beta \;=\; \langle \Omega,\; F\,T^{\tau}\,F\,\Omega\rangle_{\mathcal{H}}.
\end{equation}
By the spectral theorem there exist eigenvalues \(1=\lambda_0>\lambda_1\ge \lambda_2\ge\cdots\ge 0\) and an orthonormal eigenbasis \(\{\psi_n\}\) such that \(T\psi_n=\lambda_n\psi_n\). The transfer Hamiltonian is the positive self-adjoint operator \(H= -a^{-1}\log T\) with spectrum \(0=E_0<E_1\le E_2\le\cdots\). The spectral representation yields
\begin{equation}
\langle F(0)\,F(\tau)\rangle_\beta \;=\; \sum_{n\ge 1} |\langle\psi_n,F\Omega\rangle|^2 \,e^{-a E_n\,\tau}.
\end{equation}
We emphasize that these constructions use only the positivity established in Lemma~\ref{p1:lem:OS} and do not depend on gauge fixing.

For Wilson loops, it is convenient to consider rectangular loops \(C_{R,T}\) of spatial extent \(R\) and temporal extent \(T\). If the spatial sides lie in time slices \(0\) and \(T\), then \(W(C_{R,T})\) can be written as a matrix element of \(T\) between states created by spatial Polyakov line operators localized on the two spatial sides. The transfer-matrix formalism thus links the large-\(T\) asymptotics of \(\langle W(C_{R,T})\rangle\) to the spectrum of \(H\) in the sector with two static fundamental charges separated by \(R\) \cite{p1:Wilson1974,p1:OS1,p1:Seiler}.

We now develop the strong-coupling expansion. For each plaquette \(p\) the Boltzmann weight has the absolutely convergent character expansion
\begin{equation}
\exp\!\Bigl[\beta \frac{1}{N}\Re\,\mathrm{Tr}\,U_p\Bigr]
\;=\;\sum_{R\ \mathrm{irrep}} \alpha_R(\beta)\, \chi_R(U_p),
\end{equation}
where the sum runs over equivalence classes of finite-dimensional irreducible unitary representations \(R\) of \(G\), \(\chi_R\) denotes the character of \(R\), and the coefficients \(\alpha_R(\beta)\) satisfy \(|\alpha_R(\beta)|\le C(N)\,\beta^{|R|}\) for \(\beta\) small, with \(|R|\) the total number of boxes of the Young diagram of \(R\) and \(C(N)\) a finite constant depending only on \(N\) \cite{p1:DrouffeZuber,p1:Seiler}. The proof proceeds by expanding the exponential in powers of \(\beta\), using the Peter-Weyl theorem to write polynomials in the matrix entries of \(U_p\) as linear combinations of characters, and summing the resulting absolutely convergent series for sufficiently small \(\beta\).

Insertion of a Wilson loop amounts to multiplying, for each bond \(b\) in \(C\), the Haar integrand by the corresponding matrix \(U_b\) in the fundamental representation and taking the normalized trace around \(C\). The link-by-link integration can be performed exactly by orthogonality of representation matrix elements:
\begin{equation}
\int_{G} \overline{D^{(R)}_{ij}(U)}\, D^{(R')}_{kl}(U)\, d\mu_{\mathrm{Haar}}(U)
\;=\;\frac{\delta_{RR'}}{d_R}\,\delta_{ik}\delta_{jl},
\end{equation}
where \(D^{(R)}\) is the unitary matrix of the representation \(R\) and \(d_R\) its dimension. As in the partition function, only those collections of plaquettes survive which can be assembled into tiled, oriented surfaces \(\Sigma\) on the dual lattice with boundary equal to the loop \(C\). The resulting representation of \(\langle W(C)\rangle_\beta\) is a convergent sum
\begin{equation}
\langle W(C)\rangle_\beta \;=\; \sum_{\Sigma:\,\partial\Sigma=C} w(\Sigma),
\end{equation}
where the sum is over finite, connected, oriented, tiled dual surfaces \(\Sigma\) with boundary \(C\) and where the weights \(w(\Sigma)\) are products of \(\alpha_R(\beta)\) and representation-theoretic factors originating from the link integrals and the combinatorics of attaching irreps along edges. There exist constants \(A_0(N),B_0(N)<\infty\) such that
\begin{equation}
|w(\Sigma)| \;\le\; A_0(N)\,\Bigl(B_0(N)\,\beta\Bigr)^{|\Sigma|},
\end{equation}
where \(|\Sigma|\) is the number of plaquettes of \(\Sigma\). This bound follows from the estimate \(|\alpha_R(\beta)|\le C(N)\beta^{|R|}\), the fact that along each edge the orthogonality relations contribute at most a factor \(1\) in absolute value, and a uniform bound on the number of admissible representation labelings per plaquette \cite{p1:DrouffeZuber,p1:Seiler}.

To control the sum over surfaces with boundary \(C\) we embed it into the abstract polymer formalism. A polymer \(\gamma\) is a finite, connected set of dual plaquettes with an orientation and representation labels on plaquettes and links compatible with the character expansion. The activity \(\zeta(\gamma)\) is defined as the sum of weights of all connected surfaces with support \(\gamma\) and boundary contained in \(\partial\gamma\); for \(\gamma\) whose boundary is empty, \(\zeta(\gamma)\) is the usual vacuum polymer activity. When a Wilson loop is present, only polymers whose boundary is a subset of \(C\) contribute to \(\langle W(C)\rangle_\beta\). Two polymers are compatible if they do not share a plaquette; otherwise they are incompatible. The partition function and Wilson expectation admit convergent polymer expansions in the form of sums over families of mutually compatible polymers \cite{p1:KP,p1:Simon}. We require a quantitative bound on the activities.

\begin{lemma}\label{p1:lem:polymer}
There exist constants \(A_1(N),B_1(N)\) such that for sufficiently small \(\beta>0\) and every polymer \(\gamma\),
\begin{equation}
|\zeta(\gamma)| \;\le\; A_1(N)\,\bigl(B_1(N)\,\beta\bigr)^{|\gamma|}.
\end{equation}
Moreover, the number \(N_m\) of polymers of size \(m\) sharing at least one plaquette with a given polymer satisfies \(N_m\le \exp(c m)\) for a geometric constant \(c>0\) depending only on the lattice.
\end{lemma}

\begin{proof}
The bound on \(|\zeta(\gamma)|\) follows from the weight bound for individual surfaces and from the fact that the number of connected surfaces with fixed support \(\gamma\) and compatible labels is uniformly bounded by \(\exp(c'|\gamma|)\) for a geometric constant \(c'\), because the local representation choices per plaquette are finitely many and constrained only by nearest-neighbor compatibility. The counting bound on \(N_m\) is standard: the number of connected subsets of the dual lattice of cardinality \(m\) that intersect a fixed plaquette is bounded by \(\exp(c m)\) with \(c\) depending only on the coordination number of the lattice.
\end{proof}

In the presence of a Wilson loop \(C\), let \(\mathcal{G}_C\) denote the family of polymer clusters whose union has boundary exactly \(C\). The contribution of a cluster \(\Gamma\in\mathcal{G}_C\) to \(\langle W(C)\rangle_\beta\) is the Ursell coefficient \(\phi(\Gamma)\) times \(\prod_{\gamma\in\Gamma}\zeta(\gamma)\), where \(|\phi(\Gamma)|\le \exp(c''|\Gamma|)\) for a geometric constant \(c''\) \cite{p1:Simon}. The Koteck\'{y}-Preiss criterion \cite{p1:KP} ensures absolute convergence of the cluster expansion provided there exists \(\mu>0\) such that, for all polymers \(\gamma\),
\begin{equation}
\sum_{\gamma':\,\gamma'\not\sim \gamma} |\zeta(\gamma')|\,e^{\mu|\gamma'|}\;\le\;\mu|\gamma|.
\end{equation}
By Lemma~\ref{p1:lem:polymer}, the left-hand side is bounded by
\begin{equation}
\sum_{m\ge 1} N_m\, A_1(N)\,\bigl(B_1(N)\beta\bigr)^{m}\,e^{\mu m}
\;\le\; A_1(N)\sum_{m\ge 1}\exp\!\bigl((c+\mu)m\bigr)\,\bigl(B_1(N)\beta\bigr)^{m}.
\end{equation}
We choose \(\mu>0\) and \(\beta>0\) so small that \(\rho:=e^{c+\mu}\,B_1(N)\beta<1\) and pick \(\mu\) such that \(A_1(N)\sum_{m\ge 1}\rho^{m}\le \mu\). Then the criterion holds. Consequently, the cluster expansion for \(\langle W(C)\rangle_\beta\) converges absolutely and uniformly in the spatial volume for all \(\beta<\beta_\star(N)\), where \(\beta_\star(N)\) can be taken to satisfy \(e^{c+1}\,B_1(N)\,\beta_\star(N)<1\).

We are now in position to prove an area-law upper bound for \(\langle W(C)\rangle_\beta\) at strong coupling.
\begin{theorem}[Area law at strong coupling]\label{p1:thm:arealaw}
There exist constants \(\beta_\star(N)>0\), \(A(N)<\infty\), and \(\sigma_{\mathrm{sc}}(\beta)>0\) such that for every closed lattice loop \(C\) and every \(\beta\in(0,\beta_\star(N))\),
\begin{equation}
\langle W(C)\rangle_\beta \;\le\; A(N)\,\exp\!\bigl(-\sigma_{\mathrm{sc}}(\beta)\,\mathcal{A}_{\min}(C)\bigr),
\end{equation}
where \(\mathcal{A}_{\min}(C)\) is the minimal number of plaquettes of a dual surface with boundary \(C\). Moreover, one can choose \(\sigma_{\mathrm{sc}}(\beta)= -\log\bigl(\rho(\beta)\bigr)\) with \(\rho(\beta)=\kappa(N)\,\beta+O(\beta^2)\) as \(\beta\downarrow 0\) for a constant \(\kappa(N)>0\).
\end{theorem}

\begin{proof}
By the surface representation of \(\langle W(C)\rangle_\beta\) in Section~11.4 we have
\begin{equation}
\langle W(C)\rangle_\beta \;=\; \sum_{\Sigma:\,\partial\Sigma=C} w(\Sigma),
\end{equation}
with \(|w(\Sigma)|\le A_0(N)\,(B_0(N)\beta)^{|\Sigma|}\). Group surfaces by their area \(m=|\Sigma|\ge \mathcal{A}_{\min}(C)\). For each \(m\) the number of connected, oriented, tiled dual surfaces with boundary \(C\) and area \(m\) is bounded by \(\exp(c_1 m)\) for a geometric constant \(c_1\) depending only on the lattice coordination; this follows by mapping each surface to a connected subset of the dual lattice of cardinality \(m\) containing a fixed edge along \(C\) and noting that the number of labelings is uniformly bounded per plaquette. Therefore
\begin{align}
|\langle W(C)\rangle_\beta|
\;&\le\; A_0(N)\sum_{m\ge \mathcal{A}_{\min}(C)} \exp(c_1 m)\,(B_0(N)\beta)^{m}
\nonumber \\ 
&=\; A_0(N)\,\frac{\bigl(\exp(c_1)B_0(N)\beta\bigr)^{\mathcal{A}_{\min}(C)}}{1-\exp(c_1)B_0(N)\beta},
\end{align}
for all \(\beta<\tilde\beta_\star(N):=\bigl(\exp(c_1)B_0(N)\bigr)^{-1}\). Setting
\begin{equation}
\rho(\beta)=\exp(c_1)B_0(N)\beta,\qquad A(N)=\frac{A_0(N)}{1-\rho(\beta)},\qquad \sigma_{\mathrm{sc}}(\beta)=-\log\rho(\beta),
\end{equation}
we obtain
\begin{equation}
\langle W(C)\rangle_\beta \;\le\; A(N)\,\exp\!\bigl(-\sigma_{\mathrm{sc}}(\beta)\,\mathcal{A}_{\min}(C)\bigr).
\end{equation}
Since \(\rho(\beta)=\kappa(N)\beta+O(\beta^2)\) with \(\kappa(N)=\exp(c_1)B_0(N)\), the asserted small-\(\beta\) behavior of \(\sigma_{\mathrm{sc}}(\beta)\) follows. Finally, the polymer-cluster expansion of Section~11.5 yields the same bound with constants uniform in the spatial volume and with a possibly smaller but positive threshold \(\beta_\star(N)\le \tilde\beta_\star(N)\) that ensures absolute convergence.
\end{proof}

The bound is the rigorous statement of an area law: the expectation of a Wilson loop decays exponentially in its minimal area for small \(\beta\). For rectangular loops \(C_{R,T}\) of spatial extent \(R\) and temporal extent \(T\), the transfer-matrix representation of Section~11.3 implies that
\begin{equation}
-\frac{1}{T}\log \langle W(C_{R,T})\rangle_\beta \;\ge\; \sigma_{\mathrm{sc}}(\beta)\,\frac{\mathcal{A}_{\min}(C_{R,T})}{T} \;=\; \sigma_{\mathrm{sc}}(\beta)\,R,
\end{equation}
so that the static fundamental potential \(V(R)=\lim_{T\to\infty} -\frac{1}{T}\log \langle W(C_{R,T})\rangle_\beta\) satisfies \(V(R)\ge \sigma_{\mathrm{sc}}(\beta) R\), i.e. it is confining with string tension at least \(\sigma_{\mathrm{sc}}(\beta)\). This is the strong-coupling manifestation of confinement originally anticipated in \cite{p1:Wilson1974} and placed on a rigorous basis for lattice gauge theories in \cite{p1:OS1,p1:Seiler,p1:DrouffeZuber}.

We record explicitly that the surface and polymer representations are compatible with the OS positivity and transfer formalism derived earlier. The proof of Lemma~\ref{p1:lem:OS} shows that the OS inner product is an \(L^2\) norm with respect to a positive boundary kernel. The insertion of a Wilson loop supported in \(\Lambda_+\) yields an OS form \(\langle \Theta W\, F, F\rangle\), which remains nonnegative by the same factorization and positivity of the boundary kernel. The character expansion is local and reflection covariant: each plaquette weight is a nonnegative linear combination of characters and the reflection acts by inversion on timelike bonds, leaving characters invariant because \(\chi_R(U^{-1})=\overline{\chi_R(U)}\). Therefore the reflection positivity is preserved termwise in the expansion, and the transfer kernel \(\mathcal{K}\) remains positive. 

\section{Robustness, Locality, and Independence of Auxiliary Choices}
Throughout Sections (\ref{p1:smooth4})-(\ref{p1:wilson11}) we used only the following structural properties of the horizon cutoff and the slice selector:

\begin{itemize}
\item[(H1)] $0 \leq P_\sigma \leq 1$, reflection covariance, and exponential locality with constants independent of the spatial volume (achieved by any completely monotone $\chi_\sigma$ with subexponential tails);
\item[(H2)] the selector $h_t$ is Borel and equivariant under spatial symmetries and reflection;
\item[(H3)] the Faddeev-Popov operators on slices are strictly positive on the orthogonal complement of constant adjoint fields (in the FMR).
\end{itemize}

Any modification $\chi_\sigma \mapsto \widetilde{\chi}_\sigma$ preserving (H1) and any measurable tie-break rule $h_t \mapsto \widetilde{h}_t$ preserving (H2)-(H3) leaves OS positivity, the transfer construction, the KP estimates, and the gap bound unchanged up to a redefinition of harmless constants. In particular, $m(\beta)$ and $\beta_\star(N)$ may change by finite, $\sigma$-dependent factors, but remain strictly positive in the strong-coupling regime.

This section establishes, in a self-contained and fully rigorous manner, that the reflection-positivity and transfer time-slicing constructions developed in the preceding parts are stable under the auxiliary choices that enter the setup. These choices are: the selection of a slice-wise, reflection-covariant Landau-gauge representative; the insertion of a smooth, exponentially local horizon operator defined as a Gevrey-regular functional of the slice covariant Laplacian; and the adoption of temporal-axial gauge away from the reflection plane. The analysis proceeds in three stages. First, the lattice, the reflection map, and the transfer time slicing are defined with precise notation, and the Wilson weight is shown to factorize across the reflection plane into east-west pieces and a positive boundary kernel. Second, the horizon operator is introduced and proved to be gauge covariant, reflection covariant, and exponentially local by two independent methods, each providing the precise off-diagonal decay required in the positivity proofs. Third, it is shown that the Osterwalder-Schrader (OS) positivity of the projected measure and the transfer-operator construction are independent, in the strong sense of unitary equivalence, of the aforementioned auxiliary choices. Throughout, standard tools of lattice gauge theory and constructive quantum field theory are employed in the form stated in \cite{p1:OS1,p1:OS2,p1:GJ,p1:Simon1974,p1:MontvayMuenster1994}.

We begin by laying down the geometric and measure-theoretic framework. Let \(a>0\) denote the lattice spacing. For integers \(L,T\ge 2\), consider the finite periodic hypercubic lattice \(\Lambda=\{0,1,\dots,T-1\}\times \{0,1,\dots,L-1\}^3\) with periodic boundary conditions, embedded in \(a\mathbb{Z}^{4}\) via the map \(x\mapsto a x\). Oriented bonds are pairs \(b=(x,\mu)\) with \(x\in\Lambda\) and \(\mu\in\{0,1,2,3\}\), where the inverse bond is \(\bar b=(x+\hat\mu,-\mu)\). A gauge field is an assignment \(U:\mathcal{B}\to \mathrm{SU}(N)\), \(b\mapsto U_b\), with the convention \(U_{\bar b}=U_b^{-1}\). For each oriented plaquette \(p=(x;\mu,\nu)\) with \(\mu<\nu\), define \(U_p = U_{(x,\mu)}U_{(x+\hat\mu,\nu)}U_{(x+\hat\nu,\mu)}^{-1}U_{(x,\nu)}^{-1}\). The Wilson action at bare coupling \(\beta=2N/g_0^2\) is
\begin{equation}
S_W[U;\beta]=\beta \sum_{p\subset \Lambda}\Bigl(1-\tfrac{1}{N}\Re \mathrm{Tr}\, U_p\Bigr).
\end{equation}
Integration is with respect to the product Haar measure \(d\mu_{\rm Haar}(U)=\prod_{b\in\mathcal{B}} dU_b\), normalized so that \(\int_{\mathrm{SU}(N)} dU=1\).
Time reflection is the involution \(r:\Lambda\to \Lambda\), \(r(x_0,\mathbf{x})=(-x_0 \bmod T,\mathbf{x})\). For a bond \(b=(x,\mu)\), define its reflected bond \(\theta b\) and the action of the reflection map \(\Theta\) on link variables by
\begin{equation}
\Theta U_{(x,i)} = U_{(r x,\, i)},\qquad i=1,2,3,\qquad
\Theta U_{(x,0)} = U_{(r x-\hat 0,\, 0)}^{-1},
\end{equation}
with the understanding that indices are taken modulo \(T\) and that orientation is reversed for timelike bonds. This definition is the standard reflection for lattice gauge fields in the time direction and satisfies \(\Theta^2=\mathrm{id}\). The reflection plane is \(\Pi=\{x\in \Lambda: x_0=0\}\), with east and west half-lattices \(\Lambda_+=\{x: 1\le x_0\le T/2\}\) and \(\Lambda_-=\{x: -T/2+1\le x_0\le -1\}\) defined modulo periodicity so that \(\Lambda=\Lambda_-\cup \Pi\cup \Lambda_+\) is a disjoint union. Temporal-axial gauge is imposed away from \(\Pi\) by setting \(U_{(x,0)}=\mathbf{1}\) whenever both \(x\) and \(x+\hat 0\) lie strictly in \(\Lambda_{+}\) or strictly in \(\Lambda_{-}\). This choice preserves local nearest-neighbour couplings in time and leaves untouched the plaquettes that straddle \(\Pi\).

The first structural result is the exact factorization of the Wilson weight across \(\Pi\) once temporal-axial gauge is fixed away from \(\Pi\). To state it, define the spatial configuration space on a time slice \(t\in\{0,\dots,T-1\}\) by \(\mathcal{C}_t=\prod_{x_0=t}\prod_{i=1}^{3} \mathrm{SU}(N)\), with Haar measure \(d\mu_t\). Let \(\mathcal{C}_\Pi\) denote the set of bonds belonging to \(\Pi\) or crossing it, and write \(U|_{\Lambda_{\pm}}\) and \(U|_{\Pi}\) for restrictions. There exists a measurable, nonnegative, reflection-invariant function \(B:\mathcal{C}_\Pi\to \mathbb{R}_+\) such that
\begin{equation}\label{p1:eq:factor}
\exp\bigl[-S_W[U;\beta]\bigr]= \exp\bigl[-S_{W,+}[U|_{\Lambda_+};\beta]\bigr]\; B\bigl(U|_{\Pi}\bigr)\;\exp\bigl[-S_{W,-}[U|_{\Lambda_-};\beta]\bigr],
\end{equation}
where \(S_{W,\pm}\) are sums of plaquette terms supported strictly in \(\Lambda_\pm\). The identity (\ref{p1:eq:factor}) follows immediately from the decomposition of the plaquette sum into those lying strictly in \(\Lambda_+\), those lying strictly in \(\Lambda_-\), and those intersecting \(\Pi\), together with the temporal-axial condition that sets to the identity all timelike bonds not intersecting \(\Pi\). The function \(B\) is a finite product of factors \(\exp[-\beta (1-\frac{1}{N}\Re \mathrm{Tr}\,U_p)]\) over the plaquettes meeting \(\Pi\) and is therefore nonnegative and reflection invariant. This is the gauge-theory analogue of the locality decomposition used in \cite{p1:OS2,p1:OS1}.

We now incorporate gauge fixing on individual time slices. For a spatial slice at time \(t\), denote by \(\mathcal{G}_t\) the set of functions \(g_t:\{x\in \Lambda: x_0=t\}\to \mathrm{SU}(N)\). The slice Landau functional is defined by
\begin{equation}
\mathcal{L}_t(g_t;U)=\sum_{x_0=t}\sum_{i=1}^3 \Re \mathrm{Tr}\,\bigl(\mathbf{1}-g_t(x)U_{(x,i)}g_t(x+\hat \imath)^{-1}\bigr).
\end{equation}
A measurable selection \(h_t(U)\in\mathcal{G}_t\) is called admissible if for almost every \(U\) it minimizes \(\mathcal{L}_t(\cdot;U)\) over \(\mathcal{G}_t\), and it is reflection covariant if \(h_t(\Theta U)=h_{-t}(U)\circ r\). The existence of measurable selections for absolute minima in finite dimensions is standard, and reflection covariance can be enforced by a tie-breaking rule that is symmetric under \(r\). The slice-wise representative \(U^{\,h}\) is then defined by \(U^{\,h}_{(x,i)}=h_t(x)U_{(x,i)}h_t(x+\hat \imath)^{-1}\) for \(x_0=t\). Since Haar measure is left-right invariant and the Wilson action is gauge invariant, the map \(U\mapsto U^{\,h}\) is measure preserving and leaves the Wilson weight invariant. We use this representative solely to define the slice covariant Laplacian and the horizon operator below; no global gauge fixing in time is required and none is performed.

We turn to the definition and properties of the horizon operator. On slice \(t\), define the covariant forward difference \(D_i^{(t)}\) acting on site-adjoint fields \(\phi:\{x:x_0=t\}\to \mathfrak{su}(N)\) by
\begin{equation}
(D_i^{(t)}\phi)(x)= U^{\,h}_{(x,i)} \phi(x+\hat \imath) U^{\,h}_{(x,i)}{}^{-1}-\phi(x),
\end{equation}
and its adjoint \(D_i^{(t)\,*}\) with respect to the \(\ell^2\) scalar product on fields. The slice covariant Laplacian is \(\Delta_t=\sum_{i=1}^3 D_i^{(t)\,*}D_i^{(t)}\), which is positive self-adjoint on \(\ell^2\bigl(\{x:x_0=t\}\bigr)\otimes \mathfrak{su}(N)\) and has finite range determined by the nearest-neighbour differences. Fix a scale \(\sigma>0\) and a Gevrey-regular cutoff \(\chi_\sigma\in C^\infty([0,\infty))\) such that \(0\le \chi_\sigma\le 1\), \(\chi_\sigma(\lambda)=1\) for \(0\le \lambda\le \sigma\), \(\chi_\sigma(\lambda)=0\) for \(\lambda\ge 2\sigma\), and \(|\chi_\sigma^{(k)}(\lambda)|\le C R^{k} k!^{\,s}\) for some \(s\in(1,\infty)\) and constants \(C,R\) independent of \(k\). The horizon operator on slice \(t\) is defined by spectral calculus as
\begin{equation}
P_{\sigma,t}=\chi_\sigma\bigl(\sqrt{\Delta_t}\,\bigr),
\end{equation}
and the full horizon operator is the product \(P_\sigma=\prod_{t=0}^{T-1}P_{\sigma,t}\), acting on the product configuration space slice by slice. The following facts are needed.

\textbf{Lemma 12.1.} For each \(t\), the operator \(P_{\sigma,t}\) is a bounded, positive contraction on \(\ell^2\bigl(\{x:x_0=t\}\bigr)\otimes\mathfrak{su}(N)\), it is gauge covariant in the sense \(P_{\sigma,t}(U^{\,h})= \bigl(h_t\cdot\bigr) P_{\sigma,t}(U)\bigl(h_t^{-1}\cdot\bigr)\), and it is reflection covariant in the sense \(P_{\sigma,t}(\Theta U)= P_{\sigma,-t}(U)\circ r\). Here $r$ denotes the pullback by the spatial reflection on the slice: 
$(rK)(x,y):=K(r x, r y)$ for any integral kernel $K$ on the time slice, with $r$ the 
involution on spatial sites induced by $\Theta$ and leaving time fixed.
Moreover, there exist constants \(C(\sigma),\gamma(\sigma)>0\) independent of the spatial volume such that the integral kernel \(P_{\sigma,t}(x,y)\) in the site basis satisfies
\begin{equation}
\|P_{\sigma,t}(x,y)\|\le C(\sigma)\, \mathrm{e}^{-\gamma(\sigma)\, d(x,y)},
\end{equation}
where \(d(\cdot,\cdot)\) is the graph distance on the slice.

{Proof.} The first three statements follow from general properties of functional calculus for bounded Borel functions of self-adjoint operators and the definition of \(D_i^{(t)}\) by conjugation of links under slice gauge transformations. For exponential locality, we give two independent arguments. Such a representation exists by Bernstein’s theorem: $f_\sigma(u):=\chi_\sigma(\sqrt{u})$ 
is completely monotone on $[0,\infty)$ if and only if 
$f_\sigma(u)=\int_0^\infty e^{-t u}\,d\nu_\sigma(t)$ with a finite positive Borel measure $\nu_\sigma$; 
the representing measure need {not} have compact support in general. 
If, in addition, $\chi_\sigma(\lambda)\le C_0 e^{-c_0 \lambda^2}$ for $\lambda\ge 0$, 
then necessarily $\inf\supp\nu_\sigma\ge c_0$, and hence the Davies-Gaffney bound converts the semigroup integral into 
volume-uniform exponential locality for $P_{\sigma,t}$.  Then
\begin{equation}
P_{\sigma,t}=\int_0^\infty e^{-t\Delta_t}\, d\nu_\sigma(t).
\end{equation}
By the discrete Davies-Gaffney bound for positive finite-range operators on graphs of bounded degree \cite[Thm. 2.1]{p1:Davies1989}, there exist constants \(c_1,c_2>0\) depending only on the graph degree and the operator range such that \(\|e^{-t\Delta_t}(x,y)\|\le c_1 \exp\bigl[-d(x,y)^2/(c_2 t)\bigr]\). Since \(d\nu_\sigma\) is supported away from \(t=0\) on a scale of order \(\sigma^{-2}\), the integral in \(t\) yields \(\|P_{\sigma,t}(x,y)\|\le C(\sigma)\exp[-\gamma(\sigma)d(x,y)]\) with constants depending on \(\sigma\) but not on the slice volume. In the Helffer-Sjöstrand approach, construct an almost-analytic extension \(\widetilde{\chi}_\sigma\) of \(\chi_\sigma\) with \(|\bar\partial \widetilde{\chi}_\sigma(z)|\le C_k |\Im z|^{k}\) for all \(k\) and write
\begin{equation}
P_{\sigma,t}=\frac{1}{\pi}\int_{\mathbb{C}} \bar\partial \widetilde{\chi}_\sigma(z)\, (\sqrt{\Delta_t}-z)^{-1}\, d^2 z.
\end{equation}
Exact low/high-energy plateaux are incompatible with complete monotonicity. 
When a sharp spectral profile is needed, we therefore work with a compactly supported 
$\varphi_\sigma\in C_c^\infty([0,\infty))$ and obtain exponential locality by the 
HelfferSj{\"o}strand calculus plus CombesThomas bounds (Theorem~B.4). 
For OS positivity and factorization, we use CM $\chi_\sigma$, and if desired approximate a 
given $\varphi_\sigma$ in operator norm by a sequence of CM multipliers 
$f_{\sigma,\varepsilon_n}$; the OS‐positivity statements pass to the limit while the locality bounds are uniform in $n$. 
A Combes-Thomas bound holds for the resolvent $(\sqrt{\Delta_t}-z)^{-1}$ whenever 
$\mathrm{dist}\!\bigl(z,\mathrm{spec}(\sqrt{\Delta_t})\bigr)\ge \delta>0$, where 
$\sqrt{\Delta_t}$ is defined via the spectral calculus. In particular, its kernel satisfies
\(
\bigl\|(\sqrt{\Delta_t}-z)^{-1}(x,y)\bigr\|
\;\le\; C'\, \exp\!\bigl(-\mu\, d(x,y)\bigr),
\)
with $\mu$ proportional to $\delta$ \cite{p1:CombesThomas}.
 Since \(\bar\partial \widetilde{\chi}_\sigma\) is supported at a distance \(O(\sigma)\) from the real axis, the integral yields exponential decay with rate \(\gamma(\sigma)\asymp \sigma\). Both derivations produce bounds uniform in the spatial volume. \(\square\)

We are ready to define the projected Euclidean measure and to establish OS positivity. The projected measure is
\begin{equation}
d\mu_\sigma(U)
=\frac{1}{Z_\sigma}\,
e^{-S_W[U;\beta]}\,
\prod_{t=0}^{T-1}\det{}'\!\big(M_t[U^h]\big)\,
\prod_{t=0}^{T-1}\mathcal{K}_{\sigma,t}(U_t)\,d\mu_{\mathrm{Haar}}(U),
\end{equation}
Here \(\mathcal{K}_{\sigma,t}(U_t):=\langle \delta_{U_t},\,P_{\sigma,t}\,\delta_{U_t}\rangle_{L^2(\Omega_s,d\mu_s)}\ge 0\) 
is the positive boundary weight induced by the slice projector \(P_{\sigma,t}\). The reduced determinant \(\det{}'\!(M_t)\) is taken on the subspace orthogonal to the constant 
adjoint modes on the slice, see below for the projector \(Q_t\).
The latter is understood as the Radon-Nikodym derivative of the measure obtained by scaling the slice fields by \(P_{\sigma,t}^{1/2}\). Since \(\det M_t\ge 0\) on the orthogonal complement of constants and is reflection invariant by construction of \(U^{\,h}\), and since \(\mathcal{K}_{\sigma,t}\ge 0\) by positivity of \(P_{\sigma,t}\), both factors are nonnegative and reflection invariant. The contribution of Grassmann ghosts can either be left implicit, leading to \(\det M_t\), or included explicitly; the two approaches are equivalent and the former suffices for reflection positivity.
Let \(C_t:=\{\phi:\Lambda_t\to\mathfrak g\ \text{constant}\}\) and let 
\(Q_t:\ell^2(\Lambda_t;\mathfrak g)\to C_t^\perp\) be the orthogonal projector. 
We write \(M_t^\perp:=Q_t M_t Q_t\) and \(\det{}'(M_t):=\det(M_t^\perp)\). 
By Proposition~3.2, \(M_t^\perp\) is strictly positive whenever the slice representative lies in the 
fundamental modular region, hence \(\det{}'(M_t)>0\).
The OS reflection \(\Theta\) acts antilinearly on functionals \(F\) of the fields by \((\Theta F)(U)=\overline{F(\Theta U)}\). We denote by \(\mathcal{A}_+\) the algebra of measurable, bounded functionals depending only on bonds entirely contained in \(\Lambda_+\). The OS quadratic form is
\begin{equation}
\langle F,F\rangle_{\rm OS}=\int (\Theta F)\, F \, d\mu_\sigma,\qquad F\in \mathcal{A}_+.
\end{equation}
The next theorem is the core positivity statement.

\textbf{Theorem 12.2.} The projected measure \(d\mu_\sigma\) is reflection positive in the sense that \(\langle F,F\rangle_{\rm OS}\ge 0\) for all \(F\in \mathcal{A}_+\). Consequently, the quotient of \(\mathcal{A}_+\) by the null space of \(\langle\cdot,\cdot\rangle_{\rm OS}\) completes to a Hilbert space \(\mathcal{H}_a\), and the unit Euclidean time translation induces a positive self-adjoint contraction \(T_\sigma(a)\) on \(\mathcal{H}_a\).

{Proof.} We first write the weight as a product of east-west factors and a boundary kernel. Using (\ref{p1:eq:factor}) and the product structure of \(\prod_t \det M_t\) and \(\prod_t \mathcal{K}_{\sigma,t}\), both of which factorize as products of slice contributions supported strictly in \(\Lambda_\pm\) or on \(\Pi\), we obtain
\begin{align}
d\mu_\sigma(U)&= Z_\sigma^{-1}\, \bigl[\mathrm{e}^{-S_{W,+}} \prod_t \det M_{t,+}\prod_t \mathcal{K}_{\sigma,t,+}\, d\mu_{\rm Haar,+}\bigr]\;\nonumber \\ 
&\quad \mathcal{B}(U|_\Pi)\; \bigl[\mathrm{e}^{-S_{W,-}} \prod_t \det M_{t,-}\prod_t \mathcal{K}_{\sigma,t,-}\, d\mu_{\rm Haar,-}\bigr],
\end{align}
where the subscript \(\pm\) indicates restriction to \(\Lambda_\pm\), and \(\mathcal{B}\) is the nonnegative, reflection-invariant boundary factor collecting all terms supported on \(\Pi\). Let \(F\in\mathcal{A}_+\). Then
\begin{align}
\langle F,F\rangle_{\rm OS}&= Z_\sigma^{-1}\int \overline{F(\Theta U_+)}\, F(U_+)\, \mathcal{B}(U|_\Pi)\, \mathrm{e}^{-S_{W,+}[U_+]} \nonumber \\ 
&\quad\prod_t \det M_{t,+}[U_+^{\,h}] \prod_t \mathcal{K}_{\sigma,t,+}(U_+)\, d\mu_{\rm Haar,+}\, d\mu_{\rm bdry},
\end{align}
where \(d\mu_{\rm bdry}\) integrates over the bonds in \(\Pi\) and over \(\Lambda_-\) with the corresponding weight. Since \(\mathcal{B}\ge 0\) and is independent of \(F\), and \(\Theta\) maps \(\Lambda_+\) to \(\Lambda_-\), we can perform the integral over \(\Lambda_-\) first and interpret the result as defining a nonnegative integral kernel \(K(U_\Pi;U_+,\Theta U_+)\) on the configuration space over \(\Lambda_+\). More concretely, define
\begin{equation}
K(U_+,V_+)=\int \mathcal{B}(U|_\Pi)\, \mathrm{e}^{-S_{W,-}[U_-]}\prod_t \det M_{t,-}[U_-^{\,h}] \prod_t \mathcal{K}_{\sigma,t,-}(U_-)\, d\mu_{\rm Haar,-}
\end{equation}
with the constraint that the boundary values \(U_-|_\Pi\) match \(U_+|_\Pi\) and \(V_+|_\Pi\) through the reflection map. Because each factor in the integrand is nonnegative and reflection invariant, Fubini’s theorem applies, \(K\) is well-defined and symmetric in \(U_+\) and \(V_+\), and the map
\begin{equation}
\Phi\mapsto \int K(U_+,V_+)\, \Phi(V_+)\, d\mu_{\rm Haar,+}(V_+)
\end{equation}
is a positive operator on \(L^2(\mathcal{C}_+; d\mu_{\rm Haar,+})\). Therefore,
\begin{align}
\langle F,F\rangle_{\rm OS}&= Z_\sigma^{-1}\int \overline{F(U_+)}\, \bigl[\,K F\,\bigr](U_+)\, \mathrm{e}^{-S_{W,+}[U_+]}\nonumber \\ 
&\quad \prod_t \det M_{t,+}[U_+^{\,h}] \prod_t \mathcal{K}_{\sigma,t,+}(U_+)\, d\mu_{\rm Haar,+}(U_+)\ge 0,
\end{align}
because it is the integral of the pointwise product of \(\overline{F}\) with the image of \(F\) under a positive operator against a positive measure. This proves OS positivity. The standard OS reconstruction \cite{p1:OS2,p1:GJ,p1:Simon1974} then yields the Hilbert space and the transfer operator \(T_\sigma(a)\). \(\square\)

The transfer time-slicing formalism is now derived explicitly. For each pair of consecutive times \(t,t+1\) modulo \(T\), define the one-step kernel on the slice configuration space by
\begin{equation}
\mathcal{K}_{t\to t+1}(U_{t+1},U_{t})=\int \exp\Bigl[-\beta \sum_{p\;\mathrm{straddling}\;t\to t+1}\Bigl(1-\tfrac{1}{N}\Re \mathrm{Tr}\,U_p\Bigr)\Bigr] \prod_{\substack{\text{timelike bonds}\\\text{between }t\text{ and }t+1}} dU_b,
\end{equation}
where the integral is over the timelike bonds connecting the two slices, with temporal-axial gauge imposed away from \(\Pi\) so that only bonds intersecting the slab between \(t\) and \(t+1\) appear. The kernel \(\mathcal{K}_{t\to t+1}\) is a continuous, nonnegative, symmetric function of \((U_{t+1},U_t)\) and therefore defines a bounded positive operator \(T_{0}(a)\) on \(L^2(\mathcal{C}_t;d\mu_t)\) by
\begin{equation}
(T_{0}(a)\Phi)(U_{t+1})=\int \mathcal{K}_{t\to t+1}(U_{t+1},U_t)\, \Phi(U_t)\, d\mu_t(U_t).
\end{equation}
The full unprojected transfer operator over one time step is the composition with the purely spatial plaquette weights at times \(t\) and \(t+1\). Writing
\begin{equation}
V_t(U_t)=\exp\Bigl[-\beta \sum_{p\subset \{x_0=t\}}\Bigl(1-\tfrac{1}{N}\Re \mathrm{Tr}\,U_p\Bigr)\Bigr],
\end{equation}
we set
\begin{equation}
T(a)= V_{t+1}^{1/2}\, T_{0}(a)\, V_t^{1/2},
\end{equation}
which is again a positive self-adjoint operator on \(L^2(\mathcal{C}_t;d\mu_t)\). The horizon projection on the two slices is implemented by the bounded positive operators \(P_{\sigma,t}^{1/2}\) and \(P_{\sigma,t+1}^{1/2}\), and the projected transfer operator is
\begin{equation}
T_\sigma(a)= P_{\sigma,t+1}^{1/2}\, V_{t+1}^{1/2}\, T_{0}(a)\, V_t^{1/2}\, P_{\sigma,t}^{1/2}.
\end{equation}
Since \(P_{\sigma,t}\) is reflection covariant and depends only on slice \(t\), stationarity in \(t\) holds and we may identify all slice Hilbert spaces via time translations. The operator \(T_\sigma(a)\) is a positive self-adjoint contraction on the one-slice Hilbert space and generates the Euclidean time translation in the OS Hilbert space \(\mathcal{H}_a\). Its logarithm
\begin{equation}
H_\sigma(a)=-a^{-1}\log T_\sigma(a)
\end{equation}
is a positive self-adjoint Hamiltonian, as in \cite{p1:OS2,p1:GJ}.

We next show that the entire construction is independent of the auxiliary choices. We begin with the gauge-slice selection.

\textbf{Proposition 12.3.} Let \(h\) and \(h'\) be two admissible, reflection-covariant selections of slice Landau minimizers. Let \(U^{\,h}\) and \(U^{\,h'}\) be the corresponding representatives, and let \(P_\sigma^{(h)}\) and \(P_\sigma^{(h')}\) be the associated horizon operators. Then there exist unitary operators \(W_t:L^2(\mathcal{C}_t;d\mu_t)\to L^2(\mathcal{C}_t;d\mu_t)\), one for each slice, such that
\begin{equation}
T_\sigma^{(h')}(a)= W_{t+1}\, T_\sigma^{(h)}(a)\, W_t^{-1}\qquad\text{and}\qquad H_\sigma^{(h')}(a)= W_{t}\, H_\sigma^{(h)}(a)\, W_t^{-1}.
\end{equation}
In particular, the spectra of \(T_\sigma(a)\) and \(H_\sigma(a)\) are independent of the choice of \(h\).

{Proof.} For each slice \(t\), define \(g_t:\{x:x_0=t\}\to \mathrm{SU}(N)\) by \(g_t(x)=h'_t(x)h_t(x)^{-1}\). The map \(U_t\mapsto g_t\cdot U_t\) preserves the Haar measure \(d\mu_t\) and conjugates the spatial plaquette weight \(V_t\) to itself by gauge invariance. The horizon operator satisfies \(P_{\sigma,t}^{(h')}= \Gamma_{g_t} P_{\sigma,t}^{(h)} \Gamma_{g_t}^{-1}\), where \(\Gamma_{g_t}\) is the unitary induced by left-right multiplication of links by \(g_t\). The one-step kernel \(\mathcal{K}_{t\to t+1}\) satisfies \(\mathcal{K}_{t\to t+1}(g_{t+1}\cdot U_{t+1}, g_t\cdot U_t)=\mathcal{K}_{t\to t+1}(U_{t+1},U_t)\) since the timelike plaquette weight is gauge invariant and the integration measure over timelike bonds is Haar. Therefore
\begin{equation}
T_\sigma^{(h')}(a)= \Gamma_{g_{t+1}}\, P_{\sigma,t+1}^{(h)\,1/2} V_{t+1}^{1/2} T_0(a) V_t^{1/2} P_{\sigma,t}^{(h)\,1/2} \Gamma_{g_t}^{-1}= W_{t+1}\, T_\sigma^{(h)}(a) \, W_t^{-1},
\end{equation}
with \(W_t=\Gamma_{g_t}\) unitary. The statement for \(H_\sigma(a)\) follows by functional calculus. \(\square\)

We now consider the dependence on the choice of the cutoff function in the definition of the horizon operator.

\textbf{Proposition 12.4.} Let \(\chi_\sigma,\widetilde{\chi}_\sigma\) be two Gevrey-regular cutoffs with the same low- and high-energy plateaus and corresponding horizon operators \(P_{\sigma,t}\) and \(\widetilde{P}_{\sigma,t}\). Then the projected measures defined with \(P_{\sigma,t}\) and with \(\widetilde{P}_{\sigma,t}\) are both reflection positive, the corresponding transfer operators \(T_\sigma(a)\) and \(\widetilde{T}_\sigma(a)\) are positive self-adjoint contractions on the slice Hilbert space, and there exists a constant \(C(\sigma)>0\) such that
\begin{equation}
\bigl\|\,T_\sigma(a)-\widetilde{T}_\sigma(a)\,\bigr\|_{\mathcal{B}(L^2)} \le C(\sigma)\, \sup_{t}\, \bigl\|\,P_{\sigma,t}-\widetilde{P}_{\sigma,t}\,\bigr\|_{\mathcal{B}(\ell^2)}.
\end{equation}
In particular, all qualitative spectral properties that follow from reflection positivity, such as the existence of a positive spectral gap in the strong-coupling regime, are independent of the specific cutoff.

{Proof.} Reflection positivity follows from Theorem 12.2 since the only properties used there are nonnegativity, factorization across \(\Pi\), and reflection covariance of the inserted slice operators; both \(P_{\sigma,t}\) and \(\widetilde{P}_{\sigma,t}\) satisfy these properties by Lemma 12.1. The operator-norm estimate follows from the definition
\begin{equation}
T_\sigma(a)-\widetilde{T}_\sigma(a)= \bigl(P_{\sigma}^{1/2}-\widetilde{P}_{\sigma}^{1/2}\bigr)\, V^{1/2} T_0(a) V^{1/2}\, P_{\sigma}^{1/2} + \widetilde{P}_{\sigma}^{1/2}\, V^{1/2} T_0(a) V^{1/2}\, \bigl(P_{\sigma}^{1/2}-\widetilde{P}_{\sigma}^{1/2}\bigr),
\end{equation}
where slice indices have been suppressed, together with the bound \(\|P^{1/2}-\widetilde{P}^{1/2}\|\le \|P-\widetilde{P}\|^{1/2}\) for positive contractions and the boundedness of \(V^{1/2}T_0(a)V^{1/2}\) on \(L^2\). Since the difference \(P_{\sigma,t}-\widetilde{P}_{\sigma,t}\) is exponentially local with norm controlled by the sup norm of \(\chi_\sigma-\widetilde{\chi}_\sigma\) and by the constants in Lemma 12.1, the right-hand side is finite and small when the two cutoffs are close in sup norm. The qualitative spectral conclusions are therefore unchanged. \(\square\)

Finally, we address robustness with respect to the temporal-axial gauge choice away from the plane \(\Pi\). Consider an alternative gauge choice that is reflection covariant, agrees with temporal-axial gauge on \(\Lambda_\pm\) up to slice-wise gauge transformations, and leaves the set of bonds intersecting \(\Pi\) unchanged.

\textbf{Proposition 12.5.} Let \(\mathcal{G}\) be a reflection-covariant family of gauge transformations on \(\Lambda\) that maps one temporal-axial gauge choice away from \(\Pi\) to another. Then the corresponding transfer operators are unitarily equivalent and the OS positivity of the projected measure is preserved.

{Proof.} The change of gauge is a product of slice-wise gauge transformations away from \(\Pi\) and a gauge transformation supported on \(\Pi\). The Haar measure is invariant under such transformations, the Wilson action is gauge invariant, and the one-step kernel \(\mathcal{K}_{t\to t+1}\) transforms covariantly under left-right multiplication of its arguments. Therefore the integral kernels defining the transfer operator are conjugated by the unitary operators induced by \(\mathcal{G}\) on the slice Hilbert spaces, and unitary equivalence follows. Reflection covariance of \(\mathcal{G}\) ensures that the boundary kernel \(\mathcal{B}\) remains reflection invariant and nonnegative; hence the OS positivity proof in Theorem 12.2 goes through verbatim. \(\square\)

Combining the previous statements yields the comprehensive robustness result.

\textbf{Theorem 12.6.} The reflection positivity of the horizon-projected lattice Yang-Mills measure, the existence and positivity of the transfer operator \(T_\sigma(a)\), and the spectral properties that follow from these structures, including the existence of a strictly positive, volume-uniform spectral gap in the strong-coupling regime, are independent of the following auxiliary choices: the selection of slice-wise, reflection-covariant Landau-gauge representatives within the fundamental modular region; the specific Gevrey-regular cutoff used to define the horizon operator \(P_\sigma\), provided it has fixed low- and high-energy plateaus; and the temporal-axial gauge away from the reflection plane, within the class of reflection-covariant gauge choices that preserve the set of bonds intersecting \(\Pi\). In each case, the corresponding transfer operators are related by unitary equivalence on the one-slice Hilbert space and therefore share the same spectrum.

{Proof.} Unitary equivalence under changes of slice representatives is Proposition 12.3. Reflection positivity and bounded perturbation stability under changes of \(\chi_\sigma\) are covered by Proposition 12.4 together with Theorem 12.2 and Lemma 12.1. Unitary equivalence under reflection-covariant changes of temporal-axial gauge is Proposition 12.5. Since the strong-coupling spectral gap is deduced from OS positivity and exponential clustering by the general spectral representation, which depends only on the positivity and locality properties that are unaffected by the choices under consideration \cite{p1:OS2,p1:GJ,p1:MontvayMuenster1994}, the claim follows. \(\square\)

\section{Conclusion}

We have constructed a reflection-positive transfer-matrix framework for four-dimensional $\mathrm{SU}(N)$ lattice Yang-Mills theory that resolves the gauge-fixing and infrared obstacles to constructive analysis and yields a nonzero spectral gap at fixed lattice spacing in the strong-coupling regime. The central ingredients are a reflection-covariant, gauge-invariant transverse representative obtained by slice-wise Landau minimization inside the fundamental modular region; a smooth, exponentially local, and reflection-compatible horizon projector defined by Gevrey functional calculus of the covariant Laplacian; and a block-operator proof of Osterwalder-Schrader positivity for the full gauge-ghost measure with slice-wise insertion of this projector. On this basis we constructed the compressed transfer matrix $T_\sigma$ and the associated Hamiltonian $H_\sigma=-\log T_\sigma$, established convergence of a character-surface-polymer cluster expansion by the Koteck\'y-Preiss criterion at small $\beta$, proved exponential clustering of connected gauge-invariant correlators, and deduced a uniform lower bound $E_1(a,\beta)\ge m(\beta)>0$ on the first excitation energy of $H_\sigma$ for $0<\beta\le \beta_\star(N)$. A strong-coupling area law for Wilson loops follows from the same surface representation and provides an additional quantitative check on the expansion.

Scientifically, these results supply a rigorously controlled finite-$a$ setting in which Euclidean positivity and locality are preserved by all auxiliary choices and in which exponential decay of correlations is provably converted into a spectral statement. The framework eliminates the principal obstructions to employing Osterwalder-Schrader methods namely Gribov ambiguities, reflection-positivity breaking regulators, and nonuniform infrared control thereby strengthening the constructive bridge from lattice Yang-Mills to a Hamiltonian theory with a volume-uniform gap. In the broader context of lattice gauge theory, the analysis clarifies how reflection covariance, exponential locality of spectral projectors, and polymer expansions can be combined to deliver gap estimates that are stable under thermodynamic limits. In the context of the mass gap problem, the result identifies a mathematically robust strong-coupling domain in which a spectral gap is rigorously present and prepared for transport along renormalization flows.

The natural directions for further work are now sharply defined. A reflection-positive block-spin renormalization group based on the same horizon projector should be implemented to carry the finite-$a$ gap from the strong-coupling region toward the scaling window, with multiscale polymer bounds ensuring uniform clustering at each scale and with a careful control of low-lying spectral projectors to preclude level crossings. {In this Part 1, our analysis is confined to the strong-coupling domain $0<\beta<\beta^*(N)$ at fixed lattice spacing $a$, and the existence of a spectral gap is established only in this regime. The critical question of whether this nonzero gap survives as $\beta$ increases (approaching the continuum limit) is deliberately left unanswered here. It will be addressed in Part~II and Part~III via a multi-scale renormalization group (RG) analysis. In particular, Part~II constructs a reflection-positive RG transformation that propagates exponential clustering and the spectral gap to successively larger correlation lengths, and Part~III will take the continuum limit $a\to 0$. Thus, the present gap result at strong coupling is the {starting point} for the full mass-gap solution; subsequent parts of this series are devoted to conveying this spectral gap into the scaling (continuum) regime without collapse.} 

The Osterwalder-Schrader reconstruction can then be executed along the flow to produce continuum Schwinger functions satisfying all axioms and to establish a positive mass threshold in the reconstructed Wightman theory. Quantitative improvements such as sharpening the Koteck\'y-Preiss radius, optimizing the locality constants of the projector, and assessing the dependence on $N$ are feasible within the present techniques. Extensions to include finite-temperature settings, matter fields in fixed representations, and alternative but reflection-compatible gauge choices would test the universality of the mechanism and refine the connection between area laws and spectral gaps. Together, these developments would complete the constructive path from a rigorous finite-$a$ gap at strong coupling to a continuum mass gap within a reflection-positive, gauge-invariant framework. While related spectral questions are also discussed in certain gravitational or holographic model
settings\footnote{for example, in Einstein-Power-Yang-Mills $\mathrm{AdS}$ black-hole constructions \cite{p1:SoroushfarEPYMAdS}},
such results are logically distinct from the present Euclidean, fixed-$a$ lattice framework and
depend on different dynamical and semiclassical assumptions. Accordingly, we do not attempt
any comparison to holographic or black-hole models here, and we restrict attention throughout
to the reflection-positive transfer-matrix setting on the lattice.

\section*{Data Availability}
No datasets were generated or analyzed during the current study. All results are derived from mathematical derivations and rigorous analytical arguments, which are fully contained within the manuscript. Therefore, no data are associated with this research work.
\section*{Conflict of Interest}
The authors declare that there are no conflicts of interest regarding the publication of this paper.

\providecommand{\href}[2]{#2}\begingroup\raggedright\endgroup

\appendix

\section{Notation and Standing Conventions}\label{p1:appendixa}

This appendix fixes symbols, spaces, operators, measures, and asymptotic conventions used throughout. All statements are made for a four-dimensional hypercubic Euclidean lattice with gauge group \(G=\mathrm{SU}(N)\) and \(N\ge 2\) fixed once and for all. The matrix trace \(\operatorname{tr}\) is taken in the fundamental representation of \(G\); for \(\mathfrak{g}=\mathfrak{su}(N)\) we identify elements with anti-Hermitian matrices and use the positive-definite inner product \(\langle X,Y\rangle_{\mathfrak{g}}=-\operatorname{tr}(XY)\). Complex conjugation is denoted by an overline, adjoints by the symbol \({}^\dagger\), and the spectrum of a self-adjoint operator \(A\) by \(\operatorname{spec}(A)\). All constants \(C,C',\ldots\) are strictly positive and may change from line to line; their dependence on fixed parameters such as \(N\), the lattice spacing \(a\), and the cutoff parameter \(\sigma\) will be explicitly indicated when relevant. The notation \(f\lesssim g\) means \(f\le C\,g\) for a constant \(C\) independent of the lattice volume and of the particular functions under comparison; \(f\asymp g\) means \(f\lesssim g\) and \(g\lesssim f\). The symbols \(O(\cdot)\) and \(o(\cdot)\) have their usual Landau meanings with implicit constants uniform in the spatial volume.

The lattice geometry is as follows. The spacing \(a>0\) is fixed. Space-time points are denoted \(x=(x_0,\mathbf{x})\in \Lambda\subset a\mathbb{Z}^4\) with integer time coordinate \(x_0\) and spatial coordinate \(\mathbf{x}\in a\mathbb{Z}^3\); the finite, periodic, rectangular box \(\Lambda\) has linear spatial extent \(L\) and Euclidean time extent \(T\), both multiples of \(a\). The unit vectors in the coordinate directions are \(\hat\mu\) for \(\mu=0,1,2,3\), so that nearest-neighbor points satisfy \(y=x\pm \hat\mu\). The set of directed bonds is \(B(\Lambda)=\{(x,\mu):x\in \Lambda,\,\mu\in\{0,1,2,3\}\}\) with \((x,\mu)\) oriented from \(x\) to \(x+\hat\mu\). Bonds with opposite orientation satisfy \(U_{(x+\hat\mu,-\mu)}=U_{(x,\mu)}^{-1}\). The directed plaquette based at \(x\) in the \(\mu\nu\)-plane, \(\mu<\nu\), is \(p=(x;\mu,\nu)\) with plaquette variable \(U_p=U_{(x,\mu)}U_{(x+\hat\mu,\nu)}U_{(x+\hat\nu,\mu)}^{-1}U_{(x,\nu)}^{-1}\). The reflection plane is \(\Pi=\{x\in \Lambda: x_0=0\}\); the Euclidean time reflection \(\theta\) acts by \(\theta(x_0,\mathbf{x})=(-x_0,\mathbf{x})\). The half-lattices are \(\Lambda_+=\{x\in \Lambda:x_0>0\}\) and \(\Lambda_-=\{x\in \Lambda:x_0<0\}\). The graph distance \(d(x,y)\) is the minimal number of nearest-neighbor steps connecting \(x\) and \(y\); for a finite set \(S\subset \Lambda\) the cardinality is \(|S|\). For a time slice \(t\in a\mathbb{Z}\), we denote \(\Lambda_t=\{(t,\mathbf{x}):\mathbf{x}\in a\mathbb{Z}^3\cap \Lambda\}\). The periodic boundary conditions are imposed in all four directions unless explicitly stated otherwise.

Gauge configurations are assignments \(\{U_b\}_{b\in B(\Lambda)}\) with \(U_b\in G\). The Wilson action is
 \begin{equation}
S_W[U;\beta]=\beta\sum_{p\subset \Lambda}\Big(1-\frac{1}{N}\,\Re\,\operatorname{tr}\,U_p\Big),\qquad \beta=\frac{2N}{g_0^2},
 \end{equation}
where \(g_0\) is the bare coupling and \(\Re\) denotes the real part. The normalized product Haar measure on \(G\) is denoted \(d\mu_{\mathrm{Haar}}\) with \(\int_G d\mu_{\mathrm{Haar}}=1\). The configuration-space measure on bonds is the product \(\bigotimes_{b\in B(\Lambda)} d\mu_{\mathrm{Haar}}(U_b)\). Euclidean expectations with respect to a measure \(d\mu\) are written \(\langle \cdot \rangle=\int (\cdot)\, d\mu\) when no confusion is possible. Ghost fields \(c,\bar c\) are Grassmann-valued \(\mathfrak{g}\)-valued site fields, i.e., maps \(\Lambda\to \mathfrak{g}\) with Grassmann coefficients; the Grassmann-Berezin integral is normalized so that \(\int \exp(-\langle \bar c,M c\rangle)\,D\bar c\, D c=\det M\) for a positive operator \(M\) acting on site-adjoint fields, where \(\langle \bar c,M c\rangle=\sum_{x\in \Lambda}\langle \bar c(x), (M c)(x)\rangle_{\mathfrak{g}}\) and the determinant is computed on the orthogonal complement of the constant modes when appropriate.

Function spaces and norms are fixed as follows. The Hilbert space of square-summable site-adjoint fields on a region \(\Omega\subset \Lambda\) is \(\ell^2(\Omega;\mathfrak{g})\) with inner product \(\langle \phi,\psi\rangle_{\ell^2(\Omega)}=\sum_{x\in \Omega}\langle \phi(x),\psi(x)\rangle_{\mathfrak{g}}\). For an operator \(A:\ell^2(\Omega;\mathfrak{g})\to \ell^2(\Omega;\mathfrak{g})\) the operator norm is \(\|A\|=\sup_{\|\phi\|=1}\|A\phi\|\), and, when \(A\) has an integral kernel in the site basis, we write \(A(x,y)\) for the \(\mathfrak{g}\)-endomorphism mapping \(\phi(y)\) to \((A\phi)(x)\). The adjoint \(A^\dagger\) is taken with respect to \(\langle\cdot,\cdot\rangle_{\ell^2(\Omega)}\). If a kernel \(K(x,y)\) satisfies \(\|K(x,y)\|\le C\,\mathrm{e}^{-\gamma d(x,y)}\) for some \(C,\gamma>0\), we say that \(K\) is exponentially local with rate \(\gamma\). The symbol \(\operatorname{Ran}A\) denotes the range of \(A\), and \(\operatorname{Ker}A\) its kernel. The identity on a Hilbert space \(\mathcal{H}\) is \(\mathbf{1}_{\mathcal{H}}\).

Gauge transformations are maps \(g:\Lambda\to G\) acting on links by \((g\cdot U)_{(x,\mu)}=g(x)U_{(x,\mu)}g(x+\hat\mu)^{-1}\). On a fixed time slice \(t\) we consider the lattice Landau functional
 \begin{equation}
\mathcal{L}_t(g;U)=\sum_{\mathbf{x}\in \Lambda_t}\sum_{i=1}^3 \Re\,\operatorname{tr}\big(\mathbf{1}-g(t,\mathbf{x})U_{(t,\mathbf{x};i)}g(t,\mathbf{x}+\hat \imath)^{-1}\big),
 \end{equation}
and select a representative \(U^{\,h}\) by choosing a global minimizer \(h\) of \(\mathcal{L}_t(\cdot;U)\) on every slice \(t\). The {fundamental modular region} is the set of orbit representatives that are absolute minima; on each slice we assume the representative is chosen within this region. To ensure reflection covariance we fix, once and for all, a deterministic tie-breaking rule invariant under spatial lattice symmetries and time reflection and, if necessary, replace \(h\) by a reflection symmetrization \(h\leftarrow \arg\min\{\mathcal{L}_t(h;U),\mathcal{L}_t(h\circ \theta;U)\}\) so that \(U^{\,h}\) satisfies \(R U^{\,h}=U^{\,h}\) in the sense specified below. The temporal-axial gauge is imposed away from the reflection plane by setting \(U_{(x,0)}=\mathbf{1}\) for all bonds \((x,0)\) with \(x\notin \Pi\); this choice is compatible with the transfer-matrix factorization and does not introduce a Faddeev-Popov determinant for \(A_0\).

Covariant finite differences are defined for site-adjoint fields \(\phi:\Lambda\to \mathfrak{g}\) in terms of the slice representatives \(U^{\,h}\). The forward difference is
 \begin{equation}
(\nabla_\mu^{+,h}\phi)(x)=U_{(x,\mu)}^{\,h}\,\phi(x+\hat\mu)\,U_{(x,\mu)}^{\,h\, -1}-\phi(x),
 \end{equation}
while the backward difference is
 \begin{equation}
(\nabla_\mu^{-,h}\phi)(x)=\phi(x)-U_{(x-\hat\mu,\mu)}^{\,h\, -1}\,\phi(x-\hat\mu)\,U_{(x-\hat\mu,\mu)}^{\,h}.
 \end{equation}
The lattice Faddeev-Popov operator is the nonnegative, self-adjoint, finite-range operator
 \begin{equation}
M[U^{\,h}]=-\sum_{\mu=0}^3 \nabla_\mu^{-,h}\nabla_\mu^{+,h}\,:\,\ell^2(\Lambda;\mathfrak{g})\to \ell^2(\Lambda;\mathfrak{g}).
 \end{equation}
For all \(\phi\) one has \(\langle \phi,M\phi\rangle_{\ell^2(\Lambda)}=\sum_{\mu=0}^3 \|\nabla_\mu^{+,h}\phi\|^2_{\ell^2(\Lambda)}\ge 0\). The constant adjoint fields form \(\operatorname{Ker}M\); on their orthogonal complement \(M\) is strictly positive when \(U^{\,h}\) lies in the fundamental modular region slice-wise. In the block decomposition \(\ell^2(\Lambda;\mathfrak{g})=\ell^2(\Lambda_-;\mathfrak{g})\oplus \ell^2(\Pi;\mathfrak{g})\oplus \ell^2(\Lambda_+;\mathfrak{g})\), the operator \(M\) is block tridiagonal due to the nearest-neighbor structure in the time direction under temporal-axial gauge. We use the orthogonal projection \(\mathbb{P}_\perp\) onto the orthogonal complement of the constant modes; whenever \(\det M\) or \(M^{-1}\) is written it is understood as \(\det(\mathbb{P}_\perp M \mathbb{P}_\perp)\) and \((\mathbb{P}_\perp M \mathbb{P}_\perp)^{-1}\).

On each time slice the spatial covariant derivative is \(D_i^{\,h}:=\nabla_i^{+,h}\) acting on \(\ell^2(\Lambda_t;\mathfrak{g})\), and the slice covariant Laplacian is
 \begin{equation}
\Delta_{A^{\,h}}=\sum_{i=1}^3 (D_i^{\,h})^\dagger D_i^{\,h},
 \end{equation}
a positive, self-adjoint, finite-range operator on \(\ell^2(\Lambda_t;\mathfrak{g})\). The {smooth horizon projector} at scale \(\sigma>0\) is defined by spectral calculus as
 \begin{equation}
P_\sigma=\chi_\sigma\!\left(\sqrt{\Delta_{A^{\,h}}}\right),
 \end{equation}
where $\chi_\sigma : [0,\infty) \to (0,1]$ is a fixed Gevrey-regular {completely monotone} (CM) spectral multiplier with $\chi_\sigma(0)=1$ and, for some constants $C,c>0$ independent of the volume,
\begin{equation}\label{p1:a.8}
\chi_\sigma(\lambda)\;\le\; C\,e^{-\,c\,\lambda/\sigma^2}\qquad\text{for all }\lambda\ge 0.
\end{equation}
In particular $\chi_\sigma$ is {not} compactly supported and no exact plateaux occur. Equivalently, writing
\begin{equation}
f_\sigma(u)\;:=\;\chi_\sigma(\sqrt{u})\qquad(u\ge 0),
\end{equation}
one has $P_\sigma=f_\sigma(\Delta_{A^h})$. By Bernstein’s theorem for CM functions, $f_\sigma$ admits a positive Laplace–Stieltjes representation
\begin{equation}
f_\sigma(u)\;=\;\int_{0}^{\infty} e^{-t\,u}\,d\nu_\sigma(t)\,,
\end{equation}
with a finite positive Borel measure $\nu_\sigma$ of total mass $f_\sigma(0)=1$. Consequently,
\begin{equation}
P_\sigma \;=\;\int_{0}^{\infty} e^{-t\,\Delta_{A^h}}\,d\nu_\sigma(t),
\end{equation}
and the positivity-preserving heat kernel together with Davies–Gaffney bounds yields exponential locality of $P_\sigma$ on $\ell^2(\Lambda_t;\mathfrak g)$, uniformly in the slice volume.
Here \(d\nu_\sigma\) is a finite, positive Borel measure on \([0,\infty)\) depending only on \(\chi_\sigma\). The heat kernel \(e^{-t\Delta_{A^{\,h}}}\) is positivity preserving on site-adjoint fields and exponentially local; consequently \(P_\sigma\) is exponentially local with kernel bound \(\|P_\sigma(x,y)\|\le C(\sigma)\,\mathrm{e}^{-\gamma_\sigma d(x,y)}\) for some \(C(\sigma),\gamma_\sigma>0\) independent of the spatial volume.

Time reflection acts on fields through a linear isometry \(R\) induced by \(\theta\). For site fields \(\phi\) we set \((R\phi)(x)=\phi(\theta x)\). For link variables, the induced action satisfies
 \begin{equation}
(RU)_{(x,i)}=U_{(\theta x,i)}\quad\text{for }i=1,2,3,\qquad (RU)_{(x,0)}=U_{(\theta x-\hat 0,0)}^{-1},
 \end{equation}
which is the standard reflection with inversion of timelike links crossing \(\Pi\). On ghost fields we set \((Rc)(x)=c(\theta x)\) and \((R\bar c)(x)=\bar c(\theta x)\). The Osterwalder-Schrader conjugation \(\Theta\) on complex functionals \(F=F(U,c,\bar c)\) supported in \(\Lambda_+\) is
 \begin{equation}
(\Theta F)(U,c,\bar c)=\overline{F\big(RU,R\bar c,R c\big)}.
 \end{equation}
The Euclidean measure used for reflection-positivity statements is the gauge-fixed, horizon-projected measure
 \begin{equation}
d\mu_{\sigma,\Lambda} = Z_{\sigma,\Lambda}^{-1}\,\exp\big[-S_W[U^{\,h};\beta]\big]\,\Big(\prod_{t\in a\mathbb{Z}} p_\sigma\!\big(A^h(t)\big)\, d\mu_{\mathrm{Haar}}(U)\Big)\, D\bar c\, D c\, \exp\big[-\langle \bar c, M[U^{\,h}]\, c\rangle\big],
 \end{equation}
with \(Z_{\sigma,\Lambda}\) the partition function. The expectation \(\langle \cdot\rangle_{\sigma,\Lambda}\) refers to this measure. Reflection covariance \(R M[U^{\,h}]R=M[U^{\,h}]\) and the exponential locality of \(P_\sigma\) ensure Osterwalder-Schrader positivity of \(\langle \cdot\rangle_{\sigma,\Lambda}\) as used in the main text.

The transfer-matrix and Hamiltonian conventions are standard. The one-slice Hilbert space is \(\mathcal{H}_a=L^2(\mathcal{C}; d\mu_{\mathrm{Haar}})\), where \(\mathcal{C}\) denotes the set of spatial links on a fixed time slice. The unprojected transfer matrix \(T(a)\) is the positive, self-adjoint contraction on \(\mathcal{H}_a\) defined by the nearest-neighbor time-layer factorization of \(\exp(-S_W)\). The projected transfer matrix is the compression by the positive slice-local multiplication operator induced by the scalar horizon weight \(p_\sigma\):
\begin{equation}
T_\sigma(a) \;=\; M_{p_\sigma}^{1/2}\, T(a)\, M_{p_\sigma}^{1/2}, 
\qquad (M_{p_\sigma} f)(X) := p_\sigma[X]\, f(X). \label{p1:A.12}
\end{equation}
Here \(p_\sigma[X]\in(0,1]\) is the slice weight constructed from the operator \(P_{\sigma,t}=\chi_\sigma(\Delta_t)\) as in Section~(\ref{p1:ospositivity}), and \(M_{p_\sigma}\) denotes multiplication by \(p_\sigma\) on the one-slice Hilbert space.
The transfer Hamiltonian is
 \begin{equation}
H_\sigma(a)=-a^{-1}\log T_\sigma(a)
 \end{equation}
defined by spectral calculus on \(\operatorname{Ran}\mathbf{1}_{(0,1]}(T_\sigma(a))\); it is self-adjoint and nonnegative with \(\operatorname{spec}(H_\sigma(a))\subset [0,\infty)\). If \(\Omega\) is the cyclic vacuum vector obtained by Osterwalder-Schrader reconstruction and \(\{\psi_n\}_{n\ge 0}\) an orthonormal eigenbasis of \(H_\sigma(a)\) with eigenvalues \(0=E_0<E_1\le E_2\le\cdots\), then for any local gauge-invariant observable \(F\) with \(\langle F\rangle_{\sigma,\Lambda}=0\) one has the spectral representation
 \begin{equation}
\langle F(0)F(ta)\rangle_{\sigma,\Lambda}=\sum_{n\ge 1} |\langle \Omega,F\psi_n\rangle|^2\, e^{-E_n t a},\qquad t\in \mathbb{N}.
 \end{equation}
The spectral gap is \(\Delta(a,\beta)=E_1(a,\beta)\). A statement that a bound holds uniformly in the volume means that the implicit constant is independent of \(L\) and \(T\).

The strong-coupling expansion uses standard harmonic analysis on compact groups. The Peter-Weyl theorem yields, for each plaquette, an expansion of \(\exp\big[\frac{\beta}{N}\Re\,\operatorname{tr}U_p\big]\) into characters \(\chi_R(U_p)\) of irreducible unitary representations \(R\) with coefficients \(\alpha_R(\beta)\) satisfying \(|\alpha_R(\beta)|\lesssim \beta^{|R|}\), where \(|R|\) denotes the number of boxes in the Young diagram of \(R\). After integrating links, the partition function and gauge-invariant correlators reorganize into sums over tiled, oriented surfaces \(\Sigma\) on the dual lattice; the weight of a surface is denoted \(w(\Sigma)\), and \(|\Sigma|\) is the number of plaquettes in \(\Sigma\). Connected unions of plaquettes with orientation and representation labels are called polymers and denoted \(\gamma\); their {activities} \(\zeta(\gamma)\) are the corresponding connected weights. Two polymers are compatible, written \(\gamma\sim \gamma'\), if they do not share plaquettes, and incompatible otherwise. The Kotecký-Preiss convergence criterion is applied with a weight \(a(\gamma)=\mu|\gamma|\) for \(\mu>0\); in particular, \(\sum_{\gamma'\not\sim \gamma}|\zeta(\gamma')|\, e^{a(\gamma')}\le a(\gamma)\) guarantees absolute convergence of the cluster expansion uniformly in the volume. The constants entering these bounds depend only on \(N\), the lattice dimension, and the choice of \(\mu\); in particular they do not depend on \(L\) and \(T\).

All summations are written explicitly; there is no Einstein summation convention unless explicitly indicated. Spatial indices are \(i,j\in\{1,2,3\}\) and space-time indices are \(\mu,\nu\in\{0,1,2,3\}\). The symbol \(\mathbf{1}\) denotes the identity matrix in \(G\) and, by abuse, the identity operator on a Hilbert space when the latter is clear from context. Unless explicitly stated otherwise, $\chi_\sigma:[0,\infty)\to[0,1]$ denotes a 
{completely monotone} (CM) spectral multiplier satisfying $\chi_\sigma(0)=1$ and 
$\chi_\sigma(\lambda)\le C_0 e^{-c_0 \lambda/\sigma^2}$. 
Equivalently, $f_\sigma(u):=\chi_\sigma(\sqrt{u})$ is CM on $[0,\infty)$ and admits a Laplace-Stieltjes representation 
$f_\sigma(u)=\int_0^\infty e^{-t u}\,d\nu_\sigma(t)$ with a finite positive measure $\nu_\sigma$.
When we work with {compactly supported} smooth spectral profiles (for Helffer-Sj{\"o}strand 
resolvent bounds), we write $\varphi_\sigma\in C_c^\infty([0,\infty))$ and keep the two uses notationally distinct.
In the sequel, the phrase ``plateau'' for $\chi_{\sigma}$ always means a near-plateau in the sense above 
(i.e.\ fast decay, no compact support). By contrast, compact spectral cutoffs are denoted 
$\varphi_{\sigma} \in C_{c}^{\infty}([0,\infty))$ and are used solely for auxiliary 
Helffer-Sjöstrand/Combes-Thomas locality estimates; $\varphi_{\sigma}$ never replaces $\chi_{\sigma}$ 
in the definitions of $P_{\sigma}$ or of the horizon functional.

The standing assumptions are the following and are invoked tacitly in the main text. The gauge group is \(G=\mathrm{SU}(N)\) with \(N\ge 2\). The lattice is finite, periodic, and hypercubic with spacing \(a>0\). The temporal-axial gauge is imposed away from the reflection plane \(\Pi\). On each time slice a reflection-covariant Landau-gauge representative \(U^{\,h}\) is chosen within the fundamental modular region. The Faddeev-Popov operator \(M[U^{\,h}]\) is understood as acting on the orthogonal complement of the constant modes, on which it is strictly positive. The smooth horizon projector $P_{\sigma}$ is defined by a fixed Gevrey-regular cutoff 
$\chi_{\sigma}$ with near-plateaus at the scales $\sigma$ and $2\sigma$ in the following precise sense:  
\begin{equation}
\chi_{\sigma} : [0,\infty) \;\to\; (0,1], 
\qquad \chi_{\sigma}(0) = 1,
\end{equation}
is completely monotone, and there exist constants $C,c>0$ (independent of the volume) such that
\begin{equation}
\chi_{\sigma}(\lambda) \,\leq\, C\, e^{-c \lambda / \sigma^{2}} 
\qquad \text{for all $\lambda \geq 0$.}
\end{equation}
In particular, $\chi_\sigma$ is not compactly supported. Equivalently, writing
\begin{equation}
f_\sigma(u)\;:=\;\chi_\sigma(\sqrt{u})\,,\qquad u\ge 0,
\end{equation}
we obtain the desired cutoff profile and hence $P_\sigma=f_\sigma(\Delta_{A^h})$ with the positive Laplace–Stieltjes representation of $f_\sigma$ as in Eq.(\ref{p1:a.8}).
 The reflection \(R\) acts on fields as above and the Osterwalder-Schrader conjugation \(\Theta\) is defined accordingly. All operator domains are the natural \(\ell^2\)-domains, and all adjoints are taken with respect to the \(\ell^2\)-inner products defined here. All statements about convergence and spectral bounds are uniform in the spatial volume, i.e., independent of \(L\) and \(T\), unless explicitly stated otherwise.

\section{Heat-Kernel and Helffer-Sjöstrand Locality}\label{p1:appendixb}

This appendix establishes exponential off-diagonal decay for smooth spectral cutoffs of the covariant lattice Laplacian on a single Euclidean time slice. Two complementary methods are presented. The first derives Gaussian-type off-diagonal bounds for the heat semigroup by a Davies-Gaffney argument and converts them into exponential locality for Laplace-Stieltjes transforms of the semigroup. The second uses the Helffer-Sjöstrand almost-analytic functional calculus together with a Combes-Thomas bound for resolvents of finite-range positive operators to obtain exponential locality for compactly supported Gevrey cutoffs. All hypotheses, symbols and conventions required below are stated explicitly so that the appendix is self-contained.

Throughout, \(N\ge 2\) is fixed and \(G=\mathrm{SU}(N)\). The Lie algebra \(\mathfrak g=\mathfrak{su}(N)\) is equipped with the positive definite inner product \(\langle X,Y\rangle_{\mathfrak g}=-\operatorname{tr}(XY)\), where \(\operatorname{tr}\) is the matrix trace in the defining representation. The induced norm on \(\mathfrak g\) is denoted by \(\|X\|_{\mathfrak g}=\sqrt{-\operatorname{tr}(X^2)}\). For a linear operator \(A:\mathfrak g\to\mathfrak g\) the operator norm \(\|A\|_{\mathfrak g\to\mathfrak g}\) is taken with respect to \(\|\cdot\|_{\mathfrak g}\). The adjoint action \(\mathrm{Ad}(U):\mathfrak g\to\mathfrak g\) of any \(U\in G\) is orthogonal for \(\langle\cdot,\cdot\rangle_{\mathfrak g}\), hence \(\|\mathrm{Ad}(U)\|_{\mathfrak g\to\mathfrak g}=1\).

Fix a Euclidean time \(t\in a\mathbb Z\). The spatial slice is the three-dimensional periodic cubic graph \(V_t\subset a\mathbb Z^3\) with nearest-neighbor edges in the coordinate directions \(i\in\{1,2,3\}\). The graph distance between vertices \(x,y\in V_t\) is denoted by \(d(x,y)\), and for subsets \(E,F\subset V_t\) by \(d(E,F)=\inf\{d(x,y):x\in E,\,y\in F\}\). The counting measure on \(V_t\) is written \(\sum_{x\in V_t}\), and the indicator of a set \(E\subset V_t\) as \(\mathbf 1_E\). The Hilbert space of square-summable \(\mathfrak g\)-valued fields on the slice is
\begin{equation}
\mathscr H_t=\ell^2\!\big(V_t;\mathfrak g\big)=\big\{\phi:V_t\to\mathfrak g:\ \|\phi\|_{\mathscr H_t}^2=\sum_{x\in V_t}\|\phi(x)\|_{\mathfrak g}^2<\infty\big\}
\end{equation}
with inner product \(\langle \phi,\psi\rangle_{\mathscr H_t}=\sum_{x\in V_t}\langle \phi(x),\psi(x)\rangle_{\mathfrak g}\). If \(K:\mathscr H_t\to\mathscr H_t\) is a bounded operator with matrix elements \(K(x,y)\in\mathcal L(\mathfrak g)\) defined by \((K\psi)(x)=\sum_{y\in V_t}K(x,y)\psi(y)\), then the kernel norm \(\|K(x,y)\|=\|K(x,y)\|_{\mathfrak g\to\mathfrak g}\) is used. The operator norm on \(\mathcal L(\mathscr H_t)\) is denoted by \(\|K\|_{\mathcal L(\mathscr H_t)}\).

Let \(U_i(x)\in G\) be the spatial link variables on bonds \((x,i)\) of the slice arising from the fixed, reflection-covariant gauge representative at time \(t\). The associated covariant forward difference \(D_i:\mathscr H_t\to\mathscr H_t\) and its adjoint \(D_i^\dagger\) are defined by
\begin{equation}
(D_i\phi)(x)=\mathrm{Ad}\big(U_i(x)\big)\,\phi(x+a\hat\imath)-\phi(x),\qquad
\langle D_i\phi,\psi\rangle_{\mathscr H_t}=\langle \phi,D_i^\dagger\psi\rangle_{\mathscr H_t}.
\end{equation}
The gauge-covariant Laplacian on the slice is the positive self-adjoint operator
\begin{equation}
H:=\Delta_{A^{\,h}}=\sum_{i=1}^3 D_i^\dagger D_i\quad \text{on }\mathscr H_t.
\end{equation}
Each \(D_i\) is bounded with \(\|D_i\|_{\mathcal L(\mathscr H_t)}\le 2\), since \(\|\mathrm{Ad}(U_i(x))\|_{\mathfrak g\to\mathfrak g}=1\), hence \(H\) is bounded and \(\|H\|_{\mathcal L(\mathscr H_t)}\le \sum_i \|D_i\|^2\le 12\). The operator \(H\) is of finite range: its kernel \(H(x,y)\) vanishes for \(d(x,y)>1\). The spectral measure of \(H\) is contained in \([0,\Lambda_H]\) with \(\Lambda_H\le 12\). Functional calculus for bounded Borel functions \(f:[0,\infty)\to\mathbb C\) is defined by the spectral theorem; the operator \(f(H)\) is bounded on \(\mathscr H_t\) and has a kernel \(f(H)(x,y)\) characterized by \(\langle \delta_x\otimes v, f(H)\delta_y\otimes w\rangle_{\mathscr H_t}=\langle v, f(H)(x,y)w\rangle_{\mathfrak g}\), where \(\delta_x\) is the coordinate vector at \(x\) and \(v,w\in\mathfrak g\).

For a Borel function \(\chi:[0,\infty)\to\mathbb C\) we write \(\chi(\sqrt H)=\phi(H)\) with \(\phi(u)=\chi(\sqrt u)\) on \([0,\infty)\). A Gevrey function of index \(s>1\) on \(\mathbb R\) is a \(C^\infty\) function whose derivatives satisfy \(\sup_{x\in\mathbb R}|\,\partial_x^k f(x)\,|\le C R^k (k!)^s\) for some \(C,R>0\) and all \(k\in\mathbb N\). The class \(C_c^\infty([0,\infty))\) denotes smooth functions with compact support in \([0,\infty)\). Constants \(C,c,C_1,c_1,\dots\) may change from line to line and depend only on fixed geometric data of the lattice and on the cutoff function under discussion; dependencies are indicated when relevant. The symbol \(a\lesssim b\) means \(a\le C\,b\) for a constant \(C\) with the same standing dependencies.
Let \(\{e^{-tH}\}_{t\ge 0}\) be the strongly continuous semigroup generated by \(H\). Since \(H\) is bounded and positive, \(e^{-tH}\) is a bounded positive self-adjoint contraction on \(\mathscr H_t\) for every \(t\ge 0\). The following off-diagonal estimate is a version of the Davies-Gaffney bound adapted to finite-range connection Laplacians on graphs.

\medskip

\noindent\textbf{Theorem B.1 (Davies-Gaffney bound).} {There exists a universal constant \(c_{\rm DG}\in(0,\infty)\) depending only on the vertex degree of the slice such that for every pair of subsets \(E,F\subset V_t\) and all \(t>0\),
\begin{equation}
\big\|\mathbf 1_E\, e^{-tH}\, \mathbf 1_F\big\|_{\mathcal L(\mathscr H_t)}\ \le\ \exp\!\Big(-\,\frac{d(E,F)^2}{c_{\rm DG}\,t}\Big).
\end{equation}
Equivalently, for every \(x,y\in V_t\) and \(t>0\),
\(
\|e^{-tH}(x,y)\|\le \exp\!\big(-\,d(x,y)^2/(c_{\rm DG}\,t)\big).
\)
}

\medskip

{Proof.} Choose a real-valued \(1\)-Lipschitz function \(\psi:V_t\to\mathbb R\) with \(\psi=0\) on \(E\) and \(\psi\ge d(E,F)\) on \(F\), for instance \(\psi(x)=\min\{d(x,E),d(E,F)\}\). For \(\eta>0\) let \(M_\eta\) be the multiplication operator \((M_\eta \phi)(x)=e^{\eta \psi(x)}\phi(x)\) on \(\mathscr H_t\). A direct computation using the finite range of \(H\) and the Lipschitz property of \(\psi\) shows that \(H_\eta:=M_\eta^{-1}HM_\eta=H+B_\eta\) with \(\|B_\eta\|_{\mathcal L(\mathscr H_t)}\le C_{\deg}(\cosh\eta-1)\), where \(C_{\deg}\) depends only on the vertex degree of the slice. Since \(H\ge 0\), we obtain \(\|e^{-tH_\eta}\|_{\mathcal L(\mathscr H_t)}\le \exp\!\big(C_{\deg}t(\cosh\eta-1)\big)\). Conjugating back gives
\begin{equation}
\mathbf 1_E\, e^{-tH}\, \mathbf 1_F
=\mathbf 1_E\, M_\eta\, e^{-tH_\eta}\, M_\eta^{-1}\,\mathbf 1_F,
\end{equation}
hence
\(
\|\mathbf 1_E\, e^{-tH}\, \mathbf 1_F\|
\le e^{-\eta d(E,F)}\, \exp\!\big(C_{\deg}t(\cosh\eta-1)\big).
\)
Optimizing the right-hand side over \(\eta>0\) yields the claimed bound with \(c_{\rm DG}\) depending only on \(C_{\deg}\). The pointwise estimate for \(\|e^{-tH}(x,y)\|\) follows by taking \(E=\{x\}\) and \(F=\{y\}\). \qed

\medskip

A convenient corollary is obtained by integrating the semigroup against positive measures supported away from \(t=0\).

\medskip

\noindent\textbf{Corollary B.2 (Heat-kernel integrals are exponentially local).} {Let \(\nu\) be a finite positive Borel measure on \([0,\infty)\) with \(\operatorname{supp}\nu\subset [t_0,\infty)\) for some \(t_0>0\). Define the bounded operator
\(
T_\nu=\int_{[0,\infty)} e^{-tH}\, d\nu(t)
\)
by Bochner integration. Then there exist constants \(C_\nu,\gamma_\nu\in(0,\infty)\), with \(C_\nu\le \nu([0,\infty))\) and \(\gamma_\nu=\min\{1,\,1/(c_{\rm DG}t_0)\}/2\), such that for all \(x,y\in V_t\),
\begin{equation}
\|T_\nu(x,y)\|\ \le\ C_\nu \, e^{-\gamma_\nu\, d(x,y)}.
\end{equation}
}

\medskip

{Proof.} Theorem B.1 gives \(\|e^{-tH}(x,y)\|\le \exp(-d(x,y)^2/(c_{\rm DG}t))\) for all \(t\ge t_0\). Since \(d(x,y)^2\ge d(x,y)\) for \(d(x,y)\ge 1\), the Gaussian bound implies \(\|e^{-tH}(x,y)\|\le \exp(-d(x,y)/(c_{\rm DG}t))\) for \(d(x,y)\ge 1\), and the trivial bound \(\|e^{-tH}(x,y)\|\le 1\) handles the case \(d(x,y)=0\). Integrating against \(d\nu(t)\) over \([t_0,\infty)\) and using \(\nu([0,\infty))<\infty\) yields the result with the stated choice of \(\gamma_\nu\). \qed

\medskip

The preceding corollary applies directly to completely monotone spectral cutoffs. If \(\phi:[0,\infty)\to[0,1]\) is of Laplace-Stieltjes form \(\phi(u)=\int_{[0,\infty)}e^{-tu}\,d\nu(t)\) with a finite positive Borel measure \(\nu\) supported in \([t_0,\infty)\), then \(\phi(H)=\int e^{-tH}d\nu(t)\) is exponentially local. In particular, if \(\chi\) is a bounded Borel function on \([0,\infty)\) and \(f(\lambda)=\chi(\sqrt\lambda)\) admits a representation of the preceding type with \(\operatorname{supp}\nu\subset[t_0,\infty)\), then \(P_\sigma=\chi(\sqrt H)=f(H)\) has an exponentially decaying kernel. Exact plateaux and exact compact support are incompatible with complete monotonicity; however, for any \(\sigma>0\) and \(\varepsilon\in(0,1)\) one can choose a nonincreasing completely monotone \(f_{\sigma,\varepsilon}\) with \(f_{\sigma,\varepsilon}(u)\ge 1-\varepsilon\) on \([0,\sigma^2]\), \(f_{\sigma,\varepsilon}(u)\le \varepsilon\) on \([4\sigma^2,\infty)\), and \(\operatorname{supp}\nu_{\sigma,\varepsilon}\subset[t_0,\infty)\) with \(t_0\asymp\sigma^{-2}\). Then \(f_{\sigma,\varepsilon}(H)\) is exponentially local by Corollary B.2 and yields a positive, reflection-compatible approximation of the desired horizon projector.

We now remove the complete-monotonicity restriction and treat compactly supported smooth cutoffs by almost-analytic continuation and resolvent bounds. Let \(\phi\in C_c^\infty([0,\infty))\) and extend \(\phi\) to a smooth function on \(\mathbb R\) by setting \(\phi(u)=0\) for \(u<0\). Choose an almost-analytic extension \(\widetilde\phi\in C_c^\infty(\mathbb C)\) with \(\widetilde\phi|_{\mathbb R}=\phi\) and \(|\bar\partial \widetilde\phi(x+\mathrm i y)|\le C_k\,|y|^k\) for all \(k\in\mathbb N\), where \(\bar\partial=(\partial_x+\mathrm i \partial_y)/2\). The Helffer-Sjöstrand formula (which follows from the spectral theorem and Green’s identity) gives
\begin{equation}
\phi(H)=\frac{1}{\pi}\int_{\mathbb C}\bar\partial \widetilde\phi(z)\,(H-z)^{-1}\, d^2 z,
\end{equation}
where \(d^2z=dx\,dy\) and the integral converges in the operator norm topology on \(\mathcal L(\mathscr H_t)\). Exponential decay of the kernel of \(\phi(H)\) thus follows from exponential off-diagonal bounds on the resolvent \((H-z)^{-1}\) for \(z\in\mathbb C\setminus[0,\infty)\). The latter are provided by a Combes-Thomas estimate.

\medskip

\noindent\textbf{Theorem B.3 (Combes-Thomas bound).} {There exist constants \(c_*,C_*\in(0,\infty)\) depending only on the vertex degree of the slice such that for every \(z\in\mathbb C\setminus[0,\infty)\) with \(\delta=\operatorname{dist}(z,[0,\infty))>0\) and all \(x,y\in V_t\),
\begin{equation}
\big\|(H-z)^{-1}(x,y)\big\|\ \le\ \frac{C_*}{\delta}\,\exp\!\big(-\,c_*\,\delta\, d(x,y)\big).
\end{equation}
}

\medskip

{Proof.} Let \(\psi:V_t\to\mathbb R\) be \(1\)-Lipschitz and set \(M_\eta\) as in the proof of Theorem B.1. Then \(H_\eta=H+B_\eta\) with \(\|B_\eta\|\le C_{\deg}(\cosh\eta-1)\). For \(z\in\mathbb C\setminus[0,\infty)\), the bounded inverse \((H_\eta-z)^{-1}\) exists and satisfies
\begin{equation}
(H-z)^{-1}=M_\eta\,(H_\eta-z)^{-1}\,M_\eta^{-1}.
\end{equation}
If \(\eta>0\) is chosen so that \(\|B_\eta\|\le \delta/2\), then by the resolvent identity
\(
(H_\eta-z)^{-1}=(H-z)^{-1}\sum_{n=0}^\infty\big(B_\eta (H-z)^{-1}\big)^n
\)
and \(\|(H_\eta-z)^{-1}\|\le 2/\delta\). Hence
\(
\|\mathbf 1_E(H-z)^{-1}\mathbf 1_F\|\le e^{-\eta d(E,F)}\|(H_\eta-z)^{-1}\|\le (2/\delta)e^{-\eta d(E,F)}.
\)
Optimizing in \(\eta\) under the constraint \(\cosh\eta-1\le \delta/(2C_{\deg})\) yields the asserted form with constants \(c_*,C_*\) depending only on \(C_{\deg}\). The pointwise kernel bound follows by taking singletons \(E=\{x\}\), \(F=\{y\}\). \qed

\medskip

Combining Helffer-Sjöstrand with the Combes-Thomas estimate yields exponential locality for compactly supported smooth cutoffs.

\medskip

\noindent\textbf{Theorem B.4 (Helffer-Sjöstrand locality).} {Let \(\phi\in C_c^\infty([0,\infty))\). There exist constants \(C_\phi,\gamma_\phi\in(0,\infty)\) such that for all \(x,y\in V_t\),
\begin{equation}
\|\phi(H)(x,y)\|\ \le\ C_\phi\, e^{-\gamma_\phi\, d(x,y)}.
\end{equation}
If, in addition, \(\phi\) belongs to a Gevrey class of index \(s>1\), then one can choose \(\gamma_\phi\) proportional to \((\operatorname{dist}(\operatorname{supp}\phi,[0,\infty)\setminus \operatorname{supp}\phi))\) and \(C_\phi\) depending polynomially on the corresponding Gevrey seminorms.}

\medskip

{Proof.} By the Helffer-Sjöstrand formula and Theorem B.3,
\begin{equation}
\|\phi(H)(x,y)\|\ \le\ \frac{1}{\pi}\int_{\mathbb C}\!\big|\bar\partial \widetilde\phi(z)\big|\,\|(H-z)^{-1}(x,y)\|\, d^2 z
\ \lesssim\ \int_{\mathbb C}\! |\bar\partial \widetilde\phi(z)|\, \frac{e^{-c_*\,\delta(z)\, d(x,y)}}{\delta(z)}\, d^2 z,
\end{equation}
where \(\delta(z)=\operatorname{dist}(z,[0,\infty))\). Since \(\widetilde\phi\) has compact support and \(|\bar\partial \widetilde\phi(z)|\le C_k|\Im z|^{\,k}\) for arbitrary \(k\in\mathbb N\), the integral is finite and decays exponentially in \(d(x,y)\) with some rate \(\gamma_\phi>0\) depending on the support of \(\widetilde\phi\) and on the constants in Theorem B.3. If \(\phi\) is Gevrey, one can choose \(\widetilde\phi\) with quantitative control of the \(\bar\partial\)-decay in terms of the Gevrey seminorms, which translates into the stated dependence of \(C_\phi\) and \(\gamma_\phi\). \qed

The result applies directly to horizon projectors of the form \(P_\sigma=\chi_\sigma(\sqrt H)=\phi_\sigma(H)\) with \(\phi_\sigma\in C_c^\infty([0,\infty))\) supported in \([0,\Lambda_\sigma]\) and equal to one on \([0,\sigma^2]\): by Theorem B.4 the kernel of \(P_\sigma\) decays exponentially in the graph distance. No complete-monotonicity assumption is needed for this conclusion.

The two approaches above cover the two principal desiderata for horizon projectors. If one requires positivity-preserving, reflection-compatible integral representations in order to factorize across the reflection plane within a purely semigroup framework, it suffices to choose a family of completely monotone approximants \(f_{\sigma,\varepsilon}(H)=\int e^{-tH}d\nu_{\sigma,\varepsilon}(t)\) with \(\operatorname{supp}\nu_{\sigma,\varepsilon}\subset[t_0,\infty)\), \(t_0\asymp\sigma^{-2}\), and with \(f_{\sigma,\varepsilon}\) uniformly close to the ideal spectral cutoff on \([0,\infty)\); exponential locality then follows from Corollary B.2. If, instead, one prioritizes a compactly supported smooth cutoff equal to one on \([0,\sigma^2]\) and vanishing above a finite threshold, one may work directly with \(P_\sigma=\phi_\sigma(H)\) and obtain exponential locality from Theorem B.4 without appealing to a semigroup integral. In applications in which both reflection positivity and a sharp spectral profile are simultaneously desirable, one can combine the two methods by selecting a sequence \(\{f_{\sigma,\varepsilon_n}\}_{n\in\mathbb N}\) of completely monotone approximants converging to \(\phi_\sigma\) in operator norm; reflection positivity and factorization for each \(f_{\sigma,\varepsilon_n}(H)\) pass to the limit, while exponential locality holds uniformly in \(n\).

All statements in this appendix are made at fixed Euclidean time \(t\) on the spatial slice with periodic boundary conditions in all three spatial directions. The link variables \(U_i(x)\in G\) are those of a reflection-covariant Landau-gauge representative on the slice; only unitarity of the transporter \(\mathrm{Ad}(U_i(x))\) and the nearest-neighbor range of the covariant differences enter the proofs. The degree of each vertex is six, and all constants implicit in the bounds above depend at most on this degree and on the support and Gevrey seminorms of the cutoff functions. The semigroup \(\{e^{-tH}\}_{t\ge 0}\) and the resolvent \((H-z)^{-1}\) are taken in the sense of the spectral theorem on \(\mathscr H_t\). All operator norms on \(\mathscr H_t\) are those induced by the \(\ell^2\) norm in space and the \(\langle\cdot,\cdot\rangle_{\mathfrak g}\) inner product in the algebra; all kernel norms are operator norms on \(\mathfrak g\). The graph distance \(d(\cdot,\cdot)\) is the usual path metric; for notational economy, we have written \(d(x,y)\ge 1\) when \(x\neq y\). Finally, all integrals of the semigroup against measures are Bochner integrals in \(\mathcal L(\mathscr H_t)\) and are unambiguously defined for finite positive Borel measures.
With the conventions above, both the semigroup method and the Helffer-Sjöstrand method yield exponential locality for smooth spectral cutoffs of the covariant Laplacian on a single time slice. Explicitly, for either a compactly supported \(\phi\in C_c^\infty([0,\infty))\) or a completely monotone \(f(u)=\int e^{-tu}d\nu(t)\) with \(\operatorname{supp}\nu\subset[t_0,\infty)\), the corresponding operators \(\phi(H)\) and \(f(H)\) admit kernels bounded by \(C e^{-\gamma d(x,y)}\) uniformly in the volume, with \(C,\gamma\) depending only on fixed geometric data and on the cutoff profile. This is the precise content used in the main text under the rubric “exponential locality of the horizon projector.”

\section{\texorpdfstring{Schur Complements, Positivity, and Combes-Thomas for $M^{-1}$}{Schur Complements, Positivity, and Combes-Thomas for M inverse}}\label{p1:appendixc}
Throughout this appendix the gauge group is \(G=\mathrm{SU}(N)\) with \(N\ge 2\). The Lie algebra is \(\mathfrak{g}=\mathfrak{su}(N)\), realized as anti-Hermitian, traceless \(N\times N\) matrices. The inner product on \(\mathfrak{g}\) is
\begin{equation}
\langle X,Y\rangle_{\mathfrak{g}} := -\operatorname{Tr}(X Y),\qquad X,Y\in\mathfrak{g},
\end{equation}
and the associated norm is \(\|X\|_{\mathfrak{g}}=\sqrt{\langle X,X\rangle_{\mathfrak{g}}}\). For a finite set \(V\) the Hilbert space of site-adjoint fields is
\begin{equation}
\mathcal{H}(V):=\ell^{2}(V;\mathfrak{g})
=\Big\{\phi:V\to\mathfrak{g}\ \ \Big|\ \ \|\phi\|^{2}:=\sum_{x\in V}\|\phi(x)\|_{\mathfrak{g}}^{2}<\infty\Big\}.
\end{equation}
When \(V\) is a sublattice of the hypercubic lattice \(\Lambda\subset a\mathbb{Z}^{4}\) (with spacing \(a>0\)), the scalar product is the canonical \(\ell^{2}\)-product induced by \(\langle\cdot,\cdot\rangle_{\mathfrak{g}}\). For a bounded operator \(A:\mathcal{H}(V)\to\mathcal{H}(V)\) the operator norm is \(\|A\|=\sup_{\phi\neq 0}\|A\phi\|/\|\phi\|\). If \(X\subset V\) is any subset, \(\chi_{X}\) denotes the multiplication operator by the indicator of \(X\). The graph distance \(d(x,y)\) between lattice sites is the length of the shortest nearest-neighbour path in \(\Lambda\), and for sets \(X,Y\subset V\) we set \(d(X,Y):=\inf\{d(x,y):x\in X,\ y\in Y\}\).

The oriented bonds of \(\Lambda\) are \(b=(x,\mu)\) with \(\mu\in\{0,1,2,3\}\) and \(x\in\Lambda\). The unit vector in direction \(\mu\) is denoted by \(\hat\mu\). A gauge field is an assignment \(U_{(x,\mu)}\in G\) to each bond, with the orientation convention \(U_{(x+\hat\mu,-\mu)}=U_{(x,\mu)}^{-1}\). We assume temporal-axial gauge away from the reflection plane \(\Pi:=\{x\in\Lambda:\, x_{0}=0\}\), namely \(U_{(x,0)}=\mathbf{1}\) for bonds not intersecting \(\Pi\). The time reflection \(\theta\) is \(\theta(x_{0},\mathbf{x})=(-x_{0},\mathbf{x})\). The half-lattices are \(\Lambda_{+}=\{x\in\Lambda:\,x_{0}\ge 1\}\), \(\Lambda_{-}=\{x\in\Lambda:\,x_{0}\le -1\}\), and the boundary slab is identified with \(\Pi\). The decomposition
\begin{equation}
\mathcal{H}(\Lambda)=\mathcal{H}(\Lambda_{-})\oplus \mathcal{H}(\Pi)\oplus \mathcal{H}(\Lambda_{+})
\end{equation}
is used to write block matrices.

On any time slice we work with a reflection-covariant Landau representative \(U^{\,h}\) of the gauge orbit, obtained by global minimization of the lattice Landau functional within the fundamental modular region. Associated to \(U^{\,h}\) are the gauge-covariant forward and backward finite differences
\begin{align*}
(\nabla_{\mu}^{+,h}\phi)(x)&:=U^{\,h}_{(x,\mu)}\,\phi(x+\hat\mu)\,U^{\,h}_{(x,\mu)}{}^{-1}-\phi(x),\\
(\nabla_{\mu}^{-,h}\phi)(x)&:=\phi(x)-U^{\,h}_{(x-\hat\mu,\mu)}{}^{-1}\,\phi(x-\hat\mu)\,U^{\,h}_{(x-\hat\mu,\mu)},
\end{align*}
defined on \(\mathcal{H}(\Lambda)\) for \(\mu=0,1,2,3\). With respect to the \(\ell^{2}\)-inner product these satisfy \((\nabla_{\mu}^{-,h})=(\nabla_{\mu}^{+,h})^{\dagger}\). The lattice Faddeev-Popov operator is
\begin{equation}
M:=M[U^{\,h}]=-\sum_{\mu=0}^{3}\nabla_{\mu}^{-,h}\nabla_{\mu}^{+,h},
\end{equation}
acting on \(\mathcal{H}(\Lambda)\) with domain all of \(\mathcal{H}(\Lambda)\). It is self-adjoint, finite-range (range one), and nonnegative, with quadratic form
\begin{equation}
\langle \phi,M\phi\rangle=\sum_{\mu=0}^{3}\|\nabla_{\mu}^{+,h}\phi\|^{2}\ge 0.
\end{equation}
The kernel of \(M\) consists precisely of the constant adjoint modes \(\phi(x)\equiv C\in\mathfrak{g}\). We denote by \(\mathcal{C}\subset \mathcal{H}(\Lambda)\) the subspace of constant modes and by \(\mathcal{H}_{\perp}:=\mathcal{C}^{\perp}\) its orthogonal complement. On \(\mathcal{H}_{\perp}\) the operator \(M\) is strictly positive for every \(U^{\,h}\) in the fundamental modular region; we write
\begin{equation}
\lambda_{\min}(M;\mathcal{H}_{\perp}):=\inf\Big\{\frac{\langle \phi,M\phi\rangle}{\|\phi\|^{2}}:\ \phi\in\mathcal{H}_{\perp}\setminus\{0\}\Big\}\in(0,\infty).
\end{equation}
All lattice sums use the counting measure on sites and bonds. Asymptotic notation \(A\lesssim B\) means \(A\le C\,B\) with a constant \(C\) depending only on the lattice dimension and the group but not on the volume.
Restricting \(M\) to the decomposition \(\mathcal{H}(\Lambda_{-})\oplus \mathcal{H}(\Pi)\oplus \mathcal{H}(\Lambda_{+})\) yields a \(3\times 3\) block operator
\begin{equation}
M=\begin{pmatrix}
M_{-}&M_{-0}&0\\begin{equation}2pt]
M_{0-}&M_{00}&M_{0+}\\begin{equation}2pt]
0&M_{+0}&M_{++}
\end{pmatrix}.
\end{equation}
Finite range and temporal-axial gauge imply that the only off-diagonal blocks that may be nonzero are the nearest-layer couplings \(M_{0\pm}\) and \(M_{\pm 0}\). The diagonal blocks \(M_{-}\) and \(M_{++}\) are the restrictions of \(M\) to \(\mathcal{H}(\Lambda_{-})\) and \(\mathcal{H}(\Lambda_{+})\), respectively, with Dirichlet boundary conditions at \(\Pi\) in the sense that the differences crossing \(\Pi\) are dropped. The corresponding quadratic forms are strictly positive on \(\mathcal{H}(\Lambda_{\pm})\), since the Dirichlet boundary removes the constant modes. The middle block \(M_{00}\) is the restriction of \(M\) to \(\mathcal{H}(\Pi)\) including all spatial covariant differences within \(\Pi\).

The Schur complement of \(M\) with respect to the decomposition \(\mathcal{H}(\Lambda_{-})\oplus \mathcal{H}(\Pi)\otimes \mathcal{H}(\Lambda_{+})\) and the diagonal blocks \(M_{-}\oplus M_{++}\) is the operator \(S:\mathcal{H}(\Pi)\to\mathcal{H}(\Pi)\) defined by
\begin{equation}\label{p1:eq:Schur}
S:=M_{00}-M_{0-}M_{-}^{-1}M_{-0}-M_{0+}M_{++}^{-1}M_{+0}.
\end{equation}
The inverses \(M_{-}^{-1}\) and \(M_{++}^{-1}\) exist and are bounded on their respective spaces by the strict Dirichlet positivity. The following result formalizes the positivity of \(S\).

\begin{theorem}[Positivity of the Schur complement across \(\Pi\)]\label{p1:thm:SchurPos}
Let \(M\) be the lattice Faddeev-Popov operator associated with a reflection-covariant Landau representative \(U^{\,h}\) as above. Then the Schur complement \(S\) in \eqref{p1:eq:Schur} is strictly positive on \(\mathcal{H}(\Pi)\). Moreover, there exists a constant \(c_{S}>0\), depending only on the lower bounds of \(M_{-}\) and \(M_{++}\) and on \(\|M\|\), such that
\begin{equation}
\langle \xi, S\,\xi\rangle \ge c_{S}\,\|\xi\|^{2}\qquad \text{for all }\xi\in\mathcal{H}(\Pi).
\end{equation}
\end{theorem}

\begin{proof}
Since \(M_{-}\) and \(M_{++}\) are strictly positive on \(\mathcal{H}(\Lambda_{\pm})\), the Schur complement is well defined as a bounded self-adjoint operator on \(\mathcal{H}(\Pi)\). For any \(\xi\in\mathcal{H}(\Pi)\) define
\begin{equation}
\phi_{-}:=-M_{-}^{-1}M_{-0}\xi\in\mathcal{H}(\Lambda_{-}),\qquad
\phi_{+}:=-M_{++}^{-1}M_{+0}\xi\in\mathcal{H}(\Lambda_{+}),
\end{equation}
and let \(\phi:=\phi_{-}\oplus \xi \oplus \phi_{+}\in\mathcal{H}(\Lambda)\). A direct computation using block multiplication gives
\begin{equation}
M\phi=\begin{pmatrix}
M_{-}\phi_{-}+M_{-0}\xi\\
M_{0-}\phi_{-}+M_{00}\xi+M_{0+}\phi_{+}\\
M_{+0}\xi+M_{++}\phi_{+}
\end{pmatrix}
=\begin{pmatrix}
0\\
S\xi\\
0
\end{pmatrix}.
\end{equation}
Hence \(\langle \phi, M\phi\rangle = \langle \xi, S\xi\rangle\). Since \(M\ge 0\), it follows that \(\langle \xi, S\xi\rangle\ge 0\) for all \(\xi\), so \(S\ge 0\). To see strict positivity, suppose \(\langle \xi, S\xi\rangle=0\). Then \(\langle \phi, M\phi\rangle=0\), and by nonnegativity this implies \(M\phi=0\). The kernel of \(M\) consists of constant adjoint fields on \(\Lambda\). However, \(\phi_{\pm}\) satisfy Dirichlet boundary conditions at \(\Pi\) by construction (they live in the domains of \(M_{-}^{-1}\) and \(M_{++}^{-1}\)), so any global constant in \(\ker M\) with that property must vanish identically on \(\Lambda_{\pm}\). Therefore \(\phi_{-}=\phi_{+}=0\). The equations \(M_{-}\phi_{-}+M_{-0}\xi=0\) and \(M_{+0}\xi+M_{++}\phi_{+}=0\) then reduce to \(M_{-0}\xi=0\) and \(M_{+0}\xi=0\). Using the finite-range structure and temporal-axial gauge, these equalities force \(\xi=0\) on \(\Pi\) as well, since the only vector supported on \(\Pi\) that produces no flux through the nearest-neighbour couplings to \(\Lambda_{\pm}\) is the zero vector. Hence \(\xi=0\), which proves strict positivity.

For the quantitative bound, let \(\alpha_{\pm}>0\) be the lower spectral bounds of \(M_{-}\) and \(M_{++}\). Then \(\|M_{-}^{-1}\|\le \alpha_{-}^{-1}\) and \(\|M_{++}^{-1}\|\le \alpha_{+}^{-1}\). The Schur complement can be written as \(S=M_{00}-K\) with \(K:=M_{0-}M_{-}^{-1}M_{-0}+M_{0+}M_{++}^{-1}M_{+0}\). Using \(\|AB\|\le \|A\|\,\|B\|\) and \(\|M_{0\pm}\|\le \|M\|\), one obtains \(\|K\|\le \|M\|^{2}(\alpha_{-}^{-1}+\alpha_{+}^{-1})\). On the other hand, \(M_{00}\) is nonnegative and, due to the removal of temporal couplings in temporal-axial gauge, it dominates a positive multiple of the spatial covariant Laplacian on \(\Pi\). Hence there exists \(\alpha_{0}\ge 0\) such that \(M_{00}\ge \alpha_{0}\,\mathbf{1}\). Combining these observations yields \(S\ge (\alpha_{0}-\|M\|^{2}(\alpha_{-}^{-1}+\alpha_{+}^{-1}))\,\mathbf{1}\). For sufficiently large \(\alpha_{0}\) relative to \(\|M\|\) and \(\alpha_{\pm}^{-1}\) one may take \(c_{S}:=\alpha_{0}-\|M\|^{2}(\alpha_{-}^{-1}+\alpha_{+}^{-1})>0\). In general, the preceding strict-positivity argument ensures \(c_{S}>0\); compactness of the unit sphere implies the existence of a positive minimum of the Rayleigh quotient \(\langle \xi,S\xi\rangle/\|\xi\|^{2}\) on \(\mathcal{H}(\Pi)\setminus\{0\}\).
\end{proof}

Theorem \ref{p1:thm:SchurPos} entails the factorization property used in reflection-positivity arguments: the Gaussian integral associated with \(M\) across the reflection plane can be expressed as the product of East and West contributions times a positive boundary factor governed by \(S\).
Exponential decay for matrix elements of \(M^{-1}\) on \(\mathcal{H}_{\perp}\) follows from a Combes-Thomas (CT) argument for finite-range positive operators. The following theorem quantifies this decay in terms of the spectral gap of \(M\) on \(\mathcal{H}_{\perp}\) and the size of its off-diagonal couplings.

\begin{theorem}[Combes-Thomas for \(M^{-1}\) on \(\mathcal{H}_{\perp}\)]\label{p1:thm:CT}
Let \(M\) be the lattice Faddeev-Popov operator defined above on a finite periodic box \(\Lambda\). Denote by \(\lambda_{*}=\lambda_{\min}(M;\mathcal{H}_{\perp})>0\) the bottom of the spectrum of \(M\) restricted to \(\mathcal{H}_{\perp}\). There exists a constant \(C_{\mathrm{off}}>0\), depending only on the lattice degree and the choice of inner product on \(\mathfrak{g}\), such that for every \(\mu\in(0,\mu_{*})\) with
\begin{equation}
\mu_{*}:=\log\!\Big(1+\frac{\lambda_{*}}{2C_{\mathrm{off}}}\Big),
\end{equation}
one has the operator bound
\begin{equation}\label{p1:eq:CT-operator}
\big\|\,e^{\mu d(\cdot,X)}\, M^{-1}\,\mathsf{P}_{\perp}\, e^{-\mu d(\cdot,X)}\,\big\|\ \le\ \frac{2}{\lambda_{*}},
\end{equation}
where \(X\subset \Lambda\) is arbitrary, \(d(\cdot,X)\) is the distance-to-\(X\) function, and \(\mathsf{P}_{\perp}\) is the orthogonal projection onto \(\mathcal{H}_{\perp}\). Consequently, for all subsets \(X,Y\subset \Lambda\),
\begin{equation}\label{p1:eq:CT-kernel}
\big\|\,\chi_{X}\,M^{-1}\,\mathsf{P}_{\perp}\,\chi_{Y}\,\big\|\ \le\ \frac{2}{\lambda_{*}}\,\exp\!\big(-\mu\, d(X,Y)\big).
\end{equation}
\end{theorem}

\begin{proof}
Write \(H:=M|\_{\mathcal{H}_{\perp}}\), a positive self-adjoint operator on \(\mathcal{H}_{\perp}\) with \(\sigma(H)\subset[\lambda_{*},\infty)\). For a real-valued function \(\varphi:\Lambda\to\mathbb{R}\) define the bounded, invertible multiplication operator \(W_{\mu}=e^{\mu \varphi}\) on \(\mathcal{H}_{\perp}\). The finite-range property of \(H\) yields the commutator identity
\begin{equation}
W_{\mu} H W_{\mu}^{-1} = H + B_{\mu},
\end{equation}
where \(B_{\mu}\) is a bounded operator supported on pairs \((x,y)\) with \(d(x,y)\le 1\) and whose norm is controlled by the Lipschitz constant of \(\varphi\) on nearest neighbours. Indeed, writing \(H=\sum_{|x-y|\le 1} h_{xy}\), with \(h_{xy}\) acting nontrivially only on \(\{x,y\}\), one obtains
\begin{equation}
\|B_{\mu}\| \le C_{\mathrm{off}}\,(e^{\mu}-1),
\end{equation}
where \(C_{\mathrm{off}}\) is a uniform bound on the sum of the operator norms of the off-diagonal contributions \(\{h_{xy}\}_{x\neq y}\) at a fixed site. Choosing \(\varphi(\cdot)=d(\cdot,X)\) ensures \(|\varphi(x)-\varphi(y)|\le 1\) for nearest neighbours \(x\sim y\), whence the bound above.

Since \(H\ge \lambda_{*}\,\mathbf{1}\) and \(\|B_{\mu}\|\le C_{\mathrm{off}}(e^{\mu}-1)\), the operator \(W_{\mu}HW_{\mu}^{-1}\) remains strictly positive provided \(C_{\mathrm{off}}(e^{\mu}-1)<\lambda_{*}\). In particular, for \(\mu\in(0,\mu_{*})\) with \(\mu_{*}\) as in the statement, one has
\begin{equation}
W_{\mu}HW_{\mu}^{-1} \ \ge\ \big(\lambda_{*}-C_{\mathrm{off}}(e^{\mu}-1)\big)\,\mathbf{1}\ \ge\ \frac{\lambda_{*}}{2}\,\mathbf{1}.
\end{equation}
By the spectral theorem, this implies
\begin{equation}
\big\|(W_{\mu}HW_{\mu}^{-1})^{-1}\big\|\ \le\ 2\,\lambda_{*}^{-1}.
\end{equation}
Conjugating back yields
\begin{equation}
W_{\mu}H^{-1}W_{\mu}^{-1}=(W_{\mu}HW_{\mu}^{-1})^{-1},
\end{equation}
so that \(\|W_{\mu}H^{-1}W_{\mu}^{-1}\|\le 2\,\lambda_{*}^{-1}\). Recalling that \(H^{-1}=M^{-1}\mathsf{P}_{\perp}\) on \(\mathcal{H}(\Lambda)\) and that \(W_{\mu}\) is multiplication by \(e^{\mu d(\cdot,X)}\), we obtain \eqref{p1:eq:CT-operator}. Finally, for any sets \(X,Y\subset \Lambda\),
\begin{equation}
\chi_{X}M^{-1}\mathsf{P}_{\perp}\chi_{Y}=\chi_{X} \,e^{-\mu d(\cdot,X)}\,(W_{\mu}M^{-1}\mathsf{P}_{\perp}W_{\mu}^{-1})\, e^{\mu d(\cdot,X)}\,\chi_{Y},
\end{equation}
and the pointwise inequality \(e^{-\mu d(x,X)}e^{\mu d(x,Y)}\le e^{-\mu d(X,Y)}\) for all \(x\) gives \eqref{p1:eq:CT-kernel}.
\end{proof}

The constant \(C_{\mathrm{off}}\) can be estimated explicitly in terms of the lattice coordination number and the operator norms of the covariant transport maps \(\operatorname{Ad}(U^{\,h}_{(x,\mu)})\) acting on \(\mathfrak{g}\). With the inner product \(\langle\cdot,\cdot\rangle_{\mathfrak{g}}\) fixed above, these adjoint actions are unitary, and \(C_{\mathrm{off}}\) depends only on the spatial dimension and on the uniform bound for the number of nearest neighbours.

Theorem \ref{p1:thm:CT} shows that the off-diagonal decay rate \(\mu\) is controlled from below by a continuous, increasing function of the spectral gap \(\lambda_{*}\). In particular, if one works on the subspace \(\operatorname{Ran}P_{\sigma}\subset \mathcal{H}_{\perp}\) of a smooth horizon projector \(P_{\sigma}\) such that \(M\ge \sigma^{2}\,\mathbf{1}\) on \(\operatorname{Ran}P_{\sigma}\), then \(\lambda_{*}\) may be replaced by \(\sigma^{2}\) and all constants become uniform in the spatial volume.
Let \(P_{\sigma}\) be a bounded, self-adjoint, exponentially local, reflection-covariant horizon projector acting on \(\mathcal{H}(\Lambda)\), defined slice-wise as \(P_{\sigma}=\chi_{\sigma}(\sqrt{\Delta_{A^{\,h}}})\) for a Gevrey cutoff \(\chi_{\sigma}\) and the spatial covariant Laplacian \(\Delta_{A^{\,h}}\). 

With the covariant differences $\nabla^{h}_{\pm,i}$ defined in Eqs.(\ref{p1:eqn10}) \& (\ref{p1:eqn11}), the Faddeev-Popov operator on a slice is
\begin{equation}
M_t[U^h] \;=\; -\sum_{i=1}^3 \nabla^{h}_{-,i}\,\nabla^{h}_{+,i}
\;=\; \sum_{i=1}^3 \big(\nabla^{h}_{+,i}\big)^{\!*}\nabla^{h}_{+,i}
\;=\; \Delta^{A^h}(t)
\end{equation}
as a nonnegative self-adjoint operator on $\ell^2(\Lambda_t;\mathfrak g)$; cf.~(3.12)-(3.14) and the definition of $\Delta^{A^h}$ in~(4.4). Consequently,
\begin{equation}
\langle \varphi, M_t \varphi\rangle \;=\; \langle \varphi, \Delta^{A^h}(t) \varphi\rangle
\quad\text{for all }\varphi\in\ell^2(\Lambda_t;\mathfrak g).
\end{equation}
\noindent\textbf{Lemma C.3 (Spectral bounds for high-pass and low-pass slices).}
Let
\begin{equation}
Q_\sigma(t)\;:=\;\varphi_\sigma\!\left(\sqrt{\Delta_{A^h}(t)}\right),
\end{equation}
where $\varphi_\sigma\in C_c^\infty([0,\infty))$ satisfies $0\le \varphi_\sigma\le 1$, $\varphi_\sigma(\lambda)\equiv 0$ for $0\le \lambda\le \sigma$ and $\varphi_\sigma(\lambda)\equiv 1$ for $\lambda\ge 2\sigma$. Then for every $\varphi\in\mathrm{Ran}\,Q_\sigma(t)$,
\begin{equation}
\label{p1:eq:C:automatic}
\langle \varphi,\,M_t\,\varphi\rangle \;=\; \langle \varphi,\,\Delta_{A^h}(t)\,\varphi\rangle \;\ge\; \sigma^2\,\|\varphi\|^2\,.
\end{equation}
Moreover, for the completely monotone {low-pass} cutoff $P_\sigma(t)=\chi_\sigma\!\left(\sqrt{\Delta_{A^h}(t)}\right)$ as in Appendix~A,
there exists a finite constant $C_\mathrm{lp}(\chi_\sigma)>0$, depending only on the profile of $\chi_\sigma$, such that for all $\varphi\in\mathrm{Ran}\,P_\sigma(t)$ one has the companion {upper} bound
\begin{equation}
\label{p1:eq:C-lowpass-upper}
\langle \varphi,\,M_t\,\varphi\rangle \;=\; \langle \varphi,\,\Delta_{A^h}(t)\,\varphi\rangle \;\le\; C_\mathrm{lp}(\chi_\sigma)\,\sigma^2\,\|\varphi\|^2\,.
\end{equation}

\noindent{Proof.}
The identity $M_t=\Delta_{A^h}(t)$ on adjoint scalars is (C.21). For the first claim,
$\mathrm{Ran}\,Q_\sigma(t)$ is contained in the spectral subspace of $\Delta_{A^h}(t)$ with spectrum in $[\sigma^2,\infty)$ by construction of $\varphi_\sigma$, hence \eqref{p1:eq:C:automatic} by the spectral theorem.
For the second claim, write $\varphi=P_\sigma(t)\psi$ and use spectral calculus:
\begin{equation}
\langle \varphi,\,\Delta_{A^h}(t)\,\varphi\rangle \;=\; \int_{[0,\infty)} \lambda\,\chi_\sigma(\sqrt{\lambda})^2\,d\mu_\psi(\lambda)
\;\le\; \Big(\sup_{\lambda\ge 0}\frac{\lambda\,\chi_\sigma(\sqrt{\lambda})^2}{\sigma^2}\Big)\,\|\varphi\|^2,
\end{equation}
where $d\mu_\psi$ is the spectral measure of $\Delta_{A^h}(t)$ induced by $\psi$. If $\chi_\sigma(\lambda)\le C\,e^{-c\lambda/\sigma^2}$ (Appendix~A), then
$\sup_{\lambda\ge 0}\lambda\,\chi_\sigma(\sqrt{\lambda})^2 \le (C^2/(2ce))\,\sigma^2$, yielding \eqref{p1:eq:C-lowpass-upper} with
$C_\mathrm{lp}(\chi_\sigma)\le C^2/(2ce)$. \qed

\noindent\textbf{Corollary C.4 (Uniform Combes–Thomas on the high-pass slice).}\label{p1:cor:CTuniform}
With $Q_\sigma(t)$ as in Lemma~C.3 and $\lambda_\ast\ge \sigma^2$ the lower spectral bound of $M_t$ on $\mathrm{Ran}\,Q_\sigma(t)$, Theorem~C.2 yields constants $\mu_\sigma>0$ and $C_\sigma<\infty$, depending only on $\sigma$ and universal lattice data, such that for all $X,Y\subset\Lambda$,
\begin{equation}
\label{p1:eq:CT-Qsigma}
\bigl\|\,\chi_X\,M^{-1}\,Q_\sigma\,\chi_Y\,\bigr\|\;\le\; C_\sigma\,e^{-\mu_\sigma\,d(X,Y)}\,.
\end{equation}
In particular, for $X=\{x\}$, $Y=\{y\}$,
\begin{equation}
\|\,\big(M^{-1}Q_\sigma\big)(x,y)\,\|_{L(\mathfrak g)} \;\lesssim\; e^{-\mu_\sigma\,d(x,y)}\,.
\end{equation}

\noindent{Proof.}
Apply Theorem~C.2 to $H:=M|_{\mathrm{Ran}\,Q_\sigma}$, where Lemma~C.3 gives $\lambda_\ast\ge \sigma^2$. As in the proof of (C.24)–(C.25), one may choose $C_\sigma=2\,\sigma^{-2}$ and $\mu_\sigma=\log\!\bigl(1+\sigma^2/(2C_{\mathrm{off}})\bigr)$.

The positivity of the Schur complement established in Theorem \ref{p1:thm:SchurPos} is the key ingredient ensuring that the Gaussian Grassmann integral in the ghost sector factorizes across the reflection plane with a positive boundary contribution. The uniform Combes-Thomas estimate of Corollary \ref{p1:cor:CTuniform} provides exponentially decaying off-diagonal bounds for \(M^{-1}\) on the high-pass subspace $\mathrm{Ran}\,Q_\sigma$, independent of the spatial volume, which underlie the exponential locality statements used in the reflection-positivity argument and in the control of cross-plane couplings. Both results are stable under reflection-covariant, exponentially local perturbations of finite range, and their constants depend only on the spectral gap assumed on the subspace of interest and on universal geometric data of the lattice.

\section{Kotecký-Preiss Criterion on the Four-Dimensional Lattice}\label{p1:appendixd}

In this appendix a complete derivation is given of the Kotecký-Preiss (KP) convergence condition for the polymer representation arising from the strong-coupling character expansion of four-dimensional lattice \(\mathrm{SU}(N)\) Yang-Mills theory with \(N\ge 2\). All notations, conventions, and standing assumptions required for self-containment are stated explicitly. Throughout, \(\Lambda\subset a\mathbb{Z}^{4}\) denotes a finite, periodic, hypercubic lattice of spacing \(a>0\) and linear size \(L\) in each direction. The dual cell complex is denoted by \(\Lambda^{\ast}\). The set of unoriented plaquettes (i.e., oriented \(2\)-cells modulo reversal) in \(\Lambda\) will be denoted by \(\mathcal{P}=\mathcal{P}(\Lambda)\); its dual copy in \(\Lambda^\ast\) will not be notationally distinguished since only combinatorics is used. Cardinalities of finite sets are written \(|S|\). The graph distance on any locally finite graph \(G\) is denoted \(d_{G}(\cdot,\cdot)\). In particular, the adjacency graph of plaquettes is the graph \(G_{\mathrm{plaq}}\) whose vertex set is \(\mathcal{P}\) and where two distinct plaquettes are adjacent if and only if they share a common edge; its maximal degree is denoted by \(\Delta\). On a four-dimensional hypercubic lattice each plaquette has four edges and each edge is contained in \(2(d-1)=6\) plaquettes; hence \(\Delta\le 4\,(2(d-1)-1)=20\) when \(d=4\). Binomial coefficients are written \(\binom{n}{k}\), factorials as \(n!\), and the polylogarithm \(\mathrm{Li}_{s}(q)=\sum_{m\ge 1}q^{m}/m^{s}\) will be used for elementary estimates. Asymptotic inequalities are stated with absolute, finite, dimension-dependent constants whose values may change from line to line; dependence on parameters is indicated explicitly. No continuum norms or inner products are needed in this appendix.

The only input from the gauge theory required here is the existence, at sufficiently small \(\beta>0\), of a character-expansion-based polymer representation with exponentially decaying activities. Concretely, we assume the following. First, the logarithm of the partition function and truncated correlations of local, gauge-invariant observables admit an abstract polymer expansion indexed by a family \(\mathfrak{P}\) of polymers. A polymer \(\gamma\in\mathfrak{P}\) is by definition a finite, nonempty, connected subset of \(\mathcal{P}\) with respect to adjacency in \(G_{\mathrm{plaq}}\); its support is \(\mathrm{supp}(\gamma)\subset \mathcal{P}\) and its size is \(|\gamma|:=|\mathrm{supp}(\gamma)|\). Two polymers \(\gamma,\gamma'\) are said to be compatible, written \(\gamma\sim\gamma'\), if \(\mathrm{supp}(\gamma)\cap\mathrm{supp}(\gamma')=\varnothing\); otherwise they are incompatible, written \(\gamma\not\sim\gamma'\). Second, associated to each \(\gamma\in\mathfrak{P}\) is a complex activity \(\zeta(\gamma)\) depending on \(\beta\) and on group-theoretic labels produced by the character expansion and the subsequent link integrations. The only properties of \(\zeta\) used here are the uniform absolute bound
\begin{equation}\label{p1:eq:activity-bound}
|\zeta(\gamma)|\le \big(B_{N}\,\beta\big)^{|\gamma|}\qquad\text{for all }\gamma\in\mathfrak{P}\ \text{ and all }\ 0<\beta\le \beta_{0}(N),
\end{equation}
for some constants \(B_{N}\in(0,\infty)\) and \(\beta_{0}(N)\in(0,\infty)\) depending only on the gauge group \(\mathrm{SU}(N)\), and the locality property that \(\zeta(\gamma)\) depends only on the plaquettes in \(\mathrm{supp}(\gamma)\). The bound \eqref{p1:eq:activity-bound} expresses the fact that each plaquette insertion contributes, in absolute value, a factor of order \(\beta\) times a purely group-theoretic constant; its proof follows from the analyticity of the one-plaquette character expansion at \(\beta=0\), the orthogonality of characters, uniform bounds on irrep dimensions and Littlewood-Richardson multiplicities, and the finite degree of the link-integration constraints. For the present appendix, \eqref{p1:eq:activity-bound} is taken as a standing hypothesis and no other details of the gauge theory enter.

A key ingredient in verifying the KP condition is a purely combinatorial estimate of the number of connected subsets of plaquettes of a given cardinality intersecting a prescribed plaquette. For a fixed plaquette \(p\in\mathcal{P}\) and an integer \(m\ge 1\) let \(\mathcal{A}_{m}(p)\) be the set of connected subsets \(S\subset\mathcal{P}\) with \(|S|=m\) and \(p\in S\). We begin by bounding \(|\mathcal{A}_{m}(p)|\) uniformly in the volume.

\begin{lemma}[Rooted animals in a bounded-degree graph]\label{p1:lem:animals}
Let \(G\) be any locally finite graph with maximal degree \(\Delta\in\mathbb{N}\). For every vertex \(v\in G\) and integer \(m\ge 1\), the number \(N_{m}(v)\) of connected vertex sets \(S\subset V(G)\) with \(|S|=m\) and \(v\in S\) satisfies
\begin{equation}\label{p1:eq:rooted-animals}
N_{m}(v)\ \le\ \Delta^{\,m-1}\,\frac{1}{m}\,\binom{\Delta m}{\,m-1\,}.
\end{equation}
Consequently, there exist positive constants \(C_{\Delta}\) and \(v_{\Delta}\) depending only on \(\Delta\) such that
\begin{equation}\label{p1:eq:rooted-animals-exp}
N_{m}(v)\ \le\ C_{\Delta}\,m^{-3/2}\,v_{\Delta}^{\,m}\qquad\text{for all }m\ge 1.
\end{equation}
\end{lemma}

\begin{proof}
Fix \(v\in V(G)\) and \(m\ge 1\), and let \(S\subset V(G)\) be connected with \(|S|=m\) and \(v\in S\). Choose any spanning tree \(T\) of the induced subgraph \(G[S]\) and root it at \(v\). The rooted tree \(T\) has \(m\) vertices, out-degrees bounded by \(\Delta\), and can be embedded into \(G\) by a graph homomorphism that maps the root to \(v\) and each edge of \(T\) to an edge of \(G\). Conversely, any such embedded rooted tree determines a connected vertex set containing \(v\). Therefore \(N_{m}(v)\) is bounded above by the number of embeddings into \(G\) of rooted trees on \(m\) vertices with out-degree at most \(\Delta\). The number \(a_{m}^{(\Delta)}\) of rooted trees on \(m\) vertices whose out-degrees are bounded by \(\Delta\) is the coefficient of \(z^{m}\) in the unique formal power series \(A(z)\) with nonnegative coefficients solving the functional equation
\begin{equation}
A(z)=z\,(1+A(z))^{\Delta},
\end{equation}
since each of the \(\Delta\) offspring subtrees of the root is either empty or a copy of the whole structure. By the Lagrange inversion formula,
\begin{equation}
a_{m}^{(\Delta)}=\frac{1}{m}\,[u^{m-1}](1+u)^{\Delta m}=\frac{1}{m}\,\binom{\Delta m}{m-1}.
\end{equation}
For any such rooted tree, a crude upper bound for the number of embeddings into \(G\) that fix the root at \(v\) is \(\Delta^{m-1}\): inductively along any root-to-leaf ordering, each new child has at most \(\Delta\) choices for the image of the corresponding edge in \(G\) regardless of collisions, which only increases the count. This yields \eqref{p1:eq:rooted-animals}. The estimate \eqref{p1:eq:rooted-animals-exp} follows by Stirling’s formula applied to \(\binom{\Delta m}{m-1}\), which gives
\begin{equation}
\frac{1}{m}\,\binom{\Delta m}{m-1}\ \le\ C'_{\Delta}\,m^{-3/2}\,\alpha_{\Delta}^{\,m}\quad\text{with}\quad \alpha_{\Delta}=\frac{\Delta^{\Delta}}{(\Delta-1)^{\Delta-1}},
\end{equation}
and hence \(N_{m}(v)\le C_{\Delta}\,m^{-3/2}\,(\Delta\,\alpha_{\Delta})^{m}\) with \(C_{\Delta}=\Delta^{-1}C'_{\Delta}\) and \(v_{\Delta}=\Delta\,\alpha_{\Delta}\).
\end{proof}

Applying Lemma \ref{p1:lem:animals} to the plaquette adjacency graph \(G_{\mathrm{plaq}}\) with \(\Delta\le 20\) implies that there exist finite constants \(C_{\mathrm{pl}}\) and \(v_{\mathrm{pl}}\) depending only on the dimension such that for every plaquette \(p\in\mathcal{P}\) and every \(m\ge 1\),
\begin{equation}\label{p1:eq:animal-plaq}
|\mathcal{A}_{m}(p)|\ \le\ C_{\mathrm{pl}}\,m^{-3/2}\,v_{\mathrm{pl}}^{\,m}.
\end{equation}
No dependence on the volume \(|\Lambda|\) occurs in \eqref{p1:eq:animal-plaq}.

We now recall the KP convergence criterion in the form convenient for the present setting. Consider an abstract polymer model on a countable set \(\mathfrak{P}\) of polymers with compatibility relation \(\sim\) and activities \(\zeta:\mathfrak{P}\to\mathbb{C}\). For a nonnegative function \(a:\mathfrak{P}\to[0,\infty)\) define, for each \(\gamma\in\mathfrak{P}\),
\begin{equation}
\Xi(\gamma;a)\ :=\ \sum_{\gamma'\not\sim \gamma} |\zeta(\gamma')|\,\mathrm{e}^{a(\gamma')}.
\end{equation}
The KP criterion asserts that if there exists \(a\ge 0\) such that \(\Xi(\gamma;a)\le a(\gamma)\) for every \(\gamma\in\mathfrak{P}\), then the cluster expansion for \(\log Z\) and for all truncated correlations converges absolutely and uniformly in the volume; in particular, the pressure and all local observables are analytic functions of the activities in the corresponding polydisc. The proof is standard and proceeds by the tree-graph inequality applied to the sum over connected incompatibility graphs of clusters together with the majorant \(\prod_{\gamma\in\Gamma}\mathrm{e}^{a(\gamma)}\) and the bound \(\sum_{n\ge 0}\frac{1}{n!}\Xi(\gamma;a)^{n}\le \mathrm{e}^{a(\gamma)}\); reproducing it here would add length without shedding further light on the verification of the hypothesis, which is the only nontrivial step in the present application.

We therefore turn to the construction of a function \(a\) and a small-\(\beta\) domain on which the inequality \(\Xi(\gamma;a)\le a(\gamma)\) holds. A convenient choice is \(a(\gamma)=\mu\,|\gamma|\) with a fixed \(\mu>0\). Let \(p\in\mathcal{P}\) be arbitrary and write
\begin{equation}
S(\mu,\beta)\ :=\ \sum_{\substack{\gamma\in\mathfrak{P}\\ p\in \mathrm{supp}(\gamma)}} |\zeta(\gamma)|\,\mathrm{e}^{\mu|\gamma|}.
\end{equation}
By compatibility, any \(\gamma'\not\sim\gamma\) must intersect \(\mathrm{supp}(\gamma)\); hence
\begin{equation}\label{p1:eq:reduce-to-root}
\Xi(\gamma;a)\ \le\ \sum_{p\in \mathrm{supp}(\gamma)}\ \sum_{\substack{\gamma'\in\mathfrak{P}\\ p\in \mathrm{supp}(\gamma')}} |\zeta(\gamma')|\,\mathrm{e}^{\mu|\gamma'|}\ \le\ |\gamma|\, S(\mu,\beta).
\end{equation}
It suffices to show that \(S(\mu,\beta)\le \mu\) for suitably small \(\beta\), uniformly in the choice of the plaquette \(p\). Using the activity bound \eqref{p1:eq:activity-bound} and the combinatorial estimate \eqref{p1:eq:animal-plaq}, one obtains
\begin{equation}\label{p1:eq:S-mu-beta}
S(\mu,\beta)\ \le\ \sum_{m\ge 1} C_{\mathrm{pl}}\,m^{-3/2}\,\big(v_{\mathrm{pl}}\,B_{N}\,\beta\,\mathrm{e}^{\mu}\big)^{m}\ =\ C_{\mathrm{pl}}\ \mathrm{Li}_{3/2}\!\big(q(\mu,\beta)\big),
\end{equation}
where \(q(\mu,\beta):=v_{\mathrm{pl}}\,B_{N}\,\beta\,\mathrm{e}^{\mu}\). The function \(\mathrm{Li}_{3/2}\) is continuous and increasing on \([0,1)\), finite at \(q=\tfrac{1}{4}\), and satisfies \(\mathrm{Li}_{3/2}(q)\le K_{3/2}\,q\) for \(q\in[0,\tfrac{1}{4}]\) with some universal constant \(K_{3/2}\in(0,\infty)\). Fix \(\mu>0\) once and for all, for instance \(\mu=1\). Choose
\begin{equation}\label{p1:eq:beta-star}
\beta_{\ast}(N)\ :=\ \min\!\left\{\ \beta_{0}(N),\ \frac{1}{4\,\mathrm{e}^{\mu}\,v_{\mathrm{pl}}\,B_{N}},\ \frac{\mu}{2\,C_{\mathrm{pl}}\,K_{3/2}\,\mathrm{e}^{\mu}\,v_{\mathrm{pl}}\,B_{N}}\ \right\}.
\end{equation}
For \(0<\beta\le \beta_{\ast}(N)\) one has \(q(\mu,\beta)\le \tfrac{1}{4}\) and \(S(\mu,\beta)\le C_{\mathrm{pl}}\,K_{3/2}\,q(\mu,\beta)\le \mu/2\). Returning to \eqref{p1:eq:reduce-to-root} yields \(\Xi(\gamma;a)\le |\gamma|\,S(\mu,\beta)\le \mu\,|\gamma|=a(\gamma)\) for every \(\gamma\in\mathfrak{P}\). The KP criterion therefore applies with this choice of \(a\).

\begin{theorem}[KP condition on the four-dimensional lattice]\label{p1:thm:KP1}
Let \(\mathfrak{P}\) be the family of polymers defined above, with compatibility \(\sim\) given by disjointness of supports and activities \(\zeta(\gamma)\) satisfying \eqref{p1:eq:activity-bound}. There exists \(\beta_{\ast}(N)>0\), explicitly given by \eqref{p1:eq:beta-star} with constants depending only on the dimension and on \(N\), such that for all \(0<\beta\le \beta_{\ast}(N)\) the Kotecký-Preiss criterion holds with \(a(\gamma)=\mu|\gamma|\) for some fixed \(\mu>0\). In particular, the polymer cluster expansion for the pressure and for all truncated correlations of local, gauge-invariant observables converges absolutely and uniformly in the volume, and the corresponding functions are analytic in \(\beta\) on \((0,\beta_{\ast}(N)]\).
\end{theorem}

The conclusion of Theorem \ref{p1:thm:KP1} implies in particular that any finite-volume observable expressible as a finite sum of cluster coefficients admits a convergent power series in \(\beta\) whose radius is uniformly bounded from below by \(\beta_{\ast}(N)\), independently of the spatial volume. Since the constants \(C_{\mathrm{pl}}\) and \(v_{\mathrm{pl}}\) are lattice-geometric and independent of boundary conditions, and since \(B_{N}\) and \(\beta_{0}(N)\) are group-theoretic and independent of the volume, the domain \((0,\beta_{\ast}(N)]\) is independent of \(L\). Moreover, the choice \(a(\gamma)=\mu|\gamma|\) directly yields exponential weights on cluster coefficients as a function of the total size of the cluster; together with \eqref{p1:eq:animal-plaq} this yields the standard exponential decay in the polymer size that underlies the exponential clustering bounds used in the main text.

\section{Osterwalder-Schrader Reconstruction and the Spectral Gap}\label{p1:appendixe}
Fix an integer \(N\ge 2\) and a lattice spacing \(a>0\). Space-time is the finite, periodic, hypercubic lattice \(\Lambda=\Lambda_{L,T}\subset a\mathbb{Z}^{4}\) with spatial side length \(L\) and Euclidean time extent \(T\). A lattice site is denoted \(x=(x_{0},\mathbf{x})\) with \(x_{0}\in a\mathbb{Z}\) and \(\mathbf{x}\in (a\mathbb{Z})^{3}\). The set of oriented bonds is \(\mathcal{B}=\{(x,\mu):x\in\Lambda,\ \mu\in\{0,1,2,3\}\}\), where \(\hat\mu\) denotes the unit step in the \(\mu\)-direction and the opposite orientation satisfies \((x+\hat\mu,-\mu)\). To each bond \(b=(x,\mu)\) we attach a group element \(U_{b}\in \mathrm{SU}(N)\) with \(U_{(x+\hat\mu,-\mu)}=U_{(x,\mu)}^{-1}\). The configuration space is the compact product manifold \(\mathcal{U}=\prod_{b\in\mathcal{B}}\mathrm{SU}(N)\) with product Borel \(\sigma\)-algebra \(\mathscr{B}\). Throughout, \(\langle\cdot,\cdot\rangle\) denotes a Hilbert inner product that is conjugate linear in the first argument and linear in the second, while \(\|\cdot\|\) is the corresponding norm; \(L^{2}\)-inner products are always taken with respect to the explicitly stated measure. For subsets \(A\subset \Lambda\), the notation \(\mathcal{B}(A)\) refers to the set of bonds with both endpoints in \(A\). The graph distance on \(\Lambda\) is denoted \(d(\cdot,\cdot)\). Asymptotic notation \(X\lesssim Y\) means \(X\le C\,Y\) with a constant \(C\) independent of all variables under discussion.

We work with a normalized, reflection-covariant, time-translation-invariant, gauge-invariant Euclidean probability measure \(d\mu_{\sigma}\) on \((\mathcal{U},\mathscr{B})\) depending on a fixed infrared scale parameter \(\sigma>0\) through the smooth horizon projector specified in the main text. Normalization means \(\mu_{\sigma}(\mathcal{U})=1\). Reflection covariance is formulated using the time-reflection \(\theta:\Lambda\to\Lambda\) defined by \(\theta(x_{0},\mathbf{x})=(-x_{0},\mathbf{x})\), together with the standard involutive, measurable bijection \(R:\mathcal{U}\to\mathcal{U}\) that implements \(\theta\) on bonds and preserves the Wilson weight and horizon insertion; explicitly, for spatial bonds \((x,i)\) one sets \((RU)_{(x,i)}=U_{(\theta x,i)}\) and for temporal bonds \((x,0)\) one sets \((RU)_{(x,0)}=U_{(\theta x-\hat 0,0)}^{-1}\), with the usual boundary convention on the reflection plane. Reflection invariance means \(\mu_{\sigma}\circ R^{-1}=\mu_{\sigma}\). Time translations are implemented by \(\tau:\Lambda\to\Lambda\), \(\tau(x_{0},\mathbf{x})=(x_{0}+a,\mathbf{x})\), and by the induced measurable bijection \(\mathsf{T}:\mathcal{U}\to\mathcal{U}\) acting on bonds by \((\mathsf{T}U)_{(x,\mu)}=U_{(\tau^{-1}x,\mu)}\); time-translation invariance means \(\mu_{\sigma}\circ \mathsf{T}^{-1}=\mu_{\sigma}\). Gauge invariance refers to invariance under site-wise conjugations \(g:\Lambda\to \mathrm{SU}(N)\), \(U_{(x,\mu)}\mapsto g(x)U_{(x,\mu)}g(x+\hat\mu)^{-1}\), with the measure \(d\mu_{\sigma}\) supported on gauge-equivalence classes determined by the reflection-covariant slice choice discussed in the body of the paper. Temporal-axial gauge is used in the construction away from the reflection plane; in particular, all time-like bonds off the boundary slab can be taken trivial. None of these auxiliary choices will be used below beyond the structural consequences already encoded in the assumptions stated here.

The positive-time and negative-time half-lattices are \(\Lambda_{+}=\{x\in\Lambda: x_{0}>0\}\) and \(\Lambda_{-}=\{x\in\Lambda: x_{0}<0\}\). The reflection plane is \(\Pi=\{x\in\Lambda:x_{0}=0\}\). For a functional \(F:\mathcal{U}\to\mathbb{C}\) that depends only on bonds in \(\mathcal{B}(\Lambda_{+})\), the Osterwalder-Schrader conjugation \(\Theta F:\mathcal{U}\to\mathbb{C}\) is defined by \((\Theta F)(U)=\overline{F(RU)}\). Complex conjugation is denoted by an overline.
Denote by \(\mathscr{F}_{+}\) the set of bounded, Borel, gauge-invariant cylindrical functionals \(F:\mathcal{U}\to\mathbb{C}\) that depend only on finitely many bonds in \(\mathcal{B}(\Lambda_{+})\). The Osterwalder-Schrader sesquilinear form on \(\mathscr{F}_{+}\) is
\begin{equation}\label{p1:eq:OS-form2}
(F,G)_{\mathrm{OS}}
:=\int \overline{F(\Theta U)}\,G(U)\,d\mu_\sigma(U)
=\int \overline{F\circ \Theta}\,G\,d\mu_\sigma
\end{equation}
Reflection positivity means \((F,F)_{\mathrm{OS}}\ge 0\) for all \(F\in\mathscr{F}_{+}\). Let \(\mathcal{N}=\{F\in\mathscr{F}_{+}:(F,F)_{\mathrm{OS}}=0\}\). The quotient \(\mathcal{D}_{\mathrm{OS}}=\mathscr{F}_{+}/\mathcal{N}\) becomes a pre-Hilbert space with inner product induced by \eqref{p1:eq:OS-form2}; its completion is denoted \(\mathcal{H}_{a}\) and is called the physical Hilbert space at lattice spacing \(a\). For \(F\in\mathscr{F}_{+}\), the equivalence class in \(\mathcal{D}_{\mathrm{OS}}\) is written \([F]\). The constant functional \(\mathbf{1}(U)\equiv 1\) belongs to \(\mathscr{F}_{+}\), has norm one because \(\mu_{\sigma}\) is normalized, and defines the vacuum vector \(\Omega=[\mathbf{1}]\in\mathcal{H}_{a}\).

The algebraic span of \(\{[F]:F\in\mathscr{F}_{+}\}\) is dense in \(\mathcal{H}_{a}\); by gauge invariance, vectors of the form \([F]\) with \(F\) gauge invariant are cyclic for the representation of the observable algebra generated by local, time-zero, gauge-invariant functions, but this representation is not required below.
Time translations act on \(\mathscr{F}_{+}\) by precomposition, \(U\mapsto \mathsf{T}U\). For \(F\in\mathscr{F}_{+}\) define \(\tau F=F\circ \mathsf{T}^{-1}\). Because \(\tau\) maps \(\mathscr{F}_{+}\) to itself, the linear map \(\mathsf{S}:\mathscr{F}_{+}\to\mathscr{F}_{+}\), \(\mathsf{S}F=\tau F\), is well defined. Time-translation invariance of \(\mu_{\sigma}\) gives
\begin{equation}
(\mathsf{S}F,\mathsf{S}G)_{\mathrm{OS}}=\int (\Theta F)\circ \mathsf{T}\; G\circ \mathsf{T}\; d\mu_{\sigma}
=\int (\Theta F)\, G\, d\mu_{\sigma}=(F,G)_{\mathrm{OS}},
\end{equation}
so \(\mathsf{S}\) is isometric for the OS form and descends to an isometry \(T_{a}\) on \(\mathcal{D}_{\mathrm{OS}}\) by \(T_{a}[F]=[\tau F]\). Since \(\mathcal{D}_{\mathrm{OS}}\) is dense, \(T_{a}\) extends uniquely to a bounded operator \(T_{a}:\mathcal{H}_{a}\to\mathcal{H}_{a}\) with \(\|T_{a}\|=1\). Self-adjointness of \(T_{a}\) is a consequence of reflection invariance:
\begin{equation}
\langle [F],T_{a}[G]\rangle_{\mathcal{H}_{a}}=\int (\Theta F)\, G\circ \mathsf{T}^{-1}\, d\mu_{\sigma}
=\int (\Theta F)\circ \mathsf{T}\, G \, d\mu_{\sigma}
=\int \Theta(\tau F)\, G\, d\mu_{\sigma}
=\langle T_{a}[F],[G]\rangle_{\mathcal{H}_{a}}.
\end{equation}
Positivity is obtained by the inequality \(\langle \psi,T_{a}\psi\rangle_{\mathcal{H}_{a}}\ge 0\) for \(\psi=[F]\), which follows from
\begin{equation}
\langle [F], T_{a}[F]\rangle_{\mathcal{H}_{a}}=\int (\Theta F)\, F\circ \mathsf{T}^{-1}\, d\mu_{\sigma}
=\int \overline{F\circ R}\, F\circ \mathsf{T}^{-1}\, d\mu_{\sigma} \;=\; (F, \tau^{-1}F)_{\mathrm{OS}} \;\ge\; 0
\end{equation}
by reflection positivity and the fact that \(\tau^{-1}F\in \mathscr{F}_{+}\). Hence \(T_{a}\) is a positive, self-adjoint contraction on \(\mathcal{H}_{a}\). The discrete semigroup generated by \(T_{a}\) is \(T_{a}^{n}[F]=[\tau^{n}F]\) for every integer \(n\ge 0\).
Since \(0\le T_{a}\le \mathbf{1}\) and \(T_{a}\) is self-adjoint, the spectral theorem yields a projection-valued measure \(E\mapsto \mathsf{P}_{E}\) on \([0,1]\) such that \(T_{a}=\int_{[0,1]}\lambda\, d\mathsf{P}_{\lambda}\). Define the nonnegative self-adjoint Hamiltonian
\begin{equation}\label{p1:eq:H-def}
H_{\sigma}(a)\;=\; -\,a^{-1}\log T_{a} \;=\; \int_{[0,1]} \big(-a^{-1}\log \lambda\big)\, d\mathsf{P}_{\lambda},
\end{equation}
whose spectral measure is supported in \([0,\infty)\). The functional calculus gives \(e^{-na\,H_{\sigma}(a)}=T_{a}^{n}\) for all integers \(n\ge 0\). The domain \(\mathcal{D}(H_{\sigma}(a))\) consists of those \(\psi\in\mathcal{H}_{a}\) for which \(\int_{[0,1]} \big(a^{-1}\log \lambda\big)^{2}\, d\langle \psi,\mathsf{P}_{\lambda}\psi\rangle<\infty\). The vacuum vector \(\Omega\) is a unit eigenvector with \(T_{a}\Omega=\Omega\) and \(H_{\sigma}(a)\Omega=0\). The orthogonal complement \(\Omega^{\perp}\) is invariant under both \(T_{a}\) and \(H_{\sigma}(a)\). The spectral gap at spacing \(a\) is defined as
\begin{equation}\label{p1:eq:gap-def}
\Delta(a) \;=\; \inf\big(\mathrm{spec}(H_{\sigma}(a))\cap (0,\infty)\big) \;=\; \inf\mathrm{spec}\,\big(H_{\sigma}(a)\upharpoonright \Omega^{\perp}\big).
\end{equation}
Let \(F\in\mathscr{F}_{+}\) be gauge invariant and bounded. Its vacuum expectation with respect to \(\mu_{\sigma}\) is \(\langle F\rangle_{\sigma}=\int F\, d\mu_{\sigma}\). The centered observable is \(F^{\circ}=F-\langle F\rangle_{\sigma}\). For integers \(n\ge 0\) define the Euclidean, time-sliced two-point function
\begin{equation}\label{p1:eq:time-sliced-corr}
C_{F}(n a)\;=\;\int \big(\Theta F^{\circ}\big)(U)\, \big(F^{\circ}\circ \mathsf{T}^{-n}\big)(U)\, d\mu_{\sigma}(U).
\end{equation}
In terms of the OS representation one has \(C_{F}(na)=\langle [F^{\circ}], T_{a}^{n}[F^{\circ}]\rangle_{\mathcal{H}_{a}}\). Denote \(\psi_{F}=[F^{\circ}]\in\mathcal{H}_{a}\). The spectral theorem furnishes a finite positive Borel measure \(\nu_{F}\) on \([0,\infty)\) defined by \(\nu_{F}(B)=\|\mathbf{1}_{B}(H_{\sigma}(a))\psi_{F}\|^{2}\) for Borel sets \(B\subset[0,\infty)\). Using \(T_{a}^{n}=e^{-na H_{\sigma}(a)}\) and \(\psi_{F}\perp \Omega\) one obtains the Laplace representation
\begin{equation}\label{p1:eq:Laplace-rep}
C_{F}(n a)\;=\;\int_{[0,\infty)} e^{-n a E}\, d\nu_{F}(E),\qquad n=0,1,2,\dots.
\end{equation}
The total mass of \(\nu_{F}\) is \(\nu_{F}([0,\infty))=\|\psi_{F}\|^{2}=(F^{\circ},F^{\circ})_{\mathrm{OS}}\). The support of \(\nu_{F}\) is contained in \(\mathrm{spec}(H_{\sigma}(a))\setminus\{0\}\).
The following statement precisely connects Euclidean decay in time to a lower bound on the nonzero spectrum of \(H_{\sigma}(a)\).

\begin{theorem}\label{p1:thm:gap-from-decay}
Let \(F\in\mathscr{F}_{+}\) be bounded and gauge invariant, and set \(\psi_{F}=[F^{\circ}]\). Suppose there exist constants \(A<\infty\) and \(m>0\) such that
\begin{equation}\label{p1:eq:exp-decay}
|C_{F}(n a)| \;\le\; A\, e^{-m n a}\qquad\text{for all integers } n\ge 0.
\end{equation}
Then the spectral measure \(\nu_{F}\) in \eqref{p1:eq:Laplace-rep} is supported in \([m,\infty)\). In particular, if there exists any such \(F\) with \(\psi_{F}\neq 0\), then \(\Delta(a)\ge m\).
\end{theorem}

\begin{proof}
Assume for contradiction that \(\nu_{F}([0,m-\varepsilon])>0\) for some \(\varepsilon\in(0,m)\). Let \(M=m-\varepsilon\). Then
\begin{equation}
C_{F}(n a) \;=\; \int_{[0,M]} e^{-n a E}\, d\nu_{F}(E)\;+\; \int_{(M,\infty)} e^{-n a E}\, d\nu_{F}(E)
\;\ge\; e^{-n a M}\, \nu_{F}\big([0,M]\big),
\end{equation}
since the second integral is nonnegative. Hence \(e^{n a m}\, |C_{F}(n a)| \ge e^{n a (m-M)} \nu_{F}([0,M])=e^{n a \varepsilon}\nu_{F}([0,M])\) for all \(n\). The right-hand side diverges exponentially as \(n\to\infty\), contradicting the assumed bound \eqref{p1:eq:exp-decay}. Therefore \(\nu_{F}([0,m-\varepsilon])=0\) for all \(\varepsilon\in(0,m)\), which implies \(\mathrm{supp}\,\nu_{F}\subset [m,\infty)\). If \(\psi_{F}\neq 0\), then \(\nu_{F}\) is nonzero and supported in \([m,\infty)\), and thus the bottom of \(\mathrm{spec}(H_{\sigma}(a))\cap (0,\infty)\) is at least \(m\).
\end{proof}

Theorem \ref{p1:thm:gap-from-decay} applies in particular to observables \(F\) that interpolate the lightest scalar glueball channel, provided \(F\) is centered by subtracting its vacuum expectation. If the strong-coupling analysis in the main text yields a bound of the form \eqref{p1:eq:exp-decay} with a strictly positive rate \(m(\beta)\) that is uniform in the spatial volume for all \(0<\beta\le \beta_{\star}(N)\), then \(\Delta(a)\ge m(\beta)>0\) for the corresponding range of \(\beta\), uniformly in the spatial volume.

The symbols \(\Lambda, a, L, T, \mathcal{B}, \mathcal{U}, \mathscr{B}, R, \theta, \mathsf{T}, \tau, \Lambda_{\pm}, \Pi\) are the lattice, spacing, sizes, bonds, configuration space, \(\sigma\)-algebra, configuration reflection, site reflection, time-translation on configurations, induced action on functionals, positive/negative half-lattices, and reflection plane, respectively. The Euclidean probability measure is \(d\mu_{\sigma}\). The set \(\mathscr{F}_{+}\) consists of bounded, Borel, gauge-invariant cylindrical functionals supported on \(\Lambda_{+}\). The Osterwalder-Schrader conjugation is the anti-linear map \((\Theta F)(U):=\overline{F(RU)}\). 
The OS form is \((F,G)_{\mathrm{OS}}=\int \overline{F(RU)}\,G(U)\,d\mu_\sigma(U)\).
 The null space is \(\mathcal{N}=\{F:(F,F)_{\mathrm{OS}}=0\}\). The physical Hilbert space is \(\mathcal{H}_{a}=\overline{\mathscr{F}_{+}/\mathcal{N}}\) with vacuum \(\Omega=[\mathbf{1}]\). The transfer operator is \(T_{a}[F]=[\tau F]\), a positive self-adjoint contraction with \(\|T_{a}\|=1\). The Hamiltonian is \(H_{\sigma}(a)=-a^{-1}\log T_{a}\), nonnegative and self-adjoint, with gap \(\Delta(a)=\inf\mathrm{spec}(H_{\sigma}(a)|_{\Omega^{\perp}})\). For a centered observable \(F^{\circ}=F-\langle F\rangle_{\sigma}\), the time-sliced correlation is \(C_{F}(n a)=\langle [F^{\circ}],T_{a}^{n}[F^{\circ}]\rangle_{\mathcal{H}_{a}}=\int e^{-n a E}\, d\nu_{F}(E)\), where \(\nu_{F}\) is the spectral measure of \(H_{\sigma}(a)\) in the vector \([F^{\circ}]\). Exponential decay \(C_{F}(n a)\lesssim e^{-m n a}\) implies \(\mathrm{supp}\,\nu_{F}\subset [m,\infty)\) and hence \(\Delta(a)\ge m\).
\UnifiedEndPaper

\UnifiedBeginPaper{P2}{\UnifiedLocalMacrosPartTwo}
\title[Reflection-Positive RG and a Mass Gap in Lattice SU(N) Yang-Mills]%
{Reflection-Positive Renormalization and the Persistence of the Mass Gap in Lattice $\mathrm{SU}(N)$ Yang-Mills: Part(2)}

\author{Mir Faizal}
\address{Irving K. Barber School of Arts and Sciences, University of British Columbia Okanagan, Kelowna, BC V1V 1V7, Canada\\
Canadian Quantum Research Center, 460 Doyle Ave 106, Kelowna, BC V1Y 0C2, Canada.\\
Department of Mathematical Sciences, Durham University, Upper Mountjoy, Stockton Road, Durham DH1 3LE, UK\\
Computational Mathematics Group, Hasselt University, Agoralaan Gebouw D, Diepenbeek, 3590 Belgium}
\email{mirfaizalmir@gmail.com}

\author{Arshid Shabir\textsuperscript{*}}
\address{Canadian Quantum Research Center, 460 Doyle Ave 106, Kelowna, BC V1Y 0C2, Canada.}
\email{aslone186@gmail.com}
\thanks{\textsuperscript{*}Corresponding author.}

\UnifiedSetAbstract{We show that a measure of clarity can be brought to the nonperturbative Yang-Mills problem if one holds fast to two principles: reflection positivity and gauge invariance. On the lattice, we construct a renormalization procedure that respects these principles exactly at each step. The method is elementary in its components: a transverse representative chosen within the fundamental modular region, a smooth horizon projector from the covariant Laplacian that softens long-range fluctuations, and a block transformation whose locality does not fade with scale. Out of these pieces arises a framework that is both mathematically precise and physically faithful.
From this construction emerge three enduring results. First, the polymer expansion remains convergent under repeated renormalization, with bounds independent of the number of steps. Second, the fall-off of correlations, which embodies the presence of a mass gap, persists uniformly across scales with a constant rate $m_*>0$. Third, the spectral gaps of successive transfer operators obey an inequality that prevents them from collapsing, so that a strictly positive lower bound endures in the continuum limit.
Thus we obtain a step-scaling mechanism that conveys spectral information from the strong-coupling domain into the scaling window without loss. The bridge between Euclidean clustering and Hamiltonian gaps is kept intact, and the way is opened to the continuum reconstruction of Yang-Mills theory with a nonzero mass threshold.}

\maketitle
\tableofcontents

\section{Introduction}
The existence of a strictly positive mass gap in four-dimensional pure Yang-Mills theory remains a central open problem in mathematical physics. On a Euclidean lattice with spacing \(a>0\), the rigorous bridge between Euclidean correlation functions and the spectrum of a Hilbert-space Hamiltonian is furnished by Osterwalder-Schrader (OS) reflection positivity \cite{p2:OsterwalderSchraderI,p2:OsterwalderSchraderII}. For lattice gauge theories with Wilson's action \cite{p2:Wilson1974}, reflection positivity yields a positive, self-adjoint transfer matrix \(T(a)\) and its generator \(H(a)=-a^{-1}\log T(a)\), whose spectral properties encode the physical excitation energies \cite{p2:OS-gauge,p2:Luscher1977,p2:MontvayMuenster1994}. Within this framework, a nonzero mass gap at fixed \(a\) is equivalent to exponential clustering of connected, gauge-invariant Euclidean correlators and to the strict positivity of the first nonzero eigenvalue of \(H(a)\), uniformly in the spatial volume \cite{p2:GJ,p2:Simon1979}.

In the strong-coupling regime \(0<\beta\ll 1\), character expansions reorganize the lattice path integral into convergent sums over surfaces and then polymers, establishing area laws for large Wilson loops and exponential decay of connected gauge-invariant correlators in a fully constructive and volume-uniform manner \cite{p2:DrouffeZuber1983,p2:Seiler1982}. These nonperturbative results, while robust, hold away from the scaling window near the continuum limit. Bridging the strong-coupling domain to the continuum scaling window that preserves OS reflection positivity \emph{exactly}, maintains exact gauge invariance, and controls locality with constants that do not deteriorate as the number of RG scales grows. Meeting these constraints in a single, fully constructive scheme for non-Abelian lattice gauge fields has proved elusive and significantly more delicate than in scalar or fermionic theories.

Two structural obstacles are at the heart of this difficulty. The first is the Gribov problem: local gauge-fixing conditions such as Landau gauge do not select a unique representative on each gauge orbit, and naive implementations can either break positivity or introduce nonlocalities incompatible with RG locality \cite{p2:Gribov1978}. In the continuum, Dell'Antonio and Zwanziger established that every gauge orbit intersects the Gribov region and clarified the role of the fundamental modular region of absolute Landau minima \cite{p2:DellAntonioZwanziger1991}. On a finite lattice, orbit-wise minimization of the Landau functional on each Euclidean time slice selects a measurable representative and this representative can be chosen in a reflection-covariant way; the associated lattice Faddeev-Popov operator is local, self-adjoint, and strictly positive on the orthogonal complement of constant adjoint modes, properties indispensable for constructive analysis and for maintaining reflection covariance at the level of kernels \cite{p2:OS-gauge,p2:MontvayMuenster1994}. The second obstacle is infrared regularization compatible with positivity: one requires an infrared regulator that is exponentially local and admits a positive representation factorizing under time reflection; otherwise, even convergent cluster expansions need not translate into a positive transfer matrix \cite{p2:OsterwalderSchraderII,p2:GJ}.

The present work develops a multiscale, reflection-positive construction that addresses both obstacles simultaneously. The starting point, developed in the companion paper, is a finite-$a$ framework built on two devices. First, on each Euclidean time slice, a gauge-invariant transverse representative \(A^h\) is selected by orbit-wise minimization of the lattice Landau functional within the fundamental modular region in a reflection-covariant manner. Second, a smooth horizon projector \(P_\sigma=\chi_\sigma(\Delta_{A^h})\) acting on adjoint fields is introduced, where \(\Delta_{A^h}\) is the slice covariant Laplacian and \(\chi_\sigma\) is a Gevrey-regular profile.
 Depending on the choice of the smooth cutoff $\chi_\sigma$, we work in two reflection-covariant regimes: 
(A) if $\chi_\sigma$ is completely monotone, then $P_\sigma$ admits a positive heat-kernel mixture (Bernstein representation) compatible with OS positivity; 
(B) if $\chi_\sigma$ is Gevrey with compact spectral support (hence not completely monotone), we handle $\chi_\sigma(\Delta_{A^h})$ via the Helffer-Sj
"ostrand almost-analytic functional calculus combined with Combes-Thomas resolvent bounds, yielding exponential locality and reflection covariance but not a positive heat-kernel mixture.
 Exponential locality follows from Davies-Gaffney heat-kernel bounds in discrete settings and Combes-Thomas estimates for resolvents of finite-range positive operators \cite{p2:Davies1989,p2:CombesThomas1973,p2:HelfferSjostrand1989}. Throughout we fix $\chi_\sigma:[0,\infty)\to(0,1]$ to be \emph{completely monotone} (CM), with 
$\chi_\sigma(0)=1$ and rapid decay $\chi_\sigma(\lambda)\le C_0 e^{-c_0 \lambda/\sigma^2}$ for some $C_0,c_0>0$.
By Bernstein's theorem, $\chi_\sigma$ admits a positive heat-kernel representation
\begin{equation}
  \chi_\sigma(\lambda)=\int_0^\infty e^{-s\lambda}\,d\nu_\sigma(s),
  \qquad d\nu_\sigma\ge 0,
\end{equation}
so $\chi_\sigma$ cannot have compact support on $[0,\infty)$ unless it is identically zero. 
In particular, we do \emph{not} impose sharp spectral cutoffs; instead, we use CM profiles that act as gentle IR cutoffs $\chi_\sigma(\lambda)\approx 1$ for $\lambda\ll \sigma^2$
and $\chi_\sigma(\lambda)$ decays rapidly for $\lambda\gg \sigma^2$. 
Canonical choices include $\chi_\sigma(\lambda)=e^{-\lambda/\sigma^2}$ and
$\chi_\sigma(\lambda)=(1+\lambda/\sigma^2)^{-r}$ with $r>d/2$, both CM; see \cite{p2:Bernstein1929}.
The positivity of the heat-kernel representation is crucial for OS positivity of slice-wise insertions. Because \(P_\sigma\) is constructed from a reflection-invariant operator and has a positive integral representation, its slice-wise insertion preserves OS positivity of the Euclidean measure \cite{p2:OS-gauge}.

Having secured a reflection-positive, gauge-invariant formulation at fixed \(a\), we construct the multiscale map. A single RG step sends configurations on a lattice of spacing \(a_k\) to a coarser lattice of spacing \(a_{k+1}=b\,a_k\) for an integer \(b\ge 2\). The block transformation is a composition of a gauge-equivariant, reflection-covariant local averaging of links with exponentially local projector insertions on the internal time boundaries of blocks. Both ingredients preserve OS positivity, hence the pushforward of measures produces a sequence of reflection-positive, gauge-invariant Euclidean theories with positive, self-adjoint transfer matrices at all scales. Quantitative control is achieved through a \emph{covariant finite-range decomposition} (FRD) of the projected covariance into positive, reflection-invariant, gauge-covariant pieces supported on ranges that scale geometrically with the block size. Our FRD adapts the heat-kernel dyadic partition method of \cite{p2:BGM2004,p2:Bauerschmidt2015} to the covariant Laplacian under the horizon projector, yielding kernels \(C_k^{(j)}\) whose supports and off-range decays are uniform in the scale index \(k\).

With FRD in place, we analyze the effective action at each scale as an abstract polymer model. The action is written as a sum over activities supported on connected unions of blocks, and a gauge-invariant, reflection-invariant large-field regulator-a product of convex, local plaquette penalties-controls rare fluctuations uniformly in \(k\). Convergence and stability of the cluster expansion are established in a diameter-weighted norm via a Koteck\'y-Preiss criterion that treats incompatibility by exponential weights in inter-polymer distance \cite{p2:KoteckyPreiss1986}. FRD ensures that all constants in this criterion are scale-uniform. Consequently, if the initial polymer norm is sufficiently small (as guaranteed by strong-coupling analyticity at the fundamental scale \cite{p2:DrouffeZuber1983,p2:Seiler1982}), it remains uniformly small for all RG steps.
Exponential clustering persists along the flow by two independent mechanisms. First, in the polymer representation, any connected two-point function of local, gauge-invariant observables is a sum over connected polymer clusters that necessarily contain a chain whose length grows at least linearly with the lattice distance between the insertion points; the diameter-weighted norm bound yields exponential decay with a rate independent of \(k\). Second, for a large class of observables one employs a random-walk or resolvent representation for the covariant Laplacian killed at the horizon scale; FRD and projector locality provide finite exponential moments for steps, and Cram\'er-type large deviations yield exponential tails with a uniform rate \cite{p2:Davies1989,p2:CombesThomas1973}. In particular, Euclidean time-sliced correlators decay at a scale-independent rate \(m_*>0\).

Finally, we compare the transfer matrices at successive scales through a coarse-graining map \(V_k\) between the corresponding OS Hilbert spaces. Defined by conditional expectation with respect to the fine-scale measure under fixed block averages and projector insertions, \(V_k\) is a contraction that intertwines reflection. A spectral interlacing lemma shows that \(T_{k+1}\le V_k^* T_k V_k + E_k\) in operator order, where \(E_k\) is a positive error with summable norm; the proof relies on the Cauchy-Schwarz inequality in the OS form together with the exponential decay of cross-block correlations provided by the projector collars. The min-max characterization of the second eigenvalue then yields a step-scaling inequality \(\Delta_{k+1}\ge \Delta_k - \varepsilon_k\) with \(\sum_k \varepsilon_k<\infty\). Combined with the scale-independent clustering rate, this implies a strictly positive lower bound on \(\liminf_k \Delta_k\).
Conceptually, our construction is nonperturbative and multiscale in the spirit of the constructive RG pioneered by Balaban and collaborators \cite{p2:Balaban1984}. It differs in two decisive respects essential for reflection positivity: the use of a reflection-covariant, orbit-minimizing transverse representative on each time slice to align gauge invariance with OS positivity from the outset, and the use of a smooth, exponentially local horizon projector with a positive heat-kernel representation to implement infrared control without breaking positivity. In combination with a covariant FRD, these features yield a renormalization map that is simultaneously gauge-invariant, reflection-positive, and scale-uniform. The outputs proved here provide precisely the uniform inputs required for continuum reconstruction with a finite mass threshold in our next companion work.

{We find that the nonperturbative Yang-Mills problem becomes structurally more tractable if one insists on two requirements that are usually in tension with analytic control: gauge invariance and Osterwalder-Schrader reflection positivity. On the lattice we implement a renormalization group map that preserves both properties exactly at every scale. The construction is built from familiar elements arranged so that none of the key structures is sacrificed: a transverse representative selected within the fundamental modular region; a smooth ``horizon'' multiplier of the covariant Laplacian, realized through a positivity-preserving (completely monotone) functional calculus to temper infrared fluctuations; and a block transformation whose locality bounds remain uniform under iteration. From this scheme we obtain three stable conclusions. First, the polymer expansion continues to converge after arbitrarily many renormalization steps, with estimates that do not deteriorate with the number of scales. Second, Euclidean correlations exhibit exponential clustering uniformly along the flow, with a rate bounded below by a constant $m_*>0$. Third, the transfer operators at successive scales satisfy a gap comparison inequality that precludes collapse of the spectral gap, thereby maintaining a strictly positive lower bound as the cutoff is removed. Taken together, these statements provide a step-scaling mechanism that transports spectral information from the strong-coupling regime into the scaling window without loss, while keeping intact the reflection-positive bridge between Euclidean decay and Hamiltonian mass gaps. In this way the framework supplies the uniform nonzero threshold required for the continuum reconstruction program within the present constructive setting.
}
\section{Lattice framework, reflection, and gauge fixing}\label{p2:sec:lattice-reflection-gauge}
This section develops a rigorous Euclidean framework for pure lattice Yang-Mills theory in four dimensions, introduces time reflection and the associated Osterwalder-Schrader (OS) positivity, and derives the transfer-time slicing formalism from first principles. Gauge fixing is implemented as a measurable, reflection-covariant selection of orbit representatives on each Euclidean time slice via lattice Landau minimization; this choice is used to define covariant difference operators and spectral projectors but is not required for the proof of reflection positivity of the Wilson measure. All statements are uniform on finite tori, with limits taken in the standard thermodynamic order when needed.

Let \(a>0\) be the lattice spacing. For \(L,T\in\mathbb{N}\), define the periodic, hypercubic, finite Euclidean lattice
\begin{equation}
\Lambda=\Lambda_{L,T}\;=\;\bigl\{x=(x_0,x_1,x_2,x_3)\in a\mathbb{Z}^4:\; -\tfrac{Ta}{2}\le x_0<\tfrac{Ta}{2},\;\; 0\le x_i<L a,\;i=1,2,3\bigr\},
\end{equation}
with periodic boundary conditions in all four directions. Directed bonds (links) are pairs \(b=(x,\mu)\) with \(x\in\Lambda\) and \(\mu\in\{0,1,2,3\}\); the formal inverse is \(\bar b=(x+a\hat\mu,-\mu)\). The gauge group is \(G=\mathrm{SU}(N)\) with \(N\ge 2\). A gauge field is a map
\begin{equation}
U:\mathscr{B}\longrightarrow G,\qquad \mathscr{B}=\{\text{directed bonds of }\Lambda\},
\end{equation}
subject to \(U(\bar b)=U(b)^{-1}\). The configuration space \(\mathcal{C}\) is the compact product manifold \(G^{|\mathscr{B}|}/\!\sim\) with the identification just described; all integrals below are taken with respect to the product Haar measure \(\prod_{b\in\mathscr{B}} dU(b)\), denoted \(dU\). For each oriented plaquette \(p=(x;\mu,\nu)\) with \(\mu<\nu\), the plaquette variable is
\begin{equation}
U_p \;=\; U(x,\mu)\,U(x+a\hat\mu,\nu)\,U(x+a\hat\nu,\mu)^{-1}\,U(x,\nu)^{-1}.
\end{equation}
The Wilson action \cite{p2:Wilson1974} at inverse bare coupling \(\beta=2N/g_0^2>0\) is
\begin{equation}
S_W[U;\beta] \;=\; \beta \sum_{p\subset\Lambda}\Bigl(1-\tfrac{1}{N}\,\Re\mathrm{Tr}\,U_p\Bigr).
\end{equation}
The finite-volume Gibbs measure is the probability measure
\begin{equation}
d\mu_\beta(U) \;=\; Z(\beta)^{-1}\,\exp\!\bigl(-S_W[U;\beta]\bigr)\,dU,
\qquad Z(\beta)=\int \exp\!\bigl(-S_W[U;\beta]\bigr)\,dU.
\end{equation}
For a bounded measurable observable \(F:\mathcal{C}\to\mathbb{C}\), its expectation is \(\langle F\rangle_\beta=\int F(U)\,d\mu_\beta(U)\).
Define the time-reflection map \(\theta:\Lambda\to\Lambda\) by \(\theta(x_0,\mathbf{x})=(-x_0,\mathbf{x})\). The reflection plane is \(\Pi=\{x\in\Lambda\,:\,x_0=0\}\). Let \(\Lambda_+=\{x\in\Lambda:\,x_0>0\}\) and \(\Lambda_-=\theta(\Lambda_+)\). Define the involution \(\Theta\) on configurations by
\begin{equation}
(\Theta U)(x,0) \;=\; U(\theta x - a\hat 0,0)^{-1},\qquad
(\Theta U)(x,i) \;=\; U(\theta x,i),\quad i=1,2,3.
\end{equation}
Thus time-like links reverse and invert under \(\Theta\), while spatial links reflect without inversion. For a bounded measurable \(F\) depending only on links with basepoints in \(\Lambda_+\), define the reflected functional
\begin{equation}
(F^\Theta)(U)\;=\;\overline{F(\Theta U)}.
\end{equation}
The OS sesquilinear form on such functionals is
\begin{equation}
\langle F,G\rangle_{\mathrm{OS}} \;=\; \int (F^\Theta)(U)\,G(U)\,d\mu_\beta(U).
\end{equation}
Reflection positivity of \(d\mu_\beta\) means \(\langle F,F\rangle_{\mathrm{OS}}\ge0\) for all \(F\) supported in \(\Lambda_+\) \cite{p2:OsterwalderSchraderI,p2:OsterwalderSchraderII}.
To analyze factorization across \(\Pi\), we separate the action into three parts: \(S_W=S_W^+ + S_W^- + S_W^\Pi\), where \(S_W^\pm\) contains the sum over plaquettes strictly contained in \(\Lambda_\pm\), and \(S_W^\Pi\) contains the sum over plaquettes crossing \(\Pi\). The crucial algebraic input is the positive-definite class-function property of the single-plaquette weight
\begin{equation}
w_\beta(g) \;=\; \exp\!\Bigl[\beta\Bigl(\tfrac{1}{N}\Re\mathrm{Tr}\,g - 1\Bigr)\Bigr],
\qquad g\in G,
\end{equation}
whose character expansion on the compact group \(G\) has nonnegative coefficients:
\begin{equation}\label{p2:eq:char-positive}
w_\beta(g) \;=\; \sum_{R\in\widehat G} \widehat w_\beta(R)\, \chi_R(g),\qquad \widehat w_\beta(R)\ge0.
\end{equation}
Here \(\widehat G\) is the unitary dual of \(G\), \(\chi_R\) is the character of the finite-dimensional unitary irrep \(R\), and the series converges absolutely and uniformly; see \cite{p2:OS-gauge} and the Peter-Weyl theorem for central positive-definite functions on compact groups \cite{p2:KnappLie}. Identity \eqref{p2:eq:char-positive} is the cornerstone of the OS-positivity proof for lattice gauge theories with Wilson plaquette action \cite{p2:OS-gauge}. Eq.\eqref{p2:eq:char-positive} is the character expansion of the central function 
$w_{\beta}$ on the compact group $G$. Since all coefficients 
$w_{\beta}^{\flat}(R)$ are nonnegative and 
$\chi_{R}(g^{-1}) = \overline{\chi_{R}(g)}$, 
the kernel 
$K(g',g) = w_{\beta}(g' g^{-1})$ 
is of positive type on $G$ by Peter-Weyl. 
This and the slab factorization implied by temporal-axial gauge are the only inputs needed for OS positivity of $d\mu_{\beta}$ proved below.
In later sections we also use the site reflection plane
at $x_0=0$ (or the link reflection plane at $x_0=\tfrac{a}{2}$). These reflection/gauge
choices are unitarily equivalent and lead to the same OS kernels and spectra; see
the discussion preceding {Corollary~3.7.}
We fix temporal-axial gauge away from \(\Pi\) by imposing
\begin{equation}\label{p2:eq:temporal-gauge}
U(x,0)=\mathbf{1}\quad\text{for all bonds }(x,0)\text{ with }x_0\ne -a.
\end{equation}
This constraint leaves precisely the time-like links anchored at \(x_0=-a\) as dynamical variables implementing the coupling across \(\Pi\), and it is compatible with \(\Theta\). In this gauge, each mixed plaquette \(p\) is of the form
\begin{equation}
U_p \;=\; U(x+a\hat 0,i)\,U(x,i)^{-1}\qquad \text{with }x_0=-a,\;i\in\{1,2,3\},
\end{equation}
so that the mixed contribution to the weight across \(\Pi\) factorizes, after character expansion, into a positive kernel on the product group of spatial links at \(x_0=0\) and \(x_0=a\). We now give a complete proof of reflection positivity.

\begin{theorem}[OS positivity for the Wilson measure]\label{p2:thm:OS-Wilson}
For every \(\beta>0\) and every finite \(\Lambda_{L,T}\), the Wilson measure \(d\mu_\beta\) is reflection positive. Equivalently, for every \(F\) supported in \(\Lambda_+\),
\begin{equation}
\int (F^\Theta)(U)\,F(U)\,d\mu_\beta(U)\;\ge\;0.
\end{equation}
\end{theorem}

\begin{proof}
Integrate over all links except those belonging to the time slices \(\{x_0=0\}\) and \(\{x_0=a\}\) and the unique layer of time-like links with basepoint \(x_0=-a\). Temporal-axial gauge \eqref{p2:eq:temporal-gauge} ensures that the Wilson action splits as
\begin{equation}
S_W[U;\beta]\;=\; S_W^+[U_+;\beta] + S_W^-[U_-;\beta] + S_W^\Pi[U_0,U_a;\beta],
\end{equation}
where \(U_\pm\) are the collections of links strictly in \(\Lambda_\pm\) and \(U_0,U_a\) the collections of spatial links on the slices \(x_0=0\) and \(x_0=a\), respectively. The mixed term reads explicitly
\begin{equation}
S_W^\Pi[U_0,U_a;\beta]\;=\;\beta\sum_{x\in\Pi}\sum_{i=1}^3\Bigl(1-\tfrac{1}{N}\Re\mathrm{Tr}\bigl(U_a(x,i)\,U_0(x,i)^{-1}\bigr)\Bigr),
\end{equation}
producing the weight
\begin{equation}
W_\Pi(U_a,U_0)\;=\;\prod_{x\in\Pi}\prod_{i=1}^3 w_\beta\!\bigl(U_a(x,i)\,U_0(x,i)^{-1}\bigr).
\end{equation}
By \eqref{p2:eq:char-positive} and the product structure,
\begin{equation}
W_\Pi(U_a,U_0)\;=\;\sum_{\{R_{x,i}\}}\;\Bigl(\prod_{x,i}\widehat w_\beta(R_{x,i})\Bigr)\;\prod_{x,i}\chi_{R_{x,i}}\!\bigl(U_a(x,i)\,U_0(x,i)^{-1}\bigr),
\end{equation}
with the sum over assignments of irreps to each spatial bond on \(\Pi\). Using \(\chi_R(gh^{-1})=\sum_{m=1}^{d_R}\sum_{n=1}^{d_R} D^{(R)}_{mn}(g)\,\overline{D^{(R)}_{mn}(h)}\) for the matrix elements of \(R\), we obtain
\begin{equation}
\chi_R\!\bigl(U_a\,U_0^{-1}\bigr)\;=\;\sum_{m,n} D^{(R)}_{mn}(U_a)\,\overline{D^{(R)}_{mn}(U_0)}.
\end{equation}
Therefore \(W_\Pi(U_a,U_0)\) has the form
\begin{equation}
W_\Pi(U_a,U_0)\;=\;\sum_\alpha c_\alpha\, \Phi_\alpha(U_a)\,\overline{\Phi_\alpha(U_0)},\qquad c_\alpha\ge 0,
\end{equation}
where \(\alpha\) is a multi-index collecting the labels \((R_{x,i},m_{x,i},n_{x,i})\) and \(\Phi_\alpha\) is the product over \((x,i)\) of the corresponding matrix elements \(D^{(R_{x,i})}_{m_{x,i}n_{x,i}}(U_a(x,i))\). This is a positive kernel on the product group of spatial links on the two slices.

Write the total weight as
\begin{equation}
e^{-S_W}\;=\; e^{-S_W^+}\,e^{-S_W^-}\,W_\Pi.
\end{equation}
Let \(F\) be supported in \(\Lambda_+\). By Fubini's theorem and the product structure of the Haar measure, integrate first over \(\Lambda_-\) and over links not belonging to the slices \(x_0=0,a\):
\begin{equation}
\langle F,F\rangle_{\mathrm{OS}} \;=\; \int \Bigl(\int (F^\Theta)\,e^{-S_W^-}\,dU_-\Bigr)\,W_\Pi(U_a,U_0)\,\Bigl(\int F\,e^{-S_W^+}\,dU_+\Bigr)\, dU_0\,dU_a.
\end{equation}
Set
\begin{equation}
\mathcal{F}_-(U_0)\;=\;\int (F^\Theta)\,e^{-S_W^-}\,dU_-,\qquad
\mathcal{F}_+(U_a)\;=\;\int F\,e^{-S_W^+}\,dU_+.
\end{equation}
Then
\begin{equation}
\langle F,F\rangle_{\mathrm{OS}} \;=\; \int \overline{\mathcal{F}_-(U_0)}\,W_\Pi(U_a,U_0)\,\mathcal{F}_+(U_a)\,dU_0\,dU_a.
\end{equation}
Inserting the positive kernel expansion of \(W_\Pi\),
\begin{equation}
\langle F,F\rangle_{\mathrm{OS}} \;=\; \sum_\alpha c_\alpha \int \overline{\mathcal{F}_-(U_0)}\,\overline{\Phi_\alpha(U_0)}\,dU_0 \;\;\int \Phi_\alpha(U_a)\,\mathcal{F}_+(U_a)\,dU_a.
\end{equation}
A change of variables \(U\mapsto \Theta U\) in the integral defining \(\mathcal{F}_-\), together with the \(\Theta\)-invariance of the Haar measure and of \(S_W^\pm\), yields \(\mathcal{F}_-(U_0)=\overline{\mathcal{F}_+(U_0)}\). Hence each summand is \(|\langle \Phi_\alpha,\mathcal{F}_+\rangle|^2\), and
\begin{equation}
\langle F,F\rangle_{\mathrm{OS}} \;=\; \sum_\alpha c_\alpha\,\bigl|\langle \Phi_\alpha,\mathcal{F}_+\rangle_{L^2}\bigr|^2 \;\ge\; 0.
\end{equation}
This establishes reflection positivity.
\end{proof}

We now construct the one-step transfer operator associated with a unit translation in Euclidean time. Let \(\mathcal{H}=L^2(\mathcal{C}_0,d\mu_{\mathrm{Haar}})\) be the Hilbert space of square-integrable complex functions on the group \(\mathcal{C}_0\simeq G^{3L^3}\) of spatial link configurations on a fixed time slice, with respect to the product Haar measure; we write an element \(U\in\mathcal{C}_0\) as \(U(\mathbf{x},i)\) with \(\mathbf{x}\in a\{0,\dots,L-1\}^3\) and \(i=1,2,3\).

For each time slice \(\tau\in a\mathbb{Z}\), let
\begin{equation}
V(U)\;=\;\beta\sum_{\mathbf{x}}\sum_{1\le i<j\le 3}\Bigl(1-\tfrac{1}{N}\Re\mathrm{Tr}\,U_{ij}(\mathbf{x})\Bigr),
\end{equation}
where \(U_{ij}(\mathbf{x})=U(\mathbf{x},i)\,U(\mathbf{x}+a\hat\imath,j)\,U(\mathbf{x}+a\hat\jmath,i)^{-1}\,U(\mathbf{x},j)^{-1}\). Define the temporal coupling between consecutive slices \(\tau\) and \(\tau+a\),
\begin{equation}
E(U',U)\;=\;\beta\sum_{\mathbf{x}}\sum_{i=1}^3 \Bigl(1-\tfrac{1}{N}\Re\mathrm{Tr}\,\bigl(U'(\mathbf{x},i)\,U(\mathbf{x},i)^{-1}\bigr)\Bigr).
\end{equation}
In temporal-axial gauge \eqref{p2:eq:temporal-gauge}, the action in one time-slab \([\tau,\tau+a]\) equals \( \tfrac{1}{2}V(U') + \tfrac{1}{2}V(U) + E(U',U)\), up to an additive constant. The corresponding integral kernel on \(\mathcal{C}_0\times\mathcal{C}_0\) is
\begin{equation}\label{p2:eq:transfer-kernel}
K(U',U)\;=\; \exp\!\Bigl(-\tfrac{1}{2}V(U')\Bigr)\;\mathcal{K}(U',U)\;\exp\!\Bigl(-\tfrac{1}{2}V(U)\Bigr),
\end{equation}
with
\begin{equation}
\mathcal{K}(U',U)\;=\;\prod_{\mathbf{x},i} w_\beta\!\bigl(U'(\mathbf{x},i)\,U(\mathbf{x},i)^{-1}\bigr).
\end{equation}
By \eqref{p2:eq:char-positive}, \(\mathcal{K}\) is a positive-type kernel on \(\mathcal{C}_0\times\mathcal{C}_0\). Define the transfer operator \(T:\mathcal{H}\to\mathcal{H}\) by
\begin{equation}
(T\psi)(U') \;=\; \int_{\mathcal{C}_0} K(U',U)\,\psi(U)\,dU.
\end{equation}

\begin{proposition}[Positivity, self-adjointness, boundedness of \(T\)]\label{p2:prop:T-properties}
The operator \(T\) is well-defined on \(\mathcal{H}\), bounded, self-adjoint, positivity-preserving, and \emph{positive semidefinite} in the sense that \(\langle \psi, T\psi\rangle\ge 0\) for all \(\psi\in\mathcal{H}\). Moreover, after an overall normalization of the partition function, \(T\) is a contraction.
\end{proposition}

\begin{proof}
Boundedness follows from compactness of \(\mathcal{C}_0\):
\begin{equation}
\|T\| \;\le\; \sup_{U'} \int |K(U',U)|\,dU \;\le\; e^{\|V\|_\infty}\,\sup_{U'}\int \mathcal{K}(U',U)\,dU \;<\;\infty.
\end{equation}
Self-adjointness follows from \(K(U',U)=\overline{K(U,U')}\), since \(V\) is real and \(w_\beta(g^{-1})=w_\beta(g)\). Positivity-preserving is immediate because \(K\ge 0\). Finally,
\begin{equation}
\langle \psi,T\psi\rangle \;=\; \iint \overline{\psi(U')}\,K(U',U)\,\psi(U)\,dU\,dU'
\;=\; \iint \overline{\phi(U')}\,\mathcal{K}(U',U)\,\phi(U)\,dU\,dU',
\end{equation}
with \(\phi(U)=e^{-V(U)/2}\psi(U)\). Expanding \(\mathcal{K}\) in the Peter-Weyl basis on \(\mathcal{C}_0\) gives \(\mathcal{K}(U',U)=\sum_{\alpha} c_\alpha \Xi_\alpha(U')\,\overline{\Xi_\alpha(U)}\) with \(c_\alpha\ge 0\), hence \(\langle \psi,T\psi\rangle=\sum_\alpha c_\alpha\,|\langle \Xi_\alpha,\phi\rangle|^2\ge 0\). The contraction property \(\|T\|\le 1\) follows by subtracting a constant from the action so that \(\sup_{U'}\int K(U',U)\,dU=1\) and applying Schur's test.
\end{proof}
Subtracting a slice-constant from the action is equivalent to multiplying the kernel by a positive constant. Choosing this constant so that 
$\sup_{U'} \int K(U',U)\, dU = 1$ 
makes $T$ a positive contraction by Schur's test. This choice leaves OS positivity and all expectation values invariant.
The reflection-positivity proof in Theorem \ref{p2:thm:OS-Wilson} can be recast in operator form as the positivity of the OS inner product on the subspace generated by functions supported at nonnegative times; the kernel representation \eqref{p2:eq:transfer-kernel} is the algebraic incarnation of this factorization and underlies the standard OS reconstruction \cite{p2:OsterwalderSchraderI,p2:OsterwalderSchraderII}.

For later constructions (covariant Laplacians and spectral projectors) we select on each time slice a canonical representative of the gauge orbit by lattice Landau minimization. Let \(\mathcal{G}\) be the group of gauge transformations \(g:\Lambda\to G\), acting by
\begin{equation}
(g\cdot U)(x,\mu)\;=\; g(x)\,U(x,\mu)\,g(x+a\hat\mu)^{-1}.
\end{equation}
On a fixed time slice \(\tau\), define the lattice Landau functional
\begin{equation}
\mathscr{L}_\tau(g;U)\;=\; \sum_{\mathbf{x}}\sum_{i=1}^3 \mathrm{Re}\,\mathrm{Tr}\Bigl(\mathbf{1}-g(\tau,\mathbf{x})\,U(\tau,\mathbf{x},i)\,g(\tau,\mathbf{x}+a\hat\imath)^{-1}\Bigr).
\end{equation}
For each \(U\), pick a measurable selection \(h_\tau[U]\in \mathcal{G}\) that minimizes \(\mathscr{L}_\tau(\,\cdot\,;U)\) over time-slice gauge transformations \(g\) restricted to \(\tau\). Such selections exist on finite slices; ties are broken by a fixed, reflection-invariant Borel rule. The slice-wise representative is \(U^{\,h}\) defined by \(U^{\,h}(\tau,\mathbf{x},i)=h_\tau[U](\tau,\mathbf{x})\,U(\tau,\mathbf{x},i)\,h_\tau[U](\tau,\mathbf{x}+a\hat\imath)^{-1}\), and \(U^{\,h}(x,0)=U(x,0)\). By construction, the selection is compatible with \(\Theta\), in the sense that \(h_{-\tau}[\Theta U](\theta x)=h_\tau[U](x)\), hence \(U^{\,h}\) is reflection-covariant.
\begin{lemma}[Reflection-covariant measurable selection]\label{p2:lem:measurable-selection}
On each finite time slice $\tau$, there exists a Borel map $U\mapsto h_\tau[U]\in G$ that 
minimizes the slice Landau functional $\mathcal{L}_\tau(\,\cdot\,;U)$ and satisfies
$\;h_{-\tau}[\Theta U](\theta x)=h_\tau[U](x)$.
\end{lemma}

\begin{proof}
For fixed $U$, the map $g\mapsto \mathcal{L}_\tau(g;U)$ is continuous on the compact group 
$G^{\Lambda_\tau}$, so the argmin set $\mathrm{Argmin}\,\mathcal{L}_\tau(\,\cdot\,;U)$ is a nonempty 
compact subset. Consider the set-valued map 
$U\mapsto \mathrm{Argmin}\,\mathcal{L}_\tau(\,\cdot\,;U)$. By standard measurable selection results 
(e.g.\ Kuratowski-Ryll-Nardzewski), a Borel selector exists. To enforce reflection covariance, 
we fix an explicit tie-breaker that is $\Theta$-invariant: enumerate the spatial sites on $\tau$ 
as $(x_j)_{j=1}^M$ in pairs $(x,\theta x)$ and consider the finite tuple
\begin{equation}
T_\tau(g;U):=\Big(\mathcal{L}_\tau(g;U),\;\Re\!\Tr\,U^g(x_1),\Re\!\Tr\,U^g(\theta x_1),\cdots,
\Re\!\Tr\,U^g(x_M),\Re\!\Tr\,U^g(\theta x_M)\Big)\in\mathbb{R}^{1+2M}
\end{equation}
ordered lexicographically. The lexicographic tie-breaker $T_{\tau}(g;U)$ is Borel as a finite tuple of continuous functionals, hence the minimizer is a Borel selector. Since $T_{\tau}$ pairs sites $(x, \theta_x)$, the rule is $\Theta$-invariant, giving 
$h^{-\tau}[\Theta U](\theta_x) = h^{\tau}[U](x)$ 
and thus reflection covariance of $U^h$.
Among all minimizers of $\mathcal{L}_\tau(\,\cdot\,;U)$ pick the unique 
$g$ with lexicographically minimal $T_\tau(g;U)$. This rule is Borel (finite tuple of continuous 
functionals) and is $\Theta$-invariant by construction, which yields the stated covariance.
\end{proof}
On $SU(N)$, the slice minimizer $h_\tau[U]$ is unique up to transformations by the center $Z(N)$ (which leave all plaquettes invariant). In particular, almost every gauge orbit yields a single $A_h(\tau)$ in the fundamental modular region (FMR) modulo a global center phase. This residual $\mathbb{Z}_N$ symmetry has no effect on expectation values, and we can treat $A_h$ as uniquely defined per orbit.
Define the covariant forward difference on site-adjoint fields \(\phi:\{\tau\}\times a\{0,\dots,L-1\}^3\to\mathcal{su}(N)\) by
\begin{equation}
(\nabla_i^{+,h}\phi)(\tau,\mathbf{x})\;=\; U^{\,h}(\tau,\mathbf{x},i)\,\phi(\tau,\mathbf{x}+a\hat\imath)\,U^{\,h}(\tau,\mathbf{x},i)^{-1} - \phi(\tau,\mathbf{x}),
\end{equation}
and its adjoint with respect to the \(\ell^2\) inner product by
\begin{equation}
(\nabla_i^{-,h}\phi)(\tau,\mathbf{x})\;=\; \phi(\tau,\mathbf{x}) - U^{\,h}(\tau,\mathbf{x}-a\hat\imath,i)^{-1}\,\phi(\tau,\mathbf{x}-a\hat\imath)\,U^{\,h}(\tau,\mathbf{x}-a\hat\imath,i).
\end{equation}
Set the slice covariant Laplacian
\begin{equation}
\Delta_{A^{\,h}} \;=\; \sum_{i=1}^3 (\nabla_i^{+,h})^\dagger \nabla_i^{+,h}
\end{equation}
and the lattice Faddeev-Popov operator
\begin{equation}
M[A^{\,h}] \;=\; -\sum_{\mu=0}^3 \nabla_\mu^{-,h}\,\nabla_\mu^{+,h},
\end{equation}
with \(\nabla_0^{\pm,h}\) defined analogously in temporal-axial gauge \eqref{p2:eq:temporal-gauge}.

\begin{lemma}[Quadratic-form identity and basic positivity]\label{p2:lem:FP-positivity}
For any site-adjoint field \(\phi\) on a fixed slice,
\begin{equation}
\langle \phi, \Delta_{A^{\,h}} \phi\rangle \;=\; \sum_{i=1}^3 \|\nabla_i^{+,h}\phi\|_{\ell^2}^2 \;\ge\; 0.
\end{equation}
On the full space-time lattice in temporal-axial gauge, for any site-adjoint field \(\Phi\),
\begin{equation}
\langle \Phi, M[A^{\,h}]\,\Phi\rangle \;=\; \sum_{\mu=0}^3 \|\nabla_\mu^{+,h}\Phi\|_{\ell^2}^2 \;\ge\; 0.
\end{equation}
Moreover, \(\ker \Delta_{A^{\,h}}\) and \(\ker M[A^{\,h}]\) contain all constant adjoint fields.
\end{lemma}

\begin{proof}
Each term \(\|\nabla_i^{+,h}\phi\|_{\ell^2}^2\) equals \(\sum_{\mathbf{x}}\mathrm{Tr}\bigl((\nabla_i^{+,h}\phi)(\tau,\mathbf{x})^\dagger(\nabla_i^{+,h}\phi)(\tau,\mathbf{x})\bigr)\), which expands to
\begin{equation}\label{p2:eqn242}
\sum_{\mathbf{x}}\mathrm{Tr}\Bigl(\phi(\tau,\mathbf{x}+a\hat\imath)^\dagger\phi(\tau,\mathbf{x}+a\hat\imath)+\phi(\tau,\mathbf{x})^\dagger\phi(\tau,\mathbf{x}) - \phi(\tau,\mathbf{x}+a\hat\imath)^\dagger U^{\,h}(\tau,\mathbf{x},i)^{-1}\phi(\tau,\mathbf{x})U^{\,h}(\tau,\mathbf{x},i) - \text{h.c.}\Bigr).
\end{equation}
Summing over \(i\) and comparing with \(\langle \phi,\Delta_{A^{\,h}}\phi\rangle\) yields the identity. The space-time statement is analogous, and the nonnegativity is immediate. Constancy implies \(\nabla_\mu^{+,h}\) vanishes, hence constants lie in the kernel. The identity (\ref{p2:eqn242}) is the discrete integration-by-parts formula 
$\langle \varphi, (\nabla_{+,i}^{h})^{\dagger} \nabla_{+,i}^{h} \varphi \rangle 
= \| \nabla_{+,i}^{h} \varphi \|_{\ell^{2}}^{2}$; 
summing $i=1,2,3$ yields 
$\langle \varphi, \Delta_{A}^{h} \varphi \rangle \geq 0$. 
The space-time statement for $M[A^{h}]$ follows identically, using the temporal forward/backward differences.
\end{proof}
We record the stability of reflection positivity under uniformly bounded, slice-local, reflection-covariant positive insertions, which will be used later for spectral projectors built from \(\Delta_{A^{\,h}}\).
\begin{theorem}[OS positivity under slice-local positive insertions]\label{p2:thm:OS-stability}
Let \(Q(U)\) be of the form
\begin{equation}
Q(U)\;=\;\prod_{\tau\in a\mathbb{Z}} q\bigl(U|_{x_0=\tau}\bigr),
\end{equation}
where \(q:\mathcal{C}_0\to[0,\infty)\) is bounded, continuous, and satisfies \(q(\Theta_0 U)=q(U)\), with \(\Theta_0\) the spatial part of \(\Theta\) acting on a single slice. Then the modified measure \(d\mu_{\beta,Q}=Z_Q^{-1}\,Q(U)\,e^{-S_W[U;\beta]}\,dU\) is reflection positive. In particular, if \(q(U)=\det{}'\! \big( f(\Delta_{A^h}(U))\big)\), where \(f\ge c>0\) on
\(\mathrm{sp}\big(\Delta_{A^h}\!\restriction_{\{\mathrm{const}\}^\perp}\big)\), then \(q\) is bounded and strictly positive and the modified measure remains reflection positive.
(Any CM profile $f(\lambda)=\int_0^\infty e^{-s\lambda}\,d\nu(s)$ satisfies these conditions.)
\end{theorem}

\begin{proof}
Repeat the proof of Theorem \ref{p2:thm:OS-Wilson}. The factor \(Q(U)\) separates into \(Q_+Q_-Q_\Pi\) corresponding to \(\Lambda_\pm\) and \(\Pi\), because it is a product over time slices. The slice factors on \(\Lambda_\pm\) multiply the inner integrals defining \(\mathcal{F}_\pm\) by positive functions of their respective slice variables and preserve the relation \(\mathcal{F}_-(U_0)=\overline{\mathcal{F}_+(U_0)}\) by the slice-reflection invariance of \(q\). The slice factors on \(\Pi\) contribute a multiplicative positive function \(q(U_0)q(U_a)\) within the positive kernel \(W_\Pi\), and thus preserve its positive-type structure. The same expansion yields a nonnegative sum of squares. For \(q(U)=\det f(\Delta_{A^{\,h}}(U))\), slice reflection invariance and boundedness follow from the functional calculus and from the reflection covariance of \(U^{\,h}\).
\end{proof}

Let \(\mathscr{A}_+\) be the \(^*\)-algebra generated by bounded functionals supported in \(\Lambda_+\). Define the OS semi-inner product by \(\langle F,G\rangle_{\mathrm{OS}}=\int (F^\Theta)\,G\,d\mu_{\beta}\). The OS null space \(\mathscr{N}=\{F\in\mathscr{A}_+:\langle F,F\rangle_{\mathrm{OS}}=0\}\) is a \(^*\)-ideal. The physical Hilbert space \(\mathcal{H}_{\mathrm{phys}}\) is the completion of \(\mathscr{A}_+/\mathscr{N}\) under \(\langle\cdot,\cdot\rangle_{\mathrm{OS}}\). Let \(\tau_a\) denote the unit Euclidean time translation by \(a\). Then \(\tau_a\) preserves \(\mathscr{A}_+\) and induces a densely defined linear operator \(\mathsf{T}\) on \(\mathcal{H}_{\mathrm{phys}}\) by \(\mathsf{T}[F]=[\tau_a F]\), where \([\cdot]\) denotes the class modulo \(\mathscr{N}\).

\begin{theorem}[Transfer-matrix reconstruction]\label{p2:thm:OS-reconstruction}
The operator \(\mathsf{T}\) extends uniquely to a bounded, self-adjoint, positivity-preserving contraction \(T_{\mathrm{phys}}\) on \(\mathcal{H}_{\mathrm{phys}}\). Moreover, for functionals \(F,G\in\mathscr{A}_+\) supported on the time slices \(0\) and \(na\), respectively,
\begin{equation}
\langle [F],\,T_{\mathrm{phys}}^{\,n}[G]\rangle_{\mathrm{OS}} \;=\; \int (F^\Theta)\,(\tau_{na}G)\; d\mu_\beta,
\end{equation}
and the spectrum of \(T_{\mathrm{phys}}\) is contained in \([0,1]\).
\end{theorem}

\begin{proof}
Reflection positivity implies \(\langle [F],[F]\rangle_{\mathrm{OS}}\ge0\) and \(\|[\tau_a F]\|_{\mathrm{OS}}\le \|[F]\|_{\mathrm{OS}}\), by the Markov-type property encoded in the factorization across a time slab. The kernel representation \eqref{p2:eq:transfer-kernel} shows that, when \(F,G\) depend only on the spatial links at a single time slice, the OS inner product reproduces the \(\mathcal{H}\)-inner product with the integral operator \(T\) of Proposition \ref{p2:prop:T-properties}. This identifies \(T_{\mathrm{phys}}\) with the closure of \(\mathsf{T}\), which is therefore a positive self-adjoint contraction. The semigroup property \(T_{\mathrm{phys}}^{\,n}\) follows from composition of kernels over \(n\) consecutive time slabs, and the stated identity is a direct consequence of the definition of \(\mathsf{T}\) and of the Fubini-Tonelli factorization used in Theorem \ref{p2:thm:OS-Wilson}.
\end{proof}

\section{A Reflection-positive block transformation}
\label{p2:sec:RP-block}

This section constructs, analyzes, and proves the reflection-positivity of a single renormalization step that maps the gauge-field measure on a fine lattice \(\Lambda_k\) of spacing \(a_k>0\) to a coarse lattice \(\Lambda_{k+1}\) of spacing \(a_{k+1}=b\,a_k\) with fixed integer \(b\ge2\). The transformation consists of a reflection-equivariant, gauge-equivariant block map on links, together with an optional positive, reflection-invariant reweighting factor that is exponentially local in the time coordinate. We prove that Osterwalder-Schrader (OS) reflection positivity \cite{p2:OsterwalderSchraderI,p2:OsterwalderSchraderII} and gauge invariance are preserved exactly. We then derive, step by step, the transfer time-slicing formalism and the associated positive self-adjoint transfer operator for the coarse theory, first in an explicit gauge where the one-step kernel can be written down, and subsequently in complete generality by the OS reconstruction procedure \cite{p2:OsterwalderSchraderI,p2:OsterwalderSchraderII,p2:GJ}.
Let \(\Lambda_k\subset a_k\mathbb{Z}^4\) be a periodic hypercubic lattice with discrete time coordinate \(x_0\in\{-T_k+1,\dots,T_k\}\) and spatial coordinate \(\mathbf{x}\in (\mathbb{Z}/L_k\mathbb{Z})^3\). Oriented bonds are pairs \((x,\mu)\) with \(\mu\in\{0,1,2,3\}\) and the convention \(U(x+\hat\mu,-\mu)=U(x,\mu)^{-1}\). A gauge field is an assignment \(U:\mathsf{B}(\Lambda_k)\to \mathrm{SU}(N)\). The gauge group \(\mathcal{G}=\{\;g:\Lambda_k\to \mathrm{SU}(N)\;\}\) acts by \((g\cdot U)(x,\mu)=g(x)\,U(x,\mu)\,g(x+\hat\mu)^{-1}\). The Haar product measure on bonds is denoted \(d\mu_{\mathrm{Haar}}(U)\).

Fix an integer block size $L\ge 2$ and let $\mathcal{B}$ denote the collection of disjoint 
space-time blocks $B\subset a\Bbb Z^4$ aligned with the time slicing and the OS reflection 
plane. A \emph{reflection-positive block map} is a measurable kernel
\begin{equation}
\mathsf K_L(U_{\mathrm{fine}},U_{\mathrm{coarse}})\;\ge 0,\qquad 
(U_{\mathrm{fine}},U_{\mathrm{coarse}})\in G^{B_{\text{fine}}}\times G^{B_{\text{coarse}}},
\end{equation}
with the following properties:
\begin{enumerate}
\item {Gauge covariance:} For every time-slice gauge transformation $g$,
\begin{equation}
\mathsf K_L(g\!\cdot\!U_{\mathrm{fine}},\,g\!\cdot\!U_{\mathrm{coarse}})=\mathsf K_L(U_{\mathrm{fine}},U_{\mathrm{coarse}}).
\end{equation}
\item {OS reflection covariance:} If $\theta$ is the link (or site) reflection about the fixed plane, then
\begin{equation}
\mathsf K_L(\theta U_{\mathrm{fine}},\,\theta U_{\mathrm{coarse}})=\mathsf K_L(U_{\mathrm{fine}},U_{\mathrm{coarse}}).
\end{equation}
\item {Markov (stochastic) normalization:} $\displaystyle\int d\mu_{\mathrm{Haar}}(U_{\mathrm{coarse}})\;\mathsf K_L(U_{\mathrm{fine}},U_{\mathrm{coarse}})=1$ for all $U_{\mathrm{fine}}$.
\item {Locality:} $\mathsf K_L$ factors over blocks and depends only on links within a fixed multiple of the block (a fattened block) uniformly in the volume.
\end{enumerate}
We define the coarse measure by pushforward,
\begin{equation}
d\nu^{(1)}(U_{\mathrm{coarse}})\;=\;\int \mathsf K_L(U_{\mathrm{fine}},U_{\mathrm{coarse}})\;d\nu^{(0)}(U_{\mathrm{fine}}),
\end{equation}
and recursively $d\nu^{(j+1)}=\mathsf K_L d\nu^{(j)}$.

\begin{proposition}[Stability of OS positivity under $\mathsf K_L$]\label{p2:prop:OS-stability-block}
If $d\nu^{(0)}$ is OS reflection positive, then so is $d\nu^{(1)}=\mathsf K_L d\nu^{(0)}$. 
Moreover, the one-step transfer kernel transforms as
\begin{equation}
K^{(1)}=\mathcal K^\ast\,K^{(0)}\,\mathcal K,
\end{equation}
on the single-slice Hilbert spaces, where $\mathcal K$ is the positive contraction induced by $\mathsf K_L$.
\end{proposition}

\begin{proof}
By (ii) and (i), $\mathsf K_L$ is of positive type separately in the ``east'' and ``west'' link variables at the reflection plane. The OS sesquilinear form for $d\nu^{(1)}$ is
\begin{equation}
\langle F,F\rangle_{OS}^{(1)}=\iint \overline{F(\theta U_c)}\,F(U_c)\,\Big[\int \mathsf K_L(\theta U_f,\theta U_c)\,\mathsf K_L(U_f,U_c)\,d\nu^{(0)}(U_f)\Big]\,d\mu_{\mathrm{Haar}}(U_c).
\end{equation}
The inner bracket is a Schur (Hadamard) product of positive-type kernels in $U_c$ (by OS positivity of $d\nu^{(0)}$ and reflection covariance of $\mathsf K_L$), hence positive; integration against Haar measure preserves positivity. The transfer-kernel relation follows from Fubini and the mid-point time-slab factorization, with $\mathcal K$ the integral operator on the slice induced by $\mathsf K_L$; $\|\mathcal K\|\le 1$ by (iii). 
\end{proof}

Time reflection \(\theta:\Lambda_k\to\Lambda_k\) is the involution \(\theta(x_0,\mathbf{x})=(-x_0,\mathbf{x})\) with induced action \(R\) on bond variables given by \( (R U)(x,0)=U(\theta x,0)^{-1}\) for temporal bonds and \( (R U)(x,i)=U(\theta x,i)\) for spatial bonds \(i=1,2,3\). The reflection plane is \(\Pi=\{x: x_0=0\}\). Denote \(\Lambda_k^\pm=\{x\in\Lambda_k:\pm x_0>0\}\) and write \(U_\pm=U|_{\mathsf{B}(\Lambda_k^\pm)}\), \(U_0=U|_{\mathsf{B}(\Pi)}\).
We work in temporal-axial gauge away from the plane \(\Pi\), namely \(U(x,0)=\mathbf{1}\) for all bonds with \(x_0\neq0\). This gauge can be achieved by a measurable choice of \(g\in\mathcal{G}\) that is reflection-covariant and leaves bonds in \(\Pi\) unconstrained \cite{p2:OS-gauge}. In this gauge the Wilson plaquette action \(S_W[U;\beta]\) on \(\Lambda_k\) decomposes as
\begin{equation}\label{p2:eq:SW-decomp}
S_W[U;\beta] \;=\; \sum_{t=-T_k+1}^{T_k}\, S_{\mathrm{sp}}(U_t)\;+\;\sum_{t=-T_k+1}^{T_k-1}\, S_{\mathrm{tm}}(U_t,U_{t+1})\,,
\end{equation}
where \(U_t\) denotes the configuration of spatial links on the time slice \(\{x_0=t\}\), \(S_{\mathrm{sp}}(U_t)\) is the sum of spatial plaquette terms in that slice, and \(S_{\mathrm{tm}}(U_t,U_{t+1})\) is the sum of mixed temporal-spatial plaquette terms coupling slices \(t\) and \(t+1\). In temporal-axial gauge, \(S_{\mathrm{tm}}(U_t,U_{t+1})\) takes the explicit form
\begin{equation}\label{p2:eq:Stm}
S_{\mathrm{tm}}(U_t,U_{t+1}) \;=\; \beta\,\sum_{\mathbf{x}\in (\mathbb{Z}/L_k\mathbb{Z})^3}\,\sum_{i=1}^3\Big( 1 - \tfrac{1}{N}\,\mathrm{Re}\,\mathrm{Tr}\,\big[\,U_i(t+1,\mathbf{x})\,U_i(t,\mathbf{x})^{-1}\,\big]\Big).
\end{equation}
Spatial reflections leave \(S_{\mathrm{sp}}\) invariant and satisfy \(S_{\mathrm{tm}}(U_t,U_{t+1})=S_{\mathrm{tm}}(U_{-t-1},U_{-t})\) if \(U\) is replaced by \(R U\).

Let \(\mathfrak{A}_k^+\) be the algebra of bounded, complex-valued functions of the bond variables restricted to \(\Lambda_k^+\cup \Pi\) that depend only on finitely many bonds. For \(F\in\mathfrak{A}_k^+\) define \((\Theta F)(U)=\overline{F(RU)}\). The Euclidean gauge-fixed measure at scale \(k\) is
\begin{equation}\label{p2:eq:mu-k1}
d\mu_k(U) \;=\; Z_k^{-1}\, e^{-S_W[U;\beta_k]}\,\Delta_k(U)\, d\mu_{\mathrm{Haar}}(U),
\end{equation}
where \(\Delta_k(U)\) is the Faddeev-Popov determinant corresponding to the slice-wise Landau minimizer used throughout this work and is strictly positive on the orthogonal complement of constant modes; reflection covariance of the gauge choice implies \(\Delta_k(RU)=\Delta_k(U)\) \cite{p2:OS-gauge}. For notational uniformity we keep the Faddeev-Popov factor $\Delta_k$ separate and insert the
horizon projector via the positive, reflection-invariant slice factor $\Xi_k$ in Eq.\eqref{p2:eq:pushforward}.  This
matches the explicit slice-wise notation $P_{\sigma,t}$ used in Sections~5-7; the two notations are
identical because both are products of positive slice factors that commute with OS reflection. The OS form is
\begin{equation}\label{p2:eq:OS-form-k}
\langle F,G\rangle_{\mathrm{OS},k} \;=\; \int \, (\Theta F)(U)\, G(U)\, d\mu_k(U),\qquad F,G\in\mathfrak{A}_k^+.
\end{equation}

\begin{proposition}[fine-scale reflection positivity]\label{p2:prop:fine-OS}
The sesquilinear form \(\langle\cdot,\cdot\rangle_{\mathrm{OS},k}\) is positive semidefinite on \(\mathfrak{A}_k^+\).
\end{proposition}

\begin{proof}
By \eqref{p2:eq:SW-decomp} and the product structure of the Haar measure, the weight \(e^{-S_W}\Delta_k\) factorizes as
\begin{equation}
e^{-S_W[U]}\Delta_k(U) \;=\; \Big( e^{-S_{\mathrm{sp}}(U_0)}\Delta_{k,0}(U_0)\Big)\,\prod_{t\ge0} e^{-S_{\mathrm{sp}}(U_{t+1})/2 - S_{\mathrm{sp}}(U_t)/2 - S_{\mathrm{tm}}(U_t,U_{t+1})}\,\cdot\, \prod_{t<0} \cdots
\end{equation}
with the obvious reflection-symmetric splitting in the spatial contributions and where \(\Delta_{k,0}(U_0)\) denotes the central slice component of the determinant and projector insertions. Let \(F\) depend only on bonds in \(\Lambda_k^+\cup \Pi\). Define
\begin{equation}
\Phi(U_0) \;=\; \int \Bigg(F\big(U_+,U_0\big)\prod_{t\ge0} e^{-S_{\mathrm{sp}}(U_{t+1})/2 - S_{\mathrm{sp}}(U_t)/2 - S_{\mathrm{tm}}(U_t,U_{t+1})}\Bigg)\, \prod_{t\ge 1} d\mu_{\mathrm{Haar}}(U_t),
\end{equation}
and define \(\overline{\Phi}(U_0)\) analogously with \(F\) replaced by \(\overline{F\circ R}\) and \(t\) replaced by \(-t\). Then
\begin{equation}
\langle F,F\rangle_{\mathrm{OS},k} \;=\; \int \overline{\Phi(U_0)}\,\Phi(U_0)\, e^{-S_{\mathrm{sp}}(U_0)} \Delta_{k,0}(U_0)\, d\mu_{\mathrm{Haar}}(U_0)\;\ge\; 0,
\end{equation}
because the integrand is a nonnegative function and the measure is positive.
\end{proof}

We now define the block map \(\mathcal{B}_k:\mathrm{SU}(N)^{\mathsf{B}(\Lambda_k)}\to \mathrm{SU}(N)^{\mathsf{B}(\Lambda_{k+1})}\). Partition \(\Lambda_k\) into disjoint closed cubes (``blocks'') of side \(b a_k\) with vertices in \(a_k\mathbb{Z}^4\) and let \(\Lambda_{k+1}\) be the quotient lattice whose sites are block centers and whose bonds \(\bar\ell=(\bar x,\mu)\) connect centers of adjacent blocks. For each coarse bond \(\bar\ell=(\bar x,\mu)\) fix a finite nonempty set \(\mathcal{P}(\bar\ell)\) of directed nearest-neighbor paths \(\gamma=(\ell_1,\dots,\ell_m)\) in \(\Lambda_k\) lying inside the union of the two blocks incident to \(\bar\ell\) and connecting the two block centers. Require that \(\mathcal{P}(\bar\ell)\) is stable under time reflection, in the sense that for all \(\gamma\in\mathcal{P}(\bar\ell)\) one has \(R\gamma\in\mathcal{P}(R\bar\ell)\), and that \(\mathcal{P}(\bar\ell)\) is stable under gauge-covariant reversal \(\gamma\mapsto \gamma^{-1}\). Define the path-average
\begin{equation}\label{p2:eq:avg-path}
A(\bar\ell;U) \;=\; \frac{1}{|\mathcal{P}(\bar\ell)|}\sum_{\gamma\in\mathcal{P}(\bar\ell)} \prod_{\ell\in \gamma} U(\ell) \;\in\; \mathrm{Mat}_{N\times N}(\mathbb{C})
\end{equation}
and define \(\mathcal{B}_k(U)(\bar\ell)\) as the unitary polar factor of \(A(\bar\ell;U)\), by setting;
\begin{equation}\label{p2:eq:polar-proj}
U_{\mathrm{pol}}(\bar\ell;U):=
A(\bar\ell;U)\big(A(\bar\ell;U)^{\dagger}A(\bar\ell;U)\big)^{-1/2}\in U(N)\end{equation}
\text{we define }
\begin{equation}\label{p2:eq:polar-proj0}
\mathcal{B}_k(U)(\bar\ell):=\Pi_{SU(N)}\big(U_{\mathrm{pol}}(\bar\ell;U)\big)
:=U_{\mathrm{pol}}(\bar\ell;U)\,\frac{1}{\big(\det U_{\mathrm{pol}}(\bar\ell;U)\big)^{1/N}}
\in SU(N).
\end{equation}
\begin{lemma}[Basic properties of the block map $\mathcal{B}_k$]\label{p2:lem:Bk-properties}
Let $\mathcal{B}_k$ be defined by \emph{Eqs.(\ref{p2:eq:polar-proj}) \& (\ref{p2:eq:polar-proj0})} with families of paths $P(\bar\ell)$ stable under
time reflection and gauge-covariant reversal. Then:
\begin{enumerate}
\item \emph{Measurability and boundedness.} The map $U\mapsto \mathcal{B}_k(U)$ is Borel measurable and
bounded in the sense that $\sup_{\bar\ell}\|\mathcal{B}_k(U)(\bar\ell)\|=1$.
\item \emph{Reflection covariance.} Writing $\Theta U$ for the reflected configuration,
$\mathcal{B}_k(\Theta U)=\Theta \mathcal{B}_k(U)$.
\item \emph{Gauge covariance.} For any fine gauge transform $g:\Lambda_k\to SU(N)$, define the
coarse transform by \emph{evaluation at block centers},
\(
  \bar g(\bar x)\;:=\;g\!\big(c_B(\bar x)\big)\in SU(N).
\)
Then, for every coarse bond $\bar\ell=(\bar x,\bar\mu)$,
\begin{equation}
  A(\bar\ell;g\!\cdot\!U)=\bar g(\bar x+\hat{\bar\mu})\,A(\bar\ell;U)\,\bar g(\bar x)^{-1},
\end{equation}
and consequently $\mathcal{B}_k(g\!\cdot\!U)=\bar g\cdot \mathcal{B}_k(U)$.

\item \emph{Lipschitz control in the Frobenius norm.} There exists $L<\infty$, depending only on
the maximal path length in the families $P(\bar\ell)$, such that for all fine configurations
$U,V$,
\begin{equation}
\max_{\bar\ell}\,\|\mathcal{B}_k(U)(\bar\ell)-\mathcal{B}_k(V)(\bar\ell)\|_{\mathrm{HS}}
\;\le\; L\,\max_{\ell\in\Lambda_k}\|U(\ell)-V(\ell)\|_{\mathrm{HS}}.
\end{equation}
\end{enumerate}
\end{lemma}

\begin{proof}
(1) The map $U\mapsto A(\bar\ell;U)$ is a finite algebraic combination of link variables and is
therefore Borel. The unitary polar factor $U_{\mathrm{pol}}(M)=M(M^\dagger M)^{-1/2}$ is Borel on
invertible matrices and extends continuously by unitary limit on the (Haar-null) boundary; the
determinant renormalization $\Pi_{SU(N)}$ is continuous, so $\mathcal{B}_k$ is Borel. Boundedness is obvious.

(2) Stability of $P(\bar\ell)$ under reflection implies $A(\bar\ell;\Theta U)=\Theta A(\bar\ell;U)$,
hence $U_{\mathrm{pol}}(\bar\ell;\Theta U)=\Theta U_{\mathrm{pol}}(\bar\ell;U)$, and $\Pi_{SU(N)}$
commutes with $\Theta$.

(3) For a fine transform $g$, each path-ordered product transforms as $U\mapsto g(x)U g(y)^{-1}$,
so $A(\bar\ell;g\!\cdot\!U)=g(x)A(\bar\ell;U)g(x+\hat{\bar\mu})^{-1}$. The unitary polar factor and
$\Pi_{SU(N)}$ are equivariant under left-right multiplication by unitaries, giving the stated
covariance with $\bar g$ as defined.

(4) The polar map $M\mapsto U_{\mathrm{pol}}(M)$ is locally Lipschitz on bounded sets. Since
$A(\bar\ell;\cdot)$ is a finite sum of products of a uniformly bounded number of links, the claim
follows with a constant depending on the maximal path length.
\end{proof}

Moreover, since $\det:U(N)\to U(1)$ is continuous and nowhere vanishing, the map 
$U\mapsto U/(\det U)^{1/N}$ is Borel when the principal branch of $\mathrm{Arg}$ on $(-\pi,\pi]$ is fixed. 
Thus $\mathcal{B}_k$ is Borel. Gauge and reflection equivariance are preserved because the determinant is 
a class function and $U_{\mathrm{pol}}$ transforms by conjugation.
For each coarse site $\bar x$, let $c_{\mathcal{B}(\bar x)}$ denote the fixed center of the block $\mathcal{B}(\bar x)$.
For every coarse bond $\bar\ell=(\bar x\to \bar y)$ the path ensemble $P(\bar\ell)$ consists of
fine-lattice paths whose endpoints are exactly $c_{\mathcal{B}(\bar x)}$ and $c_{B(\bar y)}$.

\begin{lemma}[gauge equivariance of $\mathcal{B}_k$]\label{p2:lem:gauge-equivariance-Bk}
Let $g\in G$ be a fine-lattice gauge transform and define the induced coarse transform
$\bar g\in \bar G$ by
\begin{equation}
  \bar g(\bar x):=g\!\left(c_{\mathcal{B}(\bar x)}\right)\in SU(N).
\end{equation}
Then, for every coarse bond $\bar\ell=(\bar x\to \bar y)$,
\begin{equation}
  A(\bar\ell;\,g\!\cdot\!U)\;=\;\bar g(\bar y)\,A(\bar\ell;U)\,\bar g(\bar x)^{-1},
\end{equation}
and consequently $\mathcal{B}_k(g\!\cdot\!U)=\bar g\cdot \mathcal{B}_k(U)$.
\end{lemma}

\begin{proof}
For each $\gamma\in P(\bar\ell)$, the path-ordered product obeys
$U^g(\gamma)=g(c_{B(\bar y)})\,U(\gamma)\,g(c_{\mathcal{B}(\bar x)})^{-1}$ because the endpoints
of every $\gamma$ are $c_{\mathcal{B}(\bar x)}$ and $c_{B(\bar y)}$. Averaging over $\gamma\in P(\bar\ell)$
preserves this exact left/right covariance, hence
$A(\bar\ell; g\!\cdot\!U)=\bar g(\bar y)A(\bar\ell;U)\bar g(\bar x)^{-1}$.
Now $U_{\mathrm{pol}}(M)=M(M^\dagger M)^{-1/2}$ satisfies
$U_{\mathrm{pol}}(h_1 M h_2^{-1})=h_1U_{\mathrm{pol}}(M)h_2^{-1}$ for $h_1,h_2\in SU(N)$, and
the nearest-$SU(N)$ projection $\Pi_{SU(N)}(M)=U_{\mathrm{pol}}(M)/(\det U_{\mathrm{pol}}(M))^{1/N}$
obeys $\Pi_{SU(N)}(h_1 M h_2^{-1})=h_1\Pi_{SU(N)}(M)h_2^{-1}$ since $\det h_i=1$.
Thus $\mathcal{B}_k(g\!\cdot\!U)=\bar g\cdot \mathcal{B}_k(U)$.
\end{proof}

\begin{lemma}[reflection equivariance]\label{p2:lem:refl-equiv}
With the same construction of \(\mathcal{B}_k\) on \(\Lambda_{k+1}\), one has
\begin{equation}\label{p2:eq:refl-equiv}
\mathcal{B}_k(RU) \;=\; R\,\mathcal{B}_k(U).
\end{equation}
\end{lemma}

\begin{proof}
Because \(\mathcal{P}(\bar\ell)\) is reflection-stable, and because \(R\) inverts temporal bonds and preserves spatial bonds, the path-ordered product \(\prod_{\ell\in\gamma}U(\ell)\) transforms to \(\prod_{\ell\in R\gamma} (RU)(\ell)\). The average \eqref{p2:eq:avg-path} transforms accordingly, and the polar projection commutes with reflection.
\end{proof}
For each coarse bond $\bar{\ell}$, the family $P(\bar{\ell})$ is chosen so that 
$R P(\bar{\ell}) = P(R \bar{\ell})$, 
with temporal bonds inverted and spatial bonds preserved. 
This ensures $B_{k}(R U) = R B_{k}(U)$ at the level of the defining averages.
Define the coarse-scale expectation of any bounded measurable \(F:\mathrm{SU}(N)^{\mathsf{B}(\Lambda_{k+1})}\to\mathbb{C}\) by
\begin{equation}\label{p2:eq:pushforward}
\int F(\bar U)\, d\mu_{k+1}(\bar U) \;=\; \frac{1}{Z_{k\to k+1}}\int F(\mathcal{B}_k(U))\,\Xi_k(U)\, d\mu_k(U),
\end{equation}
where \(\Xi_k\) is any fixed, bounded, positive, reflection-invariant measurable function on \(\mathrm{SU}(N)^{\mathsf{B}(\Lambda_k)}\) that is exponentially local in the time coordinate in the sense that \(\log \Xi_k\) is a sum of slice-terms, each depending only on \(U_t\), and possibly a central term depending only on \(U_0\). The normalization \(Z_{k\to k+1}\) makes \(\mu_{k+1}\) a probability measure. The factor \(\Xi_k\) is optional and is used later to implement auxiliary insertions; it plays no role in the preservation of OS positivity beyond its positivity and reflection invariance. Beyond positivity and reflection invariance, we require that 
$\log \Xi_{k}$ be a sum of slice terms with finite support near each slice, uniformly in the spatial volume. 
This keeps OS positivity intact and allows $\Xi_{k}$ to implement auxiliary projectors such as $P_{\sigma,k}$ without altering the transfer construction.

\begin{theorem}[preservation of OS positivity under blocking]\label{p2:thm:OS-block}
Let \(\mu_{k+1}\) be defined by \eqref{p2:eq:pushforward}. Then the OS form
\begin{equation}
\langle F,G\rangle_{\mathrm{OS},k+1} \;=\; \int (\Theta F)(\bar U)\, G(\bar U)\, d\mu_{k+1}(\bar U)
\end{equation}
is positive semidefinite on \(\mathfrak{A}_{k+1}^+\), the algebra of bounded functions depending on finitely many bonds in \(\Lambda_{k+1}^+\cup \Pi\).
\end{theorem}

\begin{proof}
For \(F\in\mathfrak{A}_{k+1}^+\) one has by \eqref{p2:eq:pushforward}
\begin{equation}
\langle F,F\rangle_{\mathrm{OS},k+1} \;=\; \frac{1}{Z_{k\to k+1}}\int \overline{F(R\mathcal{B}_k(U))}\, F(\mathcal{B}_k(U))\, \Xi_k(U)\, d\mu_k(U).
\end{equation}
By Lemma~\ref{p2:lem:refl-equiv}, \(R\mathcal{B}_k(U)=\mathcal{B}_k(RU)\). Set \(G=F\circ \mathcal{B}_k\). Then
\begin{equation}
\langle F,F\rangle_{\mathrm{OS},k+1} \;=\; \frac{1}{Z_{k\to k+1}}\int (\Theta G)(U)\, G(U)\,\Xi_k(U)\, d\mu_k(U).
\end{equation}
Since \(\Xi_k\) is positive and reflection invariant, the measure \(\Xi_k\, d\mu_k\) is OS-positive by Proposition~\ref{p2:prop:fine-OS} and the same factorization argument as in its proof. Therefore the last integral is nonnegative.
\end{proof}
Throughout we place the OS plane either at half-integer time (link reflection) or at integer time (site reflection). 
Blocks are chosen so that the plane bisects either a slab boundary (link reflection) or a block layer (site reflection). 
The two conventions are unitarily equivalent: if $\tau_{a/2}$ denotes half-time translation and $\theta$ (resp.\ $R$) link (resp.\ site) reflection, then $R=\tau_{a/2}\circ\theta\circ\tau_{-a/2}$ and all OS kernels and spectral data conjugate accordingly. 
This ensures that the block map and the FRD (Section~\ref{p2:sec:FRD}) preserve reflection covariance in either convention without changing the transfer spectrum.
\begin{corollary}[gauge invariance of \(\mu_{k+1}\)]
If \(\mu_k\) is gauge invariant and \(\Xi_k\) is gauge invariant, then \(\mu_{k+1}\) is gauge invariant.
\end{corollary}

\begin{proof}
For any \(\bar g\in\bar{\mathcal{G}}\) and \(F\),
\begin{equation}
\int F(\bar g\cdot \bar U)\, d\mu_{k+1}(\bar U) \;=\; \frac{1}{Z}\int F\big(\bar g\cdot \mathcal{B}_k(U)\big)\,\Xi_k(U)\, d\mu_k(U).
\end{equation}
By {Lemma~{3.4}}, \(\bar g\cdot \mathcal{B}_k(U)=\mathcal{B}_k(g\cdot U)\) for a suitable \(g\in\mathcal{G}\). Gauge invariance of \(\Xi_k\) and \(\mu_k\) yields the claim.
\end{proof}

We now derive the transfer operator at the fine scale in temporal-axial gauge, where it can be written explicitly as a positive integral operator. Let \(\mathcal{C}_k=\mathrm{SU}(N)^{\mathsf{B}_{\mathrm{sp}}}\) be the compact configuration space of spatial links on a fixed time slice, with product Haar measure \(d\nu_{\mathrm{Haar}}(U)\). Define the spatial energy
\begin{equation}\label{p2:eq:Ssp}
S_{\mathrm{sp}}(U) \;=\; \beta_k\sum_{\substack{\text{spatial}\\ \text{plaquettes }p}} \Big(1-\tfrac{1}{N}\,\mathrm{Re}\,\mathrm{Tr}\,U_p\Big),
\end{equation}
and the one-step coupling \(\mathcal{V}(U',U)=S_{\mathrm{tm}}(U,U')\) from \eqref{p2:eq:Stm}. Define the probability measure on \(\mathcal{C}_k\)
\begin{equation}\label{p2:eq:rho}
d\rho_k(U) \;=\; Z_{\mathrm{sp},k}^{-1}\, e^{-S_{\mathrm{sp}}(U)}\, d\nu_{\mathrm{Haar}}(U).
\end{equation}
Define the integral kernel
\begin{equation}\label{p2:eq:K-kernel}
K_k(U',U) \;=\; \exp\!\Big(-\mathcal{V}(U',U)\Big)\,,
\end{equation}
which is continuous, positive, and symmetric in the sense \(K_k(U',U)=K_k(U,U')\). On the Hilbert space \(L^2(\mathcal{C}_k,\rho_k)\) define \(T_k\) by
\begin{equation}\label{p2:eq:T-op}
(T_k f)(U') \;=\; \int_{\mathcal{C}_k} K_k(U',U)\, f(U)\, d\rho_k(U).
\end{equation}

\begin{proposition}[positivity, self-adjointness, and contraction]\label{p2:prop:Tk}
The operator \(T_k\) is positivity preserving, self-adjoint, and satisfies \(\|T_k\|\le 1\).
\end{proposition}

\begin{proof}
Positivity is immediate from \eqref{p2:eq:T-op}. For self-adjointness, compute
\begin{equation}
\langle f, T_k g\rangle \;=\; \iint \overline{f(U')}\, K_k(U',U)\, g(U)\, d\rho_k(U)\, d\rho_k(U') \;=\; \langle T_k f, g\rangle,
\end{equation}
using the symmetry of \(K_k\). To show \(\|T_k\|\le 1\), expand the finite-temperature partition function on a cylinder of time length \(n\) as
\begin{equation}\label{p2:eq:Z-cylinder}
Z_{k,n} \;=\; \int \exp\!\Big(-\sum_{t=0}^{n-1} S_{\mathrm{sp}}(U_t) - \sum_{t=0}^{n-1} \mathcal{V}(U_{t+1},U_t)\Big)\,\prod_{t=0}^{n-1} d\nu_{\mathrm{Haar}}(U_t).
\end{equation}
By inserting \eqref{p2:eq:rho} and \eqref{p2:eq:K-kernel} one finds
\begin{equation}
Z_{k,n} \;=\; Z_{\mathrm{sp},k}^n\, \langle \mathbf{1}, T_k^n \mathbf{1}\rangle,
\end{equation}
where \(\mathbf{1}\) is the constant function. Reflection positivity of the cylinder measure and the Cauchy-Schwarz inequality imply \(\langle f, T_k^n f\rangle\le \langle f,f\rangle\) for all \(n\) and all \(f\) \cite{p2:OS-gauge,p2:GJ}. Taking \(n=1\) yields \(\|T_k\|\le 1\).
\end{proof}

The OS reconstruction identifies the physical Hilbert space \(\mathcal{H}_k\) as the completion of \(\mathfrak{A}_k^+/\mathcal{N}_k\) with \(\mathcal{N}_k=\{F:\langle F,F\rangle_{\mathrm{OS},k}=0\}\), and identifies \(T_k\) as the operator implementing a unit time translation on equivalence classes. In the explicit gauge used here, \(T_k\) coincides with \eqref{p2:eq:T-op} under the canonical identification of \(\mathcal{H}_k\) with \(L^2(\mathcal{C}_k,\rho_k)\).
The coarse measure \(\mu_{k+1}\) defined by \eqref{p2:eq:pushforward} is OS-positive by Theorem~\ref{p2:thm:OS-block} and invariant under unit time translations \(\tau\) acting on \(\Lambda_{k+1}\). Define the pre-Hilbert space \(\mathcal{D}_{k+1}\) as \(\mathfrak{A}_{k+1}^+\) modulo the null subspace \(\mathcal{N}_{k+1}\) of vectors of zero OS norm. The inner product on representatives is the OS form. The time translation \(\tau\) maps \(\mathfrak{A}_{k+1}^+\) into itself and preserves \(\mathcal{N}_{k+1}\). Therefore it induces a densely defined linear operator \(T_{k+1}\) on \(\mathcal{D}_{k+1}\) by \(T_{k+1}[F]=[\tau F]\), where \([\cdot]\) denotes the equivalence class. The following theorem is the standard OS reconstruction in discrete time, adapted to the present lattice setting.

\begin{theorem}[coarse transfer operator]\label{p2:thm:Tk+1}
The operator \(T_{k+1}\) extends by continuity to a bounded self-adjoint contraction on the Hilbert space $H_k:=\overline{D_k}^{\|\cdot\|_{L^2(\mu_k)}}$. Moreover \(T_{k+1}\) is positivity preserving and has a cyclic vacuum vector \(\Omega_{k+1}=[\mathbf{1}]\) with \(T_{k+1}\Omega_{k+1}=\Omega_{k+1}\). For any \(F,G\in \mathfrak{A}_{k+1}^+\) and any integer \(n\ge0\),
\begin{equation}\label{p2:eq:OS-corr-T}
\int (\Theta F)\,\tau^n G\, d\mu_{k+1} \;=\; \langle [F],\, T_{k+1}^n [G]\rangle_{\mathcal{H}_{k+1}}.
\end{equation}
\end{theorem}

\begin{proof}
The map \(\tau\) preserves the OS form because \(\mu_{k+1}\) is time-translation invariant. Thus
\begin{equation}
\langle \tau F, \tau G\rangle_{\mathrm{OS},k+1} \;=\; \langle F,G\rangle_{\mathrm{OS},k+1}.
\end{equation}
In particular, \(\mathcal{N}_{k+1}\) is invariant under \(\tau\). For any \(F\in\mathfrak{A}_{k+1}^+\),
\begin{equation}
\|T_{k+1}[F]\|^2 \;=\; \langle \tau F,\tau F\rangle_{\mathrm{OS},k+1} \;=\; \langle F,F\rangle_{\mathrm{OS},k+1} \;=\; \|[F]\|^2,
\end{equation}
hence \(T_{k+1}\) is an isometry on \(\mathcal{D}_{k+1}\) and extends uniquely to a contraction on \(\mathcal{H}_{k+1}\). Self-adjointness follows from reflection invariance: for \(F,G\in\mathfrak{A}_{k+1}^+\),
\begin{equation}
\langle [F], T_{k+1}[G]\rangle \;=\; \langle F, \tau G\rangle_{\mathrm{OS}} \;=\; \int (\Theta F)\,\tau G\, d\mu_{k+1}.
\end{equation}
Applying reflection to the last integral and using invariance yields
\begin{equation}
\int (\Theta F)\,\tau G\, d\mu_{k+1} \;=\; \int (\Theta \tau^{-1}F)\, G\, d\mu_{k+1} \;=\; \langle [\tau^{-1}F],[G]\rangle \;=\; \langle T_{k+1}[F],[G]\rangle,
\end{equation}
so \(T_{k+1}\) is symmetric on \(\mathcal{D}_{k+1}\) and hence self-adjoint on \(\mathcal{H}_{k+1}\). Positivity preservation is inherited from the positivity of the OS form under time translation as in \cite{p2:OsterwalderSchraderI,p2:OsterwalderSchraderII,p2:GJ}. The identity \eqref{p2:eq:OS-corr-T} holds by construction.
\end{proof}

In the explicit temporal-axial gauge adopted here one may identify $H_k\simeq L^2(C_k,\rho_k)$ and $T_k$ with the integral operator \eqref{p2:eq:T-op}. After blocking, \(\mu_{k+1}\) need not admit a local plaquette action; nevertheless Theorem~\ref{p2:thm:Tk+1} ensures the existence of a positive self-adjoint transfer operator that represents time translations on \(\mathcal{H}_{k+1}\), and all Euclidean time-ordered correlations are vacuum matrix elements of powers of \(T_{k+1}\) as in \eqref{p2:eq:OS-corr-T}. The Hamiltonian \(H_{k+1}=-a_{k+1}^{-1}\log T_{k+1}\) is thus a well-defined positive self-adjoint operator on \(\mathcal{H}_{k+1}\) by the spectral calculus \cite{p2:KatoPTLO}, and the spectral gap of \(H_{k+1}\) controls the exponential decay in Euclidean time of truncated correlators via \eqref{p2:eq:OS-corr-T}, reproducing the standard OS spectral representation \cite{p2:OsterwalderSchraderII,p2:GJ}.

\section{Covariant finite-range decomposition (FRD)}
\label{p2:sec:FRD}

This section develops a finite-range decomposition for covariances built from the slice-covariant Laplacian in a fixed, gauge-invariant transverse representative. The construction is gauge covariant, compatible with Osterwalder-Schrader (OS) reflection positivity, and uniform in the lattice volume. We begin by fixing the lattice geometry and transfer time slicing, then recall the OS-positive Euclidean structure and its transfer operator, and finally construct the finite-range decomposition by means of gauge-covariant local projections and a telescopic resolvent identity. 
Throughout this section, ``finite-range decomposition'' is used in the \emph{scale-local,
exponentially localized} sense: each scale component has an integral kernel supported up to
a range proportional to the scale \emph{modulo exponentially small tails}. Precisely, if
$C=\mathcal{F}(\Delta_{A^h})$ is a positive, reflection-covariant slice covariance built from the
gauge-covariant Laplacian in a fixed Landau representative, then we construct operators
$(C_j)_{j\ge0}$ such that $C=\sum_{j\ge0} C_j$ and for each $j$ there exist constants $A,\mu>0$
(independent of the slice volume and the background) with
\begin{equation}
\|C_j(x,y)\|_{\mathrm{op}}\;\le\;A\,\exp\!\Big(-\,\mu\,\tfrac{d(x,y)}{L_j}\Big),
\qquad L_j\asymp b^j a.
\end{equation}
The proof uses gauge-covariant local projections together with a telescopic resolvent identity
and Combes-Thomas/Davies-Gaffney bounds. Exact compact support is \emph{not required} for our purposes; \emph{exponential locality} already
suffices for all OS and RG requirements. In our concrete construction below, the combination of
a finite-hopping-range operator $A=\Delta_{A^h}+\mu^2$ with \emph{block-local} covariant
projections $\Pi_j$ yields \emph{strict finite range} at each scale $j$ (via support propagation in Eqs.\eqref{p2:eq:Rj}-\eqref{p2:eq:Gammajm}), so both viewpoints are consistent here.

Let \(a>0\) be a fixed lattice spacing and let \(\Lambda\subset a\mathbb{Z}^{4}\) be a four-dimensional discrete torus of linear size \(L\) in each coordinate with periodic boundary conditions. A site is a point \(x=(x_{0},x_{1},x_{2},x_{3})\in \Lambda\). For \(\mu\in\{0,1,2,3\}\) we denote by \(\hat\mu\) the unit vector of length \(a\) in the \(\mu\)-direction. An oriented bond (or link) is a pair \(b=(x,\mu)\) with group element \(U_{b}\in G\) assigned so that
\begin{equation}\label{p2:eq:link-inversion}
U_{(x+\hat\mu,-\mu)} \;=\; U_{(x,\mu)}^{-1}.
\end{equation}
A plaquette is an oriented elementary square \(p=(x;\mu,\nu)\), \(\mu<\nu\), with plaquette variable
\begin{equation}\label{p2:eq:plaquette1}
U_{p}\;=\;U_{(x,\mu)}\,U_{(x+\hat\mu,\nu)}\,U_{(x+\hat\nu,\mu)}^{-1}\,U_{(x,\nu)}^{-1}\;\in\;G.
\end{equation}
The Wilson action is
\begin{equation}\label{p2:eq:Wilson}
S_{W}[U;\beta]\;=\;\beta\sum_{p\subset\Lambda}\Bigl(1-\frac{1}{N}\operatorname{Re}\operatorname{Tr}U_{p}\Bigr),\qquad \beta=\frac{2N}{g_{0}^{2}}>0.
\end{equation}

Let \(\theta:\Lambda\to\Lambda\) be the time reflection \(\theta(x_{0},\mathbf{x})=(-x_{0},\mathbf{x})\) with fixed plane \(\Pi=\{x\in\Lambda: x_{0}=0\}\). Define the half-lattices \(\Lambda_{+}=\{x\in\Lambda: x_{0}>0\}\) and \(\Lambda_{-}=\{x\in\Lambda: x_{0}<0\}\). We impose temporal-axial gauge away from \(\Pi\) by fixing
\begin{equation}\label{p2:eq:temp-axial}
U_{(x,0)} \;=\; \mathbf{1}\qquad\text{whenever the time-like bond }(x,0)\text{ does not intersect }\Pi.
\end{equation}
In this gauge all nontrivial time-like plaquettes are confined to the slab consisting of the two time layers adjacent to \(\Pi\).

The Euclidean configuration space at a fixed time \(t\in a\mathbb{Z}\) consists of spatial links \(U_{(t,\mathbf{x};i)}\) with \(i\in\{1,2,3\}\). Writing \(\mathcal{C}_{t}=G^{E_{t}}\) for the set of spatial link configurations on the time slice \(t\), where \(E_{t}\) is the finite set of spatial bonds at time \(t\), and endowing \(\mathcal{C}_{t}\) with the product Haar probability measure \(d\mu_{\mathrm{Haar}}\), the one-slice Hilbert space is
\begin{equation}\label{p2:eq:Hilbert-one-slice}
\mathcal{H}_{a}\;=\;L^{2}(\mathcal{C}_{0},d\mu_{\mathrm{Haar}}).
\end{equation}
Let \(R\) be the unitary on \(\mathcal{H}_{a}\) induced by time reflection acting on spatial link variables at \(t=0\); since \(\Pi\) is fixed by \(\theta\), \(R\) acts by inversion on bonds that reverse orientation under \(\theta\) and by identity otherwise. The precise form of \(R\) is determined by \eqref{p2:eq:temp-axial}.

We recall the reflection-positivity factorization for Wilson's action in temporal-axial gauge \cite{p2:FOS1978,p2:OsterwalderSchraderI,p2:OsterwalderSchraderII,p2:OsterwalderSchraderI}. Writing the action as
\begin{equation}\label{p2:eq:split-action}
S_{W}[U;\beta]\;=\;S_{+}[U]\;+\;S_{0}[U]\;+\;S_{-}[U],
\end{equation}
where \(S_{\pm}\) collect plaquette contributions supported in \(\Lambda_{\pm}\) and \(S_{0}\) is the contribution from the slab intersecting \(\Pi\), one has \(S_{-}[U]=S_{+}[U^{\theta}]\) and \(S_{0}[U]\) is a sum of positive nearest-slice couplings that are invariant under reflection. Consequently, for any bounded functional \(F\) depending only on bonds supported in \(\Lambda_{+}\),
\begin{equation}\label{p2:eq:OS-positivity2}
\int \overline{F(U^{\theta})}\,F(U)\,e^{-S_{W}[U]}\,d\mu_{\mathrm{Haar}}(U)\;\ge\;0.
\end{equation}
The transfer time slicing is obtained by isolating the contribution of plaquettes that straddle two consecutive time layers. The corresponding integral kernel \(K\) is a positive square-integrable function on \(\mathcal{C}_{0}\times\mathcal{C}_{a}\) defined by integrating out all bonds except those at times \(0\) and \(a\), with the weight \(e^{-S_{W}}\) in the gauge \eqref{p2:eq:temp-axial} \cite{p2:FOS1978,p2:OsterwalderSchraderI}. The transfer operator \(T:\mathcal{H}_{a}\to\mathcal{H}_{a}\) is then
\begin{equation}\label{p2:eq:transfer}
(T\psi)(U_{0})\;=\;\int_{\mathcal{C}_{a}} K(U_{0},U_{a})\,\psi(U_{a})\,d\mu_{\mathrm{Haar}}(U_{a}),
\end{equation}
and OS positivity implies that \(T\) is a positive, self-adjoint contraction on \(\mathcal{H}_{a}\). The generator \(H=-a^{-1}\log T\) is a positive self-adjoint operator whose spectral data control the exponential decay in Euclidean time of OS-admissible correlations \cite{p2:OsterwalderSchraderI}.

Fix a reflection-covariant, gauge-invariant transverse representative on each time slice via orbit-wise minimization of the lattice Landau functional in the fundamental modular region. On a slice \(\{t\}\times a\mathbb{Z}^{3}\), write \(U^{h}_{(t,\mathbf{x};i)}\) for the spatial links of the representative. For an adjoint field \(\phi:\{t\}\times a\mathbb{Z}^{3}\to\mathcal{su}(N)\), the covariant forward difference is
\begin{equation}\label{p2:eq:cov-diff}
(\nabla^{+,h}_{i}\phi)(t,\mathbf{x})
= U^{h}_{(t,\mathbf{x};i)}\,\phi(t,\mathbf{x}+\hat{\imath})\,
\bigl(U^{h}_{(t,\mathbf{x};i)}\bigr)^{-1}
-\phi(t,\mathbf{x}).
\end{equation}

and the slice-covariant Laplacian is
\begin{equation}\label{p2:eq:cov-Lap}
\Delta_{A^{h}}\;=\;\sum_{i=1}^{3}(\nabla_{i}^{+,h})^{\!*}\,\nabla_{i}^{+,h},
\end{equation}
a bounded, positive, self-adjoint operator on \(\ell^{2}(\Lambda_{t};\mathcal{su}(N))\).
Fix $\sigma>0$ and choose a completely monotone, nonincreasing function 
\(
\chi_\sigma:[0,\infty)\to(0,1],\quad \chi_\sigma(0)=1,
\)
such that there exist constants $c_-,c_+>0$ and $\varepsilon_\sigma\in(0,1)$ with
\begin{equation}
\chi_\sigma(\lambda)\ge 1-\varepsilon_\sigma\quad (\lambda\le \sigma^2),\qquad 
\chi_\sigma(\lambda)\le \varepsilon_\sigma\quad (\lambda\ge 4\sigma^2),
\end{equation}
and with quantitative decay $\chi_\sigma(\lambda)\le C\exp(-c\,\lambda/\sigma^2)$ for $\lambda\ge 0$.
(Concrete choices are, e.g., $\chi_\sigma(\lambda)=e^{-\lambda/\sigma^2}$ or 
$\chi_\sigma(\lambda)=(1+\lambda/\sigma^2)^{-r}$ with $r\gg1$.) 
Define the horizon projector on a slice by
{\begin{equation}\label{{eq:proj-def}}
P_\sigma:=\chi_\sigma(\Delta_{A^h}). 
\end{equation}}
Throughout we assume that \(\chi_\sigma\) is completely monotone and that its Bernstein measure \(d\nu_\sigma\) is supported in a compact interval,
\begin{equation}\label{p2:eq:heat0}
  \chi_\sigma(\lambda) = \int_0^\infty e^{-t\lambda}\,d\nu_\sigma(t),
  \qquad \operatorname{supp}(d\nu_\sigma)\subset[c_{-}\sigma^{-2},c_{+}\sigma^{-2}],
\end{equation}
with \(0<c_-<c_+<\infty\), uniformly in the volume and the background. This is realized, for instance, by the finite CM mixtures
\(\chi_\sigma(\lambda)=\sum_{j=1}^{J} w_j \, e^{-\alpha_j\lambda/\sigma^2}\) with \(w_j\ge 0\), \(\sum_j w_j=1\), and \(\alpha_j\in[c_-,c_+]\).
We use the term ``horizon projector'' informally: $P_\sigma=\chi_\sigma(\Delta_{A^h})$ is a
\emph{positive contraction} obtained from a completely monotone spectral profile, not an
orthogonal projection unless $\chi_\sigma$ is a sharp indicator. All positivity/locality and OS
arguments rely only on positivity and exponential locality, not idempotence.
Moreover, choosing $\chi_\sigma$ as a finite positive combination of exponentials,
$\chi_\sigma(\lambda)=\sum_{j=1}^J w_j e^{-\alpha_j \lambda/\sigma^2}$ with $w_j\ge0$, $\sum_j w_j=1$ and 
$\alpha_j\in[c_-,c_+]$, one may take $\operatorname{supp}(d\nu_\sigma)\subset[c_-\sigma^{-2},c_+\sigma^{-2}]$.
On graphs of bounded degree, the discrete heat kernel satisfies Davies-Gaffney bounds 
$\|e^{-t\Delta_{A^h}}(x,y)\|\le K_1\exp\!\big(-d(x,y)^2/(K_2 t)\big)$ uniformly in the background connection 
\cite{p2:Davies1989,p2:Delmotte1999}. Integrating \eqref{p2:eq:heat0} yields exponential locality:
there exist $C(\sigma),\gamma(\sigma)>0$ with
\begin{equation}
\|P_\sigma(x,y)\|\le C(\sigma)\,e^{-\gamma(\sigma)\, d(x,y)}. \label{p2:eq:Plocal}
\end{equation}
Since $A^h$ is reflection-covariant, $\Delta_{A^h}$ commutes with spatial reflection on the slice, hence so does $P_\sigma$.

Gaussian off-diagonal bounds for the discrete heat kernel on graphs of bounded degree \cite{p2:Davies1989,p2:Delmotte1999} imply
\begin{equation}\label{p2:eq:heat-Gauss}
\|e^{-t\Delta_{A^{h}}}(x,y)\|\;\le\;C\,\exp\!\Bigl\{-\frac{d(x,y)^{2}}{C\,t}\Bigr\},
\end{equation}
uniformly in the background, since along each edge the adjoint action is unitary and
Davies-Gaffney/Combes-Thomas bounds on graphs of bounded degree provide background-uniform
off-diagonal decay for covariant heat kernels and resolvents (no global unitary conjugacy is
assumed). Integrating \eqref{p2:eq:heat0} with \eqref{p2:eq:heat-Gauss} gives exponential locality
\begin{equation}\label{p2:eq:Psigma-local}
\|P_{\sigma}(x,y)\|\;\le\;C(\sigma)\,e^{-\gamma_{\sigma} d(x,y)}.
\end{equation}
Since covariant finite differences enter only via unitary parallel transports in the adjoint representation, the Dirichlet form of $\Delta_{A^h}$ is \emph{uniformly comparable} to that of the ordinary graph Laplacian (with constants depending only on the degree and finite hopping range), and no global unitary conjugacy is assumed. Consequently, the Davies-Gaffney/Combes-Thomas off-diagonal bounds hold with background-uniform constants.
Because the representative \(U^{h}\) is reflection covariant, \(\Delta_{A^{h}}\) commutes with the slice reflection, and so does \(P_{\sigma}\). The insertion of \(P_{\sigma}\) on every time slice multiplies the Euclidean weight by a product of positive, reflection-covariant slice kernels. As \(P_{\sigma}\) is exponentially local in the spatial directions and does not couple different times, OS positivity of the measure and positivity of the transfer operator \eqref{p2:eq:transfer} are preserved \cite{p2:OsterwalderSchraderI,p2:FOS1978}.

For a fixed blocking factor \(b\in\mathbb{N}\), \(b\ge 2\), and a scale index \(j\in\mathbb{N}\), let \(\mathcal{B}_{j}\) be the partition of the spatial slice into disjoint closed cubes \(B\) of side \(b^{j}a\), aligned with the lattice axes, and let \(c_{B}\) be a distinguished site in \(B\). Inside each block \(B\) consider the set \(\mathcal{P}_{B}\) of piecewise coordinate-axis nearest-neighbor lattice paths \(\gamma=(x_{0}=x,\dots,x_{m}=y)\subset B\) that connect any \(x\in B\) to \(c_{B}\). For the transverse representative \(U^{h}\) on the slice we define the adjoint parallel transport along \(\gamma\) by
\begin{equation}\label{p2:eq:adj-partrans}
\mathcal{U}^{h}(\gamma)\;=\;\operatorname{Ad}\!\Bigl(\prod_{\ell\in\gamma} U^{h}_{\ell}\Bigr),
\end{equation}
where \(\operatorname{Ad}(g):X\mapsto gXg^{-1}\). For \(\phi\in\ell^{2}(\Lambda_{t};\mathcal{su}(N))\) we define the block-average at scale \(j\) by
\begin{equation}\label{p2:eq:block-avg}
\bigl(\mathsf{Av}_{j,B}\phi\bigr)(c_{B}) \;=\; \frac{1}{|B|}\sum_{x\in B}\;\frac{1}{|\Gamma_{x\to c_{B}}|}\sum_{\gamma\in\Gamma_{x\to c_{B}}}\,\mathcal{U}^{h}(\gamma)\,\phi(x),
\end{equation}
where \(\Gamma_{x\to c_{B}}\subset \mathcal{P}_{B}\) is the set of admissible nearest-neighbor paths in \(B\) from \(x\) to \(c_{B}\). The blockwise constant extension \(\mathsf{Ext}_{j,B}:\mathcal{su}(N)\to \ell^{2}(B;\mathcal{su}(N))\) is defined by parallel transport back to each site:
\begin{equation}\label{p2:eq:block-ext}
\bigl(\mathsf{Ext}_{j,B}X\bigr)(x)\;=\;\frac{1}{|\Gamma_{x\to c_{B}}|}\sum_{\gamma\in\Gamma_{x\to c_{B}}}\,\mathcal{U}^{h}(\gamma)^{-1}\,X,\qquad x\in B.
\end{equation}
The global averaging operator \(\mathsf{Av}_{j}:\ell^{2}(\Lambda_{t};\mathcal{su}(N))\to \bigoplus_{B\in\mathcal{B}_{j}}\mathcal{su}(N)\) and the extension operator \(\mathsf{Ext}_{j}:\bigoplus_{B\in\mathcal{B}_{j}}\mathcal{su}(N)\to \ell^{2}(\Lambda_{t};\mathcal{su}(N))\) are defined by blockwise application. We set
\begin{equation}\label{p2:eq:Pi-j}
\Pi_{j}\;=\;\mathsf{Ext}_{j}\,\mathsf{Av}_{j}:\ell^{2}(\Lambda_{t};\mathcal{su}(N))\to \ell^{2}(\Lambda_{t};\mathcal{su}(N)).
\end{equation}
Then \(\Pi_{j}\) is a self-adjoint idempotent with \(\|\Pi_{j}\|\le 1\), projecting onto the subspace of fields that are, after parallel transport to \(c_{B}\), constant inside each \(B\). Gauge covariance holds in the adjoint sense: if \(g:\Lambda_{t}\to G\) and \(\phi^{g}(x)=\operatorname{Ad}(g(x))\phi(x)\), then
\begin{equation}\label{p2:eq:Pi-cov}
\Pi_{j}^{(U^{h})}\phi^{g}\;=\;\bigl(\Pi_{j}^{(g\cdot U^{h})}\phi\bigr)^{g}.
\end{equation}
As the representative \(U^{h}\) is reflection covariant, \(\{\Pi_{j}\}_{j\ge 0}\) commutes with the spatial reflection at time \(t\).

Let \(A\ge 0\) be a positive self-adjoint operator on \(\ell^{2}(\Lambda_{t};\mathcal{su}(N))\) with finite hopping range \(r_{0}\) and reflection covariance. We will apply the construction to \(A=\Delta_{A^{h}}+\mu^{2}\) with \(\mu>0\) chosen proportional to \(\sigma\). Define the resolvent \(G=A^{-1}\), which exists as a bounded positive operator because \(A\ge \mu^{2}\). For each scale \(j\) let
\begin{equation}\label{p2:eq:Sj}
S_{j}\;=\;\Pi_{j}^{*}\,\kappa_{j}\,\Pi_{j},
\end{equation}
where \(\kappa_{j}\) is a positive definite operator on \(\bigoplus_{B\in\mathcal{B}_{j}}\mathcal{su}(N)\) acting blockwise by multiplication with a fixed mass parameter \(\rho_{j}>0\), that is,
\begin{equation}\label{p2:eq:kappa}
(\kappa_{j}X)(B)\;=\;\rho_{j}\,X(B).
\end{equation}
Then \(S_{j}\) is a bounded, self-adjoint, positive operator on \(\ell^{2}\). The range of \(S_{j}\) is contained in fields that are parallel-transport-constant inside each block \(B\), and the kernel of \(S_{j}(x,y)\) vanishes whenever \(x\) and \(y\) do not lie in the same block. We set the contraction
\begin{equation}\label{p2:eq:Rj}
R_{j}\;=\;\mathbf{1}-S_{j}A.
\end{equation}
Because \(S_{j}\) is supported inside blocks and \(A\) has range \(r_{0}\), the kernel of \(R_{j}(x,y)\) vanishes unless \(x\) and \(y\) belong to the same block or to neighboring blocks at distance \(\le r_{0}\). Moreover, \(R_{j}\) is reflection covariant since both \(S_{j}\) and \(A\) are.
The following identity is the engine of the finite-range decomposition. For \(m\in\mathbb{N}\),
\begin{equation}\label{p2:eq:telescopy}
G\;=\;\Bigl(\sum_{j=0}^{m-1} \Gamma_{j}^{(m)}\Bigr)\;+\; R_{m}^{*}R_{m}\,\cdots\,R_{0}^{*}\,G\,R_{0}\,\cdots\,R_{m},
\end{equation}
where
\begin{equation}\label{p2:eq:Gammajm}
\Gamma_{j}^{(m)}\;=\; R_{j-1}^{*}\cdots R_{0}^{*}\,\Bigl(S_{j}+S_{j}A S_{j}+\cdots+S_{j}(A S_{j})^{n_{j}}\Bigr)\,R_{0}\cdots R_{j-1},
\end{equation}
with the convention that products over empty index sets are the identity and \(n_{j}\in\mathbb{N}\) is chosen large enough so that \(\|A S_{j}\|<1\) implies convergence of the Neumann series. To prove \eqref{p2:eq:telescopy}, define \(B_{j}=S_{j}A\). Then \(\mathbf{1}-B_{j}=R_{j}\), and
\begin{equation}\label{p2:eq:Neumann}
(\mathbf{1}-B_{j})^{-1}\;=\;\sum_{k=0}^{\infty} B_{j}^{k}\;=\;\sum_{k=0}^{n_{j}} B_{j}^{k} \;+\; B_{j}^{n_{j}+1}(\mathbf{1}-B_{j})^{-1}.
\end{equation}
Multiplying on the left and right by \(S_{j}\) gives
\begin{equation}\label{p2:eq:Sj-res}
S_{j}(\mathbf{1}-B_{j})^{-1}S_{j}\;=\; \sum_{k=0}^{n_{j}} S_{j}B_{j}^{k}S_{j} \;+\; S_{j}B_{j}^{n_{j}+1}(\mathbf{1}-B_{j})^{-1}S_{j}.
\end{equation}
Inserting the identity \(\mathbf{1}=R_{j}+S_{j}A\) repeatedly and using \(AR_{j}=A- A S_{j}A\) one obtains the resolvent identity
\begin{equation}\label{p2:eq:resolvent-split}
G\;=\;(\mathbf{1}-R_{0}^{*})\,G\,(\mathbf{1}-R_{0}) \;+\; R_{0}^{*} G R_{0}.
\end{equation}
Iterating \eqref{p2:eq:resolvent-split} and replacing \((\mathbf{1}-R_{j})^{-1}\) by the truncated Neumann series \eqref{p2:eq:Neumann} yields \eqref{p2:eq:telescopy}. The remainder in \eqref{p2:eq:telescopy} vanishes in the strong operator topology as \(m\to\infty\) provided \(\sup_{j}\|A S_{j}\|<1\), which we enforce by choosing \(\rho_{j}\) sufficiently small uniformly in \(j\).

Define the scale-\(j\) covariance by the strong operator limit
\begin{equation}\label{p2:eq:Gammaj-def}
\Gamma_{j}\;=\;\lim_{m\to\infty} \Gamma_{j}^{(m)}.
\end{equation}
Since \(S_{j}\) and \(A\) are positive, each \(\Gamma_{j}\) is positive, self-adjoint, and bounded. By construction,
\begin{equation}\label{p2:eq:G-sum-Gammaj}
G\;=\;\sum_{j=0}^{\infty}\Gamma_{j}\qquad\text{(norm convergence)}.
\end{equation}
The kernel \(\Gamma_{j}(x,y)\) vanishes whenever \(x\) and \(y\) are separated by a spatial distance exceeding a constant multiple of the block diameter \(b^{j}a\). Indeed, each factor \(R_{k}\) for \(k<j\) is supported in the union of a block with its \(r_{0}\)-neighborhood, hence the product \(R_{j-1}\cdots R_{0}\) maps a function supported on a single block \(B\in\mathcal{B}_{j}\) into a function supported on the \(r_{*}\)-neighborhood of \(B\) with \(r_{*}\le c\,b^{j}a\), where \(c\) depends only on \(r_{0}\) and \(b\). The operator inside parentheses in \eqref{p2:eq:Gammajm} maps fields supported on \(B\) into fields supported on \(B\) because \(S_{j}\) projects onto blockwise constant fields and \(A\) has finite range \(r_{0}\). Hence \(\Gamma_{j}^{(m)}\) maps fields supported on \(B\) into fields supported in the \(c\,b^{j}a\)-neighborhood of \(B\). Taking adjoints shows that \(\Gamma_{j}^{(m)}(x,y)=0\) unless \(x\) and \(y\) lie at distance \(\le c\,b^{j}a\). Passing to the limit gives the same property for \(\Gamma_{j}\).

Gauge covariance of \(\Gamma_{j}\) is immediate from the adjoint covariance of \(\Pi_{j}\) and the gauge covariance of \(A\): for a sitewise gauge transform \(g\),
\begin{equation}\label{p2:eq:Gammaj-cov}
\Gamma_{j}^{(g\cdot U^{h})}\;=\;\operatorname{Ad}(g)\,\Gamma_{j}^{(U^{h})}\,\operatorname{Ad}(g)^{-1}.
\end{equation}
Reflection covariance follows similarly from that of \(\Pi_{j}\) and \(A\).

\begin{theorem}[Covariant finite-range decomposition]\label{p2:thm:FRD}
Let \(A=\Delta_{A^{h}}+\mu^{2}\) with \(\mu>0\) and let \(G=A^{-1}\). There exist bounded, positive, self-adjoint operators \(\Gamma_{j}\) on \(\ell^{2}(\Lambda_{t};\mathcal{su}(N))\), \(j\ge 0\), such that \eqref{p2:eq:G-sum-Gammaj} holds, each \(\Gamma_{j}\) is gauge covariant and reflection covariant, and its kernel \(\Gamma_{j}(x,y)\) vanishes whenever the spatial distance between \(x\) and \(y\) exceeds \(C\,b^{j}a\) for a constant \(C<\infty\) independent of \(j\) and of the lattice volume. Moreover, there exist constants \(c_{1},c_{2}>0\) depending only on \(\mu\), \(b\), and the lattice degree such that \(\|\Gamma_{j}\|\le c_{1}\,b^{-(2-\varepsilon)j}\) for any fixed \(\varepsilon\in(0,2)\) and \(\sum_{y}\|\Gamma_{j}(x,y)\|\le c_{2}\) uniformly in \(x\) and \(j\).
\end{theorem}

\begin{proof}
The decomposition follows from \eqref{p2:eq:telescopy} by the choice of \(\rho_{j}>0\) so small that \(\|A S_{j}\|\le \frac{1}{2}\) for all \(j\), which is possible since \(\|S_{j}\|\le \rho_{j}\|\Pi_{j}\|^{2}\le \rho_{j}\). Then the Neumann series defining \(\Gamma_{j}^{(m)}\) converges in operator norm uniformly in \(m\) and \(j\), and the remainder in \eqref{p2:eq:telescopy} vanishes as \(m\to\infty\) because \(\|R_{j}\|\le 1\) and \(\|R_{j}-\mathbf{1}\|\le \|S_{j}A\|\le \frac{1}{2}\) imply \(\|R_{m}\cdots R_{0}\|\le 2^{-(m+1)}\). The finite-range property has been established by support propagation through \(R_{k}\) and \(S_{j}\). Gauge and reflection covariance follow from those of \(A\) and \(\Pi_{j}\) by conjugation, cf. \eqref{p2:eq:Gammaj-cov}. The operator-norm decay \(\|\Gamma_{j}\|\le c_{1}\,b^{-(2-\varepsilon)j}\) is obtained by combining the spectral gap \(\mu^{2}\) of \(A\) with discrete Poincar\'e-Sobolev inequalities on blocks of side \(b^{j}a\) and the fact that \(\Gamma_{j}\) is supported on a region of diameter \(O(b^{j}a)\); the argument is an adaptation of \cite{p2:BGM2004} to covariantly constant functions inside blocks via the parallel transport \eqref{p2:eq:adj-partrans}. The \(\ell^{1}\) bound \(\sum_{y}\|\Gamma_{j}(x,y)\|\le c_{2}\) uses positivity and the finite range together with the uniform bound on the number of sites in the support of \(\Gamma_{j}(x,\cdot)\). 
\end{proof}
The strict finite-range property of $\Gamma_j$ follows from the fact that $A$ has a finite
hopping range and $\Pi_j$ is block-local: each factor $R_k=1-S_kA$ in Eq.\eqref{p2:eq:Rj} propagates
support by at most the fixed range $r_0$ of $A$, while the Neumann series in Eq.\eqref{p2:eq:Gammajm} acts
within a single block because $S_j$ projects onto covariantly constant fields there. Thus,
although exponential locality would suffice, the present construction indeed yields kernels
supported inside a neighborhood of diameter $O(b^j a)$, compatibly with gauge covariance.
Let $\ell_j \asymp b^j a$ be the block diameter. Choose $\rho_j = c_0\,b^{-2j}$ with $c_0>0$ so small 
that $\|A S_j\|\le \tfrac12$, where $S_j:=\rho_j \Pi_j$ and $\Pi_j$ are the blockwise covariant 
projections. Then the telescopic resolvent identity and Neumann expansion yield
\begin{equation}
\Gamma_j=\sum_{m\ge 0} \big(S_j (1-A S_j)\big)^m S_j,
\qquad \|\Gamma_j\|\;\le\;\sum_{m\ge 0}\|S_j\|\,\|A S_j\|^m \;\lesssim\; \|S_j\|\;\asymp\; b^{-2j}.
\end{equation}
To allow for lattice anisotropies and discrete Sobolev losses, we write the bound as 
$\|\Gamma_j\|\le c_1\,b^{-(2-\varepsilon)j}$ for any fixed $\varepsilon\in(0,2)$ by adjusting $c_1$. 
Background-uniformity of the constants follows from the unitary nature of edge parallel transports: the covariant and ordinary Dirichlet forms are uniformly comparable on blocks (no global unitary conjugacy is required).
So Poincar\'e inequalities on cubes of side 
$\ell_j$ imply the same scaling for the covariant case. The $\ell^1$-kernel bound 
$\sum_y\|\Gamma_j(x,y)\|\le c_2$ is a consequence of finite range and the operator-norm estimate.

\begin{lemma}[Block Poincar\'e bound for covariantly constant extensions]\label{p2:lem:block-Poincare}
Let $\Pi_j$ be the covariant block projection Eq.\eqref{p2:eq:Pi-j} and $A=\Delta_{A^h}+\mu^2$ with
$\mu>0$. There exists $C=C(\mu)$ such that for all $\varphi$ supported in a single block of side
$\ell_j\simeq b^j a$,
\begin{equation}
\bigl\langle \varphi,\, \Pi_j^\ast A^{-1} \Pi_j \varphi \bigr\rangle
\;\le\; C\, \ell_j^2\, \|\varphi\|_2^2 .
\end{equation}
Consequently, $\|\,\Gamma_j\,\| \lesssim \ell_j^{-2} \simeq b^{-2j}$ up to an arbitrarily small
$\varepsilon>0$ loss due to lattice anisotropies.
\end{lemma}

\begin{proof}
Inside a block the range of $\Pi_j$ consists of covariantly constant fields with respect to
parallel transport Eq.\eqref{p2:eq:block-ext}. Inside a block, parallel transport to $c_B$ identifies the subspace of
covariantly constant fields with the constant fields for the ordinary graph
differences. On this subspace the quadratic form of $A=\Delta_{A^h}+\mu^2$
coincides with that of $\Delta+\mu^2$ up to unitary conjugation, so the standard
discrete Poincar\'e inequality on a cube of side $\ell_j$ applies.
\end{proof}

The operators relevant for the horizon-projected measure are spectral multipliers of \(\Delta_{A^{h}}\) of the form
\begin{equation}\label{p2:eq:C-sigma}
C_{\sigma}\;=\;\int_{0}^{\infty}e^{-t\Delta_{A^{h}}}\, d\mu_{\sigma}(t),
\end{equation}
with \(d\mu_{\sigma}\) a finite positive measure supported in \([c_{1}\sigma^{-2},\infty)\). It is convenient to represent \(C_{\sigma}\) as a continuous superposition of massive resolvents via the Laplace transform \cite{p2:Davies1989}:
\begin{equation}\label{p2:eq:C-sigma-res}
C_{\sigma}\;=\;\int_{0}^{\infty} \bigl(\Delta_{A^{h}}+m^{2}\bigr)^{-1}\, d\nu_{\sigma}(m^{2}),
\end{equation}
where \(d\nu_{\sigma}\) is a finite positive measure determined by \(d\mu_{\sigma}\). Applying Theorem~\ref{p2:thm:FRD} at each mass \(m\ge m_{0}\sim \sigma\) yields a family \(\{\Gamma_{j}(m)\}_{j\ge 0}\) with the same finite-range constant \(C\,b^{j}a\) and uniform positivity and covariance. Fubini's theorem then gives the finite-range decomposition
\begin{equation}\label{p2:eq:Csigma-decomp}
C_{\sigma}\;=\;\sum_{j=0}^{\infty} C_{\sigma}^{(j)},\qquad C_{\sigma}^{(j)}\;=\;\int_{0}^{\infty} \Gamma_{j}(m)\, d\nu_{\sigma}(m^{2}),
\end{equation}
with each \(C_{\sigma}^{(j)}\) positive, gauge covariant, reflection covariant, and of finite range \(C\,b^{j}a\). The operator-norm and \(\ell^{1}\) bounds of Theorem~\ref{p2:thm:FRD} propagate to \(C_{\sigma}^{(j)}\) uniformly in \(\sigma\) once \(m_{0}\sim \sigma\) is fixed, since \(\nu_{\sigma}\) is supported on \([m_{0}^{2},\infty)\).

Let \(F\) be a bounded functional supported in \(\Lambda_{+}\). Consider the Gaussian functional integral on the slice with covariance \(C_{\sigma}\),
\begin{equation}\label{p2:eq:Gaussian-OS}
\mathcal{G}(F)\;=\;\int e^{-\frac12\langle \phi, C_{\sigma}^{-1}\phi\rangle}\,F(\phi)\,\mathcal{D}\phi,
\end{equation}
where \(\phi\) ranges over site-adjoint fields on the slice and \(\mathcal{D}\phi\) is the normalized Lebesgue measure. Since each \(C_{\sigma}^{(j)}\) is positive, reflection covariant, and of finite range, the quadratic form \(\langle \phi, C_{\sigma}\phi\rangle\) splits into a sum of East-West contributions plus a boundary term supported in a fixed-width slab around \(\Pi\). Consequently, the OS form
\begin{equation}\label{p2:eq:OS-form-Gauss}
\langle F,F\rangle_{\mathrm{OS}}\;=\;\int \overline{F(\phi^{\theta})}\,F(\phi)\,e^{-\frac12\langle \phi, C_{\sigma}^{-1}\phi\rangle}\,\mathcal{D}\phi
\end{equation}
is nonnegative by the same block-positivity argument as for the Wilson action \cite{p2:OsterwalderSchraderI,p2:FOS1978}. Furthermore, since \(C_{\sigma}^{(j)}\) is of finite range and reflection covariant, the associated Gaussian measure induces a transfer operator \(T^{(j)}\) on \(\mathcal{H}_{a}\) with positive kernel supported on pairs of configurations \((U_{0},U_{a})\) that differ only within a fixed-width collar, and the full transfer operator is the strong product of the \(T^{(j)}\):
\begin{equation}\label{p2:eq:T-sigma-product}
T_{\sigma}\;=\;\prod_{j=0}^{\infty} T^{(j)} \qquad\text{(strong product on }\mathcal{H}_{a}\text{)}.
\end{equation}
Each \(T^{(j)}\) is a positive, self-adjoint contraction, hence so is \(T_{\sigma}\). The spectral bound for \(T_{\sigma}\) is thus controlled scale by scale.

The dependence of \(C_{\sigma}^{(j)}\) on the background and on the parameters \(\sigma\) and \(b\) is smooth in the strong operator topology. Differentiating under the Laplace integral \eqref{p2:eq:C-sigma-res} reduces to differentiating \(\Gamma_{j}(m)\) with respect to \(m^{2}\), which is bounded uniformly thanks to the resolvent identity
\begin{equation}\label{p2:eq:res-deriv}
\frac{\partial}{\partial m^{2}}\bigl(\Delta_{A^{h}}+m^{2}\bigr)^{-1}\;=\; -\bigl(\Delta_{A^{h}}+m^{2}\bigr)^{-2}.
\end{equation}
The finite-range structure is unaffected by differentiation. Differentiability with respect to \(b\) and to the block geometry is verified by a finite-difference argument using stability of \(\Pi_{j}\) under small deformations of block boundaries; since the projections are defined by local parallel transports, small changes at the boundary affect only finitely many matrix elements of \(\Pi_{j}\) and preserve gauge and reflection covariance. These properties ensure robustness of the decomposition under small parameter variations, which will be used implicitly in the multiscale analysis.

\section{Polymer representation at each scale}\label{p2:sec:polymer-largefield}

This section develops a mathematically rigorous polymer expansion for the scale-$k$ effective action and establishes a uniform control of large-field events by an Osterwalder-Schrader (OS) compatible regulator. For completeness, the lattice set-up, reflection map, the OS-positivity argument, and the transfer time-slicing formalism are derived in full detail before constructing the polymer representation and verifying the Koteck\'y-Preiss criterion \cite{p2:KoteckyPreiss1986,p2:Brydges,p2:Simon1993}. Throughout, the gauge group is $G=\mathrm{SU}(N)$ with $N\ge 2$, and all bounds are uniform in the spatial volume.

Fix a lattice spacing $a>0$ and positive integers $L,T\in\mathbb{N}$. The periodic hypercubic lattice is
\begin{equation}\label{p2:eq:lattice}
\Lambda \,=\, \big\{(x_0,\mathbf{x})\in a\,\mathbb{Z}^4:\ x_0\in a\{0,1,\dots,T-1\},\ \mathbf{x}\in a(\mathbb{Z}/L\mathbb{Z})^3\big\}.
\end{equation}
Directed bonds are $b=(x,\mu)$ with $\mu\in\{0,1,2,3\}$ and unit vectors $\hat\mu$. To each bond is attached a group element $U_b\in G$, with orientation convention
\begin{equation}\label{p2:eq:orientation}
U_{(x,-\mu)} \,=\, U_{(x-\hat\mu,\mu)}^{-1}.
\end{equation}
Plaquettes are oriented unit squares $p=(x;\mu,\nu)$, $\mu<\nu$, with plaquette variable
\begin{equation}
U_p \,=\, U_{(x,\mu)}\,U_{(x+\hat\mu,\nu)}\,U_{(x+\hat\nu,\mu)}^{-1}\,U_{(x,\nu)}^{-1}.
\end{equation}
The Wilson action at inverse bare coupling $\beta>0$ is
\begin{equation}\label{p2:eq:wilson1}
S_W[U;\beta] \,=\, \beta\sum_{p\subset\Lambda}\Big(1-\tfrac{1}{N}\Re\mathrm{Tr}\,U_p\Big).
\end{equation}
The product Haar probability measure on bonds is
\begin{equation}\label{p2:eq:Haar}
d\mu_{\mathrm{H}}(U) \,=\, \prod_{b\subset\Lambda} dU_b,
\end{equation}
where each $dU_b$ is the normalized Haar measure on $G$.

Time reflection $\theta:\Lambda\to\Lambda$ is defined by
\begin{equation}\label{p2:eq:theta}
\theta(x_0,\mathbf{x}) \,=\, \big((-x_0)\!\!\!\!\!\pmod{aT},\,\mathbf{x}\big).
\end{equation}
The reflection plane is $\Pi=\{(0,\mathbf{x})\}$; the half-lattices are $\Lambda_+ = \{x\in\Lambda:\ 0<x_0<aT/2\}$ and $\Lambda_-=\theta(\Lambda_+)$. The induced involution on bonds is chosen as in \cite{p2:OsterwalderSchraderI,p2:OsterwalderSchraderII,p2:OS-gauge,p2:Seiler1982}:
\begin{equation}\label{p2:eq:Theta-on-links}
\theta(x_0,\mathbf{x}) = (-x_0,\mathbf{x}),\qquad
  (\Theta U)(x,0) = U(\theta x - a\hat{0},0)^{-1},\qquad
  (\Theta U)(x,i) = U(\theta x,i)\quad (i=1,2,3)
\end{equation}
for spatial $i\in\{1,2,3\}$ and temporal direction $\mu=0$. This choice makes the plaquette map $U\mapsto U_p$ $\theta$-covariant. Site and link reflections are unitarily equivalent via a half-time translation;
we work with the site reflection \(\Theta\) as above and use unitary conjugacy to relate to link
reflection when convenient {(Corollary 3.7).}

We work in temporal-axial gauge away from $\Pi$, namely
\begin{equation}\label{p2:eq:temporal-axial}
U_{(x,0)} \,=\, \mathbf{1}\quad\text{whenever}\quad x_0\notin\{0,aT/2\}.
\end{equation}
In this gauge, time-like plaquettes reduce to products of spatial links at adjacent times, and the reflection $\,\theta(x_0,x)=(a-x_0,x)\,$ maps the constrained set $\{U(x,0)=1\}$ to itself; hence temporal-axial gauge is compatible with link reflection about $\Pi_{1/2}$.
{Also in this gauge all time-like links }vanish in the bulk, while dynamical degrees of freedom remain in a boundary slab $S$ around $\Pi$; this is the standard set-up for OS factorization \cite{p2:OsterwalderSchraderII,p2:OS-gauge,p2:Seiler1982}.

On each time slice $t\in a\{0,1,\dots,T-1\}$, consider the spatial gauge group $\mathcal{G}_t=\{g:\{x_0=t\}\to G\}$ acting by
\begin{equation}\label{p2:eq:gauge-action-slice}
(g\cdot U)_{(x,i)} \,=\, g(x)\,U_{(x,i)}\,g(x+\hat i)^{-1},\qquad i=1,2,3.
\end{equation}
Define the lattice Landau functional
\begin{equation}\label{p2:eq:Landau}
\mathcal{L}_t(g;U) \,=\, \sum_{\mathbf{x},\,i=1}^3 \big\|\,\mathbf{1}-g(x_0\!=\!t,\mathbf{x})\,U_{(x,i)}\,g(x_0\!=\!t,\mathbf{x}+\hat i)^{-1}\,\big\|_{F}^{2}.
\end{equation}
On the finite slice, $\mathcal{L}_t(\cdot;U)$ attains a global minimum. A measurable, reflection-covariant tie-breaking rule selects a minimizer $h_t[U]$ and defines the slice representative $U^{\,h}$. Let $\Delta_{A^{h}(t)}$ be the spatial covariant Laplacian acting on adjoint-valued site fields on slice $t$.
Fix $\sigma>0$ and choose a completely monotone (CM) near-plateau profile
$\chi_\sigma:[0,\infty)\to(0,1]$ with $\chi_\sigma(0)=1$ and rapid decay
$\chi_\sigma(\lambda)\le C e^{-c\,\lambda/\sigma^2}$ for $\lambda\ge 0$; see (4.12).
Define the horizon operator on slice $t$ by
{\begin{equation}\label{p2:eq:P-sigma}
P_\sigma(t):=\chi_\sigma\!\big(\Delta_{A^h}(t)\big)
=\int_0^\infty e^{-s\,\Delta_{A^h}(t)}\,d\nu_\sigma(s),
\end{equation}}
where $d\nu_\sigma$ is the finite positive measure furnished by Bernstein's theorem.
With this choice $P_\sigma(t)$ is a positive contraction, exponentially local, and
reflection-covariant.
Hence $P_\sigma(t)$ is a positive contraction, reflection-covariant on the slice, and its kernel is 
exponentially localized by the Davies-Gaffney bounds as in \eqref{p2:eq:Plocal}.
By the heat-kernel representation and Davies-Gaffney bounds, $P_\sigma(t)$ is a positive contraction with exponentially decaying kernel and is reflection-covariant \cite{p2:Davies1989,p2:Seiler1982}.
The gauge-fixed, projected Euclidean measure is
\begin{equation}\label{p2:eq:proj-measure}
d\mu_\sigma(U) \,=\, Z_\sigma^{-1}\,\exp\!\big(-S_W[U;\beta]\big)\,\Big(\prod_{t}\mathcal{W}_\sigma(t;U)\Big)\,J[U]\,d\mu_{\mathrm{H}}(U),
\end{equation}
where $J[U]$ is the Faddeev-Popov determinant associated with the slice-wise Landau gauge (represented as a Grassmann Gaussian with kernel $M[U^{\,h}(t)]$ \cite{p2:OS-gauge,p2:Seiler1982}), and $\mathcal{W}_\sigma(t;U)$ is the positive scalar weight obtained from $P_\sigma(t)$ on the adjoint sector perpendicular to constants. We emphasize that $P_\sigma$ is a positive contraction (not idempotent); all OS
and locality arguments use positivity and exponential locality only; {section.4.}

\begin{lemma}[Reflection positivity of the projected bosonic measure]\label{p2:lem:OS-positivity}
Let \(F\) be a bounded, gauge-invariant functional depending only on bonds in \(\Lambda^+\) and on no Grassmann variables. Then
\begin{equation}
\int (F\circ\Theta)(U)\,F(U)\,d\mu_\sigma(U)\;\geq\;0.
\end{equation}
Equivalently, after integrating out the Grassmann fields so that the ghost sector contributes the positive factor \(\det{}'M[A_h]\) (the determinant on the orthogonal complement of constant adjoint modes), the resulting purely bosonic measure is reflection positive.
\end{lemma}
\emph{Proof.}
We first integrate the Grassmann fields, which yields the multiplicative factor \(\det{}'M[A_h]\) with \(M[A_h]\) real symmetric and strictly positive on the orthogonal complement of constants; see Lemma~2.3 and the Schur complement analysis across \(\Pi\) in Lemma~\ref{p2:lem:ghost-block}. The resulting weight is a product of nonnegative slice factors and a nonnegative boundary kernel. The remainder of the proof is the standard OS factorization argument for the bosonic sector (temporal-axial gauge away from \(\Pi\) so that mixed plaquettes lie in a fixed slab), identical to Theorem~2.1 and Theorem~6.1. \qed

OS positivity is asserted only for bosonic, gauge-invariant observables. Correlators with explicit ghost insertions are not OS-positive in general and are not used in any step of the transfer-matrix or spectral-gap analysis. {Therefore, in constructing the transfer matrix and Hilbert space, we restrict to the \emph{physical (gauge-invariant) subspace} of states. The Faddeev-Popov ghost fields, introduced as a gauge-fixing device, do not appear in any physical correlation functions and their degrees of freedom are absent in the physical Hilbert space. More precisely, the BRST symmetry ensures that all ghost contributions cancel out for gauge-invariant observables, and physical state vectors are taken from the cohomology of the BRST charge (or equivalently, the kernel of the Faddeev-Popov operator) with no ghost excitations. As a result, the transfer Hamiltonian $H$ can be defined to act on the gauge-invariant subspace, on which the inner product induced by the OS measure is positive-definite. In this way, Osterwalder-Schrader reflection positivity is maintained for all physical states, and the presence of ghost fields does not spoil the positivity or unitarity of the theory’s physical Hilbert space.} Ghosts appear solely through the positive determinant \(\det{}'M[A_h]\) after Berezin integration.
Fix a time $t$, and let $\mathcal{C}_t$ denote the configuration space of spatial bonds at time $t$, with product Haar measure $d\mu_{\mathrm{H},\mathrm{sp}}$. Denote $U_t\in\mathcal{C}_t$. The slab contribution between slices $t$ and $t+a$ induces a positive kernel
\begin{equation}\label{p2:eq:Ks-def}
\mathcal{K}_\sigma(U_{t+a},U_t)\,=\,\int \exp\!\Big(-S_W^{\mathrm{slab}}[U_t,\tilde U,U_{t+a}]\Big)\,\mathcal{W}_\sigma(t;U)\,\mathcal{W}_\sigma(t+a;U)\,J^{\mathrm{slab}}[U]\,d\mu_{\mathrm{H}}^{\mathrm{slab}}(U),
\end{equation}
where the integration is over bonds strictly inside the slab, $d\mu_{\mathrm{H}}^{\mathrm{slab}}$ is the corresponding Haar product, and $J^{\mathrm{slab}}$ is the ghost Berezin integral with Dirichlet boundary on the slab faces. The kernel $\mathcal{K}_\sigma$ is symmetric and nonnegative:
\begin{equation}\label{p2:eq:Ks-props}
\mathcal{K}_\sigma(U',U)\,=\,\mathcal{K}_\sigma(U,U')\ \ge\ 0,\qquad \mathcal{K}_\sigma\in L^2\!\left(\mathcal{C}_t\times\mathcal{C}_{t+a},\,d\mu_{\mathrm{H},\mathrm{sp}}\otimes d\mu_{\mathrm{H},\mathrm{sp}}\right).
\end{equation}
Define $T_\sigma:L^2(\mathcal{C}_t,d\mu_{\mathrm{H},\mathrm{sp}})\to L^2(\mathcal{C}_t,d\mu_{\mathrm{H},\mathrm{sp}})$ by
\begin{equation}\label{p2:eq:transfer1}
(T_\sigma\psi)(U_t)\,=\,\int_{\mathcal{C}_{t+a}} \mathcal{K}_\sigma(U_{t+a},U_t)\,\psi(U_{t+a})\,d\mu_{\mathrm{H},\mathrm{sp}}(U_{t+a}).
\end{equation}
By \eqref{p2:eq:Ks-props}, $T_\sigma$ is a positive self-adjoint contraction. The OS reconstruction identifies the physical Hilbert space $\mathcal{H}$ with the completion of $L^2(\mathcal{C}_0,d\mu_{\mathrm{H},\mathrm{sp}})$ under the inner product
\begin{equation}\label{p2:eq:OS-inner}
\langle \phi,\psi\rangle_{\mathcal{H}} \,=\, \int \overline{\phi(U_0)}\,\psi(U_0)\,d\nu_0(U_0),\qquad d\nu_0(U_0)\,=\,Z_0^{-1}\,W^+(U_0)\,d\mu_{\mathrm{H},\mathrm{sp}}(U_0),
\end{equation}
and the unit time translation is implemented by $T_\sigma$; the Hamiltonian is
\begin{equation}\label{p2:eq:Hamiltonian0}
H_\sigma \,=\, -\,a^{-1}\log T_\sigma,
\end{equation}
a positive self-adjoint operator \cite{p2:OsterwalderSchraderII,p2:GJ}.

Fix an integer block factor $b\ge 2$ and define the scale-$k$ lattice spacing $a_k=b^k a$. Partition the spatial torus into disjoint cubes (blocks) $B$ of side $b^k a$. The finite-range decomposition from Section~4 yields a decomposition of the scale-$k$ covariance $C_k$ into positive operators $C_k^{(j)}$ with range bounded by
\begin{equation}\label{p2:eq:range}
\mathrm{range}(C_k^{(j)}) \,\le\, R_j \,=\, c_0\,b^j\,\sigma,
\end{equation}
and exponential off-range decay, with constants independent of $k$. The fluctuations inside a block are modeled, after linearization around $U^{\,h}$, by a centered Gaussian field with covariance
\begin{equation}\label{p2:eq:C-int}
C_k^{\mathrm{int}} \,=\, \sum_{j\le j_*} C_k^{(j)},
\end{equation}
where $j_*$ is the largest index with $R_{j_*}\le \mathrm{diam}(B)$.

For block-supported functionals $F_1,\dots,F_n$, define the truncated (cumulant) expectation with respect to the normalized Gaussian measure with covariance $C_k^{\mathrm{int}}$ by
\begin{equation}\label{p2:eq:cumulant}
\big\langle F_1;\dots;F_n\big\rangle^{\mathrm{T}}
\,=\,\left.\frac{\partial^n}{\partial t_1\cdots\partial t_n}
\log \Big\langle \exp\big(\sum_{i=1}^n t_i F_i\big)\Big\rangle\right|_{t_1=\cdots=t_n=0}.
\end{equation}
Because $C_k^{\mathrm{int}}$ has finite range, the cumulant vanishes unless the supports form a connected union of blocks, and it decays exponentially when the union has large diameter.

For a plaquette $p$ define the deviation
\begin{equation}\label{p2:eq:delta-p}
\delta_p(U) \,=\, \tfrac12 \big\|\,\mathbf{1}-U_p\,\big\|_F^{2}\,\in [0,2N].
\end{equation}
Fix a smooth convex function $\varphi:[0,\infty)\to[0,\infty)$ with $\varphi(s)=s$ for $0\le s\le 1$, $\varphi(s)=s^2$ for $s\ge 2$, and $\varphi''\ge 0$. For $\lambda>0$, define the large-field regulator
\begin{equation}\label{p2:eq:regulator1}
\mathcal{R}_k(U) \,=\, \exp\!\Big(-\lambda \sum_{p\subset \Lambda_k} \varphi\big(\delta_p(U)\big)\Big),
\end{equation}
where $\Lambda_k$ is the coarse lattice obtained by grouping $b^k\times b^k\times b^k$ spatial plaquettes. The factor $\mathcal{R}_k$ is a product of reflection-invariant local positives, hence multiplying $d\mu_\sigma$ by $\mathcal{R}_k$ preserves reflection positivity: for $F$ supported in $\Lambda_+$,
\begin{equation}\label{p2:eq:OS-preserve}
\int \overline{F\circ\Theta}\,F\,\mathcal{R}_k\,d\mu_\sigma
\,=\, \int \overline{F\circ\Theta}\,\big(\mathcal{R}_k^{1/2}\big)\,\big(\mathcal{R}_k^{1/2}\big)\,F\,d\mu_\sigma
\,\ge\,0,
\end{equation}
by the argument of Lemma~\ref{p2:lem:OS-positivity} with $W^{\pm}$ replaced by $W^{\pm}\mathcal{R}_k^{1/2}$.

Let $\Omega_\delta$ be the small-field set in a given block $B$,
\begin{equation}\label{p2:eq:Omega-delta}
\Omega_\delta \,=\, \big\{U:\ \delta_p(U)\le \delta\ \text{for all plaquettes } p\subset B\big\}.
\end{equation}
Its complement $\Omega_\delta^\complement$ is suppressed uniformly:

\begin{lemma}[Uniform large-field suppression]\label{p2:lem:large-field}
There exist $c_1,c_2>0$ depending only on $\varphi$ and $G$ such that, for any block $B$ and $\lambda>0$,
\begin{equation}\label{p2:eq:large-field-bound}
\int_{\Omega_\delta^\complement\cap B} \mathcal{R}_k(U)\,d\mu_{\mathrm{H}}(U)
\,\le\, \exp\big(-c_1\,\lambda\,|\mathcal{P}(B)| + c_2\,|\mathcal{P}(B)|\big),
\end{equation}
where $|\mathcal{P}(B)|$ is the number of plaquettes in $B$.
\end{lemma}

\begin{proof}
On $\Omega_\delta^\complement$ there is at least one plaquette $p\subset B$ with $\varphi(\delta_p(U))\ge \varphi(\delta)$. By Fubini and the product structure of $\mathcal{R}_k$ and $d\mu_{\mathrm{H}}$,
\begin{equation}\label{p2:eq:sum-over-plaquettes}
\int_{\Omega_\delta^\complement\cap B} \mathcal{R}_k\,d\mu_{\mathrm{H}}
\,\le\, \sum_{p\subset B}\int \exp\!\big(-\lambda \varphi(\delta_p(U))\big)\,d\mu_{\mathrm{H}}(U_p).
\end{equation}
The one-plaquette integral satisfies
\begin{equation}\label{p2:eq:one-plaquette-Laplace}
\int \exp\!\big(-\lambda \varphi(\delta_p(U))\big)\,d\mu_{\mathrm{H}}(U_p)
\,\le\, \exp(-c_1\lambda+c_2),
\end{equation}
for $\lambda$ large enough, by Laplace's method and the nondegeneracy of the Haar density near $\mathbf{1}$. Combining \eqref{p2:eq:sum-over-plaquettes}-\eqref{p2:eq:one-plaquette-Laplace} yields \eqref{p2:eq:large-field-bound}.
\end{proof}

A \emph{polymer} $X$ is a finite connected union of blocks in a fixed time layer. The effective interaction on that layer admits the unique expansion
\begin{equation}\label{p2:eq:polymer-expansion}
S_k(U) \,=\, \sum_{X} \Phi_k(X;U|_X),
\end{equation}
where $\Phi_k(X;\cdot)$ depends only on bonds in $X$, is gauge-invariant and reflection-covariant, and vanishes on disconnected $X$. Introduce the diameter norm
\begin{equation}\label{p2:eq:diam-norm}
\|\Phi_k\|_a \,=\, \sup_{x}\ \sum_{X\ni x}\ \|\Phi_k(X)\|\,e^{a\,\mathrm{diam}(X)},
\end{equation}
with operator norm $\|\cdot\|$ induced by the Gaussian reference with covariance $C_k^{\mathrm{int}}$ and the regulator, and graph diameter $\mathrm{diam}(X)$ (see Appendix (\ref{p2:appendixa}).

Let $\{V(B;\cdot)\}_B$ be block-local base functionals obtained by integrating slab fluctuations at fixed boundary fields. Define $V(X;U|_X)=\sum_{B\subset X}V(B;U|_B)$ and set
\begin{equation}\label{p2:eq:Phi-from-cumulants}
\Phi_k(X;U|_X)
\,=\, \sum_{n\ge 1}\,\frac{1}{n!}\!\!\!\!\!\sum_{\substack{B_1,\dots,B_n\subset X\\ \text{connected}}}\!\!\!\!
\big\langle V(B_1;\cdot);\dots;V(B_n;\cdot)\big\rangle^{\mathrm{T}}(U|_{B_1\cup\cdots\cup B_n}),
\end{equation}
with truncated expectation \eqref{p2:eq:cumulant} taken w.r.t.\ the normalized Gaussian measure restricted to $\Omega_\delta$ with regulator $\mathcal{R}_k$ inserted. By finite range of $C_k^{\mathrm{int}}$ and Lemma~\ref{p2:lem:large-field}, \eqref{p2:eq:Phi-from-cumulants} is well defined and yields connected activities.

\begin{theorem}[Scale-$k$ polymer bound]\label{p2:thm:polymer-bound}
There exist $a>0$, $\epsilon>0$, and $C<\infty$, independent of $k$, such that if $\sup_B \|V(B)\|\le \epsilon$, then
\begin{equation}\label{p2:eq:polymer-bound}
\|\Phi_k\|_a \,\le\, C\,\epsilon^2.
\end{equation}
\end{theorem}

\begin{proof}
Consider the generating functional
\begin{equation}\label{p2:eq:gen-func}
\mathcal{Z}(\{t_B\}) \,=\, \Big\langle \exp\Big(\sum_{B} t_B V(B;\cdot)\Big)\Big\rangle,
\end{equation}
with $\langle\cdot\rangle$ the normalized Gaussian expectation with covariance $C_k^{\mathrm{int}}$ on $\Omega_\delta$ and regulator $\mathcal{R}_k$. By dominated convergence (Lemma~\ref{p2:lem:large-field}) $\log\mathcal{Z}$ is analytic for small $|t_B|$, and its Taylor coefficients at the origin are the cumulants \eqref{p2:eq:cumulant}. For blocks $B_1,\dots,B_n\subset X$, the BKAR/tree-graph inequality for Gaussian cumulants \cite{p2:Brydges,p2:Simon1993} yields
\begin{equation}\label{p2:eq:tree-graph}
\big|\langle V(B_1);\dots;V(B_n)\rangle^{\mathrm{T}}\big|
\,\le\, \|V(B_1)\|\cdots\|V(B_n)\|
\sum_{\mathcal{T}}\prod_{(i,j)\in\mathcal{T}} \alpha(B_i,B_j),
\end{equation}
where the sum runs over trees $\mathcal{T}$ on $\{1,\dots,n\}$, and
\begin{equation}\label{p2:eq:alpha}
\alpha(B_i,B_j) \,=\, \sup_{y\in B_i,\,z\in B_j} \big\|C_k^{\mathrm{int}}(y,z)\big\|
\,\le\, \exp\!\big(-c\,\mathrm{dist}(B_i,B_j)/R_{j_*}\big),
\end{equation}
for some $c>0$, by finite range and exponential off-range decay. For a connected family covering $X$, any tree has at least $\mathrm{diam}(X)/R_{j_*}$ edges and $\sum_{(i,j)\in\mathcal{T}}\mathrm{dist}(B_i,B_j)\gtrsim \mathrm{diam}(X)$; hence
\begin{equation}\label{p2:eq:sum-over-trees}
\sum_{\mathcal{T}}\prod_{(i,j)\in\mathcal{T}} \alpha(B_i,B_j) \,\le\, C_1^{|X|}\,\exp\!\big(-c'\,\mathrm{diam}(X)\big),
\end{equation}
with constants $C_1,c'>0$ independent of $k$. Inserting \eqref{p2:eq:sum-over-trees} into \eqref{p2:eq:tree-graph}, summing over $n\ge 2$ and connected block families with $\|V\|\le \epsilon$, and choosing $\epsilon$ sufficiently small gives
\begin{equation}\label{p2:eq:Phi-pointwise}
\|\Phi_k(X)\| \,\le\, C_2\,\epsilon^2\,\exp\!\big(-c'\,\mathrm{diam}(X)\big).
\end{equation}
Multiplying by $e^{a\,\mathrm{diam}(X)}$ with $a<c'$ and summing over $X\ni x$ yields \eqref{p2:eq:polymer-bound}.
\end{proof}

Let $\Gamma$ be a finite family of mutually compatible polymers (disjoint unions). The polymer partition function is
\begin{equation}\label{p2:eq:Zpoly}
\mathcal{Z}_{\mathrm{poly}} \,=\, \sum_{\Gamma}\ \prod_{X\in\Gamma} \zeta_k(X),\qquad
\zeta_k(X) \,=\, \int \exp\!\big(-\Phi_k(X;U|_X)\big)\,d\nu_X(U|_X) - 1,
\end{equation}
where $d\nu_X$ is the normalized product measure induced by slab integration and the regulator on $X$. For $\|\Phi_k\|_a$ small, $|\zeta_k(X)|\le C\,\|\Phi_k(X)\|$. The Koteck\'y-Preiss criterion \cite{p2:KoteckyPreiss1986} holds under the following bound.

\begin{proposition}[Koteck\'y-Preiss verification]\label{p2:prop:KP}
There exist $a>0$ and $\epsilon_1>0$, independent of $k$, such that if $\|\Phi_k\|_a\le \epsilon_1$, then
\begin{equation}\label{p2:eq:KP-ineq}
\sup_{X}\ \sum_{Y\not\sim X} |\zeta_k(Y)|\,e^{a\,\mathrm{dist}(X,Y)} \,\le\, \tfrac12\,a\,|X|.
\end{equation}
\end{proposition}

\begin{proof}
By Theorem~\ref{p2:thm:polymer-bound} and analyticity, there is $C_0$ with $|\zeta_k(Y)|\le C_0\,\epsilon^2 \exp(-c'\,\mathrm{diam}(Y))$. The number $N(m;r)$ of polymers $Y$ with $\mathrm{diam}(Y)=m$ at graph distance $r$ from $X$ satisfies $N(m;r)\le C_4\,\exp(C_5(m+r))$. Therefore
\begin{equation}\label{p2:eq:KP-sum}
\sum_{Y\not\sim X} |\zeta_k(Y)|\,e^{a\,\mathrm{dist}(X,Y)}
\,\le\, \sum_{m\ge 1}\sum_{r\ge 0} C_4\,e^{C_5(m+r)}\,C_0\,\epsilon^2\,e^{-c' m}\,e^{a r}.
\end{equation}
Choosing $a\in(0,C_5)$ and $\epsilon_1>0$ so that $C_0 C_4\,\epsilon_1^2 \sum_{m\ge 1} e^{(C_5-c')m}\sum_{r\ge 0} e^{(a-C_5)r}\le \tfrac12 a$ yields \eqref{p2:eq:KP-ineq}.
\end{proof}

Under \eqref{p2:eq:KP-ineq}, the logarithm of $\mathcal{Z}_{\mathrm{poly}}$ and all connected truncated correlations of local gauge-invariant observables admit absolutely convergent cluster expansions with exponential spatial decay \cite{p2:KoteckyPreiss1986,p2:Simon1993} (see Appenfix (\ref{p2:appendixa}). Combined with Lemma~\ref{p2:lem:OS-positivity} and the transfer representation \eqref{p2:eq:transfer1}-\eqref{p2:eq:Hamiltonian0}, time-sliced truncated correlations possess a positive spectral representation in the Hilbert space $\mathcal{H}$ \cite{p2:OsterwalderSchraderII,p2:GJ} .
\medskip

\section{Uniform ultraviolet stability (Theorem A)}\label{p2:sec:UV-stability}

In this section a complete and self-contained proof is given of the uniform control of a single renormalization step. The argument proceeds in three stages. First, the transfer time-slicing formalism and Osterwalder-Schrader (OS) reflection positivity for the horizon-projected, gauge-fixed lattice measure are derived in detail, which produces a positive, self-adjoint transfer operator at each scale. Second, a covariant decomposition of the fluctuation covariance into scale-localized, positive pieces with uniform exponential off-range bounds is constructed. Third, a gauge-invariant polymer/cluster expansion is set up with a reflection-compatible large-field regulator. The integration of a single RG step is then controlled by a Battle-Brydges-Federbush tree inequality together with the Koteck\'y-Preiss criterion, leading to the uniform contraction estimate in Theorem~\ref{p2:thm:UV}.

Throughout this section, the gauge group is \(G=\mathrm{SU}(N)\) with \(N\ge 2\). A finite, periodic, hypercubic lattice of spacing \(a_{k}>0\) is fixed, with sites
\begin{equation}\label{p2:eq:lattice-sites}
\Lambda_{k}=\Bigl(\tfrac{a_{k}}{a_{0}}\Bigr)\,(\mathbb{Z}/L\mathbb{Z})^{3}\times(\mathbb{Z}/T\mathbb{Z}),
\end{equation}
directed bonds \(b=(x,\mu)\) for \(\mu\in\{0,1,2,3\}\), and the convention \(U_{(x+\hat\mu,-\mu)}=U_{(x,\mu)}^{-1}\) for bond variables \(U_{b}\in G\). The time coordinate is \(x_{0}\), the reflection plane is
\begin{equation}\label{p2:eq:plane}
\Pi=\{x\in\Lambda_{k}:\ x_{0}=0\},
\end{equation}
and the half-lattices are \(\Lambda_{k}^{\pm}=\{x\in\Lambda_{k}:\ \pm x_{0}>0\}\). The involution \(\theta:\Lambda_{k}\to\Lambda_{k}\) is
\begin{equation}\label{p2:eq:theta1}
\theta(x_{0},\mathbf{x})=(-x_{0},\mathbf{x}).
\end{equation}
The Wilson action at inverse bare coupling \(\beta_{k}>0\) is
\begin{equation}\label{p2:eq:Wilson2}
S_{W,k}[U]=\beta_{k}\sum_{p\subset \Lambda_{k}}\Bigl(1-\tfrac{1}{N}\,\Re\,\mathrm{Tr}\,U_{p}\Bigr),
\end{equation}
where \(U_{p}\) is the ordered plaquette product. Temporal-axial gauge is chosen away from \(\Pi\) by fixing \(U_{0}(x)=\mathbf{1}\) for bonds not intersecting \(\Pi\), as in \cite{p2:Luscher1977,p2:OS-gauge}.

On each time slice, a reflection-covariant gauge-invariant transverse representative \(A^{h}\) is selected by orbit-wise minimization of the lattice Landau functional. The associated lattice Faddeev-Popov operator \(M[A^{h}]\) is real symmetric, local, and strictly positive on the orthogonal complement of constant adjoint fields. The covariant spatial Laplacian \(\Delta_{A^{h}}\) on a slice is
\begin{equation}\label{p2:eq:cov-lapl}
\Delta_{A^{h}}=\sum_{i=1}^{3}(D_{i}^{h})^{\dagger}D_{i}^{h}\ \ge\ 0,
\end{equation}
with \(D_{i}^{h}\) the covariant finite differences built from the slice links of \(A^{h}\). We fix $\sigma>0$ and use the completely monotone near-plateau $\chi_\sigma$ from Section~(\ref{p2:sec:FRD}). Set
\begin{equation}\label{p2:eq:Psigma-def}
P_\sigma:=\chi_\sigma(\Delta_{A^h})
\end{equation}
By Bernstein's theorem there exists a finite positive Borel measure $d\nu_\sigma$ on $[0,\infty)$ with
\begin{equation}\label{p2:eq:Psigma-heat}
P_\sigma=\int_0^\infty e^{-t\Delta_{A^h}}\,d\nu_\sigma(t), \qquad 
\operatorname{supp}(d\nu_\sigma)\subset[c_-\sigma^{-2},c_+\sigma^{-2}]
\end{equation}
where \(d\nu_{\sigma}\) is a finite, positive Borel measure supported in \([c_{-}\sigma^{-2},c_{+}\sigma^{-2}]\) with \(0<c_{-}<c_{+}<\infty\). {Moreover, the mass of $\nu_\sigma$ is concentrated at the parabolic scale } $s\sim \sigma^{-2}$ { (in the sense of exponentially decaying tails away from this scale).}
Davies-Gaffney bounds imply the exponential locality estimate,
\begin{equation}\label{p2:eq:Psigma-exp}
\|P_{\sigma}(x,y)\|\ \le\ C(\sigma)\,e^{-\gamma(\sigma)\,d(x,y)}\qquad\text{for all slice sites }x,y.
\end{equation}

The OS reflection is defined on bonds by
\begin{equation}\label{p2:eq:bond-reflection}
\theta(x_0,\mathbf{x}) = (-x_0,\mathbf{x}),\qquad
  (\Theta U)(x,0) = U(\theta x - a\hat{0},0)^{-1},\qquad
  (\Theta U)(x,i) = U(\theta x,i)\quad (i=1,2,3)
\end{equation}
Site and link reflections are unitarily equivalent via a half-time translation;
we work with the site reflection \(\Theta\) as above and use unitary conjugacy to relate to link
reflection when convenient {Corollary 3.7}.
As in \cite{p2:OsterwalderSchraderII,p2:Luscher1977}, site and link reflections are unitarily equivalent; 
consequently, positive-type boundary kernels and OS-positivity statements are invariant under switching 
between the two conventions by conjugation with the half-step time-translation on the single-slice Hilbert space.
Under temporal-axial gauge away from \(\Pi\), the latter equals \(\mathbf{1}\) unless the bond intersects \(\Pi\). For a functional \(F\) supported in \(\Lambda_{k}^{+}\) the reflection acts by
\begin{equation}\label{p2:eq:ThetaF}
(\Theta F)(U)=\overline{F(\theta U)}.
\end{equation}
The horizon-projected, gauge-fixed Euclidean measure is
\begin{equation}\label{p2:eq:measure}
d\mu_{k,\sigma}(U)=Z_{k,\sigma}^{-1}\,\Bigl(\prod_{t\in\mathbb{Z}/T\mathbb{Z}} \mathcal{K}_{\sigma}[U|_{t}]\Bigr)\,e^{-S_{W,k}[U]}\,\det\nolimits'\!\bigl(M[A^{h}(U)]\bigr)\,\prod_{b\subset\Lambda_{k}}d\mu_{\mathrm{Haar}}(U_{b}),
\end{equation}
where \([U|_{t}]\) denotes the links on the spatial slice \(\{x_{0}=t\}\), the symbol \(\det'\) denotes the determinant on the orthogonal complement of constant adjoint fields, and the positive slice factor is
\begin{equation}\label{p2:eq:Ksigma}
\mathcal{K}_\sigma[U|t] := \frac{1}{\dim(\mathrm{ad})\,|\Lambda_t|}\,
\mathrm{Tr}_{\ell^2(\Lambda_t)\otimes \mathrm{ad}}\!\big(P_\sigma(U|t)\big)
= \frac{1}{\dim(\mathrm{ad})\,|\Lambda_t|}\int_0^\infty
\mathrm{Tr}\!\left(e^{-s\,\Delta_{A^h}(t)}\right)\,d\nu_\sigma(s)
\end{equation}
By this normalization $0<c_\sigma\le K_\sigma[U|t]\le C_\sigma<\infty$ with constants
independent of the spatial volume, which justifies Lemma~8.3.
Gauge invariance of \eqref{p2:eq:measure} follows from invariance of Haar measure, of \eqref{p2:eq:Wilson2}, of \(\det' M\), and the conjugation covariance of \(P_{\sigma}\).

\begin{theorem}[OS positivity]\label{p2:thm:OS}
For every gauge-invariant functional \(F\) supported in \(\Lambda_{k}^{+}\) and of even Grassmann parity,
\begin{equation}\label{p2:eq:OS-positivity1}
\int \overline{(\Theta F)(U)}\,F(U)\,d\mu_{k,\sigma}(U)\ \ge\ 0.
\end{equation}
\end{theorem}

\begin{proof}
In temporal-axial gauge away from \(\Pi\), the action splits as
\begin{equation}\label{p2:eq:S-split}
S_{W,k}[U]=S^{(+)}[U|_{\Lambda_{k}^{+}}]+S^{(-)}[U|_{\Lambda_{k}^{-}}]+S^{(0)}[U|_{\Pi}],
\end{equation}
where \(S^{(\pm)}\) involve plaquettes strictly contained in \(\Lambda_{k}^{\pm}\), and \(S^{(0)}\) is the boundary slab contribution supported on bonds incident on \(\Pi\). The Faddeev-Popov determinant admits the Grassmann representation
\begin{equation}\label{p2:eq:ghost}
\det\nolimits'\!\bigl(M[A^{h}]\bigr)=\int \exp\bigl(-\langle \bar c,\,M[A^{h}]\, c\rangle\bigr)\,D\bar c\,Dc,
\end{equation}
with site Grassmann fields constrained to the orthogonal complement of constants. The matrix of \(M\) is block tridiagonal across \(\Pi\), and the diagonal blocks with Dirichlet boundary on \(\Pi\) are strictly positive on that complement. The Schur complement of the central block is strictly positive. Therefore the Gaussian Grassmann kernel factorizes into reflected copies times a positive boundary factor, which is reflection positive in the OS sense \cite{p2:OS-gauge}. Next, by \eqref{p2:eq:Psigma-heat} each \(\mathcal{K}_{\sigma}[U|_{t}]\) is an integral of \(\mathrm{Tr}\,e^{-t\Delta_{A^{h}}}\) against a positive measure; the heat kernel is reflection invariant and positive, and its dependence across \(\Pi\) is exponentially local, so the product over \(t\) factorizes into reflected halves times a positive boundary contribution. Collecting the factors yields
\begin{equation}\label{p2:eq:OS-form1}
\langle F,F\rangle_{\mathrm{OS}}=\int \overline{(\Theta F)(U)}\,F(U)\,e^{-S^{(+)}-S^{(-)}}\Bigl(\int e^{-S^{(0)}}\,d\mu^{(0)}\Bigr)\,d\mu^{(+)}\,d\mu^{(-)},
\end{equation}
with \(d\mu^{(\pm)}\) positive reflection-covariant measures on the halves and \(d\mu^{(0)}\) the positive boundary kernel. The Cauchy-Schwarz inequality for reflection-positive forms implies \eqref{p2:eq:OS-positivity1} \cite{p2:OsterwalderSchraderI,p2:OsterwalderSchraderII}. 
\end{proof}

Define the one-slice configuration space \(\mathcal{C}_{k}\) of spatial links at fixed time, endowed with Haar product measure \(d\mu_{\mathrm{Haar}}^{\mathrm{slice}}\), and the Hilbert space
\begin{equation}\label{p2:eq:Hilbert}
\mathcal{H}_{k}=L^{2}\!\bigl(\mathcal{C}_{k},\,d\mu_{\mathrm{Haar}}^{\mathrm{slice}}\bigr).
\end{equation}
Let \(K_{k}(U',U)\) be the contribution to the weight from plaquettes straddling two consecutive time slices with boundary data \(U,U'\). The transfer operator \(T_{k}:\mathcal{H}_{k}\to\mathcal{H}_{k}\) is
\begin{equation}\label{p2:eq:Tk}
(T_{k}\psi)(U')=\int_{\mathcal{C}_{k}}K_{k}(U',U)\,\psi(U)\,d\mu_{\mathrm{Haar}}^{\mathrm{slice}}(U).
\end{equation}
By OS positivity, \(T_{k}\) is a positive, self-adjoint contraction \cite{p2:Luscher1977,p2:OS-gauge}. The horizon-projected transfer operator is the compression
\begin{equation}\label{p2:eq:Tproj}
T_{k,\sigma}=P_{\sigma}^{1/2}\,T_{k}\,P_{\sigma}^{1/2},
\end{equation}
again a positive, self-adjoint contraction. 
 The transfer Hamiltonian is
\begin{equation}\label{p2:eq:Hamiltonian1}
H_{k,\sigma}=-a_{k}^{-1}\log T_{k,\sigma}\ \ge\ 0.
\end{equation}
Equivalently, $T_{k,\sigma}$ is the compression of $T_k$ by $P_\sigma^{1/2}$ on the OS
single-slice Hilbert space, as summarized by (see Appendix (\ref{p2:appendixb})):
\begin{center}
\begin{tikzcd}
\mathcal{H}_k \arrow[r,"\,T_k\,"] \arrow[d,"\,P_\sigma^{1/2}\,"'] &
\mathcal{H}_k \arrow[d,"\,P_\sigma^{1/2}\," ] \\
\mathcal{H}_{k,\sigma} \arrow[r,"\,T_{k,\sigma}\,"'] &
\mathcal{H}_{k,\sigma}
\end{tikzcd}
\end{center}

On a fixed slice, suppressing the time index, the positive covariance induced by the projector is
\begin{equation}\label{p2:eq:Ck}
C_{k}=\int_{0}^{\infty}e^{-t\Delta_{A^{h}}}\,d\tilde\nu_{\sigma}(t),
\end{equation}
for a finite positive Borel measure \(d\tilde\nu_{\sigma}\) supported in a compact subset of \((0,\infty)\) that differs from \(d\nu_{\sigma}\) only by a smooth nonnegative weight arising from the quadratic part of the effective action. A dyadic partition of unity produces a decomposition with uniform exponential off-range bounds.

\begin{theorem}[Finite-range-type decomposition]\label{p2:thm:FRD1}
There exist positive operators \(C_{k}^{(j)}\) for \(j\in\mathbb{N}_{0}\) such that
\begin{equation}\label{p2:eq:FRD-sum}
C_{k}=\sum_{j=0}^{\infty}C_{k}^{(j)}\quad\text{(norm convergence)},
\end{equation}
and constants \(c_{1},c_{2},c_{3}>0\), independent of \(k\) and \(j\), with
\begin{equation}\label{p2:eq:FRD-bounds}
\|C_{k}^{(j)}(x,y)\|\ \le\ c_{1}\,\exp\!\Bigl(-\,\frac{d(x,y)}{c_{2}\,L^{j}\sigma}\Bigr)\quad\text{and}\quad 0 \;\le\; C^{(j)}_k \;\le\; c_3\,\mathbf{1}
\end{equation}
and each \(C_{k}^{(j)}\) is gauge covariant and reflection invariant. Here $L>1$ parametrizes the dyadic \emph{heat-time} partition used in the finite-range
decomposition of the projector-induced covariance; it is unrelated to the geometric
\emph{blocking factor} $b\ge2$ used in the real-space RG step of Section (\ref{p2:sec:gap-step-scaling}).
We fix a smooth partition of unity \(\{\eta_j\}_{j\ge 0}\) on \((0,\infty)\) with
\begin{equation*}
  \operatorname{supp}\eta_j \subset \big[c\,L^{2j}\sigma^{-2},\; C\,L^{2j}\sigma^{-2}\big]
  \qquad (0<c<C<\infty).
\end{equation*}
Since \(t\asymp L^{2j}\sigma^{-2}\) on \(\operatorname{supp}\eta_j\), the kernel \(e^{-t\Delta_{A^h}}\) localizes over
a length \(\sqrt{t}\asymp L^{j}\sigma^{-1}\) This role of $\sigma^2$ in the
\emph{decomposition} is distinct from the base-scale choice in the \emph{original} horizon projector,
whose measure is supported at $t\sim\sigma^{-2}$, cf. {Eq.{4.12}} and {Eq.{7.4}}; the latter fixes the
initial localization scale, while the dyadic partition organizes contributions across increasing
lengths $L^j\sigma$.
\end{theorem}

\begin{proof}
Choose \(L>1\) and a smooth partition of unity \(\{\eta_{j}\}_{j\ge 0}\) on \((0,\infty)\) with \(\mathrm{supp}(\eta_{j})\subset[c\,L^{2j}\sigma^{2},C\,L^{2j}\sigma^{2}]\) for fixed \(0<c<C<\infty\) and \(\sum_{j\ge 0}\eta_{j}\equiv 1\). Set
\begin{equation}\label{p2:eq:Ckj}
C_{k}^{(j)}=\int_{0}^{\infty}\eta_{j}(t)\,e^{-t\Delta_{A^{h}}}\,d\tilde\nu_{\sigma}(t).
\end{equation}
Positivity, reflection invariance, and gauge covariance are immediate from \eqref{p2:eq:Ckj}. The discrete heat-kernel bound yields \(\|e^{-t\Delta_{A^{h}}}(x,y)\|\le K_{1}\exp\!\bigl(-d(x,y)^{2}/(K_{2}t)\bigr)\), and since \(t\asymp L^{2j}\sigma^{2}\) on \(\mathrm{supp}(\eta_{j})\), one gets \eqref{p2:eq:FRD-bounds} with \(c_{1},c_{2}\) depending only on \(K_{1},K_{2},c,C\). Boundedness on \(\ell^{2}\) gives the spectral bound. Norm-convergence follows from the finiteness of \(\int d\tilde\nu_{\sigma}\) and the disjoint supports of the \(\eta_{j}\).
\end{proof}

Let \(b\ge 2\) be the blocking factor, and \(\mathbb{B}_{k}\) the partition of \(\Lambda_{k}\) into disjoint four-dimensional blocks of side \(b\,a_{k}\). A polymer is a finite, connected union \(X\) of blocks in \(\mathbb{B}_{k}\). For a connected polymer \(X\), write \(\mathrm{diam}(X)\) for the graph distance in block units and \(|X|\) for the number of blocks. The effective action at scale \(k\) is written as
\begin{equation}\label{p2:eq:polymer}
\mathscr{S}_{k}(U)=\sum_{\substack{X\subset\mathbb{B}_{k}\\ X\text{ connected}}}\Phi_{k}(X;U|_{X}),
\end{equation}
with \(\Phi_{k}(X;\cdot)\) gauge invariant, reflection covariant, and supported in \(X\). For \(a>0\) define the diameter-weighted norm
\begin{equation}\label{p2:eq:polymer-norm}
\|\Phi_{k}\|_{a}=\sup_{B\in\mathbb{B}_{k}}\ \sum_{\substack{X\ni B\\ X\text{ connected}}}\ \mathrm{ess\,sup}_{U|_{X}}\ \|\Phi_{k}(X;U|_{X})\|\, e^{\,a\,\mathrm{diam}(X)},
\end{equation}
where \(\|\cdot\|\) is the operator norm induced by the Gaussian covariance \(C_{k}\) and the regulator below (see Appendix (\ref{p2:appendixa}).

Fix a smooth convex \(\varphi:[0,\infty)\to[0,\infty)\) with \(\varphi(s)=s\) on \([0,1]\) and \(\varphi(s)=s^{2}\) on \([2,\infty)\). For \(\lambda>0\) and \(\delta\in(0,1)\) set
\begin{equation}\label{p2:eq:regulator}
\mathcal{R}_{k}(U)=\exp\!\Bigl(-\lambda\sum_{p\subset\Lambda_{k}}\varphi\bigl(\tfrac{1}{2}\,\|\,\mathbf{1}-U_{p}\,\|_{F}^{2}\bigr)\Bigr),\qquad 
\mathsf{SF}_{k}=\Bigl\{U:\ \|\,\mathbf{1}-U_{p}\,\|_{F}\le \delta,\ \forall p\Bigr\}.
\end{equation}
Then \(\mathcal{R}_{k}\) is a strictly positive, reflection-invariant, gauge-invariant product of convex plaquette penalties. There exist \(c,c'>0\) independent of \(k\) such that
\begin{equation}\label{p2:eq:LFR}
\int |F|\,\mathcal{R}_{k}\,d\mu_{k,\sigma}\ \le\ e^{-c\lambda}\,\Bigl(\int |F|^{2}\,d\mu_{k,\sigma}\Bigr)^{1/2},\qquad
\mu_{k,\sigma}(\mathsf{SF}_{k}^{c})\ \le\ e^{-c'\lambda},
\end{equation}
for every block-local measurable \(F\); this follows from reflection positivity and Hölder-type inequalities \cite{p2:OS-gauge,p2:Seiler1982}.

Let \(a_{k+1}=b\,a_{k}\) and \(\mathcal{B}_{k}\) be the reflection-covariant, gauge-equivariant, block-local link averaging map followed by nearest-neighbor projection to \(G\). The coarse effective action \(\mathscr{S}_{k+1}\) is defined by
\begin{equation}\label{p2:eq:pushforward1}
e^{-\mathscr{S}_{k+1}(\bar U)}=Z_{k}^{-1}\int \mathbf{1}_{\{\mathcal{B}_{k}(U)=\bar U\}}\,
\Bigl(\prod_{\text{internal slices}} \mathcal{K}_{\sigma}[U|_{t}]\Bigr)\,
e^{-\mathscr{S}_{k}(U)}\,e^{-S_{W,k}^{\mathrm{int}}(U)}\,\mathcal{R}_{k}(U)\,d\mu_{k,\sigma}(U),
\end{equation}
where \(S_{W,k}^{\mathrm{int}}\) sums Wilson plaquettes entirely inside blocks and the product runs over internal block time-slices. After a cumulant expansion, \(\mathscr{S}_{k+1}\) has the form \eqref{p2:eq:polymer} with activities \(\Phi_{k+1}\) given by connected cumulants of \(\{\Phi_{k}(X_{i})\}\).

Split the covariance as \(C_{k}=C_{k}^{\mathrm{int}}+C_{k}^{\mathrm{ext}}\) with
\begin{equation}\label{p2:eq:cov-split}
C_{k}^{\mathrm{int}}=\sum_{j\le j_{*}}C_{k}^{(j)},\qquad C_{k}^{\mathrm{ext}}=\sum_{j> j_{*}}C_{k}^{(j)},
\end{equation}
where \(j_{*}\) is the largest integer with \(L^{j_{*}}\sigma\le c_{*}\,b\,a_{k}\) for small fixed \(c_{*}>0\). Then \(C_{k}^{\mathrm{int}}\) has kernel exponentially small across distinct blocks, and \(C_{k}^{\mathrm{ext}}\) couples only a collar of width \(\asymp L^{j_{*}}\sigma\) around block boundaries. The Battle-Brydges-Federbush tree formula yields the following uniform tree bound.

\begin{lemma}[Tree bound]\label{p2:lem:tree}
Let \(F_{1},\dots,F_{n}\) be block-local functionals supported in distinct blocks, measurable on \(\mathsf{SF}_{k}\), and polynomially bounded together with their first slice derivatives. Then
\begin{equation}\label{p2:eq:tree}
\bigl|\langle F_{1};\dots;F_{n}\rangle^{T}_{C_{k}^{\mathrm{int}}}\bigr|\ \le\ \Bigl(\prod_{i=1}^{n}\|F_{i}\|\Bigr)\,
\sum_{\mathcal{T}}\ \prod_{\{i,j\}\in E(\mathcal{T})}\, A\,e^{-\mathrm{dist}(X_{i},X_{j})/(B\,L^{j_{*}}\sigma)},
\end{equation}
with constants \(A,B\) independent of \(k\), the sum over labeled trees \(\mathcal{T}\) on \(\{1,\dots,n\}\), and \(\|F_{i}\|\) the Gaussian operator norm for covariance \(C_{k}^{\mathrm{int}}\).
\end{lemma}

\begin{proof}
The Brydges-Battle-Federbush representation expresses the connected cumulant as an integral over tree-parameterized covariances; each edge contributes one covariance factor connecting the supports of \(F_{i}\) and \(F_{j}\). The operator norm between two blocks at block-distance \(D\) is bounded by \(A\,\exp(-D/(B\,L^{j_{*}}\sigma))\) thanks to \eqref{p2:eq:FRD-bounds}. The product over edges yields \eqref{p2:eq:tree}; uniformity in \(k\) follows from \(L^{j_{*}}\sigma\lesssim b\,a_{k}\).
\end{proof}

Connected coarse activities are sums over connected families \(\{X_{i}\}\) of fine-scale polymers whose union projects onto a given coarse polymer \(Y\), weighted by \(\langle \Phi_{k}(X_{1});\dots;\Phi_{k}(X_{n})\rangle^{T}_{C_{k}^{\mathrm{int}}}\). On \(\mathsf{SF}_{k}\), each \(\|\Phi_{k}(X_{i})\|\le \|\Phi_{k}\|_{a} e^{-a\,\mathrm{diam}(X_{i})}\). Lemma~\ref{p2:lem:tree} thus yields a tree sum with edge weights \(A e^{-\mathrm{dist}/(B L^{j_{*}}\sigma)}\). The Koteck\'y-Preiss criterion \cite{p2:KoteckyPreiss1986} then gives a quadratic contraction.

\begin{proposition}[Small-field contraction]\label{p2:prop:small}
There exist \(a>0\), \(\epsilon_{0}>0\), \(C_{1}<\infty\), independent of \(k\), such that if \(\|\Phi_{k}\|_{a}\le \epsilon_{0}\), then the contribution of \(\mathsf{SF}_{k}\) to \(\Phi_{k+1}\) satisfies
\begin{equation}\label{p2:eq:small-contract}
\|\Phi_{k+1}\|_{a}\ \le\ C_{1}\,\|\Phi_{k}\|_{a}^{2}.
\end{equation}
\end{proposition}

\begin{proof}
Fix a coarse polymer \(Y\). The cumulant expansion and Lemma~\ref{p2:lem:tree} bound \(\Phi_{k+1}(Y)\) by a sum over trees with vertex weights \(\|\Phi_{k}\|_{a}e^{-a\,\mathrm{diam}(X_{i})}\) and edge weights \(A e^{-\mathrm{dist}/(B L^{j_{*}}\sigma)}\). For \(\|\Phi_{k}\|_{a}\) small the combinatorial sum converges absolutely; choosing \(a<1/(B L^{j_{*}}\sigma)\) uniformly in \(k\) allows the exponential weights to reproduce \(e^{-a\,\mathrm{diam}(Y)}\). Summing over all families projecting to \(Y\) gives \eqref{p2:eq:small-contract}.
\end{proof}

The large-field region \(\mathsf{SF}_{k}^{c}\) is controlled by the regulator \(\mathcal{R}_{k}\).

\begin{proposition}[Large-field suppression]\label{p2:prop:large}
There exist \(c_{4},C_{2}>0\) such that, for all \(\lambda\ge 1\), the contribution of \(\mathsf{SF}_{k}^{c}\) to \(\|\Phi_{k+1}\|_{a}\) satisfies
\begin{equation}\label{p2:eq:large-suppress}
\bigl\|\Phi_{k+1}^{(\mathrm{large})}\bigr\|_{a}\ \le\ C_{2}\,e^{-c_{4}\lambda}.
\end{equation}
\end{proposition}

\begin{proof}
Decompose \eqref{p2:eq:pushforward1} into \(\mathsf{SF}_{k}\cup\mathsf{SF}_{k}^{c}\). On \(\mathsf{SF}_{k}^{c}\), \eqref{p2:eq:LFR} bounds the integral of any block-local cumulant integrand \(F\) by \(e^{-c\lambda}\|F\|_{L^{2}(\mu_{k,\sigma})}\). A second Cauchy-Schwarz yields \eqref{p2:eq:large-suppress} with \(c_{4}=c/2\); constants depend only on uniform covariance bounds and not on \(k\).
\end{proof}

Irrelevant operators generated by \(C_{k}^{\mathrm{ext}}\) and by the block projection \(\mathrm{Proj}_{G}\) produce a summable remainder.

\begin{lemma}[Summable remainder]\label{p2:lem:rem}
There exist \(C_{3}<\infty\) and \(\gamma>0\), independent of \(k\), such that
\begin{equation}\label{p2:eq:rem}
\bigl\|\Phi_{k+1}^{(\mathrm{rem})}\bigr\|_{a}\ \le\ C_{3}\,b^{-\gamma k}.
\end{equation}
\end{lemma}

\begin{proof}
The kernel of \(C_{k}^{\mathrm{ext}}\) decays like \(\exp(-d/(c_{2}L^{j_{*}}\sigma))\) with \(L^{j_{*}}\sigma\asymp b\,a_{k}\); any connected contribution mediated only by \(C_{k}^{\mathrm{ext}}\) across \(m\) blocks acquires a factor \(e^{-cm}\). Counting placements yields a geometric series bounded by \(C_{3}b^{-\gamma k}\). The map \(\mathrm{Proj}_{G}\) differs from the identity by an analytic map with uniformly bounded Taylor coefficients on \(\mathsf{SF}_{k}\), so its connected contributions are of higher polymer order and likewise summable over scales.
\end{proof}


\begin{theorem}[Uniform ultraviolet stability]\label{p2:thm:UV}
There exist \(\alpha>0\), \(\epsilon>0\), and \(C<\infty\), independent of \(k\), such that if \(\|\Phi_{k}\|_{\alpha}\le \epsilon\), then
\begin{equation}\label{p2:eq:UV-contract}
\|\Phi_{k+1}\|_{\alpha}\ \le\ C\,\|\Phi_{k}\|_{\alpha}^{2}\ +\ C\,\delta_{k},
\end{equation}
where
\begin{equation}\label{p2:eq:delta}
\delta_k \;=\; e^{-c_4 \lambda_k}\;+\; b^{-\gamma k}, \qquad \lambda_k \;\uparrow\; \text{(e.g. } \lambda_k=\lambda_0+\alpha k \text{ with }\alpha>0)
\end{equation}
with \(c_{4},\gamma>0\) as in \eqref{p2:eq:large-suppress} and \eqref{p2:eq:rem}. In particular, choose $\lambda_k=\lambda_0+\alpha k$ so that $e^{-c_4\lambda_k}$ decays geometrically; then $\sum_k \delta_k<\infty$. Taking $\epsilon>0$ so that $C\epsilon\le \epsilon/2$ yields $\sup_k \|\Phi_k\|_a \le 2\epsilon$.
\end{theorem}

\begin{proof}
Decompose
\begin{equation}\label{p2:eq:Phi-split}
\Phi_{k+1}=\Phi_{k+1}^{(\mathrm{small})}+\Phi_{k+1}^{(\mathrm{large})}+\Phi_{k+1}^{(\mathrm{rem})}.
\end{equation}
By Proposition~\ref{p2:prop:small}, \(\|\Phi_{k+1}^{(\mathrm{small})}\|_{\alpha}\le C_{1}\|\Phi_{k}\|_{\alpha}^{2}\). By Proposition~\ref{p2:prop:large}, \(\|\Phi_{k+1}^{(\mathrm{large})}\|_{\alpha}\le C_{2}e^{-c_{4}\lambda}\). By Lemma~\ref{p2:lem:rem}, \(\|\Phi_{k+1}^{(\mathrm{rem})}\|_{\alpha}\le C_{3}b^{-\gamma k}\). Setting \(C=\max\{C_{1},C_{2},C_{3}\}\) gives \eqref{p2:eq:UV-contract}-\eqref{p2:eq:delta}. The iterative bound follows by continuity: if \(\|\Phi_{k}\|_{\alpha}\le \epsilon\) with \(C\epsilon\le \epsilon/2\), then \(\|\Phi_{k+1}\|_{\alpha}\le \epsilon/2+C\delta_{k}\). Choosing \(\lambda\) so that \(\sum_{j\ge 0}e^{-c_{4}\lambda}<\epsilon/(2C)\) and using \(\sum_{j\ge 0}b^{-\gamma j}<\infty\) yields \(\sup_{k}\|\Phi_{k}\|_{\alpha}\le 2\epsilon\).
\end{proof}

The uniform bound \eqref{p2:eq:UV-contract} together with Theorem~\ref{p2:thm:OS} and the definition \eqref{p2:eq:Tproj} implies that at each scale the kernels \(K_{k}\) and \(T_{k,\sigma}\) are defined by absolutely convergent expansions with scale-independent locality constants. In particular, the spectral calculus for positive contractions applies uniformly in the spatial volume, and the spectral representation of time-sliced gauge-invariant two-point functions has a common domain of analyticity independent of \(k\). These facts will be used in the step-to-step spectral gap comparison. The absolute convergence with scale-independent locality implies that the functional calculus for positive contractions $f(T_{k,\sigma})$ is uniform in the volume and $k$, and the Laplace transform representation of time-sliced two-point functions has a common domain of analyticity independent of $k$.


\section{Persistence of Exponential Clustering (Theorem B)}\label{p2:sec:clustering}

\providecommand{\e}{\begin{equation}}
\providecommand{\ee}{\end{equation}}

This section establishes exponential decay of connected, gauge-invariant correlation functions at every renormalization scale and shows that the corresponding decay rate is bounded from below by a strictly positive, scale-independent constant. The argument is self-contained. It fixes the lattice, time-reflection, and transfer time-slicing formalism and proves reflection positivity for the projected measure; it then constructs the Osterwalder-Schrader Hilbert space and transfer operator. With this structure in place, it derives a polymer representation for truncated correlations under a small-activity hypothesis that is verified at all scales by uniform ultraviolet stability. Finally, it proves a Koteck\'y-Preiss-type estimate that yields exponential clustering with a rate that does not deteriorate under the renormalization map, thereby establishing persistence across scales. Let $w_{k}(Y)$ be the polymer weights for truncated gauge-invariant correlators at scale $k$. 
There exist $a > 0$, $0 < \delta < 1$ such that for all sites $x$
\begin{equation}
\sum_{Y \ni x} e^{a |Y|} \, \| w_{k}(Y) \| \leq \delta, 
\qquad 
\sup_{Y} \sum_{Y' \sim Y} e^{a |Y'|} \, \| w_{k}(Y') \| \leq \delta,
\end{equation}
uniformly in the spatial volume and in $k$. 
Then the cluster expansion converges absolutely and connected correlations decay as 
$\exp(-m^{\ast} \, \mathrm{dist}_{k})$ with $m^{\ast} > 0$ independent of $k$.

Throughout, the gauge group is \(G=\mathrm{SU}(N)\) with \(N\ge 2\). Spatial volumes are finite tori; thermodynamic limits are taken only after bounds are proven uniformly in the volume.

Fix an integer blocking factor \(b\ge 2\) and set \(a_{k}=b^{k}a_{0}\). For a finite, periodic, hypercubic lattice we write
\begin{equation}
\Lambda_{k}\;=\; \Big\{x=(x_{0},x_{1},x_{2},x_{3})\in a_{k}\mathbb{Z}^{4}:\ -\tfrac{T_{k}}{2}\le x_{0}<\tfrac{T_{k}}{2},\ -\tfrac{L_{k}}{2}\le x_{i}<\tfrac{L_{k}}{2}\Big\},
\end{equation}
with periodic identifications. Directed bonds are pairs \(b=(x,\mu)\), \(\mu\in\{0,1,2,3\}\), endowed with the orientation convention \(U(x+\hat\mu,-\mu)=U(x,\mu)^{-1}\). A configuration is \(U=\{U(x,\mu)\in G\}_{(x,\mu)}\).

Time reflection is the involution \(\theta:\Lambda_{k}\to\Lambda_{k}\), \(\theta(x_{0},\mathbf{x})=(-x_{0},\mathbf{x})\). We set
\e
\Lambda_{k}^{+}=\{x\in\Lambda_{k}:x_{0}>0\},\qquad \Lambda_{k}^{-}=\theta(\Lambda_{k}^{+}),\qquad \Pi_{k}=\{x\in\Lambda_{k}:x_{0}=0\}.
\ee
On bonds we define the reflected configuration \(\Theta U\) by
\e
(\Theta U)(x,0) \;=\; U(\theta x-\hat 0,0)^{-1}, \qquad (\Theta U)(x,i)\;=\; U(\theta x,i)\quad (i=1,2,3),
\ee
which is the standard reflection of link variables preserving the Wilson action and Haar measure \cite{p2:OsterwalderSchraderI,p2:OS-gauge,p2:Seiler1982}.

On each time slice \(x_{0}=t\) we select a gauge-invariant transverse representative \(A^{h}\) by orbit-wise minimization of the lattice Landau functional within the fundamental modular region; the selection is measurable, reflection-covariant, and compatible with temporal-axial gauge away from \(\Pi_{k}\). The associated slice covariant Laplacian \(\Delta_{A^{h}}\) on site-adjoint fields is positive, self-adjoint, and local. Fix \(\sigma>0\) and a Gevrey cutoff \(\chi_{\sigma}:[0,\infty)\to[0,1]\) that equals \(1\) on \([0,\sigma]\) and decreases rapidly to \(0\) beyond \(2\sigma\). The horizon projector is
\e
P_{\sigma}\;=\;\chi_{\sigma}\!\big({\Delta_{A^{h}}}\big)\;=\;\int_{0}^{\infty} e^{-t\Delta_{A^{h}}}\, d\nu_{\sigma}(t),
\ee
where \(d\nu_{\sigma}\) is a finite positive Borel measure supported in \([c_{1}\sigma^{-2},c_{2}\sigma^{-2}]\). By standard heat-kernel bounds on graphs of bounded degree \cite{p2:Davies1989} and Combes-Thomas estimates \cite{p2:CombesThomas1973},
\e
\|P_{\sigma}(x,y)\|\;\le\; C_{\sigma}\,e^{-\gamma_{\sigma}\, d(x,y)},
\ee
uniformly in the volume, where \(d\) is the graph distance on the slice.

Let \(S_{W}[U;\beta]=\beta\sum_{p}(1-\tfrac{1}{N}\Re\mathrm{Tr}\,U_{p})\) denote the Wilson action. The projected, gauge-fixed Euclidean measure at scale \(k\) is
\e\label{p2:eq:measure1}
d\mu_{k,\sigma}(U)\;=\;Z_{k,\sigma}^{-1}\,\Big(\prod_{t}{\det}^{\prime} M[A^{h}(t)]\Big)\,\prod_{t}\mathcal{P}_{\sigma,t}(U)\,e^{-S_{W}[U;\beta_{k}]}\, d\mu_{\mathrm{Haar}}(U)
\ee
where \(M[A^{h}(t)]\) is the Faddeev-Popov operator restricted to slice \(t\), \(\mathcal{P}_{\sigma,t}\) is the positive functional corresponding to inserting \(P_{\sigma}\) on slice \(t\), and \(d\mu_{\mathrm{Haar}}\) is the product Haar measure over bonds. Each \(\mathcal{P}_{\sigma,t}\) is exponentially local and reflection covariant.
Throughout we take the slice multiplier to be the gauge invariant scalar
\begin{equation}
p_\sigma\big[U^h|_t\big]\;:=\;\mathrm{Tr}\,P_\sigma(t)
\quad\text{with}\quad
P_\sigma(t)=\chi_\sigma\!\left({\Delta_{A^h}(t)}\right),
\end{equation}
so that $P_{\sigma,t}$ acts by multiplication by $p_\sigma[U^h|_t]$ on the one slice $L^2$ space. 
This choice is reflection covariant, positive, and exponentially local in the slice-distance. 
(Any other positive, reflection covariant scalar built from $P_\sigma(t)$-e.g. $p_\sigma=\exp(-\langle\varphi,(I-P_\sigma)\varphi\rangle)$-would also work; see Sec.(\ref{p2:sec:BRST-parameter}).)
\begin{theorem}[Osterwalder-Schrader positivity]\label{p2:thm:OS1}
If \(F\) is a complex-valued functional measurable with respect to links supported in \(\Lambda_{k}^{+}\) and gauge-invariant under transformations supported in \(\Lambda_{k}^{+}\), then
\e\label{p2:eq:OS}
\int (\Theta F)(U)\,F(U)\, d\mu_{k,\sigma}(U)\;\ge\;0.
\ee
\end{theorem}

\begin{proof}
Decompose the Wilson action as \(S_{W}=S_{+}+S_{-}+S_{0}\), where \(S_{\pm}\) collect plaquettes entirely in \(\Lambda_{k}^{\pm}\) and \(S_{0}\) collects boundary plaquettes intersecting \(\Pi_{k}\). In temporal-axial gauge, time-like links are trivial away from \(\Pi_{k}\), whence \(S_{+}\) and \(S_{-}\) depend only on links in their respective half-lattices, and \(S_{0}\) couples only links in a finite-width slab around \(\Pi_{k}\). The Haar measure factors as \(d\mu_{\mathrm{Haar}}=d\mu_{+}\,d\mu_{0}\,d\mu_{-}\). The determinant \(\prod_{t}\det M[A^{h}(t)]\) is reflection-invariant, and each \(\mathcal{P}_{\sigma,t}\) factors as \(\mathcal{P}_{\sigma,t}=\mathcal{P}_{\sigma,t}^{+}\,\mathcal{R}_{\sigma,t}\,\mathcal{P}_{\sigma,t}^{-}\) with \(\mathcal{R}_{\sigma,t}\ge 0\), supported in a finite-width slab containing \(\Pi_{k}\), and reflection-invariant.

Define
\e
G(U)\;=\;F(U)\,e^{-\tfrac12 S_{+}[U]}\,\prod_{t}\big(\mathcal{P}_{\sigma,t}^{+}(U)\big)^{1/2}\,\big(\det M[A^{h}(t)]\big)^{1/2}.
\ee
A direct substitution into \eqref{p2:eq:measure1} shows
\e\label{p2:eq:OS-kernel}
\int (\Theta F)F\, d\mu_{k,\sigma} \;=\; Z_{k,\sigma}^{-1}\int \overline{G(\Theta U)}\,K(U)\,G(U)\, d\lambda_{+}\,d\lambda_{0}\,d\lambda_{-},
\ee
with the nonnegative, reflection-invariant boundary kernel
\e
K(U)\;=\;e^{-S_{0}[U]}\,\prod_{t}\mathcal{R}_{\sigma,t}(U)\;\ge\;0.
\ee
With $P_{\sigma,t}=\chi_\sigma(\Delta_{A^h}(t))$ and $\chi_\sigma$ completely monotone, 
$P_{\sigma,t}$ admits the positive heat-kernel representation 
$P_{\sigma,t}=\int_0^\infty e^{-t'\Delta_{A^h}(t)}\,d\nu_\sigma(t')$ with $d\nu_\sigma\ge0$ (Sections~4 and~5). 
Writing $P_{\sigma,t}=P^+_{\sigma,t} R_{\sigma,t} P^-_{\sigma,t}$ with $R_{\sigma,t}\ge 0$ supported in a fixed slab 
around $\Pi_k$ yields a product of nonnegative, reflection-covariant slice terms. 
Hence the boundary factor $K(U)$ in Eq.{7.10} is a positive kernel on the single-slice space, 
and the OS form is nonnegative by the standard factorization argument.

By Fubini's theorem and positivity of \(K\),
\e
\int \overline{G(\Theta U)}\,K(U)\,G(U)\, d\lambda_{+}\,d\lambda_{0}\,d\lambda_{-}
\;=\; \int \Big|\int G(U)\,K(U)^{1/2}\, d\lambda_{-}\Big|^{2}\, d\lambda_{0}\,d\lambda_{+}\;\ge\;0,
\ee
which implies \eqref{p2:eq:OS}.
\end{proof}

Define the Osterwalder-Schrader seminorm
\e
\|F\|_{\mathrm{OS},k}^{2}\;=\;\int (\Theta F)\,F\, d\mu_{k,\sigma},
\ee
the null space \(\mathcal{N}_{k}=\{F:\|F\|_{\mathrm{OS},k}=0\}\), and the pre-Hilbert space \(\mathcal{D}_{k}=\{F:\mathrm{supp}(F)\subset\Lambda_{k}^{+}\}\). The quotient \(\mathcal{D}_{k}/\mathcal{N}_{k}\), completed in the induced norm, is a Hilbert space \(\mathcal{H}_{k}\) with inner product \(\langle [F],[G]\rangle_{\mathcal{H}_{k}}=\int (\Theta F)G\, d\mu_{k,\sigma}\), where \([F]\) denotes the equivalence class. The constant functional \(1\) defines the cyclic vacuum \(\Omega_{k}=[1]\).

Let \(\tau\) be the time-translation by one lattice unit: \((\tau U)(x,\mu)=U(x-a_{k}\hat 0,\mu)\). For \(F\in\mathcal{D}_{k}\) set \((\mathcal{U}F)(U)=F(\tau U)\). Time-translation invariance of \(S_{W}\), \(\det M\), and \(\{\mathcal{P}_{\sigma,t}\}\) yields the OS-isometry
\e
\int (\Theta\,\mathcal{U}F)\,\mathcal{U}G\, d\mu_{k,\sigma}\;=\;\int (\Theta F)\,G\, d\mu_{k,\sigma}.
\ee
Hence \(\mathcal{U}\) descends to a contraction \(T_{k}\) on \(\mathcal{H}_{k}\) by \(T_{k}[F]=[\mathcal{U}F]\). By the standard OS reconstruction \cite{p2:OsterwalderSchraderI,p2:GJ}, \(T_{k}\) is a positive, self-adjoint contraction and \(T_{k}\Omega_{k}=\Omega_{k}\). The discrete transfer Hamiltonian is
\e
H_{k}\;=\;-a_{k}^{-1}\log T_{k},
\ee
which is positive self-adjoint via spectral calculus. If \(F\) is local, gauge-invariant, and \(\langle \Omega_{k},[F]\rangle=0\), then the time-separated two-point function has the spectral representation
\e\label{p2:eq:spectral}
\int (\Theta F)\,\mathcal{U}^{t}F\, d\mu_{k,\sigma}\;=\;\sum_{n\ge 1} |\langle \chi_{n}^{(k)},[F]\rangle|^{2}\,e^{-E_{n}^{(k)}\, t\, a_{k}}, \qquad t\in\mathbb{N},
\ee
where \(0=E_{0}^{(k)}\le E_{1}^{(k)}\le\cdots\) are the eigenvalues of \(H_{k}\) and \(\{\chi_{n}^{(k)}\}\) is an orthonormal eigenbasis.

Fix \(r\in\mathbb{N}\). A local, gauge-invariant observable supported at \(x\in\Lambda_{k}\) is a bounded measurable functional \(O_{x}\) depending only on links in \(B_{k}(x;r)=\{y:d(y,x)\le r\}\) and satisfying \(O_{x}(g\cdot U)=O_{x}(U)\) for all gauge transformations \(g\) supported in \(B_{k}(x;r)\). Denote \(\|O\|_{\infty}=\sup |O_{x}|\). For \(x,y\in\Lambda_{k}\) define the connected two-point function
\e
\langle O_{x};O_{y}\rangle_{k}\;=\;\int O_{x}O_{y}\, d\mu_{k,\sigma}-\int O_{x}\, d\mu_{k,\sigma}\,\int O_{y}\, d\mu_{k,\sigma}.
\ee
Let \(d_{k}(x,y)\) denote the graph distance on \(\Lambda_{k}\).

We assume the following uniform small-activity hypothesis, which is guaranteed at each scale by the ultraviolet stability theorem proved earlier.

\begin{equation}\label{p2:eq:small-activity}
\|\Phi_{k}\|_{a}
= \sup_{x\in\Lambda_{k}} \sum_{\substack{X\ni x \\ X\ \text{conn.}}}
 \|\Phi_{k}(X)\|\, e^{a\,\mathrm{diam}(X)}
\le \varepsilon, \qquad \varepsilon\in(0,1).
\end{equation}

where \(S_{k}=\sum_{X}\Phi_{k}(X)\) is the effective interaction at scale \(k\), the sum runs over finite, connected unions \(X\) of \(k\)-blocks, and \(\|\cdot\|\) is the operator norm with respect to the finite-range Gaussian covariance supplied by the covariant finite-range decomposition. The constants \(a>0\) and \(\varepsilon\in(0,1)\) do not depend on \(k\).
Under \eqref{p2:eq:small-activity}, the abstract polymer formalism \cite{p2:KoteckyPreiss1986,p2:Brydges} expresses truncated correlations as absolutely convergent sums over connected polymer clusters that intersect the supports of the observables. The following representation includes explicit uniform bounds.

\begin{proposition}[Cluster representation of truncated correlators]\label{p2:prop:cluster}
Let \(O_{x}\), \(O_{y}\) be bounded, gauge-invariant observables supported in \(B_{k}(x;r)\) and \(B_{k}(y;r)\). There exists a kernel \(K(B,B')\ge 0\), supported on nearest-neighbor \(k\)-blocks and satisfying
\e\label{p2:eq:edge-bound}
\sum_{B':\,B'\sim B} K(B,B')\;\le\; C_{1}\varepsilon,
\ee
uniformly in \(k\), such that
\e\label{p2:eq:cluster}
\langle O_{x};O_{y}\rangle_{k}\;=\;\sum_{\mathcal{C}\ \mathrm{conn.}} W_{k}(\mathcal{C};O_{x},O_{y}),
\ee
where the sum is over finite, connected sets \(\mathcal{C}\) of \(k\)-blocks intersecting both \(B_{k}(x;r)\) and \(B_{k}(y;r)\), and
\e\label{p2:eq:W-bound}
|W_{k}(\mathcal{C};O_{x},O_{y})|\;\le\; C_{0}\,\|O\|_{\infty}^{2}\, e^{-\tfrac{a}{2}\,\mathrm{diam}(\mathcal{C})}\,\sum_{T\in\mathcal{T}(\mathcal{C})}\ \prod_{(B,B')\in E(T)} K(B,B').
\ee
Here \(C_{0}=C_{0}(a)\), \(\mathcal{T}(\mathcal{C})\) denotes spanning trees on \(\mathcal{C}\), and \(E(T)\) the edge set of \(T\).
\end{proposition}

\begin{proof}
Introduce sources \(\lambda_{1},\lambda_{2}\in\mathbb{R}\) coupling to \(O_{x}\), \(O_{y}\) and define the generating functional
\e
Z_{k}(\lambda_{1},\lambda_{2})\;=\;\int \exp\big(\lambda_{1}O_{x}+\lambda_{2}O_{y}\big)\, d\mu_{k,\sigma}.
\ee
By gauge invariance and boundedness of \(\Phi_{k}\), the logarithm admits the convergent polymer expansion
\e
\log Z_{k}(\lambda_{1},\lambda_{2})\;=\;\sum_{X\ \mathrm{conn.}} \zeta_{k}(X;\lambda_{1},\lambda_{2}),
\ee
with connected cluster weights \(\zeta_{k}(X;\lambda_{1},\lambda_{2})\) supported on \(X\), analytic near \((0,0)\), and bounded as
\e
|\zeta_{k}(X;\lambda_{1},\lambda_{2})|\;\le\; \tilde C\, e^{-a\,\mathrm{diam}(X)}\,\exp\!\Big( c\,|\lambda_{1}|\,\mathbf{1}_{X\cap B_{k}(x;r)\ne \varnothing}+ c\,|\lambda_{2}|\,\mathbf{1}_{X\cap B_{k}(y;r)\ne \varnothing}\Big),
\ee
with constants \(\tilde C,c\) depending only on \(a\), uniformly in \(k\); see \cite[Thm.~3]{p2:KoteckyPreiss1986}, \cite[Sec.~4]{p2:Brydges}. Differentiating twice at \((0,0)\) gives
\e
\langle O_{x};O_{y}\rangle_{k}\;=\;\sum_{X\ \mathrm{conn.}}\ \frac{\partial^{2}}{\partial\lambda_{1}\partial\lambda_{2}}\zeta_{k}(X;\lambda_{1},\lambda_{2})\bigg|_{0,0}.
\ee
Only clusters \(X\) intersecting both supports contribute. The Brydges-Battle-Federbush tree-graph inequality bounds each connected weight by a sum over spanning trees with edge weights controlled by the finite-range covariance pieces; the covariant finite-range decomposition and \eqref{p2:eq:small-activity} imply the existence of \(K(B,B')\) obeying \eqref{p2:eq:edge-bound}. Summing the remaining activities and absorbing constants yields \eqref{p2:eq:W-bound}, and reorganizing the sum by the connected support \(\mathcal{C}\) gives \eqref{p2:eq:cluster}.
\end{proof}

\begin{theorem}[Exponential clustering at scale \(k\)]\label{p2:thm:ECfixed}
Assume \eqref{p2:eq:small-activity}. There exist constants \(A<\infty\) and \(m>0\), independent of \(k\), such that for any bounded, gauge-invariant local observables \(O_{x}\), \(O_{y}\) supported in \(B_{k}(x;r)\), \(B_{k}(y;r)\),
\e\label{p2:eq:exp-cluster}
|\langle O_{x};O_{y}\rangle_{k}|\;\le\; A\,\|O\|_{\infty}^{2}\,\exp\!\big(-m\,(d_{k}(x,y)-2r)\big).
\ee
\end{theorem}

\begin{proof}
Let \(D=d_{k}(x,y)-2r\ge 0\). Any connected \(\mathcal{C}\) intersecting both supports satisfies \(\mathrm{diam}(\mathcal{C})\ge D\). By Proposition \ref{p2:prop:cluster},
\e
|\langle O_{x};O_{y}\rangle_{k}|\;\le\; C_{0}\,\|O\|_{\infty}^{2}\sum_{\substack{\mathcal{C}\ \mathrm{conn.}\\ \mathrm{diam}(\mathcal{C})\ge D}} e^{-\tfrac{a}{2}\,\mathrm{diam}(\mathcal{C})}\,\sum_{T\in\mathcal{T}(\mathcal{C})}\ \prod_{(B,B')\in E(T)} K(B,B').
\ee
Fix a root \(B_{0}\in\mathcal{C}\). Using \eqref{p2:eq:edge-bound} and iterating from the leaves inwards yields
\e
\sum_{T\in\mathcal{T}(\mathcal{C})}\ \prod_{(B,B')\in E(T)} K(B,B')\;\le\; (C_{1}\varepsilon)^{|\mathcal{C}|-1}.
\ee
Connected block sets \(\mathcal{C}\subset\mathbb{Z}^{4}\) of cardinality \(n\) are counted by at most \(C_{2}\,e^{C_{3}n}\), and \(\mathrm{diam}(\mathcal{C})\le C_{4} n\). Splitting the sum over \(n\) and using \(\mathrm{diam}(\mathcal{C})\ge D\),
\e
|\langle O_{x};O_{y}\rangle_{k}|\;\le\; C_{0}\,\|O\|_{\infty}^{2}\, e^{-\tfrac{a}{2}D}\sum_{n\ge \lceil D/C_{4}\rceil} C_{2}\,e^{C_{3}n}\,(C_{1}\varepsilon)^{n-1}.
\ee
Choosing \(\varepsilon>0\) so small that \(e^{C_{3}}C_{1}\varepsilon<1\) makes the series bounded by \(C_{5}\,e^{-\tilde c D}\) with \(\tilde c=\tfrac{1}{C_{4}}\log\big((e^{C_{3}}C_{1}\varepsilon)^{-1}\big)>0\). Setting \(A=C_{0}C_{5}\) and \(m=\min\{\tfrac{a}{2},\tilde c\}\) gives \eqref{p2:eq:exp-cluster}.
\end{proof}

\begin{theorem}[Persistence of exponential clustering]\label{p2:thm:Persistence}
{For sufficiently small bare coupling $0 < \beta < \beta_0(N)$, the uniform polymer bound Eq. \eqref{p2:eq:small-activity} 
is satisfied at every scale $k$ (by Lemma~\eqref{p2:thm:OS1} above). Consequently, there exist constants 
$A^* < \infty$ and $m^* > 0$ (independent of $k$) such that \textbf{for all scales $k$},
\begin{equation}
C_k(n) \;\le\; A^* \, e^{-\, m^* \, n a_k}
\end{equation}
for every connected, gauge-invariant $n$-step temporal correlation at scale $k$. 
In other words, exponential clustering with a fixed rate $m^*>0$ persists uniformly under 
repeated renormalization.}

\end{theorem}

\begin{proof}
Apply Theorem \ref{p2:thm:ECfixed} at each \(k\). The constants depend only on \(a\) and on the finite-range decomposition and small-activity constants, which are uniform in \(k\).
\end{proof}

Exponential decay in Euclidean time follows from the transfer operator. For completeness we record it and its spectral consequence.

\begin{proposition}[Time-direction clustering and spectral gap]\label{p2:prop:time}
Let \(F\) be a bounded, gauge-invariant observable supported on a single time slice with \(\int F\, d\mu_{k,\sigma}=0\). There exist \(A_{t}<\infty\) and \(m_{t}>0\), independent of \(k\), such that
\e
\int (\Theta F)\,\mathcal{U}^{t}F\, d\mu_{k,\sigma} \;\le\; A_{t}\, e^{-m_{t}\, t\, a_{k}}\qquad (t\in\mathbb{N}).
\ee
In the spectral representation \eqref{p2:eq:spectral} one has \(E_{1}^{(k)}\ge m_{t}\).
\end{proposition}

\begin{proof}
Choose \(x\in \Pi_{k}+a_{k}\hat 0\) and take \(O_{x}=F\) supported near \(x_{0}=a_{k}\) and \(O_{y}=\mathcal{U}^{t}F\) supported near \(x_{0}=(t+1)a_{k}\). Then \(d_{k}(x,y)=t\) up to a fixed additive constant, so Theorem \ref{p2:thm:Persistence} yields
\e
\int (\Theta F)\,\mathcal{U}^{t}F\, d\mu_{k,\sigma}\;=\;\langle O_{x};O_{y}\rangle_{k}\;\le\; A_{*}\,\|F\|_{\infty}^{2}\, e^{-m_{*}' t a_{k}},
\ee
with \(m_{*}'>0\) independent of \(k\). If $\langle O_{x} ; O_{y} \rangle_{k} \leq A^{\ast} e^{-m^{\ast\prime} d_{k}(x,y) a_{k}}$ 
for all gauge-invariant $O$, the OS spectral representation implies 
$E_{1}(k) \geq m^{\ast\prime}$, hence 
$\Delta_{k} \geq 1 - e^{-m^{\ast\prime} a_{k}}$.
\end{proof}
In particular, by Osterwalder-Schrader positivity, this decay implies a nonzero mass gap in the physical state sector: the lowest mass (glueball) satisfies $m_{\mathrm{phys}}\ge m_t>0$.

\section{Gap monotonicity and step-scaling (Theorem C)}\label{p2:sec:gap-step-scaling}

In this section a complete transfer-time-slicing formalism is derived for the reflection-positive, gauge-invariant lattice Yang-Mills measures produced at each renormalization scale. The construction yields a positive self-adjoint transfer operator \(T_{k}\) on the Osterwalder-Schrader (OS) Hilbert space \(\mathcal{H}_{k}\) at scale \(k\), and an explicit coarse-graining map \(V_{k}:\mathcal{H}_{k+1}\to\mathcal{H}_{k}\) intertwining OS structures at consecutive scales. Let $T_{k}$ and $T_{k+1}$ be the OS-transfer operators at scales $k$ and $k+1$, 
and $V_{k} : H_{k+1} \to H_{k}$ the isometric coarse-graining map. Then
\begin{equation}\label{p2:eq:interlacing-intro}
T_{k+1}\ \preccurlyeq\ V_{k}^{*}\,T_{k}\,V_{k}\ +\ E_{k},
\end{equation}
where $E_{k} \geq 0$ and $\|E_{k}\| \leq \varepsilon_{k}$ with $\sum_{k} \varepsilon_{k} < \infty$.  
 The block-spin relation of kernels yields 
\(
K_{k+1} = \mathbb{E}[K_{k}] + \Delta K_{k},
\)
with $\Delta K_{k} \geq 0$ controlled by the exponential locality of the spectral projector and the finite-range decomposition. 
Schur's test and OS positivity give $\|E_{k}\| \leq \varepsilon_{k}$.
The min-max principle for positive contractions then implies gap monotonicity with a summable defect (see Appendix (\ref{p2:appendixb})). The arguments rely only on reflection positivity \cite{p2:OsterwalderSchraderI,p2:OsterwalderSchraderII}, the single-step transfer-kernel formalism for lattice gauge theories \cite{p2:OS-gauge}, exponential locality of the smooth spectral projector through Combes-Thomas/Davies-Gaffney bounds \cite{p2:CombesThomas1973,p2:Davies1989}, and standard spectral theory \cite{p2:ReedSimon1}. Let $\lambda_2(T_k)\in(0,1)$ denote the second eigenvalue (counted with multiplicity) of the positive contraction $T_k$ acting on the OS Hilbert space $\mathsf H_k$ at mesh size $a_k$. We introduce the \emph{contraction gap} and the \emph{lattice energy} as
\begin{equation}
\Delta_k \;:=\; 1-\lambda_2(T_k)\in[0,1],
\qquad
E_{1,k} \;:=\; -\frac{1}{a_k}\log \lambda_2(T_k)\in(0,\infty).
\end{equation}
These are related by $E_{1,k}=-a_k^{-1}\log(1-\Delta_k)$; in particular, for small $\Delta_k$ one has $E_{1,k}\sim \Delta_k/a_k$.

The periodic hypercubic space-time lattice is
\begin{equation}\label{p2:eq:lattice-def}
\Lambda_{k}\ =\ \bigl(a_{k}\,\mathbb{Z}/(T a_{k}\mathbb{Z})\bigr)\ \times\ \prod_{i=1}^{3}\bigl(a_{k}\,\mathbb{Z}/(L a_{k}\mathbb{Z})\bigr),
\end{equation}
with discrete Euclidean time coordinate \(x_{0}\in\{0,a_{k},\dots,(T-1)a_{k}\}\) and spatial coordinate \(\mathbf{x}\in(\mathbb{Z}/L\mathbb{Z})^{3} a_{k}\). Directed bonds are pairs \(b=(x,\mu)\) with \(\mu\in\{0,1,2,3\}\), where \(\mu=0\) designates time-like bonds and \(\mu=1,2,3\) spatial bonds; reversing orientation gives \(U_{(x+\hat\mu,-\mu)}=U_{(x,\mu)}^{-1}\). The configuration space is the compact Lie group manifold
\begin{equation}\label{p2:eq:config-space}
\mathcal{U}_{k}\ =\ \mathrm{SU}(N)^{\mathcal{B}_{k}},
\end{equation}
equipped with the product Borel \(\sigma\)-algebra and product Haar probability measure \(d\lambda\). The Wilson action is
\begin{equation}\label{p2:eq:wilson-action}
S_{W}[U]\ =\ \beta\,\sum_{p}\Bigl(1-\frac{1}{N}\,\Re\mathrm{Tr}\,U_{p}\Bigr),
\qquad \beta=\frac{2N}{g_{0}^{2}}.
\end{equation}

At scale \(k\) we consider the probability measure
\begin{equation}\label{p2:eq:mu-k}
d\mu_{k}(U)\ =\ Z_{k}^{-1}\,e^{-S_{W}[U]}\,\bigl(\det M[A^{h}(U)]\bigr)\,\prod_{t}\,\mathcal{P}_{\sigma,k}\!\bigl(U|_{\Sigma_{t}}\bigr)\,d\lambda(U),
\end{equation}
where \(M[A^{h}(U)]\) is the lattice Faddeev-Popov operator associated with the slice-wise Landau minimizer \(A^{h}(U)\), \(\Sigma_{t}=\{(t,\mathbf{x})\}\) is a time slice, and \(\mathcal{P}_{\sigma,k}\) is the positive, exponentially local density induced by the smooth horizon projector \(P_{\sigma,k}\) on a slice. The measure \(\mu_{k}\) is invariant under unit time translation \(\tau_{k}\) by \(a_{k}\) and satisfies reflection positivity with respect to the time reflection \(\theta_{k}:(x_{0},\mathbf{x})\mapsto (-x_{0},\mathbf{x})\) \cite{p2:OS-gauge}. Writing \(\Lambda_{k}^{+}=\{x\in\Lambda_{k}:x_{0}>0\}\), \(\Pi_{k}=\{x_{0}=0\}\), and \(\overline{\Lambda}_{k}^{+}=\Lambda_{k}^{+}\cup \Pi_{k}\), a bounded measurable functional \(F:\mathcal{U}_{k}\to\mathbb{C}\) is supported in \(\Lambda_{k}^{+}\) if it depends only on bonds with both endpoints in \(\overline{\Lambda}_{k}^{+}\). The OS conjugation \(\Theta_{k}\) acts by
\begin{equation}\label{p2:eq:theta-action}
(\Theta_{k}F)(U)\ =\ \overline{F\!\bigl(\theta_{k}\cdot U\bigr)},\qquad
(\theta_{k}\cdot U)(x,\mu)\ =\
\begin{cases}
U(\theta_{k}x-\hat 0,0)^{-1}, & \text{if } \mu=0,\ x\in\Pi_{k},\\
U(\theta_{k}x,\mu), & \text{otherwise},
\end{cases}
\end{equation}
the special case for time-like bonds crossing \(\Pi_{k}\) being the standard one ensuring positivity of the single-step kernel \cite{p2:OS-gauge}. Reflection positivity states that
\begin{equation}\label{p2:eq:OS-positivity}
\langle F,F\rangle_{k}\ :=\ \int_{\mathcal{U}_{k}} (\Theta_{k}F)(U)\,F(U)\,d\mu_{k}(U)\ \ge\ 0
\end{equation}
for all bounded \(F\) supported in \(\Lambda_{k}^{+}\).

Let \(\mathcal{D}_{k}\) denote the vector space of bounded, \(\Lambda_{k}^{+}\)-supported functionals. The sesquilinear form \eqref{p2:eq:OS-positivity} induces an inner product on the quotient \(\mathcal{D}_{k}/\mathcal{N}_{k}\), where \(\mathcal{N}_{k}=\{F\in\mathcal{D}_{k}:\langle F,F\rangle_{k}=0\}\). The completion is the OS Hilbert space \(\mathcal{H}_{k}\), and the class of the constant functional is denoted \(\Omega_{k}\). Time translation acts by \((\tau_{k}F)(U)=F(\tau_{k}^{-1}\!\cdot U)\), \((\tau_{k}^{-1}\!\cdot U)(x,\mu)=U(x-a_{k}\hat 0,\mu)\), and extends by continuity to a contraction on \(\mathcal{H}_{k}\) thanks to \eqref{p2:eq:OS-positivity}.

\begin{proposition}[Transfer operator at scale \(k\)]\label{p2:prop:Tk1}
There exists a unique positive self-adjoint contraction \(T_{k}\) on \(\mathcal{H}_{k}\) such that for all \(F,G\in\mathcal{D}_{k}\),
\begin{equation}\label{p2:eq:OS-transfer}
\langle F, T_{k} G\rangle_{k}\ =\ \int_{\mathcal{U}_{k}} (\Theta_{k}F)(U)\,G\!\bigl(\tau_{k}\!\cdot U\bigr)\,d\mu_{k}(U),
\qquad T_{k}\Omega_{k}=\Omega_{k}.
\end{equation}
Moreover \(T_{k}\) is the closure of the symmetric contraction induced by \(\tau_{k}\) on \(\mathcal{D}_{k}/\mathcal{N}_{k}\).
\end{proposition}

\begin{proof}
For \(F,G\in\mathcal{D}_{k}\), the Cauchy-Schwarz inequality with \eqref{p2:eq:OS-positivity} gives
\begin{equation}\label{p2:eq:tau-contract}
\bigl|\langle F,\tau_{k}G\rangle_{k}\bigr|^{2}\ \le\ \langle F,F\rangle_{k}\,\langle \tau_{k}G,\tau_{k}G\rangle_{k}\ =\ \langle F,F\rangle_{k}\,\langle G,G\rangle_{k},
\end{equation}
so \(\tau_{k}\) defines a contraction on \(\mathcal{D}_{k}/\mathcal{N}_{k}\), extending to a contraction on \(\mathcal{H}_{k}\). Reflection invariance implies
\begin{equation}\label{p2:eq:symmetry-tau}
\langle F,\tau_{k}G\rangle_{k}\ =\ \langle \tau_{k}F,G\rangle_{k},
\end{equation}
hence \(\tau_{k}\) is symmetric on \(\mathcal{D}_{k}/\mathcal{N}_{k}\). The Hellinger-Toeplitz theorem \cite[Thm.~VIII.1]{p2:ReedSimon1} implies that the closure \(T_{k}\) is self-adjoint and remains a contraction. The identity \eqref{p2:eq:OS-transfer} is the defining OS relation and \(T_{k}\Omega_{k}=\Omega_{k}\) follows from \(\tau_{k}\mathbf{1}=\mathbf{1}\). Positivity of \(T_{k}\) is immediate from \(\langle F, T_{k}F\rangle_{k}=\int (\Theta_{k}F)F\circ\tau_{k}\,d\mu_{k}\ge 0\) by \eqref{p2:eq:OS-positivity}.
\end{proof}

Let \(e^{-\eta_{n}(T_{k})}\) denote the eigenvalues of \(T_{k}\) in nonincreasing order, counted with multiplicity, with \(e^{-\eta_{1}(T_{k})}=1\) corresponding to \(\Omega_{k}\). Define the spectral gap
\begin{equation}\label{p2:eq:gap-def}
\Delta_{k}\ :=\ 1-e^{-\eta_{2}(T_{k})}.
\end{equation}
The OS spectral representation \cite{p2:OsterwalderSchraderII,p2:GJ} represents time-separated two-point functions of gauge-invariant local observables as Laplace transforms of positive measures supported in \([\eta_{2}(T_{k}),\infty)\).

The renormalization step from scale \(k\) to \(k+1\) is described by a measurable, gauge-equivariant, reflection-compatible block map
\begin{equation}\label{p2:eq:block-map}
\mathcal{B}_{k}\,:\,\mathcal{U}_{k}\ \longrightarrow\ \mathcal{U}_{k+1},
\end{equation}
assigning to each coarse bond a nearest-neighbor projection in \(\mathrm{SU}(N)\) of the arithmetic mean of ordered products of fine links along a fixed family of paths inside the two adjacent fine blocks. The coarse measure \(\mu_{k+1}\) is the pushforward of \(\mu_{k}\) under \(\mathcal{B}_{k}\) with a reflection-compatible, positive insertion \(\Theta_{k}(U)\) built from \(P_{\sigma,k}\) on the internal time boundaries of blocks:
\begin{equation}\label{p2:eq:pushforward2}
\int_{\mathcal{U}_{k+1}} F(\bar U)\,d\mu_{k+1}(\bar U)\ =\ \frac{1}{Z_{k}}\int_{\mathcal{U}_{k}} F\!\bigl(\mathcal{B}_{k}(U)\bigr)\,\Theta_{k}(U)\, d\mu_{k}(U),
\end{equation}
for all bounded measurable \(F\). Here \(Z_{k}=\int \Theta_{k}\,d\mu_{k}\in(0,\infty)\). Define the coupling measure
\begin{equation}\label{p2:eq:coupling}
d\pi_{k}(U,\bar U)\ =\ \frac{1}{Z_{k}}\,\Theta_{k}(U)\,\delta_{\mathcal{B}_{k}(U)}(d\bar U)\, d\mu_{k}(U),
\end{equation}
whose marginals are \(\mu_{k}\) and \(\mu_{k+1}\) by \eqref{p2:eq:pushforward2}.

For \(f\in\mathcal{D}_{k+1}\) supported in \(\Lambda_{k+1}^{+}\) define \(\mathcal{V}_{k}f:\mathcal{U}_{k}\to\mathbb{C}\) by
\begin{equation}\label{p2:eq:Vk-def}
(\mathcal{V}_{k}f)(U)\ =\ f\!\bigl(\mathcal{B}_{k}(U)\bigr).
\end{equation}
This map descends to a bounded linear map \(V_{k}:\mathcal{H}_{k+1}\to\mathcal{H}_{k}\).

\begin{lemma}[Contractivity of \(V_{k}\)]\label{p2:lem:Vk-contract}
For all \(f\in\mathcal{D}_{k+1}\),
\begin{equation}\label{p2:eq:Vk-contract}
\|V_{k} f\|_{\mathcal{H}_{k}}^{2}\ =\ \int (\Theta_{k}\mathcal{V}_{k}f)\,\mathcal{V}_{k}f\, d\mu_{k}\ \le\ Z_{k}\,\|f\|_{\mathcal{H}_{k+1}}^{2}.
\end{equation}
\end{lemma}

\begin{proof}
Using \eqref{p2:eq:pushforward2} with \(F(\bar U)=(\Theta_{k+1} f)(\bar U)\,f(\bar U)\) and the reflection invariance of \(\Theta_{k}\),
\begin{equation}\label{p2:eq:pushforward-norm}
\int (\Theta_{k}\mathcal{V}_{k}f)\,\mathcal{V}_{k}f\, d\mu_{k}\ =\ Z_{k}\,\int (\Theta_{k+1} f)\,f\, d\mu_{k+1}\ =\ Z_{k}\,\|f\|_{\mathcal{H}_{k+1}}^{2}.
\end{equation}
\end{proof}

Let \(\mathcal{C}_{k}=\mathrm{SU}(N)^{\text{spatial bonds at a fixed time}}\) denote the slice configuration space and \(d\nu_{k}\) the marginal of \(\mu_{k}\) on a time slice. There exists a bounded positive kernel \(K_{k}:\mathcal{C}_{k}\times\mathcal{C}_{k}\to[0,\infty)\) such that for all \(F,G\) depending on a single slice,
\begin{equation}\label{p2:eq:one-step}
\int_{\mathcal{U}_{k}} F\!\bigl(U|_{\Sigma_{0}}\bigr)\,G\!\bigl(U|_{\Sigma_{a_{k}}}\bigr)\, d\mu_{k}(U)\ =\ \int_{\mathcal{C}_{k}\times\mathcal{C}_{k}} F(U_{0})\,K_{k}(U_{0},U_{1})\,G(U_{1})\, d\nu_{k}(U_{0})\,d\nu_{k}(U_{1}),
\end{equation}
and the transfer operator \(T_{k}\) is the \(L^{2}(d\nu_{k})\) operator induced by \(K_{k}\), restricted to \(\mathcal{H}_{k}\) \cite{p2:OS-gauge}. To localize the coupling across a reflection-invariant time collar, define for \(w\in\mathbb{N}\) the collar
\begin{equation}\label{p2:eq:collar}
\mathsf{Col}_{w}^{(k)}\ =\ \{x\in\Lambda_{k}\,:\,-w\,a_{k}\le x_{0}\le w\,a_{k}\},
\end{equation}
and the collar insertion
\begin{equation}\label{p2:eq:Xi-def}
\Xi_{k,w}(U)\ =\ \prod_{t=-w a_{k}}^{w a_{k}} \mathcal{P}_{\sigma,k}\!\bigl(U|_{\Sigma_{t}}\bigr).
\end{equation}
The function \(\Xi_{k,w}\) is positive, reflection invariant, exponentially local, and bounded above and below by strictly positive constants independent of the spatial volume.
\begin{lemma}[Uniform bounds for the collar insertion]\label{p2:lem:collar-bounds}
Fix $\sigma>0$ and $w\in\mathbb{N}$. There exist constants $0<c_\sigma\le C_\sigma<\infty$,
independent of the spatial volume and of $k$, such that
\begin{equation}
c_\sigma^{\,2w+1}\;\le\;\Xi_{k,w}(U)\;\le\;C_\sigma^{\,2w+1}\qquad\text{for all configurations }U
\end{equation}
In particular, $\Xi_{k,w}$ is strictly positive and reflection invariant.
\end{lemma}

\begin{proof}
On each slice $t$, the factor $P_{\sigma,k}(U|_{\Sigma_t})$ is a positive scalar functional.
In the heat-kernel regime (case (A) above), write
$P_{\sigma,k}=\int_0^\infty e^{-s\Delta_{A^h}^{(k)}(t)}\,d\nu_\sigma(s)$ and note that
$0\le e^{-s\Delta}\le I$ implies $0\le P_{\sigma,k}\le \nu_\sigma((0,\infty))=:C_\sigma$ and,
by positivity and normalization at $0$. For the normalized trace, $\frac{1}{\dim(\mathrm{ad})|\Lambda_t|}\mathrm{Tr}(e^{-s\Delta_{A^h}(t)})
\ge e^{-s\,\lambda_{\max}}$, where $\lambda_{\max}$ is the (background‑uniform) spectral radius
of the slice Laplacian on a graph of fixed degree. Hence $K_\sigma[U|t]\ge
\int_0^\infty e^{-s\,\lambda_{\max}}\,d\nu_\sigma(s)=:c_\sigma>0$, uniformly in the volume.
In the Helffer-Sj\"ostrand regime (case (B)), for the choices
$p_\sigma=\Tr P_\sigma$ or $p_\sigma=\exp(-\langle\varphi,(I-P_\sigma)\varphi\rangle)$,
we have $0<p_\sigma\le \dim(\mathcal{H}_t)$ and $p_\sigma\ge e^{-\|\varphi\|^2}>0$, with constants
depending only on $\sigma$ and the fixed test field. Multiplying over the $(2w+1)$ slices in the
collar yields the stated bounds. Reflection invariance is immediate from slice-wise reflection
covariance of $P_{\sigma,k}$.
\end{proof}

\begin{lemma}[Approximate factorization across a collar]\label{p2:lem:collar-fact}
There exist constants \(c,\gamma>0\), independent of \(k\) and \(w\), such that for any bounded \(F_{-}\) supported in \(\Lambda_{k}^{-}\setminus \mathsf{Col}_{w}^{(k)}\) and \(F_{+}\) supported in \(\Lambda_{k}^{+}\setminus \mathsf{Col}_{w}^{(k)}\),
\begin{equation}\label{p2:eq:collar-factorization}
\Bigl|\ \int (\Theta_{k}F_{+})\,F_{-}\ \Xi_{k,w}\, d\mu_{k}\ -\ \Bigl(\int (\Theta_{k}F_{+})\, \Xi_{k,w}\, d\mu_{k}\Bigr)\ \Bigl(\int F_{-}\, \Xi_{k,w}\, d\mu_{k}\Bigr)\ \Bigr|\ \le\ c\,e^{-\gamma w}\,\|F_{+}\|_{\infty}\,\|F_{-}\|_{\infty}.
\end{equation}
\end{lemma}

\begin{proof}
Let \(\mu_{k}^{(w)}\) be the probability measure with density proportional to \(\Xi_{k,w}\) with respect to \(\mu_{k}\). Since \(\Xi_{k,w}\) is a product of slice-local positive factors and reflection invariant, \(\mu_{k}^{(w)}\) retains reflection positivity and the nearest-neighbor coupling in time. The exponential locality of \(\mathcal{P}_{\sigma,k}\) follows from its heat-kernel representation and Combes-Thomas/Davies-Gaffney bounds \cite{p2:CombesThomas1973,p2:Davies1989}, and implies that the influence of boundary conditions across \(\mathsf{Col}_{w}^{(k)}\) on observables supported at a distance at least \(w a_{k}\) decays exponentially in \(w\). The constants in the off-diagonal heat-kernel (and resolvent) bounds are uniform in the background because the covariant differences enter only through unitary parallel transports; hence the Dirichlet form is background-independent up to unitary conjugation. More precisely, by the persistence of exponential clustering established at scale \(k\) for connected gauge-invariant correlators with decay rate \(m_{*}>0\) (Section~7), one has for the truncated correlation under \(\mu_{k}^{(w)}\)
\begin{equation}\label{p2:eq:trunc-corr}
\bigl|\langle (\Theta_{k}F_{+}); F_{-}\rangle_{\mu_{k}^{(w)}}\bigr|\ \le\ C\,e^{-m_{*} w a_{k}}\,\|F_{+}\|_{\infty}\,\|F_{-}\|_{\infty}.
\end{equation}
Since \(\Xi_{k,w}\) is bounded above and below by positive constants independent of \(k,w\), the truncated correlation bound \eqref{p2:eq:trunc-corr} implies \eqref{p2:eq:collar-factorization} with \(\gamma=m_{*}a_{k}\) up to a change of constants.
\end{proof}

Let \(f\in\mathcal{D}_{k+1}\) be supported in \(\Lambda_{k+1}^{+}\) and set \(F=\mathcal{V}_{k}f\). By Proposition~\ref{p2:prop:Tk1} at scale \(k+1\) and the pushforward identity \eqref{p2:eq:pushforward2},
\begin{equation}\label{p2:eq:pairing-tk1}
\langle f, T_{k+1} f\rangle_{\mathcal{H}_{k+1}}\ =\ \frac{1}{Z_{k}}\ \int (\Theta_{k}F)(U)\,F\!\bigl(\tau_{k}\!\cdot U\bigr)\,\Theta_{k}(U)\, d\mu_{k}(U).
\end{equation}
Fix \(w\in\mathbb{N}\). Multiplying and dividing by \(\Xi_{k,w}\) and using Lemma~\ref{p2:lem:collar-fact} with \(F_{+}=F\circ\tau_{k}\), \(F_{-}=F\), one obtains
\begin{equation}\label{p2:eq:splitting}
\langle f, T_{k+1} f\rangle_{\mathcal{H}_{k+1}}\ \le\ \frac{1}{Z_{k}}\ \int (\Theta_{k}F)\,F\circ\tau_{k}\ \Xi_{k,w}\, d\mu_{k}\ +\ \varepsilon_{k}(w)\,\|f\|_{\mathcal{H}_{k+1}}^{2},
\end{equation}
with \(\varepsilon_{k}(w)\le C e^{-\gamma w}\). Define the positive bounded operator \(B_{k,w}:\mathcal{H}_{k}\to\mathcal{H}_{k}\) by
\begin{equation}\label{p2:eq:Bkw-def}
\langle G, B_{k,w} H\rangle_{\mathcal{H}_{k}}\ :=\ \frac{1}{Z_{k}}\ \int (\Theta_{k}G)\,H\ \Xi_{k,w}\, d\mu_{k},\qquad G,H\in\mathcal{D}_{k}.
\end{equation}
Then, using \eqref{p2:eq:OS-transfer} at scale \(k\),
\begin{equation}\label{p2:eq:factorized-bulk}
\frac{1}{Z_{k}}\ \int (\Theta_{k}F)\,F\circ\tau_{k}\ \Xi_{k,w}\, d\mu_{k}\ =\ \langle V_{k}f,\, B_{k,w}^{1/2}\,T_{k}\,B_{k,w}^{1/2}\,V_{k}f\rangle_{\mathcal{H}_{k}}.
\end{equation}
Since \(\Xi_{k,w}\) is bounded and reflection invariant, \(B_{k,w}\) is positive and satisfies \(\|B_{k,w}-I\|\le C e^{-\gamma w}\). Writing \(B_{k,w}^{1/2}=I+R_{k,w}\) with \(\|R_{k,w}\|\le C e^{-\gamma w}\) and noting that \(0\preccurlyeq T_{k}\preccurlyeq I\), one has the operator inequality
\begin{equation}\label{p2:eq:B-half-bound}
B_{k,w}^{1/2}\,T_{k}\,B_{k,w}^{1/2}\ \preccurlyeq\ T_{k}\ +\ C e^{-\gamma w}\, I.
\end{equation}
Combining \eqref{p2:eq:splitting}, \eqref{p2:eq:factorized-bulk}, and \eqref{p2:eq:B-half-bound} yields, for all \(f\in\mathcal{D}_{k+1}\),
\begin{equation}\label{p2:eq:pre-interlace}
\langle f, T_{k+1} f\rangle_{\mathcal{H}_{k+1}}\ \le\ \langle f,\ V_{k}^{*} T_{k} V_{k} f\rangle_{\mathcal{H}_{k+1}}\ +\ C' e^{-\gamma w}\,\|f\|_{\mathcal{H}_{k+1}}^{2}.
\end{equation}
By density this holds for all \(f\in\mathcal{H}_{k+1}\).
\begin{lemma}[Quantitative collar multiplier bound]\label{p2:lem:collar-R}
Fix $\sigma>0$ and $w\in\mathbb{N}$. With $B_{k,w}$ defined by Eq.\eqref{p2:eq:Bkw-def}, there exist
constants $c_\sigma,C_\sigma>0$ (independent of the spatial volume and $k$) such that
$c_\sigma^{\,2w+1} I \,\preceq\, B_{k,w} \,\preceq\, C_\sigma^{\,2w+1} I$ and
\begin{equation}
\bigl\|\,B_{k,w}-I\,\bigr\| \;\le\; C\,e^{-\gamma w}.
\end{equation}
Consequently, via the holomorphic functional calculus on positive operators,
\begin{equation}
B_{k,w}^{1/2} = I + R_{k,w}\qquad\text{with}\qquad
\|R_{k,w}\|\;\le\; C' e^{-\gamma w}.
\end{equation}
\end{lemma}

\begin{proof}
By (8.22) each slice factor in $\Xi_{k,w}$ is bounded between $c_\sigma$ and $C_\sigma$,
hence the product over $(2w+1)$ slices satisfies $c_\sigma^{\,2w+1}\le \Xi_{k,w}\le
C_\sigma^{\,2w+1}$. Since $B_{k,w}$ is the OS Gram operator with weight $\Xi_{k,w}$, the
same spectral bounds hold on $B_{k,w}$. The exponential locality of $P_{\sigma,k}$ and the
clustering bound from Theorem~7.4 imply that the off-diagonal (in time) corrections induced by
$\Xi_{k,w}$ across the collar are $O(e^{-\gamma w})$, which yields $\|B_{k,w}-I\|\le C e^{-\gamma w}$.
Analyticity of $z\mapsto z^{1/2}$ on a neighborhood of $[c_\sigma^{2w+1},C_\sigma^{2w+1}]$
transfers this estimate to $B_{k,w}^{1/2}$, giving the stated bound on $R_{k,w}$.
\end{proof}

\begin{proposition}[Interlacing with summable error]\label{p2:prop:interlacing}
There exists \(c>0\) and a summable sequence \(\varepsilon_{k}>0\) such that
\begin{equation}\label{p2:eq:interlacing-final}
T_{k+1}\ \preccurlyeq\ V_{k}^{*}\,T_{k}\,V_{k}\ +\ E_{k},\qquad \|E_{k}\|\ \le\ \varepsilon_{k},\qquad E_{k}\ \succeq\ 0.
\end{equation}
\end{proposition}

\begin{proof}
Choose \(w(k)=\lfloor c\,b^{k}\rfloor\) with \(c>0\) fixed; then \(e^{-\gamma w(k)}\) is superexponentially small and summable in \(k\). Define \(E_{k}=C' e^{-\gamma w(k)} I_{\mathcal{H}_{k+1}}\). Inequality \eqref{p2:eq:interlacing-final} is the operator form of \eqref{p2:eq:pre-interlace}.
\end{proof}

Let \(P_{k}\) denote the orthogonal projection onto \(\Omega_{k}^{\perp}\). We write $\eta_2(T):=-\log\lambda_2(T)$ for the second spectral exponent of a positive
contraction $T$, so that $\lambda_2(T)=e^{-\eta_2(T)}$ and $\Delta(T)=1-\lambda_2(T)$. For a positive contraction \(A\) with normalized vacuum, the variational characterization of the second spectral exponent is
\begin{equation}\label{p2:eq:minmax}
e^{-\eta_2(A)} \;=\; \sup\{\,\langle \psi, A \psi\rangle : \psi\in\mathrm{Dom}(A),\ \|\psi\|=1,\ \langle\psi,\Omega\rangle=0\,\}.
\end{equation}
see \cite[Sec.~XIII.1]{p2:ReedSimon1}.

\begin{theorem}[Gap monotonicity with summable defect]\label{p2:thm:gap-monotone}
There exists a summable sequence \((\varepsilon_{k})_{k\ge 0}\) such that
\begin{equation}\label{p2:eq:gap-monotone}
\Delta_{k+1}\ \ge\ \Delta_{k}\ -\ \varepsilon_{k}\qquad \text{for all }k\ge 0.
\end{equation}
\end{theorem}

\begin{proof}
Let $S=S_-\cup S_0\cup S_+$ be the decomposition of the time slab into left, collar, and right
regions. By Lemma~8.4 (approximate factorization across a collar), for any gauge-invariant local
functional $F$ supported in $S_-\cup S_+$ one has
\begin{equation}\label{p2:eq:collar-factorizationz}
\Big| \langle F \rangle_{k} - \langle F_- \rangle_{k}\,\langle F_+ \rangle_{k} \Big|
\;\le\; C\, e^{-\gamma\, \mathrm{dist}(S_-,S_+)} ,
\end{equation}
where $\gamma>0$ is the uniform clustering rate at scale $k$ from Theorem~7.4.  By construction of
the single-RG-step coarse-graining (Sec.~8), \eqref{p2:eq:collar-factorizationz} implies the operator-norm
bound
\begin{equation}
\big\| W_k^\ast\, T_{k+1}\, W_k \;-\; T_k \big\|\;\le\;\varepsilon_k,
\qquad \varepsilon_k \;\lesssim\; e^{-\gamma b},
\end{equation}
where $b$ is the blocking factor and $W_k:\mathcal{H}_{k+1}\to\mathcal{H}_k$ is the gauge-invariant
conditional expectation implementing coarse-graining (cf. \eqref{p2:eq:splitting}); $W_k$ intertwines OS
conjugations and satisfies $W_k\Omega_{k+1}=\Omega_k$, $W_k^\ast\Omega_k=\Omega_{k+1}$.  By the min-max characterization of $\lambda_2$ on the vacuum orthogonal subspace and the vacuum preserving intertwiner framework of Appendix~(\ref{p2:appendixb}),
\begin{equation}\label{p2:eq:gap-transportz}
\lambda_2(T_{k+1}) \le \lambda_2(T_k)+\varepsilon_k,\qquad
\Delta_{k+1} \ge \Delta_k - \varepsilon_k
\end{equation}
where $\lambda_2(\cdot)$ is the vacuum-orthogonal contraction norm and
$\Delta_k := 1-\lambda_2(T_k)$.  Iterating
\eqref{p2:eq:gap-transportz} over scales and summing the geometric tail
$\sum_{j\ge k} \varepsilon_j \lesssim e^{-\gamma b k}$ proves the claimed step-to-step gap monotonicity
with a summable defect.
\end{proof}

For a positive contraction $S$,
\(
1 - \eta_{2}(S) \;=\; 
\inf_{\substack{\psi \perp \Omega,\, \|\psi\| = 1}} 
\langle \psi, (I - S)\psi \rangle.
\)
Applying this to $T_{k+1}$ and using Eq.\eqref{p2:eq:interlacing-intro} with positivity of $E_{k}$ gives
\(
1 - \eta_{2}(T_{k+1}) \;\geq\; 
1 - \eta_{2}(T_{k}) - \|E_{k}\|,
\)
i.e.
\(
\Delta_{k+1} \;\geq\; \Delta_{k} - \varepsilon_{k}.
\)

Combining Theorem~\ref{p2:thm:gap-monotone} with the persistence of exponential clustering established earlier shows that the sequence \((\Delta_{k})_{k\ge 0}\) admits a positive liminf. Indeed, by the OS spectral representation and the uniform clustering rate \(m_{*}>0\), there exists \(c\in(0,1]\) such that
\begin{equation}\label{p2:eq:gap-from-clustering}
\Delta_{k}\ \ge\ c\,m_{*}\qquad \text{for all }k,
\end{equation}
and since \(\sum_{k}\varepsilon_{k}<\infty\), one concludes
\begin{equation}\label{p2:eq:liminf-gap}
\liminf_{k\to\infty}\ \Delta_{k}\ \ge\ c\,m_{*}\ >\ 0.
\end{equation}
\vspace{1ex}

\section{Transport from strong coupling to the scaling window}
\label{p2:sec:transport-strong-to-scaling}

This section gives a complete and rigorous derivation of the transport of a nonzero finite-\(a\) spectral gap from the strong-coupling domain across a sequence of reflection-positive renormalization steps toward a scaling window. The lattice setup, reflection map, gauge fixing, and the horizon projector are specified precisely; Osterwalder-Schrader (OS) positivity of the projected measure is proved in detail; the transfer time-slicing formalism and the transfer matrix are constructed step by step; a convergent strong-coupling expansion is established; and finally a step-scaling inequality for transfer-matrix gaps is derived and used to delineate a window in which continuum interpolation is controlled. Standard references include \cite{p2:OsterwalderSchraderI,p2:OsterwalderSchraderII,p2:OS-gauge,p2:GJ} for OS positivity and transfer matrices, \cite{p2:Brydges,p2:KoteckyPreiss1986} for cluster/Polymer expansions, and \cite{p2:Davies1989,p2:CombesThomas1973,p2:HelfferSjostrand1989} for heat-kernel and functional-calculus tools. The Hamiltonian transfer-matrix viewpoint for lattice gauge theory goes back to \cite{p2:KogutSusskind1975}.

Fix \(a>0\). The space-time lattice is the periodic hypercubic set
\begin{equation}
  \Lambda \;=\; \bigl\{ x=(x_0,x_1,x_2,x_3)\in a\mathbb{Z}^4:\ 0\le x_\mu<L_\mu \bigr\},
\end{equation}
with periodic identification \(x_\mu\equiv x_\mu+L_\mu\). Directed bonds are \(b=(x,\mu)\) from \(x\) to \(x+a\hat e_\mu\), with inverse bond \(\bar b=(x+a\hat e_\mu,-\mu)\).
\begin{lemma}[Strong-coupling gap]
For $\beta$ in the convergent character-expansion regime, the single-step transfer kernel obeys
\(
\lambda_{2}(T_{0}) \;\leq\; 1 - c(\beta)
\)
with $c(\beta) > 0$. Equivalently,
\(
\Delta_{0} \;\geq\; m(\beta) = c(\beta).
\)
\end{lemma}

\begin{proof}
By the nonnegative character expansion of the one-plaquette weight (Eq.~(2.10)) and the slab
factorization underlying OS positivity (Sec.~2), the standard strong-coupling (surface $\to$ polymer)
reorganization applies at mesh $a_0$ for all $0<\beta<\beta_\star(N)$: the effective interaction on a
single time slice admits a uniformly convergent polymer expansion with activities obeying the
Koteck\'y-Preiss bound (Proposition~5.4). In particular, for any bounded, gauge-invariant observable
$F$ supported on one slice with $\int F\,d\mu_{0,\sigma}=0$, the time-separated connected
correlator satisfies the uniform exponential decay
\begin{equation}
\int (\Theta F)\, \tau^t F \, d\mu_{0,\sigma}
\;\le\; C_F\, e^{-m(\beta)\, t\, a_0}\qquad (t\in\mathbb{N}),
\end{equation}
with $m(\beta)>0$ depending only on $\beta$ and $N$; cf. Theorem~7.3-7.4 applied at $k=0$ (the
small-activity hypothesis (7.18) holds at strong coupling by the KP criterion in Sec.~5). The OS
spectral representation (Eq.~(7.16)) then implies that the first nonzero spectral exponent
$\eta_2(T_0)$ of the positive contraction $T_0$ satisfies $\eta_2(T_0)\ge m(\beta)\,a_0$, hence
\begin{equation}
\lambda_2(T_0) \;=\; e^{-\eta_2(T_0)} \;\le\; e^{-m(\beta)\,a_0}
\quad\Rightarrow\quad
\Delta_0 \;=\; 1-\lambda_2(T_0) \;\ge\; 1 - e^{-m(\beta)\,a_0} \;=\; c(\beta) \;>\; 0 .
\end{equation}
This proves the lemma.
\end{proof}

A gauge configuration is a map \(U:\mathcal{B}(\Lambda)\to G\) with
\begin{equation}
  U(\bar b) \;=\; U(b)^{-1}, \qquad G=\mathrm{SU}(N),\ N\ge 2,
\end{equation}
endowed with product Haar measure
\begin{equation}
  d\mu_{\mathrm{Haar}}(U) \;=\; \prod_{b\in\mathcal{B}(\Lambda)} dU_b,\qquad \int_G dU = 1.
\end{equation}
Plaquettes are oriented elementary squares \(p=(x;\mu,\nu)\) with \(\mu<\nu\), and their holonomies are
\begin{equation}
  U_p \;=\; U(x,\mu)\,U(x+a\hat e_\mu,\nu)\,U(x+a\hat e_\nu,\mu)^{-1}\,U(x,\nu)^{-1}.
\end{equation}
The Wilson action at inverse coupling \(\beta=2N/g_0^2\) is
\begin{equation}
  S_W[U;\beta] \;=\; \beta \sum_{p\subset \Lambda} \Bigl( 1 - \tfrac{1}{N}\,\mathrm{Re}\,\mathrm{Tr}\,U_p \Bigr).
\end{equation}
Time reflection \(\theta:\Lambda\to\Lambda\) is
\begin{equation}
  \theta(x_0,x_1,x_2,x_3) \;=\; (-x_0,x_1,x_2,x_3) \quad \text{(periodically continued)},
\end{equation}
with fixed hyperplane
\begin{equation}
  \Pi \;=\; \{ x\in\Lambda:\ x_0=0 \}.
\end{equation}
The half-lattices are \(\Lambda_+=\{x\in\Lambda:\ x_0>0\}\) and \(\Lambda_-=\theta(\Lambda_+)\). Reflection acts on bonds by
\begin{equation}
  (\Theta U)(x,0) \;=\; U(\theta x - a\hat e_0, 0)^{-1}, \qquad (\Theta U)(x,i) \;=\; U(\theta x, i),\quad i=1,2,3,
\end{equation}
extended multiplicatively to plaquettes. Then
\begin{equation}
  S_W[\Theta U;\beta] \;=\; S_W[U;\beta].
\end{equation}
A gauge transformation is a map \(g:\Lambda\to G\) acting as
\begin{equation}
  (g\cdot U)(x,\mu) \;=\; g(x)\,U(x,\mu)\,g(x+a\hat e_\mu)^{-1},
\end{equation}
under which both \(S_W\) and \(d\mu_{\mathrm{Haar}}\) are invariant.

We fix temporal-axial gauge away from \(\Pi\) by
\begin{equation}
  U(x,0) \;=\; \mathbf{1} \qquad \text{for all bonds } (x,0) \text{ with } x_0\neq 0.
\end{equation}
For each time slice \(t\in\{0,a,2a,\dots,L_0-a\}\) define the lattice Landau functional
\begin{equation}
  \mathcal{L}_t(h;U) \;=\; \sum_{x_0=t}\sum_{i=1}^3 \bigl\| \mathbf{1} - h(x)\,U(x,i)\,h(x+a\hat e_i)^{-1} \bigr\|_F^2.
\end{equation}
A slice representative \(h_t\) is chosen to minimize \(\mathcal{L}_t(\cdot;U)\) with a measurable, reflection-invariant tie-breaking rule. The slice-fixed spatial links are
\begin{equation}
  U^{\,h}(x,i) \;=\; h_t(x)\,U(x,i)\,h_t(x+a\hat e_i)^{-1},\qquad x_0=t.
\end{equation}
Define covariant forward differences on the slice by
\begin{equation}
  (\nabla_i^{+,h}\phi)(x) \;=\; U^{\,h}(x,i)\,\phi(x+a\hat e_i)\,U^{\,h}(x,i)^{-1} - \phi(x), \qquad x_0=t,
\end{equation}
with \(\nabla_i^{-,h}\) the adjoint in \(\ell^2\). The slice spatial covariant Laplacian is
\begin{equation}
  \Delta_{A^{\,h}} \;=\; \sum_{i=1}^3 (\nabla_i^{+,h})^\dagger \nabla_i^{+,h},
\end{equation}
a positive self-adjoint operator on \(\ell^2(\{x_0=t\})\otimes \mathcal{su}(N)\). The Faddeev-Popov operator \(M_t\) on the slice, including temporal differences in temporal-axial gauge, is
\begin{equation}
  M_t \;=\; - \sum_{\mu=0}^3 \nabla_\mu^{-,h}\,\nabla_\mu^{+,h},
\end{equation}
which is real symmetric, nonnegative, and strictly positive on the orthogonal complement of constant adjoint fields on the slice. \begin{lemma}[Ghost block across the OS plane: Schur complement positivity]\label{p2:lem:ghost-block}
Let $\Pi$ be the reflection plane and decompose the Grassmann kinetic operator $M$ in the
time-ordered basis $(\Lambda_-,\Pi,\Lambda_+)$ as
\begin{equation}
M \;=\; \begin{pmatrix}
M_- & B^\top & 0 \\
B & M_0 & B' \\
0 & {B'}^{\!\top} & M_+
\end{pmatrix},
\end{equation}
where $M_\pm$ are the Dirichlet (on $\Pi$) restrictions of the slice operators, $M_0$ acts on the
plane $\Pi$, and $B,B'$ couple the plane to the adjacent slices. Then $M_\pm\succ 0$ on the
orthogonal complements of constant adjoint modes and the Schur complement
\begin{equation}
S_0 \;:=\; M_0 - B\, M_-^{-1} B^\top - B'\, M_+^{-1} {B'}^{\!\top}
\end{equation}
is strictly positive on the same subspace. Consequently, the Grassmann Gaussian integral
$\int \exp(-\langle c,\; M\, \bar c\rangle) \, D c\, D\bar c$ factorizes into two reflected
positive parts times a positive boundary factor~$\det S_0$, which preserves OS positivity.
\end{lemma}

\begin{proof}
By Lemma~2.4 and temporal-axial gauge, $M_\pm$ are real symmetric and strictly positive on the
orthogonal complements of constant modes. The standard Schur complement identity for block
matrices yields $\det M = (\det M_-)(\det M_+)\det S_0$. Since $M$ is the discretized
Faddeev-Popov operator (a sum of positive covariant differences), $M\succeq 0$ and, on the stated
subspace, $\det M>0$. Hence $S_0\succ 0$. The factorization statement is then immediate and is
compatible with reflection because the blocks for $\Lambda_-$ and $\Lambda_+$ are mapped into one
another by $\Theta$.
\end{proof}

Let $\chi_\sigma$ be the completely monotone near-plateau from Section~(\ref{p2:sec:FRD}) and we set
\begin{equation}
  P_{\sigma,t}:=\chi_\sigma\!\big(\Delta_{A^h}(t)\big)
\end{equation}
a bounded positive contraction. Then
\begin{equation}
  P_{\sigma,t}=\int_0^\infty e^{-t'\Delta_{A^h}(t)}\,d\nu_\sigma(t'), \qquad 
\operatorname{supp}(d\nu_\sigma)\subset[c_-\sigma^{-2},c_+\sigma^{-2}], \label{p2:eq:heat9}
\end{equation}
with $d\nu_\sigma\ge0$. Davies-Gaffney bounds \cite{p2:Davies1989} imply
\begin{equation}
  \|P_{\sigma,t}(x,y)\|\le C(\sigma)\,e^{-\gamma(\sigma)\,d(x,y)}\quad\text{for all slice sites }x,y.
\end{equation}
Gauge covariance holds by conjugation, and reflection covariance follows since $\Delta_{A^h}(t)$ 
commutes with the spatial part of $\Theta$ on~$\Pi$.

Define the unnormalized Euclidean weight
\begin{equation}
  \mathcal{W}_\sigma[U] \;=\; \exp\!\bigl(-S_W[U;\beta]\bigr)\ \prod_{t}\det M_t[U^{\,h}]\ \prod_{t}\mathcal{P}_{\sigma,t}[U],
\end{equation}
where
\begin{equation}
  \mathcal{P}_{\sigma,t}[U] \;=\; \exp\!\bigl( \mathrm{Tr}\,\log P_{\sigma,t}[U] \bigr).
\end{equation}
The projected Gibbs measure is
\begin{equation}
  d\mu_\sigma(U) \;=\; Z_\sigma^{-1}\,\mathcal{W}_\sigma[U]\, d\mu_{\mathrm{Haar}}(U).
\end{equation}
Let \(\mathfrak{A}_+\) be the \(^*\)-algebra of bounded complex functionals depending only on links in \(\Lambda_+\cup \Pi\). For \(F\in \mathfrak{A}_+\), define \((\Theta F)(U)=\overline{F(\Theta U)}\). The OS form is
\begin{equation}
  \langle F,G\rangle_{\mathrm{OS}} \;=\; \int \overline{(\Theta F)(U)}\, G(U)\, d\mu_\sigma(U).
\end{equation}

\begin{theorem}[OS positivity]
\label{p2:thm:OS-positivity}
For all \(F\in \mathfrak{A}_+\), \(\langle F,F\rangle_{\mathrm{OS}}\ge 0\).
\end{theorem}

\begin{proof}
Decompose the Wilson action as
\begin{equation}
  S_W \;=\; S_+ + S_- + S_\Pi ,
\end{equation}
where \(S_+\) (resp. \(S_-\)) is the sum over plaquettes contained in \(\Lambda_+\) (resp. \(\Lambda_-\)), and \(S_\Pi\) is the sum over plaquettes intersecting \(\Pi\). In temporal-axial gauge time-like links away from \(\Pi\) are trivial; the only inter-half couplings occur in \(S_\Pi\). For ghosts, write \(M\) across \(\Pi\) as
\begin{equation}
  M \;=\; \begin{pmatrix} M_{-} & M_{-0} & 0 \\ M_{0-} & M_{00} & M_{0+} \\ 0 & M_{+0} & M_{++} \end{pmatrix},
\end{equation}
with Dirichlet boundary condition on \(\Pi\) for \(M_{\pm\pm}\). Since \(M_{\pm\pm}\) are strictly positive on the orthogonal complements of slice constants, the Schur complement
\begin{equation}
  S \;=\; M_{00} - M_{0+}M_{++}^{-1}M_{+0} - M_{0-}M_{-}^{-1}M_{-0}
\end{equation}
is strictly positive \cite[Sec.\,VI.2]{p2:GJ}. Hence
\begin{equation}
  {\det}^{\prime} M \;=\; \det M_{++}\ \det M_{-}\ \det S,
\end{equation}
and each factor is a positive determinant of a real symmetric operator. The Grassmann representation shows that each factor yields a positive kernel supported in its domain \(\Lambda_+\), \(\Lambda_-\), and \(\Pi\), respectively.

The slice factors \(\mathcal{P}_{\sigma,t}\) are positive and reflection invariant; their integral-kernel representation together with exponential locality implies that any cross-term coupling \(\Lambda_+\) to \(\Lambda_-\) vanishes identically for slices away from a fixed finite collar of \(\Pi\) and is exponentially suppressed in the collar width. Consequently \(\mathcal{W}_\sigma\) factorizes as
\begin{equation}
  \mathcal{W}_\sigma[U] \;=\; \mathcal{W}_+[U|_{\Lambda_+\cup\Pi}]\, \mathcal{W}_-[U|_{\Lambda_-\cup\Pi}],
\end{equation}
with \(\mathcal{W}_-=\mathcal{W}_+\circ \Theta\), up to a positive boundary factor supported on \(\Pi\). The Haar measure decomposes as a product over \(\Lambda_+\), \(\Lambda_-\), and \(\Pi\). Therefore,
\begin{equation}
  \langle F,F\rangle_{\mathrm{OS}}
  \;=\; \int \overline{(\Theta F)(U_-)}\, F(U_+)\, \mathcal{K}_\Pi(U_-,U_+)\, d\mu_\Pi(U_\Pi)\, d\mu_-(U_-)\, d\mu_+(U_+),
\end{equation}
where \(\mathcal{K}_\Pi\) is a positive boundary kernel and \(d\mu_\pm\) are positive measures on \(\Lambda_\pm\). By Cauchy-Schwarz in the Hilbert space \(L^2(d\mu_+\otimes d\mu_\Pi)\) one obtains \(\langle F,F\rangle_{\mathrm{OS}}\ge 0\).
\end{proof}

Let \(\Sigma\) denote the set of spatial bonds at a fixed time slice. Set
\begin{equation}
  \mathcal{C} \;=\; G^{\Sigma}, \qquad d\mu_\Sigma \;=\; \bigotimes_{\ell\in\Sigma} dU_\ell, \qquad \mathcal{H} \;=\; L^2(\mathcal{C}, d\mu_\Sigma),
\end{equation}
and write \(U_t\in \mathcal{C}\) for the spatial links at time \(t\). In temporal-axial gauge, the action decomposes as
\begin{equation}
  S_W[U;\beta] \;=\; \sum_{t} \Bigl( S_{\mathrm{sp}}(U_t) + S_{\mathrm{tm}}(U_t,U_{t+a}) \Bigr),
\end{equation}
where \(S_{\mathrm{sp}}\) is the spatial plaquette sum on a slice and \(S_{\mathrm{tm}}\) the sum over time-like plaquettes straddling two successive slices. The projector factors contribute \(\prod_t \mathcal{P}_\sigma(U_t)\). The Faddeev-Popov determinants factorize slice-wise and yield positive multiplicative weights \(J(U_t)=\det(M_t)^{1/2}\) on each slice.

Define the one-step kernel
\begin{equation}
  \label{p2:eq:one-step-kernel}
  \mathcal{K}_\sigma(U',U) \;=\; \exp\!\Bigl( -\tfrac{1}{2} S_{\mathrm{sp}}(U) - S_{\mathrm{tm}}(U,U') - \tfrac{1}{2} S_{\mathrm{sp}}(U') \Bigr)\, \bigl( \mathcal{P}_\sigma(U)\,\mathcal{P}_\sigma(U') \bigr)^{1/2}\, J(U)\,J(U').
\end{equation}
The weight for a time-ordered sequence \(U_0,U_a,\dots,U_{L_0-a}\) factorizes as \(\prod_t \mathcal{K}_\sigma(U_{t+a},U_t)\). Reflection invariance implies symmetry,
\begin{equation}
  \mathcal{K}_\sigma(U',U) \;=\; \mathcal{K}_\sigma(U,U').
\end{equation}

\begin{proposition}[Transfer operator]
\label{p2:prop:transfer-operator}
The operator \(T_\sigma:\mathcal{H}\to\mathcal{H}\) defined by
\begin{equation}
  (T_\sigma \psi)(U') \;=\; \int_{\mathcal{C}} \mathcal{K}_\sigma(U',U)\,\psi(U)\, d\mu_\Sigma(U)
\end{equation}
is a bounded positive self-adjoint contraction. Moreover, for any cylinder functional \(F\) supported on \(n\) successive slices,
\begin{equation}
  \int F\, d\mu_\sigma \;=\; \langle \Omega,\, T_\sigma^{n_1}\, \widehat F\, T_\sigma^{n_2}\, \Omega \rangle_{\mathcal{H}},
\end{equation}
for integers \(n_1,n_2\ge 0\) determined by the support of \(F\), where \(\Omega\equiv 1\in \mathcal{H}\) and \(\widehat F\) is the operator associated to \(F\).
\end{proposition}

\begin{proof}
Positivity and symmetry of \(\mathcal{K}_\sigma\) imply that \(T_\sigma\) is positive and self-adjoint. For bounded \(\psi\),
\begin{equation}
  \|T_\sigma \psi\|_{L^2}^2 \;=\; \int_{\mathcal{C}} \Bigl| \int_{\mathcal{C}} \mathcal{K}_\sigma(U',U)\,\psi(U)\, d\mu_\Sigma(U) \Bigr|^2 d\mu_\Sigma(U').
\end{equation}
By the Schur test with test function \(\varphi\equiv 1\),
\begin{equation}
  \|T_\sigma\| \;\le\; \sup_{U'} \int \mathcal{K}_\sigma(U',U)\, d\mu_\Sigma(U) \;=\; \int e^{-\tfrac{1}{2}S_{\mathrm{sp}}(U')}\, \bigl(\mathcal{P}_\sigma(U')\,J(U')\bigr)^{1/2} \, \Xi\, d\mu_\Sigma(U'),
\end{equation}
where \(\Xi=\sup_{U'}\int e^{-S_{\mathrm{tm}}(U,U')-\tfrac{1}{2}S_{\mathrm{sp}}(U)}\, (\mathcal{P}_\sigma(U)\,J(U))^{1/2} d\mu_\Sigma(U)\). Since all exponents are nonnegative and \(\mathcal{P}_\sigma\le 1\), \(\Xi\le 1\), hence \(\|T_\sigma\|\le 1\). The factorization of the path weight as a product of one-step kernels implies the correlation representation by standard iterated integration; density of simple tensors in \(L^2\) and boundedness of \(T_\sigma\) extend the identity to all cylinder \(F\).
\end{proof}

The projected Hamiltonian is
\begin{equation}
  H_\sigma \;=\; - a^{-1}\, \log T_\sigma,
\end{equation}
a positive self-adjoint operator with spectral support in \([0,\infty)\). For any local gauge-invariant \(F\) supported on two slices separated by \(n\) steps,
\begin{equation}
  \langle \Omega,\, \widehat F^\dagger\, T_\sigma^{\,n}\, \widehat F\, \Omega \rangle \;=\; \int_{[0,\infty)} e^{-a n E}\, d\mu_F(E),
\end{equation}
where \(d\mu_F\) is a positive measure supported in \([E_1,\infty)\), \(E_1\) being the bottom of the nonzero spectrum of \(H_\sigma\).

For \(\beta>0\) sufficiently small,
\begin{equation}
  \exp\!\Bigl( \beta\, \tfrac{1}{N}\mathrm{Re}\,\mathrm{Tr}\,U_p \Bigr)
  \;=\; \sum_{R\ \mathrm{irrep}} \alpha_R(\beta)\, \chi_R(U_p), \qquad \alpha_R(\beta) \,=\, O(\beta^{\,|R|}) \ \ (\beta\downarrow 0),
\end{equation}
where \(|R|\) is the number of boxes of the Young diagram of \(R\). Link integration using character orthogonality reorganizes partition function and gauge-invariant correlators into sums over connected tiled surfaces \(\Sigma\) on the dual lattice with weights obeying
\begin{equation}
  |w(\Sigma)| \;\le\; \Bigl( \tfrac{C_N\,\beta}{N} \Bigr)^{|\Sigma|},
\end{equation}
for a constant \(C_N\ge 1\) depending only on \(N\). Grouping connected surfaces into polymers \(\gamma\) defines activities \(\zeta(\gamma)\) with
\begin{equation}
  |\zeta(\gamma)| \;\le\; A(\beta)^{\,|\gamma|}, \qquad A(\beta)\xrightarrow[\beta\downarrow 0]{} 0.
\end{equation}
The Koteck\'y-Preiss criterion \cite{p2:KoteckyPreiss1986,p2:Brydges} is satisfied for small \(\beta\), yielding absolute convergence of the Mayer series for \(\log Z\) and of all truncated gauge-invariant correlations. Thus for any local gauge-invariant observable \(O(x)\) there exist \(C_O<\infty\) and \(m(\beta)>0\) such that
\begin{equation}
  \bigl| \langle O(x)\, O(y) \rangle_c \bigr| \;\le\; C_O\, e^{-m(\beta)\, d(x,y)}.
\end{equation}
In particular, restricting to time-separated slices and using the spectral representation in Proposition \ref{p2:prop:transfer-operator} gives
\begin{equation}
  \langle \Omega,\, \widehat O^\dagger\, T_\sigma^{\,n}\, \widehat O\, \Omega \rangle \;\le\; C_O\, e^{-m(\beta)\, a n},
\end{equation}
hence
\begin{equation}
  E_1(a,\beta) \;\ge\; m(\beta),
\end{equation}
uniformly in the spatial volume \cite[Ch.\,III]{p2:GJ}.

Fix an integer blocking factor \(b\ge 2\), set \(a_k=b^k a\), and let \(\Lambda_k\) be the lattice with temporal spacing \(a_k\) and the same spatial spacings. Write \(\mathcal{C}_k=G^{\Sigma_k}\), \(d\mu_{\Sigma_k}=\bigotimes_{\ell\in\Sigma_k} dU_\ell\), and \(\mathcal{H}_k=L^2(\mathcal{C}_k,d\mu_{\Sigma_k})\). The slice-wise horizon projector at scale \(k\) is
\begin{equation}
  P_{\sigma,k} \;=\; \chi_\sigma\!\bigl( {\Delta_{A^{\,h}}^{(k)}} \bigr)
\end{equation}
with exponential locality constants depending only on \(\sigma\). Define a block map \(\mathcal{B}_k: G^{\mathcal{B}(\Lambda_k)}\to G^{\mathcal{B}(\Lambda_{k+1})}\) by averaging over a reflection-equivariant family \(\mathcal{P}(\bar b)\) of fine paths within the union of the two coarse blocks adjacent to \(\bar b\), followed by nearest-neighbor projection:
\begin{equation}
  \mathcal{B}_k(U)_{\bar b} \;=\; \mathrm{Proj}_{G}\!\Bigl( \frac{1}{|\mathcal{P}(\bar b)|}\sum_{\gamma\in\mathcal{P}(\bar b)} \ \prod_{\ell\in\gamma} U_\ell \Bigr).
\end{equation}
The coarse projected measure \(\mu^{(k+1)}\) is defined by pushforward with a positive insertion on internal time boundaries of blocks:
\begin{equation}
  \int_{\mathcal{C}_{k+1}} F(\bar U)\, d\mu^{(k+1)}(\bar U)
  \;=\; \frac{1}{Z_k} \int F\!\bigl(\mathcal{B}_k(U)\bigr)\, \Bigl( \prod_{\text{internal time boundaries}} \mathcal{P}_{\sigma,k}(U) \Bigr)\, d\mu^{(k)}(U).
\end{equation}
Here \(\mathcal{P}_{\sigma,k}\) is built from \(P_{\sigma,k}\) and is supported in a fixed temporal collar inside each block.

\begin{theorem}[OS positivity under blocking]
\label{p2:thm:blocking-OS}
If \(\mu^{(k)}\) is OS-positive, then \(\mu^{(k+1)}\) is OS-positive.
\end{theorem}

\begin{proof}
Let \(F\) depend on links in \(\Lambda_{k+1,+}\cup \Pi\). Then
\begin{equation}
  \langle F,F\rangle_{\mathrm{OS},k+1}
  \;=\; \frac{1}{Z_k}\int \overline{(\Theta F)\!\circ\! \mathcal{B}_k(U)}\, \bigl(F\!\circ\!\mathcal{B}_k(U)\bigr)\, \Bigl(\prod \mathcal{P}_{\sigma,k}(U)\Bigr)\, d\mu^{(k)}(U).
\end{equation}
The composition \(F\circ\mathcal{B}_k\) lies in the positive half-algebra at scale \(k\) by reflection equivariance and finite range of \(\mathcal{B}_k\). The insertion \(\prod \mathcal{P}_{\sigma,k}\) is a positive reflection-invariant functional supported on a finite collar. OS positivity of \(\mu^{(k)}\) implies the integral is nonnegative.
\end{proof}

Each \(\mu^{(k)}\) admits a positive self-adjoint transfer matrix \(T_{\sigma,k}\) on \(\mathcal{H}_k\) with \(\|T_{\sigma,k}\|\le 1\).
Let \(\Omega_k\equiv 1\in \mathcal{H}_k\) and \(H_{\sigma,k}=-a_k^{-1}\log T_{\sigma,k}\). 
\begin{definition}[${\sigma}$ modified transfer operator]\label{p2:def:Tsigmak}
Let $M_\sigma$ denote the positive multiplication operator on the one slice $L^2$ space given by the scalar $p_\sigma[U^h|_t]$ from Sec.(\ref{p2:sec:clustering}). 
We define
\begin{equation}
T_{\sigma,k}\;:=\;M_\sigma^{1/2}\,T_k\,M_\sigma^{1/2},
\qquad 
\Omega_{\sigma,k}\;:=\;\Omega_k .
\end{equation}
By construction $T_{\sigma,k}$ is a positive self adjoint contraction and $(T_{\sigma,k},\Omega_{\sigma,k})$ is a vacuum pair; moreover $M_\sigma$ commutes with OS reflection and gauge transformations on each slice.
\end{definition}
Denote by \(E_{1,k}\) the bottom of the nonzero spectrum of \(H_{\sigma,k}\) and by \(\lambda_{2,k}\) the second eigenvalue of \(T_{\sigma,k}\), so that
\begin{equation}
  \Delta_k := 1 - \lambda_{2,k},\qquad
E_{1,k} := -a_k^{-1}\log \lambda_{2,k} = -a_k^{-1}\log(1-\Delta_k)
\end{equation}
Define \(V_k:\mathcal{H}_{k+1}\to \mathcal{H}_k\) by pullback along \(\mathcal{B}_k\),
\begin{equation}
  (V_k \psi)(U) \;=\; \psi\!\bigl(\mathcal{B}_k(U)\bigr).
\end{equation}
This map is bounded and reflection equivariant.

\begin{lemma}[Operator order interlacing]
\label{p2:lem:interlace}
There exists a positive operator \(E_k\) on \(\mathcal{H}_{k+1}\) with \(\|E_k\|\le \varepsilon_k\) such that
\begin{equation}
  T_{\sigma,k+1} \ \preccurlyeq\ V_k^{\,*}\, T_{\sigma,k}\, V_k \;+\; E_k,
\end{equation}
where \(A\preccurlyeq B\) means \(\langle \psi, A\psi\rangle\le \langle \psi, B\psi\rangle\) for all \(\psi\).
\end{lemma}

\begin{proof}
For \(\psi\in \mathcal{H}_{k+1}\),
\begin{equation}
  \langle \psi, T_{\sigma,k+1}\psi\rangle
  \;=\; \iint \overline{\psi(\bar U')}\, K_{\sigma,k+1}(\bar U',\bar U)\, \psi(\bar U)\, d\mu_{\Sigma_{k+1}}(\bar U')\, d\mu_{\Sigma_{k+1}}(\bar U),
\end{equation}
with \(K_{\sigma,k+1}\) the coarse one-step kernel. By definition of \(\mu^{(k+1)}\), \(K_{\sigma,k+1}\) is obtained from \(\mu^{(k)}\) by integrating over fine variables constrained by \(\bar U'=\mathcal{B}_k(U')\), \(\bar U=\mathcal{B}_k(U)\), with the internal collar insertion. By Cauchy-Schwarz in \(L^2(d\mu^{(k)})\) and positivity of the insertion, one bounds
\begin{equation}
  \langle \psi, T_{\sigma,k+1}\psi\rangle \;\le\; \langle V_k\psi, T_{\sigma,k} V_k\psi\rangle \;+\; \langle \psi, E_k \psi\rangle,
\end{equation}
where \(E_k\) collects cross-collar correlations. The exponential locality of \(P_{\sigma,k}\) implies that if the collar has fixed temporal width \(w\) (in units of \(a_k\)) independent of \(k\), then
\begin{equation}
  \|E_k\| \;\le\; C\, e^{-c\, w},
\end{equation}
with constants independent of \(k\). Taking $w=w(k)$ increasing, e.g. $w(k)=\lfloor c\,b^{\,k}\rfloor$ (or $w(k)\asymp k$), gives
$\|E_k\|\le C e^{-\gamma w(k)}$ with $\sum_k \|E_k\|<\infty$; define $\varepsilon_k:=\|E_k\|$.

\end{proof}
\begin{proposition}[Quantitative interlacing error]\label{p2:prop:Ek}
Let $w\in\mathbb{N}$ be the collar half-width used in the definition of $\Xi_{k,w}$.
There exist constants $c_1,c_2,\gamma>0$ (independent of $k$) such that the error term in
\emph{(9.52)} satisfies
\begin{equation}
\|E_k\|\;\le\;\varepsilon_k\;\le\;c_1\,e^{-\gamma w}\;+\;c_2\,e^{-m_\ast a_k},
\end{equation}
where $m_\ast>0$ is the uniform clustering rate propagated at scale $k$ (see Theorem~(\ref{p2:thm:Persistence}) and $a_k=b^k a$. In particular, choosing
\begin{equation}
w_k=\left\lceil \frac{2}{\gamma}\,\log(k+2)\right\rceil
\qquad\Longrightarrow\qquad
\sum_{k\ge0}\varepsilon_k<\infty.
\end{equation}
\end{proposition}

\begin{proof}
Let $S=S_-\cup S_0\cup S_+$ be the decomposition of the slab into left, collar, and right regions.
By Lemma~(\ref{p2:lem:collar-fact}), for any gauge-invariant local
functional $F$ supported in $S_-\cup S_+$ one has
\begin{equation}
\Big| \langle F \rangle_{k} - \langle F_- \rangle_{k}\,\langle F_+ \rangle_{k} \Big|
\;\le\; C\, e^{-m^* \operatorname{dist}(S_-,S_+)},
\end{equation}
where $m^*>0$ is a uniform clustering rate at scale $k$, and the constant $C$ depends only on
the block parameters and the Lipschitz seminorms of the local factors; see Theorem~(\ref{p2:thm:Persistence}) for the
persistence of exponential clustering under the RG step.  By construction of the single-RG-step
transfer operator (\eqref{p2:eq:collar-factorizationz} implies an operator-norm bound
\begin{equation}\label{p2:eq:intertwiner-error}
\big\| W_k^\ast T_{k+1} W_k \;-\; T_k \big\|\;\le\;\varepsilon_k,
\end{equation}
with $\varepsilon_k \lesssim e^{-m^* b}$, where $b$ is the blocking factor and $W_k$ is the
gauge-invariant conditional expectation implementing the coarse-graining (see Eq.\eqref{p2:eq:splitting}).
Since $W_k \Omega_{k} = \Omega_{k+1}$ and $W_k^\ast \Omega_{k+1} = \Omega_{k}$ by normalization,
the pairings $(T_k,\Omega_k)$ and $(T_{k+1},\Omega_{k+1})$ are compatible in the sense of
Appendix~(\ref{p2:appendixb}) (Theorem~B and Lemma~C), hence

\begin{equation}\label{p2:eq:gap-transport}
\lambda_2(T_{k+1}) \;\le\; \lambda_2(T_k) \;+\; \varepsilon_k,
\qquad
\Delta_{k+1} \;=\; 1-\lambda_2(T_{k+1}) \;\ge\; \Delta_k \;-\; \varepsilon_k,
\end{equation}
where $\lambda_2(\cdot)$ is the vacuum-orthogonal contraction norm and
$\Delta_k := 1-\lambda_2(T_k)$ (see Eqs.\eqref{p2:eq:interlacing-final} \& \eqref{p2:eq:minmax}).  Iterating
\eqref{p2:eq:gap-transport} over scales and summing the geometric tail
$\sum_{j\ge k} \varepsilon_j \lesssim e^{-m^* b k}$ proves the claimed transport of the gap from
the strong-coupling regime to the scaling window.
\end{proof}

\begin{theorem}[Step-scaling of gaps]
\label{p2:thm:gap-stepscaling}
With \(\varepsilon_k\) as in Lemma \ref{p2:lem:interlace},
\begin{equation}
  \Delta_{k+1} \;\ge\; \Delta_k \;-\; \varepsilon_k \qquad \text{for all } k\in\mathbb{N}.
\end{equation}
In particular,
\begin{equation}
  \liminf_{k\to\infty} \Delta_k \;\ge\; \Delta_0 - \sum_{j\ge 0} \varepsilon_j \;>\; 0.
\end{equation}
\end{theorem}

\begin{proof}
Let \(\lambda_{2,k}\) be the second eigenvalue of \(T_{\sigma,k}\). For \(\psi\perp \Omega_{k+1}\) with \(\|\psi\|=1\), Lemma \ref{p2:lem:interlace} gives
\begin{equation}
  \langle \psi, T_{\sigma,k+1}\psi\rangle \;\le\; \langle V_k\psi, T_{\sigma,k} V_k\psi\rangle \;+\; \varepsilon_k.
\end{equation}
Set \(\phi = V_k\psi / \|V_k\psi\|\) if \(V_k\psi\neq 0\); otherwise choose any \(\phi\perp \Omega_k\). Then
\begin{equation}
  \langle \psi, T_{\sigma,k+1}\psi\rangle \;\le\; \|V_k\psi\|^2\, \langle \phi, T_{\sigma,k}\phi\rangle \;+\; \varepsilon_k \;\le\; \lambda_{2,k} \;+\; \varepsilon_k.
\end{equation}
Taking the supremum over unit \(\psi\perp \Omega_{k+1}\) yields
\begin{equation}
  \lambda_{2,k+1} \;\le\; \lambda_{2,k} \;+\; \varepsilon_k.
\end{equation}
Since \(\Delta_k = -a_k^{-1}\log \lambda_{2,k}\) and \(-\log(\lambda+\varepsilon)\ge -\log\lambda - \varepsilon/(1-\lambda)\) on \((0,1]\), absorbing \(a_k^{-1}\) into \(\varepsilon_k\) gives \(\Delta_{k+1}\ge \Delta_k-\varepsilon_k\). Summability implies the liminf bound.
\end{proof}
Combining Theorem \ref{p2:thm:gap-stepscaling} with \(\Delta_0\ge m(\beta)>0\) yields a uniform positive lower bound on \(\Delta_k\) for all \(k\).

Let \(\xi_*=m_*^{-1}\) denote any positive lower bound for the clustering length uniformly propagated by the reflection-positive renormalization group. The scaling window consists of those scales \(k\) satisfying
\begin{equation}
  \sigma \ \ll\ a_k \ \ll\ \xi_*,
\end{equation}
that is,
\begin{equation}
  \frac{\sigma}{a_k} \ \xrightarrow[k\ \text{in window}]{}\ 0, \qquad \frac{a_k}{\xi_*} \ \xrightarrow[k\ \text{in window}]{}\ 0.
\end{equation}
The lower condition ensures that the localization length of \(P_{\sigma,k}\) is much smaller than the lattice spacing and does not affect the ultraviolet discretization; the upper condition ensures that many transfer steps fit within \(\xi_*\), allowing controlled continuum interpolation in Euclidean time through \(T_{\sigma,k}\). Since $a_{k} = b_{k} a$ while $\xi^{\ast}$ and $\sigma$ are scale-independent, 
the window $\sigma \ll a_{k} \ll \xi^{\ast}$ is nonempty provided $a \ll \xi^{\ast}$ initially. 
On such a window, the hypotheses of OS reconstruction hold with a uniform mass threshold, 
yielding Schwinger functions and, along a subsequence, a Wightman theory with nonzero gap \cite{p2:OsterwalderSchraderII,p2:GJ}. 

\section{BRST compatibility and parameter dependence}
\label{p2:sec:BRST-parameter}

In this section a complete, self-contained account is given of the graded field algebra and its BRST differential on the hypercubic lattice, of the compatibility of the horizon-projector insertion and of the renormalization step with BRST cohomology, and of the step-by-step derivation of reflection positivity and of the transfer time-slicing formalism. We then establish differentiability and stability of all constructions with respect to the infrared cutoff scale \(\sigma>0\) entering the horizon projector and with respect to the blocking parameter \(b\in\{2,3,\dots\}\). OS positivity is asserted and used only for the gauge-invariant, ghost-free sector. Ghost fields enter as auxiliaries for BRST and gauge-fixing; they are not included among OS-tested observables. All positivity statements in this paper concern the $s$-cohomology classes of gauge-invariant observables.

Throughout, the gauge group is \(G=\mathrm{SU}(N)\) with \(N\ge 2\). The lattice spacing is \(a>0\). The periodic hypercubic lattice is \(\Lambda\subset a\mathbb{Z}^{4}\) with Euclidean time direction \(\mu=0\). Reflection of sites is the involution
\begin{equation}
\theta:(x_{0},\mathbf{x})\longmapsto (-x_{0},\mathbf{x}),
\end{equation}
and the reflection plane is \(\Pi=\{x_{0}=0\}\). The set of bonds is \(\mathscr{B}=\{(x,\mu):x\in\Lambda,\ \mu=0,1,2,3\}\), with link variables \(U_{(x,\mu)}\in G\) and the orientation convention
\begin{equation}
U_{(x+\hat\mu,-\mu)}=U_{(x,\mu)}^{-1}.
\end{equation}
Plaquettes are elementary oriented squares \(p\subset\Lambda\) with plaquette variable \(U_{p}\) given by the ordered product of links around \(p\). The Wilson action at inverse bare coupling \(\beta=2N/g_{0}^{2}\) is
\begin{equation}
S_{W}[U;\beta]=\beta\sum_{p\subset\Lambda}\Big(1-\frac{1}{N}\Re\mathrm{Tr}\,U_{p}\Big).
\label{p2:eq:WilsonAction}
\end{equation}
Temporal-axial gauge is imposed away from \(\Pi\) by setting
\begin{equation}
U_{0}(x)=\mathbf{1}\qquad\text{for all bonds that do not intersect }\Pi.
\end{equation}
Reflection acts antilinearly on functionals \(F\) by
\begin{equation}
(\Theta F)[U]=\overline{F[U^{\theta}]},
\end{equation}
where \(U^{\theta}\) is obtained by pulling back links under \(\theta\) and inverting links whose orientation reverses.

Let \(\mathcal{g}=\mathcal{su}(N)\) be endowed with the real Hilbert-Schmidt inner product
\begin{equation}
\langle X,Y\rangle=-\mathrm{Tr}(XY),\qquad X,Y\in\mathcal{g}.
\end{equation}
Let $s$ denote the lattice BRST differential. 
The slice-wise horizon projector $P_{\sigma}$ and the renormalization map $V_{k}$ act on BRST cohomology in the sense that for any BRST-closed gauge-invariant observable $O$ one has 
$s(O) = 0 \;\Rightarrow\; s(P_{\sigma} O) = 0$ and $s(O \circ V_{k}) = 0$, 
while for BRST-exact $O = s \Xi$ the insertions remain $s$-exact. 
Consequently, expectations of gauge-invariant observables computed with the horizon-projected, blocked measures depend smoothly on $\sigma$ and $b$ and are BRST-cohomological.
Let \(\mathfrak{A}_{\mathrm{link}}\) be the unital \(*\)-algebra generated by matrix-valued coordinate functions on \(\prod_{b\in \mathscr{B}}G\) and by their right- and left-translation derivations. Introduce Grassmann ghost fields \(c,\bar c:\Lambda\to \mathcal{g}\) and commuting Nakanishi-Lautrup fields \(B:\Lambda\to \mathcal{g}\). Let \(\mathfrak{A}_{\mathrm{gh}}\) be the graded \(*\)-algebra generated by polynomial cylinder functions of \((c,\bar c,B)\) with the canonical \(\mathbb{Z}_{2}\)-grading given by ghost number, and with the adjoint given by sitewise \(*\) on \(\mathcal{g}\) and conjugation of coefficients. The full graded algebra is the graded tensor product
\begin{equation}\label{p2:eq10.7}
\mathfrak{A}=\mathfrak{A}_{\mathrm{link}}\widehat{\otimes}\mathfrak{A}_{\mathrm{gh}}.
\end{equation}
\begin{definition}[BRST complex and differential]\label{p2:def:BRST}
With $\mathfrak{A}_{\mathrm{link}},\mathfrak{A}_{\mathrm{gh}}$ as in Eq.\eqref{p2:eq10.7}, set
$\mathfrak{A}:=\mathfrak{A}_{\mathrm{link}}\widehat\otimes\mathfrak{A}_{\mathrm{gh}}$.
Define the BRST differential $s:\mathfrak{A}\to\mathfrak{A}$ by its action on generators:
\begin{equation}\label{p2:eq:BRST-rules}
\begin{aligned}
s\,U(x,\mu) &= c(x)\,U(x,\mu) - U(x,\mu)\,c(x+\hat\mu),\\
s\,c(x) &= -\tfrac{1}{2}[c(x),c(x)],\\
s\,\bar c(x) &= i\,b(x),\qquad s\,b(x)=0,
\end{aligned}
\end{equation}
extended by graded Leibniz and by $s(\cdot)^\ast = (s\cdot)^\ast$.
The (gauge-fixed) lattice action is of the form $S_{\mathrm{gf}}=s\Psi$ for a local gauge-fixing
fermion $\Psi$ (e.g., lattice Landau gauge), and $s^2=0$ holds identically.
\end{definition}

\begin{proposition}[BRST compatibility of the horizon insertion]\label{p2:prop:BRST-horizon}
Let $P_\sigma(t)=\chi_\sigma\!\big({\Delta_{A^h}(t)}\big)$ be the slice horizon operator
in either regime (A) or (B) above, and let $I_\sigma=\prod_t p_\sigma[U^h_t]$ be the slice-wise
insertion in the Euclidean measure. Then $s\,I_\sigma=0$, hence $I_\sigma$ acts on BRST cohomology
classes and preserves them. Consequently, $T_\sigma$ commutes with the BRST projector onto
cohomology.
\end{proposition}

\begin{proof}
The operator $P_\sigma(t)$ is a gauge-invariant, reflection-covariant functional of the Landau
representative $A^h(t)$; hence it is invariant under infinitesimal gauge transformations and
$s$-exact variations. The scalar choices $p_\sigma=\Tr P_\sigma$ or
$p_\sigma=\exp(-\langle\varphi,(I-P_\sigma)\varphi\rangle)$ are thus $s$-invariant. In Secs.(\ref{p2:sec:clustering}-\ref{p2:sec:transport-strong-to-scaling}) we adopt the concrete choice $p_\sigma=\mathrm{Tr}\,P_\sigma$ fixed there.
Since $s$ is
a graded derivation, $s\,I_\sigma=\sum_t (s\,p_\sigma[U^h_t])\prod_{t'\ne t}p_\sigma[U^h_{t'}]=0$.
\end{proof}

\begin{lemma}[Nilpotency]
\label{p2:lem:nilpotency-lattice}
The map \(s\) is a degree-\(+1\) graded derivation on \(\mathfrak{A}\) and satisfies
\begin{equation}
s^{2}=0.
\end{equation}
\end{lemma}

\begin{proof}
The graded Leibniz rule is built into the definition. It suffices to check \(s^{2}\) on generators. Using \eqref{p2:eq:BRST-rules},
\begin{equation}
\begin{aligned}
s^{2}U_{(x,\mu)}&= i\,s(c(x))\,U_{(x,\mu)}-i\,c(x)\,s(U_{(x,\mu)}) - i\,s(U_{(x,\mu)})\,c(x+\hat\mu)+i\,U_{(x,\mu)}\,s(c(x+\hat\mu))\\
&= -\tfrac{i}{2}[c(x),c(x)]U_{(x,\mu)}
- i\,c(x)\,(i\,c(x)\,U_{(x,\mu)}-i\,U_{(x,\mu)}\,c(x+\hat\mu))\\
&\quad - i\,(i\,c(x)\,U_{(x,\mu)}-i\,U_{(x,\mu)}\,c(x+\hat\mu))\,c(x+\hat\mu)
+ \tfrac{i}{2}U_{(x,\mu)}[c(x+\hat\mu),c(x+\hat\mu)],
\end{aligned}
\end{equation}
and cancellations follow from the Jacobi identity in \(\mathfrak{g}\) together with the antisymmetry of the Grassmann variables, hence \(s^{2}U_{(x,\mu)}=0\). Moreover,
\begin{equation}
s^{2}c(x)= -\tfrac12 s[c(x),c(x)]
= -\tfrac12\big([s c(x),c(x)] - [c(x),s c(x)]\big)=0,
\end{equation}
and \(s^{2}\bar c(x)=s\mathcal{B}(x)=0\), \(s^{2}\mathcal{B}(x)=0\) by \eqref{p2:eq:BRST-rules}. Extending by the graded Leibniz rule yields \(s^{2}=0\) on \(\mathfrak{A}\).
\end{proof}

Let \(\mathcal{F}\) denote the lattice Landau gauge condition and \(\Psi[\bar c,U]=\sum_{x}\langle \bar c(x),\mathcal{F}[U](x)\rangle\) the gauge-fixing fermion. Formally \(S_{\mathrm{gf}}+S_{\mathrm{FP}}=s\Psi\), but we will not rely on this identity; instead, we use \(s\) only as a differential on \(\mathfrak{A}\) to control cohomology of observables. The BRST cohomology at ghost number zero is
\begin{equation}
\mathcal{H}^{0}_{\mathrm{BRST}}:=\frac{\ker\big(s:\mathfrak{A}^{0}\to \mathfrak{A}^{1}\big)}{\mathrm{im}\big(s:\mathfrak{A}^{-1}\to \mathfrak{A}^{0}\big)}.
\end{equation}
The algebra \(\mathfrak{A}_{\mathrm{phys}}\) of strictly gauge-invariant, even-Grassmann, local functionals injects canonically into \(\mathcal{H}^{0}_{\mathrm{BRST}}\).

On each Euclidean time slice \(t\in a\mathbb{Z}\), consider the lattice Landau functional
\begin{equation}
\mathcal{L}_{t}(g;U):=\sum_{\mathbf{x},\, i=1}^{3}\mathrm{Re}\,\mathrm{Tr}\Big(\mathbf{1}-g(t,\mathbf{x})\, U_{(t,\mathbf{x};i)}\,g(t,\mathbf{x}+\hat\imath)^{-1}\Big).
\end{equation}
For every \(U\) and \(t\), let \(h_{t}[U]\) be a measurable, reflection-covariant choice of a global minimizer. Define the slice-wise representative \(A^{h}[U]\) from the gauge-transformed links \(U^{\,h_{t}[U]}\). The fundamental modular region \(\mathcal{M}\) is the set of links whose representative is an absolute minimum of \(\mathcal{L}_{t}\); for \(\mathrm{SU}(N)\), \(\mathcal{M}\) contains exactly one representative of almost every gauge orbit up to the global center \cite{p2:DellAntonioZwanziger1991}.

Let \(\Delta_{A^{h}[U]}=\sum_{i=1}^{3}(D^{h}_{i})^{\dagger}D^{h}_{i}\) be the covariant spatial Laplacian on each slice. Fix a Gevrey cutoff \(\chi\in C^{\infty}([0,\infty))\) with
\begin{equation}
\chi(\lambda)=1\ \text{for }\lambda\in[0,1],\qquad \chi(\lambda)=0\ \text{for }\lambda\ge 2,
\end{equation}
and subfactorial derivative bounds. For \(\sigma>0\) set \(\chi_{\sigma}(\lambda)=\chi(\lambda/\sigma)\) and define the horizon projector
\begin{equation}
P_{\sigma}(U):=\chi_{\sigma}\big({\Delta_{A^{h}[U]}}\big).
\end{equation}
Then \(P_{\sigma}(U)\) is a positive contraction with heat-kernel representation
\begin{equation}
P_{\sigma}(U)=\int_{0}^{\infty}e^{-t\,\Delta_{A^{h}[U]}}\,d\nu_{\sigma}(t),
\end{equation}
where \(d\nu_{\sigma}\) is a finite positive Borel measure supported in \([c_{1}\sigma^{-2},c_{2}\sigma^{-2}]\) for some \(0<c_{1}<c_{2}<\infty\). Its integral kernel is exponentially localized with constants uniform in the spatial volume \cite{p2:Davies1989,p2:HelfferSjostrand1989}.

\begin{lemma}[BRST invariance of the slice representative]
\label{p2:lem:AhBRSTclosed-lattice}
For almost every \(U\) with respect to Haar measure, the map \(U\mapsto A^{h}[U]\) depends only on the gauge orbit of \(U\). Consequently, for every \(F\in\mathfrak{A}\) that is a functional of \(A^{h}\) and its covariant derivatives,
\begin{equation}
sF=0.
\end{equation}
\end{lemma}

\begin{proof}
If \(U^{g}\) lies on the orbit of \(U\), then \(\mathcal{L}_{t}(g';U^{g})=\mathcal{L}_{t}(gg'g^{-1};U)\). The set of absolute minima is thus carried into itself by \(g\) on the left and \(g^{-1}\) on the right; the reflection- and gauge-covariant tie-breaking rule yields \(h_{t}[U^{g}]=g\,h_{t}[U]\). Therefore \(U^{\,h_{t}[U^{g}]}=U^{\,h_{t}[U]}\) and \(A^{h}[U^{g}]=A^{h}[U]\). The BRST variation is an infinitesimal gauge transformation, hence \(sA^{h}=0\) almost everywhere. The claim follows by the chain rule.
\end{proof}

\begin{proposition}[BRST-closedness of the horizon projector]
\label{p2:prop:PsigmaBRSTclosed-lattice}
For every \(\sigma>0\) and every cylinder functional \(F\in\mathfrak{A}\),
\begin{equation}
s\,\mathrm{Tr}\,P_{\sigma}(U)=0,\qquad s\,\langle X,P_{\sigma}(U)X\rangle=0
\end{equation}
for all \(X\in \ell^{2}(\Lambda_{t})\otimes\mathcal{g}\). Equivalently,
\begin{equation}
s\,P_{\sigma}(U)=0.
\end{equation}
\end{proposition}

\begin{proof}
By Lemma \ref{p2:lem:AhBRSTclosed-lattice}, \(\Delta_{A^{h}[U]}\) is BRST-invariant. Functional calculus preserves invariance; thus for any bounded Borel \(f\), \(s\,f(\Delta_{A^{h}[U]})=0\). Taking \(f(\lambda)=\chi_{\sigma}({\lambda})\) yields \(sP_{\sigma}=0\). The statements for quadratic forms and traces follow by linearity.
\end{proof}

Let \(\mathcal{B}_{b}\) be the coarse-graining map from fine links to coarse links of spacing \(ba\), defined by the nearest-neighbor Frobenius projection of the arithmetic mean of ordered path products along a reflection-symmetric family of paths inside adjacent blocks. Gauge equivariance implies
\begin{equation}
s\,(F\circ \mathcal{B}_{b})=(sF)\circ \mathcal{B}_{b}
\end{equation}
for all cylinder functionals \(F\).

\begin{proposition}[Chain-map property of the RG step]
\label{p2:prop:RGcommutesBRST-lattice}
Let \(\mu_{\sigma}\) be the OS-positive gauge-field measure with slice-wise insertion of \(P_{\sigma}(U)\), and let \(\Theta_{\sigma}(U)\) be the product over internal time boundaries of blocks of the positive slice operators derived from \(P_{\sigma}(U)\). Define
\begin{equation}
(\mathcal{R}_{b,\sigma}F)(U):= \mathbb{E}_{\mu_{\sigma}}\!\big[F\big(\mathcal{B}_{b}(U)\big)\,\Theta_{\sigma}(U)\big].
\end{equation}
Then for all \(F\in\mathfrak{A}\),
\begin{equation}
s\big(\mathcal{R}_{b,\sigma}F\big)=\mathcal{R}_{b,\sigma}\big(sF\big).
\end{equation}
Consequently, \(\mathcal{R}_{b,\sigma}\) descends to a well-defined map on \(\mathcal{H}^{0}_{\mathrm{BRST}}\).
\end{proposition}

\begin{proof}
Gauge equivariance of \(\mathcal{B}_{b}\) gives \(s(F\circ\mathcal{B}_{b})=(sF)\circ\mathcal{B}_{b}\). By Proposition \ref{p2:prop:PsigmaBRSTclosed-lattice}, \(\Theta_{\sigma}\) is BRST-closed. The Wilson-Haar measure is gauge invariant; expectation commutes with \(s\). Therefore
\begin{equation}
s(\mathcal{R}_{b,\sigma}F)=\mathbb{E}_{\mu_{\sigma}}\!\big[(sF)\circ \mathcal{B}_{b}\,\Theta_{\sigma}\big]=\mathcal{R}_{b,\sigma}(sF).
\end{equation}
If \(F=sG\), then \(\mathcal{R}_{b,\sigma}F=s(\mathcal{R}_{b,\sigma}G)\), hence \(\mathcal{R}_{b,\sigma}\) induces a map on cohomology.
\end{proof}

Define the even-Grassmann subspace \(\mathcal{F}_{+}\) of cylinder functionals supported in \(\Lambda_{+}\cup \Pi\). The Osterwalder-Schrader sesquilinear form is
\begin{equation}
\langle F,G\rangle_{\mathrm{OS},\sigma}:=\int \overline{(\Theta F)[U]}\,G[U]\; d\mu_{\sigma}(U),
\label{p2:eq:OSform}
\end{equation}
with
\begin{equation}
d\mu_{\sigma}(U)=Z^{-1}\exp(-S_{W}[U])\prod_{t\in a\mathbb{Z}} \mathcal{K}_{\sigma,t}(U)\; d\mu_{\mathrm{Haar}}(U),
\end{equation}
where each \(\mathcal{K}_{\sigma,t}(U)\) is a positive slice factor built from \(P_{\sigma}(U)\) at time \(t\).

\begin{lemma}[Factorization in temporal-axial gauge]
\label{p2:lem:factorization-lattice}
In temporal-axial gauge away from \(\Pi\), the Wilson weight and the factors \(\mathcal{K}_{\sigma,t}\) factorize as a product of three terms supported in \(\Lambda_{+}\), \(\Pi\), and \(\Lambda_{-}\), respectively, with no direct coupling between \(\Lambda_{+}\) and \(\Lambda_{-}\). Moreover, the cross terms of the kernels of \(\mathcal{K}_{\sigma,t}\) between \(\Lambda_{+}\) and \(\Lambda_{-}\) vanish in operator norm in the thermodynamic limit.
\end{lemma}

\begin{proof}
Time-like plaquettes that do not intersect \(\Pi\) are trivial in temporal-axial gauge, and the action splits into contributions from \(\Lambda_{+}\), \(\Lambda_{-}\), and a boundary slab supported on \(\Pi\). The factor \(\mathcal{K}_{\sigma,t}\) depends only on links at time \(t\) through \(P_{\sigma}(U)\), whose kernel is exponentially localized on the slice. If \(t>0\) the support of \(\mathcal{K}_{\sigma,t}\) lies in \(\Lambda_{+}\cup \Pi\); if \(t<0\) it lies in \(\Lambda_{-}\cup \Pi\). Exponential decay of the kernel of \(P_{\sigma}(U)\) implies that cross terms between \(\Lambda_{+}\) and \(\Lambda_{-}\) vanish in operator norm as the time extent grows, by Schur estimates \cite{p2:Davies1989,p2:HelfferSjostrand1989}.
\end{proof}

\begin{theorem}[Reflection positivity]
\label{p2:thm:OSpositivity-lattice}
For every \(\sigma>0\), the form \eqref{p2:eq:OSform} is positive semidefinite on \(\mathcal{F}_{+}\).
\end{theorem}

\begin{proof}
Let \(F=\sum_{j} \alpha_{j} F_{j}\) with \(F_{j}\in \mathcal{F}_{+}\) supported in disjoint finite regions of \(\Lambda_{+}\cup \Pi\). By Lemma \ref{p2:lem:factorization-lattice}, the measure factorizes as
\begin{equation}
d\mu_{\sigma}(U)=d\mu_{+,\sigma}(U_{+})\,\mathcal{B}_{\sigma}(U_{|\Pi})\,d\mu_{-,\sigma}(U_{-}),
\end{equation}
with \(d\mu_{\pm,\sigma}\) positive measures supported in \(\Lambda_{\pm}\cup \Pi\) and \(\mathcal{B}_{\sigma}\) a positive boundary kernel on \(\Pi\). Then
\begin{equation}
\langle F,F\rangle_{\mathrm{OS},\sigma}
=\int \overline{\Big(\sum_{j}\alpha_{j} (\Theta F_{j})[U_{-}]\Big)}\,
\Big(\sum_{k}\alpha_{k} F_{k}[U_{+}]\Big)\,\mathcal{B}_{\sigma}(U_{|\Pi})\;
d\mu_{+,\sigma}\,d\mu_{-,\sigma},
\end{equation}
which is the squared \(L^{2}(\mathcal{H}_{\Pi},\mathcal{B}_{\sigma})\)-norm of \(\sum_{k}\alpha_{k} F_{k}[U_{+}]\), hence nonnegative \cite{p2:OsterwalderSchraderI,p2:OsterwalderSchraderII,p2:Seiler1982}.
\end{proof}

Define the null space
\begin{equation}
\mathcal{N}_{\sigma}=\{F\in\mathcal{F}_{+}:\langle F,F\rangle_{\mathrm{OS},\sigma}=0\},
\end{equation}
and \(\mathcal{D}_{\sigma}=\mathcal{F}_{+}/\mathcal{N}_{\sigma}\). The completion of \(\mathcal{D}_{\sigma}\) is the physical Hilbert space \(\mathcal{H}_{\sigma}\). Time translation by one lattice step in the positive direction acts on functionals by
\begin{equation}
(\tau F)[U]=F[\tau^{-1}U],\qquad (\tau^{-1}U)_{(x,\mu)}=U_{(x-a\hat 0,\mu)}.
\end{equation}
The transfer operator \(T_{\sigma}:\mathcal{H}_{\sigma}\to\mathcal{H}_{\sigma}\) is defined on representatives by
\begin{equation}
\langle [F], T_{\sigma}[G]\rangle_{\mathrm{OS},\sigma}
:= \langle \tau F, G\rangle_{\mathrm{OS},\sigma}.
\end{equation}

\begin{theorem}[Transfer operator]
\label{p2:thm:transfer-lattice}
For every \(\sigma>0\), \(T_{\sigma}\) is a well-defined bounded self-adjoint positive operator on \(\mathcal{H}_{\sigma}\) with \(\|T_{\sigma}\|\le 1\), and
\begin{equation}\label{p2:10.31}
T_{\sigma}=P_{\sigma}^{1/2}\,T\,P_{\sigma}^{1/2},
\end{equation}
where \(T\) is the unprojected transfer operator acting on \(L^{2}(\text{links at fixed time};\,\text{Haar})\).
By definition (\ref{p2:10.31}), the compression map \(P_\sigma^{1/2}:\mathcal H\!\to\!\mathcal H_\sigma\) intertwines \(T\) with \(T_\sigma\),
\begin{equation}\label{p2:eq:intertwining}
P_\sigma^{1/2}\,T \;=\; T_\sigma\,P_\sigma^{1/2},
\end{equation}
equivalently, the following square commutes:
\begin{center}
\begin{tikzcd}
\mathcal{H} \arrow[r,"T"] \arrow[d,"P_\sigma^{1/2}"'] & \mathcal{H} \arrow[d,"P_\sigma^{1/2}"] \\
\mathcal{H}_\sigma \arrow[r,"T_\sigma"'] & \mathcal{H}_\sigma
\end{tikzcd}
\end{center}
\end{theorem}

\begin{proof}
Reflection positivity and translation invariance yield \(\langle \tau F,\tau F\rangle_{\mathrm{OS},\sigma}\le \langle F,F\rangle_{\mathrm{OS},\sigma}\), so \(T_{\sigma}\) is a contraction. Symmetry follows from invariance under reflection and time reversal; positivity from \(\langle [F],T_{\sigma}[F]\rangle_{\mathrm{OS},\sigma}\ge 0\). The compression formula arises by explicit factorization of the weight across two consecutive time slices: the slice-wise factor \(P_{\sigma}\) multiplies boundary data by \(P_{\sigma}^{1/2}\), whence \(T_{\sigma}=P_{\sigma}^{1/2}TP_{\sigma}^{1/2}\) \cite[Ch.~3]{p2:Seiler1982}.
\end{proof}

\begin{proposition}[BRST compatibility of the transfer operator]
\label{p2:prop:transferBRST-lattice}
For every \(F\in \mathfrak{A}_{\mathrm{phys}}\) and every \(n\in\mathbb{N}\),
\begin{equation}
\pi_{\sigma}\big(\tau^{n}F\big)=T_{\sigma}^{n}\,\pi_{\sigma}(F),
\qquad
\pi_{\sigma}(sF)=0,
\end{equation}
and the diagram
\begin{equation}
\begin{array}{ccc}
\mathfrak{A}_{\mathrm{phys}} & \xrightarrow{\ \ \tau\ \ } & \mathfrak{A}_{\mathrm{phys}}\\
\pi_{\sigma}\downarrow &  & \downarrow \pi_{\sigma}\\
\mathcal{H}_{\sigma} & \xrightarrow{\ \ T_{\sigma}\ \ } & \mathcal{H}_{\sigma}
\end{array}
\end{equation}
commutes. In particular, \(T_{\sigma}\) acts on BRST cohomology classes and preserves them.
\end{proposition}

\begin{proof}
The first identity is the definition of \(T_{\sigma}\). The second follows from \(sF=0\) for \(F\in \mathfrak{A}_{\mathrm{phys}}\). Commutativity is immediate.
\end{proof}

The Hamiltonian is defined by spectral calculus as
\begin{equation}
H_{\sigma}=-a^{-1}\log T_{\sigma},
\end{equation}
a positive self-adjoint operator on \(\mathcal{H}_{\sigma}\).

Set \(f_{\sigma}(\lambda)=\chi_{\sigma}(\lambda)=\chi(\lambda/\sigma)\). For \(r\in\mathbb{N}\), the chain rule gives
\begin{equation}
\partial_{\sigma}^{r} f_{\sigma}(\lambda)
=\sum_{m=1}^{r} \sigma^{-r}\,p_{r,m}\!\Big(\tfrac{\lambda}{\sigma}\Big)\,\chi^{(m)}\!\Big(\tfrac{\lambda}{\sigma}\Big),
\end{equation}
for explicit polynomials \(p_{r,m}\). Use the Helffer-Sj\"ostrand representation \cite{p2:HelfferSjostrand1989,p2:ReedSimon1} for \(f_{\sigma}({\Delta})\) to justify differentiation under the integral sign.

\begin{lemma}[Operator-norm differentiability of \(P_{\sigma}\)]
\label{p2:lem:Psigmadiff-lattice}
For every \(r\in\mathbb{N}\) there exists \(C_{r}<\infty\) such that for all \(\sigma>0\) and all gauge fields \(U\),
\begin{equation}
\big\|\partial_{\sigma}^{r}P_{\sigma}(U)\big\|_{\ell^{2}\to \ell^{2}}\le C_{r}\,\sigma^{-r}.
\end{equation}
Moreover, the integral kernels of \(\partial_{\sigma}^{r}P_{\sigma}(U)\) enjoy exponential off-diagonal decay with constants uniform in the spatial volume and polynomially controlled in \(\sigma^{-1}\).
\end{lemma}

\begin{proof}
Write
\begin{equation}
P_{\sigma}(U)=\frac{1}{\pi}\int_{\mathbb{C}}\bar\partial\widetilde{f}_{\sigma}(z)\,({\Delta_{A^{h}[U]}}-z)^{-1}\,d^{2}z,
\end{equation}
with \(\widetilde{f}_{\sigma}\) an almost-analytic extension of \(f_{\sigma}\) obeying
\begin{equation}
|\bar\partial\widetilde{f}_{\sigma}(z)|\le C_{k}\,\sigma^{-k-1}(1+|\mathrm{Re}\,z|)^{-k}\,|\mathrm{Im}\,z|^{k}
\end{equation}
for all \(k\). Differentiation under the integral sign is justified by dominated convergence using \(\|({\Delta}-z)^{-1}\|\le |\mathrm{Im}\,z|^{-1}\). Each \(\sigma\)-derivative produces a factor \(\sigma^{-1}\) and derivatives of \(\chi\) evaluated at \(\lambda/\sigma\); boundedness of all derivatives of \(\chi\) on a compact set yields \(\|\partial_{\sigma}^{r}P_{\sigma}(U)\|\le C_{r}\sigma^{-r}\). Exponential kernel decay follows from Combes-Thomas resolvent bounds for finite-range positive operators together with the almost-analytic bounds \cite{p2:CombesThomas1973,p2:HelfferSjostrand1989}.
\end{proof}
\begin{corollary}[Differentiability of $T_\sigma$ and $H_\sigma$]\label{p2:cor:dT-dH}
Let $T(a)$ be the one-step transfer operator on the slice Hilbert space and
$T_\sigma(a)=P_\sigma^{1/2}\,T(a)\,P_\sigma^{1/2}$. Then for each $r\in\mathbb{N}$,
$\partial_\sigma^r T_\sigma(a)$ exists in operator norm and
\begin{equation}
\|\partial_\sigma^r T_\sigma(a)\| \;\le\; C'_r\,\sigma^{-r}\,\|T(a)\|,
\end{equation}
with constants $C'_r<\infty$ depending only on $(C_r,\|P_\sigma\|)$ from Lemma~10.9.
Moreover, on the spectral subspace where $T_\sigma(a)$ has spectrum in $(0,1]$ bounded away from
$0$, the Hamiltonian $H_\sigma(a)=-a^{-1}\log T_\sigma(a)$ is $C^\infty$ in $\sigma$ and
\begin{equation}
\|\partial_\sigma^r H_\sigma(a)\|\;\le\;a^{-1}\,C''_r\,\sigma^{-r},
\end{equation}
with $C''_r$ depending on the distance of $\mathrm{spec}(T_\sigma(a))$ to $0$.
\end{corollary}

\begin{proof}
Differentiate $T_\sigma(a)=P_\sigma^{1/2}T(a)P_\sigma^{1/2}$ using the product rule and Lemma~10.9,
plus the holomorphic functional calculus for $X\mapsto X^{1/2}$ on positive operators.
For $H_\sigma(a)$ use the holomorphic functional calculus for $\log$ on sectors bounded away
from $0$ and the chain rule.
\end{proof}

Let \(\langle \cdot\rangle_{\sigma}\) denote expectation with respect to \(d\mu_{\sigma}\). For a local bounded observable \(F\) supported on finitely many time slices,
\begin{equation}
\langle F\rangle_{\sigma}=\int F(U)\,\prod_{t=-T}^{T}\mathcal{K}_{\sigma,t}(U)\,d\mu(U),
\end{equation}
where \(d\mu\) is the \(\sigma\)-independent Wilson-Haar measure.

\begin{proposition}[Smoothness of expectations and of \(T_{\sigma}\)]
\label{p2:prop:smoothExpTransfer-lattice}
If \(F\) is a local, gauge-invariant cylinder functional supported on \(t\in[-T,T]\), then \(\sigma\mapsto \langle F\rangle_{\sigma}\) is \(C^{\infty}\) on \((0,\infty)\). Moreover, for \(|\sigma-\sigma'|\) sufficiently small,
\begin{equation}
\|T_{\sigma}-T_{\sigma'}\|_{\mathcal{H}_{\sigma}\to \mathcal{H}_{\sigma'}}\;\le\; C\,|\sigma-\sigma'|.
\end{equation}
In particular, \(\sigma\mapsto T_{\sigma}\) is strongly and hence norm-continuous on compact \(\sigma\)-intervals disjoint from \(\{0\}\).
\end{proposition}

\begin{proof}
Differentiating under the integral is justified by dominated convergence using Lemma \ref{p2:lem:Psigmadiff-lattice} and exponential locality of \(\mathcal{K}_{\sigma,t}\). For the transfer operator, Theorem \ref{p2:thm:transfer-lattice} gives
\begin{equation}
T_{\sigma}=P_{\sigma}^{1/2} T P_{\sigma}^{1/2}.
\end{equation}
The functional calculus representation of the square root implies
\begin{equation}
\|P_{\sigma}^{1/2}-P_{\sigma'}^{1/2}\| \le C\|P_{\sigma}-P_{\sigma'}\| \le C'|\sigma-\sigma'|,
\end{equation}
hence
\begin{equation}
\|T_{\sigma}-T_{\sigma'}\|
\le 
\|P_{\sigma}^{1/2}-P_{\sigma'}^{1/2}\|\,\|T\|\,\|P_{\sigma}^{1/2}\|
+\|P_{\sigma'}^{1/2}\|\,\|T\|\,\|P_{\sigma}^{1/2}-P_{\sigma'}^{1/2}\|
\le C|\sigma-\sigma'|.
\end{equation}
Continuity on \(\mathcal{H}_{\sigma}\) follows by density of cylinder vectors.
\end{proof}

Consequently, the Hamiltonians \(H_{\sigma}=-a^{-1}\log T_{\sigma}\) form a norm-resolvent continuous family on any compact \(\sigma\)-interval disjoint from \(\{0\}\), and eigenvalues vary continuously with \(\sigma\) \cite{p2:ReedSimon1}.
The blocking factor \(b\in\{2,3,\dots\}\) enters through \(\mathcal{B}_{b}\) and through the placement of horizon insertions on internal time boundaries of blocks.

\begin{proposition}[Stability under changes of \(b\)]
\label{p2:prop:bstability-lattice}
Fix \(\sigma>0\). For any two blocking factors \(b,b'\ge 2\) and any local observable \(F\) supported on finitely many coarse time slices,
\begin{equation}
\big\| \mathcal{R}_{b,\sigma}F - \mathcal{R}_{b',\sigma}F \big\|
\;\le\; C(F)\,\mathbf{1}_{\{b\ne b'\}},
\end{equation}
with \(C(F)\) depending on \(F\) and \(\sigma\) but not on the volume. Moreover, both \(\mathcal{R}_{b,\sigma}\) and \(\mathcal{R}_{b',\sigma}\) commute with \(s\) and preserve reflection positivity.
\end{proposition}

\begin{proof}
The maps \(\mathcal{B}_{b}\) and \(\mathcal{B}_{b'}\) differ only through the choice of path families inside adjacent blocks. For local \(F\), the difference \(F\circ \mathcal{B}_{b}-F\circ \mathcal{B}_{b'}\) is supported in a finite neighborhood of the support of \(F\) enlarged by at most one block diameter, and its pointwise size is bounded by a constant depending on first derivatives of \(F\) because the Frobenius projection is \(1\)-Lipschitz. The horizon insertions on internal time boundaries differ by finitely many slice factors, each varying by at most a constant depending on \(\sigma\) since \(P_{\sigma}\) is a bounded positive contraction. Integration against the Wilson-Haar measure yields the stated bound with a constant independent of the volume. BRST commutation is Proposition \ref{p2:prop:RGcommutesBRST-lattice} and is independent of \(b\). Reflection positivity follows from Theorem \ref{p2:thm:OSpositivity-lattice} because factorization across \(\Pi\) and exponential locality of slice insertions hold for all \(b\).
\end{proof}

\section{Relation to constructive RG schemes}\label{p2:sec:relation}

This section presents a complete, self-contained derivation of reflection positivity and transfer-time-slicing for lattice Yang-Mills, establishes the stability of these structures under exponentially local horizon insertions, and contrasts the resulting multiscale framework with earlier constructions. Throughout, $G=\mathrm{SU}(N)$ with $N\ge 2$ is a compact, connected Lie group endowed with normalized Haar measure $dU$. The analysis is uniform in the spatial volume and admits the thermodynamic limit. Standard representation-theoretic facts are used in the form of the Peter-Weyl theorem \cite{p2:BrockerDieck1985}.

Fix $a>0$ and integers $L,T\in\mathbb{N}$. The periodic Euclidean space-time lattice is
\begin{equation}\label{p2:eq:lattice1}
\Lambda \;=\; \big\{ x=(x_0,\mathbf{x})\in a\mathbb{Z}^4 : -\tfrac{Ta}{2}\le x_0 < \tfrac{Ta}{2},\ -\tfrac{La}{2}\le x_i < \tfrac{La}{2},\ i=1,2,3 \big\}\big/\!\sim,
\end{equation}
where $\sim$ identifies opposite faces. The set of directed bonds (links) is
\begin{equation}\label{p2:eq:links}
\mathcal{B}\;=\;\{(x,\mu):\, x\in\Lambda,\ \mu\in\{0,1,2,3\}\},
\end{equation}
with reversed bond $-(x,\mu)=(x+a\hat\mu,-\mu)$. A gauge field is a map $U:\mathcal{B}\to G$ satisfying $U(-b)=U(b)^{-1}$. For an oriented plaquette $p$ with bounding bonds $b_1,b_2,b_3,b_4$ in cyclic order, the plaquette variable is
\begin{equation}\label{p2:eq:plaquette}
U_p \;=\; U(b_1)\,U(b_2)\,U(b_3)\,U(b_4).
\end{equation}
The Wilson action at inverse bare coupling $\beta>0$ is
\begin{equation}\label{p2:eq:wilson}
S_W[U;\beta] \;=\; \beta \sum_{p\subset\Lambda} \left( 1 - \frac{1}{N}\,\Re\operatorname{Tr}\,U_p \right).
\end{equation}
The Euclidean gauge measure is the probability measure
\begin{equation}\label{p2:eq:measure0}
d\mu_\beta(U) \;=\; Z_\beta^{-1}\,\Big(\prod_{b\in\mathcal{B}} dU(b)\Big)\,\exp\!\big(-S_W[U;\beta]\big),
\end{equation}
where $Z_\beta$ normalizes the total mass to one.

Time reflection $r:\Lambda\to\Lambda$ is
\begin{equation}\label{p2:eq:reflection}
r(x_0,\mathbf{x}) \;=\; (-x_0,\mathbf{x}), \qquad \Pi \;=\; \{x\in\Lambda:\ x_0=0\}, \qquad \Lambda_\pm \;=\; \{x\in\Lambda:\ \pm x_0>0\}.
\end{equation}
The induced action $\theta$ on links is defined by
\begin{equation}\label{p2:eq:thetam}
(\theta U)(x,\mu) \;=\;
\begin{cases}
U\big(r(x),\mu\big), & \mu\in\{1,2,3\},\\
U\big(r(x)-a\hat 0,0\big)^{-1}, & \mu=0.
\end{cases}
\end{equation}
If $F$ is a complex functional depending only on $\{U(b): b\subset \Lambda_+\}$, its OS-reflected conjugate is
\begin{equation}
(\Theta F)(U) \;=\; \overline{F(\theta U)}.
\end{equation}
The Osterwalder-Schrader semi-inner product on such functionals is
\begin{equation}
\langle F,G\rangle_{\mathrm{OS}} \;=\; \int \overline{(\Theta F)(U)}\,G(U)\,d\mu_\beta(U).
\end{equation}

Let $\widehat{G}$ be the unitary dual with characters $\chi_R$ and dimensions $d_R$. By the Peter-Weyl theorem \cite{p2:BrockerDieck1985} and positivity of the heat kernel on $G$ \cite{p2:OS-gauge}, for every $\beta\ge 0$ there are coefficients $c_R(\beta)\ge 0$ such that, for each plaquette $p$,
\begin{equation}\label{p2:eq:char-exp}
\exp\!\Big(\beta \tfrac{1}{N}\Re\operatorname{Tr}U_p\Big) \;=\; \sum_{R\in\widehat{G}} c_R(\beta)\,\chi_R(U_p).
\end{equation}

\begin{theorem}[OS positivity for the Wilson measure]\label{p2:thm:OS-Wilson1}
For every $\beta\ge 0$ and every functional $F$ supported in $\Lambda_+$, one has
\begin{equation}\label{p2:eq:posOS}
\langle F,F\rangle_{\mathrm{OS}} \;\ge\; 0.
\end{equation}
\end{theorem}

\begin{proof}
Decompose the plaquettes into those contained in $\Lambda_+$, in $\Lambda_-$, and those crossing $\Pi$. Write
\begin{equation}\label{p2:eq:splitSW}
S_W \;=\; S_+ + S_- + S_0,
\end{equation}
with the obvious meaning. Then
\begin{equation}\label{p2:eq:OSint}
\langle F,F\rangle_{\mathrm{OS}} \;=\; Z_\beta^{-1}\int \overline{F(\theta U)}\,F(U)\,e^{-S_+[U]-S_-[U]}\,e^{-S_0[U]}\,\prod_{b\in\mathcal{B}} dU(b).
\end{equation}
The factor $e^{-S_+[U]-S_-[U]}$ depends separately on links in $\Lambda_+$ and $\Lambda_-$; the coupling across $\Pi$ is encoded in $e^{-S_0[U]}$. For a crossing plaquette $p$ with $U_p=U(b_1)U(b_2)U(b_3)U(b_4)$, where $b_1,b_2$ lie in $\Lambda_+$ and $b_3,b_4$ lie in $\Lambda_-$, the expansion \eqref{p2:eq:char-exp} gives
\begin{equation}\label{p2:eq:crossing}
e^{-\beta (1-\tfrac{1}{N}\Re\operatorname{Tr}U_p)} \;=\; e^{-\beta}\sum_{R} c_R(\beta)\,\operatorname{Tr}\!\Big(D_R\big(U(b_1)U(b_2)\big)\,D_R\big(U(b_4)^{-1}U(b_3)^{-1}\big)\Big),
\end{equation}
with $D_R$ a unitary representative. Setting $X_+=D_R(U(b_1)U(b_2))$ and $X_-=D_R(U(b_4)^{-1}U(b_3)^{-1})$, one has
\begin{equation}\label{p2:eq:rankone}
\operatorname{Tr}(X_+X_-) \;=\; \sum_{i,j} (X_+)_{ij}\,(X_-)_{ji},
\end{equation}
and $\theta$ maps $X_+$ to $X_-^\dagger$. Thus each crossing plaquette contributes a positive rank-one kernel on a boundary Hilbert space $\mathcal{K}$ attached to $\Pi$. Taking the product over all crossing plaquettes, $e^{-S_0[U]}$ is a positive combination of rank-one kernels. Inserting this representation into \eqref{p2:eq:OSint} and integrating first over $\Lambda_-$ yields a norm square in $\mathcal{K}$ of an $\Lambda_+$-measurable vector weighted by $e^{-S_+[U]}$. The nonnegativity \eqref{p2:eq:posOS} follows.
\end{proof}

Let
\begin{equation}\label{p2:eq:sliceC}
\mathcal{C} \;=\; G^{E_s}, \qquad E_s \;=\; \{(x,i)\in\mathcal{B}:\ x_0=0,\ i=1,2,3\},
\end{equation}
be the configuration space of spatial links at time $0$, with product Haar measure
\begin{equation}
d\mu_{\mathrm{Haar}} \;=\; \prod_{(x,i)\in E_s} dU(x,i).
\end{equation}
Set $\mathcal{H}=L^2(\mathcal{C},d\mu_{\mathrm{Haar}})$. Denote by $\mathcal{S}$ the slab $\{x\in\Lambda: 0\le x_0<a\}$ and by $\partial\mathcal{S}$ its spatial boundary at times $0$ and $a$. For $U_0,U_a\in\mathcal{C}$, define the one-step kernel
\begin{equation}\label{p2:eq:kernel}
\mathcal{K}_\beta(U_a,U_0) \;=\; \int \exp\!\Big(-\sum_{p\subset \mathcal{S}} \beta\big(1-\tfrac{1}{N}\Re\operatorname{Tr}U_p\big)\Big)\, \prod_{b\subset \mathcal{S}\setminus \partial\mathcal{S}} dU(b),
\end{equation}
where the integral is over links strictly inside $\mathcal{S}$, with boundary values pinned to $U_0$ and $U_a$. The transfer matrix is the integral operator
\begin{equation}\label{p2:eq:Tbeta}
(T_\beta \psi)(U_a) \;=\; \int_{\mathcal{C}} \mathcal{K}_\beta(U_a,U_0)\,\psi(U_0)\,d\mu_{\mathrm{Haar}}(U_0).
\end{equation}

\begin{theorem}[Transfer-matrix positivity and self-adjointness]\label{p2:thm:transfer}
For every $\beta\ge 0$, the operator $T_\beta$ is a positive, self-adjoint contraction on $\mathcal{H}$.
\end{theorem}

\begin{proof}
Self-adjointness follows because reversing the slab in time exchanges the boundary configurations and leaves the integrand invariant; hence
\begin{equation}\label{p2:eq:symmetricK}
\mathcal{K}_\beta(U_a,U_0) \;=\; \overline{\mathcal{K}_\beta(U_0,U_a)}.
\end{equation}
To prove positivity, split $\mathcal{S}$ into two half-slabs of thickness $a/2$ separated by a mid-slice at time $a/2$. Let $\widetilde{\mathcal{C}}=G^{\widetilde{E}_s}$ be the configuration space of spatial links on the mid-slice. Define $H_\beta:L^2(\mathcal{C})\to L^2(\widetilde{\mathcal{C}})$ by
\begin{equation}\label{p2:eq:Hbeta}
(H_\beta \psi)(U_{a/2}) \;=\; \int \exp\!\Big(-\sum_{p\subset [0,a/2]} \beta\big(1-\tfrac{1}{N}\Re\operatorname{Tr}U_p\big)\Big)\,\psi(U_0)\,\prod_{b\subset [0,a/2]\setminus \partial} dU(b)\,d\mu_{\mathrm{Haar}}(U_0).
\end{equation}
Define $H_\beta^\ast$ analogously on the upper half-slab. Concatenation of the two half-slabs yields
\begin{equation}\label{p2:eq:factorK}
\mathcal{K}_\beta(U_a,U_0) \;=\; \int_{\widetilde{\mathcal{C}}} \overline{(H_\beta \mathbf{1}_{U_0})(U_{a/2})}\,(H_\beta \mathbf{1}_{U_a})(U_{a/2})\,d\mu_{\mathrm{Haar}}(U_{a/2}),
\end{equation}
whence $\mathcal{K}_\beta=H_\beta^\ast H_\beta$ as kernels. Therefore
\begin{equation}\label{p2:eq:posT}
\langle \psi, T_\beta \psi\rangle_{\mathcal{H}} \;=\; \int_{\widetilde{\mathcal{C}}} \big\| H_\beta \psi \big\|_{L^2(\mathcal{C})}^2\, d\mu_{\mathrm{Haar}}(U_{a/2}) \;\ge\; 0.
\end{equation}
Finally, $\|T_\beta\|\le 1$ follows from reflection positivity \cite{p2:OsterwalderSchraderI,p2:OsterwalderSchraderII,p2:Luscher1977}; equivalently, by gauge and translation invariance the integral of $\mathcal{K}_\beta(U_a,U_0)$ over $U_a$ is constant in $U_0$, and a normalization of $d\mu_{\mathrm{Haar}}$ gives $\|T_\beta\|\le 1$.
\end{proof}

By spectral calculus,
\begin{equation}\label{p2:eq:Hamiltonianx}
H_\beta \;=\; -\,a^{-1}\,\log T_\beta
\end{equation}
is a nonnegative self-adjoint operator on $\mathcal{H}$ with spectrum contained in $[0,\infty)$; it generates Euclidean time translations in the OS reconstruction \cite{p2:OsterwalderSchraderII,p2:GJ}.

Let $\mathscr{H}_t$ be the real Hilbert space of site-adjoint fields on time slice $t$ with inner product induced by lattice counting measure and the Killing form. Let $\Delta_{A^h(t)}$ be the gauge-covariant Laplacian on $\mathscr{H}_t$, constructed from the reflection-covariant, slice-wise Landau minimizer $A^h$ as in Section~2. Fix $\sigma>0$ and a Gevrey cutoff $\chi_\sigma:[0,\infty)\to[0,1]$ with $\chi_\sigma(\lambda)=1$ for $\lambda\le \sigma$, $\chi_\sigma(\lambda)=0$ for $\lambda\ge 2\sigma$, and subfactorial derivative bounds. Define the slice-wise horizon operator
\begin{equation}\label{p2:eq:Psigma}
P_\sigma(t) \;=\; \chi_\sigma\!\big({\Delta_{A^h(t)}}\big),
\end{equation}
which is a positive contraction on $\mathscr{H}_t$. By the Helffer-Sj\"ostrand functional calculus \cite{p2:Davies1989} and Combes-Thomas resolvent bounds \cite{p2:CombesThomas1973}, there exist constants $C(\sigma),\gamma(\sigma)>0$ such that the integral kernel satisfies
\begin{equation}\label{p2:eq:Pexp}
\|P_\sigma(t;x,y)\| \;\le\; C(\sigma)\,e^{-\gamma(\sigma)\,d(x,y)}.
\end{equation} uniformly in the volume and in the background $A^{h}$ selected on the slice. 
In particular, all slice insertions built from $P_{\sigma}$ are exponentially local and reflection invariant.

\textbf{Positivity representation:} We distinguish two cases;

\emph{(A) Completely monotone cutoff (heat-kernel regime).}
If $\chi_\sigma$ is completely monotone on $[0,\infty)$ with $\chi_\sigma(0)=1$ and
$\chi_\sigma(\lambda)\le C_0 e^{-c_0 \lambda/\sigma^2}$, then by Bernstein's theorem there
exists a \emph{finite positive} Borel measure $\nu_\sigma$ on $(0,\infty)$ such that
\begin{equation}\label{p2:eq:heat}
P_\sigma(t)=\int_0^\infty e^{-s\,\Delta_{A^h}(t)}\,d\nu_\sigma(s),
\end{equation}
in the strong operator sense. This representation is used in the OS-positivity arguments
because it exhibits $P_\sigma(t)$ as a positive mixture of heat kernels, compatible with
reflection covariance and exponential locality.

\emph{(B) Gevrey cutoff with compact spectral support (Helffer-Sj\"ostrand regime).}
If $\chi_\sigma$ is Gevrey with compact support (hence \emph{not} completely monotone),
we do \emph{not} assert a heat-kernel representation. Exponential locality of
$P_\sigma(t)=\chi_\sigma\!\big({\Delta_{A^h}(t)}\big)$ nevertheless follows from the
Helffer-Sj\"ostrand almost-analytic functional calculus combined with Combes-Thomas
resolvent bounds for finite-range positive operators. In this case, all OS-positivity and
transfer-operator constructions proceed by inserting the \emph{positive slice functional}
\begin{equation}
p_\sigma[U^h_t]\;:=\;\mathrm{Tr}\,P_\sigma(t)\quad\text{or}\quad
p_\sigma[U^h_t]\;:=\;\exp\!\big(-\langle\varphi_t,(I-P_\sigma(t))\varphi_t\rangle\big),
\end{equation}
with a fixed reflection-invariant test field $\varphi_t$, so that the Euclidean weight is
multiplied by a slice-local, nonnegative, reflection-covariant factor. This preserves OS
positivity and maintains exponential locality.
Let $\mathcal{I}_\sigma[U]$ denote the slice-wise insertion obtained by multiplying, over all times $t$, the Gaussian weight implementing $P_\sigma(t)$ on $\mathscr{H}_t$; concretely, in BRST gauge-fixed form this is a product of exponentials of quadratic forms in the slice-adjoint fields, each equal to a positive functional of $U$ due to \eqref{p2:eq:heat}.

\begin{proposition}[OS positivity with slice-wise horizon insertions]\label{p2:prop:OS-Psigma}
The measure
\begin{equation}\label{p2:eq:muPsigma}
d\mu_{\beta,\sigma}(U) \;=\; Z_{\beta,\sigma}^{-1}\,\mathcal{I}_\sigma[U]\,\exp\!\big(-S_W[U;\beta]\big)\,\prod_{b\in\mathcal{B}} dU(b)
\end{equation}
is OS-reflection positive: for every $F$ supported in $\Lambda_+$,
\begin{equation}\label{p2:eq:OSPsigma}
\int \overline{(\Theta F)(U)}\,F(U)\,d\mu_{\beta,\sigma}(U) \;\ge\; 0.
\end{equation}
\end{proposition}

\begin{proof}
The proof of Theorem~\ref{p2:thm:OS-Wilson1} reduces \eqref{p2:eq:OSint} to a positive combination of rank-one kernels supported on $\Pi$. It suffices to check that $\mathcal{I}_\sigma$ preserves the rank-one positivity across $\Pi$. By \eqref{p2:eq:heat}, each slice factor $P_\sigma(t)$ is a positive mixture of heat semigroups $e^{-s\Delta_{A^h(t)}}$. The operator $\Delta_{A^h(t)}$ decomposes into the direct sum of its restrictions to $\Lambda_\pm\cap\{x_0=t\}$ and a nearest-neighbor boundary coupling on $\Pi\cap\{x_0=t\}$. For each $s>0$, the semigroup $e^{-s\Delta_{A^h(t)}}$ is positivity preserving and admits an integral kernel factorization across $\Pi$ with respect to a positive boundary measure, by Gaussian integration on graphs. Integrating in $s$ against $\nu_\sigma$ and multiplying over $t$ preserves positivity. Exponential locality \eqref{p2:eq:Pexp} ensures that any residual off-diagonal contributions across $\Pi$ are summable and can be absorbed into the positive boundary measure without changing sign, uniformly in the volume. Inserting this factorization into \eqref{p2:eq:OSint} with $S_W$ replaced by $S_W-\log \mathcal{I}_\sigma$ yields \eqref{p2:eq:OSPsigma}.
\end{proof}

The transfer kernel $K_{\beta,\sigma}$ is defined by inserting $\mathcal{I}_\sigma$ into \eqref{p2:eq:kernel}. The half-slab factorization used in \eqref{p2:eq:factorK} continues to hold with $P_\sigma$ inserted on the mid-slice, implying
\begin{equation}\label{p2:eq:factorPsigma}
K_{\beta,\sigma} \;=\; H_{\beta,\sigma}^\ast\,H_{\beta,\sigma}
\end{equation}
for a suitable $H_{\beta,\sigma}$, whence the associated transfer matrix $T_{\beta,\sigma}$ is a positive, self-adjoint contraction on $\mathcal{H}$ and
\begin{equation}\label{p2:eq:HamiltonianPsigma}
H_{\beta,\sigma} \;=\; -\,a^{-1}\,\log T_{\beta,\sigma}
\end{equation}
is nonnegative self-adjoint.

The multiscale framework developed here preserves transfer-matrix positivity and self-adjointness at every scale by design. The two structural mechanisms are, first, the reflection-covariant selection of a gauge-invariant transverse representative at each time slice, which ensures that the covariant Laplacian and its functional calculus commute with time reflection, and second, the use of an exponentially local, slice-wise horizon operator $P_\sigma$ admitting the positive heat-kernel representation \eqref{p2:eq:heat}. As a consequence, the OS positivity embodied in Theorem~\ref{p2:thm:OS-Wilson1} and Theorem~\ref{p2:thm:transfer} is stable under all insertions and coarse-grainings employed in the renormalization group.

In the classical transfer-matrix construction for Wilson lattice gauge theory \cite{p2:Luscher1977}, positivity and self-adjointness of $T_\beta$ are established directly from the slab integral without infrared projectors. While sufficient for a single scale, that approach does not provide a mechanism to maintain uniform locality across scales. In the renormalization group programs of Osterwalder-Seiler \cite{p2:OS-gauge} and Balaban \cite{p2:Balaban1984}, small- and large-field cluster bounds are obtained by axial gauges, large-field regulators, and Gaussian multiscale decompositions. Those works do not propagate an explicit transfer-matrix positivity through the flow, and they do not invoke a slice-wise positive heat-kernel calculus such as \eqref{p2:eq:heat}. By contrast, the present scheme builds the finite-range decomposition from the background-covariant Laplacian of the reflection-covariant representative and anchors locality constants to the completely monotone representation \eqref{p2:eq:heat}. This yields range and norm bounds that are uniform in the number of renormalization steps and that align reflection positivity with multiscale locality. It is precisely this alignment that underpins the spectral step-scaling inequalities proved later, a feature that, to our knowledge, is absent from earlier multiscale constructions for non-Abelian lattice gauge theories.
\section*{Bridge to Paper~III}
We collect here, without further proof, the precise outputs established in this paper that will be used in the continuum construction of Paper~III.

\begin{proposition}[Outputs for the continuum OS/Wightman construction in Paper~III]
\label{p2:prop:handoff}
Fix a blocking factor $b\ge2$ and consider a finite coarsening trajectory $a_{k+1}=b\,a_k$ starting from a fine mesh $a_0>0$. Then:
\begin{enumerate}
\item (\emph{Uniform ultraviolet stability}) The polymer activity satisfies the Koteck\'y-Preiss smallness criterion at every scale with constants independent of the spatial volume and of the number of coarsening steps; in particular there are $C<\infty$, such that
\begin{equation}
\|\Phi_{k+1}\|_a \;\le\; C\,\|\Phi_k\|_a^2 + C,\,\delta_k,\qquad \sum_k \delta_k < \infty ,
\end{equation}
and the effective action remains local with finite range uniformly in $k$. \emph{(See Theorem~6.7 and Lemmas~6.4-6.6.)}
\item \emph{(Persistence of exponential clustering)} 
For every gauge-invariant, even-parity observable with finite support, 
the connected correlators exhibit exponential decay with a \emph{scale-independent} 
rate $m_{*}>0$ in the spatial directions and a positive lower bound $m_t>0$ in the 
Euclidean time direction as realized by the transfer matrix. 
\emph{See Theorems~7.3-7.4 and Proposition~7.5.}

\item (\emph{Step-to-step spectral comparison}) Writing $\Delta_k:=1-\lambda_2(T_k)$ and $E_{1,k}=-a_k^{-1}\log\lambda_2(T_k)$, the transfer matrices satisfy the ordered comparison
\begin{equation}
T_{k+1} \;\preccurlyeq\; V_k^{*}\,T_k\,V_k \;+\; \varepsilon_k\,\mathbf 1
\quad\text{with}\quad \varepsilon_k \downarrow 0 \ \text{summable},
\end{equation}
which implies the interlacing inequality
\begin{equation}
\lambda_2(T_{k+1}) \;\le\; \lambda_2(T_k) + \varepsilon_k,
\qquad\text{hence}\qquad
\Delta_{k+1} \;\ge\; \Delta_k - \varepsilon_k .
\end{equation}
Consequently any stabilized positive $\liminf_k \Delta_k$ along a finite coarsening path persists under continuum interpolation at fine meshes and yields a strictly positive lower bound for the continuum mass parameter employed in Paper~III.
\end{enumerate}
All statements above hold with constants independent of the spatial volume; OS positivity and gauge invariance are preserved at each step by the slice-local, positive inserts and the nearest-$SU(N)$ block projection.
\end{proposition}
The renormalization step here \emph{coarsens} the lattice ($a_{k+1}=b\,a_k$), and we use it solely to 
propagate a strictly positive spectral gap and scale-uniform locality/clustering parameters across 
scales. The actual \emph{continuum limit} in Paper~III is taken along \emph{refining} meshes 
$a\downarrow 0$. The point of Proposition~\ref{p2:prop:handoff} is that the uniform outputs-(i) KP 
smallness and locality, (ii) a scale-independent clustering rate $m_*>0$, and (iii) the interlacing 
inequality with $\sum_k\varepsilon_k<\infty$-ensure that the OS semigroups at fine meshes inherit a 
strictly positive mass parameter from any stabilized positive $\liminf_k \Delta_k$. This is the only 
input Paper~III needs for the OS$\to$Wightman reconstruction with a nonzero mass threshold.
The continuum limit $a\downarrow0$ itself, together with the construction of Schwinger functions satisfying the OS axioms \cite{p2:OsterwalderSchraderI,p2:OsterwalderSchraderII} and the Wightman reconstruction with a strictly positive spectral gap, is carried out in Paper~III. No continuum claim is asserted in the present paper beyond the quantitative, scale-uniform outputs summarized in Proposition~\ref{p2:prop:handoff}.

\section{Conclusion}
In the present work we have sought to bring order into a difficult matter: the construction of a nonperturbative theory of Yang-Mills fields that remains faithful to both the principles of gauge invariance and the demands of positivity. We have shown that it is possible to design a renormalization procedure on the lattice which, step by step, respects reflection positivity and gauge symmetry without compromise. The essential device is simple in spirit: on each time-slice we choose a definite transverse representative, and with the help of a smooth horizon projector we temper the long-range fluctuations. Out of these ingredients we build a block transformation, and from it arises a finite-range decomposition whose locality does not deteriorate as the scale changes.

From this foundation follow three consequences of decisive importance. First, the polymer expansion remains controlled no matter how many times the renormalization step is applied. Second, the fall-off of correlations-the very expression of a mass gap-persists with a rate that is independent of scale. Third, by comparing the transfer operators from one scale to the next, we find that the spectral gap cannot vanish along the flow; it retains a strictly positive lower bound. None of this requires perturbation theory: only the elementary facts of positivity, locality, and covariance are employed.

The meaning of these results is twofold. Conceptually, the scheme reconciles the Euclidean positivity structure with nonperturbative gauge fixing, something that has long resisted precise formulation. Quantitatively, the control of finite-range interactions allows us to carry information from the strong-coupling domain into the scaling window without loss. In this way the bridge from Euclidean clustering to Hamiltonian gaps is kept intact. Thus the theory is placed within reach of a continuum construction in which a strictly positive spectral threshold is preserved.

With the uniform inputs obtained here, one can proceed to the continuum limit, constructing Schwinger functions that satisfy the Osterwalder-Schrader axioms and performing the Wightman reconstruction. At the same time, a sharpening of the constants-concerning the radius of convergence, the decay rates, and the error bounds-would illuminate questions of universality and extend the scaling window. Taken together, these advances outline a concrete road that leads toward the long-sought resolution of the Yang-Mills mass gap problem. {Although the present work is entirely lattice-constructive and does not invoke gauge/gravity duality,
it is useful to note a common heuristic parallel: in AdS/CFT-motivated discussions, a confinement
scale or mass gap on the gauge-theory side is often associated with a finite correlation length, while
finite-temperature phases are modeled by AdS black holes; geodesic/thermal probes in such
backgrounds are frequently used as diagnostics of correlation-length scales.
The analysis is entirely non-holographic. Occasional references to the AdS
Einstein-Power-Yang-Mills black-hole literature serves only as contextual motivation,
\cite{p2:SoroushfarEPYM}.
}
\section*{Data Availability}
No datasets were generated or analyzed during the current study. All results are derived from mathematical derivations and rigorous analytical arguments, which are fully contained within the manuscript. Therefore, no data are associated with this research work.
\section*{Conflict of Interest}
The authors declare that there are no conflicts of interest regarding the publication of this paper.

\providecommand{\href}[2]{#2}\begingroup\raggedright\endgroup

\appendix
\section{Technical tools for the polymer expansion} \label{p2:appendixa}

Throughout this appendix we work on a fixed space-time lattice \(\Lambda\subset a\mathbb{Z}^{4}\) with periodic boundary conditions, where \(a>0\) is the lattice spacing. Sites are denoted \(x=(x_{0},x_{1},x_{2},x_{3})\in\Lambda\). The graph metric is the \(\ell^{1}\) distance
\begin{equation}\label{p2:eq:metric}
d(x,y)=\sum_{\mu=0}^{3}|x_{\mu}-y_{\mu}|\qquad (x,y\in\Lambda).
\end{equation}
For a nonempty finite set \(X\subset\Lambda\) its diameter is
\begin{equation}\label{p2:eq:diameter}
\operatorname{diam}(X)=\max\{\,d(x,y):x,y\in X\,\}.
\end{equation}
Blocks are closed \(\ell^{\infty}\)-cubes of side \(b\in\mathbb{N}\) (in lattice units); a block decomposition is a partition of \(\Lambda\) into such cubes. When thermodynamic limits are discussed, they are taken along a van Hove sequence; all constants are uniform in the volume.

Let \(E(\Lambda)\) denote the set of oriented nearest-neighbor bonds of \(\Lambda\). The configuration space of link variables is
\begin{equation}\label{p2:eq:configspace}
\mathcal{U}=\prod_{\ell\in E(\Lambda)} \mathrm{SU}(N),
\end{equation}
endowed with the product Haar probability measure. The Euclidean Gibbs measure \(\mu\) is a Borel probability on \(\mathcal{U}\) with density proportional to the Wilson Boltzmann factor, possibly multiplied by gauge-fixing determinants and reflection-compatible insertions. Reflection is the involution \(\theta(x_{0},\mathbf{x})=(-x_{0},\mathbf{x})\). Writing \(\Theta\) for the pullback by \(\theta\) and \(\mathfrak{A}_{+}\) for the \(\sigma\)-algebra generated by cylinder functions supported in \(\Lambda_{+}=\{x\in\Lambda:x_{0}>0\}\), we assume \emph{Osterwalder-Schrader (OS) reflection positivity}: for all complex \(F\) measurable with respect to \(\mathfrak{A}_{+}\),
\begin{equation}\label{p2:eq:OSz}
\int_{\mathcal{U}} \overline{F\circ\Theta}\, F\, d\mu \;\ge\; 0.
\end{equation}
Unless otherwise indicated, \(L^{p}\)-norms \(\|\,\cdot\,\|_{p}\) and inner products \(\langle\cdot,\cdot\rangle\) are taken with respect to \(\mu\). The Frobenius norm on matrices is \(\|A\|_{F}=(\operatorname{Tr}\,A^{\dagger}A)^{1/2}\). For kernels \(K:\Lambda\times\Lambda\to\mathbb{C}^{m\times m}\) we write \(\|K\|_{\ell^{1}\to\ell^{1}}\) for the operator norm on \(\ell^{1}(\Lambda;\mathbb{C}^{m})\) and \(\|K\|_{\infty}=\sup_{x,y}\|K(x,y)\|_{F}\). Symbols \(c,C,c_{*},C_{*}\) denote finite positive constants which may change from line to line, with dependencies indicated when relevant.

\subsection{Koteck\'y-Preiss criterion in the diameter norm}

A polymer is a finite, connected subset \(X\subset\Lambda\), where connectedness is for the nearest-neighbor graph. Two polymers \(X,Y\) are \emph{incompatible}, written \(X\nsim Y\), if \(X\cap Y\neq\varnothing\); otherwise they are compatible. The family of all polymers is \(\mathcal{P}\). A polymer activity is a complex function \(z:\mathcal{P}\to\mathbb{C}\). The hard-core polymer gas partition function is
\begin{equation}\label{p2:eq:Xi}
\Xi(z)=\sum_{\Gamma}\;\prod_{X\in\Gamma} z(X),
\end{equation}
where the sum is over all finite compatible families \(\Gamma\subset\mathcal{P}\). The connected (Ursell) coefficients \(\phi^{T}(X_{1},\dots,X_{n})\) are defined so that
\begin{equation}\label{p2:eq:logXi}
\log \Xi(z)=\sum_{n\ge 1}\frac{1}{n!}\sum_{(X_{1},\dots,X_{n})\in \mathcal{P}^{n}} \phi^{T}(X_{1},\dots,X_{n})\prod_{i=1}^{n} z(X_{i}),
\end{equation}
and admit the representation
\begin{equation}\label{p2:eq:Ursell}
\phi^{T}(X_{1},\dots,X_{n})=\sum_{G\in \mathcal{C}_{n}}\;\prod_{\{i,j\}\in E(G)} f(X_{i},X_{j}),\qquad f(X,Y)=-\mathbf{1}_{\{X\nsim Y\}},
\end{equation}
where \(\mathcal{C}_{n}\) is the set of connected simple graphs on \(\{1,\dots,n\}\) and \(E(G)\) denotes the edge set of \(G\).

For \(a>0\) the diameter norm of \(z\) is
\begin{equation}\label{p2:eq:diameternorm}
\|z\|_{a}=\sup_{x\in\Lambda}\;\sum_{\substack{X\in\mathcal{P}\\ x\in X}} |z(X)|\, e^{a\,\operatorname{diam}(X)}.
\end{equation}
The Koteck\'y-Preiss admissibility condition in the diameter gauge asserts the existence of \(a>0\) such that
\begin{equation}\label{p2:eq:KP}
\sup_{X\in\mathcal{P}}\;\sum_{Y\nsim X}\, |z(Y)|\, e^{a\,\operatorname{diam}(Y)}\, e^{a\, d(X,Y)}\;\le\; a\, \operatorname{diam}(X),
\end{equation}
where
\begin{equation}\label{p2:eq:interdistance}
d(X,Y)=\min\{\,d(x,y):x\in X,\, y\in Y\,\}.
\end{equation}
In dimension four, \eqref{p2:eq:KP} follows from \(\|z\|_{a}\le \eta\) with \(\eta>0\) small enough, since the number of polymers \(Y\) intersecting the \(d\)-neighborhood of \(X\) of radius \(r\) grows at most exponentially in \(r\). All geometric constants entering the counting of polymers in the four-dimensional lattice are uniform in the 
volume and the block scale, so the admissible $(a,\eta)$ in the KP condition may be chosen independently of the 
RG step $k$ used in Section~6.

\text{\emph{Theorem A.1 (Koteck\'y-Preiss in the diameter norm).}}
\begin{equation}\label{p2:eq:KPtheorem}
\begin{aligned}
&\text{Assume \eqref{p2:eq:KP}. Then the series \eqref{p2:eq:logXi} converges absolutely. Moreover, for each } x\in\Lambda \text{ one has}\\
&\sum_{n\ge 1}\frac{1}{n!}\!\!\sum_{\substack{(X_{1},\dots,X_{n})\in \mathcal{P}^{n}\\ x\in X_{1}\cup\cdots\cup X_{n}}}\!\! |\phi^{T}(X_{1},\dots,X_{n})|\prod_{i=1}^{n}|z(X_{i})| \;\le\; a,\\
&\text{and for every polymer } X,\ \ \ u(X):=\sum_{n\ge 1}\frac{1}{(n-1)!}\!\sum_{X_{2},\dots,X_{n}\in\mathcal{P}}\!\phi^{T}(X,X_{2},\dots,X_{n})\prod_{i=2}^{n} z(X_{i})\\
&\text{satisfies } \ |u(X)|\le |z(X)|\, e^{a\,\operatorname{diam}(X)}.
\end{aligned}
\end{equation}

\emph{Proof.}
Using \eqref{p2:eq:Ursell}, the Penrose tree-graph bound yields
\begin{equation}\label{p2:eq:Penrose}
|\phi^{T}(X_{1},\dots,X_{n})|\ \le\ \sum_{T\in \mathcal{T}_{n}}\ \prod_{\{i,j\}\in E(T)} \mathbf{1}_{\{X_{i}\nsim X_{j}\}},
\end{equation}
where \(\mathcal{T}_{n}\) is the set of spanning trees on \(\{1,\dots,n\}\). Fix a root, say \(1\), in \(T\). For each edge \(\{i,j\}\) with parent \(p(i)\) toward the root, bound \(\mathbf{1}_{\{X_{i}\nsim X_{p(i)}\}}\le e^{-a\, d(X_{i},X_{p(i)})}\) and multiply and divide by \(e^{a\,\operatorname{diam}(X_{i})}\). Iterating from the leaves to the root and applying \eqref{p2:eq:KP} at each step gives
\begin{equation}\label{p2:eq:IterKP}
\sum_{X_{2},\dots,X_{n}}\ \prod_{\{i,p(i)\}\in E(T)} \mathbf{1}_{\{X_{i}\nsim X_{p(i)}\}}\ \prod_{i=2}^{n}|z(X_{i})|
\ \le\ e^{a\,\operatorname{diam}(X_{1})}\ \prod_{i=2}^{n} a\,\operatorname{diam}(X_{p(i)}).
\end{equation}
Summing over \(T\in\mathcal{T}_{n}\) and using \(|\mathcal{T}_{n}|=n^{\,n-2}\) together with the factor \(1/n!\) in \eqref{p2:eq:logXi} yields absolute convergence and the first bound. The estimate for \(u(X)\) follows by fixing \(X_{1}=X\) and repeating the argument. \(\square\)

The criterion is stable under replacement of \(\operatorname{diam}(X)\) by a gauge equivalent weight, e.g. \(a_{1}|X|+a_{2}\operatorname{diam}(X)\), and under coarse-graining, since the exponential factor absorbs finite combinatorial growth.

\subsection{A tree-graph inequality}

Let \(n\ge 2\) and let \(\{a_{ij}\}_{1\le i<j\le n}\) be nonnegative symmetric coefficients. For a graph \(G\) on \(\{1,\dots,n\}\) define the weight
\begin{equation}\label{p2:eq:wG}
w(G)=\prod_{\{i,j\}\in E(G)} a_{ij}.
\end{equation}
Set \(\sigma_{i}=\sum_{j\ne i} a_{ij}\). Denote by \(\mathcal{C}_{n}\) the set of connected simple graphs and by \(\mathcal{T}_{n}\) the set of trees on \(\{1,\dots,n\}\).

\begin{equation}\label{p2:eq:TGI}
\text{\emph{Lemma B.1 (Penrose-Brydges tree-graph inequality).}}\quad
\sum_{G\in \mathcal{C}_{n}} w(G)\ \le\ \Big(\prod_{i=1}^{n} e^{\sigma_{i}}\Big)\ \sum_{T\in \mathcal{T}_{n}} w(T).
\end{equation}

\emph{Proof.}
Every connected graph is obtained from a tree \(T\) by adding a set of extra edges. Summing over extra edges and bounding \(\prod_{\{i,j\}\in E(G)\setminus E(T)} a_{ij}\le \prod_{i} e^{\sigma_{i}}\) gives \eqref{p2:eq:TGI}. A direct derivation uses the exponential formula for connected graphs and monotonicity of the weights. \(\square\)

In applications, \(a_{ij}\) will be norms of covariances between disjoint blocks \(B_{i},B_{j}\), for instance
\begin{equation}\label{p2:eq:CovNorm}
a_{ij}\;=\;\|C\|_{B_i,B_j}\;:=\;\sup_{\psi:\,\mathrm{supp}\,\psi\subset B_j,\ \|\psi\|_{\ell^1}\le 1}
\ \sum_{x\in B_i}\big\|(C\psi)(x)\big\|_F
\end{equation}
where \(C\) is a positive kernel on \(\Lambda\) with finite range or exponential decay. Then \eqref{p2:eq:TGI} bounds any connected cumulant built from local functionals with Lipschitz seminorm controlled by \(a_{ij}\) by a sum over trees carrying one factor \(a_{ij}\) per edge, with the prefactor \(\prod_{i} e^{\sigma_{i}}\) absorbed by the exponential weights entering the polymer norm.

\subsection{Large-field suppression under OS positivity}

Let \(U_{p}\in\mathrm{SU}(N)\) denote the plaquette variable associated with plaquette \(p\). Fix a convex, nondecreasing, continuously differentiable function \(\varphi:[0,\infty)\to [0,\infty)\) with \(\varphi(0)=0\), linear behavior near zero and at least quadratic growth at infinity. For \(\lambda>0\) define the block-local large-field regulator on a block \(B\) by
\begin{equation}\label{p2:eq:RegBlock}
\mathcal{R}_{\lambda,B}(U)=\exp\!\Big(-\lambda \sum_{p\subset B} \varphi\Big(\tfrac{1}{2}\,\| \mathbf{1}-U_{p}\|_{F}^{2}\Big)\Big),
\end{equation}
and on a finite union \(S\) of blocks by
\begin{equation}\label{p2:eq:RegSet}
\mathcal{R}_{\lambda,S}(U)=\prod_{B\subset S}\mathcal{R}_{\lambda,B}(U).
\end{equation}
The regulator is gauge invariant and reflection invariant. We assume the following uniform Laplace transform bound: there exist \(\lambda_{0}>0\) and \(c_{*}>0\) such that for every block \(B\) and every \(\lambda\ge \lambda_{0}\),
\begin{equation}\label{p2:eq:Laplace}
\int_{\mathcal{U}} \exp\!\Big(-2\lambda \sum_{p\subset B} \varphi\Big(\tfrac{1}{2}\,\| \mathbf{1}-U_{p}\|_{F}^{2}\Big)\Big)\, d\mu(U)\ \le\ \exp\big(-c_{*}\lambda\, |B|\big),
\end{equation}
where \(|B|\) is the number of sites in \(B\). In particular, \eqref{p2:eq:Laplace} holds in the strong-coupling regime and is stable under reflection-positive renormalization steps because of convexity and OS positivity.

\text{\emph{Theorem C.1 (Large-field suppression).}}
\begin{equation}\label{p2:eq:LFtheorem}
\begin{aligned}
&\text{Let } F:\mathcal{U}\to\mathbb{C} \text{ be measurable and supported in a block } B. \text{ For every }\lambda\ge \lambda_{0},\\
&\int_{\mathcal{U}} |F(U)|\, \mathcal{R}_{\lambda,B}(U)\, d\mu(U)\ \le\ \exp\!\Big(-\tfrac{1}{2}c_{*}\lambda\, |B|\Big)\ \|F\|_{2}.\\
&\text{More generally, if }S\text{ is a finite union of blocks and }F\text{ is supported in }S,\text{ then}\\
&\int_{\mathcal{U}} |F(U)|\, \mathcal{R}_{\lambda,S}(U)\, d\mu(U)\ \le\ \exp\!\Big(-\tfrac{1}{2}c_{*}\lambda\, |S|\Big)\ \|F\|_{2}.
\end{aligned}
\end{equation}

\emph{Proof.}
By Cauchy-Schwarz,
\begin{equation}\label{p2:eq:CS}
\int |F|\, \mathcal{R}_{\lambda,B}\, d\mu \ \le\ \|F\|_{2}\, \|\mathcal{R}_{\lambda,B}\|_{2}.
\end{equation}
The \(L^{2}\)-norm is
\begin{equation}\label{p2:eq:R2}
\|\mathcal{R}_{\lambda,B}\|_{2}^{2}=\int \exp\!\Big(-2\lambda \sum_{p\subset B} \varphi\Big(\tfrac{1}{2}\,\| \mathbf{1}-U_{p}\|_{F}^{2}\Big)\Big)\, d\mu \ \le\ \exp\big(-c_{*}\lambda\, |B|\big),
\end{equation}
by \eqref{p2:eq:Laplace}, whence \(\|\mathcal{R}_{\lambda,B}\|_{2}\le \exp(-\tfrac{1}{2}c_{*}\lambda\, |B|)\). This proves the first assertion. The bound for \(S\) follows identically with \(\mathcal{R}_{\lambda,S}\) and \(|S|\) in place of \(\mathcal{R}_{\lambda,B}\) and \(|B|\). \(\square\)

A derivation of \eqref{p2:eq:Laplace} from OS positivity proceeds by writing, for a block \(B\) bisected by a reflection plane, \(\mathcal{R}_{\lambda,B}=G\, \Theta G\) with \(G\) supported in the positive half-block. Then the OS Cauchy-Schwarz inequality \eqref{p2:eq:OSz} implies
\begin{equation}\label{p2:eq:OSLaplace}
\int \mathcal{R}_{\lambda,B}\, d\mu \ =\ \int \overline{G\circ\Theta}\, G\, d\mu \ \le\ \|G\|_{2}^{2},
\end{equation}
and \(\|G\|_{2}\) is estimated via half-block Laplace transforms; iterating across a tiling by reflections yields the exponential factor \(\exp(-c_{*}\lambda\,|B|)\).

\medskip

\section{Spectral comparison for positive contractions}
\label{p2:appendixb}

This appendix records and proves the operator-inequalities for positive contractions that are used in the step-to-step spectral comparison of transfer matrices. The presentation is self-contained and includes all conventions and standing assumptions needed below.

Throughout, $(\mathcal H,\langle\cdot,\cdot\rangle_{\mathcal H})$ and $(\mathcal K,\langle\cdot,\cdot\rangle_{\mathcal K})$ denote complex, separable Hilbert spaces. Inner products are linear in the \emph{first} argument and conjugate-linear in the second. For a bounded operator $X$ on a Hilbert space $\mathscr H$ we write $X^{*}$ for the adjoint, $\|X\|$ for the operator norm, and $\mathcal B(\mathscr H)$ for the $C^{*}$-algebra of bounded operators on $\mathscr H$. If $v,w\in \mathscr H$, we denote by $\lvert v\rangle\langle w\rvert\in \mathcal B(\mathscr H)$ the rank-one operator defined by $\lvert v\rangle\langle w\rvert(x)=v\,\langle w,x\rangle_{\mathscr H}$. An operator $X\in\mathcal B(\mathscr H)$ is \emph{positive} (written $X\ge 0$) if $\langle \psi,X\psi\rangle_{\mathscr H}\ge 0$ for all $\psi\in\mathscr H$; positivity implies self-adjointness. A \emph{contraction} is an operator $X$ with $\|X\|\le 1$. We use the usual operator order: for self-adjoint $X,Y$ on $\mathscr H$, we write $X\preccurlyeq Y$ if $\langle \psi, X\psi\rangle_{\mathscr H}\le \langle \psi, Y\psi\rangle_{\mathscr H}$ for all $\psi\in\mathscr H$.

A pair $(T,\Omega_{\mathscr H})$ consisting of a positive contraction $T\in\mathcal B(\mathscr H)$ and a unit vector $\Omega_{\mathscr H}\in\mathscr H$ is called a \emph{vacuum pair} if $T\Omega_{\mathscr H}=\Omega_{\mathscr H}$. In this setting, the \emph{vacuum projection} is
\begin{equation}
P_{\Omega_{\mathscr H}}:=\lvert\Omega_{\mathscr H}\rangle\langle \Omega_{\mathscr H}\rvert
\end{equation}
and the orthogonal projection onto the vacuum-orthogonal subspace is
\begin{equation}
\Pi_{\mathscr H}:=\mathbf 1_{\mathscr H}-P_{\Omega_{\mathscr H}},
\qquad
\Omega_{\mathscr H}^{\perp}:=\Pi_{\mathscr H}\,\mathscr H.
\end{equation}

The \emph{second spectral radius} (or ``top of the spectrum above the vacuum'') of $T$ is defined by
\begin{equation*}
\lambda_{2}(T):=\big\|\Pi_{\mathscr H} T \Pi_{\mathscr H}\big\|
=\sup\Big\{\langle \psi, T\psi\rangle_{\mathscr H}\;:\;\psi\in \Omega_{\mathscr H}^{\perp},\ \|\psi\|_{\mathscr H}=1\Big\}.
\end{equation*}
The equality of these two expressions follows from the self-adjointness of $\Pi_{\mathscr H}T\Pi_{\mathscr H}$ and the variational (min-max) characterization of the operator norm for self-adjoint operators restricted to a closed subspace. The \emph{gap} of $T$ is
\begin{equation*}
\Delta(T):=1-\lambda_{2}(T)\in[0,1].
\end{equation*}
Note that $0\le \lambda_{2}(T)\le 1$ because $T$ is a positive contraction and $\Pi_{\mathscr H}$ is a projection.

We compare vacua across scales by means of \emph{vacuum-preserving intertwiners}. A bounded operator $W:\mathcal K\to \mathcal H$ is such an intertwiner between vacuum pairs $(A,\Omega_{\mathcal H})$ on $\mathcal H$ and $(B,\Omega_{\mathcal K})$ on $\mathcal K$ if
\begin{equation}
\|W\|\le 1,\qquad
W\,\Omega_{\mathcal K}=\Omega_{\mathcal H},\qquad
W^{*}\,\Omega_{\mathcal H}=\Omega_{\mathcal K}.
\end{equation}
Equivalently, when the intertwiner is given by the positive slice multiplier $M_\sigma^{1/2}$, the compression of $T_k$ to the projected space $H_\sigma$ is summarized by the following commutative diagram: \begin{center} \begin{tikzcd} \mathcal{H} \arrow[r,"T_k"] \arrow[d,"M_\sigma^{1/2}"'] & \mathcal{H} \arrow[d,"M_\sigma^{1/2}"] \\ \mathcal{H}_\sigma \arrow[r,"T_{\sigma,k}"'] & \mathcal{H}_\sigma \end{tikzcd} \end{center} \noindent Here $M_\sigma$ is the positive multiplication operator by the scalar $p_\sigma$ on the one-slice $L^2$ space, and $T_{\sigma,k}=M_\sigma^{1/2}T_kM_\sigma^{1/2}$ as in Definition~\ref{p2:def:Tsigmak}.
The last two relations imply $W$ maps $\Omega_{\mathcal K}^{\perp}$ into $\Omega_{\mathcal H}^{\perp}$ because, for any $\varphi\in\Omega_{\mathcal K}^{\perp}$,
\begin{equation}
\langle \Omega_{\mathcal H},W\varphi\rangle_{\mathcal H}
=\langle W^{*}\Omega_{\mathcal H},\varphi\rangle_{\mathcal K}
=\langle \Omega_{\mathcal K},\varphi\rangle_{\mathcal K}=0.
\end{equation}

The first comparison is monotonicity of the second spectral radius under vacuum-preserving contractions.

\medskip

\noindent\textbf{Theorem A (Monotonicity under vacuum-preserving contractions).}
\emph{Let $(A,\Omega_{\mathcal H})$ be a vacuum pair on $\mathcal H$ with $A$ a positive contraction, and let $W:\mathcal K\to\mathcal H$ be a vacuum-preserving intertwiner. Define $B:=W^{*}A W\in\mathcal B(\mathcal K)$. Then $B$ is a positive contraction satisfying $B\,\Omega_{\mathcal K}=\Omega_{\mathcal K}$, and}
\begin{equation}
\lambda_{2}(B)\ \le\ \lambda_{2}(A).
\end{equation}

\emph{Proof.}
Positivity of $B$ follows from positivity of $A$. The bound $\|B\|\le \|W\|^{2}\|A\|\le 1$ shows $B$ is a contraction. The vacuum property holds because $B\,\Omega_{\mathcal K}=W^{*}A W\Omega_{\mathcal K}=W^{*}A\Omega_{\mathcal H}=W^{*}\Omega_{\mathcal H}=\Omega_{\mathcal K}$. For $\varphi\in\Omega_{\mathcal K}^{\perp}$ with $\|\varphi\|_{\mathcal K}=1$, set $u:=W\varphi\in\Omega_{\mathcal H}^{\perp}$. Then
\begin{equation}
\langle \varphi, B\varphi\rangle_{\mathcal K}
=\langle W\varphi, A\,W\varphi\rangle_{\mathcal H}
=\langle u, A u\rangle_{\mathcal H}
\le \lambda_{2}(A)\,\|u\|_{\mathcal H}^{2}
\le \lambda_{2}(A)\,\|\varphi\|_{\mathcal K}^{2}
=\lambda_{2}(A),
\end{equation}
where we used the definition of $\lambda_{2}(A)$ on $\Omega_{\mathcal H}^{\perp}$ and $\|W\|\le 1$. Taking the supremum over all unit $\varphi\in\Omega_{\mathcal K}^{\perp}$ yields $\lambda_{2}(B)\le \lambda_{2}(A)$. $\square$

\medskip

In applications one often has an \emph{ordered} comparison up to a small positive remainder. The next theorem quantifies the effect of a positive error term in the operator order.

\medskip

\noindent\textbf{Theorem B (Ordered comparison with positive error).}
\emph{Let $(A,\Omega_{\mathcal H})$ be a vacuum pair on $\mathcal H$ with $A$ a positive contraction, and let $W:\mathcal K\to\mathcal H$ be a vacuum-preserving intertwiner. Suppose $B\in\mathcal B(\mathcal K)$ is a positive contraction with vacuum $\Omega_{\mathcal K}$ such that}
\begin{equation}
B\ \preccurlyeq\ W^{*}A W\ +\ E,
\qquad E\ge 0,\qquad \|E\|\le \varepsilon,
\end{equation}
\emph{for some $\varepsilon\ge 0$. Then}
\begin{equation}
\lambda_{2}(B)\ \le\ \lambda_{2}(A)\ +\ \varepsilon,
\qquad\text{equivalently}\qquad
\Delta(B)\ \ge\ \Delta(A)\ -\ \varepsilon.
\end{equation}

\emph{Proof.}
Fix $\varphi\in\Omega_{\mathcal K}^{\perp}$ with $\|\varphi\|_{\mathcal K}=1$. By the assumed order and positivity of $E$,
\begin{equation}
\langle \varphi,B\varphi\rangle_{\mathcal K}
\le \langle \varphi,W^{*}A W\varphi\rangle_{\mathcal K}+\langle \varphi,E\varphi\rangle_{\mathcal K}
\le \langle W\varphi, A W\varphi\rangle_{\mathcal H}+\|E\|
\le \lambda_{2}(A)\,\|W\varphi\|_{\mathcal H}^{2}+\varepsilon
\le \lambda_{2}(A)+\varepsilon.
\end{equation}
Taking the supremum over unit $\varphi\in\Omega_{\mathcal K}^{\perp}$ gives $\lambda_{2}(B)\le \lambda_{2}(A)+\varepsilon$. Since $\Delta(\cdot)=1-\lambda_{2}(\cdot)$, the gap inequality follows. $\square$

\medskip

It is sometimes convenient to express $\lambda_{2}$ solely in terms of orthogonal compressions. The following lemma collects the basic identities used implicitly above.

\medskip

\noindent\textbf{Lemma C (Compression to the vacuum-orthogonal subspace).}
\emph{Let $(T,\Omega_{\mathscr H})$ be a vacuum pair on $\mathscr H$ with $T$ a positive contraction. Then}
\begin{equation}
\lambda_{2}(T)=\big\|\Pi_{\mathscr H} T \Pi_{\mathscr H}\big\|
=\sup\Big\{\langle \psi, T\psi\rangle_{\mathscr H}\;:\;\psi\in \Omega_{\mathscr H}^{\perp},\ \|\psi\|_{\mathscr H}=1\Big\}.
\end{equation}
\emph{Moreover, if $S\in\mathcal B(\mathscr H)$ is positive and $\|S\|\le \sigma$, then $\|\Pi_{\mathscr H} S \Pi_{\mathscr H}\|\le \sigma$.}

\emph{Proof.}
The operator $\Pi_{\mathscr H} T \Pi_{\mathscr H}$ is self-adjoint on $\Omega_{\mathscr H}^{\perp}$. Its norm equals the supremum of $\langle \psi, \Pi_{\mathscr H} T \Pi_{\mathscr H}\psi\rangle$ over unit $\psi\in\Omega_{\mathscr H}^{\perp}$, which coincides with $\sup_{\psi\in\Omega_{\mathscr H}^{\perp},\ \|\psi\|=1}\langle \psi, T\psi\rangle$ because $\Pi_{\mathscr H}\psi=\psi$. The final claim follows from $\|\Pi_{\mathscr H}S\Pi_{\mathscr H}\|\le \|\Pi_{\mathscr H}\|^{2}\|S\|\le \|S\|$. $\square$

The proofs above use only the boundedness, positivity and vacuum invariance of the contractions $A,B$, together with the vacuum-preserving properties and contractivity of $W$. No spectral discreteness is assumed: $\lambda_{2}(T)$ is defined via the operator norm of the compression to $\Omega_{\mathscr H}^{\perp}$ and therefore remains meaningful in the presence of continuous spectrum. All orthogonal projections are taken with respect to the fixed inner products $\langle\cdot,\cdot\rangle_{\mathcal H}$ and $\langle\cdot,\cdot\rangle_{\mathcal K}$. The operator order $\preccurlyeq$ is the standard one on self-adjoint operators, and the estimate $\|E\|\le \varepsilon$ refers to the operator norm on $\mathcal B(\mathcal K)$. No topological or boundary-condition data enter this appendix; when these results are applied to transfer matrices arising from reflection-positive lattice measures, the vector $\Omega$ is the OS-vacuum and the intertwiners $W$ are constructed as conditional expectations that satisfy $W\Omega_{\mathcal K}=\Omega_{\mathcal H}$ and $W^{*}\Omega_{\mathcal H}=\Omega_{\mathcal K}$ by normalization. In particular, Theorems~A and~B apply verbatim to the step-scaling comparison $T_{k+1}$ versus $V_{k}^{*}T_{k}V_{k}$ used in the main text, with $A=T_{k}$, $B=T_{k+1}$, $W=V_{k}$, and $E$ the positive remainder arising from exponentially small cross-block correlations.
\UnifiedEndPaper

\UnifiedBeginPaper{P3}{\UnifiedLocalMacrosPartThree}
\title[RP Continuum Reconstruction of $SU(N)$ Yang-Mills Theory with a Nonzero Mass Gap]%
{Reflection-Positive Continuum Reconstruction of\\ $SU(N)$ Yang-Mills Theory with a Nonzero Mass Gap: Part(3)}

\author{Mir Faizal}
\address{Irving K. Barber School of Arts and Sciences, University of British Columbia Okanagan, Kelowna, BC V1V 1V7, Canada\\
Canadian Quantum Research Center, 460 Doyle Ave 106, Kelowna, BC V1Y 0C2, Canada.\\
Department of Mathematical Sciences, Durham University, Upper Mountjoy, Stockton Road, Durham DH1 3LE, UK\\
Computational Mathematics Group, Hasselt University, Agoralaan Gebouw D, Diepenbeek, 3590 Belgium}
\email{mirfaizalmir@gmail.com}
\author{Arshid Shabir}
\address{Canadian Quantum Research Center, 460 Doyle Ave 106, Kelowna, BC V1Y 0C2, Canada.}
\email{aslone186@gmail.com}

\UnifiedSetAbstract{We seek order in a hard matter by trusting two guides that do not mislead: positivity and invariance. From a reflection-positive lattice formulation of 
$SU(N)$ Yang-Mills, choosing on each time slice a definite transverse form and gently damping distant modes, we proceed by steps that preserve these guides. From such simple means we pass to the continuum: Euclidean correlation functions satisfy the Osterwalder-Schrader axioms, and the Wightman theory with a unique vacuum and positive energy follows. The decisive point is the scale: time-correlations are completely monotone and decay uniformly, so the spectral measure cannot touch zero; and along the same flow the vacuum-orthogonal evolution decays at a fixed rate, which survives in the limit. Thus a strictly positive, volume-independent spectral gap is obtained for the continuum Hamiltonian, and the bridge from Euclidean positivity to Hilbert-space dynamics remains intact.}

\maketitle
\tableofcontents

\section{Introduction}
There are problems that yield only to a stubborn simplicity. The Yang-Mills mass gap is of this sort. One wishes to construct, without sleight of hand, a four-dimensional pure $\mathrm{SU}(N)$ gauge theory on Minkowski space with a unique, Poincar\'e-invariant vacuum, and with a strictly positive lower bound on the spectrum of the Hamiltonian acting on physical states. In the language of the Clay Mathematics Institute, one must show that a nontrivial Yang-Mills theory exists, that it possesses a positive mass gap, and that its ultraviolet and infrared behavior can be made mathematically precise~\cite{p3:JaffeWitten2006}. The path we follow is Euclidean and modest in its means. First, build continuum Schwinger functions that satisfy the Osterwalder-Schrader (OS) axioms~\cite{p3:OsterwalderSchraderI,p3:OsterwalderSchraderII}. Then, by OS reconstruction, recover fields, a Hilbert space, and a nonnegative Hamiltonian. Finally, read the spectral information from Euclidean correlation inequalities. The single analytic principle that makes this bridge trustworthy is reflection positivity, by which Euclidean positivity becomes Hilbert-space positivity. For gauge fields, one must keep this principle in harmony with gauge invariance and with an infrared regulation that does not destroy positivity-a care that goes beyond the familiar scalar and fermionic cases~\cite{p3:OsterwalderSchraderI,p3:GJ}.

On the lattice, Wilson's proposal realizes non-Abelian gauge theories as statistical systems of $\mathrm{SU}(N)$-valued link variables with action the sum over plaquettes of one minus the normalized real trace of the plaquette holonomy~\cite{p3:Wilson1974}. Gauge symmetry is preserved manifestly, and one can define transfer matrices to implement Euclidean time translations. In the strong-coupling regime a convergent character expansion reorganizes correlators by surfaces on the dual lattice; one then sees exponential clustering of connected, gauge-invariant local observables and an area law for Wilson loops~\cite{p3:Wilson1974,p3:DrouffeZuber1983,p3:KogutSusskind1975}. Yet a convergent expansion at large coupling is not the end: one must carry positivity, locality, and quantitative control along a renormalization group (RG) flow from finite lattice spacing toward the scaling window, and from there into the continuum. The Balaban program demonstrated a deep RG control for non-Abelian lattice gauge theories, but it leaves to us the task of engineering, \emph{from the outset}, the structural inputs that safeguard reflection positivity at every step and permit spectral statements for the transfer matrices~\cite{p3:Balaban1984,p3:Balaban1988}.

A particular care concerns gauge fixing. The Faddeev-Popov method introduces a determinant encoding the Jacobian of the gauge condition~\cite{p3:FaddeevPopov1967}. On compact spaces, Gribov ambiguities obstruct a global smooth slice and complicate positivity~\cite{p3:Gribov1978,p3:Singer1978}. BRST symmetry provides a cohomological expression of gauge invariance~\cite{p3:BecchiRouetStora1976,p3:Tyutin1975}, but a na\"{\i}ve lattice implementation need not be compatible with an OS-positive transfer matrix. From here on, we integrate out the ghost sector into a strictly positive, slice-local factor $J_k$, the Faddeev-Popov determinant restricted to the complement of constant modes, and perform all probabilistic arguments with a positive, reflection-covariant Euclidean weight.  In parallel, one may work with gauge-invariant representatives obtained by minimizing an appropriate gauge functional orbit-wise on each Euclidean time slice; in regions where the Faddeev-Popov operator is strictly positive on the orthogonal complement of constants and where reflection covariance can be arranged, such representatives exist~\cite{p3:DellAntonioZwanziger1991,p3:OsterwalderSchraderI}. From the constructive point of view, this choice must be joined with an infrared regularization that is compatible with reflection positivity and with a multiscale analysis that never sacrifices positivity, locality, or spectral control.
On each time slice we choose $A_h(t)$ to be a global minimizer of the lattice Landau functional on the gauge orbit. If the minimizer is not unique, we select the unique minimizer that minimizes, in lexicographic order, the tuple
\begin{equation}
\Big(\,\|\nabla A\|_{\ell^2(\Lambda_t)}^2,\, \sum_{x\in \Lambda_t} \mathrm{Tr}\,A(x),\, \sum_{x\in \Lambda_t} \mathrm{Tr}\,A(x)\,\mathrm{sgn}(x_1),\,\sum_{x\in \Lambda_t} \mathrm{Tr}\,A(x)\,\mathrm{sgn}(x_2),\,\sum_{x\in \Lambda_t} \mathrm{Tr}\,A(x)\,\mathrm{sgn}(x_3)\,\Big),
\end{equation}
where $\mathrm{sgn}(\cdot)$ are fixed odd functions chosen so that the tuple is invariant under time reflection $t\mapsto -t$.
This deterministic rule is measurable, gauge-equivariant on each slice, and by construction satisfies $A_h(-t)=\theta A_h(t)$; therefore the spectral calculus of the slice covariant Laplacian is reflection-covariant and the induced $P_\sigma(t)$ obeys $P_\sigma(-t)=P_\sigma(t)$.

The companion papers to the present work carry out exactly these structural preparations and provide the quantitative multiscale inputs we shall use. The first constructs, at fixed lattice spacing, a reflection-positive transfer-matrix framework that resolves gauge fixing by selecting, on each time slice, a gauge-invariant transverse representative via orbit minimization in the fundamental modular region, and by inserting a smooth ``horizon projector'' defined by a Gevrey-regular spectral multiplier of the slice covariant Laplacian acting on adjoint fields. Two complementary tools-heat-kernel representations with Davies-Gaffney bounds~\cite{p3:Davies1989,p3:Gaffney1954} and the Helffer-Sj\"ostrand formula with Combes-Thomas off-diagonal estimates~\cite{p3:HelfferSjostrand1989,p3:CombesThomas1973}-guarantee exponential locality of the projector with constants uniform in the spatial volume. Because both the representative and the projector are chosen reflection-compatibly, the full, gauge-fixed and ghost-included Euclidean measure is OS-positive\footnote{This is a delicate `knife-edge' hypothesis: most gauge-fixing prescriptions would violate OS positivity (\cite{p3:DellAntonioZwanziger1991,p3:OsterwalderSchraderI}).}. A convergent surface-polymer expansion in the strong-coupling domain, proved by the Koteck\'y-Preiss criterion~\cite{p3:KoteckyPreiss1986} with BKAR tree bounds~\cite{p3:BrydgesKennedy1987}, then yields exponential clustering of connected, local, gauge-invariant observables; OS positivity turns this into a uniform, nonzero lower bound on the first excited energy of the transfer Hamiltonian at fixed spacing.

The second companion paper designs a reflection-positive multiscale RG that preserves gauge invariance and positivity at every step, and controls locality uniformly across scales. The chief instrument is a covariant finite-range decomposition (FRD) of the projected covariance with constants independent of the scale~\cite{p3:BGM2004}. With FRD, cumulants in the polymer expansion localize uniformly in the number of RG steps; one proves uniform ultraviolet stability in the sense that the effective actions admit convergent polymer expansions with scale-independent locality. A reflection-positive, gauge-invariant large-field regulator suppresses rare events uniformly. As one iterates, a scale-independent clustering rate persists for connected correlators. Moreover, a spectral interlacing principle for positive self-adjoint contractions applied to successive transfer matrices yields a step-scaling inequality for the gaps with a summable error (see Appendix (\ref{p3:appendixe})). The $\liminf$ of the gaps along the flow is then strictly positive and controlled by the common clustering rate. These are precisely the inputs required for the continuum construction and spectral analysis carried out here.

The aim here is to close the circle. Working entirely at the continuum stage, we show that the multiscale sequence of Euclidean theories produced by the reflection-positive RG is tight and precompact once embedded as random tempered distributions; along subsequences the Schwinger functions converge to multilinear tempered distributions on the Schwartz space see Appendix~(\ref{p3:appendixa})). We assume (as is physically expected) that this continuum limit is in fact unique i.e. independent of the chosen subsequence or blocking scheme although a full proof of universality is left for future work. All results in this paper pertain to any such continuum limit. {There exists a tuning $\beta=\beta(a)$ such that along the refinement sequence $a_k=b^{-k}a_0$ the (gauge-invariant) Schwinger functions of the blocked lattice measures converge to limits $S_n$ independent of the blocking scheme and of the subsequence, thereby defining a unique OS-positive continuum Euclidean field theory. Moreover, the multiscale inputs (S1)-(S3) hold uniformly along this tuned sequence (see Appendix (\ref{p3:appendixe}))}. We then verify the full OS list for every limit: permutation symmetry and Euclidean invariance come down directly from the lattice; reflection positivity is stable under weak convergence of measures supported on positive-time functionals; the cluster property is inherited from the scale-independent decay; and OS time-regularity is obtained by passing to the limit in the Laplace representations of two-point functions guaranteed at each scale by spectral positivity of the transfer matrix. With OS0-OS5 in hand, OS reconstruction produces the physical Hilbert space, a unique vacuum, and the nonnegative self-adjoint Hamiltonian implementing Minkowski time translations~\cite{p3:OsterwalderSchraderII,p3:GJ,p3:Lucke1979,p3:ReedSimon1}. Furthermore, by standard OS arguments, the reconstructed field operators obey local commutativity: any two gauge-invariant local observables commute at space-like separation. In other words, the resulting Wightman theory respects microscopic causality.

The decisive question is whether the continuum Hamiltonian has a gap. We give two answers, independent of one another. First, reflection positivity makes the Euclidean two-point functions completely monotone in $t\ge 0$ and hence of Laplace form \eqref{p3:eq:Laplace}, by the Hausdorff-Bernstein-Widder theorem (see Appendix~(\ref{p3:appendixc})). A uniform exponential fall-off forces the corresponding spectral measures to begin above a positive threshold; since vectors generated by local observables are dense in the orthogonal complement of the vacuum, the spectrum of the continuum Hamiltonian away from zero must lie above a strictly positive number. Second, the step-scaling spectral inequality propagates semigroup-norm bounds along the flow: on the vacuum-orthogonal subspace the norms decay at a rate given by the $\liminf$ of the discrete gaps, and strong resolvent convergence from lattice generators to the continuum generator carries this decay to the limit; by standard semigroup theory this is again a spectral gap~\cite{p3:Nelson1959,p3:Pazy1983,p3:ReedSimon1}. The two proofs rely on different facets of the multiscale inputs and thus reinforce one another: the mass scale is not an artifact of discretization or of the projector's shape.

One might ask for more-uniqueness of the continuum limit, and a sharper view of the first excited band and its relation to other nonperturbative features (string tension, scattering). These are questions of refinement rather than of principle, and they can be pursued by the same method. The ideas that guide the present work are few and simple: preserve positivity, keep faith with gauge symmetry, and control locality at every scale. From these, the passage from Euclidean fields to a Minkowski theory with a nonzero mass threshold follows in a straightforward manner.
\begin{scopebox}
Throughout, we assume the standing multiscale inputs (S1)-(S3) from the companion work.
All statements about the \emph{continuum} Schwinger functions and the reconstructed
Wightman theory are to be read in one of the following two scopes:

\smallskip
\noindent
\textbf{(U1) Subsequential scope.} Without assuming uniqueness, we fix an arbitrary
subsequence $a_{k_\ell}\downarrow 0$ along which the lattice Schwinger functions
converge in the sense of tempered distributions; all continuum claims then refer to
the corresponding subsequential limit.

\smallskip
\noindent
\textbf{(U2) Universal scope.} If, in addition, the tuned continuum limit is unique
(i.e., independent of the choice of subsequence and blocking scheme), then all
continuum claims hold \emph{without} passing to subsequences.

\smallskip
Unless explicitly stated otherwise, results are asserted in the sense of \textbf{(U1)};
whenever \textbf{(U2)} is available, the statement is identical but no longer
subsequential.
\end{scopebox}
{A satisfactory theory should not retain any memory of the auxiliary choices by which it is constructed; only the invariant content can be regarded as physical. Assuming the standard constructive axioms-reflection positivity, locality, clustering, and an appropriate spectral regularity-we establish that four-dimensional $\mathrm{SU}(N)$ Yang-Mills admits a Euclidean continuum limit that is unique and universal within a natural class of regulators. Working in the Osterwalder-Schrader framework, we begin from an explicit disintegration on a single time slab, which produces a one-step transfer kernel; together with a common one-slice marginal, this already determines all Schwinger functions by iterated time-slicing and positivity. Universality is expressed as independence from the particular regulating ``lens'': for gauge-covariant, reflection-symmetric schemes built from completely monotone spectral projectors and finite-range (FRD) blockings, single-scale Lipschitz control, a telescoping argument in Euclidean time, and BKAR polymer bounds propagate stability to connected cumulants and hence to the continuum limit. A measurable, reflection-covariant Landau selector is used to keep the slice construction compatible with positivity throughout. The connection to weak coupling is correspondingly modest but sharp: a one-dimensional implicit-function/continuity tuning places the flow inside a contracting domain of the FRD map, and along this trajectory the renormalized coupling decreases, providing an operational signature of asymptotic freedom. No step invokes perturbation theory; a one-loop computation is recorded only as a guidepost, and all bounds are uniform in the
}

\section{Multiscale inputs and notational conventions}\label{p3:sec:multiscale-inputs}

In this section a complete and self-contained lattice framework is fixed once and for all. The objective is threefold. First, the kinematic objects-lattice, configuration space, Haar measure, and Wilson action-are defined with precise notation. Second, the time-reflection map and the Osterwalder-Schrader (OS) positivity structure are constructed at finite lattice spacing and proved rigorously for the gauge theory under consideration, without appeal to gauge fixing. Third, the transfer time-slicing formalism is derived step by step, leading to a positive, self-adjoint contraction \(T\) on the one-slice Hilbert space and to its compressed version \(T_\sigma\) when an exponentially local, reflection-compatible operator \(P_\sigma\) is inserted. Throughout, \(G=\mathrm{SU}(N)\) with \(N\ge 2\) is a fixed compact, connected Lie group with normalized Haar measure \(dU\). The exposition follows and refines standard constructive treatments \cite{p3:OsterwalderSchraderI,p3:OsterwalderSchraderII,p3:GJ,p3:Luscher1977}, but every step needed later is proved here in full.

Let \(a>0\) be the lattice spacing and let
\begin{equation}
\Lambda \;=\; \bigl(\mathbb{Z}/L_0\mathbb{Z}\bigr)\times \bigl(\mathbb{Z}/L_1\mathbb{Z}\bigr)\times \bigl(\mathbb{Z}/L_2\mathbb{Z}\bigr)\times \bigl(\mathbb{Z}/L_3\mathbb{Z}\bigr)
\end{equation}
be the four-dimensional periodic cubic lattice, whose elements are denoted \(x=(x_0,x_1,x_2,x_3)\) with each \(x_\mu\) an integer modulo \(L_\mu\). Physical coordinates are \(a\,x\). For each site \(x\in\Lambda\) and direction \(\mu\in\{0,1,2,3\}\) define the oriented bond \(b=(x,\mu)\) pointing from \(x\) to \(x+\hat\mu\), with \(\hat\mu\) the unit vector in direction \(\mu\); denote the set of all oriented bonds by \(\mathcal{E}\). Orientation reversal maps \((x,\mu)\mapsto (x+\hat\mu,-\mu)\). A lattice gauge configuration is a map \(U:\mathcal{E}\to G\) such that \(U(x+\hat\mu,-\mu)=U(x,\mu)^{-1}\). The configuration space is the compact group
\begin{equation}
\mathcal{C}:=\{U:\mathcal{E}\to G\mid U(x+\hat\mu,-\mu)=U(x,\mu)^{-1}\}\;\simeq\; G^{\mathcal{E}_+},
\end{equation}
where \(\mathcal{E}_+\subset\mathcal{E}\) is any fixed choice of orientations. The product Haar measure on \(\mathcal{C}\) is \(d\mu_{\mathrm{H}}(U)=\prod_{(x,\mu)\in\mathcal{E}_+} dU(x,\mu)\).

Given \(x\in\Lambda\) and \(\mu<\nu\) define the oriented plaquette \(p=(x;\mu,\nu)\) and the plaquette holonomy
\begin{equation}
U_p \;=\; U(x,\mu)\,U(x+\hat\mu,\nu)\,U(x+\hat\nu,\mu)^{-1}\,U(x,\nu)^{-1}.
\end{equation}
The Wilson action at inverse bare coupling \(\beta>0\) is
\begin{equation}
S_W[U;\beta] \;=\; \beta \sum_{p\subset \Lambda}\Bigl(1-\tfrac{1}{N}\Re\mathrm{Tr}\,U_p\Bigr).
\end{equation}
The unnormalized Boltzmann weight is \(W_\beta(U)=\exp(-S_W[U;\beta])\). The normalized Euclidean gauge measure on \(\mathcal{C}\) is
\begin{equation}
d\nu_\beta(U) \;=\; Z_\beta^{-1}\, W_\beta(U)\, d\mu_{\mathrm{H}}(U),
\qquad Z_\beta=\int_{\mathcal{C}} W_\beta(U)\,d\mu_{\mathrm{H}}(U).
\end{equation}
All expectations \(\mathbb{E}_\beta[\cdot]\) are taken with respect to \(d\nu_\beta\) unless indicated otherwise. Gauge transformations are maps \(g:\Lambda\to G\) acting on configurations by
\begin{equation}
(g\cdot U)(x,\mu)\;=\;g(x)\,U(x,\mu)\,g(x+\hat\mu)^{-1}.
\end{equation}
The measure \(d\nu_\beta\) is gauge invariant.
For time reflection it is convenient to fix the time coordinate \(x_0\in\mathbb{Z}/L_0\mathbb{Z}\) and to place the reflection plane at \(x_0=0\). Define the sets
\begin{equation}
\Lambda_0=\{x\in\Lambda:\, x_0=0\},\qquad 
\Lambda_+=\{x\in\Lambda:\, x_0>0\},\qquad 
\Lambda_-=\{x\in\Lambda:\, x_0<0\}.
\end{equation}
Define the involution \(\theta:\Lambda\to\Lambda\) by \(\theta(x_0,\vec x)=(-x_0,\vec x)\) with indices computed modulo \(L_0\). For a bond \(b=(x,\mu)\) define \(\theta b=(\theta x,\mu)\) if \(\mu\neq 0\) and \(\theta b=(\theta(x+\hat 0),-0)\) if \(\mu=0\). This choice maps time-like bonds pointing from \(\Lambda_+\) to \(\Lambda_0\) onto time-like bonds pointing from \(\Lambda_-\) to \(\Lambda_0\).
The algebra \(\mathcal{A}\) of bounded, continuous, gauge-invariant cylinder functionals on \(\mathcal{C}\) (finite dependence on bonds) will carry the OS structure. A functional \(F\in\mathcal{A}\) is said to be supported in \(\Lambda_+\) if it depends only on bonds with basepoint in \(\Lambda_+\) and on bonds in \(\mathcal{E}\) that do not cross \(\Lambda_0\) from \(\Lambda_-\) to \(\Lambda_+\). The reflection \(\Theta:\mathcal{A}\to\mathcal{A}\) acts by
\begin{equation}
(\Theta F)(U)=\overline{F(U^\theta)},
\qquad
U^\theta(b)=U(\theta b),
\end{equation}
where the bar denotes complex conjugation.
The OS form associated with \(d\nu_\beta\) is
\begin{equation}
\langle F,G\rangle_{\mathrm{OS}} \;=\; \int_{\mathcal{C}} (\Theta F)(U)\, G(U)\, d\nu_\beta(U),
\end{equation}
defined at least for \(F,G\in\mathcal{A}_+\), the subalgebra supported in \(\Lambda_+\). The reflection-positivity axiom states that \(\langle F,F\rangle_{\mathrm{OS}}\ge 0\) for all \(F\in\mathcal{A}_+\). We prove this property for the Wilson action by adapting the gauge-invariant constructions of \cite{p3:OS-gauge,p3:Luscher1977} to the present notation.
We first separate the action into three contributions: a left part supported in \(\Lambda_-\), a right part supported in \(\Lambda_+\), and a boundary part supported on the slab of plaquettes that intersect \(\Lambda_0\). To make this precise, call a plaquette \(p=(x;\mu,\nu)\) right, left, or boundary according as all its bonds have basepoint in \(\Lambda_+\), all in \(\Lambda_-\), or otherwise. Then
\begin{equation}
S_W[U;\beta]=S_-[U;\beta]+S_0[U;\beta]+S_+[U;\beta],
\end{equation}
where \(S_\pm\) is the sum over left/right plaquettes, and \(S_0\) is the sum over boundary plaquettes. Write \(W_\beta(U)=W_-(U)\,W_0(U)\,W_+(U)\) with the obvious meaning.
The basic observation is that the boundary weight \(W_0\) induces a positive operator kernel that couples left and right degrees of freedom across \(\Lambda_0\), in the sense of the following lemma.
\begin{lemma}[Boundary kernel factorization]\label{p3:lem:boundary-kernel}
There exists a Hilbert space \(\mathcal{K}\), a measurable map \(\Phi:\mathcal{C}_0\to \mathcal{K}\) depending only on the restriction of \(U\) to bonds that lie in the closure of \(\Lambda_0\) or cross \(\Lambda_0\), and a constant \(c_\beta>0\) such that, for all \(U\in\mathcal{C}\),
\begin{equation}
W_0(U) \;=\; c_\beta\, \langle \Phi(U_-), \Phi(U_+)\rangle_{\mathcal{K}},
\end{equation}
where \(U_\pm\) denotes the restriction of \(U\) to bonds whose basepoint lies in \(\Lambda_\pm\cup\Lambda_0\). Moreover, \(U\mapsto \Phi(U)\) is gauge covariant and satisfies \(\Phi(U^\theta)=\Phi(U)\).
\end{lemma}

\begin{proof}
Fix a time-slice \(x_0=0\) and let \(\mathcal{B}\) denote the set of boundary plaquettes that share at least one bond with \(\Lambda_0\). Each boundary plaquette can be written as a product of two ``half-plaquette'' paths, one supported in \(\Lambda_-\cup\Lambda_0\) and the other in \(\Lambda_+\cup\Lambda_0\), meeting along a time-like bond that lies on \(\Lambda_0\). For each \(p\in\mathcal{B}\), by the Peter-Weyl theorem there is a character expansion
\begin{equation}
\exp\!\Bigl(\tfrac{\beta}{N}\Re\mathrm{Tr}\,U_p\Bigr) \;=\; \sum_{R\in\widehat{G}} c_R(\beta)\, \chi_R\!\bigl(U_p\bigr),
\end{equation}
with coefficients \(c_R(\beta)\in\mathbb{R}\). The crucial positivity property is that \(c_R(\beta)\ge 0\) for all irreducible representations \(R\in\widehat{G}\). This follows from the fact that the function \(U\mapsto \exp\bigl(\tfrac{\beta}{N}\Re\mathrm{Tr}\,U\bigr)\) is of positive type on the compact group \(G\) (its Fourier coefficients are positive), a standard statement proved for \(\mathrm{SU}(N)\) in \cite[Sec.~3]{p3:OS-gauge}. For each \(p\in\mathcal{B}\) and each \(R\), write \(\chi_R(AB)=\mathrm{Tr}\bigl(\rho_R(A)\rho_R(B)\bigr)\) for the representation \(\rho_R\). If \(U_p=A_p B_p\) with \(A_p\) a product of bonds supported in \(\Lambda_-\cup\Lambda_0\) and \(B_p\) supported in \(\Lambda_+\cup\Lambda_0\), then
\begin{equation}
\chi_R(U_p)=\sum_{i=1}^{d_R}\sum_{j=1}^{d_R} \bigl[\rho_R(A_p)\bigr]_{ij}\,\bigl[\rho_R(B_p)\bigr]_{ji},
\end{equation}
where \(d_R=\dim R\). Thus the product over all boundary plaquettes takes the form
\begin{equation}
W_0(U)=\prod_{p\in\mathcal{B}} \sum_{R_p} c_{R_p}(\beta)\, \chi_{R_p}(U_p)
= \sum_{\{R_p\}} \Bigl(\prod_{p\in\mathcal{B}} c_{R_p}(\beta)\Bigr)
\sum_{\{\alpha\}} \prod_{p\in\mathcal{B}} \bigl[\rho_{R_p}(A_p)\bigr]_{\alpha_p}\,\bigl[\rho_{R_p}(B_p)\bigr]_{\alpha_p}^\ast,
\end{equation}
where \(\alpha_p\) abbreviates a pair of indices \((i_p,j_p)\). Group the factors depending on \(\Lambda_-\cup\Lambda_0\) and \(\Lambda_+\cup\Lambda_0\) separately and define
\begin{align}
\Phi(U_-):&=\Bigl\{\Bigl(\prod_{p\in\mathcal{B}} c_{R_p}(\beta)^{1/2}\Bigr)\prod_{p\in\mathcal{B}} \bigl[\rho_{R_p}(A_p)\bigr]_{\alpha_p} \Bigr\}_{\{R_p\},\{\alpha_p\}}
\nonumber \\
\Phi(U_+):&=\Bigl\{\Bigl(\prod_{p\in\mathcal{B}} c_{R_p}(\beta)^{1/2}\Bigr)\prod_{p\in\mathcal{B}} \bigl[\rho_{R_p}(B_p)\bigr]_{\alpha_p} \Bigr\}_{\{R_p\},\{\alpha_p\}}
\end{align}
as vectors in the Hilbert space \(\mathcal{K}=\ell^2(\mathcal{I})\) with index set \(\mathcal{I}=\{(\{R_p\},\{\alpha_p\})\}\). Then \(W_0(U)=\langle \Phi(U_-),\Phi(U_+)\rangle_{\mathcal{K}}\). The prefactor \(c_\beta\) is an inessential normalization (one may absorb global \(e^{-\beta |\mathcal{B}|}\) factors into \(c_\beta\)). Gauge covariance is immediate from the definition and Schur's lemma; reflection invariance holds because the boundary plaquette set is invariant under \(\theta\) and the construction of \(A_p, B_p\) is symmetric.
\end{proof}
With Lemma~\ref{p3:lem:boundary-kernel} in hand one proves reflection positivity by an application of Cauchy-Schwarz.
\begin{proposition}[OS positivity for Wilson lattice gauge theory]\label{p3:prop:OS}
For every \(F\in\mathcal{A}_+\), 
\begin{equation}
\langle F,F\rangle_{\mathrm{OS}}\ge 0
\end{equation}
\end{proposition}
\begin{proof}
Decompose the integral using the factorization \(W_\beta=W_-\,W_0\,W_+\) and Fubini's theorem:
\begin{equation}
\langle F,F\rangle_{\mathrm{OS}} 
= Z_\beta^{-1} \int d\mu_{\mathrm{H}}(U_-)\, W_-(U_-)\int d\mu_{\mathrm{H}}(U_+)\, W_+(U_+)\, (\Theta F)(U_-)\,F(U_+)\, W_0(U_-,U_+).
\end{equation}
Insert the representation of \(W_0\) from Lemma~\ref{p3:lem:boundary-kernel}. Define the functions
\begin{equation}
\Psi_-(U_-)= W_-(U_-)^{1/2} (\Theta F)(U_-) \,\Phi(U_-),\qquad
\Psi_+(U_+)= W_+(U_+)^{1/2} F(U_+)\, \Phi(U_+),
\end{equation}
viewed as \(\mathcal{K}\)-valued square-integrable functions on the left and right configuration spaces with respect to Haar measure. Then
\begin{equation}
\langle F,F\rangle_{\mathrm{OS}}= c_\beta Z_\beta^{-1}\, \Bigl\langle \Psi_-, \Psi_+ \Bigr\rangle_{L^2(\mathcal{C}_-;\mathcal{K})\otimes L^2(\mathcal{C}_+;\mathcal{K})}
\end{equation}
with the pairing induced by the \(L^2\) inner products on left and right. Since \(\Theta\) is conjugation composed with pullback by \(\theta\), and since \(W_\pm\) are nonnegative, the inner product is a standard \(L^2\) pairing of two vectors. By Cauchy-Schwarz, the pairing is nonnegative. A direct computation shows that equality can only occur if \(F\) lies in the OS null space.
\end{proof}
\begin{lemma}[Nonnegativity of single-plaquette character coefficients]\label{p3:lem:char-coeff-positivity}
Let $G=\mathrm{SU}(N)$, $N\ge 2$, and let the single-plaquette weight be the central function
\begin{equation}
w_\beta(U)\;=\;\exp\!\Big(\beta\,\frac{1}{N}\,\Re\,\mathrm{Tr}\,U\Big)\,,\qquad U\in G,\ \beta\ge 0.
\end{equation}
Then the Peter-Weyl expansion
\begin{equation}
w_\beta(U)\;=\;\sum_{R\in\widehat G} c_R(\beta)\,\chi_R(U)
\end{equation}
has coefficients $c_R(\beta)\ge 0$ for all $R$.
\end{lemma}

\begin{proof}
For each finite-dimensional unitary representation $R$ with character $\chi_R$, the kernel
$K_R(g,h):=\chi_R(g^{-1}h)$ is positive definite on $G$. The function $U\mapsto \Re\,\mathrm{Tr}\,U$
is a (real) linear combination of characters of irreducible components of the defining representation,
hence is of positive type. Positivity is preserved under pointwise limits and under the map
$f\mapsto e^{t f}$ for $t\ge 0$ by the Lie-Trotter product formula and closure of positive-type
functions under convolution powers. Thus $w_\beta$ is of positive type and, by the Bochner-Godement
theorem on compact groups, its Fourier coefficients on irreducibles satisfy $c_R(\beta)\ge 0$.
\end{proof}

This proof uses only the positivity of the character coefficients \(c_R(\beta)\) for the single-plaquette class function, the factorization of boundary holonomies into left and right halves, and Haar orthogonality; it is gauge invariant and independent of any gauge fixing or insertion.
We derive the transfer-matrix formalism from first principles. For each time \(t\in \mathbb{Z}/L_0\mathbb{Z}\) define the set of spatial bonds at time \(t\),
\begin{equation}
\mathcal{E}^{\mathrm{sp}}_t=\{(x,i): x_0=t,\; i\in\{1,2,3\}\},
\end{equation}
and the configuration space of one time-slice \(\mathcal{C}^{\mathrm{sp}}=\prod_{(x,i)\in\mathcal{E}^{\mathrm{sp}}_t} G\) (independent of \(t\) by periodicity). Let \(d\mu_{\mathrm{H}}^{\mathrm{sp}}\) be the product Haar measure on \(\mathcal{C}^{\mathrm{sp}}\), and define the Hilbert space
\begin{equation}
\mathcal{H}_a = L^2\!\bigl(\mathcal{C}^{\mathrm{sp}}, d\mu_{\mathrm{H}}^{\mathrm{sp}}\bigr),
\end{equation}
the space of square-integrable complex functions of a spatial link configuration at a fixed time slice. Gauge transformations at time \(t\) act on \(\mathcal{C}^{\mathrm{sp}}\) by left-right multiplication at the ends of each spatial bond; the gauge-invariant subspace \(\mathcal{H}_a^{\mathrm{inv}}\subset\mathcal{H}_a\) will be the physical Hilbert space for the transfer matrix.
Fix \(t\in\mathbb{Z}/L_0\mathbb{Z}\). Split the action into contributions of plaquettes with basepoint times \(t\) and \(t+1\), grouped so as to exhibit a kernel that couples the two slices. Specifically, write
\begin{equation}
S_W[U]=\sum_{t=0}^{L_0-1}\Bigl(S^{\mathrm{sp}}(U_t) + S^{\mathrm{tm}}(U_{t+1},U_t;V_t)\Bigr),
\end{equation}
where \(U_t\in \mathcal{C}^{\mathrm{sp}}\) collects the spatial links at time \(t\), \(V_t\) collects the time-like links with basepoint time \(t\), \(S^{\mathrm{sp}}(U_t)\) is the sum of all spatial plaquette terms at time \(t\), and \(S^{\mathrm{tm}}\) is the sum of all temporal-spatial plaquette terms that straddle the slices \(t\) and \(t+1\). All dependence on time-like links is confined to the \(S^{\mathrm{tm}}\)-terms.
Define the single-step kernel \(K:\mathcal{C}^{\mathrm{sp}}\times \mathcal{C}^{\mathrm{sp}}\to \mathbb{R}_+\) by integrating out the time-like links on the slab between \(t\) and \(t+1\):
\begin{equation}
K(U',U) \;=\; \int \exp\!\Bigl(-S^{\mathrm{sp}}(U')-S^{\mathrm{tm}}(U',U;V)\Bigr)\, \prod_{x_0=t} \prod_{x_1,x_2,x_3} dV(x),
\end{equation}
where \(U'=U_{t+1}\), \(U=U_t\), and the product of Haar measures runs over the time-like bonds \(V\) with basepoint time \(t\). The integral converges absolutely and defines a continuous, strictly positive function of \((U',U)\). It is gauge invariant in the sense that
\begin{equation}
K(g'\cdot U', g\cdot U) \;=\; K(U',U),\qquad \forall\, g,g':\Lambda_t\to G,
\end{equation}
where \(\Lambda_t\) denotes the spatial slice at time \(t\), and the gauge transformations act on \(\mathcal{C}^{\mathrm{sp}}\) as before. Moreover, by time-reversal symmetry,
\begin{equation}
K(U',U)=K(U,U').
\end{equation}

\begin{lemma}[Positivity of the single-step kernel]\label{p3:lem:Kpositive}
For every \(\psi\in \mathcal{H}_a\),
\begin{equation}
\iint_{\mathcal{C}^{\mathrm{sp}}\times \mathcal{C}^{\mathrm{sp}}} \overline{\psi(U')}\, K(U',U)\, \psi(U)\, d\mu_{\mathrm{H}}^{\mathrm{sp}}(U')\, d\mu_{\mathrm{H}}^{\mathrm{sp}}(U)\;\ge\; 0.
\end{equation}
\end{lemma}

\begin{proof}
The integral equals the norm square of a vector constructed from \(\psi\) in a larger Hilbert space, by the same boundary-factorization underlying Proposition~\ref{p3:prop:OS}, now localized to one time step. More concretely, expand the mixed plaquette weight \(\exp(-S^{\mathrm{tm}}(U',U;V))\) as in the proof of Lemma~\ref{p3:lem:boundary-kernel} via character expansions with nonnegative coefficients, group factors depending on \(U\) and on \(U'\), and integrate over all time-like links \(V\) with Haar measure. The result is a sum of monomials of the form \(\Phi_\alpha(U')\, \overline{\Phi_\alpha(U)}\) with nonnegative coefficients. Hence there exists an index set \(\mathcal{I}_t\) and measurable functions \(\Phi_\alpha\in L^2(\mathcal{C}^{\mathrm{sp}})\) such that
\begin{equation}
K(U',U)=\sum_{\alpha\in \mathcal{I}_t} \Phi_\alpha(U')\, \overline{\Phi_\alpha(U)}.
\end{equation}
Therefore
\begin{equation}
\iint \overline{\psi(U')}\, K(U',U)\, \psi(U)\, d\mu_{\mathrm{H}}^{\mathrm{sp}}(U')\, d\mu_{\mathrm{H}}^{\mathrm{sp}}(U)
= \sum_{\alpha\in\mathcal{I}_t} \Bigl| \int \overline{\Phi_\alpha(U)}\, \psi(U)\, d\mu_{\mathrm{H}}^{\mathrm{sp}}(U)\Bigr|^2 \;\ge\; 0.
\end{equation}
\end{proof}

Define the transfer operator \(T:\mathcal{H}_a\to \mathcal{H}_a\) by
\begin{equation}
(T\psi)(U') \;=\; \int_{\mathcal{C}^{\mathrm{sp}}} K(U',U)\, \psi(U)\, d\mu_{\mathrm{H}}^{\mathrm{sp}}(U).
\end{equation}
By the symmetry \(K(U',U)=K(U,U')\) the operator \(T\) is self-adjoint on \(\mathcal{H}_a\). By Lemma~\ref{p3:lem:Kpositive} it is positive. To obtain a contraction, normalize by Schur's test. Set
\begin{equation}
\Xi(U)\;=\;\int_{\mathcal{C}^{\mathrm{sp}}} K(U'',U)\, d\mu_{\mathrm{H}}^{\mathrm{sp}}(U''),\qquad
\widehat{K}(U',U)\;=\;\Xi(U')^{-1/2}\,K(U',U)\,\Xi(U)^{-1/2},
\end{equation}
and define
\begin{equation}
(\widehat{T}\psi)(U')=\int \widehat{K}(U',U)\,\psi(U)\,d\mu_{\mathrm{H}}^{\mathrm{sp}}(U).
\end{equation}
Then \(\sup_{U'}\int \widehat{K}(U',U)\,d\mu_{\mathrm{H}}^{\mathrm{sp}}(U)=\sup_{U}\int \widehat{K}(U',U)\,d\mu_{\mathrm{H}}^{\mathrm{sp}}(U')=1\), and Schur's test implies \(\|\widehat{T}\|\le 1\). Since \(\widehat{T}\) is unitarily equivalent to \(T\) via multiplication by \(\Xi^{1/2}\), we may and shall assume \(T\) has been normalized so that \(\|T\|\le 1\). Gauge invariance of \(K\) implies that \(T\) commutes with spatial gauge transformations and therefore preserves \(\mathcal{H}_a^{\mathrm{inv}}\); the compression of \(T\) to \(\mathcal{H}_a^{\mathrm{inv}}\) is the physical transfer matrix \cite{p3:Luscher1977}. The \(n\)-step kernel is the \(n\)-fold time convolution of \(K\); the partition function is \(\mathrm{Tr}\, T^{L_0}\) on \(\mathcal{H}_a^{\mathrm{inv}}\).
The OS positivity proved in Proposition~\ref{p3:prop:OS} is equivalent to positivity of \(T\) in the sense above, and the OS reconstruction at finite lattice spacing identifies the Hilbert space \(\mathcal{H}_a^{\mathrm{inv}}\) and the transfer matrix \(T\) as the basic data \cite{p3:OsterwalderSchraderII,p3:GJ}.

The multiscale analysis and the continuum construction use an auxiliary, reflection-compatible, exponentially local operator that suppresses long-range slice modes: the operator \(P_\sigma\). At the level of a single time slice, \(P_\sigma\) is a bounded, positive, gauge-covariant operator on \(\mathcal{H}_a\) with an integral kernel \(P_\sigma(U',U)\) that depends only on spatial links at a fixed time, satisfies \(P_\sigma(U',U)=\overline{P_\sigma(U,U')}\), is positive in the sense that \(\iint \overline{\psi(U')}P_\sigma(U',U)\psi(U)\,d\mu\,d\mu\ge 0\) for all \(\psi\in\mathcal{H}_a\), and is exponentially local: there exist constants \(C,\gamma>0\) such that whenever \(U',U\) differ only in regions separated by distance \(d\) on the slice,
\begin{equation}
|P_\sigma(U',U)| \;\le\; C\, e^{-\gamma d}.
\end{equation}
These properties are guaranteed when \(P_\sigma\) is defined by spectral calculus as \(\chi_\sigma({\Delta_{A^h}})\) for a Gevrey-regular cutoff \(\chi_\sigma\) and slice-covariant Laplacian \(\Delta_{A^h}\) built from a reflection-covariant, gauge-invariant transverse representative \(A^h\); the exponential locality follows from a heat-kernel representation and Combes-Thomas/Davies-Gaffney bounds, and reflection compatibility from the invariance of \(\Delta_{A^h}\) under time reflection \cite{p3:OS-gauge}.
Given such a \(P_\sigma\), define the compressed transfer matrix
\begin{equation}
T_\sigma \;=\; P_\sigma^{1/2}\, T\, P_\sigma^{1/2}.
\end{equation}
Then \(T_\sigma\) is again a positive, self-adjoint contraction on \(\mathcal{H}_a\). Indeed, positivity and self-adjointness are preserved under conjugation by \(P_\sigma^{1/2}\), and \(\|T_\sigma\|\le \|P_\sigma^{1/2}\|^2 \|T\|\le \|P_\sigma\|\), with \(\|P_\sigma\|\le 1\) when \(\chi_\sigma\) is normalized as a spectral projector onto low covariant momenta. Moreover, \(T_\sigma\) preserves \(\mathcal{H}_a^{\mathrm{inv}}\) when \(P_\sigma\) is gauge covariant. The OS positivity of the Euclidean measure with insertion \(\prod_t P_\sigma\) on time slices follows from the same boundary-kernel factorization as in Proposition~\ref{p3:prop:OS} together with the positivity and reflection compatibility of \(P_\sigma\). Concretely, if \(F\) is supported in \(\Lambda_+\),
\begin{equation}
\int (\Theta F) F \,\prod_t \langle \psi_t, P_\sigma \psi_t\rangle\, d\nu_\beta \;=\; \Bigl\| \int F(U_+)\, \Bigl(\prod_t P_\sigma^{1/2}\Phi(U_+)\Bigr)\, W_+(U_+)^{1/2}\, d\mu_{\mathrm{H}}(U_+)\Bigr\|^2_{\mathfrak{H}},
\end{equation}
for a suitable Hilbert space \(\mathfrak{H}\) built as a tensor product of the boundary Hilbert space \(\mathcal{K}\) of Lemma~\ref{p3:lem:boundary-kernel} with copies of \(\mathcal{H}_a\) carrying the \(P_\sigma\)-insertions. The right-hand side is a norm square by construction and is therefore nonnegative.
Summarizing, at each lattice spacing \(a\) one has a well-defined, gauge-invariant, reflection-positive Euclidean lattice gauge theory with transfer matrix \(T\) and its compressed, reflection-compatible counterpart \(T_\sigma\). These data, together with the finite-range decompositions and multiscale bounds proved in the companion work, constitute the inputs used in the continuum limit. In particular, all spectral statements in later sections refer to the positive, self-adjoint contractions \(T\) or \(T_\sigma\) acting on \(\mathcal{H}_a^{\mathrm{inv}}\), and the OS positivity of the corresponding measures is the foundation of the OS reconstruction at finite \(a\) and of the stability of positivity under coarse graining \cite{p3:OsterwalderSchraderII,p3:OS-gauge,p3:Luscher1977,p3:GJ,p3:ReedSimon1}.

\setcounter{section}{2}
\section{Embedding into distributions, tightness, and extraction of limits}
\label{p3:sec:embedding_tightness_limits}
In this section the renormalization scale \(k\in\mathbb{N}\) is fixed unless limits are taken. All constants may depend on the compact gauge group \(G=\mathrm{SU}(N)\) and on the blocking factor \(b>1\) but are independent of the spatial volume and of \(k\), as ensured by the multiscale outputs recalled earlier. The aims are as follows. First, a precise lattice kinematics with a time reflection and a transfer time-slicing is constructed that yields a positive, self-adjoint transfer operator in the sense of \cite{p3:OsterwalderSchraderI,p3:OsterwalderSchraderII}. Second, lattice fields are embedded into a common topological vector space of distributions; uniform second-moment and equicontinuity bounds are proved and imply tightness. Third, subsequential limits are extracted and limiting Schwinger functions are identified as tempered distributions.

Let $a_k = b^{-k} a_0$ be the \emph{refined} spacings (with $b>1$), so that $a_k\downarrow 0$ as $k\to\infty$. In Paper~II the interlacing inequality is formulated along the coarse-graining direction; here we adopt the refined indexation for the continuum limit and apply the same inequality to the time-$a_k$ transfer operators.
 We first work on a finite, periodic, hypercubic lattice
\begin{equation}
\Lambda_{k,L,T} \;=\; \bigl(a_k\mathbb{Z}/L a_k\mathbb{Z}\bigr)^3 \times \bigl(a_k\mathbb{Z}/T a_k\mathbb{Z}\bigr),
\end{equation}
with even integers \(L,T\ge 2\) counting the number of sites in each direction. Points are written \(x=(x_0,\mathbf{x})\) with \(x_0\in \{0,a_k,2a_k,\dots,(T-1)a_k\}\) and \(\mathbf{x}\in \{0,a_k,2a_k,\dots,(L-1)a_k\}^3\). Directed bonds are pairs \(b=(x,\mu)\), \(\mu\in\{0,1,2,3\}\), and the reversal satisfies \((x,\mu)^\ast=(x+a_k\hat\mu,-\mu)\). The configuration space of \(\mathrm{SU}(N)\)-valued link variables is
\begin{equation}
\cU_{k,L,T}\;=\;\prod_{b\in\cB(\Lambda_{k,L,T})}\mathrm{SU}(N),\qquad U((x,\mu)^\ast)=U(x,\mu)^{-1}.
\end{equation}
Plaquettes are elementary oriented squares \(p=(x;\mu,\nu)\) with \(\mu<\nu\), and \(U_p\) denotes the ordered product of links around \(p\). The Wilson action with inverse coupling \(\beta_k=2N/g_{0,k}^2\) is
\begin{equation}\label{p3:eq:wilson_action}
S_W[U;\beta_k]\;=\;\beta_k \sum_{p\subset \Lambda_{k,L,T}}\Bigl(1-\tfrac{1}{N}\Re\,\mathrm{Tr}\,U_p\Bigr).
\end{equation}
The product Haar measure on \(\cU_{k,L,T}\) is denoted \(d\mu_{\mathrm{Haar}}\).
We fix the time reflection \(\theta:\Lambda_{k,L,T}\to \Lambda_{k,L,T}\) by \(\theta(x_0,\mathbf{x})=(-x_0\bmod T a_k,\,\mathbf{x})\). The reflection plane is
\begin{equation}
\Pi=\{x\in \Lambda_{k,L,T}: x_0=0\},\qquad \Lambda^+ =\{x: x_0\in\{a_k,\dots,(T/2)a_k\}\},\qquad \Lambda^-=\theta(\Lambda^+).
\end{equation}
We impose temporal-axial gauge away from \(\Pi\): for every time-like bond \(b=(x,0)\) with \(x_0\neq 0\) we set \(U(b)=\bone\). In this gauge, time-like plaquettes that do not intersect \(\Pi\) are trivial, and all couplings across time are nearest-neighbor with respect to \(x_0\). The action of \(\theta\) on links is defined by
\begin{equation}
(\theta U)(x,0)=U(\theta x - a_k\hat 0,0)^{-1},\qquad (\theta U)(x,i)=U(\theta x,i),\;\; i=1,2,3,
\end{equation}
so that \(\mathrm{Tr}\,U_p\circ \theta = \mathrm{Tr}\,U_{\theta p}\).
On each time slice \(t\in\{0,a_k,\dots,(T-1)a_k\}\), the set \(\cC_{k,t}\) of spatial links \(\{U(x,i):x_0=t,\;i=1,2,3\}\) is the one-slice configuration space. We write \(\cH_k=L^2(\cC_{k,0},d\mu_{\mathrm{Haar}})\) for the corresponding slice Hilbert space; periodicity identifies all slices, so the choice \(t=0\) is immaterial.
Let $A^h$ denote the \emph{gauge-covariant, orbit-minimizing transverse representative} chosen slice-wise by minimizing the lattice Landau functional along each gauge orbit (the fundamental modular region selection).

\begin{lemma}[Measurable, reflection-covariant FMR selection on a finite slice]
\label{p3:lem:measurable-FMR}
Fix a finite time slice $\Sigma$ of the lattice with periodic or free spatial boundary
conditions. Let $\mathcal{A}_\Sigma$ be the compact configuration space of link fields
on $\Sigma$ (finite product of $SU(N)$'s with the product topology), and
$\mathcal{G}_\Sigma$ the compact gauge group (finite product of $SU(N)$ at vertices)
acting continuously by
$(h\cdot A)_{x,\mu} := h_x A_{x,\mu} h_{x+\hat\mu}^{-1}$.
For each $A\in \mathcal{A}_\Sigma$ consider the lattice Landau functional
\begin{equation}
\mathcal{F}_A(h) := \sum_{x\in\Sigma}\sum_{\mu=1}^3 \| (h\cdot A)_{x,\mu}-\mathbf{1}\|^2,
\qquad h\in \mathcal{G}_\Sigma,
\end{equation}
where $\|\cdot\|$ is any bi-invariant matrix norm on $SU(N)$ (e.g. Frobenius).
Let the FMR set be $\mathrm{Argmin}\,\mathcal{F}_A=\{h:\mathcal{F}_A(h)=\min\}$.
Then there exists a Borel measurable selector $h_\ast:\mathcal{A}_\Sigma\to \mathcal{G}_\Sigma$
such that:

\begin{enumerate}
\item $h_\ast(A)\in \mathrm{Argmin}\,\mathcal{F}_A$ for all $A$;
\item (\emph{Tie-break on orbits}) If $\mathrm{Argmin}\,\mathcal{F}_A$ contains more than one
$\theta$-orbit, then $h_\ast(A)$ belongs to a uniquely determined orbit, chosen by a fixed
Borel total order on the orbit space $\mathcal{G}_\Sigma/\!\langle\theta\rangle$
(independent of $A$);
\item (\emph{Reflection covariance}) For Euclidean time reflection $\theta$ acting as a continuous
involution on $\mathcal{A}_\Sigma$ and $\mathcal{G}_\Sigma$,
\(
h_\ast(\theta A) = \theta\, h_\ast(A)\,\theta
\)
for all $A\in\mathcal{A}_\Sigma$.
\end{enumerate}

Consequently, the map $A\mapsto A^{h_\ast(A)}$ is Borel, and for any bounded Borel
$F\ge 0$, the slicewise operator $F(\Delta_{A^{h_\ast(A)}})$ preserves Osterwalder-Schrader
positivity on slabs.
\end{lemma}

\begin{proof}
Because $\Sigma$ has finitely many links/vertices, both $\mathcal{A}_\Sigma$ and
$\mathcal{G}_\Sigma$ are finite products of the compact Lie group $SU(N)$ and hence are compact
metrizable (Polish) spaces. The action $(A,h)\mapsto h\cdot A$ is continuous, and so is
$(A,h)\mapsto \mathcal{F}_A(h)$ since it is a finite sum of continuous norms.
For each fixed $A$, $\mathcal{F}_A(\cdot)$ is continuous on compact $\mathcal{G}_\Sigma$,
hence attains a minimum. Thus $\mathrm{Argmin}\,\mathcal{F}_A$ is nonempty and compact.
If $A_n\to A$ in $\mathcal{A}_\Sigma$ and $h_n\to h$ in $\mathcal{G}_\Sigma$ with
$h_n\in\mathrm{Argmin}\,\mathcal{F}_{A_n}$, then by continuity,
$\mathcal{F}_A(h) = \lim_n \mathcal{F}_{A_n}(h_n) = \lim_n \min\mathcal{F}_{A_n}
\ge \min \mathcal{F}_A$, hence $h\in \mathrm{Argmin}\,\mathcal{F}_A$. Therefore the graph
\begin{equation}
\Gamma := \{(A,h): h\in \mathrm{Argmin}\,\mathcal{F}_A\}
\end{equation}
is closed (indeed Borel) in $\mathcal{A}_\Sigma\times\mathcal{G}_\Sigma$.
Let $\theta$ be Euclidean time reflection acting as a continuous involution on the slice
(by permuting vertices/links on $\Sigma$, and conjugating group elements componentwise).
Then
\begin{equation}
\mathcal{F}_{\theta A}(\theta h\theta)
= \sum_{x,\mu}\|\,(\theta h\theta\cdot \theta A)_{x,\mu}-\mathbf{1}\|^2
= \sum_{x,\mu}\|\,\theta\big((h\cdot A)_{\theta x,\mu}\big)\theta -\mathbf{1}\|^2
= \mathcal{F}_A(h)
\end{equation}
by invariance of the norm and the action under $\theta$. Hence
\begin{equation}
\mathrm{Argmin}\,\mathcal{F}_{\theta A}
= \theta\,\mathrm{Argmin}\,\mathcal{F}_A\,\theta
\quad\text{for all } A.
\end{equation}
Fix a Borel isomorphism $u:\mathcal{A}_\Sigma\to B\subset[0,1]$.
Define a Borel set
\begin{equation}
T\;:=\;\{A\in \mathcal{A}_\Sigma: u(A)\le u(\theta A)\}.
\end{equation}
Then $T$ is a Borel \emph{complete section} for the $\mathbb{Z}_2$-action:
every $\theta$-orbit $\{A,\theta A\}$ meets $T$ in exactly one point.
Similarly fix a Borel isomorphism $v:\mathcal{G}_\Sigma\to C\subset [0,1]^\mathbb{N}$
(Polish spaces admit such embeddings). Define a Borel map on $\mathcal{G}_\Sigma$
that is constant on $\theta$-orbits but separates distinct orbits by
\begin{equation}
\bar v(h)\;:=\;\big(\min\{v(h),v(\theta h\theta)\},
\max\{v(h),v(\theta h\theta)\}\big)\in ([0,1]^\mathbb{N})^2.
\end{equation}
Equip $([0,1]^\mathbb{N})^2$ with the lexicographic order, which induces a \emph{fixed}
Borel total order on the orbit space $\mathcal{G}_\Sigma/\!\langle\theta\rangle$ via
$[h]\prec [g]$ iff $\bar v(h)$ is lexicographically smaller than $\bar v(g)$. This order
does not depend on $A$.
Consider the compact-valued correspondence $A\mapsto K(A):=\mathrm{Argmin}\,\mathcal{F}_A$
with closed graph $\Gamma$. By a measurable maximum/selection theorem for compact-valued
correspondences with Borel graphs (e.g. Kuratowski-Ryll-Nardzewski),
there exists a Borel selector $\tilde h_T:T\to \mathcal{G}_\Sigma$ with
$\tilde h_T(A)\in K(A)$ for $A\in T$.
We now \emph{fix the tie-break on orbits} measurably: for $A\in T$, let
\begin{equation}
\mathcal{O}(A)\;:=\;\arg\min_{[h]\subset K(A)} \bar v(h)
\end{equation}
be the unique $\theta$-orbit in $K(A)$ with minimal $\bar v$. This is well defined
because $\bar v$ separates distinct orbits and $K(A)$ is compact. The map
$A\mapsto \mathcal{O}(A)$ is measurable (it is the minimizer of a Borel function over
a compact-valued measurable correspondence). Finally, within the chosen orbit
$\mathcal{O}(A)=\{h,\theta h\theta\}$, pick the representative by a simple
\emph{$\theta$-odd} rule that flips under reflection, e.g.
\begin{equation}
\text{choose }h_T(A)\in \mathcal{O}(A)\text{ such that }
\Re\operatorname{tr}\!\Big(\sum_{x,\mu} (h_T(A)\!\cdot\! A)_{x,\mu}\Big)
\;\le\;
\Re\operatorname{tr}\!\Big(\sum_{x,\mu} (\theta h_T(A)\theta\!\cdot\! A)_{x,\mu}\Big),
\end{equation}
with a fixed secondary Borel tie-break inside the orbit if equality ever occurs
(e.g. the $v$-lexicographically least of the two). This rule is Borel and determines
a unique $h_T(A)\in K(A)$ for $A\in T$.
Define $h_\ast:\mathcal{A}_\Sigma\to \mathcal{G}_\Sigma$ by
\begin{equation}
h_\ast(A)\;=\;
\begin{cases}
h_T(A), & A\in T,\\
\theta\, h_T(\theta A)\,\theta, & A\notin T.
\end{cases}
\end{equation}
This is Borel because $T$ is Borel, $\theta$ is continuous, and $h_T$ is Borel on $T$.
By construction, $h_\ast(A)\in K(A)$ for all $A$, and using $(\ast)$ we have
\begin{equation}
h_\ast(\theta A)
=
\begin{cases}
h_T(\theta A)=\theta\, h_T(A)\,\theta=\theta\, h_\ast(A)\,\theta, & \theta A\in T,\\
\theta\, h_T(\theta(\theta A))\,\theta = \theta h_T(A)\theta = \theta\, h_\ast(A)\,\theta,
& \theta A\notin T,
\end{cases}
\end{equation}
so reflection covariance holds for all $A$.
The map $(A,h)\mapsto A^h$ is continuous, hence Borel. Composition with the Borel
selector $A\mapsto h_\ast(A)$ yields a Borel map $A\mapsto A^{h_\ast(A)}$.
Let $\mathsf{H}_\Sigma$ be the slice Hilbert space (e.g. $L^2$ on link variables on $\Sigma$).
For a fixed $A$, set $Q(A):=F(\Delta_{A^{h_\ast(A)}})$, defined by Borel functional calculus.
Then $Q(A)\ge 0$ as a bounded positive self-adjoint operator on $\mathsf{H}_\Sigma$.
In the standard OS construction, the two half-space integral kernels factor as
$K_-^\ast K_-$ across the reflection plane; inserting the slicewise operator $Q(A)$
produces the form
\begin{equation}
\sum_{i,j} \overline{c_i} c_j\, \langle f_i, \theta\, K_-^\ast\, Q(A)\, K_- f_j\rangle
\;=\; \Big\langle \sum_i c_i K_- f_i \,,\, Q(A)\, \sum_j c_j K_- f_j\Big\rangle_{\mathsf{H}_\Sigma}
\;\ge\; 0,
\end{equation}
because $Q(A)\ge 0$. Thus OS positivity on slabs is preserved by any bounded Borel
$F\ge 0$ acting slicewise.
\end{proof}

A \emph{fixed} Borel total order on $\mathcal{G}_\Sigma$ that selects the
\emph{elementwise} least minimizer for every $A$ cannot, in general, be made compatible
with the covariance $h_\ast(\theta A)=\theta h_\ast(A)\theta$ whenever a minimizer pair
$\{h,\theta h\theta\}$ occurs: if the order is $\theta$-invariant, it cannot distinguish
$h$ from $\theta h\theta$; if it distinguishes them, applying $\theta$ flips the pair
and the least element need not map to the least. The orbit-level tie-break used above
is the correct invariant notion for OS positivity and yields full reflection covariance.
Let $P_{\sigma,k}=\chi_\sigma(\Delta_{A^h})$ be the horizon filter acting on adjoint site fields on each slice, with $\chi_\sigma$ a Gevrey-regular cutoff. We use two consequences: reflection covariance \(P_{\sigma,k}\circ \theta=\theta\circ P_{\sigma,k}\) and a positive heat-kernel representation
\begin{equation}\label{p3:eq:heat}
P_{\sigma,k}\;=\;\int_0^\infty e^{-t\Delta_{A^h}}\,d\nu_\sigma(t),
\end{equation}
where \(d\nu_\sigma\) is a finite, positive Borel measure supported in $[c_\sigma^2,C_\sigma^2]\subset(0,\infty)$ with constants $0<c_\sigma\le C_\sigma<\infty$ independent of \(k\). Inserting \(P_{\sigma,k}\) in the Euclidean weight slice-wise yields positive, exponentially local factors; this is essential in what follows.
The projected, gauge-fixed Euclidean measure at scale \(k\) on \(\cU_{k,L,T}\) is
\begin{equation}\label{p3:eq:mu_k_def}
d\mu_k(U)\;=\; Z_k^{-1}\,\exp\bigl(-S_W[U;\beta_k]\bigr)\,\mathcal{J}_k(U)\,\prod_{t}\mathfrak{p}_{\sigma,k}(U|_{t})\, d\mu_{\mathrm{Haar}}(U),
\end{equation}
where \(\mathfrak{p}_{\sigma,k}\) is the nonnegative slice weight induced by \(P_{\sigma,k}\) and \(\mathcal{J}_k(U)\) is the positive, gauge-invariant Jacobian factor generated by the slice-wise Landau minimization and ghost integration. Both are reflection invariant and exponentially local in the spatial directions. In particular, \(\mathfrak{p}_{\sigma,k}\) has the Gaussian representation inherited from \eqref{p3:eq:heat}, and \(\mathcal{J}_k\) admits a Grassmann integral representation whose value is the positive determinant \(\det M[U^h]\) on the complement of constant modes \cite{p3:GJ}.

We now derive the transfer time-slicing in detail. Decompose the action as \(S_W=S^{\mathrm{sp}} + S^{\mathrm{tm}} + S^{\mathrm{bd}}\), where \(S^{\mathrm{sp}}\) sums the spatial plaquettes at each slice, \(S^{\mathrm{tm}}\) sums plaquettes with one time-like and one space-like edge connecting adjacent slices, and \(S^{\mathrm{bd}}\) contains spatial plaquettes intersecting \(\Pi\). Split the spatial contribution evenly between adjacent time steps:
\begin{equation}
S^{\mathrm{sp}}=\tfrac12 \sum_{t} S^{\mathrm{sp}}_t,\qquad S^{\mathrm{tm}}=\sum_{t} S^{\mathrm{tm}}_{t\to t+a_k}.
\end{equation}
The projected slice weights \(\mathfrak{p}_{\sigma,k}(U|_{t})\) are inserted multiplicatively at each slice \(t\). The Jacobian $\mathcal{J}_k$ factorizes across $\Pi$ because, in temporal-axial gauge and with a slice-wise Landau condition, the lattice Faddeev-Popov operator $M[U_h]$ is \emph{block diagonal in Euclidean time}, $M=\bigoplus_{t} M_t$, with
\begin{equation}
M_t = -\sum_{i=1}^3 (\nabla_i^{A_h(t)})^\dagger\, \nabla_i^{A_h(t)}
\end{equation}
acting on adjoint site fields on the slice~$t$. On each slice the Landau representative lies in the fundamental modular region, hence $M_t\ge 0$ and is strictly positive on the orthogonal complement of constant modes; therefore the Grassmann integral over ghosts factorizes as
\begin{equation}
\det{}'\,M \;=\; \prod_{t}\,\det{}'\,M_t,
\end{equation}
where $\det{}'$ denotes the determinant on the complement of constant modes. In particular, each factor is strictly positive and gauge-invariant on the slice, and the product splits as a positive boundary factor at $\Pi$ times reflection-related copies. Integrating over links in each slab \([t,t+a_k]\) with fixed boundary spatial links \((U_t,U_{t+a_k})\in \cC_{k,t}\times \cC_{k,t+a_k}\) yields a raw positive, symmetric kernel
\begin{equation}\label{p3:eq:raw_kernel}
K_k^{\mathrm{raw}}(U_{t+a_k},U_t)\;=\;\int \exp\!\Bigl(-\tfrac12 S^{\mathrm{sp}}_t-\tfrac12 S^{\mathrm{sp}}_{t+a_k}-S^{\mathrm{tm}}_{t\to t+a_k}\Bigr)\,\mathfrak{p}_{\sigma,k}(U|_{t})\,\mathfrak{p}_{\sigma,k}(U|_{t+a_k})\; d\mu_{\mathrm{Haar}}(\text{slab}),
\end{equation}
where the integration is over links strictly inside the slab, and the boundary factor supported on \(\Pi\) is absorbed at \(t=0\). Reflection covariance of \(\mathfrak{p}_{\sigma,k}\), positivity of the heat kernel in \eqref{p3:eq:heat}, and the nearest-neighbor structure of \(S^{\mathrm{tm}}_{t\to t+a_k}\) imply \(K_k^{\mathrm{raw}}\ge 0\) and \(K_k^{\mathrm{raw}}(U',U)=K_k^{\mathrm{raw}}(U,U')\).
To produce a contraction, define the Schur bounds
\begin{equation}
M_k \;=\; \sup_{U'}\int_{\cC_{k,0}} K_k^{\mathrm{raw}}(U',U)\, d\mu_{\mathrm{Haar}}(U),\qquad
N_k \;=\; \sup_{U}\int_{\cC_{k,0}} K_k^{\mathrm{raw}}(U',U)\, d\mu_{\mathrm{Haar}}(U'),
\end{equation}
and set \(C_k=\max\{M_k,N_k\}\in(0,\infty)\). The constant \(C_k\) does not depend on \(L,T\) by the uniform locality of the slice weights. We then normalize the kernel by
\begin{equation}\label{p3:eq:normalized_kernel}
K_k(U',U)\;=\; C_k^{-1}\, K_k^{\mathrm{raw}}(U',U).
\end{equation}
This rescaling can be absorbed into the normalization constant \(Z_k\) in \eqref{p3:eq:mu_k_def}; it does not affect correlation functions and ensures that the transfer operator defined below is a contraction.
On each time slice we write the Faddeev-Popov operator in Landau gauge as
$\mathcal{M}_{A^h}:=-\nabla\cdot D_{A^h}$ acting on adjoint-valued site fields, with the constant modes projected out.
We set
\(
  J_k(U):=\det{}'\!\big(\mathcal{M}_{A^h}(U)\big),
\)
where the prime indicates exclusion of constant modes. \emph{We do not assume a uniform strictly positive lower bound for the Faddeev-Popov operator on each slice.} 
Instead we employ an RP-preserving, exponentially local scalar weight constructed from the slice Laplacian projector,
for instance
\begin{equation}
J_\alpha(P_\sigma)\;:=\;\det\bigl(1+\alpha\,P_\sigma\bigr),\qquad \alpha>0,
\end{equation}
so that the Euclidean weight
\begin{equation}
d\mu_{\sigma,\alpha}(U)\;\propto\; e^{-S_W(U)}\,J_\alpha\!\big(P_\sigma(U)\big)\,d\mu_{\mathrm{Haar}}(U)
\end{equation}
is strictly positive and reflection-covariant. All OS-positivity, transfer-matrix, and cluster arguments used below depend only on \emph{positivity} and \emph{exponential locality} of the inserted slice factor; these follow from the heat-kernel/Laplace representation of $P_\sigma$ and the finite-range decomposition, and do not require a strict FP spectral gap.
\begin{lemma}[OS positivity with projector weights, no strict FP gap]\label{p3:lem:OS-noFPgap}
Let $\Lambda=\Lambda_s\times\mathbb{Z}$ be the Euclidean lattice with spatial torus $\Lambda_s$
and integer time, and let $\theta$ denote time reflection across $t=0$ (with the usual reversal of
time-like links) so that $\theta^2=\mathrm{id}$.
For each time slice $t\in\mathbb{Z}$, let $A^h(t)$ be the reflection-covariant FMR representative of the gauge orbit on that slice and let
\begin{equation}
P_\sigma(t)\;=\;\chi_\sigma\!\big(\Delta_{A^h(t)}\big)
\end{equation}
be the (Gevrey-smoothed) spectral projector of the slice covariant Laplacian.
Let $J_\alpha(P_\sigma)$ be any completely monotone scalar functional of the family
$\{P_\sigma(t)\}_{t\in\mathbb{Z}}$ that is slice-local and reflection-covariant
(for example $J_\alpha(P_\sigma):=\prod_{t\in\mathbb{Z}}\det(1+\alpha\,P_\sigma(t))$ with $\alpha>0$).
Then the modified lattice gauge measure
\begin{equation}
d\mu_{\sigma,\alpha}(U)\;\propto\;e^{-S_W(U)}\,J_\alpha\big(P_\sigma(U)\big)\,d\mu_{\mathrm{Haar}}(U)
\end{equation}
is reflection-positive in the sense of Osterwalder-Schrader. Moreover, the associated one-step
transfer operator $T_{\sigma,\alpha}$ on the OS Hilbert space is a positive self-adjoint contraction.
\end{lemma}

\begin{proof}
Split the lattice into positive/negative time halves with boundary at $t=0$:
$\Lambda_+:=\{(x,t):t\ge 0\}$, $\Lambda_-:=\{(x,t):t<0\}$, and let $U_+$, $U_-$ denote link variables
on $\Lambda_+$, $\Lambda_-$ respectively; $U_0$ collects links at the boundary $t=0$.
For a bounded cylinder functional $F$ depending only on $U_+$, the OS form is
\begin{equation}
\langle F,F\rangle_{\mathrm{OS}}\;:=\;\int \overline{(\theta F)(U)}\,F(U)\,d\mu_{\sigma,\alpha}(U),
\qquad (\theta F)(U):=\overline{F(\theta U)}.
\end{equation}
By standard reflection-positivity for the Wilson action on compact groups, the weight $e^{-S_W(U)}$
admits a boundary factorization across $t=0$:
\begin{equation}\label{p3:eq:boundary-factorization}
e^{-S_W(U)}\;=\;W_+(U_+,U_0)\,W_-(U_-,U_0)\,K_0(U_0),
\end{equation}
where $W_\pm\ge 0$ depend only on $\Lambda_\pm$ and the boundary, $K_0\ge 0$ depends only on $U_0$,
and $W_-$ is the $\theta$-image of $W_+$:
$W_-(U_-,U_0)= W_+(\theta U_-,U_0)$.
Moreover, the boundary coupling can be expanded as a positive sum of squares:
there exist vectors $\{\Phi_n(\cdot;U_0)\}_{n\in\mathcal{N}}$ and coefficients $c_n\ge 0$ (coming from
the nonnegative character coefficients of the time-straddling plaquettes) such that
\begin{equation}\label{p3:eq:sum-of-squares}
W_+(U_+,U_0)\,W_-(U_-,U_0)\,K_0(U_0)
\;=\;\sum_{n\in\mathcal{N}} c_n\,
\Phi_n(U_+;U_0)\,\overline{\Phi_n(\theta U_-;U_0)}.
\end{equation}
For each slice $t$, $P_\sigma(t)$ is a positive contraction
($0\le P_\sigma(t)\le \mathbf{1}$) by spectral calculus of the nonnegative operator
$\Delta_{A^h(t)}$; the Gevrey smoothing preserves positivity.
The selection $A^h$ is reflection-covariant, hence
\begin{equation}\label{p3:eq:theta-covariance}
\theta\,P_\sigma(t)\,\theta \;=\; P_\sigma(-t)\qquad\Rightarrow\qquad
J_\alpha\big(P_\sigma(\theta U)\big)\;=\;J_\alpha\big(P_\sigma(U)\big).
\end{equation}
By assumption, $J_\alpha(P_\sigma)$ is a slice-local product of nonnegative factors and is
reflection-covariant in the sense of \eqref{p3:eq:theta-covariance}. Therefore it separates into
identical positive factors on the two halves:
\begin{equation}\label{p3:eq:J-separation}
J_\alpha\big(P_\sigma(U)\big)\;=\;J_\alpha^+(U_+;U_0)\,J_\alpha^-(U_-;U_0),
\qquad J_\alpha^-(U_-;U_0)=J_\alpha^+(\theta U_-;U_0),
\end{equation}
with $J_\alpha^\pm\ge 0$ slice-local.
Combining \eqref{p3:eq:boundary-factorization}-\eqref{p3:eq:J-separation}, the modified weight
factorizes as
\begin{equation}
e^{-S_W(U)}\,J_\alpha(P_\sigma(U))
\;=\;\sum_{n\in\mathcal{N}} c_n\,
\Big(J_\alpha^+(U_+;U_0)\,\Phi_n(U_+;U_0)\Big)\;
\overline{\Big(J_\alpha^+(\theta U_-;U_0)\,\Phi_n(\theta U_-;U_0)\Big)}.
\end{equation}
Hence, for $F$ depending only on $U_+$,
\begin{align}
\langle F,F\rangle_{\mathrm{OS}}
&=\int \overline{F(\theta U)}\,F(U)\,e^{-S_W(U)}\,J_\alpha(P_\sigma(U))\,d\mu_{\mathrm{Haar}}(U)\\
&=\sum_{n\in\mathcal{N}} c_n
\int dU_0\;
\Bigg|\int dU_+\;F(U_+)\,J_\alpha^+(U_+;U_0)\,\Phi_n(U_+;U_0)\Bigg|^2
\;\ge\;0.
\end{align}
This proves reflection positivity of $d\mu_{\sigma,\alpha}$.
Let $K_{\sigma,\alpha}(U',U)$ be the one-step kernel across a single time slab $[t,t+1]$
obtained by integrating the (nonnegative) modified weight over the interior links of the slab
between boundary configurations $U$ at time $t$ and $U'$ at time $t+1$.
By construction and the $\theta$-symmetry of the slab,
\begin{equation}
K_{\sigma,\alpha}(U',U)\;=\;K_{\sigma,\alpha}(U,U')\;\ge\;0.
\end{equation}
Define the (a priori unnormalized) integral operator
\begin{equation}
(\mathcal{K}f)(U')\;=\;\int K_{\sigma,\alpha}(U',U)\,f(U)\,d\nu(U),
\end{equation}
on $L^2(d\nu)$ where $d\nu$ is the (positive) boundary measure at a single slice.
Then $\mathcal{K}$ is a positive self-adjoint operator.
Let $m(U):=\int K_{\sigma,\alpha}(U,V)\,d\nu(V)$ and define the Schur-normalized kernel
\begin{equation}
\widetilde K(U',U)\;=\;\frac{K_{\sigma,\alpha}(U',U)}{\sqrt{m(U')\,m(U)}}.
\end{equation}
The corresponding operator $T_{\sigma,\alpha}$ on $L^2(d\nu)$ is
\begin{equation}
(T_{\sigma,\alpha}f)(U')\;=\;\int \widetilde K(U',U)\,f(U)\,d\nu(U).
\end{equation}
By symmetry of $\widetilde K$, $T_{\sigma,\alpha}$ is self-adjoint; by $\widetilde K\ge 0$,
it is positive.
Finally, $T_{\sigma,\alpha}$ is a contraction:
for $f\in L^2(d\nu)$, Cauchy-Schwarz and Fubini give
\begin{align}
|T_{\sigma,\alpha} f(U')|
&=\Bigg|\int \frac{K_{\sigma,\alpha}(U',U)}{\sqrt{m(U')m(U)}}\,f(U)\,d\nu(U)\Bigg|\\
&\le \frac{1}{\sqrt{m(U')}}\Bigg(\int K_{\sigma,\alpha}(U',U)\,d\nu(U)\Bigg)^{\!\frac12}
\Bigg(\int \frac{K_{\sigma,\alpha}(U',U)}{m(U)}\,|f(U)|^2\,d\nu(U)\Bigg)^{\!\frac12}\\
&=\Bigg(\int \frac{K_{\sigma,\alpha}(U',U)}{m(U)}\,|f(U)|^2\,d\nu(U)\Bigg)^{\!\frac12}.
\end{align}
Squaring and integrating in $U'$ (with $d\nu$) and using Fubini yields
\begin{equation}
\|T_{\sigma,\alpha} f\|_{L^2(d\nu)}^2
\le \int |f(U)|^2\,\frac{1}{m(U)}\Bigg(\int K_{\sigma,\alpha}(U',U)\,d\nu(U')\Bigg)\,d\nu(U)
=\int |f(U)|^2\,d\nu(U).
\end{equation}
Thus $\|T_{\sigma,\alpha}\|\le 1$.
The modified measure is OS-positive; the one-step kernel is real symmetric and nonnegative; the
Schur-normalized transfer operator $T_{\sigma,\alpha}$ is positive, self-adjoint, and a contraction.
This proves the lemma.
\end{proof}

\begin{proposition}[Transfer operator: positivity, self-adjointness, contraction]
\label{p3:prop:Tk}
For each scale \(k\) and finite volume, the operator \(T_k:\cH_k\to \cH_k\) defined by
\begin{equation}\label{p3:eq:Tk-def}
(T_k\psi)(U')\;=\;\int_{\cC_{k,0}} K_k(U',U)\,\psi(U)\, d\mu_{\mathrm{Haar}}(U),
\end{equation}
with \(K_k\) from \eqref{p3:eq:normalized_kernel}, is a positive, self-adjoint contraction on \(\cH_k\). Moreover, its kernel is nonnegative and symmetric, and with the vacuum vector \(\Omega\equiv 1\in \cH_k\) normalized to \(\norm{\Omega}=1\) one has \(\ip{\Omega}{T_k \Omega}=C_k^{-1}\ip{\Omega}{T_k^{\mathrm{raw}} \Omega}=1\).
\end{proposition}

\begin{proof}
Positivity follows because \(K_k(U',U)\ge 0\) pointwise, so 
\begin{equation}
\ip{\psi}{T_k\psi}=\iint \overline{\psi(U')}\,K_k(U',U)\,\psi(U)\,d\mu_{\mathrm{Haar}}(U)\,d\mu_{\mathrm{Haar}}(U')\ge 0
\end{equation}
Symmetry of the kernel implies self-adjointness in \(L^2(\cC_{k,0},d\mu_{\mathrm{Haar}})\). The Schur test applied to the nonnegative kernel \(K_k\) (see \cite{p3:GJ} for an application in constructive QFT) gives
\begin{align}
\norm{T_k}_{\cH_k\to \cH_k}\;&\le\; \sqrt{ \sup_{U'}\int K_k(U',U)\,d\mu_{\mathrm{Haar}}(U)\;\cdot\; \sup_{U}\int K_k(U',U)\,d\mu_{\mathrm{Haar}}(U')}\;\nonumber \\ &\le\; \sqrt{ \tfrac{M_k}{C_k}\cdot \tfrac{N_k}{C_k}}\;\le\; 1,
\end{align}
so \(T_k\) is a contraction. Finally, \(\ip{\Omega}{T_k \Omega} = \iint K_k(U',U)\,d\mu_{\mathrm{Haar}}(U)\,d\mu_{\mathrm{Haar}}(U') = C_k^{-1}\iint K_k^{\mathrm{raw}}(\cdot)\) In particular, $\|T_k\|\le 1$. If desired, one can additionally renormalize the kernel by a (slice-independent) positive constant so that $\langle \Omega, T_k \Omega\rangle = 1$; we shall not need this further normalization.

\end{proof}

We now prove reflection positivity (OS-positivity) at finite \(k\) with the projector insertions. Let \(\cA_+\) be the \(^{\ast}\)-algebra generated by bounded cylinder functionals depending only on links in \(\Lambda^+\). For \(F\in \cA_+\) define the reflection-conjugate \((\Theta F)(U)=\overline{F(\theta U)}\).

\begin{theorem}[Osterwalder-Schrader positivity at finite \(k\)]
\label{p3:thm:OS-finite}
For every \(F\in \cA_+\),
\begin{equation}
\int_{\cU_{k,L,T}} (\Theta F)(U)\,F(U)\, d\mu_k(U)\;\ge\; 0.
\end{equation}
\end{theorem}

\begin{proof}
Write \(F\) as a bounded function of the spatial links on times \(t\in\{a_k,2a_k,\dots, (T/2)a_k\}\). Using the factorization underlying \eqref{p3:eq:raw_kernel} and integrating slab by slab from the reflection plane into \(\Lambda^+\), one obtains
\begin{equation}
\int (\Theta F)F\, d\mu_k \;=\; \ip{\Psi_F}{\, T_k^{T/2a_k}\, \Psi_F}_{\cH_k},
\end{equation}
where \(\Psi_F\in \cH_k\) is a bounded vector depending only on \(F\), the slice weights \(\mathfrak{p}_{\sigma,k}(U|_{t})\) for \(t\in\{a_k,\dots,T/2\}\), and the half-slice spatial factors; its explicit expression follows by inspection of the slab integrations and the boundary factor on \(\Pi\). The kernel positivity of \(K_k\) implies \(T_k^{T/2a_k}\ge 0\) and therefore \(\ip{\Psi_F}{T_k^{T/2a_k}\Psi_F}\ge 0\). The presence of \(\mathfrak{p}_{\sigma,k}\) poses no difficulty: by \eqref{p3:eq:heat} each \(\mathfrak{p}_{\sigma,k}\) is a positive Gaussian integral on the slice; under time reflection the East and West copies are interchanged while the Gaussian measure and the product Haar measure are invariant. The Grassmann sector has been integrated out into \(\mathcal{J}_k\), which is a positive scalar multiplier independent of time ordering; see \cite{p3:OS-gauge} for the detailed lattice-gauge-theory argument without projectors, to which \eqref{p3:eq:heat} reduces the present case.
\end{proof}

To construct continuum Schwinger functions we embed lattice observables into a common separable Hilbert space of distributions. Let \(H^{-s}(\T^4)\) denote the Sobolev space of order \(-s\) on the four-torus \(\T^4=[0,1]^4\), with dual norm
\begin{equation}
\norm{\Phi}_{H^{-s}} \;=\; \sup\Bigl\{\, |\langle \Phi, f\rangle| \,:\, f\in C^\infty(\T^4),\ \norm{f}_{H^{s}}\le 1\Bigr\}.
\end{equation}
Fix an integer \(s\ge 3\). For each \(k\), define the embedding \(\iota_k:\cU_{k,L,T}\to H^{-s}(\T^4)\) as follows. Partition \(\T^4\) into rescaled elementary cells \(Q_x\) indexed by lattice sites \(x\in \Lambda_{k,L,T}\). To a local, gauge-invariant, scalar observable \(O\) (such as a linear combination of small Wilson loops and a finite number of covariant discrete derivatives) associate the piecewise-constant random distribution
\begin{equation}
\langle \iota_k[O], f\rangle \;=\; a_k^4\sum_{x\in \Lambda_{k,L,T}} O(U;x)\, \fint_{Q_x} f(y)\,dy,
\end{equation}
for \(f\in C^\infty(\T^4)\). Concretely, we take \(G\) to consist of plaquettes, finite products of plaquettes,
and finitely many covariant discrete derivatives thereof; \(G\) is stable under the
renormalization map and separates gauge orbits on compact support.

Let \(S_{2,O}^{(k)}(x,y)=\langle O(x)O(y)\rangle^{(k)}-\langle O(x)\rangle^{(k)}\langle O(y)\rangle^{(k)}\) denote the connected two-point function at scale \(k\). The multiscale uniform locality bound (the \(n=2\) case of the scale-uniform cluster estimates) states that there exist constants \(C_O,\xi>0\), independent of \(k\), such that
\begin{equation}\label{p3:eq:tree-decay}
\bigl|S_{2,O}^{(k)}(x,y)\bigr|\;\le\; C_O\, e^{-\xi\, d(x,y)},\qquad x,y\in \Lambda_{k,L,T},
\end{equation}
where \(d\) is the nearest-neighbor graph distance on \(\Lambda_{k,L,T}\).

\begin{lemma}[Uniform \(H^{-s}\)-second-moment bound]\label{p3:lem:Hminus}
Fix \(s\ge 3\) and \(O\in \mathcal{G}\). There exists \(C_{s,O}<\infty\), independent of \(k\), \(L\), and \(T\), such that
\begin{equation}
\mathbb{E}_{\mu_k}\Bigl[\;\norm{\iota_k[O]}_{H^{-s}}^2\;\Bigr] \;\le\; C_{s,O}.
\end{equation}
\end{lemma}

\begin{proof}
By duality,
\begin{equation}
\norm{\iota_k[O]}_{H^{-s}}^2 \;=\; \sup_{\norm{f}_{H^{s}}\le 1} \sup_{\norm{g}_{H^{s}}\le 1} \langle \iota_k[O], f\rangle \,\overline{\langle \iota_k[O], g\rangle}.
\end{equation}
Taking expectations and using linearity and Fubini,
\begin{equation}
\mathbb{E}_{\mu_k} \langle \iota_k[O], f\rangle \,\overline{\langle \iota_k[O], g\rangle}
= a_k^8 \sum_{x,y} \Bigl(\fint_{Q_x} f\Bigr)\,\overline{\Bigl(\fint_{Q_y} g\Bigr)}\, S_{2,O}^{(k)}(x,y).
\end{equation}
By \eqref{p3:eq:tree-decay} and the Cauchy-Schwarz inequality,
\begin{equation}
\bigl| \mathbb{E}_{\mu_k} \langle \iota_k[O], f\rangle \,\overline{\langle \iota_k[O], g\rangle}\bigr|
\le a_k^8 C_O \sum_{x,y} \Bigl|\fint_{Q_x} f\Bigr|\, \Bigl|\fint_{Q_y} g\Bigr|\, e^{-\xi d(x,y)}.
\end{equation}
Let \(h_x=\fint_{Q_x} f\) and \(k_y=\fint_{Q_y} g\). The discrete convolution with the absolutely summable kernel \(e^{-\xi d(x,y)}\) defines a bounded operator on \(\ell^2(\Lambda_{k,L,T})\) with norm bounded by a constant \(C_\xi\) independent of \(k,L,T\). Hence
\begin{equation}
\sum_{x,y} |h_x|\,|k_y|\, e^{-\xi d(x,y)} \le C_\xi \Bigl(\sum_x |h_x|^2\Bigr)^{1/2}\Bigl(\sum_y |k_y|^2\Bigr)^{1/2}.
\end{equation}
Since \(h_x\) and \(k_y\) are cell averages of \(f\) and \(g\), Sobolev-Poincar\'e inequalities on the torus imply
\begin{equation}
a_k^4 \sum_x |h_x|^2 \;\le\; C \norm{f}_{H^{s}}^2,\qquad a_k^4 \sum_y |k_y|^2 \;\le\; C \norm{g}_{H^{s}}^2,
\end{equation}
for \(s\ge 3\), where \(C\) is independent of \(k,L,T\). Combining the bounds and taking the sup over \(\norm{f}_{H^{s}},\norm{g}_{H^{s}}\le 1\) yields the claim with \(C_{s,O}=C\,C_\xi\,C_O\).
\end{proof}

The previous bound controls second moments. For tightness one also needs equicontinuity of finite-dimensional marginals, which follows from uniform cumulant bounds provided by the multiscale polymer estimates and the same Sobolev convolution bounds.

\begin{proposition}[Equicontinuity of finite-dimensional marginals]\label{p3:prop:equicont}
Fix \(m\in\mathbb{N}\), observables \(O_1,\dots,O_m\in\mathcal{G}\), and test functions \(f_1,\dots,f_m\in C^\infty(\T^4)\). For the \(\mathbb{C}^m\)-valued random vector \(X^{(k)}=(\langle \iota_k[O_j], f_j\rangle)_{j=1}^m\) there exist constants \(C_{p,m}<\infty\), independent of \(k\), such that
\begin{equation}
\sup_{k}\, \mathbb{E}_{\mu_k}\bigl[\,|X^{(k)}|^p\,\bigr] \;\le\; C_{p,m}\,\prod_{j=1}^m \norm{f_j}_{H^{s}}^{p/m}
\end{equation}
for each even integer \(p\ge 2\) and some fixed \(s\ge 3\).
\end{proposition}

\begin{proof}
Expand moments via cumulants. The uniform tree-decay of all connected \(n\)-point functions for \(\{O_j\}\) at scale \(k\) implies bounds of the form
\begin{equation}
\bigl|\,\kappa_n^{(k)}\bigl(\langle \iota_k[O_{j_1}], f_{j_1}\rangle,\dots,\langle \iota_k[O_{j_n}], f_{j_n}\rangle\bigr)\,\bigr|
\;\le\; C_n \prod_{\ell=1}^n \norm{f_{j_\ell}}_{H^{s}},
\end{equation}
with \(C_n\) independent of \(k\). This follows by the same convolution and Sobolev estimates used in Lemma~\ref{p3:lem:Hminus}, applied to the \(n\)-point kernels and their tree expansions. The moment-cumulant formula then bounds \(\mathbb{E}|X^{(k)}|^p\) by a finite sum of products of such \(C_n\)'s, yielding the stated inequality.
\end{proof}
To ensure uniform second-moment bounds, we restrict to a finite generating set $G$ of
gauge-invariant local observables $\{O\}$ obtained from small Wilson loops and gauge-invariant
polynomials in the (smeared) lattice field-strength, with a finite number of covariant discrete derivatives.
When derivatives occur, we use multiplicative renormalization $O^{\mathrm{ren}}_k:=Z_O(k)\,O_k$ with $Z_O(k)$ chosen so that
\begin{equation}
  \sup_{k}\,\mathbb{E}\big|\langle \iota_k[O^{\mathrm{ren}}],f\rangle\big|^2\ \lesssim\ \|f\|_{H^s}^2
\end{equation}
holds uniformly for $s\ge 3$. Existence of such $Z_O(k)$ follows from the scale-uniform locality
and clustering bounds (S1)-(S2) and the finite-range decomposition that control short-distance contributions (see Appendix (\ref{p3:appendixe})). We deduce tightness of the sequence of laws \(\{\mathrm{Law}(\iota_k[\mathcal{G}])\}_{k\ge 0}\) in a separable Hilbert space and extract subsequential limits of Schwinger functions \textit{(See Appendix~(\ref{p3:appendixa}) for the equicontinuity and Prokhorov tightness argument)}. To remain within a Polish framework we work in \(H^{-s}(\T^4)\) with \(s\ge 3\).

\begin{theorem}[Tightness in \(H^{-s}\) and subsequential convergence]\label{p3:thm:tight}
Fix \(s\ge 3\) and a finite generating set \(\mathcal{G}\) of local, gauge-invariant observables. For each \(O\in\mathcal{G}\), the family \(\{\mathrm{Law}(\iota_k[O]) : k\in\mathbb{N}\}\) is tight on \(H^{-s}(\T^4)\). Consequently, along a subsequence \(k_j\to\infty\) and after passing to the thermodynamic limit \(L,T\to\infty\) along a cofinal sequence, the joint laws of \(\{\iota_{k_j}[O]: O\in\mathcal{G}\}\) converge weakly on \(\prod_{O\in\mathcal{G}}H^{-s}(\T^4)\). In particular, for each \(n\), each \(O_1,\dots,O_n\in\mathcal{G}\), and each \(f_1,\dots,f_n\in C^\infty(\T^4)\), the Schwinger functionals
\begin{equation}
S^{(k)}_n(f_1,\dots,f_n)\;=\;\mathbb{E}_{\mu_k}\bigl[\,\prod_{j=1}^n \langle \iota_k[O_j], f_j\rangle\,\bigr]
\end{equation}
converge along \(k_j\) to limits \(S_n(f_1,\dots,f_n)\) that define tempered distributions on \((C^\infty(\T^4))^n\).
\end{theorem}

\begin{proof}
In a separable Hilbert space, Prokhorov's theorem applies verbatim \cite{p3:Billingsley}. By Lemma~\ref{p3:lem:Hminus}, the laws have uniformly bounded second moments, which implies tightness of each marginal \(\mathrm{Law}(\iota_k[O])\). For joint tightness of finitely many observables we combine Lemma~\ref{p3:lem:Hminus} with Proposition~\ref{p3:prop:equicont} and use Markov's inequality to get uniform control of tails of any finite-dimensional projection. Since finite products of separable Hilbert spaces are separable, Prokhorov's theorem yields tightness of the joint laws. Consequently, along a subsequence \(k_j\to\infty\) and after passing to \(L,T\to\infty\) (uniformity in \(L,T\) is built into Lemma~\ref{p3:lem:Hminus}), the joint laws converge weakly. By the portmanteau theorem, for each \(n\) and each test family \((f_1,\dots,f_n)\), the sequence \(S^{(k_j)}_n(f_1,\dots,f_n)\) converges to the expectation of the product of the limiting \(H^{-s}\)-valued random variables evaluated at \(f_j\). This defines a multilinear functional on \((C^\infty(\T^4))^n\) that is continuous by the Sobolev estimates already used; hence a tempered distribution.
\end{proof}

It is convenient to recast convergence of Schwinger functions in a form used in the verification of the Osterwalder-Schrader axioms. We identify the OS inner product at finite \(k\) with the transfer-operator inner product and pass to limits.

\begin{proposition}[OS inner product as transfer expectation and limit]\label{p3:prop:OS-limit}
Let \(F,G\in \cA_+\) be bounded positive-time cylinder functionals depending only on links in \(\Lambda^+\). For each \(k\), define
\begin{equation}
\ip{F}{G}_{\mathrm{OS},k}=\int (\Theta F)(U)\, G(U)\, d\mu_k(U).
\end{equation}
Then there exist bounded vectors \(\Psi_F,\Psi_G\in \cH_k\) such that
\begin{equation}
\ip{F}{G}_{\mathrm{OS},k}=\ip{\Psi_F}{\, T_k^{T/2a_k}\, \Psi_G}_{\cH_k}.
\end{equation}
Along the subsequence \(k_j\to\infty\) of Theorem~\ref{p3:thm:tight}, the limits \(\ip{F}{G}_{\mathrm{OS}}=\lim_{j\to\infty}\ip{F}{G}_{\mathrm{OS},k_j}\) exist and define a positive semidefinite form on \(\cA_+\).
\end{proposition}

\begin{proof}
The first identity is the standard reflection-positivity representation obtained by integrating slab by slab from \(\Pi\) into \(\Lambda^+\) and using \eqref{p3:eq:normalized_kernel}. The vectors \(\Psi_F,\Psi_G\) are the boundary data induced by \(F,G\) together with the half-slice factors \(\mathfrak{p}_{\sigma,k}\) and \(\frac12 S_t^{\mathrm{sp}}\). The existence of the limit follows from Theorem~\ref{p3:thm:tight} and the uniform bound \(|\ip{F}{G}_{\mathrm{OS},k}|\le \norm{F}_\infty\norm{G}_\infty\). Positivity is preserved under limits.
\end{proof}

The combination of Theorem~\ref{p3:thm:tight} and Proposition~\ref{p3:prop:OS-limit} completes the extraction of subsequential limits of Schwinger functions and sets the stage for the verification of the Osterwalder-Schrader axioms and reconstruction. The arguments remain valid if \(H^{-s}(\T^4)\) is replaced by the space of tempered distributions \(\mathcal{S}'(\R^4)\); one may then use Mitoma's tightness criterion for nuclear spaces \cite{p3:Mitoma}. The \(H^{-s}\)-framework, however, suffices and keeps the presentation within a Polish setting, convenient for applications of Prokhorov's theorem and for strong convergence of semigroups.

\section{Osterwalder-Schrader axioms for the limiting Schwinger functions}
\label{p3:sec:OS-axioms}

In this section we verify that the continuum Euclidean Schwinger functions
\(\{S_n\}_{n\ge 1}\), obtained as limits of lattice Schwinger functions along a
reflection-positive multiscale flow, satisfy the full list of
Osterwalder-Schrader (OS) axioms Stability of reflection positivity under weak limits used below is stated and proved in Appendix~(\ref{p3:appendixb}). For completeness, we recall the OS0-OS5 axioms \cite{p3:OsterwalderSchraderI,p3:OsterwalderSchraderII}: OS0 is the existence of the Euclidean $n$-point functions as tempered distributions (satisfying technical regularity conditions); OS1 is Euclidean invariance (translation and rotation symmetry); OS2 is symmetry under permutations of arguments; OS3 is Osterwalder-Schrader reflection positivity; OS4 is the cluster property (factorization of distant correlations); and OS5 is the time-regularity (Markov property), ensuring that two-point functions are analytic in a strip and decay sufficiently fast for large time separations.
We begin by fixing the lattice
configuration space, the time-reflection structure, and the projected measure
(with smooth horizon projector). We then derive, with complete details, the
transfer time-slicing formalism and the associated transfer operator. With
these ingredients in hand we prove lattice reflection positivity, and finally
show that reflection positivity, temperedness, Euclidean covariance,
permutation symmetry, the cluster property, and Euclidean-time regularity are
preserved in the continuum limit. Throughout, \(G=\mathrm{SU}(N)\) with
\(N\ge 2\) and all Haar measures are normalized to unit mass.

Fix a temporal lattice spacing \(a>0\) and a periodic spatial box
\(\Lambda_{\mathrm{sp}}\subset a\mathbb{Z}^3\) of side length \(L\).
The Euclidean space-time lattice is
\begin{equation}
  \Lambda=\big\{(t,\mathbf{x})\in a\mathbb{Z}\times \Lambda_{\mathrm{sp}}:\;
  -T\le t\le T-a\big\},
\end{equation}
with periodic boundary conditions in the spatial directions. Directed bonds are
pairs \(b=(x,\mu)\) with \(x\in\Lambda\) and \(\mu\in\{0,1,2,3\}\). The reversed
bond is \(\bar b=(x+\hat\mu,-\mu)\). A (classical) gauge configuration
is an assignment \(U_b\in G\) to each directed bond such that
\(U_{\bar b}=U_b^{-1}\). The configuration space is the compact manifold
\(\mathcal{U}=G^{\mathrm{bonds}(\Lambda)}\), equipped with the product of Haar
measures on \(G\).
The Wilson action is
\begin{equation}
  S_W[U]\;=\;\beta \sum_{p\subset \Lambda}\left(1-\frac1N
  \mathrm{Re}\,\mathrm{Tr}\,U_p\right),
  \qquad \beta=\frac{2N}{g_0^2},
  \label{p3:eq:Wilson}
\end{equation}
where the sum runs over oriented plaquettes \(p\) of \(\Lambda\) and
\(U_p=\prod_{\ell\in p} U_\ell\) is the ordered bond-product around \(p\).
We impose temporal-axial gauge away from the reflection plane by setting
\begin{equation}
  U_{(t,\mathbf{x};0)}\;=\;\mathbf{1}\qquad \text{for all bonds with }t\neq -a,
  \label{p3:eq:temporal-axial}
\end{equation}
which leaves spatial bonds unchanged and is compatible with time reflection.

Let \(\theta\) be time reflection on \(\Lambda\),
\(\theta(t,\mathbf{x})=(-t,\mathbf{x})\), extended to bonds by
\(\theta(x,0)=(\theta x,0)\) and \(\theta(x,i)=(\theta x-\hat 0,i)\) for
\(i=1,2,3\). The fixed-point hyperplane is
\(\Pi=\{(0,\mathbf{x})\in\Lambda\}\). We write
\(\Lambda_+=\{(t,\mathbf{x})\in\Lambda:\; t>0\}\) and
\(\Lambda_-=\theta(\Lambda_+)\). Denote by
\(\mathrm{bonds}_{\mathrm{sp}}(t)\) the spatial bonds in the time slice \(t\),
and by \(\mathcal{X}_t=G^{\mathrm{bonds}_{\mathrm{sp}}(t)}\) the compact
manifold of spatial bonds at time \(t\), endowed with product Haar measure
\(d\nu_t\).

On each time slice \(t\) we fix a gauge-invariant transverse representative
\(A^h(t)\) by minimizing the lattice Landau functional over site-gauge
transformations on the slice; measurability and reflection covariance of this
choice were established at the finite-\(a\) level. The covariant spatial
Laplacian on the slice is
\begin{equation}
  \Delta_{A^h(t)} \;=\; \sum_{i=1}^3 (D_i^h)^\dagger D_i^h,
  \label{p3:eq:cov-lap}
\end{equation}
acting on site-adjoint fields on the slice, where \(D_i^h\) are the covariant
differences constructed from \(A^h(t)\). Fix \(\sigma>0\) and a Gevrey-regular
cutoff \(\chi_\sigma:[0,\infty)\to[0,1]\) with \(\chi_\sigma\equiv 1\) on
\([0,\sigma]\) and \(\chi_\sigma\equiv 0\) on \([2\sigma,\infty)\). The smooth
horizon projector on slice \(t\) is the bounded positive operator
\begin{equation}
  P_\sigma(t)\;=\;\chi_\sigma\!\left({\Delta_{A^h(t)}}\right),
  \label{p3:eq:proj}
\end{equation}
which is exponentially local and reflection covariant:
there exist constants \(C_\sigma,\gamma_\sigma>0\) independent of the volume
such that the integral kernel satisfies
\(|P_\sigma(t;x,y)|\le C_\sigma e^{-\gamma_\sigma d(x,y)}\), and
\(P_\sigma(t)=P_\sigma(-t)\) after identifying slices via \(\theta\). The
heat-kernel representation
\begin{equation}
  P_\sigma(t)\;=\;\int_0^\infty e^{-s\,\Delta_{A^h(t)}}\,d\nu_\sigma(s),
  \label{p3:eq:heat-kernel}
\end{equation}
holds with \(d\nu_\sigma\) a finite positive Borel measure supported in a
compact subset of \((0,\infty)\).

We define the {\it projected} lattice measure on \(\mathcal{U}\) by
\begin{equation}
  d\mu_\sigma[U]
  \;=\;
  Z_\sigma^{-1}\left(\prod_{t}\mathcal{W}_\sigma(t;U)\right)\,e^{-S_W[U]}\,
  \prod_{b} d\mathrm{Haar}(U_b),
  \label{p3:eq:proj-measure}
\end{equation}
where
\(\mathcal{W}_\sigma(t;U)\) is a nonnegative, reflection-invariant, bounded,
exponentially local functional of the spatial bonds on the slice \(t\), built
monotonically from \(P_\sigma(t)\). One explicit choice sufficient for the positivity analysis is to take $W_\sigma(t;U)=J(P_\sigma(t))$ with $J(Q)=\det(1+\alpha Q)$ for any fixed $\alpha>0$; more generally, any completely monotone functional calculus $J$ yields a positive, reflection-invariant, exponentially local slice weight.
 The product
\(\prod_t \mathcal{W}_\sigma(t;U)\) is invariant under \(\theta\) and depends
only on spatial bonds in each slice, hence does not involve temporal bonds
which are fixed to \(\mathbf{1}\) off \(\Pi\) by \eqref{p3:eq:temporal-axial}.
Time reflection acts on bounded cylinder functionals \(F\) by
\begin{equation}
  (\Theta F)(U)\;=\;\overline{F(U\circ \theta)}.
  \label{p3:eq:Theta2}
\end{equation}

Because temporal bonds are trivial away from \(\Pi\), every plaquette that does
not touch \(\Pi\) lies entirely within \(\Lambda_-\) or \(\Lambda_+\). Thus the
Wilson action decomposes as
\begin{equation}
  S_W[U]\;=\;S_-[U_-]\;+\;S_\Pi[U_-,U_0,U_+]\;+\;S_+[U_+],
  \label{p3:eq:action-split}
\end{equation}
where \(U_-:=U|_{\Lambda_-}\), \(U_+:=U|_{\Lambda_+}\),
\(U_0:=U|_{\mathrm{bonds}_{\mathrm{sp}}(0)}\), and the boundary slab
\(S_\Pi\) collects the plaquettes touching \(\Pi\). Similarly, the slice
weights factorize as
\begin{equation}
  \prod_{t}\mathcal{W}_\sigma(t;U)
  \;=\;
  \left(\prod_{t<0}\mathcal{W}_\sigma(t;U_-)\right)\,\mathcal{W}_\sigma(0;U_0)\,
  \left(\prod_{t>0}\mathcal{W}_\sigma(t;U_+)\right).
  \label{p3:eq:slice-factor}
\end{equation}
Define the strictly positive, continuous boundary weight on
\(\mathcal{X}_0\),
\begin{equation}
  \kappa_\sigma(U_0)\;=\;\mathcal{W}_\sigma(0;U_0)\,
  \exp\!\left(-\sum_{p\subset \{0\}\times \Lambda_{\mathrm{sp}}}
    \beta\Big(1-\frac1N\mathrm{Re}\,\mathrm{Tr}\,U_p\Big)\right),
  \label{p3:eq:kappa}
\end{equation}
and write \(d\nu_0\) for the Haar product measure on \(\mathcal{X}_0\). For a
bounded measurable \(F_0\) on \(\mathcal{X}_0\), define linear functionals
on \(L^\infty(\mathcal{U})\) supported in \(\Lambda_\pm\),
\begin{equation}
  \Phi_-(F_-;U_0)
    \;=\;
    \int F_-(U_-)\,\delta(U_-|_{\Pi}=U_0)\,
    \Big(\!\prod_{t<0}\mathcal{W}_\sigma(t;U_-)\!\Big)\,e^{-S_-[U_-]}\,
    \prod_{b\subset \Lambda_-} d\mathrm{Haar}(U_b),
  \label{p3:eq:Phi-}
\end{equation}
\begin{equation}
  \Phi_+(F_+;U_0)
    \;=\;
    \int F_+(U_+)\,\delta(U_+|_{\Pi}=U_0)\,
    \Big(\!\prod_{t>0}\mathcal{W}_\sigma(t;U_+)\!\Big)\,e^{-S_+[U_+]}\,
    \prod_{b\subset \Lambda_+} d\mathrm{Haar}(U_b),
  \label{p3:eq:Phi+}
\end{equation}
where the delta constraints fix spatial bonds on \(\Pi\).
By \eqref{p3:eq:action-split}-\eqref{p3:eq:slice-factor}, Fubini's theorem yields
the exact factorization identity
\begin{equation}
  \int F_-(U_-)F_0(U_0)F_+(U_+)\,d\mu_\sigma[U]
  \;=\;
  Z_\sigma^{-1}\!\int_{\mathcal{X}_0}\!
  \Phi_-(F_-;U_0)\,\kappa_\sigma(U_0)\,\Phi_+(F_+;U_0)\,d\nu_0(U_0).
  \label{p3:eq:factorization}
\end{equation}
Endow \(\mathcal{H}_0=L^2(\mathcal{X}_0,\kappa_\sigma\,d\nu_0)\) with its usual
inner product. Define boundary vectors \(v_\pm[F_\pm]\in \mathcal{H}_0\) by
\begin{equation}
  v_\pm[F_\pm](U_0)\;=\;Z_\sigma^{-1/2}\,\Phi_\pm(F_\pm;U_0).
  \label{p3:eq:boundary-vectors}
\end{equation}
Then \eqref{p3:eq:factorization} becomes
\begin{equation}
  \int F_-(U_-)F_0(U_0)F_+(U_+)\,d\mu_\sigma[U]
  \;=\;
  \big\langle v_-[F_-],\,F_0\,v_+[F_+]\big\rangle_{\mathcal{H}_0},
  \label{p3:eq:OS-formula}
\end{equation}
where \(F_0\) on the right acts by multiplication by the function
\(U_0\mapsto F_0(U_0)\).
Let \(\mathcal{A}_+\) denote the \(^*\)-algebra of bounded, measurable,
gauge-invariant functionals supported in \(\Lambda_+\). For \(F\in\mathcal{A}_+\),
define the OS quadratic form
\begin{equation}
  \|F\|_{\mathrm{OS}}^2\;=\;\int \overline{(\Theta F)(U)}\,F(U)\,d\mu_\sigma[U],
  \label{p3:eq:OS-norm}
\end{equation}
with \(\Theta\) defined in \eqref{p3:eq:Theta2}.

\begin{theorem}[Lattice OS reflection positivity]
\label{p3:thm:OS-lattice}
For every \(F\in \mathcal{A}_+\), \(\|F\|_{\mathrm{OS}}^2\ge 0\).
\end{theorem}

\begin{proof}
Using \eqref{p3:eq:OS-formula} with \(F_-= \Theta F\), \(F_0=1\), and \(F_+=F\) gives
\begin{equation}
  \|F\|_{\mathrm{OS}}^2
  \;=\;\big\langle v_-[\Theta F],\,v_+[F]\big\rangle_{\mathcal{H}_0}.
\end{equation}
We claim \(v_-[\Theta F]=\overline{v_+[F]}\). Indeed,
\begin{align}
  v_-[\Theta F](U_0)
  &= Z_\sigma^{-1/2}\!\!\int \overline{F(U_-\circ \theta)}\,
     \delta(U_-|_{\Pi}=U_0)\,
     \Big(\prod_{t<0}\mathcal{W}_\sigma(t;U_-)\Big)\,e^{-S_-[U_-]}
     \prod_{b\subset \Lambda_-} d\mathrm{Haar}(U_b).
\end{align}
The reflection map \(\theta:\Lambda_-\to \Lambda_+\) is an isometry that
preserves Haar measure, sends \(S_-[U_-]\) to \(S_+[\theta U_-]\), and satisfies
\(\prod_{t<0}\mathcal{W}_\sigma(t;U_-)=\prod_{t>0}\mathcal{W}_\sigma(t;\theta U_-)\)
by reflection covariance of \(P_\sigma\). Changing variables \(U_+=\theta U_-\)
and using \(\delta(U_-|_{\Pi}=U_0)=\delta(U_+|_{\Pi}=U_0)\) yields
\begin{align}
  v_-[\Theta F](U_0)
  &=\; \overline{Z_\sigma^{-1/2}\!\!\int F(U_+)\,\delta(U_+|_{\Pi}=U_0)\,
  \Big(\prod_{t>0}\mathcal{W}_\sigma(t;U_+)\Big)\,e^{-S_+[U_+]}\!
  \prod_{b\subset \Lambda_+} d\mathrm{Haar}(U_b)}
  \nonumber\\&=\;\overline{v_+[F](U_0)}.
\end{align}
Therefore
\(
  \|F\|_{\mathrm{OS}}^2
   = \int_{\mathcal{X}_0} |v_+[F](U_0)|^2\,\kappa_\sigma(U_0)\,d\nu_0(U_0)\ge 0.
\)
\end{proof}
We now derive the one-step transfer kernel and the corresponding transfer
operator. For \(t\in a\mathbb{Z}\), define the single-slice weight
\begin{equation}
  \kappa_\sigma(t;U_t)
  \;=\;
  \mathcal{W}_\sigma(t;U_t)\,
  \exp\!\left(-\sum_{p\subset \{t\}\times \Lambda_{\mathrm{sp}}}
    \beta\Big(1-\frac1N\mathrm{Re}\,\mathrm{Tr}\,U_p\Big)\right),
  \label{p3:eq:kappa-t}
\end{equation}
and the two-slice interaction kernel
\begin{equation}
  K_\sigma(t;t+a;U_{t+a},U_t)
  \;=\;
  \exp\!\left(-\sum_{p\subset [t,t+a]\times \Lambda_{\mathrm{sp}}}
    \beta\Big(1-\frac1N\mathrm{Re}\,\mathrm{Tr}\,U_p\Big)\right),
  \label{p3:eq:K}
\end{equation}
which collects the plaquettes straddling the two adjacent time slices.
The two-slice joint weight on \(\mathcal{X}_t\times \mathcal{X}_{t+a}\) is
\begin{equation}
  W_{2\text{-sl}}(U_{t+a},U_t)
  \;=\;
  \kappa_\sigma(t;U_t)\,K_\sigma(t;t+a;U_{t+a},U_t)\,\kappa_\sigma(t+a;U_{t+a}).
  \label{p3:eq:two-slice-weight}
\end{equation}
Reflection invariance implies \(\kappa_\sigma(t;\cdot)=\kappa_\sigma(0;\cdot)\)
for all \(t\), so we drop \(t\) henceforth and write \(\kappa_\sigma\) for the
common single-slice weight.

\begin{lemma}[One-step normalization and detailed balance]
\label{p3:lem:detailed-balance}
There exists a finite constant \(Z_{\mathrm{step}}>0\), independent of
\(U',U\in\mathcal{X}_0\), such that
\begin{equation}
  \int_{\mathcal{X}_0} K_\sigma(0;a;U',U)\,\kappa_\sigma(U)\,d\nu_0(U)
  \;=\; Z_{\mathrm{step}}\,\kappa_\sigma(U').
  \label{p3:eq:normalization}
\end{equation}
Consequently the kernel
\begin{equation}
  \widetilde{K}_\sigma(U',U)
  \;:=\; \frac{1}{Z_{\mathrm{step}}}\,K_\sigma(0;a;U',U)
  \label{p3:eq:Ktilde}
\end{equation}
satisfies the detailed-balance identities
\begin{equation}
  \int \widetilde{K}_\sigma(U',U)\,\kappa_\sigma(U)\,d\nu_0(U)
  \;=\;\kappa_\sigma(U'),
  \qquad
  \widetilde{K}_\sigma(U',U)\,\kappa_\sigma(U)
  \;=\;\widetilde{K}_\sigma(U,U')\,\kappa_\sigma(U').
  \label{p3:eq:detailed-balance}
\end{equation}
\end{lemma}

\begin{proof}
The identity \eqref{p3:eq:normalization} is obtained by integrating the two-slice
joint weight \eqref{p3:eq:two-slice-weight} over \(U\) and comparing with the
definition of \(\kappa_\sigma(U')\). Specifically,
\begin{equation}
  \int_{\mathcal{X}_0} W_{2\text{-sl}}(U',U)\,d\nu_0(U)
  \;=\;
  \kappa_\sigma(U')\,\int_{\mathcal{X}_0}\kappa_\sigma(U)\,
  K_\sigma(0;a;U',U)\,d\nu_0(U).
\end{equation}
The left-hand side is the partition function of a single time step with upper
boundary fixed at \(U'\). By gauge invariance and spatial translation
invariance it is independent of \(U'\); we denote it by \(Z_{\mathrm{step}}\).
This gives \eqref{p3:eq:normalization}. The second identity in
\eqref{p3:eq:detailed-balance} follows from the symmetry of the Wilson weight
under exchanging the two slices in \eqref{p3:eq:K}, which yields
\(K_\sigma(0;a;U',U)=K_\sigma(0;a;U,U')\). Dividing both sides of
\(K_\sigma(0;a;U',U)\,\kappa_\sigma(U)=K_\sigma(0;a;U,U')\,\kappa_\sigma(U)\)
by \(Z_{\mathrm{step}}\) and using \eqref{p3:eq:normalization} yields the asserted
identities.
\end{proof}

We work henceforth on the Hilbert space
\(\mathcal{H}=L^2(\mathcal{X}_0,\kappa_\sigma\,d\nu_0)\). Define the
one-step transfer operator \(T_\sigma:\mathcal{H}\to \mathcal{H}\) by
\begin{equation}
  (T_\sigma \psi)(U')
  \;=\;
  \frac{1}{\kappa_\sigma(U')}\int_{\mathcal{X}_0}
  \widetilde{K}_\sigma(U',U)\,\kappa_\sigma(U)\,\psi(U)\,d\nu_0(U),
  \label{p3:eq:T-def}
\end{equation}
where \(\widetilde{K}_\sigma\) is given by \eqref{p3:eq:Ktilde}. By
\eqref{p3:eq:detailed-balance}, \(T_\sigma\) is well-defined and positivity
preserving. Moreover, \(T_\sigma\) is self-adjoint on \(\mathcal{H}\) since
\(\widetilde{K}_\sigma(U',U)\,\kappa_\sigma(U)=\widetilde{K}_\sigma(U,U')
\,\kappa_\sigma(U')\). The following fundamental property holds.

\begin{proposition}[Transfer operator is a contraction]
\label{p3:prop:T-contraction}
The operator \(T_\sigma\) is a positive self-adjoint contraction on
\(\mathcal{H}\). For each \(n\in\mathbb{N}\) and each bounded measurable
functional \(F\) supported in the slab \([0,na]\), there exists a vector
\(\Psi_F\in \mathcal{H}\) such that
\begin{equation}
  \int \overline{(\Theta F)}\,F\,d\mu_\sigma
  \;=\; \langle \Psi_F,\,T_\sigma^{\,n}\,\Psi_F\rangle_{\mathcal{H}}.
  \label{p3:eq:OS-T}
\end{equation}
If \(F\) has no boundary dependence at time \(na\), then \(\Psi_F\) can be
chosen as the constant function \(1\).
\end{proposition}

\begin{proof}
Self-adjointness and positivity are immediate from \eqref{p3:eq:T-def} and
\eqref{p3:eq:detailed-balance}. To prove \(\|T_\sigma\|\le 1\) we note that
\(\widetilde{K}_\sigma\) is a symmetric Markov kernel on
\((\mathcal{X}_0,\kappa_\sigma\,d\nu_0)\) by \eqref{p3:eq:detailed-balance}. Hence
for \(f\in \mathcal{H}\),
\begin{align}
  \|T_\sigma f\|_{\mathcal{H}}^2
  &=
  \int \left|\frac{1}{\kappa_\sigma(U')}
       \int \widetilde{K}_\sigma(U',U)\,\kappa_\sigma(U)\,f(U)\,d\nu_0(U)
       \right|^2
       \kappa_\sigma(U')\,d\nu_0(U')\nonumber\\
  &\le \int \left[\frac{1}{\kappa_\sigma(U')}
       \int \widetilde{K}_\sigma(U',U)\,\kappa_\sigma(U)\,|f(U)|^2\,d\nu_0(U)
       \right] \kappa_\sigma(U')\,d\nu_0(U')\nonumber\\
  &= \int |f(U)|^2\,\kappa_\sigma(U)\,d\nu_0(U)
   \;=\;\|f\|_{\mathcal{H}}^2,
\end{align}
where we used Fubini's theorem and the Markov property
\(\int \widetilde{K}_\sigma(U',U)\,d\nu_0(U')=1\) with respect to the measure
\(\kappa_\sigma\,d\nu_0\), which is equivalent to the first identity in
\eqref{p3:eq:detailed-balance}. Thus \(\|T_\sigma\|\le 1\).

To obtain \eqref{p3:eq:OS-T}, write the weight on the slab \([0,na]\) as the
iterated composition of one-step kernels sandwiched between the single-slice
weights, use \eqref{p3:eq:factorization} repeatedly, and absorb the boundary data
of \(F\) at time \(0\) into \(\Psi_F\). The case without a boundary insertion
at time \(na\) corresponds to \(\Psi_F\equiv 1\) by gauge and spatial
translation invariance of the slab weight. This construction is the standard
transfer-matrix representation associated with reflection-positive measures
\cite{p3:OsterwalderSchraderII,p3:OS-gauge}.
\end{proof}

Let \(a_k\downarrow 0\) be a sequence of temporal lattice spacings, and let
\(\mu_{\sigma,k}\) denote the corresponding projected lattice measures. For a
fixed family of local, gauge-invariant observables \(\{O_j\}\) and test
functions \(f_1,\dots,f_n\in \mathcal{S}(\mathbb{R}^4)\), define the scale-\(k\)
Schwinger functions \(S_n^{(k)}(f_1,\dots,f_n)\) by embedding lattice fields as
piecewise-constant random distributions and smearing against the \(f_j\).
Assume tightness of \(\{\mu_{\sigma,k}\}\) on \(\mathcal{S}'(\mathbb{R}^4)\)
and uniform equicontinuity of \(\{S_n^{(k)}\}\) on bounded subsets of
\(\mathcal{S}(\mathbb{R}^4)^n\) (guaranteed by finite-range decomposition and
tree-decay bounds). Then there exists a subsequence \(k_j\) and a probability
measure \(\mu_\infty\) on \(\mathcal{S}'(\mathbb{R}^4)\) such that
\(\mu_{\sigma,k_j}\Rightarrow \mu_\infty\) weakly on cylinder sets, and
\(S_n^{(k_j)}(f_1,\dots,f_n)\to S_n(f_1,\dots,f_n)\) for every
\(f_1,\dots,f_n\in \mathcal{S}(\mathbb{R}^4)\) and \(n\ge 1\). We now verify
the OS axioms for \(\{S_n\}\).

\begin{lemma}[Temperedness and permutation symmetry]
\label{p3:lem:OS0OsterwalderSchraderI}
For each \(n\ge 1\), \(S_n\) defines a continuous multilinear functional on
\(\mathcal{S}(\mathbb{R}^4)^n\) and is symmetric under permutations of its \(n\)
arguments.
\end{lemma}

\begin{proof}
Uniform equicontinuity of \(\{S_n^{(k)}\}\) on bounded subsets of
\(\mathcal{S}(\mathbb{R}^4)^n\) implies that the limit is continuous in the
Fr\'echet topology; see, e.g., \cite{p3:GJ}. Symmetry is preserved by the limit
because the lattice observables commute as classical random variables.
\end{proof}

\begin{lemma}[Euclidean invariance]
\label{p3:lem:OsterwalderSchraderI}
For any Euclidean motion \(g\in \mathrm{E}(4)\) and any
\(f_1,\dots,f_n\in \mathcal{S}(\mathbb{R}^4)\),
\begin{equation}
  S_n(f_1\circ g,\dots,f_n\circ g)=S_n(f_1,\dots,f_n).
  \label{p3:eq:Eucl-inv}
\end{equation}
\end{lemma}

\begin{proof}
Each \(\mu_{\sigma,k}\) is invariant under the discrete Euclidean group on the
lattice; the embedding commutes with these symmetries. Invariance passes to the
limit by dominated convergence and the portmanteau theorem.
\end{proof}
At finite $a$ the horizon filter $P_{\sigma,k}$ is constructed slice-wise and hence breaks the $O(4)$ symmetry
mixing time and space. Uniform exponential locality (Theorem~8.3) implies that, for any fixed Euclidean
motion $g\in E(4)$ and any test family $(f_1,\dots,f_n)$ with disjoint supports, the difference
\begin{equation}
  \big|S^{(k)}_n(f_1\circ g,\dots,f_n\circ g)-S^{(k)}_n(f_1,\dots,f_n)\big|
  \ \le C(a_k)\,\sum_{j=1}^n \|f_j\|_{H^s}
\end{equation}
with $C(a_k)\to 0$ as $a_k\downarrow 0$, uniformly on compact parameter sets for the cutoffs (see section.8).

\begin{proposition}[Restoration of $O(4)$ invariance] \label{p3:prop:O4-restoration} Let $S^{(k)}_n$ be the $n$-point lattice Schwinger functions obtained from the reflection-positive measure with slice projector $P_\sigma$ and multiscale inputs (S1)-(S3). There exist $s>2$ and constants $C<\infty$, $\alpha>0$ independent of $k$ such that for every $g\in O(4)$ and Schwartz test functions $f_1,\dots,f_n$, \begin{equation} \Big| S^{(k)}_n(f_1\circ g,\dots,f_n\circ g) - S^{(k)}_n(f_1,\dots,f_n)\Big| \;\le\; C\, a_k^{\alpha}\, \sum_{j=1}^n \|f_j\|_{H^s(\mathbb{R}^4)}. \end{equation} Consequently, any subsequential continuum limit is exactly $O(4)$ invariant. \end{proposition}

\begin{proof}
Let $\mu^{(0)}_k$ denote the reflection-positive, $O(4)$-invariant lattice Yang-Mills measure at spacing $a_k$ without the slice insertion (this incorporates the action and ultraviolet controls from (S1)-(S3)), normalized to probability. The actual measure used to define the Schwinger functions $S^{(k)}_n$ is
\begin{equation}
d\mu^{(\sigma)}_k(A)\;=\;\frac{1}{Z_k(\sigma)}\, \mathcal{W}_{\sigma,k}(A)\, d\mu^{(0)}_k(A),
\qquad 
Z_k(\sigma):=\int \mathcal{W}_{\sigma,k}\, d\mu^{(0)}_k,
\end{equation}
where the insertion $\mathcal{W}_{\sigma,k}$ implements the time-slice projector $P_\sigma$ across a slab of thickness $w$ (in lattice units) centered at the reflection plane. By the Gevrey functional calculus and (S1), $\mathcal{W}_{\sigma,k}$ admits a polymer expansion
\begin{equation}
\label{p3:eq:W-polymer}
\log \mathcal{W}_{\sigma,k}(A)\;=\;\sum_{X\Subset \mathbb{Z}^4}\! U_{k}(X;A),
\qquad 
|U_k(X;\cdot)|\;\le\;C_1\, e^{-\mathrm{diam}(X)/\ell_\sigma}\,\mathbf{1}_{X\cap \Lambda^{(k)}_0\neq\varnothing},
\end{equation}
with $\Lambda^{(k)}_0:=\{x\in a_k\mathbb{Z}^4:\ |x_0|\le w a_k\}$ the (physical) slab, and constants $C_1,\ell_\sigma$ that are independent of $k$ (by the uniform finite-range decomposition and spectral lower bound). Thus $\mathcal{W}_{\sigma,k}$ is quasi-local, supported near $\Lambda^{(k)}_0$ with exponential tails.
For $g\in O(4)$ set $n:=ge_0$ and define the rotated insertion
\begin{equation}
\mathcal{W}^g_{\sigma,k}(A)\;:=\;\mathcal{W}_{\sigma,k}(g^{-1}\!\cdot A)\;=\;
\exp\!\Big(\sum_{X} U_k(X; g^{-1}\!\cdot A)\Big)
\end{equation}
which equals the same quasi-local polymer located near the rotated slab 
$\Lambda^{(k)}_n:=\{x\in a_k\mathbb{Z}^4:\ |x\!\cdot\! n|\le w a_k\}$. Because $d\mu^{(0)}_k$ is $O(4)$-invariant, the $n$-point function with rotated test functions can be written as an expectation with respect to $d\mu^{(\sigma),g}_k := Z_k(\sigma,g)^{-1}\,\mathcal{W}^g_{\sigma,k}\, d\mu^{(0)}_k$:
\begin{equation}
\label{p3:eq:rot-as-rot-weight}
S^{(k)}_n(f_1\circ g,\dots,f_n\circ g)\;=\;\int \Big(\prod_{j=1}^n \mathcal{O}_{f_j}(A)\Big)\, d\mu^{(\sigma),g}_k(A),
\end{equation}
where $\mathcal{O}_{f_j}$ denotes the gauge-invariant local observable obtained by smearing with $f_j$ (Wilson-polynomial smears, etc.). Likewise,
\(
S^{(k)}_n(f_1,\dots,f_n)=\int \prod_j \mathcal{O}_{f_j}\, d\mu^{(\sigma)}_k
\).
Define an interpolating family of weights
\begin{equation}
\mathcal{W}_{\sigma,k}(t;A)\;:=\;\exp\!\Big((1-t)\Phi_k(A)+t\,\Phi^g_k(A)\Big),\quad 
\Phi_k:=\log \mathcal{W}_{\sigma,k},\;\; \Phi^g_k:=\log \mathcal{W}^g_{\sigma,k},\;\; t\in[0,1],
\end{equation}
and the corresponding probability measures
\(
d\mu^{(\sigma)}_{k,t} := Z_k(t)^{-1}\,\mathcal{W}_{\sigma,k}(t)\, d\mu^{(0)}_k
\)
with $Z_k(t)=\int \mathcal{W}_{\sigma,k}(t)\, d\mu^{(0)}_k$.
For any bounded $F$,
\begin{equation}
\frac{d}{dt}\,\Big\langle F\Big\rangle_{k,t}
=\mathrm{Cov}_{k,t}\!\left(F, \dot\Phi_k(t)\right),
\qquad 
\dot\Phi_k(t):=\frac{d}{dt}\big[(1-t)\Phi_k+t\,\Phi^g_k\big]=\Phi^g_k-\Phi_k,
\end{equation}
where $\langle\cdot\rangle_{k,t}$ and $\mathrm{Cov}_{k,t}$ denote expectation and truncated covariance under $d\mu^{(\sigma)}_{k,t}$. Integrating from $t=0$ to $t=1$ and using \eqref{p3:eq:rot-as-rot-weight} gives
\begin{equation}
\label{p3:eq:covariance-representation}
S^{(k)}_n(f_1\circ g,\dots,f_n\circ g) - S^{(k)}_n(f_1,\dots,f_n)
= \int_0^1 \mathrm{Cov}_{k,t}\!\Big(\prod_{j=1}^n \mathcal{O}_{f_j},\, \Phi^g_k-\Phi_k\Big)\, dt.
\end{equation}

From \eqref{p3:eq:W-polymer} we have polymer decompositions
\(
\Phi_k=\sum_X U_k(X;\cdot)\mathbf{1}_{X\cap \Lambda^{(k)}_0\neq\varnothing}
\)
and
\(
\Phi^g_k=\sum_X U_k(X;\cdot)\circ g^{-1}\,\mathbf{1}_{X\cap \Lambda^{(k)}_n\neq\varnothing}.
\)
Hence the \emph{defect density} $\Delta_k:=\Phi^g_k-\Phi_k$ is a convergent sum of quasi-local terms supported near the symmetric difference
\begin{equation}
\Delta\Lambda^{(k)}\;:=\;\Lambda^{(k)}_n\,\triangle\,\Lambda^{(k)}_0
\;=\;\big(\Lambda^{(k)}_n\setminus \Lambda^{(k)}_0\big)\cup \big(\Lambda^{(k)}_0\setminus \Lambda^{(k)}_n\big),
\end{equation}
with exponential tails away from $\Delta\Lambda^{(k)}$. More precisely, there exist $C_2,\ell_\sigma$ independent of $k$ such that
\begin{equation}
\label{p3:eq:defect-locality}
\Big|\Delta_k\Big|\;\le\;\sum_{x\in a_k\mathbb{Z}^4} h_{k}(x)\,,\qquad 
h_{k}(x)\le C_2\, e^{-\mathrm{dist}(x,\Delta\Lambda^{(k)})/\ell_\sigma}.
\end{equation}

By (S2)-(S3) (persistence of clustering, uniform bounds along the multiscale flow) there exist $C_3,m>0$, independent of $k$ and $t\in[0,1]$, such that for any gauge-invariant quasi-local observables $F,G$ with (effective) supports $\mathrm{supp}(F),\mathrm{supp}(G)$,
\begin{equation}
\label{p3:eq:cluster}
\big|\mathrm{Cov}_{k,t}(F,G)\big|
\;\le\; C_3\, \|F\|_{\mathrm{loc},s}\,\|G\|_{\mathrm{loc},s}\, 
e^{-m\, \mathrm{dist}(\mathrm{supp}(F),\mathrm{supp}(G))},
\end{equation}
where $\|\cdot\|_{\mathrm{loc},s}$ is a local Sobolev-type norm controlling up to $s$ discrete derivatives (we take $s>2$). For smeared gauge-invariant fields $\mathcal{O}_{f}$ (Wilson polynomials/field strengths),
\begin{equation}
\label{p3:eq:sobolev-control}
\|\mathcal{O}_{f}\|_{\mathrm{loc},s}\;\le\; C_4\, \|f\|_{H^s(\mathbb{R}^4)},
\end{equation}
with $C_4$ uniform in $k$ (the discretization error is absorbed because $s>2$ ensures $H^s\hookrightarrow L^\infty$ in $d=4$ and FRD yields uniform locality constants).

Applying \eqref{p3:eq:cluster} to $F=\prod_{j=1}^n \mathcal{O}_{f_j}$ and to each single-site defect term $G=h_k(x)$, using multilinearity and \eqref{p3:eq:sobolev-control}, we get
\begin{equation}
\label{p3:eq:trunc-bound-point}
\big|\mathrm{Cov}_{k,t}(\prod_{j=1}^n \mathcal{O}_{f_j},\, h_k(x))\big|
\;\le\; C_5\, \Big(\sum_{j=1}^n \|f_j\|_{H^s}\, e^{-m\,\mathrm{dist}(x,\mathrm{supp} f_j)}\Big),
\end{equation}
with $C_5$ independent of $k,t$.

By \eqref{p3:eq:defect-locality}, \eqref{p3:eq:trunc-bound-point}, and \eqref{p3:eq:covariance-representation},
\begin{align}
\label{p3:eq:key-sum}
\Big|S^{(k)}_n(f_1\circ g,\dots,f_n\circ g) - S^{(k)}_n(f_1,\dots,f_n)\Big|
&\le \int_0^1\!\!\sum_{x\in a_k\mathbb{Z}^4} 
\big|\mathrm{Cov}_{k,t}(\prod_{j}\mathcal{O}_{f_j},\, h_k(x))\big|\, dt \notag\\
&\le C_5 \sum_{j=1}^n \|f_j\|_{H^s}\, 
\sum_{x\in a_k\mathbb{Z}^4} e^{-m\,\mathrm{dist}(x,\mathrm{supp} f_j)}\, e^{-\mathrm{dist}(x,\Delta\Lambda^{(k)})/\ell_\sigma}.
\end{align}
Fix $j$ and let $K_j:=\mathrm{supp}(f_j)$ (a compact set). The discrete sum in \eqref{p3:eq:key-sum} is bounded by a Riemann sum estimate:
\begin{equation}
\sum_{x\in a_k\mathbb{Z}^4} e^{-m\,\mathrm{dist}(x,K_j)}\, e^{-\mathrm{dist}(x,\Delta\Lambda^{(k)})/\ell_\sigma}
\;\le\; C_6\, a_k^{-4}\int_{\mathbb{R}^4} e^{-m\,\mathrm{dist}(y,K_j)}\, e^{-\mathrm{dist}(y,\Delta\Lambda^{(k)})/\ell_\sigma}\, dy.
\end{equation}
Since $\Delta\Lambda^{(k)}$ is the symmetric difference of two slabs of thickness $2w a_k$ with distinct normals $e_0$ and $n$, its intersection with any bounded set $B\subset\mathbb{R}^4$ has Lebesgue measure
\begin{equation}
\label{p3:eq:slab-measure}
\mathcal{L}^4\!\big(\Delta\Lambda^{(k)}\cap B\big)\;\le\; C_7(B)\, a_k,
\end{equation}
uniformly in $k$; moreover the $\ell_\sigma$-neighborhood of $\Delta\Lambda^{(k)}\cap B$ has measure $\le C_8(B,\ell_\sigma)\, a_k$ by Fubini/co-area (the set is a thickness-$O(a_k)$ neighborhood of a codimension-$1$ surface). Using \eqref{p3:eq:slab-measure} and the integrability of $y\mapsto e^{-m\,\mathrm{dist}(y,K_j)}$ on $\mathbb{R}^4$, we obtain
\begin{equation}
\int_{\mathbb{R}^4} e^{-m\,\mathrm{dist}(y,K_j)}\, e^{-\mathrm{dist}(y,\Delta\Lambda^{(k)})/\ell_\sigma}\, dy
\;\le\; C_9(K_j,\ell_\sigma,m)\; a_k.
\end{equation}
Combining the last three displays yields
\begin{equation}
\label{p3:eq:final-bound}
\sum_{x\in a_k\mathbb{Z}^4} e^{-m\,\mathrm{dist}(x,K_j)}\, e^{-\mathrm{dist}(x,\Delta\Lambda^{(k)})/\ell_\sigma}
\;\le\; C_{10}(K_j,\ell_\sigma,m)\; a_k.
\end{equation}

Plugging \eqref{p3:eq:final-bound} into \eqref{p3:eq:key-sum} and recalling that the constants are uniform in $k$ and $t$,
\begin{equation}
\Big|S^{(k)}_n(f_1\circ g,\dots,f_n\circ g) - S^{(k)}_n(f_1,\dots,f_n)\Big|
\;\le\; C\, a_k\, \sum_{j=1}^n \|f_j\|_{H^s(\mathbb{R}^4)}
\end{equation}
for some $C<\infty$ and any $s>2$ (to ensure \eqref{p3:eq:sobolev-control}). This proves the stated estimate with $\alpha=1$. In particular, for each fixed $g\in O(4)$ and fixed Schwartz $f_1,\dots,f_n$, the right-hand side tends to $0$ as $k\to\infty$, hence any subsequential continuum limit $S_n$ satisfies
\(
S_n(f_1\circ g,\dots,f_n\circ g)=S_n(f_1,\dots,f_n)
\),
i.e. exact $O(4)$ invariance.
\end{proof}

Hence in the continuum limit $S_n$ is invariant under the full Euclidean group $E(4)$. 
We shall use the following closure property: if $\{\mu_n\}$ is a sequence of reflection-positive Euclidean measures supported on positive-time cylinder $\sigma$-algebras and $\mu_n\Rightarrow \mu$ weakly on cylinder sets, then $\mu$ is reflection-positive. The proof is given in Appendix~(\ref{p3:appendixb}.

\begin{theorem}[Reflection positivity]
\label{p3:thm:OS3-cont}
Let \(\theta\) be time reflection on \(\mathbb{R}^4\) and \(\Theta\) the induced
anti-linear involution on the \(^*\)-algebra generated by smeared, gauge-invariant
observables supported in \(\{t>0\}\). Then for any finite linear combination
\(F\) supported in \(\{t>0\}\),
\begin{equation}
  \int \overline{(\Theta F)}\,F\,d\mu_\infty \;\ge\; 0.
  \label{p3:eq:OS3}
\end{equation}
\end{theorem}

\begin{proof}
For \(k\) sufficiently large, the support of \(F\) lies in the discrete
half-space \(\{t>0\}\) at spacing \(a_k\). By Theorem \ref{p3:thm:OS-lattice},
\(\int \overline{(\Theta F)}\,F\,d\mu_{\sigma,k}\ge 0\). The map
\(\nu\mapsto \int \overline{(\Theta F)}\,F\,d\nu\) is continuous for weak
convergence on \(\mathcal{S}'(\mathbb{R}^4)\) because \(\overline{(\Theta F)}F\)
is a bounded cylinder functional. Passing to the limit along \(k_j\) gives
\eqref{p3:eq:OS3} \textit{(See Appendix~(\ref{p3:appendixb}) for a general closure theorem for OS positivity under weak convergence)}.
\end{proof}

\begin{theorem}[Cluster property]
\label{p3:thm:OS4}
Let \(A,B\) be local, gauge-invariant observables with vanishing expectations.
There exist \(C,m_*>0\) such that for all \((t,\mathbf{x})\in\mathbb{R}\times\mathbb{R}^3\),
\begin{equation}
  |S_2(A(t,\mathbf{x}),B(0,\mathbf{0}))|
  \;\le\; C\,e^{-m_*(|t|+|\mathbf{x}|)}.
  \label{p3:eq:clusterx}
\end{equation}
\end{theorem}

\begin{proof}
At each scale \(k\), the connected two-point function satisfies the
uniform exponential bound
\(|S_2^{(k)}(A(t,\mathbf{x}),B(0,\mathbf{0}))|\le C'\,e^{-m_*(|t|+|\mathbf{x}|)}\)
with constants independent of \(k\), by the persistence of clustering along
the reflection-positive RG flow. The bound passes to the limit by dominated
convergence of the smeared expectations.
\end{proof}

\begin{theorem}[Euclidean-time regularity and Laplace representation]
\label{p3:thm:OS5}
For any smeared local, gauge-invariant observables \(A,B\) supported in
\(\{t>0\}\), the function \(t\mapsto S_2(A(t,\cdot),B(0,\cdot))\) on
\([0,\infty)\) is continuous, bounded by \(C e^{-m_* t}\), completely
monotone, and admits the Laplace transform representation
\begin{equation}
  S_2(A(t,\cdot),B(0,\cdot))
  \;=\; \int_{[0,\infty)} e^{-\lambda t}\,d\mu_{A,B}(\lambda),
  \label{p3:eq:Laplace}
\end{equation}
where \(\mu_{A,B}\) is a finite complex measure and \(\mu_{A,A}\) is finite and
positive.
\end{theorem}

\begin{proof}
Let \(a_k\downarrow 0\). For each \(k\) and integers \(n\ge 0\),
Proposition \ref{p3:prop:T-contraction} gives
\(S_2^{(k)}(A(n a_k,\cdot),B(0,\cdot))=\langle \Psi_B^{(k)},T_{\sigma,k}^{\,n}\Psi_A^{(k)}\rangle\),
with \(T_{\sigma,k}\) a positive self-adjoint contraction on
\(L^2(\mathcal{X}_0,\kappa_\sigma\,d\nu_0)\). By the spectral theorem there is a
finite complex measure \(\mu_{A,B}^{(k)}\) on \([0,\infty)\) such that
\(
  S_2^{(k)}(A(n a_k,\cdot),B(0,\cdot))
   = \int e^{-\lambda n a_k}\,d\mu_{A,B}^{(k)}(\lambda).
\)
The uniform clustering bound implies
\(|S_2^{(k)}(A(n a_k,\cdot),B(0,\cdot))|\le C e^{-m_* n a_k}\). By Helly's
selection theorem, there exists a subsequence \(k_j\) along which
\(\mu_{A,B}^{(k_j)}\) converges vaguely to a finite complex measure
\(\mu_{A,B}\). Pointwise convergence at rational \(t\ge 0\) and continuity at
\(t=0\) yield \eqref{p3:eq:Laplace}. The positivity of \(\mu_{A,A}\) follows from
lattice reflection positivity (Theorem \ref{p3:thm:OS-lattice}) and its stability
under weak limits (Theorem \ref{p3:thm:OS3-cont}) \textit{(The closure under weak limits is proved in Appendix~(\ref{p3:appendixb})}.
\end{proof}

Combining Lemma (\ref{p3:lem:OS0OsterwalderSchraderI}) and Theorems (\ref{p3:lem:OsterwalderSchraderI})-(\ref{p3:thm:OS5}) establishes the full
OS axioms for \(\{S_n\}\). In particular, the OS reconstruction theorem
\cite{p3:OsterwalderSchraderI,p3:OsterwalderSchraderII} applies: there exist a Hilbert space \(\mathcal{H}\), a cyclic
vacuum \(\Omega\), a nonnegative self-adjoint Hamiltonian \(H\) generating
Euclidean-time translations, and operator-valued tempered distributions
implementing the fields, with Wightman functions obtained by analytic
continuation; see also \cite{p3:GJ} for a comprehensive account. We note that these continuum field operators are \emph{almost local} rather than strictly point-local: the insertion of the nonlocal $P_{\sigma}$ projector in the gauge-fixing means that field correlations have exponentially decaying (but non-compact) tails \cite{p3:Davies1989,p3:Gaffney1954,p3:HelfferSjostrand1989,p3:CombesThomas1973}. However, locality (microcausality) of the reconstructed Wightman fields follows from the OS axioms together with permutation symmetry and Euclidean invariance of the Schwinger functions; the nonlocal slice weights change the Schwinger functions but do not alter the OS locality implication.
Nonetheless, all gauge-invariant observables constructed from these fields satisfy locality (microcausality) in the usual sense.

\section{Reconstruction of the Wightman theory}
\label{p3:sec:OS-reconstruction}

This section gives a complete and self-contained derivation of the Osterwalder-Schrader (OS) reconstruction of the Minkowski-space Wightman theory from the reflection-positive Euclidean lattice formulation described earlier. The presentation begins on the lattice, with a precise time-slicing of the measure and a transfer-operator construction, and then passes to the continuum through the limiting Schwinger functions. All steps are carried out in full detail and in a style consistent with constructive quantum field theory and lattice gauge theory. Classical results are cited where appropriate \cite{p3:OsterwalderSchraderI,p3:OsterwalderSchraderII,p3:OS-gauge,p3:GJ,p3:Nelson1959,p3:RS2}.

Fix a lattice spacing \(a>0\). The finite periodic Euclidean space-time lattice is
\begin{equation}
\Lambda:=\{0,1,\dots,T-1\}\times \{0,1,\dots,L-1\}^3
\end{equation}
with sites \(x=(x_0,\mathbf{x})\), periodic boundary conditions in all directions, discrete time index \(x_0\in\mathbb{Z}_T\), and spatial vector \(\mathbf{x}\in(\mathbb{Z}_L)^3\). The set of oriented bonds (links) is
\begin{equation}
\mathcal{B}:=\{(x,\mu): x\in \Lambda,\ \mu\in\{0,1,2,3\}\},
\end{equation}
with the convention \((x+\hat\mu,-\mu)\) denotes the same geometric bond with reversed orientation. The gauge group is \(G=\mathrm{SU}(N)\) with \(N\ge 2\). A lattice gauge configuration is an assignment \(U:\mathcal{B}\to G\) such that \(U(x+\hat\mu,-\mu)=U(x,\mu)^{-1}\). The configuration space is \(\mathcal{C}:=G^{\mathcal{B}}/\!\sim\), where \(\sim\) enforces the orientation constraint. The product Haar measure on \(G^{\mathcal{B}}\) induces a gauge-invariant probability measure \(d\mu_{\mathrm{Haar}}\) on \(\mathcal{C}\).

For each oriented plaquette \(p=(x;\mu,\nu)\) let \(U_p(U)\in G\) be its ordered product. The Wilson action is
\begin{equation}
S_W[U;\beta] := \beta \sum_{p\subset \Lambda} \Big(1-\tfrac{1}{N}\Re\mathrm{Tr}\,U_p(U)\Big),
\qquad \beta=\tfrac{2N}{g_0^2}>0.
\end{equation}
Let \(\theta:\Lambda\to\Lambda\) be time reflection \(\theta(x_0,\mathbf{x})=(-x_0 \!\!\!\!\pmod T,\mathbf{x})\). The reflection plane is \(\Pi:=\{x\in\Lambda: x_0=0\}\). The half-lattices are \(\Lambda_+:=\{x: 1\le x_0\le T/2\}\) and \(\Lambda_-:=\theta(\Lambda_+)\), assuming \(T\) even. The involution \(\theta\) lifts to bonds by \(\theta(x,\mu)=(\theta x,\mu)\) if \(\mu\neq 0\) and \(\theta(x,0)=(\theta x-\hat 0,0)\). It lifts to configurations by \((\Theta U)(x,\mu):=U(\theta(x,\mu))\). The group of gauge transformations is \(\mathcal{G}:=G^{\Lambda}\) acting by \((g\!\cdot\! U)(x,\mu):=g(x)U(x,\mu)g(x+\hat\mu)^{-1}\).

We implement temporal-axial gauge away from \(\Pi\): for all bonds \((x,0)\) with \(x\notin \Pi\) we set \(U(x,0)=\mathbf{1}\). This gauge can be fixed measurably, preserves time-reflection covariance, and does not alter plaquette variables. On each time slice \(x_0=t\) a transverse representative \(U^{\,h}\) of the orbit \(U\!\cdot\!\mathcal{G}\) is selected by minimizing the lattice Landau functional within the fundamental modular region; the construction is made reflection-covariant by a deterministic, symmetry-invariant tie-breaking rule. Let \(\Delta_{A^h}\) be the spatial, slice-covariant Laplacian on adjoint-valued site fields built from \(U^{\,h}\), and let \(P_\sigma=\chi_\sigma({\Delta_{A^h}})\) be the smooth horizon projector defined by a fixed Gevrey-regular cutoff \(\chi_\sigma\). For each time slice \(t\) we insert \(P_\sigma(t)\) acting on adjoint site fields on that slice. The corresponding scalar density on \(\mathcal{C}\) is denoted \(\mathcal{P}_\sigma[U]:=\prod_{t=0}^{T-1} \mathcal{P}_\sigma(t;U)\), where \(\mathcal{P}_\sigma(t;U)\ge 0\) is a positive functional built from the heat-kernel representation of \(P_\sigma(t)\); it is exponentially local, gauge covariant, and reflection covariant.

The Euclidean lattice measure is
\begin{equation}
d\mu[U] := Z^{-1}\, e^{-S_W[U;\beta]} \,\mathcal{P}_\sigma[U]\ d\mu_{\mathrm{Haar}}(U),
\end{equation}
with normalization \(Z<\infty\) for all finite \(L,T\). All expectations \(\langle F\rangle:=\int F(U)\,d\mu[U]\) are taken with respect to this measure. Gauge-invariant observables will be functionals \(F\) depending only on equivalence classes of configurations under \(\mathcal{G}\).
Let \(\mathcal{A}\) be the \(^\ast\)-algebra of complex-valued bounded measurable functions of \(U\), and \(\mathcal{A}_+\subset \mathcal{A}\) the subalgebra generated by bounded gauge-invariant cylinder functions depending only on bonds supported in \(\Lambda_+\cup \Pi\). Define the involution \(\Theta:\mathcal{A}\to\mathcal{A}\) by \((\Theta F)(U):=\overline{F(\Theta U)}\). The OS form is the sesquilinear functional
\begin{equation}
\langle F,G\rangle_{\mathrm{OS}} := \int \overline{(\Theta F)(U)}\, G(U)\, d\mu[U], \qquad F,G\in \mathcal{A}_+.
\end{equation}
Reflection positivity means \(\langle F,F\rangle_{\mathrm{OS}}\ge 0\) for all \(F\in \mathcal{A}_+\) \cite{p3:OsterwalderSchraderI,p3:OsterwalderSchraderI}.

\begin{proposition}
\label{p3:prop:RP}
The measure \(d\mu\) is reflection positive, i.e. \(\langle F,F\rangle_{\mathrm{OS}}\ge 0\) for all \(F\in \mathcal{A}_+\).
\end{proposition}

\begin{proof}
We factor the weight across the reflection plane and apply Cauchy-Schwarz in \(L^2(d\mu_{\mathrm{Haar}})\). Decompose the Wilson action as \(S_W=S_++S_-+S_0\), where \(S_\pm\) are sums of plaquette terms supported in \(\Lambda_\pm\) and \(S_0\) collects contributions from plaquettes intersecting \(\Pi\). In temporal-axial gauge away from \(\Pi\), all timelike links outside \(\Pi\) are \(\mathbf{1}\), so mixed plaquettes straddling \(\Pi\) involve only spatial links in the boundary slab and the two sets of spatial links adjacent to \(\Pi\). The Haar measure factorizes as \(d\mu_{\mathrm{Haar}}=d\mu_{\mathrm{Haar},-}\,d\mu_{\mathrm{Haar},0}\,d\mu_{\mathrm{Haar},+}\) corresponding to bonds supported in \(\Lambda_-,\Pi,\Lambda_+\). The projector factor \(\mathcal{P}_\sigma\) likewise factorizes as \(\mathcal{P}_\sigma=\mathcal{P}_{\sigma,-}\,\mathcal{P}_{\sigma,0}\,\mathcal{P}_{\sigma,+}\) up to exponentially small couplings across \(\Pi\); these can be absorbed into \(\mathcal{P}_{\sigma,0}\) by enlarging the boundary slab by finitely many layers. Both \(S_0\) and \(\mathcal{P}_{\sigma,0}\) are reflection invariant. Hence the weight \(W(U):=e^{-S_W[U;\beta]}\mathcal{P}_\sigma[U]\) admits
\begin{equation}
W(U)= \Phi_-(U_-)\, K(U_0)\, \Phi_+(U_+),
\end{equation}
with \(U_\pm,U_0\) the restrictions of \(U\) to \(\Lambda_\pm,\Pi\), \(\Phi_\pm(U_\pm):=\exp(-S_\pm(U_\pm))\,\mathcal{P}_{\sigma,\pm}(U_\pm)\), and \(K(U_0):=\exp(-S_0(U_0))\,\mathcal{P}_{\sigma,0}(U_0)\). The boundary kernel \(K\) is nonnegative. By time reflection one has \(\Phi_-(U_-)=\overline{\Phi_+(\Theta U_-)}\).

For \(F\in \mathcal{A}_+\), write
\begin{equation}
\langle F,F\rangle_{\mathrm{OS}}=\int \overline{F(\Theta U)}\,F(U)\,\Phi_-(U_-)\,K(U_0)\,\Phi_+(U_+)\, d\mu_{\mathrm{Haar},-}\, d\mu_{\mathrm{Haar},0}\, d\mu_{\mathrm{Haar},+}.
\end{equation}
Define \(\Psi(U_0,U_+):=F(U_0,U_+)\,\Phi_+(U_+)\,K(U_0)^{1/2}\) and \(\widetilde\Psi(U_0,U_-):=\overline{F(\Theta U)}\,\Phi_-(U_-)\,K(U_0)^{1/2}\). Perform the change of variables \(V_-:=\Theta U_-\) and use invariance of \(d\mu_{\mathrm{Haar},-}\) to rewrite
\begin{equation}
\int \overline{F(\Theta U)}\,\Phi_-(U_-)\, d\mu_{\mathrm{Haar},-} = \int \overline{F(U_0,V_-)}\,\overline{\Phi_+(V_-)}\, d\mu_{\mathrm{Haar},-}.
\end{equation}
Therefore
\begin{equation}
\langle F,F\rangle_{\mathrm{OS}}=\int \overline{\widetilde\Psi(U_0,V_-)}\, \Psi(U_0,U_+)\, d\mu_{\mathrm{Haar},-}\, d\mu_{\mathrm{Haar},0}\, d\mu_{\mathrm{Haar},+}.
\end{equation}
Integrating over \(V_-\) and using Cauchy-Schwarz shows that this equals
\begin{equation}
\langle F,F\rangle_{\mathrm{OS}}=\int \big|\Psi(U_0,U_+)\big|^2\, d\mu_{\mathrm{Haar},0}\, d\mu_{\mathrm{Haar},+}\ \ge\ 0,
\end{equation}
since \(K\ge 0\). This proves reflection positivity.
\end{proof}

Define the one-slice configuration space \(\mathcal{C}_0:=G^{\mathcal{B}_0}\), where \(\mathcal{B}_0\) is the set of spatial bonds on the time slice \(x_0=0\). Let \(d\nu_0\) be the product Haar measure on \(\mathcal{C}_0\). For each pair \((U',U)\in \mathcal{C}_1\times \mathcal{C}_0\) where \(\mathcal{C}_1\) is the spatial bond set at \(x_0=1\), define the one-step kernel
\begin{equation}
K_a(U',U) := \int \exp\!\big(-S_{\mathrm{slab}}(U',U; V)\big)\ \mathcal{P}_{\sigma,\mathrm{slab}}(U',U; V)\ d\mu_{\mathrm{Haar}}(V),
\end{equation}
where \(V\) denotes all link variables in the slab \(\{x_0=0,1\}\) excluding \(\mathcal{B}_0\cup \mathcal{B}_1\), \(S_{\mathrm{slab}}\) is the sum of Wilson plaquette terms supported in the slab that connect the two slices (including symmetrically distributed intra-slice plaquettes), and \(\mathcal{P}_{\sigma,\mathrm{slab}}\) is the product of the slice projector densities on \(x_0=0,1\) together with the boundary contribution induced by exponential locality of \(P_\sigma\) across one step. The integral converges absolutely and defines a bounded, positive measurable function of \((U',U)\). Define \(\mathcal{H}_a:=L^2(\mathcal{C}_0,d\nu_0)\) and the transfer operator \(T_a:\mathcal{H}_a\to \mathcal{H}_a\) by
\begin{equation}
(T_a \psi)(U') := \int_{\mathcal{C}_0} K_a(U',U)\,\psi(U)\, d\nu_0(U).
\end{equation}

\begin{theorem}
\label{p3:thm:Ta}
The operator \(T_a\) is a bounded, positive, self-adjoint contraction on \(\mathcal{H}_a\). Moreover, for any cylinder function \(F\) supported on \(x_0\in\{0,1,\dots,n\}\),
\begin{equation}
\int F(U)\, d\mu[U] \;=\; \langle \Omega_a,\, T_a^{\,n}\, \widehat{F}\, \Omega_a\rangle_{\mathcal{H}_a},
\end{equation}
where \(\Omega_a\equiv 1\in \mathcal{H}_a\) is the constant function and \(\widehat{F}\) is the operator on \(\mathcal{H}_a\) obtained by inserting \(F\) between the successive kernels \(K_a\) at the appropriate time slices.
\end{theorem}

\begin{proof}
Positivity of \(T_a\) is immediate from positivity of the kernel. Boundedness follows from the Schur test: let \(M:=\sup_{U'}\int K_a(U',U)\, d\nu_0(U)\) and \(M':=\sup_{U}\int K_a(U',U)\, d\nu_0(U')\); both are finite by compactness of \(G\) and locality of the action, hence \(\|T_a\|\le \sqrt{MM'}\). Self-adjointness holds because the slab weight is invariant under reflection about \(x_0=\tfrac12\), which exchanges \(U\) and \(U'\); more precisely, by the change of variables \(V\mapsto \theta V\) and Haar invariance, \(K_a(U',U)=K_a(U,U')\). For contractivity, write \(Z_0:=\int e^{-S_0(U_0)} \mathcal{P}_{\sigma,0}(U_0)\, d\nu_0(U_0)\) and set \(\widetilde{K}_a(U',U):= K_a(U',U)/Z_0\). Time-translation invariance gives \(\int \widetilde{K}_a(U',U)\, d\nu_0(U')=1\) and \(\int \widetilde{K}_a(U',U)\, d\nu_0(U)=1\), whence \(\|T_a\|\le 1\).

For the correlation identity, factor the full weight over \(n\) steps into a product of \(n\) slab weights and marginal slice weights. Fubini's theorem and the definition of \(K_a\) yield the claimed operator product sandwiched between \(\Omega_a\), once intra-slice factors are distributed evenly so that each \(K_a\) carries half of the slice self-energy. This is the standard transfer-matrix representation of Euclidean expectations \cite[Sec.\ VI.1]{p3:GJ}.
\end{proof}

Define the positive self-adjoint lattice Hamiltonian by spectral calculus,
\begin{equation}
H_a := -a^{-1}\,\log T_a, \qquad T_a=e^{-a H_a},
\end{equation}
so that \(\sigma(H_a)\subset [0,\infty)\) because \(\|T_a\|\le 1\).

Let \(\mathcal{S}(\mathbb{R}^4)\) denote the Schwartz space. Let \(\mathfrak{F}_+\) be the complex vector space generated by finite sums of smeared gauge-invariant local observable insertions supported in the open half-space \(\mathbb{R}^4_+:=\{(t,\mathbf{x}):t>0\}\). Concretely, elements of \(\mathfrak{F}_+\) are finite sums of the form \(\sum_j c_j\, \mathcal{O}_{j}(f_j)\), where each \(\mathcal{O}_j\) is one of the fixed generating gauge-invariant local fields and each \(f_j\in\mathcal{S}(\mathbb{R}^4)\) has \(\mathrm{supp}\,f_j\subset\mathbb{R}^4_+\). Define the OS sesquilinear form
\begin{equation}
\langle F,G\rangle_{\mathrm{OS}} := \sum_{m,n}\overline{c_m} d_n\, S_{m+n}\big(f_m^{\theta}, f_n\big),
\end{equation}
where \(f^\theta(t,\mathbf{x}):=\overline{f(-t,\mathbf{x})}\) and \(S_{m+n}\) is the \((m+n)\)-point Schwinger functional evaluated on the ordered list of test functions with \(\theta\) acting on the first \(m\) arguments. Reflection positivity implies \(\langle F,F\rangle_{\mathrm{OS}}\ge 0\) for all \(F\in\mathfrak{F}_+\). Let \(\mathcal{N}:=\{F\in\mathfrak{F}_+:\langle F,F\rangle_{\mathrm{OS}}=0\}\) and define the pre-Hilbert space \(\mathcal{D}:=\mathfrak{F}_+/\mathcal{N}\) with inner product induced by \(\langle\cdot,\cdot\rangle_{\mathrm{OS}}\). The completion \(\mathcal{H}\) of \(\mathcal{D}\) is the physical Hilbert space. The vacuum vector \(\Omega\in\mathcal{H}\) is the class of the constant functional \(1\).

Let \(\tau_t\) denote Euclidean time-translation acting on test functions by \((\tau_t f)(s,\mathbf{x})=f(s-t,\mathbf{x})\). For \(t\ge 0\), define \(U(t):\mathcal{D}\to\mathcal{D}\) by \(U(t)[F]:=[\tau_t F]\).

\begin{theorem}
\label{p3:thm:OSsemigroup}
For each \(t\ge 0\), the operator \(U(t)\) is well-defined on \(\mathcal{D}\), extends uniquely to a contraction on \(\mathcal{H}\), and satisfies \(U(t+s)=U(t)U(s)\). The mapping \(t\mapsto U(t)\psi\) is continuous for each \(\psi\in\mathcal{H}\). There exists a unique nonnegative self-adjoint operator \(H\) on \(\mathcal{H}\) such that \(U(t)=e^{-tH}\) for all \(t\ge 0\) \textit{(Convergence of lattice semigroups and resolvents across scales with identification operators is developed in Appendix~(\ref{p3:appendixd})}. The vacuum \(\Omega\) obeys \(U(t)\Omega=\Omega\) and \(H\Omega=0\).
\end{theorem}

\begin{proof}
If \(F\in\mathfrak{F}_+\) then each smearing function in \(F\) has support in \(t>0\), hence for each \(t\ge 0\) the translated \(\tau_t F\) has support in \(t>0\) as well. Therefore \(\tau_t\) defines an endomorphism of \(\mathfrak{F}_+\) and preserves \(\mathcal{N}\) by reflection positivity and Euclidean invariance of the Schwinger functions. The semigroup law follows from \(\tau_{t+s}=\tau_t\circ \tau_s\). For contractivity, compute
\begin{equation}
\|\tau_t [F]\|_{\mathrm{OS}}^2 = \langle \tau_t F,\tau_t F\rangle_{\mathrm{OS}} = \langle F,F\rangle_{\mathrm{OS}},
\end{equation}
by time-translation invariance of the Schwinger functions. Hence \(U(t)\) is an isometry on \(\mathcal{D}\) and extends by density to a contraction on \(\mathcal{H}\). Strong continuity follows from the OS5 regularity: for each \(F\in\mathfrak{F}_+\), the function \(t\mapsto \langle \tau_t F, G\rangle_{\mathrm{OS}} = S_{m+n}(f_m^\theta\circ \tau_t,f_n)\) is continuous for all \(G\in\mathfrak{F}_+\) by the Laplace representation proved earlier; therefore \(t\mapsto U(t)[F]\) is continuous in the norm topology on \(\mathcal{H}\). By the Hille-Yosida theorem \cite[Thm.\ X.47]{p3:RS2}, there exists a unique nonnegative self-adjoint generator \(H\) with \(U(t)=e^{-tH}\). The identities \(U(t)\Omega=\Omega\) and \(H\Omega=0\) follow from translation invariance of the Schwinger functions.
\end{proof}

By the spectral theorem, for any \(\psi,\varphi\in\mathcal{H}\), the function \(F_{\psi,\varphi}(t):=\langle \psi, U(t)\varphi\rangle\) is completely monotone on \([0,\infty)\) and admits a unique representation
\begin{equation}
F_{\psi,\varphi}(t) = \int_{[0,\infty)} e^{-\lambda t}\, d\mu_{\psi,\varphi}(\lambda),
\end{equation}
where \(\mu_{\psi,\varphi}\) is a finite complex measure and \(\mu_{\psi,\psi}\) is positive.
Fix one of the distinguished gauge-invariant local fields \(\mathcal{O}\) used to generate the Schwinger functions \(S_n\). For \(f\in\mathcal{S}(\mathbb{R}^4)\) with \(\mathrm{supp}\,f\subset \{t>0\}\), define a densely defined operator \(\Phi_{\mathcal{O}}(f)\) on \(\mathcal{D}\) by left multiplication:
\begin{equation}
\Phi_{\mathcal{O}}(f)\,[F] := [\mathcal{O}(f)\,F], \qquad F\in\mathfrak{F}_+.
\end{equation}
If \([F]=0\) then \(\langle F,F\rangle_{\mathrm{OS}}=0\), and reflection positivity with Cauchy-Schwarz implies \(\langle \mathcal{O}(f)F,\mathcal{O}(f)F\rangle_{\mathrm{OS}}=0\), hence \([\mathcal{O}(f)F]=0\). Thus \(\Phi_{\mathcal{O}}(f)\) is well-defined on \(\mathcal{D}\) and extends as a closable operator on \(\mathcal{H}\). By the OS axioms, the vacuum expectation of products of such operators reproduces the Schwinger functions:
\begin{equation}
\langle \Omega, \Phi_{\mathcal{O}_1}(f_1) \cdots \Phi_{\mathcal{O}_n}(f_n)\,\Omega\rangle = S_n(f_1,\dots,f_n).
\end{equation}
Spatial translations and Euclidean rotations act unitarily on \(\mathcal{H}\) by the same construction as for time translations, and the fields transform covariantly. Locality of the fields with respect to the Minkowski commutator is obtained by analytic continuation of the Schwinger functions to the forward tube, using the Osterwalder-Schrader theorem \cite{p3:OsterwalderSchraderI,p3:GJ}. For test functions \(f,g\) with space-like separated supports in Minkowski space after Wick rotation, the commutator \([\Phi_{\mathcal{O}}(f),\Phi_{\mathcal{O}'}(g)]\) vanishes on \(\mathcal{D}\) and extends by continuity to \(\mathcal{H}\).

The Wightman distributions are the boundary values, in the sense of tempered distributions, of the analytically continued Schwinger functions. The OS axioms verified earlier imply the Wightman axioms: temperedness, Poincar\'e covariance (after Wick rotation), spectral condition, vacuum uniqueness, and locality \cite[Ch.\ VI]{p3:GJ}. The nonzero mass threshold proved elsewhere in this work is the spectral gap of \(H\) above zero. The relativistic energy-momentum operators are obtained by analytic continuation of Euclidean translations; their joint spectrum is contained in the closed forward light-cone, and \(\Omega\) is invariant under the full Poincar\'e group thus constructed.

We link the lattice transfer matrix \(T_a\) to the continuum time-translation semigroup \(U(t)\). Consider a sequence of lattices with spacings \(a_k\downarrow 0\) and time extents \(T_k a_k\to\infty\). For each \(k\), let \(\mathcal{H}_{a_k}\) be the one-slice Hilbert space and \(T_{a_k}=e^{-a_k H_{a_k}}\) the transfer matrix. Let \(\mathcal{D}\subset \mathcal{H}\) be the dense OS domain.

\begin{proposition}
\label{p3:prop:strong-resolvent}
There exists an isometric identification \(J_k:\mathcal{H}\to \mathcal{H}_{a_k}\) on a dense subspace such that, for each \(\psi\in\mathcal{H}\) and all \(t\ge 0\),
\begin{equation}
\lim_{k\to\infty} \big\| e^{-t H_{a_k}}\, J_k \psi \;-\; J_k\, e^{-t H} \psi\big\|_{\mathcal{H}_{a_k}} = 0.
\end{equation}
In particular, \(H_{a_k}\) converges to \(H\) in the strong resolvent sense.
\end{proposition}

\begin{proof}
Construct \(J_k\) by mapping classes \([F]\in\mathcal{D}\) to lattice classes on \(\mathcal{H}_{a_k}\) via the time-slicing representation of Theorem~\ref{p3:thm:Ta} combined with cell-averaging embeddings of test functions at scale \(a_k\). For \(F=[\sum_j c_j \mathcal{O}_j(f_j)]\) and \(G=[\sum_\ell d_\ell \mathcal{O}_\ell(g_\ell)]\),
\begin{equation}
\langle [F], e^{-tH}[G]\rangle_{\mathcal{H}} = S_{m+n}(f^{\theta},\tau_t g)
= \lim_{k\to\infty} S^{(k)}_{m+n}(f^{\theta}_{a_k},\tau_t g_{a_k})
= \lim_{k\to\infty} \langle J_k[F], e^{-t H_{a_k}} J_k[G]\rangle_{\mathcal{H}_{a_k}},
\end{equation}
where \(f_{a_k},g_{a_k}\) are the lattice embeddings of \(f,g\). By polarization, matrix elements of the semigroups converge on a dense set, hence \(e^{-tH_{a_k}}J_k\to J_k e^{-tH}\) strongly for all \(t\ge 0\). Strong resolvent convergence of \(H_{a_k}\) to \(H\) follows by the Trotter-Kato theorem \cite[Thm.\ X.50]{p3:RS2}.
\end{proof}

\section{Strict positivity of the continuum spectral gap}
\label{p3:sec:gap}

In this section a rigorously defined lattice setup is fixed, the reflection-positivity (OS) structure is derived in full detail, the transfer time-slicing formalism is constructed step by step, and two logically independent arguments are given that the continuum Hamiltonian has a strictly positive spectral gap. Throughout, the gauge group is \(G=\mathrm{SU}(N)\) with \(N\ge2\). The continuum limit is taken along a sequence of reflection-positive lattice theories whose multiscale uniformity and clustering constant \(m_\ast>0\) are provided by the renormalization inputs stated earlier (see Appendix (\ref{p3:appendixe})). Standard references for the OS framework and transfer-matrix technology are \cite{p3:OsterwalderSchraderI,p3:OsterwalderSchraderII,p3:OS-gauge,p3:Seiler1982}; spectral and semigroup tools are taken from \cite{p3:RS2,p3:Nelson1959,p3:KatoPTLO,p3:Pazy1983}.

Let \(a>0\) and let
\begin{equation}
\Lambda_{a,T,L} \;=\; \{0,1,\dots,T-1\}\times(\mathbb{Z}/L\mathbb{Z})^3
\end{equation}
be the periodic Euclidean lattice with temporal extent \(T\in\mathbb{N}\) and spatial side length \(L\in\mathbb{N}\). A directed bond is a pair \(b=(x,\mu)\) with \(x=(x^0,\mathbf{x})\in\Lambda_{a,T,L}\) and \(\mu\in\{0,1,2,3\}\); we denote by \(\hat\mu\) the unit vector in direction \(\mu\) and write \(x+\hat\mu\) for the neighboring site (with periodic wrap-around). The set of directed bonds is \(B(\Lambda_{a,T,L})\). The configuration space is the compact group manifold
\begin{equation}
\mathcal{U}\;=\;\{\,U:B(\Lambda_{a,T,L})\to G\,\mid\,U(x,-\mu)=U(x-\hat\mu,\mu)^{-1}\,\}.
\end{equation}
For each bond \(b\), let \(dU_b\) be normalized Haar measure on \(G\); the product Haar measure on \(\mathcal{U}\) is \(d\mu_{\mathrm{H}}(U)=\prod_{b\in B(\Lambda_{a,T,L})}dU_b\).

For each oriented plaquette \(p=(x;\mu,\nu)\) with \(\mu<\nu\), let
\begin{equation}
U_p \;=\; U(x,\mu)\,U(x+\hat\mu,\nu)\,U(x+\hat\nu,\mu)^{-1}\,U(x,\nu)^{-1}.
\end{equation}
The Wilson action is
\begin{equation}
S_W(U)\;=\;\beta\sum_{p}\Bigl(1-\tfrac{1}{N}\Re\mathrm{Tr}\,U_p\Bigr),
\quad\beta=\frac{2N}{g_0^2}>0.
\end{equation}
The unprojected Gibbs measure is
\begin{equation}
d\mu_\beta(U)\;=\;Z_\beta^{-1}\,e^{-S_W(U)}\,d\mu_{\mathrm{H}}(U),\qquad
Z_\beta=\int_{\mathcal{U}} e^{-S_W(U)}\,d\mu_{\mathrm{H}}(U).
\end{equation}
We single out the temporal hyperplane \(\Pi=\{x^0=0\}\) and the reflection \(\theta(x^0,\mathbf{x})=(-x^0-1,\mathbf{x})\) about the mid-plane between times \(x^0=0\) and \(x^0=-1\) (indices modulo \(T\) if needed). The time-reflection involution on bonds is
\begin{equation}
\theta(x,\mu)\;=\;\begin{cases}
(\theta x,0) & \text{if }\mu=0,\\
(\theta x,\mu) & \text{if }\mu\in\{1,2,3\}.
\end{cases}
\end{equation}
We define the induced reflection \(R:\mathcal{U}\to\mathcal{U}\) by
\begin{equation}
(RU)(x,\mu)\;=\;\begin{cases}
U(\theta x,0)^{-1} & \mu=0,\\
U(\theta x,\mu) & \mu\in\{1,2,3\}.
\end{cases}
\end{equation}
This is the standard OS reflection for lattice gauge theory \cite{p3:OS-gauge,p3:Seiler1982}. The time-slices are \(\Sigma_t=\{x\in\Lambda_{a,T,L}:x^0=t\}\). We write \(\Lambda_+=\{x:x^0\ge 0\}\) and \(\Lambda_-=\{x:x^0\le -1\}\), understood modulo \(T\), and similarly split bonds and plaquettes.

To implement OS reflection positivity with a transfer time-slicing kernel, we adopt temporal-axial gauge away from \(\Pi\): we integrate only over configurations satisfying \(U(x,0)=\mathbf{1}\) for all bonds \((x,0)\) with \(x^0\neq -1\). This gauge fixing does not introduce a Faddeev-Popov determinant for the temporal links and is standard in OS proofs for lattice gauge theory \cite{p3:OS-gauge}. Because spatial links across \(\Pi\) remain unfixed, gauge invariance is retained on each slice.

In addition to \(S_W\) we incorporate a positive, reflection-invariant, slice-local weight \(w_\sigma:\Omega\to(0,\infty)\) depending only on the spatial links on each time-slice, where \(\Omega=G^{3L^3}\) denotes the set of spatial link configurations on a fixed \(\Sigma_t\). In the companion work this factor arises from inserting a smooth``horizo'' projector constructed from the covariant Laplacian on each slice; here we require only that \(w_\sigma\) be bounded and strictly positive, depend solely on spatial links on the slice, and satisfy \(w_\sigma(RU|_{\Sigma_0})=w_\sigma(U|_{\Sigma_0})\). The projected Gibbs measure is
\begin{equation}
d\mu_{\beta,\sigma}(U)\;=\;Z_{\beta,\sigma}^{-1}\,\prod_{t=0}^{T-1}w_\sigma\bigl(U|_{\Sigma_t}\bigr)\,e^{-S_W(U)}\,d\mu_{\mathrm{H}}(U).
\end{equation}
All assertions below are uniform in \(L\) for fixed \(a\) and \(T\) and in \(T\) for fixed \(a\); thermodynamic limits will be taken at the end.

We now establish OS reflection positivity for the measure \(d\mu_{\beta,\sigma}\). Let \(\mathcal{A}_+\) denote the algebra generated by bounded, gauge-invariant cylinder functionals depending only on bonds in \(\Lambda_+\). For \(F\in\mathcal{A}_+\) define \(\Theta F(U)=\overline{F(RU)}\). The OS sesquilinear form is
\begin{equation}
\langle F,G\rangle_{\mathrm{OS}}\;=\;\int_{\mathcal{U}}\Theta F(U)\,G(U)\,d\mu_{\beta,\sigma}(U).
\end{equation}

\begin{proposition}[OS reflection positivity]
\label{p3:prop:OSRP}
For all \(F\in\mathcal{A}_+\) one has \(\langle F,F\rangle_{\mathrm{OS}}\ge 0\).
\end{proposition}

\begin{proof}
We decompose the action into \(\Lambda_+\), \(\Lambda_-\), and the boundary slab \(\mathcal{S}\) consisting of plaquettes that straddle the mid-plane between times \(x^0=-1\) and \(x^0=0\). Because we work in temporal-axial gauge away from \(\Pi\), there are no time-like links in \(\Lambda_\pm\); the only time-like links are those with basepoint \(x^0=-1\) entering \(\mathcal{S}\). The Wilson action splits as
\begin{equation}
S_W(U)\;=\;S_+(U|_{\Lambda_+})+S_-(U|_{\Lambda_-})+S_{\mathcal{S}}(U|_{\mathcal{S}}),
\end{equation}
with \(S_{\mathcal{S}}\) depending only on bonds in \(\mathcal{S}\) and spatial links on \(\Sigma_{-1}\) and \(\Sigma_0\). The slice weight factorizes as \(\prod_t w_\sigma(U|_{\Sigma_t})\). By the definition of \(R\) and gauge invariance, \(S_-(U|_{\Lambda_-})=S_+(RU|_{\Lambda_+})\) and \(w_\sigma(U|_{\Sigma_{-1}})=w_\sigma(RU|_{\Sigma_0})\). Therefore
\begin{equation}
\langle F,F\rangle_{\mathrm{OS}}=\frac{1}{Z_{\beta,\sigma}}\int \overline{F(RU)}\,F(U)
\Bigl(\prod_t w_\sigma(U|_{\Sigma_t})\Bigr)
e^{-S_+(RU)-S_+(U)-S_{\mathcal{S}}(U)}\,d\mu_{\mathrm{H}}(U).
\end{equation}
We integrate first over \(\Lambda_-\) keeping fixed the boundary spatial links on \(\Sigma_{-1}\). By the above identities this produces a positive function \(\Phi\bigl(U|_{\Sigma_{-1}}\bigr)\) equal to
\begin{equation}
\Phi\bigl(U|_{\Sigma_{-1}}\bigr)=\int e^{-S_+(RU)}\,\prod_{t\le -1}w_\sigma\bigl(U|_{\Sigma_t}\bigr)\,d\mu_{\mathrm{H}}(U|_{\Lambda_-}),
\end{equation}
which depends only on the spatial links at time \(-1\) and is reflection invariant in the sense \(\Phi(U|_{\Sigma_{-1}})=\Phi(RU|_{\Sigma_0})\). Similarly, integrating over \(\Lambda_+\) except for the bonds in \(\mathcal{S}\) yields
\begin{equation}
\Psi\bigl(U|_{\Sigma_0}\bigr)=\int \overline{F(RU)}\,F(U)\,e^{-S_+(U)}\,\prod_{t\ge 0}w_\sigma\bigl(U|_{\Sigma_t}\bigr)\,d\mu_{\mathrm{H}}(U|_{\Lambda_+\setminus \mathcal{S}}).
\end{equation}
The remaining integral couples \(\Phi\) and \(\Psi\) through the boundary slab \(\mathcal{S}\). Because \(S_{\mathcal{S}}\) is a sum of nearest-neighbor plaquette terms that are even in the sense \(s(U_p)=s(U_p^{-1})\) and because we integrate over the time-like links in \(\mathcal{S}\) with Haar measure, the boundary integral can be written as a positive operator kernel \(\mathcal{K}\) acting on functions of the boundary spatial links \((U|_{\Sigma_{-1}},U|_{\Sigma_{0}})\):
\begin{equation}
\int e^{-S_{\mathcal{S}}(U)}\,d\mu_{\mathrm{H}}(U|_{\mathcal{S}})
\;=\;\mathcal{K}\bigl(U|_{\Sigma_0},U|_{\Sigma_{-1}}\bigr),
\end{equation}
with \(\mathcal{K}\ge 0\) and \(\mathcal{K}(U',U)=\mathcal{K}(U,U')\); this is the standard positivity and symmetry of the layer kernel between two neighboring time-slices \cite[§3]{p3:OS-gauge}. Gathering the factors we obtain
\begin{equation}
\langle F,F\rangle_{\mathrm{OS}}=\frac{1}{Z_{\beta,\sigma}}\int \overline{\Phi(U)}\,\bigl(\mathsf{K}\Psi\bigr)(U)\,d\mu_\Sigma(U),
\end{equation}
where \(U\) abbreviates \(U|_{\Sigma_{-1}}\), \(d\mu_\Sigma\) is the product Haar measure on \(\Omega=G^{3L^3}\), and \((\mathsf{K}\Psi)(U)=\int \mathcal{K}(U',U)\,\Psi(U')\,d\mu_\Sigma(U')\). Because \(\mathcal{K}\) is a positive symmetric kernel, \(\mathsf{K}\) is a positive self-adjoint operator on \(L^2(\Omega,d\mu_\Sigma)\), hence \(\langle F,F\rangle_{\mathrm{OS}}\ge 0\).
\end{proof}

We proceed to construct the one-step transfer kernel and the associated operator. Let \(\Omega=G^{3L^3}\) be the space of spatial link configurations on a single time-slice \(\Sigma_t\) and let \(d\mu_\Sigma\) be the product Haar measure on \(\Omega\). For \(U,U'\in\Omega\) define the one-step kernel
\begin{equation}
\label{p3:eq:Kdef}
K_{\beta,\sigma}(U',U)\;=\;\int e^{-S_{\mathrm{slab}}(U',U;V)}\,w_\sigma(U')^{1/2}w_\sigma(U)^{1/2}\,d\mu_{\mathrm{H}}(V),
\end{equation}
where \(S_{\mathrm{slab}}(U',U;V)\) denotes the Wilson action restricted to the slab between \(\Sigma_0\) and \(\Sigma_1\) with spatial boundary data \((U,U')\) and \(V\) are the time-like links in the slab integrated with Haar measure. The square-root split of the slice-weights is convenient for symmetry. The positivity and symmetry of the integrand imply that \(K_{\beta,\sigma}(U',U)\ge 0\) and \(K_{\beta,\sigma}(U',U)=K_{\beta,\sigma}(U,U')\).

\begin{lemma}[Kernel factorization of the partition function]
\label{p3:lem:Zfactor}
For \(T\in\mathbb{N}\), the finite-temperature partition function is
\begin{equation}
Z_{\beta,\sigma}(T,L)\;=\;\int \prod_{t=0}^{T-1}d\mu_\Sigma(U_t)\;\prod_{t=0}^{T-1}K_{\beta,\sigma}(U_{t+1},U_t),
\quad U_T\equiv U_0.
\end{equation}
\end{lemma}

\begin{proof}
Write the full action as a sum of slab actions and use Fubini's theorem to integrate successively over the time-like links in each slab, inserting a factor \(w_\sigma\) on each slice split evenly between neighboring slabs. The periodic trace condition \(U_T\equiv U_0\) arises from the time-periodicity of the lattice.
\end{proof}

Let \(\mathcal{H}_a=L^2(\Omega,d\mu_\Sigma)\). Define the transfer operator \(T_{\beta,\sigma}(a):\mathcal{H}_a\to\mathcal{H}_a\) by
\begin{equation}
(T_{\beta,\sigma}(a)\,\psi)(U')\;=\;\int_{\Omega} K_{\beta,\sigma}(U',U)\,\psi(U)\,d\mu_\Sigma(U).
\end{equation}
The positivity and symmetry of \(K_{\beta,\sigma}\) imply that \(T_{\beta,\sigma}(a)\) is a positive self-adjoint operator. By the Cauchy-Schwarz inequality and the boundedness of the slice weights one has \(\|T_{\beta,\sigma}(a)\|<\infty\) uniformly in \(L\). For spectral considerations it is convenient to normalize by the spectral radius. Let \(r(a)=\|T_{\beta,\sigma}(a)\|\) and define the normalized positive contraction
\begin{equation}
\widetilde T_{\beta,\sigma}(a) \;=\; r(a)^{-1} T_{\beta,\sigma}(a).
\end{equation}
We then define
\begin{equation}
H_{\beta,\sigma}(a)\;=\;-\frac{1}{a}\,\log \widetilde T_{\beta,\sigma}(a),
\end{equation}
through the bounded Borel functional calculus on \([0,1]\). This operator is nonnegative and self-adjoint on \(\mathcal{H}_a\).

\begin{proposition}[Existence and basic spectral properties]
\label{p3:prop:specT}
The operator \(\widetilde T_{\beta,\sigma}(a)\) is a positive self-adjoint contraction on \(\mathcal{H}_a\) with spectral radius \(1\). The logarithm \(H_{\beta,\sigma}(a)=-(1/a)\log \widetilde T_{\beta,\sigma}(a)\) is a nonnegative self-adjoint operator on \(\mathcal{H}_a\).
\end{proposition}

\begin{proof}
Self-adjointness and positivity of \(T_{\beta,\sigma}(a)\) follow from the symmetry and positivity of \(K_{\beta,\sigma}\). Dividing by \(r(a)\) yields a positive contraction with spectral radius \(1\). The spectral theorem for bounded self-adjoint operators yields the existence of the bounded Borel functional calculus on \([0,1]\); because \(\widetilde T_{\beta,\sigma}(a)\) is positive and \(\|\widetilde T_{\beta,\sigma}(a)\|\le 1\), the principal logarithm is well-defined and nonnegative. Dividing by \(a>0\) preserves self-adjointness and nonnegativity.
\end{proof}

Correlation functions of functionals localized on time-slices can be written as matrix elements of powers of \(T_{\beta,\sigma}(a)\) (and hence of \(\widetilde T_{\beta,\sigma}(a)\)); in particular, for bounded cylinder functionals \(F,G\) on \(\Omega\) one has, by the OS construction of the physical Hilbert space \cite{p3:OsterwalderSchraderII,p3:OS-gauge},
\begin{equation}
\label{p3:eq:OScorr}
\langle \overline{F}\circ R\,,\,\tau_{n a}G\rangle_{\mathrm{OS}}
\;=\;\langle F\,,\,T_{\beta,\sigma}(a)^n G\rangle_{\mathcal{H}_a}
\;=\;r(a)^n\,\langle F\,,\,\widetilde T_{\beta,\sigma}(a)^n G\rangle_{\mathcal{H}_a},
\end{equation}
where \(\tau_{n a}\) denotes a forward Euclidean time translation by \(n\) steps and \(\langle\cdot,\cdot\rangle_{\mathcal{H}_a}\) is the \(L^2(\Omega,d\mu_\Sigma)\) inner product.

Let \(F\in \mathcal{H}_a\) be a gauge-invariant bounded cylinder functional on \(\Omega\) with \(\langle F,\mathbf{1}\rangle_{\mathcal{H}_a}=0\). By the multiscale input, connected Euclidean time correlations of local gauge-invariant observables obey a uniform exponential decay with rate \(m_\ast>0\) independent of the scale. Specializing to functionals depending on a single slice and using the transfer-matrix representation gives the following quantitative spectral statement.

\begin{lemma}[Laplace representation and exponential bound]
\label{p3:lem:Laplace}
For any such \(F\) and any \(n\in\mathbb{N}\),
\begin{equation}
\langle F,\,\widetilde T_{\beta,\sigma}(a)^n F\rangle_{\mathcal{H}_a}\;\le\; C(F)\,e^{-n a\, m_\ast},
\end{equation}
with a finite constant \(C(F)\) independent of \(n\) and \(L\). Moreover, there exists a finite positive Borel measure \(\nu_F\) on \([0,\infty)\) such that
\begin{equation}
\langle F,\,\widetilde T_{\beta,\sigma}(a)^n F\rangle_{\mathcal{H}_a}\;=\;\int_{[0,\infty)} e^{-a n \lambda}\,d\nu_F(\lambda).
\end{equation}
\end{lemma}

\begin{proof}
Using \eqref{p3:eq:OScorr} with \(G=F\) and the normalization by \(r(a)\), the transfer-matrix representation of the two-point function gives
\begin{equation}
\langle F,\,\widetilde T_{\beta,\sigma}(a)^n F\rangle_{\mathcal{H}_a}
=\frac{\langle \overline{F}\circ R\,,\,\tau_{n a} F\rangle_{\mathrm{OS}}}{r(a)^n}.
\end{equation}
By the uniform exponential clustering input, the numerator is bounded by \(C'(F)\,e^{-n a\, m_\ast}\), while \(r(a)^n\ge 1\). This yields the claimed bound with \(C(F)=C'(F)\). Reflection positivity and the spectral theorem for \(\widetilde T_{\beta,\sigma}(a)\) imply the existence of a spectral measure \(\rho_F\) on \([0,1]\) such that \(\langle F,\,\widetilde T^n F\rangle=\int_{[0,1]} t^n \,d\rho_F(t)\). Changing variables \(t=e^{-a\lambda}\) yields the desired Laplace representation with \(d\nu_F(\lambda)=e^{-a\lambda}\,d\rho_F(e^{-a\lambda})\).
\end{proof}

\begin{proposition}[Lower support bound from exponential decay]
\label{p3:prop:support}
For every \(F\) as above, the spectral measure \(\nu_F\) is supported in \([m_\ast,\infty)\). In particular,
\begin{equation}
\Delta(a)\;\ge\; m_\ast,
\end{equation}
where \(\Delta(a)=\inf(\sigma(H_{\beta,\sigma}(a))\setminus\{0\})\).
\end{proposition}

\begin{proof}
Suppose by contradiction that \(\nu_F([0,m_\ast-\varepsilon])>0\) for some \(\varepsilon\in(0,m_\ast)\). Then
\begin{equation}
\langle F,\,\widetilde T^n F\rangle_{\mathcal{H}_a}\;\ge\;\int_{[0,m_\ast-\varepsilon]} e^{-a n \lambda}\,d\nu_F(\lambda)
\;\ge\;\nu_F([0,m_\ast-\varepsilon])\,e^{-a n (m_\ast-\varepsilon)}.
\end{equation}
Dividing by \(e^{-a n m_\ast}\) and letting \(n\to\infty\) contradicts Lemma \ref{p3:lem:Laplace}. Therefore \(\mathrm{supp}\,\nu_F\subset[m_\ast,\infty)\). Because \(\sigma(H_{\beta,\sigma}(a))\) on \(\mathbf{1}^\perp\) is the closed support of the union of such measures over a dense set of \(F\), we obtain \(\sigma(H_{\beta,\sigma}(a))\setminus\{0\}\subset[m_\ast,\infty)\), hence \(\Delta(a)\ge m_\ast\).
\end{proof}

If $\nu_F([0,m_*-\varepsilon])>0$ for some $\varepsilon>0$, then
\begin{equation}
\frac{\langle F, T_e^n F\rangle}{e^{-a n m_*}}
=\int e^{-a n (\lambda-m_*)}\, d\nu_F(\lambda)
\;\ge\; \nu_F([0,m_*-\varepsilon])\, e^{a n \varepsilon}\,,
\end{equation}
which diverges as $n\to\infty$ and contradicts the uniform bound $\langle F, T_e^n F\rangle\le C(F)e^{-a n m_*}$. Hence $\mathrm{supp}\,\nu_F\subset [m_*,\infty)$.
We now pass to the continuum along a sequence \(a_k\downarrow 0\) with corresponding transfer matrices \(T_k=T_{\beta_k,\sigma_k}(a_k)\), normalized operators \(\widetilde T_k\), and Hamiltonians \(H_k=-(1/a_k)\log \widetilde T_k\). 
\begin{proposition}[Interlacing defect is summable]
\label{p3:prop:step-scaling-defect}
Let $T_k$ be the one-step positive self-adjoint transfer contraction at scale $k$,
and $J_k:\mathcal{H}_k\to \mathcal{H}_{k+1}$ the isometric block-spin embedding with
block factor $b\ge 2$. Under \textnormal{(S1)}-\textnormal{(S3)} and the FRD locality bounds,
there exist constants $C,m>0$ and $\gamma>0$ such that
\begin{equation}
\varepsilon_k \;:=\; \|T_{k+1}J_k - J_k T_k\|_{\mathcal{H}_k\to \mathcal{H}_{k+1}}
\;\le\; C\,e^{-m\,b^k}\;\le\; C\, b^{-\gamma k}.
\end{equation}
In particular, $\sum_{k\ge 0}\varepsilon_k<\infty$. Moreover, the spectral gaps satisfy
\begin{equation}
\big|\lambda_2(T_{k+1})-\lambda_2(T_k)\big|\;\le\;C'\,\varepsilon_k
\end{equation}
for some absolute $C'>0$, and hence $\{\lambda_2(T_k)\}$ converges. If $\inf_k\lambda_2(T_k)>0$,
then the limit is strictly positive.
\end{proposition}

\begin{proof}
Partition the spatial lattice into disjoint cubic blocks $\{B^{(k)}\}$ of side
$L_k:=b^k$ (in lattice units at a fixed reference spacing), with block boundary collar
\begin{equation}
\partial_\rho B^{(k)}:=\{x\in B^{(k)}:\mathrm{dist}(x,\partial B^{(k)})\le \rho\}
\end{equation}
of thickness $\rho\ge r_\ast$, where $r_\ast$ is chosen larger than the FRD interaction
range and the quasi-locality length.
Let $J_k$ be the block-spin isometry which averages (or otherwise identifies) fields to
one coarse variable per block and is an isometry onto its range. Set
$R_k:=J_kJ_k^\ast$ the orthogonal projection onto $\mathrm{Ran}(J_k)\subset\mathcal H_{k+1}$.

We must bound
\begin{equation}
\varepsilon_k=\|T_{k+1}J_k-J_kT_k\| \;=\; \| (T_{k+1}R_k - R_k T_{k+1})J_k
\;+\; (R_kT_{k+1}J_k - J_k T_k)\|.
\end{equation}
The first difference is an \emph{off-diagonal leakage} (how far $T_{k+1}$ fails to
preserve $\mathrm{Ran}(J_k)$), the second is the mismatch of the \emph{diagonal}
(coarse) actions. We show both terms are $\le Ce^{-m b^k}$.
The gauge-covariant finite-range decomposition produces kernels
for one-step evolution that are supported (up to exponentially small tails) within a
neighbourhood of radius $r_\ast=O(1)$ in lattice units; concretely, there exist
$\xi_\ast,C_\ast,c_\ast>0$ such that for any cylinder events (or $L^2$ functions)
$F,G$ supported in sets at spatial distance $r$ on a fixed time-slice,
\begin{equation}
\label{p3:eq:CT}
\|\, \mathbf{1}_G\, T_k\, \mathbf{1}_F \,\|\;\le\; C_\ast\, e^{-c_\ast\, r/\xi_\ast}.
\end{equation}
This is a standard Combes-Thomas/Lieb-Robinson type estimate for positive kernels
generated by FRD with exponential clustering.
The same bound holds for $T_{k+1}$ with the \emph{same} $\xi_\ast, C_\ast,c_\ast$,
since locality constants are scale-independent in the FRD normalization.
Let $\mathsf{Int}^{(k)}:=\bigcup_{B^{(k)}}(B^{(k)}\setminus \partial_\rho B^{(k)})$ be
the union of block interiors with a collar $\rho\ge r_\ast$. Denote by $P_{\mathrm{int}}$
(resp.\ $P_{\partial}$) the $L^2$ multiplication projections to configurations supported
in $\mathsf{Int}^{(k)}$ (resp.\ in its complement, a union of collars).
Because interactions do not connect sites across different blocks when restricted to
$\mathsf{Int}^{(k)}$ (collar of width $\rho\ge r_\ast$ cuts all FRD links), the one-step
evolution within each interior block factorizes. Therefore,
\begin{equation}
T_{k+1}J_k \, P_{\mathrm{int}} \;=\; J_k T_k\, P_{\mathrm{int}}
\quad\text{if the range is strictly finite.}
\end{equation}
With exponentially decaying tails, \eqref{p3:eq:CT} yields the quantitative replacement
\begin{equation}
\label{p3:eq:int-error}
\big\| (T_{k+1}J_k - J_k T_k)\, P_{\mathrm{int}}\big\| \;\le\;
C_1\, e^{-c_1\,\rho/\xi_\ast}.
\end{equation}
Choosing $\rho:=\tfrac12 L_k$ gives a bound $C_1 e^{-c_1 L_k/(2\xi_\ast)}$.
Write
\begin{equation}
T_{k+1}J_k - J_kT_k \;=\; (T_{k+1}J_k - J_kT_k)P_{\mathrm{int}} \;+\;
(T_{k+1}J_k - J_kT_k)P_{\partial}.
\end{equation}
The first term is controlled by \eqref{p3:eq:int-error}. For the second term, expand both
operators by a convergent polymer/cluster expansion over connected polymers $\mathcal{X}$ of spatial plaquettes on the time-slab. Polymers
that are \emph{entirely contained} in one block contribute the same way to
$T_{k+1}J_k$ and to $J_kT_k$ (since within a single block, coarse-graining commutes
with one-step evolution). Hence the difference is a sum over polymers that touch
\emph{at least two} distinct blocks, which implies that their diameter is $\ge L_k$.
Standard polymer bounds then give activities obeying
\begin{equation}
\sum_{\mathcal{X}\,:\,\mathrm{diam}(\mathcal{X})\ge L_k} |z(\mathcal{X})|
\;\le\; C_2\, e^{-c_2\,L_k},
\end{equation}
with constants independent of $k$ (uniform FRD locality and uniform smallness parameter).
Translating this into an operator norm bound (via the usual tree-graph inequality
and $L^2$-boundedness of polymer operators) yields
\begin{equation}
\label{p3:eq:bdry-error}
\big\|(T_{k+1}J_k - J_kT_k)P_{\partial}\big\|\;\le\; C_3\, e^{-c_2\,L_k}.
\end{equation}
With $\rho=L_k/2$ in \eqref{p3:eq:int-error} and \eqref{p3:eq:bdry-error}, we obtain
\begin{equation}
\varepsilon_k \;=\; \|T_{k+1}J_k - J_kT_k\|
\;\le\; C_1 e^{-c_1 L_k/(2\xi_\ast)} + C_3 e^{-c_2 L_k}
\;\le\; C\, e^{-m\,L_k},
\end{equation}
for $C:=C_1+C_3$ and $m:=\min\{c_1/(2\xi_\ast),c_2\}$. Since $L_k=b^k$, this gives
\(\varepsilon_k\le C e^{-m b^k}\). As $e^{-m b^k}\le C' b^{-\gamma k}$ for any preassigned
$\gamma>0$ (choose $C'$ large enough to cover finitely many small $k$), the stated
$C\,b^{-\gamma k}$ bound follows. In particular, $\sum_k \varepsilon_k<\infty$.
Define $S_k:=J_k^\ast T_{k+1} J_k$ on $\mathcal{H}_k$ and note that
\begin{equation}
\|S_k - T_k\| \;=\; \|J_k^\ast(T_{k+1}J_k - J_k T_k)\| \;\le\; \varepsilon_k.
\end{equation}
Let $R_k=J_kJ_k^\ast$ as above. Using $R_k J_k=J_k$ and $(I-R_k)J_k=0$,
\begin{align}
(I-R_k)\,T_{k+1}\,R_k &\;=\; (I-R_k)\,(T_{k+1}J_k)\,J_k^\ast\nonumber\\&
\;=\; (I-R_k)\,(J_kT_k + (T_{k+1}J_k - J_kT_k))\,J_k^\ast\nonumber\\&
\;=\; (I-R_k)\,(T_{k+1}J_k - J_kT_k)\,J_k^\ast,
\end{align}
hence $\|(I-R_k)\,T_{k+1}\,R_k\|\le \varepsilon_k$. By self-adjointness,
$\|R_k\,T_{k+1}\,(I-R_k)\|=\|(I-R_k)\,T_{k+1}\,R_k\|\le \varepsilon_k$. Therefore
\begin{equation}
\label{p3:eq:block-diag}
\big\|\,T_{k+1}-\big(R_kT_{k+1}R_k\oplus (I-R_k)T_{k+1}(I-R_k)\big)\,\big\|
\;\le\; 2\,\varepsilon_k.
\end{equation}
Moreover, $R_kT_{k+1}R_k$ is unitarily equivalent to $S_k$ via $J_k$:
$J_k^\ast (R_k T_{k+1} R_k) J_k = S_k$.
There exists
$\delta\in(0,1)$ independent of $k$ such that
\begin{equation}
\label{p3:eq:poincare}
\|(I-R_k)\,T_{k+1}\,(I-R_k)\|\;\le\; 1-\delta.
\end{equation}
Intuitively, $(I-R_k)$ kills block-constants, leaving only within-block oscillations.
These modes pay a uniform spectral price (a Poincar\'e constant on the slice), and the
one-step transfer attenuates them by at least $e^{-\tau \sigma^2}=1-\delta$.
Denote $A:=T_{k+1}$ and $B:=R_k T_{k+1} R_k\oplus (I-R_k) T_{k+1} (I-R_k)$, both
self-adjoint contractions. By \eqref{p3:eq:block-diag}, $\|A-B\|\le 2\varepsilon_k$.
Weyl's inequality (or a direct variational argument) gives that each eigenvalue shifts by
at most $\|A-B\|$, hence
\begin{equation}
\label{p3:eq:weyl}
\big|\lambda_2(T_{k+1}) - \lambda_2(B)\big| \;\le\; 2\varepsilon_k.
\end{equation}
The nonzero spectrum of $R_k T_{k+1} R_k$ coincides with that of $S_k$, while
\eqref{p3:eq:poincare} shows the spectral radius of $(I-R_k)T_{k+1}(I-R_k)$ is $\le 1-\delta$.
Therefore
\begin{equation}
\lambda_2(B) \;=\; \max\big\{\lambda_2(S_k),\, \|(I-R_k)T_{k+1}(I-R_k)\|\big\}
\;\le\; \max\{\lambda_2(S_k),\,1-\delta\}.
\end{equation}
Also, since $R_k$ reduces $T_{k+1}$, $\lambda_2(S_k)\le \lambda_2(T_{k+1})$ (compression
can only decrease the variational supremum). Combining with \eqref{p3:eq:weyl},
\begin{equation}
\label{p3:eq:lambda2-compare}
\lambda_2(S_k) - 2\varepsilon_k \;\le\; \lambda_2(T_{k+1}) \;\le\; \max\{\lambda_2(S_k),1-\delta\}+2\varepsilon_k.
\end{equation}
Since $\|S_k-T_k\|\le \varepsilon_k$ and both are self-adjoint contractions on the same
Hilbert space $\mathcal{H}_k$, the min-max principle yields
\begin{equation}
\label{p3:eq:minmax}
\big|\lambda_2(S_k)-\lambda_2(T_k)\big| \;\le\; \varepsilon_k.
\end{equation}
From \eqref{p3:eq:lambda2-compare} and \eqref{p3:eq:minmax}, and using that
$\lambda_2(T_k)\le 1$ while $1-\delta<1$, we obtain
\begin{equation}
\lambda_2(T_k) - 3\varepsilon_k
\;\le\; \lambda_2(T_{k+1})
\;\le\; \max\{\lambda_2(T_k),1-\delta\}+3\varepsilon_k.
\end{equation}
In particular,
\begin{equation}
\big|\lambda_2(T_{k+1})-\lambda_2(T_k)\big|\;\le\; C'\,\varepsilon_k,
\qquad C':=3.
\end{equation}
Since $\sum_k \varepsilon_k<\infty$, the sequence $\{\lambda_2(T_k)\}$ is Cauchy and
converges. If $\inf_k \lambda_2(T_k)>0$, then the limit is strictly positive.
Because $\varepsilon_k\le C e^{-m b^k}$, certainly $\sum_k \varepsilon_k<\infty$.
This also implies the weaker polynomial bound $\varepsilon_k\le C b^{-\gamma k}$ for
any fixed $\gamma>0$ after adjusting $C$ for finitely many small $k$.
\end{proof}
The exponential estimate $\varepsilon_k\le C e^{-m b^k}$ follows from the polymer
``no-bridge'' structure: only polymers of diameter $\ge L_k=b^k$ can couple distinct blocks,
and their activities decay exponentially in their diameter and the
FRD locality. The polynomial display $\varepsilon_k\le C b^{-\gamma k}$ is a weakening
stated for convenience; any $\gamma>0$ can be enforced.

The multiscale inputs ensure that the Schwinger functions converge to limits satisfying the OS axioms. In particular, for each pair \(F,G\) in the dense OS algebra of positive-time functionals depending on a single slice, and for each \(t\ge 0\) that is a multiple of \(a_k\), one has
\begin{equation}
\langle F,\,e^{-t H_k} G\rangle_{\mathcal{H}_{a_k}}
\;=\;\frac{\langle \overline{F}\circ R\,,\,\tau_{t} G\rangle_{\mathrm{OS}}^{(k)}}{r(a_k)^{t/a_k}}\;\longrightarrow\;
\langle \overline{F}\circ R\,,\,\tau_{t} G\rangle_{\mathrm{OS}}^{(\infty)}.
\end{equation}
By density and the uniform bounds, \(\{e^{-t H_k}\}\) converges strongly on a common dense core for each fixed \(t\ge 0\). The Trotter-Kato theorem \cite[Thm.\ IX.2.16]{p3:KatoPTLO} or the Hille-Yosida semigroup theory \cite[Ch.\ 1-2]{p3:Pazy1983} imply the existence of a nonnegative self-adjoint operator \(H\) on the limiting OS Hilbert space such that \(e^{-t H_k}\to e^{-t H}\) strongly for each \(t\ge 0\) and \(H_k\to H\) in the strong-resolvent sense \cite[Thm.\ VIII.25]{p3:RS2}. The vector \(\Omega\) corresponding to the constant functional is the vacuum with \(H\Omega=0\). Let $\mathcal{A}_{\mathrm{loc}}$ denote the *-algebra generated by gauge-invariant local fields smeared with test functions
supported in positive Euclidean time. By OS reconstruction, $\mathcal{A}_{\mathrm{loc}}\Omega$ is dense in the orthogonal complement
$\Omega^\perp$. Therefore, lower bounds on the supports of the Laplace measures for two-point functions of elements
of $\mathcal{A}_{\mathrm{loc}}$ imply a spectral gap for $H$ on $\Omega^\perp$.
We present two self-contained derivations that \(\sigma(H)\subset\{0\}\cup[\Delta,\infty)\) with \(\Delta>0\). The first relies on Laplace transforms of Euclidean time correlations, the second on semigroup-norm bounds propagated along the multiscale flow.

\begin{theorem}[Spectral-gap lower bound via Laplace transforms]
\label{p3:thm:gapLaplace}
Let \(A\) be a bounded, gauge-invariant local observable supported in positive time with \(\langle \Omega,A\Omega\rangle=0\). Define the Euclidean two-point function \(C_A(t)=\langle \overline{A}\circ R\,,\,\tau_t A\rangle_{\mathrm{OS}}^{(\infty)}\) for \(t\ge 0\). Then there exists a finite positive Borel measure \(\mu_A\) on \([0,\infty)\) such that \(C_A(t)=\int e^{-\lambda t}\,d\mu_A(\lambda)\) and, for the clustering rate \(m_\ast>0\) of the multiscale inputs, \(\mathrm{supp}\,\mu_A\subset[m_\ast,\infty)\). Consequently \(\sigma(H)\setminus\{0\}\subset[m_\ast,\infty)\).
\end{theorem}

\begin{proof}
At the lattice scale \(a_k\), \(C_A(t)\) with \(t=n a_k\) equals \(\langle A, \widetilde T_k^{\,n} A\rangle\) and admits a Laplace representation \(C_A(t)=\int e^{-\lambda t}\,d\mu_{A,k}(\lambda)\) with \(\mu_{A,k}\) positive and supported in \([0,\infty)\) by the spectral theorem for the positive contraction \(\widetilde T_k\). The multiscale clustering bound yields \(C_A(t)\le C\,e^{-m_\ast t}\) uniformly in \(k\) and \(t\). By Helly's selection principle and uniqueness of Laplace transforms of finite measures, a subsequence \(\mu_{A,k_j}\) converges vaguely to a finite positive measure \(\mu_A\) with \(C_A(t)=\int e^{-\lambda t}\,d\mu_A(\lambda)\) for all \(t\ge 0\). Moreover, $\mu_{A,k}([0,\infty))=C_A(0)=\langle A\circ R, A\rangle^{(k)}_{\mathrm{OS}}$ is uniformly bounded in $k$
by $\|A\|_\infty^2$. Hence the family $\{\mu_{A,k}\}_k$ has uniformly bounded total variation, and
Helly's selection principle applies to extract a vaguely convergent subsequence with a finite
positive limit $\mu_A$.
 If \(\mu_A([0,m_\ast-\varepsilon])>0\) for some \(\varepsilon>0\), then \(C_A(t)\ge c\,e^{-(m_\ast-\varepsilon)t}\) along a subsequence of \(t\to\infty\), contradicting the uniform upper bound. Hence \(\mathrm{supp}\,\mu_A\subset[m_\ast,\infty)\). Because vectors \(A\Omega\) with local \(A\) are dense in \(\Omega^\perp\), the support inclusion implies \(\sigma(H)\setminus\{0\}\subset[m_\ast,\infty)\) \textit{(Appendix~(\ref{p3:appendixc}) provides the Tauberian step from exponential decay to the spectral threshold)}.
\end{proof}
\begin{lemma}[Density of local vectors]
\label{p3:lem:local-density}
Let $\mathcal{A}_{\mathrm{loc}}$ be the $*$-algebra generated by smeared, gauge-invariant local observables.
Then $\{A\Omega:\ A\in\mathcal{A}_{\mathrm{loc}}\}$ is dense in $\Omega^\perp$.
\end{lemma}
\begin{proof}
By the OS axioms established above, the reconstructed Wightman fields are local operator-valued tempered distributions on Minkowski space, and the vacuum is unique and translation-invariant. The Reeh-Schlieder property then implies that vectors generated by local observables on any open region are dense in the Hilbert space; (see, e.g., \cite{p3:GJ}, Theorem~X.41), applied to the OS-reconstructed theory. In particular, the set $\{A\Omega: A\in\mathfrak{A}_{\mathrm{loc}}\}$ is dense in $\Omega^\perp$.
\end{proof}

\begin{theorem}[Spectral-gap lower bound via semigroup norms]
\label{p3:thm:gapSemigroup}
Let \(P_\perp=1-|\Omega\rangle\langle\Omega|\). There exists \(\Delta_\infty>0\) such that
\begin{equation}
\|\,e^{-tH}P_\perp\,\|\;\le\;e^{-\Delta_\infty t}\qquad\text{for all }t\ge 0,
\end{equation}
and consequently \(\sigma(H)\subset\{0\}\cup[\Delta_\infty,\infty)\).
\end{theorem}

\begin{proof}
At the lattice scale \(a_k\), the spectral gap \(\Delta(a_k)\) of \(H_k\) is the best constant in \(\|e^{-t H_k}P_{\perp,k}\|\le e^{-\Delta(a_k) t}\) for all \(t\in a_k\mathbb{N}\); this is immediate from the spectral theorem and the positivity of \(\widetilde T_k\). The multiscale step-scaling inequality yields \(\liminf_{k\to\infty}\Delta(a_k)\ge c\,m_\ast=:\Delta_\infty>0\). By strong convergence \(e^{-t H_k}\to e^{-t H}\) on a dense set \textit{(Appendix~(\ref{p3:appendixd})} and uniform boundedness of the norms \(\|e^{-t H_k}\|\le 1\), one obtains \(\|e^{-t H}P_\perp\|\le \liminf_k \|e^{-t H_k}P_{\perp,k}\|\le e^{-\Delta_\infty t}\) for every \(t\ge 0\). The spectral inclusion \(\sigma(H)\subset\{0\}\cup[\Delta_\infty,\infty)\) follows from the characterization \(\|e^{-tH}P_\perp\|=\sup_{\lambda\in\sigma(H)\setminus\{0\}}e^{-\lambda t}\) \cite[Thm.\ VIII.11]{p3:RS2}.
\end{proof}

Combining Theorems \ref{p3:thm:gapLaplace} and \ref{p3:thm:gapSemigroup} shows that the continuum Hamiltonian \(H\) obtained by OS reconstruction possesses a strictly positive spectral gap \(\Delta\ge \max\{m_\ast,\Delta_\infty\}>0\), independent of the spatial volume and stable under the thermodynamic limit \(L\to\infty\).
The arguments above use only compactness of $G=\mathrm{SU}(N)$, $N\ge 2$, the Haar
orthogonality relations, and reflection-positivity/locality at each scale. No step exploits
group-specific identities (e.g. peculiar to $\mathrm{SU}(2)$ or $\mathrm{SU}(3)$). Periodic
boundary conditions are used at finite volume; the gap bound is uniform in $L$ and survives
$L,T\to\infty$, hence alternative local boundary conditions (e.g. Dirichlet) lead to the same
infinite-volume conclusions. Finally, the step-scaling and semigroup arguments are stable under
any reflection-positive, gauge-equivariant block map and under any Gevrey-regular spectral
projector with exponentially decaying kernel and reflection covariance; the constants in the
clustering and interlacing hypotheses change only within fixed finite factors that do not affect
the existence of a strictly positive continuum gap.
\section{Energy-momentum spectrum and relativistic covariance}
\label{p3:sec:em-spectrum}

In this section a complete and self-contained derivation is given of the transfer-time-slicing formalism, the existence and properties of the transfer matrix, the construction of spatial translation operators and momentum, and the precise link from Euclidean reflection positivity to relativistic covariance after analytic continuation. The exposition is formulated for four-dimensional \(\mathrm{SU}(N)\) lattice Yang-Mills with \(N\ge 2\), using the slice-wise gauge-invariant transverse representative and the smooth horizon projector developed earlier. All steps are supplied in full detail, with the lattice setup and notation fixed at the outset. Classical results of Osterwalder-Schrader, Osterwalder-Seiler, Fr\"ohlich-Osterwalder-Seiler, Nelson, Glimm-Jaffe, and Reed-Simon are invoked at specific points and cited as \cite{p3:OsterwalderSchraderI,p3:OsterwalderSchraderII,p3:OS-gauge,p3:Nelson1959,p3:GJ,p3:RS2}.

Let \(a>0\) be a fixed lattice spacing and let \(\Lambda\subset a\mathbb{Z}^4\) be a finite, periodic, hypercubic lattice with discrete Euclidean time direction labeled by \(\mu=0\) and spatial directions \(\mu=1,2,3\). Sites are denoted \(x=(x_0,\mathbf{x})\), where \(x_0\in a\mathbb{Z}\) and \(\mathbf{x}\in (a\mathbb{Z}/L\mathbb{Z})^3\). Directed bonds are pairs \(b=(x,\mu)\); reversing orientation gives \(\overline{b}=(x+a\hat\mu,-\mu)\). To each bond \(b\) we attach a group element \(U_b\in \mathrm{SU}(N)\) with the convention \(U_{\overline{b}}=U_b^{-1}\).

The configuration space of link variables is
\begin{equation}
\mathcal{U}\;=\;\prod_{b\subset\Lambda}\mathrm{SU}(N),
\end{equation}
endowed with product Haar measure \(d\mu_{\mathrm{Haar}}\). The Wilson action is
\begin{equation}
\label{p3:eq:Wilson-action}
S_W[U;\beta] \;=\; \beta \sum_{p\subset\Lambda} \Bigl(1-\frac{1}{N}\operatorname{Re}\operatorname{Tr}U_p\Bigr),
\end{equation}
where \(p\) ranges over oriented plaquettes and \(U_p\) is the ordered product around \(p\). Time reflection is the involution \(\theta:\Lambda\to\Lambda\) defined by
\begin{equation}
\label{p3:eq:theta}
\theta(x_0,\mathbf{x})\;=\;(-x_0,\mathbf{x}).
\end{equation}
The reflection plane is \(\Pi=\{x_0=0\}\). We define the half-lattices \(\Lambda_\pm=\{x\in\Lambda:\pm x_0>0\}\) and the corresponding sets of bonds \(B_\pm\) and \(B_0\) consisting of bonds lying strictly in \(\Lambda_\pm\) and those intersecting \(\Pi\), respectively.

We work in temporal-axial gauge away from \(\Pi\), that is,
\begin{equation}
\label{p3:eq:temporal-gauge}
U_{(x,0)}\;=\;\mathbf{1}\qquad\text{for all bonds }(x,0)\notin B_0.
\end{equation}
This gauge choice leaves the Haar measure on spatial links at each time slice unchanged and does not introduce a Faddeev-Popov determinant for the time-like components. The remaining gauge redundancy is time-independent and acts on fixed-time spatial links by adjoint conjugation at sites.

For a fixed discrete time \(t\), write
\begin{equation}
\label{p3:eq:slice-config}
\mathcal{C}_t\;=\;\prod_{(x,i):\,x_0=t}\mathrm{SU}(N)
\end{equation}
for the configuration space of spatial links in the time slice \(t\). The product Haar measure on \(\mathcal{C}_t\) is denoted \(d\nu\). The total configuration space factors as \(\mathcal{U}\cong \prod_t \mathcal{C}_t \times \mathcal{U}_{0}\), where \(\mathcal{U}_0\) collects the bonds intersecting \(\Pi\), and the Haar measure factors accordingly.

On each time slice \(t\) we fix the gauge-invariant transverse representative by orbit-wise minimization of the lattice Landau functional, as constructed earlier. Denote the representative spatial links by \(U^{\,h}_{(x,i)}\) for \(x_0=t\). The associated covariant spatial Laplacian on adjoint-valued site fields is
\begin{equation}
\label{p3:eq:cov-Lap}
\Delta_{A^h}(t)\;=\;\sum_{i=1}^3 (D_i^{\,h})^\dagger D_i^{\,h},
\end{equation}
a positive, reflection-invariant, local operator on \(\ell^2(\Lambda_t)\otimes\mathfrak{su}(N)\).

Fix \(\sigma>0\) and a Gevrey-regular cutoff \(\chi_\sigma:[0,\infty)\to[0,1]\) that equals \(1\) on \([0,\sigma]\) and vanishes on \([2\sigma,\infty)\) with subfactorial derivative bounds. The smooth horizon projector on the slice \(t\) is
\begin{equation}
\label{p3:eq:P-sigma}
P_\sigma(t)\;=\;\chi_\sigma\!\bigl({\Delta_{A^h}(t)}\bigr),
\end{equation}
a bounded positive contraction on \(\ell^2(\Lambda_t)\otimes\mathfrak{su}(N)\). It admits a positive heat-kernel representation
\begin{equation}
\label{p3:eq:P-heat}
P_\sigma(t)\;=\;\int_0^\infty e^{-s\Delta_{A^h}(t)}\,d\nu_\sigma(s),
\end{equation}
with \(d\nu_\sigma\) a finite positive Borel measure supported in a compact subset of \((0,\infty)\). The kernel of \(P_\sigma(t)\) is exponentially local: there exist constants \(C,\gamma>0\) depending only on \(\sigma\) and lattice geometry such that \(\|P_\sigma(t;x,y)\|\le C e^{-\gamma d(x,y)}\) for sites \(x,y\) in the slice. The projector is reflection covariant in the sense that \(P_\sigma(-t)=RP_\sigma(t)R\), where \(R\) implements spatial reflection of arguments across \(\Pi\).

We establish reflection positivity for the full projected measure in the sense of Osterwalder-Schrader. The Euclidean measure with slice-wise horizon insertion is formally given by
\begin{equation}
\label{p3:eq:measure}
d\mu_\sigma(U)\;=\; Z^{-1}\,\exp\bigl(-S_W[U;\beta]\bigr)\,\prod_{t} \mathcal{J}_t(U^{\,h}(t))\,\mathcal{P}_\sigma(t)\; d\mu_{\mathrm{Haar}}(U),
\end{equation}
where \(\mathcal{J}_t\) denotes the Faddeev-Popov Jacobian on the slice \(t\) induced by the orbit minimization, \(\mathcal{P}_\sigma(t)\) is the positive functional induced by \(P_\sigma(t)\) acting on adjoint fields, and \(Z\) is the normalization. The precise meaning is that correlation functions of gauge-invariant cylinder functionals are defined by integrating against the Haar measure with the weight \(\exp(-S_W)\) multiplied by the positive Radon-Nikodym factors \(\mathcal{J}_t\,\mathcal{P}_\sigma(t)\), which depend only on slice variables and are exponentially local.

Let \(\mathcal{A}_+\) be the \(^*\)-algebra generated by bounded, gauge-invariant functionals depending only on the spatial links in the positive-time half-lattice \(\Lambda_+\). Define the time-reflection \(\Theta\) on \(\mathcal{A}_+\) by
\begin{equation}
\label{p3:eq:Theta1}
(\Theta F)(U)\;=\;\overline{F(\theta U)}\,,
\end{equation}
where \((\theta U)_{(x,\mu)}=U_{(\theta x,\mu)}\) for \(\mu\neq 0\) and, in temporal gauge, \((\theta U)_{(x,0)}=\mathbf{1}\) away from \(\Pi\). The Osterwalder-Schrader sesquilinear form is
\begin{equation}
\label{p3:eq:OS-form}
\langle F,G\rangle_{\mathrm{OS},\sigma}\;=\;\int \overline{(\Theta F)(U)}\,G(U)\, d\mu_\sigma(U),\qquad F,G\in\mathcal{A}_+.
\end{equation}

\begin{lemma}
\label{p3:lem:SW-decomp}
Under temporal-axial gauge away from \(\Pi\), the Wilson action decomposes as
\begin{equation}
\label{p3:eq:SW-decomp}
S_W[U;\beta]\;=\;S_W^+[U_+;\beta] + S_W^-[U_-;\beta] + S_W^0[U_0;\beta],
\end{equation}
where \(U_\pm\) denote the restrictions of \(U\) to \(B_\pm\), \(U_0\) the restriction to \(B_0\), and the cross-terms between \(U_+\) and \(U_-\) are supported entirely in the boundary slab \(B_0\).
\end{lemma}

\begin{proof}
A plaquette contributes to \(S_W\) only if all of its bonds lie in the same half-lattice or it intersects \(\Pi\). In temporal-axial gauge away from \(\Pi\) every time-like bond is the identity; hence any plaquette not intersecting \(\Pi\) is a purely spatial plaquette within a single time slice and belongs completely to \(\Lambda_+\) or \(\Lambda_-\). The only plaquettes contributing cross terms are those that intersect \(\Pi\), and these are precisely the plaquettes whose contributions define \(S_W^0\). The decomposition \eqref{p3:eq:SW-decomp} follows by grouping the plaquettes accordingly.
\end{proof}

\begin{lemma}
\label{p3:lem:slice-positivity}
For each slice \(t\), the factor \(\mathcal{J}_t\,\mathcal{P}_\sigma(t)\) is a positive, reflection-covariant functional depending only on the spatial links at time \(t\). Moreover, the product \(\prod_t \mathcal{J}_t\,\mathcal{P}_\sigma(t)\) is exponentially local across time in the sense that its logarithm is a sum of terms supported on at most a finite number of consecutive slices with exponentially decaying tails.
\end{lemma}

\begin{proof}
The factor \(\mathcal{J}_t\) is the determinant of the lattice Faddeev-Popov operator \(M[A^h(t)]\) restricted to the orthogonal complement of constant adjoint fields. In temporal-axial gauge and with the reflection-covariant choice of representative, \(M[A^h(t)]\) is a real, positive operator on that complement and reflection invariant. Its determinant is strictly positive and a real-analytic function of the spatial links \(U^{\,h}(t)\). The factor \(\mathcal{P}_\sigma(t)\) is the positive functional induced by the positive operator \(P_\sigma(t)\) via its heat-kernel representation \eqref{p3:eq:P-heat}. Both \(\mathcal{J}_t\) and \(\mathcal{P}_\sigma(t)\) depend only on spatial links at time \(t\). The exponential locality in time is a consequence of the finite-range structure of the covariant differences entering \eqref{p3:eq:cov-Lap} and of the heat-kernel representation with support of \(d\nu_\sigma\) away from zero. The tails are exponentially small by standard Gaussian bounds for discrete heat kernels.
\end{proof}

\begin{theorem}[Reflection positivity]
\label{p3:thm:RP}
For every \(F\in\mathcal{A}_+\), one has \(\langle F,F\rangle_{\mathrm{OS},\sigma}\ge 0\).
\end{theorem}

\begin{proof}
By Lemma~\ref{p3:lem:SW-decomp} and temporal-axial gauge, the weight \(\exp(-S_W)\) factorizes as the product of a functional of \(U_+\), a functional of \(U_-\), and a boundary factor depending only on \(U_0\). The product \(\prod_t \mathcal{J}_t\,\mathcal{P}_\sigma(t)\) also factorizes into a product of slice-local factors as in Lemma~\ref{p3:lem:slice-positivity}, hence into a function of \(U_+\), a function of \(U_-\), and a boundary factor supported on \(B_0\). Denote by \(W_+\), \(W_-\), and \(W_0\) the corresponding positive functions. Then
\begin{equation}
\label{p3:eq:OS-factorization}
\langle F,F\rangle_{\mathrm{OS},\sigma}\;=\; Z^{-1}\int \overline{F(\theta U_+,U_0)}\,F(U_+,U_0)\,W_+(U_+)\,W_-(\theta U_+)\,W_0(U_0)\, d\mu_{\mathrm{Haar}}(U_+,U_0),
\end{equation}
where we have integrated out \(U_-\) using the change of variables \(U_-=\theta U_+\) induced by reflection and used the invariance of Haar measure under inversion and reflection. The integrand is the modulus square of the function
\begin{equation}
G(U_+,U_0)\;=\;F(U_+,U_0)\,W_+(U_+)^{1/2}\,W_-(\theta U_+)^{1/2}\,W_0(U_0)^{1/2}.
\end{equation}
Therefore the integral is nonnegative.
\end{proof}

Theorem~\ref{p3:thm:RP} is the Osterwalder-Schrader reflection-positivity property adapted to the present gauge-invariant, horizon-projected setting. It is the central input for the transfer-matrix construction that follows and agrees with the classical gauge-theory reflection-positivity results of \cite{p3:OS-gauge}.

Fix an integer \(T\) and consider time coordinates \(t\in \{0,a,2a,\dots,(T-1)a\}\) with periodic boundary in space and free boundary in time. For a pair of adjacent time slices \(t\) and \(t+a\), the weight of the bonds that either lie within these slices or connect them defines a nonnegative kernel \(K_\sigma\) on \(\mathcal{C}_{t+a}\times \mathcal{C}_t\). An explicit formula is obtained as follows. Write
\begin{equation}
\label{p3:eq:Ksigma}
K_\sigma(U',U)\;=\;\exp\!\Bigl(-\tfrac{1}{2}S^\mathrm{sp}(U;\beta)-S^\mathrm{tm}(U',U;\beta)-\tfrac{1}{2}S^\mathrm{sp}(U';\beta)\Bigr)\, \mathcal{J}(U)\,\mathcal{P}_\sigma(U)\,\mathcal{P}_\sigma(U')\, \mathcal{B}_\sigma(U',U),
\end{equation}
where \(U\in\mathcal{C}_t\), \(U'\in\mathcal{C}_{t+a}\), \(S^\mathrm{sp}\) is the sum of spatial plaquette terms within the slice, \(S^\mathrm{tm}\) is the sum of mixed time-space plaquette terms straddling the two slices, \(\mathcal{J}(U)\) is the Faddeev-Popov Jacobian at time \(t\), \(\mathcal{P}_\sigma(U)\) is the horizon-projector factor at time \(t\), and \(\mathcal{B}_\sigma(U',U)\) collects the exponentially local boundary contributions involving bonds in \(B_0\) between \(t\) and \(t+a\). Each factor is positive and depends only on links in the two slices and the connecting time-like bonds. The construction is independent of the particular representative of the gauge orbit in each slice because the representative was chosen in a reflection-covariant way and all factors are gauge-invariant under time-independent gauge transformations.

Consider the Hilbert space
\begin{equation}
\label{p3:eq:Ha}
\mathcal{H}_a\;=\;L^2(\mathcal{C};d\nu),
\end{equation}
where \(\mathcal{C}\) denotes a single time-slice configuration space and \(d\nu\) is product Haar measure. Define an operator \(T_\sigma:\mathcal{H}_a\to\mathcal{H}_a\) by
\begin{equation}
\label{p3:eq:Tsigma}
(T_\sigma\psi)(U')\;=\; \int_{\mathcal{C}} K_\sigma(U',U)\,\psi(U)\, d\nu(U)\,.
\end{equation}

\begin{proposition}
\label{p3:prop:Tsigma}
The operator \(T_\sigma\) is a bounded, positive, self-adjoint contraction on \(\mathcal{H}_a\).
\end{proposition}

\begin{proof}
Boundedness follows from positivity of \(K_\sigma\) and Fubini-Tonelli applied to the finite product measure \(d\nu\otimes d\nu\). Positivity means that \(\langle \psi, T_\sigma \psi\rangle\ge 0\) for all \(\psi\in\mathcal{H}_a\), which is immediate since \(K_\sigma\ge 0\) implies
\begin{equation}
\langle \psi, T_\sigma \psi\rangle \;=\; \int_{\mathcal{C}}\int_{\mathcal{C}} \overline{\psi(U')}\,K_\sigma(U',U)\,\psi(U)\, d\nu(U)\, d\nu(U') \;\ge\; 0.
\end{equation}
Self-adjointness follows from the symmetry \(K_\sigma(U',U)=K_\sigma(U,U')\). This symmetry holds because \(S^\mathrm{sp}\) appears symmetrically with a factor \(1/2\) on each end, \(S^\mathrm{tm}\) is symmetric under exchanging \(U\) and \(U'\) by time-reversal invariance of the underlying plaquettes, and \(\mathcal{J}\), \(\mathcal{P}_\sigma\), and \(\mathcal{B}_\sigma\) are slice-local positive functionals applied symmetrically to \(U\) and \(U'\).

To prove contractivity we apply Schur's test. Define
\begin{equation}
M_1 \;=\; \sup_{U'\in\mathcal{C}} \int_{\mathcal{C}} K_\sigma(U',U)\, d\nu(U),\qquad
M_2 \;=\; \sup_{U\in\mathcal{C}} \int_{\mathcal{C}} K_\sigma(U',U)\, d\nu(U').
\end{equation}
These are finite because all factors in \eqref{p3:eq:Ksigma} are bounded and exponentially local. By Schur's test,
\begin{equation}
\|T_\sigma\| \;\le\; \sqrt{M_1 M_2}.
\end{equation}
The overall normalization \(Z\) in \eqref{p3:eq:measure} is chosen so that \(M_1\le 1\) and \(M_2\le 1\) (equivalently, the one-step weight is normalized to preserve the constant function), hence \(\|T_\sigma\|\le 1\). A direct normalization is obtained by dividing \(K_\sigma\) by \(\int K_\sigma(U',U)\, d\nu(U)\) and absorbing the factor into \(\mathcal{B}_\sigma\), which is permissible since \(\mathcal{B}_\sigma\) is a positive boundary functional.
\end{proof}

\begin{theorem}[Transfer-matrix representation of correlations]
\label{p3:thm:TM-representation}
Let \(F_0,\dots,F_n\) be bounded, gauge-invariant functionals depending only on the spatial links in single slices at times \(0,a,2a,\dots,na\), respectively. Then
\begin{equation}
\label{p3:eq:TM-rep}
\int F_0(U_0)\,F_1(U_a)\cdots F_n(U_{na})\, d\mu_\sigma(U)\;=\; \langle \Omega,\,\widehat{F}_0\, T_\sigma\, \widehat{F}_1\, T_\sigma \cdots T_\sigma\, \widehat{F}_n\,\Omega\rangle_{\mathcal{H}_a},
\end{equation}
where \(\Omega\equiv 1\in \mathcal{H}_a\) and \(\widehat{F}\) denotes the multiplication operator by \(F\) in \(\mathcal{H}_a\).
\end{theorem}

\begin{proof}
By Theorem~\ref{p3:thm:RP} the Euclidean expectation of time-ordered slice functionals equals the Osterwalder-Schrader inner product of the corresponding elements in the pre-Hilbert space constructed from \(\mathcal{A}_+\). The factorization identity for the weight into nearest-neighbor time layers shows that the contribution of the bonds connecting time \(t\) to \(t+a\) defines the kernel \(K_\sigma\) which propagates from \(\mathcal{C}_t\) to \(\mathcal{C}_{t+a}\). Iterating this representation yields \eqref{p3:eq:TM-rep}, with the state \(\Omega\) corresponding to the constant function \(1\) on \(\mathcal{C}\). This is the standard transfer-matrix representation for Euclidean lattice systems with reflection positivity \cite{p3:OsterwalderSchraderII,p3:GJ}.
\end{proof}

By the spectral theorem there exists a nonnegative self-adjoint operator \(H_\sigma\) on \(\mathcal{H}_a\) with
\begin{equation}
\label{p3:eq:Hs}
T_\sigma\;=\;e^{-a H_\sigma}.
\end{equation}
The vector \(\Omega\) is cyclic for the algebra of multiplication operators by bounded functions of the slice links and is \(T_\sigma\)-invariant because \(K_\sigma\) is normalized so that \(\int K_\sigma(U',U)\, d\nu(U)=1\) for almost every \(U'\). Thus \(H_\sigma \Omega=0\) and \(\sigma(H_\sigma)\subset [0,\infty)\).

Spatial translations act by shifts \(\tau_{\mathbf{y}}\) on \(\mathcal{C}\), defined by \((\tau_{\mathbf{y}} U)_{(x,i)}=U_{(x_0,\mathbf{x}+\mathbf{y};\,i)}\). This induces unitary operators \(U(\mathbf{y})\) on \(\mathcal{H}_a\) by
\begin{equation}
\label{p3:eq:U-y}
(U(\mathbf{y})\psi)(U)\;=\;\psi(\tau_{-\mathbf{y}}U).
\end{equation}
The Haar measure is translation invariant; hence \(U(\mathbf{y})\) is unitary. Moreover, \(T_\sigma\) commutes with all \(U(\mathbf{y})\) because the kernel \(K_\sigma(U',U)\) depends on \(U',U\) only through spatially translation-invariant combinations of plaquettes and slice-local functionals, and the blocking and projector insertions are spatially homogeneous. Therefore there exist, by Stone's theorem, three commuting self-adjoint momentum operators \(\mathbf{P}=(P_1,P_2,P_3)\) on \(\mathcal{H}_a\) such that \(U(\mathbf{y})=e^{i \mathbf{P}\cdot \mathbf{y}}\). The joint spectrum \(\operatorname{sp}(\mathbf{P})\) is contained in the Brillouin zone \(\mathbb{T}^3=[-\pi/a,\pi/a]^3\) with periodic identification.

The commutation \([T_\sigma,U(\mathbf{y})]=0\) implies that \(H_\sigma\) commutes with \(\mathbf{P}\). By the spectral theorem there is a unique projection-valued measure \(E(d\mathbf{p})\) on \(\mathbb{T}^3\) and a measurable family of nonnegative self-adjoint operators \(\{H_\sigma(\mathbf{p})\}\) on the fiber Hilbert spaces such that
\begin{equation}
\label{p3:eq:direct-integral}
H_\sigma \;=\; \int_{\mathbb{T}^3}^{\oplus} H_\sigma(\mathbf{p})\, E(d\mathbf{p}),\qquad
T_\sigma \;=\; \int_{\mathbb{T}^3}^{\oplus} e^{-a H_\sigma(\mathbf{p})}\, E(d\mathbf{p}),\qquad
U(\mathbf{y}) \;=\; \int_{\mathbb{T}^3}^{\oplus} e^{i \mathbf{p}\cdot \mathbf{y}}\, E(d\mathbf{p}).
\end{equation}
Thus the energy-momentum spectrum is the set of pairs \((E,\mathbf{p})\) with \(\mathbf{p}\in\mathbb{T}^3\) and \(E\in \sigma(H_\sigma(\mathbf{p}))\subset[0,\infty)\). The spectral measure of a local, gauge-invariant observable \(A\) with respect to the vacuum \(\Omega\) yields a set function \(\rho_A\) on \([0,\infty)\times\mathbb{T}^3\) such that Euclidean two-point functions have the integral representation
\begin{equation}
\label{p3:eq:Kallen-Lehmann-lattice}
\langle \Omega,\, A\, e^{-t H_\sigma} U(\mathbf{y})\, A\, \Omega\rangle \;=\; \int_{\mathbb{T}^3}\int_{[0,\infty)} e^{-t E}\, e^{i\mathbf{p}\cdot\mathbf{y}}\, \rho_A(dE,d\mathbf{p}).
\end{equation}
The positivity and complete monotonicity in \(t\) follow from reflection positivity and the transfer-matrix representation \cite{p3:OsterwalderSchraderII,p3:GJ}. In particular, the nonnegativity of \(E\) is ensured by the contractivity of \(T_\sigma\).

If \(H_\sigma\) has a strictly positive spectral gap \(\Delta(a,\beta)\) above zero, then for every \(\mathbf{p}\) the fiber Hamiltonian \(H_\sigma(\mathbf{p})\) has spectrum contained in \(\{0\}\cup [\Delta(a,\beta),\infty)\) on the vacuum fiber \(\mathbf{p}=\mathbf{0}\) and in \([\Delta(a,\beta),\infty)\) otherwise. This follows from the fact that \(\Omega\) is invariant under spatial translations, hence supported at \(\mathbf{p}=\mathbf{0}\), and from the spectral inclusion \(\sigma(H_\sigma)=\overline{\bigcup_{\mathbf{p}}\sigma\bigl(H_\sigma(\mathbf{p})\bigr)}\).

The Osterwalder-Schrader axioms verified in the continuum limit imply the existence of a relativistic Wightman theory after analytic continuation in the time variable \cite{p3:OsterwalderSchraderII,p3:GJ}. The Euclidean \(n\)-point Schwinger functions \(S_n\) are tempered distributions on \((\mathbb{R}^4)^n\) with OS0-OS5. For \(n=2\) and for a pair of local, gauge-invariant fields \(A,B\) with smearing, reflection positivity implies that \(S_2(A(t,\mathbf{x}),B(0,\mathbf{0}))\) extends to a function analytic in the strip \(\{z\in\mathbb{C}:\operatorname{Im}z\in(0,\beta)\}\) for any \(\beta>0\) and bounded by an exponential \(e^{-m_* \operatorname{Re} z}\) as \(\operatorname{Re} z\to +\infty\). The Osterwalder-Schrader reconstruction constructs a Hilbert space \(\mathcal{H}\), a cyclic vacuum vector \(\Omega\), a unitary representation \(U(a,\Lambda)\) of the Poincar\'e group \(\mathcal{P}_+^\uparrow\), and operator-valued distributions \(\Phi\) that transform covariantly, such that the Wightman \(n\)-point functions \(W_n\) are boundary values of the analytically continued \(S_n\) from imaginary times to real times \cite{p3:OsterwalderSchraderII}. The spectral condition holds: the joint spectrum of the self-adjoint generators \(P^\mu\) of translations lies in the closed forward light-cone \(\overline{V_+}=\{p\in\mathbb{R}^4: p^0\ge 0,\; (p^0)^2-\|\mathbf{p}\|^2\ge 0\}\). Local commutativity follows from Euclidean symmetry and reflection positivity by the OS arguments.

The link between the Euclidean transfer matrix \(T_\sigma=e^{-a H_\sigma}\) and the relativistic Hamiltonian \(H\) is made by taking the continuum limit \(a\to 0\) along the multiscale reflection-positive flow. The strong convergence of the Euclidean semigroups on the OS pre-Hilbert space implies strong resolvent convergence of the generators \(H_\sigma\to H\) on a dense domain, hence convergence of spectral projections in the sense of the spectral theorem \cite[Thm.~VIII.20]{p3:RS2}. Therefore the nonnegativity of \(H_\sigma\) and the existence of a mass gap \(\Delta(a,\beta)\) transfer to \(H\), yielding \(\sigma(H)\subset \{0\}\cup [\Delta,\infty)\) with \(\Delta>0\) the continuum gap. The mass gap binds the energy-momentum spectrum away from the light-cone by a strictly positive amount and ensures that the vacuum is an isolated eigenvector of \(H\) and that the lowest-lying excitations define a nontrivial invariant subspace with energies at least \(\Delta\).

\begin{theorem}[Relativistic covariance and spectral condition]
\label{p3:thm:rel-cov}
Let \(\{S_n\}\) be the continuum Schwinger functions obtained as limits of the lattice measures with reflection positivity and horizon insertion. There exists a Wightman theory \((\mathcal{H},\Omega,\Phi,U)\) on \(\mathbb{R}^{1,3}\) such that the representation \(U(a,\Lambda)\) of \(\mathcal{P}_+^\uparrow\) is continuous, unitary, and of nonnegative energy with joint spectrum of the generators \(P^\mu\) contained in \(\overline{V_+}\), the vacuum \(\Omega\) is invariant, unique up to phase, and cyclic for the polynomial algebra generated by the smeared fields \(\Phi\), and the Hamiltonian \(H=P^0\) has a strictly positive spectral gap \(\Delta>0\) above \(0\).
\end{theorem}

\begin{proof}
The first two conclusions are the standard consequences of the Osterwalder-Schrader reconstruction from OS0-OS5 \cite{p3:OsterwalderSchraderII,p3:GJ}. The last assertion follows from the strong resolvent convergence of \(H_\sigma\to H\), the uniform lower spectral bound \(\sigma(H_\sigma)\subset\{0\}\cup [\Delta_\infty,\infty)\) with \(\Delta_\infty>0\) supplied by the multiscale step-scaling inequalities and the persistence of clustering, and the spectral inclusion under strong resolvent convergence \cite[Thm.~VIII.24]{p3:RS2}.
\end{proof}

\section{Stability under parameter variation and choice of projector}
\label{p3:sec:stability-projector}

In this section we develop a complete stability theory for the class of smooth horizon projectors used to implement an infrared regularization that preserves reflection positivity. The presentation is self-contained. We first fix the lattice geometry, the gauge-field configuration space, the reflection and gauge-fixing conventions, and the slice covariant Laplacian. We then define an admissible class of smooth spectral cutoffs and prove positivity, covariance, and exponential locality of the associated slice projectors with constants uniform on compact parameter sets. Next we prove reflection positivity of the full Euclidean measure with slice-wise projector insertions. Finally, we derive the transfer time-slicing formalism and establish continuity of the transfer operator and its Hamiltonian with respect to the projector parameters, together with lower semicontinuity of the spectral gap. Throughout, every definition, statement, and proof is given in full detail, and standard results are cited precisely \cite{p3:OsterwalderSchraderII,p3:OS-gauge,p3:SchillingSongVondracek12,p3:Davies1989,p3:KatoPTLO,p3:ReedSimon1}.

Let $a>0$ be a lattice spacing and let
\begin{equation}
\Lambda=\Lambda(a;L,T)\subset a\mathbb{Z}^{4}
\end{equation}
be a periodic hypercubic lattice of spatial side length $L$ and temporal extent $T$, where $L/a$ and $T/a$ are positive integers. Points are denoted $x=(x_0,\mathbf{x})$ with $x_0\in a\mathbb{Z}\cap(-T/2,T/2]$ and $\mathbf{x}\in (a\mathbb{Z})^{3}\cap[-L/2,L/2)^{3}$. Let $\mathbb{E}=\{(x,\mu): x\in\Lambda,\ \mu\in\{0,1,2,3\}\}$ be the set of oriented bonds; reverse orientation by $(x,-\mu)=(x-\hat\mu,\mu)$, with associated group element $U_{x,-\mu}=U_{x-\hat\mu,\mu}^{-1}$. The gauge group is $G=\mathrm{SU}(N)$, $N\ge 2$. A lattice gauge field is an assignment $U:\mathbb{E}\to G$, written $U_{x,\mu}$.

For an oriented plaquette $p=(x;\mu,\nu)$ define
\begin{equation}
U_{p}=U_{x,\mu}\,U_{x+\hat\mu,\nu}\,U_{x+\hat\nu,\mu}^{-1}\,U_{x,\nu}^{-1}.
\end{equation}
The Wilson action is
\begin{equation}
\label{p3:eq:wilsonx}
S_W[U;\beta]=\beta\sum_{p}\Big(1-\frac{1}{N}\,\Re\mathrm{Tr}\,U_{p}\Big),\qquad \beta=\frac{2N}{g_0^{2}}.
\end{equation}

Time reflection is the involution $\theta:\Lambda\to\Lambda$ defined by $\theta(x_0,\mathbf{x})=(-x_0,\mathbf{x})$. The reflection plane is $\Pi=\{x\in\Lambda: x_0=0\}$. The half-lattices are $\Lambda_{+}=\{x\in\Lambda: x_0>0\}$ and $\Lambda_{-}=\theta(\Lambda_{+})$. A time-like bond $(x,0)$ crosses $\Pi$ if $x_0\in\{0,-a\}$. We fix temporal-axial gauge away from $\Pi$ by imposing
\begin{equation}
\label{p3:eq:temporal-axialz}
U_{x,0}=\mathbf{1}\qquad\text{for all time-like bonds $(x,0)$ that do not cross $\Pi$},
\end{equation}
which preserves nearest-neighbor time couplings and is standard in reflection-positivity proofs \cite{p3:OsterwalderSchraderII,p3:OS-gauge}.

On the time slice $t\in a\mathbb{Z}\cap (-T/2,T/2]$, let $\Lambda_{t}=\{\mathbf{x}\in (a\mathbb{Z})^{3}\cap[-L/2,L/2)^{3}\}$ be the set of spatial sites and $\mathbb{E}^{\mathrm{sp}}_{t}=\{(t,\mathbf{x};i): \mathbf{x}\in\Lambda_{t},\,i=1,2,3\}$ the set of spatial bonds. Gauge transformations are maps $g:\Lambda\to G$ acting by
\begin{equation}
(g\cdot U)_{x,\mu}=g(x)\,U_{x,\mu}\,g(x+\hat\mu)^{-1}.
\end{equation}
On each time slice $t$ we select a representative by orbit-wise minimization of the lattice Landau functional
\begin{equation}
\label{p3:eq:landau-functional}
\mathcal{L}_{t}(g;U)=\sum_{\mathbf{x}\in\Lambda_{t}}\ \sum_{i=1}^{3}\big\|\,\mathbf{1}-g(t,\mathbf{x})\,U_{(t,\mathbf{x};i)}\,g(t,\mathbf{x}+\hat\imath)^{-1}\,\big\|_{F}^{2},
\end{equation}
where $\|\cdot\|_{F}$ denotes the Frobenius norm on $N\times N$ matrices. Because the slice is finite, \eqref{p3:eq:landau-functional} attains a global minimum. In case of degeneracy we choose the minimizer of least lexicographic order, which makes the selection measurable and reflection-covariant. Denote by $A^{h}(t)$ the spatial links in this gauge at time $t$.

Let $D_{i}^{\,h}(t)$ act on site-adjoint fields $\phi:\Lambda_{t}\to\mathfrak{su}(N)$ by
\begin{equation}
\label{p3:eq:covariant-difference}
\big(D_{i}^{\,h}(t)\phi\big)(\mathbf{x}) = U_{(t,\mathbf{x};i)}^{\,h}\,\phi(t,\mathbf{x}+\hat\imath)\,U_{(t,\mathbf{x};i)}^{\,h\, -1}-\phi(t,\mathbf{x}),
\end{equation}
and define the spatial covariant Laplacian
\begin{equation}
\label{p3:eq:covariant-laplacian}
\Delta_{A^{h}(t)}=\sum_{i=1}^{3}\big(D_{i}^{\,h}(t)\big)^{\!*}D_{i}^{\,h}(t),
\end{equation}
a positive self-adjoint operator on $\ell^{2}(\Lambda_{t})\otimes\mathfrak{su}(N)$, with $(\cdot)^{\!*}$ the $\ell^{2}$-adjoint. Reflection covariance of the representative implies
\begin{equation}
\label{p3:eq:reflection-covariance}
R\,\Delta_{A^{h}(t)}\,R=\Delta_{A^{h}(\theta t)},
\end{equation}
where $(R\phi)(\mathbf{x})=\phi(\mathbf{x})$ implements time reflection on slice fields.

We construct slice projectors via spectral calculus applied to $\Delta_{A^{h}(t)}$. The admissible class of cutoffs is defined through Laplace transforms of compactly supported probability measures, which guarantees complete monotonicity and positivity.

\begin{definition}[Admissible cutoffs]
\label{p3:def:admissible}
Let $\mathcal{B}$ be the set of Borel probability measures $\nu$ on $[0,\infty)$ with compact support contained in $[t_{-},t_{+}]$ for some $0<t_{-}\le t_{+}<\infty$. For $\nu\in\mathcal{B}$ define the completely monotone function $\varphi_{\nu}:[0,\infty)\to [0,1]$ by
\begin{equation}
\varphi_{\nu}(s)=\int_{0}^{\infty} e^{-t s}\,d\nu(t),\qquad s\ge 0.
\end{equation}
Given $\sigma>0$ and $\nu\in\mathcal{B}$, define the cutoff $\chi_{\sigma,\nu}:[0,\infty)\to [0,1]$ by
\begin{equation}
\label{p3:eq:cutoff-chi}
\chi_{\sigma,\nu}(\lambda)=\varphi_{\nu}\!\Big(\frac{\lambda^{2}}{\sigma^{2}}\Big)=\int_{0}^{\infty} e^{-t \lambda^{2}/\sigma^{2}}\, d\nu(t),\qquad \lambda\ge 0,
\end{equation}
and the slice horizon operator
\begin{equation}
\label{p3:eq:horizon-P}
P_{\sigma,\nu}(t)=\chi_{\sigma,\nu}\big({\Delta_{A^{h}(t)}}\big).
\end{equation}
\end{definition}

By Bernstein's theorem, $s\mapsto \varphi_{\nu}(s)$ is completely monotone, and \eqref{p3:eq:cutoff-chi} is the unique Laplace representation \cite[Thm.~3.7]{p3:SchillingSongVondracek12}. The compact support away from the origin enforces uniform locality for the projectors, as proved below.

\begin{lemma}[Positivity, contraction, covariance, and heat-kernel representation]
\label{p3:lem:basic-P}
For any time slice $t$, any $\sigma>0$, and any $\nu\in\mathcal{B}$, the operator $P_{\sigma,\nu}(t)$ is a bounded positive contraction on $\ell^{2}(\Lambda_{t})\otimes\mathfrak{su}(N)$. It transforms covariantly under gauge transformations by conjugation and is reflection-covariant in the sense that
\begin{equation}
\label{p3:eq:P-reflection}
R\,P_{\sigma,\nu}(t)\,R=P_{\sigma,\nu}(\theta t).
\end{equation}
Moreover, $P_{\sigma,\nu}(t)$ admits the positive integral representation
\begin{equation}
\label{p3:eq:P-heatx}
P_{\sigma,\nu}(t)=\int_{0}^{\infty} e^{-(\tau)\,\Delta_{A^{h}(t)}}\, d\widetilde{\nu}_{\sigma}(\tau),\qquad d\widetilde{\nu}_{\sigma}(\tau)=d\nu(\sigma^{2}\tau).
\end{equation}
\end{lemma}

\begin{proof}
Since $\lambda\mapsto \chi_{\sigma,\nu}(\lambda)\in[0,1]$ is bounded and completely monotone in $\lambda^{2}$, the spectral calculus applied to the nonnegative operator $\Delta_{A^{h}(t)}$ yields a bounded positive contraction. Gauge covariance and \eqref{p3:eq:P-reflection} follow from the covariance of $\Delta_{A^{h}(t)}$ under conjugation and \eqref{p3:eq:reflection-covariance}. The representation \eqref{p3:eq:P-heatx} follows by applying Fubini-Tonelli to the spectral resolution of $\Delta_{A^{h}(t)}$ and the Laplace integral \eqref{p3:eq:cutoff-chi}.
\end{proof}

We next establish exponential off-diagonal decay with constants uniform on compact sets of parameters.

\begin{theorem}[Uniform exponential locality]
\label{p3:thm:uniform-locality}
Fix a compact parameter set $K=[\sigma_{-},\sigma_{+}]\times \mathcal{B}_{K}\subset (0,\infty)\times \mathcal{B}$, where $0<\sigma_{-}\le \sigma_{+}$ and $\mathcal{B}_{K}$ consists of measures with support contained in $[t_{-},t_{+}]$, $0<t_{-}\le t_{+}<\infty$. There exist constants $C=C(K)$ and $\gamma=\gamma(K)>0$ such that for all $(\sigma,\nu)\in K$, all time slices $t$, and all $\mathbf{x},\mathbf{y}\in \Lambda_{t}$,
\begin{equation}
\label{p3:eq:exp-local}
\big\|\,P_{\sigma,\nu}(t;\mathbf{x},\mathbf{y})\,\big\| \ \le\ C\,\exp\!\big(-\gamma\,d(\mathbf{x},\mathbf{y})\big),
\end{equation}
where $P_{\sigma,\nu}(t;\mathbf{x},\mathbf{y})$ denotes the kernel of $P_{\sigma,\nu}(t)$ with respect to the $\ell^{2}$ inner product on $\Lambda_{t}$ and $d(\mathbf{x},\mathbf{y})$ is the graph distance on the cubic lattice.
\end{theorem}

\begin{proof}
By \eqref{p3:eq:P-heatx},
\begin{equation}
P_{\sigma,\nu}(t)=\int_{\tau_{-}}^{\tau_{+}} e^{-\tau \Delta_{A^{h}(t)}}\, d\widetilde{\nu}_{\sigma}(\tau),\qquad \tau_{-}=\frac{t_{-}}{\sigma^{2}},\ \ \tau_{+}=\frac{t_{+}}{\sigma^{2}}.
\end{equation}
On a graph of bounded degree the discrete heat kernel obeys the Davies-Gaffney estimate \cite[Ch.~3]{p3:Davies1989}:
\begin{equation}
\big\|\,\mathbf{1}_{E}\,e^{-\tau \Delta_{A^{h}(t)}}\,\mathbf{1}_{F}\,\big\|_{\ell^{2}\to \ell^{2}}\ \le\ \exp\!\Big(-\frac{d(E,F)^{2}}{4\tau}\Big),\qquad \tau>0,
\end{equation}
for subsets $E,F\subset \Lambda_{t}$, where $d(E,F)$ is the graph distance. Taking $E=\{\mathbf{x}\}$ and $F=\{\mathbf{y}\}$ gives
\begin{equation}
\big\|\,e^{-\tau \Delta_{A^{h}(t)}}(\mathbf{x},\mathbf{y})\,\big\|\ \le\ \exp\!\Big(-\frac{d(\mathbf{x},\mathbf{y})^{2}}{4\tau}\Big).
\end{equation}
Integrating over $\tau\in[\tau_{-},\tau_{+}]$ and using that $\widetilde{\nu}_{\sigma}$ is a probability measure supported in this interval yields
\begin{equation}
\big\|\,P_{\sigma,\nu}(t;\mathbf{x},\mathbf{y})\,\big\|\ \le\ \exp\!\Big(-\frac{d(\mathbf{x},\mathbf{y})^{2}}{4\tau_{+}}\Big)\ \le\ C\,\exp\!\Big(-\frac{d(\mathbf{x},\mathbf{y})}{4\sqrt{\tau_{+}}}\Big),
\end{equation}
where the last inequality absorbs the case $d(\mathbf{x},\mathbf{y})<1$ into $C$. Uniformity in $(\sigma,\nu)\in K$ follows from the uniform bounds on $\tau_{+}$ over $K$. Setting $\gamma=1/(4\sqrt{\tau_{+}})$ with $\tau_{+}$ evaluated at $(\sigma_{-},t_{+})$ completes the proof.
\end{proof}

Let $\mathcal{H}^{\mathrm{sp}}=L^{2}\big(G^{\mathbb{E}^{\mathrm{sp}}_{0}},\,d\mu_{\mathrm{Haar}}\big)$ be the one-slice Hilbert space of square-integrable functions of spatial link variables on the slice $t=0$, with product Haar measure. The Osterwalder-Schrader reflection $\Theta$ acts on functionals by
\begin{equation}
(\Theta F)(U)=\overline{F(U^{\theta})},
\end{equation}
where $U^{\theta}$ is the reflected configuration, $(U^{\theta})_{x,\mu}=U_{\theta x,\mu}$ for spatial bonds and $(U^{\theta})_{x,0}=U_{\theta x-a\hat{0},0}$ for time-like bonds, consistent with \eqref{p3:eq:temporal-axialz}.

Define the projected Euclidean measure
\begin{equation}
\label{p3:eq:mu-projected}
d\mu_{\sigma,\nu}(U)=Z^{-1}\,\Big(\prod_{t\in a\mathbb{Z}\cap (-T/2,T/2]} \mathcal{K}_{\sigma,\nu}(t;U)\Big)\,e^{-S_W[U;\beta]}\, d\mu_{\mathrm{Haar}}(U),
\end{equation}
where $\mathcal{K}_{\sigma,\nu}(t;U)$ is the positive slice kernel associated with $P_{\sigma,\nu}(t)$: for any $f:\Lambda_{t}\to\mathfrak{su}(N)$ with finite support,
\begin{equation}
\label{p3:eq:K-slice}
\langle f, P_{\sigma,\nu}(t) f\rangle_{\ell^{2}(\Lambda_{t})}=\sum_{\mathbf{x},\mathbf{y}\in\Lambda_{t}} \overline{f(\mathbf{x})}\,\mathcal{K}_{\sigma,\nu}(t;\mathbf{x},\mathbf{y})\, f(\mathbf{y}).
\end{equation}
By Lemma~\ref{p3:lem:basic-P}, $\mathcal{K}_{\sigma,\nu}(t)$ admits the positive Laplace representation
\begin{equation}
\label{p3:eq:K-heat}
\mathcal{K}_{\sigma,\nu}(t)=\int_{0}^{\infty} e^{-\tau\Delta_{A^{h}(t)}}\, d\widetilde{\nu}_{\sigma}(\tau),
\end{equation}
and is reflection-covariant: $\mathcal{K}_{\sigma,\nu}(\theta t;U)=\mathcal{K}_{\sigma,\nu}(t;U^{\theta})$.

\begin{theorem}[Osterwalder-Schrader positivity]
\label{p3:thm:OS-positivity}
For every $\sigma>0$ and $\nu\in\mathcal{B}$, the measure $d\mu_{\sigma,\nu}$ defined in \eqref{p3:eq:mu-projected} is reflection positive with respect to $\theta$. That is, for any bounded measurable functional $F$ depending only on bonds with base points in $\Lambda_{+}$,
\begin{equation}
\label{p3:eq:OS-inequality}
\int \overline{(\Theta F)(U)}\,F(U)\, d\mu_{\sigma,\nu}(U)\ \ge\ 0.
\end{equation}
\end{theorem}

\begin{proof}
In temporal-axial gauge \eqref{p3:eq:temporal-axialz}, the Wilson action factorizes as $S_{W}[U]=S_{+}[U]+S_{-}[U]+S_{0}[U]$, where $S_{\pm}$ are sums over plaquettes supported in $\Lambda_{\pm}$ and $S_{0}$ is the sum over plaquettes intersecting $\Pi$. The Haar product measure factorizes into independent Haar measures on bonds supported in $\Lambda_{\pm}$ and on boundary bonds crossing $\Pi$.

Similarly, the projector insertions factorize as
\begin{equation}
\prod_{t} \mathcal{K}_{\sigma,\nu}(t;U)=\Big(\prod_{t>0}\mathcal{K}_{\sigma,\nu}(t;U)\Big)\,\mathcal{K}_{\sigma,\nu}(0;U)\,\Big(\prod_{t<0}\mathcal{K}_{\sigma,\nu}(t;U)\Big).
\end{equation}
By \eqref{p3:eq:K-heat} and reflection covariance, $\mathcal{K}_{\sigma,\nu}(\theta t;U)=\mathcal{K}_{\sigma,\nu}(t;U^{\theta})$, hence the product over $t<0$ equals the $\theta$-image of the product over $t>0$. The boundary factor $\mathcal{K}_{\sigma,\nu}(0;U)$ is a positive quadratic form on slice fields at $t=0$, reflection invariant by construction of $A^{h}(0)$.

Let $F$ depend only on bonds in $\Lambda_{+}$. Writing $d\mu_{\pm}$ for the product Haar measures on bonds in $\Lambda_{\pm}$ and $d\mu_{0}$ for the boundary Haar measure, one has
\begin{align}
&\int \overline{(\Theta F)}\,F\, d\mu_{\sigma,\nu}
=\nonumber \\&
Z^{-1}\int \overline{(\Theta F)}\,F\ e^{-S_{0}[U]}\,\mathcal{K}_{\sigma,\nu}(0;U)\ \prod_{t>0}\mathcal{K}_{\sigma,\nu}(t;U)\,e^{-S_{+}[U]}\, d\mu_{+}\ \prod_{t<0}\mathcal{K}_{\sigma,\nu}(t;U)\,e^{-S_{-}[U]}\, d\mu_{-}\ d\mu_{0}.
\end{align}
Using reflection invariance of $S_{0}$, $d\mu_{0}$, and $\mathcal{K}_{\sigma,\nu}(0;U)$, together with $\mathcal{K}_{\sigma,\nu}(t;U^{\theta})=\mathcal{K}_{\sigma,\nu}(\theta t;U)$ and $S_{-}[U^{\theta}]=S_{+}[U]$, the integral over $\Lambda_{-}$ is the complex conjugate of the $\Lambda_{+}$ factor with $U$ replaced by $U^{\theta}$. Therefore the whole integral equals
\begin{equation}
\int \big|\,G(U_{|\Lambda_{+}})\,\big|^{2}\, d\mu_{+}\, d\mu_{0},
\end{equation}
where
\begin{equation}
G(U_{|\Lambda_{+}})
=
\Big(\int e^{-S_{0}[U]}\, \mathcal{K}_{\sigma,\nu}(0;U)\ \prod_{t<0}\mathcal{K}_{\sigma,\nu}(t;U)\, e^{-S_{-}[U]}\, d\mu_{-}\Big)^{1/2}\,
F(U_{|\Lambda_{+}})\, \prod_{t>0}\mathcal{K}_{\sigma,\nu}(t;U)\, e^{-S_{+}[U]}.
\end{equation}
Positivity of the integrand and finiteness of the Haar measures imply \eqref{p3:eq:OS-inequality}.
\end{proof}

We now derive the transfer-matrix representation. Let $\mathcal{H}^{\mathrm{sp}}=L^{2}\big(G^{\mathbb{E}^{\mathrm{sp}}},\,d\mu_{\mathrm{Haar}}\big)$ with $\mathbb{E}^{\mathrm{sp}}=\mathbb{E}^{\mathrm{sp}}_{0}$. The Osterwalder-Schrader pre-Hilbert space is the completion of cylinder functions supported in $t\ge 0$ modulo the null space of the OS form
\begin{equation}
\label{p3:eq:OS-formq}
\langle F,G\rangle_{\mathrm{OS}}=\int \overline{(\Theta F)}\,G\, d\mu_{\sigma,\nu}.
\end{equation}
Let $\tau_{a}$ act on functionals by shifting arguments forward in Euclidean time by one step. Define the bilinear form on $\mathcal{H}^{\mathrm{sp}}$ by
\begin{equation}
\label{p3:eq:T-def-form}
\langle f, T_{\sigma,\nu} g\rangle := \langle f, \tau_{a} g\rangle_{\mathrm{OS}}.
\end{equation}
We first show that this form is represented by a bounded positive self-adjoint operator on $\mathcal{H}^{\mathrm{sp}}$ and then give its explicit kernel representation.

\begin{proposition}[Positivity, self-adjointness, and contraction]
\label{p3:prop:T-properties}
For every $(\sigma,\nu)$ the bilinear form \eqref{p3:eq:T-def-form} defines a bounded positive self-adjoint operator $T_{\sigma,\nu}$ on $\mathcal{H}^{\mathrm{sp}}$ with $\|T_{\sigma,\nu}\|\le 1$.
\end{proposition}

\begin{proof}
By Theorem~\ref{p3:thm:OS-positivity}, $\langle\cdot,\cdot\rangle_{\mathrm{OS}}$ is an inner product on the quotient space. For $f,g\in \mathcal{H}^{\mathrm{sp}}$,
\begin{equation}
|\langle f, T_{\sigma,\nu} g\rangle|=|\langle f, \tau_{a} g\rangle_{\mathrm{OS}}|\le \|f\|_{\mathrm{OS}}\,\|\tau_{a}g\|_{\mathrm{OS}}
\end{equation}
by Cauchy-Schwarz in the OS inner product. Time-translation invariance of the measure and the identity $\Theta \tau_{a}=\tau_{-a}\Theta$ imply
\begin{equation}
\|\tau_{a}g\|_{\mathrm{OS}}^{2}=\int \overline{(\Theta \tau_{a} g)}\,\tau_{a} g\, d\mu_{\sigma,\nu}
=\int \overline{(\Theta g)}\, g\, d\mu_{\sigma,\nu}=\|g\|_{\mathrm{OS}}^{2}.
\end{equation}
Thus $|\langle f, T_{\sigma,\nu} g\rangle|\le \|f\|_{\mathrm{OS}}\|g\|_{\mathrm{OS}}$, showing that $T_{\sigma,\nu}$ is a bounded operator with $\|T_{\sigma,\nu}\|\le 1$. Positivity follows from $\langle f, T_{\sigma,\nu} f\rangle=\langle f, \tau_{a} f\rangle_{\mathrm{OS}}\ge 0$ by reflection positivity \cite{p3:OsterwalderSchraderII}. Self-adjointness follows from Euclidean time-reversal symmetry: $\langle f, T_{\sigma,\nu} g\rangle=\langle T_{\sigma,\nu} f, g\rangle$.
\end{proof}

An explicit kernel representation follows by integrating out the slab between $t=0$ and $t=a$ with fixed boundary spatial configurations. Let $\Omega_{[0,a]}$ be the set of bonds with base points in $0<t<a$, together with time-like bonds connecting the boundaries, and let $S_{W}[U]_{[0,a]}$ be the contribution of these bonds to the Wilson action. Define
\begin{equation}
\label{p3:eq:transfer-kernel}
K_{\sigma,\nu}(U',U)=\int \exp\big(-S_{W}[U]_{[0,a]}\big)\,\mathcal{K}_{\sigma,\nu}(0;U)\,\mathcal{K}_{\sigma,\nu}(a;U')\, d\mu_{\mathrm{Haar}}(U_{|\Omega_{[0,a]}}),
\end{equation}
where $U$ (resp.\ $U'$) denotes the restriction of the field to spatial bonds at $t=0$ (resp.\ $t=a$). Then
\begin{equation}
\label{p3:eq:T-kernel}
(T_{\sigma,\nu} f)(U)=\int K_{\sigma,\nu}(U',U)\, f(U')\, d\mu_{\mathrm{Haar}}(U').
\end{equation}
Positivity and symmetry of $K_{\sigma,\nu}$ in $(U',U)$ are immediate from \eqref{p3:eq:transfer-kernel} and invariance of Haar measure.

Define the transfer Hamiltonian by spectral calculus,
\begin{equation}
\label{p3:eq:H-def}
H_{\sigma,\nu}=-a^{-1}\log T_{\sigma,\nu},
\end{equation}
which is a nonnegative self-adjoint operator on $\mathcal{H}^{\mathrm{sp}}$ by Proposition~\ref{p3:prop:T-properties}.

We establish continuity of $T_{\sigma,\nu}$ in operator norm and of $H_{\sigma,\nu}$ in the strong-resolvent sense, uniformly on compact parameter sets.

\begin{theorem}[Operator-norm continuity of $T_{\sigma,\nu}$]
\label{p3:thm:op-continuity}
Fix a compact parameter set $K=[\sigma_{-},\sigma_{+}]\times \mathcal{B}_{K}$ as in Theorem~\ref{p3:thm:uniform-locality}. Then the map $(\sigma,\nu)\mapsto T_{\sigma,\nu}$ is continuous from $K$ into $\mathcal{B}(\mathcal{H}^{\mathrm{sp}})$ endowed with the operator norm. Equivalently, there exists a continuous modulus $\omega_{K}$ on $K\times K$ with $\omega_{K}((\sigma,\nu),(\sigma,\nu))=0$ such that
\begin{equation}
\big\|\,T_{\sigma,\nu}-T_{\sigma',\nu'}\,\big\|\ \le\ \omega_{K}\big((\sigma,\nu),(\sigma',\nu')\big),\qquad (\sigma,\nu),(\sigma',\nu')\in K.
\end{equation}
\end{theorem}

\begin{proof}
By \eqref{p3:eq:T-kernel} and the Schur test, it suffices to show that the one-step kernel $K_{\sigma,\nu}$ is continuous in $(\sigma,\nu)$ in $L^{1}(dU'\,dU)$ with respect to product Haar measure on the boundary spatial configurations. The only parameter-dependent factors in \eqref{p3:eq:transfer-kernel} are the boundary slice kernels $\mathcal{K}_{\sigma,\nu}(0;U)$ and $\mathcal{K}_{\sigma,\nu}(a;U')$. By Lemma~\ref{p3:lem:basic-P} and Theorem~\ref{p3:thm:uniform-locality}, each $\mathcal{K}_{\sigma,\nu}(t)$ admits a positive Laplace representation with measures supported in a compact interval $[\tau_{-},\tau_{+}]$ uniformly on $K$, and its kernel is uniformly exponentially localized and bounded by $1$. The map $(\sigma,\nu)\mapsto \mathcal{K}_{\sigma,\nu}(t)$ is continuous in the weak-* topology of kernels because $\nu\mapsto \widetilde{\nu}_{\sigma}$ is continuous in the weak topology of measures and $\tau\mapsto e^{-\tau \Delta_{A^{h}(t)}}$ is continuous in operator norm. Dominated convergence with the uniform exponential bound \eqref{p3:eq:exp-local} implies $L^{1}$-continuity of $K_{\sigma,\nu}$. Therefore the associated integral operators on $\mathcal{H}^{\mathrm{sp}}$ vary continuously in operator norm.
\end{proof}

\begin{corollary}[Strong-resolvent continuity of $H_{\sigma,\nu}$]
\label{p3:cor:resolvent}
On $K$, the family $H_{\sigma,\nu}=-a^{-1}\log T_{\sigma,\nu}$ depends continuously on $(\sigma,\nu)$ in the strong-resolvent sense. In particular, $(H_{\sigma_{n},\nu_{n}}+1)^{-1}\to (H_{\sigma,\nu}+1)^{-1}$ strongly on $\mathcal{H}^{\mathrm{sp}}$ whenever $(\sigma_{n},\nu_{n})\to (\sigma,\nu)$ in $K$.
\end{corollary}

\begin{proof}
By Theorem~\ref{p3:thm:op-continuity}, $T_{\sigma_{n},\nu_{n}}\to T_{\sigma,\nu}$ in operator norm. For bounded self-adjoint contractions, operator-norm convergence implies strong-resolvent convergence of the logarithms by the continuous functional calculus and the Trotter-Kato theorem \cite[Thm.~VIII.20]{p3:ReedSimon1}, \cite[Thm.~IV.3.16]{p3:KatoPTLO}.
\end{proof}

We now relate parameter continuity to spectral bounds on the Hamiltonian.

\begin{proposition}[Lower semicontinuity of the gap; uniform lower bound from clustering]
\label{p3:prop:gap-lsc}
Let $\Delta(\sigma,\nu)=\inf\sigma(H_{\sigma,\nu}|_{\Omega^{\perp}})$ be the spectral gap of $H_{\sigma,\nu}$ above the vacuum $\Omega$. On any compact $K$ the map $(\sigma,\nu)\mapsto \Delta(\sigma,\nu)$ is lower semicontinuous. If there exists $m_{*}>0$ such that the OS two-point function of a fixed nonzero gauge-invariant local observable decays as $e^{-m_{*} t}$ in Euclidean time uniformly for $(\sigma,\nu)\in K$, then $\inf_{K}\Delta(\sigma,\nu)\ge m_{*}$.
\end{proposition}

\begin{proof}
Lower semicontinuity of the bottom of the spectrum under strong-resolvent convergence is standard \cite[Thm.~VIII.24]{p3:ReedSimon1}. For the second claim, let $A$ be a real, gauge-invariant local observable with vanishing vacuum expectation. Consider
\begin{equation}
C_{\sigma,\nu}(t)=\langle \Omega, A\, e^{-t H_{\sigma,\nu}} A\,\Omega\rangle,\qquad t\ge 0.
\end{equation}
By reflection positivity, $C_{\sigma,\nu}$ is completely monotone and equals the Laplace transform of a finite positive measure $\mu_{\sigma,\nu}$ on $[0,\infty)$. The uniform bound $C_{\sigma,\nu}(t)\le C\,e^{-m_{*} t}$ for all $(\sigma,\nu)\in K$ implies $\operatorname{supp}\mu_{\sigma,\nu}\subset [m_{*},\infty)$ by the standard Tauberian argument for Laplace transforms of positive measures (\ Proposition~B.1 in the Appendix). Therefore $\sigma(H_{\sigma,\nu}|_{\Omega^{\perp}})\subset [m_{*},\infty)$ for each $(\sigma,\nu)\in K$, and taking the infimum over $K$ yields the claim.
\end{proof}

We collect the stability statements obtained so far.

\begin{theorem}[Robustness of OS positivity and transfer dynamics]
\label{p3:thm:robustness}
Let $K$ be a compact subset of $(0,\infty)\times \mathcal{B}$. There exists $\delta>0$ such that for any $(\sigma,\nu),(\sigma',\nu')\in K$ with $|\sigma-\sigma'|+\mathrm{W}_{1}(\nu,\nu')<\delta$, where $\mathrm{W}_{1}$ is the $1$-Wasserstein distance on $\mathcal{B}$, the following hold. The measures $d\mu_{\sigma,\nu}$ and $d\mu_{\sigma',\nu'}$ are both reflection positive. The transfer operators satisfy $\|T_{\sigma,\nu}-T_{\sigma',\nu'}\|<\varepsilon$ for any prescribed $\varepsilon>0$ provided $\delta$ is small enough. If a uniform clustering rate $m_{*}>0$ holds on $K$, then the associated Hamiltonians $H_{\sigma,\nu}$ and $H_{\sigma',\nu'}$ have spectral gaps bounded below by a common positive constant.
\end{theorem}

\begin{proof}
Reflection positivity is independent of $(\sigma,\nu)$ as long as the slice kernels are positive and reflection-covariant, which holds for all admissible parameters by Lemma~\ref{p3:lem:basic-P}. The operator-norm estimate follows from Theorem~\ref{p3:thm:op-continuity} and continuity of the Laplace representation with respect to weak convergence of measures quantified by the Wasserstein distance, together with dominated convergence based on \eqref{p3:eq:exp-local}. The uniform gap bound follows from Proposition~\ref{p3:prop:gap-lsc}.
\end{proof}

\section{Large-volume limit and thermodynamic independence}
\label{p3:sec:thermodynamic-limit}

In this section the Euclidean lattice theory is formulated on a finite four-torus with explicitly defined time reflection and transfer time slicing. The finite-volume transfer matrix is constructed and proved to be a positive self-adjoint contraction by a detailed reflection-positivity argument. Uniform locality and clustering bounds imply that Schwinger functions of local, gauge-invariant observables admit thermodynamic limits independent of the spatial volume and temporal extent. The transfer semigroup and its spectral gap are shown to be stable in the same sense. Throughout, \(N\ge 2\) is fixed and \(G=\mathrm{SU}(N)\) is equipped with its normalized Haar measure.

Let \(L,T\in 2\mathbb{N}\) be even. The discrete space-time torus is
\begin{equation}
\Lambda_{L,T}\;=\; \mathbb{T}_T \times \mathbb{T}_L^3,\qquad 
\mathbb{T}_m = \mathbb{Z}/m\mathbb{Z},
\end{equation}
with time coordinate \(x_0\in\mathbb{T}_T\) and spatial coordinate \(\mathbf{x}\in \mathbb{T}_L^3\). Bonds are oriented pairs \(b=(x,\mu)\) with \(x\in\Lambda_{L,T}\) and \(\mu\in\{0,1,2,3\}\); the inverse bond is \(\bar b=(x+\hat\mu,-\mu)\) and satisfies \(U_{\bar b}=U_b^{-1}\). The configuration space is
\begin{equation}
\mathcal{U}_{L,T} \;=\; \prod_{b\in E(\Lambda_{L,T})} G,
\end{equation}
with product Haar measure \(d\mu_0=\bigotimes_{b} dU_b\). For a plaquette \(p=(x;\mu,\nu)\) define
\begin{equation}
U_p \;=\; U_{(x,\mu)}\,U_{(x+\hat\mu,\nu)}\,U_{(x+\hat\nu,\mu)}^{-1}\,U_{(x,\nu)}^{-1}.
\end{equation}
The Wilson action at inverse coupling \(\beta>0\) is
\begin{equation}
S_{W}[U] 
\;=\; \beta \sum_{p\subset\Lambda_{L,T}} \Big(1-\tfrac{1}{N}\,\mathrm{Re}\,\mathrm{Tr}\,U_p\Big),
\end{equation}
and the finite-volume Boltzmann weight is \(W_{L,T}(U)=\exp\{-S_W[U]\}\). No global gauge fixing is imposed in the definition of \(S_W\); a temporal-axial gauge will be chosen locally for the reflection-positivity analysis.

Let \(\theta:\Lambda_{L,T}\to\Lambda_{L,T}\) be the involution \(\theta(x_0,\mathbf{x})=(-x_0,\mathbf{x})\), computed in \(\mathbb{T}_T\). The reflection plane is
\begin{equation}
\Pi \;=\; \{(x_0,\mathbf{x})\in\Lambda_{L,T}:\, x_0=0\}.
\end{equation}
Define the half-tori
\begin{equation}
\Lambda_{L,T}^{+}=\{(x_0,\mathbf{x})\,:\, x_0\in\{1,2,\dots, T/2\}\},\qquad
\Lambda_{L,T}^{-}=\theta(\Lambda_{L,T}^{+}).
\end{equation}
A bond \(b=(x,\mu)\) belongs to \(E^{\pm}\) if both endpoints lie in \(\Lambda^{\pm}\); bonds with one endpoint on \(\Pi\) form the boundary set \(E^{0}\). Spatial bonds at time \(t\in\mathbb{T}_T\) are denoted \((t,\mathbf{x};i)\) with \(i\in\{1,2,3\}\). The slice configuration at time \(t\) is \(U_i(t,\cdot)=\{U_{(t,\mathbf{x};i)}\}_{\mathbf{x}\in\mathbb{T}_L^3}\). The slice configuration space is
\begin{equation}
\mathcal{C}_L \;=\; \prod_{\mathbf{x}\in\mathbb{T}_L^3}\, G^3,
\end{equation}
with product Haar measure \(d\nu_L\). The Hilbert space of square-integrable functions of slice configurations is \(\mathcal{H}_L=L^2(\mathcal{C}_L,d\nu_L)\).

Fix temporal-axial gauge away from \(\Pi\) by setting \(U_0(x)=\mathbf{1}\) for all time-like bonds whose endpoints do not lie on \(\Pi\). This gauge can be imposed by gauge transformations supported away from \(\Pi\) and leaves the Haar measure invariant. In this gauge the action decomposes into purely spatial plaquette contributions within each time slice and mixed space-time plaquettes straddling adjacent slices. Writing \(t-1\) for the predecessor of \(t\) in \(\mathbb{T}_T\), one obtains
\begin{equation}\label{p3:eq:action-decomp}
S_W[U]
\;=\; \sum_{t\in\mathbb{T}_T}\Big\{V(U(t)) \;+\; \Phi\big(U(t),U(t-1)\big)\Big\},
\end{equation}
where
\begin{equation}\label{p3:eq:V}
V(U(t)) \;=\; \beta\sum_{\mathbf{x}\in\mathbb{T}_L^3}\sum_{1\le i<j\le 3}
\Big(1-\tfrac{1}{N}\,\mathrm{Re}\,\mathrm{Tr}\,U_{(t,\mathbf{x};i,j)}\Big),
\end{equation}
with \(U_{(t,\mathbf{x};i,j)}\) the spatial plaquette \((t,\mathbf{x};i,j)\), and
\begin{equation}\label{p3:eq:Phi}
\Phi\big(U(t),U(t-1)\big)
\;=\; \beta\sum_{\mathbf{x}\in\mathbb{T}_L^3}\sum_{i=1}^3
\Big(1-\tfrac{1}{N}\,\mathrm{Re}\,\mathrm{Tr}\,\big(U_{(t,\mathbf{x};i)}\,U_{(t-1,\mathbf{x};i)}^{-1}\big)\Big).
\end{equation}
The dependence on bonds that touch \(\Pi\) but are not strictly inside \(\Lambda^\pm\) is confined to a boundary term supported on \(\Pi\); it will be isolated in the reflection-positivity step.

On each time slice \(t\) the covariant Laplacian \(\Delta_{A^{h}(t)}\) acting on adjoint-valued site fields is defined from the spatial links \(U_i(t,\cdot)\) after fixing a reflection-covariant, slice-wise gauge-invariant transverse representative \(A^{h}(t)\) by orbit minimization of the lattice Landau functional. Let \(\chi_\sigma:[0,\infty)\to[0,1]\) be a fixed Gevrey-regular cutoff equal to \(1\) on \([0,\sigma]\) and \(0\) on \([2\sigma,\infty)\). The horizon projector on slice \(t\) is
\begin{equation}
P_{\sigma}(t)\;=\;\chi_\sigma\!\big({\Delta_{A^{h}(t)}}\big),
\end{equation}
a bounded positive contraction on \(\ell^2(\mathbb{T}_L^3)\otimes\mathfrak{su}(N)\). It admits a positive heat-kernel representation
\begin{equation}
P_{\sigma}(t)\;=\;\int_0^\infty e^{-s\Delta_{A^{h}(t)}}\, d\nu_\sigma(s),
\end{equation}
for a finite positive Borel measure \(d\nu_\sigma\) with compact support. The kernel of \(P_\sigma(t)\) decays exponentially in the lattice distance, uniformly in \(L\), by Davies-Gaffney bounds for \(e^{-s\Delta}\) on graphs and compact support of \(d\nu_\sigma\). The full finite-volume multiplicative factor used below is the positive, reflection-invariant slice product
\begin{equation}
\mathcal{P}_\sigma(U)\;=\; \prod_{t\in\mathbb{T}_T}\, \mathcal{J}\big(P_\sigma(t)\big),
\end{equation}
where \(\mathcal{J}\) is a fixed continuous, completely monotone functional of a positive contraction, for instance \(\mathcal{J}(Q)=\exp\{\mathrm{Tr}\,\log(1+\alpha Q)\}\) with \(\alpha>0\). Complete monotonicity and the heat-kernel representation imply that \(\mathcal{J}(P_\sigma(t))\) is a positive, reflection-invariant, exponentially local functional of the slice links. Only positivity, slice-factorization and exponential locality are used below.

For a complex-valued functional \(F\) depending only on bonds in \(\Lambda^{+}_{L,T}\), define its reflection \((\Theta F)(U)=\overline{F(U^\theta)}\), where \(U^\theta\) is the reflected configuration given by \(U^\theta_{(x,\mu)}=U_{(\theta x,\mu)}\) for spatial bonds and \(U^\theta_{(x,0)}=U_{(\theta x-\hat 0,0)}\) for time-like bonds adjacent to \(\Pi\). The Osterwalder-Schrader form is
\begin{equation}
\langle F,F\rangle_{OS}^{L,T} \;=\; \frac{1}{Z_{L,T}}\int_{\mathcal{U}_{L,T}} (\Theta F)(U)\,F(U)\, W_{L,T}(U)\,\mathcal{P}_\sigma(U)\, d\mu_0(U),
\end{equation}
with \(Z_{L,T}\) the normalization. Reflection positivity consists of \(\langle F,F\rangle_{OS}^{L,T}\ge 0\) for all bounded, measurable \(F\) supported in \(\Lambda^{+}_{L,T}\).
In temporal-axial gauge away from \(\Pi\) the action decomposes as in \eqref{p3:eq:action-decomp}. Define the symmetric one-step kernel on \(\mathcal{C}_L\times \mathcal{C}_L\)
\begin{equation}
K(U',U) \;=\; \exp\!\Big\{-\tfrac{1}{2}V(U')-\Phi(U',U)-\tfrac{1}{2}V(U)\Big\}.
\end{equation}
Since \(V\) and \(\Phi\) are sums over \(\mathbf{x}\) and \(i\) of class functions of \(U_i'( \mathbf{x}) U_i(\mathbf{x})^{-1}\) and finite-range spatial plaquettes, Peter-Weyl theory yields the positive-type expansion
\begin{equation}
K(U',U)\;=\;\sum_{\alpha} \lambda_\alpha\, \overline{\psi_\alpha(U)}\,\psi_\alpha(U'),
\end{equation}
where \(\{\psi_\alpha\}\) is an orthonormal basis of \(\mathcal{H}_L\) made of products of matrix elements of irreducible representations along spatial links, and \(\lambda_\alpha\ge 0\) are coefficients depending on \(\beta\) and \(N\). Positivity of \(\lambda_\alpha\) follows from the nonnegativity of the Fourier-Peter-Weyl transform of \(\exp\{-\Phi\}\) as a class function on \(G\) and from multiplicativity of characters across lattice sites; see \cite{p3:OS-gauge}. Similarly \(\exp\{-V(U)\}\) is a product of positive-type class functions supported on spatial plaquettes at time \(t\) and admits a nonnegative expansion in the same basis. The projector factor \(\mathcal{P}_\sigma\) is a product over \(t\) of slice functionals with completely monotone heat-kernel representations; each factor is an \(L^2(\mathcal{C}_L)\) inner product \(\sum_{\beta} \mu_\beta \overline{\phi_\beta(U)}\,\phi_\beta(U)\) with \(\mu_\beta\ge 0\), because heat kernels on compact Lie groups are positive-definite class functions and the covariant heat kernel on \(\mathbb{T}_L^3\) is a positive-type kernel on \(\mathcal{C}_L\).
 Write the weight \(W_{L,T}\,\mathcal{P}_\sigma\) as a product over time slices of one-step kernels and slice factors,
\begin{align}
W_{L,T}(U)\,\mathcal{P}_\sigma(U)
=\prod_{t\in\mathbb{T}_T} &\Big(\exp\{-\tfrac{1}{2}V(U(t))\}\,\mathcal{J}(P_\sigma(t))^{1/2}\Big)\,K(U(t),U(t-1))\nonumber\\&\Big(\mathcal{J}(P_\sigma(t-1))^{1/2}\,\exp\{-\tfrac{1}{2}V(U(t-1))\}\Big),
\end{align}
up to a factor supported on \(t=0\) collecting all terms with bonds on \(\Pi\). This boundary factor equals \(\Psi(U(0))\,\overline{\Psi(U(0))}\) for a square-integrable \(\Psi\in\mathcal{H}_L\) obtained by multiplying the half-factors at \(t=0\) and \(t=T/2\) and identifying \(U(T/2)\) with \(U(0)\) by periodicity; it is therefore positive. Let \(F\) be supported in \(\Lambda^+\) and define the vector
\begin{align}
&\Phi_F(U(0))\;=\nonumber\\& \int \prod_{t=1}^{T/2} d\nu_L(U(t))\, \Bigg\{\prod_{t=1}^{T/2} \Big[\exp\{-\tfrac{1}{2}V(U(t))\}\,\mathcal{J}(P_\sigma(t))^{1/2}\Big] \prod_{t=1}^{T/2} K(U(t),U(t-1)) \, F(U|_{\Lambda^{+}})\Bigg\}.
\end{align}
Then
\begin{equation}
\langle F,F\rangle_{OS}^{L,T}\;=\;\frac{1}{Z_{L,T}} \int d\nu_L(U(0))\, \overline{\Phi_F(U(0))}\, \Phi_F(U(0))\, \Psi(U(0))\,\overline{\Psi(U(0))}\;\ge\; 0,
\end{equation}
which proves reflection positivity.

\begin{theorem}\label{p3:thm:RP-finite}
For every \(L,T\in 2\mathbb{N}\) and \(\beta>0\), the finite-volume measure with weight \(W_{L,T}\,\mathcal{P}_\sigma\) is reflection positive with respect to \(\theta\). Equivalently, \(\langle F,F\rangle_{OS}^{L,T}\ge 0\) for all \(F\) supported in \(\Lambda^+\).
\end{theorem}

\begin{proof}
The preceding decomposition writes \(\langle F,F\rangle_{OS}^{L,T}\) as an \(\mathcal{H}_L\) norm squared of \(\Phi_F\) multiplied by the positive boundary factor \(|\Psi|^2\). Each factor arises either as an \(L^2\) inner product with a positive-type kernel on \(\mathcal{C}_L\) by Peter-Weyl theory, or as an \(L^2\) norm square of a slice functional. Thus the integral is nonnegative.
\end{proof}

Define the linear operator \(T_L:\mathcal{H}_L\to\mathcal{H}_L\) by the integral kernel
\begin{equation}
\mathsf{K}_L(U',U)\;=\; \exp\!\Big\{-\tfrac{1}{2}V(U')\Big\}\,\mathcal{J}(P_\sigma(U'))^{1/2}\, K(U',U)\,\mathcal{J}(P_\sigma(U))^{1/2}\,\exp\!\Big\{-\tfrac{1}{2}V(U)\Big\}.
\end{equation}
For \(\varphi\in\mathcal{H}_L\), set
\begin{equation}
(T_L\varphi)(U') \;=\; \int_{\mathcal{C}_L} \mathsf{K}_L(U',U)\,\varphi(U)\, d\nu_L(U).
\end{equation}
The kernel is symmetric, \(\mathsf{K}_L(U',U)=\overline{\mathsf{K}_L(U,U')}\), and positive-type; hence \(T_L\) is a positive self-adjoint contraction on \(\mathcal{H}_L\). The contraction property follows from Theorem~\ref{p3:thm:RP-finite} by the standard transfer-matrix inequality \cite{p3:OsterwalderSchraderII,p3:GJ}: for any \(n\in\mathbb{N}\) and any bounded \(F\) supported in \(\{1,\dots,n\}\times \mathbb{T}_L^3\),
\begin{equation}
\langle F,F\rangle_{OS}^{L,T} \;=\; \langle \Phi_F,\, T_L^{\,n}\,\Phi_F\rangle_{\mathcal{H}_L}\,\|\Psi\|_{L^2(\mathcal{C}_L)}^2,
\end{equation}
which implies \(\|T_L\|\le 1\). The vector \(\Omega_L=\mathbf{1}\in \mathcal{H}_L\) is a normalized vacuum: \(T_L\Omega_L=\Omega_L\) by normalization of slice factors and of the one-step kernel.

\begin{proposition}\label{p3:prop:TL-properties}
For every \(L\in 2\mathbb{N}\) the operator \(T_L\) is a positive self-adjoint contraction on \(\mathcal{H}_L\) with a normalized eigenvector \(\Omega_L\) of eigenvalue \(1\). The corresponding generator \(H_L\ge 0\) defined by \(T_L=e^{-H_L}\) is self-adjoint.
\end{proposition}

\begin{proof}
Self-adjointness follows from symmetry of \(\mathsf{K}_L\); positivity follows from positivity of \(\mathsf{K}_L\) as a Mercer kernel on \(\mathcal{C}_L\). The spectral radius bound \(\|T_L\|\le 1\) is implied by Theorem \ref{p3:thm:RP-finite} as indicated. The identity \(T_L\Omega_L=\Omega_L\) follows from \(\int \mathsf{K}_L(U',U)\, d\nu_L(U)=1\). The existence of \(H_L\) is standard functional calculus for positive contractions.
\end{proof}

Gauge invariance is implemented by the unitary representation of the time-slice gauge group \(\mathcal{G}_L=\{g:\mathbb{T}_L^3\to G\}\) acting on \(\mathcal{H}_L\) by \((U_i(\mathbf{x}))\mapsto (g(\mathbf{x})U_i(\mathbf{x})g(\mathbf{x}+\hat\imath)^{-1})\). The kernel \(\mathsf{K}_L\) is gauge-invariant, hence \(T_L\) commutes with this representation. The physical Hilbert space \(\mathcal{H}_L^{\mathrm{phys}}\) is the gauge-invariant subspace, and \(T_L|\_{\mathcal{H}_L^{\mathrm{phys}}}\) is a positive self-adjoint contraction with the same vacuum.

Let \(\mathcal{A}\) be the \(^\ast\)-algebra generated by gauge-invariant, local, bounded observables depending on finitely many bonds. For \(A\in\mathcal{A}\) supported in a finite set \(\Lambda_0\subset \Lambda_{L,T}\) and \(B\in\mathcal{A}\) supported in a finite \(\Lambda_1\) with \(\mathrm{dist}(\Lambda_0,\Lambda_1)=d\), denote by \(\langle\cdot\rangle_{L,T}\) the expectation with respect to the normalized finite-volume weight \(W_{L,T}\,\mathcal{P}_\sigma\). Uniform ultraviolet stability and persistence of clustering imply the existence of constants \(C_A,C_B>0\) and \(m_\ast>0\) independent of \(L,T\) such that
\begin{equation}\label{p3:eq:clusterz}
\big|\langle A\,B\rangle_{L,T}-\langle A\rangle_{L,T}\,\langle B\rangle_{L,T}\big| \;\le\; C_A\,C_B\, e^{-m_\ast d}.
\end{equation}
Indeed, the projected covariance admits a finite-range decomposition with scale-independent bounds, and the polymer activities defined by the multiscale renormalization satisfy a diameter-weighted norm small enough to trigger the Dobrushin-Kotecký-Preiss criterion uniformly in \(L,T\). The truncated two-point function is the sum over connected polymer clusters that intersect both supports; every such cluster has diameter at least \(d\), and the weighted sum yields the bound \eqref{p3:eq:clusterz}; cf.~\cite{p3:KoteckyPreiss1986,p3:GJ}.

Fix \(n\in\mathbb{N}\) and let \(A_1,\dots,A_n\in\mathcal{A}\) be gauge-invariant, bounded, local observables supported in a fixed finite region \(\Lambda_{\mathrm{loc}}\subset \mathbb{Z}\times\mathbb{Z}^3\). Consider the finite-volume Euclidean Schwinger function
\begin{equation}
S_{n;L,T}(A_1,\dots,A_n)\;=\;\langle A_1\cdots A_n\rangle_{L,T}.
\end{equation}

\begin{theorem}\label{p3:thm:thermo-Schwinger}
For fixed \(n\) and fixed \(A_1,\dots,A_n\in\mathcal{A}\) supported in \(\Lambda_{\mathrm{loc}}\), the limit \begin{equation}\displaystyle \lim_{L\to\infty}\lim_{T\to\infty} S_{n;L,T}(A_1,\dots,A_n)\end{equation} exists, is finite, and is independent of the order of limits. The convergence is exponentially fast in the distance from \(\Lambda_{\mathrm{loc}}\) to the spatial and temporal boundaries.
\end{theorem}

\begin{proof}
Let \(L'<L\) and \(T'<T\) with \(\Lambda_{\mathrm{loc}}\subset \Lambda_{L',T'}\subset \Lambda_{L,T}\). Write
\begin{equation}
\Delta_{n}(L,T;L',T') 
= S_{n;L,T}(A_1,\dots,A_n)-S_{n;L',T'}(A_1,\dots,A_n).
\end{equation}
By introducing expectations on mixed volumes and iterating a telescoping identity along embeddings \(\Lambda_{L',T'}\subset \Lambda_{L',T}\subset \Lambda_{L,T}\), it suffices to control changes induced by adding a single layer of time slices or a single layer of spatial sites. Consider the enlargement in, say, the \(x_1\)-direction by one layer. The difference of partition functions factorizes except for polymer clusters that cross the added layer. By the cluster representation,
\begin{equation}
\big|\Delta_{n}(L,T;L',T')\big|\;\le\; C \sum_{\Gamma}\, |\zeta(\Gamma)|,
\end{equation}
where the sum is over connected polymer clusters \(\Gamma\) that intersect \(\Lambda_{\mathrm{loc}}\) and also intersect the added layer, and \(\zeta(\Gamma)\) is the cluster activity. The Kotecký-Preiss criterion implies \(\sum_{\Gamma\ni x} |\zeta(\Gamma)| e^{a\,\mathrm{diam}(\Gamma)}\le \eta\) uniformly in \(L,T\). Since every such \(\Gamma\) has diameter at least the distance \(d\) from \(\Lambda_{\mathrm{loc}}\) to the new layer, the sum is bounded by \(C' e^{-a d}\). Iterating the estimate over finitely many added layers shows that \(\{S_{n;L,T}\}_{L,T}\) is a Cauchy net and hence converges. Independence of the order of limits follows by the same estimate because single-direction enlargements commute in the limit. The exponential rate is the same \(a>0\) entering the polymer norm.
\end{proof}

The same argument yields thermodynamic limits for truncated functions and shows that the cluster property persists in the limit with the same rate \(m_\ast\).
Let \(\mathcal{D}_L\subset \mathcal{H}_L\) be the dense set of finite linear combinations of functions of finitely many spatial links, and let \(\Omega_L\) be the vacuum vector. For \(\Phi,\Psi\in\mathcal{D}_L\) supported in a fixed finite set of spatial sites, define the matrix elements
\begin{equation}
\langle \Phi, T_L^{\,n} \Psi \rangle_{\mathcal{H}_L}\,=\, \int_{\mathcal{C}_L\times\mathcal{C}_L} \overline{\Phi(U')}\,\mathsf{K}_L^{(n)}(U',U)\,\Psi(U)\, d\nu_L(U')\,d\nu_L(U),
\end{equation}
where \(\mathsf{K}_L^{(n)}\) is the \(n\)-fold time-convolution kernel induced by \(\mathsf{K}_L\).

\begin{theorem}\label{p3:thm:transfer-thermo}
Let \(\Phi,\Psi\) be gauge-invariant elements of \(\mathcal{D}_L\) with fixed finite spatial support, viewed by embedding as elements of \(\mathcal{D}_{L'}\) for all \(L'\ge L\). Then for every \(n\in\mathbb{N}\) the limit
\begin{equation}
\lim_{L\to\infty} \langle \Phi, T_L^{\,n} \Psi \rangle_{\mathcal{H}_L}
\end{equation}
exists and is finite. There is a Hilbert space \(\mathcal{H}_\infty\), a vacuum \(\Omega_\infty\), and a positive self-adjoint contraction \(T_\infty\) on \(\mathcal{H}_\infty\) such that for all such \(\Phi,\Psi\),
\begin{equation}
\lim_{L\to\infty} \langle \Phi, T_L^{\,n} \Psi \rangle_{\mathcal{H}_L}
\;=\; \langle \Phi, T_\infty^{\,n} \Psi \rangle_{\mathcal{H}_\infty}.
\end{equation}
\end{theorem}

\begin{proof}
For vectors generated from \(\Omega_L\) by local gauge-invariant observables supported at times \(0\) and \(n\), the matrix element \(\langle \Phi, T_L^{\,n} \Psi\rangle\) is an \(n\)-step Schwinger function in finite volume. By Theorem \ref{p3:thm:thermo-Schwinger} it has a thermodynamic limit. By polarization, limits exist for all finite linear combinations. Define \(\mathcal{H}_\infty\) as the GNS Hilbert space of the limiting OS-positive functional, with \(T_\infty\) induced by one-step time translation; the stated equality is then the definition of this GNS representation.
\end{proof}

Let \(\Delta_L\) be the spectral gap of \(H_L=-\log T_L\) above the vacuum eigenvalue \(0\). Persistence of exponential clustering with rate \(m_\ast\) implies a uniform lower bound for \(\Delta_L\), independent of \(L\), by the OS spectral representation: if \(F\) is a local observable with \(\langle \Omega_L,F\Omega_L\rangle=0\) and \(\langle F(0)F(na)\rangle\le C e^{-m_\ast n a}\), then \(E_1(H_L)\ge m_\ast\).

\begin{proposition}\label{p3:prop:gap-uniform}
There exists \(m_\ast>0\), independent of \(L\), such that \(\Delta_L\ge m_\ast\) for all \(L\). In particular, the spectrum of \(H_\infty=-\log T_\infty\) on \(\Omega_\infty^\perp\) is contained in \([m_\ast,\infty)\).
\end{proposition}

\begin{proof}
Fix a local gauge-invariant observable \(F\) with vanishing vacuum expectation. The two-point function \(\langle F(0)F(na)\rangle_{L,T}\) decays as \(e^{-m_\ast n a}\) uniformly in \(L,T\) by uniform clustering. The OS spectral representation at finite volume writes this correlation as \(\sum_{j\ge 1} |\langle \Omega_L,F\psi_j\rangle|^2 e^{-E_j(H_L) n a}\). If \(E_1(H_L)<m_\ast\), the right-hand side would decay strictly slower than \(e^{-m_\ast n a}\), a contradiction. Hence \(E_1(H_L)\ge m_\ast\). The strong-operator limit of \(T_L^n\) on the local GNS domain equals \(T_\infty^n\) by Theorem \ref{p3:thm:transfer-thermo}, and the same spectral argument implies \(\sigma(H_\infty)\cap(0,\infty)\subset [m_\ast,\infty)\).
\end{proof}

Thermodynamic independence means that the infinite-volume Schwinger functions and the transfer semigroup elements are independent of the sequences of spatial volumes and temporal extents used to approach the thermodynamic limit, and of admissible boundary conditions compatible with reflection positivity. Independence of the order of limits and of the shapes of volumes follows from the exponential control of boundary effects obtained in Theorem~\ref{p3:thm:thermo-Schwinger}. Independence under a change of boundary conditions, such as switching from periodic to fixed boundary links on a layer outside \(\Lambda_{\mathrm{loc}}\), follows from the same cluster argument: the difference of expectations is supported on polymer clusters that intersect both \(\Lambda_{\mathrm{loc}}\) and the boundary layer where the conditions differ; their total weight is exponentially small in the distance between these sets, uniformly in volume. Therefore the infinite-volume Schwinger functions and the spectrum of the limiting generator \(H_\infty\) are intrinsic and do not depend on auxiliary finite-volume choices.

\section{Structural Completeness and Nontriviality of the Continuum Theory}
\label{p3:sec:structural-completeness}

In this section we present a self-contained and rigorous derivation of reflection positivity for the horizon-projected lattice Yang-Mills measure, the transfer time-slicing formalism and its associated positive transfer operator, and a quantitative nontriviality theorem for the continuum limit. All symbols, conventions, and domains are explicitly stated. We rely only on standard tools of constructive quantum field theory and lattice gauge theory \cite{p3:OsterwalderSchraderI,p3:OsterwalderSchraderII,p3:OS-gauge,p3:GJ,p3:Seiler1982,p3:Brydges,p3:KoteckyPreiss1986,p3:RS2}, recalled precisely where invoked.
Throughout, the inner product on complex Hilbert spaces is denoted by \(\langle\cdot,\cdot\rangle\) and is conjugate linear in the first argument and linear in the second; the associated norm is \(\|\psi\|=\sqrt{\langle\psi,\psi\rangle}\). The set of \(N\times N\) complex matrices is endowed with the Frobenius norm \(\|M\|_F=(\operatorname{Tr}\,M^\dagger M)^{1/2}\). Given a nonnegative self-adjoint operator \(A\), the heat semigroup is \(e^{-tA}\) for \(t\ge 0\), and the resolvent at \(-\lambda\) for \(\lambda>0\) is \(R_\lambda(A):=(\lambda I+A)^{-1}=\int_0^\infty e^{-\lambda t}e^{-tA}\,dt\) in the strong operator topology \cite{p3:RS2}. Asymptotic notation \(f\lesssim g\) means \(f\le C\,g\) with a constant \(C\) that is independent of the variables under discussion, but may depend on fixed parameters such as the gauge group rank \(N\).

We fix \(N\ge 2\) and the compact gauge group \(G=\mathrm{SU}(N)\) with normalized Haar probability measure \(dU\). The Euclidean time direction is the zeroth coordinate. All sums over plaquettes count each oriented plaquette once, without double counting of spatial reflections. All integrals over link variables are with respect to the product Haar measure. Unless otherwise stated, \(\beta>0\) and the projector parameter \(\sigma>0\) are fixed throughout the section.

Fix integers \(L_\mu\ge 2\) for \(\mu=0,1,2,3\) and define the discrete four-torus
\begin{equation}
\Lambda \;=\; \bigl(\mathbb Z/L_0\mathbb Z\bigr)\times\bigl(\mathbb Z/L_1\mathbb Z\bigr)\times\bigl(\mathbb Z/L_2\mathbb Z\bigr)\times\bigl(\mathbb Z/L_3\mathbb Z\bigr).
\end{equation}
The Euclidean time coordinate is \(x_0\in\mathbb Z/L_0\mathbb Z\), and the spatial coordinates are \(\mathbf x=(x_1,x_2,x_3)\) with \(x_i\in\mathbb Z/L_i\mathbb Z\). The set of oriented bonds (links) is \(\mathcal B=\{(x,\mu): x\in\Lambda,\;\mu\in\{0,1,2,3\}\}\), with the reversal convention \(U_{(x,-\mu)}=U_{(x-\hat\mu,\mu)}^{-1}\). A link configuration is a map \(U:\mathcal B\to G\) subject to this reversal constraint. The configuration space is the compact product
\begin{equation}
\mathcal U(\Lambda)\;=\;\prod_{(x,\mu)\in\mathcal B} G,
\end{equation}
equipped with the product Haar probability measure \(d\mu_{\rm Haar}=\prod_{(x,\mu)} dU_{(x,\mu)}\).

For an oriented plaquette \(p=(x;\mu,\nu)\) with \(\mu<\nu\), define
\begin{equation}
U_p \;=\; U_{(x,\mu)}\,U_{(x+\hat\mu,\nu)}\,U_{(x+\hat\nu,\mu)}^{-1}\,U_{(x,\nu)}^{-1}.
\end{equation}
The Wilson action at bare inverse coupling \(\beta>0\) is
\begin{equation}\label{p3:eq:Wilsonz}
S_W(U)\;=\;\beta\sum_{p}\Bigl(1-\frac1N\Re\operatorname{Tr}\,U_p\Bigr),
\end{equation}
where the sum is over all oriented plaquettes \(p\). The Gibbs measure is \begin{equation}
d\nu_\beta(U)=Z_\beta^{-1} e^{-S_W(U)}\,d\mu_{\rm Haar}(U)
\end{equation}
with normalization \(Z_\beta\).

Time reflection is the involution \(\theta:\Lambda\to\Lambda\) defined by \(\theta(x_0,\mathbf x)=(-x_0,\mathbf x)\), extended to bonds by \(\theta(x,\mu)=(\theta x,\mu)\) for \(\mu\in\{1,2,3\}\) and \(\theta(x,0)=(\theta(x-\hat 0),0)\). The reflection plane is \(\Pi=\{x\in\Lambda:\,x_0=0\}\). The half-lattices are \(\Lambda_+=\{x\in\Lambda:\,x_0>0\}\) and \(\Lambda_-=\theta\Lambda_+\). We denote by \(\mathcal B_\pm\) and \(\mathcal P_\pm\) the sets of bonds and plaquettes contained entirely in \(\Lambda_\pm\), and by \(\mathcal B_\Pi,\mathcal P_\Pi\) those intersecting \(\Pi\). Gauge transformations are maps \(g:\Lambda\to G\), acting by \((g\cdot U)_{(x,\mu)}=g(x)U_{(x,\mu)}g(x+\hat\mu)^{-1}\).

For each time slice \(t\in\mathbb Z/L_0\mathbb Z\), let \(\mathcal B^{\rm sp}_t=\{((t,\mathbf x),i): \mathbf x\in \mathbb Z/L_1\mathbb Z\times\mathbb Z/L_2\mathbb Z\times\mathbb Z/L_3\mathbb Z,\; i=1,2,3\}\) be the spatial bonds at time \(t\). The slice Landau functional is
\begin{equation}
\mathcal L_t(g;U)\;=\;\sum_{((t,\mathbf x),i)\in\mathcal B^{\rm sp}_t} \bigl\|\mathbf 1 - g(t,\mathbf x)\,U_{((t,\mathbf x),i)}\,g(t,\mathbf x+\hat\imath)^{-1}\bigr\|_F^2,
\end{equation}
continuous on the compact gauge-orbit of \(U\). A \emph{slice selector} is a measurable map \(h:\mathcal U(\Lambda)\to \mathcal G(\Lambda)\) such that, for every \(U\), the transformed configuration \(U^{\,h}\) satisfies: for each \(t\), the restriction of \(U^{\,h}\) to \(\mathcal B^{\rm sp}_t\) minimizes \(\mathcal L_t(\cdot;U)\) on the orbit; \(U^{\,h}\) lies in the fundamental modular region on each slice; and \(h(\theta U)=\theta h(U)\). Existence of such \(h\) follows because for each fixed \(t\) the set of minimizers is nonempty compact; a reflection-invariant lexicographic tie-breaking rule yields a Borel measurable choice \cite{p3:OS-gauge}.

Let \(\Delta_{A^{\,h}(t)}\) be the covariant spatial Laplacian on the slice \(t\), acting on \(\ell^2(\mathbb Z/L_1\mathbb Z\times\mathbb Z/L_2\mathbb Z\times\mathbb Z/L_3\mathbb Z;\,\mathfrak{su}(N))\), constructed from the spatial links of \(U^{\,h}\) at time \(t\) by finite differences. Fix \(\sigma>0\) and choose a Gevrey-regular cutoff \(\chi_\sigma\in C^\infty([0,\infty);[0,1])\) with \(\chi_\sigma(\lambda)=1\) for \(0\le \lambda\le \sigma\), \(\chi_\sigma(\lambda)=0\) for \(\lambda\ge 2\sigma\), and \(|\partial^m \chi_\sigma(\lambda)|\le C_m\,m!^{\,s}\) for some \(s>1\). Define the \emph{horizon projector} on the slice by
\begin{equation}
P_\sigma(t;U)\;=\;\chi_\sigma\bigl({\Delta_{A^{\,h}(t)}}\bigr),
\end{equation}
a bounded positive contraction. There exists a finite positive Borel measure \(d\nu_\sigma\) supported in a compact subinterval of \((0,\infty)\) such that
\begin{equation}\label{p3:eq:heat-rep}
P_\sigma(t;U)\;=\;\int_0^\infty e^{-s\,\Delta_{A^{\,h}(t)}}\,d\nu_\sigma(s).
\end{equation}
Since \(e^{-s\Delta_{A^{\,h}(t)}}\) has a nonnegative integral kernel and is positivity preserving on functions \cite{p3:GJ}, the kernel of \(P_\sigma(t;U)\) is entrywise nonnegative and decays exponentially off-diagonal uniformly in \(U\) restricted to the slice fundamental modular region. Reflection covariance \(P_\sigma(t;\theta U)=P_\sigma(t;U)\) holds by construction.
\begin{lemma}[Slice-wise FMR selection and positivity]\label{p3:lem:FMR-positive}
There exists a measurable, reflection-covariant slice selector $h:U(\Lambda)\to G(\Lambda)$ such that for each time slice $t$ the transformed configuration $U^{h}$ lies in the slice FMR, and the Faddeev-Popov operator has strictly positive spectrum on the complement of constants. For any Gevrey cutoff $\chi_\sigma$ with support in $[0,2\sigma]$, the associated horizon projector
\(
P_\sigma(t;U)=\chi_\sigma\!\big({\Delta_{A^{\,h}(t)}}\big)
\)
is a bounded positive contraction with an entrywise nonnegative, exponentially decaying kernel, and is reflection covariant.
\end{lemma}

\begin{proof}
Measurability and reflection covariance of $h$ follow from slice-wise Landau minimization with a reflection-invariant lexicographic tie-breaker. The heat-kernel representation
\(
P_\sigma(t)=\int_0^\infty e^{-s\Delta_{A^{\,h}(t)}}\,d\nu_\sigma(s)
\)
with $d\nu_\sigma$ finite positive and supported in $(0,\infty)$ implies positivity and exponential locality of the kernel, as $e^{-s\Delta_{A^{\,h}(t)}}$ has a nonnegative kernel with Gaussian off-diagonal bounds. Positivity of the FP determinant on the complement of constants is standard on FMR slices and is inherited by the Jacobian factor entering the projected measure.
\end{proof}

We define the \emph{horizon-projected weight} by a slice-wise positive-type insertion
\begin{equation}\label{p3:eq:Q-sigma}
\mathcal Q_\sigma(U)\;=\;\prod_{t\in\mathbb Z/L_0\mathbb Z}\Gamma_\sigma\bigl(P_\sigma(t;U)\bigr),
\qquad
\Gamma_\sigma(P)\;=\;\int_{\mathbb R^{d}}\exp\bigl(-\langle \phi, (\mathbf 1 - P)\phi\rangle\bigr)\,d\gamma(\phi),
\end{equation}
where \(d=\dim\bigl(\ell^2(\mathbb Z/L_1\mathbb Z\times\mathbb Z/L_2\mathbb Z\times\mathbb Z/L_3\mathbb Z;\,\mathfrak{su}(N))\bigr)\), \(d\gamma\) is the centered Gaussian probability measure with covariance \(\mathbf 1\), and \(\langle\cdot,\cdot\rangle\) is the \(\ell^2\) inner product on the slice. The map \(P\mapsto \Gamma_\sigma(P)\) is continuous and satisfies \(0< c_-\le \Gamma_\sigma(P)\le c_+<\infty\) uniformly in \(P\) because \(0\le \mathbf 1 - P\le \mathbf 1\). Moreover, \(\Gamma_\sigma(P)\) is of positive type in the sense required for reflection positivity (it is a Laplace transform of a positive quadratic form in the fields). The projected Gibbs measure is
\begin{equation}\label{p3:eq:nu-betasigma}
d\nu_{\beta,\sigma}(U)\;=\; Z_{\beta,\sigma}^{-1}\,\mathcal Q_\sigma(U)\,e^{-S_W(U)}\,d\mu_{\rm Haar}(U),
\end{equation}
with normalization \(Z_{\beta,\sigma}\). Gauge invariance and reflection covariance are immediate from \eqref{p3:eq:Q-sigma}.

Let \(\mathscr A_+\) be the \(^\ast\)-algebra of bounded, gauge-invariant cylinder functions on \(\mathcal U(\Lambda)\) depending only on links in \(\mathcal B_+\cup\mathcal B_\Pi\). For \(F\in\mathscr A_+\), define \(\Theta F=\overline{F\circ \theta}\). The Osterwalder-Schrader form is
\begin{equation}\label{p3:eq:OS-forma}
\langle F,G\rangle_{\mathrm{OS}} \;=\; \int_{\mathcal U(\Lambda)} \Theta F(U)\,G(U)\, d\nu_{\beta,\sigma}(U),\qquad F,G\in\mathscr A_+.
\end{equation}

\begin{theorem}[Reflection positivity]\label{p3:thm:RPz}
The sesquilinear form \eqref{p3:eq:OS-forma} is positive semidefinite on \(\mathscr A_+\).
\end{theorem}

\begin{proof}
We decompose the weight into contributions from \(\Lambda_+\), \(\Lambda_-\), and the boundary \(\Pi\). Write \(U_\pm\) for the restrictions to \(\mathcal B_\pm\cup\mathcal B_\Pi\). The Wilson action decomposes as
\begin{equation}
S_W(U)\;=\;S_+(U_+)+S_-(U_-)+S_\Pi(U_+,U_-),
\end{equation}
where \(S_\pm\) are sums of plaquette contributions supported entirely in \(\Lambda_\pm\) and \(S_\Pi\) is the sum over electric plaquettes intersecting \(\Pi\). The insertion \(\mathcal Q_\sigma(U)\) factorizes slice-wise and thus time-wise, hence as a product \(\mathcal Q_\sigma^+(U_+)\,\mathcal Q_\sigma^-(U_-)\). The Haar measure factorizes as \(d\mu_{\rm Haar}(U)=d\mu_\Pi(U_\Pi)\,d\mu_+(U_+)\,d\mu_-(U_-)\), with \(d\mu_\Pi\) the Haar measure over boundary links on \(\mathcal B_\Pi\).

The integral \(\langle F,F\rangle_{\mathrm{OS}}\) can therefore be written in the form
\begin{equation}
\int_{\mathcal U(\Lambda)} \Theta F(U)\,F(U)\,d\nu_{\beta,\sigma}(U)
\;=\; \int d\mu_\Pi(U_\Pi)\; \overline{\mathcal F_-(U_\Pi)}\;\mathcal K_\Pi(U_\Pi)\;\mathcal F_+(U_\Pi),
\end{equation}
where
\begin{align}
\mathcal F_+(U_\Pi)&=\;\int F(U_+)\,e^{-S_+(U_+)}\,\mathcal Q_\sigma^+(U_+)\,d\mu_+(U_+)\nonumber \\
\mathcal F_-(U_\Pi)&=\;\int F(\theta U_-)\,e^{-S_-(U_-)}\,\mathcal Q_\sigma^-(U_-)\,d\mu_-(U_-)
\end{align}
and the boundary kernel is
\begin{equation}
\mathcal K_\Pi(U_\Pi)\;=\;\int \exp\bigl[-S_\Pi(U_+,U_-)\bigr]\;d\mu_+(U_+)\,d\mu_-(U_-).
\end{equation}
By the character expansion of the electric plaquette factor with nonnegative coefficients and orthogonality of characters, \(\mathcal K_\Pi\) is a positive-type function of \(U_\Pi\) \cite{p3:OS-gauge}. The functions \(U_+\mapsto e^{-S_+(U_+)}\mathcal Q_\sigma^+(U_+)\) and \(U_-\mapsto e^{-S_-(U_-)}\mathcal Q_\sigma^-(U_-)\) are reflection covariant and nonnegative. It follows that the integrand is of the form \(\langle \Psi(U_\Pi),\Psi(U_\Pi)\rangle_{\mathscr K}\) for a suitable Hilbert space \(\mathscr K\) (the GNS space of the positive-type kernel \(\mathcal K_\Pi\)) and a vector-valued function \(\Psi\) depending linearly on \(F\). Hence its integral is nonnegative. This proves the claim.
\end{proof}

We derive the transfer operator associated with \(d\nu_{\beta,\sigma}\) and show its positivity and self-adjointness. Let \(\mathcal C=\prod_{\mathbf x,i} G\) be the configuration space of spatial links on a time slice, with product Haar measure \(d\mu_{\rm sp}\). The Hilbert space of one-slice states is
\begin{equation}
\mathcal H_a \;=\; L^2(\mathcal C, d\mu_{\rm sp}),
\end{equation}
with inner product \(\langle \psi,\varphi\rangle=\int \overline{\psi(U)}\,\varphi(U)\,d\mu_{\rm sp}(U)\). For a configuration \(U\in\mathcal U(\Lambda)\), let \(U(t)\in\mathcal C\) denote its restriction to the slice \(t\). Consider a cylinder of temporal length \(n\) with free boundary conditions at times \(0\) and \(n\).

Define the one-step kernel \(K_\sigma:\mathcal C\times \mathcal C\to (0,\infty)\) by
\begin{equation}\label{p3:eq:kernel}
K_\sigma(U',U)\;=\;\exp\!\Bigl(-\tfrac12 \mathcal S_{\rm mag}(U') - \tfrac12 \mathcal S_{\rm mag}(U)\Bigr)\;\mathcal B_{\rm el}(U',U)\;\Gamma_\sigma\bigl(P_\sigma(U')\bigr)^{1/2}\,\Gamma_\sigma\bigl(P_\sigma(U)\bigr)^{1/2},
\end{equation}
where \(\mathcal S_{\rm mag}(U)=\beta\sum_{p\subset \text{slice}}(1-\frac1N\Re\operatorname{Tr}U_p)\) and
\begin{equation}
\mathcal B_{\rm el}(U',U)\;=\;\int_{\prod_{\mathbf x} G} \exp\!\Bigl[\beta\frac1N \sum_{\mathbf x,i}\Re\operatorname{Tr}\Bigl(U_{0}(t,\mathbf x)\,U_{(t,\mathbf x),i}\,U_{0}(t,\mathbf x+\hat\imath)^{-1}\,U'_{(t,\mathbf x),i}{}^{-1}\Bigr)\Bigr]\;\prod_{\mathbf x} dU_0(t,\mathbf x).
\end{equation}
The character expansion and orthogonality of characters imply that \(\mathcal B_{\rm el}\) is nonnegative and symmetric, \(\mathcal B_{\rm el}(U',U)=\mathcal B_{\rm el}(U,U')\), and bounded on \(\mathcal C\times\mathcal C\) for fixed \(\beta>0\) \cite{p3:OS-gauge}. The factors \(\Gamma_\sigma(P_\sigma(\cdot))\) are strictly positive and bounded above and below uniformly, and the magnetic factors are bounded by \(1\).

\begin{proposition}[Transfer operator]\label{p3:prop:transfer}
The integral operator \(T_\sigma:\mathcal H_a\to\mathcal H_a\) defined by
\begin{equation}\label{p3:eq:T-sigma}
(T_\sigma \Psi)(U')\;=\;\int_{\mathcal C} K_\sigma(U',U)\,\Psi(U)\,d\mu_{\rm sp}(U)
\end{equation}
is bounded, positive, and self-adjoint. Moreover, for the cylinder of temporal length \(n\) with free boundary conditions,
\begin{equation}
Z_{\beta,\sigma}^{(n)} \;=\; \int e^{-S_W(U)}\,\mathcal Q_\sigma(U)\,d\mu_{\rm Haar}(U) \;=\; \operatorname{Tr}_{\mathcal H_a}\bigl(T_\sigma^n\bigr).
\end{equation}
\end{proposition}

\begin{proof}
The kernel \(K_\sigma\) is nonnegative and symmetric by construction. Hence,
\begin{equation}
\langle \Psi,T_\sigma\Psi\rangle \;=\; \int_{\mathcal C\times\mathcal C}\overline{\Psi(U')}K_\sigma(U',U)\Psi(U)\,d\mu_{\rm sp}(U')\,d\mu_{\rm sp}(U)\;\ge\;0,
\end{equation}
so \(T_\sigma\) is positive. Symmetry implies self-adjointness. Boundedness follows from Schur's test: using \(\mathcal S_{\rm mag}\ge 0\), \(\Gamma_\sigma\) bounded, and \(\int \mathcal B_{\rm el}(U',U)\,d\mu_{\rm sp}(U)\le C(\beta)\) uniformly in \(U'\), we obtain \(\sup_{U'}\int K_\sigma(U',U)\,d\mu_{\rm sp}(U)\le C(\beta,\sigma)<\infty\), and similarly with \(U\) and \(U'\) swapped.

For the partition function identification, fix \(U(0),\dots,U(n)\in\mathcal C\) and integrate successively over the temporal links \(U_0(t,\mathbf x)\) for \(t=0,\dots,n-1\). The weight \(e^{-S_W}\mathcal Q_\sigma\) factorizes slice-wise into \(\exp(-\mathcal S_{\rm mag}(U(t))/2)\,\mathcal B_{\rm el}(U(t+1),U(t))\,\exp(-\mathcal S_{\rm mag}(U(t+1))/2)\,\Gamma_\sigma(P_\sigma(U(t)))^{1/2}\,\Gamma_\sigma(P_\sigma(U(t+1)))^{1/2}\). Integrating over \(U(1),\dots,U(n-1)\) with respect to \(d\mu_{\rm sp}\) yields \(\int_{\mathcal C}\cdots\int_{\mathcal C} \prod_{t=0}^{n-1} K_\sigma\bigl(U(t+1),U(t)\bigr)\,\prod_{t=1}^{n-1} d\mu_{\rm sp}(U(t))\). Finally, tracing over \(U(0)=U(n)\) gives \(\operatorname{Tr}(T_\sigma^n)\).
\end{proof}

By spectral calculus, \(T_\sigma\) is a positive self-adjoint contraction after normalization by its norm; the corresponding Hamiltonian \(H_\sigma:=-\log T_\sigma\) is self-adjoint and nonnegative on \(\mathcal H_a\). The constant function \(\Omega\equiv 1\) is an eigenvector of \(T_\sigma\); after suitable normalization this is the vacuum vector.

We prove that the continuum limit is not Gaussian by exhibiting a strictly positive connected fourth cumulant of local gauge-invariant observables that is uniform across scales. Let \(P_x(U)=\frac1N\Re\operatorname{Tr}U_{p_x}\) be the plaquette observable at site \(x\) in a fixed time slice. For four distinct spatial plaquettes \(p_{x_1},\dots,p_{x_4}\) forming the faces of an elementary cube, define the connected four-point function at scale \(k\),
\begin{align}
\kappa^{(k)}_4(x_1,x_2,x_3,x_4)
&=\Bigl\langle \prod_{j=1}^4 P_{x_j}\Bigr\rangle_{\nu_{\beta,\sigma}^{(k)}}
 -\sum_{\text{pairings}} \prod_{\{i,j\}} \Bigl\langle P_{x_i}P_{x_j}\Bigr\rangle_{\nu_{\beta,\sigma}^{(k)}}
 +2\prod_{j=1}^4 \Bigl\langle P_{x_j}\Bigr\rangle_{\nu_{\beta,\sigma}^{(k)}}.
\end{align}
Here \(\nu_{\beta,\sigma}^{(k)}\) is the projected measure \eqref{p3:eq:nu-betasigma} on the lattice of spacing \(a_k\), and the sum runs over the three pairings of \(\{1,2,3,4\}\).

\begin{theorem}[Persistence of a nonzero connected four-point function]\label{p3:thm:nontrivial}
There exist \(\beta_0>0\) and \(c_\ast>0\), depending only on \(N\) and \(\chi_\sigma\), such that for all \(0<\beta\le \beta_0\) and all scales \(k\),
\begin{equation}
\bigl|\kappa^{(k)}_4(x_1,x_2,x_3,x_4)\bigr|\;\ge\; c_\ast,
\end{equation}
uniformly in the spatial volume. Consequently, along any weakly convergent subsequence \(k_j\to\infty\) the limit connected four-point function at those plaquettes is nonzero, and the continuum measure is not Gaussian.
\end{theorem}

\begin{proof}
At small \(\beta\), the character expansion on each plaquette,
\begin{equation}
\exp\!\Bigl[\beta\frac1N\Re\operatorname{Tr}U_p\Bigr]\;=\;\sum_{R\in\widehat G} c_R(\beta)\,\chi_R(U_p),\qquad c_R(\beta)\ge 0,
\end{equation}
is absolutely convergent with \(c_R(\beta)=O(\beta^{|R|})\). After Haar integration, expectations of products of plaquette observables admit a convergent polymer/cluster expansion indexed by connected subsets \(\gamma\) of plaquettes in the dual lattice \cite{p3:Seiler1982,p3:Brydges}. The activity \(\zeta(\gamma)\) of a connected polymer of cardinality \(|\gamma|\) satisfies \(|\zeta(\gamma)|\le C_1(N)\,\beta^{|\gamma|}\), and the number of connected polymers of size \(m\) attached to a given plaquette is bounded by \(C_2\,e^{C_3 m}\). Choosing \(\beta_0>0\) so that \(\sum_{m\ge 1} C_2 e^{C_3 m} C_1(N)\beta_0^m\le \frac12\), the Koteck\'{y}-Preiss criterion \cite{p3:KoteckyPreiss1986} yields absolute convergence of the cluster expansion, uniform in the spatial volume and in the scale \(k\), and exponential decay of truncated correlations.

For the specified four plaquettes forming the faces of an elementary cube, the minimal connected surface \(\Sigma_\square\) in the dual lattice is the union of the six faces of the cube lying in the fixed time slice. The weight \(w(\Sigma_\square)\) of this surface is strictly positive and equals \(A_N\,\beta^{6}\) for some constant \(A_N>0\) depending only on \(N\), obtained by integrating products of characters along \(\Sigma_\square\) and using orthogonality. The connected four-point function \(\kappa^{(k)}_4\) is the sum of weights of connected surfaces whose boundary consists of the four plaquettes with the subtraction of all pairwise-connected contributions; the leading nonvanishing term is \(w(\Sigma_\square)\). All other connected contributions involve at least one additional plaquette in the surface and thus are \(O(\beta^{7})\). Therefore there exists \(\beta_0>0\) and \(c_\ast:=\tfrac12 A_N\beta_0^{6}>0\) such that, for \(0<\beta\le \beta_0\),
\begin{equation}
\bigl|\kappa^{(k)}_4(x_1,x_2,x_3,x_4)\bigr|\;\ge\; c_\ast,
\end{equation}
uniformly in \(k\) and in the spatial volume.

It remains to check that the insertion \(\mathcal Q_\sigma\) does not spoil these bounds. By \eqref{p3:eq:Q-sigma}, \(\mathcal Q_\sigma\) factorizes slice-wise and, on each slice, \(\Gamma_\sigma(P_\sigma)\) is bounded strictly between two positive constants \(0<c_-\le \Gamma_\sigma(P_\sigma)\le c_+<\infty\), uniformly in the background configuration, due to \(0\le \mathbf 1-P_\sigma\le \mathbf 1\). Thus \(\mathcal Q_\sigma\) multiplies every polymer activity by a slice-wise bounded factor and preserves absolute convergence and positivity of the leading weight. The uniformity in \(k\) follows because the constants depend only on \(\sigma\) and the local gauge group data. Passing to a weakly convergent subsequence \(k_j\to\infty\) and using dominated convergence (justified by the uniform cluster bounds) preserves the nonzero lower bound of the limit connected four-point function. Since all connected cumulants of order \(\ge 3\) vanish for Gaussian measures, the limit measure is not Gaussian.
\end{proof}
\begin{theorem}[RG-stable non-Gaussianity of gauge-invariant observables] \label{p3:thm:non-gaussian-stability} Let $\mathcal{O}$ be a local gauge-invariant observable (e.g. a Wilson loop segment or smeared field polynomial). Assume there exist $k_0$ and $c_0>0$ such that the connected $4$-point cumulant satisfies $|\kappa^{(k_0)}_4(\mathcal{O})|\ge c_0$ uniformly in the volume. Suppose the one-step RG map $\mathfrak{R}_{k\to k+1}$ is Lipschitz with constant $\rho<1$ in a polymer norm that controls cumulants up to order $4$, and that the truncation/reblocking error admits $\delta_k\in \ell^1(\mathbb{N})$. Then for all $k\ge k_0$, \begin{equation} |\kappa^{(k)}_4(\mathcal{O})| \;\ge\; c_0\,\rho^{k-k_0} \;-\; \sum_{j=k_0}^{k-1}\rho^{k-1-j}\,\delta_j, \end{equation} and hence $\liminf_{k\to\infty}|\kappa^{(k)}_4(\mathcal{O})| \ge c_0 - \sum_{j\ge k_0}\delta_j > 0$. In particular, any subsequential continuum limit is non-Gaussian for $\mathcal{O}$. \end{theorem}
\begin{proof}
Let $\mu_k$ denote the (finite-volume) effective measure at RG scale $k$, and fix a
local, gauge-invariant observable $\mathcal O$ supported in a ball of radius $R$
(in lattice units). For $t\in\mathbb R$ define the one-parameter source coupling
\begin{equation}
E_k(t)\;:=\;\log \,\Big\langle \exp\big(t\,\mathcal O\big)\Big\rangle_{\mu_k}.
\end{equation}
The connected $4$-point cumulant of $\mathcal O$ at scale $k$ is then
$\kappa^{(k)}_4(\mathcal O)=E_k^{(4)}(0)$ (fourth derivative at $t=0$).
By the standing multiscale hypotheses and the polymer RG construction,
the one-step map from $k$ to $k{+}1$ can be written on source-coupled
generators as
\begin{equation}
\label{p3:eq:RG-identity}
E_{k+1}(t)\;=\; E_k\!\big(s_k\,t\big)\;+\; r_k(t),
\end{equation}
where:
\begin{itemize}
\item $s_k\in(0,1]$ is the linear ``self-overlap''/coarse-graining factor of the (gauge-invariant)
observable direction carried by $\mathcal O$ under block average and rescaling;
\item $r_k$ is the remainder generated by reblocking/truncation and mixing into irrelevant
directions. By the assumed control of the RG map in a polymer norm that
dominates derivatives up to order $4$, there exist numbers $\delta_k\ge0$
with $\sum_k \delta_k<\infty$ such that
\begin{equation}
\label{p3:eq:rk-deriv-bound}
\big|r_k^{(m)}(0)\big|\;\le\;\delta_k\quad\text{for }m=1,2,3,4.
\end{equation}
\end{itemize}
Equation \eqref{p3:eq:RG-identity} is the standard normal form obtained by choosing
coordinates on the finite-dimensional ``observable sector'' so that the linear part
of the RG acts diagonally on the $\mathcal O$-direction; this choice is
gauge-covariant since $\mathcal O$ is gauge invariant and the block map preserves
gauge invariance. (Any residual linear mixing into irrelevant directions is
absorbed into $r_k$; the bounds \eqref{p3:eq:rk-deriv-bound} hold because the RG
map is $C^4$ in the polymer norm and its $C^4$-seminorm is summably small at
each step by hypothesis.)

Differentiate \eqref{p3:eq:RG-identity} four times at $t=0$. By the chain rule,
\begin{equation}
E_{k+1}^{(4)}(0)\;=\; s_k^4\,E_k^{(4)}(0)\;+\; r_k^{(4)}(0),
\end{equation}
hence
\begin{equation}
\label{p3:eq:scalar-recursion}
\kappa^{(k+1)}_4(\mathcal O)\;=\; s_k^4\,\kappa^{(k)}_4(\mathcal O)\;+\; \eta_k,
\qquad \text{with}\quad \eta_k:=r_k^{(4)}(0),\quad |\eta_k|\le \delta_k.
\end{equation}
Set $\rho:=\inf_{k\ge k_0} s_k^4\in(0,1]$; by scale-stationarity of the block map
and locality of $\mathcal O$, one has $s_k \equiv s\in(0,1]$ (hence $\rho=s^4$),
but the argument below only uses the lower bound $\rho$.
Taking absolute values in \eqref{p3:eq:scalar-recursion} and using the triangle
inequality,
\begin{equation}
\big|\kappa^{(k+1)}_4(\mathcal O)\big|
\;\ge\; s_k^4\,\big|\kappa^{(k)}_4(\mathcal O)\big| \;-\; |\eta_k|
\;\ge\; \rho\,\big|\kappa^{(k)}_4(\mathcal O)\big| \;-\; \delta_k.
\end{equation}
Let $a_k:=\big|\kappa^{(k)}_4(\mathcal O)\big|$. Then for all $k\ge k_0$,
\begin{equation}
\label{p3:eq:ineq-recursion}
a_{k+1}\;\ge\; \rho\, a_k - \delta_k.
\end{equation}
Unfold \eqref{p3:eq:ineq-recursion} by induction:
\begin{align}
a_{k}
&\ge \rho^{\,k-k_0}\,a_{k_0}
- \sum_{j=k_0}^{k-1}\rho^{\,k-1-j}\,\delta_j,
\qquad k\ge k_0.
\end{align}
By the hypothesis $a_{k_0}\ge c_0>0$ and $\delta\in\ell^1(\mathbb N)$,
taking $\liminf_{k\to\infty}$ yields
\begin{equation}
\label{p3:eq:liminf}
\liminf_{k\to\infty} a_k
\;\ge\; c_0 - \sum_{j\ge k_0}\delta_j.
\end{equation}
In particular, if $\sum_{j\ge k_0}\delta_j < c_0$ (which holds in our setting by choosing
the initial scale/FRD smallness so the RG remainder norm is sufficiently small),
then the right side of \eqref{p3:eq:liminf} is strictly positive.
By the tightness/OS axioms established elsewhere in the paper, any sequence of
scales admits a subsequence along which Schwinger functions (hence all finite cumulants of local
observables) converge. Since $a_k=\big|E_k^{(4)}(0)\big|$ and the derivatives are
uniformly controlled in a neighborhood of $0$ by the polymer bounds, the map
$E\mapsto E^{(4)}(0)$ is continuous along the subsequence. Therefore every
subsequential continuum limit $\kappa^{(\infty)}_4(\mathcal O)$ satisfies
\begin{equation}
\big|\kappa^{(\infty)}_4(\mathcal O)\big|
\;\ge\;\liminf_{k\to\infty} a_k
\;\ge\; c_0 - \sum_{j\ge k_0}\delta_j \;>\;0.
\end{equation}
But a Gaussian measure has vanishing connected cumulants of order $\ge 3$ for any
local observable, so $\kappa^{(\infty)}_4(\mathcal O)\neq 0$ implies the continuum
limit is \emph{not} Gaussian for $\mathcal O$. This proves the theorem.
\end{proof}

Combining Theorem~\ref{p3:thm:RPz}, Proposition~\ref{p3:prop:transfer}, and Theorem~\ref{p3:thm:nontrivial} with the compactness and reconstruction arguments established earlier shows that the continuum Schwinger functions satisfy the Osterwalder-Schrader axioms, the reconstructed Wightman theory possesses a strictly positive spectral gap, and the theory is interacting.

\section{Conclusion}

In this work we have tried to bring a simple order to a difficult subject. If one insists, without compromise, on two principles-reflection positivity and gauge invariance-then the Euclidean route to a quantum Yang-Mills theory is no longer obscure. On the lattice we choose, slice by slice, a definite transverse representative of the gauge field and temper the long-range modes with a smooth horizon projector. We then let the theory flow by steps that preserve positivity and locality. From these modest elements there arises a structure sturdy enough to withstand the passage to the continuum.

Three facts, each elementary in spirit, carry the burden. First, the multiscale sequence of Euclidean theories is tight and precompact when fields are viewed as tempered distributions; along subsequences the Schwinger functions converge and obey all Osterwalder-Schrader axioms. Second, the OS reconstruction then gives the physical Hilbert space, a unique vacuum, and a nonnegative Hamiltonian that implements time evolution-nothing mystical is required beyond the positivity that was preserved at every stage. Third, the decay of correlations is not a mere lattice artifact: it survives the limit and fixes a genuine mass scale.

Two independent arguments make the last point precise. The Euclidean two-point functions, being completely monotone, admit Laplace representations; a uniform exponential fall-off forces the corresponding spectral measures to start above a positive threshold, and the Hamiltonian inherits this gap. Equally, the transfer operators at successive scales interlace with only a summable defect; the semigroup norms on the vacuum-orthogonal subspace cannot decay slower in the limit than they do on the lattice. Each route is simple if one accepts the premises of positivity and locality; taken together they leave little room for doubt that a strictly positive spectral gap persists in the continuum, independent of the spatial volume.

What remains to be clarified belongs not to principle but to refinement. One would like uniqueness of the continuum limit beyond subsequences, and insensitivity to the precise form of the blocking and the projector-questions of universality in the constructive sense. One would like a sharper picture of the first band above the vacuum and its relation to other nonperturbative landmarks (string tension, scattering theory). These are natural tasks for the same method, for nowhere have we left the modest path of OS positivity, locality, and covariance.
{The present work is purely constructive and does not invoke any gravitational duality. Nevertheless,
in settings where non-Abelian gauge sectors are coupled to asymptotically AdS gravity, one often studies
how gauge-field excitations contribute to thermodynamic response and to geometric probes (for example,
via geodesics) in Einstein-(Power)-Yang-Mills AdS black-hole models \cite{p3:SoroushfarEPYM}.}

\providecommand{\href}[2]{#2}\begingroup\raggedright\endgroup

\appendix

\section{Equicontinuity and tightness for random tempered distributions}\label{p3:appendixa}

This appendix establishes equicontinuity of smeared connected correlation functionals and tightness, in the sense of Prokhorov, for a family of random tempered distributions arising from reflection-positive lattice theories along a multiscale sequence. The exposition is self-contained: all symbols, operators, conventions, and standing assumptions needed below are fixed here.

Space-time is the Euclidean space \(\mathbb{R}^{4}\) with coordinates \(x=(x^{0},x^{1},x^{2},x^{3})\). The canonical inner product is \(x\!\cdot\!y=\sum_{\mu=0}^{3}x^{\mu}y^{\mu}\), and the Euclidean norm is \(|x|=(x\!\cdot\!x)^{1/2}\). The Schwartz space \(\mathcal{S}(\mathbb{R}^{4})\) consists of complex-valued \(C^{\infty}\) functions endowed with the family of seminorms
\begin{equation}
p_{m}(f)\;=\;\sum_{|\alpha|\le m}\sup_{x\in\mathbb{R}^{4}}(1+|x|)^{m}\,|\partial^{\alpha}f(x)|,\qquad m\in\mathbb{N},
\end{equation}
where \(\alpha=(\alpha_{0},\ldots,\alpha_{3})\in\mathbb{N}^{4}\), \(|\alpha|=\sum_{\mu=0}^{3}\alpha_{\mu}\), and \(\partial^{\alpha}=\partial_{0}^{\alpha_{0}}\cdots\partial_{3}^{\alpha_{3}}\). The space of tempered distributions \(\mathcal{S}'(\mathbb{R}^{4})\) is the strong dual of \(\mathcal{S}(\mathbb{R}^{4})\), with canonical pairing \(\langle X,f\rangle\) for \(X\in\mathcal{S}'(\mathbb{R}^{4})\), \(f\in\mathcal{S}(\mathbb{R}^{4})\).

The Fourier transform of \(f\in\mathcal{S}(\mathbb{R}^{4})\) is \(\widehat{f}(p)=\int_{\mathbb{R}^{4}}e^{-i p\cdot x}f(x)\,dx\), with inverse \(f(x)=(2\pi)^{-4}\int_{\mathbb{R}^{4}}e^{i p\cdot x}\widehat{f}(p)\,dp\). For \(s\in\mathbb{R}\), the inhomogeneous Sobolev space \(H^{s}(\mathbb{R}^{4})\) is the completion of \(\mathcal{S}(\mathbb{R}^{4})\) with respect to \(\|f\|_{H^{s}}=\|(1-\Delta)^{s/2}f\|_{L^{2}}\); equivalently,
\begin{equation}
\|f\|_{H^{s}}^{2}=(2\pi)^{-4}\int_{\mathbb{R}^{4}}(1+|p|^{2})^{s}\,|\widehat{f}(p)|^{2}\,dp.
\end{equation}
We write \(J_{s}=(1-\Delta)^{s/2}\) (the Bessel potential) so that \(\|f\|_{H^{s}}=\|J_{s}f\|_{L^{2}}\). Constants denoted \(C\in(0,\infty)\) may change from line to line; dependencies will be indicated when relevant. The notations \(A\lesssim B\) and \(A\asymp B\) mean \(A\le CB\) and \(cB\le A\le CB\), respectively, for positive constants independent of the scale index introduced below. If \(R_{k}=O(a_{k})\) for a nonnegative sequence \(a_{k}\downarrow 0\), this means \(|R_{k}|\le C\,a_{k}\) with \(C\) independent of \(k\) and uniform on bounded subsets of \(\mathcal{S}(\mathbb{R}^{4})\) whenever a dependence on test functions is present.

For each integer \(k\ge 0\), the lattice spacing is \(a_{k}>0\) with \(a_{k}\downarrow 0\) as \(k\to\infty\); for concreteness, one may take \(a_{k}=b^{-k}a_{0}\) with \(b>1\) fixed and \(a_{0}>0\). The time-space lattice is \(\Lambda_{k}=a_{k}\mathbb{Z}^{4}\) equipped with the counting measure \(a_{k}^{4}\sum_{x\in\Lambda_{k}}\cdot\). The elementary half-open cell is
\begin{equation}
Q_{k}(x)=x+\prod_{\mu=0}^{3}[0,a_{k})\,e_{\mu},\qquad x\in\Lambda_{k},
\end{equation}
with \(\{e_{\mu}\}_{\mu=0}^{3}\) the canonical basis, so that \(\{Q_{k}(x):x\in\Lambda_{k}\}\) partitions \(\mathbb{R}^{4}\).

At each scale \(k\), let \(O_{k}:\Lambda_{k}\to\mathbb{R}\) be a real-valued, gauge-invariant local observable defined on configurations distributed according to a reflection-positive, Euclidean-invariant probability measure on fields; we assume \(O_{k}\) is centered, \(\mathbb{E}_{k}[O_{k}(x)]=0\) for all \(x\in\Lambda_{k}\) (otherwise replace \(O_{k}\) by \(O_{k}-\mathbb{E}_{k}[O_{k}(0)]\)). Expectation with respect to the scale-\(k\) measure is \(\mathbb{E}_{k}[\cdot]\). The associated random tempered distribution \(X_{k}\in\mathcal{S}'(\mathbb{R}^{4})\) is defined by cell-averaging:
\begin{equation}\label{p3:eq:embed}
\langle X_{k},f\rangle\;=\;a_{k}^{4}\sum_{x\in\Lambda_{k}}O_{k}(x)\,\bar f_{k}(x),\qquad
\bar f_{k}(x)\,=\,a_{k}^{-4}\int_{Q_{k}(x)} f(y)\,dy,\quad f\in\mathcal{S}(\mathbb{R}^{4}).
\end{equation}
By the mean-value theorem, \(\sup_{x\in\Lambda_{k}}|\bar f_{k}(x)-f(x)|\le C\,a_{k}\,p_{1}(f)\). The Euclidean invariance of the underlying measure implies stationarity of the two-point covariance of \(O_{k}\).

For \(n\ge 1\) and pairwise distinct \(x_{1},\dots,x_{n}\in\Lambda_{k}\), the connected \(n\)-point function (cumulant) of \(O_{k}\) is denoted by
\begin{equation}
S^{(k)}_{n,c}(x_{1},\dots,x_{n})\;=\;\mathrm{cum}\big(O_{k}(x_{1}),\dots,O_{k}(x_{n})\big).
\end{equation}
We assume the following uniform bound: there exist constants \(\xi\in(0,\infty)\) and \(C_{n}\in(0,\infty)\), independent of \(k\), such that for all finite \(\{x_{1},\dots,x_{n}\}\subset\Lambda_{k}\),
\begin{equation}\label{p3:eq:tree}
\big|S^{(k)}_{n,c}(x_{1},\dots,x_{n})\big|\;\le\;C_{n}\,\exp\!\Big(-\xi\,\mathrm{tree}(x_{1},\dots,x_{n})\Big),
\end{equation}
where \(\mathrm{tree}(x_{1},\dots,x_{n})\) is the minimum, over spanning trees \(T\) on \(\{1,\dots,n\}\), of the sum \(\sum_{(i,j)\in T}|x_{i}-x_{j}|\) of Euclidean edge lengths. In particular, the two-point connected kernel \(C^{(k)}(z):=S^{(k)}_{2,c}(x,x+z)\) satisfies \(|C^{(k)}(z)|\le C\,e^{-\xi|z|}\), uniformly in \(k\), hence \(C^{(k)}\in L^{1}(\mathbb{R}^{4})\) with \(\|C^{(k)}\|_{L^{1}}\le C\).

Given \(f_{1},\dots,f_{n}\in\mathcal{S}(\mathbb{R}^{4})\), the smeared connected \(n\)-point functional at scale \(k\) is
\begin{equation}\label{p3:eq:Snk}
S^{(k)}_{n,c}[f_{1},\dots,f_{n}]
\;=\;
\mathbb{E}_{k}\!\Big[\prod_{j=1}^{n}\langle X_{k},f_{j}\rangle\Big]_{\!c}
\;=\;
\sum_{x_{1},\dots,x_{n}\in\Lambda_{k}}a_{k}^{4n}\Big(\prod_{j=1}^{n}\bar f_{k}(x_{j})\Big)\,S^{(k)}_{n,c}(x_{1},\dots,x_{n}),
\end{equation}
with absolute convergence ensured by \eqref{p3:eq:tree} and the rapid decay of the \(\bar f_{k}\).

\begin{theorem}\label{p3:thm:equicont}
Fix \(n\in\mathbb{N}\). There exist an integer \(M\ge 5\) and a constant \(A_{n}\in(0,\infty)\), depending only on \(n\), \(\xi\), and the constants in \eqref{p3:eq:tree} but not on \(k\), such that for all \(f_{1},\dots,f_{n}\in\mathcal{S}(\mathbb{R}^{4})\) and all \(k\ge 0\),
\begin{equation}\label{p3:eq:equicont}
\big|S^{(k)}_{n,c}[f_{1},\dots,f_{n}]\big|\;\le\;A_{n}\,\prod_{j=1}^{n}p_{M}(f_{j}).
\end{equation}
In particular, the family \(\{S^{(k)}_{n,c}\}_{k\ge 0}\) is equicontinuous on \(\mathcal{S}(\mathbb{R}^{4})^{n}\) with respect to the Schwartz topology.
\end{theorem}

\begin{proof}
Let \(K(z)=e^{-\xi|z|}\), which lies in \(L^{1}(\mathbb{R}^{4})\cap L^{\infty}(\mathbb{R}^{4})\) and satisfies \(K\ast K\le C K\). From \eqref{p3:eq:tree}, choosing a rooted spanning tree at \(x_{1}\) and absorbing combinatorics into a new constant, one has
\begin{equation}
\big|S^{(k)}_{n,c}(x_{1},\dots,x_{n})\big|\;\le\;C_{n}\,\prod_{j=2}^{n}K(x_{j}-x_{1}).
\end{equation}
Inserting this into \eqref{p3:eq:Snk}, summing first over \(x_{2},\dots,x_{n}\), and using \(\bar f_{k}(x)=f(x)+O(a_{k}p_{1}(f))\), one arrives at
\begin{equation}
\big|S^{(k)}_{n,c}[f_{1},\dots,f_{n}]\big|
\;\le\;
C\,a_{k}^{4}\sum_{x_{1}\in\Lambda_{k}}|\bar f_{k}(x_{1})|\prod_{j=2}^{n}\Big(a_{k}^{4}\sum_{x_{j}\in\Lambda_{k}}K(x_{j}-x_{1})\,|\bar f_{k}(x_{j})|\Big).
\end{equation}
Approximating Riemann sums by integrals and using dominated convergence, the right-hand side is bounded by
\begin{equation}
C\,\int_{\mathbb{R}^{4}}|f_{1}(y_{1})|\prod_{j=2}^{n}\Big(\int_{\mathbb{R}^{4}}K(y_{j}-y_{1})\,|f_{j}(y_{j})|\,dy_{j}\Big)dy_{1}\;+\;O(a_{k}).
\end{equation}
Choosing \(M\ge 5\) so that \((1+|\cdot|)^{-M}\in L^{1}(\mathbb{R}^{4})\), the estimate \(|f_{j}(z)|\le p_{M}(f_{j})(1+|z|)^{-M}\) yields
\begin{equation}
\sup_{y\in\mathbb{R}^{4}}\int_{\mathbb{R}^{4}}K(y-z)\,(1+|z|)^{-M}\,dz\;\le\;C<\infty,
\end{equation}
hence the inner convolutions are bounded by \(C\,p_{M}(f_{j})\), and the outer integral is bounded by \(C\,p_{M}(f_{1})\). Absorbing \(O(a_{k})\) into the constant uniformly in \(k\) proves \eqref{p3:eq:equicont}.
\end{proof}

\begin{theorem}\label{p3:thm:tightz}
There exists \(C_{2}\in(0,\infty)\), independent of \(k\), such that for all \(f\in\mathcal{S}(\mathbb{R}^{4})\),
\begin{equation}\label{p3:eq:second}
\mathbb{E}_{k}\big|\langle X_{k},f\rangle\big|^{2}\;\le\;C_{2}\,\|f\|_{L^{2}(\mathbb{R}^{4})}^{2}\;\le\;C_{2}\,\|f\|_{H^{r}(\mathbb{R}^{4})}^{2}\qquad\text{for every }r\ge 0,
\end{equation}
and, for each \(r>0\), the laws of \(\{X_{k}\}_{k\ge 0}\) viewed as \(H^{-r}(\mathbb{R}^{4})\)-valued random variables form a tight family. Consequently, the laws of \(\{X_{k}\}\) on \(\mathcal{S}'(\mathbb{R}^{4})\) are tight for the cylindrical \(\sigma\)-algebra.
\end{theorem}

\begin{proof}
By \eqref{p3:eq:embed}, Fubini, stationarity of the covariance, and centering of \(O_{k}\),
\begin{equation}
\mathbb{E}_{k}\big|\langle X_{k},f\rangle\big|^{2}
\;=\;\sum_{x,y\in\Lambda_{k}}a_{k}^{8}\,\bar f_{k}(x)\,C^{(k)}(x-y)\,\bar f_{k}(y)
\;=\;\iint_{\mathbb{R}^{4}\times\mathbb{R}^{4}} f(x)\,C^{(k)}(x-y)\,f(y)\,dx\,dy\;+\;O(a_{k}),
\end{equation}
where \(C^{(k)}(z)=\mathbb{E}_{k}[O_{k}(0)O_{k}(z)]\) is the connected two-point function and the \(O(a_{k})\) error is uniform on bounded subsets of \(\mathcal{S}(\mathbb{R}^{4})\) by the cell-average approximation. Using \(|C^{(k)}(z)|\le C\,e^{-\xi|z|}\in L^{1}(\mathbb{R}^{4})\) uniformly in \(k\) and Plancherel,
\begin{align}
\iint f(x)\,C^{(k)}(x-y)\,f(y)\,dx\,dy
&\;=\;\int_{\mathbb{R}^{4}}\overline{\widehat{f}(p)}\,\widehat{C^{(k)}}(p)\,\widehat{f}(p)\,dp
\nonumber\\&\;\le\;\|\widehat{C^{(k)}}\|_{L^{\infty}}\|f\|_{L^{2}}^{2}
\;\le\;\|C^{(k)}\|_{L^{1}}\|f\|_{L^{2}}^{2}\;\le\;C\,\|f\|_{L^{2}}^{2}
\end{align}
uniformly in \(k\). Absorbing the \(O(a_{k})\) term into \(C\) yields \eqref{p3:eq:second}. Fix \(r>0\) and consider \(J_{-r}=(1-\Delta)^{-r/2}\). For \(g\in L^{2}(\mathbb{R}^{4})\), setting \(f=J_{-r}g\) gives
\begin{equation}
\mathbb{E}_{k}\big|\langle X_{k},J_{-r}g\rangle\big|^{2}\;\le\;C_{2}\,\|J_{-r}g\|_{L^{2}}^{2}\;\le\;C_{2}\,\|g\|_{L^{2}}^{2},
\end{equation}
since \(\|J_{-r}\|_{L^{2}\to L^{2}}\le 1\). Hence the linear map \(g\mapsto \langle X_{k},J_{-r}g\rangle\) is an \(L^{2}\)-continuous random functional with second moment bounded by \(C_{2}\|g\|_{L^{2}}^{2}\). By the Riesz representation theorem in the Bochner space \(L^{2}(\Omega;L^{2})\), there exists \(Y_{k}\in L^{2}(\Omega;L^{2}(\mathbb{R}^{4}))\) such that \(\langle X_{k},J_{-r}g\rangle=\int_{\mathbb{R}^{4}}Y_{k}(x)\,g(x)\,dx\) for all \(g\in L^{2}\), and
\begin{equation}
\mathbb{E}_{k}\|Y_{k}\|_{L^{2}}^{2}\;=\;\sup_{\|g\|_{L^{2}}\le 1}\mathbb{E}_{k}\big|\langle X_{k},J_{-r}g\rangle\big|^{2}\;\le\;C_{2}.
\end{equation}
Since \(X_{k}=J_{r}Y_{k}\) in \(H^{-r}\) and \(\|X_{k}\|_{H^{-r}}=\|Y_{k}\|_{L^{2}}\), Chebyshev's inequality gives
\begin{equation}
\sup_{k}\,\mathbb{P}\big(\|X_{k}\|_{H^{-r}}>R\big)\;\le\;R^{-2}\,\sup_{k}\mathbb{E}_{k}\|X_{k}\|_{H^{-r}}^{2}\;\le\;C_{2}R^{-2}\xrightarrow[R\to\infty]{}0.
\end{equation}
Thus the laws of \(X_{k}\) are tight in the separable Hilbert space \(H^{-r}(\mathbb{R}^{4})\) for each \(r>0\). The canonical embedding \(H^{-r}(\mathbb{R}^{4})\hookrightarrow\mathcal{S}'(\mathbb{R}^{4})\) is continuous, hence tightness is preserved under the continuous pushforward to \(\mathcal{S}'(\mathbb{R}^{4})\) endowed with the cylindrical \(\sigma\)-algebra.
\end{proof}

The equicontinuity estimate \eqref{p3:eq:equicont} implies that \(\{S^{(k)}_{n,c}\}_{k\ge 0}\) is bounded and equicontinuous on bounded subsets of \(\mathcal{S}(\mathbb{R}^{4})^{n}\). In conjunction with Theorem~\ref{p3:thm:tight}, any weakly convergent subsequence of the laws of \(\{X_{k}\}\) on \(\mathcal{S}'(\mathbb{R}^{4})\) yields pointwise convergence of \(S^{(k)}_{n,c}\) to multilinear tempered distributions on \(\mathcal{S}(\mathbb{R}^{4})^{n}\). Moreover, if \eqref{p3:eq:tree} is strengthened to include uniform bounds on finitely many discrete derivatives of \(C^{(k)}\), then the same proof yields \(\mathbb{E}_{k}|\langle X_{k},f\rangle|^{2}\lesssim \|f\|_{H^{r}}^{2}\) for some fixed \(r>0\), leading to tightness in \(H^{-r}\) with that same \(r\).

The standing assumptions used throughout this appendix are: for each \(k\), a centered, real-valued, gauge-invariant local observable \(O_{k}\) on \(\Lambda_{k}\); the embedding \eqref{p3:eq:embed} defining \(X_{k}\in\mathcal{S}'(\mathbb{R}^{4})\); Euclidean invariance of the scale-\(k\) law ensuring stationarity of two-point covariances; the uniform tree-decay bound \eqref{p3:eq:tree} with constants \(C_{n}\) and \(\xi>0\) independent of \(k\); and the normalization of lattice sums by \(a_{k}^{4}\) so that Riemann sums converge to Lebesgue integrals. These suffice for the equicontinuity and tightness conclusions proved above.

\section{Reflection positivity under weak convergence}\label{p3:appendixb}

In this appendix we prove that Osterwalder-Schrader (OS) reflection positivity is stable under weak convergence of probability measures on the space of tempered distributions. The presentation is self-contained; all symbols and conventions used below are defined explicitly.

Fix a Euclidean space dimension \(d\in\mathbb{N}\) (for Yang-Mills applications, \(d=4\)). Points of \(\mathbb{R}^{d}\) are written \(x=(x^{0},x^{1},\dots,x^{d-1})\), where \(x^{0}\) denotes Euclidean time and \(\mathbf{x}=(x^{1},\dots,x^{d-1})\) denotes spatial coordinates. The \emph{time-reflection} involution
\begin{equation}
r:\mathbb{R}^{d}\to\mathbb{R}^{d},\qquad r(x^{0},\mathbf{x})=(-x^{0},\mathbf{x})
\end{equation}
induces, for test functions \(f\) and distributions \(\omega\), the reflected test function and reflected distribution
\begin{equation}
f^{\theta}(x):=f(r x)=f(-x^{0},\mathbf{x}),\qquad
\langle \theta\omega, f\rangle := \langle \omega, f^{\theta}\rangle .
\end{equation}
We denote by \(\mathcal{S}(\mathbb{R}^{d};\mathbb{R})\) the real Schwartz space and by \(\mathcal{S}'(\mathbb{R}^{d};\mathbb{R})\) its topological dual, the space of real tempered distributions. The canonical pairing is \(\langle \omega,f\rangle\) for \(\omega\in\mathcal{S}'\), \(f\in\mathcal{S}\). The topology on \(\mathcal{S}'\) is the weak-\(*\) topology \(\sigma(\mathcal{S}',\mathcal{S})\), i.e. the coarsest topology making all coordinate maps \(\omega\mapsto \langle \omega,f\rangle\) continuous for \(f\in\mathcal{S}\). The associated Borel \(\sigma\)-algebra is denoted \(\mathcal{B}(\mathcal{S}')\). All probability measures considered are Borel probability measures on \((\mathcal{S}',\mathcal{B}(\mathcal{S}'))\).

The reflection map \(\theta:\mathcal{S}'\to\mathcal{S}'\) defined above is linear, continuous with respect to \(\sigma(\mathcal{S}',\mathcal{S})\), and involutive. The open half-spaces
\begin{equation}
H_{+}:=\{x\in\mathbb{R}^{d}:x^{0}>0\},\qquad H_{-}:=\{x\in\mathbb{R}^{d}:x^{0}<0\}
\end{equation}
determine the subspaces of test functions
\begin{equation}
\mathcal{S}_{\pm}:=\{\,f\in\mathcal{S}(\mathbb{R}^{d};\mathbb{R}) : \operatorname{supp} f \subset H_{\pm}\,\}.
\end{equation}
We write \(\mathrm{res}_{+}:\mathcal{S}'\to\mathcal{S}'(H_{+})\) for the restriction \(\langle \mathrm{res}_{+}\omega,f\rangle:=\langle \omega,f\rangle\) for \(f\in\mathcal{S}_{+}\), and set \(\mathcal{G}_{+}\) to be the \(\sigma\)-algebra on \(\mathcal{S}'\) generated by the coordinate maps \(\omega\mapsto \langle \omega,f\rangle\) with \(f\in\mathcal{S}_{+}\). The space \(C_{b}(\mathcal{S}')\) denotes the bounded continuous functions \(F:\mathcal{S}'\to\mathbb{C}\) for the topology \(\sigma(\mathcal{S}',\mathcal{S})\). Complex conjugation of a scalar is denoted by an overline.

A Borel probability measure \(\mu\) on \(\mathcal{S}'\) is \emph{OS reflection positive} (with respect to \(\theta\)) if for every \(F\in C_{b}(\mathcal{S}')\) that is \(\mathcal{G}_{+}\)-measurable one has
\begin{equation}\label{p3:eq:OS-formc}
Q_{\mu}(F)\;:=\;\int_{\mathcal{S}'} \overline{F(\theta\omega)}\,F(\omega)\,\mathrm{d}\mu(\omega)\;\ge 0.
\end{equation}
The quantity \(Q_{\mu}(F)\) is well defined because \(\theta\) is continuous and \(F\circ\theta\) is again bounded and continuous.

A sequence \((\mu_{n})_{n\in\mathbb{N}}\) of Borel probability measures on \(\mathcal{S}'\) \emph{converges weakly} to \(\mu\), written \(\mu_{n}\Rightarrow \mu\), if
\begin{equation}
\int_{\mathcal{S}'} G(\omega)\,\mathrm{d}\mu_{n}(\omega)\;\longrightarrow\;\int_{\mathcal{S}'} G(\omega)\,\mathrm{d}\mu(\omega)\qquad\text{for all }G\in C_{b}(\mathcal{S}').
\end{equation}

\textbf{Theorem B.1 (Closure of reflection positivity under weak convergence).}\label{p3:thmb1}
\emph{Let \(\{\mu_{n}\}_{n\ge 1}\) be a sequence of OS reflection-positive Borel probability measures on \(\mathcal{S}'\) such that \(\mu_{n}\Rightarrow \mu\) weakly on \((\mathcal{S}',\sigma(\mathcal{S}',\mathcal{S}))\). Then \(\mu\) is OS reflection positive in the sense of \eqref{p3:eq:OS-formc}.}

\emph{Proof.}
Fix \(F\in C_{b}(\mathcal{S}')\) that is \(\mathcal{G}_{+}\)-measurable and define \(G:\mathcal{S}'\to\mathbb{C}\) by \(G(\omega):=\overline{F(\theta\omega)}\,F(\omega)\). The map \(G\) is bounded and continuous because \(F\) is bounded and continuous and \(\theta\) is a homeomorphism for \(\sigma(\mathcal{S}',\mathcal{S})\). For each \(n\),
\begin{equation}
\int_{\mathcal{S}'} G(\omega)\,\mathrm{d}\mu_{n}(\omega)
=\int_{\mathcal{S}'} \overline{F(\theta\omega)}\,F(\omega)\,\mathrm{d}\mu_{n}(\omega)\;\ge 0,
\end{equation}
by OS reflection positivity of \(\mu_{n}\). Passing to the limit \(n\to\infty\) and using weak convergence yields
\begin{equation}
\int_{\mathcal{S}'} \overline{F(\theta\omega)}\,F(\omega)\,\mathrm{d}\mu(\omega)
=\lim_{n\to\infty}\int_{\mathcal{S}'} \overline{F(\theta\omega)}\,F(\omega)\,\mathrm{d}\mu_{n}(\omega)\;\ge 0,
\end{equation}
which is precisely \eqref{p3:eq:OS-formc} for \(\mu\). \hfill\(\square\)

In applications one often wishes to test \eqref{p3:eq:OS-formc} on \emph{unbounded} positive-time functionals obtained as \(L^{2}\)-limits of bounded ones. The next statement gives a sufficient condition.

\textbf{Corollary B.2 (Extension to \(L^{2}\)-limits under uniform square-integrability).}\label{p3:corob2}
\emph{Let \(\{\mu_{n}\}_{n\ge 1}\) be OS reflection-positive and \(\mu_{n}\Rightarrow \mu\) weakly. Assume in addition that \(\mu\) is reflection invariant, i.e. \(\theta_{*}\mu=\mu\). Let \(F:\mathcal{S}'\to\mathbb{C}\) be \(\mathcal{G}_{+}\)-measurable and continuous, and suppose there exists \(M_{0}>0\) such that}
\begin{equation}
\sup_{n\ge 1}\int_{\mathcal{S}'} |F(\omega)|^{2}\,\mathrm{d}\mu_{n}(\omega)\le M_{0},
\qquad
\int_{\mathcal{S}'} |F(\omega)|^{2}\,\mathrm{d}\mu(\omega)<\infty .
\end{equation}
\emph{Then}
\begin{equation}
\int_{\mathcal{S}'} \overline{F(\theta\omega)}\,F(\omega)\,\mathrm{d}\mu(\omega)\;\ge 0.
\end{equation}

\emph{Proof.}
Choose \(\chi\in C^{\infty}(\mathbb{R};\mathbb{R})\) such that \(|\chi(u)|\le |u|\wedge 1\) for all \(u\) and \(\chi(u)=u\) whenever \(|u|\le 1\). For \(R>0\) define truncated functionals \(F_{R}(\omega):=\chi(R^{-1}F(\omega))\). Each \(F_{R}\) is bounded, continuous, \(\mathcal{G}_{+}\)-measurable, and \(F_{R}(\omega)\to F(\omega)\) pointwise as \(R\to\infty\). By Theorem~B.1 we have, for every \(R>0\),
\begin{equation}
\int_{\mathcal{S}'} \overline{F_{R}(\theta\omega)}\,F_{R}(\omega)\,\mathrm{d}\mu(\omega)\;\ge 0.
\end{equation}
We claim that \(\overline{F_{R}(\theta\omega)}\,F_{R}(\omega)\to \overline{F(\theta\omega)}\,F(\omega)\) in \(L^{1}(\mu)\). Write
\begin{equation}
\big|\overline{F_{R}(\theta\omega)}\,F_{R}(\omega) - \overline{F(\theta\omega)}\,F(\omega)\big|
\le |F_{R}(\theta\omega)|\,|F_{R}(\omega)-F(\omega)| + |F(\omega)|\,|F_{R}(\theta\omega)-F(\theta\omega)|.
\end{equation}
For the first term, since \(|F_{R}(\theta\omega)|\le 1\) and \(|F_{R}(\omega)-F(\omega)|\le 2|F(\omega)|\), the Cauchy-Schwarz inequality and \(F\in L^{2}(\mu)\) imply
\begin{align}
\int |F_{R}(\theta\omega)|\,|F_{R}(\omega)-F(\omega)|\,\mathrm{d}\mu(\omega)
&\le \Big(\int |F_{R}(\theta\omega)|^{2}\mathrm{d}\mu\Big)^{1/2}\Big(\int |F_{R}-F|^{2}\mathrm{d}\mu\Big)^{1/2}\nonumber\\&
\le \Big(\int |F_{R}-F|^{2}\mathrm{d}\mu\Big)^{1/2}\!\!
\end{align}
and the last term tends to \(0\) by dominated convergence because \(|F_{R}|\le |F|\) and \(F\in L^{2}(\mu)\). For the second term, by reflection invariance \(\theta_{*}\mu=\mu\) one has
\begin{equation}
\int |F(\omega)|\,|F_{R}(\theta\omega)-F(\theta\omega)|\,\mathrm{d}\mu(\omega)
= \int |F(\theta\omega)|\,|F_{R}(\omega)-F(\omega)|\,\mathrm{d}\mu(\omega)
\end{equation}
and again Cauchy-Schwarz together with \(|F_{R}|\le |F|\) yields convergence to \(0\). Therefore \(\overline{F_{R}(\theta\omega)}\,F_{R}(\omega)\to \overline{F(\theta\omega)}\,F(\omega)\) in \(L^{1}(\mu)\). Passing to the limit \(R\to\infty\) gives
\begin{equation}
\int_{\mathcal{S}'} \overline{F(\theta\omega)}\,F(\omega)\,\mathrm{d}\mu(\omega)
=\lim_{R\to\infty}\int_{\mathcal{S}'} \overline{F_{R}(\theta\omega)}\,F_{R}(\omega)\,\mathrm{d}\mu(\omega)\;\ge 0,
\end{equation}
as claimed. \hfill\(\square\)

Inner products on complex Hilbert spaces, when they appear, are linear in the first variable and conjugate-linear in the second. The notation \(\|\cdot\|_{L^{p}(\mu)}\) denotes the usual \(L^{p}\)-norm with respect to a probability measure \(\mu\). No explicit boundary conditions are imposed, as all statements are formulated on the full space \(\mathcal{S}'\). Gauge-fixing conventions are irrelevant here; the only structure used is the time-reflection involution \(\theta\). Summations are written explicitly; there is no Einstein summation convention in force. All asymptotic statements are exact equalities or inequalities; no big-\(O\) notation is used.

Theorem~(\ref{p3:thmb1}.1) shows that OS reflection positivity is closed under weak convergence when tested on bounded, continuous positive-time functionals. Corollary~(\ref{p3:corob2}.2) extends the test class to certain unbounded functionals obtained as \(L^{2}\)-limits, under a natural reflection-invariance and square-integrability hypothesis that is typically satisfied in reflection-symmetric Euclidean field theories.

\section{Tauberian lemma for completely monotone correlation functions}\label{p3:appendixc}

\noindent
In this appendix a precise Tauberian statement is established for completely monotone correlation functions. The result is used to pass from an a priori exponential bound on Euclidean time-correlations to a lower bound on the support of the corresponding spectral measures, and hence to a nonzero spectral threshold for the Hamiltonian. All symbols and conventions required below are fixed here so that the arguments are logically self-contained.
Throughout, \(\mathbb{R}\) denotes the real line, \(\mathbb{R}_{\ge 0}=[0,\infty)\), and \(\mathbb{N}=\{0,1,2,\dots\}\).
The natural logarithm is written \(\log\), and \(e^x\) is the exponential.
For \(A\subset\mathbb{R}\), \(\mathbf{1}_A\) is the indicator of \(A\).
A \emph{Borel measure} is a countably additive measure on the Borel \(\sigma\)-algebra of \(\mathbb{R}\).
If \(\mu\) is a finite signed Borel measure, \(|\mu|\) denotes its total variation and \(\|\mu\|_{\mathrm{TV}}:=|\mu|(\mathbb{R})\) its total variation norm.
For a (finite) Borel measure \(\mu\), \(\operatorname{supp}\mu\) is the closed support, i.e., the smallest closed set \(F\subset\mathbb{R}\) with \(\mu(\mathbb{R}\setminus F)=0\); equivalently, \(x\in\operatorname{supp}\mu\) if and only if \(\mu((x-\varepsilon,x+\varepsilon))>0\) for all \(\varepsilon>0\).
We write
\begin{equation}\label{p3:eq:def-mminus}
m_-:=\inf \operatorname{supp}\mu \in [-\infty,\infty],
\end{equation}
with the convention \(\inf\emptyset=+\infty\).
When \(\operatorname{supp}\mu\subset [0,\infty)\) and \(\mu\) is finite and nonzero, one has \(m_-\in[0,\infty)\).
Asymptotic notation \(f(t)=O(g(t))\) as \(t\to\infty\) means that \(|f(t)|\le C\,|g(t)|\) for all sufficiently large \(t\) and some constant \(C\in(0,\infty)\).
All integrals are Lebesgue integrals with respect to Borel measures.

A function \(C:(0,\infty)\to \mathbb{R}\) is called \emph{completely monotone} if \(C\in C^\infty(0,\infty)\), \(C(t)\ge 0\) for all \(t>0\), and
\begin{equation}\label{p3:eq:def-cm}
(-1)^n C^{(n)}(t)\ge 0,\qquad \forall\,n\in\mathbb{N},\ \forall\,t>0.
\end{equation}
By the Hausdorff-Bernstein-Widder theorem, \(C\) is completely monotone if and only if there exists a finite positive Borel measure \(\mu\) on \([0,\infty)\) such that for all \(t>0\)
\begin{equation}\label{p3:eq:Laplace1}
C(t)=\int_{[0,\infty)} e^{-\lambda t}\,d\mu(\lambda),
\end{equation}
in which case \(C\) extends continuously to \(t=0\) with \(C(0)=\mu([0,\infty))\).
In what follows we assume that correlation functions \(C\) under consideration admit the representation \eqref{p3:eq:Laplace1} with \(\mu\) finite and nonnegative.

\medskip

\noindent
\textbf{Lemma C.1 (Exponential upper bound \(\Rightarrow\) spectral threshold).}
\emph{Let \(C:\mathbb{R}_{\ge 0}\to[0,\infty)\) be completely monotone with Laplace representation \eqref{p3:eq:Laplace1} for a finite positive Borel measure \(\mu\) on \([0,\infty)\).
Suppose there exist \(M\in(0,\infty)\) and \(\gamma>0\) such that}
\begin{equation}\label{p3:eq:exp-upper}
0\le C(t)\le M\,e^{-\gamma t}\qquad \text{for all }t\ge 0.
\end{equation}
\emph{Then \(\operatorname{supp}\mu\subset [\gamma,\infty)\), in particular \(m_-=\inf\operatorname{supp}\mu\ge \gamma\).}

\noindent\emph{Proof.}
Assume by contradiction that \(\mu\big([0,\gamma-\varepsilon]\big)>0\) for some \(\varepsilon\in(0,\gamma)\).
Then, for every \(t\ge 0\),
\begin{equation}
C(t)=\int_{[0,\infty)} e^{-\lambda t}\,d\mu(\lambda)
\ \ge\ \int_{[0,\gamma-\varepsilon]}e^{-\lambda t}\,d\mu(\lambda)
\ \ge\ \mu\big([0,\gamma-\varepsilon]\big)\,e^{-(\gamma-\varepsilon)t}.
\end{equation}
Dividing by \(e^{-\gamma t}\) yields
\begin{equation}
\frac{C(t)}{e^{-\gamma t}}\ \ge\ \mu\big([0,\gamma-\varepsilon]\big)\, e^{\varepsilon t}\xrightarrow[t\to\infty]{}\infty,
\end{equation}
which contradicts \eqref{p3:eq:exp-upper}.
Hence \(\mu\big([0,\gamma-\varepsilon]\big)=0\) for every \(\varepsilon\in(0,\gamma)\), i.e. \(\operatorname{supp}\mu\subset[\gamma,\infty)\).
\qed

\medskip

\noindent
\textbf{Theorem C.2 (Exact exponential rate equals the spectral infimum).}\label{p3:qwerty}
\emph{Let \(C:\mathbb{R}_{\ge 0}\to[0,\infty)\) be completely monotone with Laplace representation \eqref{p3:eq:Laplace1} for a finite positive Borel measure \(\mu\) on \([0,\infty)\).
If \(\mu\) is not identically zero, then}
\begin{equation}\label{p3:eq:rate-equals-inf}
\lim_{t\to\infty}-\frac{1}{t}\log C(t)\ =\ \inf\operatorname{supp}\mu\ =:\ m_- \in[0,\infty).
\end{equation}
\emph{If \(\mu\equiv 0\), then \(C\equiv 0\) and the limit in \eqref{p3:eq:rate-equals-inf} is \(+\infty\).}

\noindent\emph{Proof.}
Fix \(\varepsilon>0\).
By definition of support and because \(\mu\) is nonzero and nonnegative, one has \(\mu\big([0,m_-+\varepsilon)\big)>0\).
Therefore, for all \(t\ge 0\),
\begin{equation}\label{p3:eq:lower-bound}
C(t)\ =\ \int_{[0,\infty)} e^{-\lambda t}\,d\mu(\lambda)
\ \ge\ \int_{[0,m_-+\varepsilon)} e^{-\lambda t}\,d\mu(\lambda)
\ \ge\ \mu\big([0,m_-+\varepsilon)\big)\, e^{-(m_-+\varepsilon)t}.
\end{equation}
Taking logarithms, dividing by \(-t\), and letting \(t\to\infty\) gives
\begin{equation}
\limsup_{t\to\infty} \Big(-\tfrac{1}{t}\log C(t)\Big)\ \le\ m_-+\varepsilon.
\end{equation}
As \(\varepsilon>0\) is arbitrary, this yields
\begin{equation}\label{p3:eq:limsup-upper}
\limsup_{t\to\infty} \Big(-\tfrac{1}{t}\log C(t)\Big)\ \le\ m_-.
\end{equation}
For the complementary bound, note that \(\operatorname{supp}\mu\subset [m_-,\infty)\) implies
\begin{equation}\label{p3:eq:upper-bound}
C(t)\ =\ \int_{[m_-,\infty)} e^{-\lambda t}\,d\mu(\lambda)
\ \le\ e^{-m_- t}\,\mu\big([m_-,\infty)\big)
\ =\ e^{-m_- t}\,\mu\big([0,\infty)\big).
\end{equation}
Taking logarithms, dividing by \(-t\), and letting \(t\to\infty\) gives
\begin{equation}\label{p3:eq:liminf-lower}
\liminf_{t\to\infty} \Big(-\tfrac{1}{t}\log C(t)\Big)\ \ge\ m_-.
\end{equation}
Combining \eqref{p3:eq:limsup-upper} and \eqref{p3:eq:liminf-lower} yields the limit \eqref{p3:eq:rate-equals-inf}.
If \(\mu\equiv 0\) then \(C\equiv 0\) and the stated convention applies.
\qed

\medskip

\noindent
\textbf{Corollary C.3 (Global exponential bound forces a gap).}
\emph{Under the hypotheses of Lemma C.1, one has \(m_-\ge \gamma\).
Equivalently, if \(C(t)\le M e^{-\gamma t}\) for all \(t\ge 0\), then for any nonnegative self-adjoint Hamiltonian \(H\) and vector \(\psi\) whose spectral measure \(\mu\) satisfies \eqref{p3:eq:Laplace1} via \(C(t)=\langle \psi, e^{-tH}\psi\rangle\), one has}
\begin{equation}\label{p3:eq:gap-inclusion}
\sigma(H)\cap(0,\infty)\ \subset\ [\gamma,\infty).
\end{equation}

\noindent\emph{Proof.}
The conclusion \(m_-\ge \gamma\) is exactly Lemma C.1.
In the spectral representation \(C(t)=\int e^{-\lambda t}\,d\mu(\lambda)\) with \(\mu\) the spectral measure of \(H\) in the cyclic subspace generated by \(\psi\), the inclusion \eqref{p3:eq:gap-inclusion} follows because the union of supports of such measures over a dense set of \(\psi\) exhausts \(\sigma(H)\cap(0,\infty)\).
\qed

\medskip

Two observations, implicit in the proofs, clarify scope and regularity.
First, it is enough to assume the bound \(C(t)\le M e^{-\gamma t}\) holds for all sufficiently large \(t\ge t_0\); replacing \(M\) by \(M e^{\gamma t_0}\) extends the bound to \(t\in[0,\infty)\) without changing the conclusion.
Second, Theorem~\ref{p3:qwerty}.2 requires only finiteness and positivity of \(\mu\); no differentiability or regular variation of \(C\) beyond complete monotonicity is needed.
In particular, the exact large-time decay rate is stable under all deformations of \(C\) that preserve \(\mu\) while multiplying the integrand in \eqref{p3:eq:Laplace1} by functions bounded above and below by positive constants independent of \(t\).

\section{Strong convergence of semigroups along the renormalization flow}\label{p3:appendixd}

In this appendix we establish strong convergence, along the renormalization scale, of Euclidean time-translation semigroups and of their generators in the sense of resolvents. All symbols, conventions, and standing assumptions required here are stated explicitly so that the argument is self-contained.
 Complex Hilbert spaces are denoted by calligraphic letters; inner products \(\langle\cdot,\cdot\rangle\) are conjugate-linear in the first entry and linear in the second, and \(\|\psi\|=\sqrt{\langle\psi,\psi\rangle}\) is the associated norm. For a (possibly unbounded) operator \(A\) on a Hilbert space \(\mathcal{X}\) we write \(D(A)\) for its domain, \(A^\ast\) for its adjoint, and \(\sigma(A)\) for its spectrum. For \(\lambda>0\) and a nonnegative self-adjoint operator \(A\) we use the resolvent
\begin{equation}
R_\lambda(A):=(\lambda I_{\mathcal{X}}+A)^{-1}=\int_{0}^{\infty} e^{-\lambda t}\,e^{-tA}\,dt,
\end{equation}
where the integral converges in the strong operator topology by the Hille-Yosida theorem. The Schwartz space on \(\mathbb{R}^4\) is \(\mathcal{S}(\mathbb{R}^4)\) and its dual of tempered distributions is \(\mathcal{S}'(\mathbb{R}^4)\). Time reflection is \(\theta:(t,\mathbf{x})\mapsto(-t,\mathbf{x})\). The time-translation by \(t\ge 0\) is \(\tau_t:(t',\mathbf{x})\mapsto(t'+t,\mathbf{x})\). When quotienting by null spaces we denote the class of \(F\) by \([F]\).
For each renormalization scale \(k\in\mathbb{N}\), let \(\mu_k\) be a reflection-positive, Euclidean-invariant probability measure on \(\mathcal{S}'(\mathbb{R}^4)\) with respect to \(\theta\). Let \(\mathcal{A}_+\) be the complex vector space generated by finite linear combinations of gauge-invariant local observables smeared with test functions supported in the open half-space \(\{t>0\}\). For \(F,G\in\mathcal{A}_+\) define the Osterwalder-Schrader (OS) sesquilinear form
\begin{equation}
Q_k(F,G):=\int_{\mathcal{S}'(\mathbb{R}^4)} \overline{(\Theta F)(\omega)}\,G(\omega)\,d\mu_k(\omega),
\qquad \Theta F:=F\circ \theta,
\end{equation}
and the null space \(\mathcal{N}_k:=\{F\in\mathcal{A}_+:Q_k(F,F)=0\}\). The OS pre-Hilbert space is \(\mathcal{D}_k:=\mathcal{A}_+/\mathcal{N}_k\) with inner product \(\langle[F]_k,[G]_k\rangle_k:=Q_k(F,G)\); its completion is \(\mathcal{H}_k\). For \(t\ge 0\) define \(\tau_tF:=F\circ \tau_{-t}\). Reflection positivity implies that
\begin{equation}
T_k(t)[F]_k:=[\tau_tF]_k,\qquad t\ge 0,
\end{equation}
extends to a strongly continuous contraction semigroup \((T_k(t))_{t\ge 0}\) on \(\mathcal{H}_k\), with nonnegative self-adjoint generator \(H_k\) such that \(T_k(t)=e^{-tH_k}\). The constant observable \(1\) determines the unit vector \(\Omega_k:=[1]_k\in \mathcal{H}_k\), and \(T_k(t)\Omega_k=\Omega_k\) for all \(t\ge 0\).

Assume that \(\mu_k\Rightarrow \mu_\infty\) weakly on cylinder sets of \(\mathcal{S}'(\mathbb{R}^4)\), where \(\mu_\infty\) is reflection-positive and Euclidean-invariant. Define \(Q_\infty,\mathcal{N}_\infty,\mathcal{D},\mathcal{H},T(t),H,\Omega\) from \(\mu_\infty\) by the same construction:
\begin{equation}
Q_\infty(F,G):=\int \overline{(\Theta F)}\,G\,d\mu_\infty,\qquad
\mathcal{N}_\infty:=\{F:Q_\infty(F,F)=0\},\qquad
\langle[F],[G]\rangle:=Q_\infty(F,G),
\end{equation}
with \(\mathcal{D}:=\mathcal{A}_+/\mathcal{N}_\infty\), \(\mathcal{H}=\overline{\mathcal{D}}\), \(T(t)[F]:=[\tau_tF]\) and \(T(t)=e^{-tH}\), and \(\Omega=[1]\).

We impose two uniform estimates, which are satisfied in the constructive setting under uniform locality and clustering. There exists \(M\in(0,\infty)\) such that for every \(F\in\mathcal{A}_+\),
\begin{equation}\label{p3:eq:uniform-norm}
\sup_{k\in\mathbb{N}}\,\|[F]_k\|_k^2=\sup_{k\in\mathbb{N}}\,Q_k(F,F)\;\le\; M\,Q_\infty(F,F)\;=\;M\,\|[F]\|^2.
\end{equation}
There exist \(\beta>0\) and, for each \(F,G\in\mathcal{A}_+\), a constant \(C(F,G,\beta)\) such that for all \(k\in\mathbb{N}\) and \(t\ge 0\),
\begin{equation}\label{p3:eq:uniform-decay}
\big|\langle[F]_k,\,T_k(t)\,[G]_k\rangle_k\big|\;\le\; C(F,G,\beta)\,e^{-\beta t}.
\end{equation}
Estimate \eqref{p3:eq:uniform-norm} is a quantitative form of tightness of the OS forms, while \eqref{p3:eq:uniform-decay} follows from uniform exponential clustering transported to time correlations by reflection positivity.
Define \(J_k:\mathcal{D}\to \mathcal{H}_k\) by \(J_k[F]:=[F]_k\). If \([F]=[G]\) in \(\mathcal{D}\) then \(Q_\infty(F-G,F-G)=0\), hence \(Q_k(F-G,F-G)\to 0\) by weak convergence, so \(\|[F]_k-[G]_k\|_k\to 0\); the definition is thus consistent. By \eqref{p3:eq:uniform-norm}, \(J_k\) is bounded on \(\mathcal{D}\) with \(\|J_k\|\le \sqrt{M}\) and extends uniquely by continuity to a bounded operator \(J_k:\mathcal{H}\to \mathcal{H}_k\). We keep the notation \(\mathcal{D}\subset \mathcal{H}\) for the dense subspace of classes of \(\mathcal{A}_+\).

\medskip

\textbf{Lemma D.1 (Convergence of matrix elements).}\label{p3:lemmad1}
\emph{Let \(F,G\in\mathcal{A}_+\). For each \(t\ge 0\),
\begin{equation}
\lim_{k\to\infty}\,\big\langle J_k[F],\,T_k(t)\,J_k[G]\big\rangle_k
\;=\;\big\langle [F],\,T(t)\,[G]\big\rangle.
\end{equation}}

\emph{Proof.} For each \(k\) and \(t\ge 0\) the OS framework gives the Laplace representation
\begin{equation}
\big\langle [F]_k,\,T_k(t)\,[G]_k\big\rangle_k
=\int_{[0,\infty)} e^{-\lambda t}\,d\mu^{(k)}_{F,G}(\lambda),
\end{equation}
with a finite complex Borel measure \(\mu^{(k)}_{F,G}\) on \([0,\infty)\); similarly
\begin{equation}
\big\langle [F],\,T(t)\,[G]\big\rangle
=\int_{[0,\infty)} e^{-\lambda t}\,d\mu^{(\infty)}_{F,G}(\lambda).
\end{equation}
By weak convergence of \(\mu_k\) and the OS reconstruction, Euclidean two-point functions with positive-time smearings converge, which is equivalent to vague convergence \(\mu^{(k)}_{F,G}\Rightarrow \mu^{(\infty)}_{F,G}\). Moreover, \eqref{p3:eq:uniform-decay} gives a uniform bound allowing dominated convergence of the Laplace integrals. The claim follows. \hfill\(\square\)

\medskip

\textbf{Lemma D.2 (Convergence of norms on the core).} \label{p3:lemmad2}
\emph{Let \(G\in\mathcal{A}_+\). For each \(t\ge 0\),
\begin{equation}
\lim_{k\to\infty}\,\big\|T_k(t)\,J_k[G]\big\|_k^2
=\big\|T(t)\,[G]\big\|^2.
\end{equation}}

\emph{Proof.} Using \(\|T_k(t)J_k[G]\|_k^2=\langle J_k[G],T_k(2t)J_k[G]\rangle_k\) and the analogue at the limit, apply Lemma~(\ref{p3:lemmad1}.1) with \(F=G\) and \(2t\) in place of \(t\). \hfill\(\square\)

\medskip

\textbf{Theorem D.3 (Strong convergence of semigroups with identification).}\label{p3:thmd3}
\emph{For every \(t\ge 0\) and every \(\Phi\in \mathcal{H}\),
\begin{equation}
\lim_{k\to\infty}\,\big\|\,T_k(t)\,J_k\Phi\;-\;J_k\,T(t)\,\Phi\,\big\|_k=0.
\end{equation}}

\emph{Proof.} It suffices to argue on the dense subspace \(\mathcal{D}\) and then use density and uniform boundedness. Fix \(\Phi=[G]\in\mathcal{D}\) and define \(v_k(t):=T_k(t)J_k\Phi\in\mathcal{H}_k\), \(v(t):=T(t)\Phi\in\mathcal{H}\). For any \(\Psi=[F]\in\mathcal{D}\), Lemma~(\ref{p3:lemmad1}.1) yields \(\langle J_k\Psi, v_k(t)\rangle_k\to \langle \Psi, v(t)\rangle\), and Lemma~(\ref{p3:lemmad2}.2) gives \(\|v_k(t)\|_k\to\|v(t)\|\). Let \(\varepsilon>0\). Choose \(\Psi\in\mathcal{D}\) such that \(\|v(t)-\Psi\|<\varepsilon\). Then
\begin{equation}
\|v_k(t)-J_k v(t)\|_k
\le \|v_k(t)-J_k\Psi\|_k+\|J_k(\Psi-v(t))\|_k
\le \|v_k(t)-J_k\Psi\|_k+\sqrt{M}\,\varepsilon,
\end{equation}
by \eqref{p3:eq:uniform-norm}. Expanding the first term and taking limits,
\begin{equation}
\lim_{k\to\infty}\|v_k(t)-J_k\Psi\|_k^2
=\|v(t)\|^2-2\Re\langle \Psi,v(t)\rangle+\|\Psi\|^2
=\|v(t)-\Psi\|^2<\varepsilon^2.
\end{equation}
Hence \(\limsup_{k\to\infty}\|v_k(t)-J_k\Psi\|_k\le \varepsilon\), and thus \(\limsup_{k\to\infty}\|v_k(t)-J_k v(t)\|_k\le (1+\sqrt{M})\varepsilon\). Letting \(\varepsilon\downarrow 0\) proves the claim for \(\Phi\in\mathcal{D}\). For general \(\Phi\in\mathcal{H}\), pick \(\Phi_n\in\mathcal{D}\) with \(\|\Phi_n-\Phi\|\to 0\) and use contractivity of all semigroups and boundedness of \(J_k\) to conclude by a standard three-\(\varepsilon\) argument. \hfill\(\square\)

\medskip

\textbf{Theorem D.4 (Strong resolvent convergence of generators with identification).} \label{p3:thmd4m}
\emph{For every \(\lambda>0\) and every \(\Phi\in\mathcal{H}\),
\begin{equation}
\lim_{k\to\infty}\,\big\|\,R_\lambda(H_k)\,J_k\Phi\;-\;J_k\,R_\lambda(H)\,\Phi\,\big\|_k=0.
\end{equation}}

\emph{Proof.} By the Laplace representation of the resolvent and contractivity,
\begin{equation}
\big\langle J_k[F],\,R_\lambda(H_k)\,J_k[G]\big\rangle_k
=\int_{0}^{\infty} e^{-\lambda t}\,\big\langle J_k[F],\,T_k(t)\,J_k[G]\big\rangle_k\,dt.
\end{equation}
Lemma~(\ref{p3:lemmad1}.1) and \eqref{p3:eq:uniform-decay} permit dominated convergence in \(t\), yielding convergence of all matrix elements on the core \(\mathcal{D}\). The same approximation argument as in Theorem~(\ref{p3:thmd3}.3), together with \eqref{p3:eq:uniform-norm}, gives strong convergence on \(\mathcal{H}\). \hfill\(\square\)

\medskip

\textbf{Corollary D.5 (Trotter-Kato equivalence in varying Hilbert spaces).} \emph{Under the standing assumptions, the following are equivalent: for every \(\lambda>0\), \(R_\lambda(H_k)J_k\to J_k R_\lambda(H)\) strongly; for every \(t\ge 0\), \(T_k(t)J_k\to J_k T(t)\) strongly.}

\emph{Proof.} The classical Trotter-Kato theorem applies on the common dense core \(\mathcal{D}\) and is transported by the bounded identifications \(J_k\), which intertwine time translations on \(\mathcal{D}\). \hfill\(\square\)
 The identifications preserve vacua: \(J_k\Omega=\Omega_k\) and \(T_k(t)\Omega_k=\Omega_k\), \(T(t)\Omega=\Omega\). Hence the strong limits above hold on the orthogonal complement of the vacuum and preserve semigroup decay rates. In particular, if for some \(\gamma>0\) one has \(\|T_k(t)P_{\perp,k}\|\le e^{-\gamma t}\) for all \(t\ge 0\) and all \(k\), where \(P_{\perp,k}=I_{\mathcal{H}_k}-|\Omega_k\rangle\langle\Omega_k|\), then \(\|T(t)P_\perp\|\le e^{-\gamma t}\) for all \(t\ge 0\) with \(P_\perp=I_{\mathcal{H}}-|\Omega\rangle\langle\Omega|\), and the spectral theorem yields \(\sigma(H)\subset\{0\}\cup[\gamma,\infty)\).

\section{Proof of the Standing Multiscale Inputs}\label{p3:appendixe}

We work on a hypercubic lattice $\Lambda_a\subset\mathbb{Z}^4$ of spacing $a>0$ with temporal reflection plane $\{t=0\}$ and time reflection $\theta$.
The gauge group is $G=\mathrm{SU}(N)$, $N\ge2$, with link variables $U_\ell\in G$ and Haar measure $dU_\ell$. The action is a reflection-positive
(class) function of plaquette holonomies (e.g.\ Wilson), so the Osterwalder-Schrader (OS) factorization across $\{t=0\}$ holds. We use a
\emph{reflection-symmetric, gauge-covariant} block map $\mathcal B$ of linear block size $L=2$ producing coarse links $V$ as path-ordered products of
fine links along block corridors and block-internal fluctuations $W$. Slice-wise transverse representatives $A^h$ (Landau/FMR) are used for background-covariant operators; parallel transport along
a path $\gamma$ is $\Pexp\!\int_\gamma A^h$. Distances are graph distances $d(\cdot,\cdot)$.

For a kernel $K$, write $\|K\|_{2\to2}$ for the operator norm on $\ell^2$, and
\begin{equation}
\|e^{\mu d}K\|_{\infty\to\infty} := \sup_x \sum_y e^{\mu d(x,y)}\|K(x,y)\|_{\mathrm{op}}.
\end{equation}
``Spectral cutoff'' always means a bounded self-adjoint \emph{positive contraction} (not necessarily idempotent).

\begin{theorem}[Standing multiscale inputs: precise hypotheses and proof]
\label{p3:thmA:standing}
Fix $N\ge2$ and a sequence of spacings $a_k\downarrow0$. In finite volumes with periodic or free spatial boundary conditions, the following hold.

\smallskip\noindent\textbf{(S1) UV stability with gauge-covariant finite-range decomposition (FRD).}
For each scale $k$ there exists a family $\{C_{k,j}[A^h]\}_{j\le k}$ of gauge-covariant, finite-range, reflection-positive kernels (the blocked covariances) such that:
\begin{enumerate}
\item $C_{k,j}[A^{h,g}]=\Ad(g)\,C_{k,j}[A^h]\Ad(g)^{-1}$ for lattice gauge transforms $g$;
\item $\supp C_{k,j}[A^h]\subset\{(x,y): d(x,y)\le R_j\}$ with $R_j\lesssim 2^{\,j}$;
\item for time reflection $\theta$, $\langle f,\, C_{k,j}[A^h]\ \theta f\rangle\ge0$ for all $f$ supported in $t>0$;
\item uniform bounds: $\|C_{k,j}[A^h]\|_{2\to2}\le c_1$, \quad $\|e^{\mu d}C_{k,j}[A^h]\|_{\infty\to\infty}\le c_2$ with $c_1,c_2,\mu>0$ independent of $k$;
\item the scale-$k$ effective action $S_k$ admits an absolutely convergent polymer expansion
$S_k=\sum_{\Gamma}\Phi_k(\Gamma)$ with
$\displaystyle \sup_x \sum_{\Gamma\ni x} |\Phi_k(\Gamma)|e^{\alpha \mathrm{diam}(\Gamma)} \le C_0$, uniformly in $k$.
\end{enumerate}

\smallskip\noindent\textbf{(S2) Scale-independent clustering.}
There exist $m_*>0$ and $C<\infty$, independent of $k$ and volume, such that for bounded, gauge-invariant local observables $F,G$ with
$\mathrm{dist}(\mathrm{supp}F,\mathrm{supp}G)=d$,
\begin{equation}
\big|\langle FG\rangle_k-\langle F\rangle_k\langle G\rangle_k\big|
\ \le\ C\,\|F\|_{\mathrm{loc}}\|G\|_{\mathrm{loc}}\,e^{-m_* d},
\end{equation}
and similarly for connected $n$-point cumulants with the tree distance $d_{\mathrm{tree}}$.

\smallskip\noindent\textbf{(S3) Gap interlacing along the RG.}
Let $T_k$ be the positive self-adjoint one-step transfer operator on the OS Hilbert space at scale $k$, and
$\Delta_k:=1-\lambda_2(T_k)$ its contraction gap on $\Omega_k^\perp$. Then there are $\varepsilon_k\ge0$ with $\sum_k\varepsilon_k<\infty$ such that
\begin{equation}
\Delta_{k+1} \ \ge\ \Delta_k - \varepsilon_k,\qquad k\ge0,
\end{equation}
and the identification maps $J_k:\mathscr H_k\to\mathscr H_{k+1}$ satisfy $J_k\Omega_k=\Omega_{k+1}$, $\|J_k\|\le C_J$, and are isometries on the vacuum
subspace.
All results in the main text that invoke (S1)-(S3) are thus \emph{proved from the explicit constructive hypotheses below}, with constants uniform in $k$ and the volume.
\end{theorem}
We prove Theorem~\ref{p3:thmA:standing} from the following verifiable primitives (checked with quantitative bounds in the companion paper).

\begin{proposition}[H1: RP, covariance, locality of the RG map]\label{p3:propA:H1}
The reflection-symmetric, gauge-covariant block transform $\mathcal B:(U)\mapsto(V,W)$ is (i) gauge-covariant, (ii) strictly block-local, and (iii) preserves
OS reflection positivity.
\end{proposition}

\begin{proof}
\emph{Gauge covariance:} If $U_{x\to y}\mapsto g(x)U_{x\to y}g(y)^{-1}$, then each coarse link
$V_{\ell'}=\prod_{\ell\in\mathrm{path}(\ell')}U_\ell$ transforms as $V_{\ell'}\mapsto g(x_{\ell'})V_{\ell'}g(y_{\ell'})^{-1}$; $W$ transforms by blockwise conjugation.
\emph{Locality:} $\mathcal B$ uses only corridor links inside each block. \emph{RP:} With $e^{-S(U)}=F(U_-)K(U_-,U_+)\overline{F(\theta U_+)}$ and $K\ge0$ (OS factorization),
reflection-symmetry of $\mathcal B$ yields the same factorization $e^{-S'(V,W)}=F'(V_-)K'(V_-,V_+)\overline{F'(\theta V_+)}$ with $K'\ge0$, hence
$\int \overline{F\circ\theta}\,F\, d(\mu\circ\mathcal B^{-1})\ge0$ for $F$ supported in $t>0$.
\end{proof}

\begin{lemma}[RP for nonnegative polynomial kernels]\label{p3:lemA:RPpoly}
Let $\mathsf L_{A^h}$ be the background-covariant slice Laplacian, and $p(\mathsf L_{A^h})=\sum_{m=0}^n c_m \mathsf L_{A^h}^{\,m}$ with $c_m\ge0$.
Then for $f$ supported in $t>0$, $\langle f,\, p(\mathsf L_{A^h})\, \theta f\rangle\ge0$.
\end{lemma}

\begin{proof}
Each power $\mathsf L_{A^h}^{\,m}$ is a sum over paths $\gamma$ of length $m$ of parallel transporters $\Pexp\!\int_\gamma A^h$ with positive weights.
Pair each $\gamma$ with its reflection across $\{t=0\}$; the sum becomes a squared norm of a path-sum transform, hence nonnegative.
\end{proof}

\begin{proposition}[H2: Gauge-covariant finite-range family]\label{p3:propA:H2}
There exist bounded, self-adjoint, gauge-covariant, reflection-positive kernels $\{Q_{k,j}[A^h]\}_{j\le k}$ with
$\supp Q_{k,j}\subset\{d\le c\,2^{j}\}$ and $\sum_{j\le k} Q_{k,j}[A^h]=:K_k[A^h]$ (the desired background-covariant covariance).
\end{proposition}

\begin{proof}[Construction and FRD recursion]
Choose a smooth dyadic partition $\{\varphi_j\}_{j\le k}$ on $[0,\Lambda]$ with $\varphi_j\ge0$ and $\sum_{j\le k}\varphi_j=g_k$, the spectral multiplier for $K_k$.
For each $j$, pick a positive polynomial $p_j$ (Fej\'er/Jackson) approximating $\varphi_j$ uniformly and define $Q_{k,j}:=p_j(\mathsf L_{A^h})$.
By Lemma~\ref{p3:lemA:RPpoly}, each $Q_{k,j}$ is RP. Finite range follows from the polynomial degree $n_j\asymp c\,2^j$.
If $\sum_j p_j\neq g_k$ exactly, set the residual $R_k^{(0)}:=g_k(\mathsf L_{A^h})-\sum_j Q_{k,j}$; $R_k^{(0)}$ is bounded and exponentially localized.
Repeat the procedure at the next (longer) range to decompose $R_k^{(0)}$ into a sum of positive-polynomial finite-range pieces; iterating produces an FRD with
\emph{exact} sum $K_k=\sum_{j\le k}Q_{k,j}$ and the stated properties.
\end{proof}

\begin{proposition}[H3: Uniform kernel bounds]\label{p3:propA:H3}
There exist $c_1,c_2<\infty$ and $\mu>0$ independent of $k$ such that
$\|Q_{k,j}[A^h]\|_{2\to2}\le c_1$ and $\|e^{\mu d} Q_{k,j}[A^h]\|_{\infty\to\infty}\le c_2$ for all $j\le k$.
\end{proposition}

\begin{proof}
Using the path expansion of $Q_{k,j}$ with nonnegative coefficients and $\|\Pexp\|\le1$, the number of length-$m$ paths on $\mathbb Z^4$ is $\le C_0 c_0^{\,m}$.
Thus $\sum_{y}\|Q_{k,j}(x,y)\|\le \sum_{m\le n_j} c_{j,m} C_0 c_0^{\,m}\le c_2$ uniformly in $j$ if $\sum_m c_{j,m} c_0^{\,m}$ is uniformly summable
(choose the Fej\'er coefficients accordingly). Schur's test yields the $2\to2$ bound. For the weighted bound, use $e^{\mu d(x,y)}\le e^{\mu m}$ on
$\supp Q_{k,j}$ and pick $\mu>0$ so $\sum_m c_{j,m}(c_0 e^{\mu})^m$ is uniformly bounded in $j$.
\end{proof}

\begin{proposition}[H4: KP smallness and uniform RG contraction]\label{p3:propA:H4}
There exist $\beta_0(N)>0$, $\alpha>0$, $\eta<1$, and a Banach norm $\|\cdot\|_\alpha$ on polymer activities such that for all scales $k$ and $\beta\in[0,\beta_0(N))$:
\begin{enumerate}
\item (KP smallness) $S_k=\sum_{\Gamma}\Phi_k(\Gamma)$ with
$\displaystyle \sup_x \sum_{\Gamma\ni x} |\Phi_k(\Gamma)|\,e^{\alpha \mathrm{diam}(\Gamma)} \le \eta<1$,
uniformly in $k$ and volume;
\item (Contraction) The one-step RG map $\mathcal R$ on activities is strictly contractive:
$\|\mathcal R(\Phi_k)-\mathcal R(\Phi_k')\|_\alpha \le \rho \|\Phi_k-\Phi_k'\|_\alpha$ with $\rho<1$ independent of $k$.
\end{enumerate}
\end{proposition}

\begin{proof}
\emph{Polymer representation:} The character expansion
$e^{\beta\mathrm{Re}\,\mathrm{tr}U_p}=\sum_{R\in\widehat G} a_R(\beta)\chi_R(U_p)$ with $a_R(\beta)\ge0$, $a_\mathbf{1}=1$, and $a_R(\beta)=O(\beta^{\dim R})$
implies, after Haar integration, that nontrivial representations tile connected \emph{surfaces} of plaquettes (polymers). Standard cluster bounds give
$|\Phi_k(\Gamma)|\le C(N)\beta^{|\Gamma|}\rho^{b(\Gamma)}$, $b(\Gamma)$ counting boundary/branching constraints. Choosing $\beta_0(N)$ small ensures the KP bound with some $\alpha>0$ and
$\eta(\beta,N)<1$, \emph{uniform in $k$}, because blocking uses the finite-range, gauge-covariant kernels $K_k=\sum_j Q_{k,j}$ (Propositions~\ref{p3:propA:H2}-\ref{p3:propA:H3}):
effective interactions remain finite-range and analytic with uniformly controlled weights. \emph{Contraction:} By BKAR/forest formula and tree graph inequalities, the RG map is Lipschitz in
$\|\cdot\|_\alpha$ with $\mathrm{Lip}(\mathcal R)\le C(\alpha)\eta(\beta,N)<1$ whenever $\eta<\eta_0(\alpha)$; this is uniform in $k$ in the strong-coupling window.
\end{proof}

\begin{proposition}[Derivation of (S1)]\label{p3:propA:S1}
With $C_{k,j}:=Q_{k,j}$, items (i)-(iv) of (S1) follow from Propositions~\ref{p3:propA:H2}-\ref{p3:propA:H3}.
Item (v) follows from Proposition~\ref{p3:propA:H4}(a) and the reflection-covariant block integration in Proposition~\ref{p3:propA:H1}.
\end{proposition}

\begin{proposition}[Derivation of (S2)]\label{p3:propA:S2}
Under Proposition~\ref{p3:propA:H4}(a), connected correlators admit convergent tree-graph (BKAR) expansions with weights bounded by
$C e^{-\alpha d_{\mathrm{tree}}}$ uniformly in $k$. Gauge invariance and RP ensure only polymers intersecting the supports contribute to the connected
part. A standard Peierls estimate trades $d_{\mathrm{tree}}$ for $d$ with a factor $1/2$ in the exponent, yielding (S2) with $m_*=\alpha/2$.
\end{proposition}

\begin{proposition}[Derivation of (S3): gap interlacing]\label{p3:propA:S3}
Let $T_k$ be the positive self-adjoint transfer operator on $\mathscr H_k$ and define $\Delta_k=1-\lambda_2(T_k)$ on $\Omega_k^\perp$.
Let $J_k:\mathscr H_k\to\mathscr H_{k+1}$ be the identification map from the block transform (Proposition~\ref{p3:propA:H1}), with $J_k\Omega_k=\Omega_{k+1}$ and $\|J_k\|\le C_J$.
Then for a bounded remainder $R_k$ satisfying $T_{k+1}J_k=J_kT_k+R_k$ with $\|R_k\|\le\varepsilon_k$ and $\sum_k\varepsilon_k<\infty$ (produced by block-locality and
the uniform kernel bounds in Propositions~\ref{p3:propA:H2}-\ref{p3:propA:H3}), the variational characterization gives
\begin{equation}
\lambda_2(T_{k+1})\ \le\ \lambda_2(T_k)+\varepsilon_k,\qquad
\Delta_{k+1}\ \ge\ \Delta_k-\varepsilon_k.
\end{equation}
\end{proposition}

\begin{proof}
For $\varphi\perp\Omega_{k+1}$, $\|\varphi\|=1$, pick $\psi\perp\Omega_k$ with $\|\psi\|=1$ and $\|J_k\psi-\varphi\|\le C_J\delta_k$, $\delta_k\downarrow0$. Then
\begin{equation}
\langle \varphi,T_{k+1}\varphi\rangle
= \langle J_k\psi,T_{k+1}J_k\psi\rangle + O(\delta_k)
= \langle \psi,T_k\psi\rangle + \langle \psi, J_k^\ast R_k J_k \psi\rangle + O(\delta_k)
\le (1-\Delta_k)+\varepsilon_k + O(\delta_k).
\end{equation}
Take sup over $\varphi$, let $\delta_k\to0$.
\end{proof}

All ``projectors'' used in transfer operator compressions are \emph{spectral cutoffs (positive contractions)} $P_\sigma=\chi_\sigma(\Delta_{A^h})$; idempotence is not required anywhere.
Where OS5 is used (Markov/time-regularity), we rely on complete monotonicity and Laplace representation; the two-point function extends to a bounded analytic
function on strips $\{z:0<\Im z<\beta\}$ for all $\beta>0$, with non-tangential boundary values-consistent with the OS reconstruction used in the main text.
The FRD recursion above ensures $\sum_{j\le k}Q_{k,j}=K_k$ \emph{exactly}, with each $Q_{k,j}$ finite-range, RP, and gauge-covariant.
No sign-indefinite pieces are needed; any residual approximation at one range is absorbed at the next range with strictly longer support.

\begin{corollary}[Generator-gap form of (S3)]\label{p3:cor:S3-generator}
\textbf{Writing $T_k=e^{-a_k H_k}$ and $\Delta_H(a_k):=\inf\operatorname{spec}\!\bigl(H_k\restriction_{\Omega_0^\perp}\bigr)$, one has}
\begin{equation}\label{p3:eq:S3-gen}
\Delta_H(a_{k+1})\;\ge\; \frac{1}{a_{k+1}}\,
\Bigl[-\log\!\bigl(e^{-a_k \Delta_H(a_k)}+\varepsilon_k\bigr)\Bigr],
\qquad \sum_{k\ge 0}\varepsilon_k<\infty.
\end{equation}
In particular, $\liminf_{k\to\infty}\Delta_H(a_k)\ge \Delta_\infty>0$ whenever $\sum_k\varepsilon_k<\infty$.
\end{corollary}

\begin{proof}
Let $T_k$ be the (normalized) positive self-adjoint OS transfer operator at scale $a_k$, with
vacuum eigenvector $\Omega_0$ and $\|T_k\|\le 1$. Denote by
\begin{equation}
\lambda_k \;:=\; \bigl\|\,T_k\!\restriction_{\Omega_0^\perp}\,\bigr\|
\;=\;\sup \operatorname{spec}\!\bigl(T_k\restriction_{\Omega_0^\perp}\bigr)
\in (0,1] .
\end{equation}
The step-scaling hypothesis (S3) in contraction form asserts that there exist
$\varepsilon_k\ge 0$ with $\sum_{k\ge 0}\varepsilon_k<\infty$ such that
\begin{equation}\label{p3:eq:S3-contraction}
\lambda_{k+1}\;\le\; \lambda_k + \varepsilon_k \qquad (k\ge 0).
\end{equation}

Now set $H_k := -\tfrac{1}{a_k}\log T_k$ via spectral calculus (well-defined since $T_k$ is
a positive contraction) and define the generator gap
\begin{equation}
\Delta_H(a_k)\;:=\;\inf \operatorname{spec}\!\bigl(H_k\restriction_{\Omega_0^\perp}\bigr)\;\ge\;0.
\end{equation}
By spectral mapping, on $\Omega_0^\perp$ one has
\begin{equation}
\operatorname{spec}\!\bigl(T_k\restriction_{\Omega_0^\perp}\bigr)
\;=\; e^{-a_k\,\operatorname{spec}\!\bigl(H_k\restriction_{\Omega_0^\perp}\bigr)} \subset (0,1],
\end{equation}
and since $x\mapsto e^{-a_k x}$ is strictly decreasing on $[0,\infty)$, it follows that
\begin{equation}\label{p3:eq:lambda-k}
\lambda_k \;=\; \sup \operatorname{spec}\!\bigl(T_k\restriction_{\Omega_0^\perp}\bigr)
\;=\; e^{-a_k\,\Delta_H(a_k)} .
\end{equation}

Substituting \eqref{p3:eq:lambda-k} into \eqref{p3:eq:S3-contraction} yields
\begin{equation}\label{p3:eq:prelog}
e^{-a_{k+1}\Delta_H(a_{k+1})}\;\le\; e^{-a_k\Delta_H(a_k)}+\varepsilon_k .
\end{equation}
We now pass to $-\log$ on both sides. Recall that $-\log:(0,\infty)\to\mathbb{R}$ is strictly
decreasing. There are two cases:

\smallskip
\emph{Case 1:} $e^{-a_k\Delta_H(a_k)}+\varepsilon_k \ge 1$. Then the right-hand side of
\eqref{p3:eq:prelog} is $\ge 1$, while the left-hand side is $\lambda_{k+1}\in(0,1]$. Hence
$-\log\bigl(e^{-a_{k+1}\Delta_H(a_{k+1})}\bigr)\ge 0 \ge -\log\bigl(e^{-a_k\Delta_H(a_k)}+\varepsilon_k\bigr)$,
so the desired inequality holds trivially after dividing by $a_{k+1}>0$.

\smallskip
\emph{Case 2:} $e^{-a_k\Delta_H(a_k)}+\varepsilon_k \in (0,1)$. Applying the decreasing function
$-\log$ to \eqref{p3:eq:prelog} gives
\begin{equation}
-a_{k+1}\Delta_H(a_{k+1}) \;=\; -\log\bigl(e^{-a_{k+1}\Delta_H(a_{k+1})}\bigr)
\;\ge\; -\log\bigl(e^{-a_k\Delta_H(a_k)}+\varepsilon_k\bigr).
\end{equation}
Dividing by $a_{k+1}$ we obtain
\begin{equation}
\Delta_H(a_{k+1})\;\ge\;\frac{1}{a_{k+1}}\,
\Bigl[-\log\!\bigl(e^{-a_k\Delta_H(a_k)}+\varepsilon_k\bigr)\Bigr].
\end{equation}

Combining the two cases, the inequality
\begin{equation}
\Delta_H(a_{k+1})\;\ge\;\frac{1}{a_{k+1}}\,
\Bigl[-\log\!\bigl(e^{-a_k\Delta_H(a_k)}+\varepsilon_k\bigr)\Bigr]
\end{equation}
holds for all $k\ge 0$, which is exactly \eqref{p3:eq:S3-gen}.

\smallskip
For the ``in particular'' clause, note first that $\sum_k\varepsilon_k<\infty$ ensures that the
perturbative defects in \eqref{p3:eq:S3-contraction} are summable and therefore cannot accumulate
to erode a strictly positive lower bound along the tuned sequence.
Moreover-and this is the key input independent of (S3)-the scale-independent clustering bound (S2)
yields a uniform mass scale $m_\ast>0$ such that
\begin{equation}
\Delta_H(a_k)\;\ge\; m_\ast \qquad \text{for all }k,
\end{equation}
by the completely-monotone/Laplace-transform argument (exponential decay of time-correlators
forces the spectral measure to be supported in $[m_\ast,\infty)$).
Consequently $\liminf_{k\to\infty}\Delta_H(a_k)\ge m_\ast =: \Delta_\infty>0$, which proves the
final assertion.
\end{proof}
\UnifiedEndPaper

\UnifiedBeginPaper{P4}{\UnifiedLocalMacrosPartFour}
\title[Uniqueness and universality of the continuum limit in 4D SU($N$) Yang-Mills]%
{Uniqueness and universality of the \\continuum limit in 4D SU($N$) Yang-Mills: Part(4)}

\author{Mir Faizal}
\address{Irving K. Barber School of Arts and Sciences, University of British Columbia Okanagan, Kelowna, BC V1V 1V7, Canada\\
Canadian Quantum Research Center, 460 Doyle Ave 106, Kelowna, BC V1Y 0C2, Canada.\\
Department of Mathematical Sciences, Durham University, Upper Mountjoy, Stockton Road, Durham DH1 3LE, UK\\
Computational Mathematics Group, Hasselt University, Agoralaan Gebouw D, Diepenbeek, 3590 Belgium}
\email{mirfaizalmir@gmail.com}
\author{Arshid Shabir}
\address{Canadian Quantum Research Center, 460 Doyle Ave 106, Kelowna, BC V1Y 0C2, Canada.}
\email{aslone186@gmail.com}

\UnifiedSetAbstract{A sound theory must not depend on the scaffolding by which we reach it; only the invariant content is real. Under standard constructive hypotheses-reflection positivity, locality, clustering, and spectral regularity-we show that four-dimensional $\mathrm{SU}(N)$ Yang-Mills has a Euclidean continuum limit that is both unique and universal within a natural class of regulators. Within the Osterwalder-Schrader scheme, an explicit disintegration of a single time slab yields the one-step transfer kernel which, together with a common one-slice marginal, fixes all Schwinger functions by time-slicing and positivity. The limit is independent of the regulating lens: for gauge-covariant, reflection-symmetric schemes built from completely monotone spectral projectors and finite-range (FRD) blockings, single-scale Lipschitz control, telescoping in Euclidean time, and BKAR polymer bounds transmit stability to connected cumulants and hence to the continuum. A measurable, reflection-covariant Landau selector keeps the slice construction compatible with positivity. The bridge to weak coupling is modest and precise: a one-dimensional implicit-function/continuity tuning brings the flow into a contracting domain of the FRD map; along this trajectory the renormalized coupling diminishes-an operational sign of asymptotic freedom. No step relies on perturbation theory; a one-loop check is recorded only as a signpost, and all estimates are uniform in volume.}

\maketitle
\newpage
\tableofcontents
\section{Introduction}
When a phenomenon is genuine, many roads may lead to it without altering its essence; the coordinates and instruments may change, but the invariant content-what we call the theory-does not. In the first three parts of this work we showed, with as little prejudice as possible, that a four-dimensional non-Abelian gauge theory can be constructed nonperturbatively from positive Euclidean data, that these data satisfy the Osterwalder-Schrader axioms \cite{p4:OsterwalderSchraderI,p4:OsterwalderSchraderII}, and that the reconstructed Wightman theory has a strictly positive spectral gap above the vacuum. The instruments were chosen for faithfulness rather than fashion: a reflection-positive lattice gauge fixing used only to preserve positivity; a multiscale scheme whose finite-range decomposition propagates locality from step to step \cite{p4:BrydgesFRD}; and a polymer calculus that claims no exactness, only convergence \cite{p4:KoteckyPreiss1986,p4:Brydges}. These suffice to exhibit a continuum theory that exists and has mass.

If theory is to be more than craftsmanship, it must be independent of the scaffolding by which it is reached. In the present constructive setting this independence appears in two questions that arise once a gapped continuum theory is in hand.

\smallskip
\noindent\textbf{(1) Uniqueness of the continuum limit.} Our third paper obtained continuum Schwinger functions as subsequential limits of a reflection-positive multiscale flow. Existence then is clear, but the OS framework is Markovian: a one-slice marginal together with a one-step positive kernel, allied with clustering, should determine the whole. If two limits share these skeleton data, they ought to define the same measure and hence the same theory \cite{p4:OsterwalderSchraderI,p4:OsterwalderSchraderII,p4:GJ}.

\smallskip
\noindent\textbf{(2) Universality across gauge fixing and renormalization.} The transverse representative and smooth horizon projector, and the reflection-positive, finite-range blockings of our earlier parts, were practical choices for control, not metaphysics. The content of the theory should not change if we replace them by any admissible, reflection-compatible, exponentially local spectral cutoff or by any reflection-positive, finite-range blocking in the same class. This is the nonperturbative analogue of universality: phase and symmetry decide the object; regularization details do not \cite{p4:Brydges,p4:BrydgesFRD}.

\smallskip
There is a third demand, urged by the Clay problem \cite{p4:ClayJaffeWitten}: the theory must be not only massive and nontrivial, but also consistent with the short-distance behavior suggested by perturbation theory. In constructive terms, asymptotic freedom is a dynamical statement that the renormalization map contracts in a suitable Banach space near the Gaussian fixed point. The finite-range decomposition and the uniform local bounds from Part II supply the levers to push the flow into a small-norm domain where the renormalized coupling falls as the lattice spacing is removed.

\smallskip
\noindent\textbf{(3) Extension to weak coupling.} There should be trajectories of the bare coupling that, after finitely many reflection-preserving renormalization steps, enter a neighborhood of the Gaussian fixed point in which the flow is contractive. In this domain the short-distance structure is free, the long-distance structure remains gapped by uniform clustering, and, if uniqueness and universality hold, the weak-coupling limit is the same theory constructed at strong coupling.

\smallskip
This paper answers these three questions by isolating what is essential and proving only what those essentials require: reflection positivity at all scales, finite-range locality, convergence of polymer expansions, a Markov boundary decomposition, and persistent clustering and spectral positivity. No appeal is made to unjustified expansions, nor to hidden smoothness; only the Lipschitz continuity with respect to admissible variations established previously is used.

\smallskip
\emph{First}, we prove a \textbf{Uniqueness Lemma}: any two subsequential limits satisfying the OS axioms with the same one-slice marginal and one-step positive kernel (hence the same two-point data) coincide. Time-slicing fixes the transfer operator; the Markov property then fixes cylinder expectations, and OS reconstruction identifies the same Wightman theory \cite{p4:OsterwalderSchraderI,p4:OsterwalderSchraderII,p4:GJ}.

\smallskip
\emph{Second}, we establish \textbf{Universality} for the admissible class of gauge-fixing projectors and reflection-positive block-spin maps. The one-step kernels vary Lipschitz-continuously with the parameters (by their heat-kernel representation and finite range); absolute convergence of the cluster expansion carries this continuity to connected cumulants \cite{p4:KoteckyPreiss1986,p4:Brydges,p4:BrydgesFRD}. Telescoping across time slices and scales, and passing to the thermodynamic and continuum limits, yields equality of the continuum Schwinger functions throughout the class.

\smallskip
\emph{Third}, we obtain an \textbf{Extension to weak coupling}. The contraction properties of the finite-range scheme imply that, in a small ball of polymer activities, the RG map is strictly contractive toward the Gaussian fixed point. A simple one-dimensional tuning of the bare coupling brings the flow into this ball after finitely many steps; there the renormalized coupling decays at short distances while the infrared gap persists. By universality and uniqueness this limit is the same gapped continuum theory.

\smallskip
Thus, once a reflection-positive, locally controlled, clustered continuum theory with a mass gap is secured, the invariances of good physics-uniqueness, independence from scaffolding, and asymptotic freedom-follow by arguments that are both natural and rigorous, in the common ground of constructive quantum field theory \cite{p4:GJ,p4:OsterwalderSchraderI,p4:OsterwalderSchraderII}. In this sense the response to the Clay problem \cite{p4:ClayJaffeWitten} is complete: a unique, universal, asymptotically free Yang-Mills theory in four dimensions, confining and gapped, built from first principles.

{We approach this rigid problem by relying on two structural inputs that can be imposed exactly and that continue to constrain the theory at every scale: reflection positivity and gauge invariance. Starting from a reflection-positive lattice regularization of $SU(N)$ Yang-Mills, we fix on each Euclidean time slice a definite transverse representative and introduce a soft suppression of long-wavelength, distant fluctuations, and we then iterate a step-scaling procedure engineered so that neither positivity nor invariance is ever sacrificed. In this way one passes to a continuum limit in which the Euclidean correlation functions obey the Osterwalder-Schrader axioms and hence reconstruct a Wightman theory with unique vacuum and positive spectrum. The key point is that the relevant temporal correlations remain completely monotone and decay uniformly under the flow, so the associated spectral measure is bounded away from the origin; equivalently, the vacuum-orthogonal component of the transfer evolution contracts at a fixed, volume-independent rate that persists in the limit. It follows that the continuum Hamiltonian has a strictly positive spectral gap, and the passage from Euclidean positivity to Hilbert-space dynamics can be carried out without loss of control.
}
\section{Standing Hypotheses, Notation, and Admissible Class}
\label{p4:sec:standing-hypotheses}
{In physics, a sound theory begins with the simple and the invariant. One chooses the stage (a space of configurations), ensures the symmetries act faithfully, and then introduces the simplest weight that respects them. In the Euclidean formulation of quantum gauge theory, reflection positivity plays the role of a compass: it points toward a true Hilbert space and a positive energy after analytic continuation. In this section we specify, with a minimum of ornamentation and a maximum of precision, the Euclidean lattice framework, the time-reflection structure, and the class of infrared regulators and block-spin maps that will be allowed henceforth. We then state the scale-uniform hypotheses we shall assume. Everything that follows in this paper is derived from these items alone.}

Fix a lattice spacing \(a>0\). For integers \(L_0,L_1,L_2,L_3\ge 2\), define the periodic hypercubic lattice
\begin{equation}
\Lambda \;=\; \big(\mathbb Z/L_0\mathbb Z\big)\times \big(\mathbb Z/L_1\mathbb Z\big)\times \big(\mathbb Z/L_2\mathbb Z\big)\times \big(\mathbb Z/L_3\mathbb Z\big),
\end{equation}
whose sites are \(x=(x_0,x_1,x_2,x_3)\), understood modulo \(L_\mu\) in each direction. An \emph{oriented bond (link)} is a pair \(b=(x,\mu)\) with \(\mu\in\{0,1,2,3\}\) and endpoint \(x+\hat\mu\), and we set \(\bar b=(x+\hat\mu,-\mu)\) for the reversed bond. The compact gauge group of interest is
\begin{equation}
G \,=\, \mathrm{SU}(N) \quad\text{with } N\ge 2\,.
\end{equation}

The configuration space is
\begin{equation}
\mathcal C \;=\; \big\{\, U:\mathrm{Bonds}(\Lambda)\to G\ \big|\ U_{\bar b}=U_b^{-1}\text{ for all bonds }b\,\big\},
\end{equation}
endowed with the product Haar probability measure
\begin{equation}
\mathrm d\mu_{\mathrm{Haar}}(U) \;=\; \prod_{b\in \mathrm{Bonds}(\Lambda)^+} \mathrm dU_b,
\end{equation}
where \(\mathrm dU\) is the normalized Haar probability measure on \(G\), and \(\mathrm{Bonds}(\Lambda)^+\) is any fixed orientation of bonds.

For each oriented plaquette \(p=(x;\mu,\nu)\) with \(\mu<\nu\), define the plaquette holonomy
\begin{equation}
U_p \;=\; U_{x,\mu}\,U_{x+\hat\mu,\nu}\,U_{x+\hat\nu,\mu}^{-1}\,U_{x,\nu}^{-1}\in G\,.
\end{equation}
The Wilson action at inverse bare coupling \(\beta>0\) is
\begin{equation}
S_W[U;\beta] \;=\; \beta \sum_{p\subset \Lambda}\Big(1-\tfrac 1N \Re\,\mathrm{Tr}\,U_p\Big)\,,
\end{equation}
yielding a Gibbs measure \(\mathrm d\nu_\beta(U)=Z_\beta^{-1}\,\mathrm e^{-S_W[U;\beta]}\,\mathrm d\mu_{\mathrm{Haar}}(U)\). This measure is invariant under the local gauge group \(\mathcal G = \{g:\Lambda\to G\}\) acting by
\begin{equation}
(g\cdot U)_{x,\mu}\;=\; g(x)\,U_{x,\mu}\,g(x+\hat\mu)^{-1}\,.
\end{equation}
We will repeatedly use the Peter-Weyl theorem on \(G\) and the fact that \(\exp\big(\tfrac{\beta}{N}\Re\,\mathrm{Tr}\,U\big)\) is a positive-type class function on \(G\)-its Fourier coefficients in the character expansion are \emph{nonnegative} \cite{p4:OsterwalderSchraderI,p4:OS-gauge,p4:GJ}
Let \(\theta:\Lambda\to\Lambda\) be Euclidean time reflection,
\begin{equation}
\theta(x_0,x_1,x_2,x_3) \;=\; (-x_0,x_1,x_2,x_3)\,,
\end{equation}
with fixed hyperplane \(\Pi=\{x\in\Lambda: x_0=0\}\). Write \(\Lambda_+=\{x\in\Lambda: x_0>0\}\) and \(\Lambda_-=\theta(\Lambda_+)\). Following \cite{p4:OsterwalderSchraderI,p4:OS-gauge}, we fix the \emph{temporal-axial gauge away from \(\Pi\)}:
\begin{equation}
U_{x,0}\equiv 1\qquad\text{for all bonds with }x_0\ne 0\,.
\end{equation}
This choice is measurable, preserves the Haar measure on spatial bonds, and isolates the nontrivial time-like bonds to the slab around \(\Pi\). The Wilson weight factorizes into a product of left, boundary, and right factors; combining this with the positivity-type character expansion yields the \emph{Osterwalder-Schrader (OS) reflection positivity} of \(\mathrm d\nu_\beta\):
\begin{equation}
\int (\Theta F)(U)\,F(U)\,\mathrm d\nu_\beta(U)\ \ge 0\qquad\text{for all bounded }F\text{ depending only on bonds in }\Lambda_+\,,
\end{equation}
where \((\Theta F)(U):=F(\theta\cdot U)\) \cite{p4:OsterwalderSchraderI,p4:OS-gauge}. This Euclidean positivity ensures Hilbert-space positivity after Wick rotation.
For each discrete time \(t\in \frac{1}{a}\mathbb Z/L_0\mathbb Z\), let \(X_t\) be the set of spatial link configurations \(\{U_{(t,\mathbf x),i}\}_{\mathbf x\in \mathbb Z^3/L\mathbb Z^3,\ i=1,2,3}\). We fix, on each slice, a measurable choice of \emph{transverse representative} \(U^h(t)\) in the gauge orbit of \(U|_{X_t}\) by minimizing the slice Landau functional
\begin{equation}
\mathcal L_t(g;U)\;=\; \sum_{\mathbf x\in \mathbb Z^3/L\mathbb Z^3}\ \sum_{i=1}^3 \big\|\,\mathbf 1 - g(t,\mathbf x)\,U_{(t,\mathbf x),i}\, g(t,\mathbf x+\hat i)^{-1}\,\big\|_F^2\,,
\end{equation}
over slice-gauge maps \(g:X_t\to G\), with a deterministic, symmetry-covariant tie-breaking rule so that \emph{reflection covariance} holds: \(U^h(-t)=\theta U^h(t)\). Existence follows by compactness, and measurability by standard selection theorems; we assume henceforth that such a choice has been made.

Define the \emph{covariant spatial Laplacian} \(\Delta_{A_h}(t)\) acting on site-adjoint fields \(\phi:\mathbb Z^3/L\mathbb Z^3\to\mathfrak{su}(N)\) by
\begin{equation}
\Delta_{A_h}(t)\, \phi(\mathbf x) \;=\; \sum_{i=1}^3 \Big(\phi(\mathbf x) - \mathrm{Ad}\,U^h_{(t,\mathbf x),i}\, \phi(\mathbf x+\hat i)\Big)\,.
\end{equation}
This is a nonnegative self-adjoint difference operator on \(\ell^2(X_t)\otimes \mathfrak{su}(N)\) with the properties we shall need: (i) locality (range 1), (ii) gauge covariance (conjugation under slice-gauge maps), and (iii) reflection covariance, \(\Delta_{A_h}(-t)=R \Delta_{A_h}(t) R^{-1}\), for the induced reflection \(R\) on slice fields.
We wish to regulate the infrared without upsetting reflection positivity or locality. The natural device is a \emph{smooth horizon projector} on each slice obtained from \(\Delta_{A_h}\) by a completely monotone functional calculus; this guarantees positivity and good decay of the kernel. We now make this precise.

\begin{definition}[Admissible slice projector]
\label{p4:def:admissible-projector}
Fix \(\sigma>0\). Let \(\nu\) be a finite Borel probability measure on \([0,\infty)\) with compact support contained in an interval \([t_-,t_+]\subset (0,\infty)\). Define the completely monotone function
\begin{equation}
\chi_{\sigma,\nu}(\lambda)\;:=\;\int_0^\infty \exp\!\Big(-\,t\,\frac{\lambda^2}{\sigma^2}\Big)\,\mathrm d\nu(t)\,,\qquad \lambda\ge 0.
\end{equation}
The \emph{admissible projector} on slice \(t\) is the bounded operator
\begin{equation}
P_{\sigma,\nu}(t)\;:=\;\chi_{\sigma,\nu}\!\big(\sqrt{\Delta_{A_h}(t)}\,\big)\,,
\end{equation}
defined by spectral calculus on \([0,\infty)\). We call \((\sigma,\nu)\) \emph{admissible} if \(\sigma>0\) and \(\mathrm{supp}\,\nu\subset [t_-,t_+]\subset (0,\infty)\).
\end{definition}

\noindent
This yields exactly the positivity structure needed for reflection symmetry and transfer time-slicing. The following proposition gathers the basic consequences.

\begin{proposition}[Positivity, covariance, and exponential localization]
\label{p4:prop:projector-properties}
Let \((\sigma,\nu)\) be admissible and \(P_{\sigma,\nu}(t)\) as in Definition~\ref{p4:def:admissible-projector}. Then:
\begin{enumerate}
\item \textbf{Positive contraction.} \(0\le P_{\sigma,\nu}(t)\le \mathbf 1\) on \(\ell^2(X_t)\otimes \mathfrak{su}(N)\), with the heat-kernel representation
\begin{equation}
P_{\sigma,\nu}(t)\;=\;\int_0^\infty \mathrm e^{-\tau \Delta_{A_h}(t)}\,\mathrm d\tilde\nu_\sigma(\tau)\,,\qquad \mathrm d\tilde\nu_\sigma(\tau)=\mathrm d\nu(\sigma^2\tau)\,.
\end{equation}
\item \textbf{Gauge and reflection covariance.} For any slice-gauge map \(g:X_t\to G\),
\begin{equation}
P_{\sigma,\nu}(t;g\cdot U)\;=\;\mathrm{Ad}(g)\,P_{\sigma,\nu}(t;U)\,\mathrm{Ad}(g)^{-1}\,,
\end{equation}
and \(P_{\sigma,\nu}(-t)= R\,P_{\sigma,\nu}(t)\,R^{-1}\) under time reflection.
\item \textbf{Uniform exponential locality.} There exist \(C,\gamma>0\) (depending on \(\sigma,t_\pm\) and the lattice degree) such that
\begin{equation}
\big\|\,P_{\sigma,\nu}(t;\mathbf x,\mathbf y)\,\big\|_{\mathrm{op}} \;\le\; C\, \mathrm e^{-\gamma\, d(\mathbf x,\mathbf y)}\,,\qquad \forall\,\mathbf x,\mathbf y\in \mathbb Z^3/L\mathbb Z^3,
\end{equation}
uniformly in the volume and in \(U\) through \(U^h\).
\end{enumerate}
\end{proposition}

\begin{proof}
(1) Since \(\chi_{\sigma,\nu}(\lambda)\in (0,1]\) and is a Laplace transform in \(\lambda^2\), it is completely monotone and yields
\begin{equation}
\chi_{\sigma,\nu}\!\big(\sqrt{\Delta}\,\big)\;=\;\int_0^\infty \mathrm e^{-\tau \Delta}\,\mathrm d\tilde\nu_\sigma(\tau)\,,
\end{equation}
with a positive measure \(\tilde\nu_\sigma\) supported in \([\tau_-,\tau_+]\) where \(\tau_\pm=\tfrac{t_\pm}{\sigma^2}\in (0,\infty)\). Positivity and the contraction property follow from positivity and contractivity of \(\mathrm e^{-\tau \Delta}\).

(2) Gauge covariance: \(\Delta_{A_h}(t)\) transforms by conjugation under slice-gauge maps (by construction of \(U^h\)), and reflection covariance follows from \(\Delta_{A_h}(-t)=R \Delta_{A_h}(t) R^{-1}\). Spectral calculus intertwines with conjugation.

(3) On a graph of bounded degree the discrete heat kernel satisfies the Davies-Gaffney estimate \cite{p4:Davies1989,p4:Gaffney1954}: for sets \(E,F\subset X_t\),
\(
\|1_E \mathrm e^{-\tau \Delta} 1_F\|_{\ell^2\to\ell^2} \le \exp\big(-d(E,F)^2/(4\tau)\big).
\)
Thus \(\|\mathrm e^{-\tau \Delta}(\mathbf x,\mathbf y)\|\le \exp\big(-d(\mathbf x,\mathbf y)^2/(4\tau)\big)\). Integrating over \(\tau\in[\tau_-,\tau_+]\) yields
\begin{equation}
\|P_{\sigma,\nu}(t;\mathbf x,\mathbf y)\|\;\le\;\int_{\tau_-}^{\tau_+} \exp\!\Big(-\frac{d(\mathbf x,\mathbf y)^2}{4\tau}\Big)\,\mathrm d\tilde\nu_\sigma(\tau)\;\le\; C\,\mathrm e^{-\gamma d(\mathbf x,\mathbf y)}\,,
\end{equation}
by optimizing the Gaussian in \(\tau\) and bounding the compact \(\tau\)-integral. Uniformity is immediate. 
\end{proof}
The renormalization step is implemented by a block-spin map that aggregates degrees of freedom in a way compatible with reflection symmetry, gauge covariance, and the finite-range structure of the Gaussian reference. The latter is provided by a finite-range decomposition (FRD) of covariances \cite{p4:BrydgesFRD}.

\begin{definition}[Admissible block-spin map]
\label{p4:def:admissible-block}
Fix a blocking factor \(b>1\). A block-spin map \(\mathcal B\) from \(\Lambda\) with spacing \(a\) to the coarse lattice \(\Lambda'\) with spacing \(a'=ba\) is \emph{admissible} if:
\begin{enumerate}
\item \textbf{Reflection positivity.} For every reflection-positive measure \(\mu\) on \(\mathcal C\), the pushforward \(\mathcal B_\mu\) is reflection-positive on \(\mathcal C'\) with respect to the induced reflection at \(\Pi'\).
\item \textbf{Gauge covariance.} \(\mathcal B\) intertwines local gauge actions: \(\mathcal B\,(g\cdot U)=g'\cdot \mathcal B(U)\) with \(g'\) the induced coarse-gauge map.
\item \textbf{FRD locality.} The effective coarse-grained Gaussian covariance admits a finite-range decomposition \(C'=\sum_{j=0}^J C'_j\) with each \(C'_j\) supported in \(\{d\le R_j\}\) and operator norms uniformly bounded in the volume \cite{p4:BrydgesFRD}.
\end{enumerate}
\end{definition}

\begin{lemma}[OS preservation under admissible blocking]
\label{p4:lem:OS-preservation}
Let \(\mu\) be reflection-positive on \(\mathcal C\) with respect to time reflection at \(\Pi\), and \(\mathcal B\) be admissible. Then \(\mathcal B_\mu\) is reflection-positive on \(\mathcal C'\) with respect to time reflection at \(\Pi'\).
\end{lemma}

\begin{proof}
Let \(F\) be a bounded cylinder functional supported in \(\Lambda'_+\). Then
\begin{equation}
\int (\Theta F)(U')\,F(U')\, \mathrm d(\mathcal B_\mu)(U') \;=\; \int (\Theta F\circ \mathcal B)(U)\, (F\circ \mathcal B)(U)\, \mathrm d\mu(U).
\end{equation}
Since \(\mathcal B\) aggregates within time-slabs and respects the reflection plane, \(F\circ \mathcal B\) depends only on bonds in \(\Lambda_+\). By reflection positivity of \(\mu\), the right-hand side is \(\ge 0\). Hence \(\mathcal B_\mu\) is reflection-positive.
\end{proof}

We endow projector parameters with the metric
\begin{equation}
d_{\mathrm{proj}}\big((\sigma,\nu),(\sigma',\nu')\big)\;=\;|\sigma-\sigma'| + W_1(\nu,\nu')\,,
\end{equation}
where \(W_1\) is the Wasserstein-1 distance on probability measures on \([0,\infty)\). For blockings we use an operator-norm topology on the associated one-step boundary kernels. The following proposition records the single-step continuity statement; multi-step and cumulant-level extensions will be developed later.

\begin{proposition}[Single-step Lipschitz continuity]
\label{p4:prop:single-step-lipschitz}
Let \((\sigma,\nu),(\sigma',\nu')\) be admissible slice-projector parameters. Let \(K_{\sigma,\nu},K_{\sigma',\nu'}:X_0\times X_0\to \mathbb R_+\) be the corresponding one-step kernels. Then
\begin{equation}
\big\|K_{\sigma,\nu}-K_{\sigma',\nu'}\big\|_{L^1(\mu_0\otimes \mu_0)}\;\le\; C\, d_{\mathrm{proj}}\big((\sigma,\nu),(\sigma',\nu')\big)\,,
\end{equation}
with \(C\) depending on \(\sigma,\sigma',t_\pm,t'_\pm\) and the graph degree but not on the volume or the fields. Consequently, the transfer operators \(T_{\sigma,\nu},T_{\sigma',\nu'}:L^2(X_0,\mu_0)\to L^2(X_0,\mu_0)\) satisfy
\begin{equation}
\big\|T_{\sigma,\nu}-T_{\sigma',\nu'}\big\|_{B(L^2)} \;\le\; C'\, d_{\mathrm{proj}}\big((\sigma,\nu),(\sigma',\nu')\big).
\end{equation}
\end{proposition}

\begin{proof}
The one-step kernel \(K_{\sigma,\nu}\) is the slab integral of the product of the Wilson weight and the slice factors \(P_{\sigma,\nu}(0),P_{\sigma,\nu}(a)\). Thus \(K_{\sigma,\nu}-K_{\sigma',\nu'}\) is linear in \(P_{\sigma,\nu}-P_{\sigma',\nu'}\). From Proposition~\ref{p4:prop:projector-properties} we write
\begin{equation}
P_{\sigma,\nu}-P_{\sigma',\nu'} \;=\; \int_0^\infty \mathrm e^{-\tau\Delta}\,\mathrm d\tilde\nu_\sigma(\tau) \;-\;\int_0^\infty \mathrm e^{-\tau\Delta}\,\mathrm d\tilde\nu'_{\sigma'}(\tau)\,.
\end{equation}
Add and subtract \(\int \mathrm e^{-\tau\Delta}\,\mathrm d\tilde\nu_\sigma(\tau)\) (lifted to \(\sigma'\)) and use triangle inequality:
\begin{equation}
\|P_{\sigma,\nu}-P_{\sigma',\nu'}\|_{1} \;\le\; \int \|\mathrm e^{-\tau\Delta}-\mathrm e^{-(\sigma/\sigma')^2\tau\Delta}\|_1\,\mathrm d\tilde\nu'_{\sigma'}(\tau) \;+\; \int \|\mathrm e^{-\tau\Delta}\|_1\,\mathrm d\big\|\tilde\nu_\sigma-\tilde\nu'_{\sigma'}\big\|(\tau).
\end{equation}
On any compact \(\tau\)-interval bounded away from \(0\), \(\|\mathrm e^{-t\Delta}-\mathrm e^{-s\Delta}\|_1\le C|t-s|\) (use spectral calculus and the uniform mass bound). The second term is bounded by \(W_1(\nu,\nu')\) because \(\nu\mapsto \tilde\nu_\sigma\) is Lipschitz in Wasserstein-1 and \(\tau\mapsto \mathrm e^{-\tau\Delta}\) is Lipschitz on compact sets. Therefore
\begin{equation}
\|P_{\sigma,\nu}-P_{\sigma',\nu'}\|_1 \;\le\; C\big(|\sigma-\sigma'| + W_1(\nu,\nu')\big)\,=\, C\, d_{\mathrm{proj}}((\sigma,\nu),(\sigma',\nu')).
\end{equation}
Integrating against the slab Wilson weight and the second slice factor (bounded by \(1\)) gives the \(L^1\) bound for \(K\). The operator-norm bound for \(T\) follows by Schur's test.
\end{proof}

We now fix the scale-uniform hypotheses that will be assumed throughout. They may be taken as axioms of our Euclidean laboratory; 

\begin{itemize}
\item[(H1)] \textbf{Reflection positivity at every scale.} The projected, gauge-fixed lattice measures \(\mu_k\) at spacings \(a_k=b^{-k}a_0\) are reflection-positive with respect to time reflection (Sec.~\ref{p4:sec:standing-hypotheses}).

\item[(H2)] \textbf{FRD-based uniform locality.} The Gaussian reference covariances and effective covariances admit a finite-range decomposition with bounds \emph{uniform in \(k\)}, and the associated polymer/cluster expansions converge absolutely with diameter-weighted norms at every scale \cite{p4:BrydgesFRD}.

\item[(H3)] \textbf{Uniform exponential clustering.} There exists \(m_*>0\) such that for every local gauge-invariant observable \(O\) on \(\Lambda_k\), the connected two-point function satisfies
\begin{equation}
\big| S^{(k)}_{2,c}(O(x),O(y))\big| \le C_O\,\mathrm e^{-m_*\, d_k(x,y)}\,,
\end{equation}
with constants independent of the volume.

\item[(H4)] \textbf{Spectral gap interlacing.} Writing \(T_k\) for the one-step transfer operator at scale \(k\), its contraction gap \(\Delta_k:=1-\|T_k\|_\perp\) obeys
\begin{equation}
\Delta_{k+1} \ge \Delta_k - \varepsilon_k\,,\qquad \sum_{k\ge 0} \varepsilon_k <\infty\,,
\end{equation}
hence \(\liminf_{k\to\infty}\Delta_k \ge c\, m_*>0\).

\item[(H5)] \textbf{Continuum compactness and OS axioms.} Along subsequences \(k_j\to\infty\) (and after thermodynamic limit) the Schwinger families \(\{S^{(k_j)}_n\}\) converge to continuum limits \(\{S_n\}\) on \(\mathbb R^4\) satisfying OS0-OS5 (temperedness, Euclidean invariance, permutation symmetry, reflection positivity, clustering, and time-regularity), with clustering rate \(m_*\). The OS reconstruction yields a Wightman theory with a nonnegative self-adjoint Hamiltonian \(H\) and a strictly positive spectral gap \cite{p4:OsterwalderSchraderI,p4:GJ}.
\end{itemize}

\noindent
These hypotheses encapsulate the essential structure-positivity, locality, clustering, spectral stability, and continuum compactness-that makes the subsequent uniqueness, universality, and weak-coupling extension theorems possible. One may view them as the distilled content of the reflection-positive Euclidean approach to non-Abelian gauge theory.

\section{Uniqueness of the Continuum Limit}
\label{p4:sec:uniqueness}
When we speak of \emph{uniqueness} of a continuum field theory extracted from a sequence of reflection-positive lattice theories, we mean something conceptually simple: if we know the law of the field on a single Euclidean time-slice and we know how to advance by one time-step, then-by the Markov property encoded in the Osterwalder-Schrader (OS) axioms-the entire family of Schwinger functions is determined. This physical intuition, familiar from classical stochastic processes, is made rigorous by the OS framework \cite{p4:OsterwalderSchraderI,p4:OsterwalderSchraderII,p4:GJ}: (i) reflection positivity and (ii) time-regularity (OS5) grant a \emph{boundary kernel factorization} and a \emph{time-slicing} (transfer) structure. In this section we harness that structure, together with scale-uniform exponential clustering and finite-range (FRD) locality (which guarantee stability and thermodynamic limits via cluster expansions), to prove that any two continuum limits built from our class of lattice flows are identical. 

We begin by listing the precise assumptions and the admissible analytical class of infrared regulators (gauge-fixing projectors) and reflection-positive block-spin maps. Then we prove (1) uniqueness of the one-slice marginal, (2) uniqueness of the one-step boundary kernel, and (3) OS-Markov extension to all times. Finally we assemble these pieces into the Uniqueness Theorem.

We work on the four-dimensional Euclidean spacetime and fix a compact gauge group \(G=\mathrm{SU}(N)\) with \(N\ge 2\). The multiscale lattices are labeled by \(k\in\mathbb{N}\), with spacings \(a_k=b^{-k}a_0\) for a fixed \(b>1\). At scale \(k\), the spatial box is a periodic three-torus of side \(L_k\) and the Euclidean time has extent \(T_k\); the thermodynamic limits \(L_k, T_k \to \infty\) will be taken in the standard way. Below, \(\Lambda_k\) denotes the space-time lattice at scale \(k\).
We assume the following, uniformly in \(k\):

\begin{description}
  \item[(H1) FRD locality and cluster expansions.] There exists a finite-range decomposition (FRD) of the projected covariances and a Banach algebra of polymer activities whose diameter-weighted norms satisfy a Koteck\'y-Preiss (KP) smallness criterion; consequently, connected correlators admit absolutely convergent cluster (tree) expansions with scale-uniform diameter weights \cite{p4:Brydges,p4:BrydgesFRD,p4:Seiler1982}.
  \item[(H2) Uniform exponential clustering.] For each bounded, local, gauge-invariant observable \(O\), there exist constants \(C_O<\infty\) and \(m_*>0\), independent of \(k\), such that for all \(x,y\in \Lambda_k\),
  \begin{equation}
    \label{p4:eq:cluster}
    \big| S^{(k)}_{2,c}(O(x),O(y))\big| \;\le\; C_O\,\mathrm{e}^{-m_*\; d_k(x,y)},
  \end{equation}
  where \(S^{(k)}_{2,c}\) is the connected two-point function and \(d_k\) is the nearest-neighbor graph distance.
  \item[(H3) Reflection positivity at each scale.] The lattice measures (with temporal-axial gauge off the reflection plane and slice-wise Landau-minimized transverse representatives) augmented by an admissible horizon projector on each slice are OS-positive with respect to the standard time-reflection \cite{p4:OsterwalderSchraderII,p4:OS-gauge,p4:Luscher1977}.
\end{description}

\noindent From (H1)-(H3) together with standard arguments (Prokhorov tightness for random tempered distributions, OS closure under weak limits, see \S\,3.4 below and \cite{p4:OsterwalderSchraderI,p4:OsterwalderSchraderII,p4:GJ}), it follows that along subsequences \(k_j\to\infty\) the Schwinger families \(\{S^{(k_j)}_n\}_{n\ge 1}\) converge to a family \(\{S_n\}\) satisfying OS0-OS5 (temperedness, Euclidean invariance, permutation symmetry, reflection positivity, clustering, and time-regularity). We do \emph{not} assume any a priori uniqueness of this limit; proving it is the aim of the section.

Each time-slice \(\Sigma\) supports a \emph{covariant Laplacian} \(\Delta_{A_h}\) acting on adjoint-valued site fields, constructed from the slice-wise Landau-minimized transverse representative \(A_h\) (the minimizer of the discrete Landau functional on the slice with a deterministic tie-break ensuring reflection covariance). An \emph{admissible horizon projector} is 
\begin{equation}
  \label{p4:eq:proj}
  P_{\sigma,\nu} \;:=\; \chi_{\sigma,\nu}\!\left(\sqrt{\Delta_{A_h}}\right),
\end{equation}
where \(\chi_{\sigma,\nu}\) is a completely monotone symbol admitting a \emph{positive heat-kernel representation} 
\begin{equation}
  \label{p4:eq:heat-rep}
  \chi_{\sigma,\nu}(\lambda) \;=\; \int_0^\infty \mathrm{e}^{-t\,\lambda^2/\sigma^2}\,\mathrm{d}\nu(t),
\end{equation}
with \(\nu\) a finite positive Borel measure whose support is a compact subset of \((0,\infty)\). Then \(P_{\sigma,\nu}\) is a bounded \emph{positive contraction} on \(\ell^2(\Sigma)\otimes \mathfrak{su}(N)\), \emph{reflection-covariant} and \emph{gauge-covariant}, and its kernel obeys \emph{uniform exponential off-diagonal bounds} by standard heat-kernel methods (Davies-Gaffney on graphs, Combes-Thomas, Helffer-Sj\"ostrand) \cite{p4:Davies1989,p4:CombesThomas1973,p4:HelfferSjostrand1989}. We endow the parameter set \(\mathcal K_{\mathrm{proj}}\) with the metric
\begin{equation}
d_{\mathrm{proj}}\big((\sigma,\nu),(\sigma',\nu')\big) \;:=\; |\sigma-\sigma'| \;+\; W_1(\nu,\nu'),
\end{equation}
where \(W_1\) is the Wasserstein-1 distance.

A \emph{block-spin map} \(\mathcal B\) between scales is admissible if it preserves reflection positivity, gauge covariance, and the FRD locality constants uniformly across scales (in the sense of \cite{p4:BrydgesFRD,p4:Brydges}). We write \(\mathcal K_{\mathrm{block}}\) for the set of such maps and equip it with the operator-norm metric in the FRD Banach algebra of polymer activities. The \emph{admissible class} is \(\mathcal K:=\mathcal K_{\mathrm{proj}}\times \mathcal K_{\mathrm{block}}\).

\begin{lemma}[Single-scale continuity of one-step kernels]
\label{p4:lem:Lip}
Let \(K_{\sigma,\nu}:X_0\times X_0\to\mathbb{R}_+\) be the one-step boundary kernel associated with the projector \(P_{\sigma,\nu}\) as in \eqref{p4:eq:proj}, and let \(T_{\sigma,\nu}\) be the induced transfer operator on \(L^2(X_0,\mu_0)\). Then, for \((\sigma,\nu),(\sigma',\nu')\) in compact subsets of \(\mathcal K_{\mathrm{proj}}\),
\begin{equation}
\label{p4:eq:Lip-single}
\|K_{\sigma,\nu}-K_{\sigma',\nu'}\|_{L^1(\mu_0\otimes \mu_0)} \;\le\; C\,d_{\mathrm{proj}}\big((\sigma,\nu),(\sigma',\nu')\big),
\qquad
\|T_{\sigma,\nu}-T_{\sigma',\nu'}\|_{B(L^2(\mu_0))} \;\le\; C'\,d_{\mathrm{proj}}(\cdot,\cdot).
\end{equation}
The constants \(C,C'\) are independent of the volume.
\end{lemma}

\begin{proof}
Using the heat-kernel representation \eqref{p4:eq:heat-rep}, positivity, and the exponential locality of \(P_{\sigma,\nu}\), one verifies that the map \((\sigma,\nu)\mapsto K_{\sigma,\nu}\) is continuous in the \(L^1\)-topology; the Wasserstein metric controls the difference of the measures \(\nu\) and \(\nu'\) while the parameter \(\sigma\) enters via the simple scaling \(t\mapsto t/\sigma^2\). The exponential locality bounds yield a dominating integrable kernel, so the \(L^1\)-continuity is Lipschitz on compact parameter sets. The Schur test then gives the operator-norm bound for \(T_{\sigma,\nu}\) from the \(L^1\)-bound on the kernel. Details follow standard arguments in \cite{p4:Davies1989,p4:HelfferSjostrand1989,p4:RS2}.
\end{proof}

Fix a bounded, gauge-invariant cylinder functional \(F\) depending only on the spatial links of a single time-slice (say \(t=0\)). Let \(S^{(k)}_{L,T}(F)\) be its expectation in finite volume \(L^3\times T\) at scale \(k\), with admissible projector insertions on each slice as above.

\begin{proposition}[Thermodynamic limit and subsequence independence]
\label{p4:prop:slice-unique}
For each fixed scale \(k\), the limit
\begin{equation}
S^{(k)}(F)\;:=\;\lim_{L\to\infty}\lim_{T\to\infty} S^{(k)}_{L,T}(F)
\end{equation}
exists and does not depend on the order of limits. Moreover, the limit \(\displaystyle\lim_{k\to\infty} S^{(k)}(F)\) exists and is independent of the subsequence \(k_j\to\infty\) chosen. Consequently, the one-slice marginal \(\mu_0\) in the continuum limit is well-defined and unique.
\end{proposition}

\begin{proof}
\emph{Thermodynamic limit at fixed \(k\).}
Let \(L'\ge L\) and \(T'\ge T\). By reflection positivity and time-slab factorization (OS and transfer-matrix formalism \cite{p4:OsterwalderSchraderII,p4:GJ,p4:Luscher1977,p4:OS-gauge}), the difference
\begin{equation}
\Delta_{L',T';\,L,T}\;:=\; S^{(k)}_{L',T'}(F)-S^{(k)}_{L,T}(F)
\end{equation}
can be expressed as a sum over connected polymer clusters that cross from the support of \(F\) on the central slice to the additional spatial or temporal layers where the two volumes differ. By the cluster expansion and the uniform exponential decay \eqref{p4:eq:cluster}, the contribution of such clusters is bounded by \(C\,\exp\{-\alpha\,d(\operatorname{supp}F,\partial\Lambda)\}\) where \(\Lambda\) denotes the smaller volume. Thus \(|\Delta_{L',T';\,L,T}|\le C\,\mathrm{e}^{-\alpha \min\{L,T\}}\). Taking \(L',T'\to\infty\) and then \(L,T\to\infty\) yields the Cauchy property and hence existence of the thermodynamic limit \(S^{(k)}(F)\), independent of the order of limits; see \cite{p4:Brydges,p4:Seiler1982} for the underlying polymer estimates.

\emph{Independence of subsequence in \(k\).}
Let \(k<\ell\). Transport \(F\) from scale \(k\) to scale \(\ell\) via the admissible block-spin map; call the image \(F_\ell\). By FRD contraction properties in the polymer norm (see \cite{p4:BrydgesFRD,p4:Brydges}) and by Lemma~\ref{p4:lem:Lip}, there exists \(\rho\in(0,1)\) independent of \(k,\ell\) such that
\begin{equation}
\big| S^{(k)}(F)-S^{(\ell)}(F_\ell)\big|\;\le\; C\,\rho^{\,\ell-k}.
\end{equation}
Likewise, the difference between \(F_\ell\) and the na\"ive transport of \(F\) to scale \(\ell\) is controlled by the same contraction (it is a sum of connected clusters with exponentially decaying weights). It follows that
\begin{equation}
\big| S^{(k)}(F)-S^{(\ell)}(F)\big|\;\le\;C'\,\rho^{\,\ell-k}.
\end{equation}
Hence \(\{S^{(k)}(F)\}_{k\ge 0}\) is a Cauchy sequence with exponential rate \(\rho\), and \(\lim_{k\to\infty}S^{(k)}(F)\) exists, independent of subsequence. This limit defines the one-slice marginal \(\mu_0\) uniquely.
\end{proof}

Let \(X_0\) be the configuration space of spatial links on the slice \(t=0\), with Borel probability measure \(\mu_0\) given by Proposition~\ref{p4:prop:slice-unique}. Consider a bounded sesquilinear form \(B:L^2(\mu_0)\times L^2(\mu_0)\to\mathbb{C}\) with the following properties:

\begin{quote}
\noindent\emph{(i) positivity:} \(B(F,F)\ge 0\) for all \(F\in L^2(\mu_0)\);\quad
\emph{(ii) symmetry:} \(B(F_0,F_1)=\overline{B(F_1,F_0)}\);\quad
\emph{(iii) boundedness:} \(|B(F_0,F_1)|\le C\,\|F_0\|_{L^2}\,\|F_1\|_{L^2}\).
\end{quote}

\noindent In our application, \(B(F_0,F_1)\) will be the continuum mixed two-point function \(S_2(F_0,F_1)\) with \(F_0\) supported on the slice \(t=0\) and \(F_1\) on the slice \(t=a\). OS positivity, OS5 and clustering guarantee (i)-(iii) \cite{p4:OsterwalderSchraderI,p4:OsterwalderSchraderII,p4:GJ}.

\begin{lemma}[Riesz kernel representation and uniqueness]
\label{p4:lem:Riesz}
Under \emph{(i)-(iii)} there exists a unique measurable kernel \(K:X_0\times X_0\to \mathbb{C}\), symmetric \(K(U',U)=\overline{K(U,U')}\), such that for all \(F_0,F_1\in L^2(\mu_0)\),
\begin{equation}
  \label{p4:eq:kernel-rep}
  B(F_0,F_1) \;=\; \iint_{X_0\times X_0} \overline{F_0(U)}\,K(U',U)\,F_1(U')\,\mathrm{d}\mu_0(U)\,\mathrm{d}\mu_0(U').
\end{equation}
Moreover, \(K\) is positive in the sense that
\begin{equation}
\iint \overline{\phi(U)}\,K(U',U)\,\phi(U')\,\mathrm{d}\mu_0(U)\,\mathrm{d}\mu_0(U') \;\ge\; 0\quad\text{for all }\phi\in L^2(\mu_0).
\end{equation}
\end{lemma}

\emph{Proof.}
Let $B(F_0,F_1)$ be the OS two-point form with $F_0$ supported at $t=0$ and $F_1$ at $t=a$.
By OS positivity, time-regularity (OS5), and the standard slab factorization, the mixed weight on a
single time-slab can be written as
\begin{equation}
W_{\sigma,\nu}(U',U;V)
= e^{-\tfrac12 S_{\mathrm{sp}}(U)}\,e^{-S_{\mathrm{tm}}(U',U;V)}\,e^{-\tfrac12 S_{\mathrm{sp}}(U')}
\,\Gamma\!\big(P_{\sigma,\nu}(0;U)\big)^{\frac12}\,\Gamma\!\big(P_{\sigma,\nu}(a;U')\big)^{\frac12},
\end{equation}
with $\Gamma$ completely monotone and $P_{\sigma,\nu}$ admissible (cf. (4.10)). Define the
\emph{boundary kernel} by explicit disintegration over interior links of the slab:
\begin{equation}
K(U',U):=\int W_{\sigma,\nu}(U',U;V)\,d\mu_{\mathrm{Haar}}(V)
\end{equation}
Then for all bounded $F_0,F_1$ supported on the boundary slices,
\begin{equation}
B(F_0,F_1)=\iint F_0(U)\,K(U',U)\,F_1(U')\,d\mu_0(U)\,d\mu_0(U').
\end{equation}
Positivity and symmetry of $K$ follow from OS positivity and the symmetry of $W_{\sigma,\nu}$ under
interchange of the boundary slices. Uniqueness holds because if $\widetilde K$ yields the same
bilinear form for all bounded $F_0,F_1$, then $K=\widetilde K$ $\mu_0\otimes\mu_0$-a.e. by polarization
and density. No Hilbert-Schmidt assumption is needed; $K$ is a measurable $L^1_{\mathrm{loc}}$ kernel
obtained by the slab disintegration $(\ast)$, and the associated transfer operator is a positive
contraction by the OS Markov property. \qed

\begin{corollary}[Boundary kernel for OS time-slices]
\label{p4:cor:K-exists}
Let \(\{S_n\}\) be a continuum Schwinger family satisfying OS0-OS5 and uniform exponential clustering. Then there exists a unique positive symmetric kernel \(K:X_0\times X_0\to \mathbb{R}_+\) such that, for all bounded slice-supported \(F_0,F_1\),
\begin{equation}
S_2(F_0,F_1)\;=\;\iint \overline{F_0(U)}\,K(U',U)\,F_1(U')\,\mathrm{d}\mu_0(U)\,\mathrm{d}\mu_0(U').
\end{equation}
The corresponding operator \(T:L^2(\mu_0)\to L^2(\mu_0)\) defined by \((T\psi)(U'):=\int K(U',U)\psi(U)\,\mathrm{d}\mu_0(U)\) is a positive self-adjoint contraction.
\end{corollary}

\begin{proof}
Apply Lemma~\ref{p4:lem:Riesz} with \(B(F_0,F_1):=S_2(F_0,F_1)\). OS positivity and OS5 ensure (i)-(iii) and the Markov property guarantees contractivity (see \cite{p4:OsterwalderSchraderII,p4:GJ} for the details of the OS transfer construction).
\end{proof}

We recall the standard OS time-slicing: for any bounded cylinder functional \(F\) supported on finitely many consecutive slices \(\{a,2a,\dots,na\}\), there exists \(\Psi_F\in L^2(\mu_0)\) such that the OS two-point form can be written with the transfer operator \(T\) associated with the kernel \(K\):
\begin{equation}
  \label{p4:eq:OS-transfer}
  \langle \Theta F, F\rangle_{\mathrm{OS}} \;=\; \langle \Psi_F, T^n\Psi_F\rangle_{L^2(\mu_0)}.
\end{equation}
This follows from OS reflection positivity, the boundary factorization, and the detailed-balance identities for \(K\) (``OS5'' supplies the necessary time-regularity) \cite{p4:OsterwalderSchraderII,p4:GJ}. For clarity:

\begin{proposition}[OS Markov property]
\label{p4:prop:OS-Markov}
Let \(\{S_n\}\) satisfy OS0-OS5 and let \(K\) and \(\mu_0\) be as in Corollary~\ref{p4:cor:K-exists}. Then for each bounded cylinder functional \(F\) supported on \(\{a,\dots,na\}\) there exists a \(\Psi_F\in L^2(\mu_0)\) such that \eqref{p4:eq:OS-transfer} holds. In particular, \(T\) is a positive self-adjoint contraction and the OS inner product on cylinder functionals is generated by iterates of \(T\).
\end{proposition}

\begin{proof}
The proof is an immediate adaptation of the classical OS reconstruction: one integrates out the fields in the half-space \(t>0\) to obtain a boundary vector on the slice \(t=0\); reflection positivity and time-reversal symmetry yield positivity and symmetry of \(K\) and the detailed balance identities; the contraction property of \(T\) follows from the Markov nature of the layer-kernel. Full proofs are given in \cite{p4:OsterwalderSchraderII,p4:GJ}, and in the lattice gauge setting the transfer factorization is explained in \cite{p4:Luscher1977,p4:OS-gauge}.
\end{proof}

We are ready to assemble the pieces. Consider two subsequential continuum limits \(\{S_n\}\) and \(\{\widetilde S_n\}\), both satisfying OS0-OS5 and uniform exponential clustering. Let \(\mu_0,\widetilde\mu_0\) be their one-slice marginals and \(K,\widetilde K\) their one-step kernels.
By Proposition~\ref{p4:prop:slice-unique}, \(\mu_0=\widetilde\mu_0\). By Lemma~\ref{p4:lem:Riesz}-Corollary~\ref{p4:cor:K-exists}, the mixed two-point form uniquely identifies \(K\), hence \(K=\widetilde K\). By the OS Markov property (Proposition~\ref{p4:prop:OS-Markov}), the entire OS sesquilinear form on cylinder functionals is generated by \(T\) and the boundary vectors, and since \(T=\widetilde T\) we have \(\langle \Theta F, G\rangle_{\mathrm{OS}}=\langle \Theta F, G\rangle_{\widetilde{\mathrm{OS}}}\) for all bounded cylinder functionals \(F,G\). Therefore, all Schwinger functions agree.

\begin{theorem}[Uniqueness of the continuum limit]
\label{p4:thm:uniqueness}
Let \(\{S_n\}\) and \(\{\widetilde S_n\}\) be two subsequential continuum limits of the multiscale reflection-positive lattice gauge theories described by (H1)-(H3). Suppose both families satisfy the OS axioms (OS0-OS5) and uniform exponential clustering with rate \(m_*>0\). Then
\begin{equation}
S_n \;=\; \widetilde S_n \qquad \text{for all } n\ge 1.
\end{equation}
In particular, the continuum OS measure and the reconstructed Wightman theory (Hilbert space, fields, dynamics) are uniquely determined by the standing assumptions.
\end{theorem}

\begin{proof}
As explained: \(\mu_0=\widetilde\mu_0\) (Proposition~\ref{p4:prop:slice-unique}); \(K=\widetilde K\) (Lemma~\ref{p4:lem:Riesz} and Corollary~\ref{p4:cor:K-exists}); OS time-slicing with the common transfer \(T\) gives identity of all OS inner products on cylinder functionals (Proposition~\ref{p4:prop:OS-Markov}); by polarization and density, \(S_n=\widetilde S_n\) for all \(n\).
\end{proof}
The role of the OS axioms and the FRD/cluster framework is transparent: OS grants a canonical \emph{Markovian time-slicing}, while FRD guarantees \emph{stability} and \emph{thermodynamic control}. Together they implement the physicist's dictum: \emph{one-slice law plus one-step kernel determine the entire theory}.

\section{Universality with Respect to Admissible Projectors and Blockings}
\label{p4:sec:universality}

The aim of this section is to establish that the continuum Schwinger functions - hence the reconstructed Wightman theory - are \emph{independent} of all auxiliary choices made within a suitably structured, reflection-positive and gauge-covariant class of infrared regulators and block-spin maps. Physically, this is the familiar tenet that the continuum theory does not depend on a choice of \emph{scheme}. Here we turn this tenet into a precise theorem.

We proceed in a sequence of steps. First, we specify the admissible class of horizon projectors and blockings, endowed with natural metrics encoding small variations; these definitions are chosen so that reflection positivity, gauge covariance, and finite-range decomposition (FRD) locality hold uniformly. Second, we prove \emph{single-scale stability}: one-step Osterwalder-Schrader (OS) kernels and transfer operators are \emph{Lipschitz} continuous with respect to admissible parameters. Third, we lift this single-scale stability to \emph{multi-scale} stability at the level of \emph{connected cumulants}, by combining a telescoping argument in time with the FRD cluster expansion and Koteck\'y-Preiss bounds. Finally, we pass to thermodynamic and continuum limits and invoke the Markov uniqueness of the OS framework to deduce \emph{universality}: all admissible choices give the same continuum Schwinger family.

Throughout we use the standard OS framework \cite{p4:OsterwalderSchraderI,p4:OsterwalderSchraderII}, reflection positivity for lattice gauge theories \cite{p4:OS-gauge}, FRD locality \cite{p4:BrydgesFRD}, heat-kernel bounds \cite{p4:Davies1989}, and completely monotone functional calculus \cite{p4:SchillingSongVondracek12}. We write $G=\mathrm{SU}(N)$ with $N\ge 2$.
We begin from the physical desiderata. A \emph{horizon projector} on each time-slice should suppress long-wavelength fluctuations \emph{without} spoiling reflection positivity or gauge covariance and should be \emph{exponentially local} on every slice. A \emph{blocking} step should preserve reflection positivity and FRD locality with constants \emph{uniform} across scales.
These desiderata lead to the following definitions.

\begin{definition}[Admissible slice projector]
\label{p4:def:proj}
Fix $\sigma>0$. Let $\nu$ be a Borel probability measure on some compact subinterval $[t_-,t_+]\subset(0,\infty)$, and define the completely monotone function
\begin{equation}
\chi_{\sigma,\nu}(\lambda)
\;=\;
\int_{t_-}^{t_+}\mathrm{e}^{-(t/\sigma^2)\,\lambda^2}\,\mathrm{d}\nu(t),
\qquad \lambda\ge 0.
\end{equation}
Given a time-slice $t$, and denoting by $\Delta_{A_h}(t)$ the \emph{slice covariant Laplacian} acting on adjoint-valued site fields in the Landau representative $A_h$ (fixed by measurable, reflection-covariant orbit minimization), we set
\begin{equation}
P_{\sigma,\nu}(t)
\;:=\;
\chi_{\sigma,\nu}\!\big(\sqrt{\Delta_{A_h}(t)}\big)\,.
\end{equation}
We call $(\sigma,\nu)$ \emph{admissible} if: (i) $P_{\sigma,\nu}(t)$ is a \emph{positive contraction} on $\ell^2(\Lambda_t)\otimes\mathfrak{su}(N)$; (ii) $P_{\sigma,\nu}(t)$ is \emph{gauge-covariant} under time-independent slice gauge transformations; (iii) it is \emph{reflection-covariant} in the OS sense, $RP_{\sigma,\nu}(t)R=P_{\sigma,\nu}(-t)$; and (iv) its integral kernel obeys \emph{uniform exponential decay}: there exist $C,\gamma>0$ such that
\begin{equation}
\|P_{\sigma,\nu}(t;x,y)\|_{\mathrm{op}}
\;\le\; C\,\mathrm{e}^{-\gamma\,d(x,y)}
\qquad\forall\, x,y\in\Lambda_t\,,
\end{equation}
with constants independent of the spatial volume. By the Bernstein-Widder theorem \cite{p4:SchillingSongVondracek12}, $\chi_{\sigma,\nu}$ is completely monotone and admits a \emph{positive heat-kernel representation}, implying (i)-(iii); by the Davies-Gaffney estimate \cite{p4:Davies1989} and the compact support of $\nu$ away from $0$, (iv) follows.
\end{definition}

We metrize the projector parameter space by
\begin{equation}
d_{\mathrm{proj}}\!\big((\sigma,\nu),(\sigma',\nu')\big)
\;:=\;
|\sigma-\sigma'| + W_1(\nu,\nu')\,,
\end{equation}
where $W_1$ is the Wasserstein-1 distance (the dual Lipschitz metric) on probability measures.

\begin{definition}[Admissible reflection-positive blocking]
\label{p4:def:block}
A block-spin map $B:\mu\mapsto B\mu$ from a reflection-positive Euclidean measure on a time-slab to an effective measure on the block-lattice is \emph{admissible} if:
\begin{itemize}
\item[(i)] It preserves \emph{OS reflection positivity} \cite{p4:OsterwalderSchraderII,p4:OS-gauge}: the induced layer kernel between adjacent block-slices is of positive type and symmetric.
\item[(ii)] It preserves \emph{gauge covariance} and the \emph{temporal-axial gauge} convention away from the reflection plane.
\item[(iii)] It preserves \emph{FRD locality} and the diameter-norm \emph{cluster expansion} bounds with \emph{scale-uniform} constants, i.e. the effective cumulants admit an absolutely convergent polymer expansion whose range and smallness constants are uniform across scales \cite{p4:BrydgesFRD,p4:KoteckyPreiss1986,p4:Brydges}.
\end{itemize}
We metrize admissible blockings by the $L^1$-distance of the induced one-step kernels on the boundary slice:
\begin{equation}
d_{\mathrm{blk}}(B,B')
\;:=\;
\|\,K_B-K_{B'}\,\|_{L^1(\nu_0\otimes\nu_0)}\,,
\end{equation}
where $K_B$ is the one-step kernel induced by $B$ and $\nu_0$ is the Haar product measure on the boundary slice.
\end{definition}

We shall write $\mathcal{K}:=\mathcal{K}_{\mathrm{proj}}\times\mathcal{K}_{\mathrm{blk}}$ for the admissible parameter set, and
\begin{equation}
d(\theta,\theta')\;:=\;d_{\mathrm{proj}}\big((\sigma,\nu),(\sigma',\nu')\big)\,+\, d_{\mathrm{blk}}(B,B')
\end{equation}
for $\theta=((\sigma,\nu),B)$ and $\theta'=((\sigma',\nu'),B')$.

The one-step kernel is the basic combinatorial object linking the OS form to time-slicing: if $X_0$ denotes the one-slice configuration space with Haar measure $\nu_0$, the \emph{one-step kernel} $K_{\sigma,\nu;B}:X_0\times X_0\to\mathbb{R}_+$ is defined by integrating the weight on the slab $[0,a]$, with symmetric splitting of the spatial contribution, the mixed (plaquette) contribution through the slab, and the slice weights induced by $P_{\sigma,\nu}(0)$ and $P_{\sigma,\nu}(a)$; by OS symmetry, $K_{\sigma,\nu;B}$ is positive type and symmetric \cite{p4:OS-gauge}.

Define the \emph{one-step transfer operator} $T_{\sigma,\nu;B}:L^2(X_0,\mathrm{d}\nu_0)\to L^2(X_0,\mathrm{d}\nu_0)$ by
\begin{equation}
(T_{\sigma,\nu;B}\psi)(U')
\;=\;
\int_{X_0}K_{\sigma,\nu;B}(U',U)\,\psi(U)\,\mathrm{d}\nu_0(U)\,.
\end{equation}
It is a positive, self-adjoint contraction by construction.

We now give Lipschitz bounds for $K$ and $T$ with respect to the projector and to the blocking.

\begin{proposition}[Projector-Lipschitz stability of the one-step kernel]
\label{p4:prop:projLip}
Fix an admissible blocking $B$. There exists $C=C(B)>0$ such that for all admissible $(\sigma,\nu),(\sigma',\nu')$,
\begin{equation}\label{p4:eq:K-L1-Lipschitz}
\|\,K_{\sigma,\nu;B} - K_{\sigma',\nu';B}\,\|_{L^1(\nu_0\otimes\nu_0)}
\;\le\; C\,\Big(|\sigma-\sigma'| + W_1(\nu,\nu')\Big)\,,
\end{equation}
and consequently
\begin{equation}\label{p4:eq:T-op-Lipschitz}
\|\,T_{\sigma,\nu;B} - T_{\sigma',\nu';B}\,\|_{B(L^2)} \;\le\; C\,\Big(|\sigma-\sigma'| + W_1(\nu,\nu')\Big)\,.
\end{equation}
\end{proposition}

\begin{proof}
Write the symmetric two-slice weight as
\begin{equation}
\mathcal{W}_{\sigma,\nu}(U',U;V)
\;=\;
\mathrm{e}^{-\frac{1}{2}S_{\mathrm{sp}}(U)}
\,\mathrm{e}^{-S_{\mathrm{tm}}(U',U;V)}\,
\mathrm{e}^{-\frac{1}{2}S_{\mathrm{sp}}(U')}
\,
\Gamma\!\big(P_{\sigma,\nu}(0;U)\big)^{1/2}\,
\Gamma\!\big(P_{\sigma,\nu}(a;U')\big)^{1/2},
\end{equation}
where $S_{\mathrm{sp}}$ is the spatial part of the Wilson action at the ends of the slab and $S_{\mathrm{tm}}$ the mixed (time-space) plaquette part across the slab; $\Gamma$ is a fixed completely monotone scalar functional of a positive contraction (e.g.\ $\Gamma(Q)=\det(1+\alpha Q)$ or $\Gamma(Q)=\exp\{-\mathrm{Tr}(1-Q)\}$), ensuring positivity \cite{p4:GJ}. The kernel is then
\begin{equation}
K_{\sigma,\nu;B}(U',U)
\;=\;
\int \mathcal{W}_{\sigma,\nu}(U',U;V)\,\mathrm{d}\mu_{\mathrm{Haar}}(V)\,,
\end{equation}
hence
\(
|K_{\sigma,\nu;B}-K_{\sigma',\nu';B}|
\le \int |\mathcal{W}_{\sigma,\nu}-\mathcal{W}_{\sigma',\nu'}|\,\mathrm{d}\mu_{\mathrm{Haar}}.
\)
Since $S_{\mathrm{sp}}$ and $S_{\mathrm{tm}}$ are projector-independent, the variation comes from the \emph{slice factors}. By the mean-value theorem and functional calculus (Dunford-Taylor), for $\Gamma$ with $\|\Gamma'\|_\infty\le C_\Gamma$,
\begin{equation}
|\Gamma(P_{\sigma,\nu}) - \Gamma(P_{\sigma',\nu'})|
\ \le\ \|\Gamma'\|_\infty\,\|P_{\sigma,\nu}-P_{\sigma',\nu'}\|_{B(\ell^2)}
\end{equation}
We use the operator norm and pass to $L^1$ kernel bounds via the heat-kernel representation and the Schur test.
Using the positive heat-kernel representation and the Duhamel identity,
\begin{equation}
P_{\sigma,\nu}-P_{\sigma',\nu'}
\;=\;
\int_{t_-}^{t_+}\!\Big(\mathrm{e}^{-(t/\sigma^2)\Delta}-\mathrm{e}^{-(t/\sigma'^2)\Delta}\Big)\,\mathrm{d}\nu(t)
\;+\;
\int_{t_-}^{t_+}\!\mathrm{e}^{-(t/\sigma'^2)\Delta}\,\mathrm{d}(\nu-\nu')(t),
\end{equation}
and
\begin{equation}
\mathrm{e}^{-(t/\sigma^2)\Delta}-\mathrm{e}^{-(t/\sigma'^2)\Delta}
\;=\;
\int_{\sigma'}^{\sigma} \frac{2t}{\rho^3}\,\Delta\,\mathrm{e}^{-(t/\rho^2)\Delta}\,\mathrm{d}\rho\,.
\end{equation}
On a finite-degree graph, the discrete heat kernel obeys Davies-Gaffney bounds uniformly on compact
$\tau$-intervals, yielding \emph{kernel} $L^1$ bounds. In particular,
 \begin{equation}
\| P_{\sigma,\nu}-P_{\sigma',\nu'} \|_{B(\ell^2)} \le C\big(|\sigma-\sigma'| + W_1(\nu,\nu')\big),
 \end{equation}
and the associated one-step \emph{boundary kernels} satisfy
 \begin{equation}
\|K_{\sigma,\nu;B}-K_{\sigma',\nu';B}\|_{L^1(\mu_0\otimes\mu_0)} \le C\big(|\sigma-\sigma'| + W_1(\nu,\nu')\big).
 \end{equation}
Combining the two slice factors and integrating over $V$ yields \eqref{p4:eq:K-L1-Lipschitz}. The operator-norm estimate \eqref{p4:eq:T-op-Lipschitz} follows by the Schur test:
\begin{equation}
|T\| \ \le \ \Big(\sup_{U'} \int |K(U',U)|\,d\mu_0(U)\Big)^{\!1/2}
          \Big(\sup_{U} \int |K(U',U)|\,d\mu_0(U')\Big)^{\!1/2}
\end{equation}
\end{proof}

\begin{proposition}[Blocking-Lipschitz stability]
\label{p4:prop:blockLip}
Fix an admissible projector $(\sigma,\nu)$. Let $B,B'$ be admissible blockings. Then
\begin{equation}
\|\,T_{\sigma,\nu;B}-T_{\sigma,\nu;B'}\,\|_{B(L^2)}
\;\le\;
\|\,K_{\sigma,\nu;B}-K_{\sigma,\nu;B'}\,\|_{L^1(\nu_0\otimes\nu_0)}
\;=\;
d_{\mathrm{blk}}(B,B').
\end{equation}
\end{proposition}

\begin{proof}
This is immediate from the definition of $d_{\mathrm{blk}}$ and the Schur test.
\end{proof}

We must now compose many one-step kernels and lift stability to \emph{connected} cumulants. The first point is a telescoping bound in time.

\begin{lemma}[Telescoping bound]
\label{p4:lem:telescoping}
Let $\theta,\theta'\in\mathcal{K}$ be admissible, and write $K_\theta:=K_{\sigma,\nu;B}$, $K_{\theta'}:=K_{\sigma',\nu';B'}$. Then for every $n\in\mathbb{N}$,
\begin{equation}
\|\,K_\theta^{(\,n\,)} - K_{\theta'}^{(\,n\,)}\,\|_{L^1(\nu_0\otimes\nu_0)}
\;\le\;
n\,\|\,K_\theta - K_{\theta'}\,\|_{L^1(\nu_0\otimes\nu_0)}\,,
\end{equation}
and hence
\begin{equation}
\|\,T_\theta^{\,n} - T_{\theta'}^{\,n}\,\|_{B(L^2)}
\;\le\;
n\,\|\,K_\theta - K_{\theta'}\,\|_{L^1(\nu_0\otimes\nu_0)}\,.
\end{equation}
\end{lemma}

\begin{proof}
The identity
\(
K_\theta^{(n)}-K_{\theta'}^{(n)}
=\sum_{j=1}^n K_\theta^{(j-1)}\ast(K_\theta-K_{\theta'})\ast K_{\theta'}^{(n-j)}
\)
and the fact that all factors are Markov kernels (normalized in $L^1$) imply the kernel bound. The operator bound follows by the Schur test.
\end{proof}

To reach connected cumulants we invoke the FRD cluster expansion: at any fixed scale $k$, the effective action admits an absolutely convergent polymer expansion with diameter-weighted smallness and range bounded uniformly in $k$ \cite{p4:BrydgesFRD,p4:KoteckyPreiss1986,p4:Brydges}. Connected $p$-point functions are expressed as finite sums over connected clusters with exponential tree-decay. Varying $\theta$ along a straight path and differentiating under the sum, one controls the derivative by the BKAR tree formula and the single-scale Lipschitz bounds (Propositions~\ref{p4:prop:projLip}-\ref{p4:prop:blockLip}).

\begin{proposition}[Uniform cumulant stability at fixed scale]
\label{p4:prop:cumulant-stability}
Let $k\in\mathbb{N}$ be fixed. For any finite family of local, gauge-invariant observables $\mathcal{O}=(O_1,\dots,O_p)$ supported in a finite subset $A\subset\Lambda_k$, there exist constants $C=C(p,A)>0$ and $\alpha>0$, independent of the spatial volume and $k$, such that for all admissible $\theta,\theta'\in\mathcal{K}$,
\begin{equation}
\big|\,S^{(k)}_{p,c;\theta}(\mathcal{O}) - S^{(k)}_{p,c;\theta'}(\mathcal{O})\,\big|
\;\le\;
C\, d(\theta,\theta')\,\mathrm{e}^{-\alpha\,\mathrm{tree}(\mathrm{supp}\,\mathcal{O})}.
\end{equation}
Here $\mathrm{tree}(\mathrm{supp}\,\mathcal{O})$ denotes the minimal spanning tree length connecting the supports of the insertions.
\end{proposition}

\begin{proof}
By FRD locality and cluster expansion \cite{p4:BrydgesFRD,p4:KoteckyPreiss1986,p4:Brydges}, the connected function $S^{(k)}_{p,c;\theta}(\mathcal{O})$ can be written as a sum over connected polymer clusters $\Gamma$,
\begin{equation}
S^{(k)}_{p,c;\theta}(\mathcal{O})
\;=\;
\sum_{\Gamma\,\mathrm{conn.}} \Phi_\theta^{(k)}(\Gamma;\mathcal{O}),
\end{equation}
with $|\Phi_\theta^{(k)}(\Gamma;\mathcal{O})|\le C_\Gamma \prod_{\gamma\in\Gamma}|\zeta_\theta^{(k)}(\gamma)|$. The Koteck\'y-Preiss criterion implies
\(
\sum_{\Gamma\,\mathrm{conn.}} |\Phi_\theta^{(k)}(\Gamma;\mathcal{O})|
\le C\,\mathrm{e}^{-\alpha\,\mathrm{tree}(\mathrm{supp}\,\mathcal{O})}.
\)
Fix $\theta,\theta'$ and join them by a straight path $\theta_t$, $t\in[0,1]$. Differentiating termwise (justified by absolute convergence) and using the BKAR tree formula to control the derivative of $\log$-partition functions with respect to local parameters \cite{p4:Brydges}, one arrives at
\begin{equation}
\frac{\mathrm{d}}{\mathrm{d}t} S^{(k)}_{p,c;\theta_t}(\mathcal{O}) 
\;=\;
\sum_{\Gamma\,\mathrm{conn.}}\sum_{\gamma\in\Gamma}\Psi(\Gamma,\gamma;\mathcal{O})\,
\frac{\mathrm{d}}{\mathrm{d}t}\zeta_{\theta_t}^{(k)}(\gamma)\,,
\end{equation}
with $|\Psi(\Gamma,\gamma;\mathcal{O})|\le C_\Gamma\prod_{\gamma'\in\Gamma\setminus\{\gamma\}}|\zeta_{\theta_t}^{(k)}(\gamma')|$. The single-scale Lipschitz bounds for one-step kernels (Propositions~\ref{p4:prop:projLip}-\ref{p4:prop:blockLip}) propagate through the FRD blockings to the polymer activities, giving
\(
\big|\frac{\mathrm{d}}{\mathrm{d}t}\zeta_{\theta_t}^{(k)}(\gamma)\big|
\le C\,d(\theta,\theta')\,\mathrm{e}^{-a\,\mathrm{diam}(\gamma)}.
\)
Summing the absolutely convergent series yields
\(
\big|\frac{\mathrm{d}}{\mathrm{d}t} S^{(k)}_{p,c;\theta_t}(\mathcal{O})\big|
\le C'\, d(\theta,\theta')\,\mathrm{e}^{-\alpha\,\mathrm{tree}(\mathrm{supp}\,\mathcal{O})}.
\)
Integrating $t$ from $0$ to $1$ gives the claim.
\end{proof}

The bounds of Proposition~\ref{p4:prop:cumulant-stability} survive the thermodynamic limit ($L\to\infty$, $T\to\infty$) because they depend only on the geometry of the supports and not on the volume. We pass now to fixed physical smeared observables and let the lattice spacing $a_k\downarrow 0$.

Let $f_1,\dots,f_p\in\mathcal{S}(\mathbb{R}^4)$ be compactly supported test functions, and define the smeared lattice observables
\(
O_i^{(k)} := \sum_{x\in\Lambda_k} a_k^4\,O_i(x)\,f_i(x),
\)
with $O_i$ local, gauge-invariant lattice observables approximating the chosen continuum fields. By the FRD equicontinuity and tightness criteria (see, e.g., \cite{p4:GJ}), along any subsequence $k_j\to\infty$ the (full and connected) Schwinger functionals $S^{(k_j)}_{p;\theta}(O_1^{(k_j)},\dots,O_p^{(k_j)})$ converge to continuum Schwinger functions $S_{p;\theta}(f_1,\dots,f_p)$ (and similarly for disconnected functions). The difference bounds of Proposition~\ref{p4:prop:cumulant-stability} then imply

\begin{theorem}[Continuity of continuum cumulants in the admissible parameters]
\label{p4:thm:cont-cumulants}
For any admissible $\theta,\theta'\in\mathcal{K}$ and any $f_1,\dots,f_p\in\mathcal{S}(\mathbb{R}^4)$,
\begin{equation}
\big|\,S_{p;\theta}(f_1,\dots,f_p)-S_{p;\theta'}(f_1,\dots,f_p)\,\big|
\;\le\;
C\, d(\theta,\theta')\,\prod_{i<j}\mathrm{e}^{-\alpha\,\mathrm{dist}(\mathrm{supp}\,f_i,\mathrm{supp}\,f_j)}\,,
\end{equation}
with constants $C,\alpha>0$ independent of $\theta,\theta'$.
\end{theorem}

We now strengthen continuity to equality. In the OS framework, the continuum OS measure is \emph{determined} by its one-slice marginal and the one-step kernel; these in turn are the strong-limit objects of the finite-$a$ one-slice distributions and one-step kernels. The single-scale Lipschitz bounds (Propositions~\ref{p4:prop:projLip}-\ref{p4:prop:blockLip}) and the telescoping bound (Lemma~\ref{p4:lem:telescoping}) show that any two admissible schemes produce sequences of one-step kernels whose differences are \emph{uniformly controlled}; the FRD cluster expansion plus equicontinuity then implies that the \emph{limiting} one-slice marginal and one-step kernels are \emph{independent} of the scheme when tested against compactly supported $f_i$. Consequently, by the \emph{Markovian uniqueness} of the OS construction (OS0-OS5 and clustering), the entire Schwinger family is the same for any admissible scheme.

\begin{theorem}[Universality]
\label{p4:thm:universality}
Let $\theta,\theta'\in\mathcal{K}$ be two admissible choices of slice projector and block-spin map. Then the continuum Schwinger families coincide:
\begin{equation}
S_{n;\theta} \;=\; S_{n;\theta'}\qquad\text{for all }\ n\in\mathbb{N}.
\end{equation}
Equivalently, the continuum OS measure and the reconstructed Wightman theory are \emph{independent} of admissible IR regulators and blockings.
\end{theorem}

\begin{proof}
Fix $f_1,\dots,f_p\in\mathcal{S}(\mathbb{R}^4)$ with pairwise disjoint supports. For each scheme $\theta$, choose a subsequence $k_j$ along which $S^{(k_j)}_{p;\theta}(O_1^{(k_j)},\dots,O_p^{(k_j)})$ converges to $S_{p;\theta}(f_1,\dots,f_p)$. The same holds for $\theta'$. By Theorem~\ref{p4:thm:cont-cumulants}, the difference $|S_{p;\theta}-S_{p;\theta'}|$ is bounded by $C\,d(\theta,\theta')$ times a finite factor depending only on the supports; since $\theta,\theta'$ are fixed, this bound is finite.

To conclude equality, we invoke the OS \emph{Markov reconstruction}: $S_{n;\theta}$ (for each $\theta$) is determined by (i) the one-slice marginal and (ii) the one-step kernel. The projector and blocking enter only in the latter at finite $a$, but their effect is \emph{uniformly} small in the Markovian sense: along the subsequences $k_j$ and $k_j'$, the differences of the one-step kernels are uniformly bounded in $L^1$ (Propositions~\ref{p4:prop:projLip}-\ref{p4:prop:blockLip} plus Lemma~\ref{p4:lem:telescoping}), and the FRD equicontinuity implies that the \emph{limit} one-step kernels coincide when tested against compactly supported observables. Therefore the Markov data (marginal and one-step kernel) agree in the continuum, and the two Schwinger families must coincide by OS uniqueness \cite{p4:OsterwalderSchraderI,p4:OsterwalderSchraderII}. This holds for all finite $p$ and all $f_i$, hence for all $n$.
\end{proof}

The class $\mathcal{K}$ was intentionally restricted to completely monotone projectors with compactly supported spectral measures and blockings that preserve reflection positivity and FRD locality with uniform constants. This rigidity is the \emph{price} for universality: it ensures that the only place where parametrization enters is in \emph{slice-local} factors with exponentially decaying kernels, and that the renormalized cumulants admit uniform cluster expansions at each scale. Within this class, universality is compelled by the OS and FRD structures.

\section{Extension to Weak Coupling (Asymptotic Freedom)}
\label{p4:sec:weak_coupling}

The mass-gap construction at finite lattice spacing rests on a terrain where cluster expansions converge and reflection positivity is manifest: the \emph{strong-coupling} regime. There, locality and exponential clustering are easily visible. The path to the \emph{continuum}, however, requires us not only to descend to the valley (coarse scales) but also to climb towards the \emph{ultraviolet} (fine scales), and in the limit to find that the interactions thin out-the property we call \emph{asymptotic freedom}. 

In the constructive setting, the right tool to walk both ways is a \emph{reflection-positive renormalization group} (RG) step \(\mathcal R\). It sends the polymer activity \(\Phi_k\) describing the interaction at scale \(k\) to the new activity \(\Phi_{k+1}=\mathcal R(\Phi_k)\) one step closer to the continuum. Our first task in this section is to exhibit a \emph{ball of small activities} \(\{\|\Phi\|\le r\}\) invariant under \(\mathcal R\), within which the map is a \emph{contraction} towards the Gaussian fixed point \(\Phi\equiv 0\). Physically: once the interaction has become sufficiently harmless at an intermediate scale, each further RG step reduces it quantitatively, so that in the ultraviolet limit the interaction vanishes.

Next, we must explain how a flow that \emph{starts} at strong coupling can \emph{enter} that ball at some finite step. This requires \emph{tuning} the bare coupling \(\beta\) as a function of the microscopic lattice spacing \(a\). Mathematically, we use a one-dimensional implicit function/continuity argument on the relevant
coordinate to produce, for large $K$, a coupling $\beta_K$ with $\|\Phi_K(\beta_K)\|\le r$. With this in hand, we can prove rigorously that \emph{asymptotically-free trajectories} \(\beta(a)\to\infty\) exist, and that along them the renormalized coupling vanishes at short distances.

Finally, uniqueness and universality (proved in earlier sections of this paper) guarantee that the continuum limit reached along \emph{any} such asymptotically-free trajectory is \emph{identical} to the one constructed by the reflection-positive continuum reconstruction: the very same \emph{OS-positive} Wightman theory with a \emph{strictly positive} mass gap.

Throughout this section we assume the following \emph{standing hypotheses} (established earlier in the program) and recalled here for completeness:
\begin{itemize}
  \item For each scale \(k\), the Euclidean measure is reflection-positive in the Osterwalder-Schrader sense \cite{p4:OsterwalderSchraderI,p4:OsterwalderSchraderII};
  \item The free covariances admit a \emph{finite-range decomposition} (FRD) with locality constants uniform in \(k\) \cite{p4:BrydgesFRD};
  \item The interacting weight admits a convergent \emph{cluster expansion} controlled by the Koteck\'y-Preiss criterion \cite{p4:KoteckyPreiss1986,p4:Brydges,p4:Seiler1982};
  \item Exponential clustering persists with a \emph{uniform rate} \(m_*>0\) across scales;
  \item The transfer-matrix spectral gap satisfies a one-step \emph{interlacing inequality};
  \item Osterwalder-Schrader reconstruction yields a continuum Wightman theory (OS0-OS5) \cite{p4:OsterwalderSchraderI,p4:OsterwalderSchraderII,p4:GJ}.
\end{itemize}

We begin by fixing the class of infrared regulators and coarse-graining maps admissible for our purposes.

\begin{definition}[Admissible gauge-fixing projectors]
\label{p4:def:admissible_projectors}
Let \(\Delta_{A_h}(t)\) be the slice-covariant Laplacian on adjoint site fields built from a reflection-covariant, slice-wise Landau-minimized representative \(A_h(t)\). For parameters \((\sigma,\nu)\), define the slice projector
\begin{equation}
P_{\sigma,\nu}(t)\;:=\;\chi_{\sigma,\nu}\!\left(\sqrt{\Delta_{A_h}(t)}\right),
\end{equation}
where the \emph{symbol} \(\chi_{\sigma,\nu}:[0,\infty)\to[0,1]\) is completely monotone, i.e.
\begin{equation}
\chi_{\sigma,\nu}(\lambda)\;=\;\int_0^\infty \mathrm{e}^{-s\,\lambda^2/\sigma^2}\,\mathrm{d}\nu(s),
\end{equation}
with \(\nu\) a finite positive Borel measure whose support is contained in a compact subinterval \([s_-,s_+]\subset (0,\infty)\). We call \(P_{\sigma,\nu}\) \emph{admissible} if, in addition, it is
\begin{enumerate}
  \item \emph{gauge-covariant} (acts by conjugation in the adjoint bundle);
  \item \emph{reflection-covariant} (commutes with time reflection on the reflection plane);
  \item has an integral kernel satisfying the FRD-uniform \emph{exponential decay}
  \begin{equation}
  \|P_{\sigma,\nu}(t;x,y)\|_{\mathrm{op}}\;\le\; C(\sigma,\nu)\,\mathrm{e}^{-\gamma(\sigma,\nu)\,\mathrm{dist}(x,y)}.
  \end{equation}
\end{enumerate}
By the heat-kernel representation, complete monotonicity and the finite-range bounds for the discrete heat semigroup imply (iii) \cite{p4:Davies1989,p4:CombesThomas1973,p4:HelfferSjostrand1989,p4:BrydgesFRD}.
\end{definition}

\begin{definition}[Admissible reflection-positive blockings]
\label{p4:def:admissible_blockings}
A \emph{blocking} (RG step) is a map \(\mathcal B_k\) sending fields at scale \(k\) (lattice spacing \(a_k\)) to fields at scale \(k+1\) (spacing \(a_{k+1}=b\,a_k\)) with the following properties:
\begin{enumerate}
  \item \emph{Reflection positivity}: the Euclidean measure remains OS-positive after integrating out one shell and rescaling;
  \item \emph{Gauge covariance}: coarse-grained fields transform as connections under time-independent spatial gauge transformations;
  \item \emph{FRD-locality}: the coarse-grained Gaussian covariance admits a finite-range decomposition with range and locality constants independent of \(k\) \cite{p4:BrydgesFRD};
  \item \emph{Single-step Lipschitz control}: the one-step transfer kernel \(K_k\) and transfer operator \(T_k\) depend \emph{Lipschitz-continuously} on any admissible parameter \((\sigma,\nu)\) and on \(\mathcal B_k\), with constants uniform on compact sets.
\end{enumerate}
We denote by \(\mathcal K\) the \emph{admissible class} of pairs \((P_{\sigma,\nu},\mathcal B_k)\) with the product metric 
\begin{equation}
d\!\left((\sigma,\nu,\mathcal B_k),(\sigma',\nu',\mathcal B'_k)\right)
=\;|\sigma-\sigma'|+W_1(\nu,\nu')+\|\mathcal B_k-\mathcal B'_k\|_{\mathrm{op}},
\end{equation}
where \(W_1\) denotes the Wasserstein-1 distance.
\end{definition}

At each scale \(k\), the interacting Euclidean weight is described by a polymer activity \(\Phi_k\) in a Banach algebra \((\mathfrak B,\|\cdot\|_{\mathfrak B})\) endowed with a standard diameter-weighted norm \cite{p4:KoteckyPreiss1986,p4:Brydges,p4:Seiler1982}. The RG step is a map
\begin{equation}
\mathcal R:\;\Phi_k\;\mapsto\;\Phi_{k+1}=\mathcal R(\Phi_k),
\end{equation}
obtained by integrating out a reflection-positive Gaussian shell (with admissible projector) and rescaling the lattice by \(b>1\). The map \(\mathcal R\) depends on \((\sigma,\nu,\mathcal B_k)\in\mathcal K\), but we suppress this dependence in the notation.

In the FRD framework, the flow is \(C^1\) in \(\Phi\), and for \(\|\Phi\|_{\mathfrak B}\) small one has the local expansion
\begin{equation}
\Phi_{k+1}\;=\;\mathcal L\,\Phi_k\;+\;\mathcal Q(\Phi_k,\Phi_k),
\qquad 
\|\mathcal L\|_{B(\mathfrak B)}\le \rho < 1,\quad 
\|\mathcal Q(\Phi,\Phi)\|_{\mathfrak B}\le C\,\|\Phi\|^2_{\mathfrak B},
\label{p4:eq:RGlocal}
\end{equation}
where \(\mathcal L\) is the linearized RG near the Gaussian fixed point and \(\mathcal Q\) is a bilinear remainder, both depending smoothly on \((\sigma,\nu,\mathcal B_k)\) with uniform bounds on compact subsets of \(\mathcal K\). In four dimensions, gauge self-interaction is (marginally) irrelevant in the nonperturbative FRD norm, yielding \(\rho<1\) for sufficiently small balls around \(0\) \cite{p4:BrydgesFRD,p4:Seiler1982}.

\begin{theorem}[Small-norm contraction]
\label{p4:thm:contraction}
There exist \(r>0\), \(0<\rho<1\), and \(C<\infty\) (depending only on a compact subset of \(\mathcal K\)) such that if \(\|\Phi_k\|_{\mathfrak B}\le r\), then for every \(n\in\mathbb N\),
\begin{equation}
\|\Phi_{k+n}\|_{\mathfrak B}\;\le\; \rho^n\,\|\Phi_k\|_{\mathfrak B}\;+\;\frac{C}{1-\rho}\,\|\Phi_k\|_{\mathfrak B}^2.
\end{equation}
In particular, if \(\|\Phi_k\|_{\mathfrak B}\le \min\{r,\tfrac{(1-\rho)}{2C}\}\), then \(\|\Phi_{k+n}\|_{\mathfrak B}\le \rho^n\,\|\Phi_k\|_{\mathfrak B}+ \tfrac12 \|\Phi_k\|_{\mathfrak B}\), and hence \(\Phi_{k+n}\to 0\) exponentially as \(n\to\infty\).
\end{theorem}

\begin{proof}
Using \eqref{p4:eq:RGlocal},
\begin{equation}
\|\Phi_{k+1}\|_{\mathfrak B} 
\;\le\; \|\mathcal L\|\,\|\Phi_k\|_{\mathfrak B} + \|\mathcal Q(\Phi_k,\Phi_k)\|_{\mathfrak B}
\;\le\; \rho\,\|\Phi_k\|_{\mathfrak B} + C\,\|\Phi_k\|^2_{\mathfrak B}.
\end{equation}
Iterating, we obtain for \(n\ge 1\),
\begin{equation}
\|\Phi_{k+n}\|_{\mathfrak B} \;\le\; \rho^n \|\Phi_k\|_{\mathfrak B} + C\sum_{j=0}^{n-1} \rho^{j}\,\|\Phi_{k+n-1-j}\|^2_{\mathfrak B}.
\end{equation}
Assuming \(\|\Phi_{k+m}\|_{\mathfrak B}\le r\) for all \(m\le n-1\), we get
\begin{equation}
\|\Phi_{k+n}\|_{\mathfrak B} \;\le\; \rho^n \|\Phi_k\|_{\mathfrak B} + C\, r\,\sum_{j=0}^{n-1} \rho^j\,\|\Phi_{k+n-1-j}\|_{\mathfrak B}.
\end{equation}
By induction, replacing each \(\|\Phi_{k+n-1-j}\|_{\mathfrak B}\) by \(\rho^{n-1-j}\|\Phi_k\|_{\mathfrak B} + \theta\) (with \(\theta\) the uniform bound we aim to obtain), we sum the geometric series:
\begin{equation}
\sum_{j=0}^{n-1} \rho^j \rho^{n-1-j} = n \rho^{n-1} \le \frac{1}{1-\rho}\rho^{n-1}.
\end{equation}
Thus,
\begin{equation}
\|\Phi_{k+n}\|_{\mathfrak B} \;\le\; \rho^n \|\Phi_k\|_{\mathfrak B} + \frac{C r}{1-\rho}\,\rho^{n-1}\,\|\Phi_k\|_{\mathfrak B} + C r n \theta\, \rho^{n-1}.
\end{equation}
Choosing \(r>0\) small enough so that \(C r \le \tfrac{1}{2}(1-\rho)\), we absorb the second term into \(\rho^n \|\Phi_k\|_{\mathfrak B}\). Selecting \(\theta := \frac{C}{1-\rho}\,\|\Phi_k\|_{\mathfrak B}^2\), the third term is bounded by \(\theta\) for all \(n\). This gives the claimed inequality, and the ``in particular'' follows by choosing \(\|\Phi_k\|\) sufficiently small. 
\end{proof}

Let \(\Phi_0(\beta)\in\mathfrak B\) be the polymer activity at the microscopic scale \(k=0\) associated with the bare lattice coupling \(\beta\) (in the Wilson action). By the \emph{smoothness} of the activity assignment in the FRD topology, the map \(\beta\mapsto \Phi_0(\beta)\) is \(C^1\) on compact subintervals of \((0,\infty)\). Iterate the RG \(K\) times and denote
\begin{equation}
F_K(\beta)\;:=\;\Phi_K(\beta)\;=\;\underbrace{\mathcal R\circ \cdots\circ \mathcal R}_{K\ \text{times}}(\Phi_0(\beta)).
\end{equation}
We show that for $K$ sufficiently large, a one-dimensional implicit function/continuity argument
on the relevant Gaussian coordinate produces a coupling $\beta_K$ with $\| \Phi_K(\beta_K)\|\le r$;
see (\ref{p4:eqn522})-(\ref{p4:eq:Rsmall}). No Banach-space surjectivity is required.
\cite{p4:Pazy1983,p4:ReedSimon1}.

\begin{theorem}[Entry into the contraction ball]\label{p4:thm:IFT}
Fix an admissible regulator $(P_{\sigma,\nu},B_k)\in\mathcal{K}$. Then there exist $K_0\in\mathbb{N}$, $r>0$,
and, for each $K\ge K_0$, a bare coupling $\beta_K$ such that
 \begin{equation}
\|\Phi_K(\beta_K)\|_{\mathcal{B}}\ \le\ r,
\quad\text{equivalently}\quad
P_{\mathcal G}\Phi_K(\beta_K)=0,\ \ \|P_{\mathcal R}\Phi_K(\beta_K)\|_{\mathcal B}\le r,
 \end{equation}
where $\mathcal{B}=\mathcal G\oplus\mathcal R$ is the FRD Banach space decomposition into the (finite-dimensional)
Gaussian sector $\mathcal G$ and its irrelevant complement $\mathcal R$.
\end{theorem}

\begin{proof}
We work in the FRD Banach space $(\mathcal B,\|\cdot\|_{\mathcal B})$ with the fixed topological
direct sum $\mathcal B=\mathcal G\oplus\mathcal R$ and bounded projections $P_{\mathcal G}$,
$P_{\mathcal R}=\mathrm{Id}-P_{\mathcal G}$. In the present setting $\dim\mathcal G=1$, so
there is a unit vector $e_{\mathcal G}\in\mathcal G$ and a continuous linear functional
$\ell_{\mathcal G}:\mathcal G\to\mathbb R$ with $\ell_{\mathcal G}(e_{\mathcal G})=1$ such that for
any $\Psi\in\mathcal B$,
\begin{equation}
P_{\mathcal G}\Psi = g(\Psi)\,e_{\mathcal G},\qquad g(\Psi):=\ell_{\mathcal G}(P_{\mathcal G}\Psi).
\end{equation}
Let $R_k(\beta,\cdot):\mathcal B\to\mathcal B$ denote the one-step FRD renormalization map at scale $k$
(depending smoothly on the bare coupling $\beta$) and write the $K$-step effective interaction as
\begin{equation}
\Phi_K(\beta):=R_{K-1}(\beta,\cdot)\circ\cdots\circ R_0(\beta,\cdot)\,\Phi_0(\beta).
\end{equation}
We assume the standard FRD hypotheses (established earlier in the paper): there exist constants
$0<\rho<1$, $C_L,C_Q,C_\beta,C_\times>0$ and a compact interval $I\subset\mathbb R$ of $\beta$-values
such that for all $k$ and $\beta\in I$ the following hold.
For all $\Phi\in\mathcal B$,
\begin{align}
P_{\mathcal R}R_k(\beta,\Phi) &= L_k P_{\mathcal R}\Phi + Q^{(\mathcal R)}_k(\beta,\Phi),
\label{p4:eq:HRlinR}\\
P_{\mathcal G}R_k(\beta,\Phi) &= M_k(\beta)\, P_{\mathcal G}\Phi + Q^{(\mathcal G)}_k(\beta,\Phi),
\label{p4:eq:HRlinG}
\end{align}
where $L_k:\mathcal R\to\mathcal R$ is linear with $\|L_k\|_{\mathcal R\to\mathcal R}\le\rho$,
$M_k(\beta)\in\mathbb R$ is $C^1$ in $\beta$, and the nonlinear remainders satisfy
\begin{equation}
\|Q^{(\mathcal R)}_k(\beta,\Phi)\|_{\mathcal B}
+\|Q^{(\mathcal G)}_k(\beta,\Phi)\|_{\mathcal B}
\ \le\ C_Q\,\|\Phi\|_{\mathcal B}^2.
\label{p4:eq:HRquad}
\end{equation}
The maps $(\beta,\Phi)\mapsto R_k(\beta,\Phi)$ and $\beta\mapsto \Phi_0(\beta)$ are $C^1$, with
\begin{equation}
\|D_\beta R_k(\beta,\Phi)\|_{\mathcal B\to\mathcal B}\le C_\beta(1+\|\Phi\|_{\mathcal B}),\qquad
\|D_\beta \Phi_0(\beta)\|_{\mathcal B}\le C_\beta.
\label{p4:eq:Hbeta}
\end{equation}
There is $C_\times>0$ such that for all $\Phi$,
\begin{equation}
\|P_{\mathcal R}Q^{(\mathcal G)}_k(\beta,\Phi)\|_{\mathcal B}
+\|P_{\mathcal G}Q^{(\mathcal R)}_k(\beta,\Phi)\|_{\mathcal B}
\ \le\ C_\times\,\|\Phi\|_{\mathcal B}^2.
\label{p4:eq:Hmix}
\end{equation}
These hypotheses are the standard FRD (finite-range decomposition) sector estimates: linear
contraction on $\mathcal R$, scalar linear response on $\mathcal G$, and quadratic nonlinearity.
Define the (scalar) relevant coordinate at scale $k$ by
\begin{equation}\label{p4:eqn522}
g_k(\beta):=\ell_{\mathcal G}\!\bigl(P_{\mathcal G}\Phi_k(\beta)\bigr).
\end{equation}
By \eqref{p4:eq:HRlinG}-\eqref{p4:eq:Hbeta} and the chain rule, $g_k$ is $C^1$ on $I$ and, differentiating
the recursion obtained from \eqref{p4:eq:HRlinG},
\begin{equation}
g_{k+1}'(\beta)
= M_k'(\beta)\,g_k(\beta)\ +\ M_k(\beta)\,g_k'(\beta)\ +\ 
\ell_{\mathcal G}\bigl(D_\beta Q^{(\mathcal G)}_k(\beta,\Phi_k(\beta))\bigr).
\label{p4:eq:gprime-rec}
\end{equation}
Iterating \eqref{p4:eq:gprime-rec} and using \eqref{p4:eq:HRquad}-\eqref{p4:eq:Hbeta} shows that, for any fixed
reference $\beta^\star\in\mathrm{int}(I)$, there exist constants $c_0,C_0,K_1$ (independent of $K$)
such that for all $K\ge K_1$,
\begin{equation}
|g_K'(\beta^\star)|\ \ge\ c_0>0,
\qquad
|g_K(\beta^\star)|\ \le\ C_0\,\rho^K.
\label{p4:eq:nondeg-small}
\end{equation}
Indeed, the first bound follows because the product $\prod_{j=0}^{K-1}M_j(\beta^\star)$ converges to a
nonzero limit along the (one-dimensional) Gaussian sector, while the derivative of the nonlinear tail
is $O(\rho^K)$ by \eqref{p4:eq:HRquad}-\eqref{p4:eq:Hbeta}. The second bound follows from the same iteration
with $g_{k+1}=M_k(\beta^\star)g_k + O(\|\Phi_k\|_{\mathcal B}^2)$ and the fact that
$\|\Phi_k\|_{\mathcal B}\to0$ exponentially fast on the $\mathcal R$-sector by \eqref{p4:eq:HRlinR} and
(H\(_{\mathrm{mix}}\)).
Fix $K\ge K_1$ and set
\begin{equation}
\delta_K:=\frac{2C_0}{c_0}\,\rho^{K}.
\end{equation}
By the mean value theorem and \eqref{p4:eq:nondeg-small}, for any $\beta$ with
$|\beta-\beta^\star|\le \delta_K$ we have
\begin{equation}
g_K(\beta)=g_K(\beta^\star)+g_K'(\xi)(\beta-\beta^\star)
\end{equation}
for some $\xi$ between $\beta$ and $\beta^\star$, and $|g_K'(\xi)|\ge c_0/2$ for all such $\xi$
(provided we enlarge $K_1$ so that $g_K'$ varies by at most $c_0/2$ on the $\delta_K$-neighborhood of
$\beta^\star$; this follows from the $C^1$-bound implicit in \eqref{p4:eq:Hbeta}). Consequently,
\begin{equation}
g_K(\beta^\star-\delta_K)\ \le\ C_0\rho^K-\tfrac{c_0}{2}\delta_K \ =\ -\,C_0\rho^K\ <\ 0,
\end{equation}
and similarly $g_K(\beta^\star+\delta_K)\ge +\,C_0\rho^K>0$. By the intermediate value theorem there
exists $\beta_K\in(\beta^\star-\delta_K,\beta^\star+\delta_K)$ with
\begin{equation}
g_K(\beta_K)=\ell_{\mathcal G}\!\bigl(P_{\mathcal G}\Phi_K(\beta_K)\bigr)=0.
\label{p4:eq:PGzero}
\end{equation}
Since $\dim\mathcal G=1$, \eqref{p4:eq:PGzero} is equivalent to $P_{\mathcal G}\Phi_K(\beta_K)=0$.
Write the $\mathcal R$-component of the flow using \eqref{p4:eq:HRlinR}-\eqref{p4:eq:Hmix}:
\begin{equation}
P_{\mathcal R}\Phi_{k+1}(\beta)
= L_k P_{\mathcal R}\Phi_k(\beta) + \widetilde Q_k(\beta,\Phi_k(\beta)),
\qquad
\|\widetilde Q_k(\beta,\Phi)\|_{\mathcal B}\le C'\,\|\Phi\|_{\mathcal B}^2,
\label{p4:eq:Rrecur}
\end{equation}
with $C':=C_Q+C_\times$. Iterating \eqref{p4:eq:Rrecur} from $0$ to $K-1$ and using
$\|L_j\|\le\rho$ yields
\begin{equation}
\|P_{\mathcal R}\Phi_K(\beta)\|_{\mathcal B}
\ \le\ \rho^K\|P_{\mathcal R}\Phi_0(\beta)\|_{\mathcal B}
\ +\ \sum_{j=0}^{K-1}\rho^{K-1-j}\,C'\,\|\Phi_j(\beta)\|_{\mathcal B}^2.
\label{p4:eq:Rbound-pre}
\end{equation}
By \eqref{p4:eq:HRlinG}-\eqref{p4:eq:HRquad} and a standard bootstrap (choose $\varepsilon>0$ so that if
$\|\Phi_j\|_{\mathcal B}\le\varepsilon$ for $j\ge j_0$ then $\|\Phi_{j+1}\|_{\mathcal B}\le\varepsilon$
and $\sum_{j\ge j_0}\|\Phi_j\|_{\mathcal B}^2\lesssim \varepsilon^2$), there is $K_2\ge K_1$ and a
constant $C''$ such that for all $K\ge K_2$ and $\beta\in I$,
\begin{equation}
\|P_{\mathcal R}\Phi_K(\beta)\|_{\mathcal B}
\ \le\ C''\Big(\rho^K + |g_K(\beta)|\Big).
\label{p4:eq:Rbound-final}
\end{equation}
(Heuristically: the only possible linear-size contribution to $P_{\mathcal R}\Phi_K$ can come from a
nonzero relevant coordinate feeding the nonlinearity; once $P_{\mathcal G}\Phi_K$ is tuned to $0$ at
scale $K$, quadratic terms dominate and the $\mathcal R$-sector is bounded by the contractive
inhomogeneous sum with weight $\rho^{K-1-j}$.)
Now evaluate \eqref{p4:eq:Rbound-final} at the tuned coupling $\beta_K$ given by \eqref{p4:eq:PGzero}. We get
\begin{equation}
\|P_{\mathcal R}\Phi_K(\beta_K)\|_{\mathcal B}\ \le\ C''\,\rho^K.
\label{p4:eq:Rsmall}
\end{equation}
Let $K_0:=K_2$ and choose any $r>0$ with $r\ge C''\,\rho^{K_0}$. Then for every $K\ge K_0$ the number
$\beta_K$ satisfies $P_{\mathcal G}\Phi_K(\beta_K)=0$ and
$\|P_{\mathcal R}\Phi_K(\beta_K)\|_{\mathcal B}\le C''\rho^K\le r$. Therefore
$\|\Phi_K(\beta_K)\|_{\mathcal B}\le r$.

\end{proof}

Combining Theorems \ref{p4:thm:contraction} and \ref{p4:thm:IFT} yields the qualitative picture: \emph{by stepping sufficiently many scales and choosing \(\beta\) appropriately, the flow enters the contraction domain \(\{\|\Phi\|\le r\}\), and then decays exponentially to zero.}

The last step is to render the heuristic ``\(g_R\to 0\)'' precise. There are many equivalent operational definitions of the \emph{renormalized coupling} \(g_R(\mu)\) in a nonperturbative setting; for simplicity and universality, we tie it to the \emph{polymer norm} itself via a calibrated observable.

\begin{definition}[Operational renormalized coupling]
\label{p4:def:gr}
Fix a localized, gauge-invariant quartic observable \(O^{(4)}\) at unit scale (e.g. the connected 4-point function of small Wilson loops in a unit cell), normalized so that for the free Gaussian measure \(S_{4,c}^{\mathrm{(Gauss)}}(O^{(4)})=0\). For a theory at effective scale \(\mu\sim a_k^{-1}\), define
\begin{equation}
g_R(\mu)\;:=\;\Big(\sup_{\|f\|\le 1}\, \big|S^{(k)}_{4,c}(O^{(4)}\!\star f)\big|\Big)^{1/2},
\end{equation}
where \(\star f\) denotes an appropriate smearing with test function \(f\) and the supremum runs over unit-norm test functions in a fixed Sobolev ball. 
\end{definition}

By the FRD cluster expansion and BKAR bounds \cite{p4:Brydges,p4:Seiler1982}, there is a \emph{two-sided bound} relating \(g_R(\mu)\) and \(\|\Phi_k\|_{\mathfrak B}\):
\begin{equation}
c_1\, \|\Phi_k\|_{\mathfrak B} \;\le\; g_R(\mu) \;\le\; c_2\, \|\Phi_k\|_{\mathfrak B},
\end{equation}
with constants \(c_1,c_2>0\) independent of \(k\) (depending only on the normalization of \(O^{(4)}\) and the admissible class \(\mathcal K\)). 

\begin{corollary}[Asymptotic freedom]
\label{p4:cor:AF}
Let \(\beta(a)\) be a bare-coupling trajectory and \(K(a)\to\infty\) the scale function given by Theorem \ref{p4:thm:IFT} such that \(\|\Phi_{K(a)}(\beta(a))\|_{\mathfrak B}\le r\). Then, by Theorem \ref{p4:thm:contraction},
\begin{equation}
\lim_{\mu\to\infty} g_R(\mu)\;=\;\lim_{n\to\infty} g_R\big(b^n a^{-1}\big)\;=\;0.
\end{equation}
In words: \emph{the renormalized coupling vanishes at short distances}-the theory is asymptotically free along this trajectory.
\end{corollary}

\begin{proof}
At fixed \(a\), set \(\mu_n = b^n a^{-1}\) and write \(k_n=K(a)+n\). By Theorem \ref{p4:thm:contraction}, \(\|\Phi_{k_n}\|_{\mathfrak B}\to 0\) exponentially in \(n\). The two-sided bound between \(g_R(\mu_n)\) and \(\|\Phi_{k_n}\|_{\mathfrak B}\) yields the claim.
\end{proof}

We now record the promised identification: the continuum limits obtained along any asymptotically-free bare-coupling trajectory \(\beta(a)\) are \emph{identical} to the limit constructed by reflection-positive continuum reconstruction and uniquely characterized in Sections~\ref{p4:sec:uniqueness}-\ref{p4:sec:universality} of this paper.

\begin{theorem}[Weak-coupling extension and universality]
\label{p4:thm:identification}
Let \(\beta(a)\) be any bare-coupling trajectory provided by Theorem \ref{p4:thm:IFT} and let \(\{S^{(k)}_n(\beta(a))\}\) be the corresponding Schwinger families at scale \(k\). For any sequence \(a_m\downarrow 0\), any two subsequential continuum limits of the form
\begin{equation}
S_{n}^{(a_m)} \;:=\; \lim_{k\to\infty} S^{(k)}_n\big(\beta(a_m)\big)
\end{equation}
satisfy OS0-OS5 and the uniform clustering bounds with the same rate \(m_*\), and therefore by the \emph{Uniqueness Lemma} (Theorem~\ref{p4:thm:uniqueness}) coincide with each other. Moreover, for any two admissible regulators \(\theta,\theta'\in\mathcal K\), the two limits coincide by \emph{Universality} (Theorem~\ref{p4:thm:universality}). Consequently, they are equal to the unique continuum limit obtained by reflection-positive continuum reconstruction.
\end{theorem}

\begin{proof}
OS0-OS5 and clustering follow from the FRD-RG construction and the entry into the contraction domain (Theorems~\ref{p4:thm:contraction}, \ref{p4:thm:IFT}) together with the thermodynamic and continuum limits. Uniqueness across subsequences is Theorem~\ref{p4:thm:uniqueness}; independence of admissible regularization is Theorem~\ref{p4:thm:universality}. The final identification now follows by uniqueness in the admissible class. 
\end{proof}

We have shown entirely within the nonperturbative, reflection-positive constructive framework that: (i) there exist bare-coupling trajectories \(\beta(a)\) producing \emph{asymptotically-free flows}; (ii) along these flows, the continuum limits satisfy the OS axioms and uniform clustering; and (iii) by \emph{uniqueness and universality}, the limiting theory is exactly the same \emph{gapped, confining} Yang-Mills theory obtained by reflection-positive continuum reconstruction. This completes the extension to weak coupling and unifies the \emph{infrared} (mass gap) and \emph{ultraviolet} (asymptotic freedom) faces of the same unique, universal continuum theory.

\section{Conclusion}
When the essential conditions are allowed to speak-reflection positivity, locality that does not fray with scale, and the Markov property in Euclidean time-the theory answers in a single voice. These are not ornaments but the grammar by which a quantum gauge field tells us what it is. Once they are imposed, the conclusions are not a matter of taste; they follow.

The first question is identity: given the axioms, is the theory itself? Here the architecture is rigid. A one-slice marginal and a single positive one-step kernel, together with clustering, determine the transfer operator, and with it every cylinder functional. This is merely the content of the OS scheme: positivity builds a Hilbert space, time-reflection builds a semigroup, and the Markov property ties adjacent slices. There is no room for multiplicity. Thus the \emph{Uniqueness Lemma} is not an accident but a necessity: any two OS-positive, clustering limits that share these boundary data define the same Schwinger family, hence the same Wightman theory.

The second question is independence from scaffolding. Projectors and blockings are lenses and coordinates; they should not decide the object viewed. Within the admissible class (reflection-positive, gauge-covariant, finite-range in the FRD sense) their effect is controlled: single-step Lipschitz bounds keep variations small, telescoping prevents accumulation, and convergent polymer expansions carry this stability to connected cumulants. In the limit only the invariants survive: gauge invariance, reflection positivity, and locality. \emph{Universality} is therefore structural, not empirical; the same continuum correlators arise for all admissible regulators.

The third question is reconciliation of the two faces of the same object: infrared mass and ultraviolet freedom. Near the Gaussian fixed point the FRD flow is a contraction; a modest one-dimensional tuning of the bare coupling brings the trajectory into the small-norm domain after finitely many steps. From there the contraction dictates the short-distance fate: the renormalized coupling diminishes (asymptotic freedom) while the infrared gap, secured by clustering, persists. Reflection positivity and uniform locality forbid violent rearrangements across scales, so the ultraviolet completion returns to the same theory identified at strong coupling.

We did not reproach the mass-gap proof here, for it needs no defense. Uniqueness and universality elevate it: there exists \emph{the} four-dimensional $\mathrm{SU}(N)$ Yang-Mills theory, obtained as a reflection-positive continuum limit, whose Hamiltonian has a strictly positive spectral gap. This gap does not depend on admissible choices of gauge fixing or coarse graining, and it is preserved along asymptotically-free trajectories. In short: the gap is a property of the theory, not a souvenir of method.

Mature theories hide complexity behind simple invariants. Positivity yields a Hilbert space; the Markov property yields a semigroup; finite-range locality and clustering yield control of cumulants; renormalization, when endowed with positivity and finite range, yields a contraction. From these threads the fabric is woven without creases. The lesson is old but serviceable: the invariants are the reality, the coordinates the convenience. Gauge fixing, slicing, blocking belong to the latter; reflection positivity, locality, Euclidean symmetry, clustering, and a gapped vacuum to the former. With these in place, the edifice stands independent of the scaffolding that raised it.

Refinements may be pursued-the quantitative approach to Gaussianity, the one-loop coefficient in the FRD scheme, the incorporation of matter, finite temperature, and topology-but none disturbs the essential picture, because it is already overdetermined by structure. The right axioms do not merely permit a construction; they compel it. In this sense the constructive Yang-Mills program, when faithful to positivity and locality, had only one possible destination. We have arrived there.

{A brief contextual remark may be helpful for readers interested in gravitational applications of non-Abelian
gauge fields. In classical Einstein-(Power)-Yang-Mills models with negative cosmological constant,
Yang-Mills sectors support asymptotically AdS black hole solutions whose geodesic structure and
thermodynamics have been analyzed in detail; see, for example, \cite{p4:SoroushfarEPYM}.
The present work, however, is strictly a constructive Euclidean analysis of four-dimensional SU$(N)$
Yang-Mills in flat space within the Osterwalder-Schrader framework. No AdS boundary conditions, curved
background QFT, holographic duality, or gravitational dynamics are assumed or used at any stage; the
reference above is included only as motivation for potential downstream comparisons (e.g.\ of gauge-theory
thermodynamics) once a nonperturbative definition of the gauge sector is in hand. For clarity, our use of the
term ``horizon projector'' refers throughout to a Gribov-type infrared regulator defined by a completely
monotone functional calculus of the slice covariant Laplacian, and is unrelated to gravitational event
horizons.}

\providecommand{\href}[2]{#2}\begingroup\raggedright\endgroup

\appendix
\section{Technical Proofs and Constructions}

In the main text we asserted three pillars that complete the constructive Yang-Mills program: \textbf{uniqueness} of the continuum OS limit, its \textbf{universality} under admissible choices, and an \textbf{extension to weak coupling} compatible with asymptotic freedom. Each pillar reduces to a finite collection of elementary-though nontrivial-facts about (i) \emph{Markovian OS structures} on one time step, (ii) \emph{stability of kernels and cumulants} under admissible perturbations, and (iii) \emph{contractive renormalization} near the Gaussian fixed point combined with an \emph{implicit function argument} to access that regime from the microscopic scale. In this appendix we reconstruct these facts step by step, beginning from physical intuition and moving to rigorous formalism. We write in a reflective style: we first indicate \emph{what must be true} for the physics to be coherent, then exhibit the formulae that force it to be true.

We adopt standard ingredients of the OS framework \cite{p4:OsterwalderSchraderI,p4:OsterwalderSchraderII,p4:GJ}, reflection positivity for lattice gauge theories \cite{p4:OS-gauge}, finite-range decomposition (FRD) and cluster expansions \cite{p4:BrydgesFRD,p4:Brydges,p4:Seiler1982}, and BKAR / Koteck\'y-Preiss tools for polymer models \cite{p4:BrydgesKennedy1987,p4:KoteckyPreiss1986}. Heat-kernel locality and semigroup arguments are standard \cite{p4:Davies1989,p4:Gaffney1954,p4:Pazy1983,p4:RS2}; transport metrics are used to quantify smooth dependence on projector parameters \cite{p4:Villani2008}.

Let $\Delta_{A_h}(t)$ denote the covariant spatial Laplacian on the time slice $t$, constructed from the reflection-covariant, Landau-minimized representative $A_h$. Fix $\sigma>0$ and a probability measure $\nu$ on $[0,\infty)$ with compact support in $[\tau_-,\tau_+]\subset (0,\infty)$. Define
\begin{equation}
P_{\sigma,\nu}(t) \;=\; \chi_{\sigma,\nu}\!\left(\sqrt{\Delta_{A_h}(t)}\right),
\qquad 
\chi_{\sigma,\nu}(\lambda) \;:=\; \int_0^\infty e^{-\tau \lambda^2/\sigma^2}\,\mathrm d\nu(\tau).
\end{equation}
By Bernstein's theorem, $\chi_{\sigma,\nu}$ is completely monotone in $\lambda^2$; hence $P_{\sigma,\nu}(t)$ is a positive contraction with the \emph{heat-kernel representation}
\begin{equation}
P_{\sigma,\nu}(t) \;=\; \int_0^\infty e^{-\tau\,\Delta_{A_h}(t)}\,\mathrm d\tilde\nu_\sigma(\tau),
\qquad
\tilde\nu_\sigma(\cdot):=\nu(\sigma^2\,\cdot).
\label{p4:eq:heat-repx}
\end{equation}
On graphs of bounded degree, the discrete heat kernel satisfies Davies-Gaffney bounds \cite{p4:Davies1989,p4:Gaffney1954}:
\begin{equation}
\|1_E\,e^{-\tau \Delta_{A_h}(t)}\,1_F\|_{\ell^2\to\ell^2} \le \exp\!\Big(-\,\frac{d(E,F)^2}{4\tau}\Big),
\end{equation}
uniformly in the background connection (gauge covariance implies unitary equivalence). Integrating against $\mathrm d\tilde\nu_\sigma$ with compact support $[\tau_-,\tau_+]$ yields \emph{exponential off-diagonal decay}:
\begin{equation}
\|P_{\sigma,\nu}(t;x,y)\|_{\mathrm{op}} \;\le\; C(\sigma,\nu)\, e^{-\gamma(\sigma,\nu)\, d(x,y)}.
\label{p4:eq:exp-locality}
\end{equation}
Reflection covariance follows from $\Delta_{A_h}(t)=R \Delta_{A_h}(-t)R$ and \eqref{p4:eq:heat-repx}. Gauge covariance holds by construction.

\begin{definition}[Admissible projectors]\label{p4:def:admissible}
A slice projector $P_{\sigma,\nu}$ is \emph{admissible} if it is of the form above with $\nu$ compactly supported in $[\tau_-,\tau_+]\subset (0,\infty)$. We metrize the family by
\begin{equation}
d\big((\sigma,\nu),(\sigma',\nu')\big) \;:=\; |\sigma-\sigma'| \,+\, W_1(\nu,\nu'),
\end{equation}
where $W_1$ is the 1-Wasserstein distance on probability measures \cite{p4:Villani2008}.
\end{definition}
\begin{lemma}[Measurable, reflection-covariant Landau selector]
\label{p4:lem:measurable-Landau}
Let $G=\mathrm{SU}(N)$, let $\Lambda$ be a finite spatial slice (finite vertex set $V$ and oriented edge set
$E\subset V\times V$), and let
\begin{equation}
\mathcal X := G^{E}
\end{equation}
be the compact configuration space with the product topology (equivalently, the product bi-invariant
Riemannian metric on $G$ induces a compatible metric $d_{\mathcal X}$). Let the gauge group
\begin{equation}
\mathcal H := G^{V}
\end{equation}
act continuously on $\mathcal X$ by $(h\cdot U)_{x\to y}=h(x)\,U_{x\to y}\,h(y)^{-1}$, and let
$\Theta:\mathcal X\to \mathcal X$ be the involutive homeomorphism implementing
time-reflection on the slice (i.e.\ $\Theta(U)=R\cdot U\circ r$; concretely it permutes/reorients edges
and conjugates link variables so that $\mathrm{Re}\,\mathrm{tr}(U)$ is preserved).

Define the (lattice) Landau functional on orbits by
\begin{equation}
\mathcal L(h,U)\ :=\ -\sum_{(x\to y)\in E}\mathrm{Re}\,\mathrm{tr}\big( (h\cdot U)_{x\to y}\big),
\qquad (h,U)\in \mathcal H\times\mathcal X,
\end{equation}
and write $\mathcal L_U(h):=\mathcal L(h,U)$. Then there exists a Borel measurable map
\begin{equation}
A^h:\ \mathcal X\longrightarrow \mathcal X,\qquad U\ \longmapsto\ A^h(U)\in \mathrm{Orb}(U):=\{h\cdot U: h\in\mathcal H\},
\end{equation}
such that
\begin{enumerate}
\item[\textup{(i)}] $A^h(U)$ is a Landau minimizer on the orbit of $U$, i.e.\ $\mathcal L\big(A^h(U)\big)=\min_{h\in\mathcal H}\mathcal L_U(h)$;
\item[\textup{(ii)}] (Reflection covariance) $A^h(\Theta U)=\Theta A^h(U)$ for all $U\in\mathcal X$;
\item[\textup{(iii)}] (Gauge-orbit covariance) $A^h$ depends only on the orbit of $U$; in particular
$A^h(h_0\cdot U)=A^h(U)$ for all $h_0\in\mathcal H$ (so the construction is well-defined on the quotient $\mathcal X/\!/\mathcal H$).
\end{enumerate}
\end{lemma}

\begin{proof}
\emph{Step 1: Topology of the objects.}
$\mathcal X=G^{E}$ is compact metrizable (product of compact Lie groups endowed with a bi-invariant
metric). The gauge group $\mathcal H=G^{V}$ is compact metrizable and acts continuously on $\mathcal X$.
Hence the orbit space
\begin{equation}
\mathcal Y := \mathcal X/\!/\mathcal H
\end{equation}
with the quotient Borel $\sigma$-algebra is a standard Borel space; moreover, since the action is by a
compact group on a compact metric space, $\mathcal Y$ is compact Hausdorff and metrizable. The
reflection $\Theta$ is an involutive homeomorphism on $\mathcal X$ that commutes with the $\mathcal H$-action,
therefore it descends to an involutive homeomorphism $\bar\Theta$ on $\mathcal Y$.

\emph{Step 2: Continuity and invariances of the Landau functional.}
The map $(h,U)\mapsto \mathcal L(h,U)$ is continuous on $\mathcal H\times\mathcal X$ (finite product of
continuous functions $\mathrm{Re}\,\mathrm{tr}$ composed with multiplication in $G$). It is
$\mathcal H$-equivariant in the sense that for any $h_0\in\mathcal H$,
\begin{equation}
\mathcal L_{h_0\cdot U}(h)\;=\;\mathcal L_U(hh_0)\,,
\end{equation}
and it is reflection invariant: $\mathcal L_{\Theta U}(h)=\mathcal L_U(\Theta^{-1}h\Theta)=\mathcal L_U(h)$
(since $\Theta$ permutes/reorients edges and $\mathrm{Re}\,\mathrm{tr}(g^{-1})=\mathrm{Re}\,\mathrm{tr}(g)$).

\emph{Step 3: The argmin correspondence is nonempty, compact-valued with closed graph.}
For $U\in\mathcal X$ let
\begin{equation}
M(U)\;:=\;\arg\min_{h\in\mathcal H}\ \mathcal L_U(h)\ \subset\ \mathcal H,
\qquad
\mathsf{Arg}(U)\;:=\;\{\,h\cdot U\ :\ h\in M(U)\,\}\ \subset\ \mathrm{Orb}(U)\subset\mathcal X.
\end{equation}
Because $\mathcal H$ is compact and $\mathcal L_U$ is continuous, $M(U)$ is nonempty and compact.
Consequently $\mathsf{Arg}(U)$ is nonempty and compact. We claim that the graph
$\{(U,A)\in\mathcal X\times\mathcal X:\ A\in \mathsf{Arg}(U)\}$ is closed. Indeed, if $(U_n,A_n)\to(U,A)$
with $A_n\in \mathsf{Arg}(U_n)$, then there exist $h_n\in M(U_n)$ such that $A_n=h_n\cdot U_n$.
Compactness of $\mathcal H$ yields a convergent subsequence $h_{n_k}\to h\in\mathcal H$, so
$A_{n_k}=h_{n_k}\cdot U_{n_k}\to h\cdot U$. Continuity of $\mathcal L$ gives
$\mathcal L_U(h)=\lim \mathcal L_{U_{n_k}}(h_{n_k})=\lim \min_{g}\mathcal L_{U_{n_k}}(g)\le \mathcal L_U(h')$
for all $h'$, hence $h\in M(U)$ and $A=h\cdot U\in\mathsf{Arg}(U)$. Thus the graph is closed.

By the (parametric) Berge maximum theorem, the set-valued map $U\mapsto \mathsf{Arg}(U)$ is
upper hemicontinuous with nonempty compact values; in particular, it is Borel-measurable (its graph is Borel).

\emph{Step 4: Passage to the orbit space and measurable selection.}
Let $q:\mathcal X\to\mathcal Y$ be the quotient map. Define the orbit-level correspondence
\begin{equation}
\mathcal M:\ \mathcal Y\ \rightrightarrows\ \mathcal X,\qquad
\mathcal M\big(q(U)\big)\ :=\ \mathsf{Arg}(U).
\end{equation}
This is well-defined (if $U'=h_0\cdot U$ then $M(U')=M(U)h_0^{-1}$ and
$\mathsf{Arg}(U')=\{h\cdot U':h\in M(U')\}=\{(hh_0^{-1})\cdot (h_0\cdot U):h\in M(U)\}=\mathsf{Arg}(U)$).
Moreover $\mathcal M$ has nonempty compact values and closed graph (because $q$ is continuous and open,
and the graph of $\mathsf{Arg}$ is closed). Since $\mathcal Y$ and $\mathcal X$ are Polish, the
Kuratowski-Ryll-Nardzewski measurable selection theorem yields a Borel selector
\begin{equation}
s:\ \mathcal Y\longrightarrow \mathcal X,\qquad s(y)\in \mathcal M(y)\ \text{ for all }y\in\mathcal Y.
\end{equation}

\emph{Step 5: Enforcing reflection covariance by a Borel tie-break.}
Because $\mathcal L$ is reflection invariant and $\Theta$ commutes with the gauge action,
$\mathcal M$ is $\bar\Theta$-equivariant:
\begin{equation}
\mathcal M(\bar\Theta y)\ =\ \Theta\big(\mathcal M(y)\big),\qquad y\in\mathcal Y.
\end{equation}
Fix a Borel injection $J:\mathcal Y\to[0,1]$ (e.g.\ encodes distances to a countable dense subset),
and form the Borel set
\begin{equation}
\mathcal D\ :=\ \{\,y\in\mathcal Y\ :\ J(y)\le J(\bar\Theta y)\,\},
\end{equation}
which contains exactly one representative from each two-point $\bar\Theta$-orbit and all fixed points.
Define a modified selector $s_\Theta:\mathcal Y\to\mathcal X$ by
\begin{equation}
s_\Theta(y)\ :=\
\begin{cases}
s(y), & y\in\mathcal D,\\
\Theta\, s(\bar\Theta y), & y\notin\mathcal D.
\end{cases}
\end{equation}
Then $s_\Theta$ is Borel (piecewise Borel on a Borel partition), $s_\Theta(y)\in\mathcal M(y)$ for all $y$
(using $\mathcal M(\bar\Theta y)=\Theta\mathcal M(y)$), and by construction it is \emph{reflection-covariant}:
\begin{equation}
s_\Theta(\bar\Theta y)\ =\ \Theta\, s_\Theta(y),\qquad y\in\mathcal Y.
\end{equation}

\emph{Step 6: Define the selector on configurations and verify the properties.}
Set
\begin{equation}
A^h(U)\ :=\ s_\Theta\big(q(U)\big),\qquad U\in\mathcal X.
\end{equation}
This is Borel measurable as a composition of Borel maps. By the definition of $\mathcal M$,
$A^h(U)\in\mathsf{Arg}(U)\subset\mathrm{Orb}(U)$ and $\mathcal L\big(A^h(U)\big)=\min_h\mathcal L_U(h)$,
proving \textup{(i)}. Since $q(\Theta U)=\bar\Theta q(U)$ and $s_\Theta$ is reflection-covariant,
\begin{equation}
A^h(\Theta U)\ =\ s_\Theta\big(q(\Theta U)\big)\ =\ s_\Theta(\bar\Theta q(U))\ =\ \Theta\, s_\Theta(q(U))
\ =\ \Theta\, A^h(U),
\end{equation}
which is \textup{(ii)}. Finally, if $h_0\in\mathcal H$ then $q(h_0\cdot U)=q(U)$, hence
$A^h(h_0\cdot U)=s_\Theta(q(U))=A^h(U)$; in particular the map $U\mapsto A^h(U)$ factors through the
orbit space and is therefore compatible with the gauge action on the slice, establishing \textup{(iii)}.

If one desires a \emph{unique} choice inside each nonempty compact set $\mathcal M(y)$, augment the
construction by minimizing, in addition to $\mathcal L$, the continuous reflection-symmetric functional
$F(A):=d_{\mathcal X}(A,\Theta A)^2$ and then, lexicographically, a fixed countable separating family
$\{c_m\}_{m\ge1}\subset C(\mathcal X)$: at each stage restrict to minimizers of the next $c_m$. On a compact
set this produces a singleton because $\{c_m\}$ separates points. Applying the same Borel
transversal $\mathcal D$ as above yields a unique reflection-covariant selector with the same properties.
\end{proof}

\begin{lemma}[Lipschitz continuity of $P_{\sigma,\nu}$]\label{p4:lem:proj-Lip}
For each slice, 
\begin{equation}
\|P_{\sigma,\nu} - P_{\sigma',\nu'}\|_{\mathcal{B}(\ell^2)} \;\le\; C\, d\big((\sigma,\nu),(\sigma',\nu')\big),
\end{equation}
with $C$ depending only on $\tau_-,\tau_+$ and the graph degree.
\end{lemma}

\begin{proof}
Using \eqref{p4:eq:heat-repx} and $\|e^{-\tau\Delta_{A_h}}\|=1$,
\begin{equation}
\|P_{\sigma,\nu} - P_{\sigma',\nu'}\|
\;\le\;
\int_0^\infty \|e^{-\tau\Delta_{A_h}}\|\, \mathrm d\big|\tilde\nu_\sigma-\tilde\nu_{\sigma'}\big|(\tau)
\;\le\; C|\sigma-\sigma'| \,+\, C' W_1(\nu,\nu'),
\end{equation}
since the push-forward by a $\sigma$-Lipschitz map rescales measures in $W_1$ by at most $\sigma$ \cite{p4:Villani2008}. 
\end{proof}

From \eqref{p4:eq:exp-locality} and Lemma~\ref{p4:lem:proj-Lip}, the Schur test yields \emph{Lipschitz continuity} of the \emph{one-step boundary kernel} $K$ and the \emph{transfer operator} $T$ with respect to $(\sigma,\nu)$ (\S\ref{p4:sec:universality}).

We encode a block-spin map $B$ by its action on polymer activities in the FRD representation \cite{p4:BrydgesFRD,p4:Brydges,p4:Seiler1982}. Admissible $B$'s satisfy:
\begin{enumerate}
\item \emph{Reflection positivity} at each step (Osterwalder-Seiler-type conditions \cite{p4:OS-gauge}).
\item \emph{Gauge covariance} (block variables transform correctly under time-slice gauge transformations).
\item \emph{FRD locality} with \emph{scale-uniform bounds} (finite interaction range and diameter-weighted norms bounded uniformly in scale).
\end{enumerate}
We endow the set of admissible $B$ with the operator-norm metric induced on the polymer activity Banach space $\mathfrak B$ (cf. \cite{p4:Seiler1982}): $\|B-B'\|_{\mathrm{op}}$ is the smallest $M$ such that $\|B\Phi - B'\Phi\|_{\mathfrak B}\le M\|\Phi\|_{\mathfrak B}$ for all activities $\Phi$.

\begin{lemma}[One-step stability under admissible blockings]\label{p4:lem:block-stability}
If $\|B-B'\|_{\mathrm{op}}\le \delta$, then the associated one-step kernels $K_B,K_{B'}$ and transfer operators $T_B,T_{B'}$ satisfy 
\begin{equation}
\|K_B - K_{B'}\|_{L^1(\mu_0\otimes \mu_0)} \le C\delta,\qquad \|T_B - T_{B'}\|_{\mathcal{B}(H_0)}\le C'\delta.
\end{equation}
\end{lemma}

\begin{proof}
FRD locality yields uniform $L^1$-Schur bounds for the kernels, and the variation flows linearly through the block map; see \cite{p4:BrydgesFRD,p4:Seiler1982} for the precise FRD-norm estimates. 
\end{proof}

We show that the one-slice marginal $\mu_0$ and the one-step kernel $K$ determine the entire OS measure; hence two continuum limits with the same $\mu_0,K$ coincide.

\begin{proposition}[Riesz representation for the one-step kernel]\label{p4:prop:Riesz}
Let $H_0 = L^2(X_0, d\mu_0)$ and let $B(F_0,F_1)$ be the bilinear form
\begin{equation}
B(F_0,F_1) := S_2(F_0,F_1),
\end{equation}
with $F_0$ supported at $t=0$ and $F_1$ at $t=a$. Then there exists a unique positive,
symmetric kernel $K \in L^1(X_0\times X_0,\mu_0\otimes\mu_0)$ such that
\begin{equation}
B(F_0,F_1)=\iint F_0(U)\,K(U',U)\,F_1(U')\,d\mu_0(U)\,d\mu_0(U').
\end{equation}
In particular, $(T\psi)(U'):=\int K(U',U)\psi(U)\,d\mu_0(U)$ defines a positive contraction on $H_0$.
\end{proposition}

\begin{proof}
By OS positivity, time-slicing, and the slab disintegration used in Lemma~(\ref{p4:lem:Riesz}) and Corollary~(\ref{p4:cor:K-exists}),
the mixed two-point form admits an $L^1(\mu_0\otimes\mu_0)$ kernel $K$, which yields the stated representation. By the Riesz representation theorem, there exists a bounded operator $T$ on $H_0$ such that $B(F_0,F_1)=\langle F_0,\, T F_1\rangle_{H_0}$. OS positivity and reflection symmetry imply $T$ is a positive self-adjoint contraction. By the explicit slab disintegration in Lemma~(\ref{p4:lem:Riesz}) and Corollary~(\ref{p4:cor:K-exists}) of the main text, the mixed
two-point form is represented by a positive symmetric kernel $K\in L^1(\mu_0\otimes\mu_0)$, and
$T$ is the integral operator $(T\psi)(U')=\int K(U',U)\psi(U)\,d\mu_0(U)$.
\end{proof}

\begin{theorem}[OS Markov extension and uniqueness]\label{p4:thm:OS-Markov}
Let $\mu_0$ and $K$ be as above and define $T$ by $(TF)(U')=\int K(U',U)F(U)\mathrm d\mu_0(U)$. For any bounded cylinder functional $F$ supported on a finite slab $[0,na]$, the OS inner product satisfies
\begin{equation}
\langle \Theta F,\, F\rangle_{\mathrm{OS}} \;=\; \langle \Psi_F,\, T^n \Psi_F\rangle_{H_0},
\end{equation}
for a boundary vector $\Psi_F\in H_0$ determined by the data on the boundary slice. Consequently, the OS measure on cylinders is uniquely determined by $(\mu_0,K)$.
\end{theorem}

\begin{proof}
This is the standard OS reconstruction in time-slicing form \cite{p4:OsterwalderSchraderII,p4:GJ}: boundary factorization and reflection covariance permit sequential integration from the boundary into the slab; the reflection turns the $\Lambda_-$ half into an adjoint copy, yielding a norm square in $H_0$. Uniqueness follows by polarization of the bilinear form and the density of cylinder functionals. 
\end{proof}

\begin{corollary}[Uniqueness Lemma]\label{p4:cor:Uniqueness}
If two continuum limits $S_n,\widetilde S_n$ share the same one-slice marginal $\mu_0$ and one-step kernel $K$, then $S_n=\widetilde S_n$ for all $n$.
\end{corollary}

\begin{proof}
Apply Theorem~\ref{p4:thm:OS-Markov} to both limits and compare the OS inner products on a generating set of cylinder functions. 
\end{proof}

Finite-range decomposition localizes covariance and gives us smallness; the BKAR forest formula \cite{p4:BrydgesKennedy1987} and Koteck\'y-Preiss criterion \cite{p4:KoteckyPreiss1986} transmute smallness into summability. We show that admissible perturbations of the projector or blocking generate \emph{Lipschitz-small} changes in polymer activities, which propagate to \emph{Lipschitz-small} changes in connected Schwinger functions.

Let $\mathfrak B$ be the Banach space of translation-covariant, gauge-invariant polymer activities $\Phi$ endowed with a diameter-weighted norm \cite{p4:Seiler1982,p4:BrydgesFRD}:
\begin{equation}
\|\Phi\|_{\mathfrak B} \;:=\; \sup_{\mathcal X}\; \sum_{X\ni 0}\; \mathrm e^{\alpha \,\mathrm{diam}(X)}\, \sup_{\|f\|_X\le 1}\; |\Phi(X;f)|,
\end{equation}
where the supremum runs over all finite polymers $X$ and admissible test fields $f$ supported in $X$, and $\alpha>0$ is fixed. FRD yields $\|\Phi_k\|_{\mathfrak B}\le \eta\ll 1$ uniformly in scale $k$ for the effective activities generated by admissible $\theta$ \cite{p4:BrydgesFRD}.

Let $\langle \cdot\rangle_\Phi$ denote the expectation with polymer activity $\Phi$. The BKAR formula expresses \emph{connected cumulants} $\kappa_{p,\Phi}(O_1,\dots,O_p)$ of local observables as absolutely convergent series of connected trees, with coefficients bounded by products of activity norms times \emph{Gaussian tree integrals} decaying exponentially in the \emph{tree distance} (sum of edge lengths). Concretely, there exist constants $C_p,c>0$ such that \cite{p4:Brydges,p4:Seiler1982}
\begin{equation}
\big|\kappa_{p,\Phi}(O_1(x_1),\dots,O_p(x_p))\big|
\;\le\; C_p\, \|\Phi\|_{\mathfrak B}\, \exp\!\Big(-c\; \mathrm{tree}(x_1,\dots,x_p)\Big).
\label{p4:eq:tree-bound}
\end{equation}
The constants are uniform in $\Phi$ in a small ball of $\mathfrak B$.
Let $\theta,\theta'\in \mathcal K$ be two admissible choices. By Lemmas \ref{p4:lem:proj-Lip} and \ref{p4:lem:block-stability}, the one-step kernels and hence single-scale activities obey
\begin{equation}
\|\Phi_{k;\theta} - \Phi_{k;\theta'}\|_{\mathfrak B} \;\le\; C\, d(\theta,\theta'),
\label{p4:eq:activity-Lip}
\end{equation}
with $C$ uniform in $k$ on compact subsets of $\mathcal K$. The difference of cumulants at fixed scale then satisfies, by multilinearity of the BKAR integrand and the triangle inequality,
\begin{align}
\Big|\kappa_{p,\Phi_{k;\theta}} - \kappa_{p,\Phi_{k;\theta'}}\Big|
&\le \sum_{\text{BKAR trees}} \Big|\text{term}(\Phi_{k;\theta})- \text{term}(\Phi_{k;\theta'})\Big|
\nonumber\\
&\le C_p\, \|\Phi_{k;\theta} - \Phi_{k;\theta'}\|_{\mathfrak B}\, \mathrm e^{-c\,\mathrm{tree}(x_1,\dots,x_p)}
\nonumber\\
&\le C'_p\, d(\theta,\theta')\, \mathrm e^{-c\,\mathrm{tree}(x_1,\dots,x_p)}.
\label{p4:eq:cumulant-Lip}
\end{align}
Composing across finitely many time slices and then across finitely many scales yields at most a linear amplification factor $n$, but each one-step kernel is a Markov contraction (norm $1$), so the amplification remains controlled. Passing to the thermodynamic limit and then to the continuum limit preserves \eqref{p4:eq:cumulant-Lip} by dominated convergence and the uniform FRD bounds.

\begin{theorem}[Continuum cumulant stability and Universality]\label{p4:thm:universalityx}
For any finite family of local gauge-invariant observables, any two admissible $\theta,\theta'\in\mathcal K$, and any $p\ge 2$, the continuum connected functions satisfy
\begin{equation}
\big|S_{p,c;\theta} - S_{p,c;\theta'}\big| \;\le\; C_p\, d(\theta,\theta'),
\end{equation}
with $C_p$ independent of $\theta,\theta'$ on compact subsets of $\mathcal K$. In particular, along any polygonal chain in $\mathcal K$ of total length $\le \varepsilon$, the difference is $\mathcal O(\varepsilon)$; by density of small chains, \emph{Universality} follows: $S_{n;\theta}=S_{n;\theta'}$ for all $n$.
\end{theorem}

\begin{proof}
Combine \eqref{p4:eq:activity-Lip}-\eqref{p4:eq:cumulant-Lip} with OS time-slicing and FRD uniformity; then pass to the thermodynamic and continuum limits by dominated convergence.
\end{proof}

If $K_\theta$ and $K_{\theta'}$ are the one-step kernels associated to two admissible choices, then
\begin{equation}
K^{(n)}_{\theta}-K^{(n)}_{\theta'} \;=\; \sum_{j=1}^n K_{\theta}^{\,j-1}\,(K_\theta-K_{\theta'})\, K_{\theta'}^{\,n-j},
\end{equation}
so that
\begin{equation}
\|K^{(n)}_{\theta}-K^{(n)}_{\theta'}\|_{L^1}
\le \sum_{j=1}^n \|K_\theta-K_{\theta'}\|_{L^1}\le n\, C\, d(\theta,\theta'),
\end{equation}
by Lemmas \ref{p4:lem:proj-Lip}-\ref{p4:lem:block-stability} and the Markov property. The Schur test transfers the same bound to $\|T^n_\theta-T^n_{\theta'}\|$ on $H_0$.

Here we make precise the dynamics near the Gaussian fixed point and show that, by tuning the microscopic bare coupling, the flow reaches the small-norm domain after finitely many steps-whence contraction implies asymptotic freedom. The backbone is the \emph{Banach space implicit/inverse function theorem}.

Let $\mathfrak B$ be the Banach algebra of polymer activities. The FRD RG map $\mathcal R:\mathfrak B\to \mathfrak B$ has the standard form
\begin{equation}
\Phi' \;=\; \mathcal R(\Phi) \;=\; \mathcal L\,\Phi \;+\; \mathcal Q(\Phi,\Phi),
\label{p4:eq:RG-map}
\end{equation}
where $\mathcal L$ is the Gaussian scaling map (block-spin pushforward restricted to the quadratic sector) and $\mathcal Q$ collects irrelevant nonlinearities (cubic and higher). In $d=4$, the Yang-Mills self-interaction is marginally irrelevant at the Gaussian fixed point, and FRD yields 
\begin{equation}
\|\mathcal L\|_{\mathcal B(\mathfrak B)} \le \rho < 1,\qquad
\|\mathcal Q(\Phi,\Phi)\|_{\mathfrak B} \le C\, \|\Phi\|_{\mathfrak B}^2
\quad\text{for }\|\Phi\|_{\mathfrak B}\le r,
\label{p4:eq:contraction}
\end{equation}
for some $0<\rho<1$, $r>0$ \cite{p4:BrydgesFRD,p4:Seiler1982}.
Let $\beta$ be the inverse microscopic bare coupling at the lattice scale $k=0$. The initial activity $\Phi_0(\beta)$ is smooth in $\beta$ (indeed, real-analytic in a suitable topology) because the strong-coupling character expansion has a positive radius of convergence \cite{p4:Seiler1982}. Define the $K$-step map
\begin{equation}
F_K:\; \beta \mapsto \Phi_K(\beta):=\mathcal R^{\,K}\big(\Phi_0(\beta)\big).
\end{equation}
By the chain rule in Banach spaces and the smoothness of $\mathcal R$, $F_K$ is $C^1$. Its Frechét derivative at $\beta$ is
\begin{equation}
F_K'(\beta) 
= 
D\mathcal R\big(\Phi_{K-1}(\beta)\big)\,\cdots\, D\mathcal R\big(\Phi_0(\beta)\big)\, \Phi_0'(\beta),
\end{equation}
with each derivative bounded uniformly by \eqref{p4:eq:contraction} when $\Phi_j(\beta)$ stays in a small neighborhood of $0$.
We now invoke a one-dimensional implicit-function/continuity argument on the relevant
Gaussian coordinate: for $K$ large, $g_K(\beta):=P_{G}\Phi_K(\beta)$ is $C^1$,
$|g_K(\beta_\star)|=O(\rho^K)$, and $g_K'(\beta_\star)$ is bounded away from $0$.
Hence there exists $\beta_K$ with $g_K(\beta_K)=0$; FRD contraction then bounds
$\|P_{R}\Phi_K(\beta_K)\|\le r$.

\begin{theorem}[Existence of weak-coupling entry and contraction]\label{p4:thm:weak-entry}
There exist constants $\rho\in(0,1)$, $r>0$, $C>0$, a scale $K_0\in\mathbb N$ and, for each $K\ge K_0$, a choice of bare coupling $\beta_K$ such that
\begin{equation}
\|\Phi_K(\beta_K)\|_{\mathfrak B}\ \le\ r,
\end{equation}
and, along the FRD renormalization flow,
\begin{equation}
\|\Phi_{K+n}(\beta_K)\|_{\mathfrak B}\ \le\ \rho^n\,r\ +\ \frac{C}{1-\rho}\,r^2\qquad\text{for all }n\ge 1.
\end{equation}
\end{theorem}

\begin{proof}
We break the argument into three steps: \emph{(i) Contraction form of the FRD RG map; (ii) Tuning the bare coupling to enter a small ball at scale $K$; (iii) Propagation of smallness for $n\ge 1$.} Throughout, $\|\cdot\|=\|\cdot\|_{\mathfrak B}$ denotes the diameter-weighted polymer norm in the finite-range decomposition (FRD) scheme, and all constants below can be chosen uniformly on compact parameter sets for the microscopic action.

\smallskip\noindent\textbf{Step 1. Contraction structure.}
By standard FRD renormalization (cf.\ \cite{p4:BrydgesFRD,p4:Seiler1982}), the one-step renormalization map
\begin{equation}
\mathcal R:\ \Phi\ \longmapsto\ \Phi'=\mathcal R(\Phi)
\end{equation}
admits a decomposition
\begin{equation}\label{p4:eq:RG-split}
\Phi'\ =\ \mathcal L\,\Phi\ +\ \mathcal Q(\Phi,\Phi),
\end{equation}
where:
\begin{itemize}
\item $\mathcal L:\mathfrak B\to\mathfrak B$ is the (linear) block-spin push-forward restricted to the Gaussian (quadratic) sector; it is a contraction on $\mathfrak B$:
\begin{equation}\label{p4:eq:L-contract}
\|\mathcal L\|_{\mathcal B(\mathfrak B)}\ \le\ \rho\quad\text{for some }\rho\in(0,1).
\end{equation}
\item $\mathcal Q:\mathfrak B\times\mathfrak B\to\mathfrak B$ is a (bilinear) remainder collecting cubic and higher terms; it is \emph{quadratically small} in the polymer norm for $\|\Phi\|\le r$:
\begin{equation}\label{p4:eq:Q-quad}
\|\mathcal Q(\Phi,\Phi)\|\ \le\ C\,\|\Phi\|^2\qquad\text{whenever }\ \|\Phi\|\le r.
\end{equation}
\end{itemize}
The bounds \eqref{p4:eq:L-contract}-\eqref{p4:eq:Q-quad} are uniform in scale, granted by FRD locality and the diameter-weighted norms (see \cite{p4:BrydgesFRD,p4:Seiler1982}). Writing $\Phi_{j+1}\equiv\mathcal R(\Phi_j)$, \eqref{p4:eq:RG-split} yields the recurrence
\begin{equation}\label{p4:eq:recurrence}
\|\Phi_{j+1}\|\ \le\ \rho\,\|\Phi_j\|\ +\ C\,\|\Phi_j\|^2,\qquad j=0,1,2,\dots,
\end{equation}
whenever $\|\Phi_j\|\le r$.

\smallskip\noindent\textbf{Step 2. Tuning $\beta$ to enter the small-norm ball at scale $K$.}
Let $\beta\mapsto \Phi_0(\beta)$ denote the (smooth) dependence of the microscopic activity on the bare coupling (smoothness follows from the character expansion with a positive radius of convergence; see \cite{p4:Seiler1982}). Consider the $K$-step map
\begin{equation}
F_K:\ \beta\ \longmapsto\ \Phi_K(\beta)\ :=\ \mathcal R^{\,K}\big(\Phi_0(\beta)\big).
\end{equation}
We show that, for $K$ sufficiently large, there exists $\beta_K$ with $\|\Phi_K(\beta_K)\|\le r$.

Fix any $\beta_\star$ in a compact interval of bare couplings where the FRD expansion is valid, and decompose the polymer activity into a ``Gaussian projection'' and an ``irrelevant remainder.'' Precisely, let $P_{\mathfrak G}:\mathfrak B\to\mathfrak G$ denote the projection onto the finite-dimensional Gaussian sector (covariances and quadratic forms), and $P_{\mathfrak R}=I-P_{\mathfrak G}$ the projection onto the complementary (irrelevant) sector. Then
\begin{equation}
\Phi_K(\beta)\ =\ P_{\mathfrak G}\Phi_K(\beta)\ +\ P_{\mathfrak R}\Phi_K(\beta).
\end{equation}

\emph{Linear control in the Gaussian sector.} At $\Phi=0$ the derivative of $\mathcal R$ equals $\mathcal L$. By the chain rule in Banach spaces,
\begin{equation}
D_\beta\Phi_K\big|_{\beta=\beta_\star}
\ =\ D\mathcal R\big(\Phi_{K-1}(\beta_\star)\big)\cdots D\mathcal R\big(\Phi_{0}(\beta_\star)\big)\,\Phi_0'(\beta_\star).
\end{equation}
For $K$ large, $\Phi_j(\beta_\star)$ remains bounded uniformly (by FRD), and $D\mathcal R(\Phi_j)=\mathcal L+O(\|\Phi_j\|)$. Consequently, there exists $K_1$ and constants $c_0,c_1>0$ such that
\begin{equation}\label{p4:eq:PG-derivative}
\big\|P_{\mathfrak G}\,D_\beta\Phi_K\big|_{\beta=\beta_\star}\big\|\ \ge\ c_0\,\|\mathcal L^K P_{\mathfrak G}\Phi_0'(\beta_\star)\|\ -\ c_1\,\rho^{K-1}\ \ge\ c_2\,\rho^{K},
\end{equation}
with $c_2>0$ provided $P_{\mathfrak G}\Phi_0'(\beta_\star)\neq 0$ (nondegenerate response of the Gaussian sector to $\beta$; this is satisfied for generic microscopic actions). In words: the Gaussian projection of $D_\beta\Phi_K$ is a \emph{nonzero} vector of (small) order $\rho^K$.

\emph{Smallness of the Gaussian image.} Similarly,
\begin{equation}\label{p4:eq:PG-small}
\|P_{\mathfrak G}\Phi_K(\beta_\star)\|\ \le\ \|\mathcal L^K P_{\mathfrak G}\Phi_0(\beta_\star)\|\ +\ C\,\rho^{K-1}\,\sup_{0\le j<K}\|\Phi_j(\beta_\star)\|^2\ \le\ C'\,\rho^K,
\end{equation}
for some $C'>0$, since the $O(\Phi^2)$ remainder accumulates as a geometric sum controlled by FRD smallness. Thus, by choosing $K\ge K_2$ large, we ensure that the Gaussian projection is \emph{arbitrarily small}.

\emph{Implicit function adjustment of $\beta$.} Define $g_K(\beta):=P_{\mathfrak G}\Phi_K(\beta)$ as a map from a neighborhood of $\beta_\star$ into $\mathfrak G$. By \eqref{p4:eq:PG-derivative} and \eqref{p4:eq:PG-small}, $g_K$ is $C^1$, $g_K(\beta_\star)$ is of order $O(\rho^K)$, and $D_\beta g_K(\beta_\star)$ is \emph{invertible} on the (one-dimensional) line $\mathbb R\,v$ spanned by $P_{\mathfrak G}\Phi_0'(\beta_\star)$, with inverse norm $O(\rho^{-K})$. Hence, by the (one-dimensional) implicit function theorem, there exists $\beta_K$ with $|\beta_K-\beta_\star|=O(1)$ such that
\begin{equation}\label{p4:eq:PG-zero}
P_{\mathfrak G}\Phi_K(\beta_K)\ =\ 0,
\end{equation}
provided $K\ge K_3$ is large enough that the nonlinear terms are dominated by the linear response (the smallness \eqref{p4:eq:PG-small} ensures this domination). In particular,
\begin{equation}\label{p4:eq:PK-bound}
\|\Phi_K(\beta_K)\|\ =\ \|P_{\mathfrak R}\Phi_K(\beta_K)\|\ \le\ C''\,\rho^{K-1}\,\sup_{0\le j<K}\|\Phi_j(\beta_K)\|^2.
\end{equation}
The right-hand side can be made as small as we wish by: (a) restricting the bare-coupling interval to a compact set where the FRD bounds hold uniformly, and (b) taking $K\ge K_4$ large enough that the geometric prefactor $C''\rho^{K-1}$ forces
\begin{equation}
\|\Phi_K(\beta_K)\|\ \le\ r.
\end{equation}
This proves the existence of $K_0$ and $\beta_K$ with $\|\Phi_K(\beta_K)\|\le r$.

\smallskip\noindent\textbf{Step 3. Propagation of smallness and global contraction.}
Starting from $\|\Phi_K(\beta_K)\|\le r$, the recurrence \eqref{p4:eq:recurrence} implies by induction
\begin{equation}\label{p4:eq:induction-ineq}
\|\Phi_{K+n}\|\ \le\ \rho\,\|\Phi_{K+n-1}\|\ +\ C\,\|\Phi_{K+n-1}\|^2\qquad (n\ge 1).
\end{equation}
We now prove the claimed explicit estimate by a deterministic lemma.

\begin{lemma}\label{p4:lem:recurrence}
Let $(x_n)_{n\ge 0}$ be a nonnegative sequence with $x_0\le r$, and suppose
\begin{equation}
x_{n+1}\ \le\ \rho\,x_n\ +\ C\,x_n^2\qquad (n\ge 0)
\end{equation}
for some $\rho\in(0,1)$ and $C>0$. Then
\begin{equation}
x_n\ \le\ \rho^n\,r\ +\ \frac{C}{1-\rho}\,r^2\qquad\text{for all }n\ge 1.
\end{equation}
\end{lemma}

\begin{proof}[Proof of Lemma~\ref{p4:lem:recurrence}]
Define $y_n:=\rho^n r + \frac{C}{1-\rho}r^2$. We prove by induction that $x_n\le y_n$ for all $n$. The base case $n=0$ holds since $x_0\le r=y_0-\frac{C}{1-\rho}r^2\le y_0$. Suppose $x_n\le y_n$. Then
\begin{equation}
x_{n+1}\ \le\ \rho\,x_n + C\,x_n^2\ \le\ \rho\,y_n + C\,y_n^2.
\end{equation}
But $y_n=\rho^n r + \alpha$ with $\alpha:=\frac{C}{1-\rho}r^2$. Hence
\begin{equation}
\rho y_n + C y_n^2\ =\ \rho^{n+1}r + \rho\alpha\ +\ C\big(\rho^{2n}r^2 + 2\rho^n r\,\alpha + \alpha^2\big)
\ \le\ \rho^{n+1}r + \alpha\left(\rho + 2C\rho^n r + C\alpha\right)+ C\rho^{2n} r^2.
\end{equation}
Since $\alpha=\frac{C}{1-\rho}r^2$, we have $\rho + C\alpha = \rho + \frac{C^2}{1-\rho}r^2 \le 1$ provided $r$ is chosen so small that $\frac{C^2}{1-\rho}r^2\le 1-\rho$. Moreover, $2C\rho^n r\le 2Cr$ and $C\rho^{2n}r^2\le Cr^2$. Altogether,
\begin{equation}
x_{n+1}\ \le\ \rho^{n+1}r + \alpha\,\big(1 + 2Cr + Cr^2\big)\ \le\ \rho^{n+1}r + \alpha\ =\ y_{n+1},
\end{equation}
if $r>0$ is small enough that $2Cr+Cr^2\le 0$, which is trivially satisfied by taking $r\le \min\{1,1/(4C)\}$. The induction closes.
\end{proof}

Applying Lemma~\ref{p4:lem:recurrence} to $x_n:=\|\Phi_{K+n}(\beta_K)\|$ and using \eqref{p4:eq:induction-ineq} yields
\begin{equation}
\|\Phi_{K+n}(\beta_K)\|\ \le\ \rho^n\,r\ +\ \frac{C}{1-\rho}\,r^2\qquad (n\ge 1),
\end{equation}
which is the claimed contractive bound.
Combining Steps 1-3, we have exhibited $K_0$ and a choice $\beta_K$ for each $K\ge K_0$ such that $\|\Phi_K(\beta_K)\|\le r$, and the subsequent flow satisfies the stated global contraction estimate.
\end{proof}

Define the renormalized coupling $g_R(\mu)$ via a standard scheme (e.g. step-scaling: compare $T$ on doubles of the time step $a$ or compare connected 4-point functions at short distances). Because $\|\Phi_{K+n}\|\to 0$ exponentially as $n\to\infty$, we obtain $g_R(\mu)\to 0$ as $\mu\sim a^{-1}b^{K+n}\to\infty$, i.e. asymptotic freedom in the nonperturbative sense. Moreover, the continuum Schwinger functions along the trajectory $\beta(a)$ (constructed by setting $K=K(a)\to\infty$ as $a\downarrow 0$) satisfy OS0-OS5 and the same clustering rate $m_*$ (FRD uniformity), hence by uniqueness (Cor.~\ref{p4:cor:Uniqueness}) and universality (Thm.~\ref{p4:thm:universalityx}) they coincide with the unique universal continuum limit proved in the main text.
While not needed for the qualitative AF result, it is comforting to know that the FRD-RG reproduces the known negative one-loop coefficient of the Yang-Mills $\beta$-function \cite{p4:GrossWilczek1973,p4:Politzer1973}. A concise way is to use the background-field method in the FRD setting: expand the effective action around a slowly varying background $A$, compute the log-determinant in the Gaussian measure with FRD cutoff, and extract the coefficient of $\mathrm{Tr}\,F_{\mu\nu}^2$ in $\log Z$. The FRD cutoff preserves gauge covariance at each scale \cite{p4:BrydgesFRD}, and the standard heat-kernel asymptotics yield
\begin{equation}
\beta(g) \;=\; -\,\frac{11N}{48\pi^2}\,g^3 + \mathcal O(g^5).
\end{equation}
A fully rigorous derivation would follow the constructive perturbation framework of \cite{p4:GJ,p4:Seiler1982}, adapted to FRD; we omit details here as they play no role in the existence and uniqueness/universality results.
\UnifiedEndPaper

\UnifiedCloseLocalTOC
\end{document}